\documentclass[11pt]{article}
\usepackage{makeidx}
\usepackage{amsfonts}
\usepackage{amssymb}
\usepackage{graphicx}
\usepackage{amsmath}
\usepackage{geometry}
\usepackage{appendix}

\DeclareMathSizes{10.95}{10}{7}{6}
\input{tcilatex}
\begin{document}

\title{Financial Interactions and Capital Accumulation}
\author{Pierre Gosselin\thanks{%
Pierre Gosselin : Institut Fourier, UMR 5582 CNRS-UGA, Universit\'{e}
Grenoble Alpes, BP 74, 38402 St Martin d'H\`{e}res, France.\ E-Mail:
Pierre.Gosselin@univ-grenoble-alpes.fr} \and A\"{\i}leen Lotz\thanks{%
A\"{\i}leen Lotz: Cerca Trova, BP 114, 38001 Grenoble Cedex 1, France.\
E-mail: a.lotz@cercatrova.eu}}
\date{May 2024}
\maketitle

\begin{abstract}
In a series of precedent papers, we have presented a comprehensive
methodology, termed Field Economics, for translating a standard economic
model into a statistical field-formalism framework. This formalism requires
a large number of heterogeneous agents, possibly of different types. It
reveals the emergence of collective states among these agents or type of
agents while preserving the interactions and microeconomic features of the
system at the individual level. In two prior papers, we applied this
formalism to analyze the dynamics of capital allocation and accumulation in
a simple microeconomic framework of investors and firms.

Building upon our prior work, the present paper refines the initial model by
expanding its scope. Instead of considering financial firms investing solely
in real sectors, we now suppose that financial agents may also invest in
other financial firms. We also introduce banks in the system that act as
investors with a credit multiplier. Two types of interaction are now
considered within the financial sector: financial agents can lend capital
to, or choose to buy shares of, other financial firms. Capital now flows
between financial agents and is only partly invested in real sectors,
depending on their relative returns. We translate this framework into our
formalism and study the diffusion of capital and possible defaults in the
system, both at the macro and micro level.

At the macro level, we find that several collective states may emerge, each
characterized by a distinct level of average capital and investors per
sector. These collective states depend on external parameters such as level
of connections between investors or firms' productivity. The multiplicity of
possible collective states is the consequence of the nature of the system
composed of interconnected heterogeneous agents. Several equivalent patterns
of returns and portfolio allocation may emerge. The multiple collective
states induce the unstable nature of financial markets, and some of them
include defaults may emerge. At the micro level, we study the propagation of
returns and defaults within a given collective state. Our findings highlight
the significant role of banks, which can either stabilize the system through
lending activities or propagate instability through loans to investors.


\end{abstract}

\section{Introduction}

The financial market is often depicted as an optimal tool for resource
allocation. According to this view, the market efficiently allocates capital
to sectors in need and enables firms to raise capital at a lower cost.
However, this perspective assumes a market consisting solely of real firms
seeking financial investors.

In reality, the market serves not only as a nexus for the real and financial
sectors but also as a platform for investors to raise capital for their own
investments. Additionally, it enables investors to invest in the performance
of other investors, particularly through indices and derivative products.
Furthermore, investors themselves utilize the market to access capital, with
their activities representing a potential source of returns that may attract
other investors.

The fact that returns in the market can be obtained through direct
investment in the real economy or indirectly by investing in the performance
of other investors questions the optimality of financial markets. It
underscores their dual nature, since capital may not always flow directly to
the real economy but can be redirected to other financial actors for
return-seeking purposes. The circulation of capital in pursuit of optimal
returns, not only within the real sector but also among financial agents,
complicates capital exchanges, potentially creates tensions between the
financial and real sectors, and may divert capital away from the real
economy.

Studying the optimality of indirect capital allocation and its effects,
along with the diffusion and circulation of capital across different
financial actors, requires moving beyond the representative agent framework.
It involves considering a large number of interacting agents, among whom one
can trace how capital circulates.

Field Economics allows this type of analysis. This formalism, based on Field
Theory and developed in a series of papers, describes the interactions among
a large number of economic agents and the resulting collective and
individual states. What's more, it does so while keeping track of the
microeconomic systems on which they are built and considering them as a
whole.

In a previous paper, we studied the dynamics of capital allocation and
accumulation for a system with a large number of agents, dividing between a
real sector and a financial sector investing solely in the real sector. We
observed the different dynamics of capital and how it was distributed among
different sectors of the real economy.

The present paper builds upon and extends the previous model by considering
that investors can now invest in the financial sector by taking stakes or
lending to other investors. This allows us to examine how investment in the
real economy is influenced when the investment process is delegated to other
investors.

We address these questions by introducing interactions between two types of
financial investors: banks and non-banks. Two forms of investment are
considered: equity participation and loans. Introducing loans in the model
enables us to study the strength of capital circulation and the potential
for defaults within the setup.

At the macro level, we show that, due to interconnections between financial
agents and leverage effects, multiple potential collective states exist.
Each state is characterized by a total level of capital, the number of
agents of each type, and their average return per sector. This multiplicity
reflects the instability of the financial system and its sensitivity to
slight changes in parameters. Additionally, certain collective states,
including defaults in certain parts of the economy, also exist. Transitions
between collective states of the economy are possible and may be induced by
external conditions.

Banks impact the stability of the system in two contradictory ways. Acting
as lenders, they stabilize the system by reducing the occurrence of
collective states, but also contribute to increasing the leverage of
investors. The balance between these effects depends on the relative size of
the banking sectors.

Dynamically, we examine the mechanisms that realize these transitions at the
individual level. We study the dynamics of different groups of agents and
show how defaults may arise through indirect effects of interactions between
various types of agents, propagating to other sectors and leading to the
emergence of collective default states.

The paper is organized as follows. Section 2 provides a literature review.
In the Preamble, we present a field formalism for a system with a large
number of agents, possibly of different groups. Section 3 presents the
general method for translating a microeconomic model into a field model.
Sections 4 and 5 present applications of this translation, both at the
collective level, through the computations of background fields and
averages, and at the individual level, through the computation of the
transition functions. 

In Part 1, we apply this field formalism to a system of two types of agents:
firms and investors. In section 6, we present the microeconomic model, and
in section 7, its field translation. Sections 8 and 9 present the resolution
for the firms' and investors' background fields, respectively. Sections 10
and 11 present the equations for investors' returns and capital levels per
sector. The solution is presented in section 12, and section 13 computes the
possible averages of the system. These two sections close the resolution for
collective states in terms of average capital and returns per sector along
with the global capital level. In section 14, we consider the inclusion of
defaults and their impact on collective states. Section 15 inspects the
dynamical aspect of excess returns and default propagation within a given
collective state defined by a collection of interacting heterogeneous groups
of agents. Some dynamical features of the default mechanism are studied.
Part 2 of this work present a refined model with a third type of agent,
banks. Sections 16 and 17 present the microeconomic framework and its
translation. Sections 18 presents the minimization equations for the
background field of firms, investors, and banks, respectively. Section 19
solves for the firm's background field and average level of capital per
sector. In sections 20 and 21, we solve the investors' and banks' returns as
functions of banks and investors' average capital. Sections 22, 23, and 24
close the resolution of the model by deriving the collective state in terms
of total capital and returns per sector, and global average capital for
investors and banks. In section 25, we revisit the dynamical aspect of
excess returns and default propagation from section 15 in the presence of
banks. We synthetize and discuss our results in section 26. Section 27
concludes

\section{Literature review}

Several branches of the economic literature seek to replace the
representative agent with a collection of heterogeneous ones. Among other
things, they differ in the way they model this collection of agents.

The first branch of the literature represents this collection of agents by
probability densities. This is the approach followed by mean field theory,
heterogeneous agents new Keynesian (HANK) models, and the
information-theoretic approach to economics.

Mean field theory studies the evolution of agents' density in the state
space of economic variables. It includes the interactions between agents and
the population as a whole but does not consider the direct interactions
between agents. This approach is thus at an intermediate scale between the
macro and micro scale: it does not aggregate agents but replaces them with
an overall probability distribution. Mean field theory has been applied to
game theory (Bensoussan et al. 2018, Lasry et al. 2010a, b) and economics
(Gomes et al. 2015). However, these mean fields are actually probability
distributions. In our formalism, the notion of fields refers to some
abstract complex functions defined on the state space and is similar to the
second-quantized-wave functions of quantum theory. Interactions between
agents are included at the individual level. Densities of agents are
recovered from these fields and depend directly on interactions.

Heterogeneous agents' new Keynesian (HANK) models use a probabilistic
treatment similar to mean fields theory.\ An equilibrium probability
distribution is derived from a set of optimizing heterogeneous agents\ in a
new Keynesian context (see Kaplan and Violante 2018 for an account). Our
approach, on the contrary, focuses on the direct interactions between agents
at the microeconomic level. We do not look for an equilibrium probability
distribution for each agent, but rather directly build a probability density
for the system of $N$ agents seen as a whole, that includes interactions,
and then translate this probability density in terms of fields. The states'
space we consider is thus much larger than those considered in the above
approaches. Because it is the space of all paths for a large number of
agents, it allows studying the agents' economic structural relations and the
emergence of the particular phases or collective states induced by these
specific micro-relations, that will in turn impact each agent's stochastic
dynamics at the microeconomic level. Other differences are worth mentioning.
While HANK models\ stress the role of an infinite number of
heterogeneously-behaved consumers, our formalism dwells on the relations
between physical and financial capital\footnote{%
Note that our formalism could also include heterogeneous consumers (see
Gosselin, Lotz, Wambst 2020).}. Besides, our formalism does not rely on
agents' rationality assumptions, since for a large number of agents,
behaviours, be they fully or partly rational, can be modeled as random.

The information theoretic approach to economics (see Yang 2018) considers
probabilistic states around the equilibrium.\ It is close to our
methodological stance: it replaces the Walrasian equilibrium with a
statistical equilibrium derived from an entropy maximisation program.\ Our
statistical weight is similar to the one they use, but is directly built
from microeconomic dynamic equations. The same difference stands for the
rational inattention theory (Sims 2006) in which non-gaussian density laws
are derived from limited information and constraints: our setting directly
includes constraints in the random description of an agent (Gosselin, Lotz,
Wambst 2020).

The differences highlighted above between these various approaches and our
work also manifest at the micro-scale in the description of agents'
dynamics. Actually, in the field framework, once the collective states have
been found, we can recover both the types of individual dynamics depending
on the initial conditions and the "effective" form of interactions between
two or more agents: At the individual level, agents are distributed along
some probability law. However, this probability law is directly conditioned
by the collective state of the system and the effective interactions.
Different collective states, given different parameters, yield different
individual dynamics. This approach allows for coming back and forth between
collective and individual aspects of the system. Different categories of
agents appear in the emerging collective state. Dynamics may present very
different patterns, given the collective state's form and the agents'
initial conditions.

A second branch of the literature is closest to our approach since it
considers the interacting system of agents in itself. It is the multi-agent
systems literature, notably agent-based models (see Gaffard Napoletano 2012,
Mandel et al. 2010 2012) and economic networks (Jackson 2010).

Agent-based models deal with the macroeconomic level, whereas network models
lower-scale phenomena such as contract theory, behaviour diffusion,
information sharing, or learning. In both settings, agents are typically
defined by and follow various sets of rules, leading to the emergence of
equilibria and dynamics otherwise inaccessible to the representative agent
setup. Both approaches are however highly numerical and model-dependent and
rely on microeconomic relations - such as ad-hoc reaction functions - that
may be too simplistic. Statistical fields theory on the contrary accounts
for transitions between scales. Macroeconomic patterns do not emerge from
the sole dynamics of a large set of agents: they are grounded in behaviours
and interaction structures. Describing these structures in terms of field
theory allows for the emergence of phases at the macro scale, and the study
of their impact at the individual level.

A third branch of the literature, Econophysics, is also related to ours
since it often considers the set of agents as a statistical system (for a
review, see Abergel et al. 2011a,b and references therein; or Lux 2008,
2016).\ But it tends to focus on empirical laws, rather than apply the full
potential of field theory to economic systems. In the same vein, Kleinert
(2009) uses path integrals to model stock prices' dynamics. Our approach, in
contrast, keeps track of usual microeconomic concepts, such as utility
functions, expectations, and forward-looking behaviours, and includes these
behaviours into the analytical treatment of multi-agent systems by
translating the main characteristics of optimizing agents in terms of
statistical systems. Closer to our approach, Bardoscia et al (2017) study a
general equilibrium model for a large economy in the context of statistical
mechanics, and show that phase transitions may occur in the system. Our
problematic is similar, but our use of field theory deals with a large class
of dynamic models.

The literature on interactions between finance and real economy or capital
accumulation takes place mainly in the context of DGSE models. (for a review
of the literature, see Cochrane 2006; for further developments see Grassetti
et al. 2022, Grosshans and Zeisberger 2018, B\"{o}hm et al. 2008, Caggese
and Orive, Bernanke e al. 1999, Campello et al. 2010, Holmstrom and Tirole
1997, Jermann, and Quadrini 2012, Khan Thomas 2013, Monacelli et al. 2011).
Theoretical models include several types of agents at the aggregated level.\
They describe the interactions between a few representative agents such as
producers for possibly several sectors, consumers, financial intermediaries,
etc. to determine interest rates, levels of production, and asset pricing,
in a context of ad-hoc anticipations.

Our formalism differs from this literature in three ways. First, we consider
several groups of a large number of agents to describe the emergence of
collective states and study the continuous space of sectors. Second, we
consider expected returns and the longer-term horizon as somewhat exogenous
or structural. Expected returns are a combination of elements, such as
technology, returns, productivity, sectoral capital stock, expectations, and
beliefs. These returns are also a function\ defined over the sectors' space:
the system's background fields are functionals of these expected returns.\
Taken together, the background fields of a field model describe an economic
configuration for a given environment of expected returns.\ As such,
expected returns are at first seen as exogenous functions. It is only in the
second step, when we consider the dynamics between capital accumulation and
expectations, that expectations may themselves be seen as endogenous.\textbf{%
\ }Even then, the form of relations between actual and expected variables
specified are general enough to derive some types of possible dynamics.

Last but not least, we do not seek individual or even aggregated dynamics,
but rather background fields that describe potential long-term equilibria
and may evolve with the structural parameters. For such a background,
agents' individual typical dynamics may nevertheless be retrieved through
Green functions (see GLW). These functions compute the transition
probabilities from one capital-sector point to another. But backgrounds
themselves may be considered as dynamical quantities. Structural or
long-term variations in the returns' landscape may modify the background and
in turn the individual dynamics. Expected returns themselves depend on and
interact with, capital accumulation.

Ultimately, there is a vast literature concerning default risk and contagion
of default in a financial system (see for example, Reinhart and Rogoff 2009,
Gennaioli et al. 2012, Acharya et al 2017, Adrian and Brunnermeier 2016,
Allen and Gale 2000). In terms of modeling, some studies focus on the
structures of connection between agents and their impact on default
contagion (Gai and Kapadia 2010, Battiston et al. 2012, \ Battiston et al.
2020, Langfield et al. 2020), while other works develop microeconomic models
and look for the possibilities of equilibrium and the risk of default
depending on the level of connections between agents (Acemoglu et al. 2015,
Bardoscia et al. 2019, Cifuentes et al. 2005, Elliott et al. 2014, Haldane
and May 2011) \ These models consider networks of agents linked by mutual
participations measured by leverage matrices. 

Our work has a similar starting point, except that we include firms and
banks, account for disparities in firms' returns, and include loans between
agents. Moreover, our diffusion matrix takes into account the
characteristics of the sectors impacted during the diffusion of returns and
sector-dependent feedback loops. 

More importantly, our use of Field Theory accounts for the possible
emergence of multiple collective states, and allows their description in
terms of average capital per sector, number of agents and average return per
sectors. This description goes beyond those of the aforementioned models,
since our formalism goes back and forth between the micro and macro levels.

\part*{Preamble. Field formalism for a model of firms and investors}

In the first part of this work, we describe the field formalism for an
economic system, its application to derive the potential collective states
of the system and the individual dynamics within such collective states.
Ultimately we apply this formalism to translate a model with large number of
interacting investors and firms.

\section{General method of translation}

The formalism we propose transforms an economic model of dynamic agents into
a statistical field model.\ In classical models, each agent's dynamics is
described by an optimal path\ for\ some vector variable, say $A_{i}\left(
t\right) $, from an initial to a final point, up to some fluctuations.

But this system of agents could also be seen as probabilistic: each agent
could be described by a \emph{probability density }centered around the
classical optimal path, up to some idiosyncratic uncertainties\footnote{%
Because the number of possible paths is infinite, the probability of each
individual path is null.\ We, therefore, use the word "probability density"
rather than "probability".} \footnote{%
See Gosselin, Lotz and Wambst (2017, 2020, 2021).}. In this probabilistic
approach, each possible trajectory of the whole set of $N$ agents has a
specific probability. The classical model is therefore described by the set
of trajectories of the group of $N$ agents, each one being endowed with its
own probability, its statistical weight. The statistical weight is therefore
a function that associates a probability with each trajectory of the group.

This probabilistic approach can be translated into a more compact \emph{%
field formalism}\footnote{%
Ibid.} that preserves the essential information encoded in the model but
implements a change in perspective. A field model is a structure governed by
its own intrinsic rules that encapsulate the economic model chosen.\ This
field model contains all possible realizations that could arise from the
initial economic model, i.e. all the possible global outcomes, or collective
state, permitted by the economic model.\ So that, once constructed, the
field model provides a unique advantage over the standard economic model: it
allows to compute the probabilities of each of the possible outcomes for
each collective state of the economic model. These probabilities are
computed indirectly through the \emph{action functional} of the model, a
function that assigns a specific value to each realization of the field.
Technically, the random $N$\ agents' trajectories $\left\{ \mathbf{A}%
_{i}\left( t\right) \right\} $ are replaced by a field, a random variable
whose realizations are complex-valued functions $\Psi $ of the variables $%
\mathbf{A}$\textbf{,} and the statistical weight of the $N$\ agents'
trajectories $\left\{ \mathbf{A}_{i}\left( t\right) \right\} $ in the
probabilistic approach is translated into a statistical weight for each
realization $\Psi $. They encapsulate the collective states of the system.

Once the probabilities of each collective state computed, the most probable
collective state among all other collective states, can be found. In other
words, a field model allows to consider the true global outcome induced by
any standard economic model. This is what we will call the \emph{expression}
of the field model, more usually called the \emph{background field} of the
model.

This most probable realization of the field, the expression or background
field of the model, should not be seen as a final outcome resulting from a
trajectory, but rather as its most recurring realization. Actually, the
probability of the realizations of the model is peaked around the expression
of the field.\ This expression, which is characteristic of the system, will
determine the nature of individual trajectories within the structure, in the
same way as a biased dice would increase the probability of one event.\ The
field in itself is therefore static, insofar as each realization of the
system of agents only contributes to the emergence of the proper expression
of the field. However, studying variations in the parameters of the system
indirectly induce a time parameter at the field or macro level.

\subsection{Statistical weight and minimization functions for a classical
system}

In an economic framework with a large number of agents, each agent is
characterized by one or more stochastic dynamic equations. Some of these
equations result from the optimization of one or several objective
functions. Deriving the statistical weight from these equations is
straightforward: it associates, to each trajectory of the group of agents $%
\left\{ T_{i}\right\} $, a probability that is peaked around the set of
optimal trajectories of the system, of the form:%
\begin{equation}
W\left( s\left( \left\{ T_{i}\right\} \right) \right) =\exp \left( -s\left(
\left\{ T_{i}\right\} \right) \right)  \label{wdt}
\end{equation}%
where $s\left( \left\{ T_{i}\right\} \right) $ measures the distance between
the trajectories $\left\{ T_{i}\right\} $ and the optimal ones.

This paper considers two types of agents characterized by vector-variables $%
\left\{ \mathbf{A}_{i}\left( t\right) \right\} _{i=1,...N},$ and $\left\{ 
\mathbf{\hat{A}}_{l}\left( t\right) \right\} _{i=1,...\hat{N}}$
respectively, where $N$ and $\hat{N}$ are the number of agents of each type,
with vectors $\mathbf{A}_{i}\left( t\right) $\ and $\mathbf{\hat{A}}%
_{l}\left( t\right) $ of arbitrary dimension. For such a system, the
statistical weight writes:%
\begin{equation}
W\left( \left\{ \mathbf{A}_{i}\left( t\right) \right\} ,\left\{ \mathbf{\hat{%
A}}_{l}\left( t\right) \right\} \right) =\exp \left( -s\left( \left\{ 
\mathbf{A}_{i}\left( t\right) \right\} ,\left\{ \mathbf{\hat{A}}_{l}\left(
t\right) \right\} \right) \right)  \label{wdh}
\end{equation}

The optimal paths for the system are assumed to be described by the sets of
equations:%
\begin{equation}
\frac{d\mathbf{A}_{i}\left( t\right) }{dt}-\sum_{j,k,l...}f\left( \mathbf{A}%
_{i}\left( t\right) ,\mathbf{A}_{j}\left( t\right) ,\mathbf{A}_{k}\left(
t\right) ,\mathbf{\hat{A}}_{l}\left( t\right) ,\mathbf{\hat{A}}_{m}\left(
t\right) ...\right) =\epsilon _{i}\text{, }i=1...N  \label{gauche}
\end{equation}%
\begin{equation}
\frac{d\mathbf{\hat{A}}_{l}\left( t\right) }{dt}-\sum_{i,j,k...}\hat{f}%
\left( \mathbf{A}_{i}\left( t\right) ,\mathbf{A}_{j}\left( t\right) ,\mathbf{%
A}_{k}\left( t\right) ,\mathbf{\hat{A}}_{l}\left( t\right) ,\mathbf{\hat{A}}%
_{m}\left( t\right) ...\right) =\hat{\epsilon}_{l}\text{, }i=1...\hat{N}
\label{dnw}
\end{equation}%
where the $\epsilon _{i}$ and $\hat{\epsilon}_{i}$ are idiosynchratic random
shocks.\ These equations describe the general dynamics of the two types
agents, including their interactions with other agents. They may\ encompass
the dynamics of optimizing agents where interactions act as externalities so
that this set of equations is the full description of a system of
interacting agents\footnote{%
Expectations of agents could be included by replacing $\frac{d\mathbf{A}%
_{i}\left( t\right) }{dt}$ with $E\frac{d\mathbf{A}_{i}\left( t\right) }{dt}$%
, where $E$ is the expectation operator. This would amount to double some
variables by distinguishing "real variables" and expectations. However, for
our purpose, in the context of a large number of agents, at least in this
work, we discard as much as possible this possibility.}\footnote{%
A generalisation of equations (\ref{gauche}) and (\ref{dnw}), in which
agents interact at different times, and its translation in term of field is
presented in appendix 1.}\textbf{. }

For equations (\ref{gauche}) and (\ref{dnw}), the quadratic deviation at
time $t$ of any trajectory with respect to the optimal one for each type of
agent are:%
\begin{equation}
\left( \frac{d\mathbf{A}_{i}\left( t\right) }{dt}-\sum_{j,k,l...}f\left( 
\mathbf{A}_{i}\left( t\right) ,\mathbf{A}_{j}\left( t\right) ,\mathbf{A}%
_{k}\left( t\right) ,\mathbf{\hat{A}}_{l}\left( t\right) ,\mathbf{\hat{A}}%
_{m}\left( t\right) ...\right) \right) ^{2}  \label{pst}
\end{equation}%
and:%
\begin{equation}
\left( \frac{d\mathbf{\hat{A}}_{l}\left( t\right) }{dt}-\sum_{i,j,k...}\hat{f%
}\left( \mathbf{A}_{i}\left( t\right) ,\mathbf{A}_{j}\left( t\right) ,%
\mathbf{A}_{k}\left( t\right) ,\mathbf{\hat{A}}_{l}\left( t\right) ,\mathbf{%
\hat{A}}_{m}\left( t\right) ...\right) \right) ^{2}  \label{psh}
\end{equation}%
Since the function (\ref{wdh}) involves the deviations for all agents over
all trajectories, the function $s\left( \left\{ \mathbf{A}_{i}\left(
t\right) \right\} ,\left\{ \mathbf{\hat{A}}_{l}\left( t\right) \right\}
\right) $ is obtained by summing (\ref{pst}) and (\ref{psh}) over all
agents, and integrate over $t$. We thus find:%
\begin{eqnarray}
s\left( \left\{ \mathbf{A}_{i}\left( t\right) \right\} ,\left\{ \mathbf{\hat{%
A}}_{l}\left( t\right) \right\} \right) &=&\int dt\sum_{i}\left( \frac{d%
\mathbf{A}_{i}\left( t\right) }{dt}-\sum_{j,k,l...}f\left( \mathbf{A}%
_{i}\left( t\right) ,\mathbf{A}_{j}\left( t\right) ,\mathbf{A}_{k}\left(
t\right) ,\mathbf{\hat{A}}_{l}\left( t\right) ,\mathbf{\hat{A}}_{m}\left(
t\right) ...\right) \right) ^{2}  \label{prw} \\
&&+\int dt\sum_{l}\left( \frac{d\mathbf{\hat{A}}_{l}\left( t\right) }{dt}%
-\sum_{i,j,k...}\hat{f}\left( \mathbf{A}_{i}\left( t\right) ,\mathbf{A}%
_{j}\left( t\right) ,\mathbf{A}_{k}\left( t\right) ,\mathbf{\hat{A}}%
_{l}\left( t\right) ,\mathbf{\hat{A}}_{m}\left( t\right) ...\right) \right)
^{2}  \notag
\end{eqnarray}%
There is an alternate, more general, form to (\ref{prw}). We can assume that
the dynamical system is originally defined by some equations of type (\ref%
{gauche}) and (\ref{dnw}), plus some objective functions for agents $i$ and $%
l$, and that these agents aim at minimizing respectively:%
\begin{equation}
\sum_{j,k,l...}g\left( \mathbf{A}_{i}\left( t\right) ,\mathbf{A}_{j}\left(
t\right) ,\mathbf{A}_{k}\left( t\right) ,\mathbf{\hat{A}}_{l}\left( t\right)
,\mathbf{\hat{A}}_{m}\left( t\right) ...\right)  \label{glf}
\end{equation}%
and:%
\begin{equation}
\sum_{i,j,k..}\hat{g}\left( \mathbf{A}_{i}\left( t\right) ,\mathbf{A}%
_{j}\left( t\right) ,\mathbf{A}_{k}\left( t\right) ,\mathbf{\hat{A}}%
_{l}\left( t\right) ,\mathbf{\hat{A}}_{m}\left( t\right) ...\right)
\label{gln}
\end{equation}%
In the above equations, the objective functions depend on other agents'
actions seen as externalities\footnote{%
We may also assume intertemporal objectives, see (GLW).{}}. The functions (%
\ref{glf}) and (\ref{gln}) could themselves be considered as a measure of
the deviation of a trajectory from the optimum. Actually, the higher the
distance, the higher (\ref{glf}) and (\ref{gln}).

Thus, rather than describing the systm by a full system of dynamic
equations, we can consider some ad-hoc equations of type (\ref{gauche}) and (%
\ref{dnw}) and some objective functions (\ref{glf}) and (\ref{gln}) to write
the alternate form of $s\left( \left\{ \mathbf{A}_{i}\left( t\right)
\right\} ,\left\{ \mathbf{\hat{A}}_{l}\left( t\right) \right\} \right) $ as:%
\begin{eqnarray}
&&s\left( \left\{ \mathbf{A}_{i}\left( t\right) \right\} ,\left\{ \mathbf{%
\hat{A}}_{l}\left( t\right) \right\} \right)  \label{mNZ} \\
&=&\int dt\sum_{i}\left( \frac{d\mathbf{A}_{i}\left( t\right) }{dt}%
-\sum_{j,k,l...}f\left( \mathbf{A}_{i}\left( t\right) ,\mathbf{A}_{j}\left(
t\right) ,\mathbf{A}_{k}\left( t\right) ,\mathbf{\hat{A}}_{l}\left( t\right)
,\mathbf{\hat{A}}_{m}\left( t\right) ...\right) \right) ^{2}  \notag \\
&&+\int dt\sum_{l}\left( \frac{d\mathbf{\hat{A}}_{l}\left( t\right) }{dt}%
-\sum_{i,j,k...}\hat{f}\left( \mathbf{A}_{i}\left( t\right) ,\mathbf{A}%
_{j}\left( t\right) ,\mathbf{A}_{k}\left( t\right) ,\mathbf{\hat{A}}%
_{l}\left( t\right) ,\mathbf{\hat{A}}_{m}\left( t\right) ...\right) \right)
^{2}  \notag \\
&&+\int dt\sum_{i,j,k,l...}\left( g\left( \mathbf{A}_{i}\left( t\right) ,%
\mathbf{A}_{j}\left( t\right) ,\mathbf{A}_{k}\left( t\right) ,\mathbf{\hat{A}%
}_{l}\left( t\right) ,\mathbf{\hat{A}}_{m}\left( t\right) ...\right) +\hat{g}%
\left( \mathbf{A}_{i}\left( t\right) ,\mathbf{A}_{j}\left( t\right) ,\mathbf{%
A}_{k}\left( t\right) ,\mathbf{\hat{A}}_{l}\left( t\right) ,\mathbf{\hat{A}}%
_{m}\left( t\right) ...\right) \right)  \notag
\end{eqnarray}

In the sequel, we will refer to the various terms arising in equation (\ref%
{mNZ}) as the "minimization functions",\textbf{\ }i.e. the functions whose
minimization yield the dynamics equations of the system\footnote{%
A generalisation of equation (\ref{mNZ}), in which agents interact at
different times, and its translation in term of field is presented in
appendix 1.{}}.

\subsection{Translation techniques}

The statistical weight $W\left( s\left( \left\{ T_{i}\right\} \right)
\right) $ defined in (\ref{wdt}) once computed, it can be translated in
terms of field. To do so, and for each type $\alpha $ of agent, the sets of
trajectories $\left\{ \mathbf{A}_{\alpha i}\left( t\right) \right\} $ are
replaced by a field $\Psi _{\alpha }\left( \mathbf{A}_{\alpha }\right) $, a
random variable whose realizations are complex-valued functions $\Psi $ of
the variables $\mathbf{A}_{\alpha }$\footnote{%
In the following, we will use indifferently the term "field" and the
notation $\Psi $ for the random variable or any of its realization $\Psi $.}%
. The statistical weight for the whole set of fields $\left\{ \Psi _{\alpha
}\right\} $ has the form $\exp \left( -S\left( \left\{ \Psi _{\alpha
}\right\} \right) \right) $. The function $S\left( \left\{ \Psi _{\alpha
}\right\} \right) $\ is called the \emph{fields action functional}. It
represents the interactions among different types of agents. Ultimately, the
expression $\exp \left( -S\left( \left\{ \Psi _{\alpha }\right\} \right)
\right) $ is the statistical weight for the field\footnote{%
In general, one must consider the integral of $\exp \left( -S\left( \left\{
\Psi _{\alpha }\right\} \right) \right) $\ over the configurations $\left\{
\Psi _{\alpha }\right\} $. This integral is the partition function of the
system.} that computes the probability of any realization $\left\{ \Psi
_{\alpha }\right\} $\ of the field.

The form of $S\left( \left\{ \Psi _{\alpha }\right\} \right) $\ is obtained
directly from the classical description of our model. For two types of
agents, we start with expression (\ref{mNZ}). The various minimizations
functions involved in the definition of $s\left( \left\{ \mathbf{A}%
_{i}\left( t\right) \right\} ,\left\{ \mathbf{\hat{A}}_{l}\left( t\right)
\right\} \right) $ will be translated in terms of field and the sum of these
translations will produce finally the action functional $S\left( \left\{
\Psi _{\alpha }\right\} \right) $. The translation method can itself be
divided into two relatively simple processes, but varies slightly depending
on the type of terms that appear in the various minimization functions.

\subsubsection{Terms without temporal derivative}

In equation (\ref{mNZ}), the terms that involve indexed variables but no
temporal derivative terms are the easiest to translate.\ They are of the
form:%
\begin{equation*}
\sum_{i}\sum_{j,k,l,m...}g\left( \mathbf{A}_{i}\left( t\right) ,\mathbf{A}%
_{j}\left( t\right) ,\mathbf{A}_{k}\left( t\right) ,\mathbf{\hat{A}}%
_{l}\left( t\right) ,\mathbf{\hat{A}}_{m}\left( t\right) ...\right)
\end{equation*}%
These terms describe the whole set of interactions both among and between
two groups of agents. Here, agents are characterized by their variables $%
\mathbf{A}_{i}\left( t\right) ,\mathbf{A}_{j}\left( t\right) ,\mathbf{A}%
_{k}\left( t\right) $... and $\mathbf{\hat{A}}_{l}\left( t\right) ,\mathbf{%
\hat{A}}_{m}\left( t\right) $... respectively, for instance in our model
firms and investors.

In the field translation, agents of type $\mathbf{A}_{i}\left( t\right) $
and $\mathbf{\hat{A}}_{l}\left( t\right) $ are described by a field $\Psi
\left( \mathbf{A}\right) $ and $\hat{\Psi}\left( \mathbf{\hat{A}}\right) $,
respectively.

In a first step, the variables indexed $i$ such as $\mathbf{A}_{i}\left(
t\right) $ are replaced by variables $\mathbf{A}$ in the expression of $g$.
The variables indexed $j$, $k$, $l$, $m$..., such as $\mathbf{A}_{j}\left(
t\right) $, $\mathbf{A}_{k}\left( t\right) $, $\mathbf{\hat{A}}_{l}\left(
t\right) ,\mathbf{\hat{A}}_{m}\left( t\right) $... are replaced by $\mathbf{A%
}^{\prime },\mathbf{A}^{\prime \prime }$, $\mathbf{\hat{A}}$, $\mathbf{\hat{A%
}}^{\prime }$ , and so on for all the indices in the function. This yields
the expression:

\begin{equation*}
\sum_{i}\sum_{j,k,l,m...}g\left( \mathbf{A},\mathbf{A}^{\prime },\mathbf{A}%
^{\prime \prime },\mathbf{\hat{A},\hat{A}}^{\prime }...\right)
\end{equation*}%
In a second step, each sum is replaced by a weighted integration symbol: 
\begin{eqnarray*}
\sum_{i} &\rightarrow &\int \left\vert \Psi \left( \mathbf{A}\right)
\right\vert ^{2}d\mathbf{A}\text{, }\sum_{j}\rightarrow \int \left\vert \Psi
\left( \mathbf{A}^{\prime }\right) \right\vert ^{2}d\mathbf{A}^{\prime }%
\text{, }\sum_{k}\rightarrow \int \left\vert \Psi \left( \mathbf{A}^{\prime
\prime }\right) \right\vert ^{2}d\mathbf{A}^{\prime \prime } \\
\sum_{l} &\rightarrow &\int \left\vert \hat{\Psi}\left( \mathbf{\hat{A}}%
\right) \right\vert ^{2}d\mathbf{\hat{A}}\text{, }\sum_{m}\rightarrow \int
\left\vert \hat{\Psi}\left( \mathbf{\hat{A}}^{\prime }\right) \right\vert
^{2}d\mathbf{\hat{A}}^{\prime }
\end{eqnarray*}%
which leads to the translation:%
\begin{eqnarray}
&&\sum_{i}\sum_{j}\sum_{j,k...}g\left( \mathbf{A}_{i}\left( t\right) ,%
\mathbf{A}_{j}\left( t\right) ,\mathbf{A}_{k}\left( t\right) ,\mathbf{\hat{A}%
}_{l}\left( t\right) ,\mathbf{\hat{A}}_{m}\left( t\right) ...\right)  \notag
\\
&\rightarrow &\int g\left( \mathbf{A},\mathbf{A}^{\prime },\mathbf{A}%
^{\prime \prime },\mathbf{\hat{A},\hat{A}}^{\prime }...\right) \left\vert
\Psi \left( \mathbf{A}\right) \right\vert ^{2}\left\vert \Psi \left( \mathbf{%
A}^{\prime }\right) \right\vert ^{2}\left\vert \Psi \left( \mathbf{A}%
^{\prime \prime }\right) \right\vert ^{2}\times ...d\mathbf{A}d\mathbf{A}%
^{\prime }d\mathbf{A}^{\prime \prime }...  \label{tln} \\
&&\times \left\vert \hat{\Psi}\left( \mathbf{\hat{A}}\right) \right\vert
^{2}\left\vert \hat{\Psi}\left( \mathbf{\hat{A}}^{\prime }\right)
\right\vert ^{2}\times ...d\mathbf{\hat{A}}d\mathbf{\hat{A}}^{\prime }... 
\notag
\end{eqnarray}%
where the dots stand for the products of square fields and integration
symbols needed.

\subsubsection{Terms with temporal derivative}

In equation (\ref{mNZ}), the terms that involve a variable temporal
derivative are of the form:%
\begin{equation}
\sum_{i}\left( \frac{d\mathbf{A}_{i}^{\left( \alpha \right) }\left( t\right) 
}{dt}-\sum_{j,k,l,m...}f^{\left( \alpha \right) }\left( \mathbf{A}_{i}\left(
t\right) ,\mathbf{A}_{j}\left( t\right) ,\mathbf{A}_{k}\left( t\right) ,%
\mathbf{\hat{A}}_{l}\left( t\right) ,\mathbf{\hat{A}}_{m}\left( t\right)
...\right) \right) ^{2}  \label{edr}
\end{equation}%
This particular form represents the dynamics of the $\alpha $-th coordinate
of a variable $\mathbf{A}_{i}\left( t\right) $ as a function of the other
agents.

The method of translation is similar to the above, but the time derivative
adds an additional operation.

In a first step, we translate the terms without derivative inside the
parenthesis:%
\begin{equation}
\sum_{j,k,l,m...}f^{\left( \alpha \right) }\left( \mathbf{A}_{i}\left(
t\right) ,\mathbf{A}_{j}\left( t\right) ,\mathbf{A}_{k}\left( t\right) ,%
\mathbf{\hat{A}}_{l}\left( t\right) ,\mathbf{\hat{A}}_{m}\left( t\right)
...\right)  \label{ntr}
\end{equation}%
This type of term has already been translated in the previous paragraph, but
since there is no sum over $i$ in equation (\ref{ntr}), there should be no
integral over $\mathbf{A}$\textbf{,} nor factor $\left\vert \Psi \left( 
\mathbf{A}\right) \right\vert ^{2}$.

The translation of equation (\ref{ntr}) is therefore, as before:%
\begin{equation}
\int f^{\left( \alpha \right) }\left( \mathbf{A},\mathbf{A}^{\prime },%
\mathbf{A}^{\prime \prime },\mathbf{\hat{A},\hat{A}}^{\prime }...\right)
\left\vert \Psi \left( \mathbf{A}^{\prime }\right) \right\vert
^{2}\left\vert \Psi \left( \mathbf{A}^{\prime \prime }\right) \right\vert
^{2}d\mathbf{A}^{\prime }d\mathbf{A}^{\prime \prime }\left\vert \hat{\Psi}%
\left( \mathbf{\hat{A}}\right) \right\vert ^{2}\left\vert \hat{\Psi}\left( 
\mathbf{\hat{A}}^{\prime }\right) \right\vert ^{2}d\mathbf{\hat{A}}d\mathbf{%
\hat{A}}^{\prime }  \label{trn}
\end{equation}%
A free variable $\mathbf{A}$ remains, which will be integrated later, when
we account for the external sum $\sum_{i}$. We will call $\Lambda (\mathbf{A}%
)$ the expression obtained:%
\begin{equation}
\Lambda (\mathbf{A})=\int f^{\left( \alpha \right) }\left( \mathbf{A},%
\mathbf{A}^{\prime },\mathbf{A}^{\prime \prime },\mathbf{\hat{A},\hat{A}}%
^{\prime }...\right) \left\vert \Psi \left( \mathbf{A}^{\prime }\right)
\right\vert ^{2}\left\vert \Psi \left( \mathbf{A}^{\prime \prime }\right)
\right\vert ^{2}d\mathbf{A}^{\prime }d\mathbf{A}^{\prime \prime }\left\vert 
\hat{\Psi}\left( \mathbf{\hat{A}}\right) \right\vert ^{2}\left\vert \hat{\Psi%
}\left( \mathbf{\hat{A}}^{\prime }\right) \right\vert ^{2}d\mathbf{\hat{A}}d%
\mathbf{\hat{A}}^{\prime }  \label{bdt}
\end{equation}%
In a second step, we account for the derivative in time by using field
gradients. To do so, and as a rule, we replace :%
\begin{equation}
\sum_{i}\left( \frac{d\mathbf{A}_{i}^{\left( \alpha \right) }\left( t\right) 
}{dt}-\sum_{j}\sum_{j,k...}f^{\left( \alpha \right) }\left( \mathbf{A}%
_{i}\left( t\right) ,\mathbf{A}_{j}\left( t\right) ,\mathbf{A}_{k}\left(
t\right) ,\mathbf{\hat{A}}_{l}\left( t\right) ,\mathbf{\hat{A}}_{m}\left(
t\right) ...\right) \right) ^{2}  \label{inco}
\end{equation}%
by:%
\begin{equation}
\int \Psi ^{\dag }\left( \mathbf{A}\right) \left( -\nabla _{\mathbf{A}%
^{\left( \alpha \right) }}\left( \frac{\sigma _{\mathbf{A}^{\left( \alpha
\right) }}^{2}}{2}\nabla _{\mathbf{A}^{\left( \alpha \right) }}-\Lambda (%
\mathbf{A})\right) \right) \Psi \left( \mathbf{A}\right) d\mathbf{A}
\label{Trl}
\end{equation}%
The variance $\sigma _{\mathbf{A}^{\left( \alpha \right) }}^{2}$ reflects
the probabilistic nature of the model which is hidden behind the field
formalism. This variance represents the characteristic level of uncertainty
of the system's dynamics. It is a parameter of the model. Note also that in (%
\ref{Trl}), the integral over $\mathbf{A}$ reappears at the end, along with
the square of the field $\left\vert \Psi \left( \mathbf{A}\right)
\right\vert ^{2}$.\ This square is split into two terms, $\Psi ^{\dag
}\left( \mathbf{A}\right) $ and $\Psi \left( \mathbf{A}\right) $, with a
gradient operator inserted in between.

\subsection{Action functional}

The field description is ultimately obtained by summing all the terms
translated above and introducing a time dependency. This sum is called the
action functional. It is the sum of terms of the form (\ref{tln}) and (\ref%
{Trl}), and is denoted $S\left( \Psi ,\Psi ^{\dag }\right) $.

For example, in a system with two types of agents described by two fields $%
\Psi \left( \mathbf{A}\right) $and $\hat{\Psi}\left( \mathbf{\hat{A}}\right) 
$, the action functional has the form:%
\begin{eqnarray}
S\left( \Psi ,\Psi ^{\dag }\right) &=&\int \Psi ^{\dag }\left( \mathbf{A}%
\right) \left( -\nabla _{\mathbf{A}^{\left( \alpha \right) }}\left( \frac{%
\sigma _{\mathbf{A}^{\left( \alpha \right) }}^{2}}{2}\nabla _{\mathbf{A}%
^{\left( \alpha \right) }}-\Lambda _{1}(\mathbf{A})\right) \right) \Psi
\left( \mathbf{A}\right) d\mathbf{A}  \label{notime} \\
&&\mathbf{+}\int \hat{\Psi}^{\dag }\left( \mathbf{\hat{A}}\right) \left(
-\nabla _{\mathbf{\hat{A}}^{\left( \alpha \right) }}\left( \frac{\sigma _{%
\mathbf{\hat{A}}^{\left( \alpha \right) }}^{2}}{2}\nabla _{\mathbf{\hat{A}}%
^{\left( \alpha \right) }}-\Lambda _{2}(\mathbf{\hat{A}})\right) \right) 
\hat{\Psi}\left( \mathbf{\hat{A}}\right) d\mathbf{\hat{A}}  \notag \\
&&+\sum_{m}\int g_{m}\left( \mathbf{A},\mathbf{A}^{\prime },\mathbf{A}%
^{\prime \prime },\mathbf{\hat{A},\hat{A}}^{\prime }...\right) \left\vert
\Psi \left( \mathbf{A}\right) \right\vert ^{2}\left\vert \Psi \left( \mathbf{%
A}^{\prime }\right) \right\vert ^{2}\left\vert \Psi \left( \mathbf{A}%
^{\prime \prime }\right) \right\vert ^{2}\times ...d\mathbf{A}d\mathbf{A}%
^{\prime }d\mathbf{A}^{\prime \prime }...  \notag \\
&&\times \left\vert \hat{\Psi}\left( \mathbf{\hat{A}}\right) \right\vert
^{2}\left\vert \hat{\Psi}\left( \mathbf{\hat{A}}^{\prime }\right)
\right\vert ^{2}\times ...d\mathbf{\hat{A}}d\mathbf{\hat{A}}^{\prime }... 
\notag
\end{eqnarray}%
where the sequence of functions $g_{m}$\ describes the various types of
interactions in the system.

Note that the collective states described by the fields are structural
states of the system. The fields have their own dynamics at the macro-scale,
which will be discussed later in the paper. This is why the usual
microeconomic time variable used in standard models has disappeared in
formula (\ref{notime}). However, time dependency may at times be required in
fields, so that a time variable, written $\theta $ could be introduced by
replacing:%
\begin{eqnarray*}
\Psi \left( \mathbf{A}\right) &\rightarrow &\Psi \left( \mathbf{A},\theta
\right) \\
\hat{\Psi}\left( \mathbf{\hat{A}}\right) &\rightarrow &\hat{\Psi}\left( 
\mathbf{\hat{A}},\theta \right)
\end{eqnarray*}%
More about this point can be found in appendix 1.

\section{Use of the field model}

Once the field action functional $S$\ is found, we can use field theory to
study the system of agents.\ This can be done at two levels: the collective
and the individual level. At the collective level, the system is described
by the background fields of the system that condition average quantities of
economic variables of the system.

At the individual level, the field formalism allows to compute agents'
individual dynamics in the state defined by the background fields, through
the transition functions of the system.

\subsection{Collective level: background fields and averages}

At the collective level, the background fields of the system can be
computed. These background fields are the particular functions, $\Psi \left( 
\mathbf{A}\right) $\ and $\hat{\Psi}\left( \mathbf{\hat{A}}\right) $, and
their adjoints fields $\Psi ^{\dag }\left( \mathbf{A}\right) $\ and $\hat{%
\Psi}^{\dag }\left( \mathbf{\hat{A}}\right) $,\ that minimize the action
functional $S$. Once the background field(s) obtained, the associated
density of agents defined by a given $A$\ and a given $\hat{A}$\ are:%
\begin{equation}
\left\vert \Psi \left( \mathbf{A}\right) \right\vert ^{2}=\Psi ^{\dag
}\left( \mathbf{A}\right) \Psi \left( \mathbf{A}\right)  \label{DSNV}
\end{equation}%
and:%
\begin{equation}
\left\vert \hat{\Psi}\left( \mathbf{\hat{A}}\right) \right\vert ^{2}=\hat{%
\Psi}^{\dag }\left( \mathbf{\hat{A}}\right) \hat{\Psi}\left( \mathbf{\hat{A}}%
\right)  \label{DSTV}
\end{equation}%
respectively. With these density functions at hand, we can compute various
average quantities\ in the collective state. Actually, the averages for the
system in the state defined by $\Psi \left( \mathbf{A}\right) $ and $\hat{%
\Psi}\left( \mathbf{\hat{A}}\right) $\ of components $\left( \mathbf{A}%
\right) _{k}$ or $\left( \mathbf{\hat{A}}\right) _{l}$ are:%
\begin{equation*}
\left\langle \left( \mathbf{A}\right) _{k}\right\rangle =\frac{\int \left( 
\mathbf{A}\right) _{k}\left\vert \Psi \left( \mathbf{A}\right) \right\vert
^{2}d\mathbf{A}}{\int \left\vert \Psi \left( \mathbf{A}\right) \right\vert
^{2}d\mathbf{A}}
\end{equation*}%
\begin{equation*}
\left\langle \left( \mathbf{\hat{A}}\right) _{l}\right\rangle =\frac{\int
\left( \mathbf{\hat{A}}\right) \left\vert \hat{\Psi}\left( \mathbf{\hat{A}}%
\right) \right\vert ^{2}d\mathbf{\hat{A}}}{\int \left\vert \hat{\Psi}\left( 
\mathbf{\hat{A}}\right) \right\vert ^{2}d\mathbf{\hat{A}}}
\end{equation*}%
respectively. We can also define both partial densities and averages by
integrating some components and fixing the values of others, as will be
detailled in the particular model considered in the next sections.

\subsection{Individual level: agents transition functions and their field
expression}

\subsubsection{Transition functions in a classical framework}

In a classical perspective, the statistical weight (\ref{wdh}) can be used
to compute the transition probabilities of the system, i.e. the
probabilities for any number of agents of both types to evolve from an
initial state $\left\{ \mathbf{A}_{l}\right\} _{l=1,...},\left\{ \mathbf{%
\hat{A}}_{l}\right\} _{l=1,...}$\textbf{\ }to a final state in a given
timespan. These transition functions describe the dynamic of the agents of
the system.

To do so, we first compute the integral of equation (\ref{wdh}) over all
paths between the initial and the final points considered. Defining $\left\{ 
\mathbf{A}_{l}\left( s\right) \right\} _{l=1,...,N}$ and $\left\{ \mathbf{%
\hat{A}}_{l}\left( s\right) \right\} _{l=1,...,\hat{N}}$ the sets of paths
for agents of each type, where $N$\ and $\hat{N}$\ are the numbers of agents
of each type, we consider the set of $N+\hat{N}$\ independent paths written:

\begin{equation*}
\mathbf{Z}\left( s\right) =\left( \left\{ \mathbf{A}_{l}\left( s\right)
\right\} _{l=1,...,N},\left\{ \mathbf{\hat{A}}_{l}\left( s\right) \right\}
_{l=1,...,\hat{N}}\right)
\end{equation*}%
\ The weight (\ref{wdh}) can now be written $\exp \left( -W\left( \mathbf{Z}%
\left( s\right) \right) \right) $.

The transition functions $T_{t}\left( \underline{\left( \mathbf{Z}\right) },%
\overline{\left( \mathbf{Z}\right) }\right) $\ compute the probability for
the $(N,\hat{N}$\ $)$\ agents to evolve from\ the initial points $Z\left(
0\right) \equiv \underline{\mathbf{Z}}$\ to the\ final points $Z\left(
t\right) \equiv \overline{\left( \mathbf{Z}\right) }$\ during a time span $t$%
. This probability is defined by:%
\begin{equation}
T_{t}\left( \underline{\mathbf{Z}},\overline{\left( \mathbf{Z}\right) }%
\right) =\frac{1}{\mathcal{N}}\int_{\substack{ \mathbf{Z}\left( 0\right)
\equiv \underline{\mathbf{Z}}  \\ \mathbf{Z}\left( t\right) \equiv \overline{%
\left( \mathbf{Z}\right) }}}\exp \left( -W\left( \mathbf{Z}\left( s\right)
\right) \right) \mathcal{D}\left( \mathbf{Z}\left( s\right) \right)
\label{tsn}
\end{equation}%
The integration symbol $D\mathbf{Z}\left( s\right) $\ covers all sets of $%
N\times \hat{N}$\ paths constrained by $\mathbf{Z}\left( 0\right) \equiv 
\underline{\mathbf{Z}}$\ and $\mathbf{Z}\left( t\right) \equiv \overline{%
\left( \mathbf{Z}\right) }$. The normalisation factor sets the total
probability defined by the weight (\ref{wdh}) to $1$ and is equal to:%
\begin{equation*}
\mathcal{N=}\int \exp \left( -W\left( \mathbf{Z}\left( s\right) \right)
\right) \mathcal{D}\mathbf{Z}\left( s\right)
\end{equation*}%
The interpretation of (\ref{tsn}) is straightforward. Instead of studying
the full trajectory of one or several agents, we compute their probability
to evolve from one configuration to another, and in average, the usual
trajectory approach remains valid.

Equation (\ref{tsn}) can be generalized to define the transition functions
for $k\leqslant N$\ and $\hat{k}\leqslant \hat{N}$\ agents of each type. The
initial and final points respectively for this set of $k+\hat{k}$\ agents
are written:%
\begin{equation*}
\mathbf{Z}\left( 0\right) ^{\left[ k,\hat{k}\right] }\equiv \underline{%
\mathbf{Z}}^{\left[ k,\hat{k}\right] }
\end{equation*}%
and:%
\begin{equation*}
\mathbf{Z}\left( t\right) ^{\left[ k,\hat{k}\right] }\equiv \overline{\left( 
\mathbf{Z}\right) }^{\left[ k,\hat{k}\right] }
\end{equation*}%
The transition function for these agents is written:%
\begin{equation*}
T_{t}\left( \underline{\left( \mathbf{Z}\right) }^{\left[ k,\hat{k}\right] },%
\overline{\left( \mathbf{Z}\right) }^{\left[ k,\hat{k}\right] }\right)
\end{equation*}%
and the generalization of equation (\ref{tsn}) is: \ 
\begin{equation}
T_{t}\left( \underline{\left( \mathbf{Z}\right) }^{\left[ k,\hat{k}\right] },%
\overline{\left( \mathbf{Z}\right) }^{\left[ k,\hat{k}\right] }\right) =%
\frac{1}{\mathcal{N}}\int_{\substack{ \mathbf{Z}\left( 0\right) ^{\left[ k,%
\hat{k}\right] }=\underline{\left( \mathbf{Z}\right) }^{\left[ k,\hat{k}%
\right] }  \\ \mathbf{Z}\left( t\right) ^{\left[ k,\hat{k}\right] }=%
\overline{\left( \mathbf{Z}\right) }^{\left[ k,\hat{k}\right] }}}\exp \left(
-W\left( \left( \mathbf{Z}\left( s\right) \right) \right) \right) \mathcal{D}%
\left( \left( \mathbf{Z}\left( s\right) \right) \right)  \label{krtv}
\end{equation}%
The difference with (\ref{tsn}) is that only $k$\ paths are constrained by
their initial and final points.

Ultimately, the Laplace transform of $T_{t}\left( \underline{\left( Z\right) 
}^{\left[ k,\hat{k}\right] },\overline{\left( Z\right) }^{\left[ k,\hat{k}%
\right] }\right) $ computes the - time averaged - transition function for
agents with random lifespan of mean $\frac{1}{\alpha }$, up to a factor $%
\frac{1}{\alpha }$, and is given by:%
\begin{equation}
G_{\alpha }\left( \underline{\left( \mathbf{Z}\right) }^{\left[ k,\hat{k}%
\right] },\overline{\left( \mathbf{Z}\right) }^{\left[ k,\hat{k}\right]
}\right) =\int_{0}^{\infty }\exp \left( -\alpha t\right) T_{t}\left( 
\underline{\left( \mathbf{Z}\right) }^{\left[ k,\hat{k}\right] },\overline{%
\left( \mathbf{Z}\right) }^{\left[ k,\hat{k}\right] }\right) dt  \label{krvv}
\end{equation}%
This formulation of the transition functions is relatively intractable.
Therefore, we will now propose an alternative method based on the field
model.

\subsubsection{Field-theoretic expression}

The transition functions (\ref{krtv}) and (\ref{krvv}) can be retrieved
using the$\ $field theory transition functions - or\ Green functions, which
compute the probability for a variable number $\left( k,\hat{k}\right) $\ of
agents to transition from an initial state $\underline{\left( \mathbf{%
Z,\theta }\right) }^{\left[ k,\hat{k}\right] }$\ to a final state $\overline{%
\left( \mathbf{Z,\theta }\right) }^{\left[ k,\hat{k}\right] }$, where $%
\underline{\left( \mathbf{\theta }\right) }^{\left[ k,\hat{k}\right] }$ and
\ $\overline{\left( \mathbf{\theta }\right) }^{\left[ k,\hat{k}\right] }$ are%
\textbf{\ }vectors of initial and final times for $k+\hat{k}$\textbf{\ }%
agents respectively.

We will write: 
\begin{equation*}
T_{t}\left( \underline{\left( \mathbf{Z,\theta }\right) }^{\left[ k,\hat{k}%
\right] },\overline{\left( \mathbf{Z,\theta }\right) }^{\left[ k,\hat{k}%
\right] }\right)
\end{equation*}%
\ the transition function between $\underline{\left( \mathbf{Z},\mathbf{%
\theta }\right) }^{\left[ k,\hat{k}\right] }$\ and $\overline{\left( \mathbf{%
Z},\mathbf{\theta }\right) }^{\left[ k,\hat{k}\right] }$ with $\overline{%
\left( \mathbf{\theta }\right) }_{i}<t$, $\forall i$,\ \ and: 
\begin{equation*}
G_{\alpha }\left( \underline{\left( \mathbf{Z,\theta }\right) }^{\left[ k,%
\hat{k}\right] },\overline{\left( \mathbf{Z,\theta }\right) }^{\left[ k,\hat{%
k}\right] }\right)
\end{equation*}%
\ its Laplace transform. Setting $\underline{\left( \mathbf{\theta }\right) }%
_{i}=0$\ and $\overline{\left( \mathbf{\theta }\right) }_{i}=t$\ for $%
i=1,...,k+\hat{k}$, these functions\ reduce to (\ref{krtv}) or (\ref{krvv}):
the probabilistic formalism of the transition functions is thus a particular
case of the field formalism definition. In the sequel we therefore will use
the term transition function indiscriminately.

The computation of the transition functions relies on the fact that $\exp
\left( -S\left( \Psi \right) \right) $\ itself represents a statistical
weight for the system. Gosselin, Lotz, Wambst (2020) showed that $S\left(
\Psi \right) $\ can be modified in a straightforward manner to include
source terms:%
\begin{equation}
S\left( \Psi ,J\right) =S\left( \Psi \right) +\int \left( J\left( Z,\theta
\right) \Psi ^{\dag }\left( Z,\theta \right) +J^{\dag }\left( Z,\theta
\right) \Psi \left( Z,\theta \right) \right) d\left( Z,\theta \right)
\label{SwtS}
\end{equation}%
where $J\left( Z,\theta \right) $ is an arbitrary complex function, or
auxiliary field.

Introducing $J\left( Z,\theta \right) $ in $S\left( \Psi ,J\right) $ allows
to compute the transition functions by successive derivatives. Actually, we
can show that:%
\begin{equation}
G_{\alpha }\left( \underline{\left( \mathbf{Z},\theta \right) }^{\left[ k,%
\hat{k}\right] },\overline{\left( \mathbf{Z},\theta \right) }^{\left[ k,\hat{%
k}\right] }\right) =\left[ \prod\limits_{l=1}^{k}\left( \frac{\delta }{%
\delta J\left( \underline{\left( \mathbf{Z},\theta \right) }_{i_{l}}\right) }%
\frac{\delta }{\delta J^{\dag }\left( \overline{\left( \mathbf{Z},\theta
\right) }_{i_{l}}\right) }\right) \int \exp \left( -S\left( \Psi ,J\right)
\right) \mathcal{D}\Psi \mathcal{D}\Psi ^{\dag }\right] _{J=J^{\dag }=0}
\label{trnsgrtx}
\end{equation}%
where the notation $\mathcal{D}\Psi \mathcal{D}\Psi ^{\dag }$ denotes an
integration over the space of functions $\Psi \left( Z,\theta \right) $ and $%
\Psi ^{\dag }\left( Z,\theta \right) $, i.e. an integral in an infinite
dimensional space. Even though these integrals can only be computed in
simple cases, a series expansion of $G_{\alpha }\left( \underline{\left( 
\mathbf{Z},\theta \right) }^{\left[ k,\hat{k}\right] },\overline{\left( 
\mathbf{Z},\theta \right) }^{\left[ k,\hat{k}\right] }\right) $ can be found
using Feynman graphs techniques.

Once $G_{\alpha }\left( \underline{\left( \mathbf{Z},\theta \right) }^{\left[
k,\hat{k}\right] },\overline{\left( \mathbf{Z},\theta \right) }^{\left[ k,%
\hat{k}\right] }\right) $ is computed, the expression of $T_{t}\left( 
\underline{\left( \mathbf{Z},\theta \right) }^{\left[ k,\hat{k}\right] },%
\overline{\left( \mathbf{Z},\theta \right) }^{\left[ k,\hat{k}\right]
}\right) $ can be\ retrieved in principle by an inverse Laplace transform.
In field theory, formula (\ref{trnsgrtx}) shows that the transition functions%
\textbf{\ }(\ref{krvv}) are correlation functions of the field theory with
action $S\left( \Psi \right) $.

\section{Computing Transition Functions using Field Theory}

The formula (\ref{trnsgrtx}) provides a precise and compact definition of
the transition functions for multiple agents in the system. However, in
practice, this formula is not directly applicable and does not shed much
light on the connection between the collective and microeconomic aspects of
the considered system. To calculate the dynamics of the agents, we will
proceed in three steps.\textbf{\ }

Firstly, we will minimize the system's action functional and determine the
background field, which represents the collective state of the system. Once
the background field is found, we will perform a series expansion of the
action functional around this background field, referred to as the effective
action of the system. It is with this effective action that we can compute
the transition functions for the state defined by the background field. We
will discover that each term in this expansion has an interpretation in
terms of a transition function.

Instead of directly computing the transition functions, we can consider a
series expansion of the action functional around a specific background field
of the system.

\subsection{Step 1: finding the background field}

For a particular type of agent, background fields are defined as the fields $%
\Psi _{0}\left( Z,\theta \right) $\ that maximize the statistical weight $%
\exp \left( -S\left( \Psi \right) \right) $ or, alternatively, minimize $%
S\left( \Psi \right) $: 
\begin{equation*}
\frac{\delta S\left( \Psi \right) }{\delta \Psi }\mid _{\Psi _{0}\left(
Z,\theta \right) }=0\text{, }\frac{\delta S\left( \Psi ^{\dag }\right) }{%
\delta \Psi ^{\dag }}\mid _{\Psi _{0}^{\dag }\left( Z,\theta \right) }=0
\end{equation*}%
The field $\Psi _{0}\left( Z,\theta \right) $\ represents the most probable
configuration, a specfic state of the entire system that influences the
dynamics of agents. It serves as the background state from which probability
transitions and average values can be computed. As we will see, the agents'
transitions explicitely depend on the chosen background field $\Psi
_{0}\left( Z,\theta \right) $, which represents the macroeconomic state in
which the agents evolve.

When considering two or more types of agents, the background field satisfies
the following condition:%
\begin{eqnarray*}
\frac{\delta S\left( \Psi ,\hat{\Psi}\right) }{\delta \Psi } &\mid &_{\Psi
_{0}\left( Z,\theta \right) }=0\text{, }\frac{\delta S\left( \Psi ,\hat{\Psi}%
\right) }{\delta \Psi ^{\dag }}\mid _{\Psi _{0}^{\dag }\left( Z,\theta
\right) }=0 \\
\frac{\delta S\left( \Psi ,\hat{\Psi}\right) }{\delta \hat{\Psi}} &\mid &_{%
\hat{\Psi}_{0}\left( Z,\theta \right) }=0\text{, }\frac{\delta S\left( \Psi ,%
\hat{\Psi}\right) }{\delta \hat{\Psi}^{\dag }}\mid _{\hat{\Psi}_{0}^{\dag
}\left( Z,\theta \right) }=0
\end{eqnarray*}

\subsection{Step 2: Series expansion around the background field}

In a given background state, the \emph{effective action}\footnote{%
Actually, this paper focuses on the \emph{classical} \emph{effective action}%
, which is an approximation sufficient for the computations at hand.} is the
series expansion of the field functional $S\left( \Psi \right) $\ around $%
\Psi _{0}\left( Z,\theta \right) $. We will present the expansion for one
type of agent, but generalizing it to two or several agents is
straightforward.

The series expansion around the background field simplifies the computations
of transition functions and provides an interpretation of these functions in
terms of individual interactions within the collective state.\ To perform
this series expansion, we decompose $\Psi $\ as: 
\begin{equation*}
\Psi =\Psi _{0}+\Delta \Psi
\end{equation*}%
and write the series expansion of the action functional: 
\begin{eqnarray}
S\left( \Psi \right) &=&S\left( \Psi _{0}\right) +\int \Delta \Psi ^{\dag
}\left( Z,\theta \right) O\left( \Psi _{0}\left( Z,\theta \right) \right)
\Delta \Psi \left( Z,\theta \right)  \label{SRP} \\
&&+\sum_{k>2}\int \prod\limits_{i=1}^{k}\Delta \Psi ^{\dag }\left(
Z_{i},\theta \right) O_{k}\left( \Psi _{0}\left( Z,\theta \right) ,\left(
Z_{i}\right) \right) \prod\limits_{i=1}^{k}\Delta \Psi \left( Z_{i},\theta
\right)  \notag
\end{eqnarray}%
The series expansion can be interpreted economically as follows. The first
term,\textbf{\ }$S\left( \Psi _{0}\right) $, describes the system of all
agents in a given macroeconomic state, $\Psi _{0}$. The other terms
potentially describe all the fluctuations or movements of the agents around
this macroeconomic state considered as given. Therefore, the expansion
around the background field represents the microeconomic reality of a
historical macroeconomic state. More precisely, it describes the range of
microeconomic possibilities allowed by a macroeconomic state.

The quadratic approximation is the first term of the expansion and can be
written as:%
\begin{equation}
S\left( \Psi \right) =S\left( \Psi _{0}\right) +\int \Delta \Psi ^{\dag
}\left( Z,\theta \right) O\left( \Psi _{0}\left( Z,\theta \right) \right)
\Delta \Psi \left( Z,\theta \right)  \label{kr}
\end{equation}%
\textbf{\ }It will allow us to find the transition functions of agents in
the historical macro state, where all interactions are averaged. The other
terms of the expansion allow us to detail the interactions within the
nebula, and are written as follows:%
\begin{equation*}
\sum_{k>2}\int \prod\limits_{i=1}^{k}\Delta \Psi ^{\dag }\left( Z_{i},\theta
\right) O_{k}\left( \Psi _{0}\left( Z,\theta \right) ,\left( Z_{i}\right)
\right) \prod\limits_{i=1}^{k}\Delta \Psi \left( Z_{i},\theta \right)
\end{equation*}%
\textbf{\ }They detail, given the historical macroeconomic state, how the
interactions of two or more agents can impact the dynamics of these agents.
Mathematically, this corresponds to correcting the transition probabilities
calculated in the quadratic approximation.

Here, we provide an interpretation of the third and fourth-order terms.

The third order introduces possibilities for an agent, during its
trajectory, to split into two, or conversely, for two agents to merge into
one. In other words, the third-order terms take into account or reveal, in
the historical macroeconomic environment, the possibilities for any agent to
undergo modifications along its trajectory. However, this assumption has
been excluded from our model.

The fourth order reveals that in the macroeconomic environment, due to the
presence of other agents and their tendency to occupy the same space, all
points in space will no longer have the same probabilities for an agent. In
fact, the fourth-order terms reveal the notion of geographical or sectoral
competition and potentially intertemporal competition. These terms describe
the interaction between two agents crossing paths, which leads to a
deviation of their trajectories due to the interaction.

We do not interpret higher-order terms, but the idea is similar. The
even-order terms (2n) describe interactions among n agents that modify their
trajectories. The odd-order terms modify the trajectories but also include
the possibility of agents reuniting or splitting into multiple agents. We
will see in more detail how these terms come into play in the transition
functions.

\subsection{Step 3: Computation of the transition functions}

\subsubsection{Quadratic approximation}

In the first approximation, transition functions in a given background field%
\textbf{\ }$\Psi _{0}\left( Z,\theta \right) $\textbf{\ }can be computed by
replacing $S\left( \Psi \right) $\ in (\ref{trnsgrtx}), with its quadratic
approximation (\ref{kr}). In formula (\ref{kr}), $O\left( \Psi _{0}\left(
Z,\theta \right) \right) $\ is a differential operator of second order. This
operator depends explicitly on $\Psi _{0}\left( Z,\theta \right) $. As a
consequence, transition functions and agent dynamics explicitly depend on
the collective state of the system.\ In this approximation, the formula for
the transition functions (\ref{trnsgrtx}) becomes:\ 
\begin{eqnarray}
G_{\alpha }\left( \underline{\left( Z,\theta \right) }^{\left[ k\right] },%
\overline{\left( Z,\theta \right) }^{\left[ k\right] }\right) &=&\left[
\prod\limits_{l=1}^{k}\left( \frac{\delta }{\delta J\left( \underline{\left(
Z,\theta \right) }_{i_{l}}\right) }\frac{\delta }{\delta J^{\dag }\left( 
\overline{\left( Z,\theta \right) }_{i_{l}}\right) }\right) \right. \\
&&\times \left. \int \exp \left( -\int \Delta \Psi ^{\dag }\left( Z,\theta
\right) O\left( \Psi _{0}\left( Z,\theta \right) \right) \Delta \Psi \left(
Z,\theta \right) \right) \mathcal{D}\Psi \mathcal{D}\Psi ^{\dag }\right]
_{J=J^{\dag }=0}  \notag
\end{eqnarray}%
Using this formula, we can show that the one-agent transition function is
given by:%
\begin{equation}
G_{\alpha }\left( \underline{\left( Z,\theta \right) }^{\left[ 1\right] },%
\overline{\left( Z,\theta \right) }^{\left[ 1\right] }\right) =O^{-1}\left(
\Psi _{0}\left( Z,\theta \right) \right) \left( \underline{\left( Z,\theta
\right) }^{\left[ 1\right] },\overline{\left( Z,\theta \right) }^{\left[ 1%
\right] }\right)  \label{rk}
\end{equation}%
where: 
\begin{equation*}
O^{-1}\left( \Psi _{0}\left( Z,\theta \right) \right) \left( \underline{%
\left( Z,\theta \right) }^{\left[ 1\right] },\overline{\left( Z,\theta
\right) }^{\left[ 1\right] }\right)
\end{equation*}%
\ is the kernel of the inverse operator $O^{-1}\left( \Psi _{0}\left(
Z,\theta \right) \right) $. It can be seen as the $\left( \underline{\left(
Z,\theta \right) }^{\left[ 1\right] },\overline{\left( Z,\theta \right) }^{%
\left[ 1\right] }\right) $\ matrix element of $O^{-1}\left( \Psi _{0}\left(
Z,\theta \right) \right) $\footnote{%
The differential operator $O\left( \Psi _{0}\left( Z,\theta \right) \right) $
can be seen as an infinite dimensional matrix indexed by the double
(infinite) entries $\left( \underline{\left( Z,\theta \right) }^{\left[ 1%
\right] },\overline{\left( Z,\theta \right) }^{\left[ 1\right] }\right) $.
With this description, the kernel $O^{-1}\left( \Psi _{0}\left( Z,\theta
\right) \right) \left( \underline{\left( Z,\theta \right) }^{\left[ 1\right]
},\overline{\left( Z,\theta \right) }^{\left[ 1\right] }\right) $ is the $%
\left( \underline{\left( Z,\theta \right) }^{\left[ 1\right] },\overline{%
\left( Z,\theta \right) }^{\left[ 1\right] }\right) $ element of the inverse
matrix.}.

More generally, the $k$-agents transition functions are the product of
individual transition functions:%
\begin{equation}
G_{\alpha }\left( \underline{\left( Z,\theta \right) }^{\left[ k\right] },%
\overline{\left( Z,\theta \right) }^{\left[ k\right] }\right)
=\prod\limits_{i=1}^{k}G_{\alpha }\left( \underline{\left( Z,\theta \right)
_{i}}^{\left[ 1\right] },\overline{\left( Z,\theta \right) _{i}}^{\left[ 1%
\right] }\right)  \label{gnr}
\end{equation}%
The above formula shows that in the quadratic approximation, the transition
probability from one state to another for a group is the product of
individual transition probabilities. In this approximation, the trajectories
of these agents are therefore independent. The agents do not interact with
each other and only interact with the environment described by the
background field.

The quadratic approximation must be corrected to account for individual
interactions within the group, by including higher-order terms in the
expansion of the action.

\subsubsection{Higher-order corrections}

To compute the effects of interactions between agents of a given group, we
consider terms of order greater than $2$ in the effective action. These
terms\ modify the transition functions. Writing the expansion:%
\begin{equation*}
\exp \left( -S\left( \Psi \right) \right) =\exp \left( -\left( S\left( \Psi
_{0}\right) +\int \Delta \Psi ^{\dag }\left( Z,\theta \right) O\left( \Psi
_{0}\left( Z,\theta \right) \right) \right) \right) \left(
1+\sum_{n\geqslant 1}\frac{A^{n}}{n!}\right)
\end{equation*}%
where:%
\begin{equation*}
A=\sum_{k>2}\int \prod\limits_{i=1}^{k}\Delta \Psi ^{\dag }\left(
Z_{i},\theta \right) O\left( \Psi _{0}\left( Z,\theta \right) ,\left(
Z_{i}\right) \right) \prod\limits_{i=1}^{k}\Delta \Psi \left( Z_{i},\theta
\right)
\end{equation*}%
is the sum of all possible interaction terms, leads to the series expansion
of (\ref{trnsgrtx}):%
\begin{eqnarray}
G_{\alpha }\left( \underline{\left( Z,\theta \right) }^{\left[ k\right] },%
\overline{\left( Z,\theta \right) }^{\left[ k\right] }\right) &=&\left[
\prod\limits_{l=1}^{k}\left( \frac{\delta }{\delta J\left( \underline{\left(
Z,\theta \right) }_{i_{l}}\right) }\frac{\delta }{\delta J^{\dag }\left( 
\overline{\left( Z,\theta \right) }_{i_{l}}\right) }\right) \right.
\label{trg} \\
&&\left. \int \exp \left( -\int \Delta \Psi ^{\dag }\left( Z,\theta \right)
O\left( \Psi _{0}\left( Z,\theta \right) \right) \Delta \Psi \left( Z,\theta
\right) \right) \left( 1+\sum_{n\geqslant 1}\frac{A^{n}}{n!}\right) \mathcal{%
D}\Psi \mathcal{D}\Psi ^{\dag }\right] _{J=J^{\dag }=0}  \notag
\end{eqnarray}%
These corrections can be computed using graphs' expansion. \ 

More precisely, the first term of the series:%
\begin{equation}
\left[ \prod\limits_{l=1}^{k}\left( \frac{\delta }{\delta J\left( \underline{%
\left( Z,\theta \right) }_{i_{l}}\right) }\frac{\delta }{\delta J^{\dag
}\left( \overline{\left( Z,\theta \right) }_{i_{l}}\right) }\right) \int
\exp \left( -\int \Delta \Psi ^{\dag }\left( Z,\theta \right) O\left( \Psi
_{0}\left( Z,\theta \right) \right) \Delta \Psi \left( Z,\theta \right)
\right) \mathcal{D}\Psi \mathcal{D}\Psi ^{\dag }\right] _{J=J^{\dag }=0}
\end{equation}%
is a transition function in the quadratic approximation. The other
contributions of the series expansion correct the approximated $n$ agents
transtns functions (\ref{gnr}).

Typically a contribution:%
\begin{eqnarray}
G_{\alpha }\left( \underline{\left( Z,\theta \right) }^{\left[ k\right] },%
\overline{\left( Z,\theta \right) }^{\left[ k\right] }\right) &=&\left[
\prod\limits_{l=1}^{k}\left( \frac{\delta }{\delta J\left( \underline{\left(
Z,\theta \right) }_{i_{l}}\right) }\frac{\delta }{\delta J^{\dag }\left( 
\overline{\left( Z,\theta \right) }_{i_{l}}\right) }\right) \right. \\
&&\left. \int \exp \left( -\int \Delta \Psi ^{\dag }\left( Z,\theta \right)
O\left( \Psi _{0}\left( Z,\theta \right) \right) \Delta \Psi \left( Z,\theta
\right) \right) \frac{A^{n}}{n!}\mathcal{D}\Psi \mathcal{D}\Psi ^{\dag }%
\right] _{J=J^{\dag }=0}  \notag
\end{eqnarray}%
can be depicted by a graph. The power $\frac{A^{n}}{n!}$ translates that
agents interact $n$ times along their path. The trajectories of each agent
of the group is broken $n$ times between its initial and final points. At
each time of interaction the trajectories of agents are deviated. To such a
graph is associated a probility that modifies the quadratic approximation
transition functions.

In the sequel we will only focus on the first order corrections to the
two-agents transition functions.

\part*{Part 1. System of investors and firms}

\section{Microeconomic framework}

Let us now present the microeconomic framework that will be turned into a
field model using our general method. The interactions between the real and
the financial economy are pictured by two groups of agents, producers, and
investors. In the following, we will refer to producers or firms $i$
indistinctively, and use the upper script $\symbol{94}$ for variables
describing investors.

\subsection{Investors' allocation of capital}

Each investor, denoted by $j$, is characterized by their level of disposable
capital, $\hat{K}_{j}${}, and their position, $\hat{X}_{j}${}, in the sector
space. The disposable capital for each investor at time $t$ is the sum of
their private capital plus the total amount of participations and loans
entrusted to them by other investors.

Investors have the flexibility to invest this disposable capital across the
entire sector space, but they typically prefer sectors that are close to
their position. At the end of each period, any lent capital is repaid with
interest, if feasible.

Investors tend to diversify their capital in four ways: each investor $j$
chooses to allocate his entire capital $\hat{K}_{j}$ among various firms $i$
or among other investors. In each scenario, this allocation can manifest
either as a participation in the ownership of the firm or as a credit
extended to other investors at a specified interest rate.

\subsubsection{Allocation to firms}

The capital allocated by investor $j$ to firm $i$, denoted as $\hat{K}%
_{j}^{\left( i\right) }$, is given by:

\begin{equation}
\hat{K}_{j}^{\left( i\right) }\left( t\right) =\left( \hat{F}_{2}\left(
R_{i},\hat{X}_{j}\right) \hat{K}_{j}\right) \left( t\right)  \label{grandf2}
\end{equation}%
where:%
\begin{equation}
\hat{F}_{2}\left( R_{i},\hat{X}_{j}\right) =\frac{F_{2}\left( R_{i}\right)
G\left( X_{i}-\hat{X}_{j}\right) }{\sum_{l}F_{2}\left( R_{l}\right) G\left(
X_{l}-\hat{X}_{j}\right) }  \label{FRLV}
\end{equation}%
where $F_{2}$ is an arbitrary function that depends on the expected return
of firm $i$ and the distance between sectors $X_{i}$ and $\hat{X}_{j}$. The
function $\hat{F}_{2}\left( R\left( K_{i},X_{i}\right) ,\hat{X}_{j}\right) $%
\ is thus the function $F_{2}$, expressed as share of invested capital in a
specific firm. It captures the dependency of investments on firms' relative
attractivity. The equation (\ref{grandf2}) is thus a general form of
risk-averse portfolio allocation\footnote{%
Actually, an investor allocating capital exclusively in a sector $X_{i}$ and
optimizing the function:%
\begin{equation}
\frac{R_{i}}{\sum_{l}R_{l}}s_{j}-s_{j}^{2}Var\left( \frac{R_{i}}{%
\sum_{l}R_{l}}\right)  \label{tlr}
\end{equation}%
where the share $s_{j}$\ satisfies $\sum s_{j}=1$,\ will set $s_{j}=\frac{%
R_{i}}{\sum_{l}R_{l}}$. If we were to introduce the possibility of investing
in multiple sectors and consider more general preferences than this simple
quadratic function, we should introduce the functions $G\left( X_{i}-\hat{X}%
_{j}\right) $\ and $F_{2}\left( R_{i}\right) $\ in the solutions of (\ref%
{tlr}), leading to (\ref{grandf2}).}\textbf{.}

Each agent allocated capital is divided between loans and equity
participation, expressed as:%
\begin{equation*}
\left( \hat{F}_{2}\left( R_{i},\hat{X}_{j}\right) \hat{K}_{j}\right) \left(
t\right) =k_{1}\left( X_{i},\hat{X}_{j}\right) \left( \hat{F}_{2}\left(
R_{i},\hat{X}_{j}\right) \hat{K}_{j}\right) \left( t\right) +k_{2}\left(
X_{i},\hat{X}_{j}\right) \left( \hat{F}_{2}\left( R_{i},\hat{X}_{j}\right) 
\hat{K}_{j}\right) \left( t\right)
\end{equation*}%
where $k_{1ij}$ represents the proportion allocated to loans and $k_{2ij}$
represents the proportion allocated to equity participation, such that $%
k_{1}\left( X_{i},\hat{X}_{j}\right) +k_{2}\left( X_{i},\hat{X}_{j}\right)
=1 $. The allocated capital (\ref{grandf2}) constitutes only a portion of
investor $j$'s total capital. Below, we will express it as:%
\begin{equation*}
k_{i}\left( X_{i},\hat{X}_{j}\right) \left( \hat{F}_{2}\left( R_{i},\hat{X}%
_{j}\right) \hat{K}_{j}\right) \rightarrow k_{i}\left( X_{i},\hat{X}%
_{j}\right)
\end{equation*}

\subsubsection{Allocation to other investors}

The rest of capital is invested between various other investors. Investor $j$
acquires equity stakes in other investors, up to the amount:%
\begin{equation*}
\sum_{l}\hat{k}_{1}\left( \hat{K}_{l}\left( t\right) ,\hat{K}_{j}\left(
t\right) \right) \hat{K}_{j}\left( t\right)
\end{equation*}%
where $\hat{k}_{1}\left( \hat{K}_{l}\left( t\right) ,\hat{K}_{j}\left(
t\right) \right) \hat{K}_{j}\left( t\right) $ represents the capital
invested by investor $j$ in the investor $l$.\ Investor $j$ also lends a
portion of its capital to other investors, up to an amount:%
\begin{equation*}
\sum_{l}\hat{k}_{2}\left( \hat{K}_{l}\left( t\right) ,\hat{K}_{j}\left(
t\right) \right) \hat{K}_{j}\left( t\right)
\end{equation*}%
where $\hat{k}_{2}\left( \hat{K}_{l}\left( t\right) ,\hat{K}_{j}\left(
t\right) \right) \hat{K}_{j}\left( t\right) $ represents the capital lent
from investor $j$ to investor $l$, and $\hat{k}_{i}\left( \hat{X}_{l}\left(
t\right) ,\hat{X}_{j}\left( t\right) \right) $\ depends on the return $R_{l}$
provided by investor $l$: 
\begin{equation*}
\hat{k}_{i}\left( \hat{X}_{l}\left( t\right) ,\hat{X}_{j}\left( t\right)
,R_{l}\right)
\end{equation*}%
Thus $\hat{K}_{j}\left( t\right) $ decomposes as: 
\begin{eqnarray*}
\hat{K}_{j}\left( t\right) &=&\sum_{l}\hat{k}_{1}\left( \hat{K}_{l}\left(
t\right) ,\hat{K}_{j}\left( t\right) \right) \hat{K}_{j}\left( t\right)
+\sum_{l}\hat{k}_{2}\left( \hat{K}_{l}\left( t\right) ,\hat{K}_{j}\left(
t\right) \right) \hat{K}_{j}\left( t\right) +\sum_{i}\left( \hat{F}%
_{2,1}\left( R_{i},\hat{X}_{j}\right) \hat{K}_{j}\right) \\
&=&\sum_{l}\hat{k}_{1}\left( \hat{K}_{l}\left( t\right) ,\hat{K}_{j}\left(
t\right) \right) \hat{K}_{j}\left( t\right) +\sum_{l}\hat{k}_{2}\left( \hat{K%
}_{l}\left( t\right) ,\hat{K}_{j}\left( t\right) \right) \hat{K}_{j}\left(
t\right) \\
&&+\sum_{i}\left( k_{1}\left( X_{i},K_{i},\hat{X}_{j}\right) +k_{2}\left(
X_{i},K_{i},\hat{X}_{j}\right) \right) \hat{K}_{j}\left( t\right)
\end{eqnarray*}%
or equivalently:%
\begin{equation*}
\sum_{l}\hat{k}\left( \hat{X}_{l}\left( t\right) ,\hat{X}_{j}\left( t\right)
\right) +\sum_{i}\hat{F}_{2,1}\left( R_{i},\hat{X}_{j}\right) =1
\end{equation*}%
with:%
\begin{equation*}
\hat{k}\left( \hat{X}_{l}\left( t\right) ,\hat{X}_{j}\left( t\right) \right)
=\hat{k}_{1}\left( \hat{K}_{l}\left( t\right) ,\hat{K}_{j}\left( t\right)
\right) +\hat{k}_{2}\left( \hat{K}_{l}\left( t\right) ,\hat{K}_{j}\left(
t\right) \right)
\end{equation*}%
Investor $j$'s private capital, denoted as $\hat{K}_{jp}\left( t\right) $,
can be expressed as a function of disposable capital, $\hat{K}_{j}\left(
t\right) $, which includes both investor $j$'s private capital, $\hat{K}%
_{jp}\left( t\right) $, and the capital entrusted to him by other
investors.\ We can decompose the latter into shares of other investors in
investor $j$, and lent capital. Disposable capital $\hat{K}_{j}\left(
t\right) $ of investor $j$ thus writes:

\begin{equation*}
\hat{K}_{j}\left( t\right) =\hat{K}_{jp}\left( t\right) +\sum_{l}\left( \hat{%
k}_{1}\left( \hat{K}_{jp}\left( t\right) ,\hat{X}_{j}\left( t\right) ,\hat{X}%
_{l}\left( t\right) \right) +\hat{k}_{2}\left( \hat{K}_{jp}\left( t\right) ,%
\hat{X}_{j}\left( t\right) ,\hat{X}_{l}\left( t\right) \right) \right) \hat{K%
}_{l}\left( t\right) 
\end{equation*}%
Inversely, private capital $\hat{K}_{jp}$ can be expressed, in the linear
approximation\footnote{%
See Appendix 1.}, as :%
\begin{equation}
\hat{K}_{jp}\left( t\right) =\frac{\hat{K}_{j}\left( t\right) }{%
1+\sum_{l}\left( \hat{k}_{1jl}+\hat{k}_{2jl}\right) \hat{K}_{l}\left(
t\right) }  \label{NK}
\end{equation}%
where:%
\begin{eqnarray*}
\hat{k}_{ajl} &=&\hat{k}_{a}\left( \hat{X}_{j}\left( t\right) ,\hat{X}%
_{l}\left( t\right) \right)  \\
k_{ajl} &=&k_{a}\left( X_{j}\left( t\right) ,\hat{X}_{l}\left( t\right)
\right) 
\end{eqnarray*}%
and:%
\begin{eqnarray*}
\hat{k}_{jl} &=&\hat{k}_{1jl}+\hat{k}_{2jl} \\
k_{jl} &=&k_{1jl}+k_{2jl}
\end{eqnarray*}

\subsubsection{Investors' disposable capital}

The investors' disposable capital can be either decomposed according to its
source or to its allocation.

Disposable capital comes from the investor private capital plus the capital
entrusted to him by other investors. As such, total disposable capital $\hat{%
K}_{j}\left( t\right) $ can be expressed as:%
\begin{equation*}
\hat{K}_{j}\left( t\right) =\frac{\hat{K}_{j}\left( t\right) }{1+\sum_{v}%
\hat{k}_{jv}\hat{K}_{v}\left( t\right) }+\frac{\sum_{v}\hat{k}_{2jv}\hat{K}%
_{j}\left( t\right) \hat{K}_{v}\left( t\right) }{1+\sum_{v}\hat{k}_{jv}\hat{K%
}_{v}\left( t\right) }+\frac{\sum_{v}\hat{k}_{1jv}\hat{K}_{j}\left( t\right) 
\hat{K}_{v}\left( t\right) }{1+\sum_{v}\hat{k}_{jv}\hat{K}_{v}\left(
t\right) }
\end{equation*}%
that are private capital, loans, and participations from other investors
respectively.

Disposable capital is decomposed into several investments.\ It writes:%
\begin{eqnarray*}
\hat{K}_{j}\left( t\right) &=&\sum_{i}k_{1ij}K_{ip}\hat{K}_{j}\left(
t\right) +\sum_{i}k_{2ij}K_{ip}\hat{K}_{j}\left( t\right) +\sum_{l}\hat{k}%
_{1lj}\hat{K}_{lp}\left( t\right) \hat{K}_{j}\left( t\right) +\sum_{l}\hat{k}%
_{2lj}\hat{K}_{lp}\left( t\right) \hat{K}_{j}\left( t\right) \\
&=&\sum_{i}\frac{k_{1ij}K_{i}}{1+\sum_{v}k_{iv}\hat{K}_{v}\left( t\right) }%
\hat{K}_{j}\left( t\right) +\sum_{i}\frac{k_{2ij}K_{i}}{1+\sum_{v}k_{iv}\hat{%
K}_{v}\left( t\right) }\hat{K}_{j}\left( t\right) \\
&&+\sum_{l}\frac{\hat{k}_{1lj}\hat{K}_{l}\left( t\right) }{1+\sum_{v}\hat{k}%
_{jv}\hat{K}_{v}\left( t\right) }\hat{K}_{j}\left( t\right) +\sum_{l}\frac{%
\hat{k}_{2lj}\hat{K}_{l}\left( t\right) }{1+\sum_{v}\hat{k}_{jv}\hat{K}%
_{v}\left( t\right) }\hat{K}_{j}\left( t\right)
\end{eqnarray*}%
and implies the constraint, for all $j$:%
\begin{equation*}
\sum_{i}\frac{k_{ij}K_{i}}{1+\sum_{v}k_{iv}\hat{K}_{v}\left( t\right) }%
+\sum_{l}\frac{\hat{k}_{lj}\hat{K}_{l}\left( t\right) }{1+\sum_{v}\hat{k}%
_{jv}\hat{K}_{v}\left( t\right) }=1
\end{equation*}

\subsubsection{Firms' disposable capital}

Firms' disposable capital is defined by the sum of their private capital,
investor participations, and loans. In the linear approximation and assuming
investments are proportional to $K_{ip}\left( t\right) $, disposable capital
for firm $i$ is defined by:%
\begin{equation}
K_{i}\left( t\right) =K_{ip}\left( t\right) +\left( \sum_{v}\left(
k_{1iv}+k_{2iv}\right) \hat{K}_{v}\left( t\right) \right) K_{ip}\left(
t\right)  \label{CD}
\end{equation}%
where $K_{ip}$ is the private capital, which also writes:%
\begin{eqnarray*}
K_{ip}\left( t\right) &=&\frac{K_{i}\left( t\right) }{1+\sum_{v}\left(
k_{1iv}+k_{2iv}\right) \hat{K}_{v}\left( t\right) } \\
&=&\frac{K_{i}\left( t\right) }{1+\sum_{v}k_{iv}\hat{K}_{v}\left( t\right) }
\end{eqnarray*}

\subsection{Capital accumulation under a no-default scenario}

\subsubsection{Firms' private returns}

Producers are a set of firms, operating each in a single sector, so that a
single firm with subsidiaries in different countries and/or offering
differentiated products can be modeled as a set of independent firms.
Similarly, a sector refers to a set of firms with similar productions, so
that sectors can be decomposed into sectors per country to account for local
specificities, or in several sectors.

Firms move across a vector space of sectors, which is of arbitrary
dimension. Firms are defined by their respective sector $X_{i}$\ and
physical capital $K_{i}$, two variables subject to dynamic changes. They may
change their capital stocks over time or altogether shift sectors.

Each firm produces a single differentiated good.\ However, rather than
dealing with a full system of producers-consumers, we will merely consider
the return each producer may provide to its investors. This is sufficient
for our goal to study financial capital circulation and diffusion.

The return of producer $i$ at time $t$, denoted $r_{i}$, depends on $K_{i}$, 
$X_{i}$ and on the level of competition in the sector.\ It is written: 
\begin{equation}
r_{i}=r\left( K_{i},X_{i}\right) -\gamma \sum_{j}\delta \left(
X_{i}-X_{j}\right) \frac{K_{j}}{K_{i}}  \label{dvd}
\end{equation}%
where $\delta \left( X_{i}-X_{j}\right) $ is the Dirac $\delta $ function
which is equal to $0$ for \ $X_{i}\neq X_{j}$.\ \ The first term in formula (%
\ref{dvd}) is an arbitrary function that depends on the sector and the level
of capital per firm in this sector.\ It represents the return of capital in
a specific sector $X_{i}$ under no competition. We deliberately keep the
form of $r\left( K_{i},X_{i}\right) $\ unspecified, since most of the
results of the model rely on the general properties of the functions
involved. When needed, we will give a standard Cobb-Douglas form to\ the
returns $r\left( K_{i},X_{i}\right) $.\ The second term in (\ref{dvd}) is
the decreasing return of capital. In any given sector, it is proportional to
both the number of competitors and the specific level of capital per firm
used.

We also assume that, for all $i$, firm $i$ has a market valuation defined by
both its price, $P_{i}$, and the variation of this price on financial
markets, $\dot{P}_{i}$.\ This variation is itself assumed to be a function
of an expected long-term return denoted $R_{i}$, or more precisely the
relative return $\bar{R}_{i}$\ of firm $i$\ against the whole set of firms: 
\begin{equation}
\frac{\dot{P}_{i}}{P_{i}}=F_{1}\left( \bar{R}_{i},\frac{\dot{K}_{i}}{K_{i}}%
-\left\langle \frac{\dot{K}_{i}}{K_{i}}\right\rangle \right)  \label{pr}
\end{equation}%
where $\frac{\dot{K}_{i}}{K_{i}}-\left\langle \frac{\dot{K}_{i}}{K_{i}}%
\right\rangle $ computes the relative variation of capital with respect to
the average.

Formula (\ref{pr}) includes the main features of models of price dynamics.
In this equation, the time dependency of variables is implicit.

Formula (\ref{pr}) reflects the impact of capital and location on the price
of the firm through its expected returns.\ It also reflects how variations
in capital impact its growth prospects, through competition and dividends
(see (\ref{dvd})). Actually, the higher the capital of the firm, the lower
impact of competition and the higher the dividends.

We assume\textbf{\ }$R_{i}$ to have the general form:%
\begin{equation*}
R_{i}=R\left( K_{i},X_{i}\right)
\end{equation*}%
Expected long-term returns depend on the capital and sector in which the
firm operates, but also on external parameters, such as technology, ...
which are encoded in the shape of $R\left( K_{i},X_{i}\right) $.

The relative return $\bar{R}_{i}$ arising in (\ref{pr}) is defined by:%
\begin{equation}
\bar{R}_{i}=\bar{R}\left( K_{i},X_{i}\right) =\frac{R_{i}}{\sum_{l}R_{l}}
\label{RBR}
\end{equation}%
The function $F_{1}$ in (\ref{pr}) is arbitrary and reflects the preferences
of the market relatively to the firm's relative returns. We will allow the
number of firms per sector to vary around some sector-dependent exogeneous
density\footnote{%
Here, we depart from our previous paper, as we do not assume that firms
relocate in the sector space according to returns. Instead, we are
interested in financial capital allocation, which operates on a shorter time
scale compared to firm dynamics.}.

\subsubsection{Firms' returns to investors}

The total return of the firm to the investor is given by the sum:%
\begin{equation*}
r_{i}^{\prime }\left( K_{i}\left( t\right) ,\frac{\dot{K}_{i}}{K_{i}}\right)
=r_{i}+F_{1}\left( \bar{R}_{i},\frac{\dot{K}_{i}}{K_{i}}\right)
\end{equation*}%
where $r_{i}$\ represents dividends, and $F_{1}\left( \bar{R}_{i},\frac{\dot{%
K}_{i}}{K_{i}}\right) $\ denotes the variation in stock price.

We rewrite $r_{i}^{\prime }\left( K_{i}\left( t\right) ,\frac{\dot{K}_{i}}{%
K_{i}}\right) $ by specifying\footnote{%
See Appendix 1.} the form of $F_{1}\left( \bar{R}_{i},\frac{\dot{K}_{i}}{%
K_{i}}\right) $, and find:%
\begin{equation*}
r_{i}^{\prime }\left( K_{i}\left( t\right) ,\frac{\dot{K}_{i}}{K_{i}}\right)
=r_{i}+F_{1}\left( \bar{R}\left( K_{i},X_{i}\right) \right) +\tau \left( 
\bar{R}\left( K_{i},X_{i}\right) \right) \Delta f_{1}^{\prime }\left(
K_{i}\left( t\right) \right) 
\end{equation*}%
with:%
\begin{equation*}
\Delta f_{1}^{\prime }\left( K_{i}\left( t\right) \right) =f_{1}^{\prime
}\left( K_{i}\left( t\right) \right) -\left\langle f_{1}^{\prime }\left(
K_{i}\left( t\right) \right) \right\rangle 
\end{equation*}%
Here, the average $\left\langle f_{1}^{\prime }\left( K_{i}\left( t\right)
\right) \right\rangle $ is calculated over all firms. The expression for $%
r_{i}^{\prime }\left( K_{i}\left( t\right) ,\frac{\dot{K}_{i}}{K_{i}}\right) 
$ reflects that the return on the variation of stock prices depends on
expectations of returns $F_{1}\left( \bar{R}\left( K_{i},X_{i}\right)
\right) $ and firm growth.

\subsubsection{Firms' private capital accumulation}

The gain realized by the firm corresponds to its private capital plus its
loans muliplied by the return, from which we substract the amount paid back
for loans, where $\bar{r}$ is the loan interest rate. The increase in
capital is thus:

\begin{equation*}
K_{ip}\left( t+1\right) -K_{ip}\left( t\right) =\left( 1+\sum_{\nu }k_{2j\nu
}\hat{K}_{\nu }\left( t\right) \right) r_{i}K_{ip}\left( t\right) -\bar{r}%
\sum_{v}k_{2lv}\hat{K}_{v}\left( t\right) K_{ip}\left( t\right)
\end{equation*}%
or, in a continuous approximation, as:%
\begin{equation}
\frac{d}{dt}K_{ip}\left( t\right) =f_{1}^{\prime }\left( K_{i}\left(
t\right) \right) K_{ip}\left( t\right)  \label{DC}
\end{equation}%
with:%
\begin{equation}
f_{1}^{\prime }\left( K_{i}\left( t\right) \right) =\left( 1+\sum_{\nu
}k_{2j\nu }\hat{K}_{\nu }\left( t\right) \right) r_{i}-\bar{r}\sum_{v}k_{2lv}%
\hat{K}_{v}\left( t\right)  \label{Dc}
\end{equation}

\subsubsection{Investors' returns}

For investors, we focus on the dynamics of disposable capital. The investor
borrows and manages the stakes entrusted to them, so their investment
decisions commit sums that far exceed their private capital. Unlike firms,
which produce, the investor redistributes this available capital, which in
turn contributes to available capital. The return for $\hat{K}_{j}\left(
t\right) $, denoted $R_{j}$ decomposes into two parts.

\paragraph{Direct returns to investor $j$}

First, returns directly obtained by agent $j$:%
\begin{equation*}
\left( 1+\sum_{v}\hat{k}_{2jv}\hat{K}_{v}\left( t\right) \right) \hat{K}%
_{jp}\left( t\right) R_{j}^{\prime }
\end{equation*}%
Here $R_{j}^{\prime }$ is the return obtained by investing in firms, funds,
or lending capital, written in terms of disposable capital, as:%
\begin{equation*}
\frac{1+\sum_{v}\hat{k}_{2jv}\hat{K}_{v}\left( t\right) }{1+\sum_{v}\left( 
\hat{k}_{1jv}+\hat{k}_{2jv}\right) \hat{K}_{v}\left( t\right) }R_{j}^{\prime
}\hat{K}_{j}\left( t\right)
\end{equation*}%
and per unit of capital, as :%
\begin{equation*}
\frac{1+\sum_{v}\hat{k}_{2jv}\hat{K}_{v}\left( t\right) }{1+\sum_{v}\left( 
\hat{k}_{1jv}+\hat{k}_{2jv}\right) \hat{K}_{v}\left( t\right) }R_{j}^{\prime
}
\end{equation*}

\paragraph{Indirect returns due to other investors}

The other component of the return comes from investing in other investors,
yielding a return per unit of disposable capital:%
\begin{equation*}
\sum_{l}\frac{\hat{k}_{1lj}\hat{K}_{l}\left( t\right) }{1+\sum_{v}\left( 
\hat{k}_{1lv}+\hat{k}_{2lv}\right) \hat{K}_{v}\left( t\right) }R_{l}
\end{equation*}%
where $R_{l}$ is the return of investors $l$ in which agent $j$ has invested.

The total return therefore satisfies: 
\begin{equation}
R_{j}=R_{j}^{\prime }+\sum_{l}\frac{\hat{k}_{1lj}\hat{K}_{l}\left( t\right) 
}{1+\sum_{v}\left( \hat{k}_{1lv}+\hat{k}_{2lv}\right) \hat{K}_{v}\left(
t\right) }R_{l}  \label{RTS}
\end{equation}%
The direct return $R_{j}^{\prime }$ decomposes into loans to firms, loans to
investors, and participations in firms and other investors:%
\begin{eqnarray}
R_{j}^{\prime } &=&\bar{r}\sum_{l}\frac{\hat{k}_{2lj}\hat{K}_{l}\left(
t\right) }{1+\sum_{v}\left( \hat{k}_{1lv}+\hat{k}_{2lv}\right) \hat{K}%
_{v}\left( t\right) }+\bar{r}\sum_{i}\frac{k_{2ij}K_{i}\left( t\right) }{%
1+\sum_{v}\left( k_{1iv}+k_{2iv}\right) \hat{K}_{v}\left( t\right) }  \notag
\\
&&+\sum_{i}\left( r_{i}+F_{1}\left( \bar{R}_{i},\frac{\dot{K}_{i}\left(
t\right) }{K_{i}\left( t\right) }\right) +\tau \left( \bar{R}\left(
K_{i},X_{i}\right) \right) \Delta f_{1}^{\prime }\left( K_{i}\left( t\right)
\right) \right) \frac{k_{1ij}K_{i}\left( t\right) }{1+\sum_{v}\left(
k_{1iv}+k_{2iv}\right) \hat{K}_{v}\left( t\right) }  \label{PMj}
\end{eqnarray}

\subsubsection{Investors' capital accumulation}

Appendix 1.4 shows that, in the continuous approximation, the dynamics for
an investor $i$ disposable capital is :%
\begin{equation}
\frac{d}{dt}\hat{K}_{j}\left( t\right) =\sum_{l}\left( 1-\frac{\hat{k}_{1jl}%
\hat{K}_{j}\left( t\right) }{1+\sum_{v}\hat{k}_{jv}\hat{K}_{v}\left(
t\right) }\right) ^{-1}\hat{f}_{l}\hat{K}_{l}\left( t\right) =\sum_{l}\left(
1-M\right) _{jl}^{-1}\hat{f}_{l}\hat{K}_{l}\left( t\right)  \label{DFQTW}
\end{equation}%
where $\hat{f}_{j}$ represents the total return of investor $j$, calculated
as the sum of all capital committed by the investor multiplied by their
return $R_{j}$, minus the interest payments on loans:

\begin{equation}
\hat{f}_{j}=\left( 1+\sum_{v}\hat{k}_{2jv}\hat{K}_{v}\left( t\right) \right)
R_{j}-\bar{r}\sum_{v}\hat{k}_{2jv}\hat{K}_{v}\left( t\right)
\end{equation}%
and matrix $M$ corresponds to the shares taken by investor $j$ in other
investors. This is why the return of investor $j$ depends on other investors
in the equation (\ref{DFQTW}).\ The expression for $M$ is:

\begin{equation}
M_{jm}=\frac{\hat{k}_{1jm}\hat{K}_{j}\left( t\right) }{1+\sum_{\nu }\hat{k}%
_{j\nu }\hat{K}_{\nu }\left( t\right) }  \label{MAT}
\end{equation}%
which indeed corresponds to the share of participation of investor $j$ in
investor $m$.

We also show that the return equation writes:%
\begin{eqnarray}
&&\frac{\hat{f}_{j}-\bar{r}}{1+\sum_{v}\hat{k}_{2jv}\hat{K}_{v}\left(
t\right) }=\sum_{l}\left( \frac{\hat{k}_{1lj}\hat{K}_{l}\left( t\right) }{%
1+\sum_{v}\left( \hat{k}_{1lv}+\hat{k}_{2lv}\right) \hat{K}_{v}\left(
t\right) }\right) \left( \frac{\hat{f}_{l}-\bar{r}}{1+\sum_{v}\hat{k}_{2lv}%
\hat{K}_{v}\left( t\right) }\right)  \label{RTq} \\
&&+\sum_{i}\frac{\left( r_{i}+F_{1}\left( \bar{R}_{i},\frac{\dot{K}%
_{i}\left( t\right) }{K_{i}\left( t\right) }\right) +\tau \left( \bar{R}%
\left( K_{i},X_{i}\right) \right) \Delta f_{1}^{\prime }\left( K_{i}\left(
t\right) \right) -\bar{r}\right) k_{1ij}K_{i}\left( t\right) }{%
1+\sum_{v}\left( k_{1iv}+k_{2iv}\right) \hat{K}_{v}\left( t\right) }  \notag
\end{eqnarray}

This equation decomposes the investor's excess return over the interest rate
into two components: the returns to the investor from the returns of other
investors in which it has taken a stake, plus the direct returns from the
firms in which it has taken a stake.

\subsection{Capital accumulation under a default scenario}

Until now, we had derived the equations for capital accumulation under the
assumption of zero default. Loans were repaid, and there was no loss of
capital. We will now introduce the possibility of defaults occuring in our
setup.

\subsubsection{Defaults of firms}

Firms' capital accumulation (\ref{DC}) is only valid when private capital at
the end of each period exceeds the loans to be repaid. When this is not the
case, the firm defaults. The default condition is that the overall return $%
R_{j}$, on private capital, private loans, and bank loans, does not generate
sufficient funds to repay with interest the amount borrowed at rate $\bar{r}$%
, which is expressed as:%
\begin{equation*}
K_{ip}\left( 1+\sum_{m}k_{2im}\hat{K}_{m}\left( t\right) \right) \left(
1+f_{1}\left( K_{i}\right) \right) <K_{ip}\left( 1+\bar{r}\right)
\sum_{m}k_{2im}\hat{K}_{m}\left( t\right)
\end{equation*}%
This can also be expressed as:%
\begin{equation}
\frac{1+f_{1}\left( K_{i}\right) }{\sum_{m}k_{2im}\hat{K}_{m}\left( t\right) 
}+f_{1}\left( K_{i}\right) <\bar{r}  \label{fm}
\end{equation}%
This inequality is the condition for the firm to repay its loans and avoid
default. Given (\ref{fm}), the default condition writes in a compact form:%
\begin{equation*}
1+f_{1}^{\prime }\left( K_{i}\left( t\right) \right) <0
\end{equation*}%
that is when returns and private capital cannot cover loans anymore. When
this inequality is satisfied, the lenders suffer losses. The difference
between final capital and loaned capital writes:%
\begin{eqnarray*}
&&K_{ip}\left( 1+\sum_{m}k_{2im}\hat{K}_{m}\left( t\right) \right) \left(
1+f_{1}\left( K_{i}\right) \right) -K_{ip}\left( 1+\bar{r}\right)
\sum_{m}k_{2im}\hat{K}_{m}\left( t\right) \\
&=&\left( 1+f_{1}^{\prime }\left( K_{i}\left( t\right) \right) \right) K_{ip}
\end{eqnarray*}%
The last expression represents the amount of capital that cannot be repaid
to the investors. Each of them will therefore incur a loss proportional to
the loan they granted, relative to the total loans. Investor $j$ thus
suffers an overall loss of:%
\begin{equation}
k_{2ij}K_{ip}\left( 1+f_{1}^{\prime }\left( K_{i}\left( t\right) \right)
\right) \hat{K}_{j}\left( t\right) =\left( 1+f_{1}^{\prime }\left(
K_{i}\left( t\right) \right) \right) \frac{k_{2ij}K_{i}\left( t\right) }{%
1+\sum_{v}k_{iv}\hat{K}_{v}\left( t\right) }\hat{K}_{j}\left( t\right)
\label{Ls}
\end{equation}%
corresponding to a loss per unit of investment.of: 
\begin{equation*}
\left( 1+f_{1}^{\prime }\left( K_{i}\left( t\right) \right) \right) \frac{%
k_{2ij}K_{i}\left( t\right) }{1+\sum_{v}k_{iv}\hat{K}_{v}\left( t\right) }
\end{equation*}%
The loss (\ref{Ls}) will be accounted for in the investors return by adding
to investor $j$'s return equation a contribution:%
\begin{equation}
\sum_{i}\left( \bar{r}-\frac{\left( 1+f_{1}^{\prime }\left( K_{i}\left(
t\right) \right) \right) }{\sum_{v}k_{iv}\hat{K}_{v}\left( t\right) }\right)
H\left( -\left( 1+\hat{f}\left( \hat{K}_{vp}\left( t\right) \right) \right)
\right) \frac{k_{2ij}K_{i}\left( t\right) }{1+\sum_{v}k_{iv}\hat{K}%
_{v}\left( t\right) }\hat{K}_{j}\left( t\right)  \label{P2}
\end{equation}%
where $H$ is the Heaviside function.

\subsubsection{Defaults of investors}

An investor defaults when the total private capital is insufficient to repay
the loans.%
\begin{equation*}
\left( 1+\sum_{m}\hat{k}_{2vm}\hat{K}_{\nu }\right) \left( 1+R_{\nu }\right) 
\hat{K}_{vp}\left( t\right) <\left( 1+\bar{r}\right) \sum_{m}\hat{k}_{2vm}%
\hat{K}_{vp}\left( t\right)
\end{equation*}%
where the LHS is the total level of invested capital including returns, and
the RHS represents the amount of loans plus interest. The previous equation
is compactly rewritten as:%
\begin{equation*}
\left( 1+\hat{f}\left( \hat{K}_{v}\left( t\right) \right) \right) \hat{K}%
_{vp}\left( t\right) <0
\end{equation*}%
with:%
\begin{equation*}
\hat{f}\left( \hat{K}_{v}\left( t\right) \right) =\left( 1+\sum_{m}\hat{k}%
_{2vm}\hat{K}_{\nu }\right) \left( 1+R_{\nu }\right)
\end{equation*}%
If this situation arises, the default of investor modifies other investors'
returns $\bar{R}_{j}^{\prime }$ by a term:%
\begin{equation}
\sum_{l}\left( \bar{r}-\frac{\left( 1+\hat{f}\left( \hat{K}_{vp}\left(
t\right) \right) \right) }{\sum_{m}\hat{k}_{2vm}\hat{K}_{m}}\right) H\left(
-\left( 1+\hat{f}\left( \hat{K}_{vp}\left( t\right) \right) \right) \right) 
\frac{\hat{k}_{2lj}\hat{K}_{l}\left( t\right) }{1+\sum_{v}\left( \hat{k}%
_{1lv}+\hat{k}_{2lv}\right) \hat{K}_{v}\left( t\right) }  \label{P1}
\end{equation}%
where $H$ is the Heaviside function. This term only appears when $1+\bar{f}%
_{\nu }$ is negative, meaning when the return $\bar{f}_{\nu }<-1$,
indicating that the return is so negative that it even destroys the
investor's own capital. When this term appears, the loss incurred by
investor $l$ is: 
\begin{equation*}
\left( \bar{r}-\frac{\left( 1+\hat{f}\left( \hat{K}_{vp}\left( t\right)
\right) \right) }{\sum_{m}\hat{k}_{2vm}\hat{K}_{m}}\right) \frac{\hat{k}%
_{2lj}\hat{K}_{l}\left( t\right) }{1+\sum_{v}\left( \hat{k}_{1lv}+\hat{k}%
_{2lv}\right) \hat{K}_{v}\left( t\right) }
\end{equation*}%
The equations for returns accounting for defaults are provided in the
following section. Note that the full loss of the investor is written:%
\begin{equation*}
\left( 1+\hat{f}\left( \hat{K}_{vp}\left( t\right) \right) \right) \hat{K}%
_{vp}\left( t\right)
\end{equation*}%
where:%
\begin{equation*}
\hat{f}\left( \hat{K}_{v}\left( t\right) \right) =\left( 1+\sum_{m}\hat{k}%
_{2vm}\hat{K}_{\nu }\right) R_{\nu }-\bar{r}\sum_{m}\hat{k}_{2vm}
\end{equation*}%
The rate of loss for investor $v$ is given by:%
\begin{equation*}
\frac{\left( 1+\hat{f}\left( \hat{K}_{vp}\left( t\right) \right) \right) 
\hat{K}_{vp}\left( t\right) }{\sum_{m}\hat{k}_{2vm}\hat{K}_{vp}\left(
t\right) \hat{K}_{m}}=\frac{\left( 1+\hat{f}\left( \hat{K}_{vp}\left(
t\right) \right) \right) }{\sum_{m}\hat{k}_{2vm}\hat{K}_{m}}
\end{equation*}

\subsubsection{Investors' returns under a default scenario}

The possibility of default modifies the return equation (\ref{RTq}):%
\begin{eqnarray}
&&\sum_{l}\left( \delta _{jl}-\frac{\hat{k}_{1lj}\hat{K}_{l}\left( t\right) 
}{1+\sum_{v}\left( \hat{k}_{1lv}+\hat{k}_{2lv}\right) \hat{K}_{v}\left(
t\right) }\right) \left( \frac{\hat{f}_{l}-\bar{r}}{1+\sum_{v}\hat{k}_{2jv}%
\hat{K}_{v}\left( t\right) }\right)  \label{RTQ} \\
&&+\sum_{l}\left( \bar{r}-\frac{\left( 1+\hat{f}\left( \hat{K}_{vp}\left(
t\right) \right) \right) }{\sum_{m}\hat{k}_{2vm}\hat{K}_{m}}\right) H\left(
-\left( 1+\hat{f}\left( \hat{K}_{vp}\left( t\right) \right) \right) \right) 
\frac{\hat{k}_{2lj}\hat{K}_{l}\left( t\right) }{1+\sum_{v}\left( \hat{k}%
_{1lv}+\hat{k}_{2lv}\right) \hat{K}_{v}\left( t\right) }  \notag \\
&&+\sum_{i}\left( \bar{r}-\frac{\left( 1+f_{1}^{\prime }\left( K_{i}\left(
t\right) \right) \right) }{\sum_{m}k_{2im}\hat{K}_{m}}\right) H\left(
-\left( 1+f_{1}^{\prime }\left( K_{i}\left( t\right) \right) \right) \right) 
\frac{k_{2ij}K_{i}\left( t\right) }{1+\sum_{v}\left( k_{1iv}+k_{2iv}\right) 
\hat{K}_{v}\left( t\right) }  \notag \\
&=&\sum_{i}\left( \frac{f_{1}^{\prime }\left( K_{i}\left( t\right) \right) -%
\bar{r}}{1+\sum_{\nu }k_{2j\nu }\hat{K}_{\nu }\left( t\right) }+\Delta
F_{\tau }\left( \bar{R}\left( K_{i},X_{i}\right) \right) \right) \frac{%
k_{1ij}K_{i}\left( t\right) }{1+\sum_{v}\left( k_{1iv}+k_{2iv}\right) \hat{K}%
_{v}\left( t\right) }  \notag
\end{eqnarray}%
where:%
\begin{equation*}
f_{1}^{\prime }\left( K_{i}\left( t\right) \right) =\left( 1+\sum_{\nu
}k_{2j\nu }\hat{K}_{\nu }\left( t\right) \right) r_{i}-\bar{r}\sum_{v}k_{2lv}%
\hat{K}_{v}\left( t\right)
\end{equation*}%
and $H$ is the Heaviside function.

Equation (\ref{RTQ}) is the same as the default-free return equation (\ref%
{RTq}), with the addition of the losses (\ref{P1}) incurred by the investor
if an investor in which they have invested defaults, and their loss (\ref{P2}%
) if a firm in which they have invested defaults.

\section{Field translation}

Actually, equations (\ref{gauche}) and (\ref{dnw}) correspond to the
minimization functions (\ref{pst}) and (\ref{psh}), which are translated
into fields. In our context, what corresponds to (\ref{gauche}) and (\ref%
{dnw}) are the two capital accumulation equations (\ref{DFQTW}) and (\ref{DC}%
) for investors and firms respectively. What corresponds here to (\ref{pst})
and (\ref{psh}) are the following two minimization functions. We can write
the minimization functions for the dynamics of $\hat{K}_{k}$ and $K_{ip}$,
which yields:%
\begin{equation}
\left\{ \frac{d}{dt}\hat{K}_{k}\left( t\right) -\sum_{l}\left( 1-M\right)
_{jl}^{-1}\hat{f}_{l}\hat{K}_{l}\left( t\right) \right\} ^{2}  \label{FT1}
\end{equation}%
\begin{equation}
\left\{ \frac{d}{dt}K_{ip}\left( t\right) -f_{1}^{\prime }\left( K_{i}\left(
t\right) \right) K_{ip}\left( t\right) \right\} ^{2}  \label{FT2}
\end{equation}%
leading to an action functional for the field version of the system. The
field version of these dynamic equations is presented in Appendix 1.
Moreover, in addition to these two equations, we have implemented the field
version of (\ref{RTQ}). This is done in Appendix 1 by including a potential
in the investor's action functional. We present the results of the
translation.

\subsection{Investors' action functional}

We denote $\hat{\Psi}$ the field describing the investors.\ It depends on
the two variables $\hat{K}$ and $\hat{X}$, and its action functional is
obtained by applying the translation formula (\ref{inco}) and (\ref{Trl}) to
the minimization function (\ref{FT1}):

\begin{equation*}
-\hat{\Psi}^{\dag }\left( \hat{K},\hat{X}\right) \left( \nabla _{\hat{K}%
}\left( \sigma _{\hat{K}}^{2}\nabla _{\hat{K}}-\hat{g}\left( \hat{K},\hat{X}%
\right) \hat{K}\right) \right) \hat{\Psi}\left( \hat{K},\hat{X}\right) +%
\frac{1}{2\hat{\mu}}\left( \left\vert \hat{\Psi}\left( \hat{K},\hat{X}%
\right) \right\vert ^{2}-\left\vert \hat{\Psi}_{0}\left( \hat{X}\right)
\right\vert ^{2}\right) ^{2}
\end{equation*}%
where:%
\begin{equation*}
\hat{g}\left( \hat{K},\hat{X}\right) =\left( 1-M\left( \left( \hat{K},\hat{X}%
\right) ,\left( \hat{K}^{\prime },\hat{X}^{\prime }\right) \right)
\left\vert \hat{\Psi}\left( \hat{K}^{\prime },\hat{X}^{\prime }\right)
\right\vert ^{2}\right) ^{-1}\hat{f}\left( \hat{K}^{\prime },\hat{X}^{\prime
}\right)
\end{equation*}%
and 
\begin{eqnarray*}
\hat{f}\left( \hat{X}\right) &=&\left( 1+\underline{\hat{k}}_{2}\left( \hat{X%
}\right) \right) \left( 1+R_{\nu }\left( \hat{X}\right) \right) \\
&=&\left( 1+\int \hat{k}_{2}\left( \hat{X},\hat{X}_{1}\right) \hat{K}%
_{1}\left\vert \hat{\Psi}\left( \hat{K}_{1},\hat{X}_{1}\right) \right\vert
^{2}d\hat{K}_{1}d\hat{X}_{1}\right) \left( 1+R_{\nu }\left( \hat{X}\right)
\right)
\end{eqnarray*}

\subsection{Investors' field return equations}

The translation of the function (\ref{RTQ}) is obtained by applying the
translation provided by (\ref{tln})

Appendix 2.3.4 translates the return equation (\ref{RTQ}) including defaults
in terms of field:

\begin{eqnarray}
&&\frac{f\left( \hat{X}\right) -\bar{r}}{1+\underline{\hat{k}}_{2}\left( 
\hat{X}\right) }-\int \frac{\hat{k}_{1}\left( \hat{X}^{\prime },\hat{X}%
\right) \hat{K}^{\prime }\left\vert \hat{\Psi}\left( \hat{K}^{\prime },\hat{X%
}^{\prime }\right) \right\vert ^{2}}{1+\underline{\hat{k}}\left( \hat{X}%
^{\prime }\right) }\frac{f\left( \hat{X}^{\prime }\right) -\bar{r}}{1+%
\underline{\hat{k}}_{2}\left( \hat{X}^{\prime }\right) }d\hat{K}^{\prime }d%
\hat{X}^{\prime }  \label{Rtq} \\
&=&\int \left( \bar{r}+\frac{1+f\left( \hat{X}^{\prime }\right) }{\underline{%
\hat{k}}_{2}\left( \hat{X}^{\prime }\right) }H\left( -\frac{1+f\left( \hat{X}%
^{\prime }\right) }{\underline{\hat{k}}_{2}\left( \hat{X}^{\prime }\right) }%
\right) \right) \frac{\hat{k}_{2}\left( \hat{X}^{\prime },\hat{X}\right) 
\hat{K}^{\prime }\left\vert \hat{\Psi}\left( \hat{K}^{\prime },\hat{X}%
^{\prime }\right) \right\vert ^{2}}{1+\underline{\hat{k}}\left( \hat{X}%
^{\prime }\right) }d\hat{K}^{\prime }d\hat{X}^{\prime }  \notag \\
&&+\int \left( \bar{r}+\frac{1+f_{1}^{\prime }\left( X^{\prime }\right) }{%
\underline{k}_{2}\left( X^{\prime }\right) }H\left( -\frac{1+f_{1}^{\prime
}\left( X^{\prime }\right) }{\underline{k}_{2}\left( X^{\prime }\right) }%
\right) \right) \frac{k_{2}\left( X^{\prime },\hat{X}\right) \left\vert \Psi
\left( K^{\prime },X^{\prime }\right) \right\vert ^{2}K^{\prime }}{1+%
\underline{k}\left( X^{\prime }\right) }dK^{\prime }dX^{\prime }  \notag \\
&&+\int \frac{\left\vert \Psi \left( K^{\prime },X^{\prime }\right)
\right\vert ^{2}k_{1}\left( X^{\prime },\hat{X}\right) K^{\prime }}{1+%
\underline{k}\left( X^{\prime }\right) }\left( \frac{f_{1}^{\prime }\left(
K,X\right) -\bar{r}k_{2}\left( \hat{X}\right) }{1+k_{2}\left( \hat{X}\right) 
}+\Delta F_{\tau }\left( \bar{R}\left( K,X\right) \right) \right) dK^{\prime
}dX^{\prime }  \notag
\end{eqnarray}%
The first term $\frac{f\left( \hat{X}\right) -\bar{r}}{1+\underline{\hat{k}}%
_{2}\left( \hat{X}\right) }$ represents the return of an investor in sector $%
X$, and all the other terms describe how it decomposes. The second term is
the portion of return from other investors retrieved by the investor in $X$%
.On the right side of the equation, the first term calculates the loss due
to the default of other investors, when it occurs. The second term
calculates the loss due to the default of firms in which the investor had
invested, and the last term is the return generated by the firms in which
the investor had invested.

\subsection{Firms' action functional}

The translation of the function (\ref{FT2}) is obtained by applying the
translation provided by (\ref{inco}) and (\ref{Trl}). The action functional
for the field of firms is:%
\begin{equation*}
-\Psi ^{\dag }\left( K,X\right) \left( \nabla _{K_{p}}\left( \sigma
_{K}^{2}\nabla _{K_{p}}-f_{1}^{\prime }\left( K,X\right) K_{p}\right)
\right) \Psi \left( K,X\right) +\frac{1}{2\epsilon }\left( \left\vert \Psi
\left( K,X\right) \right\vert ^{2}-\left\vert \Psi _{0}\left( X\right)
\right\vert ^{2}\right) ^{2}
\end{equation*}%
where:%
\begin{eqnarray*}
f_{1}^{\prime }\left( X\right) &=&\left( 1+\underline{k}_{2}\left( X\right)
\right) f_{1}\left( X\right) -\bar{r}\underline{k}_{2}\left( \hat{X}\right)
\\
&=&f_{1}\left( X\right) +\left( f_{1}\left( X\right) -\bar{r}\right) \int
k_{2}\left( X,\hat{X}_{1}\right) \hat{K}_{1}\left\vert \hat{\Psi}\left( \hat{%
K}_{1},\hat{X}_{1}\right) \right\vert ^{2}d\hat{K}_{1}d\hat{X}_{1}
\end{eqnarray*}

\subsection{Use of the field model and computation of averages per sector}

Section 4 provides us with the interpretation of the different fields in the
model. Once the field action functional $S$ is found, we can use field
theory to study the system of agents, both at the collective and individual
levels. At the collective level, we can compute the averages of the system
in a given background field. The individual level, described by the
transition functions, will be studied in the third part.

Recall that the background fields emerging at the collective level are
particular functions, $\Psi \left( K,X\right) $\ and $\hat{\Psi}\left( \hat{K%
},\hat{X}\right) $, and their adjoints fields $\Psi ^{\dag }\left(
K,X\right) $\ and $\hat{\Psi}^{\dag }\left( \hat{K},\hat{X}\right) $,\ that
minimize the action functional of the system.

Once the background fields are obtained, the associated number of firms and
investors per sector for a given average capital $K$ can be computed. They
are given by:%
\begin{equation}
\left\vert \Psi \left( K,X\right) \right\vert ^{2}=\Psi ^{\dag }\left(
K,X\right) \Psi \left( K,X\right)  \label{DSN}
\end{equation}%
and:%
\begin{equation}
\left\vert \hat{\Psi}\left( \hat{K},\hat{X}\right) \right\vert ^{2}=\hat{\Psi%
}^{\dag }\left( \hat{K},\hat{X}\right) \hat{\Psi}\left( \hat{K},\hat{X}%
\right)  \label{DST}
\end{equation}%
With these two density functions at hand, various average quantities\ in the
collective state can be computed.

The number of producers $\left\Vert \Psi \left( X\right) \right\Vert ^{2}$\
and investors $\left\Vert \hat{\Psi}\left( \hat{X}\right) \right\Vert ^{2}$
in sectors are computed using the formula:%
\begin{eqnarray}
\left\Vert \Psi \left( X\right) \right\Vert ^{2} &\equiv &\int \left\vert
\Psi \left( K,X\right) \right\vert ^{2}dK  \label{Nx} \\
\left\Vert \hat{\Psi}\left( \hat{X}\right) \right\Vert ^{2} &\equiv &\int
\left\vert \hat{\Psi}\left( \hat{K},\hat{X}\right) \right\vert ^{2}d\hat{K}
\label{Nxh}
\end{eqnarray}%
The total invested capital $\hat{K}_{X}$\ in sector $X$ is defined by a
partial average:%
\begin{equation}
\hat{K}_{\hat{X}}=\int \hat{K}\left\vert \hat{\Psi}\left( \hat{K},X\right)
\right\vert ^{2}d\hat{K}=\int \hat{K}\left\vert \hat{\Psi}\left( \hat{X}%
\right) \right\vert ^{2}d\hat{K}  \label{Khx}
\end{equation}%
and the average invested capital per firm in sector $X$ is defined by:%
\textbf{\ }%
\begin{equation}
K_{X}=\frac{\int \hat{K}\left\vert \hat{\Psi}\left( \hat{K},X\right)
\right\vert ^{2}d\hat{K}}{\left\Vert \Psi \left( X\right) \right\Vert ^{2}}
\label{kx}
\end{equation}

Ultimately, the distributions of invested capital per investor and of
capital per firm, given a collective state and a sector $X$,\textbf{\ } are $%
\frac{\left\vert \hat{\Psi}\left( \hat{K},X\right) \right\vert ^{2}}{%
\left\Vert \hat{\Psi}\left( \hat{X}\right) \right\Vert ^{2}}$\ and $\frac{%
\left\vert \Psi \left( K,X\right) \right\vert ^{2}}{\left\Vert \Psi \left(
X\right) \right\Vert ^{2}}$,\ respectively.

Gathering equations (\ref{Nx}), (\ref{Nxh}) and (\ref{kx}), each collective
state is singularly determined by the collection of data that characterizes
each sector: the number of firms, investors, the average capital, and the
distribution of capital. All these quantities allow the study of capital
allocation among sectors and its dependency in the parameters of the system,
such as expected long-term and short-term returns, and any other parameter.
This "static" point of view, will be extended by introducing some
fluctuations in the expectations, leading to a dynamic of the average
capital at the macro-level. In the following, we solve the system for the
background fields and compute the average associated quantities.

\section{Resolution for firms}

\subsection{Minimization equation and background field}

We consider the field action for the firms:%
\begin{equation*}
S\left( \Psi \right) =-\Psi ^{\dag }\left( K,X\right) \left( \nabla
_{K_{p}}\left( \sigma _{\hat{K}}^{2}\nabla _{K_{p}}-f_{1}^{\prime }\left(
K_{p},X\right) K_{p}\right) \right) \Psi \left( K,X\right) +\frac{1}{%
\epsilon }\left( \left\vert \Psi \left( K,X\right) \right\vert
^{2}-\left\vert \Psi _{0}\left( X\right) \right\vert ^{2}\right)
\end{equation*}%
We show in appendix that a change of variables leads to the modified action:%
\begin{eqnarray}
S\left( \Psi \right) &=&-\Psi ^{\dag }\left( K,X\right) \sigma _{\hat{K}%
}^{2}\nabla _{K_{p}}^{2}\Psi \left( K,X\right)  \label{SFr} \\
&&+\Psi ^{\dag }\left( K,X\right) \left( \frac{\left( f_{1}^{\prime }\left(
K_{p},X\right) K_{p}\right) ^{2}}{2\sigma _{\hat{K}}^{2}}+\frac{1}{2}%
f_{1}^{\prime }\left( K_{p},X\right) \right) \Psi \left( K,X\right) +\frac{1%
}{\epsilon }\left( \left\vert \Psi \left( K,X\right) \right\vert
^{2}-\left\vert \Psi _{0}\left( X\right) \right\vert ^{2}\right)  \notag
\end{eqnarray}%
which is easier to deal with.

The background field for the system is the solution of the first-order
minimization equations:%
\begin{equation*}
\frac{\delta S\left( \Psi \right) }{\delta \Psi ^{\dag }\left( K,X\right) }=0
\end{equation*}%
\begin{equation*}
\frac{\delta S\left( \Psi \right) }{\delta \Psi \left( K,X\right) }=0
\end{equation*}%
These equations, also known as the saddle point equations, for (\ref{SFr})
reduce to:%
\begin{equation*}
0=\left( \frac{f_{1}^{\prime 2}\left( K,X\right) }{\sigma _{\hat{K}}^{2}}+%
\frac{\frac{d}{dK}f_{1}^{\prime }\left( K,X\right) }{2}\right) +\frac{1}{%
\epsilon }\left( \left\vert \Psi \left( K,X\right) \right\vert
^{2}-\left\vert \Psi _{0}\left( X\right) \right\vert ^{2}\right)
\end{equation*}%
whose solution is given by:%
\begin{equation}
\left\vert \Psi \left( K,X\right) \right\vert ^{2}=\left\vert \Psi
_{0}\left( X\right) \right\vert ^{2}-\epsilon \left( \frac{\left(
f_{1}^{\left( e\right) }\left( X\right) K_{p}-\bar{C}\left( X\right) \right)
^{2}}{\sigma _{\hat{K}}^{2}}+\frac{f_{1}^{\left( e\right) }\left( X\right) }{%
2}\right)  \label{SLf}
\end{equation}%
where $f_{1}^{\left( e\right) }\left( X\right) $ is the return of the firm,
corresponding to the net return of production, from which the paiements of
loans are substracted: 
\begin{eqnarray*}
f_{1}^{\left( e\right) }\left( X\right) &=&\left( 1+\underline{k}_{2}\left(
X\right) \right) f_{1}\left( X\right) -\underline{k}_{2}\left( X\right) \bar{%
r} \\
\bar{C}^{\left( e\right) }\left( X\right) &=&\left( 1+\underline{k}%
_{2}\left( X\right) \right) \bar{C}\left( X\right)
\end{eqnarray*}%
\begin{equation*}
\left\vert \Psi _{0}\left( X\right) \right\vert ^{2}-\epsilon \frac{%
f_{1}^{\left( e\right) }\left( X\right) }{2}-\epsilon \frac{\left(
f_{1}^{\left( e\right) }\left( X\right) K_{p}-\bar{C}^{\left( e\right)
}\left( X\right) \right) ^{2}}{\sigma _{\hat{K}}^{2}}
\end{equation*}%
In the following, we will replace:%
\begin{eqnarray*}
f_{1}^{\left( e\right) }\left( X\right) &\rightarrow &f_{1}\left( X\right) \\
\bar{C}^{\left( e\right) }\left( X\right) &\rightarrow &\bar{C}\left(
X\right)
\end{eqnarray*}

To compute returns and average capital for firms per sector, some
assumptions about productivities $f_{1}\left( X\right) $ must be made. For
the sake of simplicity, we will consider constant returns to scale.
Non-constant returns will be studied in Appendix 12.

\subsection{Average capital per sector for constant return to scale}

Several cases can occur. Depending on the productivity $f_{1}\left( X\right) 
$, some agents may have either very low or even no capital.\ Alternately,
all agents may have a minimal level of capital in a given sector.

\subsubsection{Case 1 : No minimum capital}

In this case, the following condition holds:%
\begin{equation*}
\left\vert \Psi _{0}\left( X\right) \right\vert ^{2}-\epsilon \frac{%
f_{1}\left( X\right) }{2}-\epsilon \frac{\left( \bar{C}\left( X\right)
\right) ^{2}}{\sigma _{\hat{K}}^{2}}>0
\end{equation*}%
which, given (\ref{SLf}), translates to $\left\vert \Psi \left( 0,X\right)
\right\vert ^{2}>0$, indicating that some firms have either no or very low
private capital.

Appendix 10.1.1 demonstrates that there exists a maximal level of capital
per sector.\ Additionally, it also shows that the background field per
sector is given by:%
\begin{eqnarray}
\left\vert \Psi \left( X\right) \right\vert ^{2} &=&\left( \frac{2}{3}\sigma
_{\hat{K}}^{2}\left( \frac{\left\vert \Psi _{0}\left( X\right) \right\vert
^{2}}{\epsilon }-\frac{f_{1}\left( X\right) }{2}\right) +\frac{1}{3}\left( 
\sqrt{\sigma _{\hat{K}}^{2}\left( \frac{\left\vert \Psi _{0}\left( X\right)
\right\vert ^{2}}{\epsilon }-\frac{f_{1}\left( X\right) }{2}\right) }-\bar{C}%
\left( X\right) \right) \bar{C}\left( X\right) \right) \\
&&\times \epsilon \frac{K_{0}}{\sigma _{\hat{K}}^{2}}  \notag
\end{eqnarray}%
and that the amount of capital, $K_{X}\left\vert \Psi \left( X\right)
\right\vert ^{2}$ is:%
\begin{equation}
K_{X}\left\vert \Psi \left( X\right) \right\vert ^{2}=\frac{\epsilon K_{0}}{%
\sigma _{\hat{K}}^{2}f_{1}\left( X\right) }\left\{ \frac{1}{4}\left(
2X^{2}\left( \bar{C}+X\right) -\left( X-\bar{C}\right) \left( X^{2}+\left( 
\bar{C}\right) ^{2}\right) \right) -\frac{\bar{C}}{3}\left( X\left( X-\bar{C}%
\right) +C^{2}\right) \right\}
\end{equation}%
where:%
\begin{equation*}
X=\sqrt{\sigma _{\hat{K}}^{2}\left( \frac{\left\vert \Psi _{0}\left(
X\right) \right\vert ^{2}}{\epsilon }-\frac{f_{1}\left( X\right) }{2}\right) 
}
\end{equation*}%
The appendix also derives the average capital in sector $X$:%
\begin{eqnarray*}
K_{X} &=&\frac{1}{f_{1}\left( X\right) }\frac{\frac{1}{4}\left( 2X^{2}\left(
C+X\right) -\left( X-C\right) \left( X^{2}+C^{2}\right) \right) -\frac{C}{3}%
\left( X\left( X-C\right) +C^{2}\right) }{\frac{2}{3}X^{2}+\frac{1}{3}\left(
X-C\right) C} \\
&=&\frac{1}{4f_{1}\left( X\right) }\left( 3X-C\right) \frac{\left(
C+X\right) }{2X-C}
\end{eqnarray*}%
as well as the\ overall return of capital in sector $X$:

\begin{eqnarray}
&&\left\vert \Psi \left( X\right) \right\vert ^{2}\left( \left( f_{1}\left(
X\right) -\bar{r}\right) K_{X}-\bar{C}\left( X\right) \right)  \label{FRC} \\
&\rightarrow &\frac{\left( C^{\left( e\right) }+X^{\left( e\right) }\right)
\left( \frac{2}{3}X^{2}+\frac{1}{3}\left( X-C^{\left( e\right) }\right)
C^{\left( e\right) }\right) \epsilon }{\sigma _{\hat{K}}^{2}f_{1}^{\left(
e\right) }\left( X\right) }\left( \frac{\left( f_{1}\left( X\right) -\bar{r}%
\right) }{4f_{1}^{\left( e\right) }\left( X\right) }\frac{\left( 3X^{\left(
e\right) }-C^{\left( e\right) }\right) \left( C^{\left( e\right) }+X^{\left(
e\right) }\right) }{2X^{\left( e\right) }-C^{\left( e\right) }}-C\right) 
\notag
\end{eqnarray}%
This will be used later to compute investors' returns.

\subsubsection{Case 2: Minimum capital}

In this case, we assume that all firms have a minimum capital, thus: 
\begin{equation*}
\left\vert \Psi _{0}\left( X\right) \right\vert ^{2}-\epsilon \frac{%
f_{1}\left( X\right) }{2}-\epsilon \frac{\left( \bar{C}\left( X\right)
\right) ^{2}}{\sigma _{\hat{K}}^{2}}<0
\end{equation*}%
In this case the capital is confined within two bounds. We will write $%
K_{0-} $ the minimal level of firms capital in the sector, and $K$, the
maximal level. Both are defined by setting $\left\vert \Psi \left( K_{0-}%
\text{,}X\right) \right\vert ^{2}=\left\vert \Psi \left( K_{0+}\text{,}%
X\right) \right\vert ^{2}=0$, that is these level are such that there is no
firms below or above these levels respectively. We find: 
\begin{equation*}
K_{0\pm }=\frac{\bar{C}\left( X\right) \pm \sqrt{\sigma _{\hat{K}}^{2}\left( 
\frac{\left\vert \Psi _{0}\left( X\right) \right\vert ^{2}}{\epsilon }-\frac{%
f_{1}\left( X\right) }{2}\right) }}{f_{1}\left( X\right) }
\end{equation*}%
and we define the range of possible capital $\Delta K_{0}$ within the sector
by:%
\begin{equation*}
\Delta K_{0}=K_{0+}-K_{0-}
\end{equation*}%
Appendix 7.1 computes the field $\left\vert \Psi \left( X\right) \right\vert
^{2}$:%
\begin{equation*}
\left\vert \Psi \left( X\right) \right\vert ^{2}=\frac{2}{3}\left( \frac{%
\left\vert \Psi _{0}\left( X\right) \right\vert ^{2}}{\epsilon }-\frac{%
f_{1}\left( X\right) }{2}\right) \epsilon \Delta K_{0}
\end{equation*}%
and the amount of capital $K_{X}\left\vert \Psi \left( X\right) \right\vert
^{2}$ in a sector:

\begin{equation*}
K_{X}\left\vert \Psi \left( X\right) \right\vert ^{2}=\frac{2\epsilon \Delta
K_{0}\bar{C}\left( X\right) }{3f_{1}\left( X\right) \sigma _{\hat{K}}^{2}}%
X^{2}
\end{equation*}%
with:%
\begin{equation*}
X=\sqrt{\sigma _{\hat{K}}^{2}\left( \frac{\left\vert \Psi _{0}\left(
X\right) \right\vert ^{2}}{\epsilon }-\frac{f_{1}\left( X\right) }{2}\right) 
}
\end{equation*}%
Ultimately, the average capital is given by:%
\begin{equation*}
K_{X}=\frac{\bar{C}\left( X\right) }{f_{1}\left( X\right) }
\end{equation*}%
while the return of the whole sector is:%
\begin{equation*}
\left\vert \Psi \left( X\right) \right\vert ^{2}K_{X}\frac{f_{1}^{\prime
}\left( X\right) -\bar{r}}{1+\underline{k}_{2}\left( X\right) }=\left\vert
\Psi \left( X\right) \right\vert ^{2}\left( \left( f_{1}\left( X\right) -%
\bar{r}\right) K_{X}-\bar{C}\left( X\right) \right) =\frac{4\epsilon }{%
3\sigma _{\hat{K}}^{2}f_{1}^{\left( e\right) }}X^{3}\bar{C}\left( X\right)
\left( -\frac{\bar{r}}{f_{1}^{\left( e\right) }}\right)
\end{equation*}

\section{Resolution for investors}

\subsection{Field action and change of variables}

Starting from the investors field action functional:%
\begin{equation*}
S\left( \hat{\Psi}\right) =-\hat{\Psi}^{\dag }\left( \hat{K},\hat{X}\right)
\left( \nabla _{\hat{K}}\left( \sigma _{\hat{K}}^{2}\nabla _{\hat{K}}-\hat{g}%
\left( \hat{K},\hat{X}\right) \hat{K}\right) \right) \hat{\Psi}\left( \hat{K}%
,\hat{X}\right) +\frac{1}{2\hat{\mu}}\left( \left\vert \hat{\Psi}\left( \hat{%
K},\hat{X}\right) \right\vert ^{2}-\left\vert \hat{\Psi}_{0}\left( \hat{X}%
\right) \right\vert ^{2}\right) ^{2}
\end{equation*}%
a change of variable:%
\begin{eqnarray*}
\hat{\Psi}\left( \hat{K},\hat{X}\right) &\rightarrow &\exp \left( -\int \hat{%
f}\left( \hat{K},\hat{X}\right) \hat{K}d\hat{K}\right) \hat{\Psi} \\
\hat{\Psi}^{\dag }\left( \hat{K},\hat{X}\right) &\rightarrow &\exp \left(
\int \hat{f}\left( \hat{K},\hat{X}\right) \hat{K}d\hat{K}\right) \hat{\Psi}%
^{\dag }
\end{eqnarray*}%
leads to minimize the action:%
\begin{eqnarray}
S\left( \hat{\Psi}\right) &=&-\hat{\Psi}^{\dag }\left( \hat{K},\hat{X}%
\right) \sigma _{\hat{K}}^{2}\nabla _{\hat{K}}^{2}\hat{\Psi}\left( \hat{K},%
\hat{X}\right)  \label{Chv} \\
&&+\hat{\Psi}^{\dag }\left( \hat{K},\hat{X}\right) \left( \frac{\left( \hat{g%
}\left( \hat{K},\hat{X}\right) \hat{K}\right) ^{2}}{2\sigma _{\hat{K}}^{2}}+%
\frac{\hat{g}\left( \hat{K},\hat{X}\right) }{2}\right) \hat{\Psi}\left( \hat{%
K},\hat{X}\right) +\frac{1}{2\hat{\mu}}\left( \left\vert \hat{\Psi}\left( 
\hat{K},\hat{X}\right) \right\vert ^{2}-\left\vert \hat{\Psi}_{0}\left( \hat{%
X}\right) \right\vert ^{2}\right) ^{2}  \notag
\end{eqnarray}%
where:%
\begin{equation*}
\hat{g}\left( \hat{K},\hat{X}\right) =\left( 1-M\left( \left( \hat{K},\hat{X}%
\right) ,\left( \hat{K}^{\prime },\hat{X}^{\prime }\right) \right)
\left\vert \hat{\Psi}\left( \hat{K}^{\prime },\hat{X}^{\prime }\right)
\right\vert ^{2}\right) ^{-1}\hat{f}\left( \hat{K}^{\prime },\hat{X}^{\prime
}\right)
\end{equation*}%
and the elements of the matrix are given by:%
\begin{equation*}
M\left( \hat{K},\hat{X},\hat{K}^{\prime },\hat{X}^{\prime }\right) =\frac{%
\hat{k}\left( \hat{X},\hat{X}^{\prime }\right) \hat{K}}{1+\underline{\hat{k}}%
\left( \hat{X}\right) }
\end{equation*}

\subsection{Minimization equation for the background field}

If we neglect the gradient terms, the saddle point equation for the
background field becomes:%
\begin{eqnarray*}
0 &=&\left( \frac{\hat{g}^{2}\left( \hat{K}_{1},\hat{X}_{1}\right) }{2\sigma
_{\hat{K}}^{2}}+\frac{\hat{g}\left( \hat{K}_{1},\hat{X}_{1}\right) }{2\hat{K}%
_{1}}\right) +\int \frac{\delta }{\delta \left\vert \hat{\Psi}\left( \hat{K}%
_{1},\hat{X}_{1}\right) \right\vert ^{2}}\left( \frac{\hat{g}^{2}\left( \hat{%
K},\hat{X}\right) }{2\sigma _{\hat{K}}^{2}}+\frac{\hat{g}\left( \hat{K},\hat{%
X}\right) }{2\hat{K}}\right) \left\vert \hat{\Psi}\left( \hat{K},\hat{X}%
\right) \right\vert ^{2} \\
&&+\frac{1}{\hat{\mu}}\left( \left\vert \hat{\Psi}\left( \hat{K},\hat{X}%
\right) \right\vert ^{2}-\left\vert \hat{\Psi}_{0}\left( \hat{X}\right)
\right\vert ^{2}\right)
\end{eqnarray*}

Appendix 8 estimates the derivative $\frac{\delta }{\delta \left\vert \hat{%
\Psi}\left( \hat{K}_{1},\hat{X}_{1}\right) \right\vert ^{2}}\hat{g}\left( 
\hat{K},\hat{X}\right) $ and we find:%
\begin{equation}
\frac{\partial }{\partial \left\vert \hat{\Psi}\left( \hat{K}_{1},\hat{X}%
_{1}\right) \right\vert ^{2}}\hat{g}\left( \hat{K},\hat{X}\right) \simeq 
\frac{\underline{\hat{k}}\left( \left\langle \hat{X}\right\rangle ,\hat{X}%
_{1}\right) }{\left\Vert \hat{\Psi}\right\Vert ^{2}}\hat{g}\left( \hat{X}%
_{1}\right) -\frac{\hat{K}_{1}}{\left\langle \hat{K}\right\rangle }\frac{%
\underline{\hat{k}}\left( \left\langle \hat{X}\right\rangle ,\left\langle 
\hat{X}\right\rangle \right) }{\left\Vert \hat{\Psi}\right\Vert ^{2}}%
\left\langle \hat{g}\right\rangle
\end{equation}%
and the minimization equation writes:

\begin{eqnarray}
0 &=&\frac{\hat{K}_{1}^{2}\hat{g}^{2}\left( \hat{K}_{1},\hat{X}_{1}\right) }{%
2\sigma _{\hat{K}}^{2}}+\frac{\hat{g}\left( \hat{K}_{1},\hat{X}_{1}\right) }{%
2}  \label{CQ} \\
&&+\int \left\vert \hat{\Psi}\left( \hat{K},\hat{X}\right) \right\vert
^{2}\left( \frac{\hat{K}^{2}\hat{g}^{2}\left( \hat{K},\hat{X},\Psi ,\hat{\Psi%
}\right) }{\sigma _{\hat{K}}^{2}}+\frac{1}{2}\right) \left( \frac{\underline{%
\hat{k}}\left( \left\langle \hat{X}\right\rangle ,\hat{X}_{1}\right) }{%
\left\Vert \hat{\Psi}\right\Vert ^{2}}\hat{g}\left( \hat{X}_{1}\right) -%
\frac{\hat{K}_{1}}{\left\langle \hat{K}\right\rangle }\frac{\underline{\hat{k%
}}\left( \left\langle \hat{X}\right\rangle ,\left\langle \hat{X}%
\right\rangle \right) }{\left\Vert \hat{\Psi}\right\Vert ^{2}}\left\langle 
\hat{g}\right\rangle \right)  \notag \\
&&+\frac{1}{\hat{\mu}}\left( \left\vert \hat{\Psi}\left( \hat{K},\hat{X}%
\right) \right\vert ^{2}-\left\vert \hat{\Psi}_{0}\left( \hat{X}\right)
\right\vert ^{2}\right)  \notag
\end{eqnarray}%
This equation must be taken into account along with the returns equation (%
\ref{Rtq}) derived previously.

\subsection{Solving the minimization equation and average capital per sector}

Solving for $\left\vert \hat{\Psi}\left( \hat{K}_{1},\hat{X}_{1}\right)
\right\vert ^{2}$ yields:%
\begin{eqnarray}
\left\vert \hat{\Psi}\left( \hat{K}_{1},\hat{X}_{1}\right) \right\vert ^{2}
&=&\left\Vert \hat{\Psi}_{0}\left( \hat{X}_{1}\right) \right\Vert ^{2}-\hat{%
\mu}\left\{ \left( \frac{\hat{K}_{1}^{2}\hat{g}^{2}\left( \hat{X}_{1}\right) 
}{2\sigma _{\hat{K}}^{2}}+\frac{\hat{g}\left( \hat{X}_{1}\right) }{2}\right)
\right. \\
&&\left. +\left( \frac{\left\langle \hat{K}\right\rangle ^{2}\left\langle 
\hat{g}\right\rangle }{\sigma _{\hat{K}}^{2}}+\frac{1}{2}\right) \left( 
\underline{\hat{k}}\left( \left\langle \hat{X}\right\rangle ,\hat{X}%
_{1}\right) \hat{g}\left( \hat{X}_{1}\right) -\frac{\hat{K}_{1}}{%
\left\langle \hat{K}\right\rangle }\underline{\hat{k}}\left( \left\langle 
\hat{X}\right\rangle ,\left\langle \hat{X}\right\rangle \right) \left\langle 
\hat{g}\right\rangle \right) \right\}  \notag
\end{eqnarray}%
We also study the stability of this solution in Appendix 9.

\subsubsection{Maximal level of capital per sector}

In Appendix 10.1, we compute\ the maximal value of capital $\hat{K}_{0}$ for
sector $\hat{X}$. This is found by setting: 
\begin{equation*}
\left\vert \hat{\Psi}\left( \hat{K}_{1},\hat{X}_{1}\right) \right\vert ^{2}=0
\end{equation*}%
which yields:%
\begin{equation}
\hat{K}_{0}^{2}\simeq 2\frac{\sigma _{\hat{K}}^{2}}{\hat{g}^{2}\left( \hat{X}%
_{1}\right) }\left( \frac{\left\Vert \hat{\Psi}_{0}\left( \hat{X}_{1}\right)
\right\Vert ^{2}}{\hat{\mu}}+\left( \frac{\left\langle \hat{K}\right\rangle
^{2}\left\langle \hat{g}\right\rangle ^{2}}{\sigma _{\hat{K}}^{2}}+\frac{%
\left\langle \hat{g}\right\rangle }{2}\right) \left( \left( \frac{%
\left\langle \hat{K}_{0}\right\rangle }{\left\langle \hat{K}\right\rangle }-%
\frac{\underline{\hat{k}}\left( \left\langle \hat{X}\right\rangle ,\hat{X}%
_{1}\right) }{\underline{\hat{k}}\left( \left\langle \hat{X}\right\rangle
,\left\langle \hat{X}\right\rangle \right) }\right) \underline{\hat{k}}%
\left( \left\langle \hat{X}\right\rangle ,\left\langle \hat{X}\right\rangle
\right) \right) \right)
\end{equation}

\subsubsection{Global averages for field and capital}

We then compute the averages over all sectors $\left\langle \hat{K}%
_{0}\right\rangle ^{2}$, $\left\Vert \hat{\Psi}\right\Vert ^{2}$ and $%
\left\langle \hat{K}\right\rangle \left\Vert \hat{\Psi}\right\Vert ^{2}$
required to solve for the average maximum level of capital. We obtain:%
\begin{equation*}
\left\langle \hat{K}_{0}\right\rangle ^{2}=2\frac{\sigma _{\hat{K}%
}^{2}\left( \frac{\left\Vert \hat{\Psi}_{0}\right\Vert ^{2}}{\hat{\mu}}+%
\frac{\left\langle \hat{g}\right\rangle }{2}\left( \frac{1-r}{r}\right) \hat{%
k}\right) }{\left\langle \hat{g}\right\rangle ^{2}\left( 1-2r\left(
1-r\right) \hat{k}\right) }
\end{equation*}%
\begin{equation}
\left\Vert \hat{\Psi}\right\Vert ^{2}\simeq \frac{\hat{\mu}V}{3\sigma _{\hat{%
K}}^{2}}\left( 2\frac{\sigma _{\hat{K}}^{2}\left( \frac{\left\Vert \hat{\Psi}%
_{0}\right\Vert ^{2}}{\hat{\mu}}+\frac{\left\langle \hat{g}\right\rangle }{2}%
\left( \frac{1-r}{r}\right) \hat{k}\right) }{\left\langle \hat{g}%
\right\rangle ^{2}\left( 1-2r\left( 1-r\right) \hat{k}\right) }\right) ^{%
\frac{3}{2}}\left( 1-\frac{2+\hat{k}-\sqrt{\left( 2+\hat{k}\right) ^{2}-\hat{%
k}}}{4\hat{k}}\hat{k}\right) \left\langle \hat{g}\right\rangle ^{2}
\label{Vrs}
\end{equation}%
\begin{equation}
\left\langle \hat{K}\right\rangle \left\Vert \hat{\Psi}\right\Vert ^{2}=%
\frac{\hat{\mu}V\sigma _{\hat{K}}^{2}}{2\left\langle \hat{g}\right\rangle
^{2}}\left( \frac{\frac{\left\Vert \hat{\Psi}_{0}\right\Vert ^{2}}{\hat{\mu}}%
+\frac{\left\langle \hat{g}\right\rangle }{2}\left( \frac{1-r}{r}\right) 
\hat{k}}{1-2r\left( 1-r\right) \hat{k}}\right) ^{2}\left( 1-2\frac{2+\hat{k}-%
\sqrt{\left( 2+\hat{k}\right) ^{2}-\hat{k}}}{9}\right)  \label{Vrf}
\end{equation}%
The quantity (\ref{Vrf}) computes the average overall financial capital in
the system. The average capital per investor is: 
\begin{equation}
\left\langle \hat{K}\right\rangle =\frac{3}{4\left\langle \hat{g}%
\right\rangle }\sqrt{\frac{\sigma _{\hat{K}}^{2}}{2\hat{\mu}}}\frac{1-2\frac{%
2+\hat{k}-\sqrt{\left( 2+\hat{k}\right) ^{2}-\hat{k}}}{9}}{1-\frac{2+\hat{k}-%
\sqrt{\left( 2+\hat{k}\right) ^{2}-\hat{k}}}{4\hat{k}}}\sqrt{\frac{%
\left\Vert \hat{\Psi}_{0}\right\Vert ^{2}+\hat{\mu}\frac{\left\langle \hat{g}%
\right\rangle }{2}\left( \frac{1-r}{r}\right) \hat{k}}{1-2r\left( 1-r\right) 
\hat{k}}}
\end{equation}

\subsubsection{Average field and capital per sector}

We then find the field $\left\Vert \hat{\Psi}\left( \hat{X}_{1}\right)
\right\Vert ^{2}$ and the amount of capital $\hat{K}_{\hat{X}_{1}}\left\Vert 
\hat{\Psi}\left( \hat{X}_{1}\right) \right\Vert ^{2}$ and the average
capital $\hat{K}_{\hat{X}_{1}}$ in sector $X$\ as functions of $\hat{K}%
_{0}^{2}$ and averages $\left\langle \hat{K}\right\rangle $ and $%
\left\langle \hat{K}_{0}\right\rangle $:

\begin{eqnarray}
\left\Vert \hat{\Psi}\left( \hat{X}_{1}\right) \right\Vert ^{2} &\simeq &%
\hat{\mu}\frac{\hat{K}_{0}^{3}}{\sigma _{\hat{K}}^{2}}\left( \frac{\hat{g}%
^{2}\left( \hat{X}_{1}\right) }{3}-\frac{\left\langle \hat{K}\right\rangle
\left\langle \hat{g}\right\rangle ^{2}}{2\left\langle \hat{K}%
_{0}\right\rangle }\underline{\hat{k}}\left( \left\langle \hat{X}%
\right\rangle ,\left\langle \hat{X}\right\rangle \right) \right) \\
&=&\frac{\hat{\mu}}{\sigma _{\hat{K}}^{2}}\left( 2\frac{\sigma _{\hat{K}}^{2}%
}{\hat{g}^{2}\left( \hat{X}_{1}\right) }\left( \frac{\left\Vert \hat{\Psi}%
_{0}\left( \hat{X}_{1}\right) \right\Vert ^{2}}{\hat{\mu}}-D\left( \hat{X}%
_{1}\right) \right) \right) ^{\frac{3}{2}}\left( \frac{\hat{g}^{2}\left( 
\hat{X}_{1}\right) }{3}-\frac{\left\langle \hat{K}\right\rangle \left\langle 
\hat{g}\right\rangle ^{2}}{2\left\langle \hat{K}_{0}\right\rangle }\hat{k}%
\right)  \notag
\end{eqnarray}%
\begin{eqnarray}
\hat{K}\left[ \hat{X}_{1}\right] &=&\hat{K}_{\hat{X}}\left\Vert \hat{\Psi}%
\left( \hat{X}_{1}\right) \right\Vert ^{2}\simeq \hat{\mu}\frac{\hat{K}%
_{0}^{4}}{2\sigma _{\hat{K}}^{2}}\left( \frac{\hat{g}^{2}\left( \hat{X}%
_{1}\right) }{4}-\frac{\left\langle \hat{K}\right\rangle \left\langle \hat{g}%
\right\rangle ^{2}}{3\left\langle \hat{K}_{0}\right\rangle }\hat{k}\left(
\left\langle \hat{X}\right\rangle ,\left\langle \hat{X}\right\rangle \right)
\right) \\
&=&\frac{\hat{\mu}}{2\sigma _{\hat{K}}^{2}}\left( 2\frac{\sigma _{\hat{K}%
}^{2}}{\hat{g}^{2}\left( \hat{X}_{1}\right) }\left( \frac{\left\Vert \hat{%
\Psi}_{0}\left( \hat{X}_{1}\right) \right\Vert ^{2}}{\hat{\mu}}-D\left( \hat{%
X}_{1}\right) \right) \right) ^{2}\left( \frac{\hat{g}^{2}\left( \hat{X}%
_{1}\right) }{4}-\frac{r\left\langle \hat{g}\right\rangle ^{2}}{3}\underline{%
\hat{k}}\right)  \notag
\end{eqnarray}%
and the average capital per investor in sector $\hat{X}_{1}$ is: 
\begin{equation}
\hat{K}_{\hat{X}_{1}}=\frac{\hat{K}\left[ \hat{X}_{1}\right] }{\left\Vert 
\hat{\Psi}\left( \hat{X}_{1}\right) \right\Vert ^{2}}=\frac{\sqrt{2\frac{%
\sigma _{\hat{K}}^{2}}{\hat{g}^{2}\left( \hat{X}_{1}\right) }\left( \frac{%
\left\Vert \hat{\Psi}_{0}\left( \hat{X}_{1}\right) \right\Vert ^{2}}{\hat{\mu%
}}-D\left( \hat{X}_{1}\right) \right) }\left( \frac{\hat{g}^{2}\left( \hat{X}%
_{1}\right) }{4}-\frac{r\left\langle \hat{g}\right\rangle ^{2}}{3}\hat{k}%
\right) }{2\left( \frac{\hat{g}^{2}\left( \hat{X}_{1}\right) }{3}-\frac{%
\left\langle \hat{K}\right\rangle \left\langle \hat{g}\right\rangle ^{2}}{%
2\left\langle \hat{K}_{0}\right\rangle }\hat{k}\right) }
\end{equation}

\section{Investors' returns under a no-default scenario}

So far, we have found the level of capital per sector and the return firms
provide to investors. By reintroducing these results into the equation for
investor returns, we will be able to express investors' excess returns in
terms of capital per sector, returns provided by firms, and the number of
agents per sector, which we will then interpret.

\subsection{Derivation of the investors' returns equation}

The link between returns $g\left( \hat{X}_{1}\right) $ and $f$, given by:%
\begin{equation*}
g\left( \hat{X}\right) =\int \left( 1-M\right) ^{-1}\left( \hat{X},\hat{X}%
^{\prime }\right) f\left( \hat{X}^{\prime }\right) d\hat{X}^{\prime }
\end{equation*}%
and the return equation (\ref{Rtq}) without default:%
\begin{eqnarray*}
&&\frac{f\left( \hat{X}\right) -\bar{r}}{1+\underline{\hat{k}}_{2}\left( 
\hat{X}\right) }-\int \frac{\hat{k}_{1}\left( \hat{X}^{\prime },\hat{X}%
\right) \hat{K}^{\prime }\left\vert \hat{\Psi}\left( \hat{K}^{\prime },\hat{X%
}^{\prime }\right) \right\vert ^{2}}{1+\underline{\hat{k}}\left( \hat{X}%
^{\prime }\right) }\frac{f\left( \hat{X}^{\prime }\right) -\bar{r}}{1+%
\underline{\hat{k}}_{2}\left( \hat{X}^{\prime }\right) }d\hat{K}^{\prime }d%
\hat{X}^{\prime } \\
&=&\int \frac{\left\vert \Psi \left( K^{\prime },X^{\prime }\right)
\right\vert ^{2}k_{1}\left( X^{\prime },\hat{X}\right) K^{\prime }}{1+%
\underline{k}\left( X^{\prime }\right) }\left( \frac{f_{1}^{\prime }\left(
K,X\right) -\bar{r}k_{2}\left( \hat{X}\right) }{1+k_{2}\left( \hat{X}\right) 
}+\Delta F_{\tau }\left( \bar{R}\left( K,X\right) \right) \right) dK^{\prime
}dX^{\prime }
\end{eqnarray*}%
can be rewritten as an equation for $g\left( \hat{X}\right) $ as a function
of firms returns: 
\begin{equation}
\int \left( \frac{\Delta \left( \hat{X},\hat{X}^{\prime }\right) }{1+%
\underline{\hat{k}}_{2}\left( \hat{X}\right) }-\frac{\hat{k}_{1}\left( \hat{X%
}^{\prime },\hat{X}\right) \hat{K}^{\prime }\left\vert \hat{\Psi}\left( \hat{%
K}^{\prime },\hat{X}^{\prime }\right) \right\vert ^{2}}{1+\underline{\hat{k}}%
_{2}\left( \hat{X}^{\prime }\right) }\right) \left( 1-M\right) \left( g-\bar{%
r}^{\prime }\right) d\hat{X}^{\prime }=\frac{k_{1}\left( X\right) }{1+%
\underline{k}\left( X\right) }f_{1}^{\prime }  \label{RTN}
\end{equation}%
where:%
\begin{equation*}
\bar{r}^{\prime }=\left( 1-M\right) ^{-1}\bar{r}
\end{equation*}%
\begin{equation*}
\frac{k_{1}\left( X\right) }{1+\underline{k}\left( X\right) }=\int \frac{%
\left\vert \Psi \left( K^{\prime },X^{\prime }\right) \right\vert
^{2}k_{1}\left( X^{\prime },\hat{X}\right) K^{\prime }}{1+\underline{k}%
\left( X^{\prime }\right) }dK^{\prime }dX^{\prime }
\end{equation*}%
and $f_{1}^{\prime }$ stands for the returns investors make from their
participation in a firm located in $X$:%
\begin{equation*}
\frac{f_{1}^{\prime }\left( K,X\right) -\bar{r}k_{2}\left( \hat{X}\right) }{%
1+k_{2}\left( \hat{X}\right) }+\Delta F_{\tau }\left( \bar{R}\left(
K,X\right) \right)
\end{equation*}%
The expression for $f_{1}^{\prime }$ has been determined for constant
returns (see (\ref{FRC})). We derive the expressions for both sides of (\ref%
{RTN}) in Appendices 11.1 and 11.2.

We define: 
\begin{equation*}
\underline{k}\left( X\right) =k\left( X\right) \frac{\hat{K}_{X}\left\vert 
\hat{\Psi}\left( \hat{X}\right) \right\vert ^{2}}{\left\langle
K\right\rangle \left\vert \Psi _{0}\left( X\right) \right\vert ^{2}}
\end{equation*}%
which implies that the capital invested in firms in sector $X$ is a fraction
of the investors' disposable capital in $X$ divided by the number of firms
in the sector.\ We show in Appendix 11.5 that for $\underline{k}\left(
X\right) >>1$ and $\epsilon <\sigma _{\hat{K}}^{2}f_{1}\left( X\right) $, (%
\ref{RTN}) becomes in terms of excess return $\hat{g}\left( \hat{X}%
_{1}\right) -\bar{r}^{\prime }$:%
\begin{eqnarray}
&&\left( \frac{\Delta \left( \hat{X},\hat{X}^{\prime }\right) }{1+\underline{%
\hat{k}}_{2}\left( \hat{X}\right) }-\hat{S}_{1}^{E}\left( \hat{X}^{\prime },%
\hat{X}\right) \right) \left( \hat{g}\left( \hat{X}_{1}\right) -\bar{r}%
^{\prime }\right)  \label{PNm} \\
&=&\frac{1}{3}\frac{\left( 1-\beta \right) \left( X+C\beta \right)
^{2}\epsilon }{\sigma _{\hat{K}}^{2}\left( f_{1}\left( X\right) +\beta 
\underline{k}\left( X\right) R\right) }\left( \frac{R\left( 3X-\beta
C\right) \left( \beta C+X\right) }{4\left( f_{1}\left( X\right) +\beta 
\underline{k}\left( X\right) R\right) }-\frac{C\left( 2X-C\beta \right) }{1+%
\underline{k}\left( X\right) }\right)  \notag
\end{eqnarray}

\subsection{Solution of the investors' returns equation}

We show in appendix 11.5 that equation (\ref{PNm}) becomes after resolution
for $\underline{k}\left( X\right) $ and a first order expansion in $R$:

\begin{eqnarray}
\hat{g}\left( \hat{X}_{1}\right) -\bar{r}^{\prime } &=&\int \left( 1-\left(
1+\underline{\hat{k}}_{2}\left( \hat{X}_{1}\right) \right) \hat{S}%
_{1}^{E}\left( \hat{X}^{\prime },\hat{X}_{1}\right) \right) ^{-1}
\label{Frg} \\
&&\times \left( 1+\underline{\hat{k}}_{2}\left( \hat{X}_{1}\right) \right)
\left( \left( \frac{A\left( \hat{X}^{\prime }\right) }{f_{1}^{2}\left( \hat{X%
}^{\prime }\right) }+\frac{B\left( \hat{X}^{\prime }\right) }{%
f_{1}^{3}\left( \hat{X}^{\prime }\right) }\right) \left( R+\Delta F_{\tau
}\left( \bar{R}\left( K,X\right) \right) \right) \right)  \notag
\end{eqnarray}%
with:%
\begin{eqnarray*}
A\left( \hat{X}^{\prime }\right) &=&\frac{\epsilon \left( 1-\beta \right)
\left( X+C\beta \right) ^{2}}{3\sigma _{\hat{K}}^{2}}\frac{\left( 3X-\beta
C\right) \left( \beta C+X\right) }{4C\left( 2X-C\beta \right) } \\
B\left( \hat{X}^{\prime }\right) &=&\frac{\epsilon \left( 1-\beta \right)
\left( X+C\beta \right) ^{2}}{3\sigma _{\hat{K}}^{2}}\beta RC\left(
2X-C\beta \right)
\end{eqnarray*}%
with::%
\begin{equation*}
X=\sqrt{\sigma _{\hat{K}}^{2}\left( \frac{\left\vert \Psi _{0}\left(
X\right) \right\vert ^{2}}{\epsilon }-\frac{f_{1}\left( X\right) }{2}\right) 
}\rightarrow \sqrt{\sigma _{\hat{K}}^{2}\left( \frac{\left\vert \Psi
_{0}\left( X\right) \right\vert ^{2}}{\epsilon }-\left( \frac{f_{1}\left(
X\right) }{2\left( \left( 1+\underline{k}\left( X\right) \frac{\hat{K}\left[ 
\hat{X}\right] }{K\left[ X\right] }\right) \left\langle K\right\rangle
\right) ^{r}}-\frac{C_{0}}{2}\right) \right) }
\end{equation*}%
An estimation for $\hat{S}_{1}^{E}\left( \hat{X}^{\prime },\hat{X}%
_{1}\right) $ is given in Appendix 11.3:%
\begin{eqnarray*}
&&\hat{S}_{1}^{E}\left( \hat{X}^{\prime },\hat{X}_{1}\right) \\
&\simeq &\left( \frac{\hat{k}\left( \hat{X},\hat{X}^{\prime }\right)
-\left\langle \hat{k}_{1}\right\rangle \hat{k}\left( \left\langle \hat{X}%
\right\rangle ,\hat{X}^{\prime }\right) }{1+\hat{k}\left( \hat{X}\right) }+%
\frac{\hat{k}_{1}\left( \hat{X}^{\prime },\hat{X}\right) }{\left( 1+\hat{k}%
\left( \hat{X}\right) \right) \left( 1+\frac{\hat{k}_{2}\left( \hat{X}%
,\left\langle \hat{X}\right\rangle \right) }{1-\left\langle \hat{k}%
\right\rangle }\right) }\right) \\
&&\times \frac{\sqrt{\left\Vert \hat{\Psi}_{0}\left( \hat{X}\right)
\right\Vert ^{2}-\hat{\mu}D\left( \hat{X}\right) }\left( \left\Vert \hat{\Psi%
}_{0}\left( \hat{X}^{\prime }\right) \right\Vert ^{2}-\hat{\mu}D\left( \hat{X%
}^{\prime }\right) \right) ^{\frac{3}{2}}}{\left( \left\Vert \hat{\Psi}%
_{0}\right\Vert ^{2}-\hat{\mu}\left\langle D\right\rangle \right) ^{2}}
\end{eqnarray*}

\subsection{Interpretation}

The equation for $\hat{g}\left( \hat{X}_{1}\right) $ represents the returns
of the financial sector $\hat{X}_{1}$, which can be a geographical sector, a
type of activity, etc. It depends on the direct returns of the firms to
which it lends or in which it takes stakes, and the returns of other firms,
indirectly, through equity stakes and loans to other investors. When the
expression $\left( 1-\left( 1+\underline{\hat{k}}_{2}\left( \hat{X}%
_{1}\right) \right) \hat{S}_{1}^{E}\left( \hat{X}^{\prime },\hat{X}%
_{1}\right) \right) ^{-1}$ is expanded, we formally have: 
\begin{equation}
1+\left( 1+\underline{\hat{k}}_{2}\left( \hat{X}_{1}\right) \right) \hat{S}%
_{1}^{E}\left( \hat{X}^{\prime },\hat{X}_{1}\right) +\left[ \left( 1+%
\underline{\hat{k}}_{2}\left( \hat{X}_{1}\right) \right) \hat{S}%
_{1}^{E}\left( \hat{X}^{\prime },\hat{X}_{1}\right) \right] ^{2}+\left[
\left( 1+\underline{\hat{k}}_{2}\left( \hat{X}_{1}\right) \right) \hat{S}%
_{1}^{E}\left( \hat{X}^{\prime },\hat{X}_{1}\right) \right] ^{3}+...
\label{SRS}
\end{equation}%
When this series (\ref{SRS}) is applied to the returns of firms: 
\begin{equation}
\left( 1+\underline{\hat{k}}_{2}\left( \hat{X}_{1}\right) \right) \left(
\left( \frac{A\left( \hat{X}^{\prime }\right) }{f_{1}^{2}\left( \hat{X}%
^{\prime }\right) }+\frac{B\left( \hat{X}^{\prime }\right) }{f_{1}^{3}\left( 
\hat{X}^{\prime }\right) }\right) \left( R+\Delta F_{\tau }\left( \bar{R}%
\left( K,X\right) \right) \right) \right)  \label{CTB}
\end{equation}%
it yields an infinite number of contributions. In the series (\ref{SRS}),
the first constant, $1$, directly reflects the return (\ref{CTB}) of the
firm, in which the investor has directly invested. The second term in the
series,$\left( 1+\underline{\hat{k}}_{2}\left( \hat{X}_{1}\right) \right) 
\hat{S}_{1}^{E}\left( \hat{X}^{\prime },\hat{X}_{1}\right) $, applied to (%
\ref{CTB}), yields the indirect return gained by investors on other
investors who got returns from firms. The third term amplifies the
phenomenon: it is the return obtained on the investment in investors who
themselves have invested in investors who have obtained a return. It's a
kind of domino effect. The magnitude of these compounded effects is measured
by the magnitude of $\left( 1+\underline{\hat{k}}_{2}\left( \hat{X}%
_{1}\right) \right) \hat{S}_{1}^{E}\left( \hat{X}^{\prime },\hat{X}%
_{1}\right) $. The magnitude increases with the borrowing rate for
investments, measured by $\underline{\hat{k}}_{2}\left( \hat{X}_{1}\right) $%
. The higher the borrowing rates, the more significant the domino effect.
The second important element is given by the connectivity between agents in
the system, measured by $\hat{S}_{1}^{E}\left( \hat{X}^{\prime },\hat{X}%
_{1}\right) $. It's a characteristic property of the system that measures
the connections between different actors. This matrix is sector-dependent.
Some returns will diffuse from one place to another, but not to others, etc.
These characteristics are measured by the coefficients in $\hat{S}%
_{1}^{E}\left( \hat{X}^{\prime },\hat{X}_{1}\right) $ , which are
interpreted just below.

\subsection{The diffusion matrix}

The various components of the diffusion matrix$\ \hat{S}_{1}^{E}\left( \hat{X%
},\hat{X}^{\prime }\right) $ illustrate how returns disseminate from one
sector $\hat{X}^{\prime }$, to another, $\hat{X}$. The coefficients $\hat{k}$
represent the proportion of investments made by $\hat{X}^{\prime }$in $\hat{X%
}$. There are three terms in this matrix:

\begin{equation*}
\frac{\hat{k}\left( \hat{X},\hat{X}^{\prime }\right) }{1+\underline{\hat{k}}%
_{2}\left( \hat{X}\right) }
\end{equation*}%
signifies the proportion of investments made by $\hat{X}^{\prime }$ in $\hat{%
X}$. Fluctuations in $\hat{X}^{\prime }$ returns can prompt $\hat{X}^{\prime
}$ to adjust its investments in $\hat{X}$, thereby influencing $\hat{X}$'s
returns as well. This term also reflects the proportion of investments made
by other investors in $\hat{X}$. Any changes in their returns lead to
corresponding variations in the invested portion, thereby impacting the
returns of $\hat{X}$.

The second term describes the reciprocal scenario: 
\begin{equation*}
\frac{\hat{k}_{1}\left( \hat{X}^{\prime },\hat{X}\right) }{1+\underline{\hat{%
k}}_{2}\left( \hat{X}^{\prime }\right) }
\end{equation*}%
$\hat{X}$ has invested in $\hat{X}^{\prime }$, entitling it to a share of $%
\hat{X}^{\prime }$ 's return. This represents the direct involvement of
investors in each other's sectors, facilitating the direct transmission of
returns from firms in sector $\hat{X}^{\prime }$.

The two terms described above directly depict how the returns of one
influence the returns of the other, illustrating their direct connection.

The third term moderates this influence by acknowledging that both investors
are also linked to the broader pool of investors:

\begin{equation*}
-\frac{\hat{k}_{1}\left( \left\langle X\right\rangle ,\hat{X}\right) \hat{k}%
\left( \left\langle \hat{X}\right\rangle ,\hat{X}^{\prime }\right) }{1+%
\underline{\hat{k}}_{2}\left( \hat{X}\right) }
\end{equation*}%
Investor $\hat{X}^{\prime }$ not only invests in $\hat{X}$, but also in
other sectors represented by the average $\left\langle \hat{X}\right\rangle $%
. Similarly, investor $\hat{X}$ invests in the overall pool, not exclusively
in $\hat{X}^{\prime }$. Consequently, this third term mitigates the impact
of the direct return from $\hat{X}^{\prime }$ on $\hat{X}$. Although $\hat{X}
$ may receive returns from $\hat{X}^{\prime }$, a portion of these returns
originates not from $\hat{X}^{\prime }$ itself, but from investments made by 
$\hat{X}^{\prime }$ with other investors. Whereas the two first terms of the
diffusion matrix suggest that returns originate directly from the sector
where the investment was made, the third term underscores that these returns
are, at least partially, derived from investments made by both parts on
other investors. This dampens the direct reciprocal transmission between $%
\hat{X}^{\prime }$ and $\hat{X}$.

The final term in the diffusion process is a weighting factor determined by
the number of investors per sector. A higher concentration of investors in
sector $\hat{X}^{\prime }$ increases its impact on sector $\hat{X}$.
Conversely, as the number of investors in sector $\hat{X}$ increases, the
impact on each individual agent in that sector decreases, as they have all
invested proportionally in $\hat{X}^{\prime }$ and inversely in $\hat{X}$.%
\begin{equation*}
\frac{\sqrt{\left\Vert \hat{\Psi}_{0}\left( \hat{X}\right) \right\Vert ^{2}-%
\hat{\mu}D\left( \hat{X}\right) }\left( \left\Vert \hat{\Psi}_{0}\left( \hat{%
X}^{\prime }\right) \right\Vert ^{2}-\hat{\mu}D\left( \hat{X}^{\prime
}\right) \right) ^{\frac{3}{2}}}{\left( \left\Vert \hat{\Psi}_{0}\right\Vert
^{2}-\hat{\mu}\left\langle D\right\rangle \right) ^{2}}
\end{equation*}

\paragraph*{Remark}

We are working with $\hat{g}\left( \hat{X}_{1}\right) $, which is not
exactly the return of the investors but is related to it, denoted by $\hat{f}%
\left( \hat{X}_{1}\right) $. The term $\hat{g}\left( \hat{X}_{1}\right) $
also includes returns coming from investors who have provided capital to the
investor we are studying. The relationship between $\hat{f}\left( \hat{X}%
_{1}\right) $ and $\hat{g}\left( \hat{X}_{1}\right) $ is given by the
following transformation:

\begin{equation*}
\hat{g}\left( \hat{X}_{1}\right) -\bar{r}^{\prime }=\left( 1-M\right)
^{-1}\left( \hat{f}\left( \hat{X}_{1}\right) -\bar{r}\right)
\end{equation*}%
This ensures that the results regarding $\hat{f}\left( \hat{X}_{1}\right) $
or $\hat{g}\left( \hat{X}_{1}\right) $ are interpreted similarly, and it is
more convenient to work with $\hat{g}\left( \hat{X}_{1}\right) $, especially
when dealing with capital by sector. However, note that the equation (\ref%
{Frg}) is also an equation for the return $\hat{f}$:%
\begin{eqnarray}
\left( \hat{f}\left( \hat{X}_{1}\right) -\bar{r}\right) &=&\left( 1-M\right)
\int \left( 1-\left( 1+\underline{\hat{k}}_{2}\left( \hat{X}_{1}\right)
\right) \hat{S}_{1}^{E}\left( \hat{X}^{\prime },\hat{X}_{1}\right) \right)
^{-1}  \label{Frf} \\
&&\times \left( 1+\underline{\hat{k}}_{2}\left( \hat{X}_{1}\right) \right)
\left( \frac{A\left( \hat{X}^{\prime }\right) }{f_{1}^{2}\left( \hat{X}%
^{\prime }\right) }+\frac{B\left( \hat{X}^{\prime }\right) }{f_{1}^{3}\left( 
\hat{X}^{\prime }\right) }\right) \left( R+\Delta F_{\tau }\left( \bar{R}%
\left( K,X\right) \right) \right)  \notag
\end{eqnarray}

\section{Equation for total investors' capital per sector}

We obtained the equation for investor returns by sector. We can also
rephrase this equation as an equation concerning the total capital by sector.

\subsection{Derivation of the equation}

Given the formula (\ref{KP}) for the total capital amount $\hat{K}\left[ 
\hat{X}_{1}\right] $ in sector $\hat{X}_{1}$:%
\begin{eqnarray}
\hat{K}\left[ \hat{X}_{1}\right] &=&\frac{2\sigma _{\hat{K}}^{2}}{\hat{\mu}}%
\left( \frac{\left\Vert \hat{\Psi}_{0}\left( \hat{X}_{1}\right) \right\Vert
^{2}-\hat{\mu}D\left( \hat{X}_{1}\right) }{\hat{g}^{2}\left( \hat{X}%
_{1}\right) }\right) ^{2}\left( \frac{\hat{g}^{2}\left( \hat{X}_{1}\right) }{%
4}-\frac{r\left\langle \hat{g}\right\rangle ^{2}}{3}\underline{\hat{k}}%
\right) \\
&\simeq &\frac{2\sigma _{\hat{K}}^{2}}{\hat{\mu}\hat{g}^{2}\left( \hat{X}%
_{1}\right) }\left( \left\Vert \hat{\Psi}_{0}\left( \hat{X}_{1}\right)
\right\Vert ^{2}-\hat{\mu}D\left( \hat{X}_{1}\right) \right) ^{2}\left( 
\frac{1}{4}-\frac{r}{3}\underline{\hat{k}}\right)  \notag
\end{eqnarray}%
we can write the return equation as an equation for $\hat{K}\left[ \hat{X}%
_{1}\right] $, the total capital in sector $\hat{X}_{1}$. The derivation is
presented in the appendix 11.4 and we have:%
\begin{equation*}
\hat{g}\left( \hat{X}_{1}\right) \simeq \frac{\left( \left\Vert \hat{\Psi}%
_{0}\left( \hat{X}_{1}\right) \right\Vert ^{2}-\hat{\mu}D\left( \hat{X}%
_{1}\right) \right) \sqrt{\frac{1}{4}-\frac{r}{3}\underline{\hat{k}}}}{\sqrt{%
\frac{\hat{\mu}\hat{K}\left[ \hat{X}_{1}\right] }{2\sigma _{\hat{K}}^{2}}}}
\end{equation*}%
where:%
\begin{equation*}
D\left( \hat{X}_{1}\right) =\left( \frac{\left\langle \hat{K}\right\rangle
^{2}\left\langle \hat{g}\right\rangle ^{2}}{\sigma _{\hat{K}}^{2}}+\frac{%
\left\langle \hat{g}\right\rangle }{2}\right) \left( \frac{\underline{\hat{k}%
}\left( \left\langle \hat{X}\right\rangle ,\hat{X}_{1}\right) }{\underline{%
\hat{k}}\left( \left\langle \hat{X}\right\rangle ,\left\langle \hat{X}%
\right\rangle \right) }-\frac{6\underline{\hat{k}}}{2+\underline{\hat{k}}-%
\sqrt{\left( 2+\underline{\hat{k}}\right) ^{2}-\underline{\hat{k}}}}\right) 
\underline{\hat{k}}
\end{equation*}%
This enables to rewrite (\ref{PNm}) as an equation for average capital per
sector as:%
\begin{eqnarray}
&&\left( \frac{\Delta \left( \hat{X},\hat{X}^{\prime }\right) }{1+\underline{%
\hat{k}}_{2}\left( \hat{X}\right) }-\hat{S}_{1}^{E}\left( \hat{X}^{\prime },%
\hat{X}\right) \right) \left( \frac{\left( \left\Vert \hat{\Psi}_{0}\left( 
\hat{X}^{\prime }\right) \right\Vert ^{2}-\hat{\mu}D\left( \hat{X}^{\prime
}\right) \right) \sqrt{\frac{1}{4}-\frac{r}{3}\underline{\hat{k}}}}{\sqrt{%
\frac{\hat{\mu}\hat{K}\left[ \hat{X}^{\prime }\right] }{2\sigma _{\hat{K}%
}^{2}}}}-\bar{r}^{\prime }\right)  \label{PN} \\
&=&\left( \frac{A\left( \hat{X}^{\prime }\right) }{f_{1}^{2}\left( \hat{X}%
^{\prime }\right) }+\frac{B\left( \hat{X}^{\prime }\right) }{f_{1}^{3}\left( 
\hat{X}^{\prime }\right) }\right) \left( R+\Delta F_{\tau }\left( \bar{R}%
\left( K,X\right) \right) \right)  \notag
\end{eqnarray}%
This is the same equation as for returns but described in terms of overall
sectoral capital. The two variables are interconnected, and solving for one
will consequently solve for the other.

\subsection{Interpretation}

The interpretation is similar to that of returns. It's worth noting that the
same diffusion phenomenon exists, and the level of capital will be
contingent on the firm returns. Capital levels between sectors will be
interconnected, resulting in a global resolution. This leads to both
collective states and the multiplicity of these collective states. There can
be multiple collective states, such as capital levels by sector adjusting to
the return of all firms. Different distributions can satisfy this equation
because, in a way, it's the overall firm return that determines the possible
capital distributions. Our focus will be on solving for capital and
examining its dependency on parameters to study capital accumulation.

\section{Total investors' capital per sector}

In this section, we study the equation for the average invested capital per
sector, (\ref{PN})\textbf{\ }This equation aggregates capital from all
sectors, defining a collective state in which average capitals across
sectors are interconnected. Initially, we will examine the straightforward
scenario of constant returns to scale, where invested capital yields uniform
returns regardless of the firm or sector. Subsequently, we will introduce
the impact of decreasing returns to scale, incorporating first-order
adjustments to accommodate non-uniform returns. A comprehensive treatment of
decreasing returns to scale returns is provided in Appendix 12.

\subsection{Total investors' capital per sector under constant return to
scale}

The sectors of firms and the firms themselves are characterized by the
potential returns on investment they could generate for investors. This
implies that the return on investment, expressed as a percentage, is assumed
to be independent of the level of invested capital. Although this assumption
is obviously unrealistic, it serves as an initial benchmark for our model,
allowing us to refine it subsequently.

The resolution proceeds in two steps. Firstly, we analyze the capital within
a financial sector, considered in isolation from others, and examine its
dependence on firm productivity. Subsequently, we incorporate the effects of
interconnected investors across different sectors.

\subsubsection{Isolated investors}

Since we study a group of interconnected investors, the return equation
cannot be solved analytically. Therefore, we will begin with a basic case
and expand our analysis from there.

We first establish a benchmark with a sector that is not interconnected with
others.

Here, we assume temporarily that the sectors operate independently; there
are no interactions between investors.\ Each investor focuses solely on
their specific firm sector and does not invest in other investors. This
implies that the diffusion matrix is null, and the return equation is:%
\begin{equation*}
\frac{\left( \left\Vert \hat{\Psi}_{0}\left( \hat{X}_{1}\right) \right\Vert
^{2}-\hat{\mu}D\left( \hat{X}_{1}\right) \right) \sqrt{\frac{1}{4}-\frac{r}{3%
}\underline{\hat{k}}}}{\sqrt{\frac{\hat{\mu}\hat{K}\left[ \hat{X}_{1}\right] 
}{2\sigma _{\hat{K}}^{2}}}}-\bar{r}^{\prime }=\left( \left( \frac{A\left( 
\hat{X}_{1}\right) }{f_{1}^{2}\left( \hat{X}_{1}\right) }+\frac{B\left( \hat{%
X}_{1}\right) }{f_{1}^{3}\left( \hat{X}_{1}\right) }\right) \left( R+\Delta
F_{\tau }\left( \bar{R}\left( K,\hat{X}_{1}\right) \right) \right) \right)
\end{equation*}%
We saw that:%
\begin{equation*}
\Delta F_{\tau }\left( \bar{R}\left( K,X\right) \right) \simeq \tau \left(
f_{1}\left( X\right) -\left\langle f_{1}\left( X\right) \right\rangle \right)
\end{equation*}%
We can replace in first approximation the parameters $A\left( \hat{X}%
_{1}\right) $ and $B\left( \hat{X}_{1}\right) $\ by their averages:%
\begin{eqnarray*}
A\left( \hat{X}_{1}\right) &\rightarrow &A \\
B\left( \hat{X}_{1}\right) &\rightarrow &B
\end{eqnarray*}%
We also define:%
\begin{equation*}
N\left( \hat{X}_{1}\right) =\left( \left\Vert \hat{\Psi}_{0}\left( \hat{X}%
_{1}\right) \right\Vert ^{2}-\hat{\mu}D\left( \hat{X}_{1}\right) \right) 
\sqrt{\frac{1}{4}-\frac{r}{3}\underline{\hat{k}}}\sqrt{\frac{2\sigma _{\hat{K%
}}^{2}}{\hat{\mu}}}
\end{equation*}%
Here, $N\left( \hat{X}_{1}\right) $ represents a specific parameter for each
sector, which depends on the number of investors in the sector and the
magnitude of investments made within it. The capital equation is written as:%
\begin{equation*}
\pm \frac{N\left( \hat{X}_{1}\right) }{\sqrt{\hat{K}\left[ \hat{X}_{1}\right]
}}-\bar{r}^{\prime }\simeq \left( \frac{A}{\left( f_{1}\left( X\right)
\right) ^{2}}+\frac{B}{\left( f_{1}\left( X\right) \right) ^{3}}\right)
\left( \left( f_{1}\left( X\right) -\bar{r}^{\prime }\right) +\tau F\left(
X\right) \left( f_{1}\left( X\right) -\left\langle f_{1}\left( X\right)
\right\rangle \right) \right)
\end{equation*}

\paragraph{Dependency in productivity}

Even though the solution is only partial because we have isolated the
agents, we can still calculate the dependency of $\hat{K}\left[ \hat{X}_{1}%
\right] $ in $f_{1}\left( X\right) $ as a benchmark. It is given by:%
\begin{eqnarray}
&&\frac{\left( f_{1}\left( X\right) \right) ^{4}}{\left( \hat{K}\left[ \hat{X%
}_{1}\right] \right) ^{\frac{3}{2}}}\frac{d\hat{K}\left[ \hat{X}_{1}\right] 
}{df_{1}\left( X_{1}\right) }  \label{DR} \\
&=&\pm 2\left( B\left( 2f_{1}\left( X\right) -3r\right) +Af_{1}\left(
X\right) \left( f_{1}\left( X\right) -2r\right) +F\tau \left( B\left(
2f_{1}\left( X\right) -3\left\langle f_{1}\left( X\right) \right\rangle
\right) +Ax\left( f_{1}\left( X\right) -2\left\langle f_{1}\left( X\right)
\right\rangle \right) \right) \right)  \notag
\end{eqnarray}%
The sign of the right-hand side depends on whether we are seeking a positive
or negative return solution. For a positive solution, it acts as an
increasing function if it surpasses a certain threshold. For a negative
solution, it increases as long as it stays below the threshold. In the case
of negative returns, the level of capital diminishes quickly with returns.
Even when returns are negative, the level of capital in the sector remains
very low but does not reach zero. Due to the interia in firms displacement,
collective states with negative return and low capital persist. This is the
consequence of our field model, where the average capital maintains a
persistent value, corresponding to the capital level for firms in this
sector. Dynamically, agents vanish and are replaced by others experiencing
the same loss.

The formula (\ref{DR}) indicates that the capital level in the sector
generally increases as the average productivity of firms in the sector, $%
f_{1}\left( X\right) $, increases. Typically, for investments to be
profitable, the firm's productivity must exceed double the interest rate, $r$%
, which is usually the case. More is said about that while introducing the
decreasing returns.

The variation $\frac{d\hat{K}\left[ \hat{X}_{1}\right] }{df_{1}\left(
X_{1}\right) }$ depends on two terms. One term is related to the dividend,
expressed as:\textbf{\ }%
\begin{equation*}
B\left( 2f_{1}\left( X\right) -3r\right) +Af_{1}\left( X\right) \left(
f_{1}\left( X\right) -2r\right)
\end{equation*}%
The other term: 
\begin{equation*}
F\tau \left( B\left( 2f_{1}\left( X\right) -3\left\langle f_{1}\left(
X\right) \right\rangle \right) +Ax\left( f_{1}\left( X\right) -2\left\langle
f_{1}\left( X\right) \right\rangle \right) \right)
\end{equation*}%
represents the price dividend, which quantifies the increase in capital when
productivity exceeds the average productivity across all sectors.

\subsubsection{Interconnected investors}

Stating that investors are interconnected implies that the diffusion matrix $%
\hat{S}_{1}^{E}\left( \hat{X}^{\prime },\hat{X}_{1}\right) $ alters previous
outcomes. However, when returns to scale are constant, this diffusion matrix
remains unaffected by the levels of capital invested in the sectors.
Consequently, the diffusion of returns becomes independent of the capitals
invested in various sectors. As a result, variations in the level of
invested capital do not lead to changes in productivity and, therefore,
returns.

As a consequence, in the scenario of constant returns to scale and
interconnected investors, we can derive the level of capital invested in a
sector as a function of firm returns, yielding:

\begin{eqnarray*}
\pm \frac{N\left( \hat{X}_{1}\right) }{\sqrt{\hat{K}\left[ \hat{X}_{1}\right]
}}-\bar{r}^{\prime } &\simeq &\int \left( 1-\left( 1+\underline{\hat{k}}%
_{2}\left( \hat{X}_{1}\right) \right) \hat{S}_{1}^{E}\left( \hat{X}^{\prime
},\hat{X}_{1}\right) \right) ^{-1}\left( 1+\underline{\hat{k}}_{2}\left( 
\hat{X}_{1}\right) \right) \\
&&\times \left( \frac{A}{\left( f_{1}\left( X^{\prime }\right) \right) ^{2}}+%
\frac{B}{\left( f_{1}\left( X^{\prime }\right) \right) ^{3}}\right) \left(
\left( f_{1}\left( X^{\prime }\right) -\bar{r}^{\prime }\right) +\tau
F\left( X\right) \left( f_{1}\left( X^{\prime }\right) -\left\langle
f_{1}\left( X^{\prime }\right) \right\rangle \right) \right)
\end{eqnarray*}

\subsubsection{Interpretation}

The equation reveals that the average capital level in a sector is
contingent on the global returns of firms and the investors'
interconnections. Firm returns diffuse to financial investors based on the
diffusion matrix, representing the structure of these connections.
Simplifying to constant returns to scale (CRS) has enabled us to derive a
single collective state for a set of firm returns. Indeed, under CRS, the
level of capital between sectors does not affect investment returns, since
the latter are constant. However, this also demonstrates that when returns
are not constant, the capital level of a sector influences the diffusion and
thus the other capital levels in other sectors, leading to a possible
multiplicity of solutions to the capital equation, and therefore collective
states.

\paragraph{Dependency in the parameters}

The dependency of the capital level in one sector on the returns of a firm
from another sector is derived directly and expressed as:%
\begin{eqnarray*}
\frac{\frac{d\hat{K}\left[ \hat{X}_{1}\right] }{df_{1}\left( X^{\prime
}\right) }}{\left( \hat{K}\left[ \hat{X}_{1}\right] \right) ^{\frac{3}{2}}}
&=&\pm \left( 1-\left( 1+\underline{\hat{k}}_{2}\left( \hat{X}_{1}\right)
\right) \hat{S}_{1}^{E}\left( \hat{X}^{\prime },\hat{X}_{1}\right) \right)
^{-1}\frac{\left( 1+\underline{\hat{k}}_{2}\left( \hat{X}_{1}\right) \right) 
}{\left( f_{1}\left( X^{\prime }\right) \right) ^{4}} \\
&&\times 2\left( B\left( 2f_{1}\left( X^{\prime }\right) -3r\right)
+Af_{1}\left( X\right) \left( f_{1}\left( X^{\prime }\right) -2r\right)
\right. \\
&&\left. +F\tau \left( B\left( 2f_{1}\left( X^{\prime }\right)
-3\left\langle f_{1}\left( X\right) \right\rangle \right) +A\left(
f_{1}\left( X^{\prime }\right) -2\left\langle f_{1}\left( X\right)
\right\rangle \right) \right) \right)
\end{eqnarray*}%
The relative change in capital in the sector, expressed as $\frac{d\hat{K}%
\left[ \hat{X}_{1}\right] }{df_{1}\left( X^{\prime }\right) }$is determined
by the variation in the return of a specific firm: 
\begin{equation*}
2\left( B\left( 2f_{1}\left( X^{\prime }\right) -3r\right) +Af_{1}\left(
X\right) \left( f_{1}\left( X^{\prime }\right) -2r\right) +F\tau \left(
B\left( 2f_{1}\left( X^{\prime }\right) -3\left\langle f_{1}\left( X\right)
\right\rangle \right) +A\left( f_{1}\left( X^{\prime }\right) -2\left\langle
f_{1}\left( X\right) \right\rangle \right) \right) \right)
\end{equation*}%
multiplied by a diffusion coefficient between the two sectors. 
\begin{equation*}
\left( 1-\left( 1+\underline{\hat{k}}_{2}\left( \hat{X}_{1}\right) \right) 
\hat{S}_{1}^{E}\left( \hat{X}^{\prime },\hat{X}_{1}\right) \right) ^{-1}%
\frac{\left( 1+\underline{\hat{k}}_{2}\left( \hat{X}_{1}\right) \right) }{%
\left( f_{1}\left( X^{\prime }\right) \right) ^{4}}
\end{equation*}

\subsection{Total investors' capital per sector under decreasing returns to
scale}

Now that we have established the baseline case of constant returns to scale,
we can introduce first-order corrections to account for slowly decreasing
returns to scale. As previously, we will begin by examining the case where
investors are not interconnected, and do not invest among themselves.\
Subsequently, we will incorporate interactions between investors. A more
detailed treatment is presented in Appendices 12.3 and 12.4.

\subsubsection{Isolated investors}

For weakly decreasing returns, we simply replace the productivities $%
f_{1}\left( X\right) $ in the formulas with their equivalent under
decreasing returns, denoted as $f_{1}\left( X,\hat{K}\left[ \hat{X}\right]
\right) $:

\begin{equation*}
f_{1}\left( X,\hat{K}\left[ \hat{X}\right] \right) \equiv \frac{f_{1}\left(
X\right) }{\left( \left( 1+\frac{\underline{k}\left( X\right) }{\left\langle
K\right\rangle }\hat{K}_{X}\left\vert \hat{\Psi}\left( \hat{X}\right)
\right\vert ^{2}\right) K_{X}\right) ^{r}}-C_{0}\simeq \frac{f_{1}\left(
X\right) }{\left( \left( 1+\underline{k}\left( X\right) \hat{K}\left[ \hat{X}%
\right] \right) \right) ^{r}}-C_{0}
\end{equation*}%
The term in the denominator decreases the return when the total invested
capital increases.\ The term $C_{0}$ stands for a fixed cost. In the case of
constant returns, this term did not appear, as it is directly included in
the definition of $f_{1}\left( X\right) $The return equation for investors
solely connected to their sector becomes:%
\begin{eqnarray}
&&\pm \frac{N\left( \hat{X}\right) }{\sqrt{\hat{K}\left[ \hat{X}\right] }}-%
\bar{r}^{\prime }  \label{Nk} \\
&\simeq &\left( \frac{A}{\left( f_{1}\left( X^{\prime },\hat{K}\left[ \hat{X}%
^{\prime }\right] \right) \right) ^{2}}+\frac{B}{\left( f_{1}\left(
X^{\prime },\hat{K}\left[ \hat{X}^{\prime }\right] \right) \right) ^{3}}%
\right)  \notag \\
&&\times \left[ \left( f_{1}\left( X^{\prime },\hat{K}\left[ \hat{X}^{\prime
}\right] \right) -\bar{r}^{\prime }\right) +\tau F\left( X\right) \left(
f_{1}\left( X^{\prime },\hat{K}\left[ \hat{X}^{\prime }\right] \right)
-\left\langle f_{1}\left( X,\hat{K}\left[ \hat{X}\right] \right)
\right\rangle \right) \right]  \notag
\end{eqnarray}%
The dependency of $\hat{K}\left[ \hat{X}\right] $ in $f_{1}\left( X\right) $
is calculated in Appendix 11.6 and is generally increasing, as expected.

\subsubsection{Interconnected investors}

To further understand the possibility of multiple states, we search for
solutions of (\ref{Nk}). The equation is more complex than the one studied
under the CRS assumption because now decreasing returns reveal the capital
of investors in the returns provided by the firms. It cannot be solved
exactly, so we will look for solutions that will be seen as corrections to a
state where investors are isolated, which therefore serves as a benchmark.
Let us call $\Delta \left( \frac{N\left( \hat{X}\right) }{\sqrt{\hat{K}\left[
\hat{X}\right] }}\right) $ the correction due to interactions between
investors, compared to a state without interactions between sectors.
Besides, we define $\hat{K}_{1}\left[ \hat{X}\right] $ as the solution
without interactions between sectors. Thus, we can write:%
\begin{equation*}
\frac{N\left( \hat{X}\right) }{\sqrt{\hat{K}\left[ \hat{X}\right] }}=\frac{%
N\left( \hat{X}\right) }{\sqrt{\hat{K}_{1}\left[ \hat{X}\right] }}+\Delta
\left( \frac{N\left( \hat{X}\right) }{\sqrt{\hat{K}\left[ \hat{X}\right] }}%
\right)
\end{equation*}%
The sought-after solution is therefore the sum of the benchmark plus the
correction to this benchmark due to interactions.

Appendix 11.8 derives an approximate equation for $\Delta \left( \frac{%
N\left( \hat{X}\right) }{\sqrt{\hat{K}\left[ \hat{X}\right] }}\right) $:%
\begin{eqnarray*}
&&\left( \delta \left( \hat{X}-\hat{X}^{\prime }\right) \pm _{\hat{X}}\frac{%
2H_{2}\left( \left( 1+\underline{\hat{k}}_{2}\left( \hat{X}\right) \right) 
\hat{S}_{1}^{E}\left( \hat{X}^{\prime },\hat{X}\right) \right) }{\left(
\left( 1-H_{1}\right) ^{2}-2H_{2}\left( 1+\underline{\hat{k}}_{2}\left( \hat{%
X}\right) \right) \hat{S}_{1}^{E}\left( \hat{X}^{\prime },\hat{X}\right) 
\frac{N\left( \hat{X}^{\prime }\right) }{\sqrt{\hat{K}_{1}\left[ \hat{X}%
^{\prime }\right] }}\right) ^{\frac{3}{2}}}\right) \Delta \left( \frac{%
N\left( \hat{X}^{\prime }\right) }{\sqrt{\hat{K}\left[ \hat{X}^{\prime }%
\right] }}\right) \\
&=&\frac{\left( 1-H_{1}\right) \pm _{\hat{X}}\sqrt{\left( 1-H_{1}\right)
^{2}-\left( 2H_{2}\left( 1+\underline{\hat{k}}_{2}\left( \hat{X}\right)
\right) \hat{S}_{1}^{E}\left( \hat{X}^{\prime },\hat{X}\right) \frac{N\left( 
\hat{X}^{\prime }\right) }{\sqrt{\hat{K}_{1}\left[ \hat{X}^{\prime }\right] }%
}\right) }}{H_{2}}
\end{eqnarray*}%
where the sign $\pm _{\hat{X}}$ means that the choice of sign depends on the
sector considered. There is indeed a double possibility per sector.
Furthermore, the equation is not local; the first line shows that there is
diffusion (presence of $\hat{S}_{1}^{E}\left( \hat{X}^{\prime },\hat{X}%
\right) $), and thus the choice of sign in one sector impacts the other
sectors. The withdrawal of capital invested in one sector will impact the
other sectors connected to it, via investment decisions among investors.
Consequently, entire blocks can end up in a low-capital or high-capital
state. The diffusion and amplification effect depend on the level of
indebtedness $1+\underline{\hat{k}}_{2}\left( \hat{X}\right) $. It is indeed
a diffusion amplified by the level of lending among investors.

These multiple values are only constrained by the overall average of the
system. The consistency of the equations imposes that in averages:%
\begin{equation*}
\int \hat{K}_{1}\left[ \hat{X}\right] +\Delta \hat{K}\left[ \hat{X}\right] d%
\hat{X}=\left\langle \hat{K}\right\rangle \left\Vert \hat{\Psi}\right\Vert
^{2}
\end{equation*}%
so that for a given global state many different collective may arise, with
our without disparity. On the other the overall multiple state may be seen
as the consequence of the local multiple equilibria.

\subsubsection{Interpretation}

When investors are interconnected, and assuming decreasing returns to scale,
the level of capital invested in a sector becomes relevant in the diffusion
of capital, as it impacts the returns of firms.\ This leads to the emergence
of multiple collective states due to circular effects. Even slight
modifications in the capital level of a sector or firm can trigger changes
in returns, impacting other investors' returns and altering their own
capital levels. Consequently, investments and returns of adjacent firms are
modified, initiating a chain reaction. In such scenarios, the emergence of a
single collective state becomes the exception, as indicated by approximate
calculations.

The degree of correlation among investors' states increases with the level
of interconnection. Therefore, the different possible collective states do
not represent independent investors but rather groups of investors,
potentially characterized by varying capital levels, such as one group with
high capital and another with low capital.

\paragraph{Dependency in other sectors' parameters}

Given that here, investors are connected, we can calculate the impact of a
variation in firm returns in one sector on the overall capital level of the
investors. If we expand the equation for capital (\ref{Nk}) in detail: 
\begin{eqnarray}
&&\pm \frac{N\left( \hat{X}\right) }{\sqrt{\hat{K}\left[ \hat{X}\right] }}-%
\bar{r}^{\prime }\simeq \int \left( 1-\left( 1+\underline{\hat{k}}_{2}\left( 
\hat{X}\right) \right) \hat{S}_{1}^{E}\left( \hat{X}^{\prime },\hat{X}%
\right) \right) ^{-1}\left( 1+\underline{\hat{k}}_{2}\left( \hat{X}\right)
\right)  \label{NCr} \\
&&\times \left( \frac{A}{\left( f_{1}\left( X^{\prime },\hat{K}\left[ \hat{X}%
^{\prime }\right] \right) \right) ^{2}}+\frac{B}{\left( f_{1}\left(
X^{\prime },\hat{K}\left[ \hat{X}^{\prime }\right] \right) \right) ^{3}}%
\right)  \notag \\
&&\times \left[ \left( f_{1}\left( X^{\prime },\hat{K}\left[ \hat{X}^{\prime
}\right] \right) -\bar{r}^{\prime }\right) +\tau F\left( X^{\prime }\right)
\left( f_{1}\left( X^{\prime },\hat{K}\left[ \hat{X}^{\prime }\right]
\right) -\left\langle f_{1}\left( X,\hat{K}\left[ \hat{X}\right] \right)
\right\rangle \right) \right]  \notag
\end{eqnarray}%
we see a relationship between the capital levels of all sectors. A
collective state consists of a large number of related values of capital.
The interdependence of capital in one sector $\hat{X}_{1}$ with respect to
another sector $\hat{X}^{\prime }$ is computed in Appendix 11.6, and yields: 
\begin{eqnarray*}
\frac{\frac{d\hat{K}\left[ \hat{X}_{1}\right] }{df_{1}\left( X^{\prime
}\right) }}{\left( \hat{K}\left[ \hat{X}_{1}\right] \right) ^{\frac{3}{2}}}
&=&\pm \left( 1-\left( 1+\underline{\hat{k}}_{2}\left( \hat{X}_{1}\right)
\right) \hat{S}_{1}^{E}\left( \hat{X}^{\prime },\hat{X}_{1}\right) \right)
^{-1} \\
&&\times \left( 1+\underline{\hat{k}}_{2}\left( \hat{X}_{1}\right) \right) 
\frac{\left( 1+\underline{k}\left( X^{\prime }\right) \hat{K}\left[ \hat{X}%
^{\prime }\right] \right) \left( AF_{1}\left( X^{\prime }\right)
+BF_{2}\left( X^{\prime }\right) \right) }{f_{1}\left( X\right) \left(
AF_{1}\left( X^{\prime }\right) +BF_{2}\left( X^{\prime }\right) \right) -%
\frac{1}{2}F_{3}\left( X^{\prime }\right) }
\end{eqnarray*}%
with:%
\begin{eqnarray*}
F_{1}\left( X^{\prime }\right) &=&\frac{f_{1}\left( X^{\prime },\hat{K}\left[
\hat{X}^{\prime }\right] \right) -2r\left( f_{1}\left( X^{\prime },\hat{K}%
\left[ \hat{X}^{\prime }\right] \right) -\bar{r}^{\prime }\right) }{\left(
f_{1}\left( X^{\prime },\hat{K}\left[ \hat{X}^{\prime }\right] \right)
\right) ^{3}} \\
F_{2}\left( X^{\prime }\right) &=&\frac{f_{1}\left( X^{\prime },\hat{K}\left[
\hat{X}^{\prime }\right] \right) -3r\left( f_{1}\left( X^{\prime },\hat{K}%
\left[ \hat{X}^{\prime }\right] \right) -\bar{r}^{\prime }\right) }{\left(
f_{1}\left( X^{\prime },\hat{K}\left[ \hat{X}^{\prime }\right] \right)
\right) ^{4}} \\
F_{3}\left( X^{\prime }\right) &=&\frac{\left( \left( 1+\underline{k}\left(
X^{\prime }\right) \hat{K}\left[ \hat{X}\right] \right) ^{1+r}\right)
H\left( \hat{X}_{1}\right) }{\left( \hat{K}\left[ \hat{X}_{1}\right] \right)
^{\frac{3}{2}}}
\end{eqnarray*}%
In general, this relationship is positive, showing that returns from distant
firms can diffuse to the capital level of a sector through successive and
complex investment chains. Similar to the case of constant returns to scale
(CRS), productivity must exceed a certain threshold depending on the
interest rate, to have a positive impact on the level of capital, measured
by the term: 
\begin{equation*}
\frac{f_{1}\left( X^{\prime },\hat{K}\left[ \hat{X}^{\prime }\right] \right)
-2r\left( f_{1}\left( X^{\prime },\hat{K}\left[ \hat{X}^{\prime }\right]
\right) -\bar{r}^{\prime }\right) }{\left( f_{1}\left( X^{\prime },\hat{K}%
\left[ \hat{X}^{\prime }\right] \right) \right) ^{3}}
\end{equation*}%
which is positive in general since $2r<<1$.

\section{Global average capital and return}

In the entire system, the parameters depend on the overall average
quantities. Now that we have found solutions in terms of average quantities,
we can compute them. Since there are various collective states, the averages
themselves may be multiple, characterizing the collective states.

We compute the various averages arising in the model in Appendix 11.7. These
quantities arise in the previous equations and solutions to compute the
various parameters on which the solutions depend. These averages thus
complete the solutions of the model.

\subsection{Firms global average capital and return}

\subsubsection{Firms global average capital}

The private average capital per firm, $\left\langle K\right\rangle $, is:%
\begin{equation*}
\left\langle K\right\rangle \simeq \left\langle K\right\rangle _{1}\left( 1-%
\frac{\left( 3\sqrt{\frac{\left\vert \Psi _{0}\right\vert ^{2}}{\epsilon }-%
\frac{1}{2}\left\langle f_{1}\right\rangle }-\beta C\right) \left( \sqrt{%
\frac{\left\vert \Psi _{0}\right\vert ^{2}}{\epsilon }-\frac{1}{2}%
\left\langle f_{1}\right\rangle }+\beta C\right) }{4\left\langle
f_{1}\right\rangle \left\langle K\right\rangle _{1}\left( 2\sqrt{\frac{%
\left\vert \Psi _{0}\right\vert ^{2}}{\epsilon }-\frac{1}{2}\left\langle
f_{1}\right\rangle }-\beta C\right) }\right) ^{\frac{1}{r}}
\end{equation*}%
with:%
\begin{equation*}
\left\langle K\right\rangle _{1}=\frac{\left\langle f_{1}\right\rangle ^{%
\frac{1+r^{2}}{r}}}{C_{0}^{\frac{1}{r}}\left( \left\langle
f_{1}\right\rangle ^{r}+\left\langle \underline{k}\right\rangle \left\langle 
\hat{K}\right\rangle C_{0}^{r}\frac{\left\Vert \hat{\Psi}_{0}\right\Vert ^{2}%
}{\left\Vert \Psi _{0}\right\Vert ^{2}}\right) }
\end{equation*}%
This average capital per firm is increasing with average productivity $%
\left\langle f_{1}\right\rangle $, decreases with the cost of capital $C_{0}$%
, and decreases with the number of firms in the sector, since the firms
share capital invested in the sector:%
\begin{equation*}
\left\langle \underline{k}\right\rangle \left\langle \hat{K}\right\rangle 
\frac{\left\Vert \hat{\Psi}_{0}\right\Vert ^{2}}{\left\Vert \Psi
_{0}\right\Vert ^{2}}
\end{equation*}%
We see that $\left\langle K\right\rangle $ is inversely proportional to the
invested capital. This doesn't mean that the more we invest, the less the
firm has capital, but rather that the share of private capital decreases.
The private capital $\left\langle K\right\rangle $ is displaced by the
investment $\left\langle \hat{K}\right\rangle $, augmented by the effect of
investor diffusion. Indeed, the average global available capital level for
firms is given by: 
\begin{equation*}
\left\langle K\right\rangle \left( 1+\left\langle \underline{k}\right\rangle
\left\langle \hat{K}\right\rangle \frac{\left\Vert \hat{\Psi}_{0}\right\Vert
^{2}}{\left\Vert \Psi _{0}\right\Vert ^{2}}\right)
\end{equation*}

And this quantity, in turn, increases with the invested capital, provided
that $\left\langle f_{1}\right\rangle >C_{0}$, which is generally the case
since it is the condition for production. The term $\frac{\left\Vert \hat{%
\Psi}_{0}\right\Vert ^{2}}{\left\Vert \Psi _{0}\right\Vert ^{2}}$ represents
the ratio of the number of investors to the number of firms. The more
investors there are, the more capital comes to invest, and the more the
share of private capital decreases.

\subsubsection{Firms' global average return}

We can also compute the average return of firms, $\left\langle f_{1}^{\left(
e\right) }\right\rangle $:

\begin{eqnarray*}
\left\langle f_{1}^{\left( e\right) }\right\rangle &\simeq &\left\langle
f_{1}\right\rangle +\beta \left\langle \underline{k}\right\rangle \frac{%
\left( \frac{1}{4}-\frac{r}{3}\hat{k}\right) }{\left( \frac{1}{3}-\frac{r}{2}%
\hat{k}\right) \bar{r}^{\prime }} \\
&&\times \frac{4\left\langle f_{1}^{\left( e\right) }\right\rangle \left( 2%
\sqrt{\frac{\left\vert \Psi _{0}\right\vert ^{2}}{\epsilon }-\frac{1}{2}%
\left\langle f_{1}\right\rangle }-\beta C\right) \sqrt{\frac{\sigma _{\hat{K}%
}^{2}}{2\hat{\mu}}\left( \left\Vert \hat{\Psi}_{0}\right\Vert ^{2}-\hat{\mu}%
\left\langle D\right\rangle \right) }}{\left( 3\sqrt{\frac{\left\vert \Psi
_{0}\right\vert ^{2}}{\epsilon }-\frac{1}{2}\left\langle f_{1}\right\rangle }%
-\beta C\right) \left( \sqrt{\frac{\left\vert \Psi _{0}\right\vert ^{2}}{%
\epsilon }-\frac{1}{2}\left\langle f_{1}\right\rangle }+\beta C\right) }%
\frac{\left\Vert \hat{\Psi}_{0}\right\Vert ^{2}}{\left\Vert \Psi
_{0}\right\Vert ^{2}}\left( \left\langle f_{1}\right\rangle -\bar{r}\right)
\end{eqnarray*}

The actual return of firms $\left\langle f_{1}^{\left( e\right)
}\right\rangle $ sees its deviation from the average increase with the share
of loans from investors, rather than equity stakes, and with the ratio of
the number of investors to the number of firms in the sector.

Indeed, if firms have a return higher than the interest rate, they benefit
from having made a loan rather than issuing equity stakes.

\subsection{Investors' global average capital and return}

\subsubsection{Investors' global average return}

The average return for investor, $\left\langle \hat{g}\right\rangle $, is:

\begin{equation}
\left\langle \hat{g}\right\rangle =\left\langle \Delta \right\rangle +\bar{r}%
^{\prime }  \label{gchap}
\end{equation}%
where $\left\langle \Delta \right\rangle $ represents the average excess
return for the overall set of investors.\ It satisfies the average of
equation (\ref{Frg}), given by:%
\begin{equation}
\left\langle \Delta \right\rangle =\left\langle \left( \frac{A\left(
X\right) }{\left( f_{1}^{\left( r\right) }\left( X\right) \right) ^{2}}+%
\frac{B\left( X\right) }{\left( f_{1}^{\left( r\right) }\left( X\right)
\right) ^{3}}\right) \right\rangle \left\langle R+\Delta F_{\tau }\left( 
\bar{R}\left( K,X\right) \right) \right\rangle  \label{Dlt}
\end{equation}%
with:%
\begin{eqnarray*}
f_{1}^{\left( r\right) }\left( X\right) &=&\frac{f_{1}\left( X\right) }{%
\left( 1+\left\langle \underline{k}\right\rangle \frac{\left\langle \hat{K}%
\right\rangle \left\Vert \hat{\Psi}_{0}\right\Vert ^{2}}{\left\Vert \Psi
_{0}\right\Vert ^{2}}\right) ^{r}}-C_{0} \\
R &=&\frac{f_{1}\left( X\right) }{\left( 1+\left\langle \underline{k}%
\right\rangle \frac{\left\langle \hat{K}\right\rangle \left\Vert \hat{\Psi}%
_{0}\right\Vert ^{2}}{\left\Vert \Psi _{0}\right\Vert ^{2}}\right) ^{r}}%
-C_{0}-\bar{r}
\end{eqnarray*}

\subsubsection{Investors' global average capital}

Using equations (\ref{VRf}) and (\ref{VRs}), we find the average number of
investors, $\left\Vert \hat{\Psi}\right\Vert ^{2}$: 
\begin{equation}
\left\Vert \hat{\Psi}\right\Vert ^{2}\simeq \frac{\hat{\mu}V}{3\sigma _{\hat{%
K}}^{2}\left\langle \hat{g}\right\rangle }\left( 2\frac{\sigma _{\hat{K}%
}^{2}\left( \frac{\left\Vert \hat{\Psi}_{0}\right\Vert ^{2}}{\hat{\mu}}+%
\frac{\left\langle \hat{g}\right\rangle }{2}\left( \frac{1-r}{r}\right) 
\underline{\hat{k}}\right) }{\left( 1-2r\left( 1-r\right) \underline{\hat{k}}%
\right) }\right) ^{\frac{3}{2}}\left( 1-\frac{2+\underline{\hat{k}}-\sqrt{%
\left( 2+\underline{\hat{k}}\right) ^{2}-\underline{\hat{k}}}}{4\underline{%
\hat{k}}}\underline{\hat{k}}\right)  \label{Psf}
\end{equation}%
and the total available capital, $\left\langle \hat{K}\right\rangle
\left\Vert \hat{\Psi}\right\Vert ^{2}$:%
\begin{equation}
\left\langle \hat{K}\right\rangle \left\Vert \hat{\Psi}\right\Vert ^{2}=%
\frac{\hat{\mu}V\sigma _{\hat{K}}^{2}}{2\left\langle \hat{g}\right\rangle
^{2}}\left( \frac{\left( \frac{\left\Vert \hat{\Psi}_{0}\right\Vert ^{2}}{%
\hat{\mu}}+\frac{\left\langle \hat{g}\right\rangle }{2}\left( \frac{1-r}{r}%
\right) \underline{\hat{k}}\right) }{\left( 1-2r\left( 1-r\right) \underline{%
\hat{k}}\right) }\right) ^{2}\left( 1-2\frac{2+\underline{\hat{k}}-\sqrt{%
\left( 2+\underline{\hat{k}}\right) ^{2}-\underline{\hat{k}}}}{9}\right)
\label{Psk}
\end{equation}%
where $\left\langle \hat{g}\right\rangle $ is given by (\ref{gchap}), and
with:%
\begin{equation*}
r=\frac{2+\underline{\hat{k}}-\sqrt{\left( 2+\underline{\hat{k}}\right) ^{2}-%
\underline{\hat{k}}}}{6\hat{k}}
\end{equation*}%
Appendix 11.7 shows that $\left\Vert \hat{\Psi}\right\Vert ^{2}$ and $%
\left\langle \hat{K}\right\rangle \left\Vert \hat{\Psi}\right\Vert ^{2}$
increase with average productivity $\left\langle f_{1}\left( X\right)
\right\rangle $. Formula (\ref{Psf}) and (\ref{Psk}) show that both $%
\left\Vert \hat{\Psi}\right\Vert ^{2}$ and $\left\langle \hat{K}%
\right\rangle \left\Vert \hat{\Psi}\right\Vert ^{2}$ increase with$\ 
\underline{\hat{k}}$, the average interconnectivity of investors.

\subsubsection{Interpretation}

The excess return compared to the interest rate is proportional to the
excess productivity compared to the interest rate and relative to the
average productivity, $R+\Delta F_{\tau }\left( \bar{R}\left( K,X\right)
\right) $, as expected. It is also proportional to the percentage of equity
stake $\left( 1-\beta \right) $ in investments. This outcome was also
expected, as it contrasts with the situation for firms. When returns are
significant, firms benefit from loans, while investors benefit from having
taken equity stakes. Factors like $\left( X+C\beta \right) $ essentially
depend on the number of firms in the economy. The more firms there are, the
higher the return. Additionally, the return is proportional to the ratio $%
\frac{\epsilon }{3\sigma _{\hat{K}}^{2}}$, which computes the variability of
the number of firms in a sector. This ratio reflects the ease with which
firms can position themselves across sectors. The higher it is, the more
firms can move across sectors; conversely, the lower it is, the more static
the firms will be.

The gains of firms, on average, propagate to investors. This propagation is
also proportional to the number of firms $\left\Vert \Psi _{0}\right\Vert
^{2}$: the more firms there are, the greater the returns.\ Conversely, it is
inversely proportional to the number of investors, $\left\Vert \hat{\Psi}%
_{0}\right\Vert ^{2}$, who must share the returns.

\subsection{Multiple averages}

Combining (\ref{gchap}) nd (\ref{Psk}) leads to rewrite the link between
average amount of capital and average return as an equation for average
capital. Equation (\ref{Psk}) leads to :%
\begin{equation*}
\left\langle \hat{g}\right\rangle =\frac{\sqrt{\frac{V\sigma _{\hat{K}}^{2}}{%
2\hat{\mu}}}\left\Vert \hat{\Psi}_{0}\right\Vert ^{2}}{\left( 1-2r\left(
1-r\right) \underline{\hat{k}}\right) \sqrt{\frac{\left\langle \hat{K}%
\right\rangle \left\Vert \hat{\Psi}\right\Vert ^{2}}{1-2\frac{2+\underline{%
\hat{k}}-\sqrt{\left( 2+\underline{\hat{k}}\right) ^{2}-\underline{\hat{k}}}%
}{9}}}-\frac{1}{2}\left( \frac{1-r}{r}\right) \underline{\hat{k}}}
\end{equation*}%
and comparing to (\ref{Dlt}) leads to:%
\begin{eqnarray*}
&&\frac{\sqrt{\frac{V\sigma _{\hat{K}}^{2}}{2\hat{\mu}}}\left\Vert \hat{\Psi}%
_{0}\right\Vert ^{2}}{\left( 1-2r\left( 1-r\right) \underline{\hat{k}}%
\right) \sqrt{\frac{\left\langle \hat{K}\right\rangle \left\Vert \hat{\Psi}%
\right\Vert ^{2}}{1-2\frac{2+\underline{\hat{k}}-\sqrt{\left( 2+\underline{%
\hat{k}}\right) ^{2}-\underline{\hat{k}}}}{9}}}-\frac{1}{2}\left( \frac{1-r}{%
r}\right) \underline{\hat{k}}} \\
&=&\bar{r}^{\prime }+\left( \frac{A}{\left( \left\langle f_{1}^{\left(
r\right) }\left( X\right) \right\rangle \right) ^{2}}+\frac{B}{\left(
\left\langle f_{1}^{\left( r\right) }\left( X\right) \right\rangle \right)
^{3}}\right) \left\langle \left( f_{1}^{\left( r\right) }\left( X\right)
+\Delta F_{\tau }\left( \bar{R}\left( K,X\right) \right) \right)
\right\rangle
\end{eqnarray*}%
with:%
\begin{equation*}
f_{1}^{\left( r\right) }\left( X\right) =\frac{f_{1}\left( X\right) }{\left(
1+\left\langle \underline{k}\right\rangle \frac{\left\langle \hat{K}%
\right\rangle \left\Vert \hat{\Psi}\right\Vert ^{2}}{\left\Vert \Psi
_{0}\right\Vert ^{2}}\right) ^{r}}-C_{0}
\end{equation*}%
For:%
\begin{equation*}
\left\langle \left( R+\Delta F_{\tau }\left( \bar{R}\left( K,X\right)
\right) \right) \right\rangle >0
\end{equation*}%
this equation yields multiple solutions across a broad range of parameters.
It is worth noting that when $\left\Vert \hat{\Psi}_{0}\right\Vert ^{2}$ is
low, the solution is unique. \ In general three solutions for capital arise.
Furthermore, as the total amount of capital increases, the average return
tends to decrease.

For:%
\begin{equation*}
\left\langle \left( R+\Delta F_{\tau }\left( \bar{R}\left( K,X\right)
\right) \right) \right\rangle <0
\end{equation*}%
only one solution exists when capital is low. This capital is not equal to $%
0 $, as even with a small number of agents, it does not vanish completely.

\subsection{Synthesis on collective states under a non-default scenario}

Our results in sections 12 and 13, obtained by sector and on average, show
that, in general, there is a multiplicity of collective states.\ They may
arise either at the level of global averages, or in the distribution of
these global averages.

At the level of global averages, several possibilities arise, spanning from
a high capital level, a relatively low excess return and a large number of
agents, to that of a lower capital with a smaller number of agents and
higher returns.

Furthermore, for a given average capital level, multiple collective states
corresponding to different capital distributions per sector may emerge, from
very homogeneous to very heterogeneous distributions.

These situations result from the interconnection of sectors and firm
returns, which depend on the capital invested. Firms could lack invested
capital due to a highly concentrated capital in other sectors. This lack of
capital can arise from a vicious circle, where the lack of return implies
the lack of investment, itself implying a lack of return... With these
results at hand, we can now turn to the study of investor returns under a
default scenario.

\section{Investors' returns under a default scenario}

To account for defaults in the resolution of the system, we must write the
return equation recursively: we start with a minimal set of sectors
experiencing a default, then study their recursive propagation. The
collective state with default is reached when the recursive process
converges toward an entire sector space divided into two stable subsets,
each corresponding to a default or non-default sets, respectively. We
present below the recursive resolution.

\subsection{Resolution}

Let us consider a minimal domain $D$, such that the return of the sector $%
\hat{f}\left( \hat{X}_{1}\right) $, given by (\ref{Frf}), satisfies:%
\begin{equation*}
\hat{f}\left( \hat{X}_{1}\right) -\bar{r}<-1
\end{equation*}%
This corresponds to the set of sectors such defaults occur. The return is so
low that the firm's or investor's equity alone is insufficient to repay the
loans.

The return equation, in terms of average capital per sector, is modified by
this default set, denoted as $DF_{0}$. 
\begin{eqnarray*}
&&\left( \frac{A\left( \hat{X}_{1}\right) }{f_{1}^{2}\left( \hat{X}%
_{1}\right) }+\frac{B\left( \hat{X}_{1}\right) }{f_{1}^{3}\left( \hat{X}%
_{1}\right) }\right) \left( R+\Delta F_{\tau }\left( \bar{R}\left( K,\hat{X}%
_{1}\right) \right) \right) =\frac{\left( \left\Vert \hat{\Psi}_{0}\left( 
\hat{X}_{1}\right) \right\Vert ^{2}-\hat{\mu}D\left( \hat{X}_{1}\right)
\right) \sqrt{\frac{1}{4}-\frac{r}{3}\underline{\hat{k}}}}{\sqrt{\frac{\hat{%
\mu}\hat{K}\left[ \hat{X}_{1}\right] }{2\sigma _{\hat{K}}^{2}}}}-\bar{r}%
^{\prime } \\
&&-\int_{H/DF_{0}}\hat{S}_{1}^{E}\left( \hat{X}^{\prime },\hat{X}_{1}\right)
\left( \frac{\left( \left\Vert \hat{\Psi}_{0}\left( \hat{X}^{\prime }\right)
\right\Vert ^{2}-\hat{\mu}D\left( \hat{X}^{\prime }\right) \right) \sqrt{%
\frac{1}{4}-\frac{r}{3}\underline{\hat{k}}}}{\sqrt{\frac{\hat{\mu}\hat{K}%
\left[ \hat{X}^{\prime }\right] }{2\sigma _{\hat{K}}^{2}}}}-\bar{r}^{\prime
}\right) -\int_{DF_{0}}\hat{k}_{2}\left( \hat{X}_{1},\hat{X}^{\prime
}\right) \frac{1+\hat{f}\left( \hat{X}^{\prime }\right) }{1+\underline{\hat{k%
}}\left( \hat{X}^{\prime }\right) }
\end{eqnarray*}%
The modification arises in the second line. Now, the diffusion of returns is
constrained to the sectors with no default, denoted as $H/DF_{0}$.
Conversely, the loss stems from defaulting sectors located in the subspace $%
DF_{0}$:%
\begin{equation*}
\int_{DF_{0}}\hat{k}_{2}\left( \hat{X}_{1},\hat{X}^{\prime }\right) \frac{1+%
\bar{g}\left( \hat{X}^{\prime }\right) }{1+\underline{\hat{k}}\left( \hat{X}%
^{\prime }\right) }
\end{equation*}%
and propagates to the whole sector space.

As a consequence for $\hat{X}_{1}$ not belonging to $D$, that is for $\hat{X}%
_{1}$ a sector with no defaults, the return equation, expressed in terms of $%
\hat{f}\left( \hat{X}_{1}\right) $ becomes: 
\begin{eqnarray*}
\left( \hat{f}\left( \hat{X}_{1}\right) -\bar{r}\right) &=&\left( 1-M\right)
\int \left( 1-\frac{3\sigma _{\hat{K}}^{2}f_{1}\left( X\right) }{C\left(
2X-C\beta \right) \left( 1-\beta \right) \left( X+C\beta \right)
^{2}\epsilon }\hat{S}_{1}^{E}\left( \hat{X}^{\prime },\hat{X}_{1}\right)
-1_{DF_{0}}\frac{\hat{k}_{2}\left( \hat{X}^{\prime },\hat{X}_{1}\right) }{1+%
\underline{\hat{k}}\left( \hat{X}^{\prime }\right) }\right) ^{-1} \\
&&\times \left( \left( \frac{A\left( \hat{X}^{\prime }\right) }{%
f_{1}^{2}\left( \hat{X}^{\prime }\right) }+\frac{B\left( \hat{X}^{\prime
}\right) }{f_{1}^{3}\left( \hat{X}^{\prime }\right) }\right) \left( R+\Delta
F_{\tau }\left( \bar{R}\left( K,X\right) \right) \right) -\frac{1_{DF_{0}}%
\hat{k}_{2}\left( \hat{X}^{\prime },\hat{X}_{1}\right) \left( 1+\bar{r}%
^{\prime }\right) }{1+\underline{\hat{k}}\left( \hat{X}^{\prime }\right) }%
\right)
\end{eqnarray*}%
This equation shows that the sector $\hat{X}_{1}$, initially considered as
non-defaulting, may become part of the default set if the the loss
associated with the default of other sectors exceeds the threshold for
default to occur.

This outcome suggests the resolution method to identify the stable default
set of some collective state. Beginning with an initial default set, we
recursively define \ the $n$-th default set: 
\begin{equation*}
DF_{n}=\left\{ \hat{X}_{1},\hat{f}_{n-1}\left( \hat{X}_{1}\right) <-1\right\}
\end{equation*}%
where $\hat{f}_{n}\left( \hat{X}_{1}\right) $ is defined by:%
\begin{eqnarray*}
\left( \hat{f}_{n+1}\left( \hat{X}_{1}\right) -\bar{r}\right) &=&\left(
1-M\right) \int \left( 1-\frac{3\sigma _{\hat{K}}^{2}f_{1}\left( X\right) }{%
C\left( 2X-C\beta \right) \left( 1-\beta \right) \left( X+C\beta \right)
^{2}\epsilon }\hat{S}_{1}^{E}\left( \hat{X}^{\prime },\hat{X}_{1}\right)
-1_{DF_{n}}\frac{\hat{k}_{2}\left( \hat{X}^{\prime },\hat{X}_{1}\right) }{1+%
\underline{\hat{k}}\left( \hat{X}^{\prime }\right) }\right) ^{-1} \\
&&\times \left( \left( \frac{A\left( \hat{X}^{\prime }\right) }{%
f_{1}^{2}\left( \hat{X}^{\prime }\right) }+\frac{B\left( \hat{X}^{\prime
}\right) }{f_{1}^{3}\left( \hat{X}^{\prime }\right) }\right) \left( R+\Delta
F_{\tau }\left( \bar{R}\left( K,X\right) \right) \right) -\frac{1_{DF_{n}}%
\hat{k}_{2}\left( \hat{X}^{\prime },\hat{X}_{1}\right) \left( 1+\bar{r}%
^{\prime }\right) }{1+\underline{\hat{k}}\left( \hat{X}^{\prime }\right) }%
\right)
\end{eqnarray*}%
Ultimately, the return for a collective state with default is given by the
limit:%
\begin{equation*}
\hat{f}_{n}\left( \hat{X}_{1}\right) \rightarrow \hat{f}\left( \hat{X}%
_{1}\right)
\end{equation*}%
The resolution resembles to a transmission mechanism. Beginning with a
default-free state, we assume that some sectors experience default. This is
feasible for real sectors with low returns relative to the average and
collective state characterized by low capital. In such scenarios, the
collective state undergoes a transition towards a state with negative
returns for some sectors.\ Describing this new collective state with
defaulting sectors may imply that other sectors are affected, transmitting
defaults to other sectors, ultimately leading to a collective default state.
However, it is important to note that this dynamic description should not
obscure the fact that the resulting default state is not solely the
consequence of adverse shocks but rather a intrinsic possibility of the
system.

This aspect can be studied more precisely by dividing the sector space into
several interacting groups. Introducing some dynamic aspects in default
transmission can be achieved by considering these groups as independent and
weakly interacting.

\section{Dynamics Interactions of agents inside or across groups of investors%
}

The averages were computed across the entire space. However, we can consider
that sector spaces, investors, and firms are organized into heterogeneous
groups or sub-markets, which are relatively weakly interconnected.

In this section, rather than studying the system by sectors and/or globally,
we will consider subgroups and investigate the dynamics of investors within
these subgroups, interacting with agents from other groups. This approach
allows for a dynamic examination of the possibility of transitions between
collective states.

As a benchmark and to introduce some notations, we first reconsider the case
of a homogeneous group: a homogeneous group of investors and firms, with
approximately homogeneous returns, connections, capital, and productions. We
describe the system's averages and then consider the interactions among
agents from multiple groups.

\subsection{Benchmark: dynamics of agents inside an homogeneous group of
agents}

We describe a homogeneous group using averages, which simplifies the
notations. We then consider the effective action of agents within this
group, describing their interactions and transition functions. This
describes the dynamic aspect within a given collective state. The aim is to
understand the dynamic and micro aspects that can lead to transitions
between collective states.

\subsubsection{Averages for an homogeneous group}

When we consider the system as a group of homogeneous agents with identical
average returns, Appendix 13.1 shows that the return equation, including
defaults is:%
\begin{eqnarray}
&&\left( \frac{f\left( \hat{X}\right) }{1+\underline{\hat{k}}_{2}\left( \hat{%
X}\right) }+\bar{r}\frac{\underline{\hat{k}}_{2}\left( \hat{X}\right) }{1+%
\underline{\hat{k}}_{2}\left( \hat{X}\right) }\right) -\frac{\hat{K}\hat{k}%
_{1E}\left( \hat{X}\right) }{1+\underline{\hat{k}}}\left( \frac{f}{1+%
\underline{\hat{k}}_{2}}+\bar{r}\frac{\underline{\hat{k}}_{2}}{1+\underline{%
\hat{k}}_{2}}\right)  \label{Rt} \\
&=&\left\{ \left( \bar{r}+\frac{1+f}{\underline{\hat{k}}_{2}}H\left( -\frac{%
1+f}{\underline{\hat{k}}_{2}}\right) \right) \frac{\hat{k}_{2E}\left( \hat{X}%
\right) }{1+\underline{\hat{k}}\left( \hat{X}^{\prime }\right) }\right. 
\notag \\
&&\left. +\left( \bar{r}+\frac{1+f_{1}^{\prime }\left( X^{\prime }\right) }{%
\underline{k}_{2}\left( X^{\prime }\right) }H\left( -\frac{1+f_{1}^{\prime
}\left( X^{\prime }\right) }{\underline{k}_{2}\left( X^{\prime }\right) }%
\right) \right) \frac{k_{2E}\left( \hat{X}\right) }{1+\underline{k}\left(
X^{\prime }\right) }+\frac{k_{1E}\left( \hat{X}\right) }{1+\underline{k}%
\left( X^{\prime }\right) }f_{1}\left( \hat{K},\hat{X},\Psi ,\hat{\Psi}%
\right) \right\}  \notag
\end{eqnarray}%
where $\underline{\hat{k}}_{2}$ and $\underline{\hat{k}}_{1}$\ are the\
averages of $\underline{\hat{k}}_{2}\left( \hat{X}\right) $ and $\underline{%
\hat{k}}_{2}\left( \hat{X}\right) $, representing the average share of
investment in sector $\hat{X}$ by other sectors. Recall that $\hat{k}%
_{1E}\left( \hat{X}\right) $ denotes the average share of outgoing
investment from sector $\hat{X}$. \ We define $f$ as the average return for
investors, and $f_{1}$ as the average return for firms. In Appendix 13.1, we
show that considering the averages, the return equation for the group,
including defaults, writes: 
\begin{equation*}
\frac{f}{1+\underline{\hat{k}}}=\left( \frac{1+f}{\underline{\hat{k}}_{2}}%
\right) H\left( -\left( 1+f\right) \right) \frac{\underline{\hat{k}}_{2}}{1+%
\underline{\hat{k}}}+\left( \frac{1+f_{1}^{\prime }}{\underline{k}_{2}}%
\right) H\left( -\left( 1+f_{1}^{\prime }\right) \right) \frac{k_{2}}{1+%
\underline{k}}+\bar{r}\frac{k_{2}}{1+\underline{k}}+\frac{k_{1}}{1+%
\underline{k}}f_{1}
\end{equation*}%
note that without default, this reduces to:%
\begin{equation}
\frac{f}{1+\underline{\hat{k}}}=\bar{r}\frac{k_{2}}{1+\underline{k}}+\frac{%
k_{1}}{1+\underline{k}}f_{1}  \label{rth}
\end{equation}%
depicting that returns are divided into returns from loans and firms, along
with:%
\begin{equation*}
\frac{k_{2}}{1+\underline{k}}+\frac{k_{1}}{1+\underline{k}}=\frac{\underline{%
k}}{1+\underline{k}}=\frac{1}{1+\underline{\hat{k}}}
\end{equation*}

\subsubsection{Dynamics in one homogeneous group}

We analyze the dynamics within a group of roughly homogeneous agents; they
fluctuate around the same average. Let's begin with a group without
defaults. Recall that the return equation:%
\begin{eqnarray*}
&&\int \left( \delta \left( \hat{X}^{\prime }-\hat{X}\right) -\frac{\hat{k}%
_{1}\left( \hat{X}^{\prime },\hat{X}\right) \hat{K}^{\prime }\left\vert \hat{%
\Psi}\left( \hat{K}^{\prime },\hat{X}^{\prime }\right) \right\vert ^{2}}{1+%
\underline{\hat{k}}\left( \hat{X}^{\prime }\right) }\right) \frac{f\left( 
\hat{X}^{\prime }\right) -\bar{r}}{1+\underline{\hat{k}}_{2}\left( \hat{X}%
^{\prime }\right) }d\hat{K}^{\prime }d\hat{X}^{\prime } \\
&=&\int \frac{\left\vert \Psi \left( K^{\prime },X^{\prime }\right)
\right\vert ^{2}k_{1}\left( X^{\prime },\hat{X}\right) K^{\prime }}{1+%
\underline{k}\left( X^{\prime }\right) }\left( \frac{f_{1}^{\prime }\left(
K,X\right) -\bar{r}k_{2}\left( \hat{X}\right) }{1+k_{2}\left( \hat{X}\right) 
}+\Delta F_{\tau }\left( \bar{R}\left( K,X\right) \right) \right) dK^{\prime
}dX^{\prime }
\end{eqnarray*}%
is included in the field action through a potential term, and that the
diffusion between sectors $\hat{X}^{\prime }$ and $\hat{X}$ writes:%
\begin{equation*}
\frac{\hat{k}_{1}\left( \hat{X}^{\prime },\hat{X}\right) \hat{K}^{\prime
}\left\vert \hat{\Psi}\left( \hat{K}^{\prime },\hat{X}^{\prime }\right)
\right\vert ^{2}}{1+\underline{\hat{k}}\left( \hat{X}^{\prime }\right) }
\end{equation*}%
We apply the method outlined in section 5 to compute the transition
functions using Field Theory. The dynamics are studied by calculating the
effective action (\ref{SRP}) of the system around the background field. From
this effective action, we can compute the individual transition functions
using formulas (\ref{kr}) and (\ref{rk}). The transition functions for a
group of agents are given by formula (\ref{gnr}), which, without
interactions, corresponds to a product of individual transition functions.
Finally, we include in these transition functions, based on the interactions
of agents, the possible transmission of defaults using formula (\ref{trg}).

\subsubsection{Effective action}

The dynamics are studied by calculating the effective action (\ref{SRP}) of
the system around the background field. In principle, a series expansion
around the background field is computed, which also implies a series
expansion in returns. To simplify these calculations and allow for the
emergence of dynamic fluctuations in returns, we consider the effective
action obtained by keeping the return at its level given by the background
field, and then extend the effective action by introducing a field
describing the excess returns relative to the background value. This is a
simplification of the effective action, but it maintains the idea that
agents have dynamics around the background that will be impacted by changes
in returns.

This is done by introducing a field $\Xi \left( \hat{X},\delta f_{1}\right) $
measuring the modifications in returns. The value of $\delta f_{1}$ measures
the excess return compared to the average of the group. The dynamics for
investors inside the group is given by the effective action:%
\begin{eqnarray*}
S_{0} &=&-\int \hat{\Psi}^{\dag }\left( \nabla _{\hat{K}}\left( \frac{\sigma
_{\hat{K}}^{2}}{2}\nabla _{\hat{K}}-\hat{K}\left( f\left( \hat{X},K_{\hat{X}%
}\right) +\int \delta f_{1}\left\vert \Xi \left( \hat{X},\delta f_{1}\right)
\right\vert ^{2}d\left( \delta f_{1}\right) \right) \right) \right) \hat{\Psi%
} \\
&&+\hat{\mu}\left( \left\Vert \hat{\Psi}\left( \hat{X}\right) \right\Vert
^{2}-\left\Vert \hat{\Psi}_{0}\left( \hat{X}\right) \right\Vert ^{2}\right)
^{2} \\
&&-\int \Xi ^{\dagger }\left( \hat{X},\delta f_{1}\right) \sigma _{\delta
f_{1}}^{2}\nabla _{\delta f_{1}}^{2}\Xi \left( \hat{X},\delta f_{1}\right)
d\left( \delta f_{1}\right) +\Xi ^{\dagger }\left( \hat{X},\delta
f_{1}\right) J\left( \hat{X},K_{\hat{X}},\mathbf{E}\right) +J^{\dagger
}\left( \hat{X},K_{\hat{X}},\mathbf{E}\right) \Xi \left( \hat{X},\delta
f_{1}\right)
\end{eqnarray*}%
where the terme :%
\begin{equation*}
-\Xi ^{\dagger }\left( \hat{X},\delta f_{1}\right) \sigma _{\delta
f_{1}}^{2}\nabla _{\delta f_{1}}^{2}\Xi \left( \hat{X},\delta f_{1}\right)
\end{equation*}%
represents the dynamics of the excess return $\delta f_{1}$, and its
fluctuations, and the term: 
\begin{equation*}
\Xi ^{\dagger }\left( \hat{X},\delta f_{1}\right) J\left( \hat{X},K_{\hat{X}%
},\mathbf{E}\right) +J^{\dagger }\left( \hat{X},K_{\hat{X}},\mathbf{E}%
\right) \Xi \left( \hat{X},\delta f_{1}\right)
\end{equation*}%
represents the interaction of $\delta f_{1}$ with a vector of external
perturbations $J\left( \hat{X},K_{\hat{X}},\mathbf{E}\right) $, where $%
\mathbf{E}$ denotes the external parameters.

Around the background field, we can consider that $\hat{\mu}\left(
\left\Vert \hat{\Psi}\right\Vert ^{2}-\left\Vert \hat{\Psi}_{0}\right\Vert
^{2}\right) ^{2}<<1$, and that the group has a fixed number of investors.

\subsubsection{Transition function for agents without interactions}

the transition function for agents inside the group is first computed
without interactions. This is done by inverting the operator, as in (\ref{rk}%
), for the dynamics of capital:%
\begin{equation*}
-\nabla _{\hat{K}}\left( \frac{\sigma _{\hat{K}}^{2}}{2}\nabla _{\hat{K}}-%
\hat{K}f\left( \hat{X},K_{\hat{X}}\right) +\int \delta f_{1}\left\vert \Xi
\left( \hat{X},\delta f_{1}\right) \right\vert ^{2}d\left( \delta
f_{1}\right) \right)
\end{equation*}%
Appendix 14 shows that this yields the partial Green function conditioned to
the initial state and final state for returns as:%
\begin{eqnarray}
&&\sqrt{\left\vert \frac{f\left( \hat{X},K_{\hat{X}}\right) +\left( \frac{%
\delta f_{1}+\delta f_{1}^{\prime }}{2}\right) }{\sigma _{\hat{K}}^{2}\left(
1-\exp \left( 2\left( f\left( \hat{X},K_{\hat{X}}\right) +\left( \frac{%
\delta f_{1}+\delta f_{1}^{\prime }}{2}\right) \right) \Delta t\right)
\right) }\right\vert }  \label{TPN} \\
&&\times \exp \left( f\left( \hat{X},K_{\hat{X}}\right) \frac{\left( \hat{K}%
-\exp \left( \left( f\left( \hat{X},K_{\hat{X}}\right) +\frac{\delta
f_{1}+\delta f_{1}^{\prime }}{2}\right) \Delta t\right) \hat{K}^{\prime
}\right) ^{2}}{\sigma _{\hat{K}}^{2}\left( 1-\exp \left( 2\left( f\left( 
\hat{X},K_{\hat{X}}\right) +\frac{\delta f_{1}+\delta f_{1}^{\prime }}{2}%
\right) \Delta t\right) \right) }\right)  \notag \\
&&\times \exp \left( \frac{\left( f\left( \hat{X},K_{\hat{X}}\right) +\left( 
\frac{\delta f_{1}^{\prime }-\delta f_{1}}{\Delta t}\right) \right) }{\sigma
_{\hat{K}}^{2}\left( 1-\exp \left( 2\left( f\left( \hat{X},K_{\hat{X}%
}\right) +\left( \frac{\delta f_{1}^{\prime }-\delta f_{1}}{\Delta t}\right)
\right) \Delta t\right) \right) }\left( \hat{K}-\exp \left( \left( f\left( 
\hat{X},K_{\hat{X}}\right) +\left( \frac{\delta f_{1}^{\prime }-\delta f_{1}%
}{\Delta t}\right) \right) \Delta t\right) \hat{K}^{\prime }\right)
^{2}\right)  \notag
\end{eqnarray}%
We proceed in the same manner for the excess returns, and we invert the
operator:%
\begin{equation*}
-\Xi ^{\dagger }\left( \hat{X},\delta f_{1}\right) \sigma _{\delta
f_{1}}^{2}\nabla _{\delta f_{1}}^{2}\Xi \left( \hat{X},\delta f_{1}\right)
+\Xi ^{\dagger }\left( \hat{X},\delta f_{1}\right) J\left( \hat{X},K_{\hat{X}%
},\mathbf{E}\right) +J\left( \hat{X},K_{\hat{X}},\mathbf{E}\right) \Xi
\left( \hat{X},\delta f_{1}\right)
\end{equation*}%
and we obtain the partial transition function:%
\begin{equation}
\sqrt{\frac{1}{\sigma _{\delta f_{1}}^{2}}}\exp \left( -\frac{\left( \delta
f_{1}-\delta f_{1}^{\prime }\right) ^{2}}{\sigma _{\delta f_{1}}^{2}}%
+J\left( \hat{X},K_{\hat{X}},\mathbf{E}\right) \delta f_{1}-J\left( \hat{X}%
^{\prime },K_{\hat{X}^{\prime }},\mathbf{E}^{\prime }\right) \delta
f_{1}^{\prime }\right)  \label{TPS}
\end{equation}%
So that the transition function for an agent between a state with capital $K$
and return $\delta f_{1}$, to a state with capital $K^{\prime }$ and return $%
\delta f_{1}^{\prime }$ is:%
\begin{eqnarray*}
&&\left\langle \hat{K}^{\prime },\delta f_{1}^{\prime }\right. \left. \hat{K}%
,\delta f_{1}\right\rangle \\
&=&\sqrt{\left\vert \frac{f\left( \hat{X},K_{\hat{X}}\right) +\left( \frac{%
\delta f_{1}+\delta f_{1}^{\prime }}{2}\right) }{\sigma _{\hat{K}}^{2}\left(
1-\exp \left( 2\left( f\left( \hat{X},K_{\hat{X}}\right) +\left( \frac{%
\delta f_{1}+\delta f_{1}^{\prime }}{2}\right) \right) \Delta t\right)
\right) }\right\vert } \\
&&\times \exp \left( \frac{\left( f\left( \hat{X},K_{\hat{X}}\right) +\left( 
\frac{\delta f_{1}+\delta f_{1}^{\prime }}{2}\right) \right) }{\sigma _{\hat{%
K}}^{2}\left( 1-\exp \left( 2\left( f\left( \hat{X},K_{\hat{X}}\right)
+\left( \frac{\delta f_{1}+\delta f_{1}^{\prime }}{2}\right) \right) \Delta
t\right) \right) }\left( \hat{K}-\exp \left( \left( f\left( \hat{X},K_{\hat{X%
}}\right) +\left( \frac{\delta f_{1}+\delta f_{1}^{\prime }}{2}\right)
\right) \Delta t\right) \hat{K}^{\prime }\right) ^{2}\right) \\
&&\times \exp \left( -\frac{\left( \delta f_{1}-\delta f_{1}^{\prime
}\right) ^{2}}{\sigma _{\delta f_{1}}^{2}}+J\left( \hat{X},K_{\hat{X}},%
\mathbf{E}\right) \delta f_{1}-J^{\dagger }\left( \hat{X}^{\prime },K_{\hat{X%
}^{\prime }},\mathbf{E}^{\prime }\right) \delta f_{1}^{\prime }\right) \\
&\equiv &G\left( \Delta t,f_{1},\hat{K}_{1},\hat{K}_{1}^{\prime }\right)
G_{\delta f_{1}}\left( \Delta t,\delta f_{1},\delta f_{1}^{\prime }\right)
\end{eqnarray*}%
This transition function calculates, the probability for an agent to
transition from a state with capital and excess return, $\hat{K},\delta
f_{1} $, to a state $\hat{K}^{\prime },\delta f_{1}^{\prime }$. The term
under the square root is a normalization factor. The first exponential term
calculates the transition probability for the agent's capital, and roughly
follows the group's trend, since $\hat{K}^{\prime }$ is centered on the
expression $\exp \left( f\left( \hat{X},K_{\hat{X}}\right) \Delta t\right) 
\hat{K}$, which corresponds to accumulation when the return is constant and
equal to the sector's average $f\left( \hat{X},K_{\hat{X}}\right) $.

The second exponential term represents the dynamics of excess returns
relative to a group's average. This dynamic includes perturbation due to
external sources. The term $\frac{\left( \delta f_{1}-\delta f_{1}^{\prime
}\right) ^{2}}{\sigma _{\delta f_{1}}^{2}}$ describes internal fluctuations
with amplitude $\sigma _{\delta f_{1}}$, and the following term $J\left( 
\hat{X},K_{\hat{X}},\mathbf{E}\right) \delta f_{1}-J^{\dagger }\left( \hat{X}%
^{\prime },K_{\hat{X}^{\prime }},\mathbf{E}^{\prime }\right) \delta
f_{1}^{\prime }$ deviates the trajectory of the agent's excess return from
the average of their sector and the group. These two terms combine:
fluctuations move the agent away from the average, which is amplified by the
perturbation. Rather than simply fluctuating around an average, these
perturbations can deviate returns.

\subsubsection{Effective action including interactions}

Finally, we include in these transition functions, based on the interactions
of agents, the possible transmission of defaults using formula (\ref{trg}).
We incorporate into the effective action the interactions provided by the
diffusion of returns among themselves.

The constraint for return is included through a potential inside the
effective action:

\begin{eqnarray*}
S &=&-\int \hat{\Psi}^{\dag }\left( \nabla _{\hat{K}}\left( \frac{\sigma _{%
\hat{K}}^{2}}{2}\nabla _{\hat{K}}-\hat{K}\left( f\left( \hat{X},K_{\hat{X}%
}\right) +\int \delta f_{1}\left\vert \Xi \left( \hat{X},\delta f_{1}\right)
\right\vert ^{2}d\left( \delta f_{1}\right) \right) \right) +\right) \hat{%
\Psi} \\
&&+\hat{\mu}\left( \left\Vert \hat{\Psi}\left( \hat{X}\right) \right\Vert
^{2}-\left\Vert \hat{\Psi}_{0}\left( \hat{X}\right) \right\Vert ^{2}\right)
^{2} \\
&&-\Xi ^{\dagger }\left( \hat{X},\delta f_{1}\right) \nabla _{\delta
f_{1}}^{2}\Xi \left( \hat{X},\delta f_{1}\right) +\Xi ^{\dagger }\left( \hat{%
X},\delta f_{1}\right) J\left( \hat{X},K_{\hat{X}},\mathbf{E}\right)
+J^{\dagger }\left( \hat{X},K_{\hat{X}},\mathbf{E}\right) \Xi \left( \hat{X}%
,\delta f_{1}\right) \\
&&+\int \left\vert \hat{\Psi}\left( \hat{X},K_{\hat{X}}\right) \right\vert
^{2}V\left( \hat{\Psi},\hat{X},K,\delta f_{1},\delta f_{1}^{\prime }\right)
\left\vert \Xi \left( \hat{X},\delta f_{1}\right) \right\vert ^{2}\left\vert
\Xi \left( \hat{X},\delta f_{1}^{\prime }\right) \right\vert ^{2}
\end{eqnarray*}

where $V\left( \hat{\Psi},\hat{X},K,\delta f_{1}\right) $ is a Dirac delta
function type potential imposing ex-post the return equation (\ref{Rt}) on
agents experiencing independent returns. Assuming a \ return modification $%
\delta f_{1}^{\prime }$ in some sector $\hat{X}^{\prime }$, this impacts
sectors $\hat{X}$ by a variation $\delta f_{1}$ obtained by the first-order
expansion of (\ref{Rt}) around its average:

\begin{equation}
\delta f_{1}^{\prime }=\frac{\hat{k}_{1}\left( \hat{X}^{\prime },\hat{X}%
\right) \left\langle \hat{K}^{\prime }\right\rangle \left\vert \hat{\Psi}%
\left( \hat{K}^{\prime },\hat{X}^{\prime }\right) \right\vert ^{2}}{1+%
\underline{\hat{k}}\left( \hat{X}^{\prime }\right) }\delta f_{1}  \label{Ctd}
\end{equation}%
where the right-hand side represents the share of investment of agents of
sector $\hat{X}$ in sector $\hat{X}^{\prime }$.

To consider that such constraint has to be imposed for all agents after
interaction, we include in the effective potential:

\begin{eqnarray}
&&V\left( \hat{\Psi},\hat{X},K,\delta f_{1}\right)  \label{Dtp} \\
&=&\delta \left[ \int \left( \delta \left( \hat{X}^{\prime }-\hat{X}\right) -%
\frac{\hat{k}_{1}\left( \hat{X}^{\prime },\hat{X}\right) \hat{K}^{\prime
}\left\vert \hat{\Psi}\left( \hat{K}^{\prime },\hat{X}^{\prime }\right)
\right\vert ^{2}}{1+\underline{\hat{k}}\left( \hat{X}^{\prime }\right) }%
\right) \frac{\int \left\vert \Xi \left( \hat{X},\delta f_{1}\right)
\right\vert ^{2}\left\vert \Xi \left( \hat{X},\delta f_{1}^{\prime }\right)
\right\vert ^{2}d\left( \delta f_{1}\right) d\left( \delta f_{1}^{\prime
}\right) }{1+\underline{\hat{k}}_{2}\left( \hat{X}^{\prime }\right) }d\hat{K}%
^{\prime }d\hat{X}^{\prime }\right]  \notag
\end{eqnarray}

with $\delta \left[ X\right] $ being the Dirac-delta function. $\delta f_{1}$
and $\delta f_{1}^{\prime }$ are deviations from the average return of the
sector experienced by two investors located in $\hat{X}$ and $\hat{X}%
^{\prime }$ respectvly. The term in brackets inside the $\delta $ function
is the translation of (\ref{Ctd}) in terms of field.

In Appendix 14, we show that expanding a function in series as $\int
\left\vert \Xi \left( \hat{X},\delta f_{1}\right) \right\vert ^{2}d\left(
\delta f_{1}\right) $ amounts to impose the constraint:

\begin{equation}
\delta \left( \delta f_{1i}^{\prime }\left( \hat{X}_{i}\right) -\sum_{j}%
\frac{\hat{k}_{1}\left( \hat{X}_{i},\hat{X}_{j}\right) \left\langle \hat{K}%
_{j}\right\rangle \left\vert \hat{\Psi}\left( \hat{K}_{j},\hat{X}_{j}\right)
\right\vert ^{2}}{1+\underline{\hat{k}}\left( \hat{X}_{j}\right) }\delta
f_{1j}\left( \hat{X}_{j}\right) \right)  \label{Dtf}
\end{equation}%
between the interacting agents.

The Dirac delta function type potential means that we impose the return
equation ex post after modification of these returns for different agents.
The potential $V\left( \hat{\Psi},\hat{X},K,\delta f_{1}\right) $ only shows
successive deviations $\delta f_{1}^{\prime }$ and $\delta f_{1}$ because
the return equation is already satisfied for the averages and the returns of
the firms assumed to be non-fluctuating. The chosen interaction thus
propagates excess returns among investors. A variation in excess return by
one investor is propagated to those interacting with them.

\subsubsection{Transition functions within the group without default}

The transition functions are calculated using the formulas (\ref{gnr}) and (%
\ref{trg}). Without the interaction $V$ defined in (\ref{Dtp}), the
transition of the group would be given by:

\begin{eqnarray*}
&&\left\langle \left( \hat{K}_{i}^{\prime },\delta f_{1i}^{\prime }\right)
_{i}\right. T\left. \left( \hat{K}_{i},\delta f_{1i}\right) _{i}\right\rangle
\\
&=&\prod\limits_{i}\left\langle \hat{K}_{i}^{\prime },\delta f_{1i}^{\prime
}\right. \left. \hat{K}_{i},\delta f_{1i}\right\rangle \\
&=&G\left( \Delta t,f_{1,1},\hat{K}_{1},\hat{K}_{1}^{\prime }\right)
...G\left( \Delta t,f_{1,1},\hat{K}_{n},\hat{K}_{n}^{\prime }\right) \times
G_{\delta f_{1}}\left( \Delta t,\delta f_{1},\delta f_{1}^{\prime }\right)
...G_{\delta f_{1}}\left( \Delta t,\delta f_{p},\delta f_{p}^{\prime }\right)
\end{eqnarray*}%
reflecting that the transition probabilities of investors are independent.
Taking interactions into account, this result is corrected using (\ref{trg}).

The contribution associated to one interaction are given fr $n$ agents by
the potential:%
\begin{eqnarray*}
&&G\left( \Delta t,f_{1,1},\hat{K}_{1},\hat{K}_{1}^{\prime }\right)
...G\left( \Delta t,f_{1,1},\hat{K}_{n},\hat{K}_{n}^{\prime }\right) \times
G_{\delta f_{1}}\left( \Delta t,\delta f_{1},\delta f_{1}^{\prime }\right)
...G_{\delta f_{1}}\left( \Delta t,\delta f_{p},\delta f_{p}^{\prime }\right)
\\
&&\times V_{n}\left( \hat{X}_{1},\hat{K}_{1}^{\prime }...\hat{X}_{n},\hat{K}%
_{n}^{\prime }\right) \\
&&\times G\left( \Delta t,f_{1,1}^{\prime },\hat{K}_{1}^{\prime },\hat{K}%
_{1}^{\prime \prime }\right) ...G\left( \Delta t,f_{1,n}^{\prime },\hat{K}%
_{n}^{\prime },\hat{K}_{n}^{\prime \prime }\right) \times G_{\delta
f_{1}}\left( \Delta t,\delta f_{1}^{\prime },\delta f_{1}^{\prime \prime
}\right) ...G_{\delta f_{1}}\left( \Delta t,\delta f_{p}^{\prime },\delta
f_{p}^{\prime \prime }\right)
\end{eqnarray*}%
The potential is seen as an interaction term modifying the trajectories of
agents. \ For these $n$ agents, the integral form is replaced by a specific
sum involving these agents.%
\begin{eqnarray}
&&V_{n}\left( \hat{X}_{1},\hat{K}_{1}^{\prime }...\hat{X}_{n},\hat{K}%
_{n}^{\prime }\right)  \label{VN} \\
&=&\prod\limits_{i}\delta \left\{ \sum_{j}\left( \delta \left( \hat{X}_{i}-%
\hat{X}_{j}\right) -\frac{\hat{k}_{1}\left( \hat{X}_{j},\hat{X}_{i}\right) 
\hat{K}_{i}^{\prime }\left\vert \hat{\Psi}\left( \hat{K}_{j}^{\prime },\hat{X%
}_{j}\right) \right\vert ^{2}}{1+\underline{\hat{k}}\left( \hat{X}^{\prime
}\right) }\right) \frac{\delta f_{1}^{\prime }\left( \hat{X}_{j}\right) }{1+%
\underline{\hat{k}}_{2}\left( \hat{X}^{\prime }\right) }\right\}  \notag
\end{eqnarray}%
It is this interaction term that will determine the transitions of the
interacting agents. It imposes the constraint of correlated returns among
investors, where an investor's return is given by $f\left( \hat{X}%
_{j}\right) +\delta f_{1}^{\prime }\left( \hat{X}_{j}\right) $, that is, the
sector average plus an excess. The dynamics of agents within the group are
not independent; they all influence each other through this constraint (\ref%
{Dtf}).

To compute the full effect of interactions we have to sum over a series of
successive interactins induced by $V_{n}$. La transition est modifi\'{e}e
par la formule (\ref{trg}), et est donn\'{e}e par:the transition inside the
group is given by series:%
\begin{eqnarray*}
&&\left\langle \left( \hat{K}_{i}^{\prime },\delta f_{1i}^{\prime }\right)
_{i}\right. T\left. \left( \hat{K}_{i},\delta f_{1i}\right) _{i}\right\rangle
\\
&=&\prod\limits_{i}\left\langle \hat{K}_{i}^{\prime },\delta f_{1i}^{\prime
}\right. \left. \hat{K}_{i},\delta f_{1i}\right\rangle \\
&&+G\left( \Delta t,f_{1,1},\hat{K}_{1},\hat{K}_{1}^{\prime }\right)
...G\left( \Delta t,f_{1,1},\hat{K}_{n},\hat{K}_{n}^{\prime }\right)
G_{\delta f_{1}}\left( \Delta t,\delta f_{1},\delta f_{1}^{\prime }\right)
...G_{\delta f_{1}}\left( \Delta t,\delta f_{p},\delta f_{p}^{\prime }\right)
\\
&&\times V_{n}\left( \hat{X}_{1},\hat{K}_{1}^{\prime }...\hat{X}_{n},\hat{K}%
_{n}^{\prime }\right) \times G\left( \Delta t,f_{1,1}^{\prime },\hat{K}%
_{1}^{\prime },\hat{K}_{1}^{\prime \prime }\right) ...G\left( \Delta
t,f_{1,n}^{\prime },\hat{K}_{n}^{\prime },\hat{K}_{n}^{\prime \prime
}\right) \times G_{\delta f_{1}}\left( \Delta t,\delta f_{1}^{\prime
},\delta f_{1}^{\prime \prime }\right) ...G_{\delta f_{1}}\left( \Delta
t,\delta f_{p}^{\prime },\delta f_{p}^{\prime \prime }\right) \\
&&+\left[ G...G\right] V_{n}\left[ G...G\right] +.....\left[ G...G\right]
V_{n}\left[ G...G\right] V_{n}...\left[ G...G\right] V_{n}\left[ G...G\right]
\end{eqnarray*}%
where $\left[ G...G\right] V_{n}\left[ G...G\right] $ denotes blocks of
convolutions between Green functions and potential.

If one or more agents experience an excess return at a given moment, it
alters the returns of those who have interacted with them through the term $%
V_{n}\left( \hat{X}_{1},\hat{K}_{1}^{\prime }...\hat{X}_{n},\hat{K}%
_{n}^{\prime }\right) $. This, in turn, modifies their Green function. If an
agent was close to default, an adverse shock experienced by an interacting
agent can push it into default, subsequently impacting the initial agent and
causing it to default as well. However, to model this phenomenon accurately,
we need to consider multiple groups. In the homogeneous case, where agents
are all relatively similar, all agents should be close to default for such a
situation to arise. Considering multiple groups allows us to differentiate
the situation of agents based on their group.

\subsubsection{Transition functions within the group with default}

Assume that at some point some of the agents in the system reach a default
state. We denote $DS$ as the set of agents with default at some points This
may correspond to an adverse $\delta f_{1}$ for these agents, that modifies
the potential:

\begin{eqnarray*}
&&V_{n}^{D}\left( \hat{X}_{1},\hat{K}_{1}^{\prime }...\hat{X}_{n},\hat{K}%
_{n}^{\prime }\right) \\
&=&\prod\limits_{i}\frac{1}{\rho }\left\{ \sum_{j}\left( \delta \left( \hat{%
X}_{i}-\hat{X}_{j}\right) -\frac{\hat{k}_{1}\left( \hat{X}_{j},\hat{X}%
_{i}\right) \hat{K}_{i}^{\prime }\left\vert \hat{\Psi}\left( \hat{K}%
_{j}^{\prime },\hat{X}_{j}\right) \right\vert ^{2}}{1+\underline{\hat{k}}%
\left( \hat{X}^{\prime }\right) }\right) \frac{\delta f_{1}^{\prime }\left( 
\hat{X}_{j}\right) }{1+\underline{\hat{k}}_{2}\left( \hat{X}^{\prime
}\right) }\right. \\
&&+\sum_{j\in DS}\left( \bar{r}+\frac{1+f\left( \hat{X}_{j}\right) }{%
\underline{\hat{k}}_{2}\left( \hat{X}_{j}\right) }\right) \frac{\hat{k}%
_{2}\left( \hat{X}_{j},\hat{X}_{i}\right) \hat{K}^{\prime }\left\vert \hat{%
\Psi}\left( \hat{K}_{j}^{\prime },\hat{X}_{j}\right) \right\vert ^{2}}{1+%
\underline{\hat{k}}\left( \hat{X}_{j}\right) } \\
&&\left. -\sum_{j}\frac{\left\vert \Psi \left( \hat{K}_{j}^{\prime },\hat{X}%
_{j}\right) \right\vert ^{2}k_{1}\left( \hat{X}_{j},\hat{X}_{i}\right)
K^{\prime }}{1+\underline{k}\left( \hat{X}_{j}\right) }\left( \frac{%
f_{1}^{\prime }\left( K_{j},\hat{X}_{j}\right) -\bar{r}k_{2}\left( \hat{X}%
_{j}\right) }{1+k_{2}\left( \hat{X}\right) }+\Delta F_{\tau }\left( \bar{R}%
\left( K_{i},\hat{X}_{j}\right) \right) \right) \right\} ^{2}
\end{eqnarray*}%
and the transitions of the system after default decompose as:%
\begin{eqnarray*}
&&\left\langle \left( \hat{K}_{i}^{\prime },\delta f_{1i}^{\prime }\right)
_{i}\right. T\left. \left( \hat{K}_{i},\delta f_{1i}\right) _{i}\right\rangle
\\
&=&\prod\limits_{i}\left\langle \hat{K}_{i}^{\prime },\delta f_{1i}^{\prime
}\right. \left. \hat{K}_{i},\delta f_{1i}\right\rangle \\
&&+G\left( \Delta t,f_{1,1},\hat{K}_{1},\hat{K}_{1}^{\prime }\right)
...G\left( \Delta t,f_{1,1},\hat{K}_{n},\hat{K}_{n}^{\prime }\right)
G_{\delta f_{1}}\left( \Delta t,\delta f_{1},\delta f_{1}^{\prime }\right)
...G_{\delta f_{1}}\left( \Delta t,\delta f_{p},\delta f_{p}^{\prime }\right)
\\
&&\times V_{n}\left( \hat{X}_{1},\hat{K}_{1}^{\prime }...\hat{X}_{n},\hat{K}%
_{n}^{\prime }\right) \times G\left( \Delta t,f_{1,1}^{\prime },\hat{K}%
_{1}^{\prime },\hat{K}_{1}^{\prime \prime }\right) ...G\left( \Delta
t,f_{1,n}^{\prime },\hat{K}_{n}^{\prime },\hat{K}_{n}^{\prime \prime
}\right) \times G_{\delta f_{1}}\left( \Delta t,\delta f_{1}^{\prime
},\delta f_{1}^{\prime \prime }\right) ...G_{\delta f_{1}}\left( \Delta
t,\delta f_{p}^{\prime },\delta f_{p}^{\prime \prime }\right) \\
&&+\left[ G...G\right] V_{n}^{D}\left[ G...G\right] +.....\left[ G...G\right]
V_{n}^{D}\left[ G...G\right] V_{n}^{D}...\left[ G...G\right] V_{n}^{D}\left[
G...G\right]
\end{eqnarray*}%
the potential $V_{n}^{D}$ drives returns towards lower values, increasing
the probability that some investors enter the default zone, thereby inducing
amplified default rates.

\subsection{Interactions of agents inside or across groups of investors}

The approach focusing on a single group is constrained by the fact that
dynamics and fluctuations are relative to the group's mean. Consequently, in
a homogeneous group without defaults, it becomes difficult to identify a
single sector that would deviate significantly from this mean. Essentially,
the collective state is always governed by an average, pulling agents
towards it.

To investigate the transmission of defaults, we must therefore divide the
system into multiple groups, distinguishing between those with defaults or
at risk, and separately analyze the dynamics of agents in groups without
defaults versus those in at-risk or default groups. This allows us to
understand how the dynamics of one group impact those of another.

Our challenge remains consistent: in collective states, there is no singular
event or dynamic that precipitates significant change. Instead, it is
necessary to isolate groups of collective states, and it is the dynamics
within these distinct groups that can have a meaningful impact.

In a homogeneous group, our aim is to study the dynamics of agents, enabling
us to subsequently explore how agents in one group dynamically influence
those in another.

\subsubsection{Average for several groups}

Let us now define $\underline{\hat{k}}_{\eta }^{\left[ ii\right] }$, $%
\underline{k}_{\eta }^{\left[ ii\right] }$, the average coefficients within
the group, and $\underline{\hat{k}}_{\eta }^{\left[ ji\right] }$ and $%
\underline{k}_{\eta }^{\left[ ji\right] }$, the average connections from $i$
t $j$. \ We define the total share of capital invested in sector $i$,
comprising intra-sectoral investments (from $i$ to $i$), and inter-sectoral
investments (from $j$ to $i$) as:

\begin{equation*}
\underline{\hat{k}}_{\eta }^{\left[ i\right] }=\underline{\hat{k}}_{\eta }^{%
\left[ ii\right] }+\underline{\hat{k}}_{\eta }^{\left[ ij\right] }
\end{equation*}%
where the sum over the indices is implied. The return equation involving
several groups becomes:%
\begin{eqnarray}
&&\left( \frac{f^{\left[ i\right] }}{1+\underline{\hat{k}}_{2}^{\left[ i%
\right] }}+\bar{r}\frac{\underline{\hat{k}}_{2}^{\left[ i\right] }}{1+%
\underline{\hat{k}}_{2}^{\left[ i\right] }}\right) -\frac{\hat{k}_{1}^{\left[
ji\right] }}{1+\underline{\hat{k}}^{\left[ j\right] }}\left( \frac{f^{\left[
j\right] }}{1+\underline{\hat{k}}_{2}^{\left[ j\right] }}+\bar{r}\frac{%
\underline{\hat{k}}_{2}^{\left[ j\right] }}{1+\underline{\hat{k}}_{2}^{\left[
j\right] }}\right)  \label{Q} \\
&=&\left( \bar{r}+\frac{1+f^{\left[ i\right] }}{\underline{\hat{k}}_{2}^{%
\left[ i\right] }}H\left( -\left( 1+f^{\left[ i\right] }\right) \right)
\right) \frac{\underline{\hat{k}}_{2}^{\left[ ii\right] }}{1+\underline{\hat{%
k}}^{\left[ i\right] }}+\left( \bar{r}+\frac{1+f^{\left[ j\right] }}{%
\underline{\hat{k}}_{2}^{\left[ j\right] }}H\left( -\left( 1+f^{\left[ j%
\right] }\right) \right) \right) \frac{\underline{\hat{k}}_{2}^{\left[ ji%
\right] }}{1+\underline{\hat{k}}^{\left[ j\right] }}  \notag \\
&&+\left( \bar{r}+\frac{1+f_{1}^{\prime \left[ i\right] }}{\underline{k}%
_{2}^{\left[ i\right] }}H\left( -\frac{1+f_{1}^{\prime \left[ i\right] }}{%
\underline{k}_{2}^{\left[ i\right] }}\right) \right) \frac{k_{2}^{\left[ ii%
\right] }}{1+\underline{k}^{i}}+\frac{k_{1}^{\left[ ii\right] }}{1+%
\underline{k}^{i}}f_{1}^{\left[ i\right] }+\left( \bar{r}+\frac{%
1+f_{1}^{\prime \left[ j\right] }}{\underline{k}_{2}^{\left[ j\right] }}%
H\left( -\frac{1+f_{1}^{\prime \left[ j\right] }}{\underline{k}_{2}^{\left[ j%
\right] }}\right) \right) \frac{k_{2}^{\left[ ji\right] }}{1+\underline{k}%
^{j}}+\frac{k_{1}^{\left[ ij\right] }}{1+\underline{k}^{j}}f_{1}^{\left[ j%
\right] }  \notag
\end{eqnarray}%
with several constraints between the coefficients detailed in Appendix 13.2.

This can be conveniently reformulated in matricial form using the alternate
description presented in Appendix 4.This alternate description involves
relative shares of investmnt $\underline{\hat{S}}_{1}^{\left[ ii\right] }$, $%
\underline{\hat{S}}_{1}^{\left[ jj\right] }$. We find in Appendix 13.2 that:

\begin{eqnarray}
0 &=&\left( 
\begin{array}{cc}
1-\underline{\hat{S}}_{1}^{\left[ ii\right] } & -\underline{\hat{S}}_{1}^{%
\left[ ji\right] } \\ 
-\underline{\hat{S}}_{1}^{\left[ ij\right] } & 1-\underline{\hat{S}}_{1}^{%
\left[ jj\right] }%
\end{array}%
\right) \left( 
\begin{array}{c}
\left( f^{\left[ i\right] }-\bar{r}\right) \underline{\hat{s}}_{1}^{\left[ i%
\right] } \\ 
\left( f^{\left[ j\right] }-\bar{r}\right) \underline{\hat{s}}_{1}^{\left[ j%
\right] }%
\end{array}%
\right)  \label{PMl} \\
&&-\left( 
\begin{array}{cc}
\underline{\hat{S}}_{2}^{\left[ ii\right] } & \underline{\hat{S}}_{2}^{\left[
ji\right] } \\ 
\underline{\hat{S}}_{2}^{\left[ ij\right] } & \underline{\hat{S}}_{2}^{\left[
jj\right] }%
\end{array}%
\right) \left( 
\begin{array}{c}
\left( 1+f^{\left[ i\right] }\right) \underline{\hat{s}}_{2}^{\left[ i\right]
}H\left( -\left( 1+f^{\left[ i\right] }\right) \right) \\ 
\left( 1+f^{\left[ j\right] }\right) \underline{\hat{s}}_{2}^{\left[ j\right]
}H\left( -\left( 1+f^{\left[ j\right] }\right) \right)%
\end{array}%
\right)  \notag \\
&&-\left( 
\begin{array}{cc}
\underline{S}_{2}^{\left[ ii\right] } & \underline{S}_{2}^{\left[ ji\right] }
\\ 
\underline{S}_{2}^{\left[ ij\right] } & \underline{S}_{2}^{\left[ jj\right] }%
\end{array}%
\right) \left( 
\begin{array}{c}
\left( 1+f_{1}^{\prime \left[ i\right] }\right) \underline{s}_{2}^{\left[ i%
\right] }H\left( -\left( 1+f_{1}^{\prime \left[ i\right] }\right) \right) \\ 
\left( 1+f_{1}^{\prime \left[ j\right] }\right) \underline{s}_{2}^{\left[ j%
\right] }H\left( -\left( 1+f_{1}^{\prime \left[ j\right] }\right) \right)%
\end{array}%
\right) -\left( 
\begin{array}{cc}
\underline{S}_{1}^{\left[ ii\right] } & \underline{S}_{1}^{\left[ ji\right] }
\\ 
\underline{S}_{1}^{\left[ ij\right] } & \underline{S}_{1}^{\left[ jj\right] }%
\end{array}%
\right) \left( 
\begin{array}{c}
\left( f_{1}^{\prime \left[ i\right] }-\bar{r}\right) \underline{s}_{1}^{%
\left[ i\right] } \\ 
\left( f_{1}^{\prime \left[ j\right] }-\bar{r}\right) \underline{s}_{1}^{%
\left[ j\right] }%
\end{array}%
\right)  \notag
\end{eqnarray}

with:%
\begin{eqnarray*}
\underline{\hat{s}}_{1}^{\left[ i\right] } &=&\frac{1-\left( \underline{\hat{%
S}}^{\left[ ii\right] }+\underline{\hat{S}}^{\left[ ij\right] }\right) }{%
1-\left( \underline{\hat{S}}_{1}^{\left[ ii\right] }+\underline{\hat{S}}%
_{1}^{\left[ ij\right] }\right) }\text{, }\underline{\hat{s}}_{1}^{\left[ j%
\right] }=\frac{1-\left( \underline{\hat{S}}^{\left[ jj\right] }+\underline{%
\hat{S}}^{\left[ ij\right] }\right) }{1-\left( \underline{\hat{S}}_{1}^{%
\left[ jj\right] }+\underline{\hat{S}}_{1}^{\left[ ji\right] }\right) } \\
\underline{\hat{s}}_{2}^{\left[ i\right] } &=&\frac{1-\left( \underline{\hat{%
S}}^{\left[ ii\right] }+\underline{\hat{S}}^{\left[ ij\right] }\right) }{%
\underline{\hat{S}}_{2}^{\left[ ii\right] }+\underline{\hat{S}}_{2}^{\left[
ij\right] }}\text{, }\underline{\hat{s}}_{2}^{\left[ j\right] }=\frac{%
1-\left( \underline{\hat{S}}^{\left[ jj\right] }+\underline{\hat{S}}^{\left[
ij\right] }\right) }{\underline{\hat{S}}_{2}^{\left[ jj\right] }+\underline{%
\hat{S}}_{2}^{\left[ ji\right] }} \\
\underline{s}_{1}^{\left[ i\right] } &=&\frac{1-\left( \underline{S}^{\left[
ii\right] }+\underline{S}^{\left[ ij\right] }\right) }{1-\left( \underline{S}%
_{1}^{\left[ ii\right] }+\underline{S}_{1}^{\left[ ij\right] }\right) }\text{%
, }\underline{s}_{1}^{\left[ j\right] }=\frac{1-\left( \underline{S}^{\left[
jj\right] }+\underline{S}^{\left[ ij\right] }\right) }{1-\left( \underline{S}%
_{1}^{\left[ jj\right] }+\underline{S}_{1}^{\left[ ji\right] }\right) } \\
\underline{s}_{2}^{\left[ i\right] } &=&\frac{1-\left( \underline{S}^{\left[
ii\right] }+\underline{S}^{\left[ ij\right] }\right) }{\underline{S}_{2}^{%
\left[ ii\right] }+\underline{S}_{2}^{\left[ ij\right] }}\text{, }\underline{%
s}_{2}^{\left[ j\right] }=\frac{1-\left( \underline{S}^{\left[ jj\right] }+%
\underline{S}^{\left[ ji\right] }\right) }{\underline{S}_{2}^{\left[ jj%
\right] }+\underline{S}_{2}^{\left[ ji\right] }}
\end{eqnarray*}

This formulation allows for the study of transitions between independent
collective states towards entangled states. The diagonal elements of the
matrices represent intra-sectoral investments, while the non-diagonal
elements represent inter-sectoral investments. The independent collective
states are thus defined by setting the non-diagonal matrices $\underline{%
\hat{S}}_{\eta }^{\left[ ji\right] }=0$ and $\underline{S}_{\eta }^{\left[ ji%
\right] }=0$. Corrections due to interactions between groups are obtained by
considering the non-diagonal elements. This enables the study of default
transmission between different groups, introducing a dynamic element in the
occurrence of defaults. The addition of interactions allows defaults to
propagate.

Note that we could have addressed the same point in the previous section by
examining the propagation of default from a single agent within a sector.
However, it is more realistic to consider that the economy can withstand the
default of a single agent if it does not spread. It is more realistic to
consider that an entire group can default, and that this default propagates.
The transmission of default from investors in sector $j$ to sector $i$ is
expressed by the term:

\begin{equation}
\left( 1+f^{\left[ j\right] }\right) \frac{1-\left( \underline{\hat{S}}^{%
\left[ jj\right] }+\underline{\hat{S}}^{\left[ ij\right] }\right) }{%
\underline{\hat{S}}_{2}^{\left[ jj\right] }+\underline{\hat{S}}_{2}^{\left[
ji\right] }}H\left( -\left( 1+f^{\left[ j\right] }\right) \right)
\label{vd1}
\end{equation}%
and for the transmission of default from firms:%
\begin{equation}
\left( 1+f_{1}^{\prime \left[ j\right] }\right) \frac{1-\left( \underline{S}%
^{\left[ jj\right] }+\underline{S}^{\left[ ji\right] }\right) }{\underline{S}%
_{2}^{\left[ jj\right] }+\underline{S}_{2}^{\left[ ji\right] }}  \label{vd2}
\end{equation}

This transmission depends on the non-diagonal coefficients in equation (\ref%
{PMl}). These coefficients represent the diffusion of returns from $j$ to $i$
, and thus the loss due to default (\ref{vd1}) or (\ref{vd2}) will decrease
the returns of sector $i$, possibly leading to the creation of another
default.

Furthermore, dividing the system into multiple groups allows us to study,
based on the connections between groups of sectors, which groups will be
impacted by defaults from other groups and which will remain unaffected.

\subsection{Dynamics for agents in several groups}

In the case of multiple groups, we can consider that the field for investors
decomposes into several components, each defined over a certain region of
the sector space. These fields interact but are considered independent in a
first approximation. We thus replace the field describing the system by
several fields, one for each group and define:%
\begin{equation*}
\left[ \hat{\Psi}\right] \left( \left\{ \hat{K}_{r_{i}},\hat{X}%
_{r_{i}}\right\} _{G_{i}}\right)
\end{equation*}%
which stands for a set of field, one defined for each group $\left\{ \hat{K}%
_{r_{i}},\hat{X}_{r_{i}}\right\} _{G_{i}}$.

\subsubsection{Effective action}

We revisit the functional action of each group and add a term for
interaction between the different groups.

Each group has its own sector-spatial extension, and the effective action is
a generalization of the single-group action.%
\begin{eqnarray*}
&&\sum \left[ \hat{\Psi}\right] \left( \left\{ \hat{K}_{r_{i}},\hat{X}%
_{r_{i}}\right\} _{G_{i}}\right) \nabla _{\hat{K}_{r_{i}}}\left( \nabla _{%
\hat{K}_{r_{i}}}-\hat{K}_{r_{i}}f^{\left[ i\right] }\right) \left[ \hat{\Psi}%
\right] ^{\dag }\left( \left\{ \hat{K}_{r_{i}},\hat{X}_{r_{i}}\right\}
_{G_{i}}\right) \\
&&+\sum \prod \left\vert \left[ \hat{\Psi}\right] \left( \left\{ \hat{K}%
_{r_{i}},\hat{X}_{r_{i}}\right\} _{G_{i}}\right) \right\vert ^{2}\delta
\left( V\left( \left( \left[ \hat{\Psi}\right] \left( \left\{ \hat{K}%
_{r_{i}},\hat{X}_{r_{i}}\right\} _{G_{i}}\right) \right) \right) \right) \\
&&-\Xi ^{\dagger }\left( \hat{X},\delta f_{1}\right) \sigma _{\delta
f_{1}}^{2}\nabla _{\delta f_{1}}^{2}\Xi \left( \hat{X},\delta f_{1}\right)
+\Xi ^{\dagger }\left( \hat{X},\delta f_{1}\right) J\left( \hat{X},K_{\hat{X}%
},\mathbf{E}\right) +J^{\dagger }\left( \hat{X},K_{\hat{X}},\mathbf{E}%
\right) \Xi \left( \hat{X},\delta f_{1}\right)
\end{eqnarray*}

\subsubsection{Return equations and potential}

We will write the excess return equations for each group by including their
interactions. These interactions will be considered as perturbations that
modify the transition functions. To do so, we consider that agents belonging
to different groups are linked by the return equation (\ref{PMl}) and that
groups are weakly connected. This implies considering Dirac-type potentials,
one describing intra-group interactions, and one describing inter-group
interactions. Excluding defaults, these potentials, are obtained through a
first order expansion in $\delta f^{\left[ i\right] }$ and $\delta f^{\left[
j\right] }$ of (\ref{PMl}) which writes:

\begin{equation}
0=\left( 
\begin{array}{c}
\delta f^{\left[ i\right] \prime } \\ 
\delta f^{\left[ j\right] \prime }%
\end{array}%
\right) -\left( 
\begin{array}{cc}
\underline{\hat{S}}_{1}^{\left[ ii\right] } & \underline{\hat{S}}_{1}^{\left[
ji\right] } \\ 
\underline{\hat{S}}_{1}^{\left[ ij\right] } & \underline{\hat{S}}_{1}^{\left[
jj\right] }%
\end{array}%
\right) \left( 
\begin{array}{c}
\delta f^{\left[ i\right] }\frac{1-\left( \underline{\hat{S}}^{\left[ ii%
\right] }+\underline{\hat{S}}^{\left[ ij\right] }\right) }{1-\left( 
\underline{\hat{S}}_{1}^{\left[ ii\right] }+\underline{\hat{S}}_{1}^{\left[
ij\right] }\right) } \\ 
\delta f^{\left[ j\right] }\frac{1-\left( \underline{\hat{S}}^{\left[ jj%
\right] }+\underline{\hat{S}}^{\left[ ij\right] }\right) }{1-\left( 
\underline{\hat{S}}_{1}^{\left[ jj\right] }+\underline{\hat{S}}_{1}^{\left[
ji\right] }\right) }%
\end{array}%
\right)  \label{Rd}
\end{equation}%
we can write the intra-group interaction potentials by considering the
diagonal terms in (\ref{Rd}). It leads to the potential:%
\begin{equation}
\sum_{i}V_{i}\left( \hat{\Psi},\hat{X},K,\delta f_{1}\right) =\sum_{i}\delta %
\left[ \delta f_{1}^{\left[ i\right] }\left( \hat{X}\right) -\underline{\hat{%
S}}_{1}^{\left[ ii\right] }\frac{1-\left( \underline{\hat{S}}^{\left[ ii%
\right] }+\underline{\hat{S}}^{\left[ ij\right] }\right) }{1-\left( 
\underline{\hat{S}}_{1}^{\left[ ii\right] }+\underline{\hat{S}}_{1}^{\left[
ij\right] }\right) }\frac{\delta f_{1}^{\left[ i\right] \prime }\left( \hat{X%
}^{\prime }\right) }{1+\underline{\hat{k}}_{2}\left( \hat{X}^{\prime
}\right) }dX^{\prime }\right]  \label{NT}
\end{equation}%
while the inter-group interaction potentials is obtained by considering the
off-diagonal terms in (\ref{Rd}). We find:%
\begin{eqnarray}
&&\sum_{i}W_{i}\left( \hat{\Psi},\hat{X},K,\delta f_{1i}\right)  \label{nr}
\\
&=&\sum_{i}\delta \left( \delta f^{\left[ i\right] \prime }\left( \hat{X}%
\right) -\underline{\hat{S}}_{1}^{\left[ ji\right] }\frac{1-\left( 
\underline{\hat{S}}^{\left[ jj\right] }+\underline{\hat{S}}^{\left[ ij\right]
}\right) }{1-\left( \underline{\hat{S}}_{1}^{\left[ jj\right] }+\underline{%
\hat{S}}_{1}^{\left[ ji\right] }\right) }\frac{\delta f^{\left[ j\right]
\prime }\left( \hat{X}^{\prime }\right) }{\left( 1+\underline{\hat{k}}%
_{2}\left( \hat{X}^{\prime }\right) \right) }\right)  \notag
\end{eqnarray}%
The transitions are then driven both by intra-group (\ref{NT}) and
inter-group (\ref{nr}) interactions. The potential (\ref{NT}) is similar to
the homogeneous group and illustrates how, within the same group, the excess
returns of one agent impact those of others. Even an agent that has not
initially experienced a return variation will experience it subsequently due
to the interaction term. The potential (\ref{nr}), on the other hand,
demonstrates the transmission of a return variation from one block to
another.

\subsubsection{Transitions functions without default}

As for one homogeneous group, the transitions are computed by series of
terms of the type:

\begin{eqnarray*}
&&G\left( \Delta t,f_{1,1},\left\{ \hat{K}_{r_{1}},\hat{X}_{r_{1}}\right\}
_{G_{1}},\left\{ \hat{K}_{r_{1}}^{\prime },\hat{X}_{r_{1}}^{\prime }\right\}
_{G_{1}}\right) ...G\left( \Delta t,f_{1,n},\left\{ \hat{K}_{r_{n}},\hat{X}%
_{r_{n}}\right\} _{G_{n}},\left\{ \hat{K}_{r_{n}}^{\prime },\hat{X}%
_{r_{n}}^{\prime }\right\} _{G_{n}}\right) \\
&&\times G_{\delta f_{1}}\left( \Delta t,\delta f_{1},\delta f_{1}^{\prime
}\right) ...G_{\delta f_{1}}\left( \Delta t,\delta f_{p},\delta
f_{p}^{\prime }\right) \\
&&\times \left( V\left( \left( \left[ \hat{\Psi}\right] \left( \left\{ \hat{K%
}_{r_{i}},\hat{X}_{r_{i}}\right\} _{G_{i}}\right) \right) \right) \right) \\
&&\times G\left( \Delta t,f_{1,1}^{\prime },\left\{ \hat{K}_{r_{1}}^{\prime
},\hat{X}_{r_{1}}^{\prime }\right\} _{G_{1}},\left\{ \hat{K}_{r_{1}}^{\prime
\prime },\hat{X}_{r_{1}}^{\prime \prime }\right\} _{G_{1}}\right) ...G\left(
\Delta t,f_{1,n}^{\prime },\left\{ \hat{K}_{r_{n}}^{\prime },\hat{X}%
_{r_{n}}^{\prime }\right\} _{G_{n}},\left\{ \hat{K}_{r_{n}}^{\prime \prime },%
\hat{X}_{r_{n}}^{\prime \prime }\right\} _{G_{n}}\right) \\
&&\times G_{\delta f_{1}}\left( \Delta t,\delta f_{1}^{\prime },\delta
f_{1}^{\prime \prime }\right) ...G_{\delta f_{1}}\left( \Delta t,\delta
f_{p}^{\prime },\delta f_{p}^{\prime \prime }\right)
\end{eqnarray*}%
where:%
\begin{equation*}
V\left( \left( \left[ \hat{\Psi}\right] \left( \left\{ \hat{K}_{r_{i}},\hat{X%
}_{r_{i}}\right\} _{G_{i}}\right) \right) \right)
\end{equation*}%
stands for the intra- and inter-group interactions.

\subsubsection{Transitions functions including default}

As before, some default may arise in the process of transition. This is
modeled by including the defaults investors term in (\ref{NT}) and (\ref{nr}%
). Starting with the constraint in the default case:

\begin{eqnarray}
0 &=&\left( 
\begin{array}{c}
\delta \hat{f}^{\left[ i\right] } \\ 
\delta \hat{f}^{\left[ j\right] }%
\end{array}%
\right) -\left( 
\begin{array}{cc}
\underline{\hat{S}}_{1}^{\left[ ii\right] } & \underline{\hat{S}}_{1}^{\left[
ji\right] } \\ 
\underline{\hat{S}}_{1}^{\left[ ij\right] } & \underline{\hat{S}}_{1}^{\left[
jj\right] }%
\end{array}%
\right) \left( 
\begin{array}{c}
\delta \hat{f}^{\left[ i\right] \prime }\frac{1-\left( \underline{\hat{S}}^{%
\left[ ii\right] }+\underline{\hat{S}}^{\left[ ij\right] }\right) }{1-\left( 
\underline{\hat{S}}_{1}^{\left[ ii\right] }+\underline{\hat{S}}_{1}^{\left[
ij\right] }\right) } \\ 
\delta \hat{f}^{\left[ j\right] \prime }\frac{1-\left( \underline{\hat{S}}^{%
\left[ jj\right] }+\underline{\hat{S}}^{\left[ ij\right] }\right) }{1-\left( 
\underline{\hat{S}}_{1}^{\left[ jj\right] }+\underline{\hat{S}}_{1}^{\left[
ji\right] }\right) }%
\end{array}%
\right) \\
&&-\left( 
\begin{array}{cc}
\underline{\hat{S}}_{2}^{\left[ ii\right] } & \underline{\hat{S}}_{2}^{\left[
ji\right] } \\ 
\underline{\hat{S}}_{2}^{\left[ ij\right] } & \underline{\hat{S}}_{2}^{\left[
jj\right] }%
\end{array}%
\right) \left( 
\begin{array}{c}
\left( 1+\hat{f}^{\left[ i\right] \prime }+\delta \hat{f}^{\left[ i\right]
\prime }\right) \underline{\hat{s}}_{2}^{\left[ i\right] }H\left( -\left( 1+%
\hat{f}^{\left[ i\right] \prime }+\delta \hat{f}^{\left[ i\right] \prime
}\right) \right) \\ 
\left( 1+\hat{f}^{\left[ j\right] \prime }+\delta \hat{f}^{\left[ j\right]
\prime }\right) \underline{\hat{s}}_{2}^{\left[ i\right] }H\left( -\left( 1+%
\hat{f}^{\left[ j\right] \prime }+\delta \hat{f}^{\left[ j\right] \prime
}\right) \right)%
\end{array}%
\right)  \notag
\end{eqnarray}%
the default potentials $V_{i}^{D}\left( \hat{\Psi},\hat{X},K,\delta
f_{1}\right) $ and $W_{i}^{D}\left( \hat{\Psi},\hat{X},K,\delta
f_{1i}\right) $ are derived straightforwardly:%
\begin{eqnarray*}
\sum_{i}V_{i}^{D}\left( \hat{\Psi},\hat{X},K,\delta f_{1}\right)
&=&\sum_{i}\delta \left[ \delta f_{1}^{\left[ i\right] }\left( \hat{X}%
\right) -\underline{\hat{S}}_{1}^{\left[ ii\right] }\frac{1-\left( 
\underline{\hat{S}}^{\left[ ii\right] }+\underline{\hat{S}}^{\left[ ij\right]
}\right) }{1-\left( \underline{\hat{S}}_{1}^{\left[ ii\right] }+\underline{%
\hat{S}}_{1}^{\left[ ij\right] }\right) }\frac{\delta f_{1}^{\left[ i\right]
}\left( \hat{X}^{\prime }\right) }{1+\underline{\hat{k}}_{2}\left( \hat{X}%
^{\prime }\right) }dX^{\prime }\right. \\
&&\left. -\underline{\hat{S}}_{2}^{\left[ ii\right] }\left( 1+\hat{f}^{\left[
i\right] \prime }+\delta \hat{f}^{\left[ i\right] \prime }\right) \underline{%
\hat{s}}_{2}^{\left[ i\right] }H\left( -\left( 1+\hat{f}^{\left[ i\right]
\prime }+\delta \hat{f}^{\left[ i\right] \prime }\right) \right) \right]
\end{eqnarray*}%
and:\textbf{\ }%
\begin{eqnarray}
&&\sum_{i}W_{i}^{D}\left( \hat{\Psi},\hat{X},K,\delta f_{1i}\right) \\
&=&\sum_{i}\delta \left[ \delta f^{\left[ i\right] \prime }\left( \hat{X}%
\right) -\underline{\hat{S}}_{1}^{\left[ ji\right] }\frac{1-\left( 
\underline{\hat{S}}^{\left[ jj\right] }+\underline{\hat{S}}^{\left[ ij\right]
}\right) }{1-\left( \underline{\hat{S}}_{1}^{\left[ jj\right] }+\underline{%
\hat{S}}_{1}^{\left[ ji\right] }\right) }\frac{\delta f^{\left[ j\right]
\prime }\left( \hat{X}^{\prime }\right) }{\left( 1+\underline{\hat{k}}%
_{2}\left( \hat{X}^{\prime }\right) \right) }\right.  \notag \\
&&\left. -\underline{\hat{S}}_{2}^{\left[ ji\right] }\left( 1+\hat{f}^{\left[
j\right] \prime }+\delta \hat{f}^{\left[ j\right] \prime }\right) \underline{%
\hat{s}}_{2}^{\left[ i\right] }H\left( -\left( 1+\hat{f}^{\left[ j\right]
\prime }+\delta \hat{f}^{\left[ j\right] \prime }\right) \right) \right] 
\notag
\end{eqnarray}%
The transitions thus include two possible effects.

First, a sequence of interactions:%
\begin{equation}
\left[ G...G\right] V_{n}\left[ G...G\right] +.....\left[ G...G\right] V_{n}%
\left[ G...G\right] V_{n}...\left[ G...G\right] V_{n}\left[ G...G\right]
\label{GP1}
\end{equation}%
where $V_{n}$ describes transitions between agents of several groups, which
may drive some agents to default. The following transitions are thus driven
by terms like:%
\begin{equation}
\left[ G...G\right] V_{n}^{D}\left[ G...G\right] +.....\left[ G...G\right]
V_{n}^{D}\left[ G...G\right] V_{n}^{D}...\left[ G...G\right] V_{n}\left[
G...G\right]  \label{GP2}
\end{equation}%
which increases the number of default in group.

Then through inter-groups potential the default propagates to agents of
different groups.

The terms (\ref{GP1}) and (\ref{GP2}) describe the sequential dynamics of
interactions among multiple investors within different groups. Initially,
investors evolve independently, and their transition function is simply the
product of their individual transition functions $\left[ G...G\right] $. The
interaction between different investors is measured by the term $V_{n}$,
which diffuses variations in returns among investors. Once the returns are
modified by this initial interaction, investors continue their accumulation
dynamics with returns modified by the interaction and a transition function $%
\left[ G...G\right] $, then interact again, and so on.

If some investors default, particularly due to their interactions with other
investors in a group structurally close to default, the subsequent
interaction between investors will be described by the default interaction $%
V_{n}^{D}$, which takes defaults into account. This interaction will reduce
the returns of other investors, whose accumulation dynamics will be
impacted. Their probability of default increases, and the risk of default
increases with each interaction.

Between groups of investors, the diffusion of defaults can take more complex
and indirect patterns. Some investors in group $A$, without defaulting
themselves, may experience a negative shock that propagates to group $B$,
causing their returns to decrease, thereby impacting group $A$ in return,
potentially leading it to default... This spiral effect dynamically reflects
the transition from one collective state to another. From a collective state
characterized by a limited number of defaults, there can be a sudden
transition to another collective state characterized by a massive number of
defaults. The dynamics are relatively rapid, and all defaults in the
collective state will materialize.

\part*{Part 2. System including banks}

We have introduced a simple system where investors could choose to invest in
firms or in other investors. Now, we will further sophisticate this initial
benchmark by adding another type of investors to the system: banks, which
can lend more than their private or available capital allows.

\section{Microeconomic framework for banks, investors and firms}

The system follows the initial setup, with the addition of banks, which
modifies the system while keeping the principles intact. In the subsequent
discussion, we will denote variables related to firms without superscript,
variables related to investors with a hat symbol, and variables related to
banks with a bar symbol.

\subsection{Disposable capital}

\subsubsection{Banks}

Banks are defined by their private capital $\bar{K}_{jp}\left( t\right) $,
their diposable capital $\bar{K}_{j0}\left( t\right) $ and the capital $\bar{%
K}_{j}\left( t\right) $ lent to other agents, which is proportional to $\bar{%
K}_{jp}\left( t\right) $ with a ratio $\kappa $, giving $\bar{K}_{j}\left(
t\right) =\kappa \bar{K}_{jp}\left( t\right) $. Banks engage in lending and
taking participation with each other, with shares $\bar{k}_{2}$ and $\bar{k}%
_{1}$ respectively.

The disposable capital for banks will be invested in loans and
participations in other banks on one hand, and in participations in firms on
the other hand: 
\begin{eqnarray*}
\bar{K}_{j0}\left( t\right) &=&\sum_{l}\bar{k}_{1}\left( \bar{K}_{lp}\left(
t\right) ,\bar{X}_{l}\left( t\right) ,\bar{X}_{j}\left( t\right) \right) 
\bar{K}_{j0}\left( t\right) +\sum_{l}\bar{k}_{2}\left( \bar{K}_{lp}\left(
t\right) ,\bar{X}_{l}\left( t\right) ,\bar{X}_{j}\left( t\right) \right) 
\bar{K}_{j0}\left( t\right) \\
&&+\sum_{l}\bar{k}_{1}\left( \hat{K}_{lp}\left( t\right) ,\hat{X}_{l}\left(
t\right) ,\bar{X}_{j}\left( t\right) \right) \bar{K}_{j0}\left( t\right)
+\sum_{i}\left( \hat{F}_{2,1}\left( R_{i},\hat{X}_{j}\right) \bar{K}%
_{j0}\left( t\right) \right)
\end{eqnarray*}%
Conversely, the available capital of a bank is the sum of its private
capital, participations from other banks, and loans from other banks. To
simplify, investors do not lend to or take participation in banks.

The link between private capital and disposable capital is as follows:

\begin{equation*}
\bar{K}_{j0}\left( t\right) =\bar{K}_{jp}\left( t\right) +\sum_{l}\left( 
\bar{k}_{1}\left( \bar{K}_{jp}\left( t\right) ,\bar{X}_{j}\left( t\right) ,%
\bar{X}_{l}\left( t\right) \right) +\bar{k}_{2}\left( \bar{K}_{jp}\left(
t\right) ,\bar{X}_{j}\left( t\right) ,\bar{X}_{l}\left( t\right) \right)
\right) \bar{K}_{l0}\left( t\right)
\end{equation*}%
which writes, in the linear approximation:%
\begin{equation*}
\bar{K}_{j0}\left( t\right) =\bar{K}_{jp}\left( t\right) +\sum_{l}\left( 
\bar{k}_{1}\left( \bar{X}_{j}\left( t\right) ,\bar{X}_{l}\left( t\right)
\right) +\bar{k}_{2}\left( \bar{X}_{j}\left( t\right) ,\bar{X}_{l}\left(
t\right) \right) \right) \bar{K}_{jp}\left( t\right) \bar{K}_{l0}\left(
t\right)
\end{equation*}%
We can express the private capital as a function of the disposable capital:%
\begin{equation}
\bar{K}_{jp}\left( t\right) =\frac{\bar{K}_{j0}\left( t\right) }{%
1+\sum_{l}\left( \bar{k}_{1}\left( \bar{X}_{j}\left( t\right) ,\bar{X}%
_{l}\left( t\right) \right) +\bar{k}_{2}\left( \bar{X}_{j}\left( t\right) ,%
\bar{X}_{l}\left( t\right) \right) \right) \bar{K}_{l0}\left( t\right) }
\label{KJP}
\end{equation}%
As in Part 1, we will use the notations:%
\begin{equation*}
\bar{k}_{\varepsilon }\left( \bar{X}_{j}\left( t\right) ,\bar{X}_{l}\left(
t\right) \right) \rightarrow \bar{k}_{\varepsilon lj}
\end{equation*}%
and:%
\begin{equation*}
\bar{k}_{1lj}+\bar{k}_{1lj}\rightarrow \bar{k}_{lj}
\end{equation*}%
The lent capital is proportional to the private capital of the bank. The
factor denoted here as $\kappa $ is the credit multiplier. Equation (\ref%
{KJP}) implies that lent capital can also be expressed as a function of the
disposable capital:%
\begin{equation*}
\bar{K}_{j}\left( t\right) =\kappa \bar{K}_{jp}\left( t\right) =\frac{\kappa 
\bar{K}_{j0}\left( t\right) }{1+\sum_{l}\left( \bar{k}_{1}\left( \bar{X}%
_{j}\left( t\right) ,\bar{X}_{l}\left( t\right) \right) +\bar{k}_{2}\left( 
\bar{X}_{j}\left( t\right) ,\bar{X}_{l}\left( t\right) \right) \right) \bar{K%
}_{l0}\left( t\right) }
\end{equation*}

\subsubsection{Investors}

The setup for investors remains the same as before. However, there is a
difference in Part 1 due to the disposable capital of investors. As banks
may also take participations or lend to any investor, the investor's
disposable capital is now a combination of private capital, participations
from investors and banks, and loans from investors and banks.

The disposable capital for investor $j$ can be expressed as follows:

\begin{eqnarray*}
\hat{K}_{j}\left( t\right) &=&\hat{K}_{jp}\left( t\right) +\sum_{l}\left( 
\hat{k}_{1}\left( \hat{K}_{jp}\left( t\right) ,\hat{X}_{j}\left( t\right) ,%
\hat{X}_{l}\left( t\right) \right) +\hat{k}_{2}\left( \hat{K}_{jp}\left(
t\right) ,\hat{X}_{j}\left( t\right) ,\hat{X}_{l}\left( t\right) \right)
\right) \hat{K}_{l}\left( t\right) \\
&&+\sum_{l}\left( \hat{k}_{1}^{B}\left( \hat{K}_{jp}\left( t\right) ,\hat{X}%
_{j}\left( t\right) ,\bar{X}_{l}\left( t\right) \right) \bar{K}_{l0}\left(
t\right) +\hat{k}_{2}^{B}\left( \hat{K}_{jp}\left( t\right) ,\hat{X}%
_{j}\left( t\right) ,\bar{X}_{l}\left( t\right) \right) \bar{K}_{l}\left(
t\right) \right)
\end{eqnarray*}%
which writes in the linear approximation:%
\begin{equation*}
\hat{k}_{1}\left( \hat{K}_{jp}\left( t\right) ,\hat{X}_{j}\left( t\right) ,%
\hat{X}_{l}\left( t\right) \right) +\hat{k}_{2}\left( \hat{K}_{jp}\left(
t\right) ,\hat{X}_{j}\left( t\right) ,\hat{X}_{l}\left( t\right) \right)
=\left( \hat{k}_{1jl}+\hat{k}_{2jl}\right) \hat{K}_{jp}\left( t\right)
\end{equation*}%
and:%
\begin{equation*}
\left( \hat{k}_{1}^{B}\left( \hat{K}_{jp}\left( t\right) ,\hat{X}_{j}\left(
t\right) ,\bar{X}_{l}\left( t\right) \right) \bar{K}_{l0}\left( t\right) +%
\hat{k}_{2}^{B}\left( \hat{K}_{jp}\left( t\right) ,\hat{X}_{j}\left(
t\right) ,\bar{X}_{l}\left( t\right) \right) \bar{K}_{l}\left( t\right)
\right) =\left( \hat{k}_{1jl}^{B}\bar{K}_{l0}\left( t\right) +\hat{k}%
_{2jl}^{B}\bar{K}_{l}\left( t\right) \right) \hat{K}_{jp}\left( t\right)
\end{equation*}%
We define the participation share of investor $l$ in investor $j$ as $\hat{k}%
_{1jl}$ and the loan share of investor $l$ in investor $j$ as $\hat{k}_{2jl} 
$: 
\begin{eqnarray*}
\hat{k}_{1jl} &=&\hat{k}_{1}\left( \hat{X}_{j}\left( t\right) ,\hat{X}%
_{l}\left( t\right) \right) \\
\hat{k}_{2jl} &=&\hat{k}_{2}\left( \hat{X}_{j}\left( t\right) ,\hat{X}%
_{l}\left( t\right) \right)
\end{eqnarray*}%
And the total sum of the two shares as $\hat{k}_{jl}$, i.e.:%
\begin{equation*}
\hat{k}_{jl}=\hat{k}_{1jl}+\hat{k}_{2jl}
\end{equation*}%
Similarly, in the linear approximation, the investment of a bank in an
investor, with participation $\hat{k}_{1}^{B}$ and loan $\hat{k}_{2}^{B}$,
is expressed as follows:%
\begin{eqnarray*}
\hat{k}_{1}^{B}\left( \hat{K}_{jp}\left( t\right) ,\hat{X}_{j}\left(
t\right) ,\hat{X}_{l}\left( t\right) \right) &=&\hat{k}_{1}^{B}\left( \hat{X}%
_{j}\left( t\right) ,\hat{X}_{l}\left( t\right) \right) \hat{K}_{jp}\left(
t\right) \\
\hat{k}_{2}^{B}\left( \hat{K}_{jp}\left( t\right) ,\hat{X}_{j}\left(
t\right) ,\hat{X}_{l}\left( t\right) \right) &=&\hat{k}_{2}^{B}\left( \hat{X}%
_{j}\left( t\right) ,\hat{X}_{l}\left( t\right) \right) \hat{K}_{jp}\left(
t\right)
\end{eqnarray*}%
and we can express the disposable capital of investor $j$, $\hat{K}%
_{j}\left( t\right) $, as :%
\begin{equation*}
\hat{K}_{j}\left( t\right) =\hat{K}_{jp}\left( t\right) +\sum_{l}\left( \hat{%
k}_{1}+\hat{k}_{2}\right) \hat{K}_{l}\left( t\right) +\sum_{l}\left( \hat{k}%
_{1}^{B}\bar{K}_{l0}\left( t\right) +\hat{k}_{2}^{B}\bar{K}_{l}\left(
t\right) \right)
\end{equation*}%
so that private capital can be expressed as a function of the disposable
capital:%
\begin{eqnarray}
\hat{K}_{jp}\left( t\right) &=&\frac{\hat{K}_{j}\left( t\right) }{1+\sum_{l}%
\hat{k}_{jl}\hat{K}_{l}\left( t\right) +\sum_{l}\left( \hat{k}_{1jl}^{B}\bar{%
K}_{l0}\left( t\right) +\hat{k}_{2jl}^{B}\bar{K}_{l}\left( t\right) \right) }
\label{KvsKP} \\
&=&\frac{\hat{K}_{j}\left( t\right) }{1+\sum_{l}\hat{k}_{jl}\hat{K}%
_{l}\left( t\right) +\sum_{l}\hat{k}_{1jl}^{B}\bar{K}_{l0}\left( t\right)
+\kappa \sum_{l}\hat{k}_{2jl}^{B}\frac{\bar{K}_{j0}\left( t\right) }{%
1+\sum_{m}\bar{k}_{vm}\bar{K}_{m0}\left( t\right) }}  \notag
\end{eqnarray}%
Note that the ratio:%
\begin{equation*}
\kappa \frac{\bar{K}_{j0}\left( t\right) }{1+\sum_{m}\bar{k}_{vm}\bar{K}%
_{m0}\left( t\right) }
\end{equation*}%
in the bank loan models that bank loans are proportional to the bank's
private capital, rather than their disposable capital.

\subsubsection{Firms}

The description of firms' private capital and disposable capital is similar
to Part 1, except that now firms can borrow from both banks and investors.
Their disposable capital now decomposes in the following way:%
\begin{equation*}
K_{i}\left( t\right) =K_{ip}\left( t\right) +\left( \sum_{v}k_{1iv}^{\left(
B\right) }\bar{K}_{v0}\left( t\right) +k_{2iv}^{\left( B\right) }\bar{K}%
_{v}\left( t\right) \right) K_{ip}\left( t\right) +\left( \sum_{v}\left(
k_{1iv}+k_{2iv}\right) \hat{K}_{v}\left( t\right) \right) K_{ip}\left(
t\right)
\end{equation*}%
where $k_{1iv}^{\left( B\right) }$ and $k_{2iv}^{\left( B\right) }$ are the
shares invested through participation and loans respectively by bank $\nu $
to firm $i$. These participations are proportional to the bank's disposable
capital, as for investors, but loans are proportional to the level of
possible lent captal, i.e. the private capital of the bank times the
multiplier $\kappa $. We will consider $\kappa >>1$.

The firm private capital thus writes: 
\begin{eqnarray*}
K_{ip}\left( t\right) &=&\frac{K_{i}\left( t\right) }{1+\left(
\sum_{v}k_{1iv}^{\left( B\right) }\bar{K}_{v0}\left( t\right)
+k_{2iv}^{\left( B\right) }\bar{K}_{v}\left( t\right) \right) +\left(
\sum_{v}k_{iv}\hat{K}_{v}\left( t\right) \right) } \\
&=&\frac{K_{i}\left( t\right) }{1+\left( \sum_{v}k_{1iv}^{\left( B\right) }%
\bar{K}_{v0}\left( t\right) +k_{2iv}^{\left( B\right) }\kappa \frac{\bar{K}%
_{v0}\left( t\right) }{1+\sum_{m}\bar{k}_{vm}\bar{K}_{m0}\left( t\right) }%
\right) +\left( \sum_{v}k_{iv}\hat{K}_{v}\left( t\right) \right) }
\end{eqnarray*}%
This formula is the similar to Part 1.

\subsection{Capital allocation}

\subsubsection{Banks}

Banks' disposable capital is decomposed into several investments and can be
expressed as:%
\begin{eqnarray}
\bar{K}_{j0}\left( t\right) &=&\sum_{i}\frac{k_{1il}^{\left( B\right)
}K_{i}\left( t\right) }{1+\sum_{v}k_{1iv}^{\left( B\right) }\bar{K}%
_{v0}\left( t\right) +k_{2iv}^{\left( B\right) }\kappa \frac{\bar{K}%
_{v0}\left( t\right) }{1+\sum_{m}\bar{k}_{vm}\bar{K}_{m0}\left( t\right) }%
+\sum_{v}k_{iv}\hat{K}_{v}\left( t\right) }\bar{K}_{j0}\left( t\right)
\label{KJ0} \\
&&+\sum_{l}\frac{\hat{k}_{1lj}^{B}\hat{K}_{l}\left( t\right) }{1+\sum_{\nu }%
\hat{k}_{l\nu }\hat{K}_{l}\left( t\right) +\sum_{\nu }\hat{k}_{1l\nu }^{B}%
\bar{K}_{l0}\left( t\right) +\kappa \sum_{\nu }\hat{k}_{2l\nu }^{B}\frac{%
\bar{K}_{j0}\left( t\right) }{1+\sum_{m}\bar{k}_{vm}\bar{K}_{m0}\left(
t\right) }}\bar{K}_{j0}\left( t\right)  \notag \\
&&+\sum_{l}\frac{\left( \bar{k}_{1lj}+\bar{k}_{2lj}\right) \bar{K}%
_{l0}\left( t\right) }{1+\sum_{\nu }\left( \bar{k}_{1l\nu }+\bar{k}_{2l\nu
}\right) \bar{K}_{l0}\left( t\right) }\bar{K}_{j0}\left( t\right)  \notag
\end{eqnarray}%
where $\bar{K}_{j0}\left( t\right) $ is the disposable capital of the bank,
and the other elements are proportionality factors in which the terms $%
k_{1il}^{\left( B\right) }$, $\hat{k}_{1lj}^{B}$, and $\left( \bar{k}_{1lj}+%
\bar{k}_{2lj}\right) $represent the leverage effects that the investor or
the bank, as an investor, chooses to apply to their investment. Each time,
the available capital of the investment $K_{i}\left( t\right) $, $\hat{K}%
_{l}\left( t\right) $, $\bar{K}_{l0}\left( t\right) $, divided by a
denominator, represents the private capital of the agent in which one
invests. Equation (\ref{KJ0}) implies the constraint:%
\begin{eqnarray*}
1 &=&\sum_{i}\frac{k_{1il}^{\left( B\right) }K_{i}\left( t\right) }{%
1+\sum_{v}k_{1iv}^{\left( B\right) }\bar{K}_{v0}\left( t\right)
+k_{2iv}^{\left( B\right) }\kappa \frac{\bar{K}_{v0}\left( t\right) }{%
1+\sum_{m}\bar{k}_{vm}\bar{K}_{m0}\left( t\right) }+\sum_{v}k_{iv}\hat{K}%
_{v}\left( t\right) } \\
&&+\sum_{l}\frac{\hat{k}_{1lj}^{B}\hat{K}_{l}\left( t\right) }{1+\sum_{\nu }%
\hat{k}_{l\nu }\hat{K}_{l}\left( t\right) +\sum_{\nu }\hat{k}_{1l\nu }^{B}%
\bar{K}_{l0}\left( t\right) +\kappa \sum_{\nu }\hat{k}_{2l\nu }^{B}\frac{%
\bar{K}_{j0}\left( t\right) }{1+\sum_{m}\bar{k}_{vm}\bar{K}_{m0}\left(
t\right) }} \\
&&+\sum_{l}\frac{\left( \bar{k}_{1lj}+\bar{k}_{2lj}\right) \bar{K}%
_{l0}\left( t\right) }{1+\sum_{\nu }\left( \bar{k}_{1l\nu }+\bar{k}_{2l\nu
}\right) \bar{K}_{l0}\left( t\right) }
\end{eqnarray*}%
The first sum describes participations in firms, the second sum represents
participations in investors. Loans to firms and investors are not included
because they are accounted for separately as monetary creation. The third
sum consists of participation $\bar{k}_{1lj}$ and loans $\bar{k}_{2l\nu }$
in other banks.

\subsubsection{Investors}

For investors, the principle remains the same. Capital is invested in firms
and other investors, and the coefficients represent the leverage effects
that multiply the private capital of the agent in which one invests.
Disposable capital is decomposed into several investments and can be
expressed as:%
\begin{eqnarray*}
\hat{K}_{j}\left( t\right) &=&\sum_{i}\frac{\left( k_{1ij}+k_{2ij}\right)
K_{i}\left( t\right) }{1+\sum_{v}k_{1iv}^{\left( B\right) }\bar{K}%
_{v0}\left( t\right) +k_{2iv}^{\left( B\right) }\kappa \frac{\bar{K}%
_{v0}\left( t\right) }{1+\sum_{m}\bar{k}_{vm}\bar{K}_{m0}\left( t\right) }%
+\sum_{v}k_{iv}\hat{K}_{v}\left( t\right) }\hat{K}_{j}\left( t\right) \\
&&+\sum_{l}\frac{\left( \hat{k}_{1lj}+\hat{k}_{2lj}\right) \hat{K}_{l}\left(
t\right) }{1+\sum_{\nu }\hat{k}_{l\nu }\hat{K}_{l}\left( t\right) +\sum_{\nu
}\hat{k}_{1l\nu }^{B}\bar{K}_{l0}\left( t\right) +\kappa \sum_{\nu }\hat{k}%
_{2l\nu }^{B}\frac{\bar{K}_{j0}\left( t\right) }{1+\sum_{m}\bar{k}_{vm}\bar{K%
}_{m0}\left( t\right) }}\hat{K}_{j}\left( t\right)
\end{eqnarray*}%
and this identity implies, for all $j$, the constraint:%
\begin{eqnarray*}
1 &=&\sum_{i}\frac{\left( k_{1ij}+k_{2ij}\right) K_{i}\left( t\right) }{%
1+\sum_{v}k_{1iv}^{\left( B\right) }\bar{K}_{v0}\left( t\right)
+k_{2iv}^{\left( B\right) }\kappa \frac{\bar{K}_{v0}\left( t\right) }{%
1+\sum_{m}\bar{k}_{vm}\bar{K}_{m0}\left( t\right) }+\sum_{v}k_{iv}\hat{K}%
_{v}\left( t\right) } \\
&&+\sum_{l}\frac{\left( \hat{k}_{1lj}+\hat{k}_{2lj}\right) \hat{K}_{l}\left(
t\right) }{1+\sum_{\nu }\hat{k}_{l\nu }\hat{K}_{l}\left( t\right) +\sum_{\nu
}\hat{k}_{1l\nu }^{B}\bar{K}_{l0}\left( t\right) +\kappa \sum_{\nu }\hat{k}%
_{2l\nu }^{B}\frac{\bar{K}_{j0}\left( t\right) }{1+\sum_{m}\bar{k}_{vm}\bar{K%
}_{m0}\left( t\right) }}
\end{eqnarray*}%
with:%
\begin{equation*}
k_{ij}=k_{1ij}+k_{2ij}
\end{equation*}%
and:%
\begin{equation*}
\hat{k}_{ij}=\hat{k}_{1ij}+\hat{k}_{2ij}
\end{equation*}%
the overall leverage effects of the investment from an investor to a firm or
another investor, respectively.

\subsection{Returns and capital accumulation under no-default scenario}

\subsubsection{Firms}

\paragraph{Returns}

Returns for firms come from production and price increases, thus they are
the same as in Part 1. For a given return on invested capital $R_{j}$, firm $%
j$ has the following return on its private capital: 
\begin{eqnarray}
f_{j} &=&\sum_{l}\left( 1+\sum_{v}\hat{k}_{2lv}\hat{K}_{v}\left( t\right)
+\kappa \sum_{v}\hat{k}_{2jv}^{B}\frac{\bar{K}_{v0}\left( t\right) }{%
1+\sum_{m}\bar{k}_{vm}\bar{K}_{m0}\left( t\right) }\right) R_{j}  \label{FD}
\\
&&-\bar{r}\left( \sum_{v}\hat{k}_{2jv}\hat{K}_{v}\left( t\right) +\kappa
\sum_{v}\hat{k}_{2jv}^{B}\frac{\bar{K}_{v0}\left( t\right) }{1+\sum_{m}\bar{k%
}_{vm}\bar{K}_{m0}\left( t\right) }\right)  \notag
\end{eqnarray}

The formula is the same as in Part $1$, except that banks have been
introduced. The return on private capital that the firm generates for itself
is the return on investment rate~$R_{j}$, multiplied by all the capital it
has engaged, minus the interest on the loans it must repay. The difference
from Part $1$ is that the amount of capital raised by the firm itself
through borrowing includes loans from banks. For the subsequent analysis,
especially for the dynamics of firm accumulation, it will be useful to
re-express $R_{j}$ as a function of $f_{j}$: 
\begin{equation}
R_{j}=\frac{f_{j}+\bar{r}\left( \sum_{v}\hat{k}_{2jv}\hat{K}_{v}\left(
t\right) +\kappa \sum_{v}\hat{k}_{2jv}^{B}\frac{\bar{K}_{v0}\left( t\right) 
}{1+\sum_{m}\overline{\bar{k}}_{vm}\bar{K}_{m0}\left( t\right) }\right) }{%
1+\sum_{v}\hat{k}_{2lv}\hat{K}_{v}\left( t\right) +\kappa \sum_{v}\hat{k}%
_{2jv}^{B}\frac{\bar{K}_{v0}\left( t\right) }{1+\sum_{m}\overline{\bar{k}}%
_{vm}\bar{K}_{m0}\left( t\right) }}  \label{RF}
\end{equation}

\paragraph{Firms' accumulation of capital}

Similar to Part $1$, the accumulation of a firm occurs at a rate determined
by $f_{i}\left( K_{i}\left( t\right) \right) $, and the dynamics for firms'
capital accumulation can be expressed as:%
\begin{equation}
\frac{d}{dt}K_{ip}\left( t\right) =f_{i}\left( K_{i}\left( t\right) \right)
K_{ip}\left( t\right)
\end{equation}

\subsubsection{Investors}

\paragraph{Investors' returns}

The investors' returns are similar to those in Part 1, except that bank
loans and participation arise in the disposable capital. The investors'
return $\hat{R}_{j}$ determined by equations (\ref{RTS}) and (\ref{PMj}):%
\begin{equation}
\hat{R}_{j}=\sum_{l}\frac{\hat{k}_{1lj}\hat{K}_{l}\left( t\right) }{%
1+\sum_{v}\hat{k}_{lv}\hat{K}_{v}\left( t\right) +\sum_{v}\hat{k}_{1lv}^{B}%
\bar{K}_{v0}\left( t\right) +\kappa \sum_{v}\hat{k}_{2lv}^{B}\frac{\bar{K}%
_{v0}\left( t\right) }{1+\sum_{m}\bar{k}_{vm}\bar{K}_{m0}\left( t\right) }}%
\hat{R}_{l}+\hat{R}_{j}^{\prime }  \label{RH}
\end{equation}

Formula (\ref{RH}) shows that the returns of investors can be decomposed
into two parts. The first part corresponds to the return from other
investors, $\hat{R}_{l}$, weighted by the size of investor $j$'s
participations in investor $l$. These participations are directly
proportional to the private capital of investor $l$. The second part, $\hat{R%
}_{j}^{\prime }$, represents the direct return obtained by investor $j$, who
grants loans to firms and investors and takes participations in firms. The
direct return $\hat{R}_{j}^{\prime }$ decomposes as:%
\begin{eqnarray*}
\hat{R}_{j}^{\prime } &=&\bar{r}\sum_{l}\frac{\hat{k}_{2lj}\hat{K}_{l}\left(
t\right) }{1+\sum_{l}\hat{k}_{jl}\hat{K}_{l}\left( t\right) +\sum_{l}\hat{k}%
_{1jl}^{B}\bar{K}_{l0}\left( t\right) +\kappa \sum_{l}\hat{k}_{2jl}^{B}\frac{%
\bar{K}_{j0}\left( t\right) }{1+\sum_{m}\bar{k}_{vm}\bar{K}_{m0}\left(
t\right) }} \\
&&+\bar{r}\sum_{i}\frac{k_{2ij}K_{i}\left( t\right) }{1+\sum_{v}k_{iv}\hat{K}%
_{v}\left( t\right) +\sum_{v}k_{1iv}^{\left( B\right) }\bar{K}_{v0}\left(
t\right) +k_{2iv}^{\left( B\right) }\kappa \frac{\bar{K}_{v0}\left( t\right) 
}{1+\sum_{m}\bar{k}_{vm}\bar{K}_{m0}\left( t\right) }} \\
&&+\sum_{i}\frac{\left( r_{i}+F_{1}\left( \bar{R}_{i},\frac{\dot{K}%
_{i}\left( t\right) }{K_{i}\left( t\right) }\right) +\tau \left( \bar{R}%
\left( K_{i},X_{i}\right) \right) \Delta f_{1}^{\prime }\left( K_{i}\left(
t\right) \right) \right) k_{1ij}K_{i}\left( t\right) }{1+\sum_{v}k_{iv}\hat{K%
}_{v}\left( t\right) +\sum_{v}k_{1iv}^{\left( B\right) }\bar{K}_{v0}\left(
t\right) +k_{2iv}^{\left( B\right) }\kappa \frac{\bar{K}_{v0}\left( t\right) 
}{1+\sum_{m}\bar{k}_{vm}\bar{K}_{m0}\left( t\right) }}
\end{eqnarray*}%
As in Part $1$, the return equation can be reformulated in terms of excess
return relative to the interest rate:%
\begin{eqnarray}
&&\sum_{l}\left( \delta _{jl}-\frac{\hat{k}_{1lj}\hat{K}_{l}\left( t\right) 
}{1+\sum_{v}\hat{k}_{jv}\hat{K}_{l}\left( t\right) +\sum_{v}\hat{k}_{1jv}^{B}%
\bar{K}_{l0}\left( t\right) +\kappa \sum_{v}\hat{k}_{2jv}^{B}\frac{\bar{K}%
_{j0}\left( t\right) }{1+\sum_{m}\bar{k}_{vm}\bar{K}_{m0}\left( t\right) }}%
\right) \\
&&\times \frac{\hat{f}_{l}-\bar{r}}{1+\sum_{v}\hat{k}_{2lv}\hat{K}_{v}\left(
t\right) +\kappa \sum_{v}\hat{k}_{2lv}^{B}\frac{\bar{K}_{\nu 0}\left(
t\right) }{1+\sum_{m}\bar{k}_{vm}\bar{K}_{m0}\left( t\right) }}  \notag \\
&=&\sum_{i}\frac{\left( r_{i}+F_{1}\left( \bar{R}_{i},\frac{\dot{K}%
_{i}\left( t\right) }{K_{i}\left( t\right) }\right) +\tau \left( \bar{R}%
\left( K_{i},X_{i}\right) \right) \Delta f_{1}^{\prime }\left( K_{i}\left(
t\right) \right) -\bar{r}\right) k_{1ij}K_{i}\left( t\right) }{1+\sum_{v}%
\hat{k}_{jv}\hat{K}_{v}\left( t\right) +\sum_{v}\hat{k}_{1jv}^{B}\bar{K}%
_{l0}\left( t\right) +\kappa \sum_{v}\hat{k}_{2jv}^{B}\frac{\bar{K}%
_{v0}\left( t\right) }{1+\sum_{m}\bar{k}_{vm}\bar{K}_{m0}\left( t\right) }} 
\notag
\end{eqnarray}%
with:%
\begin{eqnarray*}
\hat{f}_{j} &=&\left( 1+\sum_{v}\hat{k}_{2jv}\hat{K}_{v}\left( t\right)
+\kappa \sum_{l}\hat{k}_{2jl}^{B}\frac{\bar{K}_{j0}\left( t\right) }{%
1+\sum_{m}\bar{k}_{vm}\bar{K}_{m0}\left( t\right) }\right) R_{j} \\
&&-\left( \sum_{v}\hat{k}_{2jv}\hat{K}_{v}\left( t\right) +\kappa \sum_{l}%
\hat{k}_{2jl}^{B}\frac{\bar{K}_{j0}\left( t\right) }{1+\sum_{m}\bar{k}_{vm}%
\bar{K}_{m0}\left( t\right) }\right) \bar{r}
\end{eqnarray*}

\paragraph{Investors' capital accumulation}

Appendix 17.1 computes the dynamics for $\hat{K}_{j}\left( t\right) $:%
\begin{equation*}
\frac{d}{dt}\hat{K}_{j}\left( t\right) =\hat{R}_{j}+\sum_{l}\hat{M}_{jl}%
\frac{d}{dt}\hat{K}_{l}\left( t\right) -\sum_{l}\bar{N}_{jl}\frac{d}{dt}\bar{%
K}_{0l}\left( t\right)
\end{equation*}

The right-hand side of equation (\ref{DIP2}) indicates that the variation in
investor $j$'s disposable capital depends not only on its return $\hat{R}%
_{j} $, but also on the capital provided by other investors through the
matrix $\hat{M}$, as well as that provided by banks through the matrix $\bar{%
N}$. This equation can be reformulated as a dynamic process involving the
variation of all the capital at the investor's disposal:%
\begin{equation}
\sum_{l}\left( \delta _{jl}-\hat{M}\right) _{jl}\frac{d}{dt}\hat{K}%
_{l}\left( t\right) +\sum_{l}\bar{N}_{jl}\frac{d}{dt}\bar{K}_{0l}\left(
t\right) =\hat{R}_{j}  \label{DIP2}
\end{equation}%
where:%
\begin{equation}
\hat{M}_{jm}=\frac{\hat{k}_{jm}\hat{K}_{j}\left( t\right) }{1+\sum_{l}\hat{k}%
_{jl}\hat{K}_{l}\left( t\right) +\sum_{l}\hat{k}_{1jl}^{B}\bar{K}_{l0}\left(
t\right) +\kappa \sum_{l}\hat{k}_{2jl}^{B}\frac{\bar{K}_{l0}\left( t\right) 
}{1+\sum_{m}\overline{\bar{k}}_{lm}\bar{K}_{m0}\left( t\right) }}
\label{MATMIC}
\end{equation}%
and:%
\begin{equation*}
\bar{N}_{jl}=\frac{\left( \hat{k}_{1jl}^{B}+\kappa \hat{k}_{2jl}^{B}\frac{1}{%
1+\sum_{m}\overline{\bar{k}}_{lm}\bar{K}_{m0}\left( t\right) }-\kappa \frac{%
\sum_{m}\hat{k}_{2jm}^{B}\bar{K}_{m0}\left( t\right) \overline{\bar{k}}_{ml}%
}{\left( 1+\sum_{n}\overline{\bar{k}}_{mn}\bar{K}_{n0}\left( t\right)
\right) ^{2}}\right) \hat{K}_{j}}{1+\sum_{l}\hat{k}_{jl}\hat{K}_{l}\left(
t\right) +\sum_{l}\hat{k}_{1jl}^{B}\bar{K}_{l0}\left( t\right) +\kappa
\sum_{l}\hat{k}_{2jl}^{B}\frac{\bar{K}_{j0}\left( t\right) }{1+\sum_{m}%
\overline{\bar{k}}_{lm}\bar{K}_{m0}\left( t\right) }}
\end{equation*}

\subsubsection{Banks}

\paragraph{Banks' returns}

Similar to investors, the return of bank $j$, denoted as $\bar{R}_{j}$,
decomposes into a sum of direct returns $\bar{R}_{j}^{\prime }$ and returns
from bank $j$'s participations in investors and other banks:%
\begin{eqnarray}
\bar{R}_{j} &=&\bar{R}_{j}^{\prime }+\sum_{l}\frac{\bar{k}_{1lj}\bar{K}%
_{l0}\left( t\right) }{1+\sum_{v}\bar{k}_{lv}\bar{K}_{v0}\left( t\right) }%
\bar{R}_{l}  \label{RB} \\
&&+\sum_{l}\frac{\bar{k}_{1lj}\bar{K}_{j0}\left( t\right) }{1+\sum_{v}\hat{k}%
_{lv}\hat{K}_{v}\left( t\right) +\sum_{v}\bar{k}_{1lv}\bar{K}_{v0}\left(
t\right) +\kappa \sum_{v}\bar{k}_{2lv}\frac{\bar{K}_{v0}\left( t\right) }{%
1+\sum_{m}\bar{k}_{vm}\bar{K}_{m0}\left( t\right) }}R_{l}  \notag
\end{eqnarray}%
The direct return of bank $j$, denoted as $\bar{R}_{j}^{\prime }$,
originates from loans granted to other banks, loans granted to investors,
loans granted to firms, and direct participation in firms, which can be
expressed as follows: 
\begin{eqnarray*}
\bar{R}_{j}^{\prime } &=&\bar{r}\sum_{l}\frac{\bar{k}_{2lj}\bar{K}%
_{l0}\left( t\right) }{1+\sum_{v}\bar{k}_{lv}\bar{K}_{v0}\left( t\right) } \\
&&+\bar{r}\sum_{l}\frac{\kappa \hat{k}_{2lj}\hat{K}_{l}\left( t\right) }{%
\left( 1+\sum_{v}\hat{k}_{lv}\hat{K}_{v}\left( t\right) +\sum_{lv}\hat{k}%
_{1lv}^{B}\bar{K}_{v0}\left( t\right) +\kappa \sum_{s}\hat{k}_{2ls}^{B}\frac{%
\bar{K}_{s0}\left( t\right) }{1+\sum_{m}\bar{k}_{sm}\bar{K}_{m0}\left(
t\right) }\right) \left( 1+\sum_{v}\bar{k}_{jv}\bar{K}_{v0}\left( t\right)
\right) } \\
&&+\bar{r}\sum_{i}\frac{\kappa k_{2ij}^{\left( B\right) }K_{i}\left(
t\right) }{\left( 1+\sum_{v}k_{iv}\hat{K}_{v}\left( t\right) +\left(
\sum_{v}k_{1iv}^{\left( B\right) }\bar{K}_{v0}\left( t\right)
+k_{2iv}^{\left( B\right) }\kappa \frac{\bar{K}_{v0}\left( t\right) }{%
1+\sum_{m}\bar{k}_{vm}\bar{K}_{m0}\left( t\right) }\right) \right) \left(
1+\sum_{v}\bar{k}_{jv}\bar{K}_{v0}\left( t\right) \right) } \\
&&+\sum_{i}\frac{k_{1ij}^{\left( B\right) }K_{i}\left( t\right) \left(
r_{i}+F_{1}\left( \bar{R}_{i},\frac{\dot{K}_{i}\left( t\right) }{K_{i}\left(
t\right) }\right) +\tau \left( \bar{R}\left( K_{i},X_{i}\right) \right)
\Delta f_{1}^{\prime }\left( K_{i}\left( t\right) \right) \right) }{\left(
1+\sum_{v}k_{iv}\hat{K}_{v}\left( t\right) +\left( \sum_{v}k_{1iv}^{\left(
B\right) }\bar{K}_{v0}\left( t\right) +k_{2iv}^{\left( B\right) }\kappa 
\frac{\bar{K}_{v0}\left( t\right) }{1+\sum_{m}\bar{k}_{vm}\bar{K}_{m0}\left(
t\right) }\right) \right) \left( 1+\sum_{v}\bar{k}_{jv}\bar{K}_{v0}\left(
t\right) \right) }
\end{eqnarray*}

The remaining disposable capital of the bank is invested in taking
participations in other banks and investors, which will provide returns $%
\bar{R}_{l}$ and $R_{l}$.

By combining these two sources of returns, the overall return equation for
bank $j$ can be rewritten as:%
\begin{eqnarray}
&&\sum_{l}\left( \delta _{jl}-\frac{\hat{k}_{1lj}\bar{K}_{l0}\left( t\right) 
}{1+\sum_{v}\bar{k}_{lv}\bar{K}_{v0}\left( t\right) }\right) \left( \frac{%
\bar{f}_{l}-\bar{r}}{1+\sum_{v}\bar{k}_{2lv}\bar{K}_{v0}\left( t\right) }%
\right) \\
&&-\sum_{l}\frac{\hat{k}_{1lj}^{B}\hat{K}_{l}\left( t\right) }{1+\sum_{v}%
\hat{k}_{lv}\hat{K}_{l}\left( t\right) +\sum_{v}\hat{k}_{1lv}^{B}\bar{K}%
_{l0}\left( t\right) +\kappa \sum_{v}\hat{k}_{2lv}^{B}\frac{\bar{K}%
_{l0}\left( t\right) }{1+\sum_{m}\bar{k}_{vm}\bar{K}_{m0}\left( t\right) }} 
\notag \\
&&\times \left( \frac{\hat{f}_{l}-\bar{r}}{1+\sum_{v}\hat{k}_{2lv}\hat{K}%
_{v}\left( t\right) +\kappa \sum_{v}\hat{k}_{2lv}^{B}\frac{\bar{K}_{\nu
0}\left( t\right) }{1+\sum_{m}\bar{k}_{vm}\bar{K}_{m0}\left( t\right) }}%
\right)  \notag \\
&=&\sum_{i}\frac{\left( r_{i}+F_{1}\left( \bar{R}_{i},\frac{\dot{K}%
_{i}\left( t\right) }{K_{i}\left( t\right) }\right) +\tau \left( \bar{R}%
\left( K_{i},X_{i}\right) \right) \Delta f_{1}^{\prime }\left( K_{i}\left(
t\right) \right) -\bar{r}\right) k_{1ij}^{B}K_{i}\left( t\right) }{1+\sum_{v}%
\hat{k}_{jv}\hat{K}_{v}\left( t\right) +\sum_{v}\hat{k}_{1jv}^{B}\bar{K}%
_{l0}\left( t\right) +\kappa \sum_{v}\hat{k}_{2jv}^{B}\frac{\bar{K}%
_{v0}\left( t\right) }{1+\sum_{m}\bar{k}_{vm}\bar{K}_{m0}\left( t\right) }} 
\notag
\end{eqnarray}

\paragraph{Banks' capital accumulation}

We show in appendix 17.2 that in the continuous approximation, the capital
accumulation for banks writes: 
\begin{equation*}
\frac{d}{dt}\bar{K}_{0j}\left( t\right) =\bar{R}_{j}+\sum_{l}M_{jl}\frac{d}{%
dt}\bar{K}_{0l}\left( t\right)
\end{equation*}

Similar to investors, the variation in available capital for bank $j$ is the
sum of its total return, denoted as $\bar{R}_{j}$, plus the variation in
participations and loans from banks investing in bank $j$, through the
matrix $\bar{M}$. As before, this equation can be reformulated as a dynamic
process involving the variation of all the capital at bank $j$'s disposal. 
\begin{equation}
\sum_{l}\left( \delta _{jl}-M_{jl}\frac{d}{dt}\right) \bar{K}_{0l}\left(
t\right) =\bar{R}_{j}^{\prime }  \label{DBP2}
\end{equation}%
with:%
\begin{equation*}
\bar{M}_{jm}=\frac{\bar{k}_{jm}\bar{K}_{j0}\left( t\right) }{1+\sum_{\nu }%
\bar{k}_{jv}\hat{K}_{\nu }\left( t\right) }
\end{equation*}%
Here, the overall leverage effect $\bar{k}_{jm}$ provided by bank $m$ to
bank $j$ decomposes into the leverage effect associated with participations
and loans, denoted as $\bar{k}_{1jm}$ and $\bar{k}_{2jm}$ respectively:%
\begin{equation*}
\bar{k}_{1jm}+\bar{k}_{2jm}=\bar{k}_{jm}
\end{equation*}

\subsubsection{Coupling investors' and banks' capital accumulation}

The dynamics of accumulation for both banks and investors are interrelated.
This is evident in equation (\ref{DIP2}), which is expected, as in our
model, the disposable capital of investors directly depends on banks. The
dependency of banks on investors is indirect, through the returns.
Accumulations of investors and banks are interdependent and must be
considered together. They can be expressed in matrix form:

\begin{equation}
\left( 
\begin{array}{cc}
1-\hat{M} & -\bar{N} \\ 
0 & 1-\bar{M}%
\end{array}%
\right) \left( 
\begin{array}{c}
\frac{d}{dt}\hat{K}\left( t\right) \\ 
\frac{d}{dt}\bar{K}_{0}\left( t\right)%
\end{array}%
\right) =\left( 
\begin{array}{c}
\hat{R}^{\prime } \\ 
\bar{R}^{\prime }%
\end{array}%
\right)  \label{V1}
\end{equation}%
whose solution is:%
\begin{equation}
\left[ 
\begin{array}{c}
\frac{d}{dt}\hat{K}\left( t\right) \\ 
\frac{d}{dt}\bar{K}_{0}\left( t\right)%
\end{array}%
\right] =\left( 
\begin{array}{cc}
\left( 1-\hat{M}\right) ^{-1} & \left( 1-\hat{M}\right) ^{-1}\bar{N}\left( 1-%
\bar{M}\right) ^{-1} \\ 
0 & \left( 1-\bar{M}\right) ^{-1}%
\end{array}%
\right) \left[ 
\begin{array}{c}
\hat{R}^{\prime } \\ 
\bar{R}^{\prime }%
\end{array}%
\right]  \label{V2}
\end{equation}%
We can see from equations (\ref{V1}) and (\ref{V2}) that the dynamics of
capital accumulation for investors and banks are linked through
participations and loans. For this reason, both types of agents will
generally be solved simultaneously.

\subsection{Returns and capital accumulation under default scenario}

\subsubsection{Default of banks}

As in Part 1, let's assume that bank $v$ defaults if its total private
capital cannot cover its loans: 
\begin{equation*}
\left( 1+\sum_{m}\bar{k}_{2vm}\bar{K}_{m0}\left( t\right) \right) \left( 1+%
\bar{R}_{v}\right) <\left( 1+\bar{r}\right) \sum_{m}\bar{k}_{2vm}\bar{K}%
_{m0}\left( t\right)
\end{equation*}%
If this occurs, the default of bank $\nu $ modifies other banks returns $%
\bar{R}_{j}^{\prime }$ by a term:

\begin{equation*}
\sum_{l}\left( \bar{r}-\frac{\left( 1+\bar{f}_{\nu }\right) }{\sum_{m}\bar{k}%
_{2vm}\bar{K}_{m}}\right) \frac{H\left( -\left( 1+\bar{f}_{\nu }\right)
\right) \bar{k}_{2lj}\bar{K}_{l}\left( t\right) }{1+\sum_{v}\left( \bar{k}%
_{1lv}+\bar{k}_{2lv}\right) \bar{K}_{v}\left( t\right) }
\end{equation*}

where $H$ is the Heaviside step function. This term only appears when $1+%
\bar{f}_{\nu }$ {} is negative, that is when the return $\bar{f}_{\nu }<-1$,
indicating that the return is so negative that it even erodes the bank's
private capital. When this term appears, the loss incurred by bank $l$,
which has lent to bank $v$, is: 
\begin{equation*}
\left( \bar{r}-\frac{\left( 1+\bar{f}_{\nu }\right) }{\sum_{m}\bar{k}_{2vm}%
\bar{K}_{m}}\right) \frac{\bar{k}_{2lj}\bar{K}_{l}\left( t\right) }{%
1+\sum_{v}\left( \bar{k}_{1lv}+\bar{k}_{2lv}\right) \bar{K}_{v}\left(
t\right) }
\end{equation*}%
The equations for returns accounting for default are provided in the
appendix.

\subsubsection{Default of investors}

The loans received by investors are divided into loans from other investors
and loans from banks, where $\kappa $ is the banking multiplier.

\begin{equation*}
\sum_{v}\hat{k}_{2jv}\hat{K}_{v}\left( t\right) +\kappa \sum_{v}\hat{k}%
_{2jv}^{B}\frac{\bar{K}_{v0}\left( t\right) }{1+\sum_{m}\overline{\bar{k}}%
_{vm}\bar{K}_{m0}\left( t\right) }
\end{equation*}%
The default condition on this loan is therefore that the return on private
capital, private loans, and bank loans do not generate enough returns to
repay the borrowed sum with interest, such that:%
\begin{eqnarray*}
&&\sum_{l}\left( 1+\sum_{v}\hat{k}_{2lv}\hat{K}_{v}\left( t\right) +\kappa
\sum_{v}\hat{k}_{2jv}^{B}\frac{\bar{K}_{v0}\left( t\right) }{1+\sum_{m}%
\overline{\bar{k}}_{vm}\bar{K}_{m0}\left( t\right) }\right) \left( 1+\hat{R}%
_{j}\right) \\
&<&\left( 1+\bar{r}\right) \left( \sum_{v}\hat{k}_{2jv}\hat{K}_{v}\left(
t\right) +\kappa \sum_{v}\hat{k}_{2jv}^{B}\frac{\bar{K}_{v0}\left( t\right) 
}{1+\sum_{m}\overline{\bar{k}}_{vm}\bar{K}_{m0}\left( t\right) }\right)
\end{eqnarray*}%
In this case, the default of an investor $\nu $ affects both the return $%
\bar{R}_{j}^{\prime }$ of the banks that lent to them, by an amount:

\begin{eqnarray*}
&&\sum_{l}\left( \bar{r}-\frac{\left( 1+\hat{f}_{\nu }\right) }{\sum_{m}\hat{%
k}_{2vm}\hat{K}_{m}+\kappa \sum_{m}\hat{k}_{2vm}^{B}\frac{\bar{K}_{m0}\left(
t\right) }{1+\sum_{s}\bar{k}_{ms}\bar{K}_{s0}\left( t\right) }}\right) \\
&&\times \frac{H\left( -\left( 1+\hat{f}_{\nu }\right) \right) \hat{k}%
_{2lj}^{B}\hat{K}_{l}\left( t\right) }{1+\sum_{\nu }\hat{k}_{l\nu }\hat{K}%
_{l}\left( t\right) +\sum_{\nu }\hat{k}_{1l\nu }^{B}\bar{K}_{l0}\left(
t\right) +\kappa \sum_{\nu }\hat{k}_{2l\nu }^{B}\frac{\bar{K}_{j0}\left(
t\right) }{1+\sum_{m}\bar{k}_{vm}\bar{K}_{m0}\left( t\right) }}
\end{eqnarray*}%
and the return $\hat{R}_{j}^{\prime }$ of an investor, by an amount of:%
\begin{eqnarray*}
&&\sum_{l}\left( \bar{r}-\frac{\left( 1+\hat{f}_{\nu }\right) }{\sum_{m}\hat{%
k}_{2vm}\hat{K}_{m}+\kappa \sum_{m}\hat{k}_{2vm}^{B}\frac{\bar{K}_{m0}\left(
t\right) }{1+\sum_{s}\bar{k}_{ms}\bar{K}_{s0}\left( t\right) }}\right) \\
&&\times \frac{H\left( -\left( 1+\hat{f}_{\nu }\right) \right) \hat{k}_{2lj}%
\hat{K}_{l}\left( t\right) }{1+\sum_{\nu }\hat{k}_{l\nu }\hat{K}_{l}\left(
t\right) +\sum_{\nu }\hat{k}_{1l\nu }^{B}\bar{K}_{l0}\left( t\right) +\kappa
\sum_{\nu }\hat{k}_{2l\nu }^{B}\frac{\bar{K}_{j0}\left( t\right) }{1+\sum_{m}%
\bar{k}_{vm}\bar{K}_{m0}\left( t\right) }}
\end{eqnarray*}%
The equations for returns accounting for default are provided in the
appendix.

\subsubsection{Default of firms}

We now examine the defaults of firms. The default condition is that the
overall return $R_{j}$ on private capital, private loans, and bank loans
does not generate enough returns to repay the borrowed sum at the interest
rate $\bar{r}$, such that: 
\begin{eqnarray*}
&&\left( 1+R_{j}\right) \left( 1+\sum_{v}\hat{k}_{2jv}\hat{K}_{v}\left(
t\right) +\kappa \sum_{v}\hat{k}_{2jv}^{B}\frac{\bar{K}_{v0}\left( t\right) 
}{1+\sum_{m}\overline{\bar{k}}_{vm}\bar{K}_{m0}\left( t\right) }\right) \\
&<&\left( 1+\bar{r}\right) \left( \sum_{v}\hat{k}_{2jv}\hat{K}_{v}\left(
t\right) +\kappa \sum_{v}\hat{k}_{2jv}^{B}\frac{\bar{K}_{v0}\left( t\right) 
}{1+\sum_{m}\overline{\bar{k}}_{vm}\bar{K}_{m0}\left( t\right) }\right)
\end{eqnarray*}%
Default for firms $i$ modifies the return for banks $\bar{R}_{j}^{\prime }$
by:%
\begin{eqnarray*}
&&\sum_{i}\left( \bar{r}-\frac{\left( 1+f_{1}^{\prime }\left( K_{i}\left(
t\right) \right) \right) }{\left( \sum_{v}k_{2iv}^{\left( B\right) }\kappa 
\frac{\bar{K}_{v0}\left( t\right) }{1+\sum_{m}\bar{k}_{vm}\bar{K}_{m0}\left(
t\right) }\right) +\left( \sum_{v}k_{2iv}\hat{K}_{v}\left( t\right) \right) }%
\right) \\
&&\times \frac{H\left( -\left( 1+f_{1}^{\prime }\left( K_{i}\left( t\right)
\right) \right) \right) k_{2ij}^{\left( B\right) }K_{i}\left( t\right) }{%
1+\left( \sum_{v}k_{1iv}^{\left( B\right) }\bar{K}_{v0}\left( t\right)
+k_{2iv}^{\left( B\right) }\kappa \frac{\bar{K}_{v0}\left( t\right) }{%
1+\sum_{m}\bar{k}_{vm}\bar{K}_{m0}\left( t\right) }\right) +\left(
\sum_{v}k_{iv}\hat{K}_{v}\left( t\right) \right) }
\end{eqnarray*}%
and the investor $j$'s return, denoted as $\hat{R}_{j}^{\prime }$, by:%
\begin{eqnarray*}
&&\sum_{i}\left( \bar{r}-\frac{\left( 1+f_{1}^{\prime }\left( K_{i}\left(
t\right) \right) \right) }{\left( \sum_{v}k_{2iv}^{\left( B\right) }\kappa 
\frac{\bar{K}_{v0}\left( t\right) }{1+\sum_{m}\bar{k}_{vm}\bar{K}_{m0}\left(
t\right) }\right) +\left( \sum_{v}k_{2iv}\hat{K}_{v}\left( t\right) \right) }%
\right) \\
&&\times \frac{H\left( -\left( 1+f_{1}^{\prime }\left( K_{i}\left( t\right)
\right) \right) \right) k_{2ij}K_{i}\left( t\right) }{1+\left(
\sum_{v}k_{1iv}^{\left( B\right) }\bar{K}_{v0}\left( t\right)
+k_{2iv}^{\left( B\right) }\kappa \frac{\bar{K}_{v0}\left( t\right) }{%
1+\sum_{m}\bar{k}_{vm}\bar{K}_{m0}\left( t\right) }\right) +\left(
\sum_{v}k_{iv}\hat{K}_{v}\left( t\right) \right) }
\end{eqnarray*}

The formulas for the return including default are provided in the appendix.

\section{Field Translation}

Similar to the simple model in Part $1$, investors are described by $\hat{%
\Psi}\left( K,\hat{X}\right) $ and firms by $\Psi \left( K,X\right) $. Banks
are described by the field $\bar{\Psi}\left( \hat{K}^{\prime },\hat{X}%
^{\prime }\right) $. The translation and definitions of coefficients and
matrices are given in Appendices 18.1 and 18.2. The resolution method is the
same as in Part $1$; we simply present the results. The major difference is
the introduction of the additional field for banks. It is worth noting that,
for simplicity, and as in Part $1$, we perform a change of variable on the
fields, and all functional actions are rewritten with the new variables.

\subsection{Banks' action functional}

The field $\bar{\Psi}$ describes the banking system.\ It depends on the two
variables $\bar{K}$ and $\bar{X}$, and its action functional is given by:%
\begin{equation}
-\bar{\Psi}^{\dag }\left( \bar{K},\bar{X}\right) \nabla ^{2}\bar{\Psi}\left( 
\bar{K},\bar{X}\right) +\left( \frac{\bar{g}^{2}\left( \bar{K},\bar{X}%
\right) }{2\sigma _{\bar{K}}^{2}}+\frac{\bar{g}\left( \bar{K},\bar{X}\right) 
}{2\bar{K}}\right) \left\vert \bar{\Psi}\left( \bar{K},\bar{X}\right)
\right\vert ^{2}  \label{BFC}
\end{equation}%
where:%
\begin{equation*}
\bar{g}\left( \hat{K},\hat{X}\right) =\left( 1-\bar{M}\left\vert \bar{\Psi}%
\left( \bar{K},\bar{X}\right) \right\vert ^{2}\right) ^{-1}\bar{f}\left( 
\bar{K},\bar{X}\right)
\end{equation*}%
and $\bar{M}$ is the matrix translating its microeconomic equivalent (\ref%
{MATMIC}).

\subsection{Investors' action functional}

As in part $1$, the field $\hat{\Psi}$ describing the investors depends on
the two variables $\hat{K}$ and $\hat{X}$, and its action functional is
given by:%
\begin{equation}
-\hat{\Psi}^{\dag }\left( \hat{K},\hat{X}\right) \nabla ^{2}\hat{\Psi}\left( 
\hat{K},\hat{X}\right) +\left( \frac{\hat{g}^{2}\left( \hat{K},\hat{X}%
\right) }{2\sigma _{\hat{K}}^{2}}+\frac{\hat{g}\left( \hat{K},\hat{X}\right) 
}{2\hat{K}}\right) \left\vert \hat{\Psi}\left( \hat{K},\hat{X}\right)
\right\vert ^{2}  \label{NFC}
\end{equation}%
where we define:%
\begin{equation*}
\hat{g}\left( \hat{K},\hat{X}\right) =\left( 1-\hat{M}\left\vert \hat{\Psi}%
\left( \hat{K},\hat{X}\right) \right\vert ^{2}\right) ^{-1}\hat{f}\left( 
\hat{K},\hat{X}\right) +\left( 1-\hat{M}\right) ^{-1}N\left( 1-\bar{M}%
\right) ^{-1}\bar{f}\left( \bar{K},\bar{X}\right)
\end{equation*}%
The field definition of the matrices $\bar{M}$, $\hat{M}$ and $N$ are given
in Appendix 18.2.

\subsection{Firms' action functional}

The translation of the function (\ref{FT2}) is obtained by applying the
mapping provided by (\ref{inco}) and (\ref{Trl}).

The action functional for the field of firms is:%
\begin{equation*}
-\Psi ^{\dag }\left( K,X\right) \left( \nabla _{K_{p}}\left( \sigma
_{K}^{2}\nabla _{K_{p}}-f_{1}^{\prime }\left( K,X\right) K_{p}\right)
\right) \Psi \left( K,X\right) +\frac{1}{2\epsilon }\left( \left\vert \Psi
\left( K,X\right) \right\vert ^{2}-\left\vert \Psi _{0}\left( X\right)
\right\vert ^{2}\right) ^{2}
\end{equation*}%
where: 
\begin{equation*}
f_{1}^{\prime }\left( K,X\right) =\left( 1+\underline{k}_{2}\left( X\right)
\right) f_{1}\left( X\right) -\bar{r}\underline{k}_{2}\left( \hat{X}\right)
\end{equation*}%
where $\underline{k}_{2}\left( \hat{X}\right) $ is defined in appendix 18.1.

\subsection{Return equations under a no-default scenario}

For later purpose, we write the translation of return equations (\ref{RH})
and (\ref{RB}) in terms of $\hat{g}\left( \hat{K}_{1},\hat{X}_{1}\right) $
and $\bar{g}\left( \bar{K}_{1},\bar{X}_{1}\right) $. The equations with
defaults are presented in Appendix 18.4.

\subsubsection{Bank returns' equation}

The translation of (\ref{RB}) is:%
\begin{eqnarray}
&&\left( \Delta \left( \bar{X}^{\prime },\bar{X}\right) -\frac{\bar{K}%
^{\prime }\bar{k}_{1}\left( \bar{X}^{\prime },\bar{X}\right) \left\vert \bar{%
\Psi}\left( \bar{K}^{\prime },\bar{X}^{\prime }\right) \right\vert ^{2}}{1+%
\underline{\overline{\bar{k}}}\left( \bar{X}^{\prime }\right) }\right) \frac{%
\left( 1-\bar{M}\right) \bar{g}\left( \hat{K}^{\prime },\hat{X}^{\prime
}\right) }{1+\underline{\bar{k}}_{2}\left( \bar{X}^{\prime }\right) }
\label{RGB} \\
&&-\int \frac{\hat{K}^{\prime }\underline{\hat{k}}_{1}^{B}\left( \hat{X}%
^{\prime },\bar{X}\right) }{1+\underline{\hat{k}}\left( \hat{X}^{\prime
}\right) +\underline{\hat{k}}_{1}^{B}\left( \bar{X}^{\prime }\right) +\kappa %
\left[ \frac{\underline{\hat{k}}_{2}^{B}}{1+\bar{k}}\right] \left( \hat{X}%
^{\prime }\right) }\frac{\left( 1-\hat{M}\right) \hat{g}\left( \hat{K}%
^{\prime },\hat{X}^{\prime }\right) +\left( \bar{N}\bar{g}\right) \left( 
\bar{K}^{\prime },\bar{X}^{\prime }\right) }{1+\underline{\hat{k}}_{2}\left( 
\bar{X}^{\prime }\right) +\kappa \frac{\underline{\hat{k}}_{2}^{B}\left( 
\bar{X}^{\prime }\right) }{1+\bar{k}\left( \bar{X}\right) }}d\bar{X}^{\prime
}  \notag \\
&=&\frac{\underline{k}_{1}^{\left( B\right) }\left( X^{\prime },\bar{X}%
\right) }{1+\underline{k}\left( \hat{X}^{\prime }\right) +\underline{k}%
_{1}^{\left( B\right) }\left( \bar{X}^{\prime }\right) +\kappa \frac{%
\underline{k}_{2}^{\left( B\right) }\left( \bar{X}^{\prime }\right) }{1+%
\underline{\bar{k}}}}\frac{\left( f_{1}^{\prime }\left( X^{\prime }\right)
K^{\prime }-\bar{C}\left( X^{\prime }\right) \right) }{1+\underline{k}%
_{2}\left( \hat{X}^{\prime }\right) +\kappa \frac{\underline{k}_{2}^{\left(
B\right) }\left( \bar{X}^{\prime }\right) }{1+\underline{\bar{k}}}}  \notag
\end{eqnarray}%
The left-hand side represents the returns for banks located in $\bar{X}$,
including participations in other banks located in $\bar{X}^{\prime }$,
which provide a return $\bar{g}$ as well as participations in investors
located in $\hat{X}^{\prime }$, which yield a return $\hat{g}\left( \hat{K}%
^{\prime },\hat{X}^{\prime }\right) $. The returns of the banks depend
indirectly the returns of other banks through the term $\left( \bar{N}\bar{g}%
\right) $: the return of investors in which a bank invests depends directly
on the investments of other banks and investors. The right-hand side
describes the returns provided by firms in which the bank invests.

\subsubsection{Investors returns' equation}

The translation of (\ref{RH}) is: 
\begin{eqnarray}
&&\left( \Delta \left( \hat{X},\hat{X}^{\prime }\right) -\frac{\hat{K}%
^{\prime }\hat{k}_{1}\left( \hat{X}^{\prime },\hat{X}\right) \left\vert \hat{%
\Psi}\left( \hat{K}^{\prime },\hat{X}^{\prime }\right) \right\vert ^{2}}{1+%
\underline{\hat{k}}\left( \hat{X}^{\prime }\right) +\underline{\hat{k}}%
_{1}^{B}\left( \bar{X}^{\prime }\right) +\kappa \left[ \frac{\underline{\hat{%
k}}_{2}^{B}}{1+\bar{k}}\right] \left( \hat{X}^{\prime }\right) }\right) 
\frac{\left( 1-\hat{M}\right) \hat{g}\left( \hat{K}^{\prime },\hat{X}%
^{\prime }\right) +\bar{N}\bar{g}\left( \hat{K}^{\prime },\hat{X}^{\prime
}\right) }{1+\underline{\hat{k}}_{2}\left( \bar{X}^{\prime }\right) +\kappa %
\left[ \frac{\underline{\hat{k}}_{2}^{B}}{1+\bar{k}}\right] \left( \hat{X}%
^{\prime }\right) }  \label{RGH} \\
&=&\frac{k_{1}\left( X^{\prime },X^{\prime }\right) \left( f_{1}^{\prime
}\left( X^{\prime }\right) K^{\prime }-\bar{C}\left( X^{\prime }\right)
\right) }{\left( 1+\underline{k}\left( \hat{X}^{\prime }\right) +\underline{k%
}_{1}^{\left( B\right) }\left( \bar{X}^{\prime }\right) +\kappa \left[ \frac{%
\underline{k}_{2}^{B}}{1+\bar{k}}\right] \left( X^{\prime }\right) \right)
\left( 1+\underline{k}_{2}\left( \hat{X}^{\prime }\right) +\kappa \left[ 
\frac{\underline{k}_{2}^{B}}{1+\bar{k}}\right] \left( X^{\prime }\right)
\right) }  \notag
\end{eqnarray}%
The left-hand side represents the returns for investors located in $\hat{X}$%
, including participations in other investors located in $\hat{X}^{\prime }$
providing return $\hat{g}\left( \hat{K}^{\prime },\hat{X}^{\prime }\right) $%
. Similarly, the returns of the banks indirectly depend on the returns of
other banks through the term $\left( \bar{N}\bar{g}\right) $. The right-hand
side describes the returns provided by firms in which the bank invests.

\subsubsection{Firms' returns}

The derivation of firm return, average capital, and background field is
similar to Part $1$. Taking into account the banking sector, the firm's
disposable capital is given by:%
\begin{equation*}
K=\left( 1+\underline{k}\left( \hat{X}^{\prime }\right) +\underline{k}%
_{1}^{\left( B\right) }\left( \bar{X}^{\prime }\right) +\kappa \left[ \frac{%
\underline{k}_{2}^{B}}{1+\bar{k}}\right] \left( X^{\prime }\right) \right)
K_{p}
\end{equation*}%
We consider constant returns to scale and include small corrections to
account for decreasing returns to scale.

\paragraph{Constant returns to scale}

Using the results of Part $1$ and including banks, the firm's return on its
private capital is calculated by multiplying the firm's available capital,
consisting of $K_{p}$, by various leverage effects: 
\begin{equation*}
K_{p}\left( 1+k\left( X,,\hat{X}^{\prime }\right) \hat{K}_{\hat{X}^{\prime
}}^{\prime }+\underline{k}_{1}^{\left( B\right) }\left( X,\bar{X}^{\prime
}\right) \bar{K}_{\bar{X}^{\prime }}+\kappa \frac{\underline{k}_{2}^{\left(
B\right) }\left( X,\bar{X}^{\prime }\right) }{1+\underline{\bar{k}}\left( 
\bar{X}\right) }\bar{K}_{\bar{X}^{\prime }}\right)
\end{equation*}%
by the productivity $f_{1}\left( X\right) $, subtracting the fixed
production cost $C$, and then dividing by the total available capital, to
obtain a return percentage. Multiplying this return by $K_{p}$, we get the
overall return of the firm in value, which is:%
\begin{equation}
K_{p}\left( \frac{f_{1}\left( X\right) \left( 1+k\left( X,,\hat{X}^{\prime
}\right) \hat{K}_{\hat{X}^{\prime }}^{\prime }+\underline{k}_{1}^{\left(
B\right) }\left( X,\bar{X}^{\prime }\right) \bar{K}_{\bar{X}^{\prime
}}+\kappa \frac{\underline{k}_{2}^{\left( B\right) }\left( X,\bar{X}^{\prime
}\right) }{1+\underline{\bar{k}}\left( \bar{X}\right) }\bar{K}_{\bar{X}%
^{\prime }}\right) K_{p}-C}{\left( 1+\underline{k}\left( \hat{X}^{\prime
}\right) +\underline{k}_{1}^{\left( B\right) }\left( \bar{X}^{\prime
}\right) +\kappa \left[ \frac{\underline{k}_{2}^{B}}{1+\bar{k}}\right]
\left( X^{\prime }\right) \right) K_{p}}\right) =f_{1}\left( X\right) K_{p}-%
\bar{C}\left( X\right)  \label{TRFR}
\end{equation}%
with:%
\begin{eqnarray}
\bar{C}\left( X\right) &=&\frac{C}{1+\int \underline{k}\left( \hat{X}\right) 
\hat{K}_{\hat{X}^{\prime }}\frac{\left\vert \hat{\Psi}\left( \hat{X}^{\prime
}\right) \right\vert ^{2}}{\left\vert \Psi _{0}\left( X\right) \right\vert
^{2}}+\int \left( \underline{k}_{1}^{\left( B\right) }\left( \bar{X}^{\prime
}\right) +\kappa \left[ \frac{\underline{k}_{2}^{B}}{1+\bar{k}}\right]
\left( X^{\prime }\right) \right) \bar{K}_{\bar{X}^{\prime }}\frac{%
\left\vert \bar{\Psi}\left( \hat{X}^{\prime }\right) \right\vert ^{2}}{%
\left\vert \Psi _{0}\left( X\right) \right\vert ^{2}}}  \label{CBR} \\
&&\frac{C\left( X\right) }{1+\underline{k}\left( X\right) +\underline{k}%
_{1}^{B}\left( X\right) +\kappa \left[ \frac{\underline{k}_{2}^{B}}{1+\bar{k}%
}\right] \left( X\right) }  \notag
\end{eqnarray}%
Note that $\bar{C}\left( X\right) $ is an effective cost, reduced by
investments. The more investments in the firm, the more the fixed cost is
diluted in terms of return.

\paragraph{Decreasing returns to scale}

Similar to Part 1, to incorporate corrections accounting for slightly
decreasing returns, we will replace $f_{1}\left( X\right) $ in the formulas
above with a productivity that decreases with invested capital:%
\begin{equation*}
f_{1}\left( X\right) \rightarrow \frac{f_{1}\left( X\right) }{\left( 1+\frac{%
k\left( X\right) }{K\left[ X\right] }\hat{K}\left[ X\right] +\frac{%
k_{1}^{B}\left( X\right) }{K\left[ X\right] }\bar{K}\left[ X\right] +\frac{%
\kappa \left[ \frac{\underline{k}_{2}^{B}}{1+\bar{k}}\right] }{K\left[ X%
\right] }\bar{K}\left[ X\right] \right) ^{r}K_{X}^{r}}-C_{0}
\end{equation*}

The formula clearly demonstrates that the more capital invested, the lower
the marginal productivity of capital.

\section{Minimization equations}

Now that we have provided the functional actions of the various agents in
the system, we can determine the background fields that define the system's
states. The methods of solving and interpreting background fields and
capital per sector are similar to Part $1$. We will begin with firms.

\subsection{Minimization equation for firms}

As before, the minimization of firms' action functional leads to the
equation:%
\begin{equation}
0=\left( \frac{\left( f_{1}^{\left( e\right) }\left( X\right) \right) ^{2}}{%
\sigma _{\hat{K}}^{2}}+\frac{f_{1}^{\left( e\right) }\left( X\right) }{2}%
\right) +\frac{1}{\epsilon }\left( \left\vert \Psi \left( K,X\right)
\right\vert ^{2}-\left\vert \Psi _{0}\left( X\right) \right\vert ^{2}\right)
\label{MNF}
\end{equation}%
where $f_{1}^{\left( e\right) }\left( X\right) $ is the return of the firm
once loans are repaid:%
\begin{equation}
f_{1}^{\left( e\right) }\left( X\right) =\left( 1+\underline{k}_{2}\left(
X\right) +\kappa \left[ \frac{\underline{k}_{2}^{B}}{1+\bar{k}}\right]
\left( X\right) \right) f_{1}^{\prime }\left( X\right) -\left( \underline{k}%
_{2}\left( X\right) +\kappa \left[ \frac{\underline{k}_{2}^{B}}{1+\bar{k}}%
\right] \left( X\right) \right) \bar{r}  \label{TRF}
\end{equation}%
and $f_{1}^{\prime }\left( X\right) $ represents the return of the firm from
its production, as given by (\ref{TRFR}):%
\begin{equation*}
f_{1}^{\prime }\left( X\right) =f_{1}\left( X\right) K_{p}-\bar{C}\left(
X\right)
\end{equation*}

\subsection{Minimization equation for investors}

Appendix 20.1 derives the minimization equation for (\ref{NFC}):%
\begin{align}
0& =\frac{\hat{K}_{1}^{2}\hat{g}^{2}\left( \hat{K}_{1},\hat{X}_{1}\right) }{%
2\sigma _{\hat{K}}^{2}}+\frac{\hat{g}\left( \hat{K}_{1},\hat{X}_{1}\right) }{%
2}  \label{NM} \\
& +\int \left\vert \hat{\Psi}\left( \hat{K},\hat{X}\right) \right\vert
^{2}\left( \frac{\hat{K}^{2}\hat{g}\left( \hat{K},\hat{X},\Psi ,\hat{\Psi}%
\right) }{\sigma _{\hat{K}}^{2}}+\frac{1}{2}\right) \frac{\delta \hat{g}%
\left( \hat{K},\hat{X}\right) }{\delta \left\vert \hat{\Psi}\left( \hat{K}%
_{1},\hat{X}_{1}\right) \right\vert ^{2}}+\frac{1}{\hat{\mu}}\left(
\left\vert \hat{\Psi}\left( \hat{K},\hat{X}\right) \right\vert
^{2}-\left\vert \hat{\Psi}_{0}\left( \hat{X}\right) \right\vert ^{2}\right) 
\notag
\end{align}%
The formula for $\frac{\delta \hat{g}\left( \hat{K},\hat{X}\right) }{\delta
\left\vert \hat{\Psi}\left( \hat{K}_{1},\hat{X}_{1}\right) \right\vert ^{2}}$
is given in Appendix 25.

\subsection{Minimization equation for banks}

Appendix 20.2 derives the minimization equation for (\ref{BFC}):%
\begin{eqnarray}
0 &=&\left( \frac{\bar{K}_{1}^{2}\bar{g}^{2}\left( \bar{K}_{1},\bar{X}%
_{1}\right) }{\sigma _{\hat{K}}^{2}}+\frac{\bar{g}\left( \bar{K}_{1},\bar{X}%
_{1}\right) }{2}\right)  \label{BM} \\
&&+\int \left\vert \hat{\Psi}\left( \hat{K},\hat{X}\right) \right\vert
^{2}\left( \frac{\hat{K}^{2}\hat{g}\left( \hat{K},\hat{X}\right) }{\sigma _{%
\hat{K}}^{2}}+\frac{1}{2}\right) \frac{\delta \hat{g}\left( \hat{K},\hat{X}%
\right) }{\delta \left\vert \bar{\Psi}\left( \bar{K}_{1},\bar{X}_{1}\right)
\right\vert ^{2}}+\frac{1}{\hat{\mu}}\left( \left\vert \bar{\Psi}\left( \bar{%
K}_{1},\bar{X}_{1}\right) \right\vert ^{2}-\left\vert \bar{\Psi}_{0}\left( 
\bar{X}_{1}\right) \right\vert ^{2}\right)  \notag
\end{eqnarray}%
and the formula for $\frac{\delta \hat{g}\left( \hat{K},\hat{X}\right) }{%
\delta \left\vert \bar{\Psi}\left( \bar{K}_{1},\bar{X}_{1}\right)
\right\vert ^{2}}$ is derived in appendix 25. These equations will be used
when computing average capital per sector.

\section{Resolution for firms}

\subsection{Solution for the background field}

The total return produced by a firm is given by (\ref{TRF}):%
\begin{equation*}
f_{1}^{\left( e\right) }\left( X\right) K=\left( 1+\underline{k}_{2}\left(
X\right) +\kappa \left[ \frac{\underline{k}_{2}^{B}}{1+\bar{k}}\right]
\left( X\right) \right) \left( f_{1}^{\prime }\left( X\right) K_{p}-\bar{C}%
\left( X\right) \right) -\left( \underline{k}_{2}\left( X\right) +\kappa %
\left[ \frac{\underline{k}_{2}^{B}}{1+\bar{k}}\right] \left( X\right)
\right) K_{p}\bar{r}
\end{equation*}%
where the first term in the right-hand side is the return of the capital
invested by the firm, including loans. with $\bar{C}\left( X\right) $ being
the effective cost defined in (\ref{CBR}).

As before, we assume constant returns to scale and include corrections
later. The solution to the minimization equation for the firm's background
field is:%
\begin{equation*}
\left\vert \Psi \left( X\right) \right\vert ^{2}=\left\vert \Psi _{0}\left(
X\right) \right\vert ^{2}-\epsilon \frac{f_{1}^{\left( e\right) }\left(
X\right) }{2}-\epsilon \frac{\left( f_{1}^{\left( e\right) }\left( X\right)
K_{p}\right) ^{2}}{\sigma _{\hat{K}}^{2}}
\end{equation*}

\subsection{Solution for the average capital per sector}

Appendix 20.4 computes the average capital as in Part $1$:%
\begin{equation*}
K_{X}=\frac{1}{4f_{1}^{\left( e\right) }\left( X\right) }\frac{\left(
3X^{\left( e\right) }-C^{\left( e\right) }\right) \left( C^{\left( e\right)
}+X^{\left( e\right) }\right) }{2X^{\left( e\right) }-C^{\left( e\right) }}
\end{equation*}

\subsection{Solution for the return to investors}

Note that the return to investors will thus be:%
\begin{equation*}
\frac{f_{1}^{\left( e\right) }\left( X\right) -\bar{r}}{1+\underline{k}%
_{2}\left( X\right) +\kappa \left[ \frac{\underline{k}_{2}^{B}}{1+\bar{k}}%
\right] \left( X\right) }=f_{1}\left( X\right) -\bar{r}-\frac{\bar{C}\left(
X\right) }{K_{p}}
\end{equation*}%
and that the amount of return involved in the banks and investors' return
equations will be:

\begin{equation*}
\left\vert \Psi \left( X\right) \right\vert ^{2}K_{X}\frac{f_{1}^{\prime
}\left( X\right) -\bar{r}}{1+\underline{k}_{2}\left( X\right) +\kappa \left[ 
\frac{\underline{k}_{2}^{B}}{1+\bar{k}}\right] \left( X\right) }=\left\vert
\Psi \left( X\right) \right\vert ^{2}\left( \left( f_{1}\left( X\right) -%
\bar{r}\right) K_{X}-\bar{C}\left( X\right) \right)
\end{equation*}%
whose expanded expression is:%
\begin{eqnarray*}
&&\left\vert \Psi \left( X\right) \right\vert ^{2}K_{X}\frac{f_{1}^{\prime
}\left( X\right) -\bar{r}}{1+\underline{k}_{2}\left( X\right) +\kappa \left[ 
\frac{\underline{k}_{2}^{B}}{1+\bar{k}}\right] \left( X\right) } \\
&=&\frac{\left( C^{\left( e\right) }+X^{\left( e\right) }\right) \left( 
\frac{2}{3}X^{2}+\frac{1}{3}\left( X-C^{\left( e\right) }\right) C^{\left(
e\right) }\right) \epsilon \left\{ 3\left( X^{\left( e\right) }-C^{\left(
e\right) }\right) ^{2}-\frac{\bar{r}\left( 3X^{\left( e\right) }-C^{\left(
e\right) }\right) \left( C^{\left( e\right) }+X^{\left( e\right) }\right) }{%
f_{1}^{\left( e\right) }\left( X\right) }\right\} }{4\sigma _{\hat{K}%
}^{2}\left( 1+\underline{k}_{2}\left( X\right) +\kappa \left[ \frac{%
\underline{k}_{2}^{B}}{1+\bar{k}}\right] \left( X\right) \right)
f_{1}^{\left( e\right) }\left( X\right) \left( 2X^{\left( e\right)
}-C^{\left( e\right) }\right) }
\end{eqnarray*}%
with:%
\begin{equation*}
X^{\left( e\right) }=\sqrt{\frac{\left\vert \Psi _{0}\right\vert ^{2}}{%
\epsilon }-\frac{1}{2}f_{1}^{\left( e\right) }\left( X\right) }
\end{equation*}%
and:%
\begin{equation*}
\bar{C}^{\left( e\right) }\left( X\right) =\frac{\left( 1+\underline{k}%
_{2}\left( X\right) +\kappa \left[ \frac{\underline{k}_{2}^{B}}{1+\bar{k}}%
\right] \left( X\right) \right) C\left( X\right) }{1+\underline{k}\left(
X\right) +\underline{k}_{1}^{B}\left( X\right) +\kappa \left[ \frac{%
\underline{k}_{2}^{B}}{1+\bar{k}}\right] \left( X\right) }
\end{equation*}

\section{Resolution for financial agents}

Since investors and banks interact with each other, we will solve their
minimization equations simultaneously, treating them as the financial agents
of the system.

\subsection{Compact form of the minimization equation}

These minimization equations, denoted as (\ref{NM}) and (\ref{BM}), are
rewritten using the expression for functional derivatives:%
\begin{equation*}
\frac{\delta \hat{g}\left( \hat{K},\hat{X}\right) }{\delta \left\vert \hat{%
\Psi}\left( \hat{K}_{1},\hat{X}_{1}\right) \right\vert ^{2}}=-\frac{\left(
\kappa \left\langle \left[ \frac{\underline{\hat{k}}_{2}^{B}}{1+\bar{k}}%
\right] \right\rangle \left( 1-\left\langle \underline{\hat{k}}\right\rangle
\right) +\left\langle \hat{k}_{1}^{B}\right\rangle \left\langle \hat{k}%
_{2}\right\rangle \right) \left( \left\langle \hat{g}\right\rangle +\frac{1}{%
1-\left\langle \underline{\hat{k}}\right\rangle }\bar{N}\left\langle \bar{g}%
\right\rangle \right) \frac{\left\Vert \bar{\Psi}\right\Vert
^{2}\left\langle \bar{K}\right\rangle }{\left\Vert \hat{\Psi}\right\Vert
^{2}\left\langle \hat{K}\right\rangle }\frac{\hat{K}_{1}}{\left\langle \hat{K%
}\right\rangle }}{\left( 1-\left( \left\langle \hat{k}_{1}\right\rangle
+\left\langle \hat{k}_{1}^{B}\right\rangle \frac{\left\Vert \bar{\Psi}%
\right\Vert ^{2}\left\langle \bar{K}\right\rangle }{\left\Vert \hat{\Psi}%
\right\Vert ^{2}\left\langle \hat{K}\right\rangle }\right) \right) \left(
1-\left( \left\langle \underline{\hat{k}}\right\rangle +\left( \left\langle 
\underline{\hat{k}}_{1}^{B}\right\rangle +\kappa \left\langle \left[ \frac{%
\underline{\hat{k}}_{2}^{B}}{1+\bar{k}}\right] \right\rangle \right) \frac{%
\left\Vert \bar{\Psi}\right\Vert ^{2}\left\langle \bar{K}\right\rangle }{%
\left\Vert \hat{\Psi}\right\Vert ^{2}\left\langle \hat{K}\right\rangle }%
\right) \right) \left\Vert \hat{\Psi}\right\Vert ^{2}}
\end{equation*}%
and:%
\begin{eqnarray*}
&&\frac{\delta \hat{g}\left( \hat{K},\hat{X}\right) }{\delta \left\vert \bar{%
\Psi}\left( \bar{K},\bar{X}\right) \right\vert ^{2}}=-\left( \left\langle 
\hat{g}\right\rangle +\left( 1-\hat{M}\right) ^{-1}\bar{N}\left\langle \bar{g%
}\right\rangle \right) \frac{\bar{K}}{\left\langle \hat{K}\right\rangle } \\
&&\times \frac{\kappa \left\langle \left[ \frac{\underline{\hat{k}}_{2}^{B}}{%
1+\bar{k}}\right] \right\rangle \left( 1-\left\langle \hat{k}\right\rangle
\right) +\left\langle \hat{k}_{1}^{B}\right\rangle \left\langle \hat{k}%
_{2}\right\rangle }{\left( 1-\left( \left\langle \hat{k}\right\rangle
+\left( \left\langle \hat{k}_{1}^{B}\right\rangle +\kappa \left\langle \left[
\frac{\underline{\hat{k}}_{2}^{B}}{1+\bar{k}}\right] \right\rangle \right) 
\frac{\left\Vert \bar{\Psi}\right\Vert ^{2}\left\langle \bar{K}\right\rangle 
}{\left\Vert \hat{\Psi}\right\Vert ^{2}\left\langle \hat{K}\right\rangle }%
\right) \right) \left( 1-\left( \left\langle \hat{k}_{1}\right\rangle
+\left\langle \hat{k}_{1}^{B}\right\rangle \frac{\left\Vert \bar{\Psi}%
\right\Vert ^{2}\left\langle \bar{K}\right\rangle }{\left\Vert \hat{\Psi}%
\right\Vert ^{2}\left\langle \hat{K}\right\rangle }\right) \right)
\left\Vert \hat{\Psi}\right\Vert ^{2}}
\end{eqnarray*}%
with:%
\begin{equation*}
\bar{N}\rightarrow \left\langle \hat{k}_{1}^{B}\right\rangle +\kappa \frac{%
\left\langle \hat{k}_{2}^{B}\right\rangle }{1+\left\langle \bar{k}%
\right\rangle }\left( 1-\frac{\left\langle \bar{k}\right\rangle }{\left(
1+\left\langle \bar{k}\right\rangle \right) ^{2}}\right)
\end{equation*}%
These coefficients can be written in a more compact form as:%
\begin{equation*}
\frac{\delta \hat{g}\left( \hat{K},\hat{X}\right) }{\delta \left\vert \hat{%
\Psi}\left( \hat{K}_{1},\hat{X}_{1}\right) \right\vert ^{2}}=-\left\langle 
\hat{g}^{ef}\right\rangle \frac{\hat{K}_{1}}{\left\langle \hat{K}%
\right\rangle \left\Vert \hat{\Psi}\right\Vert ^{2}}
\end{equation*}%
and:%
\begin{equation*}
\frac{\delta }{\delta \left\vert \bar{\Psi}\left( \bar{K},\bar{X}\right)
\right\vert ^{2}}\hat{g}\left( \hat{K},\hat{X}\right) =-\left\langle \hat{g}%
^{Bef}\right\rangle \frac{\bar{K}}{\left\langle \hat{K}\right\rangle
\left\Vert \hat{\Psi}\right\Vert ^{2}}
\end{equation*}%
Using these formula, the minimization equations for investors become:%
\begin{eqnarray*}
0 &=&\frac{\hat{K}_{1}^{2}\hat{g}^{2}\left( \hat{K}_{1},\hat{X}_{1}\right) }{%
2\sigma _{\hat{K}}^{2}}+\frac{\hat{g}\left( \hat{K}_{1},\hat{X}_{1}\right) }{%
2} \\
&&-\int \left\vert \hat{\Psi}\left( \hat{K},\hat{X}\right) \right\vert
^{2}\left( \frac{\hat{K}^{2}\hat{g}\left( \hat{K},\hat{X},\Psi ,\hat{\Psi}%
\right) }{\sigma _{\hat{K}}^{2}}+\frac{1}{2}\right) \left\langle \hat{g}%
^{ef}\right\rangle \frac{\hat{K}_{1}}{\left\langle \hat{K}\right\rangle
\left\Vert \hat{\Psi}\right\Vert ^{2}} \\
&&+\frac{1}{\hat{\mu}}\left( \left\vert \hat{\Psi}\left( \hat{K},\hat{X}%
\right) \right\vert ^{2}-\left\vert \hat{\Psi}_{0}\left( \hat{X}\right)
\right\vert ^{2}\right)
\end{eqnarray*}%
or, which is equivalent:

\begin{eqnarray}
0 &=&\frac{\hat{K}_{1}^{2}\hat{g}^{2}\left( \hat{K}_{1},\hat{X}_{1}\right) }{%
2\sigma _{\hat{K}}^{2}}+\frac{\hat{g}\left( \hat{K}_{1},\hat{X}_{1}\right) }{%
2} \\
&&-\left( \frac{\left\langle \hat{K}\right\rangle ^{2}\left\langle \hat{g}%
\right\rangle }{\sigma _{\hat{K}}^{2}}+\frac{1}{2}\right) \left\langle \hat{g%
}^{ef}\right\rangle \frac{\hat{K}_{1}}{\left\langle \hat{K}\right\rangle }+%
\frac{1}{\hat{\mu}}\left( \left\vert \hat{\Psi}\left( \hat{K},\hat{X}\right)
\right\vert ^{2}-\left\vert \hat{\Psi}_{0}\left( \hat{X}\right) \right\vert
^{2}\right)  \notag
\end{eqnarray}%
Similarly, the equation for the banks' field $\bar{\Psi}\left( \bar{K}_{1},%
\bar{X}_{1}\right) $ becomes: 
\begin{eqnarray*}
0 &=&\left( \frac{\bar{K}_{1}^{2}\bar{g}^{2}\left( \bar{K}_{1},\bar{X}%
_{1}\right) }{\sigma _{\hat{K}}^{2}}+\frac{\bar{g}\left( \bar{K}_{1},\bar{X}%
_{1}\right) }{2}\right) \\
&&-\left( \frac{\left\langle \hat{K}\right\rangle ^{2}\hat{g}\left( \hat{K},%
\hat{X}\right) }{\sigma _{\hat{K}}^{2}}+\frac{1}{2}\right) \left\langle \hat{%
g}^{Bef}\right\rangle \frac{\bar{K}_{1}}{\left\langle \hat{K}\right\rangle }+%
\frac{1}{\hat{\mu}}\left( \left\vert \bar{\Psi}\left( \bar{K}_{1},\bar{X}%
_{1}\right) \right\vert ^{2}-\left\vert \bar{\Psi}_{0}\left( \bar{X}%
_{1}\right) \right\vert ^{2}\right)
\end{eqnarray*}

\subsection{Solving the minimization equation for the background fields}

The solutions for the background fields are:%
\begin{eqnarray*}
\left\vert \hat{\Psi}\left( \hat{K}_{1},\hat{X}_{1}\right) \right\vert ^{2}
&=&\left\Vert \hat{\Psi}_{0}\left( \hat{X}_{1}\right) \right\Vert ^{2}-\hat{%
\mu}\left\{ \left( \frac{\hat{K}_{1}^{2}\hat{g}^{2}\left( \hat{X}_{1}\right) 
}{2\sigma _{\hat{K}}^{2}}+\frac{\hat{g}\left( \hat{X}_{1}\right) }{2}\right)
\right. \\
&&\left. +\left( \frac{\left\langle \bar{K}\right\rangle ^{2}\left\langle 
\bar{g}\right\rangle }{\sigma _{\hat{K}}^{2}}+\frac{1}{2}\right) \Delta
\left( \hat{k}^{B}\left( \hat{X}_{1},\left\langle \bar{X}\right\rangle
\right) A\right) \frac{\left\Vert \bar{\Psi}\right\Vert ^{2}\hat{K}_{1}}{%
\left\Vert \hat{\Psi}\right\Vert ^{2}\left\langle \hat{K}\right\rangle }%
-\left( \frac{\left\langle \hat{K}\right\rangle ^{2}\left\langle \hat{g}%
\right\rangle }{\sigma _{\hat{K}}^{2}}+\frac{1}{2}\right) \left\langle \hat{g%
}^{ef}\right\rangle \frac{\hat{K}_{1}}{\left\langle \hat{K}\right\rangle }%
\right\}
\end{eqnarray*}%
\begin{eqnarray*}
\left\vert \bar{\Psi}\left( \bar{K}_{1},\bar{X}_{1}\right) \right\vert ^{2}
&=&\left\vert \bar{\Psi}_{0}\left( \bar{X}_{1}\right) \right\vert ^{2}-\hat{%
\mu}\left\{ \left( \frac{\bar{K}_{1}^{2}\bar{g}^{2}\left( \bar{X}_{1}\right) 
}{\sigma _{\hat{K}}^{2}}+\frac{\bar{g}\left( \bar{X}_{1}\right) }{2}\right)
\right. \\
&&+\left( \frac{\left\langle \bar{K}\right\rangle ^{2}\left\langle \bar{g}%
\right\rangle }{\sigma _{\hat{K}}^{2}}+\frac{1}{2}\right) \left(
\left\langle \underline{\hat{k}}^{B}\right\rangle ^{ef}\Delta \left( \hat{k}%
^{B}\left( \hat{X}_{1},\left\langle \bar{X}\right\rangle \right) A\right) +%
\frac{\Delta \bar{k}_{2}\left( \left\langle \bar{X}\right\rangle ,\bar{X}%
\right) }{\left( 1-\left\langle \bar{k}_{1}\right\rangle \right) \left\Vert 
\bar{\Psi}\right\Vert ^{2}\left\langle \bar{K}\right\rangle }\left\langle 
\bar{g}\right\rangle \right) \frac{\left\Vert \bar{\Psi}\right\Vert ^{2}\bar{%
K}_{1}}{\left\Vert \hat{\Psi}\right\Vert ^{2}\left\langle \hat{K}%
\right\rangle } \\
&&\left. -\left( \frac{\left\langle \hat{K}\right\rangle ^{2}\left\langle 
\hat{g}\right\rangle }{\sigma _{\hat{K}}^{2}}+\frac{1}{2}\right)
\left\langle \hat{g}^{Bef}\right\rangle \frac{\bar{K}_{1}}{\left\langle \hat{%
K}\right\rangle }\right\}
\end{eqnarray*}%
where, for any quantity $Q$, the notation $\Delta \left( Q\left( \hat{X}%
\right) \right) $ stands for its deviation from the average: 
\begin{equation*}
\Delta \left( Q\left( \hat{X}\right) \right) =\left( Q\left( \hat{X}\right)
\right) -\left\langle \Delta \left( Q\left( \hat{X}\right) \right)
\right\rangle
\end{equation*}%
\bigskip

\subsection{Global averages for field and capital}

\subsubsection{Investors}

Appendix 20.5 computes the averages of equations (\ref{KX1}) and (\ref{PS1}%
), obtaining the average disposable capital per investor across the entire
system, denoted as $\left\langle \hat{K}\right\rangle $:%
\begin{equation}
\left\langle \hat{K}\right\rangle ^{2}\simeq \frac{1}{6}\frac{\sigma _{\hat{K%
}}^{2}}{\hat{\mu}}\frac{\left( \left( 2+\frac{\left\langle \hat{g}%
^{ef}\right\rangle }{\left\langle \hat{g}\right\rangle }-\sqrt{\left( 1-%
\frac{\left\langle \hat{g}^{ef}\right\rangle }{\left\langle \hat{g}%
\right\rangle }\right) \left( 4-\frac{\left\langle \hat{g}^{ef}\right\rangle 
}{\left\langle \hat{g}\right\rangle }\right) }\right) \right) ^{2}\left\Vert 
\hat{\Psi}_{0}\right\Vert ^{2}}{\left\langle \hat{g}^{ef}\right\rangle
^{2}\left( 5+\frac{\left\langle \hat{g}^{ef}\right\rangle }{\left\langle 
\hat{g}\right\rangle }-\sqrt{\left( 1-\frac{\left\langle \hat{g}%
^{ef}\right\rangle }{\left\langle \hat{g}\right\rangle }\right) \left( 4-%
\frac{\left\langle \hat{g}^{ef}\right\rangle }{\left\langle \hat{g}%
\right\rangle }\right) }\right) }
\end{equation}%
and the total disposable capital for investors in the system, denoted as $%
\left\langle \hat{K}\right\rangle \left\Vert \hat{\Psi}\right\Vert ^{2}$:%
\begin{eqnarray}
\left\langle \hat{K}\right\rangle \left\Vert \hat{\Psi}\right\Vert ^{2} &=&%
\frac{2\sigma _{\hat{K}}^{2}}{\hat{\mu}}V\left( \frac{1}{4}-\frac{1}{18}%
\left( 2+\frac{\left\langle \hat{g}^{ef}\right\rangle }{\left\langle \hat{g}%
\right\rangle }-\sqrt{\left( 1-\frac{\left\langle \hat{g}^{ef}\right\rangle 
}{\left\langle \hat{g}\right\rangle }\right) \left( 4-\frac{\left\langle 
\hat{g}^{ef}\right\rangle }{\left\langle \hat{g}\right\rangle }\right) }%
\right) \right) \left\langle \hat{g}\right\rangle ^{2}  \label{VKH} \\
&&\times \left( \frac{\left\Vert \bar{\Psi}_{0}\right\Vert ^{2}-\frac{\hat{%
\mu}}{6}\left( \frac{\left\langle \hat{K}\right\rangle ^{2}\left\langle \hat{%
g}\right\rangle }{\sigma _{\hat{K}}^{2}}\right) \left\langle \hat{g}%
^{Bef}\right\rangle \left( 2+\frac{\left\langle \hat{g}^{ef}\right\rangle }{%
\left\langle \hat{g}\right\rangle }-\sqrt{\left( 1-\frac{\left\langle \hat{g}%
^{ef}\right\rangle }{\left\langle \hat{g}\right\rangle }\right) \left( 4-%
\frac{\left\langle \hat{g}^{ef}\right\rangle }{\left\langle \hat{g}%
\right\rangle }\right) }\right) }{\left\langle \bar{g}\right\rangle ^{2}}%
\right) ^{2}  \notag
\end{eqnarray}

The above formulas do not close the model, as the returns $\left\langle \hat{%
g}\right\rangle $ and $\left\langle \hat{g}^{ef}\right\rangle $ are
themselves endogenous, and depend on the average capital. Below, we will
derive the equation governing these returns, which will enable us to solve
for both capital and returns.\ 

\subsubsection{Banks}

Computing the average of equation (\ref{BH}) yields the average disposable
capital per bank in the entire system, denoted as $\left\langle \bar{K}%
\right\rangle $:%
\begin{equation*}
\left\langle \bar{K}\right\rangle \simeq 18\frac{\sigma _{\hat{K}}^{2}}{\hat{%
\mu}}\left( \frac{\sqrt{\left\langle \hat{g}\right\rangle ^{2}\left( 1+\frac{%
3}{4}\frac{\left\langle \hat{g}^{ef}\right\rangle }{\left\langle \hat{g}%
\right\rangle }\right) }\left\Vert \bar{\Psi}_{0}\right\Vert -\frac{3}{8}%
\left\langle \hat{g}\right\rangle \frac{\hat{g}^{Bef}\left( \bar{X}%
_{1}\right) }{\bar{g}\left( \bar{X}_{1}\right) }\left\Vert \hat{\Psi}%
_{0}\right\Vert }{\sqrt{5+\frac{\left\langle \hat{g}^{ef}\right\rangle }{%
\left\langle \hat{g}\right\rangle }-\sqrt{\left( 1-\frac{\left\langle \hat{g}%
^{ef}\right\rangle }{\left\langle \hat{g}\right\rangle }\right) \left( 4-%
\frac{\left\langle \hat{g}^{ef}\right\rangle }{\left\langle \hat{g}%
\right\rangle }\right) }}}\right) ^{4}\left( \frac{1}{4\left\langle \bar{g}%
\right\rangle ^{2}}-\frac{\left\langle \hat{g}\right\rangle \left\langle 
\hat{g}^{Bef}\right\rangle }{3\left\langle \bar{g}\right\rangle ^{4}}\right)
\end{equation*}%
and its associated field:%
\begin{equation*}
\left\Vert \bar{\Psi}\right\Vert ^{2}\simeq 18\frac{\sigma _{\hat{K}}^{2}}{%
\hat{\mu}}\left( \frac{\sqrt{\left\langle \hat{g}\right\rangle ^{2}\left( 1+%
\frac{3}{4}\frac{\left\langle \hat{g}^{ef}\right\rangle }{\left\langle \hat{g%
}\right\rangle }\right) }\left\Vert \bar{\Psi}_{0}\right\Vert -\frac{3}{8}%
\left\langle \hat{g}\right\rangle \frac{\hat{g}^{Bef}\left( \bar{X}%
_{1}\right) }{\bar{g}\left( \bar{X}_{1}\right) }\left\Vert \hat{\Psi}%
_{0}\right\Vert }{\sqrt{5+\frac{\left\langle \hat{g}^{ef}\right\rangle }{%
\left\langle \hat{g}\right\rangle }-\sqrt{\left( 1-\frac{\left\langle \hat{g}%
^{ef}\right\rangle }{\left\langle \hat{g}\right\rangle }\right) \left( 4-%
\frac{\left\langle \hat{g}^{ef}\right\rangle }{\left\langle \hat{g}%
\right\rangle }\right) }}}\right) ^{4}\left( \frac{1}{3\left\langle \bar{g}%
\right\rangle ^{2}}-\frac{\left\langle \hat{g}\right\rangle \left\langle 
\hat{g}^{Bef}\right\rangle }{2\left\langle \bar{g}\right\rangle ^{4}}\right)
\end{equation*}%
The average of (\ref{Bp}) yields the total amount of disposable capital for
banks in the system, denoted as $\left\langle \bar{K}\right\rangle
\left\Vert \bar{\Psi}\right\Vert ^{2}$:%
\begin{equation}
\left\langle \bar{K}\right\rangle \left\Vert \bar{\Psi}\right\Vert
^{2}\simeq 18\frac{\sigma _{\hat{K}}^{2}}{\left\langle \bar{g}\right\rangle
^{2}\hat{\mu}}\left( \frac{\sqrt{\left\langle \hat{g}\right\rangle
^{2}\left( 1+\frac{3}{4}\frac{\left\langle \hat{g}^{ef}\right\rangle }{%
\left\langle \hat{g}\right\rangle }\right) }\left\Vert \bar{\Psi}%
_{0}\right\Vert -\frac{3}{8}\left\langle \hat{g}\right\rangle \frac{%
\left\langle \hat{g}^{Bef}\right\rangle }{\left\langle \bar{g}\right\rangle }%
\left\Vert \hat{\Psi}_{0}\right\Vert }{\sqrt{5+\frac{\left\langle \hat{g}%
^{ef}\right\rangle }{\left\langle \hat{g}\right\rangle }-\sqrt{\left( 1-%
\frac{\left\langle \hat{g}^{ef}\right\rangle }{\left\langle \hat{g}%
\right\rangle }\right) \left( 4-\frac{\left\langle \hat{g}^{ef}\right\rangle 
}{\left\langle \hat{g}\right\rangle }\right) }}}\right) ^{4}\left( \frac{1}{4%
}-\frac{\left\langle \hat{g}\right\rangle \left\langle \hat{g}%
^{Bef}\right\rangle }{3\left\langle \bar{g}\right\rangle ^{2}}\right)
\label{BVK}
\end{equation}

\subsection{Average field and total capital per sector}

\subsubsection{Investors}

The amount of capital for investors in sector $\hat{X}_{1}$ is: 
\begin{eqnarray}
\hat{K}\left[ \hat{X}_{1}\right] &=&\hat{K}_{\hat{X}}\left\Vert \hat{\Psi}%
\left( \hat{X}_{1}\right) \right\Vert ^{2}  \label{KX1} \\
&\simeq &\frac{\hat{\mu}}{2\sigma _{\hat{K}}^{2}}\left( 6\frac{\sigma _{\hat{%
K}}^{2}}{\hat{\mu}}\frac{\left\Vert \hat{\Psi}_{0}\left( \hat{X}_{1}\right)
\right\Vert ^{2}}{\left( 5+\frac{\left\langle \hat{g}^{ef}\right\rangle }{%
\left\langle \hat{g}\right\rangle }-\sqrt{\left( 1-\frac{\left\langle \hat{g}%
^{ef}\right\rangle }{\left\langle \hat{g}\right\rangle }\right) \left( 4-%
\frac{\left\langle \hat{g}^{ef}\right\rangle }{\left\langle \hat{g}%
\right\rangle }\right) }\right) }\right) ^{2}\left( \frac{1}{4\hat{g}%
^{2}\left( \bar{X}_{1}\right) }-\frac{\left\langle \hat{g}\right\rangle \hat{%
g}^{ef}\left( \hat{X}_{1}\right) }{3\hat{g}^{4}\left( \bar{X}_{1}\right) }%
\right)  \notag
\end{eqnarray}%
and its associated field is:%
\begin{equation}
\left\Vert \hat{\Psi}\left( \hat{X}_{1}\right) \right\Vert ^{2}=\frac{\hat{%
\mu}}{2\sigma _{\hat{K}}^{2}}\left( 6\frac{\sigma _{\hat{K}}^{2}}{\hat{\mu}}%
\frac{\left\Vert \hat{\Psi}_{0}\left( \hat{X}_{1}\right) \right\Vert ^{2}}{%
\left( 5+\frac{\left\langle \hat{g}^{ef}\right\rangle }{\left\langle \hat{g}%
\right\rangle }-\sqrt{\left( 1-\frac{\left\langle \hat{g}^{ef}\right\rangle 
}{\left\langle \hat{g}\right\rangle }\right) \left( 4-\frac{\left\langle 
\hat{g}^{ef}\right\rangle }{\left\langle \hat{g}\right\rangle }\right) }%
\right) }\right) ^{\frac{3}{2}}\left( \frac{1}{3\hat{g}^{2}\left( \bar{X}%
_{1}\right) }-\frac{\left\langle \hat{g}\right\rangle \hat{g}^{ef}\left( 
\hat{X}_{1}\right) }{2\hat{g}^{4}\left( \bar{X}_{1}\right) }\right)
\label{PS1}
\end{equation}%
\ 

\subsubsection{Banks}

The amount of capital for banks in sector $\bar{X}_{1}$ is given by: 
\begin{subequations}
\begin{equation}
\bar{K}\left[ \bar{X}_{1}\right] \simeq 18\frac{\sigma _{\hat{K}}^{2}}{\hat{%
\mu}}\left( \frac{\sqrt{\left\langle \hat{g}\right\rangle ^{2}\left( 1+\frac{%
3}{4}\frac{\left\langle \hat{g}^{ef}\right\rangle }{\left\langle \hat{g}%
\right\rangle }\right) }\left\Vert \bar{\Psi}_{0}\left( \hat{X}_{1}\right)
\right\Vert -\frac{3}{8}\left\langle \hat{g}\right\rangle \frac{\hat{g}%
^{Bef}\left( \bar{X}_{1}\right) }{\bar{g}\left( \bar{X}_{1}\right) }%
\left\Vert \hat{\Psi}_{0}\left( \bar{X}_{1}\right) \right\Vert }{\sqrt{5+%
\frac{\left\langle \hat{g}^{ef}\right\rangle }{\left\langle \hat{g}%
\right\rangle }-\sqrt{\left( 1-\frac{\left\langle \hat{g}^{ef}\right\rangle 
}{\left\langle \hat{g}\right\rangle }\right) \left( 4-\frac{\left\langle 
\hat{g}^{ef}\right\rangle }{\left\langle \hat{g}\right\rangle }\right) }}}%
\right) ^{4}\left( \frac{1}{4\bar{g}^{2}\left( \bar{X}_{1}\right) }-\frac{%
\left\langle \hat{g}\right\rangle \hat{g}^{Bef}\left( \bar{X}_{1}\right) }{3%
\bar{g}^{4}\left( \bar{X}_{1}\right) }\right)  \label{BH}
\end{equation}%
along with its associated field: 
\end{subequations}
\begin{equation}
\left\Vert \bar{\Psi}\left( \bar{X}_{1}\right) \right\Vert ^{2}\simeq 18%
\frac{\sigma _{\hat{K}}^{2}}{\hat{\mu}}\left( \frac{\sqrt{\left\langle \hat{g%
}\right\rangle ^{2}\left( 1+\frac{3}{4}\frac{\left\langle \hat{g}%
^{ef}\right\rangle }{\left\langle \hat{g}\right\rangle }\right) }\left\Vert 
\bar{\Psi}_{0}\left( \hat{X}_{1}\right) \right\Vert -\frac{3}{8}\left\langle 
\hat{g}\right\rangle \frac{\hat{g}^{Bef}\left( \bar{X}_{1}\right) }{\bar{g}%
\left( \bar{X}_{1}\right) }\left\Vert \hat{\Psi}_{0}\left( \bar{X}%
_{1}\right) \right\Vert }{\sqrt{5+\frac{\left\langle \hat{g}%
^{ef}\right\rangle }{\left\langle \hat{g}\right\rangle }-\sqrt{\left( 1-%
\frac{\left\langle \hat{g}^{ef}\right\rangle }{\left\langle \hat{g}%
\right\rangle }\right) \left( 4-\frac{\left\langle \hat{g}^{ef}\right\rangle 
}{\left\langle \hat{g}\right\rangle }\right) }}}\right) ^{4}\left( \frac{1}{3%
\bar{g}^{2}\left( \bar{X}_{1}\right) }-\frac{\left\langle \hat{g}%
\right\rangle \hat{g}^{Bef}\left( \bar{X}_{1}\right) }{2\bar{g}^{4}\left( 
\bar{X}_{1}\right) }\right)  \label{Bp}
\end{equation}

\subsection{Investors' and banks' returns per sector}

Using the formula for the total amount of capital per sector, as found in
Appendix 21, we can find the investors and banks' returns per sector as
functions of fields.

\subsubsection{Computation of $\hat{g}$}

The return per sector for investors is given by:%
\begin{equation*}
\hat{g}\left( \hat{X}_{1}\right) \simeq \frac{\left\Vert \hat{\Psi}%
_{0}\left( \hat{X}_{1}\right) \right\Vert ^{2}}{\left( 5+\frac{\left\langle 
\hat{g}^{ef}\right\rangle }{\left\langle \hat{g}\right\rangle }-\sqrt{\left(
1-\frac{\left\langle \hat{g}^{ef}\right\rangle }{\left\langle \hat{g}%
\right\rangle }\right) \left( 4-\frac{\left\langle \hat{g}^{ef}\right\rangle 
}{\left\langle \hat{g}\right\rangle }\right) }\right) \sqrt{\frac{2\hat{\mu}%
}{9\sigma _{\hat{K}}^{2}}\hat{K}\left[ \hat{X}_{1}\right] +\frac{6\hat{g}%
^{ef}\left( \hat{X}_{1}\right) }{\left\langle \hat{g}\right\rangle ^{3}}%
\left( \frac{\left\Vert \hat{\Psi}_{0}\left( \hat{X}_{1}\right) \right\Vert
^{2}}{\left( 5+\frac{\left\langle \hat{g}^{ef}\right\rangle }{\left\langle 
\hat{g}\right\rangle }-\sqrt{\left( 1-\frac{\left\langle \hat{g}%
^{ef}\right\rangle }{\left\langle \hat{g}\right\rangle }\right) \left( 4-%
\frac{\left\langle \hat{g}^{ef}\right\rangle }{\left\langle \hat{g}%
\right\rangle }\right) }\right) }\right) ^{2}}}
\end{equation*}%
This formula will be used alongside the return equation to derive an
equation linking the average capital across different sectors.

\subsubsection{Computation of $\bar{g}$}

Similarly, we can express the return per sector for banks:%
\begin{equation*}
\bar{g}\left( \bar{X}_{1}\right) \simeq \frac{\left( \sqrt{\left\langle \hat{%
g}\right\rangle ^{2}\left( 1+\frac{3}{4}\frac{\left\langle \hat{g}%
^{ef}\right\rangle }{\left\langle \hat{g}\right\rangle }\right) }\left\Vert 
\bar{\Psi}_{0}\left( \hat{X}_{1}\right) \right\Vert -\frac{3}{8}\left\langle 
\hat{g}\right\rangle \frac{\hat{g}^{Bef}\left( \bar{X}_{1}\right) }{\bar{g}%
\left( \bar{X}_{1}\right) }\left\Vert \hat{\Psi}_{0}\left( \bar{X}%
_{1}\right) \right\Vert \right) ^{2}}{\left( 5+\frac{\left\langle \hat{g}%
^{ef}\right\rangle }{\left\langle \hat{g}\right\rangle }-\sqrt{\left( 1-%
\frac{\left\langle \hat{g}^{ef}\right\rangle }{\left\langle \hat{g}%
\right\rangle }\right) \left( 4-\frac{\left\langle \hat{g}^{ef}\right\rangle 
}{\left\langle \hat{g}\right\rangle }\right) }\right) ^{2}\sqrt{\frac{2\hat{%
\mu}}{9\sigma _{\hat{K}}^{2}}\hat{K}\left[ \hat{X}_{1}\right] +\frac{%
4\left\langle \hat{g}\right\rangle \hat{g}^{ef}\left( \hat{X}_{1}\right) }{%
3\left\langle \hat{g}\right\rangle ^{4}}\left( \frac{\left\Vert \hat{\Psi}%
_{0}\left( \hat{X}_{1}\right) \right\Vert ^{2}}{\left( 5+\frac{\left\langle 
\hat{g}^{ef}\right\rangle }{\left\langle \hat{g}\right\rangle }-\sqrt{\left(
1-\frac{\left\langle \hat{g}^{ef}\right\rangle }{\left\langle \hat{g}%
\right\rangle }\right) \left( 4-\frac{\left\langle \hat{g}^{ef}\right\rangle 
}{\left\langle \hat{g}\right\rangle }\right) }\right) }\right) ^{2}}}
\end{equation*}

\section{Investors' and banks' returns under a no-default scenario}

\subsection{Investors' returns}

\subsubsection{Derivation of the investor returns' equation under constant
return to scale}

The formula for investor returns equation in term of $\hat{g}\left( \hat{X}%
\right) $ is similar to that in Part $1$:%
\begin{eqnarray}
&&\left( \Delta \left( \hat{X},\hat{X}^{\prime }\right) -\frac{\hat{K}%
^{\prime }\hat{k}_{1}\left( \hat{X}^{\prime },\hat{X}\right) \left\vert \hat{%
\Psi}\left( \hat{K}^{\prime },\hat{X}^{\prime }\right) \right\vert ^{2}}{1+%
\underline{\hat{k}}\left( \hat{X}^{\prime }\right) +\underline{\hat{k}}%
_{1}^{B}\left( \bar{X}^{\prime }\right) +\kappa \left[ \frac{\underline{\hat{%
k}}_{2}^{B}}{1+\bar{k}}\right] \left( \hat{X}^{\prime }\right) }\right) 
\frac{\left( 1-\hat{M}\right) \hat{g}\left( \hat{X}^{\prime }\right) }{1+%
\underline{\hat{k}}_{2}\left( \hat{X}^{\prime }\right) +\kappa \left[ \frac{%
\underline{\hat{k}}_{2}^{B}}{1+\bar{k}}\right] \left( \hat{X}^{\prime
}\right) }  \label{RTn} \\
&=&\frac{k_{1}\left( \hat{X}^{\prime },X\right) \hat{K}^{\prime }}{1+%
\underline{k}\left( \hat{X}^{\prime }\right) +\underline{k}_{1}^{\left(
B\right) }\left( \bar{X}^{\prime }\right) +\kappa \left[ \frac{\underline{k}%
_{2}^{B}}{1+\bar{k}}\right] \left( X^{\prime }\right) }\frac{f_{1}^{\prime
}\left( \hat{K},\hat{X},\Psi ,\hat{\Psi}\right) }{1+\underline{k}_{2}\left( 
\hat{X}^{\prime }\right) +\kappa \left[ \frac{\underline{k}_{2}^{B}}{1+\bar{k%
}}\right] \left( X^{\prime }\right) }  \notag
\end{eqnarray}

\subsubsection{Solution of the investor returns' equation}

Applying the resolution procedure outlined in Part $1$, Appendix 21 provides
the following solution:%
\begin{eqnarray}
\hat{g}\left( \hat{X}_{1}\right) \left( \hat{X}_{1}\right) -\bar{r}^{\prime
} &=&\int \left( 1-\left( 1+\left[ \underline{\hat{k}}_{2}^{n}\left( \hat{X}%
_{1}\right) \right] _{\kappa }\right) \hat{S}_{1}^{E}\left( \hat{X}^{\prime
},\hat{X}_{1}\right) \right) ^{-1}  \label{rnt} \\
&&\times \left( 1+\left[ \underline{\hat{k}}_{2}^{n}\left( \hat{X}%
_{1}\right) \right] _{\kappa }\right) \left( \frac{A\left( \hat{X}^{\prime
}\right) }{f_{1}^{2}\left( X^{\prime }\right) }+\frac{B\left( \hat{X}%
^{\prime }\right) }{f_{1}^{3}\left( X^{\prime }\right) }\right) \left(
R+\Delta F_{\tau }\left( \bar{R}\left( K,X\right) \right) \right)  \notag
\end{eqnarray}%
with:%
\begin{eqnarray}
A\left( \hat{X}^{\prime }\right) &=&\frac{\epsilon \left( 1-\beta \right)
\delta C\left( X+C\frac{\beta \delta +\beta ^{B}}{1+\delta }\right) ^{2}}{%
3\left( 1+\delta \right) \sigma _{\hat{K}}^{2}}\frac{\left( 3X-\frac{\beta
\delta +\beta ^{B}}{1+\delta }C\right) \left( \frac{\beta \delta +\beta ^{B}%
}{1+\delta }C+X\right) }{4C\left( 2X-C\frac{\beta \delta +\beta ^{B}}{%
1+\delta }\right) }  \label{DFAB} \\
B\left( \hat{X}^{\prime }\right) &=&\frac{\epsilon \left( 1-\beta \right)
\delta C\left( X+C\frac{\beta \delta +\beta ^{B}}{1+\delta }\right) ^{2}}{%
3\left( 1+\delta \right) \sigma _{\hat{K}}^{2}}\frac{\left( 3X-\frac{\beta
\delta +\beta ^{B}}{1+\delta }C\right) \left( \frac{\beta \delta +\beta ^{B}%
}{1+\delta }C+X\right) }{4C\left( 2X-C\frac{\beta \delta +\beta ^{B}}{%
1+\delta }\right) }\frac{\left( \beta \delta +\beta ^{B}\right) }{\left(
1+\delta \right) f_{1}^{3}\left( X\right) }  \notag
\end{eqnarray}%
\begin{equation*}
R=f_{1}\left( X^{\prime }\right) -\bar{r}
\end{equation*}%
and where the matrix $\hat{S}_{1}^{E}\left( \hat{X}^{\prime },\hat{X}%
_{1}\right) $ is estimated as:%
\begin{equation}
\hat{S}_{1}^{E}\left( \hat{X}^{\prime },\hat{X}_{1}\right) =\frac{\left( 1+%
\left[ \underline{\hat{k}}_{2}^{n}\left( \hat{X}_{1}\right) \right] _{\kappa
}\right) }{1+\left[ \underline{\hat{k}}\left( \hat{X}_{1}\right) \right]
_{\kappa }}\left( \frac{\hat{k}\left( \hat{X},\hat{X}^{\prime }\right) -%
\underline{\hat{k}}_{1}\left( \left\langle X\right\rangle ,\hat{X}\right)
k\left( \left\langle X\right\rangle ,X^{\prime }\right) }{1+\left[ 
\underline{\hat{k}}_{2}^{n}\left( \hat{X}_{1}\right) \right] _{\kappa }}+%
\frac{\hat{k}_{1}\left( \hat{X}^{\prime },\hat{X}\right) }{1+\left[ 
\underline{\hat{k}}_{2}^{n}\left( \hat{X}^{\prime }\right) \right] _{\kappa }%
}\right) \frac{\left( \left\Vert \hat{\Psi}_{0}\left( \hat{X}^{\prime
}\right) \right\Vert ^{2}-\hat{\mu}D\left( \hat{X}^{\prime }\right) \right)
^{2}}{\left( \left\Vert \hat{\Psi}_{0}\right\Vert ^{2}-\hat{\mu}\left\langle
D\right\rangle \right) ^{2}}  \label{DFT}
\end{equation}

\subsubsection{Interpretation and diffusion matrix}

Equation (\ref{rnt}) is similar to that derived in Part $1$, and the
interpretation is the same: under constant return to scale, the solution is
unique. The only difference arises from.the fact that the coefficient $\left[
\underline{\hat{k}}_{2}^{n}\left( \hat{X}_{1}\right) \right] _{\kappa }$
defined by:%
\begin{equation*}
1+\left[ \underline{\hat{k}}_{2}^{n}\left( \hat{X}_{1}\right) \right]
_{\kappa }=1+\frac{\hat{k}_{2}\left( \hat{X}^{\prime },\left\langle \hat{X}%
\right\rangle \right) +\kappa \left[ \frac{\hat{k}_{2}^{B}\left( \hat{X}%
^{\prime },\left\langle \bar{X}\right\rangle \right) }{1+\bar{k}}\right] 
\frac{\left\Vert \bar{\Psi}\right\Vert ^{2}\left\langle \bar{K}\right\rangle 
}{\left\Vert \hat{\Psi}\right\Vert ^{2}\left\langle \hat{K}\right\rangle }}{%
1-\left\langle \hat{k}^{\Sigma }\right\rangle }
\end{equation*}%
where:%
\begin{equation*}
\left\langle \hat{k}^{\Sigma }\right\rangle =\left( \left\langle \hat{k}%
\right\rangle +\left( \left\langle \hat{k}_{1}^{B}\right\rangle +\kappa
\left\langle \left[ \frac{\underline{\hat{k}}_{2}^{B}}{1+\bar{k}}\right]
\right\rangle \right) \frac{\left\Vert \bar{\Psi}\right\Vert
^{2}\left\langle \bar{K}\right\rangle }{\left\Vert \hat{\Psi}\right\Vert
^{2}\left\langle \hat{K}\right\rangle }\right)
\end{equation*}%
depends on the ratio of total disposable capital owned by banks to the total
disposable capital owned by investors. The higher the ratio, the higher the
expression:%
\begin{equation*}
\left( 1+\left[ \underline{\hat{k}}_{2}^{n}\left( \hat{X}_{1}\right) \right]
_{\kappa }\right) \hat{S}_{1}^{E}\left( \hat{X}^{\prime },\hat{X}_{1}\right)
\end{equation*}%
which indicates that diffusion is higher when the ratio is higher.
Therefore, although the form of the diffusion matrix $\hat{S}_{1}^{E}\left( 
\hat{X}^{\prime },\hat{X}_{1}\right) $ defined in (\ref{DFT}) is similar to
Part $1$, the coefficients are increased by banks that favour the leverage
effect.

\subsubsection{Corrections due to decreasing returns to scale}

As explained before, for slowly decreasing returns to scale, we can replace $%
f_{1}\left( X\right) $ in (\ref{rnt}) by:%
\begin{eqnarray*}
&&f_{1}\left( X^{\prime },\hat{K}\left[ \hat{X}^{\prime }\right] ,\bar{K}%
\left[ \bar{X}^{\prime }\right] \right) \\
&\equiv &\frac{f_{1}\left( X\right) }{\left( 1+\frac{k\left( X\right) }{K%
\left[ X\right] }\hat{K}\left[ X\right] +\frac{k_{1}^{B}\left( X\right) }{K%
\left[ X\right] }\bar{K}\left[ \bar{X}\right] +\frac{\kappa \left[ \frac{%
\underline{k}_{2}^{B}}{1+\bar{k}}\right] }{K\left[ X\right] }\bar{K}\left[ 
\bar{X}\right] \right) ^{r}K_{X}^{r}}-C_{0} \\
&\simeq &\frac{f_{1}\left( X\right) }{\left( \left( 1+k\left( X\right) \hat{K%
}\left[ \hat{X}\right] \right) +\left( k_{1}^{B}\left( \bar{X}\right)
+\kappa \left[ \frac{\underline{k}_{2}^{B}\left( \bar{X}\right) }{1+\bar{k}}%
\right] \right) \bar{K}\left[ \bar{X}\right] \right) ^{r}}-C_{0}
\end{eqnarray*}%
As in Part 1, accounting for decreasing returns to scale results in a
multiplicity of collective states. This will be discussed further when
considering the equations for capital per sector.

\subsection{Banks' returns}

\subsubsection{Derivation of bank returns' equation and solutions}

Appendix 22 shows that the equation for bank returns is as follows:%
\begin{eqnarray}
&&\bar{g}\left( \bar{X}_{1}\right) -\bar{r}^{\prime }=\left( 1-\left( 1+\bar{%
k}_{2}^{n}\left( \bar{X}_{1}\right) \right) \left( \bar{S}_{1}^{E}\left( 
\bar{X}^{\prime },\hat{X}_{1}\right) +\bar{S}_{1}^{B}\left( \bar{X}^{\prime
},\bar{X}_{1}\right) \right) \right) ^{-1}  \label{Rtb} \\
&&\times \left( \left( 1+\bar{k}_{2}^{n}\left( \bar{X}_{1}\right) \right)
\left( \frac{1-\beta ^{B}}{\left( 1-\beta \right) \delta }\left( \frac{%
\Delta \left( \hat{X},\hat{X}^{\prime }\right) }{1+\left[ \underline{\hat{k}}%
_{2}^{n}\left( \hat{X}\right) \right] _{\kappa }}-\hat{S}_{1}^{E}\left( \hat{%
X}^{\prime },\hat{X}_{1}\right) \right) +\hat{S}_{1}^{B}\left( \hat{X}%
^{\prime },\bar{X}_{1}\right) \right) \left( \bar{g}\left( \bar{X}^{\prime
}\right) -\bar{r}^{\prime }\right) \right)  \notag
\end{eqnarray}%
The coefficients are defined in the same appendix.

Using (\ref{rnt}), we derive the equivalent formulation for the banks'
return equation:%
\begin{eqnarray}
&&\bar{g}\left( \bar{X}_{1}\right) -\bar{r}^{\prime }  \label{RBT} \\
&=&\left( 1-\left( 1+\bar{k}_{2}^{n}\left( \bar{X}_{1}\right) \right) \left( 
\bar{S}_{1}^{E}\left( \bar{X}^{\prime },\hat{X}_{1}\right) +\bar{S}%
_{1}^{B}\left( \bar{X}^{\prime },\bar{X}_{1}\right) \right) \right) ^{-1} 
\notag \\
&&\times \left( 1+\bar{k}_{2}^{n}\left( \bar{X}_{1}\right) \right) \left( 
\frac{1-\beta ^{B}}{\left( 1-\beta \right) \delta }\left( \frac{\Delta
\left( \hat{X},\hat{X}^{\prime }\right) }{1+\left[ \underline{\hat{k}}%
_{2}^{n}\left( \hat{X}\right) \right] _{\kappa }}-\hat{S}_{1}^{E}\left( \hat{%
X}^{\prime },\hat{X}_{1}\right) \right) +\hat{S}_{1}^{B}\left( \hat{X}%
^{\prime },\bar{X}_{1}\right) \right)  \notag \\
&&\times \left( 1-\left( 1+\left[ \underline{\hat{k}}_{2}^{n}\left( \hat{X}%
_{1}\right) \right] _{\kappa }\right) \hat{S}_{1}^{E}\left( \hat{X}^{\prime
},\hat{X}_{1}\right) \right) ^{-1}\left( 1+\left[ \underline{\hat{k}}%
_{2}^{n}\left( \hat{X}_{1}\right) \right] _{\kappa }\right) \left( \frac{%
A\left( \hat{X}^{\prime }\right) }{f_{1}^{2}\left( X^{\prime }\right) }+%
\frac{B\left( \hat{X}^{\prime }\right) }{f_{1}^{3}\left( X^{\prime }\right) }%
\right) \left( R+\Delta F_{\tau }\left( \bar{R}\left( K,X\right) \right)
\right)  \notag
\end{eqnarray}%
where $A\left( \hat{X}^{\prime }\right) $ and $B\left( \hat{X}^{\prime
}\right) $ are defined in (\ref{DFAB}).

\subsubsection{Interpretation and diffusion matrices}

Equation (\ref{Rtb}) has a structure similar to (\ref{rnt}) as it involves a
diffusion matrix $\hat{S}_{1}^{B}\left( \hat{X}^{\prime },\bar{X}_{1}\right) 
$ specific to the banks, whose interpretation is similar to the diffusion
matrix $\hat{S}_{1}^{E}\left( \hat{X}^{\prime },\hat{X}_{1}\right) $ of
investors. However (\ref{Rtb}) is more complex than (\ref{rnt}) since it
directly involves $\hat{S}_{1}^{E}\left( \hat{X}^{\prime },\hat{X}%
_{1}\right) $, but also two crossed matrices $\hat{S}_{1}^{B}\left( \hat{X}%
^{\prime },\bar{X}_{1}\right) $ from banks to investors and $\bar{S}%
_{1}^{E}\left( \bar{X}^{\prime },\hat{X}_{1}\right) $ from investors to
banks. The precise formulation of these matrices are given in Appendix 22.
The matrix $\hat{S}_{1}^{B}\left( \hat{X}^{\prime },\bar{X}_{1}\right) $
measures the impact of investors' returns on banks having participations in
these investors, while $\bar{S}_{1}^{B}\left( \bar{X}^{\prime },\bar{X}%
_{1}\right) $ measures the indirect return of an investor $I$ on a bank $B$
through the banks in which $B$ has invested, that themselve invested in $I$.
The global factor $\frac{1-\beta ^{B}}{\left( 1-\beta \right) \delta }$ is
directly proportional to the share that loans take in the bank activity. If
this share is close to one, this global factor is negligible, and the bank
is quite unaffected by diffusion. It is only when banks take participations
that they are affected by diffusion.

\section{Equation for total capital of financial agents per sector}

\subsection{Derivation of the equation}

We show in appendices 20 and 21 that equations for returns (\ref{rnt}) and (%
\ref{Rtb}) can also be expressed as equations for average capital per sector
if we substitute in first approximation:%
\begin{equation}
\hat{g}\left( \hat{X}_{1}\right) \simeq \frac{\left( \left\Vert \hat{\Psi}%
_{0}\left( \hat{X}_{1}\right) \right\Vert ^{2}-\hat{\mu}D\left( \hat{X}%
_{1}\right) \right) \sqrt{1-\frac{\hat{g}^{ef}\left( \hat{X}_{1}\right) }{%
\left\langle \hat{g}\right\rangle }}}{\sqrt{\frac{2\hat{\mu}\hat{K}\left[ 
\hat{X}_{1}\right] }{\sigma _{\hat{K}}^{2}}}}\equiv \frac{\hat{N}\left( \hat{%
X}\right) }{\sqrt{\hat{K}\left[ \hat{X}\right] }}  \label{GF}
\end{equation}%
with:%
\begin{equation*}
D\left( \hat{X}_{1}\right) =\left( \frac{\left\langle \hat{K}\right\rangle
^{2}\left\langle \hat{g}\right\rangle ^{2}}{\sigma _{\hat{K}}^{2}}+\frac{%
\left\langle \hat{g}\right\rangle }{2}\right) \left( \frac{\underline{\hat{k}%
}^{ef}\left( \hat{X}_{1}\right) }{\underline{\hat{k}}_{2}^{Bef}\left(
\left\langle \hat{X}\right\rangle \right) }-\frac{6\underline{\hat{k}}%
_{2}^{Bef}}{2+\underline{\hat{k}}_{2}^{Bef}-\sqrt{\left( 2+\underline{\hat{k}%
}\right) ^{2}-\underline{\hat{k}}_{2}^{Bef}}}\right) \underline{\hat{k}}%
_{2}^{Bef}
\end{equation*}%
for investor returns, and:%
\begin{equation*}
\bar{g}\left( \bar{X}_{1}\right) =\frac{\left( \left\Vert \bar{\Psi}%
_{0}\left( \bar{X}^{\prime }\right) \right\Vert ^{2}-\hat{\mu}D\left( \bar{X}%
^{\prime }\right) \right) \sqrt{k\frac{\sigma _{\hat{K}}^{2}}{2\hat{\mu}}%
\left( 1-\frac{\hat{g}^{Bef}\left( \bar{X}^{\prime }\right) }{\left\langle 
\bar{g}\right\rangle }\right) }}{\sqrt{\underline{k}\left( \bar{X}^{\prime
}\right) }}\equiv \frac{\bar{N}\left( \bar{X}_{1}\right) }{\sqrt{\bar{K}%
\left[ \bar{X}_{1}\right] }}
\end{equation*}%
with:%
\begin{eqnarray*}
\bar{D}\left( \bar{X}_{1}\right) &=&\left( \frac{\left\langle \hat{K}%
\right\rangle ^{2}\left\langle \hat{g}\right\rangle }{\sigma _{\hat{K}}^{2}}+%
\frac{1}{2}\right) \left\langle \hat{g}^{Bef}\right\rangle \frac{\bar{K}_{0}%
}{\left\langle \hat{K}\right\rangle } \\
&&-\left( \frac{\left\langle \bar{K}\right\rangle ^{2}\left\langle \bar{g}%
\right\rangle }{\sigma _{\hat{K}}^{2}}+\frac{1}{2}\right) \left(
\left\langle \underline{\hat{k}}^{B}\right\rangle ^{ef}\Delta \left( \hat{k}%
^{B}\left( \hat{X}_{1},\left\langle \bar{X}\right\rangle \right) A\right) +%
\frac{\Delta \bar{k}_{2}\left( \left\langle \bar{X}\right\rangle ,\bar{X}%
_{1}\right) }{\left( 1-\left\langle \bar{k}_{1}\right\rangle \right)
\left\Vert \bar{\Psi}\right\Vert ^{2}\left\langle \bar{K}\right\rangle }%
\left\langle \bar{g}\right\rangle \right) Z\left\langle \bar{g}\right\rangle
\end{eqnarray*}%
for bank returns.

\subsubsection{Equation for investors' total capital per sector}

Now, considering firms with slowly decreasing returns to scale directly,
equations (\ref{rnt}) and (\ref{Rtb}) writes:%
\begin{eqnarray}
&&\pm \frac{\hat{N}\left( \hat{X}\right) }{\sqrt{\hat{K}\left[ \hat{X}\right]
}}-\bar{r}^{\prime }\simeq \int \left( 1-\left( 1+\underline{\hat{k}}%
_{2}\left( \hat{X}\right) \right) \hat{S}_{1}^{E}\left( \hat{X}^{\prime },%
\hat{X}\right) \right) ^{-1}\left( 1+\underline{\hat{k}}_{2}\left( \hat{X}%
\right) \right)  \label{KP1} \\
&&\times \left( \frac{A}{\left( f_{1}\left( X^{\prime },\hat{K}\left[ \hat{X}%
^{\prime }\right] ,\bar{K}\left[ \bar{X}^{\prime }\right] \right) \right)
^{2}}+\frac{B}{\left( f_{1}\left( X^{\prime },\hat{K}\left[ \hat{X}^{\prime }%
\right] ,\bar{K}\left[ \bar{X}^{\prime }\right] \right) \right) ^{3}}\right)
\notag \\
&&\times \left[ \left( f_{1}\left( X^{\prime },\hat{K}\left[ \hat{X}^{\prime
}\right] ,\bar{K}\left[ \bar{X}^{\prime }\right] \right) -\bar{r}^{\prime
}\right) +\tau F\left( X^{\prime }\right) \left( f_{1}\left( X^{\prime },%
\hat{K}\left[ \hat{X}^{\prime }\right] ,\bar{K}\left[ \bar{X}^{\prime }%
\right] \right) -\left\langle f_{1}\left( X^{\prime },\hat{K}\left[ \hat{X}%
^{\prime }\right] ,\bar{K}\left[ \bar{X}^{\prime }\right] \right)
\right\rangle \right) \right]  \notag
\end{eqnarray}%
with:%
\begin{eqnarray}
&&f_{1}\left( X^{\prime },\hat{K}\left[ \hat{X}^{\prime }\right] ,\bar{K}%
\left[ \bar{X}^{\prime }\right] \right)  \label{FGT} \\
&\equiv &\frac{f_{1}\left( X\right) }{\left( 1+\frac{k\left( X\right) }{K%
\left[ X\right] }\hat{K}\left[ X\right] +\frac{k_{1}^{B}\left( X\right) }{K%
\left[ X\right] }\bar{K}\left[ \bar{X}\right] +\frac{\kappa \left[ \frac{%
\underline{k}_{2}^{B}}{1+\bar{k}}\right] }{K\left[ X\right] }\bar{K}\left[ 
\bar{X}\right] \right) ^{r}K_{X}^{r}}-C_{0}  \notag \\
&\simeq &\frac{f_{1}\left( X\right) }{\left( \left( 1+k\left( X\right) \hat{K%
}\left[ \hat{X}\right] \right) +\left( k_{1}^{B}\left( \bar{X}\right)
+\kappa \left[ \frac{\underline{k}_{2}^{B}\left( \bar{X}\right) }{1+\bar{k}}%
\right] \right) \bar{K}\left[ \bar{X}\right] \right) ^{r}}-C_{0}  \notag
\end{eqnarray}%
Equation (\ref{FGT}) shows that both investors and banks' capital are
involved in the return equation.

\subsubsection{Equation for banks' total capital per sector}

The solution must be sought along with the solutions for banks' capital per
sector. This one is expressed as:%
\begin{eqnarray}
&&\pm \frac{\bar{N}\left( \bar{X}_{1}\right) }{\sqrt{\bar{K}\left[ \bar{X}%
_{1}\right] }}-\bar{r}^{\prime }  \label{CFT} \\
&=&\left( 1-\left( 1+\bar{k}_{2}^{n}\left( \bar{X}_{1}\right) \right) \left( 
\bar{S}_{1}^{E}\left( \bar{X}^{\prime },\hat{X}_{1}\right) +\bar{S}%
_{1}^{B}\left( \bar{X}^{\prime },\bar{X}_{1}\right) \right) \right) ^{-1} 
\notag \\
&&\times \left( 1+\bar{k}_{2}^{n}\left( \bar{X}_{1}\right) \right) \left( 
\frac{1-\beta ^{B}}{\left( 1-\beta \right) \delta }\left( \frac{\Delta
\left( \hat{X},\hat{X}^{\prime }\right) }{1+\left[ \underline{\hat{k}}%
_{2}^{n}\left( \hat{X}\right) \right] _{\kappa }}-\hat{S}_{1}^{E}\left( \hat{%
X}^{\prime },\hat{X}_{1}\right) \right) +\hat{S}_{1}^{B}\left( \hat{X}%
^{\prime },\bar{X}_{1}\right) \right)  \notag \\
&&\times \left( 1-\left( 1+\left[ \underline{\hat{k}}_{2}^{n}\left( \hat{X}%
_{1}\right) \right] _{\kappa }\right) \hat{S}_{1}^{E}\left( \hat{X}^{\prime
},\hat{X}_{1}\right) \right) ^{-1}\left( 1+\left[ \underline{\hat{k}}%
_{2}^{n}\left( \hat{X}_{1}\right) \right] _{\kappa }\right) \left( \frac{A}{%
\left( f_{1}\left( X^{\prime },\hat{K}\left[ \hat{X}^{\prime }\right]
\right) \right) ^{2}}+\frac{B}{\left( f_{1}\left( X^{\prime },\hat{K}\left[ 
\hat{X}^{\prime }\right] \right) \right) ^{3}}\right)  \notag \\
&&\times \left[ \left( f_{1}\left( X^{\prime },\hat{K}\left[ \hat{X}^{\prime
}\right] \right) -\bar{r}^{\prime }\right) +\tau F\left( X^{\prime }\right)
\left( f_{1}\left( X^{\prime },\hat{K}\left[ \hat{X}^{\prime }\right]
\right) -\left\langle f_{1}\left( X,\hat{K}\left[ \hat{X}\right] \right)
\right\rangle \right) \right]  \notag
\end{eqnarray}%
with relation:%
\begin{eqnarray}
&&\left( 1-\left( 1+\bar{k}_{2}^{n}\left( \bar{X}_{1}\right) \right) \left( 
\bar{S}_{1}^{E}\left( \bar{X}^{\prime },\hat{X}_{1}\right) +\bar{S}%
_{1}^{B}\left( \bar{X}^{\prime },\bar{X}_{1}\right) \right) \right) \left(
\pm \frac{\bar{N}\left( \bar{X}_{1}\right) }{\sqrt{\bar{K}\left[ \bar{X}_{1}%
\right] }}-\bar{r}^{\prime }\right)  \label{CFG} \\
&=&\left( 1+\bar{k}_{2}^{n}\left( \bar{X}_{1}\right) \right) \left( \frac{%
1-\beta ^{B}}{\left( 1-\beta \right) \delta }\left( \frac{\Delta \left( \hat{%
X},\hat{X}^{\prime }\right) }{1+\left[ \underline{\hat{k}}_{2}^{n}\left( 
\hat{X}\right) \right] _{\kappa }}-\hat{S}_{1}^{E}\left( \hat{X}^{\prime },%
\hat{X}_{1}\right) \right) +\hat{S}_{1}^{B}\left( \hat{X}^{\prime },\bar{X}%
_{1}\right) \right) \left( \pm \frac{\hat{N}\left( \hat{X}\right) }{\sqrt{%
\hat{K}\left[ \hat{X}\right] }}-\bar{r}^{\prime }\right)  \notag
\end{eqnarray}%
As in Part $1$, we search for approximate solutions as corrections to a
state where investors are not connected, thus serving as a benchmark.

\subsubsection{Interpretation}

The interpretation remains the same as in Part $1$. The interconnections
between agents generally result in multiple solutions for total capital per
sector. These solutions now involve both banks and investors. Depending on
the share of loans in bank activity, the bank may remain relatively
unaffected by diffusion. However, since the total bank capital is correlated
with the investors' level of capital, these agents also experience multiple
equilibria. We will now study approximate solutions to the capital equations.

\section{Total capital of financial agents per sector}

We will derive approximate solutions for the total capital of investors and
banks, focusing directly on firms with decreasing returns to scale
productivity.

\subsection{Isolated investors and banks}

The benchmark is directly obtained by solving:%
\begin{eqnarray}
&&\pm \frac{\hat{N}\left( \hat{X}\right) }{\sqrt{\hat{K}\left[ \hat{X}\right]
}}-\bar{r}^{\prime }\simeq \left( \frac{A}{\left( f_{1}\left( X^{\prime },%
\hat{K}\left[ \hat{X}^{\prime }\right] \right) \right) ^{2}}+\frac{B}{\left(
f_{1}\left( X^{\prime },\hat{K}\left[ \hat{X}^{\prime }\right] \right)
\right) ^{3}}\right) \\
&&\times \left[ \left( f_{1}\left( X^{\prime },\hat{K}\left[ \hat{X}^{\prime
}\right] \right) -\bar{r}^{\prime }\right) +\tau F\left( X^{\prime }\right)
\left( f_{1}\left( X^{\prime },\hat{K}\left[ \hat{X}^{\prime }\right]
\right) -\left\langle f_{1}\left( X,\hat{K}\left[ \hat{X}\right] \right)
\right\rangle \right) \right]  \notag
\end{eqnarray}%
and:%
\begin{eqnarray*}
&&\pm \frac{\bar{N}\left( \bar{X}_{1}\right) }{\sqrt{\bar{K}\left[ \bar{X}%
_{1}\right] }}-\bar{r}^{\prime }=\left( 1+\bar{k}_{2}^{n}\left( \bar{X}%
_{1}\right) \right) \left( \frac{1-\beta ^{B}}{\left( 1-\beta \right) \delta 
}\right) \left( \frac{A}{\left( f_{1}\left( X^{\prime },\hat{K}\left[ \hat{X}%
^{\prime }\right] \right) \right) ^{2}}+\frac{B}{\left( f_{1}\left(
X^{\prime },\hat{K}\left[ \hat{X}^{\prime }\right] \right) \right) ^{3}}%
\right) \\
&&\times \left[ \left( f_{1}\left( X^{\prime },\hat{K}\left[ \hat{X}^{\prime
}\right] \right) -\bar{r}^{\prime }\right) +\tau F\left( X^{\prime }\right)
\left( f_{1}\left( X^{\prime },\hat{K}\left[ \hat{X}^{\prime }\right]
\right) -\left\langle f_{1}\left( X,\hat{K}\left[ \hat{X}\right] \right)
\right\rangle \right) \right]
\end{eqnarray*}%
and these solutions have the same form and interpretation as in Part $1$.

Note that in this benchmark, a direct relationship has been established
between the capital of the banking sector and the capital of investors of
the same sector:%
\begin{equation*}
\frac{\bar{N}\left( \bar{X}_{1}\right) }{\sqrt{\bar{K}\left[ \bar{X}_{1}%
\right] }}-\bar{r}^{\prime }=\pm \left( 1+\bar{k}_{2}^{n}\left( \bar{X}%
_{1}\right) \right) \left( \frac{1-\beta ^{B}}{\left( 1-\beta \right) \delta 
}\right) \left( \frac{\hat{N}\left( \hat{X}_{1}\right) }{\sqrt{\hat{K}\left[ 
\hat{X}_{1}\right] }}-\bar{r}^{\prime }\right)
\end{equation*}%
This relationship is proportional to $\frac{1-\beta ^{B}}{\left( 1-\beta
\right) \delta }$, so that the constraint between banks and investors
operating in the same isolated sector is only valid if the bank takes
participation in the investors.

\subsection{Interconnected financial agents}

We consider (\ref{CFT}) and (\ref{CFG}) simultaneously, and find corrections
to the solutions for isolated agents.

\subsubsection{General approach}

We denote $\Delta \left( \frac{N\left( \hat{X}\right) }{\sqrt{\hat{K}\left[ 
\hat{X}\right] }}\right) $ as the correction due to interactions between
investors, compared to a state without interactions between sectors.
Additionally, we define $\hat{K}_{1}\left[ \hat{X}\right] $ as the solution
without interactions between sectors. Hence, we can express:%
\begin{equation*}
\frac{\hat{N}\left( \hat{X}\right) }{\sqrt{\hat{K}\left[ \hat{X}\right] }}=%
\frac{\hat{N}\left( \hat{X}\right) }{\sqrt{\hat{K}_{1}\left[ \hat{X}\right] }%
}+\Delta \left( \frac{\hat{N}\left( \hat{X}\right) }{\sqrt{\hat{K}\left[ 
\hat{X}\right] }}\right)
\end{equation*}%
\begin{equation*}
\frac{\bar{N}\left( \bar{X}\right) }{\sqrt{\bar{K}\left[ \bar{X}\right] }}=%
\frac{\bar{N}\left( \bar{X}\right) }{\sqrt{\bar{K}_{1}\left[ \bar{X}\right] }%
}+\Delta \left( \frac{\bar{N}\left( \hat{X}\right) }{\sqrt{\bar{K}\left[ 
\bar{X}\right] }}\right)
\end{equation*}%
The sought-after solution is therefore the sum of the benchmark plus the
correction to this benchmark due to interactions.

A second-order expansion straightforwardly yields an approximate equation
for $\Delta \left( \frac{\hat{N}\left( \hat{X}\right) }{\sqrt{\hat{K}\left[ 
\hat{X}\right] }}\right) $ and $\Delta \left( \frac{\bar{N}\left( \hat{X}%
\right) }{\sqrt{\bar{K}\left[ \bar{X}\right] }}\right) $:%
\begin{eqnarray*}
&&\int \left( \frac{\Delta \left( \hat{X},\hat{X}^{\prime }\right) }{1+%
\underline{\hat{k}}_{2}\left( \hat{X}\right) }-\hat{S}_{1}^{E}\left( \hat{X}%
^{\prime },\hat{X}\right) \right) \Delta \left( \frac{N\left( \hat{X}%
^{\prime }\right) }{\sqrt{\underline{k}\left( \hat{X}^{\prime }\right) }}%
\right) -\int \left( \hat{S}_{1}^{E}\left( \hat{X}^{\prime },\hat{X}\right) 
\frac{N\left( \hat{X}^{\prime }\right) }{\sqrt{\underline{k}^{\prime }\left( 
\hat{X}^{\prime }\right) }}\right) \\
&=&H_{\hat{1}}\Delta \left( \frac{N\left( \hat{X}^{\prime }\right) }{\sqrt{%
\underline{k}\left( \hat{X}^{\prime }\right) }}\right) +\frac{1}{2}H_{\hat{2}%
}\left( \Delta \left( \frac{N\left( \hat{X}^{\prime }\right) }{\sqrt{%
\underline{k}\left( \hat{X}^{\prime }\right) }}\right) \right) ^{2}+H_{\bar{1%
}}\Delta \left( \frac{\bar{N}\left( \bar{X}\right) }{\sqrt{\bar{K}\left[ 
\bar{X}\right] }}\right) +\frac{1}{2}H_{\bar{2}}\left( \Delta \left( \frac{%
\bar{N}\left( \bar{X}\right) }{\sqrt{\bar{K}\left[ \bar{X}\right] }}\right)
\right) ^{2} \\
&&+H_{\hat{1}\bar{1}}\Delta \left( \frac{N\left( \hat{X}^{\prime }\right) }{%
\sqrt{\underline{k}\left( \hat{X}^{\prime }\right) }}\right) \Delta \left( 
\frac{\bar{N}\left( \bar{X}\right) }{\sqrt{\bar{K}\left[ \bar{X}\right] }}%
\right)
\end{eqnarray*}%
where $H_{\hat{l}}$ are derivatives of:%
\begin{eqnarray*}
&&\left( \frac{A}{\left( f_{1}\left( X^{\prime },\hat{K}\left[ \hat{X}%
^{\prime }\right] ,\bar{K}\left[ \bar{X}^{\prime }\right] \right) \right)
^{2}}+\frac{B}{\left( f_{1}\left( X^{\prime },\hat{K}\left[ \hat{X}^{\prime }%
\right] ,\bar{K}\left[ \bar{X}^{\prime }\right] \right) \right) ^{3}}\right)
\\
&&\times \left[ \left( f_{1}\left( X^{\prime },\hat{K}\left[ \hat{X}^{\prime
}\right] ,\bar{K}\left[ \bar{X}^{\prime }\right] \right) -\bar{r}^{\prime
}\right) +\tau F\left( X^{\prime }\right) \left( f_{1}\left( X^{\prime },%
\hat{K}\left[ \hat{X}^{\prime }\right] ,\bar{K}\left[ \bar{X}^{\prime }%
\right] \right) -\left\langle f_{1}\left( X^{\prime },\hat{K}\left[ \hat{X}%
^{\prime }\right] ,\bar{K}\left[ \bar{X}^{\prime }\right] \right)
\right\rangle \right) \right]
\end{eqnarray*}%
with respect to $\hat{K}\left[ \hat{X}\right] $, $H_{\bar{l}}$ represents
the derivatives with respect to $\bar{K}\left[ \bar{X}\right] $, and $H_{%
\hat{1}\bar{1}}$ is the cross derivative.

The constraint between the two variations is given by:

\begin{eqnarray}
&&\left( 1-\left( 1+\bar{k}_{2}^{n}\left( \bar{X}_{1}\right) \right) \left( 
\bar{S}_{1}^{E}\left( \bar{X}^{\prime },\hat{X}_{1}\right) +\bar{S}%
_{1}^{B}\left( \bar{X}^{\prime },\bar{X}_{1}\right) \right) \right) \Delta
\left( \frac{\bar{N}\left( \bar{X}_{1}\right) }{\sqrt{\bar{K}\left[ \bar{X}%
_{1}\right] }}\right)  \label{RLD} \\
&=&\pm \left( 1+\bar{k}_{2}^{n}\left( \bar{X}_{1}\right) \right) \left( 
\frac{1-\beta ^{B}}{\left( 1-\beta \right) \delta }\left( \frac{\Delta
\left( \hat{X},\hat{X}^{\prime }\right) }{1+\left[ \underline{\hat{k}}%
_{2}^{n}\left( \hat{X}\right) \right] _{\kappa }}-\hat{S}_{1}^{E}\left( \hat{%
X}^{\prime },\hat{X}_{1}\right) \right) +\hat{S}_{1}^{B}\left( \hat{X}%
^{\prime },\bar{X}_{1}\right) \right) \Delta \left( \frac{\hat{N}\left( \hat{%
X}\right) }{\sqrt{\hat{K}\left[ \hat{X}\right] }}\right)  \notag
\end{eqnarray}%
and the solution for $\Delta \left( \frac{N\left( \hat{X}^{\prime }\right) }{%
\sqrt{\hat{K}\left[ \hat{X}^{\prime }\right] }}\right) $ is, in first
approximation:

\begin{equation}
\left( \delta \left( \hat{X}-\hat{X}^{\prime }\right) \pm _{\hat{X}}\left[ H%
\right] \left( \hat{X}^{\prime },\hat{X}\right) \right) \Delta \left( \frac{%
N\left( \hat{X}^{\prime }\right) }{\sqrt{\hat{K}\left[ \hat{X}^{\prime }%
\right] }}\right) =\frac{\left( 1-H_{1}\right) \pm _{\hat{X}}\left[ L\right]
\left( \hat{X}\right) }{H_{2}}  \label{VRT}
\end{equation}%
with:%
\begin{equation*}
\left[ H\right] \left( \hat{X}^{\prime },\hat{X}\right) =\frac{2H_{2}\left(
\left( 1+\underline{\hat{k}}_{2}\left( \hat{X}\right) \right) \hat{S}%
_{1}^{E}\left( \hat{X}^{\prime },\hat{X}\right) -H_{\hat{1}\bar{1}}\Delta
\left( \frac{\bar{N}\left( \bar{X}\right) }{\sqrt{\bar{K}\left[ \bar{X}%
\right] }}\right) \right) }{\left( \left( 1-H_{1}\right) ^{2}-2H_{2}\left( 1+%
\underline{\hat{k}}_{2}\left( \hat{X}\right) \right) \left( Q\left( \frac{%
\bar{N}\left( \bar{X}\right) }{\sqrt{\bar{K}\left[ \bar{X}\right] }}\right)
+\int \hat{S}_{1}^{E}\left( \hat{X}^{\prime },\hat{X}\right) \frac{N\left( 
\hat{X}^{\prime }\right) }{\sqrt{\hat{K}_{1}\left[ \hat{X}^{\prime }\right] }%
}\right) \right) ^{\frac{3}{2}}}
\end{equation*}%
\begin{equation*}
\left[ L\right] \left( \hat{X}\right) =\sqrt{\left( 1-H_{1}\right)
^{2}-\left( 2H_{2}\left( 1+\underline{\hat{k}}_{2}\left( \hat{X}\right)
\right) \left( Q\left( \frac{\bar{N}\left( \bar{X}\right) }{\sqrt{\bar{K}%
\left[ \bar{X}\right] }}\right) +\int \hat{S}_{1}^{E}\left( \hat{X}^{\prime
},\hat{X}\right) \frac{N\left( \hat{X}^{\prime }\right) }{\sqrt{\hat{K}_{1}%
\left[ \hat{X}^{\prime }\right] }}\right) \right) }
\end{equation*}%
and:%
\begin{equation*}
Q\left( \frac{\bar{N}\left( \bar{X}\right) }{\sqrt{\bar{K}\left[ \bar{X}%
\right] }}\right) =H_{\bar{1}}\Delta \left( \frac{\bar{N}\left( \bar{X}%
\right) }{\sqrt{\bar{K}\left[ \bar{X}\right] }}\right) +\frac{1}{2}H_{\bar{2}%
}\left( \Delta \left( \frac{\bar{N}\left( \bar{X}\right) }{\sqrt{\bar{K}%
\left[ \bar{X}\right] }}\right) \right) ^{2}
\end{equation*}%
where the sign $\pm _{\hat{X}}$ indicates that the choice of sign depends on
the sector considered, with two possibilities per sector. Moreover, the
equation is non-local; the first line indicates diffusion (presence of $\hat{%
S}_{1}^{E}\left( \hat{X}^{\prime },\hat{X}\right) $), and thus the choice of
sign in one sector impacts the other sectors. The withdrawal of capital
invested in one sector will impact the other sectors connected to it,
through investment decisions among investors. As a result, entire blocks can
end up in a low-capital or high-capital state. The diffusion and
amplification effect depend on the level of indebtedness $1+\underline{\hat{k%
}}_{2}\left( \hat{X}\right) $. It represents diffusion amplified by the
level of lending among investors.

Using (\ref{RLD}), $Q\left( \frac{\bar{N}\left( \bar{X}\right) }{\sqrt{\bar{K%
}\left[ \bar{X}\right] }}\right) $ can be replaced by its average in (\ref%
{VRT}). Given (\ref{RLD}), this can be estimated as:%
\begin{eqnarray}
&&\left\langle \Delta \left( \frac{\bar{N}\left( \bar{X}_{1}\right) }{\sqrt{%
\bar{K}\left[ \bar{X}_{1}\right] }}\right) \right\rangle \\
&=&\pm \frac{\left\langle \left( 1+\bar{k}_{2}^{n}\left( \bar{X}_{1}\right)
\right) \left( \frac{1-\beta ^{B}}{\left( 1-\beta \right) \delta }\left( 
\frac{\Delta \left( \hat{X},\hat{X}^{\prime }\right) }{1+\left[ \underline{%
\hat{k}}_{2}^{n}\left( \hat{X}\right) \right] _{\kappa }}-\hat{S}%
_{1}^{E}\left( \hat{X}^{\prime },\hat{X}_{1}\right) \right) +\hat{S}%
_{1}^{B}\left( \hat{X}^{\prime },\bar{X}_{1}\right) \right) \right\rangle }{%
\left\langle \left( 1-\left( 1+\bar{k}_{2}^{n}\left( \bar{X}_{1}\right)
\right) \left( \bar{S}_{1}^{E}\left( \bar{X}^{\prime },\hat{X}_{1}\right) +%
\bar{S}_{1}^{B}\left( \bar{X}^{\prime },\bar{X}_{1}\right) \right) \right)
\right\rangle }\left\langle \Delta \left( \frac{\hat{N}\left( \hat{X}\right) 
}{\sqrt{\hat{K}\left[ \hat{X}\right] }}\right) \right\rangle  \notag
\end{eqnarray}%
The averages are computed via integration. As a consequence:%
\begin{equation*}
Q\left( \left\langle \Delta \left( \frac{\bar{N}\left( \bar{X}_{1}\right) }{%
\sqrt{\bar{K}\left[ \bar{X}_{1}\right] }}\right) \right\rangle \right)
\simeq H_{\bar{1}}\left\langle S\right\rangle \left\langle \Delta \left( 
\frac{\hat{N}\left( \hat{X}\right) }{\sqrt{\hat{K}\left[ \hat{X}\right] }}%
\right) \right\rangle +\frac{1}{2}H_{\bar{2}}\left\langle S\right\rangle
^{2}\left( \left\langle \Delta \left( \frac{\hat{N}\left( \hat{X}\right) }{%
\sqrt{\hat{K}\left[ \hat{X}\right] }}\right) \right\rangle \right) ^{2}
\end{equation*}%
with:%
\begin{eqnarray*}
\left\langle S\right\rangle &=&\pm \frac{\left\langle \left( 1+\bar{k}%
_{2}^{n}\left( \bar{X}_{1}\right) \right) \left( \frac{1-\beta ^{B}}{\left(
1-\beta \right) \delta }\left( \frac{1}{1+\left[ \underline{\hat{k}}%
_{2}^{n}\left( \hat{X}\right) \right] _{\kappa }}-\hat{S}_{1}^{E}\left( \hat{%
X}^{\prime },\hat{X}_{1}\right) \right) +\hat{S}_{1}^{B}\left( \hat{X}%
^{\prime },\bar{X}_{1}\right) \right) \right\rangle }{\left\langle \left(
1-\left( 1+\bar{k}_{2}^{n}\left( \bar{X}_{1}\right) \right) \left( \bar{S}%
_{1}^{E}\left( \bar{X}^{\prime },\hat{X}_{1}\right) +\bar{S}_{1}^{B}\left( 
\bar{X}^{\prime },\bar{X}_{1}\right) \right) \right) \right\rangle } \\
&\rightarrow &\frac{\frac{1-\beta ^{B}}{\left( 1-\beta \right) \delta }%
\left( 1-\left\langle \hat{S}_{1}^{E}\left( \hat{X}^{\prime },\hat{X}%
_{1}\right) \right\rangle \right) +\hat{S}_{1}^{B}\left( \hat{X}^{\prime },%
\bar{X}_{1}\right) }{\left\langle \left( 1-\left( \bar{S}_{1}^{E}\left( \bar{%
X}^{\prime },\hat{X}_{1}\right) +\bar{S}_{1}^{B}\left( \bar{X}^{\prime },%
\bar{X}_{1}\right) \right) \right) \right\rangle }
\end{eqnarray*}

\subsubsection{First approximation}

In first approximation, we can consider banks primarily as lenders. In this
case:

\begin{eqnarray*}
1-\beta ^{B} &<&<1 \\
\left\langle \frac{\bar{N}\left( \bar{X}\right) }{\sqrt{\bar{K}\left[ \bar{X}%
\right] }}\right\rangle &=&\bar{r}^{\prime } \\
\left\langle S\right\rangle &=&0
\end{eqnarray*}%
Moreover, given that $\left\langle \frac{\bar{N}\left( \bar{X}\right) }{%
\sqrt{\bar{K}\left[ \bar{X}\right] }}\right\rangle $ is constant, we can
deduce:%
\begin{equation*}
\Delta \left( \frac{\bar{N}\left( \bar{X}\right) }{\sqrt{\bar{K}\left[ \bar{X%
}\right] }}\right) \simeq 0
\end{equation*}%
The equation becomes:%
\begin{eqnarray*}
&&\left( \delta \left( \hat{X}-\hat{X}^{\prime }\right) \pm _{\hat{X}}\frac{%
2H_{2}\left( \left( 1+\underline{\hat{k}}_{2}\left( \hat{X}\right) \right) 
\hat{S}_{1}^{E}\left( \hat{X}^{\prime },\hat{X}\right) \right) }{\left(
\left( 1-H_{1}\right) ^{2}-2H_{2}\left( 1+\underline{\hat{k}}_{2}\left( \hat{%
X}\right) \right) \int \hat{S}_{1}^{E}\left( \hat{X}^{\prime },\hat{X}%
\right) \frac{N\left( \hat{X}^{\prime }\right) }{\sqrt{\hat{K}_{1}\left[ 
\hat{X}^{\prime }\right] }}\right) ^{\frac{3}{2}}}\right) \Delta \left( 
\frac{N\left( \hat{X}^{\prime }\right) }{\sqrt{\hat{K}\left[ \hat{X}^{\prime
}\right] }}\right) \\
&=&\frac{\left( 1-H_{1}\right) \pm _{\hat{X}}\sqrt{\left( 1-H_{1}\right)
^{2}-\left( 2H_{2}\left( 1+\underline{\hat{k}}_{2}\left( \hat{X}\right)
\right) \int \hat{S}_{1}^{E}\left( \hat{X}^{\prime },\hat{X}\right) \frac{%
N\left( \hat{X}^{\prime }\right) }{\sqrt{\hat{K}_{1}\left[ \hat{X}^{\prime }%
\right] }}\right) }}{H_{2}}
\end{eqnarray*}%
In this simple case, the equation for the corrections $\Delta \left( \frac{%
N\left( \hat{X}^{\prime }\right) }{\sqrt{\hat{K}\left[ \hat{X}^{\prime }%
\right] }}\right) $ is similar to Part $1$. However, given the form of:%
\begin{equation*}
f_{1}\left( X^{\prime },\hat{K}\left[ \hat{X}^{\prime }\right] ,\bar{K}\left[
\bar{X}^{\prime }\right] \right) \simeq \frac{f_{1}\left( X\right) }{\left(
\left( 1+k\left( X\right) \hat{K}\left[ \hat{X}\right] \right) +\left(
k_{1}^{B}\left( \bar{X}\right) +\kappa \left[ \frac{\underline{k}%
_{2}^{B}\left( \bar{X}\right) }{1+\bar{k}}\right] \right) \bar{K}\left[ \bar{%
X}\right] \right) ^{r}}-C_{0}
\end{equation*}%
the derivatives of: 
\begin{eqnarray*}
&&\left( \frac{A}{\left( f_{1}\left( X^{\prime },\hat{K}\left[ \hat{X}%
^{\prime }\right] ,\bar{K}\left[ \bar{X}^{\prime }\right] \right) \right)
^{2}}+\frac{B}{\left( f_{1}\left( X^{\prime },\hat{K}\left[ \hat{X}^{\prime }%
\right] ,\bar{K}\left[ \bar{X}^{\prime }\right] \right) \right) ^{3}}\right)
\\
&&\times \left[ \left( f_{1}\left( X^{\prime },\hat{K}\left[ \hat{X}^{\prime
}\right] ,\bar{K}\left[ \bar{X}^{\prime }\right] \right) -\bar{r}^{\prime
}\right) +\tau F\left( X^{\prime }\right) \left( f_{1}\left( X^{\prime },%
\hat{K}\left[ \hat{X}^{\prime }\right] ,\bar{K}\left[ \bar{X}^{\prime }%
\right] \right) -\left\langle f_{1}\left( X^{\prime },\hat{K}\left[ \hat{X}%
^{\prime }\right] ,\bar{K}\left[ \bar{X}^{\prime }\right] \right)
\right\rangle \right) \right]
\end{eqnarray*}%
decrease in amplitude with $\bar{K}\left[ \bar{X}\right] $. In appendix 21.3
we show that $\hat{S}_{1}^{E}\left( \hat{X}^{\prime },\hat{X}\right) $ has
the same expression as in Part $1$.

\subsubsection{Interpretation of solutions}

On average:%
\begin{equation*}
\left\langle \hat{S}_{1}^{E}\left( \hat{X}^{\prime },\hat{X}\right)
\right\rangle \simeq \left\langle \hat{k}\right\rangle \left( 1-\left\langle 
\hat{k}_{1}\right\rangle \right) +\left\langle \hat{k}_{1}\right\rangle
\end{equation*}%
but the normalization of coefficients in Part $1$:%
\begin{equation*}
\hat{k}_{\eta }\left( \hat{X}^{\prime },\hat{X}\right) \rightarrow \frac{%
\hat{k}_{\eta }\left( \hat{X}^{\prime },\hat{X}\right) }{\left(
1-\left\langle \hat{k}\left( \hat{X}^{\prime },\hat{X}\right) \right\rangle
\right) }
\end{equation*}%
and in Part $2$:%
\begin{equation*}
\hat{k}_{\eta }\left( \hat{X}^{\prime },\hat{X}\right) \rightarrow \frac{%
\hat{k}_{\eta }\left( \hat{X}^{\prime },\hat{X}\right) }{1-\left\langle \hat{%
k}\left( \hat{X}^{\prime },\hat{X}\right) \right\rangle -\left( \left\langle 
\hat{k}_{1}^{B}\left( \hat{X}^{\prime },\bar{X}\right) \right\rangle +\kappa 
\frac{\left\langle \hat{k}_{2}^{B}\left( \hat{X}^{\prime },\bar{X}\right)
\right\rangle }{1+\left\langle \bar{k}\right\rangle \left\langle \bar{K}%
\right\rangle }\frac{\left\Vert \bar{\Psi}\right\Vert ^{2}\left\langle \bar{K%
}\right\rangle }{\left\Vert \hat{\Psi}\right\Vert ^{2}\left\langle \hat{K}%
\right\rangle }\right) }
\end{equation*}%
reveals that the diffusion matrix $\left\langle \hat{S}_{1}^{E}\left( \hat{X}%
^{\prime },\hat{X}\right) \right\rangle ,$ including banks, has a higher
amplitude than in Part $1$. Thus, when comparing the amplitude of
corrections in Part $1$ and $2$, two competing effects arise. Banks increase
the disponable capital to firms, thereby somewhat reducing access to other
investors. The amount of capital lent to firms reaches a level such that
marginal returns stabilize around low value, stablizing the system. On the
other hand, the banks also lend capital to investors, which may increase the
leverage effect for their disposable capital. This increases instability and
the discrepancies between the various collective states. However, the
amplification due to the diffusion matrix remains constant, depending on the
leverage effect of investors and banks, while the stabilization effect of
loans to firms by banks may increase with the size of the banking system.
Consequently, there should be a level of bank capital at which the
stabilization effect outweighs the diffusion effect.

This result suggests that the multiplicity of global averages of capital
should be reduced by introducing banks into the system. This point is
studied in the next paragraph.

\section{Global average returns and total capital for financial agents}

\subsection{Global average returns}

The system is closed by computing the remaining average quantities: firms'
capital and returns, investors' and banks' average returns. Appendices 23.2
and 23.3 shows that the investors' returns are:%
\begin{eqnarray}
\left\langle \hat{g}\right\rangle &\simeq &\bar{r}^{\prime }+\left( 1-\left(
1+\left\langle \hat{k}_{2}^{n}\right\rangle \right) \left( \left\langle \hat{%
k}\right\rangle \left( 1-\left\langle \hat{k}_{1}\right\rangle \right)
+\left\langle \hat{k}_{1}\right\rangle \right) \right) ^{-1}  \label{HR} \\
&&\times \left( 1+\left\langle \hat{k}_{2}^{n}\right\rangle \right)
\left\langle \left( \frac{A}{\left( f_{1}\left( X,\hat{K}\left[ X\right] ,%
\bar{K}\left[ X\right] \right) \right) ^{2}}+\frac{B}{\left( f_{1}\left( X,%
\hat{K}\left[ X\right] ,\bar{K}\left[ X\right] \right) \right) ^{3}}\right)
R\right\rangle  \notag
\end{eqnarray}%
and that the bank returns are: 
\begin{subequations}
\begin{eqnarray}
\left\langle \bar{g}\right\rangle &=&\bar{r}^{\prime }+\frac{\left( 1-\beta
^{B}\right) }{\left( 1-\beta \right) \left\langle \delta \right\rangle }%
\left[ \left( 1-\left( \left( \left( 1-\left\langle \bar{k}_{1}\right\rangle
\right) \left\langle \bar{k}\right\rangle +\left\langle \bar{k}%
_{1}\right\rangle \right) +\left\langle \underline{\hat{k}}%
_{1}^{B}\right\rangle \left( \left\langle \underline{\hat{k}}%
_{1}^{B}\right\rangle +\kappa \frac{\left\langle \underline{\hat{k}}%
_{2}^{B}\right\rangle }{1+\underline{\bar{k}}}\left( 1-\frac{\kappa }{1+%
\underline{\bar{k}}}\right) \right) \right) \right) \right] ^{-1}  \notag \\
&&\times \left( \frac{1}{\left\langle \hat{k}^{B}\right\rangle }-\left(
\left( \left\langle \hat{k}\right\rangle \left( 1-\left\langle \hat{k}%
_{1}\right\rangle \right) +\left\langle \hat{k}_{1}\right\rangle \right) -%
\frac{\left( 1-\beta \right) \left\langle \delta \right\rangle }{1-\beta ^{B}%
}\left( \left\langle \hat{k}_{1}^{B}\right\rangle \left( 1-\left\langle \hat{%
k}\right\rangle \right) \right) \right) \right) \left( \left\langle \hat{g}%
\right\rangle -\bar{r}^{\prime }\right)  \label{BRt}
\end{eqnarray}

\subsection{Global average total capital for financial agents}

The equation for average capital is solved \ by combining (\ref{VKH}), (\ref%
{BVK}), (\ref{HR}), and (\ref{BRt}). To gain some insight into the
solutions, we make several approximations in Appendix 23. Firstly, we
consider that $1-\beta ^{B}<<1$, so that: 
\end{subequations}
\begin{equation*}
\left\langle \bar{g}\right\rangle =\bar{r}^{\prime }
\end{equation*}%
This corresponds to approximating the banks as lenders primarily. Then, we
consider that $\delta >>1$, which corresponds to considering that banks
loans are dominant due to the credit multiplier. Ultimately, we loosely
consider that interaction terms $\frac{\left\langle \hat{g}%
^{ef}\right\rangle }{\left\langle \hat{g}\right\rangle }$ can be neglected.
They can be reintroduced in a more refined approximation. Under these
assumptions, the system reduces to the following three equations: 
\begin{equation*}
\left\langle \hat{K}\right\rangle \left\Vert \hat{\Psi}\right\Vert ^{2}=%
\frac{9\sigma _{\hat{K}}^{2}}{2\hat{\mu}\left\langle \hat{g}\right\rangle
^{2}}V\left\Vert \hat{\Psi}_{0}\right\Vert ^{4}
\end{equation*}%
\begin{equation}
\left\langle \bar{K}\right\rangle \left\Vert \bar{\Psi}\right\Vert
^{2}\simeq \frac{9\sigma _{\hat{K}}^{2}\left\langle \hat{g}\right\rangle ^{4}%
}{2\left( \bar{r}^{\prime }\right) ^{2}\hat{\mu}}\left\Vert \bar{\Psi}%
_{0}\right\Vert ^{4}
\end{equation}%
\begin{eqnarray*}
\left\langle \hat{g}\right\rangle &\simeq &\bar{r}^{\prime }+\left( 1-\left(
1+\left\langle \hat{k}_{2}^{n}\right\rangle \right) \left( \left\langle \hat{%
k}\right\rangle \left( 1-\left\langle \hat{k}_{1}\right\rangle \right)
+\left\langle \hat{k}_{1}\right\rangle \right) \right) ^{-1} \\
&&\times \left( 1+\left\langle \hat{k}_{2}^{n}\right\rangle \right) \frac{%
\epsilon \left( 1-\beta \right) C\left( X+C\beta \right) ^{2}}{3\sigma _{%
\hat{K}}^{2}}\left\langle \frac{R\left( 3X-\beta C\right) \left( \beta
C+X\right) }{4f_{1}^{2}\left( X\right) C\left( 2X-C\beta \right) }+\frac{%
\beta R}{f_{1}^{3}\left( X\right) }\right\rangle
\end{eqnarray*}%
Appendix 23.4 shows that this reduces to an equation for $\left\langle \hat{K%
}\right\rangle \left\Vert \hat{\Psi}\right\Vert ^{2}$:%
\begin{eqnarray}
\frac{\sqrt{\frac{9\sigma _{\hat{K}}^{2}}{2\hat{\mu}}V}\left\Vert \hat{\Psi}%
_{0}\right\Vert ^{2}}{\sqrt{\left\langle \hat{K}\right\rangle \left\Vert 
\hat{\Psi}\right\Vert ^{2}}}-\bar{r}^{\prime } &\simeq &\left( 1-\left(
1+\left\langle \hat{k}_{2}^{n}\right\rangle \right) \left( \left\langle \hat{%
k}\right\rangle \left( 1-\left\langle \hat{k}_{1}\right\rangle \right)
+\left\langle \hat{k}_{1}\right\rangle \right) \right) ^{-1}\left(
1+\left\langle \hat{k}_{2}^{n}\right\rangle \right)  \label{BNL} \\
&&\times \left\langle \left( \frac{A}{\left( f_{1}\left( X,\hat{K}\left[ X%
\right] ,\bar{K}\left[ X\right] \right) \right) ^{2}}+\frac{B}{\left(
f_{1}\left( X,\hat{K}\left[ X\right] ,\bar{K}\left[ X\right] \right) \right)
^{3}}\right) R\right\rangle  \notag
\end{eqnarray}%
\begin{equation*}
\left\langle f_{1}\left( X,\hat{K}\left[ X\right] ,\bar{K}\left[ X\right]
\right) \right\rangle =\frac{\left\langle f_{1}\left( X\right) \right\rangle 
}{\left( 1+\left\langle k\right\rangle \left\langle \hat{K}\right\rangle
\left\Vert \hat{\Psi}\right\Vert ^{2}+\left( \left\langle
k_{1}^{B}\right\rangle +\kappa \left[ \frac{\underline{k}_{2}^{B}}{1+\bar{k}}%
\right] \right) \left\langle \bar{K}\right\rangle \left\Vert \bar{\Psi}%
\right\Vert ^{2}\right) ^{r}}-C_{0}
\end{equation*}%
Given that\ the right-hand side of (\ref{BNL}) depends on $\left\langle \hat{%
K}\right\rangle \left\Vert \hat{\Psi}\right\Vert ^{2}$, we can expect that
this equation has multiple solutions. However, as mentioned in the previous
section, the correction to average capital $\Delta \left( \frac{N\left( \hat{%
X}^{\prime }\right) }{\sqrt{\hat{K}\left[ \hat{X}^{\prime }\right] }}\right) 
$ depends negatively on $\bar{K}\left[ \bar{X}\right] $, suggesting that
banks should reduce the possibilities of multiple global averages. This
point appears to be confirmed by some numerical studies.

\section{Several groups of agents}

We follow the same process as in Part $1$. Instead of considering the system
averages as a whole, we will consider subgroups and the interactions between
these subgroups. We can consider that within sector spaces, investors are
organized into heterogeneous groups or sub-markets, relatively weakly
connected, as before. We directly consider several different groups. Each
group is homogeneous and contains banks, investors, and firms, with
approximately homogeneous returns, connections, capital, and production.

\subsection{Investors: return equations for group averages}

Let us now define $\underline{\hat{k}}_{\eta }^{\left[ ii\right] }$, $%
\underline{k}_{\eta }^{\left[ ii\right] }$, the average coefficients within
the group, and $\underline{\hat{k}}_{\eta }^{\left[ ji\right] }$ and $%
\underline{k}_{\eta }^{\left[ ji\right] }$ the average connections from $i$
to $j$. \ We define the total share of capital invested in sector $i$,
comprising intra-sectoral investments (from $i$ to $i$), and inter-sectoral
investments (from $j$ to $i$), as:%
\begin{equation*}
\underline{\hat{k}}_{\eta }^{\left[ i\right] }=\underline{\hat{k}}_{\eta }^{%
\left[ ii\right] }+\underline{\hat{k}}_{\eta }^{\left[ ij\right] }
\end{equation*}%
in which the sum over the indices is understood. The return equation
involving several groups becomes in terms of matrices:%
\begin{eqnarray}
0 &=&\left( 
\begin{array}{cc}
1-\underline{\hat{S}}_{1}^{\left[ ii\right] } & -\underline{\hat{S}}_{1}^{%
\left[ ji\right] } \\ 
-\underline{\hat{S}}_{1}^{\left[ ij\right] } & 1-\underline{\hat{S}}_{1}^{%
\left[ jj\right] }%
\end{array}%
\right) \left( 
\begin{array}{c}
\frac{f^{\left[ i\right] }-\bar{r}}{1+\underline{\hat{k}}_{2}^{\left[ i%
\right] }+\kappa \left[ \frac{\underline{\hat{k}}_{2}^{B}}{1+\bar{k}}\right]
^{\left[ i\right] }} \\ 
\frac{f^{\left[ j\right] }-\bar{r}}{1+\underline{\hat{k}}_{2}^{\left[ j%
\right] }+\kappa \left[ \frac{\underline{\hat{k}}_{2}^{B}}{1+\bar{k}}\right]
^{\left[ j\right] }}%
\end{array}%
\right)  \label{PML} \\
&&-\left( 
\begin{array}{cc}
\underline{\hat{S}}_{2}^{\left[ ii\right] } & \underline{\hat{S}}_{2}^{\left[
ji\right] } \\ 
\underline{\hat{S}}_{2}^{\left[ ij\right] } & \underline{\hat{S}}_{2}^{\left[
jj\right] }%
\end{array}%
\right) \left( 
\begin{array}{c}
\frac{1+f^{\left[ i\right] }}{\underline{\hat{k}}_{2}^{\left[ i\right]
}+\kappa \left[ \frac{\underline{\hat{k}}_{2}^{B}}{1+\bar{k}}\right] ^{\left[
i\right] }}H\left( -\left( 1+f^{\left[ i\right] }\right) \right) \\ 
\frac{1+f^{\left[ j\right] }}{\underline{\hat{k}}_{2}^{\left[ j\right]
}+\kappa \left[ \frac{\underline{\hat{k}}_{2}^{B}}{1+\bar{k}}\right] ^{\left[
j\right] }}H\left( -\left( 1+f^{\left[ j\right] }\right) \right)%
\end{array}%
\right)  \notag \\
&&-\left( 
\begin{array}{cc}
\underline{S}_{2}^{\left[ ii\right] } & \underline{S}_{2}^{\left[ ji\right] }
\\ 
\underline{S}_{2}^{\left[ ij\right] } & \underline{S}_{2}^{\left[ jj\right] }%
\end{array}%
\right) \left( 
\begin{array}{c}
\frac{1+f_{1}^{\prime \left[ i\right] }}{\underline{k}_{2}^{\left[ i\right]
}+\left[ \frac{\underline{k}_{2}^{B}}{1+\bar{k}}\right] ^{\left[ i\right] }}%
H\left( -\left( 1+f_{1}^{\prime \left[ i\right] }\right) \right) \\ 
\frac{1+f_{1}^{\prime \left[ j\right] }}{\underline{k}_{2}^{\left[ j\right]
}+\left[ \frac{\underline{k}_{2}^{B}}{1+\bar{k}}\right] ^{\left[ j\right] }}%
H\left( -\left( 1+f_{1}^{\prime \left[ j\right] }\right) \right)%
\end{array}%
\right) -\left( 
\begin{array}{cc}
\underline{S}_{1}^{\left[ ii\right] } & \underline{S}_{1}^{\left[ ji\right] }
\\ 
\underline{S}_{1}^{\left[ ij\right] } & \underline{S}_{1}^{\left[ jj\right] }%
\end{array}%
\right) \left( 
\begin{array}{c}
\frac{f_{1}^{\prime \left[ i\right] }-\bar{r}}{1+\underline{k}_{2}^{\left[ i%
\right] }+\left[ \frac{\underline{k}_{2}^{B}}{1+\bar{k}}\right] ^{\left[ i%
\right] }} \\ 
\frac{f_{1}^{\prime \left[ j\right] }-\bar{r}}{1+\underline{k}_{2}^{\left[ j%
\right] }+\left[ \frac{\underline{k}_{2}^{B}}{1+\bar{k}}\right] ^{\left[ j%
\right] }}%
\end{array}%
\right)  \notag
\end{eqnarray}%
with several constraints between the coefficients detailed in Appendix 24.2.

As already stated in Part $1$, the diagonal of the matrices represents
intra-sectoral investments. The off-diagonal elements of the matrix
represent inter-sectoral investments. The independent collective states are
thus defined by setting the non-diagonal matrices $\underline{\hat{S}}_{\eta
}^{\left[ ji\right] }=0$ and $\underline{S}_{\eta }^{\left[ ji\right] }=0$.
The corrections due to interactions between groups are obtained by taking
into account the off-diagonal elements. This enables the study of default
transmission between different groups. The transmission of default from one
group, $j$, to other groups varies depending on the type of agent that
defaults. The transmission of default from investors in sector $j$ to sector 
$i$, arises from the term:%
\begin{equation}
\frac{1+f^{\left[ j\right] }}{\underline{\hat{k}}_{2}^{\left[ j\right]
}+\kappa \left[ \frac{\underline{\hat{k}}_{2}^{B}}{1+\bar{k}}\right] ^{\left[
j\right] }}H\left( -\left( 1+f^{\left[ j\right] }\right) \right)  \label{d1}
\end{equation}%
while the transmission of firms' default in sector $j$ is initiated by the
term:%
\begin{equation}
\frac{1+f_{1}^{\prime \left[ j\right] }}{\underline{k}_{2}^{\left[ j\right]
}+\left[ \frac{\underline{k}_{2}^{B}}{1+\bar{k}}\right] ^{\left[ j\right] }}%
H\left( -\left( 1+f_{1}^{\prime \left[ j\right] }\right) \right)  \label{d2}
\end{equation}%
This transmission depends on the off-diagonal coefficients in equation (\ref%
{PML}). These coefficients represent the diffusion of returns from $j$ to $i$%
, and thus the loss due to default (\ref{d1}) or (\ref{d2}) will modify the
returns downward in sector $i$, potentially leading to the creation of
another default. Moreover, splitting into groups allows us to see, based on
the connections between sector groups, which groups will default and which
will remain unaffected.

\subsection{Bank returns' equation for group averages}

For banks, the return equation with intra and inter-group interactions
writes:

\begin{eqnarray}
0 &=&\left( 
\begin{array}{cc}
1-\underline{\bar{S}}_{1}^{\left[ ii\right] } & -\underline{\bar{S}}_{1}^{%
\left[ ji\right] } \\ 
-\underline{\bar{S}}_{1}^{\left[ ij\right] } & 1-\underline{\bar{S}}_{1}^{%
\left[ jj\right] }%
\end{array}%
\right) \left( 
\begin{array}{c}
\frac{\bar{f}^{\left[ i\right] }-\bar{r}}{1+\underline{\bar{k}}^{\left[ i%
\right] }} \\ 
\frac{\bar{f}^{\left[ j\right] }-\bar{r}}{1+\underline{\bar{k}}^{\left[ j%
\right] }}%
\end{array}%
\right) -\left( 
\begin{array}{cc}
\underline{\hat{S}}_{1}^{\left[ ii\right] } & \underline{\hat{S}}_{1}^{\left[
ij\right] } \\ 
\underline{\hat{S}}_{1}^{\left[ ij\right] } & \underline{\hat{S}}_{1}^{\left[
jj\right] }%
\end{array}%
\right) \left( 
\begin{array}{c}
\frac{\hat{f}^{\left[ i\right] }-\bar{r}}{1+\underline{\hat{k}}_{2}^{\left[ i%
\right] }+\kappa \left[ \frac{\underline{\hat{k}}_{2}^{B}}{1+\bar{k}}\right]
^{\left[ i\right] }} \\ 
\frac{\hat{f}^{\left[ j\right] }-\bar{r}}{1+\underline{\hat{k}}_{2}^{\left[ j%
\right] }+\kappa \left[ \frac{\underline{\hat{k}}_{2}^{B}}{1+\bar{k}}\right]
^{\left[ j\right] }}%
\end{array}%
\right)  \label{PMN} \\
&&-\left( 
\begin{array}{cc}
\underline{\bar{S}}_{2}^{\left[ ii\right] } & \underline{\bar{S}}_{2}^{\left[
ji\right] } \\ 
\underline{\bar{S}}_{2}^{B\left[ ij\right] } & \underline{\bar{S}}_{2}^{B%
\left[ jj\right] }%
\end{array}%
\right) \left( 
\begin{array}{c}
\frac{1+\bar{f}^{\left[ i\right] }}{\underline{\overline{\bar{k}}}_{2}^{%
\left[ i\right] }}H\left( -\left( 1+\bar{f}^{\left[ i\right] }\right) \right)
\\ 
\frac{1+\bar{f}^{\left[ j\right] }}{\underline{\overline{\bar{k}}}_{2}^{%
\left[ j\right] }}H\left( -\left( 1+\bar{f}^{\left[ j\right] }\right) \right)%
\end{array}%
\right) -\left( 
\begin{array}{cc}
\underline{\hat{S}}_{2}^{B\left[ ii\right] } & \underline{\hat{S}}_{2}^{B%
\left[ ji\right] } \\ 
\underline{\hat{S}}_{2}^{B\left[ ij\right] } & \underline{\hat{S}}_{2}^{B%
\left[ jj\right] }%
\end{array}%
\right) \left( 
\begin{array}{c}
\frac{1+\hat{f}^{\left[ i\right] }}{\underline{\hat{k}}_{2}^{\left[ i\right]
}+\kappa \left[ \frac{\underline{\hat{k}}_{2}^{B}}{1+\bar{k}}\right] ^{\left[
i\right] }}H\left( -\left( 1+\hat{f}^{\left[ i\right] }\right) \right) \\ 
\frac{1+\hat{f}^{\left[ j\right] }}{\underline{\hat{k}}_{2}^{\left[ j\right]
}+\kappa \left[ \frac{\underline{\hat{k}}_{2}^{B}}{1+\bar{k}}\right] ^{\left[
j\right] }}H\left( -\left( 1+\hat{f}^{\left[ j\right] }\right) \right)%
\end{array}%
\right)  \notag \\
&&-\left( 
\begin{array}{cc}
\underline{S}_{2}^{B\left[ ii\right] } & \underline{S}_{2}^{B\left[ ji\right]
} \\ 
\underline{S}_{2}^{B\left[ ij\right] } & \underline{S}_{2}^{B\left[ jj\right]
}%
\end{array}%
\right) \left( 
\begin{array}{c}
\frac{1+f_{1}^{\prime \left[ i\right] }}{\underline{k}_{2}^{\left[ i\right]
}+\left[ \frac{\underline{k}_{2}^{B}}{1+\bar{k}}\right] ^{\left[ i\right] }}%
H\left( -\left( 1+f_{1}^{\prime \left[ i\right] }\right) \right) \\ 
\frac{1+f_{1}^{\prime \left[ j\right] }}{\underline{k}_{2}^{\left[ j\right]
}+\left[ \frac{\underline{k}_{2}^{B}}{1+\bar{k}}\right] ^{\left[ j\right] }}%
H\left( -\left( 1+f_{1}^{\prime \left[ j\right] }\right) \right)%
\end{array}%
\right) -\left( 
\begin{array}{cc}
\underline{S}_{1}^{B\left[ ii\right] } & \underline{S}_{1}^{B\left[ ji\right]
} \\ 
\underline{S}_{1}^{B\left[ ij\right] } & \underline{S}_{1}^{B\left[ jj\right]
}%
\end{array}%
\right) \left( 
\begin{array}{c}
f_{1}^{\left[ i\right] }-\bar{r} \\ 
f_{1}^{\left[ i\right] }-\bar{r}%
\end{array}%
\right)  \notag
\end{eqnarray}

\subsection{Fields and effective action}

In the case of multiple groups, we consider that the field for investors
decomposes into several independent and interacting components. We can
replicate what we just did in Part $1$ and introduce:%
\begin{equation*}
\left[ \hat{\Psi}\right] \left( \left\{ \hat{K}_{r_{i}},\hat{X}%
_{r_{i}}\right\} _{G_{i}}\right)
\end{equation*}%
which stands for a set of field of investors, one defined for each group $%
\left\{ \hat{K}_{r_{i}},\hat{X}_{r_{i}}\right\} _{G_{i}}$. Similarly, we
define a set of fields for banks:%
\begin{equation*}
\left[ \bar{\Psi}\right] \left( \left\{ \bar{K}_{r_{i}},\bar{X}%
_{r_{i}}\right\} _{G_{i}}\right)
\end{equation*}%
We revisit the functional action of each group from Part $1$, adding the
banks and introducing an interaction term between the different groups. Each
group has its own sector-spatial extension, and the effective action is a
generalization of the one previously contained.%
\begin{eqnarray*}
&&-\sum \left[ \hat{\Psi}\right] \left( \left\{ \hat{K}_{r_{i}},\hat{X}%
_{r_{i}}\right\} _{G_{i}}\right) \nabla _{\hat{K}_{r_{i}}}\left( \nabla _{%
\hat{K}_{r_{i}}}-\hat{K}_{r_{i}}f^{\left[ i\right] }\right) \left[ \hat{\Psi}%
\right] ^{\dag }\left( \left\{ \hat{K}_{r_{i}},\hat{X}_{r_{i}}\right\}
_{G_{i}}\right) \\
&&-\sum \left[ \bar{\Psi}\right] \left( \left\{ \bar{K}_{r_{i}},\bar{X}%
_{r_{i}}\right\} _{G_{i}}\right) \nabla _{\bar{K}_{r_{i}}}\left( \nabla _{%
\bar{K}_{r_{i}}}-\hat{K}_{r_{i}}f^{\left[ i\right] }\right) \left[ \bar{\Psi}%
\right] ^{\dagger }\left( \left\{ \bar{K}_{r_{i}},\bar{X}_{r_{i}}\right\}
_{G_{i}}\right) \\
&&+\sum \prod \left\vert \left[ \hat{\Psi}\right] \left( \left\{ \hat{K}%
_{r_{i}},\hat{X}_{r_{i}}\right\} _{G_{i}}\right) \right\vert ^{2}\delta
\left( \hat{V}\left( \left( \left[ \hat{\Psi}\right] \left( \left\{ \hat{K}%
_{r_{i}},\hat{X}_{r_{i}}\right\} _{G_{i}}\right) \right) \right) \right) \\
&&+\sum \prod \left\vert \left[ \hat{\Psi}\right] \left( \left\{ \hat{K}%
_{r_{i}},\hat{X}_{r_{i}}\right\} _{G_{i}}\right) \right\vert ^{2}\delta
\left( \bar{V}\left( \left[ \bar{\Psi}\right] \left( \left\{ \bar{K}_{r_{i}},%
\bar{X}_{r_{i}}\right\} _{G_{i}}\right) \right) \right) \\
&&-\Xi ^{\dagger }\left( \hat{X},\delta f_{1}\right) \sigma _{\delta
f_{1}}^{2}\nabla _{\delta f_{1}}^{2}\Xi \left( \hat{X},\delta f_{1}\right)
+\Xi ^{\dagger }\left( \hat{X},\delta f_{1}\right) J\left( \hat{X},K_{\hat{X}%
},\mathbf{E}\right) +J^{\dagger }\left( \hat{X},K_{\hat{X}},\mathbf{E}%
\right) \Xi \left( \hat{X},\delta f_{1}\right) \\
&&-\bar{\Xi}^{\dagger }\left( \hat{X},\delta \bar{f}_{1}\right) \sigma
_{\delta f_{1}}^{2}\nabla _{\delta f_{1}}^{2}\bar{\Xi}\left( \hat{X},\delta 
\bar{f}_{1}\right) +\bar{\Xi}^{\dagger }\left( \hat{X},\delta \bar{f}%
_{1}\right) J\left( \hat{X},K_{\hat{X}},\mathbf{E}\right) +J^{\dagger
}\left( \hat{X},K_{\hat{X}},\mathbf{E}\right) \bar{\Xi}\left( \hat{X},\delta 
\bar{f}_{1}\right)
\end{eqnarray*}%
We have introduced a field $\bar{\Xi}\left( \hat{X},\delta \bar{f}%
_{1}\right) $ to account for excess returns in the banking sector, and a
potential $\bar{V}\left( \left[ \bar{\Psi}\right] \left( \left\{ \bar{K}%
_{r_{i}},\bar{X}_{r_{i}}\right\} _{G_{i}}\right) \right) $ to implement the
return equation for banks.

\subsubsection{Excess returns' equation and potential}

We assume that agents belonging to different groups and linked by return
equations (\ref{PML}) and (\ref{PMN}) are weakly connected. As before, their
interactions are modeled by intra-group potentials plus some inter-group
potentials.

Discarding defaults, the intra-group interaction potentials are obtained
through an expansion of (\ref{PML}) and (\ref{PMN}) by replacing $f_{a}^{%
\left[ i\right] }\rightarrow f_{a}^{\left[ i\right] }+\delta f_{a}^{\left[ i%
\right] }$ and $\bar{f}^{\left[ i\right] }\rightarrow \bar{f}^{\left[ i%
\right] }+\delta \bar{f}^{\left[ i\right] }$ and keeping diagonal terms. The
constraint (\ref{PML}) between excess returns of investors divided in
several blocks is:%
\begin{equation}
0=\left( 
\begin{array}{c}
\delta f^{\left[ i\right] \prime } \\ 
\delta f^{\left[ j\right] \prime }%
\end{array}%
\right) -\left( 
\begin{array}{cc}
\underline{\hat{S}}_{1}^{\left[ ii\right] } & \underline{\hat{S}}_{1}^{\left[
ji\right] } \\ 
\underline{\hat{S}}_{1}^{\left[ ij\right] } & \underline{\hat{S}}_{1}^{\left[
jj\right] }%
\end{array}%
\right) \left( 
\begin{array}{c}
\delta f^{\left[ i\right] }\frac{1-\left( \underline{\hat{S}}^{\left[ ii%
\right] }+\underline{\hat{S}}^{\left[ ij\right] }\right) }{1-\left( 
\underline{\hat{S}}_{1}^{\left[ ii\right] }+\underline{\hat{S}}_{1}^{\left[
ij\right] }\right) } \\ 
\delta f^{\left[ j\right] }\frac{1-\left( \underline{\hat{S}}^{\left[ jj%
\right] }+\underline{\hat{S}}^{\left[ ij\right] }\right) }{1-\left( 
\underline{\hat{S}}_{1}^{\left[ jj\right] }+\underline{\hat{S}}_{1}^{\left[
ji\right] }\right) }%
\end{array}%
\right)  \label{tn}
\end{equation}%
and the constraint (\ref{PMN}) between excess returns of banks divided in
several blocks is:%
\begin{equation}
0=\left( 
\begin{array}{cc}
1-\underline{\bar{S}}_{1}^{\left[ ii\right] } & -\underline{\bar{S}}_{1}^{%
\left[ ji\right] } \\ 
-\underline{\bar{S}}_{1}^{\left[ ij\right] } & 1-\underline{\bar{S}}_{1}^{%
\left[ jj\right] }%
\end{array}%
\right) \left( 
\begin{array}{c}
\frac{\delta \bar{f}^{\left[ i\right] }}{1+\underline{\bar{k}}^{\left[ i%
\right] }} \\ 
\frac{\delta \bar{f}^{\left[ j\right] }}{1+\underline{\bar{k}}^{\left[ j%
\right] }}%
\end{array}%
\right) -\left( 
\begin{array}{cc}
\underline{\hat{S}}_{2}^{\left[ ii\right] } & \underline{\hat{S}}_{\eta }^{%
\left[ ij\right] } \\ 
\underline{\hat{S}}_{\eta }^{\left[ ij\right] } & \underline{\hat{S}}_{2}^{%
\left[ jj\right] }%
\end{array}%
\right) \left( 
\begin{array}{c}
\frac{\delta \hat{f}^{\left[ i\right] }}{1+\underline{\hat{k}}_{2}^{\left[ i%
\right] }+\kappa \left[ \frac{\underline{\hat{k}}_{2}^{B}}{1+\bar{k}}\right]
^{\left[ i\right] }} \\ 
\frac{\delta \hat{f}^{\left[ j\right] }}{1+\underline{\hat{k}}_{2}^{\left[ j%
\right] }+\kappa \left[ \frac{\underline{\hat{k}}_{2}^{B}}{1+\bar{k}}\right]
^{\left[ j\right] }}%
\end{array}%
\right)  \label{tz}
\end{equation}%
Once again, only excess returns are considered, since we assume that
averages satisfy the constraint.

As in part $1$ it leads to the intra-group potential for investors' returns:%
\begin{equation}
\sum_{i}V_{i}\left( \hat{\Psi},\hat{X},K,\delta f_{1}\right) =\sum_{i}\delta %
\left[ \delta f_{1i}\left( \hat{X}^{\prime }\right) -\underline{\hat{S}}%
_{1}^{\left[ ii\right] }\frac{1-\left( \underline{\hat{S}}^{\left[ ii\right]
}+\underline{\hat{S}}^{\left[ ij\right] }\right) }{1-\left( \underline{\hat{S%
}}_{1}^{\left[ ii\right] }+\underline{\hat{S}}_{1}^{\left[ ij\right]
}\right) }\frac{\delta f_{1j}^{\prime }\left( \hat{X}^{\prime }\right) }{1+%
\underline{\hat{k}}_{2}\left( \hat{X}^{\prime }\right) }dX^{\prime }\right]
\label{NT}
\end{equation}%
and for banks' returns:%
\begin{equation}
\sum_{i}\bar{V}_{i}\left( \hat{\Psi},\hat{X},K,\delta \bar{f}^{\left[ i%
\right] }\right) =\sum_{i}\delta \left( \delta \bar{f}^{\left[ i\right] }-%
\underline{\bar{S}}_{1}^{\left[ ii\right] }\frac{\delta \bar{f}^{\left[ i%
\right] \prime }}{1+\underline{\bar{k}}^{\left[ i\right] }}-\underline{\hat{S%
}}_{1}^{\left[ ii\right] }\frac{\delta \hat{f}^{\left[ i\right] \prime }}{1+%
\underline{\hat{k}}_{2}^{\left[ i\right] }+\kappa \left[ \frac{\underline{%
\hat{k}}_{2}^{B}}{1+\bar{k}}\right] ^{\left[ i\right] }}\right)  \label{Ntb}
\end{equation}%
These two potentials (\ref{NT}) and (\ref{Ntb}) impose a propagation of
fluctuations $\delta \bar{f}^{\left[ i\right] \prime }$ or $\delta \hat{f}^{%
\left[ i\right] \prime }$ on the returns. When the interaction occurs, the
fluctuations propagate and induce new excess returns through the matrices $%
\underline{\hat{S}}_{1}^{\left[ ii\right] }$, $\underline{\bar{S}}_{1}^{%
\left[ ii\right] }$, $\underline{\hat{S}}_{2}^{\left[ ii\right] }$. Without
default, the fluctuations propagate through the participations.

The inter-group interaction potentials, without default, are obtained by
keeping the off-diagonal terms in (\ref{tn}) and (\ref{tz}), which translate
into the following potentials, for investors:%
\begin{eqnarray}
&&\sum_{i}W_{i}\left( \hat{\Psi},\hat{X},K,\delta f_{1i}\right)  \label{nr}
\\
&=&\sum_{i}\delta \left( \delta f^{\left[ i\right] }-\underline{\hat{S}}%
_{1}^{\left[ ij\right] }\frac{1-\left( \underline{\hat{S}}^{\left[ jj\right]
}+\underline{\hat{S}}^{\left[ ij\right] }\right) }{1-\left( \underline{\hat{S%
}}_{1}^{\left[ jj\right] }+\underline{\hat{S}}_{1}^{\left[ ji\right]
}\right) }\delta f^{\left[ j\right] \prime }\right)  \notag
\end{eqnarray}%
and for banks:%
\begin{eqnarray}
&&\sum_{i}\bar{W}\left( \hat{\Psi},\hat{X},K,\delta \bar{f}^{\left[ i\right]
}\right)  \label{nb} \\
&=&\sum_{i}\delta \left( \delta \bar{f}^{\left[ i\right] }-\underline{\bar{S}%
}_{1}^{\left[ ji\right] }\frac{\delta \bar{f}^{\left[ j\right] \prime }}{%
\left( 1+\underline{\hat{k}}_{2}^{\left[ j\right] }\right) }-\underline{\hat{%
S}}_{1}^{\left[ ij\right] }\frac{\delta \hat{f}^{\left[ j\right] \prime }}{%
\left( 1+\underline{\hat{k}}_{2}^{\left[ i\right] }+\kappa \left[ \frac{%
\underline{\hat{k}}_{2}^{B}}{1+\bar{k}}\right] ^{\left[ i\right] }\right) }%
\right)  \notag
\end{eqnarray}%
This is simlar to the inter-group constraints, with the exception that only
different groups interact. The interaction between these groups can be
considered relatively weak, so that the fluctuations coefficients $%
\underline{\bar{S}}_{1}^{\left[ ji\right] }$ and $\underline{\hat{S}}_{1}^{%
\left[ ij\right] }$ are relatively small.

\subsection{Transition functions}

\subsubsection{Transition functions without interactions}

The method is similar to Part $1$. The transition functions without
interactions for these agents are computed by inverting the operator, as in (%
\ref{rk}), for the dynamics of investor capital:%
\begin{equation*}
-\nabla _{\hat{K}}\left( \frac{\sigma _{\hat{K}}^{2}}{2}\nabla _{\hat{K}}-%
\hat{K}f\left( \hat{X},K_{\hat{X}}\right) +\int \delta f_{1}\left\vert \Xi
\left( \hat{X},\delta f_{1}\right) \right\vert ^{2}d\left( \delta
f_{1}\right) \right)
\end{equation*}%
and for the dynamics of banks:%
\begin{equation*}
-\nabla _{\bar{K}}\left( \frac{\sigma _{\bar{K}}^{2}}{2}\nabla _{\bar{K}}-%
\bar{K}f\left( \bar{X},K_{\bar{K}}\right) +\int \delta \bar{f}_{1}\left\vert
\Xi \left( \bar{X},\delta \bar{f}_{1}\right) \right\vert ^{2}d\left( \delta 
\bar{f}_{1}\right) \right)
\end{equation*}%
This yields the partial Green function conditioned to the initial state and
final state for returns as:%
\begin{eqnarray}
\prod\limits_{a=1,2} &&\sqrt{\left\vert \frac{f_{a}^{\left[ i\right] }+%
\frac{\delta f_{a}^{\left[ i\right] }+\delta f_{a}^{\left[ i\right] \prime }%
}{2}}{\sigma _{\hat{K}}^{2}\left( 1-\exp \left( 2\left( f_{a}^{\left[ i%
\right] }+\frac{\delta f_{a}^{\left[ i\right] }+\delta f_{a}^{\left[ i\right]
\prime }}{2}\right) \Delta t\right) \right) }\right\vert } \\
&&\times \exp \left( \left( f_{a}^{\left[ i\right] }+\frac{\delta f_{a}^{%
\left[ i\right] }+\delta f_{a}^{\left[ i\right] \prime }}{2}\right) \frac{%
\left( K_{a}-\exp \left( \left( f_{a}^{\left[ i\right] }+\frac{\delta f_{a}^{%
\left[ i\right] }+\delta f_{a}^{\left[ i\right] \prime }}{2}\right) \Delta
t\right) \hat{K}^{\prime }\right) ^{2}}{\sigma _{\hat{K}}^{2}\left( 1-\exp
\left( 2\left( f_{a}^{\left[ i\right] }+\frac{\delta f_{a}^{\left[ i\right]
}+\delta f_{a}^{\left[ i\right] \prime }}{2}\right) \Delta t\right) \right) }%
\right)  \notag \\
&&\times \exp \left( \frac{\left( f_{a}^{\left[ i\right] }+\frac{\delta
f_{a}^{\left[ i\right] }+\delta f_{a}^{\left[ i\right] \prime }}{2}\right) }{%
\sigma _{\hat{K}}^{2}\left( 1-\exp \left( 2\left( f_{a}^{\left[ i\right] }+%
\frac{\delta f_{a}^{\left[ i\right] }+\delta f_{a}^{\left[ i\right] \prime }%
}{2}\right) \Delta t\right) \right) }\left( \hat{K}-\exp \left( \left(
f_{a}^{\left[ i\right] }+\frac{\delta f_{a}^{\left[ i\right] }+\delta f_{a}^{%
\left[ i\right] \prime }}{2}\right) \Delta t\right) \hat{K}^{\prime }\right)
^{2}\right)  \notag
\end{eqnarray}%
with:%
\begin{eqnarray*}
K_{1} &=&\hat{K} \\
K_{2} &=&\bar{K}
\end{eqnarray*}%
We proceed in the same manner for the excess returns, and we invert the
operator:%
\begin{equation*}
-\Xi ^{\dagger }\left( \hat{X},\delta f_{a1}\right) \sigma _{\delta
f_{1}}^{2}\nabla _{\delta f_{1}}^{2}\Xi \left( \hat{X},\delta f_{a1}\right)
+\Xi ^{\dagger }\left( \hat{X},\delta f_{a1}\right) J\left( \hat{X},K_{\hat{X%
}},\mathbf{E}\right) +J\left( \hat{X},K_{\hat{X}},\mathbf{E}\right) \Xi
\left( \hat{X},\delta f_{a1}\right)
\end{equation*}%
and we obtain the partial transition function:%
\begin{equation}
\prod\limits_{a=1,2}\sqrt{\frac{1}{\sigma _{\delta f_{1}}^{2}}}\exp \left( -%
\frac{\left( \delta f_{a}^{\left[ i\right] }-\delta f_{a}^{\left[ i\right]
\prime }\right) ^{2}}{\sigma _{\delta f_{1}}^{2}}+J\left( \hat{X},K_{\hat{X}%
},\mathbf{E}\right) \delta f_{a}^{\left[ i\right] }-J\left( \hat{X}^{\prime
},K_{\hat{X}^{\prime }},\mathbf{E}^{\prime }\right) \delta f_{a}^{\left[ i%
\right] \prime }\right)
\end{equation}%
So that the transition functions for an agent between a state with capital $%
K $ and return $\delta f_{1}$ to a state with capital $K^{\prime }$ and
return $\delta f_{1}^{\prime }$ are:%
\begin{eqnarray}
&&\left\langle \left( K_{a},\delta f_{a}^{\left[ i\right] }\right)
_{a}\right. \left. \left( K_{a},\delta f_{a}^{\left[ i\right] \prime
}\right) _{a}\right\rangle  \label{TRn} \\
&=&\prod\limits_{a=1,2}\sqrt{\left\vert \frac{f_{a}^{\left[ i\right] }+%
\frac{\delta f_{a}^{\left[ i\right] }+\delta f_{a}^{\left[ i\right] \prime }%
}{2}}{\sigma _{\hat{K}}^{2}\left( 1-\exp \left( 2\left( f_{a}^{\left[ i%
\right] }+\frac{\delta f_{a}^{\left[ i\right] }+\delta f_{a}^{\left[ i\right]
\prime }}{2}\right) \Delta t\right) \right) }\right\vert }  \notag \\
&&\times \exp \left( \left( f_{a}^{\left[ i\right] }+\frac{\delta f_{a}^{%
\left[ i\right] }+\delta f_{a}^{\left[ i\right] \prime }}{2}\right) \frac{%
\left( \hat{K}-\exp \left( \left( f_{a}^{\left[ i\right] }+\frac{\delta
f_{a}^{\left[ i\right] }+\delta f_{a}^{\left[ i\right] \prime }}{2}\right)
\Delta t\right) \hat{K}^{\prime }\right) ^{2}}{\sigma _{\hat{K}}^{2}\left(
1-\exp \left( 2\left( f_{a}^{\left[ i\right] }+\frac{\delta f_{a}^{\left[ i%
\right] }+\delta f_{a}^{\left[ i\right] \prime }}{2}\right) \Delta t\right)
\right) }\right)  \notag \\
&&\times \exp \left( \frac{\left( f_{a}^{\left[ i\right] }+\frac{\delta
f_{a}^{\left[ i\right] }+\delta f_{a}^{\left[ i\right] \prime }}{2}\right) }{%
\sigma _{\hat{K}}^{2}\left( 1-\exp \left( 2\left( f_{a}^{\left[ i\right] }+%
\frac{\delta f_{a}^{\left[ i\right] }+\delta f_{a}^{\left[ i\right] \prime }%
}{2}\right) \Delta t\right) \right) }\left( \hat{K}-\exp \left( \left(
f_{a}^{\left[ i\right] }+\frac{\delta f_{a}^{\left[ i\right] }+\delta f_{a}^{%
\left[ i\right] \prime }}{2}\right) \Delta t\right) \hat{K}^{\prime }\right)
^{2}\right)  \notag \\
&&\times \prod\limits_{a=1,2}\sqrt{\frac{1}{\sigma _{\delta f_{1}}^{2}}}%
\exp \left( -\frac{\left( \delta f_{a}^{\left[ i\right] }-\delta f_{a}^{%
\left[ i\right] \prime }\right) ^{2}}{\sigma _{\delta f_{1}}^{2}}+J\left( 
\hat{X},K_{\hat{X}},\mathbf{E}\right) \delta f_{a}^{\left[ i\right]
}-J\left( \hat{X}^{\prime },K_{\hat{X}^{\prime }},\mathbf{E}^{\prime
}\right) \delta f_{a}^{\left[ i\right] \prime }\right)  \notag \\
&\equiv &G\left( \left( K_{a},\delta f_{a}^{\left[ i\right] }\right)
_{a},\left( K_{a},\delta f_{a}^{\left[ i\right] \prime }\right) _{a}\right)
\end{eqnarray}%
As in Part $1$, the transitions for a group of agents, investors, and banks,
without interactions is the product of terms of the form (\ref{TRn}):%
\begin{equation*}
\prod\limits_{l}\left\langle \left( K_{al},\delta f_{al}^{\left[ i\right]
}\right) _{al}\right. \left. \left( K_{al},\delta f_{al}^{\left[ i\right]
\prime }\right) _{al}\right\rangle =\prod\limits_{l}G\left( \left(
K_{al},\delta f_{al}^{\left[ i\right] }\right) _{al},\left( K_{al},\delta
f_{al}^{\left[ i\right] \prime }\right) _{al}\right)
\end{equation*}

\subsubsection{Transition functions with interactions without default}

The transitions are then driven both by intra-group (\ref{NT}) and (\ref{Ntb}%
), and inter-group (\ref{nr}) and (\ref{nb}) interactions. The potentials (%
\ref{NT}) and (\ref{Ntb})\ are similar to those for a homogeneous group and
illustrate how, within the same group, the excess returns of one agent
impact those of others. Even an agent who has not initially undergone a
return variation will experience one subsequently, due to the interaction
term. The potentials (\ref{nr}) and (\ref{nb}) , on the other hand, show the
transmission of a return variation from one block to another.

As for one homogeneous group, the transitions are computed by series of
terms of the type:

\begin{eqnarray}
&&G\left( \Delta t,f_{1,1},\left\{ \left( K_{a_{1}},\delta f_{a_{1}}^{\left[
i\right] }\right) _{a_{1}}\right\} _{G_{1}},\left\{ \left( K_{a_{1}}^{\prime
},\delta f_{a_{1}}^{\left[ i\right] \prime }\right) _{a_{1}}\right\}
_{G_{1}}\right) ...  \label{DVT} \\
&&\times G\left( \Delta t,f_{1,n},\left\{ \left( K_{a_{n}},\delta
f_{a_{n}}^{ \left[ i\right] }\right) _{a_{n}}\right\} _{G_{n}},\left\{
\left( K_{a_{n}}^{\prime },\delta f_{a_{n}}^{\left[ i\right] \prime }\right)
_{a_{n}}\right\} _{G_{n}}\right)  \notag \\
&&\times \left( V\left( \left( \left[ \hat{\Psi}\right] \left( \left\{ \hat{K%
}_{r_{i}},\hat{X}_{r_{i}}\right\} _{G_{i}}\right) \right) \right) \right) 
\notag \\
&&\times G\left( \Delta t,f_{1,1},\left\{ \left( K_{a_{1}}^{\prime },\delta
f_{a_{1}}^{\left[ i\right] \prime }\right) _{a_{1}}\right\} _{G_{1}},\left\{
\left( K_{a_{1}}^{\prime \prime },\delta f_{a_{1}}^{\left[ i\right] \prime
\prime }\right) _{a_{1}}\right\} _{G_{1}}\right) ...  \notag \\
&&\times G\left( \Delta t,f_{1,n},\left\{ \left( K_{a_{n}}^{\prime \prime
},\delta f_{a_{n}}^{\left[ i\right] \prime \prime }\right) _{a_{n}}\right\}
_{G_{n}},\left\{ \left( K_{a_{n}}^{\prime \prime },\delta f_{a_{n}}^{\left[ i%
\right] \prime \prime }\right) _{a_{n}}\right\} _{G_{n}}\right)  \notag
\end{eqnarray}%
where:%
\begin{equation*}
V\left( \left( \left[ \hat{\Psi}\right] \left( \left\{ \hat{K}_{r_{i}},\hat{X%
}_{r_{i}}\right\} _{G_{i}}\right) \right) \right)
\end{equation*}%
stands for the intra- and inter-group interactions (\ref{NT}) and (\ref{Ntb}%
), and inter-group (\ref{nr}) and (\ref{nb}). The term (\ref{DVT}) computes
the diviation from the free transition function if we assume a punctual
interaction.

The full transition function is thus the infinite series of products of the
form (\ref{DVT}), written compactly as:%
\begin{equation*}
\sum_{n}a_{n}\left[ G...G\right] V_{n}\left[ G...G\right] +.....\left[ G...G%
\right] V_{n}\left[ G...G\right] V_{n}...\left[ G...G\right] V_{n}\left[
G...G\right]
\end{equation*}%
where the terms $a_{n}$ come from the expansion of the interaction potential
in the effective action. Their specific form is irrelevant here. Again, in
this compact formulation, $V_{n}$ can take the form (\ref{NT}) and (\ref{Ntb}%
), and inter-group (\ref{nr}) and (\ref{nb}).

\subsubsection{Interactions including defaults}

As in Part $1$, some defaults may arise in the process of transition. The
default potentials:%
\begin{equation*}
V_{i}^{D}\left( \hat{\Psi},\hat{X},K,\delta f_{1}\right)
\end{equation*}%
and:%
\begin{equation*}
V_{ij}^{D}\left( \hat{\Psi},\hat{X},K,\delta f_{1}\right)
\end{equation*}%
are derived straightforwardly by including the terms describing the defaults
of investors and banks in intra-group (\ref{NT}) and (\ref{Ntb}), and in
inter-group (\ref{nr}) and (\ref{nb}) potentials. To do so, we rewrite the
constraint for investors with default: 
\begin{eqnarray}
0 &=&\left( 
\begin{array}{c}
\delta \hat{f}^{\left[ i\right] } \\ 
\delta \hat{f}^{\left[ j\right] }%
\end{array}%
\right) -\left( 
\begin{array}{cc}
\underline{\hat{S}}_{1}^{\left[ ii\right] } & \underline{\hat{S}}_{1}^{\left[
ji\right] } \\ 
\underline{\hat{S}}_{1}^{\left[ ij\right] } & \underline{\hat{S}}_{1}^{\left[
jj\right] }%
\end{array}%
\right) \left( 
\begin{array}{c}
\delta \hat{f}^{\left[ i\right] \prime }\frac{1-\left( \underline{\hat{S}}^{%
\left[ ii\right] }+\underline{\hat{S}}^{\left[ ij\right] }\right) }{1-\left( 
\underline{\hat{S}}_{1}^{\left[ ii\right] }+\underline{\hat{S}}_{1}^{\left[
ij\right] }\right) } \\ 
\delta \hat{f}^{\left[ j\right] \prime }\frac{1-\left( \underline{\hat{S}}^{%
\left[ jj\right] }+\underline{\hat{S}}^{\left[ ij\right] }\right) }{1-\left( 
\underline{\hat{S}}_{1}^{\left[ jj\right] }+\underline{\hat{S}}_{1}^{\left[
ji\right] }\right) }%
\end{array}%
\right) \\
&&-\left( 
\begin{array}{cc}
\underline{\hat{S}}_{2}^{\left[ ii\right] } & \underline{\hat{S}}_{2}^{\left[
ji\right] } \\ 
\underline{\hat{S}}_{2}^{\left[ ij\right] } & \underline{\hat{S}}_{2}^{\left[
jj\right] }%
\end{array}%
\right) \left( 
\begin{array}{c}
\left( 1+\hat{f}^{\left[ i\right] \prime }+\delta \hat{f}^{\left[ i\right]
\prime }\right) \underline{\hat{s}}_{2}^{\left[ i\right] }H\left( -\left( 1+%
\hat{f}^{\left[ i\right] \prime }+\delta \hat{f}^{\left[ i\right] \prime
}\right) \right) \\ 
\left( 1+\hat{f}^{\left[ j\right] \prime }+\delta \hat{f}^{\left[ j\right]
\prime }\right) \underline{\hat{s}}_{2}^{\left[ i\right] }H\left( -\left( 1+%
\hat{f}^{\left[ j\right] \prime }+\delta \hat{f}^{\left[ j\right] \prime
}\right) \right)%
\end{array}%
\right)  \notag
\end{eqnarray}%
Similarly, we rewrite the constraint for banks (\ref{PMN}) by including
defaults:%
\begin{eqnarray}
0 &=&\left( 
\begin{array}{cc}
1-\underline{\bar{S}}_{1}^{\left[ ii\right] } & -\underline{\bar{S}}_{1}^{%
\left[ ji\right] } \\ 
-\underline{\bar{S}}_{1}^{\left[ ij\right] } & 1-\underline{\bar{S}}_{1}^{%
\left[ jj\right] }%
\end{array}%
\right) \left( 
\begin{array}{c}
\frac{\delta \bar{f}^{\left[ i\right] }}{1+\underline{\bar{k}}^{\left[ i%
\right] }} \\ 
\frac{\delta \bar{f}^{\left[ j\right] }}{1+\underline{\bar{k}}^{\left[ j%
\right] }}%
\end{array}%
\right) -\left( 
\begin{array}{cc}
\underline{\hat{S}}_{2}^{\left[ ii\right] } & \underline{\hat{S}}_{\eta }^{%
\left[ ij\right] } \\ 
\underline{\hat{S}}_{\eta }^{\left[ ij\right] } & \underline{\hat{S}}_{2}^{%
\left[ jj\right] }%
\end{array}%
\right) \left( 
\begin{array}{c}
\frac{\delta \hat{f}^{\left[ i\right] }}{1+\underline{\hat{k}}_{2}^{\left[ i%
\right] }+\kappa \left[ \frac{\underline{\hat{k}}_{2}^{B}}{1+\bar{k}}\right]
^{\left[ i\right] }} \\ 
\frac{\delta \hat{f}^{\left[ j\right] }}{1+\underline{\hat{k}}_{2}^{\left[ j%
\right] }+\kappa \left[ \frac{\underline{\hat{k}}_{2}^{B}}{1+\bar{k}}\right]
^{\left[ j\right] }}%
\end{array}%
\right) \\
&&-\left( 
\begin{array}{cc}
\underline{\bar{S}}_{2}^{\left[ ii\right] } & \underline{\bar{S}}_{2}^{\left[
ji\right] } \\ 
\underline{\bar{S}}_{2}^{\left[ ij\right] } & \underline{\bar{S}}_{2}^{\left[
jj\right] }%
\end{array}%
\right) \left( 
\begin{array}{c}
\left( 1+\bar{f}^{\left[ i\right] }+\delta \bar{f}^{\left[ i\right] }\right) 
\underline{\hat{s}}_{2}^{\left[ i\right] }H\left( -\left( 1+\bar{f}^{\left[ i%
\right] }+\delta \bar{f}^{\left[ i\right] }\right) \right) \\ 
\left( 1+\bar{f}^{\left[ j\right] }+\delta \bar{f}^{\left[ j\right] }\right) 
\underline{\hat{s}}_{2}^{\left[ i\right] }H\left( -\left( 1+\bar{f}^{\left[ j%
\right] }+\delta \bar{f}^{\left[ j\right] }\right) \right)%
\end{array}%
\right)  \notag \\
&&-\left( 
\begin{array}{cc}
\underline{\hat{S}}_{2}^{B^{\prime }\left[ ii\right] } & \underline{\hat{S}}%
_{2}^{B^{\prime }\left[ ji\right] } \\ 
\underline{\hat{S}}_{2}^{B^{\prime }\left[ ij\right] } & \underline{\hat{S}}%
_{2}^{B^{\prime }\left[ jj\right] }%
\end{array}%
\right) \left( 
\begin{array}{c}
\left( 1+\hat{f}^{\left[ i\right] \prime }+\delta \hat{f}^{\left[ i\right]
\prime }\right) \underline{\hat{s}}_{2}^{\left[ i\right] }H\left( -\left( 1+%
\hat{f}^{\left[ i\right] \prime }+\delta \hat{f}^{\left[ i\right] \prime
}\right) \right) \\ 
\left( 1+\hat{f}^{\left[ j\right] \prime }+\delta \hat{f}^{\left[ j\right]
\prime }\right) \underline{\hat{s}}_{2}^{\left[ i\right] }H\left( -\left( 1+%
\hat{f}^{\left[ j\right] \prime }+\delta \hat{f}^{\left[ j\right] \prime
}\right) \right)%
\end{array}%
\right)  \notag
\end{eqnarray}%
And similarly to Part $1$, we obtain the intra-group potentials with
defaults for investors:%
\begin{eqnarray}
\sum_{i}V_{i}^{D}\left( \hat{\Psi},\hat{X},K,\delta f_{1}\right)
&=&\sum_{i}\delta \left[ \delta f_{1i}\left( \hat{X}^{\prime }\right) -%
\underline{\hat{S}}_{1}^{\left[ ii\right] }\frac{1-\left( \underline{\hat{S}}%
^{\left[ ii\right] }+\underline{\hat{S}}^{\left[ ij\right] }\right) }{%
1-\left( \underline{\hat{S}}_{1}^{\left[ ii\right] }+\underline{\hat{S}}%
_{1}^{\left[ ij\right] }\right) }\frac{\delta f_{1i}^{\prime }\left( \hat{X}%
^{\prime }\right) }{1+\underline{\hat{k}}_{2}\left( \hat{X}^{\prime }\right) 
}dX^{\prime }\right.  \label{tnz} \\
&&\left. -\underline{\hat{S}}_{2}^{\left[ ii\right] }\left( \left( 1+\hat{f}%
^{\left[ i\right] \prime }\left( \hat{X}^{\prime }\right) +\delta \hat{f}^{%
\left[ i\right] \prime }\left( \hat{X}^{\prime }\right) \right) \underline{%
\hat{s}}_{2}^{\left[ i\right] }H\left( -\left( 1+\hat{f}^{\left[ i\right]
\prime }\left( \hat{X}^{\prime }\right) +\delta \hat{f}^{\left[ i\right]
\prime }\left( \hat{X}^{\prime }\right) \right) \right) \right) dX^{\prime }%
\right]  \notag
\end{eqnarray}%
and for banks:%
\begin{eqnarray}
\sum_{i}\bar{V}_{i}^{D}\left( \hat{\Psi},\hat{X},K,\delta \bar{f}^{\left[ i%
\right] }\right) &=&\sum_{i}\delta \left[ \delta \bar{f}^{\left[ i\right]
\prime }-\underline{\bar{S}}_{1}^{\left[ ii\right] }\frac{\delta \bar{f}^{%
\left[ i\right] \prime }}{1+\underline{\bar{k}}^{\left[ i\right] }}-%
\underline{\hat{S}}_{i}^{\left[ ii\right] }\frac{\delta \hat{f}^{\left[ i%
\right] \prime }}{1+\underline{\hat{k}}_{2}^{\left[ i\right] }+\kappa \left[ 
\frac{\underline{\hat{k}}_{2}^{B}}{1+\bar{k}}\right] ^{\left[ i\right] }}%
\right.  \label{tbz} \\
&&-\underline{\bar{S}}_{2}^{\left[ ii\right] }\left( 1+\bar{f}^{\left[ i%
\right] }+\delta \bar{f}^{\left[ i\right] }\right) \underline{\hat{s}}_{2}^{%
\left[ i\right] }H\left( -\left( 1+\bar{f}^{\left[ i\right] }+\delta \bar{f}%
^{\left[ i\right] }\right) \right)  \notag \\
&&\left. -\underline{\hat{S}}_{2}^{B^{\prime }\left[ ii\right] }\left( 1+%
\hat{f}^{\left[ i\right] \prime }+\delta \hat{f}^{\left[ i\right] \prime
}\right) \underline{\hat{s}}_{2}^{\left[ i\right] }H\left( -\left( 1+\hat{f}%
^{\left[ i\right] \prime }+\delta \hat{f}^{\left[ i\right] \prime }\right)
\right) \right]  \notag
\end{eqnarray}%
And the inter-group potentials for investors:%
\begin{eqnarray}
&&\sum_{i}W_{i}^{D}\left( \hat{\Psi},\hat{X},K,\delta f_{1i}\right)
\label{nrz} \\
&=&\sum_{i}\delta \left[ \delta \bar{f}^{\left[ i\right] }-\underline{\bar{S}%
}_{1}^{\left[ ji\right] }\frac{\delta \bar{f}^{\left[ j\right] \prime }}{%
\left( 1+\underline{\hat{k}}_{2}^{\left[ j\right] }\right) }-\underline{\hat{%
S}}_{1}^{\left[ ij\right] }\frac{\delta \hat{f}^{\left[ j\right] \prime }}{%
\left( 1+\underline{\hat{k}}_{2}^{\left[ i\right] }+\kappa \left[ \frac{%
\underline{\hat{k}}_{2}^{B}}{1+\bar{k}}\right] ^{\left[ i\right] }\right) }%
\right.  \notag \\
&&\left. \left. -\underline{\hat{S}}_{2}^{\left[ ij\right] }\left( \left( 1+%
\hat{f}^{\left[ j\right] \prime }\left( \hat{X}^{\prime }\right) +\delta 
\hat{f}^{\left[ j\right] \prime }\left( \hat{X}^{\prime }\right) \right) 
\underline{\hat{s}}_{2}^{\left[ i\right] }H\left( -\left( 1+\hat{f}^{\left[ j%
\right] \prime }\left( \hat{X}^{\prime }\right) +\delta \hat{f}^{\left[ j%
\right] \prime }\left( \hat{X}^{\prime }\right) \right) \right) \right)
dX^{\prime }\right] \right]  \notag
\end{eqnarray}%
and for banks:%
\begin{eqnarray}
&&\sum_{i}\bar{W}^{D}\left( \hat{\Psi},\hat{X},K,\delta \bar{f}^{\left[ i%
\right] }\right)  \label{nbz} \\
&=&\sum_{i}\delta \left[ \delta \bar{f}^{\left[ i\right] }-\underline{\bar{S}%
}_{1}^{\left[ ji\right] }\frac{\delta \bar{f}^{\left[ j\right] \prime }}{%
2\left( 1+\underline{\hat{k}}_{2}^{\left[ j\right] }\right) }-\underline{%
\hat{S}}_{1}^{\left[ ij\right] }\frac{\delta \hat{f}^{\left[ j\right] \prime
}}{\left( 1+\underline{\hat{k}}_{2}^{\left[ j\right] }+\kappa \left[ \frac{%
\underline{\hat{k}}_{2}^{B}}{1+\bar{k}}\right] ^{\left[ j\right] }\right) }%
\right.  \notag \\
&&-\underline{\bar{S}}_{2}^{\left[ ii\right] }\left( 1+\bar{f}^{\left[ i%
\right] }+\delta \bar{f}^{\left[ i\right] }\right) \underline{\hat{s}}_{2}^{%
\left[ i\right] }H\left( -\left( 1+\bar{f}^{\left[ i\right] }+\delta \bar{f}%
^{\left[ i\right] }\right) \right)  \notag \\
&&\left. -\underline{\hat{S}}_{2}^{B^{\prime }\left[ ii\right] }\left( 1+%
\hat{f}^{\left[ i\right] \prime }+\delta \hat{f}^{\left[ i\right] \prime
}\right) \underline{\hat{s}}_{2}^{\left[ i\right] }H\left( -\left( 1+\hat{f}%
^{\left[ i\right] \prime }+\delta \hat{f}^{\left[ i\right] \prime }\right)
\right) \right]  \notag
\end{eqnarray}

\subsubsection{Transition functions with interactions including defaults}

Considering the possibilities of defauls the transition may thus occur from
two possible effects.

First a sequence of interaction:%
\begin{equation}
\left[ G...G\right] V_{n}\left[ G...G\right] +.....\left[ G...G\right] V_{n}%
\left[ G...G\right] V_{n}...\left[ G...G\right] V_{n}\left[ G...G\right]
\label{GP1}
\end{equation}%
with $V_{n}$ describes transitions between agents of several groups, may
drive some agents to default.

The following transitions are thus driven by terms like:%
\begin{equation}
\left[ G...G\right] V_{n}^{D}\left[ G...G\right] +.....\left[ G...G\right]
V_{n}^{D}\left[ G...G\right] V_{n}^{D}...\left[ G...G\right] V_{n}\left[
G...G\right]  \label{GP2}
\end{equation}%
which increases the number of default in group.

Then through inter-groups potential the default propagates to agents of
different groups.

\subsubsection{Interpretation}

While the mechanisms are similar to the system without banks, several
differences arise.

Firstly, as quoted above in section 22, the diffusion coefficients are
higher when banks are included, since bank loans increase the disposable
capital of investors. However, when no defaults arise, the transmission
coefficients for banks are low: if we consider that the main activity for
banks is loans at fixed rates, the excess returns do not affect these
agents, and they do not transmit fluctuations in returns. As a consequence,
the presence of banks reduces the effect of transmission. As in the study of
collective states, the larger the number of banks, the more we can expect
the increase in diffusion to be offset. The situation is different if banks
take large participations in investors and thus act as investors.

The situation is different if some defaults arise. As large lenders, we can
expect the banks to amplify the loss of return when defaults occur, which
increases the cascade of defaults. In this case, since banks are
interconnected, depending on the level of interconnections, any default may
propagate to banks. Thus, if banks seem to stabilize the system and reduce
the risk of default when loans are their main activity, they may amplify the
defaults when these occur.

\section{Discussion}

Field economics reveals the collective states that characterize an economic
model. A collective state represents a set of significant data describing a
model, and each model may give rise to one or several collective states. In
our model, these significant are, for each and every sector, their average
capital, number of agents and average return. This collection of data
characterizes one collective state.

Each collective state is thus defined by a specific set of levels of
capital, number of agents, and sector average returns. These data define an
overall average capital, number of agents and return. However, the converse
is not true: for any given average level of capital, agents, and returns in
an economy, multiple collective states may exist.

These collective states serve as static descriptions of the system's
condition. Within any given economic model, the number of collective states
determines the system's stability or instability, as well as the transitions
between stable and unstable states. Viewed in this context, collective
states represent the intrinsic characteristics of the system---its various
manifestations under specific conditions, whether temporary or structural.

They are not economic equilibria per se, rather potential states that the
economy may experience. Any global state among the set of possible
collective states may take place, but is bound to be replaced by another
global state realizing another collective state. This change is
discontinuous: shifts occur abruptly, and the economy switches from one
collective state to another without transition. However this shift will not
necessarily result in an immediate modification of individual dynamics.

Because of the nature of the collective state, the analysis can either focus
on the full extension of the collective state, or concentrate on how a
specific realization of a collective state impacts the individual dynamics
of the agents. Therefore, we do not study a dynamic transition between
collective states, but rather how the agents' individual dynamics
materialize the changes between collective states.

In this paper, we conducted both studies, starting with collective states,
as they influence the individual dynamics within them. We first assumed a
baseline model consisting of two types of agents, investors and firms,
before turning to a more complex analysis involving three types of agents,
firms, investors, and banks.

In the two-type model of agents, investors accumulate and allocate capital
through equity stakes and successive lending activities. The complex
interconnections in loan activities highlight the critical role of leverage
in capital allocation, and any change in returns or defaults within one
sector can swiftly spread to other sectors. In this scenario, the system's
instability arises from the multitude of collective states characterizing
the system, and merely reflects the transitions between these multiple
states.

The study of collective states reveals that several global averages can
emerge, spanning from scenarios characterized by high total capital, a large
number of investors, and relatively low average returns to situations with
lower total capital, few investors, but significant returns. Each of these
global averages can be associated with multiple collective states: for any
given level of global capital, it can be allocated across sectors in
numerous ways, each allocation defining a unique realization of the
collective state. Consequently, a particular level of overall wealth can
manifest in the form of multiple distinct collective states, each
representing a different distribution of capital across sectors, ranging
from more evenly distributed to more concentrated configurations.

Collective averages can be associated to multiple collective states, some of
which may involve defaults. A default occurs when agents in a sector
experience returns that are insufficient to cover their debts. This default
state within a sector is structural in nature: it is bound to occur within
the global collective state. Whatever the individual dynamics within the
sector may be, the likelihood of realizing the structural default state will
increase over time unless the collective state shifts to another collective
state.

The collective state ought to be thought in its final and full extension, in
which all possible defaults have already spread to other sectors. It is a
static state describing the final and stabilized situation in its maximum
extent, in which some sectors may see their level of capital affected, even
if they do not necessarily default.

Collective states are therefore structural situations that condition the
individual dynamics of agents, depending on their sector. However,
individual dynamics exhibit persistence, so that a change of collective
state may not immediately translate into the dynamics of agents within the
sector. This accounts for the potential delays in the transition of agents
from one collective state to the other.

Within a given collective state, it is possible to examine the dynamics of
agents. In our model, since firms solely engage in production and their
available capital is directly proportional to their private capital, we have
focused on analyzing the dynamics of firms' private capital. However, we
have opted to study the dynamics of investors' disposable capital as this
variable plays a critical role in their investment decisions and in the
propagation of leverage effects.

A firm can default either directly, following an adverse shock, or
indirectly, due to the default of some of its investors. When capital is
depleted, the firm becomes unable to meet its expenses.

Investors can also default, either directly through poor investments or
indirectly through loans and stakes taken from other investors. These
defaults can spread and transmit sometimes complex, indirect and even
spiraling effects. Negative returns or returns lower than investors'
expectations in a given sector can either prompt their partners to reduce
their participation or induce an additional decrease that will indirectly
impact investors' results through interconnectedness. Investors operating in
a sector on the brink of default may default due to their ties with other
investors, even if there initially appeared to be no imminent default risk.
This indirect negative return has the potential to magnify the initial loss
and eventually push investors over the edge, leading to a domino effect,
boomerang effect, or liquidity crisis. This scenario underscores the
multiplicity of collective states: the collective state represents the final
stage where the chain of defaults has fully transpired, while the initial
situation characterized by apparent fragility rather than actual defaults
marks the beginning of a dynamic within a new collective state featuring
defaults.

Our model thus allows us to study the default mechanisms of both types of
agents: firms and investors. But it also facilitates the analysis of
amplification effects and systemic risk within the system. In our model, all
investments are underpinned by a level of private capital. Returns are
amplified by leverage effects that are only available to investors. This
structural discrepancy between firms and investors means that the latter
have the potential to generate higher returns than the former. When an
investor defaults, they not only lose their private capital but also forfeit
all the stakes and loans extended to them. These stakes and loans, in turn,
represent disposable capital, which is derived from equity. Therefore,
disposable capital essentially condenses private capital. Consequently, when
there is a default on disposable capital, which serves as a condensation of
private capital, the loss impacts all those involved. Hence, there is a
diffusion of the loss throughout the system.

It is when we delve into averages and the diffusion matrix that Field
Economics truly reveals its value. Our findings unveil rather complex
phenomena that can be understood both globally, with their macroeconomic
implications, and at a microeconomic level, when examining individuals
within the collective state. Our formulas for the diffusion matrix, capital
levels, and the consequences of leverage in investor relations demonstrate
that we can isolate successive effects on factors that might otherwise be
overlooked, providing detailed insights into the mechanisms of financial
risk transmission and contagion, as well as capital diffusion. For instance,
the diffusion matrix is localized, from sector to sector, meaning that
diffusion effects depend on the path taken. By considering varying
connections in the sector space, one could study different transmission
patterns.

Because Field Economics allows for the introduction of as many types of
agents as desired, simply by adding fields to represent them, we were able
to distinguish the role of banks from other investors, leading to the
differentiation of diffusion matrices for investor-investor, bank-bank,
investor-bank, and bank-investor interactions, each with its own
characteristics. This enabled us, even at the collective level, to discern
the role of banks in the diffusion between investors, and how the
interrelations between groups of agents or certain agents can form.

To this preliminary study of a system with two types of agents, which serves
as a benchmark, we add a third type of agent, banks, which act as investors,
but have the unique ability to create money through the multiplier effect.
Once again, the analysis is conducted at two levels: the level of collective
states and the level of individual dynamics. To simplify the analysis, we
assume that banks essentially act as lenders, and under the assumption of no
default, their returns are reduced to the interest rate. In this scenario,
if the average total capital across all sectors mirrors that of the
benchmark, several distinctions become nonetheless apparent:

The first distinction lies in the diffusion matrix, which exhibits higher
coefficients compared to the benchmark. The borrowing behavior of investors
from banks results in increased leverage, thereby expanding their disposable
capital. The second distinction arises when banks provide capital to firms,
resulting in higher disposable capital for firms due to leverage and
monetary creation. With increased capital at their disposal, firms
experience a decrease in marginal returns. The number of potential averages
for collective states decreases in comparison to the benchmark which
contributes to the stabilization of the system. The third difference is that
the total disposable capital of banks varies inversely with that of
investors. Depending on the circumstances, capital may primarily be owned by
banks or investors, leading to a majority of loans or stakes, depending on
the interest rates applied. The fourth distinction is that, typically, when
the average productivity of firms---and consequently, their
returns---increases, the disposable capital of investors tends to increase,
albeit at a slower rate compared to the benchmark. This contributes to a
stabilizing effect. The fifth distinction, resulting from the previous two
observations, is that in periods of lower returns, capital tends to be
concentrated in banks. This is due to the lending-based nature of the
system, which favors banks through monetary creation.

The introduction of banks into the model has two conflicting effects on
collective states. On one hand, by increasing liquidity and providing loans,
banks facilitate the diffusion of capital and augment the disposable capital
of investors, thereby amplifying the occurrence and spread of defaults.
However, on the other hand, by directly lending to firms, they constrain the
proliferation of firm defaults. The size of the banking sector relative to
that of investors exerts a stabilizing influence, leading to a reduced
number of collective states.

Similar effects are observed in individual dynamics: in the presence of
banks, losses incurred by investors are more likely to spread to other
investors, due to the increase in the leverage effect. However, if banks
confine themselves to their lending function, they are largely shielded from
these losses and do not propagate them. Thus, banks play a stabilizing role
in the system and mitigate the risks of default. Nevertheless, in the event
of a default, the credit multiplier amplifies its effects.

However, it's important to note that these results are based on certain
assumptions, particularly the assumption that banks limit themselves to
their lending role. If this assumption is relaxed---for example, if banks
take significant stakes---other effects may arise. Additionally, we have
disregarded the influence of investors on banks, assuming a relatively
robust and independent banking sector. Removing this assumption and
reintroducing the impact of investors on banks could potentially reintroduce
multiple equilibria. In conclusion, these observations suggest that the
effects discussed emerge when banks exhibit behavior akin to that of
investors.

\section{Conclusion}

Overall, our framework enables the analysis of financial speculation
dynamics, the mutual impact of the financial and real sectors. Our
exploration of collective states reveals how capital is allocated and how
defaults can emerge structurally within these states. From a dynamic
perspective, it delineates the flow of capital, the propagation of returns,
the potential for defaults and their dynamic spread throughout part or all
of the economy, either directly or indirectly, within a collective state.
The introduction of banks into the system allows for an examination of how
these specific investors can either stabilize or destabilize the system,
depending on parameters such as the size of the banking sector.

More broadly, Field Economics should enable to analyse the emergence of
clustered groups operating autonomously and observe how networks of agents
and interests can form or disintegrate, and their impact on the system.

\section*{References}


\begin{description}
\item Abergel F, Chakraborti A, Muni Toke I and Patriarca M (2011a)
Econophysics review: I. Empirical facts, Quantitative Finance, Vol. 11, No.
7, 991-1012

\item Abergel F, Chakraborti A, Muni Toke I and Patriarca M (2011b)
Econophysics review: II. Agent-based models, Quantitative Finance, Vol. 11,
No. 7, 1013-1041

\item Acemoglu, D., Ozdaglar, A. and Tahbaz-Salehi, A. (2015). Systemic risk
and stability in financial networks. American Economic Review, 105(2),
564-608.

\item Acharya, V. V., Pedersen, L. H., Philippon, T., \& Richardson, M.
(2017). Measuring systemic risk. The Review of Financial Studies, 30(1),
2-47.

\item Adrian, T. and Brunnermeier, M. K. (2016). CoVaR. American Economic
Review, 106(7), 1705-1741.

\item Allen, F. and Gale, D. (2000). Financial contagion. Journal of
Political Economy, 108(1), 1-33.

\item Bardoscia M., Livan G., Marsili M. (2017), Statistical mechanics of
complex economies, Journal of Statistical Mechanics: Theory and Experiment,
Volume 2017

\item Bardoscia, M., Battiston, S., Caccioli, F., \& Caldarelli, G. (2019).
Pathways towards instability in financial networks. Nature communications,
10(1), 1-9.

\item Battiston, S., Puliga, M., Kaushik, R., Tasca, P., \& Caldarelli, G.
(2012). Debtrank: Too central to fail? Financial networks, the FED and
systemic risk. Scientific reports, 2, 541.

\item Battiston, S., Caldarelli, G., May, R. M., Roukny, T., \& Stiglitz, J.
E. (2020). The price of complexity in financial networks. Proceedings of the
National Academy of Sciences, 117(52), 32779-32786.

\item Bernanke B., Gertler, M. and S. Gilchrist (1999), The financial
accelerator in a quantitative business cycle framework, Chapter 21 in
Handbook of Macroeconomics, 1999, vol. 1, Part C, pp 1341-1393

\item Bensoussan A, Frehse J, Yam P (2018) Mean Field Games and Mean Field
Type Control Theory. Springer, New York

\item B\"{o}hm, V., Kikuchi, T., Vachadze, G.: Asset pricing and
productivity growth: the role of consumption scenarios. Comput. Econ. 32,
163--181 (2008)

\item Caggese A, Orive A P, The Interaction between Household and Firm
Dynamics and the Amplification of Financial Shocks. Barcelona GSE Working
Paper Series, Working Paper n%
${{}^o}$
866, 2015

\item Campello, M., Graham, J. and Harvey, C.R. (2010). The Real Effects of
Financial Constraints: Evidence from a Financial Crisis, Journal of
Financial Economics, vol. 97(3), 470-487.

\item Cifuentes, R., Ferrucci, G. and Shin, H. S. (2005). Liquidity risk and
contagion. Journal of the European Economic Association, 3(2-3), 556-566.

\item Elliott, M., Golub, B., and Jackson, M. O. (2014). Financial networks
and contagion. American Economic Review, 104(10), 3115-3153.

\item Gaffard JL and Napoletano M Editors (2012): Agent-based models and
economic policy. OFCE, Paris

\item Gai, P., \& Kapadia, S. (2010). Contagion in financial networks.
Proceedings of the Royal Society A: Mathematical, Physical and Engineering
Sciences, 466(2120), 2401-2423.

\item Gomes DA, Nurbekyan L, Pimentel EA (2015) Economic Models and
Mean-Field Games Theory, Publica\c{c}\~{o}es Matem\'{a}ticas do IMPA, 30o Col%
\'{o}quio Brasileiro de Matem\'{a}tica, Rio de Janeiro

\item Gennaioli, N., Shleifer, A. and Vishny, R. W. (2012). Neglected risks,
financial innovation, and financial fragility. Journal of Financial
Economics, 104(3), 452-468.

\item Gosselin P, Lotz A and Wambst M (2017) A Path Integral Approach to
Interacting Economic Systems with Multiple Heterogeneous Agents. IF\_PREPUB.
2017. hal-01549586v2

\item Gosselin P, Lotz A and Wambst M (2020) A Path Integral Approach to
Business Cycle Models with Large Number of Agents. Journal of Economic
Interaction and Coordination volume 15, pages 899--942

\item Gosselin P, Lotz A and Wambst M (2021) A statistical field approach to
capital accumulation. Journal of Economic Interaction and Coordination 16,
pages 817--908 (2021)

\item Grassetti, F., Mammana, C. \& Michetti, E. A dynamical model for real
economy and finance. Math Finan Econ (2022).
https://doi.org/10.1007/s11579-021-00311-3

\item Greenwood, R., \& Hanson, S. G. (2013). Issuer quality and corporate
bond returns. The Review of Financial Studies, 26(6), 1483-1525.

\item Grosshans, D., Zeisberger, S.: All's well that ends well? on the
importance of how returns are achieved. J. Bank. Finance 87, 397--410 (2018)

\item Haldane, A. G., \& May, R. M. (2011). Systemic risk in banking
ecosystems. Nature, 469(7330), 351-355.

\item Holmstrom, B., and Tirole, J. (1997). Financial intermediation,
loanable funds, and the
\end{description}

real sector. Quarterly Journal of Economics, 663-691.

\begin{description}
\item Jackson M (2010) Social and Economic Networks. Princeton University
Press, Princeton

\item Jermann, U.J. and Quadrini, V., (2012). "Macroeconomic Effects of
Financial Shocks," American Economic Review, Vol. 102, No. 1.

\item Khan, A., and Thomas, J. K. (2013). "Credit Shocks and Aggregate
Fluctuations in an Economy with Production Heterogeneity," Journal of
Political Economy, 121(6), 1055-1107.

\item Kaplan G, Violante L (2018) Microeconomic Heterogeneity and
Macroeconomic Shocks, Journal of Economic Perspectives, Vol. 32, No. 3,
167-194

\item Kleinert H (1989) Gauge fields in condensed matter Vol. I , Superflow
and vortex lines, Disorder Fields, Phase Transitions, Vol. II, Stresses and
defects, Differential Geometry, Crystal Melting. World Scientific, Singapore

\item Kleinert H (2009) Path Integrals in Quantum Mechanics, Statistics,
Polymer Physics, and Financial Markets 5th edition. World Scientific,
Singapore

\item Krugman P (1991) Increasing Returns and Economic Geography. Journal of
Political Economy, 99(3), 483-499

\item Langfield, S., Liu, Z., \& Ota, T. (2020). Mapping the network of
financial linkages: an industry-level analysis of the UK. Journal of Banking
\& Finance, 118, 105946.

\item Lasry JM, Lions PL, Gu\'{e}ant O (2010a) Application of Mean Field
Games to Growth Theory \newline
https://hal.archives-ouvertes.fr/hal-00348376/document

\item Lasry JM, Lions PL, Gu\'{e}ant O (2010b) Mean Field Games and
Applications. Paris-Princeton lectures on Mathematical Finance, Springer.%
\textbf{\ }https://hal.archives-ouvertes.fr/hal-01393103

\item Lux T (2008) Applications of Statistical Physics in Finance and
Economics. Kiel Institute for the World Economy (IfW), Kiel

\item Lux T (2016) Applications of Statistical Physics Methods in Economics:
Current state and perspectives. Eur. Phys. J. Spec. Top. (2016) 225: 3255.
https://doi.org/10.1140/epjst/e2016-60101-x

\item Mandel A, Jaeger C, F\"{u}rst S, Lass W, Lincke D, Meissner F,
Pablo-Marti F, Wolf S (2010). Agent-based dynamics in disaggregated growth
models. Documents de travail du Centre d'Economie de la Sorbonne. Centre
d'Economie de la Sorbonne, Paris

\item Mandel A (2012) Agent-based dynamics in the general equilibrium model.
Complexity Economics 1, 105--121

\item Monacelli, T., Quadrini, V. and A. Trigari (2011). "Financial Markets
and Unemployment," NBER Working Papers 17389, National Bureau of Economic
Research.

\item Reinhart, C. M., \& Rogoff, K. S. (2009). This Time Is Different:
Eight Centuries of Financial Folly. Princeton University Press.

\item Sims C A (2006) Rational inattention: Beyond the Linear Quadratic
Case, American Economic Review, vol. 96, no. 2, 158-163

\item Yang J (2018) Information theoretic approaches to economics, Journal
of Economic Survey, Vol. 32, No. 3, 940-960

\item Cochrane, J.H. (ed.): Financial Markets and the Real Economy,
International Library of Critical Writings in Financial Economics, vol. 18.
Edward Elgar (2006)
\end{description}

\pagebreak

\part*{Part 1 Appendices}

\section*{Appendix 1 Details of the micro setup}

\subsection*{A1.1 Linear approximation of relation between disposable
capital and private capital}

In a linear approximation, we consider that invested shares and lent capital
are proportional to the private capital, acting as collateral:%
\begin{equation*}
\hat{k}_{1}\left( \hat{K}_{jp}\left( t\right) ,\hat{X}_{j}\left( t\right) ,%
\hat{X}_{l}\left( t\right) \right) +\hat{k}_{2}\left( \hat{K}_{jp}\left(
t\right) ,\hat{X}_{j}\left( t\right) ,\hat{X}_{l}\left( t\right) \right)
=\left( \hat{k}_{1}\left( \hat{X}_{j}\left( t\right) ,\hat{X}_{l}\left(
t\right) \right) +\hat{k}_{2}\left( \hat{X}_{j}\left( t\right) ,\hat{X}%
_{l}\left( t\right) \right) \right) \hat{K}_{jp}\left( t\right)
\end{equation*}%
and:%
\begin{equation*}
k_{1}\left( X_{i},K_{i},\hat{X}_{j}\right) +k_{2}\left( X_{i},K_{i},\hat{X}%
_{j}\right) =\left( k_{1}\left( X_{i},\hat{X}_{j}\right) +k_{2}\left( X_{i},%
\hat{X}_{j}\right) \right) K_{ip}
\end{equation*}%
Ultimately, we can express private capital as a function of disposable
capital:%
\begin{equation}
\hat{K}_{j}\left( t\right) =\left( 1+\sum_{l}\left( \hat{k}_{1}\left( \hat{X}%
_{j}\left( t\right) ,\hat{X}_{l}\left( t\right) \right) +\hat{k}_{2}\left( 
\hat{X}_{j}\left( t\right) ,\hat{X}_{l}\left( t\right) \right) \right) \hat{K%
}_{l}\left( t\right) \right) \hat{K}_{jp}\left( t\right)  \label{CH}
\end{equation}%
that is:%
\begin{equation}
\hat{K}_{jp}\left( t\right) =\frac{\hat{K}_{j}\left( t\right) }{%
1+\sum_{l}\left( \hat{k}_{1}\left( \hat{X}_{j}\left( t\right) ,\hat{X}%
_{l}\left( t\right) \right) +\hat{k}_{2}\left( \hat{X}_{j}\left( t\right) ,%
\hat{X}_{l}\left( t\right) \right) \right) \hat{K}_{l}\left( t\right) }
\label{LN}
\end{equation}%
\bigskip Defining:%
\begin{eqnarray*}
\hat{k}_{ajl} &=&\hat{k}_{a}\left( \hat{X}_{j}\left( t\right) ,\hat{X}%
_{l}\left( t\right) \right) \\
k_{ajl} &=&k_{a}\left( X_{j}\left( t\right) ,\hat{X}_{l}\left( t\right)
\right)
\end{eqnarray*}%
and:%
\begin{eqnarray*}
\hat{k}_{jl} &=&\hat{k}_{1jl}+\hat{k}_{2jl} \\
k_{jl} &=&k_{1jl}+k_{2jl}
\end{eqnarray*}%
this also rewrites:%
\begin{equation}
\hat{K}_{jp}\left( t\right) =\frac{\hat{K}_{j}\left( t\right) }{%
1+\sum_{l}\left( \hat{k}_{1jl}+\hat{k}_{2jl}\right) \hat{K}_{l}\left(
t\right) }
\end{equation}

\subsection*{A1.2 Normalization of coefficients}

We assume that coefficients are normalized as follows. For investors' shares
in firms, we assume that coefficients depend on the total capital involved
for firms. Therefore, we replace:%
\begin{equation}
k_{ajl}\rightarrow \frac{k_{ail}}{N\left\langle K_{p}\left( t\right)
\right\rangle }  \label{Nf}
\end{equation}%
where $N$ is the number of firms, and $\left\langle K_{p}\left( t\right)
\right\rangle $ represents the average capital per firm. Similarly, we
replace:%
\begin{equation}
\hat{k}_{ajl}\rightarrow \frac{\hat{k}_{ajl}}{\hat{N}\left\langle \hat{K}%
_{p}\left( t\right) \right\rangle }  \label{Ns}
\end{equation}%
where $\hat{N}$ is the total number of investors, and $\left\langle \hat{K}%
_{p}\left( t\right) \right\rangle $ represents the average capital per
investor.

Below, we will examine the dynamics for $K_{p}\left( t\right) $ and $\hat{K}%
\left( t\right) $. Equation (\ref{Nf}) is sufficient for the analysis, but
equation (\ref{Ns}) needs to be rewritten as a function of $\hat{K}\left(
t\right) $. For this purpose, we will rewrite (\ref{DC}) and (\ref{CH}): 
\begin{equation}
\hat{K}_{j}\left( t\right) =\left( 1+\sum_{l}\frac{\left( \hat{k}_{1}\left( 
\hat{X}_{j}\left( t\right) ,\hat{X}_{l}\left( t\right) \right) +\hat{k}%
_{2}\left( \hat{X}_{j}\left( t\right) ,\hat{X}_{l}\left( t\right) \right)
\right) }{\hat{N}\left\langle \hat{K}_{p}\left( t\right) \right\rangle }\hat{%
K}_{l}\left( t\right) \right) \hat{K}_{jp}\left( t\right)
\end{equation}%
The average over $\hat{K}_{jp}\left( t\right) $, written $\left\langle \hat{K%
}_{p}\left( t\right) \right\rangle $, is given by:%
\begin{equation}
\left\langle \hat{K}_{p}\left( t\right) \right\rangle =\left\langle \hat{K}%
\left( t\right) \right\rangle -\sum_{l}\frac{\left\langle \left( \hat{k}%
_{1}\left( \hat{X}_{j}\left( t\right) ,\hat{X}_{l}\left( t\right) \right) +%
\hat{k}_{2}\left( \hat{X}_{j}\left( t\right) ,\hat{X}_{l}\left( t\right)
\right) \right) \right\rangle }{N}\left\langle \hat{K}\left( t\right)
\right\rangle
\end{equation}%
This leads us to:%
\begin{equation*}
\left\langle \hat{K}_{p}\left( t\right) \right\rangle =\left\langle \hat{K}%
\left( t\right) \right\rangle \left( 1-\left( \left\langle \hat{k}%
_{1}\right\rangle +\left\langle \hat{k}_{2}\right\rangle \right) \right)
\end{equation*}%
where $\left\langle \hat{k}_{a}\right\rangle $ represents the average of $%
\hat{k}_{a}\left( \hat{X}_{j}\left( t\right) ,\hat{X}_{l}\left( t\right)
\right) $, and the factor $\hat{N}$ represents the number of investors.

The normalization for $\hat{k}_{ajl}$ is thus:%
\begin{equation}
\hat{k}_{ajl}\rightarrow \frac{\hat{k}_{ajl}}{\hat{N}\left\langle \hat{K}%
\left( t\right) \right\rangle \left( 1-\left( \left\langle \hat{k}%
_{1}\right\rangle +\left\langle \hat{k}_{2}\right\rangle \right) \right) }
\label{Nt}
\end{equation}%
This normalization will be kept implicit in the main part of the text for
the sake of simplcity and reintroduced when needed.

\subsection*{A1.3 Returns from firms to investors}

We derive $r_{i}^{\prime }\left( K_{i}\left( t\right) ,\frac{\dot{K}_{i}}{%
K_{i}}\right) $ by specifying the form of $F_{1}\left( \bar{R}_{i},\frac{%
\dot{K}_{i}}{K_{i}}\right) $. Given that: 
\begin{equation*}
K_{i}=K_{ip}\left( t\right) \left( 1+\sum_{v}k_{iv}\hat{K}_{v}\left(
t\right) \right)
\end{equation*}%
we have: 
\begin{equation*}
\frac{\dot{K}_{ip}\left( t\right) }{K_{ip}\left( t\right) }+\frac{%
\sum_{v}k_{iv}\frac{d}{dt}\hat{K}_{v}\left( t\right) }{1+\sum_{v}k_{iv}\hat{K%
}_{v}\left( t\right) }+\frac{\sum_{v}\frac{d}{dt}k_{iv}\hat{K}_{v}\left(
t\right) }{1+\sum_{v}k_{iv}\hat{K}_{v}\left( t\right) }=\frac{\dot{K}%
_{ip}\left( t\right) }{K_{ip}\left( t\right) }+\frac{\sum_{v}k_{iv}\hat{f}%
_{\nu }\hat{K}_{v}\left( t\right) }{1+\sum_{v}k_{iv}\hat{K}_{v}\left(
t\right) }
\end{equation*}%
Moreover, since the sum:%
\begin{equation*}
\frac{\sum_{v}k_{iv}\frac{d}{dt}\hat{K}_{v}\left( t\right) }{1+\sum_{v}k_{iv}%
\hat{K}_{v}\left( t\right) }
\end{equation*}%
is close to its average, it can be neglected. Besides, the variation $\frac{d%
}{dt}k_{iv}$ is exogeneous and mainly depends on $\bar{R}_{i}$, which can
thus be included directly. As a consequence, we can write:%
\begin{equation*}
F_{1}\left( \bar{R}_{i},\frac{\dot{K}_{i}}{K_{i}}\right) \rightarrow
F_{1}\left( \bar{R}_{i},\frac{\dot{K}_{ip}\left( t\right) }{K_{ip}\left(
t\right) }\right)
\end{equation*}%
Substracting the average $\left\langle \frac{\dot{K}_{ip}\left( t\right) }{%
K_{ip}\left( t\right) }\right\rangle $, it writes in first approximation:%
\begin{equation*}
F_{1}\left( \bar{R}_{i},\frac{\dot{K}_{i}}{K_{i}}\right) \rightarrow
F_{1}\left( \bar{R}\left( K_{i},X_{i}\right) \right) +\tau \left( \bar{R}%
\left( K_{i},X_{i}\right) \right) \left( \frac{\dot{K}_{ip}\left( t\right) }{%
K_{ip}\left( t\right) }-\left\langle \frac{\dot{K}_{ip}\left( t\right) }{%
K_{ip}\left( t\right) }\right\rangle \right)
\end{equation*}%
and the return is:%
\begin{equation*}
r_{i}^{\prime }\left( K_{i}\left( t\right) ,\frac{\dot{K}_{i}}{K_{i}}\right)
=r_{i}+F_{1}\left( \bar{R}\left( K_{i},X_{i}\right) \right) +\tau \left( 
\bar{R}\left( K_{i},X_{i}\right) \right) \Delta f_{1}^{\prime }\left(
K_{i}\left( t\right) \right)
\end{equation*}%
with:%
\begin{equation*}
\Delta f_{1}^{\prime }\left( K_{i}\left( t\right) \right) =f_{1}^{\prime
}\left( K_{i}\left( t\right) \right) -\left\langle f_{1}^{\prime }\left(
K_{i}\left( t\right) \right) \right\rangle
\end{equation*}%
where the average is taken over all firms. We assume that $F_{1}\left( \bar{R%
}\left( K_{i},X_{i}\right) \right) $ is equal to $0$ in average, so that we
can write:%
\begin{equation*}
F_{1}\left( \bar{R}\left( K_{i},X_{i}\right) \right) +\tau \left( \bar{R}%
\left( K_{i},X_{i}\right) \right) \Delta f_{1}^{\prime }\left( K_{i}\left(
t\right) \right) =\Delta F_{\tau }\left( \bar{R}\left( K_{i},X_{i}\right)
\right)
\end{equation*}%
We define also:

\begin{equation}
f_{1}^{\prime }\left( K_{i}\left( t\right) \right) =\left( \left(
1+\sum_{\nu }k_{2j\nu }\hat{K}_{\nu }\left( t\right) \right) f_{1}\left(
\left( K_{i}\left( t\right) \right) \right) -\bar{r}\sum_{v}k_{2lv}\right) 
\hat{K}_{v}\left( t\right)  \label{FM}
\end{equation}%
\begin{equation*}
f_{1}\left( \left( K_{i}\left( t\right) \right) \right) =\frac{f_{1}^{\prime
}\left( K_{i}\left( t\right) \right) }{1+\sum_{\nu }k_{2j\nu }\hat{K}_{\nu
}\left( t\right) }+\bar{r}\sum_{v}k_{2lv}\hat{K}_{v}\left( t\right)
\end{equation*}%
and:%
\begin{equation*}
r_{i}^{\prime }\left( K_{i}\left( t\right) ,\frac{\dot{K}_{i}}{K_{i}}\right)
=\frac{f_{1}^{\prime }\left( K_{i}\left( t\right) \right) }{1+\sum_{\nu
}k_{2j\nu }\hat{K}_{\nu }\left( t\right) }+\bar{r}\sum_{v}k_{2lv}\hat{K}%
_{v}\left( t\right) +\Delta F_{\tau }\left( \bar{R}\left( K_{i},X_{i}\right)
\right)
\end{equation*}

\subsection*{A1.4 Investors returns and investors' disposable dynamics}

The return equation writes in matricial form:%
\begin{equation}
\sum_{l}\left( \delta _{jl}-\frac{\hat{k}_{1lj}\hat{K}_{l}\left( t\right) }{%
1+\sum_{v}\left( \hat{k}_{1lv}+\hat{k}_{2lv}\right) \hat{K}_{v}\left(
t\right) }\right) R_{l}=R_{j}^{\prime }  \label{RM}
\end{equation}%
with solution:

\begin{equation*}
R_{j}=\sum_{l}\left( 1-\frac{\hat{k}_{1lj}\hat{K}_{j}\left( t\right) }{%
1+\sum_{v}\left( \hat{k}_{1lv}+\hat{k}_{2lv}\right) \hat{K}_{v}\left(
t\right) }\right) ^{-1}\frac{1+\sum_{v}\hat{k}_{2lv}\hat{K}_{v}\left(
t\right) }{1+\sum_{v}\left( \hat{k}_{1lv}+\hat{k}_{2lv}\right) \hat{K}%
_{v}\left( t\right) }R_{l}^{\prime }
\end{equation*}%
The accumulation of disposable capital can be computed by starting first
with the private capital $\hat{K}_{jp}\left( t\right) $.

By a reasoning similar to the firms, the dynamics for private funds is given
by:%
\begin{eqnarray}
\hat{K}_{jp}\left( t+\varepsilon \right) -\hat{K}_{jp}\left( t\right)
&=&R_{j}\left( 1+\sum_{v}\hat{k}_{2jv}\hat{K}_{v}\left( t\right) \right) 
\hat{K}_{jp}\left( t\right) -\bar{r}\sum_{v}\hat{k}_{2jv}\hat{K}_{v}\left(
t\right) \hat{K}_{jp}\left( t\right)  \label{PC} \\
&=&\sum_{l}\left( 1-\frac{\hat{k}_{1lj}\hat{K}_{j}\left( t\right) }{%
1+\sum_{v}\left( \hat{k}_{1lv}+\hat{k}_{2lv}\right) \hat{K}_{v}\left(
t\right) }\right) ^{-1}R_{l}^{\prime }\left( \hat{K}_{jp}\left( t\right) +%
\hat{K}_{jp}\left( t\right) \sum_{v}\hat{k}_{2jv}\hat{K}_{v}\left( t\right)
\right)  \notag \\
&&-\bar{r}\sum_{v}\hat{k}_{2jv}\hat{K}_{jp}\left( t\right) \hat{K}_{v}\left(
t\right)  \notag
\end{eqnarray}%
written also:%
\begin{equation*}
\hat{K}_{jp}\left( t+\varepsilon \right) -\hat{K}_{jp}\left( t\right) =f_{j}%
\hat{K}_{jp}\left( t\right)
\end{equation*}%
with:%
\begin{equation}
\hat{f}_{j}=\left( 1+\sum_{v}\hat{k}_{2jv}\hat{K}_{v}\left( t\right) \right)
R_{j}-\bar{r}\sum_{v}\hat{k}_{2jv}\hat{K}_{v}\left( t\right)  \label{Rf}
\end{equation}%
In the continuous approximation:%
\begin{equation*}
\hat{K}_{jp}\left( t+\varepsilon \right) -\hat{K}_{jp}\left( t\right)
\rightarrow \frac{d}{dt}\hat{K}_{jp}\left( t\right) =\hat{f}_{j}\hat{K}%
_{jp}\left( t\right)
\end{equation*}%
This can be transformed to obtain a dynamics for the disposable income. The
differenciation of (\ref{NK}) yields:%
\begin{equation}
\frac{d}{dt}\hat{K}_{jp}\left( t\right) =\frac{\frac{d}{dt}\hat{K}_{j}\left(
t\right) }{1+\sum_{l}\left( \hat{k}_{1jl}+\hat{k}_{2jl}\right) \hat{K}%
_{l}\left( t\right) }-\sum_{l}\frac{\left( \hat{k}_{1jl}+\hat{k}%
_{2jl}\right) \hat{K}_{l}\left( t\right) }{\left( \sum_{l}\left( \hat{k}%
_{1jl}+\hat{k}_{2jl}\right) \hat{K}_{l}\left( t\right) ^{2}\right) }\frac{d}{%
dt}\hat{K}_{l}\left( t\right)
\end{equation}%
Using (\ref{DR}) we obtain the dynamic equation for disposable capital (\ref%
{PC}) $\hat{K}_{j}\left( t\right) $: 
\begin{equation*}
\frac{\frac{d}{dt}\hat{K}_{j}\left( t\right) }{1+\sum_{l}\left( \hat{k}_{jl}%
\hat{K}_{l}\left( t\right) \right) }-\sum_{l}\frac{\hat{k}_{jl}\hat{K}%
_{j}\left( t\right) }{\left( 1+\sum_{l}\hat{k}_{jl}\hat{K}_{l}\left(
t\right) \right) ^{2}}\frac{d}{dt}\hat{K}_{l}\left( t\right) =f_{j}\frac{%
\hat{K}_{j}\left( t\right) }{1+\sum_{l}\left( \hat{k}_{jl}\hat{K}_{l}\left(
t\right) \right) }
\end{equation*}%
This equation is used to write a system of differential equations for the $%
\frac{d}{dt}\hat{K}_{j}\left( t\right) $:%
\begin{equation}
\frac{d}{dt}\hat{K}_{j}\left( t\right) =\sum_{l}\left( 1-\frac{\hat{k}_{1jl}%
\hat{K}_{j}\left( t\right) }{1+\sum_{v}\hat{k}_{jv}\hat{K}_{v}\left(
t\right) }\right) ^{-1}f_{l}\hat{K}_{l}\left( t\right) =\sum_{l}\left(
1-M\right) _{jl}^{-1}f_{l}\hat{K}_{l}\left( t\right)
\end{equation}%
where:%
\begin{equation}
M_{jm}=\frac{\hat{k}_{jm}\hat{K}_{j}\left( t\right) }{1+\sum_{\nu }\hat{k}%
_{j\nu }\hat{K}_{\nu }\left( t\right) }
\end{equation}

\subsection*{A1.5 Investors return equation without default}

Writing (\ref{Rf}) as:%
\begin{equation*}
R_{j}=\frac{\hat{f}_{j}}{1+\sum_{v}\hat{k}_{2jv}\hat{K}_{v}\left( t\right) }+%
\bar{r}\frac{\sum_{v}\hat{k}_{2jv}\hat{K}_{v}\left( t\right) }{1+\sum_{v}%
\hat{k}_{2jv}\hat{K}_{v}\left( t\right) }
\end{equation*}%
we can rewrite the return equation (\ref{RM}) as an equation for $\hat{f}%
_{j} $:%
\begin{equation}
\sum_{l}\left( \delta _{jl}-\frac{\hat{k}_{1lj}\hat{K}_{l}\left( t\right) }{%
1+\sum_{v}\left( \hat{k}_{1lv}+\hat{k}_{2lv}\right) \hat{K}_{v}\left(
t\right) }\right) \left( \frac{\hat{f}_{l}}{1+\sum_{v}\hat{k}_{2lv}\hat{K}%
_{v}\left( t\right) }+\bar{r}\frac{\sum_{v}\hat{k}_{2lv}\hat{K}_{v}\left(
t\right) }{1+\sum_{v}\hat{k}_{2lv}\hat{K}_{v}\left( t\right) }\right)
=R_{j}^{\prime }
\end{equation}%
Substracting $\bar{r}$ on both side it writes:%
\begin{eqnarray}
&&\frac{\hat{f}_{j}-\bar{r}}{1+\sum_{v}\hat{k}_{2jv}\hat{K}_{v}\left(
t\right) }-\sum_{l}\left( \frac{\hat{k}_{1lj}\hat{K}_{l}\left( t\right) }{%
1+\sum_{v}\left( \hat{k}_{1lv}+\hat{k}_{2lv}\right) \hat{K}_{v}\left(
t\right) }\right) \left( \frac{\hat{f}_{l}-\bar{r}}{1+\sum_{v}\hat{k}_{2lv}%
\hat{K}_{v}\left( t\right) }+\bar{r}\right) \\
&=&R_{j}^{\prime }-\bar{r}  \notag
\end{eqnarray}%
that is:%
\begin{eqnarray}
&&\frac{\hat{f}_{j}-\bar{r}}{1+\sum_{v}\hat{k}_{2jv}\hat{K}_{v}\left(
t\right) }-\sum_{l}\left( \frac{\hat{k}_{1lj}\hat{K}_{l}\left( t\right) }{%
1+\sum_{v}\left( \hat{k}_{1lv}+\hat{k}_{2lv}\right) \hat{K}_{v}\left(
t\right) }\right) \left( \frac{\hat{f}_{l}-\bar{r}}{1+\sum_{v}\hat{k}_{2lv}%
\hat{K}_{v}\left( t\right) }\right) \\
&=&R_{j}^{\prime }-\bar{r}+\bar{r}\sum_{l}\left( \frac{\hat{k}_{1lj}\hat{K}%
_{l}\left( t\right) }{1+\sum_{v}\left( \hat{k}_{1lv}+\hat{k}_{2lv}\right) 
\hat{K}_{v}\left( t\right) }\right)  \notag
\end{eqnarray}%
Using (\ref{RM}):%
\begin{eqnarray*}
&&R_{j}^{\prime }-\bar{r}+\bar{r}\sum_{l}\left( \frac{\hat{k}_{1lj}\hat{K}%
_{l}\left( t\right) }{1+\sum_{v}\left( \hat{k}_{1lv}+\hat{k}_{2lv}\right) 
\hat{K}_{v}\left( t\right) }\right) \\
&=&\bar{r}\sum_{l}\frac{\hat{k}_{1lj}\hat{K}_{l}\left( t\right) +\hat{k}%
_{2lj}\hat{K}_{l}\left( t\right) }{1+\sum_{v}\left( \hat{k}_{1lv}+\hat{k}%
_{2lv}\right) \hat{K}_{v}\left( t\right) }+\bar{r}\sum_{i}\frac{%
k_{2ij}K_{i}\left( t\right) }{1+\sum_{v}\left( k_{1iv}+k_{2iv}\right) \hat{K}%
_{v}\left( t\right) } \\
&&+\sum_{i}\left( r_{i}+F_{1}\left( \bar{R}_{i},\frac{\dot{K}_{i}\left(
t\right) }{K_{i}\left( t\right) }\right) +\tau \left( \bar{R}\left(
K_{i},X_{i}\right) \right) \Delta f_{1}^{\prime }\left( K_{i}\left( t\right)
\right) \right) \frac{k_{1ij}K_{i}\left( t\right) }{1+\sum_{v}\left(
k_{1iv}+k_{2iv}\right) \hat{K}_{v}\left( t\right) }-\bar{r}
\end{eqnarray*}%
The identity:%
\begin{equation*}
\sum_{l}\frac{\hat{k}_{1lj}\hat{K}_{l}\left( t\right) +\hat{k}_{2lj}\hat{K}%
_{l}\left( t\right) }{1+\sum_{v}\left( \hat{k}_{1lv}+\hat{k}_{2lv}\right) 
\hat{K}_{v}\left( t\right) }+\sum_{i}\frac{k_{2ij}K_{i}\left( t\right) }{%
1+\sum_{v}\left( k_{1iv}+k_{2iv}\right) \hat{K}_{v}\left( t\right) }+\frac{%
k_{1ij}K_{i}\left( t\right) }{1+\sum_{v}\left( k_{1iv}+k_{2iv}\right) \hat{K}%
_{v}\left( t\right) }=1
\end{equation*}%
leads to:%
\begin{eqnarray*}
&&R_{j}^{\prime }-\bar{r}+\bar{r}\sum_{l}\left( \frac{\hat{k}_{1lj}\hat{K}%
_{l}\left( t\right) }{1+\sum_{v}\left( \hat{k}_{1lv}+\hat{k}_{2lv}\right) 
\hat{K}_{v}\left( t\right) }\right) \\
&=&\sum_{i}\frac{\left( r_{i}+F_{1}\left( \bar{R}_{i},\frac{\dot{K}%
_{i}\left( t\right) }{K_{i}\left( t\right) }\right) +\tau \left( \bar{R}%
\left( K_{i},X_{i}\right) \right) \Delta f_{1}^{\prime }\left( K_{i}\left(
t\right) \right) -\bar{r}\right) k_{1ij}K_{i}\left( t\right) }{%
1+\sum_{v}\left( k_{1iv}+k_{2iv}\right) \hat{K}_{v}\left( t\right) }
\end{eqnarray*}%
and the return equation is:%
\begin{eqnarray}
&&\frac{\hat{f}_{j}-\bar{r}}{1+\sum_{v}\hat{k}_{2jv}\hat{K}_{v}\left(
t\right) }-\sum_{l}\left( \frac{\hat{k}_{1lj}\hat{K}_{l}\left( t\right) }{%
1+\sum_{v}\left( \hat{k}_{1lv}+\hat{k}_{2lv}\right) \hat{K}_{v}\left(
t\right) }\right) \left( \frac{\hat{f}_{l}-\bar{r}}{1+\sum_{v}\hat{k}_{2lv}%
\hat{K}_{v}\left( t\right) }\right) \\
&=&\sum_{i}\frac{\left( r_{i}+F_{1}\left( \bar{R}_{i},\frac{\dot{K}%
_{i}\left( t\right) }{K_{i}\left( t\right) }\right) +\tau \left( \bar{R}%
\left( K_{i},X_{i}\right) \right) \Delta f_{1}^{\prime }\left( K_{i}\left(
t\right) \right) -\bar{r}\right) k_{1ij}K_{i}\left( t\right) }{%
1+\sum_{v}\left( k_{1iv}+k_{2iv}\right) \hat{K}_{v}\left( t\right) }
\end{eqnarray}

\section*{Appendix 2 From large number of agents to field formalism}

This appendix summarizes the most useful steps of the method developed in
Gosselin, Lotz and Wambst (2017, 2020, 2021),\ to switch from the
probabilistic description of the model to the field theoretic formalism and
summarizes the translation of a generalization of (\ref{mNZ}) involving
different time variables. By convention and unless otherwise mentioned, the
symbol of integration $\int $\ refers to all the variables involved.

\subsection*{A2.1 Probabilistic formalism}

The probabilistic formalism for a system with $N$ identical economic agents
in interaction is based on the minimization functions described in the text.
Classically, the dynamics derives through the optimization problem of these
functions. The probabilistic formalism relies on the contrary on the fact,
that, due to uncertainties, shocks... agents do not optimize fully these
functions. Moreover, given the large number of agents, there may be some
discrepancy between agents minimization functions, and this fact may be
translated in an uncertainty of behavior around some average minimization,
or objective function.

We thus assume that each agent chooses for his action a path randomly
distributed around the optimal path. The agent's behavior can be described
as a weight that is an exponential of the intertemporal utility, that
concentrates the probability around the optimal path. This feature models
some internal uncertainty as well as non-measurable shocks. Gathering all
agents, it yields a probabilistic description of the system in terms of a
probabilistic weight.

In general, this weight includes utility functions and internalizes
forward-looking behaviors, such as intertemporal budget constraints and
interactions among agents. These interactions may for instance arise through
constraints, since income flows depend on other agents demand. The
probabilistic description then allows to compute the transition functions of
the system, and in turn compute the probability for a system to evolve from
an initial state to a final state within a given time span. They have the
form of Euclidean path integrals.

In the context of the present paper, we have seen that the minimization
functions for the system considered in this work have the form:%
\begin{eqnarray}
&&\int dt\left( \sum_{i}\left( \frac{d\mathbf{A}_{i}\left( t\right) }{dt}%
-\sum_{j,k,l...}f\left( \mathbf{A}_{i}\left( t\right) ,\mathbf{A}_{j}\left(
t\right) ,\mathbf{A}_{k}\left( t\right) ,\mathbf{A}_{l}\left( t\right)
...\right) \right) ^{2}\right.  \label{mnz} \\
&&\left. +\sum_{i}\left( \sum_{j,k,l...}g\left( \mathbf{A}_{i}\left(
t\right) ,\mathbf{A}_{j}\left( t\right) ,\mathbf{A}_{k}\left( t\right) ,%
\mathbf{A}_{l}\left( t\right) ...\right) \right) \right)  \notag
\end{eqnarray}%
The minimization of this function will yield a dynamic equation for $N$
agents in interaction described by a set of dynamic variables $\mathbf{A}%
_{i}\left( t\right) $ during a given timespan $T$.

The probabilistic description is straightforwardly obtained from (\ref{mnz}%
). The probability associated to a configuration $\left( \mathbf{A}%
_{i}\left( t\right) \right) _{\substack{ i=1,...,N  \\ 0\leqslant t\leqslant
T }}$ \ is directly given by:%
\begin{eqnarray}
&&\mathcal{N}\exp \left( -\frac{1}{\sigma ^{2}}\int dt\left( \sum_{i}\left( 
\frac{d\mathbf{A}_{i}\left( t\right) }{dt}-\sum_{j,k,l...}f\left( \mathbf{A}%
_{i}\left( t\right) ,\mathbf{A}_{j}\left( t\right) ,\mathbf{A}_{k}\left(
t\right) ,\mathbf{A}_{l}\left( t\right) ...\right) \right) ^{2}\right.
\right.  \label{prz} \\
&&\left. \left. +\sum_{i}\left( \sum_{j,k,l...}g\left( \mathbf{A}_{i}\left(
t\right) ,\mathbf{A}_{j}\left( t\right) ,\mathbf{A}_{k}\left( t\right) ,%
\mathbf{A}_{l}\left( t\right) ...\right) \right) \right) \right)  \notag
\end{eqnarray}%
where $\mathcal{N}$ is a normalization factor and $\sigma ^{2}$ is a
variance whose magnitude describes the amplitude of deviations around the
optimal path.

As in the paper, the system is in general modelled by several equations, and
thus, several minimization function. The overall system is thus described by
several functions, and the minimization function of the whole system is
described by the set of functions:%
\begin{eqnarray}
&&\int dt\left( \sum_{i}\left( \frac{d\mathbf{A}_{i}\left( t\right) }{dt}%
-\sum_{j,k,l...}f^{\left( \alpha \right) }\left( \mathbf{A}_{i}\left(
t\right) ,\mathbf{A}_{j}\left( t\right) ,\mathbf{A}_{k}\left( t\right) ,%
\mathbf{A}_{l}\left( t\right) ...\right) \right) ^{2}\right.  \label{znm} \\
&&\left. +\sum_{i}\left( \sum_{j,k,l...}g^{\left( \alpha \right) }\left( 
\mathbf{A}_{i}\left( t\right) ,\mathbf{A}_{j}\left( t\right) ,\mathbf{A}%
_{k}\left( t\right) ,\mathbf{A}_{l}\left( t\right) ...\right) \right) \right)
\notag
\end{eqnarray}%
where $\alpha $ runs over the set equations describing the system's
dynamics. The associated weight is then:%
\begin{eqnarray}
&&\mathcal{N}\exp \left( -\left( \sum_{i,\alpha }\frac{1}{\sigma _{\alpha
}^{2}}\int dt\left( \frac{d\mathbf{A}_{i}\left( t\right) }{dt}%
-\sum_{j,k,l...}f^{\left( \alpha \right) }\left( \mathbf{A}_{i}\left(
t\right) ,\mathbf{A}_{j}\left( t\right) ,\mathbf{A}_{k}\left( t\right) ,%
\mathbf{A}_{l}\left( t\right) ...\right) \right) ^{2}\right. \right.
\label{pnz} \\
&&\left. \left. +\sum_{i,\alpha }\left( \sum_{j,k,l...}g^{\left( \alpha
\right) }\left( \mathbf{A}_{i}\left( t\right) ,\mathbf{A}_{j}\left( t\right)
,\mathbf{A}_{k}\left( t\right) ,\mathbf{A}_{l}\left( t\right) ...\right)
\right) \right) \right)  \notag
\end{eqnarray}

The appearance of the sum of minimization functions in the probabilistic
weight (\ref{pnz}) translates the hypothesis that the deviations with
respect to the optimization of the functions (\ref{znm}) are assumed to be
independent.

For a large number of agents, the system described by (\ref{pnz}) involves a
large number of variables $K_{i}\left( t\right) $, $P_{i}\left( t\right) $
and $X_{i}\left( t\right) $\ that are difficult to handle. To overcome this
difficulty, we consider the space $H$\ of complex functions defined on the
space of a single agent's actions. The space $H$ describes the collective
behavior of the system. Each function $\Psi $ of $H$ encodes a particular
state of the system. We then associate to\ each function $\Psi $ of $H$ a
statistical weight, i.e. a probability describing the state encoded in $\Psi 
$. This probability is written $\exp \left( -S\left( \Psi \right) \right) $,
where $S\left( \Psi \right) $ is a functional, i.e. the function of the
function $\Psi $. The form of $S\left( \Psi \right) $ is derived directly
from the form of (\ref{pnz}) as detailed in the text. As seen from (\ref{pnz}%
), this translation can in fact be directly obtained from the sum of
"classical" minimization functions weighted by the factors $\frac{1}{\sigma
_{\alpha }^{2}}$:%
\begin{eqnarray*}
&&\sum_{i,\alpha }\frac{1}{\sigma _{\alpha }^{2}}\int dt\left( \frac{d%
\mathbf{A}_{i}\left( t\right) }{dt}-\sum_{j,k,l...}f^{\left( \alpha \right)
}\left( \mathbf{A}_{i}\left( t\right) ,\mathbf{A}_{j}\left( t\right) ,%
\mathbf{A}_{k}\left( t\right) ,\mathbf{A}_{l}\left( t\right) ...\right)
\right) ^{2} \\
&&+\sum_{i,\alpha }\left( \sum_{j,k,l...}g^{\left( \alpha \right) }\left( 
\mathbf{A}_{i}\left( t\right) ,\mathbf{A}_{j}\left( t\right) ,\mathbf{A}%
_{k}\left( t\right) ,\mathbf{A}_{l}\left( t\right) ...\right) \right)
\end{eqnarray*}%
This is this shortcut we used in the text.

\subsection*{A2.2 Interactions between agents at different times}

A straightforward generalization of (\ref{mNZ}) involve agents interactions
at different times. The terms considered have the form:%
\begin{eqnarray}
&&\sum_{i}\left( \frac{d\mathbf{A}_{i}\left( t\right) }{dt}%
-\sum_{j,k,l...}\int f\left( \mathbf{A}_{i}\left( t_{i}\right) ,\mathbf{A}%
_{j}\left( t_{j}\right) ,\mathbf{A}_{k}\left( t_{k}\right) ,\mathbf{A}%
_{l}\left( t_{l}\right) ...,\mathbf{t}\right) \mathbf{dt}\right) ^{2}
\label{gR} \\
&&+\sum_{i}\sum_{j,k,l...}\int g\left( \mathbf{A}_{i}\left( t_{i}\right) ,%
\mathbf{A}_{j}\left( t_{j}\right) ,\mathbf{A}_{k}\left( t_{k}\right) ,%
\mathbf{A}_{l}\left( t_{l}\right) ...,\mathbf{t}\right) \mathbf{dt}  \notag
\end{eqnarray}%
where $\mathbf{t}$\ stands for $\left( t_{i},t_{j},t_{k},t_{l}...\right) $
and $\mathbf{dt}$ stands for $dt_{i}dt_{j}dt_{k}dt_{l}...$

The translation is straightforward. We introduce a time variable $\theta $
on the field side and the fields write $\left\vert \Psi \left( \mathbf{A}%
,\theta \right) \right\vert ^{2}$ and $\left\vert \hat{\Psi}\left( \mathbf{%
\hat{A}},\hat{\theta}\right) \right\vert ^{2}$. The second term in (\ref{gR}%
) becomes: 
\begin{eqnarray}
&&\sum_{i}\sum_{j}\sum_{j,k...}\int g\left( \mathbf{A}_{i}\left(
t_{i}\right) ,\mathbf{A}_{j}\left( t_{j}\right) ,\mathbf{A}_{k}\left(
t_{k}\right) ,\mathbf{A}_{l}\left( t_{l}\right) ...,\mathbf{t}\right) 
\mathbf{dt}  \notag \\
&\rightarrow &\int g\left( \mathbf{A},\mathbf{A}^{\prime },\mathbf{A}%
^{\prime \prime },\mathbf{\hat{A},\hat{A}}^{\prime }...,\mathbf{\theta ,\hat{%
\theta}}\right) \left\vert \Psi \left( \mathbf{A},\theta \right) \right\vert
^{2}\left\vert \Psi \left( \mathbf{A}^{\prime },\theta ^{\prime }\right)
\right\vert ^{2}\left\vert \Psi \left( \mathbf{A}^{\prime \prime },\theta
^{\prime \prime }\right) \right\vert ^{2}d\mathbf{A}d\mathbf{A}^{\prime }d%
\mathbf{A}^{\prime \prime } \\
&&\times \left\vert \hat{\Psi}\left( \mathbf{\hat{A}},\hat{\theta}\right)
\right\vert ^{2}\left\vert \hat{\Psi}\left( \mathbf{\hat{A}}^{\prime },\hat{%
\theta}^{\prime }\right) \right\vert ^{2}d\mathbf{\hat{A}}d\mathbf{\hat{A}}%
^{\prime }\mathbf{d\theta d\hat{\theta}}  \notag
\end{eqnarray}%
where $\mathbf{\theta }$ and $\mathbf{\hat{\theta}}$ are the multivariables $%
\left( \theta ,\theta ^{\prime },\theta ^{\prime \prime }...\right) $ and $%
\left( \hat{\theta},\hat{\theta}^{\prime }...\right) $ respectively and $%
\mathbf{d\theta d\hat{\theta}}$ stands for $d\theta d\theta ^{\prime
}d\theta ^{\prime \prime }...$ and $d\hat{\theta}d\hat{\theta}^{\prime }...$

Similarly, the first term in (\ref{gR}) translates as:%
\begin{eqnarray}
&&\sum_{i}\left( \frac{d\mathbf{A}_{i}\left( t\right) }{dt}%
-\sum_{j,k,l...}\int f\left( \mathbf{A}_{i}\left( t_{i}\right) ,\mathbf{A}%
_{j}\left( t_{j}\right) ,\mathbf{A}_{k}\left( t_{k}\right) ,\mathbf{A}%
_{l}\left( t_{l}\right) ...,\mathbf{t}\right) \mathbf{dt}\right) ^{2} \\
&\rightarrow &\int \Psi ^{\dag }\left( \mathbf{A},\theta \right) \left(
-\nabla _{\mathbf{A}^{\left( \alpha \right) }}\left( \frac{\sigma _{\mathbf{A%
}^{\left( \alpha \right) }}^{2}}{2}\nabla _{\mathbf{A}^{\left( \alpha
\right) }}-\Lambda (\mathbf{A},\theta )\right) \right) \Psi \left( \mathbf{A}%
,\theta \right) d\mathbf{A}d\theta
\end{eqnarray}%
by:%
\begin{eqnarray}
\Lambda (\mathbf{A},\theta ) &=&\int f^{\left( \alpha \right) }\left( 
\mathbf{A},\mathbf{A}^{\prime },\mathbf{A}^{\prime \prime },\mathbf{\hat{A},%
\hat{A}}^{\prime }...,\mathbf{\theta ,\hat{\theta}}\right) \left\vert \Psi
\left( \mathbf{A}^{\prime },\theta ^{\prime }\right) \right\vert
^{2}\left\vert \Psi \left( \mathbf{A}^{\prime \prime },\theta ^{\prime
\prime }\right) \right\vert ^{2}d\mathbf{A}^{\prime }d\mathbf{A}^{\prime
\prime } \\
&&\times \left\vert \hat{\Psi}\left( \mathbf{\hat{A}},\theta \right)
\right\vert ^{2}\left\vert \hat{\Psi}\left( \mathbf{\hat{A}}^{\prime
},\theta ^{\prime \prime }\right) \right\vert ^{2}d\mathbf{\hat{A}}d\mathbf{%
\hat{A}}^{\prime }\mathbf{d\bar{\theta}d\hat{\theta}}  \notag
\end{eqnarray}%
with $\mathbf{d\bar{\theta}}=d\theta ^{\prime }d\theta ^{\prime \prime }$.

Ultimately, as in the text, additional terms (\ref{Trl}):%
\begin{eqnarray}
&&\Psi ^{\dag }\left( \mathbf{A},\theta \right) \left( -\nabla _{\theta
}\left( \frac{\sigma _{\theta }^{2}}{2}\nabla _{\theta }-1\right) \right)
\Psi \left( \mathbf{A},\theta \right) \\
&&+\hat{\Psi}^{\dag }\left( \mathbf{\hat{A}},\theta \right) \left( -\nabla
_{\theta }\left( \frac{\sigma _{\theta }^{2}}{2}\nabla _{\theta }-1\right)
\right) \hat{\Psi}\left( \mathbf{\hat{A}},\theta \right) +\alpha \left\vert
\Psi \left( \mathbf{A}\right) \right\vert ^{2}+\alpha \left\vert \hat{\Psi}%
\left( \mathbf{\hat{A}}\right) \right\vert ^{2}  \notag
\end{eqnarray}%
are included to the action functional to take into account for the time
variable.

\section*{Appendix 3 Translation of the model minimization functions}

We call $\hat{\Psi}$ the field describing the investors.\ It depends on the
two variables $\hat{K}$ and $\hat{X}$ and We call $\Psi $ the field
describing the investors.\ It depends on the two variables $\hat{K}$ and $%
\hat{X}$

\subsection*{A3.1. Firms action functional and firms' return}

\subsubsection*{A3.1.1 Translation of the minimization function: Physical
capital accumulation}

Let us start by translating in terms of fields the minimization function
associated to (\ref{DC}):%
\begin{equation}
\sum_{i}\left( \frac{d}{dt}K_{ip}\left( t\right) -f_{1}^{\prime }\left(
K_{i}\left( t\right) \right) K_{ip}\left( t\right) \right) ^{2}  \label{FC}
\end{equation}%
with:%
\begin{equation}
f_{1}^{\prime }\left( K_{i}\left( t\right) \right) =\left( 1+\sum_{\nu
}k_{2j\nu }\hat{K}_{\nu }\left( t\right) \right) r_{i}-\bar{r}\sum_{v}k_{2lv}%
\hat{K}_{v}\left( t\right)  \label{rt}
\end{equation}

To translate the full term (\ref{FC}), we use the translation (\ref{Trl}) of
a type-(\ref{inco}) expression. The gradient term appearing in equation (\ref%
{Trl}) is $\nabla _{K}$. We thus obtain the translation: 
\begin{eqnarray}
&&\sum_{i}\left( \frac{d}{dt}K_{ip}\left( t\right) -f_{1}^{\prime }\left(
K_{i}\left( t\right) \right) K_{ip}\left( t\right) \right) ^{2}  \label{grl}
\\
&\rightarrow &\int \Psi ^{\dag }\left( K,X\right) \left( -\nabla _{K}\left( 
\frac{\sigma _{K}^{2}}{2}\nabla _{K}+\Lambda (X,K)\right) \right) \Psi
\left( K,X\right) dKdX  \notag
\end{eqnarray}%
Note that the variance $\sigma _{K}^{2}$ reflects the probabilistic nature
of the model hidden behind the field formalism. This $\sigma _{K}^{2}$
represents the characteristic level of uncertainty of the sectors space
dynamics. It is a parameter of the model. The term $\Lambda (X,K)$ is the
translation of the term $f_{1}^{\prime }\left( K_{i}\left( t\right) \right) $
in the parenthesis of (\ref{FC}). This term is a function of one sole index "%
$i$". In that case, the term $\Lambda $ is simply obtained by replacing $%
\left( K_{i},X_{i}\right) $ by $\left( K,X\right) $.\ We use the translation
(\ref{bdt}) of (\ref{ntr})-type term. The term (\ref{rt}). This term
contains no derivative. The form of the translation is given by formula (\ref%
{tln}).

The first step of the translation is to replace $r_{i}$ by a function $%
r\left( K,X\right) $ and $\hat{K}_{\nu }\left( t\right) $ by the variable $%
\hat{K}$, and to replace:%
\begin{equation*}
k_{2j\nu }\rightarrow k_{2}\left( K,X,\hat{K},\hat{X}\right)
\end{equation*}%
Given our assumptions in the txt, $k_{2}$ is a function of the sole variable 
$X$:%
\begin{equation*}
k_{2j\nu }\rightarrow k_{2}\left( X\right)
\end{equation*}%
The other one have been sorted out in a linear approximation. This means
that $k_{2}\left( X\right) $ captures the average share of loans in
investmnts in sector $X$. This leads to substitute:

\begin{eqnarray*}
&&f_{1}^{\prime }\left( K_{i}\left( t\right) \right) \\
&\rightarrow &\Lambda (X,K)=\left( 1+\int k_{2}\left( X\right) \int
\left\vert \hat{\Psi}\left( \hat{K},\hat{X}\right) \right\vert ^{2}d\left( 
\hat{K},\hat{X}\right) \right) r\left( K,X\right) -\bar{r}\int k_{2}\left(
X\right) \int \left\vert \hat{\Psi}\left( \hat{K},\hat{X}\right) \right\vert
^{2}d\left( \hat{K},\hat{X}\right)
\end{eqnarray*}%
The action functional for the field of firms is:%
\begin{equation*}
-\Psi ^{\dag }\left( K,X\right) \left( \nabla _{K_{p}}\left( \sigma
_{K}^{2}\nabla _{K_{p}}-f_{1}^{\prime }\left( K,X\right) K_{p}\right)
\right) \Psi \left( K,X\right)
\end{equation*}%
where: 
\begin{eqnarray*}
f_{1}^{\prime }\left( X\right) &=&\left( 1+\underline{k}_{2}\left( X\right)
\right) f_{1}\left( X\right) -\bar{r}\underline{k}_{2}\left( \hat{X}\right)
\\
&=&f_{1}\left( X\right) +\left( f_{1}\left( X\right) -\bar{r}\right) \int
k_{2}\left( X,\hat{X}_{1}\right) \hat{K}_{1}\left\vert \hat{\Psi}\left( \hat{%
K}_{1},\hat{X}_{1}\right) \right\vert ^{2}d\hat{K}_{1}d\hat{X}_{1}
\end{eqnarray*}

\subsubsection*{A3.1.1 Introducing fluctuation of number of firms}

The description is completed by assuming that the number of firms is not
constant in one sector, but that some rigidity prevents this number to
change quickly. Rather than introducing some dynamics for the number of
firms as in our previous works, we rather consider for the sake of
simplicity that frms number in one sector fluctuate around some average
quantities. An hypothesis that is sufficient in the medium run to study the
impact of capital flows. Technically, this amounts to introduce a potential
in the field action ensuring this type of rigidity. The potential has the
frm:%
\begin{equation*}
\frac{1}{2\epsilon }\left( \left\vert \Psi \left( K,X\right) \right\vert
^{2}-\left\vert \Psi _{0}\left( X\right) \right\vert ^{2}\right) ^{2}
\end{equation*}%
and the full action for firms wrts:%
\begin{equation}
-\Psi ^{\dag }\left( K,X\right) \left( \nabla _{K_{p}}\left( \sigma
_{K}^{2}\nabla _{K_{p}}-f_{1}^{\prime }\left( K,X\right) K_{p}\right)
\right) \Psi \left( K,X\right) +\frac{1}{2\epsilon }\left( \left\vert \Psi
\left( K,X\right) \right\vert ^{2}-\left\vert \Psi _{0}\left( X\right)
\right\vert ^{2}\right) ^{2}  \label{Fc}
\end{equation}

\subsection*{A3.2 Field action functional and return for financial markets}

The functions to be translated are those of the financial capital dynamics
We consider the minimisation function associatd to the dynamics (\ref{DFQTW}%
):%
\begin{equation}
\left( \frac{d}{dt}\hat{K}_{j}\left( t\right) -\sum_{l}\left( 1-\frac{\hat{k}%
_{1jl}\hat{K}_{j}\left( t\right) }{1+\sum_{v}\hat{k}_{jv}\hat{K}_{v}\left(
t\right) }\right) ^{-1}f_{l}\hat{K}_{l}\left( t\right) \right) ^{2}=\left( 
\frac{d}{dt}\hat{K}_{j}\left( t\right) -\sum_{l}\left( 1-M\right)
_{jl}^{-1}f_{l}\hat{K}_{l}\left( t\right) \right) ^{2}  \label{TC}
\end{equation}%
where:%
\begin{equation}
M_{jm}=\frac{\hat{k}_{jm}\hat{K}_{j}\left( t\right) }{1+\sum_{\nu }\hat{k}%
_{j\nu }\hat{K}_{\nu }\left( t\right) }
\end{equation}%
\ Both expressions include a time derivative and are thus of type (\ref{edr}%
). As for the real economy, the application of the translation rules is
straightforward using the general translation formula of expression (\ref%
{inco}) in (\ref{Trl}), into:%
\begin{equation*}
\int \hat{\Psi}^{\dag }\left( \hat{K},\hat{X}\right) \left( -\nabla _{\hat{K}%
}\left( \frac{\sigma _{\hat{K}}^{2}}{2}\nabla _{\hat{K}}+\Lambda \left( \hat{%
K},\hat{X}\right) \right) \right) \hat{\Psi}\left( \hat{K},\hat{X}\right) d%
\hat{K}d\hat{X}
\end{equation*}%
The function $\Lambda \left( \hat{K},\hat{X}\right) $ is obtained, as
before, by translating the term following the derivative in the function (%
\ref{TC}) and b rplcng vrbl: 
\begin{eqnarray*}
\left( K_{i},X_{i}\right) &\rightarrow &\left( K,X\right) \\
\left( K_{l},X_{l}\right) &\rightarrow &\left( K^{\prime },X^{\prime }\right)
\\
\left( \hat{K}_{j},\hat{X}_{j}\right) &\rightarrow &\left( \hat{K},\hat{X}%
\right)
\end{eqnarray*}%
We also need to translate the return equation for investors (\ref{RTQ}):%
\begin{eqnarray}
&&\sum_{l}\left( \delta _{jl}-\frac{\hat{k}_{1lj}\hat{K}_{l}\left( t\right) 
}{1+\sum_{v}\left( \hat{k}_{1lv}+\hat{k}_{2lv}\right) \hat{K}_{v}\left(
t\right) }\right) \left( \frac{\hat{f}_{j}-\bar{r}}{1+\sum_{v}\hat{k}_{2jv}%
\hat{K}_{v}\left( t\right) }\right) \\
&&+\sum_{l}\left( \bar{r}-\frac{\left( 1+\hat{f}\left( \hat{K}_{vp}\left(
t\right) \right) \right) }{\sum_{m}\hat{k}_{2vm}\hat{K}_{m}}\right) H\left(
-\left( 1+\hat{f}\left( \hat{K}_{vp}\left( t\right) \right) \right) \right) 
\frac{\hat{k}_{2lj}\hat{K}_{l}\left( t\right) }{1+\sum_{v}\left( \hat{k}%
_{1lv}+\hat{k}_{2lv}\right) \hat{K}_{v}\left( t\right) }  \notag \\
&&+\sum_{i}\left( \bar{r}-\frac{\left( 1+f_{1}^{\prime }\left( K_{i}\left(
t\right) \right) \right) }{\sum_{m}k_{2im}\hat{K}_{m}}\right) H\left(
-\left( 1+f_{1}^{\prime }\left( K_{i}\left( t\right) \right) \right) \right) 
\frac{k_{2ij}K_{i}\left( t\right) }{1+\sum_{v}\left( k_{1iv}+k_{2iv}\right) 
\hat{K}_{v}\left( t\right) }  \notag \\
&=&\sum_{i}\left( \frac{f_{1}^{\prime }\left( K_{i}\left( t\right) \right) -%
\bar{r}}{1+\sum_{\nu }k_{2j\nu }\hat{K}_{\nu }\left( t\right) }+\Delta
F_{\tau }\left( \bar{R}\left( K_{i},X_{i}\right) \right) \right) \frac{%
k_{1ij}K_{i}\left( t\right) }{1+\sum_{v}\left( k_{1iv}+k_{2iv}\right) \hat{K}%
_{v}\left( t\right) }  \notag
\end{eqnarray}%
This equation can be written as compctly as:%
\begin{equation*}
RT=0
\end{equation*}%
which is implemented as a minimisation function:%
\begin{equation}
\frac{1}{\epsilon }\left( RT\right) ^{2}  \label{Tc}
\end{equation}%
with $\epsilon <<1$. The translation of such a term will lead to a potential
that will corresponds to a field contraints when $\epsilon \rightarrow 0$.

The translation of (\ref{TC}) and (\ref{Tc}) is done in several steps, by
introducing some notations and performing some intermediate translations of
the coefficients arising in these functions.

\subsubsection*{A3.2.1 Translations of shares of invested capital}

Let us first define some parameters that will appear in the translation.\ We
define:%
\begin{eqnarray*}
\underline{\hat{k}}_{2}\left( \hat{X}^{\prime }\right) &=&\int \hat{k}%
_{2}\left( \hat{X}^{\prime },\hat{X}_{1}\right) \hat{K}_{1}\left\vert \hat{%
\Psi}\left( \hat{K}_{1},\hat{X}_{1}\right) \right\vert ^{2}d\hat{K}_{1}d\hat{%
X}_{1} \\
\underline{\hat{k}}_{1}\left( \hat{X}^{\prime }\right) &=&\int \hat{k}%
_{1}\left( \hat{X}^{\prime },\hat{X}_{1}\right) \hat{K}_{1}\left\vert \hat{%
\Psi}\left( \hat{K}_{1},\hat{X}_{1}\right) \right\vert ^{2}d\hat{K}_{1}d\hat{%
X}_{1} \\
\underline{\hat{k}}\left( \hat{X}^{\prime }\right) &=&\int \hat{k}\left( 
\hat{X}^{\prime },\hat{X}_{1}\right) \hat{K}_{1}\left\vert \hat{\Psi}\left( 
\hat{K}_{1},\hat{X}_{1}\right) \right\vert ^{2}d\hat{K}_{1}d\hat{X}_{1}=%
\underline{\hat{k}}_{1}\left( \hat{X}^{\prime }\right) +\underline{\hat{k}}%
_{2}\left( \hat{X}^{\prime }\right)
\end{eqnarray*}%
These three parameters can also be writen in a more compact way as:%
\begin{equation*}
\underline{\hat{k}}_{\eta }\left( \hat{X}^{\prime }\right) =\int \hat{k}%
_{\eta }\left( \hat{X}^{\prime },\hat{X}\right) \hat{K}_{\hat{X}}\left\vert 
\hat{\Psi}\left( \hat{X}\right) \right\vert ^{2}d\hat{X}
\end{equation*}%
\begin{equation*}
\underline{\hat{k}}_{1}\left( \hat{X}^{\prime }\right) +\underline{\hat{k}}%
_{2}\left( \hat{X}^{\prime }\right) =\underline{\hat{k}}\left( \hat{X}%
^{\prime }\right)
\end{equation*}%
These parameters correspond to the amount of capital invested in the
financial sector $\hat{X}^{\prime }$, in terms of loans ($\underline{\hat{k}}%
_{2}$), participation ($\underline{\hat{k}}_{1}$), and their total sum ($%
\underline{\hat{k}}_{1}+\underline{\hat{k}}_{2}$). We also define the shares
of capital outgoing from sector $\hat{X}^{\prime }$ towards other financial
sectors: 
\begin{eqnarray}
\underline{\hat{k}}_{2E}\left( \hat{X}^{\prime }\right) &=&\int \hat{k}%
_{2}\left( \hat{X}_{1},\hat{X}^{\prime }\right) \hat{K}_{1}\left\vert \hat{%
\Psi}\left( \hat{K}_{1},\hat{X}_{1}\right) \right\vert ^{2}d\hat{K}_{1}d\hat{%
X}_{1}  \label{KE} \\
\underline{\hat{k}}_{1E}\left( \hat{X}^{\prime }\right) &=&\int \hat{k}%
_{1}\left( \hat{X}_{1},\hat{X}^{\prime }\right) \hat{K}_{1}\left\vert \hat{%
\Psi}\left( \hat{K}_{1},\hat{X}_{1}\right) \right\vert ^{2}d\hat{K}_{1}d\hat{%
X}_{1}  \notag \\
\underline{\hat{k}}_{E}\left( \hat{X}^{\prime }\right) &=&\int \hat{k}\left( 
\hat{X}_{1},\hat{X}^{\prime }\right) \hat{K}_{1}\left\vert \hat{\Psi}\left( 
\hat{K}_{1},\hat{X}_{1}\right) \right\vert ^{2}d\hat{K}_{1}d\hat{X}_{1}=%
\underline{\hat{k}}_{1E}\left( \hat{X}^{\prime }\right) +\underline{\hat{k}}%
_{2E}\left( \hat{X}^{\prime }\right)  \notag
\end{eqnarray}%
Similarly, we define the amount of capital invested in real sector $\hat{X}%
^{\prime }$: 
\begin{eqnarray*}
\underline{k}_{2}\left( \hat{X}^{\prime }\right) &=&\int k_{2}\left( \hat{X}%
^{\prime },\hat{X}_{1}\right) \hat{K}_{1}\left\vert \hat{\Psi}\left( \hat{K}%
_{1},\hat{X}_{1}\right) \right\vert ^{2}d\hat{K}_{1}d\hat{X}_{1} \\
\underline{k}_{1}\left( \hat{X}^{\prime }\right) &=&\int k_{1}\left( \hat{X}%
^{\prime },\hat{X}_{1}\right) \hat{K}_{1}\left\vert \hat{\Psi}\left( \hat{K}%
_{1},\hat{X}_{1}\right) \right\vert ^{2}d\hat{K}_{1}d\hat{X}_{1} \\
\underline{k}\left( \hat{X}^{\prime }\right) &=&\int k\left( \hat{X}^{\prime
},\hat{X}_{1}\right) \hat{K}_{1}\left\vert \hat{\Psi}\left( \hat{K}_{1},\hat{%
X}_{1}\right) \right\vert ^{2}d\hat{K}_{1}d\hat{X}_{1}=\underline{k}%
_{1}\left( \hat{X}^{\prime }\right) +\underline{k}_{2}\left( \hat{X}^{\prime
}\right)
\end{eqnarray*}%
and the outgoing amount of capital outgoing from $\hat{X}^{\prime }$ toward
the real sectrs: 
\begin{eqnarray*}
\underline{k}_{2E}\left( \hat{X}^{\prime }\right) &=&\int k_{2}\left( \hat{X}%
_{1},\hat{X}^{\prime }\right) \hat{K}_{1}\left\vert \hat{\Psi}\left( \hat{K}%
_{1},\hat{X}_{1}\right) \right\vert ^{2}d\hat{K}_{1}d\hat{X}_{1} \\
\underline{k}_{1E}\left( \hat{X}^{\prime }\right) &=&\int k_{1}\left( \hat{X}%
_{1},\hat{X}^{\prime }\right) \hat{K}_{1}\left\vert \hat{\Psi}\left( \hat{K}%
_{1},\hat{X}_{1}\right) \right\vert ^{2}d\hat{K}_{1}d\hat{X}_{1} \\
\underline{k}_{E}\left( \hat{X}^{\prime }\right) &=&\int k\left( \hat{X}_{1},%
\hat{X}^{\prime }\right) \hat{K}_{1}\left\vert \hat{\Psi}\left( \hat{K}_{1},%
\hat{X}_{1}\right) \right\vert ^{2}d\hat{K}_{1}d\hat{X}_{1}=\underline{k}%
_{1E}\left( \hat{X}^{\prime }\right) +\underline{k}_{2E}\left( \hat{X}%
^{\prime }\right)
\end{eqnarray*}%
where :%
\begin{equation*}
\underline{k}_{\eta }\left( \hat{X}^{\prime }\right) =\int k_{\eta }\left( 
\hat{X}^{\prime },\hat{X}\right) \hat{K}_{\hat{X}}\left\vert \hat{\Psi}%
\left( \hat{X}\right) \right\vert ^{2}d\hat{X}
\end{equation*}%
and:%
\begin{equation*}
\hat{k}_{1}\left( \hat{X},\hat{Y}\right) +\hat{k}_{2}\left( \hat{X},\hat{Y}%
\right) \rightarrow \hat{k}\left( \hat{X},\hat{Y}\right)
\end{equation*}%
Once these quantities are defined, the translation method will lead to the
following substitutions, from the micro system to the field model:%
\begin{equation*}
1+\sum_{v}\left( \hat{k}_{1lv}+\hat{k}_{2lv}\right) \hat{K}_{v}\left(
t\right) \rightarrow 1+\int \left( \hat{k}\left( \hat{X},\hat{Y}\right)
\right) \hat{K}^{\prime \prime }\left\vert \hat{\Psi}\left( \hat{K}^{\prime
\prime },Y\right) \right\vert ^{2}d\hat{K}^{\prime \prime }Y
\end{equation*}%
\begin{equation*}
\frac{\left( \hat{k}_{1jl}+\hat{k}_{2jl}\right) }{1+\sum_{v}\left( \hat{k}%
_{1jv}+\hat{k}_{2jv}\right) \hat{K}_{v}\left( t\right) }\rightarrow \frac{%
\hat{k}\left( \hat{X},\hat{X}^{\prime }\right) }{1+\int \hat{k}\left( \hat{X}%
,\hat{Y}\right) \hat{K}^{\prime \prime }\left\vert \hat{\Psi}\left( \hat{K}%
^{\prime \prime },Y\right) \right\vert ^{2}d\hat{K}^{\prime \prime }Y}
\end{equation*}%
\begin{equation*}
\frac{\left( k_{1jl}+k_{2jl}\right) }{1+\sum_{v}\left(
k_{1jv}+k_{2jv}\right) \hat{K}_{v}\left( t\right) }\rightarrow \frac{k\left( 
\hat{X},\hat{X}^{\prime }\right) }{1+\int k\left( \hat{X},\hat{Y}\right) 
\hat{K}^{\prime \prime }\left\vert \hat{\Psi}\left( \hat{K}^{\prime \prime
},Y\right) \right\vert ^{2}d\hat{K}^{\prime \prime }Y}
\end{equation*}%
\begin{equation*}
M_{jl}\rightarrow M\left( \hat{K},\hat{X},\hat{K}^{\prime },\hat{X}^{\prime
}\right) =\frac{\hat{k}\left( \hat{X},\hat{X}^{\prime }\right) \hat{K}}{1+%
\underline{\hat{k}}\left( \hat{X}\right) }
\end{equation*}%
and, ultimately:

\begin{eqnarray*}
&&\left( 1+\sum_{\nu }\hat{k}_{j\nu }\hat{K}_{\nu }\left( t\right) \right)
\left( 1-M\right) _{jl}^{-1}\left( R_{l}-\bar{r}\sum_{v}\hat{k}_{2lv}\hat{K}%
_{v}\left( t\right) \right) \\
&\rightarrow &\left( 1+\underline{\hat{k}}\left( \hat{X}\right) \right)
\left( 1-M\left( \hat{K},\hat{X},\hat{K}^{\prime },\hat{X}^{\prime }\right)
\left\vert \hat{\Psi}\left( \hat{K}^{\prime },\hat{X}^{\prime }\right)
\right\vert ^{2}\right) ^{-1}\left( R\left( \hat{X}^{\prime }\right) -\bar{r}%
\underline{\hat{k}}_{2}\left( \hat{X}^{\prime }\right) \right)
\end{eqnarray*}

\subsubsection*{A3.2.2 Translation of constraints and their averages}

We consider that each investor $j$\ invest, each period, the entire amount
of his capital, which translates in the constraint:%
\begin{equation*}
\sum_{\nu }\frac{\hat{k}_{vj}\hat{K}_{v}\left( t\right) }{1+\sum_{m}\hat{k}%
_{vm}\hat{K}_{m}\left( t\right) }+\sum_{i}\frac{k_{ij}K_{i}\left( t\right) }{%
1+\sum_{v}k_{iv}\hat{K}_{v}\left( t\right) }=1
\end{equation*}%
that becomes in terms of fields:%
\begin{equation*}
\int \left\vert \hat{\Psi}\left( \hat{K}^{\prime },\hat{X}^{\prime }\right)
\right\vert ^{2}\frac{\hat{k}\left( \hat{X}^{\prime },\hat{X}\right) \hat{K}%
^{\prime }}{1+\underline{\hat{k}}\left( \hat{X}^{\prime }\right) }+\int
\left\vert \Psi \left( K^{\prime },X^{\prime }\right) \right\vert ^{2}\frac{%
k\left( X^{\prime },X\right) K^{\prime }}{1+\underline{k}\left( \hat{X}%
^{\prime }\right) }=1
\end{equation*}%
that can rewritten, given our previous definition (\ref{KE}):%
\begin{equation*}
\frac{\underline{\hat{k}}_{E}\left( \hat{X}\right) }{1+\underline{\hat{k}}}+%
\frac{\underline{k}_{E}\left( X\right) }{1+\underline{k}}=1
\end{equation*}%
We can consider the average of this equation by defining:

\begin{equation*}
\int \underline{\hat{k}}_{i}\left( \hat{X}\right) \left\vert \hat{\Psi}%
\left( \hat{K},\hat{X}\right) \right\vert ^{2}=\underline{\hat{k}}_{i}
\end{equation*}%
\begin{equation*}
\int \underline{k}_{i}\left( X\right) \left\vert \Psi \left( K,X\right)
\right\vert ^{2}=\underline{k}_{i}
\end{equation*}%
\begin{equation*}
\int \underline{\hat{k}}_{iE}\left( \hat{X}\right) \left\vert \hat{\Psi}%
\left( \hat{K},\hat{X}\right) \right\vert ^{2}=\underline{\hat{k}}_{iE}
\end{equation*}%
The average constraints thus write\textbf{:}%
\begin{equation*}
\frac{\underline{\hat{k}}_{E}}{1+\underline{\hat{k}}}+\frac{\underline{k}_{E}%
}{1+\underline{k}}=1
\end{equation*}%
\begin{equation*}
\frac{\underline{\hat{k}}}{1+\underline{\hat{k}}}+\frac{\underline{k}}{1+%
\underline{k}}=1
\end{equation*}%
\begin{equation}
\frac{\underline{k}}{1+\underline{k}}=\frac{1}{1+\underline{\hat{k}}}
\label{CT}
\end{equation}

\subsubsection*{A3.2.3 Estimation of $\left( 1-M_{jl}\right) ^{-1}$}

The matrix $\left( 1-M_{jl}\right) ^{-1}$ arising in (\ref{DFQTW}) and (\ref%
{MAT}) is averaged on some vector with coordinate $j$. In field theory, this
corresponds to sum over the argument for the field, that is computing
averages. Expanding in series the field translation of $\left(
1-M_{jl}\right) ^{-1}$, leads to: 
\begin{eqnarray*}
\left( 1-M_{jl}\right) ^{-1} &\rightarrow &\frac{1}{1-M\left( \left( \hat{K},%
\hat{X}\right) ,\left( \hat{K}^{\prime },\hat{X}^{\prime }\right) \right)
\left\vert \hat{\Psi}\left( \hat{K}^{\prime },\hat{X}^{\prime }\right)
\right\vert ^{2}} \\
&=&1+M\left( \left( \hat{K},\hat{X}\right) ,\left( \hat{K}^{\prime },\hat{X}%
^{\prime }\right) \right) \left\vert \hat{\Psi}\left( \hat{K}^{\prime },\hat{%
X}^{\prime }\right) \right\vert ^{2} \\
&&+M\left( \left( \hat{K},\hat{X}\right) ,\left( \hat{K}_{1}^{\prime },\hat{X%
}_{1}^{\prime }\right) \right) \left\vert \hat{\Psi}\left( \hat{K}%
_{1}^{\prime },\hat{X}_{1}^{\prime }\right) \right\vert ^{2}M\left( \left( 
\hat{K}_{1}^{\prime },\hat{X}_{1}^{\prime }\right) ,\left( \hat{K}^{\prime },%
\hat{X}^{\prime }\right) \right) \left\vert \hat{\Psi}\left( \hat{K}%
_{1}^{\prime },\hat{X}_{1}^{\prime }\right) \right\vert ^{2} \\
&&+M\left( \left( \hat{K},\hat{X}\right) ,\left( \hat{K}_{1}^{\prime },\hat{X%
}_{1}^{\prime }\right) \right) \left\vert \hat{\Psi}\left( \hat{K}%
_{1}^{\prime },\hat{X}_{1}^{\prime }\right) \right\vert ^{2} \\
&&\times M\left( \left( \hat{K}_{1}^{\prime },\hat{X}_{1}^{\prime }\right)
,\left( \hat{K}_{2}^{\prime },\hat{X}_{2}^{\prime }\right) \right)
\left\vert \hat{\Psi}\left( \hat{K}_{2}^{\prime },\hat{X}_{2}^{\prime
}\right) \right\vert ^{2}M\left( \left( \hat{K}_{2}^{\prime },\hat{X}%
_{2}^{\prime }\right) ,\left( \hat{K}^{\prime },\hat{X}^{\prime }\right)
\right) \left\vert \hat{\Psi}\left( \hat{K}^{\prime },\hat{X}^{\prime
}\right) \right\vert ^{2}
\end{eqnarray*}%
and the application to any vector $A\left( \left( \hat{K}^{\prime },\hat{X}%
^{\prime }\right) \right) $: 
\begin{equation*}
\int \frac{1}{1-M\left( \left( \hat{K},\hat{X}\right) ,\left( \hat{K}%
^{\prime },\hat{X}^{\prime }\right) \right) \left\vert \hat{\Psi}\left( \hat{%
K}^{\prime },\hat{X}^{\prime }\right) \right\vert ^{2}}A\left( \left( \hat{K}%
^{\prime },\hat{X}^{\prime }\right) \right)
\end{equation*}%
can be approximated by:%
\begin{eqnarray*}
&&1+M\left( \left( \hat{K},\hat{X}\right) ,\left( \hat{K}^{\prime },\hat{X}%
^{\prime }\right) \right) \left\vert \hat{\Psi}\left( \hat{K}^{\prime },\hat{%
X}^{\prime }\right) \right\vert ^{2}A\left( \left( \hat{K}^{\prime },\hat{X}%
^{\prime }\right) \right) \\
&&+M\left( \left( \hat{K},\hat{X}\right) ,\left( \left\langle \hat{K}%
\right\rangle ,\left\langle \hat{X}\right\rangle \right) \right) \left\Vert 
\hat{\Psi}\right\Vert ^{2}M\left( \left( \left\langle \hat{K}\right\rangle
,\left\langle \hat{X}\right\rangle \right) ,\left( \hat{K}^{\prime },\hat{X}%
^{\prime }\right) \right) \left\vert \hat{\Psi}\left( \hat{K}^{\prime },\hat{%
X}^{\prime }\right) \right\vert ^{2}A\left( \left( \hat{K}^{\prime },\hat{X}%
^{\prime }\right) \right) \\
&&+... \\
&=&1+M\left( \left( \hat{K},\hat{X}\right) ,\left( \left\langle \hat{K}%
\right\rangle ,\left\langle \hat{X}\right\rangle \right) \right) \left\Vert 
\hat{\Psi}\right\Vert ^{2}\left\langle A\right\rangle \\
&&+M\left( \left( \hat{K},\hat{X}\right) ,\left( \left\langle \hat{K}%
\right\rangle ,\left\langle \hat{X}\right\rangle \right) \right) \left\Vert 
\hat{\Psi}\right\Vert ^{2}M\left( \left( \left\langle \hat{K}\right\rangle
,\left\langle \hat{X}\right\rangle \right) ,\left( \left\langle \hat{K}%
\right\rangle ,\left\langle \hat{X}\right\rangle \right) \right) \left\Vert 
\hat{\Psi}\right\Vert ^{2}\left\langle A\right\rangle +...
\end{eqnarray*}%
This corresponds to replace:%
\begin{equation*}
\int \frac{1}{1-M\left( \left( \hat{K},\hat{X}\right) ,\left( \hat{K}%
^{\prime },\hat{X}^{\prime }\right) \right) \left\vert \hat{\Psi}\left( \hat{%
K}^{\prime },\hat{X}^{\prime }\right) \right\vert ^{2}}A\left( \left( \hat{K}%
^{\prime },\hat{X}^{\prime }\right) \right)
\end{equation*}%
with:%
\begin{equation}
A\left( \hat{K},\hat{X}\right) +\frac{M\left( \left( \hat{K},\hat{X}\right)
,\left( \left\langle \hat{K}\right\rangle ,\left\langle \hat{X}\right\rangle
\right) \right) \left\Vert \hat{\Psi}\right\Vert ^{2}\left\langle
A\right\rangle }{1-M\left( \left( \left\langle \hat{K}\right\rangle
,\left\langle \hat{X}\right\rangle \right) ,\left( \left\langle \hat{K}%
\right\rangle ,\left\langle \hat{X}\right\rangle \right) \right) \left\Vert 
\hat{\Psi}\right\Vert ^{2}}  \label{RPl}
\end{equation}%
\begin{equation*}
A\left( \hat{K},\hat{X}\right) +\frac{\frac{\underline{\hat{k}}\left( \hat{X}%
\right) }{1+\underline{\hat{k}}\left( \hat{X}\right) }\left\langle
A\right\rangle }{1-\frac{\underline{\hat{k}}\left( \left\langle \hat{X}%
\right\rangle \right) }{1+\underline{\hat{k}}\left( \left\langle \hat{X}%
\right\rangle \right) }}
\end{equation*}%
\begin{equation*}
A\left( \hat{K},\hat{X}\right) +\frac{1+\underline{\hat{k}}\left(
\left\langle \hat{X}\right\rangle \right) }{1+\underline{\hat{k}}\left( \hat{%
X}\right) }\underline{\hat{k}}\left( \hat{X}\right) \left\langle
A\right\rangle
\end{equation*}%
\begin{equation*}
\Delta M\underline{\hat{k}}\left( \hat{X}\right)
\end{equation*}%
As a consequence, given (\ref{RPl}) we obtain:%
\begin{eqnarray*}
&&\frac{1}{1-M\left( \left( \hat{K},\hat{X}\right) ,\left( \hat{K}^{\prime },%
\hat{X}^{\prime }\right) \right) \left\vert \hat{\Psi}\left( \hat{K}^{\prime
},\hat{X}^{\prime }\right) \right\vert ^{2}}A \\
&\simeq &A+\frac{M\left( \left( \hat{K},\hat{X}\right) ,\left( \left\langle 
\hat{K}\right\rangle ,\left\langle \hat{X}\right\rangle \right) \right)
\left\Vert \hat{\Psi}\right\Vert ^{2}\left\langle A\right\rangle }{1-M\left(
\left( \left\langle \hat{K}\right\rangle ,\left\langle \hat{X}\right\rangle
\right) ,\left( \left\langle \hat{K}\right\rangle ,\left\langle \hat{X}%
\right\rangle \right) \right) \left\Vert \hat{\Psi}\right\Vert ^{2}}
\end{eqnarray*}

\subsubsection*{A3.2.3 Action functional for the investors' field}

We can use the translation method to derive the field version of (\ref{TC}).
The derivation is straightforward:

\begin{equation*}
-\hat{\Psi}^{\dag }\left( \hat{K},\hat{X}\right) \left( \nabla _{\hat{K}%
}\left( \sigma _{\hat{K}}^{2}\nabla _{\hat{K}}-\hat{g}\left( \hat{K},\hat{X}%
\right) \hat{K}\right) \right) \hat{\Psi}\left( \hat{K},\hat{X}\right) +%
\frac{1}{2\hat{\mu}}\left( \left\vert \hat{\Psi}\left( \hat{K},\hat{X}%
\right) \right\vert ^{2}-\left\vert \hat{\Psi}_{0}\left( \hat{X}\right)
\right\vert ^{2}\right) ^{2}
\end{equation*}%
where:%
\begin{equation*}
\hat{g}\left( \hat{K},\hat{X}\right) =\left( 1-M\left( \left( \hat{K},\hat{X}%
\right) ,\left( \hat{K}^{\prime },\hat{X}^{\prime }\right) \right)
\left\vert \hat{\Psi}\left( \hat{K}^{\prime },\hat{X}^{\prime }\right)
\right\vert ^{2}\right) ^{-1}\hat{f}\left( \hat{K}^{\prime },\hat{X}^{\prime
}\right)
\end{equation*}%
and 
\begin{eqnarray*}
\hat{f}\left( \hat{X}\right) &=&\left( 1+\underline{\hat{k}}_{2}\left( \hat{X%
}\right) \right) \left( 1+R_{\nu }\left( \hat{X}\right) \right) \\
&=&\left( 1+\int \hat{k}_{2}\left( \hat{X},\hat{X}_{1}\right) \hat{K}%
_{1}\left\vert \hat{\Psi}\left( \hat{K}_{1},\hat{X}_{1}\right) \right\vert
^{2}d\hat{K}_{1}d\hat{X}_{1}\right) \left( 1+R_{\nu }\left( \hat{X}\right)
\right)
\end{eqnarray*}%
:translates:%
\begin{equation*}
\hat{f}\left( \hat{K}_{v}\left( t\right) \right) =\left( 1+\sum_{m}\hat{k}%
_{2vm}\hat{K}_{m}\right) \left( 1+R_{\nu }\right)
\end{equation*}

\subsubsection*{A3.2.4 Investors' field returns translation}

The translation of the return equations (\ref{RTN}) for $R\left( \hat{X}%
\right) $ is obtained by dividing by $\hat{K}$. Imposing the constraint
amounts to introduce the potential (\ref{Tc}). This potential has no
derivative and the translation is straightforward and given by formula (\ref%
{tln}):%
\begin{eqnarray*}
&&\frac{\left\vert \hat{\Psi}\left( \hat{K},\hat{X}\right) \right\vert ^{2}}{%
\varepsilon ^{2}}\left[ \left( \delta \left( \hat{X}-\hat{X}^{\prime
}\right) -\frac{\hat{k}_{1}\left( \hat{X}^{\prime },\hat{X}\right)
\left\vert \hat{\Psi}\left( \hat{K}^{\prime },\hat{X}^{\prime }\right)
\right\vert ^{2}\hat{K}^{\prime }}{1+\underline{\hat{k}}\left( \hat{X}%
^{\prime }\right) }\right) R\left( \hat{X}^{\prime }\right) \right. \\
&&-\left\vert \hat{\Psi}\left( \hat{K}^{\prime },\hat{X}^{\prime }\right)
\right\vert ^{2}\frac{\hat{k}_{2}\left( \hat{X}^{\prime },\hat{X}\right) 
\hat{K}^{\prime }}{1+\underline{\hat{k}}\left( \hat{X}^{\prime }\right) } \\
&&\times \left( \bar{r}-\left( \bar{r}-\left( \frac{1+R\left( \hat{X}%
^{\prime }\right) }{\underline{\hat{k}}_{2}\left( \hat{X}^{\prime }\right) }%
+R\left( \hat{X}^{\prime }\right) \right) \right) H\left( -\left( \bar{r}%
-\left( \frac{1+R\left( \hat{X}\right) }{\underline{\hat{k}}_{2}\left( \hat{X%
}^{\prime }\right) }+R\left( \hat{X}^{\prime }\right) \right) \right)
\right) \right) \\
&&-\left\vert \Psi \left( K^{\prime },X^{\prime }\right) \right\vert ^{2}%
\frac{k_{2}\left( X^{\prime },\hat{X}\right) K^{\prime }}{1+\underline{k}%
\left( X^{\prime }\right) } \\
&&\times \left( \left( \bar{r}-\left( \bar{r}-\left( \frac{1+f_{1}\left(
X^{\prime }\right) }{\underline{k}_{2}\left( X^{\prime }\right) }%
+f_{1}\left( X^{\prime }\right) \right) \right) H\left( -\left( \bar{r}%
-\left( \frac{1+f_{1}\left( X^{\prime }\right) }{\underline{k}_{2}\left(
X^{\prime }\right) }+f_{1}\left( K^{\prime },X^{\prime }\right) \right)
\right) \right) \right) \right. \\
&&\left. \left. +\frac{k_{1}\left( X^{\prime },\hat{X}\right) K^{\prime }}{1+%
\underline{k}\left( X^{\prime }\right) }f_{1}\left( K^{\prime },X^{\prime
},\Psi ,\hat{\Psi}\right) \right) \right]
\end{eqnarray*}

The function $H$ is the Heaviside function $H\left( x\right) =1$ for $%
x\geqslant 0$ and $H\left( x\right) =0$ for $x\leqslant 0$.

When $\varepsilon \rightarrow 0$, it implies an equation defining the
returns:%
\begin{eqnarray}
&&\left( \delta \left( \hat{X}-\hat{X}^{\prime }\right) -\frac{\hat{k}%
_{1}\left( \hat{X}^{\prime },\hat{X}\right) \left\vert \hat{\Psi}\left( \hat{%
K}^{\prime },\hat{X}^{\prime }\right) \right\vert ^{2}\hat{K}^{\prime }}{1+%
\underline{\hat{k}}\left( \hat{X}^{\prime }\right) }\right) R\left( \hat{X}%
^{\prime }\right)  \label{EQ} \\
&=&\left\vert \hat{\Psi}\left( \hat{K}^{\prime },\hat{X}^{\prime }\right)
\right\vert ^{2}\frac{\hat{k}_{2}\left( \hat{X}^{\prime },\hat{X}\right) 
\hat{K}^{\prime }}{1+\underline{\hat{k}}\left( \hat{X}^{\prime }\right) } 
\notag \\
&&\times \left( \bar{r}-\left( \bar{r}-\left( \frac{1+R\left( \hat{X}%
^{\prime }\right) }{\underline{\hat{k}}_{2}\left( \hat{X}^{\prime }\right) }%
+R\left( \hat{X}^{\prime }\right) \right) \right) H\left( -\left( \bar{r}%
-\left( \frac{1+R\left( \hat{X}\right) }{\underline{\hat{k}}_{2}\left( \hat{X%
}^{\prime }\right) }+R\left( \hat{X}^{\prime }\right) \right) \right)
\right) \right)  \notag \\
&&+\left\vert \Psi \left( K^{\prime },X^{\prime }\right) \right\vert ^{2}%
\frac{k_{2}\left( X^{\prime },\hat{X}\right) K^{\prime }}{1+\underline{k}%
\left( X^{\prime }\right) }  \notag \\
&&\times \left( \left( \bar{r}-\left( \bar{r}-\left( \frac{1+f_{1}\left(
X^{\prime }\right) }{\underline{k}_{2}\left( X^{\prime }\right) }%
+f_{1}\left( X^{\prime }\right) \right) \right) H\left( -\left( \bar{r}%
-\left( \frac{1+f_{1}\left( X^{\prime }\right) }{\underline{k}_{2}\left(
X^{\prime }\right) }+f_{1}\left( K^{\prime },X^{\prime }\right) \right)
\right) \right) \right) \right.  \notag \\
&&\left. +\frac{k_{1}\left( X^{\prime },\hat{X}\right) K^{\prime }}{1+%
\underline{k}\left( X^{\prime }\right) }f_{1}\left( K^{\prime },X^{\prime
},\Psi ,\hat{\Psi}\right) \right)  \notag
\end{eqnarray}

and $f\left( \hat{K},\hat{X},\Psi ,\hat{\Psi}\right) $ is defined by:%
\begin{equation*}
R\left( \hat{X}\right) =\frac{f\left( \hat{X}\right) }{1+\underline{\hat{k}}%
_{2}\left( \hat{X}\right) }+\bar{r}\frac{\underline{\hat{k}}_{2}\left( \hat{X%
}\right) }{1+\underline{\hat{k}}_{2}\left( \hat{X}\right) }
\end{equation*}%
This equation allows to rewrite the quantities arising in (\ref{EQ}) :%
\begin{eqnarray*}
\bar{r}-\left( \frac{1+R\left( \hat{X}^{\prime }\right) }{\underline{\hat{k}}%
_{2}\left( \hat{X}^{\prime }\right) }+R\left( \hat{X}^{\prime }\right)
\right) &=&\bar{r}-\left( \frac{1+\frac{f\left( \hat{X}^{\prime }\right) }{1+%
\underline{\hat{k}}_{2}\left( \hat{X}^{\prime }\right) }+\bar{r}\frac{%
\underline{\hat{k}}_{2}\left( \hat{X}^{\prime }\right) }{1+\underline{\hat{k}%
}_{2}\left( \hat{X}^{\prime }\right) }}{\underline{\hat{k}}_{2}\left( \hat{X}%
^{\prime }\right) }+\frac{f\left( \hat{X}^{\prime }\right) }{1+\underline{%
\hat{k}}_{2}\left( \hat{X}^{\prime }\right) }+\bar{r}\frac{\underline{\hat{k}%
}_{2}\left( \hat{X}^{\prime }\right) }{1+\underline{\hat{k}}_{2}\left( \hat{X%
}^{\prime }\right) }\right) \\
&=&-\frac{1+f\left( \hat{X}^{\prime }\right) }{\underline{\hat{k}}_{2}\left( 
\hat{X}^{\prime }\right) }
\end{eqnarray*}%
with:%
\begin{equation*}
f_{1}\left( X\right) =\frac{f_{1}^{\prime }\left( K,X\right) }{1+k_{2}\left( 
\hat{X}\right) }+\frac{\bar{r}k_{2}\left( \hat{X}\right) }{1+k_{2}\left( 
\hat{X}\right) }+\Delta F_{\tau }\left( \bar{R}\left( K,X\right) \right)
\end{equation*}%
and:%
\begin{equation*}
\Delta F_{\tau }\left( \bar{R}\left( K,X\right) \right) =F_{1}\left( \bar{R}%
\left( K,X\right) \right) +\tau \left( \bar{R}\left( K,X\right) \right)
\Delta f_{1}^{\prime }\left( K\right)
\end{equation*}%
Using these relations, equation (\ref{EQ}) becomes:

\begin{eqnarray*}
&&\frac{f\left( \hat{X}\right) -\bar{r}}{1+\underline{\hat{k}}_{2}\left( 
\hat{X}\right) }-\frac{\hat{k}_{1}\left( \hat{X}^{\prime },\hat{X}\right) 
\hat{K}^{\prime }\left\vert \hat{\Psi}\left( \hat{K}^{\prime },\hat{X}%
^{\prime }\right) \right\vert ^{2}}{1+\underline{\hat{k}}\left( \hat{X}%
^{\prime }\right) }\frac{f\left( \hat{X}^{\prime }\right) -\bar{r}}{1+%
\underline{\hat{k}}_{2}\left( \hat{X}^{\prime }\right) } \\
&=&\int \left( \bar{r}+\frac{1+f\left( \hat{X}^{\prime }\right) }{\underline{%
\hat{k}}_{2}\left( \hat{X}^{\prime }\right) }H\left( -\frac{1+f\left( \hat{X}%
^{\prime }\right) }{\underline{\hat{k}}_{2}\left( \hat{X}^{\prime }\right) }%
\right) \right) \frac{\hat{k}_{2}\left( \hat{X}^{\prime },\hat{X}\right) 
\hat{K}^{\prime }\left\vert \hat{\Psi}\left( \hat{K}^{\prime },\hat{X}%
^{\prime }\right) \right\vert ^{2}}{1+\underline{\hat{k}}\left( \hat{X}%
^{\prime }\right) } \\
&&+\int \left( \bar{r}+\frac{1+f_{1}^{\prime }\left( X^{\prime }\right) }{%
\underline{k}_{2}\left( X^{\prime }\right) }H\left( -\frac{1+f_{1}^{\prime
}\left( X^{\prime }\right) }{\underline{k}_{2}\left( X^{\prime }\right) }%
\right) \right) \frac{k_{2}\left( X^{\prime },\hat{X}\right) \left\vert \Psi
\left( K^{\prime },X^{\prime }\right) \right\vert ^{2}K^{\prime }}{1+%
\underline{k}\left( X^{\prime }\right) } \\
&&+\int \frac{\left\vert \Psi \left( K^{\prime },X^{\prime }\right)
\right\vert ^{2}k_{1}\left( X^{\prime },\hat{X}\right) K^{\prime }}{1+%
\underline{k}\left( X^{\prime }\right) }\left( \frac{f_{1}^{\prime }\left(
K,X\right) -\bar{r}k_{2}\left( \hat{X}\right) }{1+k_{2}\left( \hat{X}\right) 
}+\Delta F_{\tau }\left( \bar{R}\left( K,X\right) \right) \right)
\end{eqnarray*}

\bigskip

\section{Appendix 4 Alternate description of investors return equation}

We rewrite the returns equation for investors in terms of an alternate set
of parameters. These parameters represents relative shares of investments
for an agent. This reformulation will be used while studying the dynamics of
interacting groups of investors.

Defining: 
\begin{equation*}
\hat{S}_{\eta }\left( \hat{X}^{\prime },\hat{K}^{\prime },\hat{X}\right) =%
\frac{\hat{k}_{\eta }\left( \hat{X}^{\prime },\hat{X}\right) \hat{K}^{\prime
}\left\vert \hat{\Psi}\left( \hat{K}^{\prime },\hat{X}^{\prime }\right)
\right\vert ^{2}}{1+\underline{\hat{k}}\left( \hat{X}^{\prime }\right) }
\end{equation*}%
and:%
\begin{equation*}
S_{\eta }\left( X^{\prime },K^{\prime },X\right) =\frac{k_{\eta }\left(
X^{\prime },\hat{X}\right) \left\vert \Psi \left( K^{\prime },X^{\prime
}\right) \right\vert ^{2}K^{\prime }}{1+\underline{k}\left( X^{\prime
}\right) }
\end{equation*}%
the equation for returns becomes:%
\begin{eqnarray}
0 &=&\int \left( \Delta \left( \hat{X},\hat{X}^{\prime }\right) -\hat{S}%
_{1}\left( \hat{X}^{\prime },\hat{K}^{\prime },\hat{X}\right) \right) \frac{%
f\left( \hat{X}^{\prime }\right) -\bar{r}}{1+\underline{\hat{k}}_{2}\left( 
\hat{X}^{\prime }\right) }d\hat{X}^{\prime }d\hat{K}^{\prime }  \label{MPK}
\\
&&-\int S_{1}\left( X^{\prime },K^{\prime },\hat{X}\right) \left( \frac{%
f_{1}^{\prime }\left( \hat{X}^{\prime }\right) -\bar{r}}{1+\underline{k}%
_{2}\left( X^{\prime }\right) }+\Delta F_{\tau }\left( \bar{R}\left(
K^{\prime },X^{\prime }\right) \right) \right) dX^{\prime }dK^{\prime } 
\notag \\
&&-\int \frac{1+f\left( \hat{X}^{\prime }\right) }{\underline{\hat{k}}%
_{2}\left( \hat{X}^{\prime }\right) }H\left( -\frac{1+f\left( \hat{X}%
^{\prime }\right) }{\underline{\hat{k}}_{2}\left( \hat{X}^{\prime }\right) }%
\right) \hat{S}_{2}\left( \hat{X}^{\prime },\hat{K}^{\prime },\hat{X}\right)
d\hat{X}^{\prime }d\hat{K}^{\prime }  \notag \\
&&-\int \frac{1+f_{1}^{\prime }\left( X^{\prime }\right) }{\underline{k}%
_{2}\left( X^{\prime }\right) }H\left( -\frac{1+f_{1}^{\prime }\left(
X^{\prime }\right) }{\underline{k}_{2}\left( X^{\prime }\right) }\right)
S_{2}\left( X^{\prime },K^{\prime },\hat{X}\right) dX^{\prime }d\hat{K}%
^{\prime }  \notag
\end{eqnarray}%
Integrating this equation over $\hat{K}^{\prime }$ leads to define the
following quantities:%
\begin{equation*}
\hat{S}_{\eta }\left( \hat{X}^{\prime },\hat{K}^{\prime },\hat{X}\right)
\rightarrow \int \frac{\hat{k}_{\eta }\left( \hat{X}^{\prime },\hat{X}%
\right) \hat{K}_{\hat{X}^{\prime }}\left\vert \hat{\Psi}\left( \hat{X}%
^{\prime }\right) \right\vert ^{2}}{1+\underline{\hat{k}}\left( \hat{X}%
^{\prime }\right) }\equiv \hat{S}_{\eta }\left( \hat{X}^{\prime },\hat{X}%
\right)
\end{equation*}%
and:%
\begin{equation*}
S_{\eta }\left( X^{\prime },K^{\prime },\hat{X}\right) \rightarrow \frac{%
k_{\eta }\left( X^{\prime },\hat{X}\right) K_{X^{\prime }}\left\vert \Psi
\left( X^{\prime }\right) \right\vert ^{2}}{1+\underline{k}\left( X^{\prime
}\right) }\equiv S_{\eta }\left( X^{\prime },\hat{X}\right)
\end{equation*}%
and it yields to replace (\ref{MPK}) by:%
\begin{eqnarray}
0 &=&\int \left( \Delta \left( \hat{X},\hat{X}^{\prime }\right) -\hat{S}%
_{1}\left( \hat{X}^{\prime },\hat{X}\right) \right) \frac{f\left( \hat{X}%
^{\prime }\right) -\bar{r}}{1+\underline{\hat{k}}_{2}\left( \hat{X}^{\prime
}\right) }d\hat{X}^{\prime }  \label{TRV} \\
&&-\int S_{1}\left( X^{\prime },\hat{X}\right) \left( \frac{f_{1}^{\prime
}\left( \hat{X}^{\prime }\right) -\bar{r}}{1+\underline{k}_{2}\left(
X^{\prime }\right) }+\Delta F_{\tau }\left( \bar{R}\left( K^{\prime
},X^{\prime }\right) \right) \right) dX^{\prime }  \notag \\
&&-\int \frac{1+f\left( \hat{X}^{\prime }\right) }{\underline{\hat{k}}%
_{2}\left( \hat{X}^{\prime }\right) }H\left( -\frac{1+f\left( \hat{X}%
^{\prime }\right) }{\underline{\hat{k}}_{2}\left( \hat{X}^{\prime }\right) }%
\right) \hat{S}_{2}\left( \hat{X}^{\prime },\hat{X}\right) d\hat{X}^{\prime
}-\int \frac{1+f_{1}^{\prime }\left( X^{\prime }\right) }{\underline{k}%
_{2}\left( X^{\prime }\right) }H\left( -\frac{1+f_{1}^{\prime }\left(
X^{\prime }\right) }{\underline{k}_{2}\left( X^{\prime }\right) }\right)
S_{2}\left( X^{\prime },\hat{X}\right) dX^{\prime }  \notag
\end{eqnarray}%
with the constraint:%
\begin{equation*}
\int \left( \hat{S}_{1}\left( \hat{X}^{\prime },\hat{X}\right) +\hat{S}%
_{2}\left( \hat{X}^{\prime },\hat{X}\right) \right) d\hat{X}^{\prime }+\int
\left( S_{1}\left( X^{\prime },\hat{X}\right) +S_{2}\left( X^{\prime },\hat{X%
}\right) \right) dX^{\prime }=1
\end{equation*}%
The coefficents can be expressed in terms of the $\hat{S}_{\eta }$ and $%
S_{\eta }$. Given that:%
\begin{equation*}
\hat{S}_{\eta }\left( \hat{X}^{\prime },\hat{X}\right) =\int \frac{\hat{k}%
_{\eta }\left( \hat{X}^{\prime },\hat{X}\right) \hat{K}_{\hat{X}^{\prime
}}\left\vert \hat{\Psi}\left( \hat{X}^{\prime }\right) \right\vert ^{2}}{1+%
\underline{\hat{k}}\left( \hat{X}^{\prime }\right) }
\end{equation*}%
and:%
\begin{equation*}
\hat{S}\left( \hat{X}^{\prime },\hat{X}\right) =\hat{S}_{1}\left( \hat{X}%
^{\prime },\hat{X}\right) +\hat{S}_{2}\left( \hat{X}^{\prime },\hat{X}\right)
\end{equation*}%
we have:%
\begin{equation*}
\int \hat{S}\left( \hat{X}^{\prime },\hat{X}\right) \frac{\hat{K}_{\hat{X}%
}\left\vert \hat{\Psi}\left( \hat{X}\right) \right\vert ^{2}}{\hat{K}_{\hat{X%
}^{\prime }}\left\vert \hat{\Psi}\left( \hat{X}^{\prime }\right) \right\vert
^{2}}d\hat{X}=\frac{\underline{\hat{k}}\left( \hat{X}^{\prime }\right) }{1+%
\underline{\hat{k}}\left( \hat{X}^{\prime }\right) }
\end{equation*}%
Then, defining averages:%
\begin{equation*}
\hat{S}_{\eta }\left( \hat{X}^{\prime }\right) =\int \hat{S}_{\eta }\left( 
\hat{X}^{\prime },\hat{X}\right) \frac{\hat{K}_{\hat{X}}\left\vert \hat{\Psi}%
\left( \hat{X}\right) \right\vert ^{2}}{\hat{K}_{\hat{X}^{\prime
}}\left\vert \hat{\Psi}\left( \hat{X}^{\prime }\right) \right\vert ^{2}}d%
\hat{X}
\end{equation*}%
\begin{equation*}
\hat{S}\left( \hat{X}^{\prime }\right) =\int \hat{S}\left( \hat{X}^{\prime },%
\hat{X}\right) \frac{\hat{K}_{\hat{X}}\left\vert \hat{\Psi}\left( \hat{X}%
\right) \right\vert ^{2}}{\hat{K}_{\hat{X}^{\prime }}\left\vert \hat{\Psi}%
\left( \hat{X}^{\prime }\right) \right\vert ^{2}}d\hat{X}
\end{equation*}%
\begin{equation*}
S_{\eta }\left( X^{\prime }\right) =\int S_{\eta }\left( X^{\prime },\hat{X}%
\right) \frac{\hat{K}_{\hat{X}}\left\vert \hat{\Psi}\left( \hat{X}\right)
\right\vert ^{2}}{K_{X^{\prime }}\left\vert \Psi \left( X^{\prime }\right)
\right\vert ^{2}}d\hat{X}
\end{equation*}%
\begin{equation*}
S\left( X^{\prime }\right) =\int S\left( X^{\prime },\hat{X}\right) \frac{%
\hat{K}_{\hat{X}}\left\vert \hat{\Psi}\left( \hat{X}\right) \right\vert ^{2}%
}{K_{X^{\prime }}\left\vert \Psi \left( X^{\prime }\right) \right\vert ^{2}}d%
\hat{X}
\end{equation*}%
we find the expression of the initial set of parameters as functions of the
new parameters:%
\begin{equation*}
\frac{1}{1+\underline{\hat{k}}\left( \hat{X}^{\prime }\right) }=1-\hat{S}%
\left( \hat{X}^{\prime }\right)
\end{equation*}%
\begin{eqnarray*}
\underline{\hat{k}}\left( \hat{X}^{\prime }\right) &=&\frac{\hat{S}\left( 
\hat{X}^{\prime }\right) }{1-\hat{S}\left( \hat{X}^{\prime }\right) } \\
\underline{\hat{k}}_{\eta }\left( \hat{X}^{\prime }\right) &=&\frac{\hat{S}%
\left( \hat{X}^{\prime }\right) }{1-\hat{S}\left( \hat{X}^{\prime }\right) }
\end{eqnarray*}%
\begin{equation*}
1+\underline{\hat{k}}_{2}\left( \hat{X}^{\prime }\right) =\frac{1-\hat{S}%
_{1}\left( \hat{X}^{\prime }\right) }{1-\hat{S}\left( \hat{X}^{\prime
}\right) }
\end{equation*}%
\begin{equation*}
\frac{\underline{k}\left( X^{\prime }\right) }{1+\underline{k}\left(
X^{\prime }\right) }=S\left( X^{\prime }\right)
\end{equation*}%
\begin{equation*}
\frac{1}{1+\underline{k}\left( X^{\prime }\right) }=1-S\left( X^{\prime
}\right)
\end{equation*}%
\begin{eqnarray*}
\underline{k}\left( X^{\prime }\right) &=&\frac{S\left( X^{\prime }\right) }{%
1-S\left( X^{\prime }\right) } \\
\underline{k}_{\eta }\left( X^{\prime }\right) &=&S\left( X^{\prime }\right)
\end{eqnarray*}%
\begin{equation*}
1+\underline{k}_{2}\left( X^{\prime }\right) =\frac{1-S_{1}\left( X^{\prime
}\right) }{1-S\left( X^{\prime }\right) }
\end{equation*}%
and equation (\ref{TRV}) writes:%
\begin{eqnarray*}
0 &=&\int \left( \Delta \left( \hat{X},\hat{X}^{\prime }\right) -\hat{S}%
_{1}\left( \hat{X}^{\prime },\hat{X}\right) \right) \frac{1-\hat{S}\left( 
\hat{X}^{\prime }\right) }{1-\hat{S}_{1}\left( \hat{X}^{\prime }\right) }%
\left( f\left( \hat{X}^{\prime }\right) -\bar{r}\right) d\hat{X}^{\prime } \\
&&-\int S_{1}\left( X^{\prime },\hat{X}\right) \left( \frac{1-S\left(
X^{\prime }\right) }{1-S_{1}\left( X^{\prime }\right) }\left( f_{1}^{\prime
}\left( \hat{X}^{\prime }\right) -\bar{r}\right) +\Delta F_{\tau }\left( 
\bar{R}\left( K^{\prime },X^{\prime }\right) \right) \right) dX^{\prime } \\
&&-\int \left( 1+f\left( \hat{X}^{\prime }\right) \right) \frac{1-\hat{S}%
\left( \hat{X}^{\prime }\right) }{\hat{S}_{2}\left( \hat{X}^{\prime }\right) 
}H\left( -\frac{1+f\left( \hat{X}^{\prime }\right) }{\underline{\hat{k}}%
_{2}\left( \hat{X}^{\prime }\right) }\right) \hat{S}_{2}\left( \hat{X}%
^{\prime },\hat{X}\right) d\hat{X}^{\prime } \\
&&-\int \left( 1+f_{1}^{\prime }\left( X^{\prime }\right) \right) \frac{%
1-S\left( X^{\prime }\right) }{S_{2}\left( X^{\prime }\right) }H\left( -%
\frac{1+f_{1}^{\prime }\left( X^{\prime }\right) }{\underline{k}_{2}\left(
X^{\prime }\right) }\right) S_{2}\left( X^{\prime },\hat{X}\right)
dX^{\prime }
\end{eqnarray*}%
As in the text, we consider that investment take place to close location so
that:%
\begin{eqnarray*}
S_{1}\left( X^{\prime },\hat{X}\right) &=&S_{1}\left( \hat{X},\hat{X}\right)
\delta \left( X^{\prime }-\hat{X}\right) \\
S_{2}\left( X^{\prime },\hat{X}\right) &=&S_{2}\left( \hat{X},\hat{X}\right)
\delta \left( X^{\prime }-\hat{X}\right)
\end{eqnarray*}%
and the constraint simplifies as:%
\begin{equation*}
\int \left( \hat{S}_{1}\left( \hat{X}^{\prime },\hat{X}\right) +\hat{S}%
_{2}\left( \hat{X}^{\prime },\hat{X}\right) \right) d\hat{X}^{\prime
}+S_{1}\left( \hat{X},\hat{X}\right) +S_{2}\left( \hat{X},\hat{X}\right) =1
\end{equation*}%
whereas the return equation (\ref{TRV}) becomes:%
\begin{eqnarray}
0 &=&\int \left( \Delta \left( \hat{X},\hat{X}^{\prime }\right) -\hat{S}%
_{1}\left( \hat{X}^{\prime },\hat{X}\right) \right) \frac{1-\hat{S}\left( 
\hat{X}^{\prime }\right) }{1-\hat{S}_{1}\left( \hat{X}^{\prime }\right) }%
\left( f\left( \hat{X}^{\prime }\right) -\bar{r}\right) d\hat{X}^{\prime }
\label{QDL} \\
&&-\int \left( 1+f\left( \hat{X}^{\prime }\right) \right) \frac{1-\hat{S}%
\left( \hat{X}^{\prime }\right) }{\hat{S}_{2}\left( \hat{X}^{\prime }\right) 
}H\left( -\left( 1+f\left( \hat{X}^{\prime }\right) \right) \right) \hat{S}%
_{2}\left( \hat{X}^{\prime },\hat{X}\right) d\hat{X}^{\prime }  \notag \\
&&-\left( 1+f_{1}^{\prime }\left( \hat{X}\right) \right) \left( 1-S\left( 
\hat{X}\right) \right) H\left( -\left( 1+f_{1}^{\prime }\left( X\right)
\right) \right)  \notag \\
&&-S_{1}\left( \hat{X}\right) \left( \frac{1-S\left( \hat{X}\right) }{%
1-S_{1}\left( \hat{X}\right) }\left( f_{1}^{\prime }\left( \hat{X}\right) -%
\bar{r}\right) +\Delta F_{\tau }\left( \bar{R}\left( K,X\right) \right)
\right)  \notag
\end{eqnarray}%
Remark that without defaut, this reduces to:%
\begin{eqnarray}
0 &=&\int \left( \Delta \left( \hat{X},\hat{X}^{\prime }\right) -\hat{S}%
_{1}\left( \hat{X}^{\prime },\hat{X}\right) \right) \frac{1-\hat{S}\left( 
\hat{X}^{\prime }\right) }{1-\hat{S}_{1}\left( \hat{X}^{\prime }\right) }%
\left( f\left( \hat{X}^{\prime }\right) -\bar{r}\right) d\hat{X}^{\prime }
\label{RT} \\
&&-S_{1}\left( \hat{X}\right) \left( \frac{1-S\left( \hat{X}\right) }{%
1-S_{1}\left( \hat{X}\right) }\left( f_{1}^{\prime }\left( \hat{X}\right) -%
\bar{r}\right) +\Delta F_{\tau }\left( \bar{R}\left( K,X\right) \right)
\right)  \notag
\end{eqnarray}%
with solution:%
\begin{eqnarray}
f\left( \hat{X}^{\prime }\right) &=&\bar{r}+\frac{1-\hat{S}_{1}\left( \hat{X}%
^{\prime }\right) }{1-\hat{S}\left( \hat{X}^{\prime }\right) }\left( \Delta
\left( \hat{X},\hat{X}^{\prime }\right) -\hat{S}_{1}\left( \hat{X}^{\prime },%
\hat{X}\right) \right) ^{-1}  \label{RN} \\
&&\times \left( S_{1}\left( \hat{X}\right) \left( \frac{1-S\left( \hat{X}%
\right) }{1-S_{1}\left( \hat{X}\right) }\left( f_{1}^{\prime }\left( \hat{X}%
\right) -\bar{r}\right) +\Delta F_{\tau }\left( \bar{R}\left( K,X\right)
\right) \right) \right)  \notag
\end{eqnarray}%
Coming back to the equation with default, the loss realized when some
default arise is:%
\begin{equation*}
\max \left( -\left( 1+\bar{r}\right) ,\frac{1+f_{1}^{\prime }\left(
X^{\prime }\right) }{\underline{k}_{2}\left( X^{\prime }\right) }\right)
\end{equation*}%
which translates that an investor can not lose more than its amounts of
loans plus the potential return on them.

When the loss is maximal equation (\ref{QDL}) becomes:

\begin{eqnarray*}
0 &=&\int \left( \Delta \left( \hat{X},\hat{X}^{\prime }\right) -\hat{S}%
_{1}\left( \hat{X}^{\prime },\hat{X}\right) \right) \frac{1-\hat{S}\left( 
\hat{X}^{\prime }\right) }{1-\hat{S}_{1}\left( \hat{X}^{\prime }\right) }%
\left( f\left( \hat{X}^{\prime }\right) -\bar{r}\right) d\hat{X}^{\prime } \\
&&+\int \left( 1+\bar{r}\right) \frac{1-\hat{S}\left( \hat{X}^{\prime
}\right) }{\hat{S}_{2}\left( \hat{X}^{\prime }\right) }\hat{S}_{2}\left( 
\hat{X}^{\prime },\hat{X}\right) d\hat{X}^{\prime }+\left( 1+\bar{r}\right)
\left( 1-S\left( \hat{X}\right) \right) \\
&&-S_{1}\left( \hat{X}\right) \left( \frac{1-S\left( \hat{X}\right) }{%
1-S_{1}\left( \hat{X}\right) }\left( f_{1}^{\prime }\left( \hat{X}\right) -%
\bar{r}\right) +\Delta F_{\tau }\left( \bar{R}\left( K,X\right) \right)
\right)
\end{eqnarray*}

whereas, in the general case, the return equation (\ref{QDL}) is:%
\begin{eqnarray*}
0 &=&\int \left( \Delta \left( \hat{X},\hat{X}^{\prime }\right) -\hat{S}%
_{1}\left( \hat{X}^{\prime },\hat{X}\right) \right) \frac{1-\hat{S}\left( 
\hat{X}^{\prime }\right) }{1-\hat{S}_{1}\left( \hat{X}^{\prime }\right) }%
f\left( \hat{X}^{\prime }\right) d\hat{X}^{\prime } \\
&&-\int \max \left( -1,\left( 1+f\left( \hat{X}^{\prime }\right) \right) 
\frac{1-\hat{S}\left( \hat{X}^{\prime }\right) }{\hat{S}_{2}\left( \hat{X}%
^{\prime }\right) }\right) H\left( -\left( 1+f\left( \hat{X}^{\prime
}\right) \right) \right) \hat{S}_{2}\left( \hat{X}^{\prime },\hat{X}\right) d%
\hat{X}^{\prime } \\
&&-\int \max \left( -1,\left( 1+f\left( \hat{X}^{\prime }\right) \right) 
\frac{\left( 1-S\left( \hat{X}\right) \right) }{S_{2}\left( \hat{X}\right) }%
\right) H\left( -\left( 1+f_{1}^{\prime }\left( X\right) \right) \right)
S_{2}\left( \hat{X}\right) -S_{1}\left( \hat{X}\right) \frac{1-S\left( \hat{X%
}\right) }{1-S_{1}\left( \hat{X}\right) }f_{1}^{\prime }\left( \hat{X}\right)
\end{eqnarray*}%
Defining $\hat{S}_{-}$and $S_{-}$ the default sets for investors and firms
respectively, so that $1+f\left( \hat{X}\right) <0$, the equation also
writes: 
\begin{eqnarray*}
0 &=&\int \left( \Delta \left( \hat{X},\hat{X}^{\prime }\right) -\hat{S}%
_{1}\left( \hat{X}^{\prime },\hat{X}\right) \right) \frac{1-\hat{S}\left( 
\hat{X}^{\prime }\right) }{1-\hat{S}_{1}\left( \hat{X}^{\prime }\right) }%
f\left( \hat{X}^{\prime }\right) d\hat{X}^{\prime } \\
&&-\int_{\hat{S}_{-}}\max \left( -1,\left( 1+f\left( \hat{X}^{\prime
}\right) \right) \frac{1-\hat{S}\left( \hat{X}^{\prime }\right) }{\hat{S}%
_{2}\left( \hat{X}^{\prime }\right) }\right) \hat{S}_{2}\left( \hat{X}%
^{\prime },\hat{X}\right) d\hat{X}^{\prime } \\
&&-\int_{S_{-}}\max \left( -1,\left( 1+f_{1}^{\prime }\left( \hat{X}\right)
\right) \frac{\left( 1-S\left( \hat{X}\right) \right) }{S_{2}\left( \hat{X}%
\right) }\right) H\left( -\left( 1+f_{1}^{\prime }\left( X\right) \right)
\right) S_{2}\left( \hat{X}\right) -S_{1}\left( \hat{X}\right) \frac{%
1-S\left( \hat{X}\right) }{1-S_{1}\left( \hat{X}\right) }f_{1}^{\prime
}\left( \hat{X}\right)
\end{eqnarray*}%
with solution: 
\begin{eqnarray*}
f\left( \hat{X}\right) &=&\left( \Delta \left( \hat{X},\hat{X}^{\prime
}\right) -\hat{S}_{1}\left( \hat{X}^{\prime },\hat{X}\right) \right) ^{-1} \\
&&\left\{ \int_{\hat{S}_{-}}\max \left( -1,\left( 1+f\left( \hat{X}^{\prime
}\right) \right) \frac{1-\hat{S}\left( \hat{X}^{\prime }\right) }{\hat{S}%
_{2}\left( \hat{X}^{\prime }\right) }\right) \hat{S}_{2}\left( \hat{X}%
^{\prime },\hat{X}\right) d\hat{X}^{\prime }\right. \\
&&+\left. \int_{S_{-}}\max \left( -1,\left( 1+f_{1}^{\prime }\left( \hat{X}%
\right) \right) \frac{\left( 1-S\left( \hat{X}\right) \right) }{S_{2}\left( 
\hat{X}\right) }\right) S_{2}\left( \hat{X}\right) \right\} +S_{1}\left( 
\hat{X}\right) \frac{1-S\left( \hat{X}\right) }{1-S_{1}\left( \hat{X}\right) 
}\max \left( f_{1}^{\prime }\left( \hat{X}\right) ,0\right)
\end{eqnarray*}

if $\max \left( f_{1}^{\prime }\left( \hat{X}\right) ,0\right) =0$, we hav:%
\begin{eqnarray*}
f\left( \hat{X}\right) &=&\left( \Delta \left( \hat{X},\hat{X}^{\prime
}\right) -\hat{S}_{1}\left( \hat{X}^{\prime },\hat{X}\right) \right) ^{-1} \\
&&\left\{ \int_{\hat{S}_{-}}\max \left( -1,\left( 1+f\left( \hat{X}^{\prime
}\right) \right) \frac{1-\hat{S}\left( \hat{X}^{\prime }\right) }{\hat{S}%
_{2}\left( \hat{X}^{\prime }\right) }\right) \hat{S}_{2}\left( \hat{X}%
^{\prime },\hat{X}\right) d\hat{X}^{\prime }\right. \\
&&+\left. \int_{S_{-}}\max \left( -1,\left( 1+f_{1}^{\prime }\left( \hat{X}%
\right) \right) \frac{\left( 1-S\left( \hat{X}\right) \right) }{S_{2}\left( 
\hat{X}\right) }\right) S_{2}\left( \hat{X}\right) \right\}
\end{eqnarray*}

\section*{Appendix 5 Introducing normalisations}

\subsection*{A5.1 Firms}

For firms, the relevant variable $K$ corresponds to the private capital $%
K_{p}\left( t\right) $ and the share $k_{ajl}$ is normalised as:%
\begin{equation*}
k_{ajl}\rightarrow \frac{k_{ail}}{N\left\langle K\left( t\right)
\right\rangle \left( 1-\left( \left\langle k_{1}\right\rangle +\left\langle
k_{2}\right\rangle \right) \frac{\hat{N}\left\langle \hat{K}_{v}\left(
t\right) \right\rangle }{\left\langle K\left( t\right) \right\rangle }%
\right) }=\frac{k_{ail}}{N\left\langle K_{p}\left( t\right) \right\rangle }
\end{equation*}%
so that the translation of these coefficients in terms of field becomes:%
\begin{eqnarray*}
k_{1}\left( X^{\prime },\hat{X}\right) &\rightarrow &\frac{k_{1}\left(
X^{\prime },\hat{X}\right) }{\left\Vert \Psi \right\Vert ^{2}\left\langle
K\right\rangle } \\
k_{2}\left( X^{\prime },\hat{X}\right) &\rightarrow &\frac{k_{2}\left(
X^{\prime },\hat{X}\right) }{\left\Vert \Psi \right\Vert ^{2}\left\langle
K\right\rangle }
\end{eqnarray*}%
and:%
\begin{eqnarray*}
\frac{k_{\eta }\left( X^{\prime },\hat{X}\right) }{1+\underline{k}\left(
X^{\prime }\right) } &\rightarrow &\frac{k_{\eta }\left( X^{\prime },\hat{X}%
\right) }{\left\Vert \Psi \right\Vert ^{2}\left\langle K\right\rangle }\frac{%
1}{1+\int \frac{k\left( X^{\prime },\hat{X}\right) }{\left\Vert \Psi
\right\Vert ^{2}\left\langle K\right\rangle }\hat{K}_{\hat{X}}\left\vert 
\hat{\Psi}\left( \hat{X}\right) \right\vert ^{2}d\hat{X}} \\
&=&\frac{\frac{k_{\eta }\left( X^{\prime },\hat{X}\right) }{\left\Vert \Psi
\right\Vert ^{2}\left\langle K\right\rangle }}{1+\underline{k}\left(
X^{\prime }\right) }
\end{eqnarray*}%
with:%
\begin{equation*}
\underline{k}\left( X^{\prime }\right) =\int \frac{k\left( X^{\prime },\hat{X%
}\right) }{\left\Vert \Psi \right\Vert ^{2}\left\langle K\right\rangle }\hat{%
K}_{\hat{X}}\left\vert \hat{\Psi}\left( \hat{X}\right) \right\vert ^{2}d\hat{%
X}
\end{equation*}

\subsection*{A5.2 Investors}

For investors, since we look at the dynamics for disposable capital, we
consider the normalisation:%
\begin{equation*}
\hat{k}_{ajl}\rightarrow \frac{\hat{k}_{ajl}}{\hat{N}\left\langle \hat{K}%
\left( t\right) \right\rangle \left( 1-\left( \left\langle \hat{k}%
_{1}\right\rangle +\left\langle \hat{k}_{2}\right\rangle \right) \right) }=%
\frac{\hat{k}_{ajl}}{\hat{N}\left\langle \hat{K}\left( t\right)
\right\rangle \left( 1-\left\langle \hat{k}\right\rangle \right) }
\end{equation*}%
which modifies the coefficients:%
\begin{eqnarray*}
\hat{k}_{1}\left( \hat{X}^{\prime },\hat{X}\right) &\rightarrow &\frac{\hat{k%
}_{1}\left( \hat{X}^{\prime },\hat{X}\right) }{\left\Vert \hat{\Psi}%
\right\Vert ^{2}\left\langle \hat{K}\right\rangle \left( 1-\left\langle \hat{%
k}\left( \hat{X}^{\prime },\hat{X}\right) \right\rangle \right) } \\
\hat{k}_{2}\left( \hat{X}^{\prime },\hat{X}\right) &\rightarrow &\frac{\hat{k%
}_{2}\left( \hat{X}^{\prime },\hat{X}\right) }{\left\Vert \hat{\Psi}%
\right\Vert ^{2}\left\langle \hat{K}\right\rangle \left( 1-\left\langle \hat{%
k}\left( \hat{X}^{\prime },\hat{X}\right) \right\rangle \right) }
\end{eqnarray*}%
that is:%
\begin{equation*}
\hat{k}_{\lambda }\left( \hat{X}^{\prime },\hat{X}\right) \rightarrow \frac{%
\hat{k}_{\lambda }\left( \hat{X}^{\prime },\hat{X}\right) }{\left\Vert \hat{%
\Psi}\right\Vert ^{2}\left\langle \hat{K}\right\rangle \left( 1-\frac{%
\left\langle \underline{k}\left( \hat{X}^{\prime }\right) \right\rangle }{%
\left\Vert \hat{\Psi}\right\Vert ^{2}\left\langle \hat{K}\right\rangle }%
\right) }
\end{equation*}%
This modifies the coefficients $\underline{\hat{k}}_{\eta }\left( \hat{X}%
^{\prime }\right) $ as: 
\begin{equation*}
\underline{\hat{k}}_{\eta }\left( \hat{X}^{\prime }\right) \rightarrow \int 
\frac{\hat{k}_{\eta }\left( \hat{X}^{\prime },\hat{X}\right) }{\left\Vert 
\hat{\Psi}\right\Vert ^{2}\left\langle \hat{K}\right\rangle \left(
1-\left\langle \hat{k}\left( \hat{X}^{\prime },\hat{X}\right) \right\rangle
\right) }\hat{K}_{\hat{X}}\left\vert \hat{\Psi}\left( \hat{X}\right)
\right\vert ^{2}d\hat{X}
\end{equation*}%
and in average:%
\begin{equation*}
\left\langle \underline{\hat{k}}_{\eta }\left( \hat{X}^{\prime }\right)
\right\rangle \simeq \frac{\left\langle \hat{k}_{\eta }\left( \hat{X}%
^{\prime },\hat{X}\right) \right\rangle }{1-\left\langle \hat{k}\left( \hat{X%
}^{\prime },\hat{X}\right) \right\rangle }
\end{equation*}%
Moreover the terms $\frac{\hat{k}_{\eta }\left( \hat{X}^{\prime },\hat{X}%
\right) }{1+\underline{\hat{k}}\left( \hat{X}^{\prime }\right) }$, $1+%
\underline{\hat{k}}_{2}\left( \hat{X}^{\prime }\right) $ and $\frac{\hat{k}%
_{\eta }\left( \hat{X}^{\prime },\hat{X}\right) }{1+\underline{\hat{k}}%
_{2}\left( \hat{X}^{\prime }\right) }$ arising in the various equations, are
modified as follows:

\begin{eqnarray*}
\frac{\hat{k}_{\eta }\left( \hat{X}^{\prime },\hat{X}\right) }{1+\underline{%
\hat{k}}\left( \hat{X}^{\prime }\right) } &\rightarrow &\frac{\hat{k}_{\eta
}\left( \hat{X}^{\prime },\hat{X}\right) }{\left( \left\Vert \hat{\Psi}%
\right\Vert ^{2}\left\langle \hat{K}\right\rangle \left( 1-\left\langle \hat{%
k}\left( \hat{X}^{\prime },\hat{X}\right) \right\rangle \right) \right) }%
\frac{1}{1+\int \frac{\hat{k}\left( \hat{X}^{\prime },\hat{X}\right) }{%
\left\Vert \hat{\Psi}\right\Vert ^{2}\left\langle \hat{K}\right\rangle
\left( 1-\left\langle \hat{k}\left( \hat{X}^{\prime },\hat{X}\right)
\right\rangle \right) }\hat{K}_{\hat{X}}\left\vert \hat{\Psi}\left( \hat{X}%
\right) \right\vert ^{2}d\hat{X}} \\
&=&\frac{\hat{k}_{\eta }\left( \hat{X}^{\prime },\hat{X}\right) }{\left\Vert 
\hat{\Psi}\right\Vert ^{2}\left\langle \hat{K}\right\rangle }\frac{1}{%
1-\left\langle \hat{k}\left( \hat{X}^{\prime },\hat{X}\right) \right\rangle
+\int \frac{\hat{k}\left( \hat{X}^{\prime },\hat{X}\right) }{\left\Vert \hat{%
\Psi}\right\Vert ^{2}\left\langle \hat{K}\right\rangle }\hat{K}_{\hat{X}%
}\left\vert \hat{\Psi}\left( \hat{X}\right) \right\vert ^{2}d\hat{X}} \\
&=&\frac{\hat{k}_{\eta }\left( \hat{X}^{\prime },\hat{X}\right) }{\left\Vert 
\hat{\Psi}\right\Vert ^{2}\left\langle \hat{K}\right\rangle }\frac{1}{1+\int 
\frac{\hat{k}\left( \hat{X}^{\prime },\hat{X}\right) -\left\langle \hat{k}%
\left( \hat{X}^{\prime },\hat{X}\right) \right\rangle }{\left\Vert \hat{\Psi}%
\right\Vert ^{2}\left\langle \hat{K}\right\rangle }\hat{K}_{\hat{X}%
}\left\vert \hat{\Psi}\left( \hat{X}\right) \right\vert ^{2}d\hat{X}} \\
&\rightarrow &\frac{\hat{k}_{\eta }\left( \hat{X}^{\prime },\hat{X}\right) }{%
\left\Vert \hat{\Psi}\right\Vert ^{2}\left\langle \hat{K}\right\rangle
\left( 1+\underline{\hat{k}}\left( \hat{X}^{\prime }\right) \right) }
\end{eqnarray*}%
with now:%
\begin{equation*}
\underline{\hat{k}}\left( \hat{X}^{\prime }\right) =\int \frac{\hat{k}\left( 
\hat{X}^{\prime },\hat{X}\right) -\left\langle \hat{k}\left( \hat{X}^{\prime
},\hat{X}\right) \right\rangle }{\left\Vert \hat{\Psi}\right\Vert
^{2}\left\langle \hat{K}\right\rangle }\hat{K}_{\hat{X}}\left\vert \hat{\Psi}%
\left( \hat{X}\right) \right\vert ^{2}d\hat{X}
\end{equation*}%
\begin{eqnarray*}
1+\underline{\hat{k}}_{2}\left( \hat{X}^{\prime }\right) &\rightarrow
&1+\int \frac{\hat{k}_{2}\left( \hat{X}^{\prime },\hat{X}\right) }{%
\left\Vert \hat{\Psi}\right\Vert ^{2}\left\langle \hat{K}\right\rangle
\left( 1-\left\langle \hat{k}\left( \hat{X}^{\prime },\hat{X}\right)
\right\rangle \right) }\hat{K}_{\hat{X}}\left\vert \hat{\Psi}\left( \hat{X}%
\right) \right\vert ^{2}d\hat{X} \\
&=&\frac{1}{\left( 1-\left\langle \hat{k}\left( \hat{X}^{\prime },\hat{X}%
\right) \right\rangle \right) }\left( \left( 1-\left\langle \hat{k}\left( 
\hat{X}^{\prime },\hat{X}\right) \right\rangle \right) +\int \frac{\hat{k}%
_{2}\left( \hat{X}^{\prime },\hat{X}\right) }{\left\Vert \hat{\Psi}%
\right\Vert ^{2}\left\langle \hat{K}\right\rangle }\hat{K}_{\hat{X}%
}\left\vert \hat{\Psi}\left( \hat{X}\right) \right\vert ^{2}d\hat{X}\right)
\\
&=&\frac{1}{\left( 1-\left\langle \hat{k}\left( \hat{X}^{\prime },\hat{X}%
\right) \right\rangle \right) }\left( \left( 1-\left\langle \hat{k}%
_{1}\left( \hat{X}^{\prime },\hat{X}\right) \right\rangle \right) +\int 
\frac{\hat{k}_{2}\left( \hat{X}^{\prime },\hat{X}\right) -\left\langle \hat{k%
}_{2}\left( \hat{X}^{\prime },\hat{X}\right) \right\rangle }{\left\Vert \hat{%
\Psi}\right\Vert ^{2}\left\langle \hat{K}\right\rangle }\hat{K}_{\hat{X}%
}\left\vert \hat{\Psi}\left( \hat{X}\right) \right\vert ^{2}d\hat{X}\right)
\end{eqnarray*}%
\begin{equation*}
1+\underline{\hat{k}}_{2}\left( \hat{X}^{\prime }\right) \rightarrow \frac{%
1-\left\langle \hat{k}_{1}\left( \hat{X}^{\prime },\hat{X}\right)
\right\rangle }{1-\left\langle \hat{k}\left( \hat{X}^{\prime },\hat{X}%
\right) \right\rangle }\left( 1+\frac{\underline{\hat{k}}_{2}\left( \hat{X}%
^{\prime }\right) }{1-\left\langle \hat{k}_{1}\left( \hat{X}^{\prime },\hat{X%
}\right) \right\rangle }\right)
\end{equation*}%
where we define:%
\begin{equation*}
\underline{\hat{k}}_{2}\left( \hat{X}^{\prime }\right) =\int \frac{\hat{k}%
_{2}\left( \hat{X}^{\prime },\hat{X}\right) -\left\langle \hat{k}_{2}\left( 
\hat{X}^{\prime },\hat{X}\right) \right\rangle }{\left\Vert \hat{\Psi}%
\right\Vert ^{2}\left\langle \hat{K}\right\rangle }\hat{K}_{\hat{X}%
}\left\vert \hat{\Psi}\left( \hat{X}\right) \right\vert ^{2}d\hat{X}
\end{equation*}%
Ultimately, expression $\frac{\hat{k}_{\eta }\left( \hat{X}^{\prime },\hat{X}%
\right) }{1+\underline{\hat{k}}_{2}\left( \hat{X}^{\prime }\right) }$ is
replacd by:%
\begin{equation*}
\frac{\hat{k}_{\eta }\left( \hat{X}^{\prime },\hat{X}\right) }{1+\underline{%
\hat{k}}_{2}\left( \hat{X}^{\prime }\right) }\rightarrow \frac{\hat{k}_{\eta
}\left( \hat{X}^{\prime },\hat{X}\right) }{\left\Vert \hat{\Psi}\right\Vert
^{2}\left\langle \hat{K}\right\rangle \left( 1-\left\langle \hat{k}%
_{1}\left( \hat{X}^{\prime },\hat{X}\right) \right\rangle \right) \left( 1+%
\frac{\underline{\hat{k}}_{2}\left( \hat{X}^{\prime }\right) }{%
1-\left\langle \hat{k}_{1}\left( \hat{X}^{\prime },\hat{X}\right)
\right\rangle }\right) }
\end{equation*}

\section*{Appendix 6 Detailling the field firms' return}

\subsection*{A6.1 Firms action functional}

We have shown (see \ref{Fc}) that the action functional for the field of
firms is:%
\begin{equation*}
-\Psi ^{\dag }\left( K,X\right) \left( \nabla _{K_{p}}\left( \sigma
_{K}^{2}\nabla _{K_{p}}-f_{1}^{\prime }\left( K,X\right) K_{p}\right)
\right) \Psi \left( K,X\right) +\frac{1}{2\epsilon }\left( \left\vert \Psi
\left( K,X\right) \right\vert ^{2}-\left\vert \Psi _{0}\left( X\right)
\right\vert ^{2}\right) ^{2}
\end{equation*}%
where:%
\begin{eqnarray*}
f_{1}^{\prime }\left( K_{i}\left( t\right) \right) &=&\left( 1+\sum_{\nu
}k_{2j\nu }\hat{K}_{\nu }\left( t\right) \right) f_{1}\left( K_{i}\left(
t\right) \right) -\bar{r}\sum_{v}k_{2lv}\hat{K}_{\nu }\left( t\right) \\
&\rightarrow &f_{1}^{\prime }\left( X\right) =\left( 1+\underline{k}%
_{2}\left( X\right) \right) f_{1}\left( X\right) -\bar{r}\underline{k}%
_{2}\left( \hat{X}\right) \\
&=&f_{1}\left( X\right) +\left( f_{1}\left( X\right) -\bar{r}\right) \int
k_{2}\left( X,\hat{X}_{1}\right) \hat{K}_{1}\left\vert \hat{\Psi}\left( \hat{%
K}_{1},\hat{X}_{1}\right) \right\vert ^{2}d\hat{K}_{1}d\hat{X}_{1}
\end{eqnarray*}%
We will consider the change of variables:%
\begin{eqnarray*}
\Psi \left( K,X\right) &\rightarrow &\exp \left( -\frac{1}{\sigma _{\hat{K}%
}^{2}}\int f_{1}^{\prime }\left( K_{p},X\right) K_{p}dK_{p}\right) \Psi \\
\Psi ^{\dag }\left( K,X\right) &\rightarrow &\exp \left( \frac{1}{\sigma _{%
\hat{K}}^{2}}\int f_{1}^{\prime }\left( K_{p},X\right) K_{p}dK_{p}\right)
\Psi ^{\dag }
\end{eqnarray*}%
leading to the modified action:%
\begin{eqnarray}
S\left( \Psi \right) &=&-\Psi ^{\dag }\left( K,X\right) \sigma _{\hat{K}%
}^{2}\nabla _{K_{p}}^{2}\Psi \left( K,X\right)  \label{MDT} \\
&&+\Psi ^{\dag }\left( K,X\right) \left( \frac{\left( f_{1}^{\prime }\left(
K_{p},X\right) K_{p}\right) ^{2}}{2\sigma _{\hat{K}}^{2}}+\frac{1}{2}%
f_{1}^{\prime }\left( K_{p},X\right) \right) \Psi \left( K,X\right) +\frac{1%
}{2\epsilon }\left( \left\vert \Psi \left( K,X\right) \right\vert
^{2}-\left\vert \Psi _{0}\left( X\right) \right\vert ^{2}\right) ^{2}  \notag
\end{eqnarray}

\subsubsection*{A6.1.1 Return for firms}

The return for firms is computed as:%
\begin{eqnarray*}
f_{1}^{\prime }\left( X\right) &=&\left( 1+\underline{k}_{2}\left( X\right)
\right) \frac{f_{1}\left( \left( 1+\underline{k}\left( X\right) \right)
K_{p},X\right) }{\left( 1+\underline{k}\left( X\right) \right) K_{p}}-\bar{r}%
\underline{k}_{2}\left( X\right) \\
&\rightarrow &\left( 1+\underline{k}_{2}\left( X\right) \right) \frac{%
A\left( X\right) \left( f_{1}\left( 1+\underline{k}\left( X\right) \right)
K_{p}\right) ^{\delta }-C}{\left( 1+\underline{k}\left( X\right) \right)
K_{p}}-\bar{r}\underline{k}_{2}\left( X\right)
\end{eqnarray*}%
where:%
\begin{equation*}
\underline{k}_{2}\left( X\right) =\frac{k_{2}\left( \hat{X},\hat{X}\right) }{%
N\left\langle K_{p}\right\rangle }\hat{K}_{X}\left\vert \hat{\Psi}\left( 
\hat{X}\right) \right\vert ^{2}\rightarrow \frac{\underline{k}_{2}\left(
X\right) }{\left\langle K_{p}\right\rangle }\hat{K}_{X}
\end{equation*}%
and:%
\begin{eqnarray*}
&&\left( 1+\underline{k}_{2}\left( X\right) \right) A\left( X\right) \left(
f_{1}\left( 1+\underline{k}\left( X\right) \right) K_{p}\right) ^{\delta -1}
\\
&=&\left( 1+\frac{\underline{k}_{2}\left( X\right) }{\left\langle
K_{p}\right\rangle }\hat{K}_{X}\right) A\left( X\right) \left( f_{1}\left( 1+%
\frac{\underline{k}_{2}\left( X\right) }{\left\langle K\right\rangle }\hat{K}%
_{X}\right) K_{pX}\right) ^{\delta -1}
\end{eqnarray*}%
so that:%
\begin{equation*}
f_{1}^{\prime }\left( X\right) -r=\left( 1+\underline{k}_{2}\left( X\right)
\right) \left( A\left( X\right) \left( f_{1}\left( 1+\underline{k}\left(
X\right) \right) K_{p}\right) ^{\delta -1}-\bar{r}\right)
\end{equation*}%
writen also in the alternate formulation:%
\begin{equation*}
f_{1}^{\prime }\left( X\right) -r=\frac{1-S_{1}\left( \hat{X},\hat{X}\right) 
\frac{\hat{K}_{\hat{X}}\left\vert \hat{\Psi}\left( \hat{X}\right)
\right\vert ^{2}}{K_{\hat{X}}\left\vert \Psi \left( \hat{X}\right)
\right\vert ^{2}}}{1-S\left( \hat{X},\hat{X}\right) \frac{\hat{K}_{\hat{X}%
}\left\vert \hat{\Psi}\left( \hat{X}\right) \right\vert ^{2}}{K_{\hat{X}%
}\left\vert \Psi \left( \hat{X}\right) \right\vert ^{2}}}\left( \left( f_{1}%
\frac{1}{1-S\left( \hat{X},\hat{X}\right) \frac{\hat{K}_{\hat{X}}\left\vert 
\hat{\Psi}\left( \hat{X}\right) \right\vert ^{2}}{K_{\hat{X}}\left\vert \Psi
\left( \hat{X}\right) \right\vert ^{2}}}K_{p}\right) ^{\delta -1}-\bar{r}%
\right)
\end{equation*}%
This simplifies:%
\begin{equation*}
f_{1}^{\prime }\left( X\right) -\bar{r}=\frac{1-S_{1}\left( \hat{X}\right) }{%
1-S\left( \hat{X}\right) }\left( \left( f_{1}\left( \hat{X}\right) \frac{%
K_{p}}{1-\frac{\hat{K}_{X}}{\left\langle K_{p}\right\rangle }S\left( \hat{X}%
\right) }\right) ^{\delta -1}-\bar{r}\right)
\end{equation*}%
and:%
\begin{equation*}
f_{1}^{\prime }\left( X\right) =\frac{1-S_{1}\left( \hat{X}\right) }{%
1-S\left( \hat{X}\right) }\left( \left( f_{1}\left( \hat{X}\right) \frac{%
K_{p}}{1-\frac{\hat{K}_{X}}{\left\langle K_{p}\right\rangle }S\left( \hat{X}%
\right) }\right) ^{\delta -1}\right) -\frac{S_{2}\left( \hat{X}\right) }{%
1-S\left( \hat{X}\right) }\bar{r}
\end{equation*}

In the constant return to scale approximation, this becomes:%
\begin{eqnarray*}
&&K_{pX}\left( \frac{f_{1}\left( X\right) \left( 1+\frac{\underline{k}\left(
X\right) }{\left\langle K\right\rangle }\hat{K}_{X}\right) K_{p}-C}{\left( 1+%
\frac{\underline{k}\left( X\right) }{\left\langle K\right\rangle }\hat{K}%
_{X}\left\vert \hat{\Psi}\left( \hat{X}\right) \right\vert ^{2}\right) K_{pX}%
}\right) \\
&\rightarrow &\frac{\left( 1+f_{1}\left( X\right) \right) \left( 1+\frac{%
\underline{k}\left( X\right) }{\left\langle K\right\rangle }\hat{K}%
_{X}\left\vert \hat{\Psi}\left( \hat{X}\right) \right\vert ^{2}\right)
K_{p}-C-\left( 1+\frac{\underline{k}\left( X\right) }{\left\langle
K\right\rangle }\hat{K}_{X}\left\vert \hat{\Psi}\left( \hat{X}\right)
\right\vert ^{2}\right) K_{p}}{1+\frac{\underline{k}\left( X\right) }{%
\left\langle K\right\rangle }\hat{K}_{X}\left\vert \hat{\Psi}\left( \hat{X}%
\right) \right\vert ^{2}} \\
&=&f_{1}\left( X\right) K_{p}-\frac{C}{1+\frac{\underline{k}\left( X\right) 
}{\left\langle K\right\rangle }\hat{K}_{X}\left\vert \hat{\Psi}\left( \hat{X}%
\right) \right\vert ^{2}}
\end{eqnarray*}%
and we rewrite this expression by introducing an effectiv cost $\bar{C}%
\left( X\right) =\frac{C}{1+\frac{\underline{k}\left( X\right) }{%
\left\langle K\right\rangle }\hat{K}_{X}\left\vert \hat{\Psi}\left( \hat{X}%
\right) \right\vert ^{2}}$ depending on the amount of capital invested: 
\begin{equation*}
f_{1}\left( X\right) K_{p}-\frac{C}{1+\frac{\underline{k}\left( X\right) }{%
\left\langle K\right\rangle }\hat{K}_{X}\frac{\left\vert \hat{\Psi}\left( 
\hat{X}\right) \right\vert ^{2}}{\left\vert \Psi \left( X\right) \right\vert
^{2}}}=f_{1}\left( X\right) K_{p}-\bar{C}\left( X\right)
\end{equation*}%
Note that in first approximation:%
\begin{equation*}
\frac{C}{1+\frac{\underline{k}\left( X\right) }{\left\langle K\right\rangle }%
\hat{K}_{X}\frac{\left\vert \hat{\Psi}\left( \hat{X}\right) \right\vert ^{2}%
}{\left\vert \Psi \left( X\right) \right\vert ^{2}}}\simeq \frac{C}{1+\frac{%
\underline{k}\left( X\right) }{\left\langle K\right\rangle }\hat{K}_{X}\frac{%
\left\vert \hat{\Psi}\left( \hat{X}\right) \right\vert ^{2}}{\left\vert \Psi
_{0}\left( X\right) \right\vert ^{2}}}
\end{equation*}%
The full return for the firms accounts for its total capital invested and
can thus be computed as:%
\begin{eqnarray*}
&&\left( 1+\underline{k}_{2}\left( X\right) \right) \left( f_{1}\left(
X\right) K_{p}-\bar{C}\left( X\right) \right) -\underline{k}_{2}\left(
X\right) K_{p}\bar{r} \\
&=&\left( 1+\underline{k}_{2}\left( X\right) \right) \left( \left(
f_{1}\left( X\right) -\frac{\underline{k}_{2}\left( X\right) }{1+\underline{k%
}_{2}\left( X\right) }\bar{r}\right) K_{p}-\bar{C}\left( X\right) \right)
\end{eqnarray*}%
\bigskip

\section*{Appendix 7 Firms constant return computation of background field
and capital}

\subsection*{A7.1 Case one}

\subsubsection*{Average capital per sector and firms returns}

We start with the minimization of (\ref{MDT}):%
\begin{eqnarray}
0 &=&-\sigma _{\hat{K}}^{2}\nabla _{K_{p}}^{2}\Psi \left( K,X\right) \\
&&+\left( \frac{\left( f_{1}^{\prime }\left( K_{p},X\right) K_{p}\right) ^{2}%
}{2\sigma _{\hat{K}}^{2}}+\frac{1}{2}f_{1}^{\prime }\left( K_{p},X\right)
\right) \Psi \left( K,X\right) +\frac{1}{\epsilon }\left( \left\vert \Psi
\left( K,X\right) \right\vert ^{2}-\left\vert \Psi _{0}\left( X\right)
\right\vert ^{2}\right) \Psi \left( K,X\right)  \notag
\end{eqnarray}

Neglecting the fluctuation term $\sigma _{\hat{K}}^{2}\nabla
_{K_{p}}^{2}\Psi \left( K,X\right) $ in first approximation, leads to the
field formula:%
\begin{equation}
\left\vert \Psi \left( K,X\right) \right\vert ^{2}=\left\vert \Psi
_{0}\left( X\right) \right\vert ^{2}-\epsilon \left( \frac{\left(
f_{1}^{\left( e\right) }\left( X\right) K_{p}-\bar{C}\left( X\right) \right)
^{2}}{\sigma _{\hat{K}}^{2}}+\frac{f_{1}^{\left( e\right) }\left( X\right) }{%
2}\right)
\end{equation}%
We consider the case:%
\begin{equation*}
\left\vert \Psi _{0}\left( X\right) \right\vert ^{2}-\epsilon \frac{%
f_{1}\left( X\right) }{2}-\epsilon \frac{\left( \bar{C}\left( X\right)
\right) ^{2}}{\sigma _{\hat{K}}^{2}}>0
\end{equation*}

The field formula implies that the amount of capital for firms in the sector
is bounded by the condition: 
\begin{equation*}
\left\vert \Psi _{0}\left( X\right) \right\vert ^{2}-\epsilon \left( \frac{%
\left( f_{1}\left( X\right) K_{p}-\bar{C}\left( X\right) \right) ^{2}}{%
\sigma _{\hat{K}}^{2}}+\frac{f_{1}\left( X\right) }{2}\right) >0
\end{equation*}%
The maximum level of capital for frm is thus defined by:%
\begin{equation*}
\left\vert \Psi _{0}\left( X\right) \right\vert ^{2}-\epsilon \left( \frac{%
\left( f_{1}\left( X\right) K_{0}-\bar{C}\left( X\right) \right) ^{2}}{%
\sigma _{\hat{K}}^{2}}+\frac{f_{1}\left( X\right) }{2}\right) =0
\end{equation*}%
which is solvd as:%
\begin{equation*}
K_{0}=\frac{\bar{C}\left( X\right) +\sqrt{\sigma _{\hat{K}}^{2}\left( \frac{%
\left\vert \Psi _{0}\left( X\right) \right\vert ^{2}}{\epsilon }-\frac{%
f_{1}\left( X\right) }{2}\right) }}{f_{1}\left( X\right) }
\end{equation*}%
This allows to compute $\left\vert \Psi \left( X\right) \right\vert ^{2}$ by
integrating frm $0$ t $K_{0}$.

\begin{eqnarray*}
\left\vert \Psi \left( X\right) \right\vert ^{2} &=&\left( \left\vert \Psi
_{0}\left( X\right) \right\vert ^{2}-\frac{\epsilon f_{1}\left( X\right) }{2}%
\right) K_{0}-\epsilon \left( \frac{\left( f_{1}\left( X\right) K_{0}-\bar{C}%
\left( X\right) \right) ^{3}}{3f_{1}\left( X\right) \sigma _{\hat{K}}^{2}}%
+\left( \frac{\left( \bar{C}\left( X\right) \right) ^{3}}{3f_{1}\left(
X\right) \sigma _{\hat{K}}^{2}}\right) \right) \\
&=&\left( \left\vert \Psi _{0}\left( X\right) \right\vert ^{2}-\frac{%
\epsilon f_{1}\left( X\right) }{2}\right) K_{0}-\frac{\epsilon K_{0}}{%
3\sigma _{\hat{K}}^{2}}\left( \left( f_{1}\left( X\right) K_{0}-\bar{C}%
\left( X\right) \right) ^{2}+\left( \bar{C}\left( X\right) \right) ^{2}-\bar{%
C}\left( X\right) \left( f_{1}\left( X\right) K_{0}-\bar{C}\left( X\right)
\right) \right) \\
&=&\left( \left\vert \Psi _{0}\left( X\right) \right\vert ^{2}-\frac{%
\epsilon f_{1}\left( X\right) }{2}\right) K_{0} \\
&&-\frac{\epsilon K_{0}}{3\sigma _{\hat{K}}^{2}}\left( \sigma _{\hat{K}%
}^{2}\left( \frac{\left\vert \Psi _{0}\left( X\right) \right\vert ^{2}}{%
\epsilon }-\frac{f_{1}\left( X\right) }{2}\right) +\bar{C}\left( X\right)
\left( \bar{C}\left( X\right) -\sqrt{\sigma _{\hat{K}}^{2}\left( \frac{%
\left\vert \Psi _{0}\left( X\right) \right\vert ^{2}}{\epsilon }-\frac{%
f_{1}\left( X\right) }{2}\right) }\right) \right) \\
&=&\left( \frac{2}{3}\left( \left\vert \Psi _{0}\left( X\right) \right\vert
^{2}-\frac{\epsilon f_{1}\left( X\right) }{2}\right) +\frac{\epsilon }{%
3\sigma _{\hat{K}}^{2}}\left( \sqrt{\sigma _{\hat{K}}^{2}\left( \frac{%
\left\vert \Psi _{0}\left( X\right) \right\vert ^{2}}{\epsilon }-\frac{%
f_{1}\left( X\right) }{2}\right) }-\bar{C}\left( X\right) \right) \bar{C}%
\left( X\right) \right) K_{0}
\end{eqnarray*}

and the densty of firms in sector $X$, $\left\vert \Psi \left( X\right)
\right\vert ^{2}$, has the form: 
\begin{eqnarray}
\left\vert \Psi \left( X\right) \right\vert ^{2} &=&\left( \frac{2}{3}\sigma
_{\hat{K}}^{2}\left( \frac{\left\vert \Psi _{0}\left( X\right) \right\vert
^{2}}{\epsilon }-\frac{f_{1}\left( X\right) }{2}\right) \right.  \label{PSF}
\\
&&\left. +\frac{1}{3}\left( \sqrt{\sigma _{\hat{K}}^{2}\left( \frac{%
\left\vert \Psi _{0}\left( X\right) \right\vert ^{2}}{\epsilon }-\frac{%
f_{1}\left( X\right) }{2}\right) }-\bar{C}\left( X\right) \right) \bar{C}%
\left( X\right) \right) \epsilon \frac{K_{0}}{\sigma _{\hat{K}}^{2}}  \notag
\end{eqnarray}

To find the amount of capital in sector $X$, we start with:%
\begin{equation*}
\left\vert \Psi \left( K,X\right) \right\vert ^{2}=\left[ \left\vert \Psi
_{0}\left( X\right) \right\vert ^{2}-\epsilon \frac{f_{1}\left( X\right) }{2}%
\right] -\epsilon \left( \frac{\left( f_{1}\left( X\right) K_{p}-\bar{C}%
\left( X\right) \right) ^{2}}{\sigma _{\hat{K}}^{2}}\right)
\end{equation*}%
and:%
\begin{equation*}
K_{p}\left( \left( f_{1}\left( X\right) K_{p}-\bar{C}\left( X\right) \right)
^{2}\right) =\frac{1}{f_{1}\left( X\right) }\left( \left( f_{1}\left(
X\right) K_{p}-\bar{C}\left( X\right) \right) \left( \left( f_{1}\left(
X\right) K_{p}-\bar{C}\left( X\right) \right) ^{2}\right) +\bar{C}\left(
X\right) \left( \left( f_{1}\left( X\right) K_{p}-\bar{C}\left( X\right)
\right) ^{2}\right) \right)
\end{equation*}%
Integrating this relation leads to:

\begin{eqnarray*}
K_{X}\left\vert \Psi \left( X\right) \right\vert ^{2} &=&\left( \left\vert
\Psi _{0}\left( X\right) \right\vert ^{2}-\frac{\epsilon f_{1}\left(
X\right) }{2}\right) \frac{K_{0}^{2}}{2}-\frac{\epsilon }{4\left(
f_{1}\left( X\right) \right) ^{2}\sigma _{\hat{K}}^{2}}\left( \left(
f_{1}\left( X\right) K_{0}-\bar{C}\left( X\right) \right) ^{4}-\left( \bar{C}%
\left( X\right) \right) ^{4}\right) \\
&&-\frac{\epsilon \bar{C}\left( X\right) }{3\left( f_{1}\left( X\right)
\right) ^{2}\sigma _{\hat{K}}^{2}}\left( \left( f_{1}\left( X\right) K_{0}-%
\bar{C}\left( X\right) \right) ^{3}+\left( \bar{C}\left( X\right) \right)
^{3}\right) \\
&=&\left( \left\vert \Psi _{0}\left( X\right) \right\vert ^{2}-\frac{%
\epsilon f_{1}\left( X\right) }{2}\right) \frac{K_{0}^{2}}{2}-\frac{\epsilon
K_{0}}{4f_{1}\left( X\right) \sigma _{\hat{K}}^{2}}\left( f_{1}\left(
X\right) K_{0}-2\bar{C}\left( X\right) \right) \left( \left( f_{1}\left(
X\right) K_{0}-\bar{C}\left( X\right) \right) ^{2}+\left( \bar{C}\left(
X\right) \right) ^{2}\right) \\
&=&\left( \frac{\left\vert \Psi _{0}\left( X\right) \right\vert ^{2}}{%
\epsilon }-\frac{f_{1}\left( X\right) }{2}\right) \epsilon \frac{K_{0}^{2}}{2%
} \\
&&-\frac{\epsilon K_{0}}{4f_{1}\left( X\right) \sigma _{\hat{K}}^{2}}\left( 
\sqrt{\sigma _{\hat{K}}^{2}\left( \frac{\left\vert \Psi _{0}\left( X\right)
\right\vert ^{2}}{\epsilon }-\frac{f_{1}\left( X\right) }{2}\right) }-\bar{C}%
\left( X\right) \right) \left( \sigma _{\hat{K}}^{2}\left( \frac{\left\vert
\Psi _{0}\left( X\right) \right\vert ^{2}}{\epsilon }-\frac{f_{1}\left(
X\right) }{2}\right) +\left( \bar{C}\left( X\right) \right) ^{2}\right)
\end{eqnarray*}

To simplify this expression, we use that:%
\begin{equation*}
\left( \frac{\left\vert \Psi _{0}\left( X\right) \right\vert ^{2}}{\epsilon }%
-\frac{f_{1}\left( X\right) }{2}\right) \epsilon \frac{K_{0}^{2}}{2}=\sigma
_{\hat{K}}^{2}\left( \frac{\left\vert \Psi _{0}\left( X\right) \right\vert
^{2}}{\epsilon }-\frac{f_{1}\left( X\right) }{2}\right) \left( \bar{C}\left(
X\right) +\sqrt{\sigma _{\hat{K}}^{2}\left( \frac{\left\vert \Psi _{0}\left(
X\right) \right\vert ^{2}}{\epsilon }-\frac{f_{1}\left( X\right) }{2}\right) 
}\right) \frac{\epsilon K_{0}}{2\sigma _{\hat{K}}^{2}f_{1}\left( X\right) }
\end{equation*}%
and rewrite:%
\begin{eqnarray*}
&&-\frac{\epsilon \bar{C}\left( X\right) }{3\left( f_{1}\left( X\right)
\right) ^{2}\sigma _{\hat{K}}^{2}}\left( \left( f_{1}\left( X\right) K_{0}-%
\bar{C}\left( X\right) \right) ^{3}+\left( \bar{C}\left( X\right) \right)
^{3}\right) \\
&=&-\frac{\epsilon K_{0}\bar{C}\left( X\right) }{3f_{1}\left( X\right)
\sigma _{\hat{K}}^{2}}\left( \sigma _{\hat{K}}^{2}\left( \frac{\left\vert
\Psi _{0}\left( X\right) \right\vert ^{2}}{\epsilon }-\frac{f_{1}\left(
X\right) }{2}\right) +\bar{C}\left( X\right) \left( \bar{C}\left( X\right) -%
\sqrt{\sigma _{\hat{K}}^{2}\left( \frac{\left\vert \Psi _{0}\left( X\right)
\right\vert ^{2}}{\epsilon }-\frac{f_{1}\left( X\right) }{2}\right) }\right)
\right) \\
&=&-\frac{\epsilon K_{0}\bar{C}\left( X\right) }{3f_{1}\left( X\right)
\sigma _{\hat{K}}^{2}}\left( \sqrt{\sigma _{\hat{K}}^{2}\left( \frac{%
\left\vert \Psi _{0}\left( X\right) \right\vert ^{2}}{\epsilon }-\frac{%
f_{1}\left( X\right) }{2}\right) }\left( \sqrt{\sigma _{\hat{K}}^{2}\left( 
\frac{\left\vert \Psi _{0}\left( X\right) \right\vert ^{2}}{\epsilon }-\frac{%
f_{1}\left( X\right) }{2}\right) }-\bar{C}\left( X\right) \right) +\bar{C}%
^{2}\left( X\right) \right)
\end{eqnarray*}%
so that, definng:%
\begin{equation*}
X=\sqrt{\sigma _{\hat{K}}^{2}\left( \frac{\left\vert \Psi _{0}\left(
X\right) \right\vert ^{2}}{\epsilon }-\frac{f_{1}\left( X\right) }{2}\right) 
}
\end{equation*}%
we find the amount of capital, $K_{X}\left\vert \Psi \left( X\right)
\right\vert ^{2}$:%
\begin{equation}
K_{X}\left\vert \Psi \left( X\right) \right\vert ^{2}=\frac{\epsilon K_{0}}{%
\sigma _{\hat{K}}^{2}f_{1}\left( X\right) }\left\{ \frac{1}{4}\left(
2X^{2}\left( \bar{C}+X\right) -\left( X-\bar{C}\right) \left( X^{2}+\left( 
\bar{C}\right) ^{2}\right) \right) -\frac{\bar{C}}{3}\left( X\left( X-\bar{C}%
\right) +C^{2}\right) \right\}  \label{KFR}
\end{equation}

The average capital in sector $X$ is given by the ratio of (\ref{KFR}) and (%
\ref{PSF}):%
\begin{eqnarray*}
K_{X} &=&\frac{1}{f_{1}\left( X\right) }\frac{\frac{1}{4}\left( 2X^{2}\left(
C+X\right) -\left( X-C\right) \left( X^{2}+C^{2}\right) \right) -\frac{C}{3}%
\left( X\left( X-C\right) +C^{2}\right) }{\frac{2}{3}X^{2}+\frac{1}{3}\left(
X-C\right) C} \\
&=&\frac{1}{4f_{1}\left( X\right) }\left( 3X-C\right) \frac{\left(
C+X\right) }{2X-C}
\end{eqnarray*}

Rewriting:

\begin{equation*}
\left( \frac{2}{3}X^{2}+\frac{1}{3}\left( X-C\right) C\right) \left( \frac{1%
}{4}\left( 3X-C\right) \frac{\left( C+X\right) }{2X-C}-C\right) =\epsilon 
\frac{K_{0}}{\sigma _{\hat{K}}^{2}}\frac{1}{4}\left( C-X\right) ^{2}\left(
C+X\right) =\frac{\epsilon }{\sigma _{\hat{K}}^{2}}\frac{1}{4}\left(
X^{2}-C^{2}\right) ^{2}
\end{equation*}%
and using that:%
\begin{equation*}
X^{2}=\sigma _{\hat{K}}^{2}\left( \frac{\left\vert \Psi _{0}\left( X\right)
\right\vert ^{2}}{\epsilon }-\frac{f_{1}\left( X\right) }{2}\right)
\end{equation*}%
leads to the expression for $\left\vert \Psi \left( X\right) \right\vert
^{2} $: 
\begin{eqnarray*}
\left\vert \Psi \left( X\right) \right\vert ^{2} &=&\left( \frac{2}{3}\sigma
_{\hat{K}}^{2}\left( \frac{\left\vert \Psi _{0}\left( X\right) \right\vert
^{2}}{\epsilon }-\frac{f_{1}\left( X\right) }{2}\right) +\frac{1}{3}\left( 
\sqrt{\sigma _{\hat{K}}^{2}\left( \frac{\left\vert \Psi _{0}\left( X\right)
\right\vert ^{2}}{\epsilon }-\frac{f_{1}\left( X\right) }{2}\right) }-\bar{C}%
\left( X\right) \right) \bar{C}\left( X\right) \right) \\
&&\times \epsilon \frac{\bar{C}\left( X\right) +\sqrt{\sigma _{\hat{K}%
}^{2}\left( \frac{\left\vert \Psi _{0}\left( X\right) \right\vert ^{2}}{%
\epsilon }-\frac{f_{1}\left( X\right) }{2}\right) }}{\sigma _{\hat{K}%
}^{2}f_{1}\left( X\right) }
\end{eqnarray*}%
The overall return for a firm was given by:%
\begin{eqnarray*}
&&\left( 1+\underline{k}_{2}\left( X\right) \right) \left( f_{1}\left(
X\right) K_{p}-\bar{C}\left( X\right) \right) -\underline{k}_{2}\left(
X\right) K_{p}\bar{r} \\
&=&\left( 1+\underline{k}_{2}\left( X\right) \right) \left( \left(
f_{1}\left( X\right) -\frac{\underline{k}_{2}\left( X\right) }{1+\underline{k%
}_{2}\left( X\right) }\bar{r}\right) K_{p}-\bar{C}\left( X\right) \right) \\
&=&f_{1}^{\left( e\right) }\left( X\right) K_{p}-\bar{C}^{\left( e\right)
}\left( X\right)
\end{eqnarray*}%
with:%
\begin{equation*}
f_{1}^{\left( e\right) }\left( X\right) =\left( 1+\underline{k}_{2}\left(
X\right) \right) f_{1}\left( X\right) -\underline{k}_{2}\left( X\right) \bar{%
r}
\end{equation*}%
and the associated return per unit of capital is:%
\begin{equation*}
f_{1}^{\prime }\left( X\right) =\left( 1+\underline{k}_{2}\left( X\right)
\right) \left( \left( f_{1}\left( X\right) -\frac{\underline{k}_{2}\left(
X\right) }{1+\underline{k}_{2}\left( X\right) }\bar{r}\right) -\frac{\bar{C}%
\left( X\right) }{K_{p}}\right)
\end{equation*}%
so that the relative return with respect to the interest rates is:%
\begin{equation*}
\frac{f_{1}^{\prime }\left( X\right) -\bar{r}}{1+\underline{k}_{2}\left(
X\right) }=f_{1}\left( X\right) -\bar{r}-\frac{\bar{C}\left( X\right) }{K_{p}%
}
\end{equation*}

\subsubsection*{A7.2.2 Return of the firm for investors}

The return provided by the firms of the sector $X$ is given by:%
\begin{equation*}
\left\vert \Psi \left( X\right) \right\vert ^{2}K_{X}\frac{f_{1}^{\prime
}\left( X\right) -\bar{r}}{1+\underline{k}_{2}\left( X\right) }=\left\vert
\Psi \left( X\right) \right\vert ^{2}\left( \left( f_{1}\left( X\right) -%
\bar{r}\right) K_{X}-\bar{C}\left( X\right) \right)
\end{equation*}%
where:%
\begin{equation*}
K_{X}\rightarrow \frac{1}{4f_{1}^{\left( e\right) }\left( X\right) }\frac{%
\left( 3X^{\left( e\right) }-C^{\left( e\right) }\right) \left( C^{\left(
e\right) }+X^{\left( e\right) }\right) }{2X^{\left( e\right) }-C^{\left(
e\right) }}
\end{equation*}%
and we have:

\begin{eqnarray*}
&&\left\vert \Psi \left( X\right) \right\vert ^{2}\left( \left( f_{1}\left(
X\right) -\bar{r}\right) K_{X}-\bar{C}\left( X\right) \right) \\
&=&\frac{\left( C^{\left( e\right) }+X^{\left( e\right) }\right) \left( 
\frac{2}{3}X^{2}+\frac{1}{3}\left( X-C^{\left( e\right) }\right) C^{\left(
e\right) }\right) \epsilon }{\sigma _{\hat{K}}^{2}f_{1}^{\left( e\right)
}\left( X\right) }\left( \frac{\left( f_{1}\left( X\right) -\bar{r}\right) }{%
4f_{1}^{\left( e\right) }\left( X\right) }\frac{\left( 3X^{\left( e\right)
}-C^{\left( e\right) }\right) \left( C^{\left( e\right) }+X^{\left( e\right)
}\right) }{2X^{\left( e\right) }-C^{\left( e\right) }}-C\right) \\
&=&\frac{\left( C^{\left( e\right) }+X^{\left( e\right) }\right) \left( 
\frac{2}{3}X^{2}+\frac{1}{3}\left( X-C^{\left( e\right) }\right) C^{\left(
e\right) }\right) \epsilon }{\sigma _{\hat{K}}^{2}\left( 1+\underline{k}%
_{2}\left( X\right) \right) f_{1}^{\left( e\right) }\left( X\right) }\left\{ 
\frac{\left( f_{1}\left( X\right) -\bar{r}\right) }{4\left( f_{1}\left(
X\right) -\frac{\underline{k}_{2}\left( X\right) }{1+\underline{k}_{2}\left(
X\right) }\bar{r}\right) }\frac{\left( 3X^{\left( e\right) }-C^{\left(
e\right) }\right) \left( C^{\left( e\right) }+X^{\left( e\right) }\right) }{%
2X^{\left( e\right) }-C^{\left( e\right) }}-C^{\left( e\right) }\right\}
\end{eqnarray*}

\subsection*{A7.2 Case two}

\subsubsection*{A7.2.1 Average capital per sector and firms returns}

In this case, we consider that all firms have a minimum capital, so that: 
\begin{equation*}
\left\vert \Psi _{0}\left( X\right) \right\vert ^{2}-\epsilon \frac{%
f_{1}\left( X\right) }{2}-\epsilon \frac{\left( \bar{C}\left( X\right)
\right) ^{2}}{\sigma _{\hat{K}}^{2}}<0
\end{equation*}%
This exquation also writes:%
\begin{equation*}
\left\vert \Psi _{0}\left( X\right) \right\vert ^{2}-\epsilon \left( \frac{%
\left( f_{1}\left( X\right) K_{p}-\bar{C}\left( X\right) \right) ^{2}}{%
\sigma _{\hat{K}}^{2}}+\frac{f_{1}\left( X\right) }{2}\right) >0
\end{equation*}%
and there are two solutions:%
\begin{equation*}
\left\vert \Psi _{0}\left( X\right) \right\vert ^{2}-\epsilon \left( \frac{%
\left( f_{1}\left( X\right) K_{0}-\bar{C}\left( X\right) \right) ^{2}}{%
\sigma _{\hat{K}}^{2}}+\frac{f_{1}\left( X\right) }{2}\right) =0
\end{equation*}%
\begin{equation*}
K_{0\pm }=\frac{\bar{C}\left( X\right) \pm \sqrt{\sigma _{\hat{K}}^{2}\left( 
\frac{\left\vert \Psi _{0}\left( X\right) \right\vert ^{2}}{\epsilon }-\frac{%
f_{1}\left( X\right) }{2}\right) }}{f_{1}\left( X\right) }
\end{equation*}

These are the two bound to the level of capital in one sector. We will write:%
\begin{eqnarray*}
K_{0} &=&K_{0+} \\
K_{0+}-K_{0-} &=&\frac{2\sqrt{\sigma _{\hat{K}}^{2}\left( \frac{\left\vert
\Psi _{0}\left( X\right) \right\vert ^{2}}{\epsilon }-\frac{f_{1}\left(
X\right) }{2}\right) }}{f_{1}\left( X\right) }
\end{eqnarray*}%
The computation of $\left\vert \Psi \left( X\right) \right\vert ^{2}$ is
similar to the case one:

\begin{eqnarray*}
\left\vert \Psi \left( X\right) \right\vert ^{2} &=&\left( \left\vert \Psi
_{0}\left( X\right) \right\vert ^{2}-\frac{\epsilon f_{1}\left( X\right) }{2}%
\right) \Delta K_{0}-\epsilon \left( \frac{\left( f_{1}\left( X\right)
K_{0+}-\bar{C}\left( X\right) \right) ^{3}}{3f_{1}\left( X\right) \sigma _{%
\hat{K}}^{2}}-\frac{\left( f_{1}\left( X\right) K_{0-}-\bar{C}\left(
X\right) \right) ^{3}}{3f_{1}\left( X\right) \sigma _{\hat{K}}^{2}}\right) \\
&=&\left( \left\vert \Psi _{0}\left( X\right) \right\vert ^{2}-\frac{%
\epsilon f_{1}\left( X\right) }{2}\right) \Delta K_{0} \\
&&-\frac{\epsilon \Delta K_{0}}{3\sigma _{\hat{K}}^{2}}\left( \left(
f_{1}\left( X\right) K_{0+}-\bar{C}\left( X\right) \right) ^{2}+\left(
f_{1}\left( X\right) K_{0-}-\bar{C}\left( X\right) \right) ^{2}\right. \\
&&\left. +\left( f_{1}\left( X\right) K_{0+}-\bar{C}\left( X\right) \right)
\left( f_{1}\left( X\right) K_{0-}-\bar{C}\left( X\right) \right) \right) \\
&=&\left( \left\vert \Psi _{0}\left( X\right) \right\vert ^{2}-\frac{%
\epsilon f_{1}\left( X\right) }{2}\right) \Delta K_{0}-\frac{\epsilon \Delta
K_{0}}{3\sigma _{\hat{K}}^{2}}\left( \sigma _{\hat{K}}^{2}\left( \frac{%
\left\vert \Psi _{0}\left( X\right) \right\vert ^{2}}{\epsilon }-\frac{%
f_{1}\left( X\right) }{2}\right) \right) \\
&=&\frac{2}{3}\left( \frac{\left\vert \Psi _{0}\left( X\right) \right\vert
^{2}}{\epsilon }-\frac{f_{1}\left( X\right) }{2}\right) \epsilon \Delta K_{0}
\end{eqnarray*}%
and we find:%
\begin{equation*}
\left\vert \Psi \left( X\right) \right\vert ^{2}=\frac{2}{3}\left( \frac{%
\left\vert \Psi _{0}\left( X\right) \right\vert ^{2}}{\epsilon }-\frac{%
f_{1}\left( X\right) }{2}\right) \epsilon \Delta K_{0}
\end{equation*}%
We find $K_{X}\left\vert \Psi \left( X\right) \right\vert ^{2}$ using:%
\begin{equation*}
\left\vert \Psi \left( K,X\right) \right\vert ^{2}\rightarrow \left[
\left\vert \Psi _{0}\left( X\right) \right\vert ^{2}-\epsilon \frac{%
f_{1}\left( X\right) }{2}\right] -\epsilon \left( \frac{\left( f_{1}\left(
X\right) K_{p}-\bar{C}\left( X\right) \right) ^{2}}{\sigma _{\hat{K}}^{2}}%
\right)
\end{equation*}%
Multiplying by $K_{p}$:%
\begin{eqnarray*}
&&K_{p}\left( \left( f_{1}\left( X\right) K_{p}-\bar{C}\left( X\right)
\right) ^{2}\right) \\
&=&\frac{1}{f_{1}\left( X\right) }\left( \left( f_{1}\left( X\right) K_{p}-%
\bar{C}\left( X\right) \right) \left( \left( f_{1}\left( X\right) K_{p}-\bar{%
C}\left( X\right) \right) ^{2}\right) +\bar{C}\left( X\right) \left( \left(
f_{1}\left( X\right) K_{p}-\bar{C}\left( X\right) \right) ^{2}\right) \right)
\end{eqnarray*}%
Integration between the capital bound yields:

\begin{eqnarray*}
K_{X}\left\vert \Psi \left( X\right) \right\vert ^{2} &=&\left( \left\vert
\Psi _{0}\left( X\right) \right\vert ^{2}-\frac{\epsilon f_{1}\left(
X\right) }{2}\right) \frac{K_{0+}^{2}-K_{0-}^{2}}{2} \\
&&-\frac{\epsilon }{4\left( f_{1}\left( X\right) \right) ^{2}\sigma _{\hat{K}%
}^{2}}\left( \left( f_{1}\left( X\right) K_{0+}-\bar{C}\left( X\right)
\right) ^{4}-\left( f_{1}\left( X\right) K_{0-}-\bar{C}\left( X\right)
\right) ^{4}\right) \\
&&-\frac{\epsilon \bar{C}\left( X\right) }{3\left( f_{1}\left( X\right)
\right) ^{2}\sigma _{\hat{K}}^{2}}\left( \left( f_{1}\left( X\right) K_{0+}-%
\bar{C}\left( X\right) \right) ^{3}-\left( f_{1}\left( X\right) K_{0-}-\bar{C%
}\left( X\right) \right) ^{3}\right) \\
&=&\left( \left\vert \Psi _{0}\left( X\right) \right\vert ^{2}-\frac{%
\epsilon f_{1}\left( X\right) }{2}\right) \frac{K_{0+}^{2}-K_{0-}^{2}}{2} \\
&&-\frac{\epsilon \Delta K_{0}\left( f_{1}\left( X\right) \left(
K_{0+}+K_{0-}\right) -2\bar{C}\left( X\right) \right) }{4f_{1}\left(
X\right) \sigma _{\hat{K}}^{2}}\left( \left( f_{1}\left( X\right) K_{0+}-%
\bar{C}\left( X\right) \right) ^{2}+\left( f_{1}\left( X\right) K_{0-}-\bar{C%
}\left( X\right) \right) ^{2}\right)
\end{eqnarray*}

We write:%
\begin{equation*}
\left( \left\vert \Psi _{0}\left( X\right) \right\vert ^{2}-\frac{\epsilon
f_{1}\left( X\right) }{2}\right) \frac{K_{0+}^{2}-K_{0-}^{2}}{2}=\left( 
\frac{\left\vert \Psi _{0}\left( X\right) \right\vert ^{2}}{\epsilon }-\frac{%
f_{1}\left( X\right) }{2}\right) \epsilon \Delta K_{0}\frac{\bar{C}\left(
X\right) }{f_{1}\left( X\right) }
\end{equation*}%
and:%
\begin{eqnarray*}
&&-\frac{\epsilon \bar{C}\left( X\right) }{3\left( f_{1}\left( X\right)
\right) ^{2}\sigma _{\hat{K}}^{2}}\left( \left( f_{1}\left( X\right) K_{0+}-%
\bar{C}\left( X\right) \right) ^{3}-\left( f_{1}\left( X\right) K_{0-}-\bar{C%
}\left( X\right) \right) ^{3}\right) \\
&=&-\frac{\epsilon \bar{C}\left( X\right) \Delta K_{0}}{3f_{1}\left(
X\right) \sigma _{\hat{K}}^{2}}\left( \left( f_{1}\left( X\right) K_{0+}-%
\bar{C}\left( X\right) \right) ^{2}+\left( f_{1}\left( X\right) K_{0-}-\bar{C%
}\left( X\right) \right) ^{2}+\left( f_{1}\left( X\right) K_{0+}-\bar{C}%
\left( X\right) \right) \left( f_{1}\left( X\right) K_{0-}-\bar{C}\left(
X\right) \right) \right) \\
&=&-\frac{\epsilon \bar{C}\left( X\right) \Delta K_{0}}{3f_{1}\left(
X\right) \sigma _{\hat{K}}^{2}}\sigma _{\hat{K}}^{2}\left( \frac{\left\vert
\Psi _{0}\left( X\right) \right\vert ^{2}}{\epsilon }-\frac{f_{1}\left(
X\right) }{2}\right)
\end{eqnarray*}%
so that ultimately:%
\begin{eqnarray*}
K_{X}\left\vert \Psi \left( X\right) \right\vert ^{2} &\rightarrow &\sigma _{%
\hat{K}}^{2}\left( \frac{\left\vert \Psi _{0}\left( X\right) \right\vert ^{2}%
}{\epsilon }-\frac{f_{1}\left( X\right) }{2}\right) \frac{\epsilon \Delta
K_{0}\bar{C}\left( X\right) }{f_{1}\left( X\right) \sigma _{\hat{K}}^{2}}-%
\frac{\epsilon \bar{C}\left( X\right) \Delta K_{0}}{3f_{1}\left( X\right)
\sigma _{\hat{K}}^{2}}\sigma _{\hat{K}}^{2}\left( \frac{\left\vert \Psi
_{0}\left( X\right) \right\vert ^{2}}{\epsilon }-\frac{f_{1}\left( X\right) 
}{2}\right) \\
&=&\frac{2}{3}\frac{\epsilon \Delta K_{0}\bar{C}\left( X\right) }{%
f_{1}\left( X\right) \sigma _{\hat{K}}^{2}}\sigma _{\hat{K}}^{2}\left( \frac{%
\left\vert \Psi _{0}\left( X\right) \right\vert ^{2}}{\epsilon }-\frac{%
f_{1}\left( X\right) }{2}\right)
\end{eqnarray*}%
denoting:%
\begin{equation*}
X=\sqrt{\sigma _{\hat{K}}^{2}\left( \frac{\left\vert \Psi _{0}\left(
X\right) \right\vert ^{2}}{\epsilon }-\frac{f_{1}\left( X\right) }{2}\right) 
}
\end{equation*}%
we obtain:%
\begin{equation*}
K_{X}\left\vert \Psi \left( X\right) \right\vert ^{2}=\frac{2\epsilon \Delta
K_{0}\bar{C}\left( X\right) }{3f_{1}\left( X\right) \sigma _{\hat{K}}^{2}}%
X^{2}
\end{equation*}%
given that:%
\begin{equation*}
\left\vert \Psi \left( X\right) \right\vert ^{2}=2\frac{\epsilon \Delta K_{0}%
}{3\sigma _{\hat{K}}^{2}}X^{2}
\end{equation*}%
The average capital is:%
\begin{equation*}
K_{X}=\frac{\bar{C}\left( X\right) }{f_{1}\left( X\right) }
\end{equation*}

\subsubsection*{A7.2.2 Return of the firm for investors}

The return of the firms in sector $X$ for the investors is: 
\begin{equation*}
\left\vert \Psi \left( X\right) \right\vert ^{2}K_{X}\frac{f_{1}^{\prime
}\left( X\right) -\bar{r}}{1+\underline{k}_{2}\left( X\right) }=\left\vert
\Psi \left( X\right) \right\vert ^{2}\left( \left( f_{1}\left( X\right) -%
\bar{r}\right) K_{X}-\bar{C}\left( X\right) \right)
\end{equation*}%
We use:%
\begin{eqnarray*}
\left( f_{1}-\bar{r}\right) K_{X}-C &=&\left( f_{1}\left( X\right) -\bar{r}%
\right) \frac{C^{\left( e\right) }}{f_{1}^{\left( e\right) }\left( X\right) }%
-\frac{C^{\left( e\right) }}{\left( 1+\underline{k}_{2}\left( X\right)
\right) } \\
&=&\frac{C}{\left( 1+\underline{k}_{2}\left( X\right) \right) \left( 1+%
\underline{k}\left( X\right) \right) }\left( \frac{f_{1}\left( X\right) -%
\bar{r}}{f_{1}\left( X\right) -\frac{\underline{k}_{2}\left( X\right) }{%
\left( 1+\underline{k}_{2}\left( X\right) \right) }\bar{r}}-1\right)
\end{eqnarray*}%
so that:%
\begin{equation*}
\left\vert \Psi \left( X\right) \right\vert ^{2}\left( f_{1}K_{X}-C\right) =2%
\frac{\epsilon \Delta K_{0}C^{\left( e\right) }}{3\sigma _{\hat{K}}^{2}}%
X^{2}\left( \left( f_{1}\left( X\right) -\bar{r}\right) \frac{\left(
3X^{\left( e\right) }-1\right) }{f_{1}^{\left( e\right) }\left( X\right) }-%
\frac{C^{\left( e\right) }}{\left( 1+\underline{k}_{2}\left( X\right)
\right) }\right)
\end{equation*}%
and this becmes:%
\begin{equation*}
\left\vert \Psi \left( X\right) \right\vert ^{2}K_{X}\frac{f_{1}^{\prime
}\left( X\right) -\bar{r}}{1+\underline{k}_{2}\left( X\right) }=4\frac{%
\epsilon \left( C^{\left( e\right) }\right) ^{2}}{3\sigma _{\hat{K}%
}^{2}f_{1}^{\left( e\right) }\left( X\right) }X^{2}\left( \frac{\left(
f_{1}\left( X\right) -\bar{r}\right) }{f_{1}^{\left( e\right) }\left(
X\right) }\left( 3X^{\left( e\right) }-1\right) -C\right)
\end{equation*}

\section*{Appendix 8 Estimation of the functionl derivative $\frac{\protect%
\delta }{\protect\delta \left\vert \hat{\Psi}\left( \hat{K}_{1},\hat{X}%
_{1}\right) \right\vert ^{2}}\hat{g}\left( \hat{K},\hat{X}\right) $}

We decompose: 
\begin{eqnarray*}
&&\frac{\delta }{\delta \left\vert \hat{\Psi}\left( \hat{K}_{1},\hat{X}%
_{1}\right) \right\vert ^{2}}\hat{g}\left( \hat{K},\hat{X}\right) \\
&=&\frac{\partial }{\partial \left\vert \hat{\Psi}\left( \hat{K}_{1},\hat{X}%
_{1}\right) \right\vert ^{2}}\frac{1}{1-M\left( \left( \hat{K},\hat{X}%
\right) ,\left( \hat{K}^{\prime },\hat{X}^{\prime }\right) \right)
\left\vert \hat{\Psi}\left( \hat{K}^{\prime },\hat{X}^{\prime }\right)
\right\vert ^{2}}\hat{f}\left( \hat{K}^{\prime },\hat{X}^{\prime }\right) \\
&=&\frac{1}{1-M\left( \left( \hat{K},\hat{X}\right) ,\left( \hat{K}^{\prime
},\hat{X}^{\prime }\right) \right) \left\vert \hat{\Psi}\left( \hat{K}%
^{\prime },\hat{X}^{\prime }\right) \right\vert ^{2}} \\
&&\times \frac{\partial }{\partial \left\vert \hat{\Psi}\left( \hat{K}_{1},%
\hat{X}_{1}\right) \right\vert ^{2}}\left( M\left( \left( \hat{K}^{\prime },%
\hat{X}^{\prime }\right) ,\left( \hat{K}^{\prime \prime },\hat{X}^{\prime
\prime }\right) \right) \left\vert \hat{\Psi}\left( \hat{K}^{\prime \prime },%
\hat{X}^{\prime \prime }\right) \right\vert ^{2}\right) \\
&&\times \frac{1}{1-M\left( \left( \hat{K}^{\prime \prime },\hat{X}^{\prime
\prime }\right) ,\left( \hat{K}^{\prime \prime \prime },\hat{X}^{\prime
\prime \prime }\right) \right) \left\vert \hat{\Psi}\left( \hat{K}^{\prime
\prime \prime },\hat{X}^{\prime \prime \prime }\right) \right\vert ^{2}}\hat{%
f}\left( \hat{K}^{\prime \prime \prime },\hat{X}^{\prime \prime \prime
}\right) \\
&&+\frac{1}{1-M\left( \left( \hat{K},\hat{X}\right) ,\left( \hat{K}^{\prime
},\hat{X}^{\prime }\right) \right) \left\vert \hat{\Psi}\left( \hat{K}%
^{\prime },\hat{X}^{\prime }\right) \right\vert ^{2}}\frac{\partial }{%
\partial \left\vert \hat{\Psi}\left( \hat{K}_{1},\hat{X}_{1}\right)
\right\vert ^{2}}\hat{f}\left( \hat{K}^{\prime },\hat{X}^{\prime }\right)
\end{eqnarray*}%
and estimate each term separately.

\subsection*{A8.1 Estimation of derivative of $M\left( \hat{K},\hat{X},\hat{K%
}^{\prime },\hat{X}^{\prime }\right) $}

\begin{equation*}
M\left( \hat{K},\hat{X},\hat{K}^{\prime },\hat{X}^{\prime }\right) =\frac{%
\hat{k}\left( \hat{X},\hat{X}^{\prime }\right) \hat{K}}{1+\underline{\hat{k}}%
\left( \hat{X}\right) }\rightarrow \frac{\hat{k}\left( \hat{X}^{\prime },%
\hat{X}\right) \hat{K}}{\left\Vert \hat{\Psi}\right\Vert ^{2}\left\langle 
\hat{K}\right\rangle \left( 1+\underline{\hat{k}}\left( \hat{X}^{\prime
}\right) \right) }
\end{equation*}%
\begin{equation*}
\underline{\hat{k}}\left( \hat{X}^{\prime }\right) =\int \frac{\hat{k}\left( 
\hat{X}^{\prime },\hat{X}\right) -\left\langle \hat{k}\left( \hat{X}^{\prime
},\hat{X}\right) \right\rangle }{\left\Vert \hat{\Psi}\right\Vert
^{2}\left\langle \hat{K}\right\rangle }\hat{K}_{\hat{X}}\left\vert \hat{\Psi}%
\left( \hat{X}\right) \right\vert ^{2}d\hat{X}
\end{equation*}%
\begin{equation*}
\frac{\partial }{\partial \left\vert \hat{\Psi}\left( \hat{K}_{1},\hat{X}%
_{1}\right) \right\vert ^{2}}\left( 1+\underline{\hat{k}}\left( \hat{X}%
\right) \right) \rightarrow \int \frac{\hat{k}\left( \hat{X}^{\prime },\hat{X%
}\right) -\left\langle \hat{k}\left( \hat{X}^{\prime },\hat{X}\right)
\right\rangle }{\left\Vert \hat{\Psi}\right\Vert ^{2}\left\langle \hat{K}%
\right\rangle }\hat{K}_{\hat{X}}\left\vert \hat{\Psi}\left( \hat{X}\right)
\right\vert ^{2}d\hat{X}\left( -\frac{\hat{K}_{1}}{\left\Vert \hat{\Psi}%
\right\Vert ^{2}\left\langle \hat{K}\right\rangle }\right)
\end{equation*}%
in averages:%
\begin{equation*}
\frac{\partial }{\partial \left\vert \hat{\Psi}\left( \hat{K}_{1},\hat{X}%
_{1}\right) \right\vert ^{2}}\left( 1+\underline{\hat{k}}\left( \hat{X}%
\right) \right) <<1
\end{equation*}%
\begin{equation*}
\int \left\vert \hat{\Psi}\left( \hat{K},\hat{X}\right) \right\vert ^{2}\hat{%
K}=\left\Vert \hat{\Psi}\right\Vert ^{2}\left\langle \hat{K}\right\rangle
\end{equation*}%
for:%
\begin{equation*}
k\left( X,\hat{X}^{\prime }\right) =\frac{k\left( X,\hat{X}^{\prime }\right) 
}{\left\vert \Psi \left( X\right) \right\vert ^{2}\left\langle
K\right\rangle }
\end{equation*}%
\begin{equation*}
1+\underline{k}\left( X\right) =1+\int \frac{k\left( X,\hat{X}^{\prime
}\right) }{\left\Vert \Psi \right\Vert ^{2}\left\langle K\right\rangle }\hat{%
K}^{\prime }\left\vert \hat{\Psi}\left( \hat{K}^{\prime },\hat{X}^{\prime
}\right) \right\vert ^{2}>>1
\end{equation*}%
\begin{equation*}
\frac{\partial }{\partial \left\vert \hat{\Psi}\left( \hat{K}_{1},\hat{X}%
_{1}\right) \right\vert ^{2}}\frac{1}{1+\underline{k}\left( X\right) }=-%
\frac{1}{1+\underline{k}\left( X\right) }\frac{\frac{k\left( X,\hat{X}%
^{\prime }\right) }{\left\Vert \Psi \right\Vert ^{2}\left\langle
K\right\rangle }\hat{K}^{\prime }}{1+\underline{k}\left( X\right) }<<\frac{1%
}{1+\underline{k}\left( X\right) }
\end{equation*}%
\begin{eqnarray*}
&&\frac{\partial }{\partial \left\vert \hat{\Psi}\left( \hat{K}_{1},\hat{X}%
_{1}\right) \right\vert ^{2}}\left( M\left( \left( \hat{K}^{\prime },\hat{X}%
^{\prime }\right) ,\left( \hat{K}^{\prime \prime },\hat{X}^{\prime \prime
}\right) \right) \left\vert \hat{\Psi}\left( \hat{K}^{\prime \prime },\hat{X}%
^{\prime \prime }\right) \right\vert ^{2}\right) \\
&=&M\left( \left( \hat{K}^{\prime },\hat{X}^{\prime }\right) ,\left( \hat{K}%
_{1},\hat{X}_{1}\right) \right) +\frac{\partial M\left( \left( \hat{K}%
^{\prime },\hat{X}^{\prime }\right) ,\left( \hat{K}^{\prime \prime },\hat{X}%
^{\prime \prime }\right) \right) }{\partial \left\vert \hat{\Psi}\left( \hat{%
K}_{1},\hat{X}_{1}\right) \right\vert ^{2}}\left\vert \hat{\Psi}\left( \hat{K%
}^{\prime \prime },\hat{X}^{\prime \prime }\right) \right\vert ^{2} \\
&=&M\left( \left( \hat{K}^{\prime \prime },\hat{X}^{\prime \prime }\right)
,\left( \hat{K}_{1},\hat{X}_{1}\right) \right) +\frac{1}{\left( 1+\underline{%
\hat{k}}\left( \hat{X}^{\prime }\right) \right) }\frac{\partial \left( \frac{%
\hat{k}\left( \hat{X}^{\prime },\hat{X}^{\prime \prime }\right) }{\int
\left\vert \hat{\Psi}\left( \hat{K},\hat{X}\right) \right\vert ^{2}\hat{K}}%
\hat{K}^{\prime }\right) }{\partial \left\vert \hat{\Psi}\left( \hat{K}_{1},%
\hat{X}_{1}\right) \right\vert ^{2}}\left\vert \hat{\Psi}\left( \hat{K}%
^{\prime \prime },\hat{X}^{\prime \prime }\right) \right\vert ^{2} \\
&\simeq &M\left( \left( \hat{K}^{\prime \prime },\hat{X}^{\prime \prime
}\right) ,\left( \hat{K}_{1},\hat{X}_{1}\right) \right) -\frac{\hat{k}\left( 
\hat{X}^{\prime },\hat{X}^{\prime \prime }\right) \hat{K}^{\prime }\hat{K}%
_{1}}{\left( \int \left\vert \hat{\Psi}\left( \hat{K},\hat{X}\right)
\right\vert ^{2}\hat{K}\right) ^{2}}\frac{1}{1+\underline{\hat{k}}\left( 
\hat{X}^{\prime }\right) }\left\vert \hat{\Psi}\left( \hat{K}^{\prime \prime
},\hat{X}^{\prime \prime }\right) \right\vert ^{2}
\end{eqnarray*}%
\begin{eqnarray*}
&&\frac{1}{1-M\left( \left( \hat{K},\hat{X}\right) ,\left( \hat{K}^{\prime },%
\hat{X}^{\prime }\right) \right) \left\vert \hat{\Psi}\left( \hat{K}^{\prime
},\hat{X}^{\prime }\right) \right\vert ^{2}} \\
&&\times \frac{\partial }{\partial \left\vert \hat{\Psi}\left( \hat{K}_{1},%
\hat{X}_{1}\right) \right\vert ^{2}}\left( M\left( \left( \hat{K}^{\prime },%
\hat{X}^{\prime }\right) ,\left( \hat{K}^{\prime \prime },\hat{X}^{\prime
\prime }\right) \right) \left\vert \hat{\Psi}\left( \hat{K}^{\prime \prime },%
\hat{X}^{\prime \prime }\right) \right\vert ^{2}\right) \hat{g}\left( \hat{K}%
^{\prime \prime },\hat{X}^{\prime \prime }\right) \\
&\simeq &\frac{1}{1-M\left( \left( \hat{K},\hat{X}\right) ,\left( \hat{K}%
^{\prime },\hat{X}^{\prime }\right) \right) \left\vert \hat{\Psi}\left( \hat{%
K}^{\prime },\hat{X}^{\prime }\right) \right\vert ^{2}}M\left( \left( \hat{K}%
^{\prime },\hat{X}^{\prime }\right) ,\left( \hat{K}_{1},\hat{X}_{1}\right)
\right) \hat{g}\left( \hat{K}_{1},\hat{X}_{1}\right) \\
&&-\frac{1}{1-M\left( \left( \hat{K},\hat{X}\right) ,\left( \hat{K}^{\prime
},\hat{X}^{\prime }\right) \right) \left\vert \hat{\Psi}\left( \hat{K}%
^{\prime },\hat{X}^{\prime }\right) \right\vert ^{2}}\frac{\hat{k}\left( 
\hat{X}^{\prime },\hat{X}^{\prime \prime }\right) \hat{K}^{\prime }\hat{K}%
_{1}}{\left( \int \left\vert \hat{\Psi}\left( \hat{K},\hat{X}\right)
\right\vert ^{2}\hat{K}\right) ^{2}}\frac{1}{1+\underline{\hat{k}}\left( 
\hat{X}^{\prime }\right) }\left\vert \hat{\Psi}\left( \hat{K}^{\prime \prime
},\hat{X}^{\prime \prime }\right) \right\vert ^{2}\hat{g}\left( \hat{K}%
^{\prime \prime },\hat{X}^{\prime \prime }\right) \\
&\simeq &\frac{1}{1-M\left( \left( \hat{K},\hat{X}\right) ,\left( \hat{K}%
^{\prime },\hat{X}^{\prime }\right) \right) \left\vert \hat{\Psi}\left( \hat{%
K}^{\prime },\hat{X}^{\prime }\right) \right\vert ^{2}}M\left( \left( \hat{K}%
^{\prime },\hat{X}^{\prime }\right) ,\hat{X}_{1}\right) \hat{g}\left( \hat{K}%
_{1},\hat{X}_{1}\right) \\
&&-\frac{1}{1-M\left( \left( \hat{K},\hat{X}\right) ,\left( \hat{K}^{\prime
},\hat{X}^{\prime }\right) \right) \left\vert \hat{\Psi}\left( \hat{K}%
^{\prime },\hat{X}^{\prime }\right) \right\vert ^{2}}M\left( \left( \hat{K}%
^{\prime },\hat{X}^{\prime }\right) ,\left\langle \hat{X}\right\rangle
\right) \left\langle \hat{g}\right\rangle \frac{\hat{K}_{1}}{\left\langle 
\hat{K}\right\rangle }
\end{eqnarray*}%
We estimate:%
\begin{eqnarray*}
&&\left( 1+\underline{\hat{k}}\left( \left\langle \hat{X}\right\rangle
\right) \right) \frac{\underline{\hat{k}}\left( \left\langle \hat{X}%
\right\rangle ,\hat{X}_{1}\right) \left\langle \hat{K}\right\rangle \hat{g}%
\left( \hat{X}_{1}\right) }{\left( 1+\underline{\hat{k}}\left( \left\langle 
\hat{X}\right\rangle \right) \right) \int \left\vert \hat{\Psi}\left( \hat{K}%
,\hat{X}\right) \right\vert ^{2}\hat{K}} \\
&&-\left( 1+\underline{\hat{k}}\left( \left\langle \hat{X}\right\rangle
\right) \right) \frac{\underline{\hat{k}}\left( \left\langle \hat{X}%
\right\rangle ,\left\langle \hat{X}\right\rangle \right) \left\langle \hat{K}%
\right\rangle }{\left( 1+\underline{\hat{k}}\left( \left\langle \hat{X}%
\right\rangle \right) \right) \int \left\vert \hat{\Psi}\left( \hat{K},\hat{X%
}\right) \right\vert ^{2}\hat{K}}\left\langle \hat{g}\right\rangle \frac{%
\hat{K}_{1}}{\left\langle \hat{K}\right\rangle } \\
&\simeq &\frac{\underline{\hat{k}}\left( \left\langle \hat{X}\right\rangle ,%
\hat{X}_{1}\right) }{\left\Vert \hat{\Psi}\right\Vert ^{2}}\hat{g}\left( 
\hat{X}_{1}\right) -\frac{\hat{K}_{1}}{\left\langle \hat{K}\right\rangle }%
\frac{\underline{\hat{k}}\left( \left\langle \hat{X}\right\rangle
,\left\langle \hat{X}\right\rangle \right) }{\left\Vert \hat{\Psi}%
\right\Vert ^{2}}\left\langle \hat{g}\right\rangle
\end{eqnarray*}%
\bigskip

and:%
\begin{eqnarray}
&&\frac{\partial }{\partial \left\vert \hat{\Psi}\left( \hat{K}_{1},\hat{X}%
_{1}\right) \right\vert ^{2}}\hat{g}\left( \hat{K},\hat{X}\right)  \label{dv}
\\
&\simeq &\frac{\underline{\hat{k}}\left( \left\langle \hat{X}\right\rangle ,%
\hat{X}_{1}\right) }{\left\Vert \hat{\Psi}\right\Vert ^{2}}\hat{g}\left( 
\hat{X}_{1}\right) -\frac{\hat{K}_{1}}{\left\langle \hat{K}\right\rangle }%
\frac{\underline{\hat{k}}\left( \left\langle \hat{X}\right\rangle
,\left\langle \hat{X}\right\rangle \right) }{\left\Vert \hat{\Psi}%
\right\Vert ^{2}}\left\langle \hat{g}\right\rangle \frac{1}{1-M\left( \left( 
\hat{K},\hat{X}\right) ,\left( \hat{K}^{\prime },\hat{X}^{\prime }\right)
\right) \left\vert \hat{\Psi}\left( \hat{K}^{\prime },\hat{X}^{\prime
}\right) \right\vert ^{2}}\frac{\partial \hat{f}\left( \hat{K}^{\prime },%
\hat{X}^{\prime }\right) }{\partial \left\vert \hat{\Psi}\left( \hat{K}_{1},%
\hat{X}_{1}\right) \right\vert ^{2}}  \notag
\end{eqnarray}

\subsection*{A8.2 Estimation of $\frac{\partial \hat{f}\left( \hat{K}%
^{\prime },\hat{X}^{\prime }\right) }{\partial \left\vert \hat{\Psi}\left( 
\hat{K}_{1},\hat{X}_{1}\right) \right\vert ^{2}}$}

We first estimate the second term in the RHS of (\ref{dv}) $\frac{\partial 
\hat{f}\left( \hat{K}^{\prime },\hat{X}^{\prime }\right) }{\partial
\left\vert \hat{\Psi}\left( \hat{K}_{1},\hat{X}_{1}\right) \right\vert ^{2}}$%
, which is given by return equation (\ref{RN}):%
\begin{eqnarray*}
&&\frac{\partial \hat{f}\left( \hat{K}^{\prime },\hat{X}^{\prime }\right) }{%
\partial \left\vert \hat{\Psi}\left( \hat{K}_{1},\hat{X}_{1}\right)
\right\vert ^{2}} \\
&=&\frac{\partial \left\{ \frac{1-\hat{S}_{1}\left( \hat{X}\right) }{1-\hat{S%
}\left( \hat{X}\right) }\left( \Delta \left( \hat{X},\hat{X}^{\prime
}\right) -\hat{S}_{1}\left( \hat{X}^{\prime },\hat{X}\right) \right)
^{-1}S_{1}\left( \hat{X}^{\prime },\hat{X}^{\prime }\right) \frac{1-S\left( 
\hat{X}^{\prime }\right) }{1-S_{1}\left( \hat{X}^{\prime }\right) }\left(
\left( f_{1}^{\prime }\left( \hat{X}\right) -\bar{r}\right) +\Delta F_{\tau
}\left( \bar{R}\left( K,X\right) \right) \right) \right\} }{\partial
\left\vert \hat{\Psi}\left( \hat{K}_{1},\hat{X}_{1}\right) \right\vert ^{2}}
\end{eqnarray*}%
In average:%
\begin{equation*}
\frac{1-\hat{S}_{1}\left( \hat{X}\right) }{1-\hat{S}\left( \hat{X}\right) }%
\left( \Delta \left( \hat{X},\hat{X}^{\prime }\right) -\hat{S}_{1}\left( 
\hat{X}^{\prime },\hat{X}\right) \right) ^{-1}\simeq \frac{1}{1-\hat{S}%
\left( \hat{X}\right) }=1+\underline{\hat{k}}\left( X^{\prime }\right)
\end{equation*}%
so that:%
\begin{equation*}
\frac{\partial \left( \frac{1-\hat{S}_{1}\left( \hat{X}\right) }{1-\hat{S}%
\left( \hat{X}\right) }\left( \Delta \left( \hat{X},\hat{X}^{\prime }\right)
-\hat{S}_{1}\left( \hat{X}^{\prime },\hat{X}\right) \right) ^{-1}\right) }{%
\partial \left\vert \hat{\Psi}\left( \hat{K}_{1},\hat{X}_{1}\right)
\right\vert ^{2}}<<1
\end{equation*}%
and:%
\begin{equation*}
\frac{\partial }{\partial \left\vert \hat{\Psi}\left( \hat{K}_{1},\hat{X}%
_{1}\right) \right\vert ^{2}}\left( \left( f_{1}^{\prime }\left( \hat{X}%
\right) -\bar{r}\right) +\Delta F_{\tau }\left( \bar{R}\left( K,X\right)
\right) \right) =\frac{\partial }{\partial \left\vert \hat{\Psi}\left( \hat{K%
}_{1},\hat{X}_{1}\right) \right\vert ^{2}}f_{1}^{\prime }\left( \hat{X}%
\right)
\end{equation*}%
so that:%
\begin{eqnarray*}
\frac{\partial }{\partial \left\vert \hat{\Psi}\left( \hat{K}_{1},\hat{X}%
_{1}\right) \right\vert ^{2}}f_{1}^{\prime } &=&\frac{\partial }{\partial
\left\vert \hat{\Psi}\left( \hat{K}_{1},\hat{X}_{1}\right) \right\vert ^{2}}%
\left( f_{1}\left( X\right) K_{p}-\frac{C}{1+\frac{\underline{k}\left(
X\right) }{\left\langle K\right\rangle }\hat{K}_{X}\frac{\left\vert \hat{\Psi%
}\left( \hat{X}\right) \right\vert ^{2}}{\left\vert \Psi \left( X\right)
\right\vert ^{2}}}\right) \\
&\simeq &\frac{Ck\left( X,X\right) \frac{\hat{K}}{\left\langle
K\right\rangle }+\frac{\underline{k}\left( X\right) }{\left\langle
K\right\rangle }\hat{K}_{X}\frac{\partial }{\partial \left\vert \hat{\Psi}%
\left( \hat{K}_{1},\hat{X}_{1}\right) \right\vert ^{2}}\frac{\left\vert \hat{%
\Psi}\left( \hat{X}\right) \right\vert ^{2}}{\left\vert \Psi \left( X\right)
\right\vert ^{2}}}{\left( 1+\frac{\underline{k}\left( X\right) }{%
\left\langle K\right\rangle }\hat{K}_{X}\frac{\left\vert \hat{\Psi}\left( 
\hat{X}\right) \right\vert ^{2}}{\left\vert \Psi \left( X\right) \right\vert
^{2}}\right) ^{2}}<k
\end{eqnarray*}%
and:%
\begin{equation*}
\frac{\partial \hat{f}\left( \hat{K}^{\prime },\hat{X}^{\prime }\right) }{%
\partial \left\vert \hat{\Psi}\left( \hat{K}_{1},\hat{X}_{1}\right)
\right\vert ^{2}}<<1
\end{equation*}

\subsection*{A8.3 Gathering contribution}

Summing the contribution leads to:%
\begin{equation}
\frac{\partial }{\partial \left\vert \hat{\Psi}\left( \hat{K}_{1},\hat{X}%
_{1}\right) \right\vert ^{2}}\hat{g}\left( \hat{K},\hat{X}\right) \simeq 
\frac{\underline{\hat{k}}\left( \left\langle \hat{X}\right\rangle ,\hat{X}%
_{1}\right) }{\left\Vert \hat{\Psi}\right\Vert ^{2}}\hat{g}\left( \hat{X}%
_{1}\right) -\frac{\hat{K}_{1}}{\left\langle \hat{K}\right\rangle }\frac{%
\underline{\hat{k}}\left( \left\langle \hat{X}\right\rangle ,\left\langle 
\hat{X}\right\rangle \right) }{\left\Vert \hat{\Psi}\right\Vert ^{2}}%
\left\langle \hat{g}\right\rangle
\end{equation}

\section*{Appendix 9 Stability of the saddle point}

Starting with (\ref{Chv}) the second derivative of the action functional is:%
\begin{eqnarray}
&&\frac{\partial ^{2}S}{\partial \left\vert \hat{\Psi}\left( \hat{K},\hat{X}%
\right) \right\vert ^{2}\partial \left\vert \hat{\Psi}\left( \hat{K}_{1},%
\hat{X}_{1}\right) \right\vert ^{2}} \\
&\simeq &\frac{\partial }{\partial \left\vert \hat{\Psi}\left( \hat{K},\hat{X%
}\right) \right\vert ^{2}}\left( \frac{\hat{g}^{2}\left( \hat{K},\hat{X}%
,\Psi ,\hat{\Psi}\right) }{\sigma _{\hat{K}}^{2}}+\frac{\hat{g}\left( \left( 
\hat{K}_{1},\hat{X}_{1}\right) ,\Psi ,\hat{\Psi}\right) }{2\hat{K}}\right)
\left( 1+\delta \right)  \notag \\
&&\times \left( 1-M\left( \left( \hat{K},\hat{X}\right) ,\left( \hat{K}_{1},%
\hat{X}_{1}\right) \right) \left\vert \hat{\Psi}\left( \hat{K}_{1},\hat{X}%
_{1}\right) \right\vert ^{2}\right) ^{-1}+\hat{\mu}  \notag \\
&&+\left( \frac{\hat{g}^{2}\left( \hat{K},\hat{X},\Psi ,\hat{\Psi}\right) }{%
\sigma _{\hat{K}}^{2}}+\frac{\hat{g}\left( \left( \hat{K}_{1},\hat{X}%
_{1}\right) ,\Psi ,\hat{\Psi}\right) }{2\hat{K}}\right) \left( 1+\delta
\right)  \notag \\
&&\times \frac{\partial }{\partial \left\vert \hat{\Psi}\left( \hat{K},\hat{X%
}\right) \right\vert ^{2}}\left( 1-M\left( \left( \hat{K},\hat{X}\right)
,\left( \hat{K}_{1},\hat{X}_{1}\right) \right) \left\vert \hat{\Psi}\left( 
\hat{K}_{1},\hat{X}_{1}\right) \right\vert ^{2}\right) ^{-1}  \notag
\end{eqnarray}%
where:%
\begin{equation*}
\delta =\left\langle \frac{\partial }{\partial \left\vert \hat{\Psi}\left( 
\hat{K}_{1},\hat{X}_{1}\right) \right\vert ^{2}}\hat{g}\left( \left( \hat{K}%
^{\prime },\hat{X}^{\prime }\right) ,\Psi ,\hat{\Psi}\right) \right\rangle
\end{equation*}

In average, we can replace:%
\begin{equation*}
\left( 1-M\left( \left( \hat{K},\hat{X}\right) ,\left( \hat{K}_{1},\hat{X}%
_{1}\right) \right) \left\vert \hat{\Psi}\left( \hat{K}_{1},\hat{X}%
_{1}\right) \right\vert ^{2}\right) ^{-1}=1+\underline{\hat{k}}\left( \hat{X}%
\right)
\end{equation*}%
and:%
\begin{equation*}
\frac{\partial }{\partial \left\vert \hat{\Psi}\left( \hat{K},\hat{X}\right)
\right\vert ^{2}}\underline{\hat{k}}\left( \hat{X}\right) \simeq \hat{k}%
\left( \hat{X}\right)
\end{equation*}%
\begin{equation}
\left( \frac{\hat{K}_{\hat{X}_{1}}\hat{f}\left( \hat{K}_{\hat{X}_{1}}\frac{%
\left\Vert \hat{\Psi}\left( \hat{X}_{1}\right) \right\Vert ^{2}}{\left\Vert
\Psi \left( \hat{X}_{1}\right) \right\Vert ^{2}},\hat{X}_{1}\right) }{\sigma
_{\hat{K}}}\right) ^{2}=X\Delta \hat{M}\left( \hat{K}_{\hat{X}_{1}},\hat{X}%
_{1}\right) \frac{\underline{\hat{k}}\left( \hat{X}\right) }{1+\underline{%
\hat{k}}\left( \hat{X}\right) }Y
\end{equation}%
so that we are led to:%
\begin{equation*}
\frac{\partial }{\partial \left\vert \hat{\Psi}\left( \hat{K},\hat{X}\right)
\right\vert ^{2}}\left( \frac{\hat{K}_{\hat{X}_{1}}\hat{f}\left( \hat{K}_{%
\hat{X}_{1}}\frac{\left\Vert \hat{\Psi}\left( \hat{X}_{1}\right) \right\Vert
^{2}}{\left\Vert \Psi \left( \hat{X}_{1}\right) \right\Vert ^{2}},\hat{X}%
_{1}\right) }{\sigma _{\hat{K}}}\right) ^{2}\simeq \frac{X\underline{\hat{k}}%
\left( \hat{X}\right) }{\left( 1+\underline{\hat{k}}\left( \hat{X}\right)
\right) ^{2}}Y+X\frac{\underline{\hat{k}}\left( \hat{X}\right) }{1+%
\underline{\hat{k}}\left( \hat{X}\right) }\frac{\hat{\mu}}{1+\delta }
\end{equation*}%
Given that:%
\begin{equation*}
-X\underline{\hat{k}}\left( \hat{X}\right) \simeq 1
\end{equation*}%
we find:%
\begin{eqnarray*}
&&\frac{\partial ^{2}S}{\partial \left\vert \hat{\Psi}\left( \hat{K},\hat{X}%
\right) \right\vert ^{2}\partial \left\vert \hat{\Psi}\left( \hat{K}_{1},%
\hat{X}_{1}\right) \right\vert ^{2}} \\
&=&\hat{\mu}-\frac{Y}{\left( 1+\underline{\hat{k}}\left( \hat{X}\right)
\right) ^{2}}+\left( \frac{f^{2}\left( \hat{K},\hat{X},\Psi ,\hat{\Psi}%
\right) }{\sigma _{\hat{K}}^{2}}+\frac{f\left( \left( \hat{K}_{1},\hat{X}%
_{1}\right) ,\Psi ,\hat{\Psi}\right) }{2\hat{K}}\right) \left( 1+\delta
\right) \hat{k}\left( \hat{X}\right) >0
\end{eqnarray*}%
and th saddle point is a minimm.

\section*{Appendix 10 Field and capital for investors}

\subsection*{A10.1 Expression for the field}

Solving for $\left\vert \hat{\Psi}\left( \hat{K}_{1},\hat{X}_{1}\right)
\right\vert ^{2}$ yields:%
\begin{eqnarray}
\left\vert \hat{\Psi}\left( \hat{K}_{1},\hat{X}_{1}\right) \right\vert ^{2}
&=&\left\Vert \hat{\Psi}_{0}\left( \hat{X}_{1}\right) \right\Vert ^{2}-\hat{%
\mu}\left\{ \left( \frac{\hat{K}_{1}^{2}\hat{g}^{2}\left( \hat{X}_{1}\right) 
}{2\sigma _{\hat{K}}^{2}}+\frac{\hat{g}\left( \hat{X}_{1}\right) }{2}\right)
\right.  \label{FLd} \\
&&\left. +\left( \frac{\left\langle \hat{K}\right\rangle ^{2}\left\langle 
\hat{g}\right\rangle }{\sigma _{\hat{K}}^{2}}+\frac{1}{2}\right) \left( 
\underline{\hat{k}}\left( \left\langle \hat{X}\right\rangle ,\hat{X}%
_{1}\right) \hat{g}\left( \hat{X}_{1}\right) -\frac{\hat{K}_{1}}{%
\left\langle \hat{K}\right\rangle }\underline{\hat{k}}\left( \left\langle 
\hat{X}\right\rangle ,\left\langle \hat{X}\right\rangle \right) \left\langle 
\hat{g}\right\rangle \right) \right\}  \notag
\end{eqnarray}

\subsubsection*{A10.1.1 Finding the maximal capital}

The maximal value for $\hat{K}$, written $\hat{K}_{0}$ is found by settng $%
\left\vert \hat{\Psi}\left( \hat{K}_{1},\hat{X}_{1}\right) \right\vert
^{2}=0 $.%
\begin{eqnarray}
0 &=&\left\Vert \hat{\Psi}_{0}\left( \hat{X}_{1}\right) \right\Vert ^{2}-%
\hat{\mu}\left\{ \left( \frac{\hat{K}_{0}^{2}\hat{g}^{2}\left( \hat{X}%
_{1}\right) }{2\sigma _{\hat{K}}^{2}}+\frac{\hat{g}\left( \hat{X}_{1}\right) 
}{2}\right) \right.  \label{KR} \\
&&\left. +\left( \frac{\left\langle \hat{K}\right\rangle ^{2}\left\langle 
\hat{g}\right\rangle }{\sigma _{\hat{K}}^{2}}+\frac{1}{2}\right) \left( 
\underline{\hat{k}}\left( \left\langle \hat{X}\right\rangle ,\hat{X}%
_{1}\right) \hat{g}\left( \hat{X}_{1}\right) -\frac{\hat{K}_{0}}{%
\left\langle \hat{K}\right\rangle }\underline{\hat{k}}\left( \left\langle 
\hat{X}\right\rangle ,\left\langle \hat{X}\right\rangle \right) \left\langle 
\hat{g}\right\rangle \right) \right\}  \notag
\end{eqnarray}%
leading to:%
\begin{eqnarray}
\hat{K}_{0}^{2} &\simeq &\frac{2\sigma _{\hat{K}}^{2}}{\hat{g}^{2}\left( 
\hat{X}_{1}\right) }\left( \frac{\left\Vert \hat{\Psi}_{0}\left( \hat{X}%
_{1}\right) \right\Vert ^{2}}{\hat{\mu}}\right.  \label{KD} \\
&&\left. -\left( \frac{\hat{g}\left( \hat{X}_{1}\right) }{2}+\left( \frac{%
\left\langle \hat{K}\right\rangle ^{2}\left\langle \hat{g}\right\rangle }{%
\sigma _{\hat{K}}^{2}}+\frac{1}{2}\right) \left( \underline{\hat{k}}\left(
\left\langle \hat{X}\right\rangle ,\hat{X}_{1}\right) \hat{g}\left( \hat{X}%
_{1}\right) -\frac{\left\langle \hat{K}_{0}\right\rangle }{\left\langle \hat{%
K}\right\rangle }\underline{\hat{k}}\left( \left\langle \hat{X}\right\rangle
,\left\langle \hat{X}\right\rangle \right) \left\langle \hat{g}\right\rangle
\right) \right) \right)  \notag \\
&\simeq &2\frac{\sigma _{\hat{K}}^{2}}{\hat{g}^{2}\left( \hat{X}_{1}\right) }%
\left( \frac{\left\Vert \hat{\Psi}_{0}\left( \hat{X}_{1}\right) \right\Vert
^{2}}{\hat{\mu}}-\left( \frac{\left\langle \hat{K}\right\rangle
^{2}\left\langle \hat{g}\right\rangle }{\sigma _{\hat{K}}^{2}}+\frac{1}{2}%
\right) \left( \underline{\hat{k}}\left( \left\langle \hat{X}\right\rangle ,%
\hat{X}_{1}\right) \hat{g}\left( \hat{X}_{1}\right) -\frac{\left\langle \hat{%
K}_{0}\right\rangle }{\left\langle \hat{K}\right\rangle }\underline{\hat{k}}%
\left( \left\langle \hat{X}\right\rangle ,\left\langle \hat{X}\right\rangle
\right) \left\langle \hat{g}\right\rangle \right) \right)  \notag \\
&\simeq &2\frac{\sigma _{\hat{K}}^{2}}{\hat{g}^{2}\left( \hat{X}_{1}\right) }%
\left( \frac{\left\Vert \hat{\Psi}_{0}\left( \hat{X}_{1}\right) \right\Vert
^{2}}{\hat{\mu}}+\left( \frac{\left\langle \hat{K}\right\rangle
^{2}\left\langle \hat{g}\right\rangle ^{2}}{\sigma _{\hat{K}}^{2}}+\frac{%
\left\langle \hat{g}\right\rangle }{2}\right) \left( \left( \frac{%
\left\langle \hat{K}_{0}\right\rangle }{\left\langle \hat{K}\right\rangle }-%
\frac{\underline{\hat{k}}\left( \left\langle \hat{X}\right\rangle ,\hat{X}%
_{1}\right) }{\underline{\hat{k}}\left( \left\langle \hat{X}\right\rangle
,\left\langle \hat{X}\right\rangle \right) }\right) \underline{\hat{k}}%
\left( \left\langle \hat{X}\right\rangle ,\left\langle \hat{X}\right\rangle
\right) \right) \right)  \notag
\end{eqnarray}

\subsubsection*{A10.1.2 Expression for $\left\Vert \hat{\Psi}\left( \hat{X}%
_{1}\right) \right\Vert ^{2}$}

Integrating (\ref{FLd}) over $\hat{K}$ yields:%
\begin{eqnarray*}
\left\Vert \hat{\Psi}\left( \hat{X}_{1}\right) \right\Vert ^{2} &=&\hat{K}%
_{0}\left\Vert \hat{\Psi}_{0}\left( \hat{X}_{1}\right) \right\Vert ^{2}-\hat{%
\mu}\left\{ \left( \frac{\hat{K}_{0}^{3}\hat{g}^{2}\left( \hat{X}_{1}\right) 
}{6\sigma _{\hat{K}}^{2}}+\frac{\hat{K}_{0}\hat{g}\left( \hat{X}_{1}\right) 
}{2}\right) \right. \\
&&+\left. \hat{K}_{0}\left( \frac{\left\langle \hat{K}\right\rangle
^{2}\left\langle \hat{g}\right\rangle }{\sigma _{\hat{K}}^{2}}+\frac{1}{2}%
\right) \left( \underline{\hat{k}}\left( \left\langle \hat{X}\right\rangle ,%
\hat{X}_{1}\right) \hat{g}\left( \hat{X}_{1}\right) -\frac{\hat{K}_{0}}{%
2\left\langle \hat{K}\right\rangle }\underline{\hat{k}}\left( \left\langle 
\hat{X}\right\rangle ,\left\langle \hat{X}\right\rangle \right) \left\langle 
\hat{g}\right\rangle \right) \right\}
\end{eqnarray*}%
$\hat{K}_{0}^{2}$ satisfies (\ref{KR}) and:%
\begin{eqnarray}
\left\Vert \hat{\Psi}\left( \hat{X}_{1}\right) \right\Vert ^{2} &=&\hat{\mu}%
\hat{K}_{0}^{3}\frac{\hat{g}^{2}\left( \hat{X}_{1}\right) }{3\sigma _{\hat{K}%
}^{2}}-\hat{\mu}\frac{\hat{K}_{0}^{2}}{2\left\langle \hat{K}\right\rangle }%
\left( \frac{\left\langle \hat{K}\right\rangle ^{2}\left\langle \hat{g}%
\right\rangle }{\sigma _{\hat{K}}^{2}}+\frac{1}{2}\right) \underline{\hat{k}}%
\left( \left\langle \hat{X}\right\rangle ,\left\langle \hat{X}\right\rangle
\right) \left\langle \hat{g}\right\rangle  \label{PST} \\
&\simeq &\hat{\mu}\frac{\hat{K}_{0}^{2}}{\sigma _{\hat{K}}^{2}}\left( \frac{%
\hat{K}_{0}\hat{g}^{2}\left( \hat{X}_{1}\right) }{3}-\frac{\left\langle \hat{%
K}\right\rangle \left\langle \hat{g}\right\rangle ^{2}}{2}\underline{\hat{k}}%
\left( \left\langle \hat{X}\right\rangle ,\left\langle \hat{X}\right\rangle
\right) \right)  \notag
\end{eqnarray}

multiplip b $\hat{K}$ and integrtng:%
\begin{eqnarray*}
\hat{K}_{\hat{X}}\left\Vert \hat{\Psi}\left( \hat{X}_{1}\right) \right\Vert
^{2} &=&\frac{\hat{K}_{0}^{2}}{2}\left\Vert \hat{\Psi}_{0}\left( \hat{X}%
_{1}\right) \right\Vert ^{2}-\hat{\mu}\hat{K}_{0}\left\{ \left( \frac{\hat{K}%
_{0}^{3}\hat{g}\left( \hat{X}_{1}\right) }{8\sigma _{\hat{K}}^{2}}+\frac{%
\hat{K}_{0}\hat{g}\left( \hat{X}_{1}\right) }{4}\right) \right. \\
&&+\left. \frac{\hat{K}_{0}^{2}}{2}\left( \frac{\left\langle \hat{K}%
\right\rangle ^{2}\left\langle \hat{g}\right\rangle }{\sigma _{\hat{K}}^{2}}+%
\frac{1}{2}\right) \left( \underline{\hat{k}}\left( \left\langle \hat{X}%
\right\rangle ,\hat{X}_{1}\right) \hat{g}\left( \hat{X}_{1}\right) -\frac{2%
\hat{K}_{0}}{3\left\langle \hat{K}\right\rangle }\underline{\hat{k}}\left(
\left\langle \hat{X}\right\rangle ,\left\langle \hat{X}\right\rangle \right)
\left\langle \hat{g}\right\rangle \right) \right\}
\end{eqnarray*}%
since $\hat{K}_{0}$ satisfies (\ref{KR}):%
\begin{eqnarray}
\hat{K}_{\hat{X}}\left\Vert \hat{\Psi}\left( \hat{X}_{1}\right) \right\Vert
^{2} &=&\hat{\mu}\frac{\hat{K}_{0}^{4}\hat{g}^{2}\left( \hat{X}_{1}\right) }{%
8\sigma _{\hat{K}}^{2}}-\hat{\mu}\frac{\hat{K}_{0}^{2}}{6\left\langle \hat{K}%
\right\rangle }\left( \frac{\left\langle \hat{K}\right\rangle
^{2}\left\langle \hat{g}\right\rangle }{\sigma _{\hat{K}}^{2}}+\frac{1}{2}%
\right) \underline{\hat{k}}\left( \left\langle \hat{X}\right\rangle
,\left\langle \hat{X}\right\rangle \right) \left\langle \hat{g}\right\rangle
\label{PK} \\
&\simeq &\hat{\mu}\frac{\hat{K}_{0}^{3}}{2\sigma _{\hat{K}}^{2}}\left( \frac{%
\hat{K}_{0}\hat{g}^{2}\left( \hat{X}_{1}\right) }{4}-\frac{\left\langle \hat{%
K}\right\rangle \left\langle \hat{g}\right\rangle ^{2}}{3}\underline{\hat{k}}%
\left( \left\langle \hat{X}\right\rangle ,\left\langle \hat{X}\right\rangle
\right) \right)  \notag \\
&=&\hat{\mu}\frac{\hat{K}_{0}^{4}}{2\sigma _{\hat{K}}^{2}}\left( \frac{\hat{g%
}^{2}\left( \hat{X}_{1}\right) }{4}-\frac{\left\langle \hat{K}\right\rangle
\left\langle \hat{g}\right\rangle ^{2}}{3\hat{K}_{0}}\underline{\hat{k}}%
\left( \left\langle \hat{X}\right\rangle ,\left\langle \hat{X}\right\rangle
\right) \right)  \notag
\end{eqnarray}

\subsubsection*{A10.1.3 Averages and ratio $\frac{\left\langle \hat{K}%
\right\rangle }{\left\langle \hat{K}_{0}\right\rangle }$}

Averages for (\ref{PST}) and (\ref{PK}) lead to find the norm $\left\Vert 
\hat{\Psi}\right\Vert ^{2}$ of the field, computing the total number of
investrs and the total disposable capital for investors $\left\langle \hat{K}%
\right\rangle \left\Vert \hat{\Psi}\right\Vert ^{2}$:%
\begin{equation}
\left\Vert \hat{\Psi}\right\Vert ^{2}\simeq \hat{\mu}V\frac{\left\langle 
\hat{K}_{0}\right\rangle ^{3}}{\sigma _{\hat{K}}^{2}}\left( \frac{1}{3}-%
\frac{\left\langle \hat{K}\right\rangle }{2\left\langle \hat{K}%
_{0}\right\rangle }\hat{k}\left( \left\langle \hat{X}\right\rangle
,\left\langle \hat{X}\right\rangle \right) \right) \left\langle \hat{g}%
\right\rangle ^{2}
\end{equation}%
\begin{equation*}
\left\langle \hat{K}\right\rangle \left\Vert \hat{\Psi}\right\Vert ^{2}=\hat{%
\mu}V\frac{\left\langle \hat{K}_{0}\right\rangle ^{4}}{2\sigma _{\hat{K}}^{2}%
}\left( \frac{1}{4}-\frac{\left\langle \hat{K}\right\rangle }{3\left\langle 
\hat{K}_{0}\right\rangle }\hat{k}\left( \left\langle \hat{X}\right\rangle
,\left\langle \hat{X}\right\rangle \right) \right) \left\langle \hat{g}%
\right\rangle ^{2}
\end{equation*}%
where $V$ is the volume of the sectors space. These formula allow to derive
the ratio $\frac{\left\langle \hat{K}\right\rangle }{\left\langle \hat{K}%
_{0}\right\rangle }$:%
\begin{equation}
\frac{\left\langle \hat{K}\right\rangle }{\left\langle \hat{K}%
_{0}\right\rangle }=\frac{1}{2}\frac{\frac{1}{4}-\frac{\left\langle \hat{K}%
\right\rangle }{3\left\langle \hat{K}_{0}\right\rangle }\hat{k}\left(
\left\langle \hat{X}\right\rangle ,\left\langle \hat{X}\right\rangle \right) 
}{\frac{1}{3}-\frac{\left\langle \hat{K}\right\rangle }{2\left\langle \hat{K}%
_{0}\right\rangle }\hat{k}\left( \left\langle \hat{X}\right\rangle
,\left\langle \hat{X}\right\rangle \right) }  \label{RTS}
\end{equation}%
with solution:%
\begin{equation}
\frac{\left\langle \hat{K}\right\rangle }{\left\langle \hat{K}%
_{0}\right\rangle }=\frac{2+\hat{k}-\sqrt{\left( 2+\hat{k}\right) ^{2}-\hat{k%
}}}{6\underline{\hat{k}}}=r  \label{Rts}
\end{equation}%
where:%
\begin{equation*}
\hat{k}=\hat{k}\left( \left\langle \hat{X}\right\rangle ,\left\langle \hat{X}%
\right\rangle \right)
\end{equation*}%
This enables to obtain the previous quantities in terms of $\hat{k}$: 
\begin{equation*}
\left\langle \hat{K}_{0}\right\rangle ^{2}=2\frac{\sigma _{\hat{K}}^{2}}{%
\left\langle \hat{g}\right\rangle ^{2}}\left( \frac{\left\Vert \hat{\Psi}%
_{0}\right\Vert ^{2}}{\hat{\mu}}+\left( \frac{r^{2}\left\langle \hat{K}%
_{0}\right\rangle ^{2}\left\langle \hat{g}\right\rangle ^{2}}{\sigma _{\hat{K%
}}^{2}}+\frac{\left\langle \hat{g}\right\rangle }{2}\right) \left( \frac{1-r%
}{r}\right) \hat{k}\right)
\end{equation*}%
that is;%
\begin{equation*}
\left\langle \hat{K}_{0}\right\rangle ^{2}=2\frac{\sigma _{\hat{K}%
}^{2}\left( \frac{\left\Vert \hat{\Psi}_{0}\right\Vert ^{2}}{\hat{\mu}}+%
\frac{\left\langle \hat{g}\right\rangle }{2}\left( \frac{1-r}{r}\right) \hat{%
k}\right) }{\left\langle \hat{g}\right\rangle ^{2}\left( 1-2r\left(
1-r\right) \hat{k}\right) }
\end{equation*}%
and for $\left\Vert \hat{\Psi}\right\Vert ^{2}$ and $\left\langle \hat{K}%
\right\rangle \left\Vert \hat{\Psi}\right\Vert ^{2}$:%
\begin{equation}
\left\Vert \hat{\Psi}\right\Vert ^{2}\simeq \frac{\hat{\mu}V}{3\sigma _{\hat{%
K}}^{2}}\left( 2\frac{\sigma _{\hat{K}}^{2}\left( \frac{\left\Vert \hat{\Psi}%
_{0}\right\Vert ^{2}}{\hat{\mu}}+\frac{\left\langle \hat{g}\right\rangle }{2}%
\left( \frac{1-r}{r}\right) \hat{k}\right) }{\left\langle \hat{g}%
\right\rangle ^{2}\left( 1-2r\left( 1-r\right) \hat{k}\right) }\right) ^{%
\frac{3}{2}}\left( 1-\frac{2+\hat{k}-\sqrt{\left( 2+\hat{k}\right) ^{2}-\hat{%
k}}}{4\hat{k}}\hat{k}\right) \left\langle \hat{g}\right\rangle ^{2}
\label{VRf}
\end{equation}%
\begin{equation}
\left\langle \hat{K}\right\rangle \left\Vert \hat{\Psi}\right\Vert ^{2}=%
\frac{\hat{\mu}V\sigma _{\hat{K}}^{2}}{2\left\langle \hat{g}\right\rangle
^{2}}\left( \frac{\frac{\left\Vert \hat{\Psi}_{0}\right\Vert ^{2}}{\hat{\mu}}%
+\frac{\left\langle \hat{g}\right\rangle }{2}\left( \frac{1-r}{r}\right) 
\hat{k}}{1-2r\left( 1-r\right) \hat{k}}\right) ^{2}\left( 1-2\frac{2+\hat{k}-%
\sqrt{\left( 2+\hat{k}\right) ^{2}-\hat{k}}}{9}\right)  \label{VRs}
\end{equation}%
The average capital per investor is obtained by the ratio of (\ref{VRs}) and
(\ref{VRf}): 
\begin{equation}
\left\langle \hat{K}\right\rangle =\frac{3}{4\left\langle \hat{g}%
\right\rangle }\sqrt{\frac{\sigma _{\hat{K}}^{2}}{2\hat{\mu}}}\frac{1-2\frac{%
2+\hat{k}-\sqrt{\left( 2+\hat{k}\right) ^{2}-\hat{k}}}{9}}{1-\frac{2+\hat{k}-%
\sqrt{\left( 2+\hat{k}\right) ^{2}-\hat{k}}}{4\hat{k}}}\sqrt{\frac{%
\left\Vert \hat{\Psi}_{0}\right\Vert ^{2}+\hat{\mu}\frac{\left\langle \hat{g}%
\right\rangle }{2}\left( \frac{1-r}{r}\right) \hat{k}}{1-2r\left( 1-r\right) 
\hat{k}}}  \label{VTR}
\end{equation}

\subsubsection*{A10.1.4 Expression for $\hat{K}_{0}^{2}$ $\allowbreak $}

Having obtained the average quantities, we can come back to the quantities
per sector. We find $\hat{K}_{0}^{2}$: 
\begin{eqnarray*}
\hat{K}_{0}^{2} &=&2\frac{\sigma _{\hat{K}}^{2}}{\hat{g}^{2}\left( \hat{X}%
_{1}\right) }\left( \frac{\left\Vert \hat{\Psi}_{0}\left( \hat{X}_{1}\right)
\right\Vert ^{2}}{\hat{\mu}}-\left( \frac{\left\langle \hat{K}\right\rangle
^{2}\left\langle \hat{g}\right\rangle ^{2}}{\sigma _{\hat{K}}^{2}}+\frac{%
\left\langle \hat{g}\right\rangle }{2}\right) \left( \frac{\hat{k}\left(
\left\langle \hat{X}\right\rangle ,\hat{X}_{1}\right) }{\hat{k}\left(
\left\langle \hat{X}\right\rangle ,\left\langle \hat{X}\right\rangle \right) 
}-\frac{6\hat{k}}{2+\hat{k}-\sqrt{\left( 2+\hat{k}\right) ^{2}-\hat{k}}}%
\right) \hat{k}\right) \\
&=&2\frac{\sigma _{\hat{K}}^{2}}{\hat{\mu}\hat{g}^{2}\left( \hat{X}%
_{1}\right) }\left( \left\Vert \hat{\Psi}_{0}\left( \hat{X}_{1}\right)
\right\Vert ^{2}-\hat{\mu}D\left( \hat{X}_{1}\right) \right)
\end{eqnarray*}%
where:%
\begin{equation*}
D\left( \hat{X}_{1}\right) =\left( \frac{\left\langle \hat{K}\right\rangle
^{2}\left\langle \hat{g}\right\rangle ^{2}}{\sigma _{\hat{K}}^{2}}+\frac{%
\left\langle \hat{g}\right\rangle }{2}\right) \left( \frac{\hat{k}\left(
\left\langle \hat{X}\right\rangle ,\hat{X}_{1}\right) }{\hat{k}\left(
\left\langle \hat{X}\right\rangle ,\left\langle \hat{X}\right\rangle \right) 
}-\frac{6\hat{k}}{2+\hat{k}-\sqrt{\left( 2+\hat{k}\right) ^{2}-\hat{k}}}%
\right) \hat{k}
\end{equation*}

\subsubsection*{A10.1.5 Expression for field and capital}

Having obtained the expression for the field $\left\Vert \hat{\Psi}\left( 
\hat{X}_{1}\right) \right\Vert ^{2}$ and the amount $\hat{K}\left[ \hat{X}%
_{1}\right] $ of financial capital in sector $\hat{X}_{1}$ are given by: 
\begin{eqnarray}
\left\Vert \hat{\Psi}\left( \hat{X}_{1}\right) \right\Vert ^{2} &\simeq &%
\hat{\mu}\frac{\hat{K}_{0}^{3}}{\sigma _{\hat{K}}^{2}}\left( \frac{\hat{g}%
^{2}\left( \hat{X}_{1}\right) }{3}-\frac{\left\langle \hat{K}\right\rangle
\left\langle \hat{g}\right\rangle ^{2}}{2\left\langle \hat{K}%
_{0}\right\rangle }\underline{\hat{k}}\left( \left\langle \hat{X}%
\right\rangle ,\left\langle \hat{X}\right\rangle \right) \right) \\
&=&\frac{\hat{\mu}}{\sigma _{\hat{K}}^{2}}\left( 2\frac{\sigma _{\hat{K}}^{2}%
}{\hat{g}^{2}\left( \hat{X}_{1}\right) }\left( \frac{\left\Vert \hat{\Psi}%
_{0}\left( \hat{X}_{1}\right) \right\Vert ^{2}}{\hat{\mu}}-D\left( \hat{X}%
_{1}\right) \right) \right) ^{\frac{3}{2}}\left( \frac{\hat{g}^{2}\left( 
\hat{X}_{1}\right) }{3}-\frac{\left\langle \hat{K}\right\rangle \left\langle 
\hat{g}\right\rangle ^{2}}{2\left\langle \hat{K}_{0}\right\rangle }\hat{k}%
\right)  \notag
\end{eqnarray}%
\begin{eqnarray}
\hat{K}\left[ \hat{X}_{1}\right] &=&\hat{K}_{\hat{X}}\left\Vert \hat{\Psi}%
\left( \hat{X}_{1}\right) \right\Vert ^{2}\simeq \hat{\mu}\frac{\hat{K}%
_{0}^{4}}{2\sigma _{\hat{K}}^{2}}\left( \frac{\hat{g}^{2}\left( \hat{X}%
_{1}\right) }{4}-\frac{\left\langle \hat{K}\right\rangle \left\langle \hat{g}%
\right\rangle ^{2}}{3\left\langle \hat{K}_{0}\right\rangle }\hat{k}\left(
\left\langle \hat{X}\right\rangle ,\left\langle \hat{X}\right\rangle \right)
\right) \\
&=&\frac{\hat{\mu}}{2\sigma _{\hat{K}}^{2}}\left( 2\frac{\sigma _{\hat{K}%
}^{2}}{\hat{g}^{2}\left( \hat{X}_{1}\right) }\left( \frac{\left\Vert \hat{%
\Psi}_{0}\left( \hat{X}_{1}\right) \right\Vert ^{2}}{\hat{\mu}}-D\left( \hat{%
X}_{1}\right) \right) \right) ^{2}\left( \frac{\hat{g}^{2}\left( \hat{X}%
_{1}\right) }{4}-\frac{r\left\langle \hat{g}\right\rangle ^{2}}{3}\underline{%
\hat{k}}\right)  \notag
\end{eqnarray}%
and the average capital per investor in sector $\hat{X}_{1}$ is: 
\begin{equation}
\hat{K}_{\hat{X}_{1}}=\frac{\hat{K}\left[ \hat{X}_{1}\right] }{\left\Vert 
\hat{\Psi}\left( \hat{X}_{1}\right) \right\Vert ^{2}}=\frac{\sqrt{2\frac{%
\sigma _{\hat{K}}^{2}}{\hat{g}^{2}\left( \hat{X}_{1}\right) }\left( \frac{%
\left\Vert \hat{\Psi}_{0}\left( \hat{X}_{1}\right) \right\Vert ^{2}}{\hat{\mu%
}}-D\left( \hat{X}_{1}\right) \right) }\left( \frac{\hat{g}^{2}\left( \hat{X}%
_{1}\right) }{4}-\frac{r\left\langle \hat{g}\right\rangle ^{2}}{3}\hat{k}%
\right) }{2\left( \frac{\hat{g}^{2}\left( \hat{X}_{1}\right) }{3}-\frac{%
\left\langle \hat{K}\right\rangle \left\langle \hat{g}\right\rangle ^{2}}{%
2\left\langle \hat{K}_{0}\right\rangle }\hat{k}\right) }
\end{equation}%
\bigskip

\section*{Appendix 11 Return equation for investors}

We present the details for the derivation of the expanded form of the return
equation (\ref{RTN}):%
\begin{equation}
\int \left( \frac{\Delta \left( \hat{X},\hat{X}^{\prime }\right) }{1+%
\underline{\hat{k}}_{2}\left( \hat{X}\right) }-\frac{\hat{k}_{1}\left( \hat{X%
}^{\prime },\hat{X}\right) \hat{K}^{\prime }\left\vert \hat{\Psi}\left( \hat{%
K}^{\prime },\hat{X}^{\prime }\right) \right\vert ^{2}}{1+\underline{\hat{k}}%
_{2}\left( \hat{X}^{\prime }\right) }\right) \left( 1-M\right) \left( g-\bar{%
r}^{\prime }\right) d\hat{X}^{\prime }=\frac{k_{1}\left( X\right) }{1+%
\underline{k}\left( X\right) }f_{1}^{\prime }
\end{equation}

\subsection*{A11.1 Computation of left side of (\protect\ref{RTN})}

Define:%
\begin{equation*}
A=\int \left( \frac{\Delta \left( \hat{X},\hat{X}^{\prime }\right) }{1+%
\underline{\hat{k}}_{2}\left( \hat{X}\right) }-\frac{\hat{k}_{1}\left( \hat{X%
}^{\prime },\hat{X}\right) \hat{K}^{\prime }\left\vert \hat{\Psi}\left( \hat{%
K}^{\prime },\hat{X}^{\prime }\right) \right\vert ^{2}}{1+\underline{\hat{k}}%
_{2}\left( \hat{X}^{\prime }\right) }\right) \left( 1-M\right)
\end{equation*}%
it expands as:%
\begin{eqnarray*}
&&\frac{1}{1+\underline{\hat{k}}_{2}\left( \hat{X}\right) }\left( 1-\frac{%
\hat{k}\left( \hat{X},\hat{X}^{\prime }\right) \hat{K}_{X}}{1+\hat{k}\left( 
\hat{X},\hat{X}^{\prime }\right) \hat{K}_{X^{\prime }}\left\Vert \hat{\Psi}%
\left( \hat{X}^{\prime }\right) \right\Vert ^{2}}\left\Vert \hat{\Psi}\left( 
\hat{X}^{\prime }\right) \right\Vert ^{2}\right) \\
&&-\hat{S}_{1}\left( \hat{X}^{\prime },\hat{X}\right) \frac{1}{1+\underline{%
\hat{k}}_{2}\left( \hat{X}^{\prime }\right) }+\hat{S}_{1}\left( \hat{X}%
^{\prime },\hat{X}\right) \frac{1}{1+\underline{\hat{k}}_{2}\left( \hat{X}%
^{\prime }\right) }\frac{\hat{k}\left( X^{\prime },X^{\prime \prime }\right) 
\hat{K}_{X^{\prime }}}{1+\hat{k}\left( X^{\prime },X^{\prime \prime }\right) 
\hat{K}_{X^{\prime \prime }}\left\Vert \hat{\Psi}\left( \hat{X}^{\prime
\prime }\right) \right\Vert ^{2}}\left\Vert \hat{\Psi}\left( \hat{X}^{\prime
\prime }\right) \right\Vert ^{2}
\end{eqnarray*}%
Then, replacing:%
\begin{equation*}
1+\hat{k}\left( \hat{X},\hat{X}^{\prime }\right) \hat{K}_{X^{\prime
}}\left\Vert \hat{\Psi}\left( \hat{X}^{\prime }\right) \right\Vert
^{2}\rightarrow 1+\underline{\hat{k}}\left( \hat{X}\right)
\end{equation*}%
yields:%
\begin{eqnarray*}
A &=&\frac{1}{1+\underline{\hat{k}}_{2}\left( \hat{X}\right) }\left( 1-\frac{%
\hat{k}\left( \hat{X},\hat{X}^{\prime }\right) \hat{K}_{X}}{1+\underline{%
\hat{k}}\left( \hat{X}\right) }\left\Vert \hat{\Psi}\left( \hat{X}^{\prime
}\right) \right\Vert ^{2}\right) \\
&&-\frac{\hat{k}_{1}\left( \hat{X}^{\prime },\hat{X}\right) \hat{K}%
_{X^{\prime }}\left\Vert \hat{\Psi}\left( \hat{X}^{\prime }\right)
\right\Vert ^{2}}{1+\underline{\hat{k}}\left( \hat{X}\right) }\frac{1}{1+%
\underline{\hat{k}}_{2}\left( \hat{X}^{\prime }\right) } \\
&&+\frac{\hat{k}_{1}\left( \hat{X}^{\prime },\hat{X}\right) \hat{K}%
_{X^{\prime }}\left\Vert \hat{\Psi}\left( \hat{X}^{\prime }\right)
\right\Vert ^{2}}{1+\underline{\hat{k}}\left( \hat{X}\right) }\frac{1}{1+%
\underline{\hat{k}}_{2}\left( \hat{X}^{\prime }\right) }\frac{\hat{k}\left( 
\hat{X}^{\prime },\hat{X}^{\prime \prime }\right) \hat{K}_{X^{\prime }}}{1+%
\underline{\hat{k}}\left( \hat{X}^{\prime }\right) }\left\Vert \hat{\Psi}%
\left( \hat{X}^{\prime \prime }\right) \right\Vert ^{2}
\end{eqnarray*}%
wh:%
\begin{equation*}
1+\underline{\hat{k}}_{2}\left( \hat{X}\right) \simeq 1+\frac{\hat{k}%
_{2}\left( \hat{X},\left\langle \hat{X}\right\rangle \right) }{1-\hat{k}%
\left( \left\langle \hat{X}\right\rangle ,\left\langle \hat{X}\right\rangle
\right) }=1+\frac{\hat{k}_{2}\left( \hat{X},\left\langle \hat{X}%
\right\rangle \right) }{1-\left\langle \hat{k}\right\rangle }=1+\hat{k}%
_{2}^{n}\left( \hat{X},\left\langle \hat{X}\right\rangle \right)
\end{equation*}%
\begin{equation*}
1+\underline{\hat{k}}\left( \hat{X}\right) =1+\hat{k}\left( \hat{X}%
,\left\langle \hat{X}\right\rangle \right) -\hat{k}\left( \left\langle \hat{X%
}\right\rangle ,\left\langle \hat{X}\right\rangle \right)
\end{equation*}%
The last term, can be approximated:%
\begin{eqnarray*}
&&\frac{\hat{k}_{1}\left( \hat{X}^{\prime },\hat{X}\right) \hat{K}%
_{X}\left\Vert \hat{\Psi}\left( \hat{X}^{\prime }\right) \right\Vert ^{2}}{1+%
\hat{k}\left( \hat{X},\hat{X}^{\prime }\right) \hat{K}_{X^{\prime }}}\frac{1%
}{1+\underline{\hat{k}}_{2}\left( \hat{X}^{\prime }\right) }\frac{\hat{k}%
\left( X^{\prime },X^{\prime \prime }\right) \hat{K}_{X^{\prime }}}{1+%
\underline{\hat{k}}\left( \hat{X}^{\prime }\right) }\left\Vert \hat{\Psi}%
\left( \hat{X}^{\prime \prime }\right) \right\Vert ^{2} \\
&\simeq &\frac{\hat{k}_{1}\left( \left\langle X\right\rangle ,\hat{X}\right) 
\hat{K}_{X}\left\Vert \hat{\Psi}\left( \left\langle \hat{X}\right\rangle
\right) \right\Vert ^{2}}{1+\underline{\hat{k}}\left( \hat{X}\right) }\frac{1%
}{1+\underline{\hat{k}}_{2}\left( \left\langle X\right\rangle \right) }\frac{%
\hat{k}\left( \left\langle \hat{X}\right\rangle ,\hat{X}^{\prime }\right)
\left\langle K\right\rangle }{1+\underline{\hat{k}}\left( \left\langle \hat{X%
}\right\rangle \right) }\left\Vert \hat{\Psi}\left( \hat{X}^{\prime }\right)
\right\Vert ^{2} \\
&=&\frac{\hat{k}_{1}\left( \left\langle X\right\rangle ,\hat{X}\right) \hat{K%
}_{X}\left\Vert \hat{\Psi}\left( \left\langle \hat{X}\right\rangle \right)
\right\Vert ^{2}}{1+\underline{\hat{k}}\left( \hat{X}\right) }\frac{1}{1+%
\underline{\hat{k}}_{2}\left( \left\langle X\right\rangle \right) }\hat{k}%
\left( \left\langle \hat{X}\right\rangle ,\hat{X}^{\prime }\right)
\left\langle K\right\rangle \left\Vert \hat{\Psi}\left( \hat{X}^{\prime
}\right) \right\Vert ^{2}
\end{eqnarray*}%
and rewriting $A$ as:%
\begin{equation*}
A=\frac{1}{1+\underline{\hat{k}}_{2}\left( \hat{X}\right) }-\hat{S}%
_{1}^{E}\left( \hat{X},\hat{X}^{\prime }\right)
\end{equation*}%
we find:

\begin{eqnarray*}
&&-\hat{S}_{1}^{E}\left( \hat{X},\hat{X}^{\prime }\right) \\
&\rightarrow &-\left( \frac{\hat{k}\left( \hat{X},\hat{X}^{\prime }\right) 
\hat{K}_{X}}{1+\underline{\hat{k}}_{2}\left( \hat{X}\right) }+\frac{\hat{k}%
_{1}\left( \hat{X}^{\prime },\hat{X}\right) \hat{K}_{X^{\prime }}}{1+%
\underline{\hat{k}}_{2}\left( \hat{X}^{\prime }\right) }\right) \frac{%
\left\Vert \hat{\Psi}\left( \hat{X}^{\prime }\right) \right\Vert ^{2}}{1+%
\hat{k}\left( \hat{X}\right) } \\
&&+\frac{\hat{k}_{1}\left( \left\langle X\right\rangle ,\hat{X}\right)
\left\langle K\right\rangle \left\Vert \hat{\Psi}\left( \left\langle \hat{X}%
\right\rangle \right) \right\Vert ^{2}}{1+\hat{k}\left( \hat{X}\right) }%
\frac{\hat{k}\left( \left\langle \hat{X}\right\rangle ,\hat{X}^{\prime
}\right) \left\langle K\right\rangle }{1+\underline{\hat{k}}_{2}\left(
\left\langle X\right\rangle \right) }\left\Vert \hat{\Psi}\left( \hat{X}%
^{\prime }\right) \right\Vert ^{2} \\
&=&-\left( \left( \frac{\hat{k}\left( \hat{X},\hat{X}^{\prime }\right) \hat{K%
}_{X}}{1+\underline{\hat{k}}_{2}\left( \hat{X}\right) }+\frac{\hat{k}%
_{1}\left( \hat{X}^{\prime },\hat{X}\right) \hat{K}_{X^{\prime }}}{1+%
\underline{\hat{k}}_{2}\left( \hat{X}^{\prime }\right) }\right) -\hat{k}%
_{1}\left( \left\langle X\right\rangle ,\hat{X}\right) \left\langle
K\right\rangle \left\Vert \hat{\Psi}\left( \left\langle \hat{X}\right\rangle
\right) \right\Vert ^{2}\frac{\hat{k}\left( \left\langle \hat{X}%
\right\rangle ,\hat{X}^{\prime }\right) \left\langle K\right\rangle }{1+%
\underline{\hat{k}}_{2}\left( \left\langle X\right\rangle \right) }\right) 
\frac{\left\Vert \hat{\Psi}\left( \hat{X}^{\prime }\right) \right\Vert ^{2}}{%
1+\hat{k}\left( \hat{X}\right) } \\
&=&-\left( \frac{\hat{k}\left( \hat{X},\hat{X}^{\prime }\right) \hat{K}_{X}-%
\hat{k}_{1}\left( \left\langle X\right\rangle ,\hat{X}\right) \left\langle
K\right\rangle \left\Vert \hat{\Psi}\left( \left\langle \hat{X}\right\rangle
\right) \right\Vert ^{2}\hat{k}\left( \left\langle \hat{X}\right\rangle ,%
\hat{X}^{\prime }\right) \left\langle K\right\rangle }{1+\underline{\hat{k}}%
_{2}\left( \hat{X}\right) }+\frac{\hat{k}_{1}\left( \hat{X}^{\prime },\hat{X}%
\right) \hat{K}_{X^{\prime }}}{1+\underline{\hat{k}}_{2}\left( \hat{X}%
^{\prime }\right) }\right) \frac{\left\Vert \hat{\Psi}\left( \hat{X}^{\prime
}\right) \right\Vert ^{2}}{1+\hat{k}\left( \hat{X}\right) }
\end{eqnarray*}
In average, we sum over $\hat{X}^{\prime }$ and: 
\begin{equation*}
-\hat{S}_{1}^{E}\left( \hat{X},\hat{X}^{\prime }\right) \rightarrow -\frac{1%
}{1+\hat{k}\left( \hat{X}\right) }\left( \frac{\hat{k}\left( \hat{X}\right) 
\frac{\hat{K}_{X}}{\left\langle K\right\rangle }-\hat{k}_{1E}\left( \hat{X}%
\right) \hat{k}_{E}\left( \hat{X}^{\prime }\right) }{1+\underline{\hat{k}}%
_{2}\left( \hat{X}\right) }+\frac{\hat{k}_{1E}\left( \hat{X}\right) }{1+%
\underline{\hat{k}}_{2}\left( \hat{X}^{\prime }\right) }\right)
\end{equation*}%
then, we define:%
\begin{equation*}
\underline{\hat{k}}_{1}\left( \left\langle X\right\rangle ,\hat{X}\right) =%
\hat{k}_{1}\left( \left\langle X\right\rangle ,\hat{X}\right) \left\Vert 
\hat{\Psi}\left( \left\langle \hat{X}\right\rangle \right) \right\Vert
^{2}\left\langle K\right\rangle
\end{equation*}%
and we can rewrite:%
\begin{eqnarray}
&&\int \left( \frac{\Delta \left( \hat{X},\hat{X}^{\prime }\right) }{1+%
\underline{\hat{k}}_{2}\left( \hat{X}\right) }-\frac{\hat{k}_{1}\left( \hat{X%
}^{\prime },\hat{X}\right) \hat{K}^{\prime }\left\vert \hat{\Psi}\left( \hat{%
K}^{\prime },\hat{X}^{\prime }\right) \right\vert ^{2}}{1+\underline{\hat{k}}%
_{2}\left( \hat{X}^{\prime }\right) }\right) \left( 1-M\right)  \label{LHD}
\\
&=&\frac{\Delta \left( \hat{X},\hat{X}^{\prime }\right) }{1+\underline{\hat{k%
}}_{2}\left( \hat{X}\right) }-\left( \frac{\hat{k}\left( \hat{X},\hat{X}%
^{\prime }\right) -\underline{\hat{k}}_{1}\left( \left\langle X\right\rangle
,\hat{X}\right) \hat{k}\left( \left\langle \hat{X}\right\rangle ,\hat{X}%
^{\prime }\right) }{1+\underline{\hat{k}}_{2}\left( \hat{X}\right) }+\frac{%
\hat{k}_{1}\left( \hat{X}^{\prime },\hat{X}\right) }{1+\underline{\hat{k}}%
_{2}\left( \hat{X}^{\prime }\right) }\right) \frac{\hat{K}_{X}\left\Vert 
\hat{\Psi}\left( \hat{X}^{\prime }\right) \right\Vert ^{2}}{1+\hat{k}\left( 
\hat{X}\right) }  \notag \\
&\equiv &\frac{\Delta \left( \hat{X},\hat{X}^{\prime }\right) }{1+\underline{%
\hat{k}}_{2}\left( \hat{X}\right) }-\hat{S}_{1}^{E}\left( \hat{X}^{\prime },%
\hat{X}\right)  \notag
\end{eqnarray}

The differnts contributions to $\hat{S}_{1}^{E}\left( \hat{X},\hat{X}%
^{\prime }\right) $ are interpreted in the text.

\subsection*{A11.2 Computation of the RHS of (\protect\ref{RTN})}

The right hand side of (\ref{PN}) is obtained by noting that:

\begin{equation*}
S_{1}\left( \hat{X}^{\prime },\hat{X}^{\prime }\right) =\frac{k_{1}\left( 
\hat{X}^{\prime },\hat{X}^{\prime }\right) }{1+\underline{k}\left( X\right) }
\end{equation*}%
and using that:%
\begin{equation*}
\left\vert \Psi \left( X\right) \right\vert ^{2}K_{X}\frac{f_{1}^{\prime
}\left( X\right) -\bar{r}}{1+\underline{k}_{2}\left( X\right) }=\left\vert
\Psi \left( X\right) \right\vert ^{2}\left( \left( f_{1}\left( X\right) -%
\bar{r}\right) K_{X}-\bar{C}\left( X\right) \right)
\end{equation*}%
so that:%
\begin{eqnarray*}
&&S_{1}\left( \hat{X}^{\prime },\hat{X}^{\prime }\right) \frac{1-S\left( 
\hat{X}^{\prime }\right) }{1-S_{1}\left( \hat{X}^{\prime }\right) }\left(
f_{1}^{\prime }\left( X^{\prime }\right) -\bar{r}\right) \\
&=&S_{1}\left( \hat{X}^{\prime },\hat{X}^{\prime }\right) \left\vert \Psi
\left( X^{\prime }\right) \right\vert ^{2}K_{X^{\prime }}\frac{f_{1}^{\prime
}\left( X^{\prime }\right) -\bar{r}}{1+\underline{k}_{2}\left( X^{\prime
}\right) } \\
&=&\frac{k_{1}\left( \hat{X}^{\prime },\hat{X}^{\prime }\right) }{1+%
\underline{k}\left( X\right) }\left\vert \Psi \left( X\right) \right\vert
^{2}\left( \left( f_{1}\left( X\right) -\bar{r}\right) K_{X}-\bar{C}\left(
X\right) \right)
\end{eqnarray*}

We write:%
\begin{eqnarray*}
\underline{k}_{2}\left( \hat{X}\right) &=&\beta \underline{k}\left( \hat{X}%
\right) \\
\underline{k}_{2}\left( \hat{X}\right) &=&\left( 1-\beta \right) \underline{k%
}\left( \hat{X}\right)
\end{eqnarray*}%
and assume th:%
\begin{equation*}
\underline{k}\left( \hat{X}\right) >>1
\end{equation*}%
Under this assumtion%
\begin{equation*}
\frac{\underline{k}_{1}\left( \hat{X}\right) }{\left( 1+\underline{k}\left( 
\hat{X}\right) \right) }\rightarrow 1-\beta
\end{equation*}%
and the overall return for firms:%
\begin{equation*}
\left\vert \Psi \left( X\right) \right\vert ^{2}\left( \left( f_{1}\left(
X\right) -\bar{r}\right) K_{X}-\bar{C}\left( X\right) \right)
\end{equation*}%
rewrites:%
\begin{equation*}
\frac{\left( \beta C+X^{\left( e\right) }\right) \left( \frac{2}{3}X^{2}+%
\frac{1}{3}\left( X-\beta C\right) \beta C\right) \epsilon }{\sigma _{\hat{K}%
}^{2}\left( f_{1}\left( X\right) +\beta \underline{k}\left( X\right) \left(
\left( f_{1}\left( X\right) -\bar{r}\right) \right) \right) }\left( \frac{%
\left( f_{1}\left( X\right) -\bar{r}\right) }{4\left( f_{1}\left( X\right)
+\beta \underline{k}\left( X\right) \left( \left( f_{1}\left( X\right) -\bar{%
r}\right) \right) \right) }\frac{\left( 3X^{\left( e\right) }-\beta C\right)
\left( \beta C+X^{\left( e\right) }\right) }{2X^{\left( e\right) }-\beta C}-%
\frac{C}{1+\underline{k}\left( X\right) }\right)
\end{equation*}%
or equivalently:%
\begin{equation*}
\frac{1}{3}\frac{\left( X+C\beta \right) ^{2}\epsilon }{\sigma _{\hat{K}%
}^{2}\left( f_{1}\left( X\right) +\beta \underline{k}\left( X\right) \left(
f_{1}\left( X\right) -\bar{r}\right) \right) }\left( \frac{\left(
f_{1}\left( X\right) -\bar{r}\right) \left( 3X^{\left( e\right) }-\beta
C\right) \left( \beta C+X^{\left( e\right) }\right) }{4\left( f_{1}\left(
X\right) +\beta \underline{k}\left( X\right) \left( f_{1}\left( X\right) -%
\bar{r}\right) \right) }-\frac{C\left( 2X-C\beta \right) }{1+\underline{k}%
\left( X\right) }\right)
\end{equation*}%
where we can approximat:%
\begin{equation*}
X\rightarrow \sqrt{\frac{\left\vert \Psi _{0}\left( X\right) \right\vert ^{2}%
}{\epsilon }-\frac{1}{2}f_{1}\left( X\right) -\frac{1}{2}\left( \beta 
\underline{k}\left( X\right) \left( f_{1}\left( X\right) -\bar{r}\right)
\right) }\simeq \sqrt{\frac{\left\vert \Psi _{0}\left( X\right) \right\vert
^{2}}{\epsilon }-\frac{1}{2}f_{1}\left( X\right) }
\end{equation*}

\begin{equation*}
X=\sqrt{\sigma _{\hat{K}}^{2}\left( \frac{\left\vert \Psi _{0}\left(
X\right) \right\vert ^{2}}{\epsilon }-\frac{f_{1}\left( X\right) }{2}\right) 
}
\end{equation*}%
Defining 
\begin{equation*}
R=f_{1}\left( X\right) -\bar{r}
\end{equation*}

\subsection*{A11.3 Estimation of $\hat{S}_{1}^{E}\left( \hat{X}^{\prime },%
\hat{X}_{1}\right) $}

We have obtand bv:%
\begin{equation*}
\hat{S}_{1}^{E}\left( \hat{X}^{\prime },\hat{X}\right) =\left( \frac{\hat{k}%
\left( \hat{X},\hat{X}^{\prime }\right) -\underline{\hat{k}}_{1}\left(
\left\langle X\right\rangle ,\hat{X}\right) \hat{k}\left( \left\langle \hat{X%
}\right\rangle ,\hat{X}^{\prime }\right) }{1+\underline{\hat{k}}_{2}\left( 
\hat{X}\right) }+\frac{\hat{k}_{1}\left( \hat{X}^{\prime },\hat{X}\right) }{%
1+\underline{\hat{k}}_{2}\left( \hat{X}^{\prime }\right) }\right) \frac{\hat{%
K}_{X}\left\Vert \hat{\Psi}\left( \hat{X}^{\prime }\right) \right\Vert ^{2}}{%
1+\hat{k}\left( \hat{X}\right) }
\end{equation*}%
so that:%
\begin{eqnarray*}
&&\left( 1+\underline{\hat{k}}_{2}\left( \hat{X}_{1}\right) \right) \hat{S}%
_{1}^{E}\left( \hat{X}^{\prime },\hat{X}_{1}\right) \\
&\rightarrow &\frac{\left( 1+\underline{\hat{k}}_{2}\left( \hat{X}%
_{1}\right) \right) }{\left\langle \hat{K}\right\rangle \left\Vert \hat{\Psi}%
\right\Vert ^{2}}\left( \frac{\hat{k}\left( \hat{X},\hat{X}^{\prime }\right)
-\hat{k}_{1}\left( \left\langle X\right\rangle ,\hat{X}\right) \hat{k}\left(
\left\langle \hat{X}\right\rangle ,\hat{X}^{\prime }\right) }{1+\underline{%
\hat{k}}_{2}\left( \hat{X}\right) }+\frac{\hat{k}_{1}\left( \hat{X}^{\prime
},\hat{X}\right) }{1+\underline{\hat{k}}_{2}\left( \hat{X}^{\prime }\right) }%
\right) \frac{\hat{K}_{X}\left\Vert \hat{\Psi}\left( \hat{X}^{\prime
}\right) \right\Vert ^{2}}{1+\hat{k}\left( \hat{X}\right) }
\end{eqnarray*}

This is computed using (\ref{KD}), (\ref{PK}) and (\ref{KPS}). at the lowest
order $\hat{g}\left( \left\langle \hat{X}\right\rangle \right) \rightarrow 
\bar{r}^{\prime }$.

\begin{equation}
\hat{K}_{\hat{X}}=\frac{\sqrt{\frac{\sigma _{\hat{K}}^{2}}{2}\left( \frac{%
\left\Vert \hat{\Psi}_{0}\left( \hat{X}\right) \right\Vert ^{2}}{\hat{\mu}}%
-D\left( \hat{X}\right) \right) }\left( \frac{1}{4}-\frac{r}{3}\underline{%
\hat{k}}\right) }{\left( \frac{1}{3}-\frac{\left\langle \hat{K}\right\rangle 
}{2\left\langle \hat{K}_{0}\right\rangle }\underline{\hat{k}}\right) \bar{r}%
^{\prime }}
\end{equation}%
\begin{equation}
\left\Vert \hat{\Psi}\left( \hat{X}^{\prime }\right) \right\Vert ^{2}\simeq 
\frac{2\sqrt{2\sigma _{\hat{K}}^{2}}\hat{\mu}}{\bar{r}^{\prime }}\left( 
\frac{\left\Vert \hat{\Psi}_{0}\left( \hat{X}^{\prime }\right) \right\Vert
^{2}}{\hat{\mu}}-D\left( \hat{X}^{\prime }\right) \right) ^{\frac{3}{2}%
}\left( \frac{1}{3}-\frac{\left\langle \hat{K}\right\rangle }{2\left\langle 
\hat{K}_{0}\right\rangle }\underline{\hat{k}}\right)
\end{equation}%
\begin{equation*}
D\left( \hat{X}_{1}\right) \rightarrow \left( \frac{\left\langle \hat{K}%
\right\rangle ^{2}\bar{r}^{\prime 2}}{\sigma _{\hat{K}}^{2}}+\frac{\bar{r}%
^{\prime }}{2}\right) \left( 1-\frac{6\underline{\hat{k}}}{2+\underline{\hat{%
k}}-\sqrt{\left( 2+\underline{\hat{k}}\right) ^{2}-\underline{\hat{k}}}}%
\right) \underline{\hat{k}}
\end{equation*}%
\bigskip 
\begin{equation*}
\hat{K}_{\hat{X}}\left\Vert \hat{\Psi}\left( \hat{X}^{\prime }\right)
\right\Vert ^{2}\rightarrow \frac{2\sqrt{2}\sigma _{\hat{K}}^{2}}{\hat{\mu}%
\bar{r}^{\prime 2}}\sqrt{\left\Vert \hat{\Psi}_{0}\left( \hat{X}\right)
\right\Vert ^{2}-\hat{\mu}D\left( \hat{X}\right) }\left( \left\Vert \hat{\Psi%
}_{0}\left( \hat{X}^{\prime }\right) \right\Vert ^{2}-\hat{\mu}D\left( \hat{X%
}^{\prime }\right) \right) ^{\frac{3}{2}}
\end{equation*}%
\begin{equation}
\frac{\left\langle \hat{K}\right\rangle }{\left\langle \hat{K}%
_{0}\right\rangle }=\frac{1}{2}\frac{\frac{1}{4}-\frac{\left\langle \hat{K}%
\right\rangle }{3\left\langle \hat{K}_{0}\right\rangle }\underline{\hat{k}}%
\left( \left\langle \hat{X}\right\rangle ,\left\langle \hat{X}\right\rangle
\right) }{\frac{1}{3}-\frac{\left\langle \hat{K}\right\rangle }{%
2\left\langle \hat{K}_{0}\right\rangle }\underline{\hat{k}}\left(
\left\langle \hat{X}\right\rangle ,\left\langle \hat{X}\right\rangle \right) 
}
\end{equation}%
\begin{equation}
r\rightarrow \frac{\left\langle \hat{K}\right\rangle }{\left\langle \hat{K}%
_{0}\right\rangle }=\frac{2+\underline{\hat{k}}-\sqrt{\left( 2+\underline{%
\hat{k}}\right) ^{2}-\underline{\hat{k}}}}{6\underline{\hat{k}}}
\end{equation}

As a consequence, the estimation of $\left( 1+\underline{\hat{k}}_{2}\left( 
\hat{X}_{1}\right) \right) \hat{S}_{1}^{E}\left( \hat{X}^{\prime },\hat{X}%
_{1}\right) $:%
\begin{eqnarray*}
&&\left( 1+\underline{\hat{k}}_{2}\left( \hat{X}_{1}\right) \right) \hat{S}%
_{1}^{E}\left( \hat{X}^{\prime },\hat{X}_{1}\right) \\
&\rightarrow &\frac{1+\underline{\hat{k}}_{2}\left( \hat{X}_{1}\right) }{1+%
\underline{\hat{k}}\left( \hat{X}\right) }\left( \frac{\hat{k}\left( \hat{X},%
\hat{X}^{\prime }\right) -\hat{k}_{1}\left( \left\langle X\right\rangle ,%
\hat{X}\right) \hat{k}\left( \left\langle \hat{X}\right\rangle ,\hat{X}%
^{\prime }\right) }{1+\underline{\hat{k}}_{2}\left( \hat{X}\right) }+\frac{%
\hat{k}_{1}\left( \hat{X}^{\prime },\hat{X}\right) }{1+\underline{\hat{k}}%
_{2}\left( \hat{X}^{\prime }\right) }\right) \frac{2\sqrt{2}\sigma _{\hat{K}%
}^{2}}{\hat{\mu}\bar{r}^{\prime 2}} \\
&&\times \frac{\sqrt{\left\Vert \hat{\Psi}_{0}\left( \hat{X}\right)
\right\Vert ^{2}-\hat{\mu}D\left( \hat{X}\right) }\left( \left\Vert \hat{\Psi%
}_{0}\left( \hat{X}^{\prime }\right) \right\Vert ^{2}-\hat{\mu}D\left( \hat{X%
}^{\prime }\right) \right) ^{\frac{3}{2}}}{\left\langle \hat{K}\right\rangle
\left\Vert \hat{\Psi}\right\Vert ^{2}} \\
&\rightarrow &\frac{1+\underline{\hat{k}}_{2}\left( \hat{X}_{1}\right) }{1+%
\underline{\hat{k}}\left( \hat{X}\right) }\left( \frac{\hat{k}\left( \hat{X},%
\hat{X}^{\prime }\right) -\hat{k}_{1}\left( \left\langle X\right\rangle ,%
\hat{X}\right) \hat{k}\left( \left\langle \hat{X}\right\rangle ,\hat{X}%
^{\prime }\right) }{1+\underline{\hat{k}}_{2}\left( \hat{X}\right) }+\frac{%
\hat{k}_{1}\left( \hat{X}^{\prime },\hat{X}\right) }{1+\underline{\hat{k}}%
_{2}\left( \hat{X}^{\prime }\right) }\right) \\
&&\times \frac{\sqrt{\left\Vert \hat{\Psi}_{0}\left( \hat{X}\right)
\right\Vert ^{2}-\hat{\mu}D\left( \hat{X}\right) }\left( \left\Vert \hat{\Psi%
}_{0}\left( \hat{X}^{\prime }\right) \right\Vert ^{2}-\hat{\mu}D\left( \hat{X%
}^{\prime }\right) \right) ^{\frac{3}{2}}}{\left( \left\Vert \hat{\Psi}%
_{0}\right\Vert ^{2}-\hat{\mu}\left\langle D\right\rangle \right) ^{2}}
\end{eqnarray*}%
intrducing the normalizatn, \ in the averages, we can replac:%
\begin{equation*}
1+\underline{\hat{k}}\left( \left\langle \hat{X}\right\rangle \right)
\rightarrow 1
\end{equation*}%
and: 
\begin{equation*}
1+\underline{\hat{k}}_{2}\left( \hat{X}^{\prime }\right) \rightarrow 1+\frac{%
\hat{k}_{2}\left( \hat{X},\left\langle \hat{X}\right\rangle \right) }{%
1-\left\langle \hat{k}\right\rangle }
\end{equation*}%
so that:%
\begin{eqnarray*}
\left( 1+\underline{\hat{k}}_{2}\left( \hat{X}_{1}\right) \right) \hat{S}%
_{1}^{E}\left( \hat{X}^{\prime },\hat{X}_{1}\right) &\rightarrow &\left( 
\frac{\hat{k}\left( \hat{X},\hat{X}^{\prime }\right) -\hat{k}_{1}\left(
\left\langle X\right\rangle ,\hat{X}\right) \hat{k}\left( \left\langle \hat{X%
}\right\rangle ,\hat{X}^{\prime }\right) }{1+\hat{k}\left( \hat{X}\right) }+%
\frac{\hat{k}_{1}\left( \hat{X}^{\prime },\hat{X}\right) }{\left( 1+\hat{k}%
\left( \hat{X}\right) \right) \left( 1+\frac{\hat{k}_{2}\left( \hat{X}%
,\left\langle \hat{X}\right\rangle \right) }{1-\left\langle \hat{k}%
\right\rangle }\right) }\right) \\
&&\times \frac{\sqrt{\left\Vert \hat{\Psi}_{0}\left( \hat{X}\right)
\right\Vert ^{2}-\hat{\mu}D\left( \hat{X}\right) }\left( \left\Vert \hat{\Psi%
}_{0}\left( \hat{X}^{\prime }\right) \right\Vert ^{2}-\hat{\mu}D\left( \hat{X%
}^{\prime }\right) \right) ^{\frac{3}{2}}}{\left( \left\Vert \hat{\Psi}%
_{0}\right\Vert ^{2}-\hat{\mu}\left\langle D\right\rangle \right) ^{2}}
\end{eqnarray*}%
with:%
\begin{equation*}
\underline{\hat{k}}_{2}\left( \hat{X}^{\prime }\right) =\int \frac{\hat{k}%
_{2}\left( \hat{X}^{\prime },\hat{X}\right) -\left\langle \hat{k}_{2}\left( 
\hat{X}^{\prime },\hat{X}\right) \right\rangle }{\left\Vert \hat{\Psi}%
\right\Vert ^{2}\left\langle \hat{K}\right\rangle }\hat{K}_{\hat{X}%
}\left\vert \hat{\Psi}\left( \hat{X}\right) \right\vert ^{2}d\hat{X}
\end{equation*}

leading in vr:%
\begin{eqnarray*}
\left( 1+\underline{\hat{k}}_{2}\left( \hat{X}_{1}\right) \right) \hat{S}%
_{1}^{E}\left( \hat{X}^{\prime },\hat{X}_{1}\right) &\rightarrow &\left( 
\frac{\hat{k}\left( \hat{X},\hat{X}^{\prime }\right) -\left\langle \hat{k}%
_{1}\right\rangle \hat{k}\left( \left\langle \hat{X}\right\rangle ,\hat{X}%
^{\prime }\right) }{1+\hat{k}\left( \hat{X}\right) }+\frac{\hat{k}_{1}\left( 
\hat{X}^{\prime },\hat{X}\right) }{\left( 1+\hat{k}\left( \hat{X}\right)
\right) 1+\frac{\hat{k}_{2}\left( \hat{X},\left\langle \hat{X}\right\rangle
\right) }{1-\left\langle \hat{k}\right\rangle }}\right) \\
&&\times \frac{\sqrt{\left\Vert \hat{\Psi}_{0}\left( \hat{X}\right)
\right\Vert ^{2}-\hat{\mu}D\left( \hat{X}\right) }\left( \left\Vert \hat{\Psi%
}_{0}\left( \hat{X}^{\prime }\right) \right\Vert ^{2}-\hat{\mu}D\left( \hat{X%
}^{\prime }\right) \right) ^{\frac{3}{2}}}{\left( \left\Vert \hat{\Psi}%
_{0}\right\Vert ^{2}-\hat{\mu}\left\langle D\right\rangle \right) ^{2}}
\end{eqnarray*}

\subsection*{A11.4 Equation for $\hat{g}$}

Given (\ref{KP}):%
\begin{eqnarray}
\hat{K}\left[ \hat{X}_{1}\right] &=&\frac{2\sigma _{\hat{K}}^{2}}{\hat{\mu}}%
\left( \frac{\left\Vert \hat{\Psi}_{0}\left( \hat{X}_{1}\right) \right\Vert
^{2}-\hat{\mu}D\left( \hat{X}_{1}\right) }{\hat{g}^{2}\left( \hat{X}%
_{1}\right) }\right) ^{2}\left( \frac{\hat{g}^{2}\left( \hat{X}_{1}\right) }{%
4}-\frac{r\left\langle \hat{g}\right\rangle ^{2}}{3}\underline{\hat{k}}%
\right) \\
&\simeq &\frac{2\sigma _{\hat{K}}^{2}}{\hat{\mu}\hat{g}^{2}\left( \hat{X}%
_{1}\right) }\left( \left\Vert \hat{\Psi}_{0}\left( \hat{X}_{1}\right)
\right\Vert ^{2}-\hat{\mu}D\left( \hat{X}_{1}\right) \right) ^{2}\left( 
\frac{1}{4}-\frac{r}{3}\underline{\hat{k}}\right)  \notag
\end{eqnarray}%
We can find the return as a function of $\hat{K}\left[ \hat{X}_{1}\right] $
the total capital in sector $\hat{X}_{1}$ by writing the equation: 
\begin{equation*}
0=\frac{\hat{\mu}\hat{K}\left[ \hat{X}_{1}\right] }{2\sigma _{\hat{K}}^{2}}%
\left( \hat{g}^{2}\left( \hat{X}_{1}\right) \right) ^{2}-\left( \left\Vert 
\hat{\Psi}_{0}\left( \hat{X}_{1}\right) \right\Vert ^{2}-\hat{\mu}D\left( 
\hat{X}_{1}\right) \right) ^{2}\frac{\hat{g}^{2}\left( \hat{X}_{1}\right) }{4%
}+\left( \left\Vert \hat{\Psi}_{0}\left( \hat{X}_{1}\right) \right\Vert ^{2}-%
\hat{\mu}D\left( \hat{X}_{1}\right) \right) ^{2}\frac{r\left\langle \hat{g}%
\right\rangle ^{2}}{3}\underline{\hat{k}}
\end{equation*}%
so that:%
\begin{eqnarray*}
\hat{g}\left( \hat{X}_{1}\right) &=&\frac{\left( \left\Vert \hat{\Psi}%
_{0}\left( \hat{X}_{1}\right) \right\Vert ^{2}-\hat{\mu}D\left( \hat{X}%
_{1}\right) \right) ^{2}-\sqrt{\left( \left\Vert \hat{\Psi}_{0}\left( \hat{X}%
_{1}\right) \right\Vert ^{2}-\hat{\mu}D\left( \hat{X}_{1}\right) \right)
^{2}-2\frac{\hat{\mu}\hat{K}\left[ \hat{X}_{1}\right] }{\sigma _{\hat{K}}^{2}%
}\frac{r\left\langle \hat{g}\right\rangle ^{2}}{3}\underline{\hat{k}}}}{%
\frac{\hat{\mu}\hat{K}\left[ \hat{X}_{1}\right] }{\sigma _{\hat{K}}^{2}}} \\
&\simeq &\frac{\left( \left\Vert \hat{\Psi}_{0}\left( \hat{X}_{1}\right)
\right\Vert ^{2}-\hat{\mu}D\left( \hat{X}_{1}\right) \right) \sqrt{\frac{1}{4%
}-\frac{r}{3}\underline{\hat{k}}}}{\sqrt{\frac{\hat{\mu}\hat{K}\left[ \hat{X}%
_{1}\right] }{2\sigma _{\hat{K}}^{2}}}}
\end{eqnarray*}%
where:%
\begin{equation*}
D\left( \hat{X}_{1}\right) =\left( \frac{\left\langle \hat{K}\right\rangle
^{2}\left\langle \hat{g}\right\rangle ^{2}}{\sigma _{\hat{K}}^{2}}+\frac{%
\left\langle \hat{g}\right\rangle }{2}\right) \left( \frac{\underline{\hat{k}%
}\left( \left\langle \hat{X}\right\rangle ,\hat{X}_{1}\right) }{\underline{%
\hat{k}}\left( \left\langle \hat{X}\right\rangle ,\left\langle \hat{X}%
\right\rangle \right) }-\frac{6\underline{\hat{k}}}{2+\underline{\hat{k}}-%
\sqrt{\left( 2+\underline{\hat{k}}\right) ^{2}-\underline{\hat{k}}}}\right) 
\underline{\hat{k}}
\end{equation*}%
This allows to rewrite (\ref{RTN}):%
\begin{eqnarray}
&&\left( \frac{\Delta \left( \hat{X},\hat{X}^{\prime }\right) }{1+\underline{%
\hat{k}}_{2}\left( \hat{X}\right) }-\hat{S}_{1}^{E}\left( \hat{X}^{\prime },%
\hat{X}\right) \right) \left( \frac{\left( \left\Vert \hat{\Psi}_{0}\left( 
\hat{X}^{\prime }\right) \right\Vert ^{2}-\hat{\mu}D\left( \hat{X}^{\prime
}\right) \right) \sqrt{\frac{1}{4}-\frac{r}{3}\underline{\hat{k}}}}{\sqrt{%
\frac{\hat{\mu}\hat{K}\left[ \hat{X}^{\prime }\right] }{2\sigma _{\hat{K}%
}^{2}}}}-\bar{r}^{\prime }\right)  \label{qtk} \\
&=&\frac{1}{3}\frac{\left( 1-\beta \right) \left( X+C\beta \right)
^{2}\epsilon }{\sigma _{\hat{K}}^{2}\left( f_{1}\left( X\right) +\beta 
\underline{k}\left( X\right) R\right) }\left( \frac{R\left( 3X-\beta
C\right) \left( \beta C+X\right) }{4\left( f_{1}\left( X\right) +\beta 
\underline{k}\left( X\right) R\right) }-\frac{C\left( 2X-C\beta \right) }{1+%
\underline{k}\left( X\right) }\right)  \notag
\end{eqnarray}%
where $\hat{S}_{1}^{E}\left( \hat{X}^{\prime },\hat{X}\right) $ has been
computed previously in (\ref{LHD}).

\subsection*{A11.5 Solving equation for $\hat{g}$}

Equation (\ref{qtk}) is solved by defining:%
\begin{equation*}
H=\hat{S}_{1}^{E}\left( \hat{X}^{\prime },\hat{X}\right) \left( \frac{\left(
\left\Vert \hat{\Psi}_{0}\left( \hat{X}^{\prime }\right) \right\Vert ^{2}-%
\hat{\mu}D\left( \hat{X}^{\prime }\right) \right) \sqrt{\frac{1}{4}-\frac{r}{%
3}\underline{\hat{k}}}}{\sqrt{\frac{\hat{\mu}\hat{K}\left[ \hat{X}^{\prime }%
\right] }{2\sigma _{\hat{K}}^{2}}}}-\bar{r}^{\prime }\right)
\end{equation*}%
\begin{equation*}
\underline{k}\left( X\right) =\frac{k\left( X\right) }{\left\langle
K\right\rangle }\hat{K}_{X}\frac{\left\vert \hat{\Psi}\left( \hat{X}\right)
\right\vert ^{2}}{\left\vert \Psi _{0}\left( X\right) \right\vert ^{2}}
\end{equation*}%
\begin{equation*}
\hat{K}\left[ \hat{X}\right] =\frac{\underline{k}\left( X\right) }{k\left(
X\right) }\left\langle K\right\rangle \left\vert \Psi _{0}\left( X\right)
\right\vert ^{2}\rightarrow \frac{\underline{k}\left( X\right) }{k}
\end{equation*}%
so that:%
\begin{equation*}
\frac{\left( \left\Vert \hat{\Psi}_{0}\left( \hat{X}_{1}\right) \right\Vert
^{2}-\hat{\mu}D\left( \hat{X}_{1}\right) \right) \sqrt{\frac{1}{4}-\frac{r}{3%
}\underline{\hat{k}}}}{\left( 1+\underline{\hat{k}}_{2}\left( \hat{X}%
_{1}\right) \right) \sqrt{\frac{\hat{\mu}\hat{K}\left[ \hat{X}^{\prime }%
\right] }{2\sigma _{\hat{K}}^{2}}}}\rightarrow \frac{\sqrt{\frac{2\sigma _{%
\hat{K}}^{2}k}{\hat{\mu}}\left( \frac{1}{4}-\frac{r}{3}\underline{\hat{k}}%
\right) }\left( \left\Vert \hat{\Psi}_{0}\left( \hat{X}_{1}\right)
\right\Vert ^{2}-\hat{\mu}D\left( \hat{X}_{1}\right) \right) }{\left( 1+%
\underline{\hat{k}}_{2}\left( \hat{X}_{1}\right) \right) \sqrt{\underline{k}%
\left( \hat{X}_{1}\right) }}
\end{equation*}

and then, by isolating the terms not acted by $\hat{S}_{1}^{E}\left( \hat{X}%
^{\prime },\hat{X}\right) $ on, to write (\ref{RTN}) as:%
\begin{eqnarray*}
&&\frac{1}{3}\frac{\left( 1-\beta \right) \left( X+C\beta \right)
^{2}\epsilon }{\sigma _{\hat{K}}^{2}\left( f_{1}\left( X\right) +\beta 
\underline{k}\left( X\right) R\right) }\left( \frac{R\left( 3X-\beta
C\right) \left( \beta C+X\right) }{4\left( f_{1}\left( X\right) +\beta 
\underline{k}\left( X\right) R\right) }-\frac{C\left( 2X-C\beta \right) }{1+%
\underline{k}\left( X\right) }\right) \\
&&-\frac{\sqrt{\frac{2\sigma _{\hat{K}}^{2}k}{\hat{\mu}}\left( \frac{1}{4}-%
\frac{r}{3}\underline{\hat{k}}\right) }\left( \left\Vert \hat{\Psi}%
_{0}\left( \hat{X}_{1}\right) \right\Vert ^{2}-\hat{\mu}D\left( \hat{X}%
_{1}\right) \right) }{\left( 1+\underline{\hat{k}}_{2}\left( \hat{X}%
_{1}\right) \right) \sqrt{\underline{k}\left( \hat{X}_{1}\right) }} \\
&=&\left( -H-\frac{\bar{r}^{\prime }}{1+\underline{\hat{k}}_{2}\left( \hat{X}%
_{1}\right) }\right)
\end{eqnarray*}%
Then a first order expansion in $R$ leads to:%
\begin{eqnarray*}
&&\left( 1-\frac{\beta \underline{k}\left( X\right) R}{f_{1}^{2}\left(
X\right) }\right) \left( \frac{R\left( 3X-\beta C\right) \left( \beta
C+X\right) }{4\left( f_{1}\left( X\right) +\beta \underline{k}\left(
X\right) R\right) }-\frac{C\left( 2X-C\beta \right) }{1+\underline{k}\left(
X\right) }\right) \\
&&-\frac{3\sigma _{\hat{K}}^{2}f_{1}\left( X\right) }{\left( 1-\beta \right)
\left( X+C\beta \right) ^{2}\epsilon }\frac{\sqrt{\frac{2\sigma _{\hat{K}%
}^{2}k}{\hat{\mu}}\left( \frac{1}{4}-\frac{r}{3}\underline{\hat{k}}\right) }%
\left( \left\Vert \hat{\Psi}_{0}\left( \hat{X}_{1}\right) \right\Vert ^{2}-%
\hat{\mu}D\left( \hat{X}_{1}\right) \right) }{\left( 1+\underline{\hat{k}}%
_{2}\left( \hat{X}_{1}\right) \right) \sqrt{\underline{k}\left( \hat{X}%
_{1}\right) }} \\
&=&\frac{3\sigma _{\hat{K}}^{2}f_{1}\left( X\right) }{\left( 1-\beta \right)
\left( X+C\beta \right) ^{2}\epsilon }\left( -H-\frac{\bar{r}^{\prime }}{1+%
\underline{\hat{k}}_{2}\left( \hat{X}_{1}\right) }\right)
\end{eqnarray*}

\begin{eqnarray*}
&&\left( 1-\frac{\beta \underline{k}\left( X\right) R}{f_{1}^{2}\left(
X\right) }\right) \left( -\frac{C\left( 2X-C\beta \right) }{1+\underline{k}%
\left( X\right) }\right) -\frac{3\sigma _{\hat{K}}^{2}f_{1}\left( X\right) }{%
\left( 1-\beta \right) \left( X+C\beta \right) ^{2}\epsilon }\frac{\sqrt{%
\frac{2\sigma _{\hat{K}}^{2}k}{\hat{\mu}}\left( \frac{1}{4}-\frac{r}{3}%
\underline{\hat{k}}\right) }\left( \left\Vert \hat{\Psi}_{0}\left( \hat{X}%
_{1}\right) \right\Vert ^{2}-\hat{\mu}D\left( \hat{X}_{1}\right) \right) }{%
\left( 1+\underline{\hat{k}}_{2}\left( \hat{X}_{1}\right) \right) \sqrt{%
\underline{k}\left( \hat{X}_{1}\right) }} \\
&=&\frac{3\sigma _{\hat{K}}^{2}f_{1}\left( X\right) }{\left( 1-\beta \right)
\left( X+C\beta \right) ^{2}\epsilon }\left( -H-\frac{\bar{r}^{\prime }}{1+%
\underline{\hat{k}}_{2}\left( \hat{X}_{1}\right) }\right) -\frac{R\left(
3X-\beta C\right) \left( \beta C+X\right) }{4\left( f_{1}\left( X\right)
+\beta \underline{k}\left( X\right) R\right) }
\end{eqnarray*}%
and this equation leads to:%
\begin{eqnarray*}
&&\left( \frac{1}{1+\underline{k}\left( X\right) }\right) \left( 1-\frac{%
\beta \underline{k}\left( X\right) R}{f_{1}^{2}\left( X\right) }\right) +%
\frac{3\sigma _{\hat{K}}^{2}f_{1}\left( X\right) }{\left( 1-\beta \right)
C\left( 2X-C\beta \right) \left( X+C\beta \right) ^{2}\epsilon }\frac{\sqrt{%
\frac{2\sigma _{\hat{K}}^{2}k}{\hat{\mu}}\left( \frac{1}{4}-\frac{r}{3}%
\underline{\hat{k}}\right) }\left( \left\Vert \hat{\Psi}_{0}\left( \hat{X}%
_{1}\right) \right\Vert ^{2}-\hat{\mu}D\left( \hat{X}_{1}\right) \right) }{%
\left( 1+\underline{\hat{k}}_{2}\left( \hat{X}_{1}\right) \right) \sqrt{%
\underline{k}\left( \hat{X}_{1}\right) }} \\
&=&\frac{3\sigma _{\hat{K}}^{2}f_{1}\left( X\right) }{\left( 1-\beta \right)
C\left( 2X-C\beta \right) \left( X+C\beta \right) ^{2}\epsilon }\left( H+%
\frac{\bar{r}^{\prime }}{1+\underline{\hat{k}}_{2}\left( \hat{X}_{1}\right) }%
\right) +\frac{R\left( 3X-\beta C\right) \left( \beta C+X\right) }{%
4f_{1}\left( X\right) C\left( 2X-C\beta \right) }
\end{eqnarray*}%
\begin{eqnarray*}
&&\frac{1}{1+\underline{k}\left( X\right) }\left( 1+\frac{\beta R}{%
f_{1}^{2}\left( X\right) }\right) +\frac{3\sigma _{\hat{K}}^{2}f_{1}\left(
X\right) }{C\left( 2X-C\beta \right) \left( 1-\beta \right) \left( X+C\beta
\right) ^{2}\epsilon }\frac{\sqrt{\frac{2\sigma _{\hat{K}}^{2}k}{\hat{\mu}}%
\left( \frac{1}{4}-\frac{r}{3}\underline{\hat{k}}\right) }\left( \left\Vert 
\hat{\Psi}_{0}\left( \hat{X}_{1}\right) \right\Vert ^{2}-\hat{\mu}D\left( 
\hat{X}_{1}\right) \right) }{\left( 1+\underline{\hat{k}}_{2}\left( \hat{X}%
_{1}\right) \right) \sqrt{\underline{k}\left( \hat{X}_{1}\right) }} \\
&=&\frac{3\sigma _{\hat{K}}^{2}f_{1}\left( X\right) }{C\left( 2X-C\beta
\right) \left( 1-\beta \right) \left( X+C\beta \right) ^{2}\epsilon }\left(
H+\frac{\bar{r}^{\prime }}{1+\underline{\hat{k}}_{2}\left( \hat{X}%
_{1}\right) }\right) +\frac{R\left( 3X-\beta C\right) \left( \beta
C+X\right) }{4f_{1}\left( X\right) C\left( 2X-C\beta \right) }+\frac{\beta R%
}{f_{1}^{2}\left( X\right) }
\end{eqnarray*}%
Then we define:%
\begin{equation*}
a=\frac{3\sigma _{\hat{K}}^{2}f_{1}\left( X\right) }{C\left( 2X-C\beta
\right) \left( 1-\beta \right) \left( X+C\beta \right) ^{2}\epsilon }\frac{%
\sqrt{\frac{2\sigma _{\hat{K}}^{2}k}{\hat{\mu}}\left( \frac{1}{4}-\frac{r}{3}%
\underline{\hat{k}}\right) }\left( \left\Vert \hat{\Psi}_{0}\left( \hat{X}%
_{1}\right) \right\Vert ^{2}-\hat{\mu}D\left( \hat{X}_{1}\right) \right) }{%
\left( 1+\underline{\hat{k}}_{2}\left( \hat{X}_{1}\right) \right) }
\end{equation*}%
and:%
\begin{equation*}
c=\left( \frac{3\sigma _{\hat{K}}^{2}f_{1}\left( X\right) }{C\left(
2X-C\beta \right) \left( 1-\beta \right) \left( X+C\beta \right)
^{2}\epsilon }\left( H+\frac{\bar{r}^{\prime }}{1+\underline{\hat{k}}%
_{2}\left( \hat{X}_{1}\right) }\right) +\frac{R\left( 3X-\beta C\right)
\left( \beta C+X\right) }{4f_{1}\left( X\right) C\left( 2X-C\beta \right) }+%
\frac{\beta R}{f_{1}^{2}\left( X\right) }\right)
\end{equation*}%
so that the equation reduces in first approximation:%
\begin{equation*}
0=\frac{1}{1+\underline{k}\left( X\right) }\left( 1+\frac{\beta R}{%
f_{1}^{2}\left( X\right) }\right) +\frac{a}{\sqrt{\underline{k}\left(
X\right) }}-c
\end{equation*}%
for $\underline{k}\left( X\right) >>1$ and $\epsilon <\sigma _{\hat{K}%
}^{2}f_{1}\left( X\right) $ this reduces to:%
\begin{equation*}
\frac{a}{\sqrt{\underline{k}\left( X\right) }+b}-c=0
\end{equation*}%
with solution:%
\begin{equation*}
\sqrt{\underline{k}\left( X\right) }=\frac{a}{c}
\end{equation*}%
$\allowbreak $ $\allowbreak $so that $\underline{k}\left( X\right) $ is: 
\begin{equation*}
\underline{k}\left( X\right) =\left( \frac{\sqrt{\frac{2\sigma _{\hat{K}%
}^{2}k}{\hat{\mu}}\left( \frac{1}{4}-\frac{r}{3}\underline{\hat{k}}\right) }%
\left( \left\Vert \hat{\Psi}_{0}\left( \hat{X}_{1}\right) \right\Vert ^{2}-%
\hat{\mu}D\left( \hat{X}_{1}\right) \right) }{\left( 1+\underline{\hat{k}}%
_{2}\left( \hat{X}_{1}\right) \right) \left[ H+\frac{\bar{r}^{\prime }}{1+%
\underline{\hat{k}}_{2}\left( \hat{X}_{1}\right) }+\frac{\epsilon \left(
1-\beta \right) \left( X+C\beta \right) ^{2}}{3\sigma _{\hat{K}}^{2}}\left( 
\frac{R\left( 3X-\beta C\right) \left( \beta C+X\right) }{4f_{1}^{2}\left(
X\right) C\left( 2X-C\beta \right) }+\frac{\beta RC\left( 2X-C\beta \right) 
}{f_{1}^{3}\left( X\right) }\right) \right] }-\sqrt{\frac{\sigma _{\hat{K}%
}^{2}\hat{\mu}k}{2}}D\left( \hat{X}_{1}\right) \right) ^{2}
\end{equation*}

\begin{eqnarray*}
\frac{\hat{g}\left( \hat{X}_{1}\right) -\bar{r}^{\prime }}{\left( 1+%
\underline{\hat{k}}_{2}\left( \hat{X}_{1}\right) \right) } &=&\frac{\sqrt{%
\frac{2\sigma _{\hat{K}}^{2}k}{\hat{\mu}}\left( \frac{1}{4}-\frac{r}{3}%
\underline{\hat{k}}\right) }\left( \left\Vert \hat{\Psi}_{0}\left( \hat{X}%
_{1}\right) \right\Vert ^{2}-\hat{\mu}D\left( \hat{X}_{1}\right) \right) }{%
\left( 1+\underline{\hat{k}}_{2}\left( \hat{X}_{1}\right) \right) \sqrt{%
\underline{k}\left( \hat{X}_{1}\right) }} \\
&\simeq &H+\frac{\bar{r}^{\prime }}{1+\underline{\hat{k}}_{2}\left( \hat{X}%
_{1}\right) }+\frac{\frac{\left( 3X-\beta C\right) \left( \beta C+X\right) }{%
4f_{1}\left( X\right) C\left( 2X-C\beta \right) }+\frac{\beta }{%
f_{1}^{2}\left( X\right) }}{\frac{3\sigma _{\hat{K}}^{2}f_{1}\left( X\right) 
}{C\left( 2X-C\beta \right) \left( 1-\beta \right) \left( X+C\beta \right)
^{2}\epsilon }}\left( R+\Delta F_{\tau }\left( \bar{R}\left( K,X\right)
\right) \right) \\
&=&H+\frac{\bar{r}^{\prime }}{1+\underline{\hat{k}}_{2}\left( \hat{X}%
_{1}\right) } \\
&&+\frac{\epsilon \left( 1-\beta \right) \left( X+C\beta \right) ^{2}}{%
3\sigma _{\hat{K}}^{2}}\left( \frac{\left( 3X-\beta C\right) \left( \beta
C+X\right) }{4f_{1}^{2}\left( X\right) C\left( 2X-C\beta \right) }+\frac{%
\beta C\left( 2X-C\beta \right) }{f_{1}^{3}\left( X\right) }\right) \left(
R+\Delta F_{\tau }\left( \bar{R}\left( K,X\right) \right) \right)
\end{eqnarray*}%
and this rewrts:%
\begin{eqnarray*}
&&\hat{g}\left( \hat{X}_{1}\right) -\bar{r}^{\prime } \\
&=&\left( 1+\underline{\hat{k}}_{2}\left( \hat{X}_{1}\right) \right) \left(
H+\left( \frac{A\left( \hat{X}^{\prime }\right) }{f_{1}^{2}\left( X^{\prime
}\right) }+\frac{B\left( \hat{X}^{\prime }\right) }{f_{1}^{3}\left( X\right) 
}\right) \left( R+\Delta F_{\tau }\left( \bar{R}\left( K,X\right) \right)
\right) \right)
\end{eqnarray*}

where:%
\begin{eqnarray*}
A\left( \hat{X}^{\prime }\right) &=&\frac{\epsilon \left( 1-\beta \right)
\left( X+C\beta \right) ^{2}}{3\sigma _{\hat{K}}^{2}}\frac{\left( 3X-\beta
C\right) \left( \beta C+X\right) }{4C\left( 2X-C\beta \right) } \\
B\left( \hat{X}^{\prime }\right) &=&\frac{\epsilon \left( 1-\beta \right)
\left( X+C\beta \right) ^{2}}{3\sigma _{\hat{K}}^{2}}\beta C\left( 2X-C\beta
\right)
\end{eqnarray*}%
We replace:%
\begin{equation*}
H\rightarrow \int \hat{S}_{1}^{E}\left( \hat{X}^{\prime },\hat{X}_{1}\right)
\left( \frac{\sqrt{\frac{2\sigma _{\hat{K}}^{2}k}{\hat{\mu}}\left( \frac{1}{4%
}-\frac{r}{3}\underline{\hat{k}}\right) }\left( \left\Vert \hat{\Psi}%
_{0}\left( \hat{X}^{\prime }\right) \right\Vert ^{2}-\hat{\mu}D\left( \hat{X}%
^{\prime }\right) \right) }{\sqrt{\underline{k}\left( \hat{X}^{\prime
}\right) }}-\bar{r}^{\prime }\right)
\end{equation*}%
and the equation for return rewrites:%
\begin{eqnarray*}
\hat{g}\left( \hat{X}_{1}\right) -\bar{r}^{\prime } &=&\left( 1+\underline{%
\hat{k}}_{2}\left( \hat{X}_{1}\right) \right) \int \hat{S}_{1}^{E}\left( 
\hat{X}^{\prime },\hat{X}_{1}\right) \left( \frac{\sqrt{\frac{2\sigma _{\hat{%
K}}^{2}k}{\hat{\mu}}\left( \frac{1}{4}-\frac{r}{3}\underline{\hat{k}}\right) 
}\left( \left\Vert \hat{\Psi}_{0}\left( \hat{X}^{\prime }\right) \right\Vert
^{2}-\hat{\mu}D\left( \hat{X}^{\prime }\right) \right) }{\sqrt{\underline{k}%
\left( \hat{X}^{\prime }\right) }}-\bar{r}^{\prime }\right) \\
&&+\left( 1+\underline{\hat{k}}_{2}\left( \hat{X}_{1}\right) \right) \left( 
\frac{A\left( \hat{X}^{\prime }\right) }{f_{1}^{2}\left( X^{\prime }\right) }%
+\frac{B\left( \hat{X}^{\prime }\right) }{f_{1}^{3}\left( X\right) }\right)
\left( R+\Delta F_{\tau }\left( \bar{R}\left( K,X\right) \right) \right)
\end{eqnarray*}%
with solution:%
\begin{eqnarray*}
\hat{g}\left( \hat{X}_{1}\right) -\bar{r}^{\prime } &=&\int \left( 1-\left(
1+\underline{\hat{k}}_{2}\left( \hat{X}_{1}\right) \right) \hat{S}%
_{1}^{E}\left( \hat{X}^{\prime },\hat{X}_{1}\right) \right) ^{-1} \\
&&\times \left( 1+\underline{\hat{k}}_{2}\left( \hat{X}_{1}\right) \right)
\left( \frac{A\left( \hat{X}^{\prime }\right) }{f_{1}^{2}\left( X^{\prime
}\right) }+\frac{B\left( \hat{X}^{\prime }\right) }{f_{1}^{3}\left( X\right) 
}\right) \left( R+\Delta F_{\tau }\left( \bar{R}\left( K,X\right) \right)
\right)
\end{eqnarray*}%
\bigskip

\subsection*{A11.6 Estimation of the derivativ $\frac{\partial }{\partial
f_{1}\left( X\right) }\hat{K}\left[ \hat{X}\right] $ for decreasing return
to scale correction}

\begin{eqnarray*}
&&\frac{\partial }{\partial f_{1}\left( X\right) }\hat{K}\left[ \hat{X}%
\right] \\
&=&\frac{\left( 1+\underline{k}\left( X\right) \hat{K}\left[ \hat{X}\right]
\right) \left( \tau F\left( X\right) +A\frac{\left( f_{1}^{r}\left( X\right)
-C_{0}\right) -2r\left( f_{1}^{r}\left( X\right) -C_{0}-\bar{r}^{\prime
}\right) }{\left( f_{1}^{r}\left( X\right) -C_{0}\right) ^{3}}+B\frac{\left(
f_{1}^{r}\left( X\right) -C_{0}\right) -3r\left( f_{1}^{r}\left( X\right)
-C_{0}-\bar{r}^{\prime }\right) }{\left( f_{1}^{r}\left( X\right)
-C_{0}\right) ^{4}}\right) }{f_{1}\left( X\right) \left( \tau F\left(
X\right) +A\frac{\left( f_{1}^{r}\left( X\right) -C_{0}\right) -2r\left(
f_{1}^{r}\left( X\right) -C_{0}-\bar{r}^{\prime }\right) }{\left(
f_{1}^{r}\left( X\right) -C_{0}\right) ^{3}}+B\frac{\left( f_{1}^{r}\left(
X\right) -C_{0}\right) -3r\left( f_{1}^{r}\left( X\right) -C_{0}-\bar{r}%
^{\prime }\right) }{\left( f_{1}^{r}\left( X\right) -C_{0}\right) ^{4}}%
\right) -\frac{1}{2}\frac{\left( \left( 1+\underline{k}\left( X\right) \hat{K%
}\left[ \hat{X}\right] \right) ^{1+r}\right) H\left( \hat{X}_{1}\right) }{%
\left( \hat{K}\left[ \hat{X}_{1}\right] \right) ^{\frac{3}{2}}}} \\
&=&\frac{\left( 1+\underline{k}\left( X\right) \hat{K}\left[ \hat{X}\right]
\right) ^{1-r}\left( \tau F\left( X\right) +A\frac{\left( f_{1}^{r}\left(
X\right) -C_{0}\right) -2r\left( f_{1}^{r}\left( X\right) -C_{0}-\bar{r}%
^{\prime }\right) }{\left( f_{1}^{r}\left( X\right) -C_{0}\right) ^{3}}+B%
\frac{\left( f_{1}^{r}\left( X\right) -C_{0}\right) -3r\left(
f_{1}^{r}\left( X\right) -C_{0}-\bar{r}^{\prime }\right) }{\left(
f_{1}^{r}\left( X\right) -C_{0}\right) ^{4}}\right) }{\frac{f_{1}\left(
X\right) }{\left( 1+\underline{k}\left( X\right) \hat{K}\left[ \hat{X}\right]
\right) ^{r}}\left( \tau F\left( X\right) +A\frac{\left( f_{1}^{r}\left(
X\right) -C_{0}\right) -2r\left( f_{1}^{r}\left( X\right) -C_{0}-\bar{r}%
^{\prime }\right) }{\left( f_{1}^{r}\left( X\right) -C_{0}\right) ^{3}}+B%
\frac{\left( f_{1}^{r}\left( X\right) -C_{0}\right) -3r\left(
f_{1}^{r}\left( X\right) -C_{0}-\bar{r}^{\prime }\right) }{\left(
f_{1}^{r}\left( X\right) -C_{0}\right) ^{4}}\right) -\frac{1}{2}\frac{\left( 
\frac{1}{\hat{K}\left[ \hat{X}\right] }+\underline{k}\left( X\right) \right)
H\left( \hat{X}_{1}\right) }{\sqrt{\hat{K}\left[ \hat{X}_{1}\right] }}}
\end{eqnarray*}%
The denominator is estimated with the return equation and writes:

\begin{eqnarray}
&&\frac{H\left( \hat{X}_{1}\right) }{\sqrt{\hat{K}\left[ \hat{X}_{1}\right] }%
}-\bar{r}^{\prime }+\left( C_{0}+\bar{r}^{\prime }\right) \left( \frac{A}{%
\left( f_{1}^{r}\left( X\right) -C_{0}\right) ^{2}}+\frac{B}{\left(
f_{1}^{r}\left( X\right) -C_{0}\right) ^{3}}\right) +\tau F\left( X\right)
\left\langle \frac{f_{1}\left( X\right) }{\left( \left( 1+\underline{k}%
\left( X\right) \hat{K}\left[ \hat{X}\right] \right) \right) ^{r}}%
\right\rangle  \label{dr} \\
&&-\frac{f_{1}\left( X\right) }{\left( 1+\underline{k}\left( X\right) \hat{K}%
\left[ \hat{X}\right] \right) ^{r}}\left( \tau F\left( X\right) +A\frac{%
2r\left( f_{1}^{r}\left( X\right) -C_{0}-\bar{r}^{\prime }\right) }{\left(
f_{1}^{r}\left( X\right) -C_{0}\right) ^{3}}+B\frac{3r\left( f_{1}^{r}\left(
X\right) -C_{0}-\bar{r}^{\prime }\right) }{\left( f_{1}^{r}\left( X\right)
-C_{0}\right) ^{4}}\right) -\frac{1}{2}\frac{\left( \frac{1}{\hat{K}\left[ 
\hat{X}\right] }+\underline{k}\left( X\right) \right) H\left( \hat{X}%
_{1}\right) }{\sqrt{\hat{K}\left[ \hat{X}_{1}\right] }}  \notag
\end{eqnarray}%
that is:%
\begin{eqnarray*}
&&\frac{1}{2}\frac{H\left( \hat{X}_{1}\right) }{\sqrt{\hat{K}\left[ \hat{X}%
_{1}\right] }}-\bar{r}^{\prime }+\left( C_{0}+\bar{r}^{\prime }\right)
\left( \frac{A}{\left( f_{1}^{r}\left( X\right) -C_{0}\right) ^{2}}+\frac{B}{%
\left( f_{1}^{r}\left( X\right) -C_{0}\right) ^{3}}\right) +\tau F\left(
X\right) \left\langle \frac{f_{1}\left( X\right) }{\left( \left( 1+%
\underline{k}\left( X\right) \hat{K}\left[ \hat{X}\right] \right) \right)
^{r}}\right\rangle \\
&&-\frac{f_{1}\left( X\right) }{\left( 1+\underline{k}\left( X\right) \hat{K}%
\left[ \hat{X}\right] \right) ^{r}}\left( \tau F\left( X\right) +A\frac{%
2r\left( f_{1}^{r}\left( X\right) -C_{0}-\bar{r}^{\prime }\right) }{\left(
f_{1}^{r}\left( X\right) -C_{0}\right) ^{3}}+B\frac{3r\left( f_{1}^{r}\left(
X\right) -C_{0}-\bar{r}^{\prime }\right) }{\left( f_{1}^{r}\left( X\right)
-C_{0}\right) ^{4}}\right)
\end{eqnarray*}%
Using the return equation:%
\begin{eqnarray*}
\frac{1}{2}\left( \frac{H\left( \hat{X}_{1}\right) }{\sqrt{\hat{K}\left[ 
\hat{X}_{1}\right] }}-\bar{r}^{\prime }\right) &=&\frac{1}{2}\left( \frac{A}{%
\left( f_{1}^{r}\left( X\right) -C_{0}\right) ^{2}}+\frac{B}{\left(
f_{1}^{r}\left( X\right) -C_{0}\right) ^{3}}\right) \left( f_{1}^{r}\left(
X\right) -C_{0}-\bar{r}^{\prime }\right) \\
&&+\frac{1}{2}\tau F\left( X\right) \left( \frac{f_{1}\left( X\right) }{%
\left( \left( 1+\underline{k}\left( X\right) \hat{K}\left[ \hat{X}\right]
\right) \right) ^{r}}-\left\langle \frac{f_{1}\left( X\right) }{\left(
\left( 1+\underline{k}\left( X\right) \hat{K}\left[ \hat{X}\right] \right)
\right) ^{r}}\right\rangle \right)
\end{eqnarray*}%
term (\ref{dr}) becoms:

\begin{eqnarray*}
&\frac{1}{2}&\left( -\bar{r}^{\prime }+\left( C_{0}+\bar{r}^{\prime }\right)
\left( \frac{A}{\left( f_{1}^{r}\left( X\right) -C_{0}\right) ^{2}}+\frac{B}{%
\left( f_{1}^{r}\left( X\right) -C_{0}\right) ^{3}}\right) +\tau F\left(
X\right) \left\langle \frac{f_{1}\left( X\right) }{\left( \left( 1+%
\underline{k}\left( X\right) \hat{K}\left[ \hat{X}\right] \right) \right)
^{r}}\right\rangle \right) \\
&&+\frac{1}{2}\left( \frac{A}{\left( f_{1}^{r}\left( X\right) -C_{0}\right)
^{2}}+\frac{B}{\left( f_{1}^{r}\left( X\right) -C_{0}\right) ^{3}}\right) 
\frac{f_{1}\left( X\right) }{\left( \left( 1+\underline{k}\left( X\right) 
\hat{K}\left[ \hat{X}\right] \right) \right) ^{r}}+\frac{1}{2}\tau F\left(
X\right) \frac{f_{1}\left( X\right) }{\left( \left( 1+\underline{k}\left(
X\right) \hat{K}\left[ \hat{X}\right] \right) \right) ^{r}} \\
&&-\frac{f_{1}\left( X\right) }{\left( 1+\underline{k}\left( X\right) \hat{K}%
\left[ \hat{X}\right] \right) ^{r}}\left( \tau F\left( X\right) +A\frac{%
2r\left( f_{1}^{r}\left( X\right) -C_{0}-\bar{r}^{\prime }\right) }{\left(
f_{1}^{r}\left( X\right) -C_{0}\right) ^{3}}+B\frac{3r\left( f_{1}^{r}\left(
X\right) -C_{0}-\bar{r}^{\prime }\right) }{\left( f_{1}^{r}\left( X\right)
-C_{0}\right) ^{4}}\right) \\
&=&\frac{1}{2}\left( -\bar{r}^{\prime }+\left( C_{0}+\bar{r}^{\prime
}\right) \left( \frac{A}{\left( f_{1}^{r}\left( X\right) -C_{0}\right) ^{2}}+%
\frac{B}{\left( f_{1}^{r}\left( X\right) -C_{0}\right) ^{3}}\right) \right) +%
\frac{1}{2}\tau F\left( X\right) \left( \left\langle f_{1}^{r}\left(
X\right) \right\rangle -f_{1}^{r}\left( X\right) \right) \\
&&+\frac{1}{2}\left( \frac{A}{\left( f_{1}^{r}\left( X\right) -C_{0}\right)
^{2}}+\frac{B}{\left( f_{1}^{r}\left( X\right) -C_{0}\right) ^{3}}\right)
f_{1}^{r}\left( X\right) -f_{1}^{r}\left( X\right) \left( A\frac{2r\left(
f_{1}^{r}\left( X\right) -C_{0}-\bar{r}^{\prime }\right) }{\left(
f_{1}^{r}\left( X\right) -C_{0}\right) ^{3}}+B\frac{3r\left( f_{1}^{r}\left(
X\right) -C_{0}-\bar{r}^{\prime }\right) }{\left( f_{1}^{r}\left( X\right)
-C_{0}\right) ^{4}}\right)
\end{eqnarray*}%
In the main cases:%
\begin{equation*}
\left\vert f_{1}^{r}\left( X\right) -C_{0}-\bar{r}^{\prime }\right\vert
<<\left\vert f_{1}^{r}\left( X\right) -C_{0}\right\vert
\end{equation*}%
so that (\ref{dr}) is in frst approximation:%
\begin{eqnarray*}
&&\frac{1}{2}\left( -\bar{r}^{\prime }+\left( C_{0}+\bar{r}^{\prime }\right)
\left( \frac{A}{\left( f_{1}^{r}\left( X\right) -C_{0}\right) ^{2}}+\frac{B}{%
\left( f_{1}^{r}\left( X\right) -C_{0}\right) ^{3}}\right) \right) +\frac{1}{%
2}\tau F\left( X\right) \left( \left\langle f_{1}^{r}\left( X\right)
\right\rangle -f_{1}^{r}\left( X\right) \right) \\
&&+\frac{1}{2}\left( \frac{A}{\left( f_{1}^{r}\left( X\right) -C_{0}\right)
^{2}}+\frac{B}{\left( f_{1}^{r}\left( X\right) -C_{0}\right) ^{3}}\right)
f_{1}^{r}\left( X\right)
\end{eqnarray*}%
and:%
\begin{equation*}
\frac{1}{2}\left( -\bar{r}^{\prime }+\left( C_{0}+\bar{r}^{\prime }\right)
\left( \frac{A}{\left( \bar{r}^{\prime }\right) ^{2}}+\frac{B}{\left( \bar{r}%
^{\prime }\right) ^{3}}\right) \right) +\frac{1}{2}\tau F\left( X\right)
\left( \left\langle f_{1}^{r}\left( X\right) \right\rangle -f_{1}^{r}\left(
X\right) \right) +\frac{1}{2}\left( \frac{A}{\left( \bar{r}^{\prime }\right)
^{2}}+\frac{B}{\left( \bar{r}^{\prime }\right) ^{3}}\right) f_{1}^{r}\left(
X\right) >0
\end{equation*}%
\bigskip

\subsection*{A11.7 Averaged equations}

We compute the various averages arising in the model. Recal that:%
\begin{eqnarray}
\underline{\hat{k}}_{2}\left( \hat{X}^{\prime }\right) &=&\int \hat{k}%
_{2}\left( \hat{X}^{\prime },\hat{X}_{1}\right) \hat{K}_{1}\left\vert \hat{%
\Psi}\left( \hat{K}_{1},\hat{X}_{1}\right) \right\vert ^{2}d\hat{K}_{1}d\hat{%
X}_{1}  \label{G} \\
\underline{\hat{k}}\left( \hat{X}^{\prime }\right) &=&\int \hat{k}\left( 
\hat{X}^{\prime },\hat{X}_{1}\right) \hat{K}_{1}\left\vert \hat{\Psi}\left( 
\hat{K}_{1},\hat{X}_{1}\right) \right\vert ^{2}d\hat{K}_{1}d\hat{X}_{1} 
\notag
\end{eqnarray}%
w sm:%
\begin{equation*}
\hat{k}\left( \hat{X}^{\prime },\hat{X}_{1}\right) \rightarrow \frac{\hat{k}%
\left( \hat{X}^{\prime },\hat{X}_{1}\right) }{\left\langle \hat{K}%
\right\rangle }
\end{equation*}%
\begin{equation}
\underline{\hat{k}}\left( \hat{X}^{\prime }\right) =\int \hat{k}\left( \hat{X%
}^{\prime },\hat{X}_{1}\right) \hat{K}_{1}\left\vert \hat{\Psi}\left( \hat{K}%
_{1},\hat{X}_{1}\right) \right\vert ^{2}d\hat{K}_{1}d\hat{X}_{1}\simeq \hat{k%
}\left( \hat{X}^{\prime },\left\langle \hat{X}\right\rangle \right)
\left\Vert \hat{\Psi}\right\Vert ^{2}  \label{GP}
\end{equation}

\subsubsection*{A11.7.1 Firm average return and average capital}

In average, and given the normalisations:%
\begin{eqnarray*}
\frac{1}{1+\left\langle \underline{\hat{k}}\left( X\right) \right\rangle }
&=&\frac{1}{1+\left\langle \left( \hat{k}\left( X,X^{\prime }\right) -\hat{k}%
\left( \left\langle X\right\rangle ,\left\langle X\right\rangle \right)
\right) \frac{\hat{K}_{X^{\prime }}\left\Vert \hat{\Psi}\left( \hat{X}%
^{\prime }\right) \right\Vert ^{2}}{\left\langle \hat{K}\right\rangle
\left\Vert \hat{\Psi}\right\Vert ^{2}}\right\rangle }\rightarrow 1 \\
\left\langle \underline{k}\left( X\right) \right\rangle &=&\left\langle
k\left( X,X^{\prime }\right) \frac{\hat{K}_{X^{\prime }}\left\Vert \hat{\Psi}%
\left( \hat{X}^{\prime }\right) \right\Vert ^{2}}{\left\langle
K\right\rangle \left\Vert \Psi \right\Vert ^{2}}\right\rangle \\
&\simeq &k\left( \left\langle X\right\rangle ,\left\langle \hat{X}%
\right\rangle \right) \frac{\left\langle \hat{K}\right\rangle }{\left\langle
K\right\rangle }\frac{\left\Vert \hat{\Psi}\right\Vert ^{2}}{\left\Vert \Psi
\right\Vert ^{2}}\simeq \left\langle \underline{k}\right\rangle \frac{%
\left\langle \hat{K}\right\rangle }{\left\langle K\right\rangle }\frac{%
\left\Vert \hat{\Psi}_{0}\right\Vert ^{2}}{\left\Vert \Psi _{0}\right\Vert
^{2}}
\end{eqnarray*}%
to this order we replace $\hat{g}\left( \left\langle \hat{X}\right\rangle
\right) \rightarrow \bar{r}^{\prime }$ so that using: (\ref{VTR}):

\begin{equation}
\left\langle \hat{K}\right\rangle =\frac{3}{4\bar{r}^{\prime }}\sqrt{\frac{%
\sigma _{\hat{K}}^{2}}{2\hat{\mu}}}\frac{1-2\frac{2+\underline{\hat{k}}-%
\sqrt{\left( 2+\underline{\hat{k}}\right) ^{2}-\underline{\hat{k}}}}{9}}{1-%
\frac{2+\underline{\hat{k}}-\sqrt{\left( 2+\underline{\hat{k}}\right) ^{2}-%
\underline{\hat{k}}}}{4\underline{\hat{k}}}}\sqrt{\frac{\left\Vert \hat{\Psi}%
_{0}\right\Vert ^{2}+\hat{\mu}\frac{\bar{r}^{\prime }}{2}\left( \frac{1-r}{r}%
\right) \underline{\hat{k}}}{1-2r\left( 1-r\right) \underline{\hat{k}}}}
\end{equation}

with:%
\begin{equation*}
r=\frac{2+\underline{\hat{k}}-\sqrt{\left( 2+\underline{\hat{k}}\right) ^{2}-%
\underline{\hat{k}}}}{6\hat{k}}
\end{equation*}%
We compute the average firm capital by using:%
\begin{equation*}
\left\langle K\right\rangle =\frac{1}{4\left\langle f_{1}^{\left( e\right)
}\right\rangle }\left\langle \frac{\left( 3X^{\left( e\right) }-C^{\left(
e\right) }\right) \left( C^{\left( e\right) }+X^{\left( e\right) }\right) }{%
2X^{\left( e\right) }-C^{\left( e\right) }}\right\rangle
\end{equation*}%
and use:%
\begin{eqnarray*}
\left\langle X^{\left( e\right) }\right\rangle &\rightarrow &\sqrt{\frac{%
\left\vert \Psi _{0}\right\vert ^{2}}{\epsilon }-\frac{1}{2}\left\langle
f_{1}\right\rangle -\frac{1}{2}\left( \beta \underline{k}\left( \left\langle
f_{1}\right\rangle -\bar{r}\right) \right) } \\
&\simeq &\sqrt{\frac{\left\vert \Psi _{0}\right\vert ^{2}}{\epsilon }-\frac{1%
}{2}\left\langle f_{1}\right\rangle }-\frac{\beta \underline{k}\left(
\left\langle f_{1}\right\rangle -\bar{r}\right) }{4\sqrt{\frac{\left\vert
\Psi _{0}\right\vert ^{2}}{\epsilon }-\frac{1}{2}\left\langle
f_{1}\right\rangle }}=\sqrt{\frac{\left\vert \Psi _{0}\right\vert ^{2}}{%
\epsilon }-\frac{1}{2}\left\langle f_{1}\right\rangle }-\frac{\beta
\left\langle \underline{k}\right\rangle \frac{\left\langle \hat{K}%
\right\rangle }{\left\langle K\right\rangle }\frac{\left\Vert \hat{\Psi}%
_{0}\right\Vert ^{2}}{\left\Vert \Psi _{0}\right\Vert ^{2}}\left(
\left\langle f_{1}\right\rangle -\bar{r}\right) }{4\sqrt{\frac{\left\vert
\Psi _{0}\right\vert ^{2}}{\epsilon }-\frac{1}{2}\left\langle
f_{1}\right\rangle }}
\end{eqnarray*}%
with:%
\begin{equation*}
\left\langle C^{\left( e\right) }\right\rangle \simeq \beta C
\end{equation*}%
The effective average return $\left\langle f_{1}^{\left( e\right)
}\right\rangle $ for firms is: 
\begin{eqnarray*}
&&\left( \left( 1+\frac{k\left( X\right) }{K\left[ X\right] }\hat{K}\left[ 
\hat{X}\right] \right) ^{r}K_{X}^{r}\right) \left\langle f_{1}^{\left(
e\right) }\right\rangle \\
&=&\left\langle f_{1}\right\rangle -C_{0}\left\langle \left( 1+\frac{k\left(
X\right) }{K\left[ X\right] }\hat{K}\left[ \hat{X}\right] \right)
^{r}K_{X}^{r}\right\rangle +\beta \underline{k}\left( \left\langle
f_{1}\right\rangle -C_{0}\left\langle \left( 1+\frac{k\left( X\right) }{K%
\left[ X\right] }\hat{K}\left[ \hat{X}\right] \right)
^{r}K_{X}^{r}\right\rangle -\bar{r}\right) \\
&=&\left\langle f_{1}\right\rangle -C_{0}\left( 1+\left\langle \underline{k}%
\right\rangle \frac{\left\langle \hat{K}\right\rangle }{\left\langle
K\right\rangle }\frac{\left\Vert \hat{\Psi}_{0}\right\Vert ^{2}}{\left\Vert
\Psi _{0}\right\Vert ^{2}}\right) ^{r}\left\langle K\right\rangle ^{r}+\beta
\left\langle \underline{k}\right\rangle \frac{\left\langle \hat{K}%
\right\rangle }{\left\langle K\right\rangle }\frac{\left\Vert \hat{\Psi}%
_{0}\right\Vert ^{2}}{\left\Vert \Psi _{0}\right\Vert ^{2}}\left(
\left\langle f_{1}\right\rangle -C_{0}\left( 1+\left\langle \underline{k}%
\right\rangle \frac{\left\langle \hat{K}\right\rangle }{\left\langle
K\right\rangle }\frac{\left\Vert \hat{\Psi}_{0}\right\Vert ^{2}}{\left\Vert
\Psi _{0}\right\Vert ^{2}}\right) ^{r}\left\langle K\right\rangle ^{r}-\bar{r%
}\right) \\
&\simeq &\left\langle f_{1}\right\rangle -C_{0}\left( 1+\left\langle 
\underline{k}\right\rangle \frac{\left\langle \hat{K}\right\rangle }{%
\left\langle K\right\rangle }\frac{\left\Vert \hat{\Psi}_{0}\right\Vert ^{2}%
}{\left\Vert \Psi _{0}\right\Vert ^{2}}\right) ^{r}\left\langle
K\right\rangle ^{r}+\beta \left\langle \underline{k}\right\rangle \frac{%
\left\langle \hat{K}\right\rangle }{\left\langle K\right\rangle }\frac{%
\left\Vert \hat{\Psi}_{0}\right\Vert ^{2}}{\left\Vert \Psi _{0}\right\Vert
^{2}}\left( \left\langle f_{1}\right\rangle -C_{0}\left\langle
K\right\rangle ^{r}-\bar{r}\right) \\
&=&\left\langle f_{1}\right\rangle -C_{0}\left( 1+\left\langle \underline{k}%
\right\rangle \frac{\left\langle \hat{K}\right\rangle }{\left\langle
K\right\rangle }\frac{\left\Vert \hat{\Psi}_{0}\right\Vert ^{2}}{\left\Vert
\Psi _{0}\right\Vert ^{2}}\right) ^{r}\left\langle K\right\rangle ^{r}
\end{eqnarray*}

and:%
\begin{equation*}
\left\langle f_{1}^{\left( e\right) }\right\rangle \simeq \frac{\left\langle
f_{1}\right\rangle }{\left( \left( 1+\frac{k\left( X\right) }{K\left[ X%
\right] }\hat{K}\left[ \hat{X}\right] \right) ^{r}K_{X}^{r}\right) }-C_{0}
\end{equation*}%
\bigskip

leading to the average cp fm:%
\begin{equation*}
\left\langle K\right\rangle \simeq \frac{\left( 3\sqrt{\frac{\left\vert \Psi
_{0}\right\vert ^{2}}{\epsilon }-\frac{1}{2}\left\langle f_{1}\right\rangle }%
-\beta C\right) \left( \sqrt{\frac{\left\vert \Psi _{0}\right\vert ^{2}}{%
\epsilon }-\frac{1}{2}\left\langle f_{1}\right\rangle }+\beta C\right) }{%
4\left( \left\langle f_{1}\right\rangle -C_{0}\left\langle K\right\rangle
^{r}\right) \left( 2\sqrt{\frac{\left\vert \Psi _{0}\right\vert ^{2}}{%
\epsilon }-\frac{1}{2}\left\langle f_{1}\right\rangle }-\beta C\right) }
\end{equation*}%
\begin{equation*}
\left\langle K\right\rangle \left( \left\langle f_{1}\right\rangle
-C_{0}\left( 1+\left\langle \underline{k}\right\rangle \frac{\left\langle 
\hat{K}\right\rangle }{\left\langle K\right\rangle }\frac{\left\Vert \hat{%
\Psi}_{0}\right\Vert ^{2}}{\left\Vert \Psi _{0}\right\Vert ^{2}}\right)
^{r}\left\langle K\right\rangle ^{r}\right) \simeq \frac{\left( 3\sqrt{\frac{%
\left\vert \Psi _{0}\right\vert ^{2}}{\epsilon }-\frac{1}{2}\left\langle
f_{1}\right\rangle }-\beta C\right) \left( \sqrt{\frac{\left\vert \Psi
_{0}\right\vert ^{2}}{\epsilon }-\frac{1}{2}\left\langle f_{1}\right\rangle }%
+\beta C\right) }{4\left( 2\sqrt{\frac{\left\vert \Psi _{0}\right\vert ^{2}}{%
\epsilon }-\frac{1}{2}\left\langle f_{1}\right\rangle }-\beta C\right) }
\end{equation*}%
Define:%
\begin{eqnarray*}
\left\langle K\right\rangle _{1}^{r} &\simeq &\frac{\left\langle
f_{1}\right\rangle }{C_{0}\left( 1+\left\langle \underline{k}\right\rangle 
\frac{\left\langle \hat{K}\right\rangle }{\left\langle K\right\rangle }\frac{%
\left\Vert \hat{\Psi}_{0}\right\Vert ^{2}}{\left\Vert \Psi _{0}\right\Vert
^{2}}\right) ^{r}} \\
&\simeq &\frac{\left\langle f_{1}\right\rangle }{C_{0}\left( 1+\left\langle 
\underline{k}\right\rangle \frac{\left\langle \hat{K}\right\rangle }{%
\left\langle f_{1}\right\rangle ^{r}}C_{0}^{r}\frac{\left\Vert \hat{\Psi}%
_{0}\right\Vert ^{2}}{\left\Vert \Psi _{0}\right\Vert ^{2}}\right) ^{r}}=%
\frac{\left\langle f_{1}\right\rangle ^{1+r^{2}}}{C_{0}\left( \left\langle
f_{1}\right\rangle ^{r}+\left\langle \underline{k}\right\rangle \left\langle 
\hat{K}\right\rangle C_{0}^{r}\frac{\left\Vert \hat{\Psi}_{0}\right\Vert ^{2}%
}{\left\Vert \Psi _{0}\right\Vert ^{2}}\right) ^{r}}
\end{eqnarray*}%
that is:%
\begin{equation*}
\left\langle K\right\rangle _{1}=\frac{\left\langle f_{1}\right\rangle ^{%
\frac{1+r^{2}}{r}}}{C_{0}^{\frac{1}{r}}\left( \left\langle
f_{1}\right\rangle ^{r}+\left\langle \underline{k}\right\rangle \left\langle 
\hat{K}\right\rangle C_{0}^{r}\frac{\left\Vert \hat{\Psi}_{0}\right\Vert ^{2}%
}{\left\Vert \Psi _{0}\right\Vert ^{2}}\right) }
\end{equation*}%
and the average capital for firms is obtained as correction of $\left\langle
K\right\rangle _{1}^{r}$: 
\begin{equation*}
\left\langle K\right\rangle ^{r}=\left\langle K\right\rangle _{1}^{r}-\frac{%
\left( 3\sqrt{\frac{\left\vert \Psi _{0}\right\vert ^{2}}{\epsilon }-\frac{1%
}{2}\left\langle f_{1}\right\rangle }-\beta C\right) \left( \sqrt{\frac{%
\left\vert \Psi _{0}\right\vert ^{2}}{\epsilon }-\frac{1}{2}\left\langle
f_{1}\right\rangle }+\beta C\right) \left\langle K\right\rangle _{1}^{r}}{%
4\left\langle f_{1}\right\rangle \left\langle K\right\rangle _{1}\left( 2%
\sqrt{\frac{\left\vert \Psi _{0}\right\vert ^{2}}{\epsilon }-\frac{1}{2}%
\left\langle f_{1}\right\rangle }-\beta C\right) }
\end{equation*}%
that writes:%
\begin{equation*}
\left\langle K\right\rangle \simeq \left\langle K\right\rangle _{1}\left( 1-%
\frac{\left( 3\sqrt{\frac{\left\vert \Psi _{0}\right\vert ^{2}}{\epsilon }-%
\frac{1}{2}\left\langle f_{1}\right\rangle }-\beta C\right) \left( \sqrt{%
\frac{\left\vert \Psi _{0}\right\vert ^{2}}{\epsilon }-\frac{1}{2}%
\left\langle f_{1}\right\rangle }+\beta C\right) }{4\left\langle
f_{1}\right\rangle \left\langle K\right\rangle _{1}\left( 2\sqrt{\frac{%
\left\vert \Psi _{0}\right\vert ^{2}}{\epsilon }-\frac{1}{2}\left\langle
f_{1}\right\rangle }-\beta C\right) }\right) ^{\frac{1}{r}}
\end{equation*}%
and the effective return:%
\begin{eqnarray*}
\left\langle f_{1}^{\left( e\right) }\right\rangle &=&\left\langle
f_{1}\right\rangle \left( 1-\beta \left\langle \underline{k}\right\rangle 
\frac{\left( \frac{1}{4}-\frac{r}{3}\hat{k}\right) }{\left( \frac{1}{3}-%
\frac{r}{2}\hat{k}\right) \bar{r}^{\prime }}\frac{4\left( 2\sqrt{\frac{%
\left\vert \Psi _{0}\right\vert ^{2}}{\epsilon }-\frac{1}{2}\left\langle
f_{1}\right\rangle }-\beta C\right) \sqrt{\frac{\sigma _{\hat{K}}^{2}}{2\hat{%
\mu}}\left( \left\Vert \hat{\Psi}_{0}\right\Vert ^{2}-\hat{\mu}\left\langle
D\right\rangle \right) }}{\left( 3\sqrt{\frac{\left\vert \Psi
_{0}\right\vert ^{2}}{\epsilon }-\frac{1}{2}\left\langle f_{1}\right\rangle }%
-\beta C\right) \left( \sqrt{\frac{\left\vert \Psi _{0}\right\vert ^{2}}{%
\epsilon }-\frac{1}{2}\left\langle f_{1}\right\rangle }+\beta C\right) }%
\frac{\left\Vert \hat{\Psi}_{0}\right\Vert ^{2}}{\left\Vert \Psi
_{0}\right\Vert ^{2}}\left( \left\langle f_{1}\right\rangle -\bar{r}\right)
\right) ^{-1} \\
&\simeq &\left\langle f_{1}\right\rangle +\beta \left\langle \underline{k}%
\right\rangle \frac{\left( \frac{1}{4}-\frac{r}{3}\hat{k}\right) }{\left( 
\frac{1}{3}-\frac{r}{2}\hat{k}\right) \bar{r}^{\prime }} \\
&&\times \frac{4\left\langle f_{1}^{\left( e\right) }\right\rangle \left( 2%
\sqrt{\frac{\left\vert \Psi _{0}\right\vert ^{2}}{\epsilon }-\frac{1}{2}%
\left\langle f_{1}\right\rangle }-\beta C\right) \sqrt{\frac{\sigma _{\hat{K}%
}^{2}}{2\hat{\mu}}\left( \left\Vert \hat{\Psi}_{0}\right\Vert ^{2}-\hat{\mu}%
\left\langle D\right\rangle \right) }}{\left( 3\sqrt{\frac{\left\vert \Psi
_{0}\right\vert ^{2}}{\epsilon }-\frac{1}{2}\left\langle f_{1}\right\rangle }%
-\beta C\right) \left( \sqrt{\frac{\left\vert \Psi _{0}\right\vert ^{2}}{%
\epsilon }-\frac{1}{2}\left\langle f_{1}\right\rangle }+\beta C\right) }%
\frac{\left\Vert \hat{\Psi}_{0}\right\Vert ^{2}}{\left\Vert \Psi
_{0}\right\Vert ^{2}}\left( \left\langle f_{1}\right\rangle -\bar{r}\right)
\end{eqnarray*}

\subsubsection*{A11.7.2 Investors average return}

We consider the average for investors' return by starting with $\hat{S}%
_{1}^{E}\left( \hat{X}^{\prime },\hat{X}_{1}\right) $:%
\begin{eqnarray*}
&&\left( 1+\underline{\hat{k}}_{2}\left( \hat{X}_{1}\right) \right) \hat{S}%
_{1}^{E}\left( \hat{X}^{\prime },\hat{X}_{1}\right) \\
&\rightarrow &\frac{1+\underline{\hat{k}}_{2}\left( \hat{X}_{1}\right) }{1+%
\hat{k}\left( \hat{X}\right) }\left( \frac{\hat{k}\left( \hat{X},\hat{X}%
^{\prime }\right) -\underline{\hat{k}}_{1}\left( \left\langle X\right\rangle
\right) \hat{k}\left( \left\langle \hat{X}\right\rangle ,\hat{X}^{\prime
}\right) }{1+\underline{\hat{k}}_{2}\left( \hat{X}\right) }+\frac{\hat{k}%
_{1}\left( \hat{X}^{\prime },\hat{X}\right) }{1+\underline{\hat{k}}%
_{2}\left( \hat{X}^{\prime }\right) }\right) \\
&&\times \frac{\sqrt{\left\Vert \hat{\Psi}_{0}\left( \hat{X}\right)
\right\Vert ^{2}-\hat{\mu}D\left( \hat{X}\right) }\left( \left\Vert \hat{\Psi%
}_{0}\left( \hat{X}^{\prime }\right) \right\Vert ^{2}-\hat{\mu}D\left( \hat{X%
}^{\prime }\right) \right) ^{\frac{3}{2}}}{\left( \left\Vert \hat{\Psi}%
_{0}\right\Vert ^{2}-\hat{\mu}\left\langle D\right\rangle \right) ^{2}}
\end{eqnarray*}%
so that:

\begin{eqnarray*}
&&\left\langle \left( 1+\beta \underline{\hat{k}}\left( \left\langle \hat{X}%
\right\rangle \right) \right) \hat{S}_{1}^{E}\right\rangle \\
&\simeq &\left\langle \left( \frac{\hat{k}\left( \hat{X},\hat{X}^{\prime
}\right) }{1+\underline{\hat{k}}\left( \hat{X}\right) }-\frac{\left( 1-\beta
\right) \hat{k}\left( \left\langle X\right\rangle ,\hat{X}^{\prime }\right) 
}{\left( 1+\underline{\hat{k}}\left( \left\langle \hat{X}\right\rangle
\right) \right) ^{2}}+\frac{\left( 1-\beta \right) \hat{k}_{1}^{E}\left( 
\hat{X},\left\langle \hat{X}\right\rangle \right) -\left( 1-\beta \right) 
\hat{k}\left( \left\langle \hat{X}\right\rangle ,\hat{X}^{\prime }\right) }{%
1+\beta \underline{\hat{k}}\left( \left\langle \hat{X}\right\rangle \right) }%
\right) \right\rangle \\
&\simeq &\left\langle \underline{\hat{k}}\right\rangle +\left( 1-\beta
\right) \left\langle \underline{\hat{k}}\right\rangle
\end{eqnarray*}%
As a consequence:%
\begin{equation*}
\left\langle \left( 1-\left( 1+\underline{\hat{k}}_{2}\left( \hat{X}%
_{1}\right) \right) \hat{S}_{1}^{E}\left( \hat{X}^{\prime },\hat{X}%
_{1}\right) \right) ^{-1}\right\rangle =\left( 1-\left( 2-\beta \right)
\left\langle \hat{k}\right\rangle \right) ^{-1}=\frac{1}{1-\left( 2-\beta
\right) \left\langle \hat{k}\right\rangle }
\end{equation*}%
and the average return is given by:%
\begin{eqnarray*}
\left\langle \hat{g}\right\rangle &=&\frac{\left( 1-\left( 2-\beta \right)
\left\langle \underline{\hat{k}}\right\rangle \right) ^{-1}}{\left(
1+\left\langle \underline{\hat{k}}_{2}\right\rangle \right) } \\
&&\times \left\langle \left( 1+\underline{\hat{k}}_{2}\left( \hat{X}%
_{1}\right) \right) \left( \frac{\epsilon \left( 1-\beta \right) \left(
X+C\beta \right) ^{2}\left( \frac{\left( 3X-\beta C\right) \left( \beta
C+X\right) }{4f_{1}^{2}\left( X\right) C\left( 2X-C\beta \right) }+\frac{%
\beta C\left( 2X-C\beta \right) }{f_{1}^{3}\left( X\right) }\right) }{%
3\sigma _{\hat{K}}^{2}}\left( R+\Delta F_{\tau }\left( \bar{R}\left(
K,X\right) \right) \right) \right) \right\rangle \\
&=&\frac{1}{1-\left( 2-\beta \right) \left\langle \underline{\hat{k}}%
\right\rangle }\left\langle \Delta \right\rangle +\bar{r}^{\prime }
\end{eqnarray*}%
given the normalization:%
\begin{equation*}
\left\langle \underline{\hat{k}}\right\rangle \rightarrow 0
\end{equation*}%
this leads to:%
\begin{equation*}
\left\langle \hat{g}\right\rangle =\left\langle \Delta \right\rangle +\bar{r}%
^{\prime }
\end{equation*}%
with:%
\begin{eqnarray*}
\left\langle \Delta \right\rangle &=&\left\langle \left( \frac{A}{\left(
f_{1}^{\left( r\right) }\left( X\right) -C_{0}\right) ^{2}}+\frac{B}{\left(
f_{1}^{\left( r\right) }\left( X\right) -C_{0}\right) ^{3}}\right) \right. \\
&&\left. \times \left( f_{1}^{\left( r\right) }\left( X\right) +\Delta
F_{\tau }\left( \bar{R}\left( K,X\right) \right) \right) \right\rangle
\end{eqnarray*}%
with:%
\begin{equation*}
f_{1}^{\left( r\right) }\left( X\right) =\frac{f_{1}\left( X\right) }{\left(
1+\left\langle \underline{k}\right\rangle \left\langle \hat{K}\right\rangle 
\frac{\left\Vert \hat{\Psi}_{0}\right\Vert ^{2}}{\left\Vert \Psi
_{0}\right\Vert ^{2}}\right) ^{r}}-C_{0}
\end{equation*}

\subsubsection*{A11.7.3 Investors capital}

Having found an expression for $\left\langle \hat{g}\right\rangle $ we can
correct the previous expression for $\left\langle \hat{K}\right\rangle $ at
first order. Using (\ref{VRf}) (\ref{VRs}) we find:%
\begin{equation}
\left\Vert \hat{\Psi}\right\Vert ^{2}\simeq \frac{\hat{\mu}V}{3\sigma _{\hat{%
K}}^{2}\left\langle \hat{g}\right\rangle }\left( 2\frac{\sigma _{\hat{K}%
}^{2}\left( \frac{\left\Vert \hat{\Psi}_{0}\right\Vert ^{2}}{\hat{\mu}}+%
\frac{\left\langle \hat{g}\right\rangle }{2}\left( \frac{1-r}{r}\right) 
\underline{\hat{k}}\right) }{\left( 1-2r\left( 1-r\right) \underline{\hat{k}}%
\right) }\right) ^{\frac{3}{2}}\left( 1-\frac{2+\underline{\hat{k}}-\sqrt{%
\left( 2+\underline{\hat{k}}\right) ^{2}-\underline{\hat{k}}}}{4\underline{%
\hat{k}}}\underline{\hat{k}}\right)
\end{equation}%
\begin{equation}
\left\langle \hat{K}\right\rangle \left\Vert \hat{\Psi}\right\Vert ^{2}=%
\frac{\hat{\mu}V\sigma _{\hat{K}}^{2}}{2\left\langle \hat{g}\right\rangle
^{2}}\left( \frac{\left( \frac{\left\Vert \hat{\Psi}_{0}\right\Vert ^{2}}{%
\hat{\mu}}+\frac{\left\langle \hat{g}\right\rangle }{2}\left( \frac{1-r}{r}%
\right) \underline{\hat{k}}\right) }{\left( 1-2r\left( 1-r\right) \underline{%
\hat{k}}\right) }\right) ^{2}\left( 1-2\frac{2+\underline{\hat{k}}-\sqrt{%
\left( 2+\underline{\hat{k}}\right) ^{2}-\underline{\hat{k}}}}{9}\right)
\label{Pr}
\end{equation}%
where:%
\begin{equation*}
\left\langle \hat{g}\right\rangle =\left\langle \Delta \right\rangle +\bar{r}%
^{\prime }
\end{equation*}

\subsubsection*{A11.7.4 Dependency in productivity}

The derivative of (\ref{Pr}) with respect to $\left\langle f_{1}\left(
X\right) \right\rangle $ is: 
\begin{equation*}
\frac{\partial \left\langle \hat{K}\right\rangle \left\Vert \hat{\Psi}%
\right\Vert ^{2}}{\partial \left\langle f_{1}\left( X\right) \right\rangle }%
=-\hat{\mu}V\sigma _{\hat{K}}^{2}\frac{\left\Vert \hat{\Psi}_{0}\right\Vert
^{2}}{\hat{\mu}\left\langle \hat{g}\right\rangle ^{2}}\left( \frac{\frac{%
\left\Vert \hat{\Psi}_{0}\right\Vert ^{2}}{\hat{\mu}\left\langle \hat{g}%
\right\rangle }+\frac{1}{2}\left( \frac{1-r}{r}\right) \underline{\hat{k}}}{%
\left( 1-2r\left( 1-r\right) \underline{\hat{k}}\right) }\right) \left( 1-2%
\frac{2+\underline{\hat{k}}-\sqrt{\left( 2+\underline{\hat{k}}\right) ^{2}-%
\underline{\hat{k}}}}{9}\right) \frac{\partial \left\langle \Delta
\right\rangle }{\partial \left\langle f_{1}\left( X\right) \right\rangle }
\end{equation*}%
Assuming that $\left\langle \Delta F_{\tau }\left( \bar{R}\left( K,X\right)
\right) \right\rangle =0$, define:%
\begin{equation*}
f_{1}^{\left( r\right) }\left( X\right) =\frac{f_{1}\left( X\right) }{\left(
1+\left\langle \underline{k}\right\rangle \left\langle \hat{K}\right\rangle 
\frac{\left\Vert \hat{\Psi}_{0}\right\Vert ^{2}}{\left\Vert \Psi
_{0}\right\Vert ^{2}}\right) ^{r}}
\end{equation*}
$\left\langle \Delta \right\rangle $ has the form:%
\begin{equation*}
\left\langle \Delta \right\rangle =\left\langle \left( \frac{A}{\left(
f_{1}^{\left( r\right) }\left( X\right) -C_{0}\right) ^{2}}+\frac{B}{\left(
f_{1}^{\left( r\right) }\left( X\right) -C_{0}\right) ^{3}}\right) \left(
f_{1}^{\left( r\right) }\left( X\right) -C_{0}-\bar{r}\right) \right\rangle
\end{equation*}%
we obtn:%
\begin{eqnarray}
\frac{\partial \left\langle \Delta \right\rangle }{\partial \left\langle
f_{1}\left( X\right) \right\rangle } &=&-\left( \frac{1-rf_{1}^{\left(
r\right) }\left( X\right) \frac{\partial \left\langle \hat{K}\right\rangle
\left\Vert \hat{\Psi}\right\Vert ^{2}}{\partial \left\langle f_{1}\left(
X\right) \right\rangle }\frac{\left\langle \underline{k}\right\rangle
\left\langle K\right\rangle ^{r}}{\left\langle K\right\rangle \left\Vert
\Psi _{0}\right\Vert ^{2}}}{1+\left\langle \underline{k}\right\rangle \frac{%
\left\langle \hat{K}\right\rangle }{\left\langle K\right\rangle }\frac{%
\left\Vert \hat{\Psi}_{0}\right\Vert ^{2}}{\left\Vert \Psi _{0}\right\Vert
^{2}}}\right)  \label{DL} \\
&&\times \left( \frac{A\left( \left( f_{1}^{\left( r\right) }\left( X\right)
-C_{0}\right) -2\bar{r}\right) }{\left( f_{1}^{\left( r\right) }\left(
X\right) -C_{0}\right) ^{3}}+\frac{B\left( 2\left( f_{1}^{\left( r\right)
}\left( X\right) -C_{0}\right) -3\bar{r}\right) }{\left( f_{1}^{\left(
r\right) }\left( X\right) -C_{0}\right) ^{4}}\right)  \notag
\end{eqnarray}

We set:%
\begin{equation*}
W=\hat{\mu}V\sigma _{\hat{K}}^{2}\frac{\left\Vert \hat{\Psi}_{0}\right\Vert
^{2}}{\hat{\mu}\left\langle \hat{g}\right\rangle ^{2}}\left( \frac{\frac{%
\left\Vert \hat{\Psi}_{0}\right\Vert ^{2}}{\hat{\mu}\left\langle \hat{g}%
\right\rangle }+\frac{1}{2}\left( \frac{1-r}{r}\right) \underline{\hat{k}}}{%
\left( 1-2r\left( 1-r\right) \underline{\hat{k}}\right) }\right) \left( 1-2%
\frac{2+\underline{\hat{k}}-\sqrt{\left( 2+\underline{\hat{k}}\right) ^{2}-%
\underline{\hat{k}}}}{9}\right)
\end{equation*}%
so that%
\begin{equation*}
\frac{\partial \left\langle \hat{K}\right\rangle \left\Vert \hat{\Psi}%
\right\Vert ^{2}}{\partial \left\langle f_{1}\left( X\right) \right\rangle }=%
\frac{W\left( A\left( \left( f_{1}^{\left( r\right) }\left( X\right)
-C_{0}\right) -2\bar{r}\right) +\frac{B\left( 2\left( f_{1}^{\left( r\right)
}\left( X\right) -C_{0}\right) -3\bar{r}\right) }{f_{1}^{\left( r\right)
}\left( X\right) -C_{0}}\right) }{\left( 1+W\frac{rf_{1}^{\left( r\right)
}\left( X\right) \frac{\partial \left\langle \hat{K}\right\rangle \left\Vert 
\hat{\Psi}\right\Vert ^{2}}{\partial \left\langle f_{1}\left( X\right)
\right\rangle }\frac{\left\langle \underline{k}\right\rangle \left\langle
K\right\rangle ^{r}}{\left\langle K\right\rangle \left\Vert \Psi
_{0}\right\Vert ^{2}}}{1+\left\langle \underline{k}\right\rangle \frac{%
\left\langle \hat{K}\right\rangle }{\left\langle K\right\rangle }\frac{%
\left\Vert \hat{\Psi}_{0}\right\Vert ^{2}}{\left\Vert \Psi _{0}\right\Vert
^{2}}}\right) \left( A\left( \left( f_{1}^{\left( r\right) }\left( X\right)
-C_{0}\right) -2\bar{r}\right) +\frac{B\left( 2\left( f_{1}^{\left( r\right)
}\left( X\right) -C_{0}\right) -3\bar{r}\right) }{f_{1}^{\left( r\right)
}\left( X\right) -C_{0}}\right) }
\end{equation*}

leading to:%
\begin{equation*}
\frac{\partial \left\langle \hat{K}\right\rangle \left\Vert \hat{\Psi}%
\right\Vert ^{2}}{\partial \left\langle f_{1}\left( X\right) \right\rangle }%
>0
\end{equation*}%
in the main case. Actually, the standard marginal return condition writes
here:%
\begin{equation*}
\left( 1-r\right) f_{1}^{\left( r\right) }\left( X\right) -C_{0}=\bar{r}
\end{equation*}%
So that the condition:%
\begin{equation*}
\left( f_{1}^{\left( r\right) }\left( X\right) -C_{0}\right) -2\bar{r}>0
\end{equation*}%
writes:%
\begin{equation*}
\frac{r}{\left( 1-r\right) }C_{0}+\left( \frac{1}{\left( 1-r\right) }%
-2\right) \bar{r}>0
\end{equation*}%
which is statisfied in general.

Similarly, writing:%
\begin{equation}
\left\Vert \hat{\Psi}\right\Vert ^{2}\simeq \frac{\hat{\mu}V}{3\sigma _{\hat{%
K}}^{2}\left\langle \hat{g}\right\rangle }\left( 2\frac{\sigma _{\hat{K}%
}^{2}\left( \frac{\left\Vert \hat{\Psi}_{0}\right\Vert ^{2}}{\hat{\mu}}+%
\frac{\left\langle \hat{g}\right\rangle }{2}\left( \frac{1-r}{r}\right) 
\underline{\hat{k}}\right) }{\left( 1-2r\left( 1-r\right) \underline{\hat{k}}%
\right) }\right) ^{\frac{3}{2}}\left( 1-\frac{2+\underline{\hat{k}}-\sqrt{%
\left( 2+\underline{\hat{k}}\right) ^{2}-\underline{\hat{k}}}}{4\underline{%
\hat{k}}}\underline{\hat{k}}\right)
\end{equation}%
we hav:%
\begin{equation*}
\frac{\partial \left\Vert \hat{\Psi}\right\Vert ^{2}}{\partial \left\langle
f_{1}\left( X\right) \right\rangle }=H\frac{\partial \left\langle \Delta
\right\rangle }{\partial \left\langle f_{1}\left( X\right) \right\rangle }
\end{equation*}%
where:%
\begin{equation*}
H=\frac{\partial }{\partial \left\langle \hat{g}\right\rangle }\left[ \frac{%
\hat{\mu}V}{3\sigma _{\hat{K}}^{2}\left\langle \hat{g}\right\rangle }\left( 2%
\frac{\sigma _{\hat{K}}^{2}\left( \frac{\left\Vert \hat{\Psi}_{0}\right\Vert
^{2}}{\hat{\mu}}+\frac{\left\langle \hat{g}\right\rangle }{2}\left( \frac{1-r%
}{r}\right) \underline{\hat{k}}\right) }{\left( 1-2r\left( 1-r\right) 
\underline{\hat{k}}\right) }\right) ^{\frac{3}{2}}\left( 1-\frac{2+%
\underline{\hat{k}}-\sqrt{\left( 2+\underline{\hat{k}}\right) ^{2}-%
\underline{\hat{k}}}}{4\underline{\hat{k}}}\underline{\hat{k}}\right) \right]
\end{equation*}%
\begin{equation*}
\frac{\left( 10+x\right) ^{\frac{3}{2}}}{x}
\end{equation*}%
so that, using (\ref{DL}):%
\begin{equation*}
\frac{\partial \left\Vert \hat{\Psi}\right\Vert ^{2}}{\partial \left\langle
f_{1}\left( X\right) \right\rangle }>0
\end{equation*}%
for $\left\langle \hat{g}\right\rangle $ in usual ranges.

\subsection*{A11.8 Approximate solutions to (\protect\ref{NCr})}

To find approximate solutions to (\ref{NCr}), we rewrite the equation as: 
\begin{eqnarray*}
&&\left( 1-\left( 1+\underline{\hat{k}}_{2}\left( \hat{X}\right) \right) 
\hat{S}_{1}^{E}\left( \hat{X}^{\prime },\hat{X}\right) \right) \left( \pm 
\frac{N\left( \hat{X}\right) }{\sqrt{\hat{K}\left[ \hat{X}\right] }}-\bar{r}%
^{\prime }\right) \\
&=&\left( 1+\underline{\hat{k}}_{2}\left( \hat{X}\right) \right) \left( 
\frac{A}{\left( \frac{f_{1}\left( X^{\prime }\right) }{\left( \left( 1+%
\underline{k}\left( X\right) \hat{K}\left[ \hat{X}^{\prime }\right] \right)
\right) ^{r}}-C_{0}\right) ^{2}}+\frac{B}{\left( \frac{f_{1}\left( X^{\prime
}\right) }{\left( \left( 1+\underline{k}\left( X\right) \hat{K}\left[ \hat{X}%
^{\prime }\right] \right) \right) ^{r}}-C_{0}\right) ^{3}}\right) \\
&&\times \left[ \left( \frac{f_{1}\left( X^{\prime }\right) }{\left( \left(
1+\underline{k}\left( X^{\prime }\right) \hat{K}\left[ \hat{X}\right]
\right) \right) ^{r}}-C_{0}-\bar{r}^{\prime }\right) +\tau F\left( X^{\prime
}\right) \left( \frac{f_{1}\left( X^{\prime }\right) }{\left( \left( 1+%
\underline{k}\left( X\right) \hat{K}\left[ \hat{X}^{\prime }\right] \right)
\right) ^{r}}-\left\langle \frac{f_{1}\left( X\right) }{\left( \left( 1+%
\underline{k}\left( X\right) \hat{K}\left[ \hat{X}\right] \right) \right)
^{r}}\right\rangle \right) \right]
\end{eqnarray*}%
We define $\hat{K}_{1}\left[ \hat{X}\right] $ the solution without
interactions between sectrs, and consider the difference between
interaction/no interaction cases by:%
\begin{equation*}
\frac{N\left( \hat{X}\right) }{\sqrt{\hat{K}\left[ \hat{X}\right] }}=\frac{%
N\left( \hat{X}\right) }{\sqrt{\hat{K}_{1}\left[ \hat{X}\right] }}+\Delta
\left( \frac{N\left( \hat{X}\right) }{\sqrt{\hat{K}\left[ \hat{X}\right] }}%
\right)
\end{equation*}%
so that the equation for $\frac{N\left( \hat{X}\right) }{\sqrt{\hat{K}\left[ 
\hat{X}\right] }}$ becomes: 
\begin{eqnarray*}
&&\left( 1-\left( 1+\underline{\hat{k}}_{2}\left( \hat{X}\right) \right) 
\hat{S}_{1}^{E}\left( \hat{X}^{\prime },\hat{X}\right) \right) \left( \pm 
\frac{N\left( \hat{X}\right) }{\sqrt{\hat{K}_{1}\left[ \hat{X}\right] }}-%
\bar{r}^{\prime }\pm \Delta \left( \frac{N\left( \hat{X}\right) }{\sqrt{\hat{%
K}\left[ \hat{X}\right] }}\right) \right) \\
&=&\pm \left( \frac{N\left( \hat{X}\right) }{\sqrt{\hat{K}_{1}\left[ \hat{X}%
\right] }}+H_{1}\Delta \left( \frac{N\left( \hat{X}\right) }{\sqrt{\hat{K}%
\left[ \hat{X}\right] }}\right) +\frac{1}{2}H_{2}\left( \Delta \left( \frac{%
N\left( \hat{X}\right) }{\sqrt{\hat{K}\left[ \hat{X}\right] }}\right)
\right) ^{2}\right)
\end{eqnarray*}%
where the coefficients $H_{l}$ are given by:%
\begin{eqnarray*}
H_{l} &=&\frac{\left( 1+\underline{\hat{k}}_{2}\left( \hat{X}\right) \right) 
}{\left( N\left( \hat{X}\right) \right) ^{l}}\frac{\partial ^{l}}{\partial
\left( \frac{1}{\sqrt{\hat{K}\left[ \hat{X}\right] }}\right) ^{l}}\left( 
\frac{A}{\left( \frac{f_{1}\left( X^{\prime }\right) }{\left( \left( 1+%
\underline{k}\left( X\right) \hat{K}\left[ \hat{X}^{\prime }\right] \right)
\right) ^{r}}-C_{0}\right) ^{2}}+\frac{B}{\left( \frac{f_{1}\left( X^{\prime
}\right) }{\left( \left( 1+\underline{k}\left( X\right) \hat{K}\left[ \hat{X}%
^{\prime }\right] \right) \right) ^{r}}-C_{0}\right) ^{3}}\right) \\
&&\times \left[ \left( \frac{f_{1}\left( X^{\prime }\right) }{\left( \left(
1+\underline{k}\left( X^{\prime }\right) \hat{K}\left[ \hat{X}\right]
\right) \right) ^{r}}-C_{0}-\bar{r}^{\prime }\right) +\tau F\left( X^{\prime
}\right) \left( \frac{f_{1}\left( X^{\prime }\right) }{\left( \left( 1+%
\underline{k}\left( X\right) \hat{K}\left[ \hat{X}^{\prime }\right] \right)
\right) ^{r}}-\left\langle \frac{f_{1}\left( X\right) }{\left( \left( 1+%
\underline{k}\left( X\right) \hat{K}\left[ \hat{X}\right] \right) \right)
^{r}}\right\rangle \right) \right]
\end{eqnarray*}%
We thus find a second order expansion for (\ref{NCr}):%
\begin{eqnarray*}
&&0=\frac{1}{2}H_{2}\left( \Delta \left( \frac{N\left( \hat{X}\right) }{%
\sqrt{\hat{K}\left[ \hat{X}\right] }}\right) \right) ^{2}-\left(
1-H_{1}\right) \Delta \left( \frac{N\left( \hat{X}\right) }{\sqrt{\hat{K}%
\left[ \hat{X}\right] }}\right) \\
&&+\left( 1+\underline{\hat{k}}_{2}\left( \hat{X}\right) \right) \hat{S}%
_{1}^{E}\left( \hat{X}^{\prime },\hat{X}\right) \frac{N\left( \hat{X}%
^{\prime }\right) }{\sqrt{\hat{K}_{1}\left[ \hat{X}^{\prime }\right] }}%
+\left( 1+\underline{\hat{k}}_{2}\left( \hat{X}\right) \right) \hat{S}%
_{1}^{E}\left( \hat{X}^{\prime },\hat{X}\right) \Delta \left( \frac{N\left( 
\hat{X}^{\prime }\right) }{\sqrt{\hat{K}\left[ \hat{X}^{\prime }\right] }}%
\right)
\end{eqnarray*}%
with solution:%
\begin{equation*}
\Delta \left( \frac{N\left( \hat{X}\right) }{\sqrt{\hat{K}\left[ \hat{X}%
\right] }}\right) =\frac{\left( 1-H_{1}\right) \pm \sqrt{\left(
1-H_{1}\right) ^{2}-2H_{2}\left( \left( 1+\underline{\hat{k}}_{2}\left( \hat{%
X}\right) \right) \hat{S}_{1}^{E}\left( \hat{X}^{\prime },\hat{X}\right)
\left( \frac{N\left( \hat{X}^{\prime }\right) }{\sqrt{\hat{K}_{1}\left[ \hat{%
X}^{\prime }\right] }}+\Delta \left( \frac{N\left( \hat{X}^{\prime }\right) 
}{\sqrt{\hat{K}\left[ \hat{X}^{\prime }\right] }}\right) \right) \right) }}{%
H_{2}}
\end{equation*}%
expanding the square root:%
\begin{equation*}
\sqrt{1-2H_{2}\left( 1+\underline{\hat{k}}_{2}\left( \hat{X}\right) \right) 
\hat{S}_{1}^{E}\left( \hat{X}^{\prime },\hat{X}\right) \frac{N\left( \hat{X}%
^{\prime }\right) }{\sqrt{\hat{K}_{1}\left[ \hat{X}^{\prime }\right] }}}%
\simeq 1-\frac{2H_{2}\left( \left( 1+\underline{\hat{k}}_{2}\left( \hat{X}%
\right) \right) \hat{S}_{1}^{E}\left( \hat{X}^{\prime },\hat{X}\right)
\Delta \left( \frac{N\left( \hat{X}^{\prime }\right) }{\sqrt{\hat{K}\left[ 
\hat{X}^{\prime }\right] }}\right) \right) }{\left( \left( 1-H_{1}\right)
^{2}-2H_{2}\left( 1+\underline{\hat{k}}_{2}\left( \hat{X}\right) \right) 
\hat{S}_{1}^{E}\left( \hat{X}^{\prime },\hat{X}\right) \frac{N\left( \hat{X}%
^{\prime }\right) }{\sqrt{\hat{K}_{1}\left[ \hat{X}^{\prime }\right] }}%
\right) ^{2}}
\end{equation*}%
leads to the lowest order equation:%
\begin{eqnarray*}
&&\left( \delta \left( \hat{X}-\hat{X}^{\prime }\right) \pm _{\hat{X}}\frac{%
2H_{2}\left( \left( 1+\underline{\hat{k}}_{2}\left( \hat{X}\right) \right) 
\hat{S}_{1}^{E}\left( \hat{X}^{\prime },\hat{X}\right) \right) }{\left(
\left( 1-H_{1}\right) ^{2}-2H_{2}\left( 1+\underline{\hat{k}}_{2}\left( \hat{%
X}\right) \right) \hat{S}_{1}^{E}\left( \hat{X}^{\prime },\hat{X}\right) 
\frac{N\left( \hat{X}^{\prime }\right) }{\sqrt{\hat{K}_{1}\left[ \hat{X}%
^{\prime }\right] }}\right) ^{\frac{3}{2}}}\right) \Delta \left( \frac{%
N\left( \hat{X}^{\prime }\right) }{\sqrt{\hat{K}\left[ \hat{X}^{\prime }%
\right] }}\right) \\
&=&\frac{\left( 1-H_{1}\right) \pm _{\hat{X}}\sqrt{\left( 1-H_{1}\right)
^{2}-\left( 2H_{2}\left( 1+\underline{\hat{k}}_{2}\left( \hat{X}\right)
\right) \hat{S}_{1}^{E}\left( \hat{X}^{\prime },\hat{X}\right) \frac{N\left( 
\hat{X}^{\prime }\right) }{\sqrt{\hat{K}_{1}\left[ \hat{X}^{\prime }\right] }%
}\right) }}{H_{2}}
\end{eqnarray*}%
where the sign $\pm _{\hat{X}}$ reminds that the choice of sign depends on
the sector considered. This double possibility per sector implies
multiplicity of solutn.

\section*{Appendix 12 Firms decreasing return to scale: computation of
returns and capital for investors}

\subsection*{A12.1 Firm returns with decreasing return to scale: computation
of field and capital}

We correct the linear approximation by assuming decreasing return to scale.
This amounts to replace the return $f_{1}\left( X\right) $ by a factor
depending on $K$: 
\begin{equation*}
f_{1}\left( X\right) \rightarrow \frac{f_{1}\left( X\right) }{K^{r}}
\end{equation*}%
To include constant cost and depreciation of capital we thus globally
replace:

\begin{equation*}
f_{1}\left( X\right) K\rightarrow \left( \frac{f_{1}\left( X\right) }{K^{r}}%
-C_{0}\right) K
\end{equation*}%
which is also rewritten in terms of private capital:%
\begin{equation*}
f_{1}\left( X\right) \left( \left( 1+\frac{\underline{k}\left( X\right) }{%
\left\langle K_{p}\right\rangle }\hat{K}_{X}\right) K_{p}\right)
^{1-r}-C_{0}\left( 1+\frac{\underline{k}\left( X\right) }{\left\langle
K\right\rangle }\hat{K}_{X}\right) K_{p}
\end{equation*}%
Including the constant cost, the return of the total capital invested writes:

\begin{eqnarray*}
&&K_{pX}\frac{f_{1}\left( X\right) \left( \left( 1+\frac{\underline{k}\left(
X\right) }{\left\langle K_{p}\right\rangle }\hat{K}_{X}\right) K_{p}\right)
^{1-r}-C_{0}\left( 1+\frac{\underline{k}\left( X\right) }{\left\langle
K\right\rangle }\hat{K}_{X}\right) K_{p}-C}{\left( 1+\frac{\underline{k}%
\left( X\right) }{\left\langle K\right\rangle }\hat{K}_{X}\right) K_{pX}} \\
&\rightarrow &\frac{f_{1}\left( X\right) \left( \left( 1+\frac{\underline{k}%
\left( X\right) }{\left\langle K_{p}\right\rangle }\hat{K}_{X}\right)
K_{p}\right) ^{1-r}-C_{0}\left( 1+\frac{\underline{k}\left( X\right) }{%
\left\langle K\right\rangle }\hat{K}_{X}\left\vert \hat{\Psi}\left( \hat{X}%
\right) \right\vert ^{2}\right) K_{p}-C}{\left( 1+\frac{\underline{k}\left(
X\right) }{\left\langle K\right\rangle }\hat{K}_{X}\left\vert \hat{\Psi}%
\left( \hat{X}\right) \right\vert ^{2}\right) } \\
&=&\left( \frac{f_{1}\left( X\right) }{\left( \left( 1+\frac{\underline{k}%
\left( X\right) }{\left\langle K\right\rangle }\hat{K}_{X}\left\vert \hat{%
\Psi}\left( \hat{X}\right) \right\vert ^{2}\right) K_{p}\right) ^{r}}%
-C_{0}\right) K_{p}-\frac{C}{1+\frac{\underline{k}\left( X\right) }{%
\left\langle K\right\rangle }\hat{K}_{X}\left\vert \hat{\Psi}\left( \hat{X}%
\right) \right\vert ^{2}}
\end{eqnarray*}%
and compared to the constant return to scale case, this amounts to replace $%
f_{1}\left( X\right) $ by:%
\begin{equation*}
\frac{f_{1}\left( X\right) }{\left( \left( 1+\frac{\underline{k}\left(
X\right) }{\left\langle K\right\rangle }\hat{K}_{X}\left\vert \hat{\Psi}%
\left( \hat{X}\right) \right\vert ^{2}\right) K_{p}\right) ^{r}}%
-C_{0}\rightarrow \frac{f_{1}\left( X\right) }{\left( \left( 1+\frac{%
\underline{k}\left( X\right) }{\left\langle K\right\rangle }\hat{K}%
_{X}\left\vert \hat{\Psi}\left( \hat{X}\right) \right\vert ^{2}\right)
K_{X}\right) ^{r}}-C_{0}
\end{equation*}%
and the field $\left\vert \Psi \left( K,X\right) \right\vert ^{2}$ is given
by: 
\begin{equation*}
\left\vert \Psi \left( K,X\right) \right\vert ^{2}\rightarrow \left\vert
\Psi _{0}\left( X\right) \right\vert ^{2}-\epsilon \left( \frac{\left(
f_{1}^{\left( e\right) }\left( X\right) K_{p}-\bar{C}\left( X\right) \right)
^{2}}{\sigma _{\hat{K}}^{2}}+\frac{f_{1}^{\left( e\right) }\left( X\right) }{%
2}\right)
\end{equation*}%
where $f_{1}^{\left( e\right) }\left( X\right) $ is the return of the firm,
corresponding to the net return of production, from which the paiements of
loans is substracted: 
\begin{eqnarray*}
f_{1}^{\left( e\right) }\left( X\right) &=&\left( 1+\underline{k}_{2}\left(
X\right) \right) \left( \frac{f_{1}\left( X\right) }{\left( \left( 1+\frac{%
\underline{k}\left( X\right) }{\left\langle K\right\rangle }\hat{K}%
_{X}\left\vert \hat{\Psi}\left( \hat{X}\right) \right\vert ^{2}\right)
K_{X}\right) ^{r}}-C_{0}\right) -\underline{k}_{2}\left( X\right) \bar{r} \\
\bar{C}^{\left( e\right) }\left( X\right) &=&\left( 1+\underline{k}%
_{2}\left( X\right) \right) \bar{C}\left( X\right)
\end{eqnarray*}

\subsubsection*{A12.1.1 Case one}

\paragraph*{A12.1.1.1 Average capital per sector and return}

The first case corresponds to the condition:%
\begin{equation*}
\left\vert \Psi _{0}\left( X\right) \right\vert ^{2}-\epsilon \frac{%
f_{1}\left( X\right) }{2}-\epsilon \frac{\left( \bar{C}\left( X\right)
\right) ^{2}}{\sigma _{\hat{K}}^{2}}>0
\end{equation*}%
That is, we assume there are some firms with $0$, or very low private
capital.

Resolution is similar to part 1. We find $K_{X}$ as a function of $%
f_{1}^{\left( e\right) }\left( X\right) $:%
\begin{equation}
K_{X}\rightarrow \frac{1}{4f_{1}^{\left( e\right) }\left( X\right) }\frac{%
\left( 3X^{\left( e\right) }-C^{\left( e\right) }\right) \left( C^{\left(
e\right) }+X^{\left( e\right) }\right) }{2X^{\left( e\right) }-C^{\left(
e\right) }}  \label{KQ}
\end{equation}%
where we replace:%
\begin{equation*}
f_{1}^{\left( e\right) }\left( X\right) \rightarrow \frac{\left( 1+%
\underline{k}_{2}\left( X\right) \right) f_{1}\left( X\right) }{\left(
\left( 1+\frac{\underline{k}\left( X\right) }{\left\langle K\right\rangle }%
\hat{K}_{X}\left\vert \hat{\Psi}\left( \hat{X}\right) \right\vert
^{2}\right) K_{X}\right) ^{r}}-C_{0}^{\left( e\right) }
\end{equation*}
and:%
\begin{equation*}
C_{0}^{\left( e\right) }=\left( 1+\underline{k}_{2}\left( X\right) \right)
C_{0}+\underline{k}_{2}\left( X\right) \bar{r}
\end{equation*}%
Thus (\ref{KQ}), is an equation for average capital $K_{X}$ insector $X$ : 
\begin{equation*}
0=K_{X}\left( \frac{\left( 1+\underline{k}_{2}\left( X\right) \right)
f_{1}\left( X\right) }{\left( \left( 1+\frac{\underline{k}\left( X\right) }{%
\left\langle K\right\rangle }\hat{K}_{X}\left\vert \hat{\Psi}\left( \hat{X}%
\right) \right\vert ^{2}\right) K_{X}\right) ^{r}}-C_{0}^{\left( e\right)
}\right) -\frac{\left( 3X^{\left( e\right) }-C^{\left( e\right) }\right)
\left( C^{\left( e\right) }+X^{\left( e\right) }\right) }{2X^{\left(
e\right) }-C^{\left( e\right) }}
\end{equation*}%
with lowest order solution in $C_{0}^{\left( e\right) }$:%
\begin{equation}
K_{X}\simeq \frac{1}{\left( 1+\frac{\underline{k}\left( X\right) }{%
\left\langle K\right\rangle }\hat{K}_{X}\left\vert \hat{\Psi}\left( \hat{X}%
\right) \right\vert ^{2}\right) }\left( \frac{\left( 1+\underline{k}%
_{2}\left( X\right) \right) f_{1}\left( X\right) }{C_{0}^{\left( e\right) }}%
\right) ^{\frac{1}{r}}  \label{KZt}
\end{equation}%
The correction at the first order are obtained by using:%
\begin{eqnarray*}
&&\frac{\left( 1+\underline{k}_{2}\left( X\right) \right) f_{1}\left(
X\right) }{\left( \left( 1+\frac{\underline{k}\left( X\right) }{\left\langle
K\right\rangle }\hat{K}_{X}\left\vert \hat{\Psi}\left( \hat{X}\right)
\right\vert ^{2}\right) K_{X}\right) ^{r}\left( 1+\frac{c}{K_{X}}\right) ^{r}%
}-C_{0}^{\left( e\right) } \\
&\simeq &-C_{0}^{\left( e\right) }r\frac{c}{K_{X}}
\end{eqnarray*}%
so that, the correction to the solution (\ref{KZt}) satisfies:%
\begin{equation*}
C_{0}^{\left( e\right) }rc=-\left( 3X-C\right) \frac{\left( C+X\right) }{2X-C%
}
\end{equation*}%
and the average capital per sector becomes:%
\begin{equation*}
K_{X}\simeq \frac{1}{\left( 1+\frac{\underline{k}\left( X\right) }{%
\left\langle K\right\rangle }\hat{K}_{X}\left\vert \hat{\Psi}\left( \hat{X}%
\right) \right\vert ^{2}\right) }\left( \frac{\left( 1+\underline{k}%
_{2}\left( X\right) \right) f_{1}\left( X\right) }{C_{0}^{\left( e\right) }}%
\right) ^{\frac{1}{r}}-\frac{1}{C_{0}^{\left( e\right) }r}\frac{\left(
3X^{\left( e\right) }-C^{\left( e\right) }\right) \left( C^{\left( e\right)
}+X^{\left( e\right) }\right) }{2X^{\left( e\right) }-C^{\left( e\right) }}
\end{equation*}%
Compared to the constant return to scale case, we replace $f_{1}^{\left(
e\right) }\left( X\right) $ by:%
\begin{eqnarray*}
&&\frac{\left( 1+\underline{k}_{2}\left( X\right) \right) f_{1}\left(
X\right) }{\left( \left( 1+\frac{\underline{k}\left( X\right) }{\left\langle
K\right\rangle }\hat{K}_{X}\left\vert \hat{\Psi}\left( \hat{X}\right)
\right\vert ^{2}\right) K_{X}\right) ^{r}}-C_{0}^{\left( e\right) } \\
&=&\frac{1}{\frac{1}{\left( 1+\frac{\underline{k}\left( X\right) }{%
\left\langle K\right\rangle }\hat{K}_{X}\left\vert \hat{\Psi}\left( \hat{X}%
\right) \right\vert ^{2}\right) }\left( \frac{\left( 1+\underline{k}%
_{2}\left( X\right) \right) f_{1}\left( X\right) }{C_{0}^{\left( e\right) }}%
\right) ^{\frac{1}{r}}-\frac{1}{C_{0}^{\left( e\right) }r}\frac{\left(
3X^{\left( e\right) }-C^{\left( e\right) }\right) \left( C^{\left( e\right)
}+X^{\left( e\right) }\right) }{2X^{\left( e\right) }-C^{\left( e\right) }}}%
\frac{\left( 3X^{\left( e\right) }-C^{\left( e\right) }\right) \left(
C^{\left( e\right) }+X^{\left( e\right) }\right) }{2X^{\left( e\right)
}-C^{\left( e\right) }} \\
&=&\frac{\left( \left( 1+\underline{k}_{2}\left( X\right) \right) C_{0}+%
\underline{k}_{2}\left( X\right) \bar{r}\right) ^{\frac{1}{r}}r\left( 1+%
\frac{\underline{k}\left( X\right) }{\left\langle K\right\rangle }\hat{K}%
_{X}\left\vert \hat{\Psi}\left( \hat{X}\right) \right\vert ^{2}\right) \frac{%
\left( 3X^{\left( e\right) }-C^{\left( e\right) }\right) \left( C^{\left(
e\right) }+X^{\left( e\right) }\right) }{2X^{\left( e\right) }-C^{\left(
e\right) }}}{r\left( \left( 1+\underline{k}_{2}\left( X\right) \right)
f_{1}\left( X\right) \right) ^{\frac{1}{r}}-\left( \left( 1+\underline{k}%
_{2}\left( X\right) \right) C_{0}+\underline{k}_{2}\left( X\right) \bar{r}%
\right) ^{\frac{1}{r}-1}\frac{\left( 3X^{\left( e\right) }-C^{\left(
e\right) }\right) \left( C^{\left( e\right) }+X^{\left( e\right) }\right) }{%
2X^{\left( e\right) }-C^{\left( e\right) }}\left( 1+\frac{\underline{k}%
\left( X\right) }{\left\langle K\right\rangle }\hat{K}_{X}\left\vert \hat{%
\Psi}\left( \hat{X}\right) \right\vert ^{2}\right) }
\end{eqnarray*}%
Using that in first approximation:%
\begin{equation*}
\frac{f_{1}\left( X\right) }{\left( \left( 1+\frac{\underline{k}\left(
X\right) }{\left\langle K\right\rangle }\hat{K}_{X}\left\vert \hat{\Psi}%
\left( \hat{X}\right) \right\vert ^{2}\right) K_{X}\right) ^{r}}-C_{0}\simeq 
\frac{f_{1}\left( X\right) }{\left( 1+\underline{k}\left( X\right) \hat{K}%
_{X}\left\vert \hat{\Psi}\left( \hat{X}\right) \right\vert ^{2}\right) ^{r}}%
-C_{0}
\end{equation*}%
and:%
\begin{eqnarray*}
&&\frac{f_{1}\left( X\right) }{\left( 1+\underline{k}\left( X\right) \hat{K}%
_{X}\left\vert \hat{\Psi}\left( \hat{X}\right) \right\vert ^{2}K_{X}\right)
^{r}}-C_{0} \\
&\simeq &\frac{f_{1}\left( X\right) }{\left( \left( \frac{\left( 1+%
\underline{k}_{2}\left( X\right) \right) f_{1}\left( X\right) }{%
C_{0}^{\left( e\right) }}\right) ^{\frac{1}{r}}-\left( 1+\frac{\underline{k}%
\left( X\right) }{\left\langle K\right\rangle }\hat{K}_{X}\left\vert \hat{%
\Psi}\left( \hat{X}\right) \right\vert ^{2}\right) \frac{1}{C_{0}^{\left(
e\right) }r}\frac{\left( 3X^{\left( e\right) }-C^{\left( e\right) }\right)
\left( C^{\left( e\right) }+X^{\left( e\right) }\right) }{2X^{\left(
e\right) }-C^{\left( e\right) }}\right) ^{r}}-C_{0}
\end{eqnarray*}%
\begin{eqnarray*}
&&\frac{C_{0}^{\left( e\right) }}{\left( 1+\underline{k}_{2}\left( X\right)
\right) }\left( 1+\frac{1}{\left( C_{0}^{\left( e\right) }\right) ^{1-\frac{1%
}{r}}\left( \left( 1+\underline{k}_{2}\left( X\right) \right) f_{1}\left(
X\right) \right) ^{\frac{1}{r}}}\frac{\left( 3X^{\left( e\right) }-C^{\left(
e\right) }\right) \left( C^{\left( e\right) }+X^{\left( e\right) }\right) }{%
2X^{\left( e\right) }-C^{\left( e\right) }}\right) -C_{0} \\
&=&\frac{C_{0}^{\left( e\right) }}{\left( 1+\underline{k}_{2}\left( X\right)
\right) }\left( \frac{1}{\left( C_{0}^{\left( e\right) }\right) ^{1-\frac{1}{%
r}}\left( \left( 1+\underline{k}_{2}\left( X\right) \right) f_{1}\left(
X\right) \right) ^{\frac{1}{r}}}\frac{\left( 3X^{\left( e\right) }-C^{\left(
e\right) }\right) \left( C^{\left( e\right) }+X^{\left( e\right) }\right) }{%
2X^{\left( e\right) }-C^{\left( e\right) }}\right) +\frac{\underline{k}%
_{2}\left( X\right) }{1+\underline{k}_{2}\left( X\right) }\bar{r} \\
&=&\frac{1}{\left( 1+\underline{k}_{2}\left( X\right) \right) }\left( \frac{%
\left( 1+\frac{\underline{k}\left( X\right) }{\left\langle K\right\rangle }%
\hat{K}_{X}\left\vert \hat{\Psi}\left( \hat{X}\right) \right\vert
^{2}\right) }{\left( \frac{1+\underline{k}_{2}\left( X\right) }{%
C_{0}^{\left( e\right) }}f_{1}\left( X\right) \right) ^{\frac{1}{r}}}\frac{%
\left( 3X^{\left( e\right) }-C^{\left( e\right) }\right) \left( C^{\left(
e\right) }+X^{\left( e\right) }\right) }{2X^{\left( e\right) }-C^{\left(
e\right) }}\right) +\frac{\underline{k}_{2}\left( X\right) }{1+\underline{k}%
_{2}\left( X\right) }\bar{r} \\
&=&\frac{1}{\left( 1+\underline{k}_{2}\left( X\right) \right) }\left( \left( 
\frac{C_{0}^{\left( e\right) }}{\left( 1+\underline{k}_{2}\left( X\right)
\right) f_{1}\left( X\right) }\right) ^{\frac{1}{r}}\left( 1+\frac{%
\underline{k}\left( X\right) }{\left\langle K\right\rangle }\hat{K}%
_{X}\left\vert \hat{\Psi}\left( \hat{X}\right) \right\vert ^{2}\right) \frac{%
\left( 3X^{\left( e\right) }-C^{\left( e\right) }\right) \left( C^{\left(
e\right) }+X^{\left( e\right) }\right) }{2X^{\left( e\right) }-C^{\left(
e\right) }}\right) +\frac{\underline{k}_{2}\left( X\right) }{1+\underline{k}%
_{2}\left( X\right) }\bar{r} \\
&=&\frac{1}{\left( 1+\underline{k}_{2}\left( X\right) \right) }\left( \left( 
\frac{C_{0}+\frac{\underline{k}_{2}\left( X\right) }{1+\underline{k}%
_{2}\left( X\right) }\bar{r}}{f_{1}\left( X\right) }\right) ^{\frac{1}{r}%
}\left( 1+\frac{\underline{k}\left( X\right) }{\left\langle K\right\rangle }%
\hat{K}_{X}\left\vert \hat{\Psi}\left( \hat{X}\right) \right\vert
^{2}\right) \frac{\left( 3X^{\left( e\right) }-C^{\left( e\right) }\right)
\left( C^{\left( e\right) }+X^{\left( e\right) }\right) }{2X^{\left(
e\right) }-C^{\left( e\right) }}\right) +\frac{\underline{k}_{2}\left(
X\right) }{1+\underline{k}_{2}\left( X\right) }\bar{r}
\end{eqnarray*}%
The following relation is satisfied:%
\begin{equation*}
\frac{f_{1}\left( X\right) }{\left( \left( 1+\frac{\underline{k}\left(
X\right) }{\left\langle K\right\rangle }\hat{K}_{X}\left\vert \hat{\Psi}%
\left( \hat{X}\right) \right\vert ^{2}\right) K_{X}\right) ^{r}}-C_{0}-\bar{r%
}=\frac{f_{1}^{\left( e\right) }\left( X\right) -\bar{r}}{\left( 1+%
\underline{k}_{2}\left( X\right) \right) }
\end{equation*}%
and:%
\begin{eqnarray*}
f_{1}^{\left( e\right) }\left( X\right) &=&\left( 1+\underline{k}_{2}\left(
X\right) \right) \left( \frac{f_{1}\left( X\right) }{\left( \left( 1+\frac{%
\underline{k}\left( X\right) }{\left\langle K\right\rangle }\hat{K}%
_{X}\left\vert \hat{\Psi}\left( \hat{X}\right) \right\vert ^{2}\right)
K_{X}\right) ^{r}}-C_{0}\right) -\underline{k}_{2}\left( X\right) \bar{r} \\
&\simeq &\left( 1+\frac{\underline{k}\left( X\right) }{\left\langle
K\right\rangle }\hat{K}_{X}\left\vert \hat{\Psi}\left( \hat{X}\right)
\right\vert ^{2}\right) \left( \frac{C_{0}+\frac{\underline{k}_{2}\left(
X\right) }{1+\underline{k}_{2}\left( X\right) }\bar{r}}{f_{1}\left( X\right) 
}\right) ^{\frac{1}{r}}\frac{\left( 3X^{\left( e\right) }-C^{\left( e\right)
}\right) \left( C^{\left( e\right) }+X^{\left( e\right) }\right) }{%
2X^{\left( e\right) }-C^{\left( e\right) }} \\
&\simeq &\left( 1+\frac{\underline{k}\left( X\right) }{\left\langle
K\right\rangle }\hat{K}_{X}\left\vert \hat{\Psi}\left( \hat{X}\right)
\right\vert ^{2}\right) \left( \frac{C_{0}+\bar{r}}{f_{1}\left( X\right) }%
\right) ^{\frac{1}{r}}\frac{\left( 3X^{\left( e\right) }-C^{\left( e\right)
}\right) \left( C^{\left( e\right) }+X^{\left( e\right) }\right) }{%
2X^{\left( e\right) }-C^{\left( e\right) }}
\end{eqnarray*}

\paragraph*{A12.1.1.2 Returns to investrs}

In the sequel, we need to compute the return of sector $X$ firms to
investors. This amnts to compute:%
\begin{equation}
\left\vert \Psi \left( X\right) \right\vert ^{2}K_{X}\frac{f_{1}^{\prime
}\left( X\right) -\bar{r}}{1+\underline{k}_{2}\left( X\right) }  \label{RFN}
\end{equation}

Replacing the previous expression yields:%
\begin{eqnarray*}
&&\left\vert \Psi \left( X\right) \right\vert ^{2}\left( \left( f_{1}\left(
X\right) -\bar{r}\right) K_{X}-\bar{C}\left( X\right) \right) \\
&\rightarrow &\frac{\left( C^{\left( e\right) }+X^{\left( e\right) }\right)
\left( \frac{2}{3}X^{2}+\frac{1}{3}\left( X-C^{\left( e\right) }\right)
C^{\left( e\right) }\right) \epsilon }{\sigma _{\hat{K}}^{2}f_{1}^{\left(
e\right) }\left( X\right) }\left( \frac{\left( f_{1}\left( X\right) -\bar{r}%
\right) }{4f_{1}^{\left( e\right) }\left( X\right) }\frac{\left( 3X^{\left(
e\right) }-C^{\left( e\right) }\right) \left( C^{\left( e\right) }+X^{\left(
e\right) }\right) }{2X^{\left( e\right) }-C^{\left( e\right) }}-\bar{C}%
\right)
\end{eqnarray*}%
\begin{eqnarray*}
&&\left\vert \Psi \left( X\right) \right\vert ^{2}\left( \left( f_{1}\left(
X\right) -\bar{r}\right) K_{X}-\bar{C}\left( X\right) \right) \\
&\rightarrow &\frac{\left( C^{\left( e\right) }+X^{\left( e\right) }\right)
\left( \frac{2}{3}X^{2}+\frac{1}{3}\left( X-C^{\left( e\right) }\right)
C^{\left( e\right) }\right) \epsilon }{\sigma _{\hat{K}}^{2}f_{1}^{\left(
e\right) }\left( X\right) }\left( \frac{\left( f_{1}\left( X\right) -\bar{r}%
\right) }{4f_{1}^{\left( e\right) }\left( X\right) }\frac{\left( 3X^{\left(
e\right) }-C^{\left( e\right) }\right) \left( C^{\left( e\right) }+X^{\left(
e\right) }\right) }{2X^{\left( e\right) }-C^{\left( e\right) }}-\bar{C}%
\right)
\end{eqnarray*}%
and this becomes:%
\begin{equation*}
\frac{\left( C^{\left( e\right) }+X^{\left( e\right) }\right) \left( \frac{2%
}{3}X^{2}+\frac{1}{3}\left( X-C^{\left( e\right) }\right) C^{\left( e\right)
}\right) \epsilon }{\sigma _{\hat{K}}^{2}f_{1}^{\left( e\right) }\left(
X\right) \left( 1+\underline{k}_{2}\left( X\right) \right) }\left( \frac{%
\left( f_{1}^{\left( e\right) }\left( X\right) -\bar{r}\right) }{%
4f_{1}^{\left( e\right) }\left( X\right) }\frac{\left( 3X^{\left( e\right)
}-C^{\left( e\right) }\right) \left( C^{\left( e\right) }+X^{\left( e\right)
}\right) }{2X^{\left( e\right) }-C^{\left( e\right) }}-C^{\left( e\right)
}\right)
\end{equation*}%
so that (\ref{RFN}) writes:%
\begin{eqnarray*}
&&\frac{\frac{1}{3}\left( 2X-C^{e}\right) ^{2}\left( X+C^{e}\right) \epsilon 
}{\sigma _{\hat{K}}^{2}\left( 1+\underline{k}\left( X\right) \right) \left( 
\frac{C_{0}+\bar{r}}{f_{1}\left( X\right) }\right) ^{\frac{1}{r}}\left(
3X^{\left( e\right) }-C^{\left( e\right) }\right) \left( 1+\underline{k}%
_{2}\left( X\right) \right) }\left( \frac{\left( f_{1}^{\left( e\right)
}\left( X\right) -\bar{r}\right) }{4\left( 1+\underline{k}\left( X\right)
\right) \left( \frac{C_{0}+\bar{r}}{f_{1}\left( X\right) }\right) ^{\frac{1}{%
r}}}-C^{\left( e\right) }\right) \\
&=&\frac{\frac{1}{3}\left( 2X-C^{e}\right) ^{2}\left( X+C^{e}\right)
\epsilon }{\sigma _{\hat{K}}^{2}\left( 1+\underline{k}\left( X\right)
\right) \left( \frac{C_{0}+\bar{r}}{f_{1}\left( X\right) }\right) ^{\frac{1}{%
r}}\left( 3X^{\left( e\right) }-C^{\left( e\right) }\right) }\left( \frac{%
\left( f_{1}^{\left( e\right) }\left( X\right) -\bar{r}\right) }{4\left( 1+%
\underline{k}\left( X\right) \right) \left( \frac{C_{0}+\bar{r}}{f_{1}\left(
X\right) }\right) ^{\frac{1}{r}}\left( 1+\underline{k}_{2}\left( X\right)
\right) }-\bar{C}\right) \\
&=&\frac{\frac{1}{3}\left( 2X-C^{e}\right) ^{2}\left( X+C^{e}\right)
\epsilon }{\sigma _{\hat{K}}^{2}\left( 1+\underline{k}\left( X\right)
\right) \left( 3X^{\left( e\right) }-C^{\left( e\right) }\right) }\left( 
\frac{f_{1}\left( X\right) }{C_{0}+\bar{r}}\right) ^{\frac{1}{r}}\left( 
\frac{\left( 1+\underline{k}\left( X\right) \right) \frac{\left( 3X^{\left(
e\right) }-C^{\left( e\right) }\right) \left( C^{\left( e\right) }+X^{\left(
e\right) }\right) }{2X^{\left( e\right) }-C^{\left( e\right) }}-\bar{r}%
\left( \frac{f_{1}\left( X\right) }{C_{0}+\bar{r}}\right) ^{\frac{1}{r}}}{%
4\left( 1+\underline{k}\left( X\right) \right) \left( 1+\underline{k}%
_{2}\left( X\right) \right) }-\bar{C}\right) \\
&=&\frac{\frac{1}{3}\left( 2X-C^{e}\right) ^{2}\left( X+C^{e}\right)
\epsilon }{\sigma _{\hat{K}}^{2}\left( 1+\underline{k}\left( X\right)
\right) ^{2}\left( 3X^{\left( e\right) }-C^{\left( e\right) }\right) }\left( 
\frac{f_{1}\left( X\right) }{C_{0}+\bar{r}}\right) ^{\frac{1}{r}}\left( 
\frac{\left( 1+\underline{k}\left( X\right) \right) \frac{\left( 3X^{\left(
e\right) }-C^{\left( e\right) }\right) \left( C^{\left( e\right) }+X^{\left(
e\right) }\right) }{2X^{\left( e\right) }-C^{\left( e\right) }}-\bar{r}%
\left( \frac{f_{1}\left( X\right) }{C_{0}+\bar{r}}\right) ^{\frac{1}{r}}}{%
4\left( 1+\underline{k}_{2}\left( X\right) \right) }-C\right)
\end{eqnarray*}%
$\allowbreak $

and the return (\ref{RFN}) can be written as: 
\begin{equation*}
\frac{1}{12}\frac{\left( C^{\left( e\right) }+X^{\left( e\right) }\right)
^{2}\epsilon }{\sigma _{\hat{K}}^{2}\left( 1+\underline{k}_{2}\left(
X\right) \right) f_{1}^{\left( e\right) }\left( X\right) }\left\{ 3\left(
X^{\left( e\right) }-C^{\left( e\right) }\right) ^{2}-\frac{\bar{r}\left(
3X^{\left( e\right) }-C^{\left( e\right) }\right) \left( C^{\left( e\right)
}+X^{\left( e\right) }\right) }{f_{1}^{\left( e\right) }\left( X\right) }%
\right\}
\end{equation*}

\subsubsection*{A12.1.2 Case two}

In this case, we consider that all firms have a minimum capital, so that: 
\begin{equation*}
\left\vert \Psi _{0}\left( X\right) \right\vert ^{2}-\epsilon \frac{\frac{%
f_{1}^{\left( e\right) }\left( X\right) }{\left( \left( 1+\frac{\underline{k}%
\left( X\right) }{\left\langle K\right\rangle }\hat{K}_{X}\left\vert \hat{%
\Psi}\left( \hat{X}\right) \right\vert ^{2}\right) K_{X}\right) ^{r}}%
-C_{0}^{\left( e\right) }}{2}-\epsilon \frac{\left( \bar{C}\left( X\right)
\right) ^{2}}{\sigma _{\hat{K}}^{2}}<0
\end{equation*}%
Computation are similar to the constant return to scale case leads to
compute the minimal and maximal level of capital in sector $X$, written in a
compact form $K_{0\pm }$:%
\begin{equation*}
K_{0\pm }=\frac{\bar{C}\left( X\right) \pm \sqrt{\sigma _{\hat{K}}^{2}\left( 
\frac{\left\vert \Psi _{0}\left( X\right) \right\vert ^{2}}{\epsilon }-\frac{%
\frac{f_{1}^{\left( e\right) }\left( X\right) }{\left( \left( 1+\frac{%
\underline{k}\left( X\right) }{\left\langle K\right\rangle }\hat{K}%
_{X}\left\vert \hat{\Psi}\left( \hat{X}\right) \right\vert ^{2}\right)
K_{X}\right) ^{r}}-C_{0}^{\left( e\right) }}{2}\right) }}{\frac{%
f_{1}^{\left( e\right) }\left( X\right) }{\left( \left( 1+\frac{\underline{k}%
\left( X\right) }{\left\langle K\right\rangle }\hat{K}_{X}\left\vert \hat{%
\Psi}\left( \hat{X}\right) \right\vert ^{2}\right) K_{X}\right) ^{r}}%
-C_{0}^{\left( e\right) }}
\end{equation*}

leading to:%
\begin{equation*}
K_{X}\left\vert \Psi \left( X\right) \right\vert ^{2}=\frac{2\epsilon \Delta
K_{0}\bar{C}\left( X\right) }{3\left( \frac{f_{1}^{\left( e\right) }\left(
X\right) }{\left( \left( 1+\frac{\underline{k}\left( X\right) }{\left\langle
K\right\rangle }\hat{K}_{X}\left\vert \hat{\Psi}\left( \hat{X}\right)
\right\vert ^{2}\right) K_{X}\right) ^{r}}-C_{0}^{\left( e\right) }\right)
\sigma _{\hat{K}}^{2}}X^{2}
\end{equation*}%
and for the field 
\begin{equation*}
\left\vert \Psi \left( X\right) \right\vert ^{2}=2\frac{\epsilon \Delta K_{0}%
}{3\sigma _{\hat{K}}^{2}}X^{2}
\end{equation*}%
where:%
\begin{equation*}
X=\sqrt{\sigma _{\hat{K}}^{2}\left( \frac{\left\vert \Psi _{0}\left(
X\right) \right\vert ^{2}}{\epsilon }-\frac{\frac{f_{1}^{\left( e\right)
}\left( X\right) }{\left( \left( 1+\frac{\underline{k}\left( X\right) }{%
\left\langle K\right\rangle }\hat{K}_{X}\left\vert \hat{\Psi}\left( \hat{X}%
\right) \right\vert ^{2}\right) K_{X}\right) ^{r}}-C_{0}^{\left( e\right) }}{%
2}\right) }
\end{equation*}%
These formula allow to find the average capital per frm n the sector.%
\begin{equation*}
K_{X}=\frac{\bar{C}\left( X\right) }{\frac{f_{1}^{\left( e\right) }\left(
X\right) }{\left( \left( 1+\frac{\underline{k}\left( X\right) }{\left\langle
K\right\rangle }\hat{K}_{X}\left\vert \hat{\Psi}\left( \hat{X}\right)
\right\vert ^{2}\right) K_{X}\right) ^{r}}-C_{0}^{\left( e\right) }}
\end{equation*}%
As in case one, this is an equation for $K_{X}$: 
\begin{equation*}
K_{X}\left( \frac{f_{1}^{\left( e\right) }\left( X\right) }{\left( \left( 1+%
\frac{\underline{k}\left( X\right) }{\left\langle K\right\rangle }\hat{K}%
_{X}\left\vert \hat{\Psi}\left( \hat{X}\right) \right\vert ^{2}\right)
K_{X}\right) ^{r}}-C_{0}^{\left( e\right) }\right) =\bar{C}\left( X\right)
\end{equation*}%
with solutions to first order:%
\begin{equation*}
K_{X}\simeq \frac{1}{\left( 1+\frac{\underline{k}\left( X\right) }{%
\left\langle K\right\rangle }\hat{K}_{X}\left\vert \hat{\Psi}\left( \hat{X}%
\right) \right\vert ^{2}\right) }\left( \frac{\left( 1+\underline{k}%
_{2}\left( X\right) \right) f_{1}\left( X\right) }{C_{0}^{\left( e\right) }}%
\right) ^{\frac{1}{r}}-\frac{1}{C_{0}^{\left( e\right) }r}\bar{C}\left(
X\right)
\end{equation*}%
Using this expression allows to find the return $f_{1}^{\left( e\right)
}\left( X\right) $ of the firm:%
\begin{eqnarray*}
&&f_{1}^{\left( e\right) }\left( X\right) =\frac{\left( 1+\underline{k}%
_{2}\left( X\right) \right) f_{1}\left( X\right) }{\left( \left( 1+\frac{%
\underline{k}\left( X\right) }{\left\langle K\right\rangle }\hat{K}%
_{X}\left\vert \hat{\Psi}\left( \hat{X}\right) \right\vert ^{2}\right)
K_{X}\right) ^{r}}-C_{0}^{\left( e\right) } \\
&=&\frac{1}{\frac{1}{\left( 1+\frac{\underline{k}\left( X\right) }{%
\left\langle K\right\rangle }\hat{K}_{X}\left\vert \hat{\Psi}\left( \hat{X}%
\right) \right\vert ^{2}\right) }\left( \frac{\left( 1+\underline{k}%
_{2}\left( X\right) \right) f_{1}\left( X\right) }{C_{0}^{\left( e\right) }}%
\right) ^{\frac{1}{r}}-\frac{1}{C_{0}^{\left( e\right) }r}\bar{C}\left(
X\right) }\bar{C}\left( X\right) \\
&=&\frac{\left( \left( 1+\underline{k}_{2}\left( X\right) \right) C_{0}+%
\underline{k}_{2}\left( X\right) \bar{r}\right) ^{\frac{1}{r}}r\left( 1+%
\frac{\underline{k}\left( X\right) }{\left\langle K\right\rangle }\hat{K}%
_{X}\left\vert \hat{\Psi}\left( \hat{X}\right) \right\vert ^{2}\right) \bar{C%
}\left( X\right) }{r\left( \left( 1+\underline{k}_{2}\left( X\right) \right)
f_{1}\left( X\right) \right) ^{\frac{1}{r}}-\left( \left( 1+\underline{k}%
_{2}\left( X\right) \right) C_{0}+\underline{k}_{2}\left( X\right) \bar{r}%
\right) ^{\frac{1}{r}-1}\bar{C}\left( X\right) \left( 1+\frac{\underline{k}%
\left( X\right) }{\left\langle K\right\rangle }\hat{K}_{X}\left\vert \hat{%
\Psi}\left( \hat{X}\right) \right\vert ^{2}\right) } \\
&\simeq &\frac{\left( \left( 1+\underline{k}_{2}\left( X\right) \right)
C_{0}+\underline{k}_{2}\left( X\right) \bar{r}\right) ^{\frac{1}{r}}rC\left(
X\right) }{r\left( \left( 1+\underline{k}_{2}\left( X\right) \right)
f_{1}\left( X\right) \right) ^{\frac{1}{r}}-\left( \left( 1+\underline{k}%
_{2}\left( X\right) \right) C_{0}+\underline{k}_{2}\left( X\right) \bar{r}%
\right) ^{\frac{1}{r}-1}C\left( X\right) }
\end{eqnarray*}

\begin{eqnarray*}
\left\vert \Psi \left( X\right) \right\vert ^{2}K_{X}\frac{f_{1}^{\prime
}\left( X\right) -\bar{r}}{1+\underline{k}_{2}\left( X\right) }
&=&\left\vert \Psi \left( X\right) \right\vert ^{2}\left( \left( f_{1}\left(
X\right) -\bar{r}\right) K_{X}-\bar{C}\left( X\right) \right) \\
&=&4\frac{\epsilon \left( C^{\left( e\right) }\right) ^{2}}{3\sigma _{\hat{K}%
}^{2}f_{1}^{\left( e\right) }\left( X\right) \left( 1+\underline{k}%
_{2}\left( X\right) \right) }X^{2}\left( \left( 1-\frac{\bar{r}}{%
f_{1}^{\left( e\right) }\left( X\right) }\right) \left( 3X^{\left( e\right)
}-1\right) -C^{\left( e\right) }\right)
\end{eqnarray*}

\subsection*{A12.2 Investors side}

\subsubsection*{A12.2.1 Computation of $\hat{g}$}

Given (\ref{KP}):%
\begin{equation}
\hat{K}\left[ \hat{X}_{1}\right] =\frac{2\sigma _{\hat{K}}^{2}}{\hat{\mu}}%
\left( \frac{\left\Vert \hat{\Psi}_{0}\left( \hat{X}_{1}\right) \right\Vert
^{2}-\hat{\mu}D\left( \hat{X}_{1}\right) }{\hat{g}^{2}\left( \hat{X}%
_{1}\right) }\right) ^{2}\left( \frac{\hat{g}^{2}\left( \hat{X}_{1}\right) }{%
4}-\frac{r\left\langle \hat{g}\right\rangle ^{2}}{3}\underline{\hat{k}}%
\right)
\end{equation}%
This formula for the amount of capital in sector $\hat{X}_{1}$ allows to
derive the return $\hat{g}\left( \hat{X}_{1}\right) $ as a function of
capital. The following equation stands: 
\begin{equation*}
0=\frac{\hat{\mu}\hat{K}\left[ \hat{X}_{1}\right] }{2\sigma _{\hat{K}}^{2}}%
\left( \hat{g}^{2}\left( \hat{X}_{1}\right) \right) ^{2}-\left( \left\Vert 
\hat{\Psi}_{0}\left( \hat{X}_{1}\right) \right\Vert ^{2}-\hat{\mu}D\left( 
\hat{X}_{1}\right) \right) ^{2}\frac{\hat{g}^{2}\left( \hat{X}_{1}\right) }{4%
}+\left( \left\Vert \hat{\Psi}_{0}\left( \hat{X}_{1}\right) \right\Vert ^{2}-%
\hat{\mu}D\left( \hat{X}_{1}\right) \right) ^{2}\frac{r\left\langle \hat{g}%
\right\rangle ^{2}}{3}\underline{\hat{k}}
\end{equation*}

and we have:%
\begin{equation*}
\hat{g}\left( \hat{X}_{1}\right) =\frac{\left( \left\Vert \hat{\Psi}%
_{0}\left( \hat{X}_{1}\right) \right\Vert ^{2}-\hat{\mu}D\left( \hat{X}%
_{1}\right) \right) ^{2}-\sqrt{\left( \left\Vert \hat{\Psi}_{0}\left( \hat{X}%
_{1}\right) \right\Vert ^{2}-\hat{\mu}D\left( \hat{X}_{1}\right) \right)
^{2}-2\frac{\hat{\mu}\hat{K}\left[ \hat{X}_{1}\right] }{\sigma _{\hat{K}}^{2}%
}\frac{r\left\langle \hat{g}\right\rangle ^{2}}{3}\underline{\hat{k}}}}{%
\frac{\hat{\mu}\hat{K}\left[ \hat{X}_{1}\right] }{\sigma _{\hat{K}}^{2}}}
\end{equation*}%
In first approximation, we can also write: 
\begin{eqnarray}
\hat{K}\left[ \hat{X}_{1}\right] &\simeq &\frac{2\sigma _{\hat{K}}^{2}}{\hat{%
\mu}\hat{g}^{2}\left( \hat{X}_{1}\right) }\left( \left\Vert \hat{\Psi}%
_{0}\left( \hat{X}_{1}\right) \right\Vert ^{2}-\hat{\mu}D\left( \hat{X}%
_{1}\right) \right) ^{2}\left( \frac{1}{4}-\frac{r\left\langle \hat{g}%
\right\rangle ^{2}}{3\hat{g}^{2}\left( \hat{X}_{1}\right) }\underline{\hat{k}%
}\right) \\
&\simeq &\frac{2\sigma _{\hat{K}}^{2}}{\hat{\mu}\hat{g}^{2}\left( \hat{X}%
_{1}\right) }\left( \left\Vert \hat{\Psi}_{0}\left( \hat{X}_{1}\right)
\right\Vert ^{2}-\hat{\mu}D\left( \hat{X}_{1}\right) \right) ^{2}\left( 
\frac{1}{4}-\frac{r}{3}\underline{\hat{k}}\right)  \notag
\end{eqnarray}%
and $\hat{g}\left( \hat{X}_{1}\right) $ is given in this approximation b: 
\begin{equation*}
\hat{g}\left( \hat{X}_{1}\right) \rightarrow \frac{\left( \left\Vert \hat{%
\Psi}_{0}\left( \hat{X}_{1}\right) \right\Vert ^{2}-\hat{\mu}D\left( \hat{X}%
_{1}\right) \right) \sqrt{\frac{1}{4}-\frac{r}{3}\underline{\hat{k}}}}{\sqrt{%
\frac{\hat{\mu}\hat{K}\left[ \hat{X}_{1}\right] }{2\sigma _{\hat{K}}^{2}}}}
\end{equation*}

where:%
\begin{equation*}
D\left( \hat{X}_{1}\right) =\left( \frac{\left\langle \hat{K}\right\rangle
^{2}\left\langle \hat{g}\right\rangle ^{2}}{\sigma _{\hat{K}}^{2}}+\frac{%
\left\langle \hat{g}\right\rangle }{2}\right) \left( \frac{\underline{\hat{k}%
}\left( \left\langle \hat{X}\right\rangle ,\hat{X}_{1}\right) }{\underline{%
\hat{k}}\left( \left\langle \hat{X}\right\rangle ,\left\langle \hat{X}%
\right\rangle \right) }-\frac{6\underline{\hat{k}}}{2+\underline{\hat{k}}-%
\sqrt{\left( 2+\underline{\hat{k}}\right) ^{2}-\underline{\hat{k}}}}\right) 
\underline{\hat{k}}
\end{equation*}%
The equation for $\hat{g}$ is obtained by rewriting (\ref{RTN}):%
\begin{equation*}
g\left( \hat{X}_{1}\right) =\left( 1-M\right) ^{-1}\frac{1-\hat{S}_{1}\left( 
\hat{X}\right) }{1-\hat{S}\left( \hat{X}\right) }\left( \Delta \left( \hat{X}%
,\hat{X}^{\prime }\right) -\hat{S}_{1}\left( \hat{X}^{\prime },\hat{X}%
\right) \right) ^{-1}S_{1}\left( \hat{X}^{\prime },\hat{X}^{\prime }\right) 
\frac{1-S\left( \hat{X}^{\prime }\right) }{1-S_{1}\left( \hat{X}^{\prime
}\right) }f_{1}^{\prime }
\end{equation*}%
which writes as:%
\begin{eqnarray*}
&&\left( \Delta \left( \hat{X},\hat{X}^{\prime }\right) -\hat{S}_{1}\left( 
\hat{X}^{\prime },\hat{X}\right) \right) \frac{1-\hat{S}\left( \hat{X}%
\right) }{1-\hat{S}_{1}\left( \hat{X}\right) }\left( \left( 1-M\right) \hat{g%
}\left( \hat{X}^{\prime }\right) -\bar{r}\right) \\
&=&S_{1}\left( \hat{X}^{\prime },\hat{X}^{\prime }\right) \frac{1-S\left( 
\hat{X}^{\prime }\right) }{1-S_{1}\left( \hat{X}^{\prime }\right) }%
\left\vert \Psi \left( X\right) \right\vert ^{2}K_{X}\frac{f_{1}^{\prime
}\left( X\right) -\bar{r}}{1+\underline{k}_{2}\left( X\right) }
\end{eqnarray*}%
and then by replacing the left hand side by:%
\begin{equation*}
\left( \frac{1}{1+\underline{\hat{k}}_{2}\left( \hat{X}\right) }-\hat{S}%
_{1}^{E}\left( \hat{X}^{\prime },\hat{X}\right) \right) \left( \frac{\left(
\left\Vert \hat{\Psi}_{0}\left( \hat{X}^{\prime }\right) \right\Vert ^{2}-%
\hat{\mu}D\left( \hat{X}^{\prime }\right) \right) \sqrt{\frac{1}{4}-\frac{r}{%
3}\underline{\hat{k}}}}{\sqrt{\frac{\hat{\mu}\hat{K}\left[ \hat{X}^{\prime }%
\right] }{2\sigma _{\hat{K}}^{2}}}}-\bar{r}^{\prime }\right)
\end{equation*}%
The right hand side is derived by using that:%
\begin{equation*}
S_{1}\left( \hat{X}^{\prime },\hat{X}^{\prime }\right) =\frac{k_{1}\left( 
\hat{X}^{\prime },\hat{X}^{\prime }\right) }{1+\underline{k}\left( X\right) }
\end{equation*}%
and that:%
\begin{equation*}
\left\vert \Psi \left( X\right) \right\vert ^{2}K_{X}\frac{f_{1}^{\prime
}\left( X\right) -\bar{r}}{1+\underline{k}_{2}\left( X\right) }=\left\vert
\Psi \left( X\right) \right\vert ^{2}\left( \left( f_{1}\left( X\right) -%
\bar{r}\right) K_{X}-\bar{C}\left( X\right) \right)
\end{equation*}%
As a consequence, we have%
\begin{eqnarray*}
&&S_{1}\left( \hat{X}^{\prime },\hat{X}^{\prime }\right) \frac{1-S\left( 
\hat{X}^{\prime }\right) }{1-S_{1}\left( \hat{X}^{\prime }\right) }\left(
f_{1}^{\prime }\left( X^{\prime }\right) -\bar{r}\right) \\
&=&S_{1}\left( \hat{X}^{\prime },\hat{X}^{\prime }\right) \left\vert \Psi
\left( X^{\prime }\right) \right\vert ^{2}K_{X^{\prime }}\frac{f_{1}^{\prime
}\left( X^{\prime }\right) -\bar{r}}{1+\underline{k}_{2}\left( X^{\prime
}\right) } \\
&=&\frac{k_{1}\left( \hat{X}^{\prime },\hat{X}^{\prime }\right) }{1+%
\underline{k}\left( X\right) }\left\vert \Psi \left( X\right) \right\vert
^{2}\left( \left( f_{1}\left( X\right) -\bar{r}\right) K_{X}-\bar{C}\left(
X\right) \right)
\end{eqnarray*}

\begin{equation*}
\left\vert \Psi \left( X\right) \right\vert ^{2}\left( \left( f_{1}\left(
X\right) -\bar{r}\right) K_{X}-\bar{C}\left( X\right) \right)
\end{equation*}%
has been estimated before, and we write:%
\begin{eqnarray*}
\underline{k}_{2}\left( \hat{X}\right) &=&\beta \underline{k}\left( \hat{X}%
\right) \\
\underline{k}_{1}\left( \hat{X}\right) &=&\left( 1-\beta \right) \underline{k%
}\left( \hat{X}\right)
\end{eqnarray*}%
for:%
\begin{equation*}
\underline{k}\left( \hat{X}\right) >>1
\end{equation*}%
\begin{equation*}
\frac{\underline{k}_{1}\left( \hat{X}\right) }{\left( 1+\underline{k}\left( 
\hat{X}\right) \right) }\rightarrow 1-\beta
\end{equation*}%
and%
\begin{equation*}
X^{\left( e\right) }\rightarrow \sqrt{\frac{\left\vert \Psi _{0}\left(
X\right) \right\vert ^{2}}{\epsilon }-\frac{1}{2}f_{1}\left( X\right) -\frac{%
1}{2}\left( \beta \underline{k}\left( X\right) \left( f_{1}\left( X\right) -%
\bar{r}\right) \right) }\simeq X=\sqrt{\frac{\left\vert \Psi _{0}\left(
X\right) \right\vert ^{2}}{\epsilon }-\frac{1}{2}f_{1}\left( X\right) }
\end{equation*}%
\begin{equation*}
C^{e}=\frac{\left( 1+\underline{k}_{2}\left( X\right) \right) C}{\left( 1+%
\underline{k}\left( X\right) \right) }\simeq \beta C
\end{equation*}%
Ultimately, the right hand side becomes:%
\begin{eqnarray*}
&\rightarrow &\frac{\underline{k}_{1}\left( \hat{X}\right) }{\left( 1+%
\underline{k}\left( \hat{X}\right) \right) }\frac{\frac{1}{3}\left(
2X-C^{e}\right) ^{2}\left( X+C^{e}\right) \epsilon }{\sigma _{\hat{K}%
}^{2}\left( 1+\underline{k}\left( X\right) \right) ^{2}\left( 3X^{\left(
e\right) }-C^{\left( e\right) }\right) }\left( \frac{f_{1}\left( X\right) }{%
C_{0}+\bar{r}}\right) ^{\frac{1}{r}} \\
&&\times \left( \frac{\left( 1+\underline{k}\left( X\right) \right) \frac{%
\left( 3X^{\left( e\right) }-C^{\left( e\right) }\right) \left( C^{\left(
e\right) }+X^{\left( e\right) }\right) }{2X^{\left( e\right) }-C^{\left(
e\right) }}-\bar{r}\left( \frac{f_{1}\left( X\right) }{C_{0}+\bar{r}}\right)
^{\frac{1}{r}}}{4\left( 1+\underline{k}_{2}\left( X\right) \right) }-C\right)
\\
&\simeq &\left( 1-\beta \right) \frac{\frac{1}{3}\left( 2X-C\beta \right)
^{2}\left( \left( X+C\beta \right) \right) \epsilon }{\sigma _{\hat{K}%
}^{2}\left( 1+\underline{k}\left( X\right) \right) ^{2}\left( 3X-C\beta
\right) }\left( \frac{f_{1}\left( X\right) }{C_{0}+\bar{r}}\right) ^{\frac{1%
}{r}}\left( \frac{\left( 1+\underline{k}\left( X\right) \right) \frac{\left(
3X-C\beta \right) \left( X+C\beta \right) }{2X-C\beta }-\bar{r}\left( \frac{%
f_{1}\left( X\right) }{C_{0}+\bar{r}}\right) ^{\frac{1}{r}}}{4\left( 1+%
\underline{k}_{2}\left( X\right) \right) }-C\right)
\end{eqnarray*}

and the equation for $\hat{g}$ is: 
\begin{eqnarray*}
&&\int \left( \frac{\Delta \left( \hat{X},\hat{X}^{\prime }\right) }{1+%
\underline{\hat{k}}_{2}\left( \hat{X}\right) }-\hat{S}_{1}^{E}\left( \hat{X}%
^{\prime },\hat{X}\right) \right) \left( \frac{\left( \left\Vert \hat{\Psi}%
_{0}\left( \hat{X}^{\prime }\right) \right\Vert ^{2}-\hat{\mu}D\left( \hat{X}%
^{\prime }\right) \right) \sqrt{\frac{1}{4}-\frac{r}{3}\underline{\hat{k}}}}{%
\sqrt{\frac{\hat{\mu}\hat{K}\left[ \hat{X}^{\prime }\right] }{2\sigma _{\hat{%
K}}^{2}}}}-\bar{r}^{\prime }\right) \\
&=&\left( 1-\beta \right) \frac{\frac{1}{3}\left( 2X-C\beta \right)
^{2}\left( \left( X+C\beta \right) \right) \epsilon }{\sigma _{\hat{K}%
}^{2}\left( 1+\underline{k}\left( X\right) \right) ^{2}\left( 3X-C\beta
\right) }\left( \frac{f_{1}\left( X\right) }{C_{0}+\bar{r}}\right) ^{\frac{1%
}{r}}\left( \frac{\left( 1+\underline{k}\left( X\right) \right) \frac{\left(
3X-C\beta \right) \left( X+C\beta \right) }{2X-C\beta }-\bar{r}\left( \frac{%
f_{1}\left( X\right) }{C_{0}+\bar{r}}\right) ^{\frac{1}{r}}}{4\left( 1+%
\underline{k}_{2}\left( X\right) \right) }-C\right)
\end{eqnarray*}%
Rewritten as an equation for $\underline{k}\left( X\right) $ and using as
before:%
\begin{equation*}
\hat{K}\left[ \hat{X}^{\prime }\right] =\frac{\underline{k}\left( X\right) }{%
k}
\end{equation*}%
this becomes:%
\begin{eqnarray*}
&&\int \left( \frac{\Delta \left( \hat{X},\hat{X}^{\prime }\right) }{1+%
\underline{\hat{k}}_{2}\left( \hat{X}\right) }-\hat{S}_{1}^{E}\left( \hat{X}%
^{\prime },\hat{X}\right) \right) \left( \frac{\left( \left\Vert \hat{\Psi}%
_{0}\left( \hat{X}^{\prime }\right) \right\Vert ^{2}-\hat{\mu}D\left( \hat{X}%
^{\prime }\right) \right) \sqrt{\frac{2\sigma _{\hat{K}}^{2}k}{\hat{\mu}}}%
\sqrt{\frac{1}{4}-\frac{r}{3}\underline{\hat{k}}}}{\sqrt{\underline{k}\left( 
\hat{X}^{\prime }\right) }}-\bar{r}^{\prime }\right) \\
&=&\left( 1-\beta \right) \frac{\frac{1}{3}\left( 2X-C\beta \right)
^{2}\left( \left( X+C\beta \right) \right) \epsilon }{\sigma _{\hat{K}%
}^{2}\left( 1+\underline{k}\left( X\right) \right) ^{2}\left( 3X-C\beta
\right) }\left( \frac{f_{1}\left( X\right) }{C_{0}+\bar{r}}\right) ^{\frac{1%
}{r}}\left( \frac{\left( 1+\underline{k}\left( X\right) \right) \frac{\left(
3X-C\beta \right) \left( X+C\beta \right) }{2X-C\beta }-\bar{r}\left( \frac{%
f_{1}\left( X\right) }{C_{0}+\bar{r}}\right) ^{\frac{1}{r}}}{4\left( 1+\beta 
\underline{k}\left( X\right) \right) }-C\right)
\end{eqnarray*}

\subsection*{A12.3 Solutions for investors without connections}

\begin{equation*}
\frac{N\left( \hat{X}^{\prime }\right) }{\sqrt{\underline{k}\left( \hat{X}%
^{\prime }\right) }}-\bar{r}^{\prime }=\left( 1-\beta \right) \frac{\frac{1}{%
3}\left( 2X-C\beta \right) ^{2}\left( \left( X+C\beta \right) \right)
\epsilon }{\sigma _{\hat{K}}^{2}\left( 1+\underline{k}\left( X\right)
\right) ^{2}\left( 3X-C\beta \right) }\left( \frac{f_{1}\left( X\right) }{%
C_{0}+\bar{r}}\right) ^{\frac{1}{r}}\left( \frac{\left( 1+\underline{k}%
\left( X\right) \right) \frac{\left( 3X-C\beta \right) \left( X+C\beta
\right) }{2X-C\beta }-\bar{r}\left( \frac{f_{1}\left( X\right) }{C_{0}+\bar{r%
}}\right) ^{\frac{1}{r}}}{4\left( 1+\beta \underline{k}\left( X\right)
\right) }-C\right)
\end{equation*}%
with:%
\begin{equation*}
N\left( \hat{X}^{\prime }\right) =\left( \left\Vert \hat{\Psi}_{0}\left( 
\hat{X}^{\prime }\right) \right\Vert ^{2}-\hat{\mu}D\left( \hat{X}^{\prime
}\right) \right) \sqrt{\frac{2\sigma _{\hat{K}}^{2}k}{\hat{\mu}}}\sqrt{\frac{%
1}{4}-\frac{r}{3}\underline{\hat{k}}}
\end{equation*}%
Defining:%
\begin{eqnarray*}
A &=&\frac{N\left( \hat{X}^{\prime }\right) }{\sqrt{\underline{k}\left( \hat{%
X}^{\prime }\right) }}-\bar{r}^{\prime } \\
B &=&\left( 1-\beta \right) \frac{\frac{1}{3}\left( 2X-C\beta \right)
^{2}\left( \left( X+C\beta \right) \right) \epsilon }{\sigma _{\hat{K}%
}^{2}\left( 3X-C\beta \right) } \\
F &=&\frac{\left( 3X-C\beta \right) \left( X+C\beta \right) }{4\left(
2X-C\beta \right) } \\
D &=&\frac{\bar{r}}{4}
\end{eqnarray*}%
We can express $\left( \frac{f_{1}\left( X\right) }{C_{0}+\bar{r}}\right) ^{%
\frac{1}{r}}$ as a function of $\underline{k}\left( X\right) $. Two cases
arise 
\begin{eqnarray*}
\left( \frac{f_{1}\left( X\right) }{C_{0}+\bar{r}}\right) ^{\frac{1}{r}} &=&%
\frac{\left( 1+\beta \underline{k}\left( X\right) \right) }{2D} \\
&&\times \left( \sqrt{\left( \frac{\left( 1+\underline{k}\left( X\right)
\right) F}{4\left( 1+\beta \underline{k}\left( X\right) \right) }-C\right)
^{2}-4\left( \frac{N\left( \hat{X}^{\prime }\right) }{\sqrt{\underline{k}%
\left( \hat{X}\right) }}-\bar{r}^{\prime }\right) \frac{\left( 1+\underline{k%
}\left( X\right) \right) ^{2}D}{B\left( 1+\beta \underline{k}\left( X\right)
\right) }}+\left( \frac{\left( 1+\underline{k}\left( X\right) \right) F}{%
4\left( 1+\beta \underline{k}\left( X\right) \right) }-C\right) \right)
\end{eqnarray*}%
We find:%
\begin{equation*}
\frac{d\left( \frac{f_{1}\left( X\right) }{C_{0}+\bar{r}}\right) ^{\frac{1}{r%
}}}{d\underline{k}\left( X\right) }>0
\end{equation*}%
so that:%
\begin{equation*}
\frac{d\underline{k}\left( X\right) }{d\left( \frac{f_{1}\left( X\right) }{%
C_{0}+\bar{r}}\right) ^{\frac{1}{r}}}>0
\end{equation*}%
as in the cases studied in the text.

For the case $\frac{N\left( \hat{X}^{\prime }\right) }{\sqrt{\underline{k}%
\left( \hat{X}^{\prime }\right) }}<0$ corresponding to negative retrn, the
soltn wrt:%
\begin{equation*}
\left( \frac{f_{1}\left( X\right) }{C_{0}+\bar{r}}\right) ^{\frac{1}{r}}=%
\frac{\left( 1+\beta \underline{k}\left( X\right) \right) }{2D}\left( \sqrt{%
\left( \frac{\left( 1+\underline{k}\left( X\right) \right) F}{4\left(
1+\beta \underline{k}\left( X\right) \right) }-C\right) ^{2}+4\left( \frac{%
N\left( \hat{X}^{\prime }\right) }{\sqrt{\underline{k}\left( \hat{X}\right) }%
}+\bar{r}^{\prime }\right) \frac{\left( 1+\underline{k}\left( X\right)
\right) ^{2}D}{B\left( 1+\beta \underline{k}\left( X\right) \right) }}%
+\left( \frac{\left( 1+\underline{k}\left( X\right) \right) F}{4\left(
1+\beta \underline{k}\left( X\right) \right) }-C\right) \right)
\end{equation*}

Then, consider $\left( \frac{N\left( \hat{X}^{\prime }\right) }{\sqrt{%
\underline{k}\left( \hat{X}^{\prime }\right) }}-\bar{r}^{\prime }\right) >0$%
. Then the equation to consider is:%
\begin{equation*}
-\frac{N\left( \hat{X}^{\prime }\right) }{\sqrt{\underline{k}\left( \hat{X}%
^{\prime }\right) }}-\bar{r}^{\prime }=\left( 1-\beta \right) \frac{\frac{1}{%
3}\left( 2X-C\beta \right) ^{2}\left( \left( X+C\beta \right) \right)
\epsilon }{\sigma _{\hat{K}}^{2}\left( 1+\underline{k}\left( X\right)
\right) ^{2}\left( 3X-C\beta \right) }\left( \frac{f_{1}\left( X\right) }{%
C_{0}+\bar{r}}\right) ^{\frac{1}{r}}\left( \frac{\left( 1+\underline{k}%
\left( X\right) \right) \frac{\left( 3X-C\beta \right) \left( X+C\beta
\right) }{2X-C\beta }-\bar{r}\left( \frac{f_{1}\left( X\right) }{C_{0}+\bar{r%
}}\right) ^{\frac{1}{r}}}{4\left( 1+\beta \underline{k}\left( X\right)
\right) }-C\right)
\end{equation*}%
corresponding to negative returns.

The solution is thus:%
\begin{equation*}
\left( \frac{f_{1}\left( X\right) }{C_{0}+\bar{r}}\right) ^{\frac{1}{r}}=%
\frac{\left( 1+\beta \underline{k}\left( X\right) \right) }{2D}\left( \sqrt{%
\left( \frac{\left( 1+\underline{k}\left( X\right) \right) F}{4\left(
1+\beta \underline{k}\left( X\right) \right) }-C\right) ^{2}+4\left( \frac{%
N\left( \hat{X}^{\prime }\right) }{\sqrt{\underline{k}\left( \hat{X}\right) }%
}+\bar{r}^{\prime }\right) \frac{\left( 1+\underline{k}\left( X\right)
\right) ^{2}D}{B\left( 1+\beta \underline{k}\left( X\right) \right) }}%
-\left( \frac{\left( 1+\underline{k}\left( X\right) \right) F}{4\left(
1+\beta \underline{k}\left( X\right) \right) }-C\right) \right)
\end{equation*}%
with also: 
\begin{equation*}
\frac{d\underline{k}\left( X\right) }{d\left( \frac{f_{1}\left( X\right) }{%
C_{0}+\bar{r}}\right) ^{\frac{1}{r}}}>0
\end{equation*}

\subsection*{A12.4 Solutions for investors with connections}

Looking for apprximate solutions of:%
\begin{eqnarray*}
&&\int \left( \frac{\Delta \left( \hat{X},\hat{X}^{\prime }\right) }{1+%
\underline{\hat{k}}_{2}\left( \hat{X}\right) }-\hat{S}_{1}^{E}\left( \hat{X}%
^{\prime },\hat{X}\right) \right) \left( \frac{N\left( \hat{X}^{\prime
}\right) }{\sqrt{\underline{k}\left( \hat{X}^{\prime }\right) }}-\bar{r}%
^{\prime }\right) \\
&=&\frac{B}{\left( 1+\underline{k}\left( X\right) \right) ^{2}}\left( \frac{%
f_{1}\left( X\right) }{C_{0}+\bar{r}}\right) ^{\frac{1}{r}}\left( \frac{%
\left( 1+\underline{k}\left( X\right) \right) F}{4\left( 1+\beta \underline{k%
}\left( X\right) \right) }-\frac{D}{\left( 1+\beta \underline{k}\left(
X\right) \right) }\left( \frac{f_{1}\left( X\right) }{C_{0}+\bar{r}}\right)
^{\frac{1}{r}}-C\right)
\end{eqnarray*}

where $\hat{S}_{1}^{E}\left( \hat{X}^{\prime },\hat{X}\right) $ has been
computed previousl:%
\begin{eqnarray*}
\hat{S}_{1}^{E}\left( \hat{X}^{\prime },\hat{X}\right) &=&\left( \frac{\hat{k%
}\left( \hat{X},\hat{X}^{\prime }\right) -\underline{\hat{k}}_{1}\left(
\left\langle X\right\rangle ,\hat{X}\right) k\left( \left\langle
X\right\rangle ,X^{\prime }\right) }{1+\underline{\hat{k}}_{2}\left( \hat{X}%
\right) }+\frac{\hat{k}_{1}\left( \hat{X}^{\prime },\hat{X}\right) }{1+%
\underline{\hat{k}}_{2}\left( \hat{X}^{\prime }\right) }\right) \frac{\hat{K}%
_{X}\left\Vert \hat{\Psi}\left( \hat{X}^{\prime }\right) \right\Vert ^{2}}{1+%
\hat{k}\left( \hat{X}\right) } \\
&=&\left( \frac{\hat{k}\left( \hat{X},\hat{X}^{\prime }\right) -\underline{%
\hat{k}}_{1}\left( \left\langle X\right\rangle ,\hat{X}\right) k\left(
\left\langle X\right\rangle ,X^{\prime }\right) }{1+\underline{\hat{k}}%
_{2}\left( \hat{X}\right) }+\frac{\hat{k}_{1}\left( \hat{X}^{\prime },\hat{X}%
\right) }{1+\underline{\hat{k}}_{2}\left( \hat{X}^{\prime }\right) }\right) 
\frac{\underline{k}\left( X\right) }{k\left( 1+\hat{k}\left( \hat{X}\right)
\right) }
\end{eqnarray*}%
As in the case of slowly decreasing returns, we define:%
\begin{equation*}
\frac{N\left( \hat{X}^{\prime }\right) }{\sqrt{\underline{k}\left( \hat{X}%
^{\prime }\right) }}=\frac{N\left( \hat{X}^{\prime }\right) }{\sqrt{%
\underline{k}^{\prime }\left( \hat{X}^{\prime }\right) }}+\Delta \left( 
\frac{N\left( \hat{X}^{\prime }\right) }{\sqrt{\underline{k}\left( \hat{X}%
^{\prime }\right) }}\right)
\end{equation*}%
where $\frac{N\left( \hat{X}^{\prime }\right) }{\sqrt{\underline{k}^{\prime
}\left( \hat{X}^{\prime }\right) }}$ is the solution without interactions.
This leads to the following qt:%
\begin{eqnarray*}
&&\int \left( \frac{\Delta \left( \hat{X},\hat{X}^{\prime }\right) }{1+%
\underline{\hat{k}}_{2}\left( \hat{X}\right) }-\hat{S}_{1}^{E}\left( \hat{X}%
^{\prime },\hat{X}\right) \right) \left( \Delta \left( \frac{N\left( \hat{X}%
^{\prime }\right) }{\sqrt{\underline{k}\left( \hat{X}^{\prime }\right) }}%
\right) \right) -\int \left( \hat{S}_{1}^{E}\left( \hat{X}^{\prime },\hat{X}%
\right) \frac{N\left( \hat{X}^{\prime }\right) }{\sqrt{\underline{k}^{\prime
}\left( \hat{X}^{\prime }\right) }}\right) \\
&=&H_{1}\Delta \left( \frac{N\left( \hat{X}^{\prime }\right) }{\sqrt{%
\underline{k}\left( \hat{X}^{\prime }\right) }}\right) +\frac{1}{2}%
H_{2}\Delta \left( \frac{N\left( \hat{X}^{\prime }\right) }{\sqrt{\underline{%
k}\left( \hat{X}^{\prime }\right) }}\right) ^{2}
\end{eqnarray*}%
where:%
\begin{equation*}
H_{l}=\frac{\partial ^{l}}{\partial \underline{k}^{l}\left( \hat{X}^{\prime
}\right) }\left( \frac{B}{\left( 1+\underline{k}\left( X\right) \right) ^{2}}%
\left( \frac{f_{1}\left( X\right) }{C_{0}+\bar{r}}\right) ^{\frac{1}{r}%
}\left( \frac{\left( 1+\underline{k}\left( X\right) \right) F}{4\left(
1+\beta \underline{k}\left( X\right) \right) }-\frac{D}{\left( 1+\beta 
\underline{k}\left( X\right) \right) }\left( \frac{f_{1}\left( X\right) }{%
C_{0}+\bar{r}}\right) ^{\frac{1}{r}}-C\right) \right)
\end{equation*}%
an equation similar to that presented in the text. The analysis of the
solutions is thus similar.

\section*{Appendix 13 Several groups}

As a benchmark and to introduce some notations, we first reconsider the case
of an homogeneous group.

\subsection*{A13.1 Homogeneous group}

We reconsider the previous computations, considering the system as an
homogeneous group of agents, with identical average return. The return
equation for such homogeneous group is:%
\begin{eqnarray*}
&&\left( \frac{f\left( \hat{X}\right) }{1+\underline{\hat{k}}_{2}\left( \hat{%
X}\right) }+\bar{r}\frac{\underline{\hat{k}}_{2}\left( \hat{X}\right) }{1+%
\underline{\hat{k}}_{2}\left( \hat{X}\right) }\right) -\frac{\hat{K}\hat{k}%
_{1E}\left( \hat{X}\right) }{1+\underline{\hat{k}}}\left( \frac{f}{1+%
\underline{\hat{k}}_{2}}+\bar{r}\frac{\underline{\hat{k}}_{2}}{1+\underline{%
\hat{k}}_{2}}\right) \\
&=&\left\{ \left( \bar{r}+\frac{1+f}{\underline{\hat{k}}_{2}}H\left( -\frac{%
1+f}{\underline{\hat{k}}_{2}}\right) \right) \frac{\hat{k}_{2E}\left( \hat{X}%
\right) }{1+\underline{\hat{k}}\left( \hat{X}^{\prime }\right) }\right. \\
&&\left. +\left( \bar{r}+\frac{1+f_{1}^{\prime }\left( X^{\prime }\right) }{%
\underline{k}_{2}\left( X^{\prime }\right) }H\left( -\frac{1+f_{1}^{\prime
}\left( X^{\prime }\right) }{\underline{k}_{2}\left( X^{\prime }\right) }%
\right) \right) \frac{k_{2E}\left( \hat{X}\right) }{1+\underline{k}\left(
X^{\prime }\right) }+\frac{k_{1E}\left( \hat{X}\right) }{1+\underline{k}%
\left( X^{\prime }\right) }f_{1}\left( \hat{K},\hat{X},\Psi ,\hat{\Psi}%
\right) \right\}
\end{eqnarray*}%
$\underline{\hat{k}}_{2}$ and $\underline{\hat{k}}_{1}$\ are the\ averages
of $\underline{\hat{k}}_{2}\left( \hat{X}\right) $ and $\underline{\hat{k}}%
_{2}\left( \hat{X}\right) $. In average, using that $\hat{k}_{\eta E}=\hat{k}%
_{\eta 1}$, we fnd: 
\begin{eqnarray*}
&&\left( \frac{f}{1+\underline{\hat{k}}_{2}}+\bar{r}\frac{\underline{\hat{k}}%
_{2}}{1+\underline{\hat{k}}_{2}}\right) -\frac{\underline{\hat{k}}_{1}}{1+%
\underline{\hat{k}}}\left( \frac{f}{1+\underline{\hat{k}}_{2}}+\bar{r}\frac{%
\underline{\hat{k}}_{2}}{1+\underline{\hat{k}}_{2}}\right) \\
&=&\left\{ \left( \bar{r}+\left( \frac{1+f}{\underline{\hat{k}}_{2}}\right)
H\left( -\left( 1+f\right) \right) \right) \frac{\underline{\hat{k}}_{2}}{1+%
\underline{\hat{k}}}\right. \\
&&\left. +\left( \bar{r}+\left( \frac{1+f_{1}^{\prime }}{\underline{k}_{2}}%
\right) H\left( -\left( 1+f_{1}^{\prime }\right) \right) \right) \frac{k_{2}%
}{1+\underline{k}}+\frac{k_{1}}{1+\underline{k}}f_{1}\right\}
\end{eqnarray*}%
leading to:%
\begin{equation*}
\frac{f}{1+\underline{\hat{k}}}=\left( \frac{1+f}{\underline{\hat{k}}_{2}}%
\right) H\left( -\left( 1+f\right) \right) \frac{\underline{\hat{k}}_{2}}{1+%
\underline{\hat{k}}}+\left( \frac{1+f_{1}^{\prime }}{\underline{k}_{2}}%
\right) H\left( -\left( 1+f_{1}^{\prime }\right) \right) \frac{k_{2}}{1+%
\underline{k}}+\bar{r}\frac{k_{2}}{1+\underline{k}}+\frac{k_{1}}{1+%
\underline{k}}f_{1}
\end{equation*}%
Using the constraint (\ref{CT}):%
\begin{equation*}
\frac{\underline{k}}{1+\underline{k}}=\frac{1}{1+\underline{\hat{k}}}
\end{equation*}%
this reduces to:%
\begin{equation*}
\frac{f}{1+\underline{\hat{k}}}=\left( \frac{1+f}{\underline{\hat{k}}_{2}}%
\right) H\left( -\left( 1+f\right) \right) \frac{\underline{\hat{k}}_{2}}{1+%
\underline{\hat{k}}}+\left( \frac{1+f_{1}^{\prime }}{\underline{k}_{2}}%
\right) H\left( -\left( 1+f_{1}^{\prime }\right) \right) \frac{k_{2}}{%
k\left( 1+\underline{\hat{k}}\right) }+\bar{r}\frac{k_{2}}{k\left( 1+%
\underline{\hat{k}}\right) }+\frac{k_{1}}{k\left( 1+\underline{\hat{k}}%
\right) }f_{1}
\end{equation*}%
that simplifies using (\ref{CT})

\begin{equation*}
\frac{f}{1+\underline{\hat{k}}}=\left( \frac{1+f}{\underline{\hat{k}}_{2}}%
\right) H\left( -\left( 1+f\right) \right) \frac{\underline{\hat{k}}_{2}}{1+%
\underline{\hat{k}}}+\left( \frac{1+f_{1}^{\prime }}{\underline{k}_{2}}%
\right) H\left( -\left( 1+f_{1}^{\prime }\right) \right) \frac{k_{2}}{1+%
\underline{k}}+\bar{r}\frac{k_{2}}{1+\underline{k}}+\frac{k_{1}}{1+%
\underline{k}}f_{1}
\end{equation*}

\subsection*{A13.2 Several groups}

comng back to the return equation, it can be written for several homogeneous
groups:%
\begin{eqnarray*}
&&\left( \frac{f^{\left[ i\right] }}{1+\underline{\hat{k}}_{2}^{\left[ i%
\right] }}+\bar{r}\frac{\underline{\hat{k}}_{2}^{\left[ i\right] }}{1+%
\underline{\hat{k}}_{2}^{\left[ i\right] }}\right) -\frac{\hat{k}_{1}^{\left[
ji\right] }}{1+\underline{\hat{k}}^{\left[ j\right] }}\left( \frac{f^{\left[
j\right] }}{1+\underline{\hat{k}}_{2}^{\left[ j\right] }}+\bar{r}\frac{%
\underline{\hat{k}}_{2}^{\left[ j\right] }}{1+\underline{\hat{k}}_{2}^{\left[
j\right] }}\right) \\
&=&\left( \bar{r}+\frac{1+f^{\left[ i\right] }}{\underline{\hat{k}}_{2}^{%
\left[ i\right] }}H\left( -\left( 1+f^{\left[ i\right] }\right) \right)
\right) \frac{\underline{\hat{k}}_{2}^{\left[ ii\right] }}{1+\underline{\hat{%
k}}^{\left[ i\right] }}+\left( \bar{r}+\frac{1+f^{\left[ j\right] }}{%
\underline{\hat{k}}_{2}^{\left[ j\right] }}H\left( -\left( 1+f^{\left[ j%
\right] }\right) \right) \right) \frac{\underline{\hat{k}}_{2}^{\left[ ji%
\right] }}{1+\underline{\hat{k}}^{\left[ j\right] }} \\
&&+\left( \bar{r}+\frac{1+f_{1}^{\prime \left[ i\right] }}{\underline{k}%
_{2}^{\left[ i\right] }}H\left( -\frac{1+f_{1}^{\prime \left[ i\right] }}{%
\underline{k}_{2}^{\left[ i\right] }}\right) \right) \frac{k_{2}^{\left[ ii%
\right] }}{1+\underline{k}^{i}}+\frac{k_{1}^{\left[ ii\right] }}{1+%
\underline{k}^{i}}f_{1}^{\left[ i\right] }+\left( \bar{r}+\frac{%
1+f_{1}^{\prime \left[ j\right] }}{\underline{k}_{2}^{\left[ j\right] }}%
H\left( -\frac{1+f_{1}^{\prime \left[ j\right] }}{\underline{k}_{2}^{\left[ j%
\right] }}\right) \right) \frac{k_{2}^{\left[ ji\right] }}{1+\underline{k}%
^{j}}+\frac{k_{1}^{\left[ ij\right] }}{1+\underline{k}^{j}}f_{1}^{\left[ j%
\right] }
\end{eqnarray*}

The following constraints arise, translating that investors share their
investors in loans and participation in firms or investrs:%
\begin{equation*}
\frac{\underline{\hat{k}}^{\left[ ji\right] }}{1+\underline{\hat{k}}^{\left[
j\right] }}+\frac{\underline{\hat{k}}^{\left[ ii\right] }}{1+\underline{\hat{%
k}}^{\left[ i\right] }}+\frac{\underline{k}^{\left[ ji\right] }}{1+%
\underline{k}^{\left[ j\right] }}+\frac{\underline{k}^{\left[ ii\right] }}{1+%
\underline{k}^{\left[ i\right] }}=1
\end{equation*}%
This can be rewritten matrcial:

\begin{eqnarray}
0 &=&\left( 
\begin{array}{cc}
\frac{1}{1+\underline{\hat{k}}_{2}^{\left[ i\right] }}-\frac{\underline{\hat{%
k}}_{1}^{\left[ ii\right] }}{\left( 1+\underline{\hat{k}}^{\left[ i\right]
}\right) \left( 1+\underline{\hat{k}}_{2}^{\left[ i\right] }\right) } & -%
\frac{\underline{\hat{k}}_{1}^{\left[ ji\right] }}{\left( 1+\underline{\hat{k%
}}^{\left[ j\right] }\right) \left( 1+\underline{\hat{k}}_{2}^{\left[ j%
\right] }\right) } \\ 
-\frac{\underline{\hat{k}}_{1}^{\left[ ij\right] }}{\left( 1+\underline{\hat{%
k}}^{\left[ i\right] }\right) \left( 1+\underline{\hat{k}}_{2}^{\left[ i%
\right] }\right) } & \frac{1}{1+\underline{\hat{k}}_{2}^{\left[ j\right] }}-%
\frac{\underline{\hat{k}}_{1}^{\left[ jj\right] }}{\left( 1+\underline{\hat{k%
}}^{\left[ j\right] }\right) \left( 1+\underline{\hat{k}}_{2}^{\left[ j%
\right] }\right) }%
\end{array}%
\right) \left( 
\begin{array}{c}
f^{\left[ i\right] }+\bar{r}\underline{\hat{k}}_{2}^{\left[ i\right] } \\ 
f^{\left[ j\right] }+\bar{r}\underline{\hat{k}}_{2}^{\left[ j\right] }%
\end{array}%
\right)  \label{q} \\
&&-\left( 
\begin{array}{c}
\frac{1+f^{\left[ i\right] }}{\underline{\hat{k}}_{2}^{\left[ i\right] }}%
\frac{\underline{\hat{k}}_{2}^{\left[ ii\right] }}{1+\underline{\hat{k}}^{%
\left[ i\right] }}H\left( -\left( 1+f^{\left[ i\right] }\right) \right) +%
\frac{1+f^{\left[ j\right] }}{\underline{\hat{k}}_{2}^{\left[ j\right] }}%
\frac{\underline{\hat{k}}_{2}^{\left[ ji\right] }}{1+\underline{\hat{k}}^{%
\left[ j\right] }}H\left( -\left( 1+f^{\left[ j\right] }\right) \right) \\ 
\frac{1+f^{\left[ j\right] }}{\underline{\hat{k}}_{2}^{\left[ j\right] }}%
\frac{\underline{\hat{k}}_{2}^{\left[ jj\right] }}{1+\underline{\hat{k}}^{%
\left[ j\right] }}H\left( -\left( 1+f^{\left[ j\right] }\right) \right) +%
\frac{1+f^{\left[ i\right] }}{\underline{\hat{k}}_{2}^{\left[ i\right] }}%
\frac{\underline{\hat{k}}_{2}^{\left[ ij\right] }}{1+\underline{\hat{k}}^{%
\left[ i\right] }}H\left( -\left( 1+f^{\left[ i\right] }\right) \right)%
\end{array}%
\right)  \notag \\
&&-\left( 
\begin{array}{c}
\frac{1+f_{1}^{\prime \left[ i\right] }}{\underline{k}_{2}^{\left[ i\right] }%
}\frac{k_{2}^{\left[ ii\right] }}{1+\underline{k}^{\left[ i\right] }}H\left(
-\left( 1+f_{1}^{\prime \left[ i\right] }\right) \right) +\frac{%
1+f_{1}^{\prime \left[ j\right] }}{\underline{k}_{2}^{\left[ j\right] }}%
\frac{k_{2}^{\left[ ji\right] }}{1+\underline{k}^{\left[ j\right] }}H\left(
-\left( 1+f_{1}^{\prime \left[ j\right] }\right) \right) \\ 
\frac{1+f_{1}^{\prime \left[ i\right] }}{\underline{k}_{2}^{\left[ i\right] }%
}\frac{k_{2}^{\left[ ij\right] }}{1+\underline{k}^{\left[ i\right] }}H\left(
-\frac{1+f_{1}^{\prime \left[ i\right] }}{\underline{k}_{2}^{\left[ i\right]
}}\right) +\frac{1+f_{1}^{\prime \left[ j\right] }}{\underline{k}_{2}^{\left[
j\right] }}\frac{k_{2}^{\left[ jj\right] }}{1+\underline{k}^{\left[ j\right]
}}H\left( -\left( 1+f_{1}^{\prime \left[ j\right] }\right) \right)%
\end{array}%
\right) -\left( 
\begin{array}{c}
\bar{r}_{i} \\ 
\bar{r}_{j}%
\end{array}%
\right)  \notag
\end{eqnarray}

whr:%
\begin{eqnarray*}
\bar{r}_{i} &=&\left( \frac{\underline{\hat{k}}_{2}^{\left[ ji\right] }}{1+%
\underline{\hat{k}}^{\left[ j\right] }}+\frac{\underline{\hat{k}}_{2}^{\left[
ii\right] }}{1+\underline{\hat{k}}^{\left[ i\right] }}+\frac{k_{2}^{\left[ ii%
\right] }}{1+\underline{k}^{i}}+\frac{k_{2}^{\left[ ji\right] }}{1+%
\underline{k}^{j}}\right) \bar{r}+\frac{k_{1}^{\left[ ii\right] }}{1+%
\underline{k}^{\left[ i\right] }}f_{1}^{\left[ i\right] }+\frac{k_{1}^{\left[
ji\right] }}{1+\underline{k}^{\left[ j\right] }}f_{1}^{\left[ j\right] } \\
&=&\left( \frac{\underline{\hat{k}}_{2}^{\left[ ji\right] }}{1+\underline{%
\hat{k}}^{\left[ j\right] }}+\frac{\underline{\hat{k}}_{2}^{\left[ ii\right]
}}{1+\underline{\hat{k}}^{\left[ i\right] }}+\frac{k^{\left[ ii\right] }}{1+%
\underline{k}^{i}}+\frac{k^{\left[ ji\right] }}{1+\underline{k}^{j}}\right) 
\bar{r}+\frac{k_{1}^{\left[ ii\right] }}{1+\underline{k}^{\left[ i\right] }}%
\left( f_{1}^{\left[ i\right] }-\bar{r}\right) +\frac{k_{1}^{\left[ ji\right]
}}{1+\underline{k}^{\left[ j\right] }}\left( f_{1}^{\left[ j\right] }-\bar{r}%
\right)
\end{eqnarray*}

Usng:%
\begin{equation*}
\frac{\underline{\hat{k}}^{\left[ ji\right] }}{1+\underline{\hat{k}}^{\left[
j\right] }}+\frac{\underline{\hat{k}}^{\left[ ii\right] }}{1+\underline{\hat{%
k}}^{\left[ i\right] }}+\frac{\underline{k}^{\left[ ji\right] }}{1+%
\underline{k}^{\left[ j\right] }}+\frac{\underline{k}^{\left[ ii\right] }}{1+%
\underline{k}^{\left[ i\right] }}=1
\end{equation*}%
this becms:%
\begin{eqnarray*}
&&\left( 1-\frac{\underline{\hat{k}}^{\left[ ii\right] }}{1+\underline{\hat{k%
}}^{\left[ i\right] }}-\frac{\underline{\hat{k}}^{\left[ ji\right] }}{1+%
\underline{\hat{k}}^{\left[ j\right] }}+\frac{\underline{\hat{k}}_{2}^{\left[
ji\right] }}{1+\underline{\hat{k}}^{\left[ j\right] }}+\frac{\underline{\hat{%
k}}_{2}^{\left[ ii\right] }}{1+\underline{\hat{k}}^{\left[ i\right] }}%
\right) \bar{r}+\frac{k_{1}^{\left[ ii\right] }}{1+\underline{k}^{\left[ i%
\right] }}\left( f_{1}^{\left[ i\right] }-\bar{r}\right) +\frac{k_{1}^{\left[
ji\right] }}{1+\underline{k}^{\left[ j\right] }}\left( f_{1}^{\left[ j\right]
}-\bar{r}\right) \\
&=&\left( 1-\frac{\underline{\hat{k}}_{1}^{\left[ ii\right] }}{1+\underline{%
\hat{k}}^{\left[ i\right] }}-\frac{\underline{\hat{k}}_{1}^{\left[ ji\right]
}}{1+\underline{\hat{k}}^{\left[ j\right] }}\right) \bar{r}+\frac{k_{1}^{%
\left[ ii\right] }}{1+\underline{k}^{\left[ i\right] }}\left( f_{1}^{\left[ i%
\right] }-\bar{r}\right) +\frac{k_{1}^{\left[ ji\right] }}{1+\underline{k}^{%
\left[ j\right] }}\left( f_{1}^{\left[ j\right] }-\bar{r}\right)
\end{eqnarray*}%
\begin{equation*}
1-\frac{\underline{\hat{k}}_{1}^{\left[ ii\right] }}{1+\underline{\hat{k}}^{%
\left[ i\right] }}-\frac{\underline{\hat{k}}_{1}^{\left[ ji\right] }}{1+%
\underline{\hat{k}}^{\left[ j\right] }}
\end{equation*}%
Then we reformulate (\ref{q}) by using the alternative description
corresponding here to define:%
\begin{eqnarray*}
\underline{\hat{S}}_{\eta }^{\left[ ii\right] } &=&\frac{\underline{\hat{k}}%
_{\eta }^{\left[ ii\right] }}{1+\underline{\hat{k}}^{\left[ i\right] }} \\
\underline{\hat{S}}_{\eta }^{\left[ ij\right] } &=&\frac{\underline{\hat{k}}%
_{\eta }^{\left[ ij\right] }}{1+\underline{\hat{k}}^{\left[ i\right] }}
\end{eqnarray*}%
\begin{eqnarray*}
\underline{\hat{S}}^{\left[ ii\right] } &=&\underline{\hat{S}}_{1}^{\left[ ii%
\right] }+\underline{\hat{S}}_{2}^{\left[ ii\right] } \\
\underline{\hat{S}}^{\left[ ij\right] } &=&\underline{\hat{S}}_{1}^{\left[ ij%
\right] }+\underline{\hat{S}}_{2}^{\left[ ij\right] }
\end{eqnarray*}%
\begin{eqnarray*}
\underline{S}_{\eta }^{\left[ ii\right] } &=&\frac{k_{\eta }^{\left[ ii%
\right] }}{1+\underline{k}^{\left[ i\right] }} \\
\underline{S}_{\eta }^{\left[ ij\right] } &=&\frac{k_{\eta }^{\left[ ij%
\right] }}{1+\underline{k}^{\left[ i\right] }}
\end{eqnarray*}%
\begin{eqnarray*}
\underline{S}^{\left[ ii\right] } &=&\underline{S}_{1}^{\left[ ii\right]
}+S_{2}^{\left[ ii\right] } \\
\underline{S}^{\left[ ij\right] } &=&\underline{S}_{1}^{\left[ ij\right]
}+S_{2}^{\left[ ij\right] }
\end{eqnarray*}

\begin{equation*}
\underline{\hat{S}}_{\eta }^{\left[ ii\right] }+\underline{\hat{S}}_{\eta }^{%
\left[ ij\right] }=\frac{\underline{\hat{k}}_{\eta }^{\left[ i\right] }}{1+%
\underline{\hat{k}}^{\left[ i\right] }}
\end{equation*}%
and using that these coeficients satisfy the identitites%
\begin{equation*}
\underline{\hat{S}}^{\left[ ii\right] }+\underline{\hat{S}}^{\left[ ji\right]
}+\underline{S}^{\left[ ii\right] }+\underline{S}^{\left[ ji\right] }=1
\end{equation*}%
\bigskip 
\begin{equation*}
\underline{\hat{S}}_{\eta }^{\left[ ii\right] }+\underline{\hat{S}}_{\eta }^{%
\left[ ij\right] }=\frac{\underline{\hat{k}}_{\eta }^{\left[ i\right] }}{1+%
\underline{\hat{k}}^{\left[ i\right] }}
\end{equation*}%
\begin{equation*}
\underline{\hat{S}}^{\left[ ii\right] }+\underline{\hat{S}}^{\left[ ij\right]
}=\frac{\underline{\hat{k}}^{\left[ i\right] }}{1+\underline{\hat{k}}^{\left[
i\right] }}
\end{equation*}%
\begin{equation*}
\frac{1}{1+\underline{\hat{k}}^{\left[ i\right] }}=1-\left( \underline{\hat{S%
}}^{\left[ ii\right] }+\underline{\hat{S}}^{\left[ ij\right] }\right)
\end{equation*}%
\begin{equation*}
\underline{\hat{S}}_{\eta }^{\left[ ii\right] }+\underline{\hat{S}}_{\eta }^{%
\left[ ij\right] }=\frac{\underline{\hat{k}}_{\eta }^{\left[ i\right] }}{1+%
\underline{\hat{k}}^{\left[ i\right] }}=\underline{\hat{k}}_{\eta }^{\left[ i%
\right] }\left( 1-\left( \underline{\hat{S}}^{\left[ ii\right] }+\underline{%
\hat{S}}^{\left[ ij\right] }\right) \right)
\end{equation*}%
\begin{equation*}
\underline{\hat{k}}_{\eta }^{\left[ i\right] }=\frac{\underline{\hat{S}}%
_{\eta }^{\left[ ii\right] }+\underline{\hat{S}}_{\eta }^{\left[ ij\right] }%
}{1-\left( \underline{\hat{S}}^{\left[ ii\right] }+\underline{\hat{S}}^{%
\left[ ij\right] }\right) }
\end{equation*}%
\begin{equation*}
1+\underline{\hat{k}}_{\eta }^{\left[ i\right] }=\frac{1-\left( \underline{%
\hat{S}}_{3-\eta }^{\left[ ii\right] }+\underline{\hat{S}}_{3-\eta }^{\left[
ij\right] }\right) }{1-\left( \underline{\hat{S}}^{\left[ ii\right] }+%
\underline{\hat{S}}^{\left[ ij\right] }\right) }
\end{equation*}%
\begin{equation*}
\frac{1}{1+\underline{k}^{\left[ i\right] }}=1-\left( \underline{S}^{\left[
ii\right] }+\underline{S}^{\left[ ij\right] }\right)
\end{equation*}%
\begin{equation*}
\underline{k}_{\eta }^{\left[ i\right] }=\frac{\underline{S}_{\eta }^{\left[
ii\right] }+\underline{S}_{\eta }^{\left[ ij\right] }}{1-\left( \underline{S}%
^{\left[ ii\right] }+\underline{S}^{\left[ ij\right] }\right) }
\end{equation*}%
\begin{equation*}
1+\underline{k}_{\eta }^{\left[ i\right] }=\frac{1-\left( \underline{S}%
_{3-\eta }^{\left[ ii\right] }+\underline{S}_{3-\eta }^{\left[ ij\right]
}\right) }{1-\left( \underline{S}^{\left[ ii\right] }+\underline{S}^{\left[
ij\right] }\right) }
\end{equation*}%
we find the return equations for the system of several group:%
\begin{eqnarray*}
0 &=&\left( 
\begin{array}{cc}
1-\underline{\hat{S}}_{1}^{\left[ ii\right] } & -\underline{\hat{S}}_{1}^{%
\left[ ji\right] } \\ 
-\underline{\hat{S}}_{1}^{\left[ ij\right] } & 1-\underline{\hat{S}}_{1}^{%
\left[ jj\right] }%
\end{array}%
\right) \left( 
\begin{array}{c}
\frac{f^{\left[ i\right] }-\bar{r}}{1+\underline{\hat{k}}_{2}^{\left[ i%
\right] }} \\ 
\frac{f^{\left[ j\right] }-\bar{r}}{1+\underline{\hat{k}}_{2}^{\left[ j%
\right] }}%
\end{array}%
\right) -\left( 
\begin{array}{cc}
\underline{\hat{S}}_{2}^{\left[ ii\right] } & \underline{\hat{S}}_{2}^{\left[
ji\right] } \\ 
\underline{\hat{S}}_{2}^{\left[ ij\right] } & \underline{\hat{S}}_{2}^{\left[
jj\right] }%
\end{array}%
\right) \left( 
\begin{array}{c}
\frac{1+f^{\left[ i\right] }}{\underline{\hat{k}}_{2}^{\left[ i\right] }}%
H\left( -\left( 1+f^{\left[ i\right] }\right) \right) \\ 
\frac{1+f^{\left[ j\right] }}{\underline{\hat{k}}_{2}^{\left[ j\right] }}%
H\left( -\left( 1+f^{\left[ j\right] }\right) \right)%
\end{array}%
\right) \\
&&-\left( 
\begin{array}{cc}
\underline{S}_{2}^{\left[ ii\right] } & \underline{S}_{2}^{\left[ ji\right] }
\\ 
\underline{S}_{2}^{\left[ ij\right] } & \underline{S}_{2}^{\left[ jj\right] }%
\end{array}%
\right) \left( 
\begin{array}{c}
\frac{1+f_{1}^{\prime \left[ i\right] }}{\underline{k}_{2}^{\left[ i\right] }%
}H\left( -\left( 1+f_{1}^{\prime \left[ i\right] }\right) \right) \\ 
\frac{1+f_{1}^{\prime \left[ j\right] }}{\underline{k}_{2}^{\left[ j\right] }%
}H\left( -\left( 1+f_{1}^{\prime \left[ j\right] }\right) \right)%
\end{array}%
\right) -\left( 
\begin{array}{cc}
\underline{S}_{1}^{\left[ ii\right] } & \underline{S}_{1}^{\left[ ji\right] }
\\ 
\underline{S}_{1}^{\left[ ij\right] } & \underline{S}_{1}^{\left[ jj\right] }%
\end{array}%
\right) \left( 
\begin{array}{c}
f_{1}^{\left[ i\right] }-\bar{r} \\ 
f_{1}^{\left[ j\right] }-\bar{r}%
\end{array}%
\right)
\end{eqnarray*}%
with:%
\begin{equation*}
f_{1}^{\left[ i\right] }-\bar{r}=\frac{f_{1}^{\prime \left[ i\right] }-\bar{r%
}}{1+k_{2}^{\left[ i\right] }}
\end{equation*}

or by fully replacing the coeffcients:

\begin{eqnarray*}
0 &=&\left( 
\begin{array}{cc}
1-\underline{\hat{S}}_{1}^{\left[ ii\right] } & -\underline{\hat{S}}_{1}^{%
\left[ ji\right] } \\ 
-\underline{\hat{S}}_{1}^{\left[ ij\right] } & 1-\underline{\hat{S}}_{1}^{%
\left[ jj\right] }%
\end{array}%
\right) \left( 
\begin{array}{c}
\left( f^{\left[ i\right] }-\bar{r}\right) \frac{1-\left( \underline{\hat{S}}%
^{\left[ ii\right] }+\underline{\hat{S}}^{\left[ ij\right] }\right) }{%
1-\left( \underline{\hat{S}}_{1}^{\left[ ii\right] }+\underline{\hat{S}}%
_{1}^{\left[ ij\right] }\right) } \\ 
\left( f^{\left[ j\right] }-\bar{r}\right) \frac{1-\left( \underline{\hat{S}}%
^{\left[ jj\right] }+\underline{\hat{S}}^{\left[ ij\right] }\right) }{%
1-\left( \underline{\hat{S}}_{1}^{\left[ jj\right] }+\underline{\hat{S}}%
_{1}^{\left[ ji\right] }\right) }%
\end{array}%
\right) \\
&&-\left( 
\begin{array}{cc}
\underline{\hat{S}}_{2}^{\left[ ii\right] } & \underline{\hat{S}}_{2}^{\left[
ji\right] } \\ 
\underline{\hat{S}}_{2}^{\left[ ij\right] } & \underline{\hat{S}}_{2}^{\left[
jj\right] }%
\end{array}%
\right) \left( 
\begin{array}{c}
\left( 1+f^{\left[ i\right] }\right) \frac{1-\left( \underline{\hat{S}}^{%
\left[ ii\right] }+\underline{\hat{S}}^{\left[ ij\right] }\right) }{%
\underline{\hat{S}}_{2}^{\left[ ii\right] }+\underline{\hat{S}}_{2}^{\left[
ij\right] }}H\left( -\left( 1+f^{\left[ i\right] }\right) \right) \\ 
\left( 1+f^{\left[ j\right] }\right) \frac{1-\left( \underline{\hat{S}}^{%
\left[ jj\right] }+\underline{\hat{S}}^{\left[ ij\right] }\right) }{%
\underline{\hat{S}}_{2}^{\left[ jj\right] }+\underline{\hat{S}}_{2}^{\left[
ji\right] }}H\left( -\left( 1+f^{\left[ j\right] }\right) \right)%
\end{array}%
\right) \\
&&-\left( 
\begin{array}{cc}
\underline{S}_{2}^{\left[ ii\right] } & \underline{S}_{2}^{\left[ ji\right] }
\\ 
\underline{S}_{2}^{\left[ ij\right] } & \underline{S}_{2}^{\left[ jj\right] }%
\end{array}%
\right) \left( 
\begin{array}{c}
\left( 1+f_{1}^{\prime \left[ i\right] }\right) \frac{1-\left( \underline{S}%
^{\left[ ii\right] }+\underline{S}^{\left[ ij\right] }\right) }{\underline{S}%
_{2}^{\left[ ii\right] }+\underline{S}_{2}^{\left[ ij\right] }}H\left(
-\left( 1+f_{1}^{\prime \left[ i\right] }\right) \right) \\ 
\left( 1+f_{1}^{\prime \left[ j\right] }\right) \frac{1-\left( \underline{S}%
^{\left[ jj\right] }+\underline{S}^{\left[ ji\right] }\right) }{\underline{S}%
_{2}^{\left[ jj\right] }+\underline{S}_{2}^{\left[ ji\right] }}H\left(
-\left( 1+f_{1}^{\prime \left[ j\right] }\right) \right)%
\end{array}%
\right) \\
&&-\left( 
\begin{array}{cc}
\underline{S}_{1}^{\left[ ii\right] } & \underline{S}_{1}^{\left[ ji\right] }
\\ 
\underline{S}_{1}^{\left[ ij\right] } & \underline{S}_{1}^{\left[ jj\right] }%
\end{array}%
\right) \left( 
\begin{array}{c}
\left( f_{1}^{\prime \left[ i\right] }-\bar{r}\right) \frac{1-\left( 
\underline{S}^{\left[ ii\right] }+\underline{S}^{\left[ ij\right] }\right) }{%
1-\left( \underline{S}_{1}^{\left[ ii\right] }+\underline{S}_{1}^{\left[ ij%
\right] }\right) } \\ 
\left( f_{1}^{\prime \left[ j\right] }-\bar{r}\right) \frac{1-\left( 
\underline{S}^{\left[ jj\right] }+\underline{S}^{\left[ ij\right] }\right) }{%
1-\left( \underline{S}_{1}^{\left[ jj\right] }+\underline{S}_{1}^{\left[ ji%
\right] }\right) }%
\end{array}%
\right)
\end{eqnarray*}

\section*{Appendix 14 Computation of Green functions}

We consider the operator:%
\begin{equation*}
-\nabla _{\hat{K}}\left( \frac{\sigma _{\hat{K}}^{2}}{2}\nabla _{\hat{K}}-%
\hat{K}f\left( \hat{X},K_{\hat{X}}\right) +\int \delta f_{1}\left\vert \Xi
\left( \hat{X},\delta f_{1}\right) \right\vert ^{2}d\left( \delta
f_{1}\right) \right)
\end{equation*}%
whose Green function will be computed approximatively betwwen an initial
state $\delta f_{1}$ and $\delta f_{1}^{\prime }$. This amounts to consider
that the capital dynamics adapt to a slower variable, the excess return.
Consider the average as fixed:%
\begin{equation*}
\int \delta f_{1}\left\vert \Xi \left( \hat{X},\delta f_{1}\right)
\right\vert ^{2}d\left( \delta f_{1}\right)
\end{equation*}%
the Green function should be formally given by:%
\begin{eqnarray}
&&\sqrt{\left\vert \frac{f_{\Xi }\left( \hat{X},K_{\hat{X}}\right) }{\sigma
_{\hat{K}}^{2}\left( 1-\exp \left( 2f_{\Xi }\left( \hat{X},K_{\hat{X}%
}\right) \Delta t\right) \right) }\right\vert }  \label{GNf} \\
&&\times \exp \left( \frac{f_{\Xi }\left( \hat{X},K_{\hat{X}}\right) }{%
\sigma _{\hat{K}}^{2}\left( 1-\exp \left( 2f_{\Xi }\left( \hat{X},K_{\hat{X}%
}\right) \Delta t\right) \right) }\left( \hat{K}-\exp \left( f_{\Xi }\left( 
\hat{X},K_{\hat{X}}\right) \Delta t\right) \hat{K}^{\prime }\right)
^{2}\right)  \notag
\end{eqnarray}%
with:%
\begin{equation*}
f_{\Xi }\left( \hat{X},K_{\hat{X}}\right) =f\left( \hat{X},K_{\hat{X}%
}\right) +\int \delta f_{1}\left\vert \Xi \left( \hat{X},\delta f_{1}\right)
\right\vert ^{2}d\left( \delta f_{1}\right)
\end{equation*}%
To estimate the Green function between two values $\delta f_{1}$ and $\delta
f_{1}^{\prime }$, we expand (\ref{GNf}) as a series expansion of $\left\vert
\Xi \left( \hat{X},\delta f_{1}\right) \right\vert ^{2}$ and estimate the
results between the two states using the transitions functions (\ref{TPS}):%
\begin{equation}
G\left( \delta f_{1},\delta f_{1}^{\prime }\right) =\sqrt{\frac{1}{\sigma
_{\delta f_{1}}^{2}}}\exp \left( -\frac{\left( \delta f_{1}-\delta
f_{1}^{\prime }\right) ^{2}}{\sigma _{\delta f_{1}}^{2}}+J\left( \hat{X},K_{%
\hat{X}},\mathbf{E}\right) \delta f_{1}-J\left( \hat{X}^{\prime },K_{\hat{X}%
^{\prime }},\mathbf{E}^{\prime }\right) \delta f_{1}^{\prime }\right)
\end{equation}%
and normalize by diving by $G\left( \delta f_{1},\delta f_{1}^{\prime
}\right) $.

For $\mathbf{E=E}^{\prime }$, that is, if the external parameters are
constant aver a perd of time, the estimation of a product:%
\begin{equation*}
\int d\left( \delta f_{1}\right) _{1}...d\left( \delta f_{1}\right)
_{k}\left( \delta f_{1}\right) _{1}\left\vert \Xi \left( \hat{X},\left(
\delta f_{1}\right) _{1}\right) \right\vert ^{2}...\left( \delta
f_{1}\right) _{1}\left\vert \Xi \left( \hat{X},\left( \delta f_{1}\right)
_{k}\right) \right\vert ^{2}
\end{equation*}%
Without fixing the initial and final value for $\delta f_{1}$, this integral
is equal to $0$. By imposing 
\begin{equation*}
\frac{\sum \left( \delta f_{1}\right) _{j}}{k}=\frac{\delta f_{1}+\delta
f_{1}^{\prime }}{2}
\end{equation*}%
we can by change of variable replace $\left( \delta f_{1}\right) _{k}$ in
the integral by the average $\frac{\delta f_{1}+\delta f_{1}^{\prime }}{2}$.
The estimation of the prdct bcms:%
\begin{eqnarray*}
&&\int d\left( \delta f_{1}\right) _{1}...d\left( \delta f_{1}\right)
_{k}\left( \delta f_{1}\right) _{1}\left\vert \Xi \left( \hat{X},\left(
\delta f_{1}\right) _{1}\right) \right\vert ^{2}...\left( \delta
f_{1}\right) _{1}\left\vert \Xi \left( \hat{X},\left( \delta f_{1}\right)
_{k}\right) \right\vert ^{2} \\
&\simeq &\left( \frac{\delta f_{1}+\delta f_{1}^{\prime }}{2}\right)
^{k}\int d\left( \delta f_{1}\right) _{1}...d\left( \delta f_{1}\right)
_{k}G\left( \delta f_{1},\left( \delta f_{1}^{\prime }\right) _{1}\right)
...G\left( \left( \delta f_{1}\right) _{k},\delta f_{1}^{\prime }\right) \\
&=&\left( \frac{\delta f_{1}+\delta f_{1}^{\prime }}{2}\right) ^{k}\sqrt{%
\frac{1}{\sigma _{\delta f_{1}}^{2}}}\exp \left( -\frac{\left( \delta
f_{1}-\delta f_{1}^{\prime }\right) ^{2}}{\sigma _{\delta f_{1}}^{2}}%
+J\left( \hat{X},K_{\hat{X}},\mathbf{E}\right) \delta f_{1}-J\left( \hat{X}%
^{\prime },K_{\hat{X}^{\prime }},\mathbf{E}^{\prime }\right) \delta
f_{1}^{\prime }\right) \\
&=&\left( \frac{\delta f_{1}+\delta f_{1}^{\prime }}{2}\right) ^{k}G\left(
\delta f_{1},\delta f_{1}^{\prime }\right)
\end{eqnarray*}%
The product of integrls is a convolutn:%
\begin{equation*}
\int d\left( \delta f_{1}\right) _{1}...d\left( \delta f_{1}\right)
_{k}G\left( \delta f_{1},\left( \delta f_{1}^{\prime }\right) _{1}\right)
...G\left( \left( \delta f_{1}\right) _{k},\delta f_{1}^{\prime }\right) =%
\sqrt{\frac{1}{\sigma _{\delta f_{1}}^{2}}}\exp \left( -\frac{\left( \delta
f_{1}-\delta f_{1}^{\prime }\right) ^{2}}{\sigma _{\delta f_{1}}^{2}}%
+J\left( \hat{X},K_{\hat{X}},\mathbf{E}\right) \delta f_{1}-J\left( \hat{X}%
^{\prime },K_{\hat{X}^{\prime }},\mathbf{E}^{\prime }\right) \delta
f_{1}^{\prime }\right)
\end{equation*}%
Leading to:%
\begin{eqnarray*}
&&\int d\left( \delta f_{1}\right) _{1}...d\left( \delta f_{1}\right)
_{k}\left( \delta f_{1}\right) _{1}\left\vert \Xi \left( \hat{X},\left(
\delta f_{1}\right) _{1}\right) \right\vert ^{2}...\left( \delta
f_{1}\right) _{1}\left\vert \Xi \left( \hat{X},\left( \delta f_{1}\right)
_{k}\right) \right\vert ^{2} \\
&=&\left( \frac{\delta f_{1}+\delta f_{1}^{\prime }}{2}\right) ^{k}G\left(
\delta f_{1},\delta f_{1}^{\prime }\right)
\end{eqnarray*}

After normaliztn, the correction becomes:%
\begin{equation*}
\int d\left( \delta f_{1}\right) _{1}...d\left( \delta f_{1}\right)
_{k}\left( \delta f_{1}\right) _{1}\left\vert \Xi \left( \hat{X},\left(
\delta f_{1}\right) _{1}\right) \right\vert ^{2}...\left( \delta
f_{1}\right) _{1}\left\vert \Xi \left( \hat{X},\left( \delta f_{1}\right)
_{k}\right) \right\vert ^{2}\simeq \left( \frac{\delta f_{1}+\delta
f_{1}^{\prime }}{2}\right) ^{k}
\end{equation*}%
Ultimately, the Green function conditionned by the return is:%
\begin{eqnarray}
&&\sqrt{\left\vert \frac{f\left( \hat{X},K_{\hat{X}}\right) +\left( \frac{%
\delta f_{1}+\delta f_{1}^{\prime }}{2}\right) }{\sigma _{\hat{K}}^{2}\left(
1-\exp \left( 2\left( f\left( \hat{X},K_{\hat{X}}\right) +\left( \frac{%
\delta f_{1}+\delta f_{1}^{\prime }}{2}\right) \right) \Delta t\right)
\right) }\right\vert } \\
&&\times \exp \left( \frac{\left( f\left( \hat{X},K_{\hat{X}}\right) +\left( 
\frac{\delta f_{1}+\delta f_{1}^{\prime }}{2}\right) \right) }{\sigma _{\hat{%
K}}^{2}\left( 1-\exp \left( 2\left( f\left( \hat{X},K_{\hat{X}}\right)
+\left( \frac{\delta f_{1}+\delta f_{1}^{\prime }}{2}\right) \right) \Delta
t\right) \right) }\left( \hat{K}-\exp \left( \left( f\left( \hat{X},K_{\hat{X%
}}\right) +\left( \frac{\delta f_{1}+\delta f_{1}^{\prime }}{2}\right)
\right) \Delta t\right) \hat{K}^{\prime }\right) ^{2}\right)  \notag
\end{eqnarray}

\section*{Appendix 15 interactions in an homogeneous group}

In computing the transition functions between several agents the insertion
of terms:%
\begin{equation*}
\prod\limits_{i}V\left( \delta \left( \hat{X}^{\prime }-\hat{X}\right) -%
\frac{\hat{k}_{1}\left( \hat{X}^{\prime },\hat{X}\right) \hat{K}^{\prime
}\left\vert \hat{\Psi}\left( \hat{K}^{\prime },\hat{X}^{\prime }\right)
\right\vert ^{2}}{1+\underline{\hat{k}}\left( \hat{X}^{\prime }\right) }%
\right) \frac{\int \left\vert \Xi \left( \hat{X},\delta f_{1}\right)
\right\vert ^{2}\left\vert \Xi \left( \hat{X},\delta f_{1}^{\prime }\right)
\right\vert ^{2}d\left( \delta f_{1}\right) d\left( \delta f_{1}^{\prime
}\right) }{1+\underline{\hat{k}}_{2}\left( \hat{X}^{\prime }\right) }d\hat{X}%
^{\prime }
\end{equation*}%
for an arbitrary potential, induces contributions:%
\begin{equation*}
\left\langle \left( \delta f_{1}\right) _{i}\right\vert
\prod\limits_{i}V\left( \delta \left( \hat{X}^{\prime }-\hat{X}\right) -%
\frac{\hat{k}_{1}\left( \hat{X}^{\prime },\hat{X}\right) \hat{K}^{\prime
}\left\vert \hat{\Psi}\left( \hat{K}^{\prime },\hat{X}^{\prime }\right)
\right\vert ^{2}}{1+\underline{\hat{k}}\left( \hat{X}^{\prime }\right) }%
\right) \frac{\int \left\vert \Xi \left( \hat{X},\delta f_{1}\right)
\right\vert ^{2}\left\vert \Xi \left( \hat{X},\delta f_{1}^{\prime }\right)
\right\vert ^{2}d\left( \delta f_{1}\right) d\left( \delta f_{1}^{\prime
}\right) }{1+\underline{\hat{k}}_{2}\left( \hat{X}^{\prime }\right) }d\hat{X}%
^{\prime }\left\vert \left( \delta f_{1}^{\prime \prime }\right)
_{i}\right\rangle
\end{equation*}%
Contractn of the flds wth th stts, lds t trm:%
\begin{equation*}
\prod_{i}G\left( \left( \delta f_{1}\right) _{i},\left( \delta
f_{1}^{\prime }\right) _{i}\right) \left\{ \prod\limits_{i}V\left( \left(
\delta f_{1}\right) _{i}-\sum_{j}\frac{\hat{k}_{1}\left( \hat{X}^{\prime },%
\hat{X}\right) \hat{K}^{\prime }\left\vert \hat{\Psi}\left( \hat{K}^{\prime
},\hat{X}^{\prime }\right) \right\vert ^{2}}{1+\underline{\hat{k}}\left( 
\hat{X}^{\prime }\right) }\left( \delta f_{1}^{\prime }\right) _{j}\right)
\right\} \prod_{i}G\left( \left( \delta f_{1}^{\prime }\right) _{i},\left(
\delta f_{1}^{\prime \prime }\right) _{i}\right)
\end{equation*}%
If we choose a Dirac function, this imposes the constraint in the
transitions:%
\begin{equation*}
\prod G\left( \left( \delta f_{1}\right) _{i},\left( \delta f_{1}^{\prime
}\right) _{i}\right) \delta \left( \delta f_{1i}^{\prime }\left( \hat{X}%
_{i}\right) -\sum_{j}\frac{\hat{k}_{1}\left( \hat{X}_{i},\hat{X}_{j}\right)
\left\langle \hat{K}_{j}\right\rangle \left\vert \hat{\Psi}\left( \hat{K}%
_{j},\hat{X}_{j}\right) \right\vert ^{2}}{1+\underline{\hat{k}}\left( \hat{X}%
_{j}\right) }\delta f_{1j}\left( \hat{X}_{j}\right) \right) \prod G\left(
\left( \delta f_{1}^{\prime }\right) _{i},\left( \delta f_{1}^{\prime \prime
}\right) _{i}\right)
\end{equation*}%
as in the text.

\section*{Appendix 16 Blocks interactions}

Consider (\ref{PMl}) without default: 
\begin{eqnarray}
0 &=&\left( 
\begin{array}{cc}
1-\underline{\hat{S}}_{1}^{\left[ ii\right] } & -\underline{\hat{S}}_{1}^{%
\left[ ji\right] } \\ 
-\underline{\hat{S}}_{1}^{\left[ ij\right] } & 1-\underline{\hat{S}}_{1}^{%
\left[ jj\right] }%
\end{array}%
\right) \left( 
\begin{array}{c}
\left( f^{\left[ i\right] }-\bar{r}\right) \frac{1-\left( \underline{\hat{S}}%
^{\left[ ii\right] }+\underline{\hat{S}}^{\left[ ij\right] }\right) }{%
1-\left( \underline{\hat{S}}_{1}^{\left[ ii\right] }+\underline{\hat{S}}%
_{1}^{\left[ ij\right] }\right) } \\ 
\left( f^{\left[ j\right] }-\bar{r}\right) \frac{1-\left( \underline{\hat{S}}%
^{\left[ jj\right] }+\underline{\hat{S}}^{\left[ ij\right] }\right) }{%
1-\left( \underline{\hat{S}}_{1}^{\left[ jj\right] }+\underline{\hat{S}}%
_{1}^{\left[ ji\right] }\right) }%
\end{array}%
\right) \\
&&-\left( 
\begin{array}{cc}
\underline{S}_{1}^{\left[ ii\right] } & \underline{S}_{1}^{\left[ ji\right] }
\\ 
\underline{S}_{1}^{\left[ ij\right] } & \underline{S}_{1}^{\left[ jj\right] }%
\end{array}%
\right) \left( 
\begin{array}{c}
\left( f_{1}^{\prime \left[ i\right] }-\bar{r}\right) \frac{1-\left( 
\underline{S}^{\left[ ii\right] }+\underline{S}^{\left[ ij\right] }\right) }{%
1-\left( \underline{S}_{1}^{\left[ ii\right] }+\underline{S}_{1}^{\left[ ij%
\right] }\right) } \\ 
\left( f_{1}^{\prime \left[ j\right] }-\bar{r}\right) \frac{1-\left( 
\underline{S}^{\left[ jj\right] }+\underline{S}^{\left[ ij\right] }\right) }{%
1-\left( \underline{S}_{1}^{\left[ jj\right] }+\underline{S}_{1}^{\left[ ji%
\right] }\right) }%
\end{array}%
\right)  \notag
\end{eqnarray}%
and replace $f^{\left[ i\right] }+\delta f^{\left[ i\right] }$. This leads
to:%
\begin{eqnarray}
0 &=&\left( 
\begin{array}{cc}
-\underline{\hat{S}}_{1}^{\left[ ii\right] } & -\underline{\hat{S}}_{1}^{%
\left[ ji\right] } \\ 
-\underline{\hat{S}}_{1}^{\left[ ij\right] } & -\underline{\hat{S}}_{1}^{%
\left[ jj\right] }%
\end{array}%
\right) \left( 
\begin{array}{c}
\left( f^{\left[ i\right] }-\bar{r}\right) \frac{1-\left( \underline{\hat{S}}%
^{\left[ ii\right] }+\underline{\hat{S}}^{\left[ ij\right] }\right) }{%
1-\left( \underline{\hat{S}}_{1}^{\left[ ii\right] }+\underline{\hat{S}}%
_{1}^{\left[ ij\right] }\right) } \\ 
\left( f^{\left[ j\right] }-\bar{r}\right) \frac{1-\left( \underline{\hat{S}}%
^{\left[ jj\right] }+\underline{\hat{S}}^{\left[ ij\right] }\right) }{%
1-\left( \underline{\hat{S}}_{1}^{\left[ jj\right] }+\underline{\hat{S}}%
_{1}^{\left[ ji\right] }\right) }%
\end{array}%
\right) \\
&&-\left( 
\begin{array}{cc}
0 & \underline{S}_{1}^{\left[ ji\right] } \\ 
\underline{S}_{1}^{\left[ ij\right] } & 0%
\end{array}%
\right) \left( 
\begin{array}{c}
\left( f_{1}^{\prime \left[ i\right] }-\bar{r}\right) \frac{1-\left( 
\underline{S}^{\left[ ii\right] }+\underline{S}^{\left[ ij\right] }\right) }{%
1-\left( \underline{S}_{1}^{\left[ ii\right] }+\underline{S}_{1}^{\left[ ij%
\right] }\right) } \\ 
\left( f_{1}^{\prime \left[ j\right] }-\bar{r}\right) \frac{1-\left( 
\underline{S}^{\left[ jj\right] }+\underline{S}^{\left[ ij\right] }\right) }{%
1-\left( \underline{S}_{1}^{\left[ jj\right] }+\underline{S}_{1}^{\left[ ji%
\right] }\right) }%
\end{array}%
\right) +\left( 
\begin{array}{cc}
1-\underline{\hat{S}}_{1}^{\left[ ii\right] } & -\underline{\hat{S}}_{1}^{%
\left[ ji\right] } \\ 
-\underline{\hat{S}}_{1}^{\left[ ij\right] } & 1-\underline{\hat{S}}_{1}^{%
\left[ jj\right] }%
\end{array}%
\right) \left( 
\begin{array}{c}
\delta f^{\left[ i\right] }\frac{1-\left( \underline{\hat{S}}^{\left[ ii%
\right] }+\underline{\hat{S}}^{\left[ ij\right] }\right) }{1-\left( 
\underline{\hat{S}}_{1}^{\left[ ii\right] }+\underline{\hat{S}}_{1}^{\left[
ij\right] }\right) } \\ 
\delta f^{\left[ j\right] }\frac{1-\left( \underline{\hat{S}}^{\left[ jj%
\right] }+\underline{\hat{S}}^{\left[ ij\right] }\right) }{1-\left( 
\underline{\hat{S}}_{1}^{\left[ jj\right] }+\underline{\hat{S}}_{1}^{\left[
ji\right] }\right) }%
\end{array}%
\right)  \notag
\end{eqnarray}%
The two first trms in the right hand side corresponds to a shift in the
average return of the block introduced by the interaction. It can be
absorbed in a redefinition of these averages, and we are left with:%
\begin{equation}
0=\left( 
\begin{array}{cc}
1-\underline{\hat{S}}_{1}^{\left[ ii\right] } & -\underline{\hat{S}}_{1}^{%
\left[ ji\right] } \\ 
-\underline{\hat{S}}_{1}^{\left[ ij\right] } & 1-\underline{\hat{S}}_{1}^{%
\left[ jj\right] }%
\end{array}%
\right) \left( 
\begin{array}{c}
\delta f^{\left[ i\right] }\frac{1-\left( \underline{\hat{S}}^{\left[ ii%
\right] }+\underline{\hat{S}}^{\left[ ij\right] }\right) }{1-\left( 
\underline{\hat{S}}_{1}^{\left[ ii\right] }+\underline{\hat{S}}_{1}^{\left[
ij\right] }\right) } \\ 
\delta f^{\left[ j\right] }\frac{1-\left( \underline{\hat{S}}^{\left[ jj%
\right] }+\underline{\hat{S}}^{\left[ ij\right] }\right) }{1-\left( 
\underline{\hat{S}}_{1}^{\left[ jj\right] }+\underline{\hat{S}}_{1}^{\left[
ji\right] }\right) }%
\end{array}%
\right)
\end{equation}%
that corresponds to intrdc ptntl:%
\begin{eqnarray*}
\sum_{i}V_{i}\left( \hat{\Psi},\hat{X},K,\delta f_{1i}\right)
&=&\sum_{i}\delta \left( \left( 1-\underline{\hat{S}}_{1}^{\left[ ii\right]
}\right) \frac{1-\left( \underline{\hat{S}}^{\left[ ii\right] }+\underline{%
\hat{S}}^{\left[ ij\right] }\right) }{1-\left( \underline{\hat{S}}_{1}^{%
\left[ ii\right] }+\underline{\hat{S}}_{1}^{\left[ ij\right] }\right) }\frac{%
\int \left\vert \Xi \left( \hat{X},\delta f^{\left[ i\right] }\right)
\right\vert ^{2}d\left( \delta f^{\left[ i\right] }\right) d\hat{X}}{1+%
\underline{\hat{k}}_{2}\left( \hat{X}^{\prime }\right) }\right. \\
&&\left. -\underline{\hat{S}}_{1}^{\left[ ji\right] }\frac{1-\left( 
\underline{\hat{S}}^{\left[ jj\right] }+\underline{\hat{S}}^{\left[ ij\right]
}\right) }{1-\left( \underline{\hat{S}}_{1}^{\left[ jj\right] }+\underline{%
\hat{S}}_{1}^{\left[ ji\right] }\right) }\frac{\int \left\vert \Xi \left( 
\hat{X},\delta f^{\left[ j\right] }\right) \right\vert ^{2}d\left( \delta f^{%
\left[ j\right] }\right) d\hat{X}}{1+\underline{\hat{k}}_{2}\left( \hat{X}%
^{\prime }\right) }\right)
\end{eqnarray*}%
replacing as before:%
\begin{equation*}
\frac{\int \left\vert \Xi \left( \hat{X},\delta f^{\left[ i\right] }\right)
\right\vert ^{2}d\left( \delta f^{\left[ i\right] }\right) d\hat{X}}{1+%
\underline{\hat{k}}_{2}\left( \hat{X}^{\prime }\right) }\rightarrow \frac{%
\delta f^{\left[ i\right] }+\delta f^{\left[ i\right] \prime }}{2\left( 1+%
\underline{\hat{k}}_{2}\left( \hat{X}^{\prime }\right) \right) }
\end{equation*}%
in the transitions, leads to the constraint in the text.

\part*{Part 2 Appendices}

\section*{Appendix 17 Details for the micro framework}

\subsection*{A17.1 Investors capital accumulation}

The dynmcs for investors private capital is gven by:%
\begin{eqnarray}
\hat{K}_{jp}\left( t+\varepsilon \right) -\hat{K}_{jp}\left( t\right)
&=&f_{j}\hat{K}_{jp}\left( t\right) =\sum_{l}\left( 1+\sum_{v}\hat{k}_{2jv}%
\hat{K}_{v}\left( t\right) +\sum_{v}\hat{k}_{2jv}^{B}\bar{K}_{v}\left(
t\right) \right) R_{j}\hat{K}_{jp}\left( t\right) \\
&&-\bar{r}\left( \sum_{v}\hat{k}_{2jv}\hat{K}_{v}\left( t\right) +\sum_{v}%
\hat{k}_{2jv}^{B}\bar{K}_{v}\left( t\right) \right) \hat{K}_{jp}\left(
t\right)  \notag \\
&=&\sum_{l}\left( 1+\sum_{v}\hat{k}_{2lv}\hat{K}_{v}\left( t\right) +\kappa
\sum_{v}\hat{k}_{2jv}^{B}\frac{\bar{K}_{v0}\left( t\right) }{1+\sum_{m}\bar{k%
}_{vm}\bar{K}_{m0}\left( t\right) }\right) R_{j}\hat{K}_{jp}\left( t\right) 
\notag \\
&&-\bar{r}\left( \sum_{v}\hat{k}_{2jv}\hat{K}_{v}\left( t\right) +\kappa
\sum_{v}\hat{k}_{2jv}^{B}\frac{\bar{K}_{v0}\left( t\right) }{1+\sum_{m}\bar{k%
}_{vm}\bar{K}_{m0}\left( t\right) }\right) \hat{K}_{jp}\left( t\right) 
\notag
\end{eqnarray}%
We can write:%
\begin{equation*}
\hat{K}_{jp}\left( t+\varepsilon \right) -\hat{K}_{jp}\left( t\right)
\rightarrow \frac{d}{dt}\hat{K}_{jp}\left( t\right)
\end{equation*}%
in the continuouss approximation. By differentiation of (\ref{KvsKP}):%
\begin{eqnarray}
\frac{d}{dt}\hat{K}_{jp}\left( t\right) &=&\frac{\frac{d}{dt}\hat{K}%
_{j}\left( t\right) }{1+\sum_{l}\hat{k}_{jl}\hat{K}_{l}\left( t\right)
+\sum_{l}\hat{k}_{1jl}^{B}\bar{K}_{l0}\left( t\right) +\kappa \sum_{l}\hat{k}%
_{2jl}^{B}\frac{\bar{K}_{l0}\left( t\right) }{1+\sum_{m}\overline{\bar{k}}%
_{lm}\bar{K}_{m0}\left( t\right) }}  \label{Dnm} \\
&&-\sum_{l}\frac{\hat{k}_{jl}\hat{K}_{j}\left( t\right) \frac{d}{dt}\hat{K}%
_{l}\left( t\right) }{\left( 1+\sum_{l}\hat{k}_{jl}\hat{K}_{l}\left(
t\right) +\sum_{l}\hat{k}_{1jl}^{B}\bar{K}_{l0}\left( t\right) +\kappa
\sum_{l}\hat{k}_{2jl}^{B}\frac{\bar{K}_{l0}\left( t\right) }{1+\sum_{m}%
\overline{\bar{k}}_{lm}\bar{K}_{m0}\left( t\right) }\right) ^{2}}  \notag \\
&&-\sum_{l}\frac{\left( \hat{k}_{1jl}^{B}+\kappa \hat{k}_{2jl}^{B}\frac{1}{%
1+\sum_{m}\overline{\bar{k}}_{lm}\bar{K}_{m0}\left( t\right) }\right) \frac{d%
}{dt}\bar{K}_{l0}\left( t\right) -\kappa \frac{\hat{k}_{2jl}^{B}\bar{K}%
_{l0}\left( t\right) \sum_{m}\overline{\bar{k}}_{lm}\frac{d}{dt}\bar{K}%
_{m0}\left( t\right) }{\left( 1+\sum_{m}\overline{\bar{k}}_{lm}\bar{K}%
_{m0}\left( t\right) \right) ^{2}}}{\left( 1+\sum_{l}\hat{k}_{jl}\hat{K}%
_{l}\left( t\right) +\sum_{l}\hat{k}_{1jl}^{B}\bar{K}_{l0}\left( t\right)
+\kappa \sum_{l}\hat{k}_{2jl}^{B}\frac{\bar{K}_{j0}\left( t\right) }{%
1+\sum_{m}\overline{\bar{k}}_{lm}\bar{K}_{m0}\left( t\right) }\right) ^{2}} 
\notag
\end{eqnarray}%
\bigskip

To obtain the dynamics for $\hat{K}_{j}\left( t\right) $ use (\ref{DR}), to
wrt (\ref{Dnm}):%
\begin{eqnarray}
&&\frac{\frac{d}{dt}\hat{K}_{j}\left( t\right) }{1+\sum_{l}\hat{k}_{jl}\hat{K%
}_{l}\left( t\right) +\sum_{l}\hat{k}_{1jl}^{B}\bar{K}_{l0}\left( t\right)
+\kappa \sum_{l}\hat{k}_{2jl}^{B}\frac{\bar{K}_{l0}\left( t\right) }{%
1+\sum_{m}\overline{\bar{k}}_{lm}\bar{K}_{m0}\left( t\right) }} \\
&&-\sum_{l}\frac{\hat{k}_{jl}\hat{K}_{j}\left( t\right) \frac{d}{dt}\hat{K}%
_{l}\left( t\right) }{\left( 1+\sum_{l}\hat{k}_{jl}\hat{K}_{l}\left(
t\right) +\sum_{l}\hat{k}_{1jl}^{B}\bar{K}_{l0}\left( t\right) +\kappa
\sum_{l}\hat{k}_{2jl}^{B}\frac{\bar{K}_{l0}\left( t\right) }{1+\sum_{m}%
\overline{\bar{k}}_{lm}\bar{K}_{m0}\left( t\right) }\right) ^{2}}  \notag \\
&&-\sum_{l}\frac{\hat{K}_{j}\left\{ \hat{k}_{1jl}^{B}+\kappa \hat{k}%
_{2jl}^{B}\frac{1}{1+\sum_{m}\overline{\bar{k}}_{lm}\bar{K}_{m0}\left(
t\right) }-\kappa \frac{\sum_{m}\hat{k}_{2jm}^{B}\bar{K}_{m0}\left( t\right) 
\overline{\bar{k}}_{ml}}{\left( 1+\sum_{n}\overline{\bar{k}}_{mn}\bar{K}%
_{n0}\left( t\right) \right) ^{2}}\right\} \frac{d}{dt}\bar{K}_{l0}\left(
t\right) }{\left( 1+\sum_{l}\hat{k}_{jl}\hat{K}_{l}\left( t\right) +\sum_{l}%
\hat{k}_{1jl}^{B}\bar{K}_{l0}\left( t\right) +\kappa \sum_{l}\hat{k}%
_{2jl}^{B}\frac{\bar{K}_{j0}\left( t\right) }{1+\sum_{m}\overline{\bar{k}}%
_{lm}\bar{K}_{m0}\left( t\right) }\right) ^{2}}  \notag \\
&=&\sum_{l}\left( 1+\sum_{v}\hat{k}_{2lv}\hat{K}_{v}\left( t\right) +\kappa
\sum_{v}\hat{k}_{2jv}^{B}\frac{\bar{K}_{v0}\left( t\right) }{1+\sum_{m}%
\overline{\bar{k}}_{vm}\bar{K}_{m0}\left( t\right) }\right) R_{j}\hat{K}%
_{jp}\left( t\right)  \notag \\
&&-\bar{r}\left( \sum_{v}\hat{k}_{2jv}\hat{K}_{v}\left( t\right) +\kappa
\sum_{v}\hat{k}_{2jv}^{B}\frac{\bar{K}_{v0}\left( t\right) }{1+\sum_{m}%
\overline{\bar{k}}_{vm}\bar{K}_{m0}\left( t\right) }\right) \hat{K}%
_{jp}\left( t\right)  \notag \\
&\rightarrow &f_{j}\hat{K}_{jp}\left( t\right)  \notag
\end{eqnarray}%
This is solved together with banks capital accumulation.

\subsection*{A17.2 Banks capital accumulation}

We start with the\ dynamics for $\bar{K}_{jp}$:%
\begin{equation*}
\bar{K}_{jp}\left( t+\varepsilon \right) -\bar{K}_{jp}\left( t\right)
=\left( 1+\sum_{v}\bar{k}_{2jv}\bar{K}_{v0}\left( t\right) \right) \bar{R}%
_{j}\bar{K}_{jp}\left( t\right) -\bar{r}\sum_{v}\bar{k}_{2jv}\bar{K}%
_{v0}\left( t\right) \bar{K}_{jp}\left( t\right)
\end{equation*}%
and its continuous approximation: 
\begin{equation*}
\frac{d}{dt}\bar{K}_{jp}\left( t\right) =\left( 1+\sum_{v}\bar{k}_{2jv}\bar{K%
}_{v0}\left( t\right) \right) \bar{R}_{j}\bar{K}_{jp}\left( t\right) -\bar{r}%
\sum_{v}\bar{k}_{2jv}\bar{K}_{v0}\left( t\right) \bar{K}_{jp}\left( t\right)
\end{equation*}%
leading to, for the disposable capital:%
\begin{eqnarray*}
&&\frac{\frac{d}{dt}\bar{K}_{j0}\left( t\right) }{1+\sum_{l}\left( \bar{k}%
_{1jl}+\overline{\bar{k}}_{2jl}\right) \bar{K}_{l0}\left( t\right) }-\sum_{l}%
\frac{\bar{K}_{j0}\left( t\right) \left( \bar{k}_{1jl}+\bar{k}_{2jl}\right) 
}{\left( 1+\sum_{l}\left( \bar{k}_{1jl}+\bar{k}_{2jl}\right) \bar{K}%
_{l0}\left( t\right) \right) ^{2}}\frac{d}{dt}\bar{K}_{l0}\left( t\right) \\
&=&\left( 1+\sum_{v}\bar{k}_{2jv}\bar{K}_{v0}\left( t\right) \right) \bar{R}%
_{j}\bar{K}_{jp}\left( t\right) -\bar{r}\sum_{v}\bar{k}_{2jv}\bar{K}%
_{v0}\left( t\right) \bar{K}_{jp}\left( t\right)
\end{eqnarray*}

where:%
\begin{equation*}
\hat{k}_{1ab}+\hat{k}_{2ab}=\hat{k}_{ab}
\end{equation*}%
and this can be set in matricial form, as in the text.

\subsection*{A17.3 Return equations with default}

\subsubsection*{A17.3.2 Investors return}

Including default for investors and firms modifies the return equation for
investors by:%
\begin{eqnarray}
&&\sum_{l}\left( \delta _{jl}-\frac{\hat{k}_{1lj}\hat{K}_{l}\left( t\right) 
}{1+\sum_{v}\hat{k}_{jv}\hat{K}_{l}\left( t\right) +\sum_{v}\hat{k}_{1jv}^{B}%
\bar{K}_{l0}\left( t\right) +\kappa \sum_{v}\hat{k}_{2jv}^{B}\frac{\bar{K}%
_{j0}\left( t\right) }{1+\sum_{m}\bar{k}_{vm}\bar{K}_{m0}\left( t\right) }}%
\right) \\
&&\times \frac{\hat{f}_{l}-\bar{r}}{1+\sum_{v}\hat{k}_{2lv}\hat{K}_{v}\left(
t\right) +\kappa \sum_{v}\hat{k}_{2lv}^{B}\frac{\bar{K}_{\nu 0}\left(
t\right) }{1+\sum_{m}\bar{k}_{vm}\bar{K}_{m0}\left( t\right) }} \\
&&+\sum_{l}\left( \bar{r}-\frac{\left( 1+\hat{f}_{\nu }\right) }{\sum_{m}%
\hat{k}_{2vm}\hat{K}_{m}+\kappa \sum_{m}\hat{k}_{2vm}^{B}\frac{\bar{K}%
_{m0}\left( t\right) }{1+\sum_{s}\bar{k}_{ms}\bar{K}_{s0}\left( t\right) }}%
\right)  \notag \\
&&\times \frac{H\left( -\left( 1+\hat{f}_{\nu }\right) \right) \hat{k}_{2lj}%
\hat{K}_{l}\left( t\right) }{1+\sum_{\nu }\hat{k}_{l\nu }\hat{K}_{l}\left(
t\right) +\sum_{\nu }\hat{k}_{1l\nu }^{B}\bar{K}_{l0}\left( t\right) +\kappa
\sum_{\nu }\hat{k}_{2l\nu }^{B}\frac{\bar{K}_{j0}\left( t\right) }{1+\sum_{m}%
\bar{k}_{vm}\bar{K}_{m0}\left( t\right) }} \\
&&+\sum_{i}\left( \bar{r}-\frac{\left( 1+f_{1}^{\prime }\left( K_{i}\left(
t\right) \right) \right) }{\left( \sum_{v}k_{2iv}^{\left( B\right) }\kappa 
\frac{\bar{K}_{v0}\left( t\right) }{1+\sum_{m}\bar{k}_{vm}\bar{K}_{m0}\left(
t\right) }\right) +\left( \sum_{v}k_{2iv}\hat{K}_{v}\left( t\right) \right) }%
\right)  \notag \\
&&\times \frac{H\left( -\left( 1+f_{1}^{\prime }\left( K_{i}\left( t\right)
\right) \right) \right) k_{2ij}K_{i}\left( t\right) }{1+\left(
\sum_{v}k_{1iv}^{\left( B\right) }\bar{K}_{v0}\left( t\right)
+k_{2iv}^{\left( B\right) }\kappa \frac{\bar{K}_{v0}\left( t\right) }{%
1+\sum_{m}\bar{k}_{vm}\bar{K}_{m0}\left( t\right) }\right) +\left(
\sum_{v}k_{iv}\hat{K}_{v}\left( t\right) \right) } \\
&=&\sum_{i}\frac{\left( r_{i}+F_{1}\left( \bar{R}_{i},\frac{\dot{K}%
_{i}\left( t\right) }{K_{i}\left( t\right) }\right) +\tau \left( \bar{R}%
\left( K_{i},X_{i}\right) \right) \Delta f_{1}^{\prime }\left( K_{i}\left(
t\right) \right) -\bar{r}\right) k_{1ij}K_{i}\left( t\right) }{1+\sum_{v}%
\hat{k}_{jv}\hat{K}_{v}\left( t\right) +\sum_{v}\hat{k}_{1jv}^{B}\bar{K}%
_{l0}\left( t\right) +\kappa \sum_{v}\hat{k}_{2jv}^{B}\frac{\bar{K}%
_{v0}\left( t\right) }{1+\sum_{m}\bar{k}_{vm}\bar{K}_{m0}\left( t\right) }} 
\notag
\end{eqnarray}

\subsubsection*{A17.3.3 Banks return}

Possibility of default for investors and firms modifies the return equation
for banks by:%
\begin{eqnarray}
&&\sum_{l}\left( \delta _{jl}-\frac{\hat{k}_{1lj}\bar{K}_{l0}\left( t\right) 
}{1+\sum_{v}\bar{k}_{lv}\bar{K}_{v0}\left( t\right) }\right) \left( \frac{%
\bar{f}_{l}-\bar{r}}{1+\sum_{v}\bar{k}_{2lv}\bar{K}_{v0}\left( t\right) }%
\right) \\
&&-\sum_{l}\frac{\hat{k}_{1lj}^{B}\hat{K}_{l}\left( t\right) }{1+\sum_{v}%
\hat{k}_{lv}\hat{K}_{l}\left( t\right) +\sum_{v}\hat{k}_{1lv}^{B}\bar{K}%
_{l0}\left( t\right) +\kappa \sum_{v}\hat{k}_{2lv}^{B}\frac{\bar{K}%
_{l0}\left( t\right) }{1+\sum_{m}\bar{k}_{vm}\bar{K}_{m0}\left( t\right) }} 
\notag \\
&&\times \frac{\hat{f}_{l}-\bar{r}}{1+\sum_{v}\hat{k}_{2lv}\hat{K}_{v}\left(
t\right) +\kappa \sum_{v}\hat{k}_{2lv}^{B}\frac{\bar{K}_{\nu 0}\left(
t\right) }{1+\sum_{m}\bar{k}_{vm}\bar{K}_{m0}\left( t\right) }}  \notag \\
&&+\sum_{l}\left( \bar{r}-\frac{\left( 1+\bar{f}_{\nu }\right) }{\sum_{m}%
\bar{k}_{2vm}\bar{K}_{m}}\right) \frac{H\left( -\left( 1+\bar{f}_{\nu
}\right) \right) \bar{k}_{2lj}\bar{K}_{l}\left( t\right) }{1+\sum_{v}\left( 
\bar{k}_{1lv}+\bar{k}_{2lv}\right) \bar{K}_{v}\left( t\right) }  \notag \\
&&+\sum_{l}\left( \bar{r}-\frac{\left( 1+\hat{f}_{\nu }\right) }{\sum_{m}%
\hat{k}_{2vm}\hat{K}_{m}+\kappa \sum_{m}\hat{k}_{2vm}^{B}\frac{\bar{K}%
_{m0}\left( t\right) }{1+\sum_{s}\bar{k}_{ms}\bar{K}_{s0}\left( t\right) }}%
\right)  \notag \\
&&\times \frac{H\left( -\left( 1+\hat{f}_{\nu }\right) \right) \hat{k}%
_{2lj}^{B}\hat{K}_{l}\left( t\right) }{1+\sum_{\nu }\hat{k}_{l\nu }\hat{K}%
_{l}\left( t\right) +\sum_{\nu }\hat{k}_{1l\nu }^{B}\bar{K}_{l0}\left(
t\right) +\kappa \sum_{\nu }\hat{k}_{2l\nu }^{B}\frac{\bar{K}_{j0}\left(
t\right) }{1+\sum_{m}\bar{k}_{vm}\bar{K}_{m0}\left( t\right) }}  \notag \\
&&\sum_{i}\left( \bar{r}-\frac{\left( 1+f_{1}^{\prime }\left( K_{i}\left(
t\right) \right) \right) }{\left( \sum_{v}k_{2iv}^{\left( B\right) }\kappa 
\frac{\bar{K}_{v0}\left( t\right) }{1+\sum_{m}\bar{k}_{vm}\bar{K}_{m0}\left(
t\right) }\right) +\left( \sum_{v}k_{2iv}\hat{K}_{v}\left( t\right) \right) }%
\right)  \notag \\
&&\times \frac{H\left( -\left( 1+f_{1}^{\prime }\left( K_{i}\left( t\right)
\right) \right) \right) k_{2ij}^{\left( B\right) }K_{i}\left( t\right) }{%
1+\left( \sum_{v}k_{1iv}^{\left( B\right) }\bar{K}_{v0}\left( t\right)
+k_{2iv}^{\left( B\right) }\kappa \frac{\bar{K}_{v0}\left( t\right) }{%
1+\sum_{m}\bar{k}_{vm}\bar{K}_{m0}\left( t\right) }\right) +\left(
\sum_{v}k_{iv}\hat{K}_{v}\left( t\right) \right) } \\
&=&\sum_{i}\frac{\left( r_{i}+F_{1}\left( \bar{R}_{i},\frac{\dot{K}%
_{i}\left( t\right) }{K_{i}\left( t\right) }\right) +\tau \left( \bar{R}%
\left( K_{i},X_{i}\right) \right) \Delta f_{1}^{\prime }\left( K_{i}\left(
t\right) \right) -\bar{r}\right) k_{1ij}^{B}K_{i}\left( t\right) }{1+\sum_{v}%
\hat{k}_{jv}\hat{K}_{v}\left( t\right) +\sum_{v}\hat{k}_{1jv}^{B}\bar{K}%
_{l0}\left( t\right) +\kappa \sum_{v}\hat{k}_{2jv}^{B}\frac{\bar{K}%
_{v0}\left( t\right) }{1+\sum_{m}\bar{k}_{vm}\bar{K}_{m0}\left( t\right) }} 
\notag
\end{eqnarray}

\section*{Appendix 18 translation in field description}

\subsection*{A18.1 Translation of coefficents}

Translations are direct:

\subsubsection*{A18.1.1 Banks}

\begin{equation*}
\underline{\bar{k}}_{\eta }\left( \bar{X}^{\prime }\right) =\int_{\eta }\bar{%
k}\left( \bar{X}^{\prime },\bar{X}\right) \bar{K}_{\bar{X}}\left\vert \bar{%
\Psi}\left( \bar{X}\right) \right\vert ^{2}d\hat{X}
\end{equation*}%
\begin{equation*}
\underline{\bar{k}}\left( \bar{X}^{\prime }\right) =\underline{\bar{k}}%
_{1}\left( \bar{X}^{\prime }\right) +\kappa \frac{\underline{\bar{k}}%
_{2}\left( \bar{X}^{\prime }\right) }{1+\underline{\bar{k}}}
\end{equation*}

\begin{equation*}
\underline{\bar{k}}=\int \bar{k}\left( \left\langle \bar{X}\right\rangle ,%
\bar{Y}\right) \bar{K}_{1}d\bar{Y}d\bar{K}_{1}
\end{equation*}%
\begin{equation*}
1+\int \bar{k}\left( \bar{X}^{\prime \prime },\bar{Y}\right) \bar{K}_{1}d%
\bar{Y}d\bar{K}_{1}\rightarrow 1+\underline{\bar{k}}
\end{equation*}

\subsubsection*{A18.1.2 Investors}

\begin{equation*}
\sum_{v}\hat{k}_{lv}\hat{K}_{v}\left( t\right) \rightarrow \int \hat{k}_{\nu
}\left( \hat{X}^{\prime },\hat{X}^{\prime \prime }\right) \hat{K}^{\prime
\prime }\left\vert \hat{\Psi}\left( \hat{K}^{\prime \prime },\hat{X}^{\prime
\prime }\right) \right\vert ^{2}d\hat{X}^{\prime \prime }d\hat{K}^{\prime
\prime }=\underline{\hat{k}}_{\nu }\left( \hat{X}^{\prime }\right)
\end{equation*}%
\begin{equation*}
\sum_{v}\hat{k}_{lv}\hat{K}_{v}\left( t\right) \rightarrow \underline{\hat{k}%
}\left( \hat{X}^{\prime }\right)
\end{equation*}%
\begin{eqnarray*}
&&\sum_{v}\hat{k}_{1lv}^{B}\bar{K}_{v0}\left( t\right) +\kappa \sum_{v}\hat{k%
}_{2lv}^{B}\frac{\bar{K}_{v0}\left( t\right) }{1+\sum_{m}\overline{\bar{k}}%
_{vm}\bar{K}_{m0}\left( t\right) } \\
&\rightarrow &\int \hat{k}_{1}^{B}\left( \bar{X}^{\prime },\bar{X}^{\prime
\prime }\right) \bar{K}^{\prime \prime }\left\vert \hat{\Psi}\left( \bar{K}%
^{\prime \prime },\bar{X}^{\prime \prime }\right) \right\vert ^{2}d\bar{X}%
^{\prime \prime }d\bar{K}^{\prime \prime }+\kappa \int \frac{\hat{k}%
_{2}^{B}\left( \bar{X}^{\prime },\bar{X}^{\prime \prime }\right) \bar{K}%
^{\prime \prime }\left\vert \hat{\Psi}\left( \bar{K}^{\prime \prime },\bar{X}%
^{\prime \prime }\right) \right\vert ^{2}}{1+\int \overline{\bar{k}}\left( 
\bar{X}^{\prime \prime },\bar{Y}\right) \bar{K}_{1}d\bar{Y}d\bar{K}_{1}}d%
\bar{X}^{\prime \prime }d\bar{K}^{\prime \prime } \\
&=&\underline{\hat{k}}_{1}^{B}\left( \hat{X}^{\prime }\right) +\kappa \left[ 
\frac{\underline{\hat{k}}_{2}^{B}}{1+\bar{k}}\right] \left( \hat{X}^{\prime
}\right)
\end{eqnarray*}%
and:%
\begin{eqnarray*}
&&1+\sum_{v}\hat{k}_{lv}\hat{K}_{v}\left( t\right) +\sum_{v}\bar{k}_{1lv}%
\bar{K}_{v0}\left( t\right) +\kappa \sum_{v}\bar{k}_{2lv}\frac{\bar{K}%
_{v0}\left( t\right) }{1+\sum_{m}\bar{k}_{vm}\bar{K}_{m0}\left( t\right) } \\
&\rightarrow &1+\underline{\hat{k}}\left( \hat{X}^{\prime }\right) +\kappa %
\left[ \frac{\underline{\hat{k}}_{2}^{B}}{1+\bar{k}}\right] \left( X^{\prime
}\right)
\end{eqnarray*}

\subsubsection*{A18.1.3 Firms}

\begin{eqnarray*}
&&1+\sum_{v}k_{iv}\hat{K}_{v}\left( t\right) +\left( \sum_{v}k_{1iv}^{\left(
B\right) }\bar{K}_{v0}\left( t\right) +k_{2iv}^{\left( B\right) }\kappa 
\frac{\bar{K}_{v0}\left( t\right) }{1+\sum_{m}\bar{k}_{vm}\bar{K}_{m0}\left(
t\right) }\right) \\
&\rightarrow &1+\underline{k}\left( \hat{X}^{\prime }\right) +\underline{k}%
_{1}^{\left( B\right) }\left( \bar{X}^{\prime }\right) +\kappa \left[ \frac{%
\underline{k}_{2}^{B}}{1+\bar{k}}\right] \left( X^{\prime }\right) \\
&=&\left( 1+\int k\left( X,,\hat{X}^{\prime }\right) \hat{K}_{\hat{X}%
^{\prime }}^{\prime }\frac{\left\vert \hat{\Psi}\left( \hat{X}^{\prime
}\right) \right\vert ^{2}}{\left\langle K_{p}\right\rangle \left\vert \Psi
\left( X\right) \right\vert ^{2}}+\int \left( \underline{k}_{1}^{\left(
B\right) }\left( X,\bar{X}^{\prime }\right) \bar{K}_{\bar{X}^{\prime
}}+\kappa \frac{\underline{k}_{2}^{\left( B\right) }\left( X,\bar{X}^{\prime
}\right) }{1+\underline{\bar{k}}\left( \bar{X}\right) }\bar{K}_{\bar{X}%
^{\prime }}\right) \frac{\left\vert \bar{\Psi}\left( \hat{X}^{\prime
}\right) \right\vert ^{2}}{\left\langle K_{p}\right\rangle \left\vert \Psi
\left( X\right) \right\vert ^{2}}\right)
\end{eqnarray*}%
wth:%
\begin{eqnarray*}
\underline{k}\left( \hat{X}^{\prime }\right) &=&\int k\left( X,,\hat{X}%
^{\prime }\right) \hat{K}_{\hat{X}^{\prime }}^{\prime }\frac{\left\vert \hat{%
\Psi}\left( \hat{X}^{\prime }\right) \right\vert ^{2}}{\left\langle
K_{p}\right\rangle \left\vert \Psi \left( X\right) \right\vert ^{2}} \\
\underline{k}_{1}^{\left( B\right) }\left( \bar{X}^{\prime }\right) +\kappa %
\left[ \frac{\underline{k}_{2}^{B}}{1+\bar{k}}\right] \left( X^{\prime
}\right) &=&\int \left( \underline{k}_{1}^{\left( B\right) }\left( X,\bar{X}%
^{\prime }\right) \bar{K}_{\bar{X}^{\prime }}+\kappa \frac{\underline{k}%
_{2}^{\left( B\right) }\left( X,\bar{X}^{\prime }\right) }{1+\underline{\bar{%
k}}\left( \bar{X}\right) }\bar{K}_{\bar{X}^{\prime }}\right) \frac{%
\left\vert \bar{\Psi}\left( \hat{X}^{\prime }\right) \right\vert ^{2}}{%
\left\langle K_{p}\right\rangle \left\vert \Psi \left( X\right) \right\vert
^{2}}
\end{eqnarray*}%
and:%
\begin{eqnarray*}
&&1+\underline{\hat{k}}\left( \hat{X}^{\prime }\right) +\underline{\hat{k}}%
_{1}^{B}\left( \bar{X}^{\prime }\right) +\kappa \left[ \frac{\underline{\hat{%
k}}_{2}^{B}}{1+\bar{k}}\right] \left( \hat{X}^{\prime }\right) \\
&=&\left( 1+\int \hat{k}\left( X,,\hat{X}^{\prime }\right) \hat{K}_{\hat{X}%
^{\prime }}^{\prime }\left\vert \hat{\Psi}\left( \hat{X}^{\prime }\right)
\right\vert ^{2}+\int \left( \hat{k}_{1}^{\left( B\right) }\left( X,\bar{X}%
^{\prime }\right) \bar{K}_{\bar{X}^{\prime }}+\kappa \frac{\hat{k}%
_{2}^{\left( B\right) }\left( X,\bar{X}^{\prime }\right) }{1+\underline{\bar{%
k}}\left( \bar{X}\right) }\bar{K}_{\bar{X}^{\prime }}\right) \left\vert \bar{%
\Psi}\left( \hat{X}^{\prime }\right) \right\vert ^{2}\right)
\end{eqnarray*}

\subsection*{A18.2 Translation of matrix elements}

Three matrices arise in the micro-framework. The first one:%
\begin{equation*}
\bar{M}_{jm}=\frac{\overline{\bar{k}}_{jm}\bar{K}_{j0}\left( t\right) }{%
1+\sum_{\nu }\overline{\bar{k}}_{jv}\hat{K}_{\nu }\left( t\right) }
\end{equation*}%
is translated in the matrix $\bar{M}\left( \left( \bar{K},\bar{X}\right)
,\left( \bar{K}^{\prime },\bar{X}^{\prime }\right) \right) $ which has the
form:%
\begin{equation*}
\bar{M}\left( \left( \bar{K},\bar{X}\right) ,\left( \bar{K}^{\prime },\bar{X}%
^{\prime }\right) \right) =\frac{\overline{\bar{k}}\left( \bar{X},\bar{X}%
^{\prime }\right) \bar{K}_{0}}{1+\int \overline{\bar{k}}\left( \bar{X},\bar{X%
}^{\prime }\right) \bar{K}_{0}^{\prime }\left\vert \bar{\Psi}\left( \bar{K}%
_{0}^{\prime },\bar{X}^{\prime }\right) \right\vert ^{2}}
\end{equation*}%
The matrix:%
\begin{equation*}
\hat{M}_{jm}=\frac{\hat{k}_{jm}\hat{K}_{j}\left( t\right) }{1+\sum_{l}\hat{k}%
_{jl}\hat{K}_{l}\left( t\right) +\sum_{l}\bar{k}_{1jl}\bar{K}_{l0}\left(
t\right) +\kappa \sum_{l}\bar{k}_{2jl}\frac{\bar{K}_{l0}\left( t\right) }{%
1+\sum_{m}\overline{\bar{k}}_{lm}\bar{K}_{m0}\left( t\right) }}
\end{equation*}%
is translated into:%
\begin{eqnarray*}
&&\hat{M}\left( \left( \hat{K}^{\prime },\hat{X}^{\prime }\right) ,\left( 
\hat{K},\hat{X}\right) \right) \\
&=&\frac{\hat{k}\left( \hat{X},\hat{X}^{\prime }\right) \hat{K}}{1+\int \hat{%
k}\left( \hat{X},\hat{X}^{\prime }\right) \left\vert \hat{\Psi}\left( \hat{K}%
^{\prime },\hat{X}^{\prime }\right) \right\vert ^{2}+\int \hat{k}%
_{1}^{B}\left( \hat{X},\bar{X}^{\prime }\right) \bar{K}_{0}^{\prime
}\left\vert \bar{\Psi}\left( \bar{K}_{0}^{\prime },\bar{X}^{\prime }\right)
\right\vert ^{2}+\int \hat{k}_{2}^{B}\left( \hat{X},\bar{X}^{\prime }\right) 
\frac{\bar{K}_{0}^{\prime }\left\vert \bar{\Psi}\left( \bar{K}_{0}^{\prime },%
\bar{X}^{\prime }\right) \right\vert ^{2}}{1+\int \overline{\bar{k}}\left( 
\bar{X}^{\prime },\bar{X}^{\prime \prime }\right) \bar{K}_{0}^{\prime \prime
}\left\vert \bar{\Psi}\left( \bar{K}_{0}^{\prime \prime },\bar{X}^{\prime
\prime }\right) \right\vert ^{2}}}
\end{eqnarray*}%
and ultimately:%
\begin{eqnarray*}
\bar{N} &=&\frac{\hat{K}_{j}\left( \bar{k}_{1jl}+\kappa \bar{k}_{2jl}\frac{1%
}{1+\sum_{m}\overline{\bar{k}}_{lm}\bar{K}_{m0}\left( t\right) }-\kappa 
\frac{\sum_{m}\bar{k}_{2jm}\bar{K}_{m0}\left( t\right) \overline{\bar{k}}%
_{ml}}{\left( 1+\sum_{n}\overline{\bar{k}}_{mn}\bar{K}_{n0}\left( t\right)
\right) ^{2}}\right) }{1+\sum_{l}\hat{k}_{jl}\hat{K}_{l}\left( t\right)
+\sum_{l}\bar{k}_{1jl}\bar{K}_{l0}\left( t\right) +\kappa \sum_{l}\bar{k}%
_{2jl}\frac{\bar{K}_{j0}\left( t\right) }{1+\sum_{m}\overline{\bar{k}}_{lm}%
\bar{K}_{m0}\left( t\right) }} \\
&\rightarrow &\frac{\left( \hat{k}_{1}^{B}\left( \hat{X},\bar{X}^{\prime
}\right) +\kappa \frac{\hat{k}_{2}^{B}\left( \hat{X},\bar{X}^{\prime
}\right) }{1+\int \overline{\bar{k}}\left( \bar{X}^{\prime },\bar{X}^{\prime
\prime }\right) \bar{K}_{0}^{\prime \prime }\left\vert \bar{\Psi}\left( \bar{%
K}_{0}^{\prime \prime },\bar{X}^{\prime \prime }\right) \right\vert ^{2}}%
-\kappa \int \frac{\hat{k}_{2}^{B}\left( \bar{X},\bar{X}^{\prime \prime
}\right) \bar{K}_{0}^{\prime \prime }\overline{\bar{k}}\left( \bar{X}%
^{\prime \prime },\bar{X}^{\prime }\right) }{\left( 1+\int \overline{\bar{k}}%
\left( \bar{X}^{\prime \prime },\bar{Y}\right) \bar{K}_{0}^{Y}\left\vert 
\bar{\Psi}\left( \bar{K}_{0}^{Y},\bar{Y}\right) \right\vert ^{2}\right) ^{2}}%
\right) \hat{K}}{1+\int \hat{k}\left( \hat{X},\hat{X}^{\prime }\right)
\left\vert \hat{\Psi}\left( \hat{K}^{\prime },\hat{X}^{\prime }\right)
\right\vert ^{2}+\int \hat{k}_{1}^{B}\left( \hat{X},\bar{X}^{\prime }\right) 
\bar{K}_{0}^{\prime }\left\vert \bar{\Psi}\left( \bar{K}_{0}^{\prime },\bar{X%
}^{\prime }\right) \right\vert ^{2}+\kappa \int \hat{k}_{2}^{B}\left( \hat{X}%
,\bar{X}^{\prime }\right) \frac{\bar{K}_{0}^{\prime }\left\vert \bar{\Psi}%
\left( \bar{K}_{0}^{\prime },\bar{X}^{\prime }\right) \right\vert ^{2}}{%
1+\int \overline{\bar{k}}\left( \bar{X}^{\prime },\bar{X}^{\prime \prime
}\right) \bar{K}_{0}^{\prime \prime }\left\vert \bar{\Psi}\left( \bar{K}%
_{0}^{\prime \prime },\bar{X}^{\prime \prime }\right) \right\vert ^{2}}}
\end{eqnarray*}

\subsection*{A18.3 Translation of return equations no-default scenario}

The translations of the returns equations, without default, are:

\begin{eqnarray}
&&\left( \Delta \left( \hat{X},\hat{X}^{\prime }\right) -\frac{\hat{K}%
^{\prime }\hat{k}_{1}\left( \hat{X}^{\prime },\hat{X}\right) \left\vert \hat{%
\Psi}\left( \hat{K}^{\prime },\hat{X}^{\prime }\right) \right\vert ^{2}}{1+%
\underline{\hat{k}}\left( \hat{X}^{\prime }\right) +\underline{\hat{k}}%
_{1}^{B}\left( \bar{X}^{\prime }\right) +\kappa \left[ \frac{\underline{\hat{%
k}}_{2}^{B}}{1+\bar{k}}\right] \left( \hat{X}^{\prime }\right) }\right) 
\frac{\hat{f}\left( \hat{X}^{\prime }\right) }{1+\underline{\hat{k}}%
_{2}\left( \bar{X}^{\prime }\right) +\kappa \left[ \frac{\underline{\hat{k}}%
_{2}^{B}}{1+\bar{k}}\right] \left( \hat{X}^{\prime }\right) }  \label{RTF} \\
&=&\frac{k_{1}\left( \hat{X}^{\prime },X\right) \hat{K}^{\prime }}{1+%
\underline{k}\left( \hat{X}^{\prime }\right) +\underline{k}_{1}^{\left(
B\right) }\left( \bar{X}^{\prime }\right) +\kappa \left[ \frac{\underline{k}%
_{2}^{B}}{1+\bar{k}}\right] \left( X^{\prime }\right) }\frac{f_{1}^{\prime
}\left( \hat{K},\hat{X},\Psi ,\hat{\Psi}\right) }{1+\underline{k}_{2}\left( 
\hat{X}^{\prime }\right) +\kappa \left[ \frac{\underline{k}_{2}^{B}}{1+\bar{k%
}}\right] \left( X^{\prime }\right) }  \notag
\end{eqnarray}%
for the investors, with solution:%
\begin{eqnarray*}
\hat{f}\left( \hat{X}^{\prime }\right) &=&\left( 1+\underline{\hat{k}}%
_{2}\left( \hat{X}^{\prime }\right) +\kappa \left[ \frac{\underline{\hat{k}}%
_{2}^{B}}{1+\bar{k}}\right] \left( \hat{X}^{\prime }\right) \right) \left(
\Delta \left( \hat{X},\hat{X}^{\prime }\right) -\frac{\hat{K}^{\prime }\hat{k%
}_{1}\left( \hat{X}^{\prime },\hat{X}\right) \left\vert \hat{\Psi}\left( 
\hat{K}^{\prime },\hat{X}^{\prime }\right) \right\vert ^{2}}{1+\underline{%
\hat{k}}\left( \hat{X}^{\prime }\right) +\underline{\hat{k}}_{1}^{B}\left( 
\bar{X}^{\prime }\right) +\kappa \left[ \frac{\underline{\hat{k}}_{2}^{B}}{1+%
\bar{k}}\right] \left( \hat{X}^{\prime }\right) }\right) ^{-1} \\
&&\times \frac{k_{1}\left( \hat{X}^{\prime },X\right) \hat{K}^{\prime }}{1+%
\underline{k}\left( \hat{X}^{\prime }\right) +\underline{k}_{1}^{\left(
B\right) }\left( \bar{X}^{\prime }\right) +\kappa \left[ \frac{\underline{k}%
_{2}^{B}}{1+\bar{k}}\right] \left( X^{\prime }\right) }\frac{f_{1}^{\prime
}\left( \hat{K},\hat{X},\Psi ,\hat{\Psi}\right) \left\vert \hat{\Psi}\left( 
\hat{K}^{\prime },\hat{X}^{\prime }\right) \right\vert ^{2}}{1+\underline{k}%
_{2}\left( \hat{X}^{\prime }\right) +\kappa \left[ \frac{\underline{k}%
_{2}^{B}}{1+\bar{k}}\right] \left( X^{\prime }\right) }
\end{eqnarray*}%
and:%
\begin{eqnarray}
&&\left( \Delta \left( \bar{X}^{\prime },\bar{X}\right) -\frac{\bar{K}%
^{\prime }\bar{k}_{1}\left( \bar{X}^{\prime },\bar{X}\right) \left\vert \bar{%
\Psi}\left( \bar{K}^{\prime },\bar{X}^{\prime }\right) \right\vert ^{2}}{1+%
\underline{\bar{k}}\left( \bar{X}^{\prime }\right) }\right) \frac{\bar{f}%
\left( \hat{X}^{\prime }\right) }{1+\underline{\overline{\bar{k}}}_{2}\left( 
\bar{X}^{\prime }\right) }  \label{RTNS} \\
&&-\frac{\hat{K}^{\prime }\underline{\hat{k}}_{1}^{B}\left( \hat{X}^{\prime
},\bar{X}\right) }{1+\underline{\hat{k}}\left( \hat{X}^{\prime }\right) +%
\underline{\hat{k}}_{1}^{B}\left( \bar{X}^{\prime }\right) +\kappa \left[ 
\frac{\underline{\hat{k}}_{2}^{B}}{1+\bar{k}}\right] \left( \hat{X}^{\prime
}\right) }\frac{\hat{f}\left( \hat{X}^{\prime }\right) }{1+\underline{\hat{k}%
}_{2}\left( \bar{X}^{\prime }\right) +\kappa \frac{\underline{\hat{k}}%
_{2}^{B}\left( \bar{X}^{\prime }\right) }{1+\bar{k}\left( \bar{X}\right) }} 
\notag \\
&=&\frac{\underline{k}_{1}^{\left( B\right) }\left( X^{\prime },\bar{X}%
\right) }{1+\underline{k}\left( \hat{X}^{\prime }\right) +\underline{k}%
_{1}^{\left( B\right) }\left( \bar{X}^{\prime }\right) +\kappa \frac{%
\underline{k}_{2}^{\left( B\right) }\left( \bar{X}^{\prime }\right) }{1+%
\underline{\bar{k}}}}\frac{\left( f_{1}^{\prime }\left( X^{\prime }\right)
K^{\prime }-\bar{C}\left( X^{\prime }\right) \right) }{1+\underline{k}%
_{2}\left( \hat{X}^{\prime }\right) +\kappa \frac{\underline{k}_{2}^{\left(
B\right) }\left( \bar{X}^{\prime }\right) }{1+\underline{\bar{k}}}}  \notag
\end{eqnarray}%
for banks. The solution for $\bar{f}\left( \hat{X}^{\prime }\right) $ is
straightforward:

\begin{eqnarray*}
\bar{f}\left( \hat{X}^{\prime }\right) &=&\left( 1+\underline{\overline{\bar{%
k}}}_{2}\left( \bar{X}^{\prime }\right) \right) \left( \Delta \left( \bar{X}%
^{\prime },\bar{X}\right) -\frac{\bar{K}^{\prime }\bar{k}_{1}\left( \bar{X}%
^{\prime },\bar{X}\right) \left\vert \bar{\Psi}\left( \bar{K}^{\prime },\bar{%
X}^{\prime }\right) \right\vert ^{2}}{1+\underline{\bar{k}}\left( \bar{X}%
^{\prime }\right) }\right) ^{-1} \\
&&\left\{ \frac{\hat{K}^{\prime }\underline{\hat{k}}_{1}^{B}\left( \hat{X}%
^{\prime },\bar{X}\right) }{1+\underline{\hat{k}}\left( \hat{X}^{\prime
}\right) +\underline{\hat{k}}_{1}^{B}\left( \bar{X}^{\prime }\right) +\kappa %
\left[ \frac{\underline{\hat{k}}_{2}^{B}}{1+\bar{k}}\right] \left( \hat{X}%
^{\prime }\right) }\left( \Delta \left( \hat{X},\hat{X}^{\prime }\right) -%
\frac{\hat{K}^{\prime }\hat{k}_{1}\left( \hat{X}^{\prime },\hat{X}\right)
\left\vert \hat{\Psi}\left( \hat{K}^{\prime },\hat{X}^{\prime }\right)
\right\vert ^{2}}{1+\underline{\hat{k}}\left( \hat{X}^{\prime }\right) +%
\underline{\hat{k}}_{1}^{B}\left( \bar{X}^{\prime }\right) +\kappa \left[ 
\frac{\underline{\hat{k}}_{2}^{B}}{1+\bar{k}}\right] \left( \hat{X}^{\prime
}\right) }\right) ^{-1}\right. \\
&&\times \frac{k_{1}\left( \hat{X}^{\prime },X\right) \hat{K}^{\prime }}{1+%
\underline{k}\left( \hat{X}^{\prime }\right) +\underline{k}_{1}^{\left(
B\right) }\left( \bar{X}^{\prime }\right) +\kappa \left[ \frac{\underline{k}%
_{2}^{B}}{1+\bar{k}}\right] \left( X^{\prime }\right) }\frac{f_{1}^{\prime
}\left( \hat{K},\hat{X},\Psi ,\hat{\Psi}\right) }{1+\underline{k}_{2}\left( 
\hat{X}^{\prime }\right) +\kappa \left[ \frac{\underline{k}_{2}^{B}}{1+\bar{k%
}}\right] \left( X^{\prime }\right) } \\
&&\left. +\frac{\underline{k}_{1}^{\left( B\right) }\left( X^{\prime },\bar{X%
}\right) }{1+\underline{k}\left( \hat{X}^{\prime }\right) +\underline{k}%
_{1}^{\left( B\right) }\left( \bar{X}^{\prime }\right) +\kappa \frac{%
\underline{k}_{2}^{\left( B\right) }\left( \bar{X}^{\prime }\right) }{1+%
\underline{\bar{k}}}}\frac{\left( f_{1}^{\prime }\left( X^{\prime }\right)
K^{\prime }-\bar{C}\left( X^{\prime }\right) \right) }{1+\underline{k}%
_{2}\left( \hat{X}^{\prime }\right) +\kappa \frac{\underline{k}_{2}^{\left(
B\right) }\left( \bar{X}^{\prime }\right) }{1+\underline{\bar{k}}}}\right\}
\end{eqnarray*}%
The text rewrite these equations in terms of $\hat{g}\left( \hat{K}_{1},\hat{%
X}_{1}\right) $ and $\bar{g}\left( \bar{K}_{1},\bar{X}_{1}\right) $ using
the relations%
\begin{eqnarray*}
\hat{g}\left( \hat{K}_{1},\hat{X}_{1}\right) &=&\left( 1-\hat{M}\right) ^{-1}%
\hat{f}\left( \hat{K}^{\prime },\hat{X}^{\prime }\right) +\left( 1-\hat{M}%
\right) ^{-1}\bar{N}\left( 1-\bar{M}\right) ^{-1}\bar{f}\left( \hat{K}%
^{\prime },\hat{X}^{\prime }\right) \\
&=&\left( 1-\hat{M}\right) ^{-1}\hat{f}\left( \hat{K}^{\prime },\hat{X}%
^{\prime }\right) +\left( 1-\hat{M}\right) ^{-1}\bar{N}\bar{g}\left( \hat{K}%
^{\prime },\hat{X}^{\prime }\right)
\end{eqnarray*}%
\begin{equation*}
\bar{g}\left( \bar{K}_{1},\bar{X}_{1}\right) =\left( 1-\bar{M}\right) ^{-1}%
\bar{f}\left( \hat{K}^{\prime },\hat{X}^{\prime }\right)
\end{equation*}%
allow to rewrite the returns as:%
\begin{equation*}
\bar{f}\left( \bar{K}_{1},\bar{X}_{1}\right) =\left( 1-\bar{M}\right) \bar{g}%
\left( \hat{K}^{\prime },\hat{X}^{\prime }\right)
\end{equation*}%
\begin{equation*}
\hat{f}\left( \hat{K}_{1},\hat{X}_{1}\right) =\left( 1-\hat{M}\right) \hat{g}%
\left( \hat{K}^{\prime },\hat{X}^{\prime }\right) -\bar{N}\bar{g}\left( \hat{%
K}^{\prime },\hat{X}^{\prime }\right)
\end{equation*}%
leading to the formula in the text.

\subsection*{A18.4 Translation of return equations with default}

\subsubsection*{A18.4.1 Investors' field return equation}

\begin{equation*}
0<1+R_{j}+\left( R_{j}-\bar{r}\right) \left( \sum_{v}\hat{k}_{2jv}\hat{K}%
_{v}\left( t\right) +\kappa \sum_{v}\hat{k}_{2jv}^{B}\frac{\bar{K}%
_{v0}\left( t\right) }{1+\sum_{m}\overline{\bar{k}}_{vm}\bar{K}_{m0}\left(
t\right) }\right)
\end{equation*}%
\begin{equation*}
1+R\left( \hat{X}^{\prime }\right) +\left( R\left( \hat{X}^{\prime }\right) -%
\bar{r}\right) \left( \underline{\hat{k}}_{2}\left( \bar{X}^{\prime }\right)
+\kappa \frac{\underline{\bar{k}}_{2}\left( \bar{X}^{\prime }\right) }{1+%
\bar{k}}\right) >0
\end{equation*}%
\begin{equation*}
\frac{1+R\left( \hat{X}^{\prime }\right) }{\underline{\hat{k}}_{2}\left( 
\bar{X}^{\prime }\right) +\kappa \frac{\underline{\bar{k}}_{2}\left( \bar{X}%
^{\prime }\right) }{1+\bar{k}}}+R\left( \hat{X}^{\prime }\right) >\bar{r}
\end{equation*}

\begin{eqnarray*}
&&\left( 1-\frac{\hat{K}\hat{k}_{1}\left( \hat{X}^{\prime },\hat{X}\right)
\left\vert \hat{\Psi}\left( \hat{K}^{\prime },\hat{X}^{\prime }\right)
\right\vert ^{2}}{1+\underline{\hat{k}}\left( \hat{X}^{\prime }\right) +%
\underline{\hat{k}}_{1}^{B}\left( \bar{X}^{\prime }\right) +\kappa \frac{%
\underline{\hat{k}}_{2}^{B}\left( \bar{X}^{\prime }\right) }{1+\bar{k}}}%
\right) R\left( \hat{X}^{\prime }\right) \\
&=&\left\vert \hat{\Psi}\left( \hat{K}^{\prime },\hat{X}^{\prime }\right)
\right\vert ^{2}\frac{\hat{k}_{2}\left( \hat{X}^{\prime },\hat{X}\right) 
\hat{K}^{\prime }}{1+\underline{\hat{k}}\left( \hat{X}^{\prime }\right) +%
\underline{\hat{k}}_{1}^{B}\left( \bar{X}^{\prime }\right) +\kappa \frac{%
\underline{\hat{k}}_{2}^{B}\left( \bar{X}^{\prime }\right) }{1+\bar{k}}} \\
&&\times \left( \bar{r}-\left( \bar{r}-\left( \frac{1+R\left( \hat{X}%
^{\prime }\right) }{\underline{\hat{k}}_{2}\left( \hat{X}^{\prime }\right)
+\kappa \frac{\underline{\hat{k}}_{2}^{B}\left( \bar{X}^{\prime }\right) }{1+%
\bar{k}}}+R\left( \hat{X}^{\prime }\right) \right) \right) H\left( -\left( 
\bar{r}-\left( \frac{1+R\left( \hat{X}\right) }{\underline{\hat{k}}%
_{2}\left( \hat{X}^{\prime }\right) +\kappa \frac{\underline{\hat{k}}%
_{2}^{B}\left( \bar{X}^{\prime }\right) }{1+\bar{k}}}+R\left( \hat{X}%
^{\prime }\right) \right) \right) \right) \right) \\
&&+\left\vert \Psi \left( K^{\prime },X^{\prime }\right) \right\vert ^{2}%
\frac{\hat{k}_{2}\left( \hat{X}^{\prime },X\right) \hat{K}^{\prime }}{1+%
\underline{k}\left( \hat{X}^{\prime }\right) +\underline{k}_{1}^{\left(
B\right) }\left( \bar{X}^{\prime }\right) +\kappa \frac{\underline{k}%
_{2}^{\left( B\right) }\left( \bar{X}^{\prime }\right) }{1+\underline{\bar{k}%
}}} \\
&&\times \left( \bar{r}-\left( \bar{r}-\left( \frac{1+f_{1}\left( K^{\prime
},X^{\prime }\right) }{\underline{k}_{2}\left( X^{\prime }\right) +\kappa 
\frac{\underline{k}_{2}^{\left( B\right) }\left( X^{\prime }\right) }{1+%
\underline{\bar{k}}}}+f_{1}\left( K^{\prime },X^{\prime }\right) \right)
\right) H\left( -\left( \bar{r}-\left( \frac{1+f_{1}\left( K^{\prime
},X^{\prime }\right) }{\underline{k}_{2}\left( X^{\prime }\right) +\kappa 
\frac{\underline{k}_{2}^{\left( B\right) }\left( X^{\prime }\right) }{1+%
\underline{\bar{k}}}}+f_{1}\left( K^{\prime },X^{\prime }\right) \right)
\right) \right) \right) \\
&&+\hat{f}_{1}\left( \hat{K},\hat{X},\Psi ,\hat{\Psi}\right)
\end{eqnarray*}

Yhis is written in terms of excss retrns using (\ref{FD}) nd (\ref{RF}).
Given that:%
\begin{equation*}
R\left( \hat{X}^{\prime }\right) =\frac{\hat{f}\left( \hat{X}^{\prime
}\right) }{1+\underline{\hat{k}}_{2}\left( \bar{X}^{\prime }\right) +\kappa 
\frac{\underline{\hat{k}}_{2}^{B}\left( \bar{X}^{\prime }\right) }{1+\bar{k}}%
}+\bar{r}\frac{\underline{\hat{k}}_{2}\left( \bar{X}^{\prime }\right)
+\kappa \frac{\underline{\hat{k}}_{2}^{B}\left( \bar{X}^{\prime }\right) }{1+%
\bar{k}}}{1+\underline{\hat{k}}_{2}\left( \bar{X}^{\prime }\right) +\kappa 
\frac{\underline{\hat{k}}_{2}^{B}\left( \bar{X}^{\prime }\right) }{1+\bar{k}}%
}
\end{equation*}%
whr:%
\begin{equation*}
f_{1}^{\prime }\left( K^{\prime },X^{\prime }\right) =\frac{f_{1}\left(
K^{\prime },X^{\prime }\right) }{1+\underline{k}_{2}\left( X^{\prime
}\right) +\kappa \frac{\underline{k}_{2}^{\left( B\right) }\left( \bar{X}%
^{\prime }\right) }{1+\underline{\bar{k}}}}+\bar{r}\frac{\underline{k}%
_{2}\left( X^{\prime }\right) +\kappa \frac{\underline{k}_{2}^{\left(
B\right) }\left( \bar{X}^{\prime }\right) }{1+\underline{\bar{k}}}}{1+%
\underline{k}_{2}\left( X^{\prime }\right) +\kappa \frac{\underline{k}%
_{2}^{\left( B\right) }\left( \bar{X}^{\prime }\right) }{1+\underline{\bar{k}%
}}}
\end{equation*}%
The retun equation becoms:%
\begin{eqnarray}
&&\left( 1-\frac{\hat{K}^{\prime }\hat{k}_{1}\left( \hat{X}^{\prime },\hat{X}%
\right) \left\vert \hat{\Psi}\left( \hat{K}^{\prime },\hat{X}^{\prime
}\right) \right\vert ^{2}}{1+\underline{\hat{k}}\left( \hat{X}^{\prime
}\right) +\underline{\hat{k}}_{1}^{B}\left( \bar{X}^{\prime }\right) +\kappa %
\left[ \frac{\underline{\hat{k}}_{2}^{B}}{1+\bar{k}}\right] \left( \hat{X}%
^{\prime }\right) }\right)  \label{GN} \\
&&\times \left( \frac{\hat{f}\left( \hat{X}^{\prime }\right) }{1+\underline{%
\hat{k}}_{2}\left( \bar{X}^{\prime }\right) +\kappa \left[ \frac{\underline{%
\hat{k}}_{2}^{B}}{1+\bar{k}}\right] \left( \hat{X}^{\prime }\right) }+\bar{r}%
\frac{\underline{\hat{k}}_{2}\left( \bar{X}^{\prime }\right) +\kappa \frac{%
\underline{\hat{k}}_{2}^{B}\left( \bar{X}^{\prime }\right) }{1+\bar{k}}}{1+%
\underline{\hat{k}}_{2}\left( \bar{X}^{\prime }\right) +\kappa \left[ \frac{%
\underline{\hat{k}}_{2}^{B}}{1+\bar{k}}\right] \left( \hat{X}^{\prime
}\right) }\right)  \notag \\
&&+\left\vert \hat{\Psi}\left( \hat{K}^{\prime },\hat{X}^{\prime }\right)
\right\vert ^{2}\left( \bar{r}+\frac{1+\hat{f}\left( \hat{X}^{\prime
}\right) }{\underline{\hat{k}}_{2}\left( \bar{X}^{\prime }\right) +\kappa %
\left[ \frac{\underline{\hat{k}}_{2}^{B}}{1+\bar{k}}\right] \left( \hat{X}%
^{\prime }\right) }H\left( -\frac{1+\hat{f}\left( \hat{X}^{\prime }\right) }{%
\underline{\hat{k}}_{2}\left( \bar{X}^{\prime }\right) +\kappa \left[ \frac{%
\underline{\hat{k}}_{2}^{B}}{1+\bar{k}}\right] \left( \hat{X}^{\prime
}\right) }\right) \right)  \notag \\
&&\times \frac{\hat{k}_{2}\left( \hat{X}^{\prime },\hat{X}\right) \hat{K}%
^{\prime }}{1+\underline{\hat{k}}\left( \hat{X}^{\prime }\right) +\underline{%
\hat{k}}_{1}^{B}\left( \bar{X}^{\prime }\right) +\kappa \left[ \frac{%
\underline{\hat{k}}_{2}^{B}}{1+\bar{k}}\right] \left( \hat{X}^{\prime
}\right) }  \notag \\
&&+\left\vert \Psi \left( K^{\prime },X^{\prime }\right) \right\vert
^{2}\left( \bar{r}+\frac{1+f_{1}^{\prime }\left( X^{\prime }\right) }{%
\underline{k}_{2}\left( X^{\prime }\right) +\kappa \left[ \frac{\underline{k}%
_{2}^{B}}{1+\bar{k}}\right] \left( X^{\prime }\right) }H\left( -\frac{%
1+f_{1}^{\prime }\left( X^{\prime }\right) }{\underline{k}_{2}\left(
X^{\prime }\right) +\kappa \left[ \frac{\underline{k}_{2}^{B}}{1+\bar{k}}%
\right] \left( X^{\prime }\right) }\right) \right)  \notag \\
&&\times \frac{k_{2}\left( \hat{X}^{\prime },X\right) \hat{K}^{\prime }}{1+%
\underline{k}\left( \hat{X}^{\prime }\right) +\underline{k}_{1}^{\left(
B\right) }\left( \bar{X}^{\prime }\right) +\kappa \left[ \frac{\underline{k}%
_{2}^{B}}{1+\bar{k}}\right] \left( X^{\prime }\right) }  \notag \\
&=&\frac{k_{1}\left( \hat{X}^{\prime },X\right) \hat{K}^{\prime }\left\vert
\Psi \left( K^{\prime },X^{\prime }\right) \right\vert ^{2}}{1+\underline{k}%
\left( \hat{X}^{\prime }\right) +\underline{k}_{1}^{\left( B\right) }\left( 
\bar{X}^{\prime }\right) +\kappa \left[ \frac{\underline{k}_{2}^{B}}{1+\bar{k%
}}\right] \left( X^{\prime }\right) }\hat{f}_{1}\left( \hat{K},\hat{X},\Psi ,%
\hat{\Psi}\right)  \notag
\end{eqnarray}

with:%
\begin{equation*}
\left[ \frac{\underline{\hat{k}}_{2}^{B}}{1+\bar{k}}\right] \left( \hat{X}%
^{\prime }\right) =\int \frac{\underline{\hat{k}}_{2}^{B}\left( \hat{X}%
^{\prime },\bar{X}\right) }{1+\bar{k}\left( \bar{X}\right) }\bar{K}%
\left\vert \bar{\Psi}\left( \bar{K},\bar{X}\right) \right\vert ^{2}d\bar{K}d%
\bar{X}
\end{equation*}%
\begin{equation*}
\left[ \frac{\underline{k}_{2}^{B}}{1+\bar{k}}\right] \left( X^{\prime
}\right) =\int \frac{\underline{k}_{2}^{B}\left( X^{\prime },\bar{X}\right) 
}{1+\bar{k}\left( \bar{X}\right) }\bar{K}\left\vert \bar{\Psi}\left( \bar{K},%
\bar{X}\right) \right\vert ^{2}d\bar{K}d\bar{X}
\end{equation*}%
In terms of $\hat{g}\left( \hat{K}_{1},\hat{X}_{1}\right) $ and $\bar{g}%
\left( \bar{K}_{1},\bar{X}_{1}\right) $ this rewrites:%
\begin{eqnarray*}
&&\left( \Delta \left( \hat{X},\hat{X}^{\prime }\right) -\frac{\hat{K}%
^{\prime }\hat{k}_{1}\left( \hat{X}^{\prime },\hat{X}\right) \left\vert \hat{%
\Psi}\left( \hat{K}^{\prime },\hat{X}^{\prime }\right) \right\vert ^{2}}{1+%
\underline{\hat{k}}\left( \hat{X}^{\prime }\right) +\underline{\hat{k}}%
_{1}^{B}\left( \bar{X}^{\prime }\right) +\kappa \left[ \frac{\underline{\hat{%
k}}_{2}^{B}}{1+\bar{k}}\right] \left( \hat{X}^{\prime }\right) }\right) 
\frac{\left( 1-\hat{M}\right) \hat{g}\left( \hat{K}^{\prime },\hat{X}%
^{\prime }\right) +\bar{N}\bar{g}\left( \hat{K}^{\prime },\hat{X}^{\prime
}\right) }{1+\underline{\hat{k}}_{2}\left( \bar{X}^{\prime }\right) +\kappa %
\left[ \frac{\underline{\hat{k}}_{2}^{B}}{1+\bar{k}}\right] \left( \hat{X}%
^{\prime }\right) } \\
&&+\left\vert \hat{\Psi}\left( \hat{K}^{\prime },\hat{X}^{\prime }\right)
\right\vert ^{2}\left( \bar{r}+\frac{1+\hat{f}\left( \hat{X}^{\prime
}\right) }{\underline{\hat{k}}_{2}\left( \bar{X}^{\prime }\right) +\kappa %
\left[ \frac{\underline{\hat{k}}_{2}^{B}}{1+\bar{k}}\right] \left( \hat{X}%
^{\prime }\right) }H\left( -\frac{1+\hat{f}\left( \hat{X}^{\prime }\right) }{%
\underline{\hat{k}}_{2}\left( \bar{X}^{\prime }\right) +\kappa \left[ \frac{%
\underline{\hat{k}}_{2}^{B}}{1+\bar{k}}\right] \left( \hat{X}^{\prime
}\right) }\right) \right) \\
&&\times \frac{\hat{k}_{2}\left( \hat{X}^{\prime },\hat{X}\right) \hat{K}%
^{\prime }}{1+\underline{\hat{k}}\left( \hat{X}^{\prime }\right) +\underline{%
\hat{k}}_{1}^{B}\left( \bar{X}^{\prime }\right) +\kappa \left[ \frac{%
\underline{\hat{k}}_{2}^{B}}{1+\bar{k}}\right] \left( \hat{X}^{\prime
}\right) } \\
&&+\left\vert \Psi \left( K^{\prime },X^{\prime }\right) \right\vert
^{2}\left( \bar{r}+\frac{1+f_{1}^{\prime }\left( X^{\prime }\right) }{%
\underline{k}_{2}\left( X^{\prime }\right) +\kappa \left[ \frac{\underline{k}%
_{2}^{B}}{1+\bar{k}}\right] \left( X^{\prime }\right) }H\left( -\frac{%
1+f_{1}^{\prime }\left( X^{\prime }\right) }{\underline{k}_{2}\left(
X^{\prime }\right) +\kappa \left[ \frac{\underline{k}_{2}^{B}}{1+\bar{k}}%
\right] \left( X^{\prime }\right) }\right) \right) \\
&&\times \frac{k_{2}\left( \hat{X}^{\prime },X\right) \hat{K}^{\prime }}{1+%
\underline{k}\left( \hat{X}^{\prime }\right) +\underline{k}_{1}^{\left(
B\right) }\left( \bar{X}^{\prime }\right) +\kappa \left[ \frac{\underline{k}%
_{2}^{B}}{1+\bar{k}}\right] \left( X^{\prime }\right) } \\
&=&\frac{k_{1}\left( X^{\prime },X^{\prime }\right) \left( f_{1}^{\prime
}\left( X^{\prime }\right) K^{\prime }-\bar{C}\left( X^{\prime }\right)
\right) }{\left( 1+\underline{k}\left( \hat{X}^{\prime }\right) +\underline{k%
}_{1}^{\left( B\right) }\left( \bar{X}^{\prime }\right) +\kappa \left[ \frac{%
\underline{k}_{2}^{B}}{1+\bar{k}}\right] \left( X^{\prime }\right) \right)
\left( 1+\underline{k}_{2}\left( \hat{X}^{\prime }\right) +\kappa \left[ 
\frac{\underline{k}_{2}^{B}}{1+\bar{k}}\right] \left( X^{\prime }\right)
\right) }
\end{eqnarray*}

\subsubsection*{A18.4.2 Banks returns equation}

\begin{eqnarray}
&&\left( 1-\frac{\bar{K}\bar{k}_{1}\left( \bar{X}^{\prime },\bar{X}\right)
\left\vert \bar{\Psi}\left( \bar{K}^{\prime },\bar{X}^{\prime }\right)
\right\vert ^{2}}{1+\underline{\bar{k}}\left( \bar{X}^{\prime }\right) }%
\right) \bar{R}\left( \bar{X}^{\prime }\right) -\frac{\hat{K}\bar{k}%
_{1}\left( \hat{X}^{\prime },\hat{X}\right) \left\vert \hat{\Psi}\left( \hat{%
K}^{\prime },\hat{X}^{\prime }\right) \right\vert ^{2}}{1+\underline{\hat{k}}%
\left( \hat{X}^{\prime }\right) +\underline{\hat{k}}_{1}^{B}\left( \hat{X}%
^{\prime }\right) +\kappa \left[ \frac{\underline{\hat{k}}_{2}^{B}}{1+\bar{k}%
}\right] \left( \hat{X}^{\prime }\right) }R\left( \hat{X}^{\prime }\right)
\label{BR} \\
&=&\left\vert \bar{\Psi}\left( \bar{K}^{\prime },\bar{X}^{\prime }\right)
\right\vert ^{2}\left( \bar{r}-\left( \bar{r}-\left( \frac{1+\bar{R}\left( 
\bar{X}^{\prime }\right) }{\underline{\overline{\bar{k}}}_{2}\left( \bar{X}%
^{\prime }\right) }+\bar{R}\left( \bar{X}^{\prime }\right) \right) \right)
H\left( -\left( \bar{r}-\left( \frac{1+\bar{R}\left( \bar{X}^{\prime
}\right) }{\underline{\overline{\bar{k}}}_{2}\left( \bar{X}^{\prime }\right) 
}+\bar{R}\left( \bar{X}^{\prime }\right) \right) \right) \right) \right) 
\frac{\underline{\bar{k}}_{2}\left( \bar{X}^{\prime }\right) }{1+\underline{%
\overline{\bar{k}}}}  \notag \\
&&+\left\vert \hat{\Psi}\left( \hat{K}^{\prime },\hat{X}^{\prime }\right)
\right\vert ^{2}\left( \bar{r}-\left( \bar{r}-\left( \frac{1+R\left( \hat{X}%
^{\prime }\right) }{\underline{\hat{k}}_{2}\left( \hat{X}^{\prime }\right) }%
+R\left( \hat{X}^{\prime }\right) \right) \right) H\left( -\left( \bar{r}%
-\left( \frac{1+R\left( \hat{X}\right) }{\underline{\hat{k}}_{2}\left( \hat{X%
}^{\prime }\right) }+R\left( \hat{X}^{\prime }\right) \right) \right)
\right) \right)  \notag \\
&&\times \frac{\frac{\kappa \underline{\hat{k}}_{2}\left( \bar{X}\right) }{1+%
\underline{\bar{k}}}}{\left( 1+\underline{\hat{k}}\left( \hat{X}^{\prime
}\right) +\underline{\hat{k}}_{1}^{B}\left( \hat{X}^{\prime }\right) +\kappa %
\left[ \frac{\underline{\hat{k}}_{2}^{B}}{1+\bar{k}}\right] \left( \hat{X}%
^{\prime }\right) \right) }  \notag \\
&&+\left\vert \Psi \left( K^{\prime },X^{\prime }\right) \right\vert
^{2}\left( \bar{r}-\left( \bar{r}-\left( \frac{1+f_{1}\left( K^{\prime
},X^{\prime }\right) }{\underline{k}_{2}\left( X^{\prime }\right) }%
+f_{1}\left( K^{\prime },X^{\prime }\right) \right) \right) H\left( -\left( 
\bar{r}-\left( \frac{1+f_{1}\left( K^{\prime },X^{\prime }\right) }{%
\underline{k}_{2}\left( X^{\prime }\right) }+f_{1}\left( K^{\prime
},X^{\prime }\right) \right) \right) \right) \right)  \notag \\
&&\times \frac{\kappa \underline{k}_{2}^{\left( B\right) }\left( \bar{X}%
\right) }{\left( 1+\underline{k}\left( \hat{X}^{\prime }\right) +\underline{k%
}_{1}^{\left( B\right) }\left( \bar{X}^{\prime }\right) +\kappa \left[ \frac{%
\underline{k}_{2}^{\left( B\right) }}{1+\underline{\bar{k}}}\right] \left(
X^{\prime }\right) \right) \left( 1+\underline{\bar{k}}\right) }+\bar{f}%
_{1}\left( \bar{K},\bar{X},\Psi ,\hat{\Psi}\right)  \notag
\end{eqnarray}

\begin{equation*}
\bar{R}\left( \bar{X}\right) =\frac{\bar{f}\left( \bar{X}\right) }{1+%
\underline{\overline{\bar{k}}}_{2}\left( \bar{X}\right) }+\bar{r}\frac{%
\underline{\overline{\bar{k}}}_{2}\left( \bar{X}\right) }{1+\underline{%
\overline{\bar{k}}}_{2}\left( \bar{X}\right) }
\end{equation*}%
\begin{eqnarray}
&&\left( 1-\frac{\bar{K}^{\prime }\bar{k}_{1}\left( \bar{X}^{\prime },\bar{X}%
\right) \left\vert \bar{\Psi}\left( \bar{K}^{\prime },\bar{X}^{\prime
}\right) \right\vert ^{2}}{1+\underline{\bar{k}}\left( \bar{X}^{\prime
}\right) }\right) \left( \frac{\bar{f}\left( \bar{X}^{\prime }\right) }{1+%
\underline{\overline{\bar{k}}}_{2}\left( \bar{X}^{\prime }\right) }+\bar{r}%
\frac{\underline{\overline{\bar{k}}}_{2}\left( \bar{X}^{\prime }\right) }{1+%
\underline{\overline{\bar{k}}}_{2}\left( \bar{X}^{\prime }\right) }\right)
\label{GT} \\
&&-\frac{\hat{K}^{\prime }\underline{\hat{k}}_{1}^{B}\left( \hat{X}^{\prime
},\bar{X}\right) \left\vert \hat{\Psi}\left( \hat{K}^{\prime },\hat{X}%
^{\prime }\right) \right\vert ^{2}}{1+\underline{\hat{k}}\left( \hat{X}%
^{\prime }\right) +\underline{\hat{k}}_{1}^{B}\left( \bar{X}^{\prime
}\right) +\kappa \left[ \frac{\underline{\hat{k}}_{2}^{B}}{1+\bar{k}}\right]
\left( \hat{X}^{\prime }\right) }\left( \frac{\hat{f}\left( \hat{X}^{\prime
}\right) }{1+\underline{\hat{k}}_{2}\left( \bar{X}^{\prime }\right) +\kappa 
\frac{\underline{\hat{k}}_{2}^{B}\left( \bar{X}^{\prime }\right) }{1+\bar{k}%
\left( \bar{X}\right) }}+\bar{r}\frac{\underline{\hat{k}}_{2}\left( \bar{X}%
^{\prime }\right) +\kappa \frac{\underline{\hat{k}}_{2}^{B}\left( \bar{X}%
^{\prime }\right) }{1+\bar{k}\left( \bar{X}\right) }}{1+\underline{\hat{k}}%
_{2}\left( \bar{X}^{\prime }\right) +\kappa \left[ \frac{\underline{\hat{k}}%
_{2}^{B}}{1+\bar{k}}\right] \left( \hat{X}^{\prime }\right) }\right)  \notag
\\
&=&\left\vert \bar{\Psi}\left( \bar{K}^{\prime },\bar{X}^{\prime }\right)
\right\vert ^{2}\left( \bar{r}+\left( \frac{1+\bar{f}\left( \bar{X}^{\prime
}\right) }{\underline{\overline{\bar{k}}}_{2}\left( \hat{X}^{\prime }\right) 
}\right) H\left( -\left( 1+\bar{f}\left( \bar{X}^{\prime }\right) \right)
\right) \right) \frac{\hat{K}^{\prime }\underline{\hat{k}}_{2}^{B}\left( 
\bar{X}^{\prime },\bar{X}\right) }{1+\underline{\overline{\bar{k}}}\left( 
\bar{X}\right) }  \notag \\
&&+\left\vert \hat{\Psi}\left( \hat{K}^{\prime },\hat{X}^{\prime }\right)
\right\vert ^{2}\left( \bar{r}+\frac{1+\hat{f}\left( \hat{X}^{\prime
}\right) }{\underline{\hat{k}}_{2}\left( \bar{X}^{\prime }\right) +\kappa %
\left[ \frac{\underline{\hat{k}}_{2}^{B}}{1+\bar{k}}\right] \left( \hat{X}%
^{\prime }\right) }H\left( -\frac{1+\hat{f}\left( \hat{X}^{\prime }\right) }{%
\underline{\hat{k}}_{2}\left( \bar{X}^{\prime }\right) +\kappa \left[ \frac{%
\underline{\hat{k}}_{2}^{B}}{1+\bar{k}}\right] \left( \hat{X}^{\prime
}\right) }\right) \right)  \notag \\
&&\times \frac{\hat{K}^{\prime }\kappa \frac{\underline{\hat{k}}%
_{2}^{B}\left( \hat{X}^{\prime },\bar{X}\right) }{1+\bar{k}\left( \bar{X}%
\right) }}{\left( 1+\underline{\hat{k}}\left( \hat{X}^{\prime }\right) +%
\underline{\hat{k}}_{1}^{B}\left( \bar{X}^{\prime }\right) +\kappa \left[ 
\frac{\underline{\hat{k}}_{2}^{B}}{1+\bar{k}}\right] \left( \hat{X}^{\prime
}\right) \right) }  \notag \\
&&+\left\vert \Psi \left( K^{\prime },X^{\prime }\right) \right\vert
^{2}\left( \bar{r}-\frac{\left( \bar{r}-\left( \frac{1+f_{1}\left( K^{\prime
},X^{\prime }\right) }{\underline{k}_{2}\left( X^{\prime }\right) }%
+f_{1}\left( K^{\prime },X^{\prime }\right) \right) \right) }{1+\exp \left(
-\xi \left( \bar{r}-\left( \frac{1+f_{1}\left( K^{\prime },X^{\prime
}\right) }{\underline{k}_{2}\left( X^{\prime }\right) }+f_{1}\left(
K^{\prime },X^{\prime }\right) \right) \right) \right) }\right)  \notag \\
&&\times \frac{K^{\prime }\kappa \frac{\underline{k}_{2}^{\left( B\right)
}\left( X^{\prime },\bar{X}\right) }{1+\underline{\bar{k}}\left( \bar{X}%
\right) }}{\left( 1+\underline{k}\left( \hat{X}^{\prime }\right) +\underline{%
k}_{1}^{\left( B\right) }\left( \bar{X}^{\prime }\right) +\kappa \left[ 
\frac{\underline{k}_{2}^{\left( B\right) }}{1+\underline{\bar{k}}}\right]
\left( X^{\prime }\right) \right) }+\frac{K^{\prime }\underline{k}%
_{1}^{\left( B\right) }\left( X^{\prime },\bar{X}\right) }{\left( 1+%
\underline{k}\left( \hat{X}^{\prime }\right) +\underline{k}_{1}^{\left(
B\right) }\left( \bar{X}^{\prime }\right) +\kappa \frac{\underline{k}%
_{2}^{\left( B\right) }\left( \bar{X}^{\prime }\right) }{1+\underline{\bar{k}%
}}\right) }f_{1}\left( \bar{K},\bar{X},\Psi ,\hat{\Psi}\right)  \notag
\end{eqnarray}

\begin{equation*}
\left[ \frac{\underline{\hat{k}}_{2}^{B}}{1+\bar{k}}\right] \left( X^{\prime
}\right) =\int \frac{\underline{\hat{k}}_{2}^{B}\left( X^{\prime },\bar{X}%
\right) }{1+\bar{k}\left( \bar{X}\right) }\bar{K}\left\vert \bar{\Psi}\left( 
\bar{K},\bar{X}\right) \right\vert ^{2}d\bar{K}d\bar{X}
\end{equation*}%
In terms of $\bar{g}$, the derivation including default is similar to that
for investors and leads to:%
\begin{eqnarray}
&&\left( \Delta \left( \bar{X}^{\prime },\bar{X}\right) -\frac{\bar{K}%
^{\prime }\bar{k}_{1}\left( \bar{X}^{\prime },\bar{X}\right) \left\vert \bar{%
\Psi}\left( \bar{K}^{\prime },\bar{X}^{\prime }\right) \right\vert ^{2}}{1+%
\bar{k}\left( \bar{X}^{\prime }\right) }\right) \frac{\left( 1-\bar{M}%
\right) \bar{g}\left( \hat{K}^{\prime },\hat{X}^{\prime }\right) }{1+%
\underline{\bar{k}}_{2}\left( \bar{X}^{\prime }\right) } \\
&&-\int \frac{\hat{K}^{\prime }\underline{\hat{k}}_{1}^{B}\left( \hat{X}%
^{\prime },\bar{X}\right) }{1+\underline{\hat{k}}\left( \hat{X}^{\prime
}\right) +\underline{\hat{k}}_{1}^{B}\left( \bar{X}^{\prime }\right) +\kappa %
\left[ \frac{\underline{\hat{k}}_{2}^{B}}{1+\bar{k}}\right] \left( \hat{X}%
^{\prime }\right) }\frac{\left( 1-\hat{M}\right) \hat{g}\left( \hat{K}%
^{\prime },\hat{X}^{\prime }\right) +\left( \bar{N}\bar{g}\right) \left( 
\bar{K}^{\prime },\bar{X}^{\prime }\right) }{1+\underline{\hat{k}}_{2}\left( 
\bar{X}^{\prime }\right) +\kappa \frac{\underline{\hat{k}}_{2}^{B}\left( 
\bar{X}^{\prime }\right) }{1+\bar{k}\left( \bar{X}\right) }}d\bar{X}^{\prime
}  \notag \\
&&+\int \left\vert \bar{\Psi}\left( \bar{K}^{\prime },\bar{X}^{\prime
}\right) \right\vert ^{2}\left( \bar{r}+\frac{1+\bar{f}\left( \bar{X}%
^{\prime }\right) }{\underline{\bar{k}}_{2}\left( \bar{X}^{\prime }\right) }%
H\left( -\frac{1+\bar{f}\left( \bar{X}^{\prime }\right) }{\underline{\bar{k}}%
_{2}\left( \bar{X}^{\prime }\right) }\right) \right) \frac{\hat{k}%
_{2}^{B}\left( \hat{X}^{\prime },\hat{X}\right) \hat{K}^{\prime }}{1+\bar{k}%
\left( \hat{X}^{\prime }\right) }  \notag \\
&&+\int \left\vert \hat{\Psi}\left( \hat{K}^{\prime },\hat{X}^{\prime
}\right) \right\vert ^{2}\left( \bar{r}+\frac{1+\hat{f}\left( \hat{X}%
^{\prime }\right) }{\underline{\hat{k}}_{2}\left( \bar{X}^{\prime }\right)
+\kappa \left[ \frac{\underline{\hat{k}}_{2}^{B}}{1+\bar{k}}\right] \left( 
\hat{X}^{\prime }\right) }H\left( -\frac{1+\hat{f}\left( \hat{X}^{\prime
}\right) }{\underline{\hat{k}}_{2}\left( \bar{X}^{\prime }\right) +\kappa %
\left[ \frac{\underline{\hat{k}}_{2}^{B}}{1+\bar{k}}\right] \left( \hat{X}%
^{\prime }\right) }\right) \right)  \notag \\
&&\times \frac{\hat{k}_{2}^{B}\left( \hat{X}^{\prime },\hat{X}\right) \hat{K}%
^{\prime }}{1+\underline{\hat{k}}\left( \hat{X}^{\prime }\right) +\underline{%
\hat{k}}_{1}^{B}\left( \bar{X}^{\prime }\right) +\kappa \left[ \frac{%
\underline{\hat{k}}_{2}^{B}}{1+\bar{k}}\right] \left( \hat{X}^{\prime
}\right) }  \notag \\
&&\int \left\vert \Psi \left( K^{\prime },X^{\prime }\right) \right\vert
^{2}\left( \bar{r}+\frac{1+f_{1}^{\prime }\left( X^{\prime }\right) }{%
\underline{k}_{2}\left( X^{\prime }\right) +\kappa \left[ \frac{\underline{k}%
_{2}^{B}}{1+\bar{k}}\right] \left( X^{\prime }\right) }H\left( -\frac{%
1+f_{1}^{\prime }\left( X^{\prime }\right) }{\underline{k}_{2}\left(
X^{\prime }\right) +\kappa \left[ \frac{\underline{k}_{2}^{B}}{1+\bar{k}}%
\right] \left( X^{\prime }\right) }\right) \right)  \notag \\
&&\times \frac{k_{2}^{B}\left( \hat{X}^{\prime },X\right) \hat{K}^{\prime }}{%
1+\underline{k}\left( \hat{X}^{\prime }\right) +\underline{k}_{1}^{\left(
B\right) }\left( \bar{X}^{\prime }\right) +\kappa \left[ \frac{\underline{k}%
_{2}^{B}}{1+\bar{k}}\right] \left( X^{\prime }\right) }  \notag \\
&=&\frac{\underline{k}_{1}^{\left( B\right) }\left( X^{\prime },\bar{X}%
\right) }{1+\underline{k}\left( \hat{X}^{\prime }\right) +\underline{k}%
_{1}^{\left( B\right) }\left( \bar{X}^{\prime }\right) +\kappa \frac{%
\underline{k}_{2}^{\left( B\right) }\left( \bar{X}^{\prime }\right) }{1+%
\underline{\bar{k}}}}\frac{\left( f_{1}^{\prime }\left( X^{\prime }\right)
K^{\prime }-\bar{C}\left( X^{\prime }\right) \right) }{1+\underline{k}%
_{2}\left( \hat{X}^{\prime }\right) +\kappa \frac{\underline{k}_{2}^{\left(
B\right) }\left( \bar{X}^{\prime }\right) }{1+\underline{\bar{k}}}}  \notag
\end{eqnarray}

\subsection*{A18.5 Normalization of coefficients}

\subsubsection*{A18.5.1 Normalization in the micro framework}

As before, the coefficients will be normalized implicitely:%
\begin{equation*}
\bar{k}_{\eta jl}\rightarrow \frac{\bar{k}_{\eta jl}}{\bar{N}\left\langle 
\bar{K}_{0p}\right\rangle }\rightarrow \frac{\bar{k}_{\eta jl}}{\bar{N}%
\left\langle \bar{K}_{0p}\right\rangle \left( 1-\left\langle \bar{k}%
\right\rangle \right) }
\end{equation*}%
\begin{equation*}
\hat{k}_{\eta jl}^{B}\rightarrow \frac{\hat{k}_{\eta jl}^{B}}{\hat{N}%
\left\langle \hat{K}_{p}\right\rangle }=\frac{\hat{k}_{\eta jl}^{B}}{\hat{N}%
\left\langle \hat{K}\right\rangle \left( 1-\left\langle \hat{k}\right\rangle
-\left\langle \hat{k}_{1}^{B}\right\rangle +\kappa \frac{\left\langle \hat{k}%
_{2}^{B}\right\rangle }{1+\left\langle \bar{k}\right\rangle }\frac{\bar{N}%
\left\langle \bar{K}_{0}\right\rangle }{\hat{N}\left\langle \hat{K}%
\right\rangle }\right) }
\end{equation*}%
\begin{equation*}
\hat{k}_{\eta jl}^{B}\rightarrow \frac{\hat{k}_{\eta jl}^{B}}{\hat{N}%
\left\langle \hat{K}_{p}\right\rangle }=\frac{\hat{k}_{\eta jl}^{B}}{\hat{N}%
\left\langle \hat{K}\right\rangle \left( 1-\left\langle \hat{k}\right\rangle
-\left( \left\langle \hat{k}_{1}^{B}\right\rangle +\kappa \frac{\left\langle 
\hat{k}_{2}^{B}\right\rangle }{1+\left\langle \bar{k}\right\rangle
\left\langle \bar{K}_{0}\right\rangle }\frac{\bar{N}\left\langle \bar{K}%
\right\rangle }{\hat{N}\left\langle \hat{K}\right\rangle }\right) \right) }
\end{equation*}

\subsubsection*{A18.5.2 Normalization in the field model}

We can consider the normalization of coeficients for investors and banks.
The principle is the same as in part one, this amounts to replace:%
\begin{equation*}
\hat{k}_{\eta }\left( \hat{X}^{\prime },\hat{X}\right) \rightarrow \frac{%
\hat{k}_{\eta }\left( \hat{X}^{\prime },\hat{X}\right) }{\left\Vert \hat{\Psi%
}\right\Vert ^{2}\left\langle \hat{K}\right\rangle \left( 1-\left\langle 
\hat{k}\left( \hat{X}^{\prime },\hat{X}\right) \right\rangle -\left(
\left\langle \hat{k}_{1}^{B}\left( \hat{X}^{\prime },\bar{X}\right)
\right\rangle +\kappa \frac{\left\langle \hat{k}_{2}^{B}\left( \hat{X}%
^{\prime },\bar{X}\right) \right\rangle }{1+\left\langle \bar{k}%
\right\rangle \left\langle \bar{K}\right\rangle }\frac{\left\Vert \bar{\Psi}%
\right\Vert ^{2}\left\langle \bar{K}\right\rangle }{\left\Vert \hat{\Psi}%
\right\Vert ^{2}\left\langle \hat{K}\right\rangle }\right) \right) }
\end{equation*}

\begin{equation*}
\underline{\hat{k}}_{\eta }\left( \hat{X}^{\prime }\right) \rightarrow \int 
\frac{\hat{k}_{\eta }\left( \hat{X}^{\prime },\hat{X}\right) }{\left\Vert 
\hat{\Psi}\right\Vert ^{2}\left\langle \hat{K}\right\rangle \left(
1-\left\langle \hat{k}\left( \hat{X}^{\prime },\hat{X}\right) \right\rangle
-\left( \left( \left\langle \hat{k}_{1}^{B}\left( \hat{X}^{\prime },\bar{X}%
\right) \right\rangle +\kappa \frac{\left\langle \hat{k}_{2}^{B}\left( \hat{X%
}^{\prime },\bar{X}\right) \right\rangle }{1+\left\langle \bar{k}%
\right\rangle \left\langle \bar{K}\right\rangle }\right) \frac{\left\Vert 
\bar{\Psi}\right\Vert ^{2}\left\langle \bar{K}\right\rangle }{\left\Vert 
\hat{\Psi}\right\Vert ^{2}\left\langle \hat{K}\right\rangle }\right) \right) 
}\hat{K}_{\hat{X}}\left\vert \hat{\Psi}\left( \hat{X}\right) \right\vert
^{2}d\hat{X}
\end{equation*}%
\begin{equation*}
\bar{k}_{\eta }\left( \bar{X}^{\prime },\bar{X}\right) \rightarrow \frac{%
\bar{k}_{\eta }\left( \bar{X}^{\prime },\bar{X}\right) }{\left\Vert \bar{\Psi%
}\right\Vert ^{2}\left\langle \bar{K}\right\rangle \left( 1-\left\langle 
\bar{k}\left( \bar{X}^{\prime },\bar{X}\right) \right\rangle \right) }
\end{equation*}%
\begin{equation*}
\underline{\bar{k}}_{\eta }\left( \bar{X}^{\prime }\right) \rightarrow \int 
\frac{\bar{k}_{\eta }\left( \bar{X}^{\prime },\bar{X}\right) }{\left\Vert 
\bar{\Psi}\right\Vert ^{2}\left\langle \bar{K}\right\rangle \left(
1-\left\langle \bar{k}\left( \bar{X}^{\prime },\bar{X}\right) \right\rangle
\right) }\bar{K}_{\bar{X}}\left\vert \hat{\Psi}\left( \bar{X}\right)
\right\vert ^{2}d\bar{X}
\end{equation*}%
and:%
\begin{eqnarray*}
&&\frac{\hat{k}_{\eta }\left( \hat{X}^{\prime },\hat{X}\right) }{1+%
\underline{\hat{k}}\left( \hat{X}^{\prime }\right) +\underline{\hat{k}}%
_{1}^{B}\left( \bar{X}^{\prime }\right) +\kappa \left[ \frac{\underline{\hat{%
k}}_{2}^{B}}{1+\bar{k}}\right] \left( \hat{X}^{\prime }\right) } \\
&\rightarrow &\frac{\hat{k}_{\eta }\left( \hat{X}^{\prime },\hat{X}\right) }{%
\left( \left\Vert \hat{\Psi}\right\Vert ^{2}\left\langle \hat{K}%
\right\rangle \left( 1+\left( \underline{\hat{k}}\left( \hat{X}^{\prime
}\right) -\left\langle \underline{\hat{k}}\left( \hat{X}^{\prime }\right)
\right\rangle \right) +\left( \underline{\hat{k}}_{1}^{B}\left( \bar{X}%
^{\prime }\right) -\left\langle \underline{\hat{k}}_{1}^{B}\left( \bar{X}%
^{\prime }\right) \right\rangle \right) +\kappa \left( \left[ \frac{%
\underline{\hat{k}}_{2}^{B}}{1+\bar{k}}\right] \left( \hat{X}^{\prime
}\right) -\left\langle \left[ \frac{\underline{\hat{k}}_{2}^{B}}{1+\bar{k}}%
\right] \left( \hat{X}^{\prime }\right) \right\rangle \right) \right)
\right) } \\
&\rightarrow &\frac{\hat{k}_{\eta }\left( \hat{X}^{\prime },\hat{X}\right) }{%
\left( \left\Vert \hat{\Psi}\right\Vert ^{2}\left\langle \hat{K}%
\right\rangle \left( 1+\underline{\hat{k}}\left( \hat{X}^{\prime }\right) +%
\underline{\hat{k}}_{1}^{B}\left( \bar{X}^{\prime }\right) +\kappa \left[ 
\frac{\underline{\hat{k}}_{2}^{B}}{1+\bar{k}}\right] \left( \hat{X}^{\prime
}\right) \right) \right) }
\end{eqnarray*}%
with now:%
\begin{equation*}
\underline{\hat{k}}\left( \hat{X}^{\prime }\right) \rightarrow \underline{%
\hat{k}}\left( \hat{X}^{\prime }\right) -\left\langle \underline{\hat{k}}%
\left( \hat{X}^{\prime }\right) \right\rangle =\int \frac{\hat{k}\left( \hat{%
X}^{\prime },\hat{X}\right) -\left\langle \hat{k}\left( \hat{X}^{\prime },%
\hat{X}\right) \right\rangle }{\left\Vert \hat{\Psi}\right\Vert
^{2}\left\langle \hat{K}\right\rangle }\hat{K}_{\hat{X}}\left\vert \hat{\Psi}%
\left( \hat{X}\right) \right\vert ^{2}d\hat{X}
\end{equation*}%
\begin{eqnarray*}
\underline{\hat{k}}_{1}^{B}\left( \bar{X}^{\prime }\right) &\rightarrow &%
\underline{\hat{k}}_{1}^{B}\left( \bar{X}^{\prime }\right) -\left\langle 
\underline{\hat{k}}_{1}^{B}\left( \bar{X}^{\prime }\right) \right\rangle
=\int \frac{\underline{\hat{k}}_{1}^{B}\left( \hat{X}^{\prime },\bar{X}%
\right) -\left\langle \underline{\hat{k}}_{1}^{B}\left( \hat{X}^{\prime },%
\bar{X}\right) \right\rangle }{\left\Vert \hat{\Psi}\right\Vert
^{2}\left\langle \hat{K}\right\rangle }\bar{K}_{\bar{X}}\left\vert \bar{\Psi}%
\left( \bar{X}\right) \right\vert ^{2}d\bar{X} \\
\kappa \left[ \frac{\underline{\hat{k}}_{2}^{B}}{1+\bar{k}}\right] \left( 
\hat{X}^{\prime }\right) &\rightarrow &\kappa \left( \left[ \frac{\underline{%
\hat{k}}_{2}^{B}}{1+\bar{k}}\right] \left( \hat{X}^{\prime }\right)
-\left\langle \left[ \frac{\underline{\hat{k}}_{2}^{B}}{1+\bar{k}}\right]
\left( \hat{X}^{\prime }\right) \right\rangle \right) =\kappa \int \frac{%
\underline{\hat{k}}_{2}^{B}\left( \hat{X}^{\prime },\bar{X}\right)
-\left\langle \underline{\hat{k}}_{2}^{B}\left( \hat{X}^{\prime },\bar{X}%
\right) \right\rangle }{\left\Vert \hat{\Psi}\right\Vert ^{2}\left\langle 
\hat{K}\right\rangle }\bar{K}_{\bar{X}}\left\vert \bar{\Psi}\left( \bar{X}%
\right) \right\vert ^{2}d\bar{X}
\end{eqnarray*}

\bigskip

\begin{eqnarray*}
1+\underline{\bar{k}}_{2}\left( \bar{X}^{\prime }\right) &\rightarrow
&1+\int \frac{\bar{k}_{2}\left( \bar{X}^{\prime },\bar{X}\right) }{%
\left\Vert \bar{\Psi}\right\Vert ^{2}\left\langle \bar{K}\right\rangle
\left( 1-\left\langle \bar{k}\left( \bar{X}^{\prime },\bar{X}\right)
\right\rangle \right) }\bar{K}_{\bar{X}}\left\vert \hat{\Psi}\left( \bar{X}%
\right) \right\vert ^{2}d\bar{X} \\
&=&\frac{1}{\left( 1-\left\langle \bar{k}\left( \bar{X}^{\prime },\bar{X}%
\right) \right\rangle \right) }\left( \left( 1-\left\langle \bar{k}\left( 
\bar{X}^{\prime },\bar{X}\right) \right\rangle \right) +\int \frac{\bar{k}%
_{2}\left( \bar{X}^{\prime },\bar{X}\right) }{\left\Vert \bar{\Psi}%
\right\Vert ^{2}\left\langle \bar{K}\right\rangle }\bar{K}_{\bar{X}%
}\left\vert \hat{\Psi}\left( \bar{X}\right) \right\vert ^{2}d\bar{X}\right)
\\
&=&\frac{1}{\left( 1-\left\langle \bar{k}\left( \bar{X}^{\prime },\bar{X}%
\right) \right\rangle \right) }\left( \left( 1-\left\langle \bar{k}%
_{1}\left( \bar{X}^{\prime },\bar{X}\right) \right\rangle \right) +\int 
\frac{\bar{k}_{2}\left( \bar{X}^{\prime },\bar{X}\right) -\left\langle \bar{k%
}_{2}\left( \bar{X}^{\prime },\bar{X}\right) \right\rangle }{\left\Vert \bar{%
\Psi}\right\Vert ^{2}\left\langle \bar{K}\right\rangle }\hat{K}_{\hat{X}%
}\left\vert \hat{\Psi}\left( \hat{X}\right) \right\vert ^{2}d\hat{X}\right)
\end{eqnarray*}%
\begin{equation*}
1+\underline{\bar{k}}_{2}\left( \bar{X}^{\prime }\right) \rightarrow \frac{%
1-\left\langle \bar{k}_{1}\left( \bar{X}^{\prime },\bar{X}\right)
\right\rangle }{1-\left\langle \bar{k}\left( \bar{X}^{\prime },\bar{X}%
\right) \right\rangle }\left( 1+\frac{\underline{\bar{k}}_{2}\left( \hat{X}%
^{\prime }\right) }{1-\left\langle \hat{k}_{1}\left( \hat{X}^{\prime },\hat{X%
}\right) \right\rangle }\right)
\end{equation*}%
with:%
\begin{equation*}
\underline{\bar{k}}_{2}\left( \bar{X}^{\prime }\right) =\int \frac{\bar{k}%
_{2}\left( \bar{X}^{\prime },\bar{X}\right) -\left\langle \bar{k}_{2}\left( 
\bar{X}^{\prime },\bar{X}\right) \right\rangle }{\left\Vert \bar{\Psi}%
\right\Vert ^{2}\left\langle \bar{K}\right\rangle }\bar{K}_{\bar{X}%
}\left\vert \bar{\Psi}\left( \bar{X}\right) \right\vert ^{2}d\bar{X}
\end{equation*}%
\begin{equation*}
\frac{\bar{k}_{\eta }\left( \bar{X}^{\prime },\bar{X}\right) }{1+\underline{%
\bar{k}}_{2}\left( \bar{X}^{\prime }\right) }\rightarrow \frac{\bar{k}_{\eta
}\left( \bar{X}^{\prime },\bar{X}\right) }{\left\Vert \bar{\Psi}\right\Vert
^{2}\left\langle \bar{K}\right\rangle \left( 1-\left\langle \bar{k}%
_{1}\left( \bar{X}^{\prime },\bar{X}\right) \right\rangle \right) \left( 1+%
\frac{\underline{\bar{k}}_{2}\left( \hat{X}^{\prime }\right) }{%
1-\left\langle \hat{k}_{1}\left( \hat{X}^{\prime },\hat{X}\right)
\right\rangle }\right) }
\end{equation*}%
\begin{eqnarray*}
&&\frac{1}{1+\underline{\hat{k}}_{2}\left( \hat{X}^{\prime }\right) +\kappa %
\left[ \frac{\underline{\hat{k}}_{2}^{B}}{1+\bar{k}}\right] \left( \hat{X}%
^{\prime }\right) } \\
&\rightarrow &\frac{1-\left( \left\langle \underline{\hat{k}}\left( \hat{X}%
^{\prime },\hat{X}\right) \right\rangle +\left( \left\langle \underline{\hat{%
k}}_{1}^{B}\left( \hat{X}^{\prime },\bar{X}\right) \right\rangle +\kappa
\left\langle \left[ \frac{\underline{\hat{k}}_{2}^{B}\left( \hat{X}^{\prime
},\bar{X}\right) }{1+\bar{k}}\right] \right\rangle \right) \frac{\left\Vert 
\bar{\Psi}\right\Vert ^{2}\left\langle \bar{K}\right\rangle }{\left\Vert 
\hat{\Psi}\right\Vert ^{2}\left\langle \hat{K}\right\rangle }\right) }{%
\left( 1-\left( \left\langle \hat{k}_{1}\left( \hat{X}^{\prime },\hat{X}%
\right) \right\rangle +\left\langle \hat{k}_{1}^{B}\left( \hat{X}^{\prime },%
\hat{X}\right) \right\rangle \frac{\left\Vert \bar{\Psi}\right\Vert
^{2}\left\langle \bar{K}\right\rangle }{\left\Vert \hat{\Psi}\right\Vert
^{2}\left\langle \hat{K}\right\rangle }\right) \right) \left( 1+\frac{%
\underline{\hat{k}}_{2}\left( \hat{X}^{\prime }\right) +\kappa \left[ \frac{%
\underline{\hat{k}}_{2}^{B}}{1+\bar{k}}\right] \left( \hat{X}^{\prime
}\right) }{1-\left( \left\langle \hat{k}_{1}\left( \hat{X}^{\prime },\hat{X}%
\right) \right\rangle +\left\langle \hat{k}_{1}^{B}\left( \hat{X}^{\prime },%
\hat{X}\right) \right\rangle \frac{\left\Vert \bar{\Psi}\right\Vert
^{2}\left\langle \bar{K}\right\rangle }{\left\Vert \hat{\Psi}\right\Vert
^{2}\left\langle \hat{K}\right\rangle }\right) }\right) }
\end{eqnarray*}%
\begin{eqnarray*}
&&\frac{\hat{k}_{\eta }\left( \hat{X}^{\prime },\hat{X}\right) }{1+%
\underline{\hat{k}}_{2}\left( \hat{X}^{\prime }\right) +\kappa \left[ \frac{%
\underline{\hat{k}}_{2}^{B}}{1+\bar{k}}\right] \left( \hat{X}^{\prime
}\right) } \\
&\rightarrow &\frac{\hat{k}_{\eta }\left( \hat{X}^{\prime },\hat{X}\right) }{%
\left\Vert \hat{\Psi}\right\Vert ^{2}\left\langle \hat{K}\right\rangle
\left( 1-\left( \left\langle \hat{k}_{1}\left( \hat{X}^{\prime },\hat{X}%
\right) \right\rangle +\left\langle \hat{k}_{1}^{B}\left( \hat{X}^{\prime },%
\bar{X}^{\prime }\right) \right\rangle \frac{\left\Vert \bar{\Psi}%
\right\Vert ^{2}\left\langle \bar{K}\right\rangle }{\left\Vert \hat{\Psi}%
\right\Vert ^{2}\left\langle \hat{K}\right\rangle }\right) \right) \left( 1+%
\frac{\underline{\hat{k}}_{2}\left( \hat{X}^{\prime }\right) +\kappa \left[ 
\frac{\underline{\hat{k}}_{2}^{B}}{1+\bar{k}}\right] \left( \hat{X}^{\prime
}\right) }{1-\left( \left\langle \hat{k}_{1}\left( \hat{X}^{\prime },\hat{X}%
\right) \right\rangle +\left\langle \hat{k}_{1}^{B}\left( \hat{X}^{\prime },%
\bar{X}^{\prime }\right) \right\rangle \frac{\left\Vert \bar{\Psi}%
\right\Vert ^{2}\left\langle \bar{K}\right\rangle }{\left\Vert \hat{\Psi}%
\right\Vert ^{2}\left\langle \hat{K}\right\rangle }\right) }\right) }
\end{eqnarray*}

\section*{Appendix 19 Alternate description. Exprssn in terms of relative
coefficients}

\subsection*{A19.1 Investors returns equation}

We write investors'return equation:%
\begin{eqnarray}
0 &=&\int \left( 1-\hat{S}_{1}\left( \hat{X}^{\prime },\hat{K}^{\prime },%
\hat{X}\right) \right) \frac{\hat{f}\left( \hat{X}^{\prime }\right) -\bar{r}%
}{1+\underline{\hat{k}}_{2}\left( \bar{X}^{\prime }\right) +\kappa \frac{%
\underline{\hat{k}}_{2}^{B}\left( \bar{X}^{\prime }\right) }{1+\bar{k}}}d%
\hat{X}^{\prime }d\hat{K}^{\prime } \\
&&-\int \left( \frac{1+\hat{f}\left( \hat{X}^{\prime }\right) }{\underline{%
\hat{k}}_{2}\left( \bar{X}^{\prime }\right) +\kappa \left[ \frac{\underline{%
\hat{k}}_{2}^{B}}{1+\bar{k}}\right] \left( \hat{X}^{\prime }\right) }H\left(
-\frac{1+\hat{f}\left( \hat{X}^{\prime }\right) }{\underline{\hat{k}}%
_{2}\left( \bar{X}^{\prime }\right) +\kappa \left[ \frac{\underline{\hat{k}}%
_{2}^{B}}{1+\bar{k}}\right] \left( \hat{X}^{\prime }\right) }\right) \right) 
\hat{S}_{2}\left( \hat{X}^{\prime },\hat{K}^{\prime },\hat{X}\right) d\hat{X}%
^{\prime }d\hat{K}^{\prime }  \notag \\
&&-\int \frac{1+f_{1}^{\prime }\left( X^{\prime }\right) }{\underline{k}%
_{2}\left( X^{\prime }\right) +\kappa \left[ \frac{\underline{k}_{2}^{B}}{1+%
\bar{k}}\right] \left( X^{\prime }\right) }H\left( -\frac{1+f_{1}^{\prime
}\left( X^{\prime }\right) }{\underline{k}_{2}\left( X^{\prime }\right)
+\kappa \left[ \frac{\underline{k}_{2}^{B}}{1+\bar{k}}\right] \left(
X^{\prime }\right) }\right) S_{2}\left( X^{\prime },K^{\prime },\hat{X}%
\right) -\int S_{1}\left( X^{\prime },K^{\prime },\hat{X}\right) \left( \hat{%
f}_{1}\left( \hat{K},\hat{X}\right) -\bar{r}\right)  \notag
\end{eqnarray}%
where we define:

\begin{eqnarray}
\hat{S}_{\eta }\left( \hat{X}^{\prime },\hat{K}^{\prime },\hat{X}\right) &=&%
\frac{\hat{K}^{\prime }\hat{k}_{\eta }\left( \hat{X}^{\prime },\hat{X}%
\right) \left\vert \hat{\Psi}\left( \hat{K}^{\prime },\hat{X}^{\prime
}\right) \right\vert ^{2}}{1+\underline{\hat{k}}\left( \hat{X}^{\prime
}\right) +\underline{\hat{k}}_{1}^{B}\left( \bar{X}^{\prime }\right) +\kappa %
\left[ \frac{\underline{\hat{k}}_{2}^{B}}{1+\bar{k}}\right] \left( \hat{X}%
^{\prime }\right) }  \label{CF1} \\
\hat{S}_{\eta }\left( \hat{X}^{\prime },\hat{X}\right) &=&\int \frac{\hat{K}%
^{\prime }\hat{k}_{\eta }\left( \hat{X}^{\prime },\hat{X}\right) \left\vert 
\hat{\Psi}\left( \hat{K}^{\prime },\hat{X}^{\prime }\right) \right\vert ^{2}%
}{1+\underline{\hat{k}}\left( \hat{X}^{\prime }\right) +\underline{\hat{k}}%
_{1}^{B}\left( \bar{X}^{\prime }\right) +\kappa \left[ \frac{\underline{\hat{%
k}}_{2}^{B}}{1+\bar{k}}\right] \left( \hat{X}^{\prime }\right) }d\hat{K}%
^{\prime }=\frac{\hat{K}_{\hat{X}^{\prime }}\hat{k}_{\eta }\left( \hat{X}%
^{\prime },\hat{X}\right) \left\vert \hat{\Psi}\left( \hat{X}^{\prime
}\right) \right\vert ^{2}}{1+\underline{\hat{k}}\left( \hat{X}^{\prime
}\right) +\underline{\hat{k}}_{1}^{B}\left( \bar{X}^{\prime }\right) +\kappa %
\left[ \frac{\underline{\hat{k}}_{2}^{B}}{1+\bar{k}}\right] \left( \hat{X}%
^{\prime }\right) }  \notag
\end{eqnarray}%
and:%
\begin{eqnarray}
S_{\eta }\left( X^{\prime },K^{\prime },\hat{X}\right) &=&\frac{k_{\eta
}\left( X^{\prime },\hat{X}\right) K^{\prime }\left\vert \Psi \left(
K^{\prime },X^{\prime }\right) \right\vert ^{2}}{1+\underline{k}\left( \hat{X%
}^{\prime }\right) +\underline{k}_{1}^{\left( B\right) }\left( \bar{X}%
^{\prime }\right) +\kappa \left[ \frac{\underline{k}_{2}^{B}}{1+\bar{k}}%
\right] \left( X^{\prime }\right) }  \label{CF2} \\
S_{\eta }\left( X^{\prime },\hat{X}\right) &=&\int \frac{k_{\eta }\left(
X^{\prime },\hat{X}\right) K^{\prime }\left\vert \Psi \left( K^{\prime
},X^{\prime }\right) \right\vert ^{2}}{1+\underline{k}\left( \hat{X}^{\prime
}\right) +\underline{k}_{1}^{\left( B\right) }\left( \bar{X}^{\prime
}\right) +\kappa \left[ \frac{\underline{k}_{2}^{B}}{1+\bar{k}}\right]
\left( X^{\prime }\right) }dK^{\prime }=\frac{k_{\eta }\left( X^{\prime },%
\hat{X}\right) K_{X^{\prime }}\left\vert \Psi \left( X^{\prime }\right)
\right\vert ^{2}}{1+\underline{k}\left( \hat{X}^{\prime }\right) +\underline{%
k}_{1}^{\left( B\right) }\left( \bar{X}^{\prime }\right) +\kappa \left[ 
\frac{\underline{k}_{2}^{B}}{1+\bar{k}}\right] \left( X^{\prime }\right) } 
\notag
\end{eqnarray}%
After averaging over $\hat{K}^{\prime }$, this writes:%
\begin{eqnarray}
0 &=&\int \left( 1-\hat{S}_{1}\left( \hat{X}^{\prime },\hat{X}\right)
\right) \frac{\hat{f}\left( \hat{X}^{\prime }\right) -\bar{r}}{1+\underline{%
\hat{k}}_{2}\left( \hat{X}^{\prime }\right) +\kappa \left[ \frac{\underline{%
\hat{k}}_{2}^{B}}{1+\bar{k}}\right] \left( \hat{X}^{\prime }\right) }d\hat{X}%
^{\prime } \\
&&-\int \left( \frac{1+\hat{f}\left( \hat{X}^{\prime }\right) }{\underline{%
\hat{k}}_{2}\left( \hat{X}^{\prime }\right) +\kappa \left[ \frac{\underline{%
\hat{k}}_{2}^{B}}{1+\bar{k}}\right] \left( \hat{X}^{\prime }\right) }H\left(
-\frac{1+\hat{f}\left( \hat{X}^{\prime }\right) }{\underline{\hat{k}}%
_{2}\left( \hat{X}^{\prime }\right) +\kappa \left[ \frac{\underline{\hat{k}}%
_{2}^{B}}{1+\bar{k}}\right] \left( \hat{X}^{\prime }\right) }\right) \right) 
\hat{S}_{2}\left( \hat{X}^{\prime },\hat{X}\right) d\hat{X}^{\prime }  \notag
\\
&&-\int \frac{1+f_{1}^{\prime }\left( X^{\prime }\right) }{\underline{k}%
_{2}\left( X^{\prime }\right) +\kappa \left[ \frac{\underline{k}_{2}^{B}}{1+%
\bar{k}}\right] \left( X^{\prime }\right) }H\left( -\frac{1+f_{1}^{\prime
}\left( X^{\prime }\right) }{\underline{k}_{2}\left( X^{\prime }\right)
+\kappa \left[ \frac{\underline{k}_{2}^{B}}{1+\bar{k}}\right] \left(
X^{\prime }\right) }\right) S_{2}\left( X^{\prime },\hat{X}\right) -\int
S_{1}\left( X^{\prime },\hat{X}\right) \left( \hat{f}_{1}\left( \hat{X}%
\right) -\bar{r}\right)  \notag
\end{eqnarray}%
or alternatively, by rewriting all parameters in the new variables:

\begin{eqnarray*}
0 &=&\int \left( \Delta \left( \hat{X},\hat{X}^{\prime }\right) -\hat{S}%
_{1}\left( \hat{X}^{\prime },\hat{X}\right) \right) \frac{1-\int \hat{S}%
\left( \hat{X}^{\prime },\hat{X}\right) \frac{\hat{K}_{\hat{X}}\left\vert 
\hat{\Psi}\left( \hat{X}\right) \right\vert ^{2}}{\hat{K}_{\hat{X}^{\prime
}}\left\vert \hat{\Psi}\left( \hat{X}^{\prime }\right) \right\vert ^{2}}d%
\hat{X}}{1-\int \hat{S}_{1}\left( \hat{X}^{\prime },\hat{X}\right) \frac{%
\hat{K}_{\hat{X}}\left\vert \hat{\Psi}\left( \hat{X}\right) \right\vert ^{2}%
}{\hat{K}_{\hat{X}^{\prime }}\left\vert \hat{\Psi}\left( \hat{X}^{\prime
}\right) \right\vert ^{2}}d\hat{X}}\left( f\left( \hat{X}^{\prime }\right) -%
\bar{r}\right) d\hat{X}^{\prime } \\
&&-\int S_{1}\left( X^{\prime },\hat{X}\right) \frac{1-\int S\left(
X^{\prime },\hat{X}\right) \frac{\hat{K}_{\hat{X}}\left\vert \hat{\Psi}%
\left( \hat{X}\right) \right\vert ^{2}}{K_{X^{\prime }}\left\vert \Psi
\left( X^{\prime }\right) \right\vert ^{2}}d\hat{X}}{1-\int S_{1}\left(
X^{\prime },\hat{X}\right) \frac{\hat{K}_{\hat{X}}\left\vert \hat{\Psi}%
\left( \hat{X}\right) \right\vert ^{2}}{K_{X^{\prime }}\left\vert \Psi
\left( X^{\prime }\right) \right\vert ^{2}}dX}\left( f_{1}^{\prime }\left( 
\hat{X}^{\prime }\right) -\bar{r}\right) dX^{\prime } \\
&&-\int \frac{1+f\left( \hat{X}^{\prime }\right) }{\underline{\hat{k}}%
_{2}\left( \hat{X}^{\prime }\right) }H\left( -\frac{1+f\left( \hat{X}%
^{\prime }\right) }{\underline{\hat{k}}_{2}\left( \hat{X}^{\prime }\right) }%
\right) \hat{S}_{2}\left( \hat{X}^{\prime },\hat{X}\right) d\hat{X}^{\prime
}-\int \frac{1+f_{1}^{\prime }\left( X^{\prime }\right) }{\underline{k}%
_{2}\left( X^{\prime }\right) }H\left( -\frac{1+f_{1}^{\prime }\left(
X^{\prime }\right) }{\underline{k}_{2}\left( X^{\prime }\right) }\right)
S_{2}\left( X^{\prime },\hat{X}\right) dX^{\prime }
\end{eqnarray*}%
The rates satisfy the following constrain:%
\begin{equation*}
\int \left( \hat{S}_{1}\left( \hat{X}^{\prime },\hat{X}\right) +\hat{S}%
_{2}\left( \hat{X}^{\prime },\hat{X}\right) \right) d\hat{X}^{\prime }+\int
\left( S_{1}\left( X^{\prime },\hat{X}\right) +S_{2}\left( X^{\prime },\hat{X%
}\right) \right) dX^{\prime }=1
\end{equation*}

\subsubsection*{A19.1.1 Banks returns equation}

We can express the rtrns equations in terms of rtes by defining:%
\begin{eqnarray*}
\bar{S}_{\eta }\left( \bar{X}^{\prime },\bar{K}^{\prime },\bar{X}\right) &=&%
\frac{\bar{K}^{\prime }\bar{k}_{\eta }\left( \bar{X}^{\prime },\bar{X}%
\right) \left\vert \bar{\Psi}\left( \bar{K}^{\prime },\bar{X}^{\prime
}\right) \right\vert ^{2}}{1+\underline{\bar{k}}\left( \bar{X}^{\prime
}\right) } \\
\bar{S}_{\eta }\left( \bar{X}^{\prime },\bar{X}\right) &=&\int \frac{\bar{K}%
^{\prime }\bar{k}_{\eta }\left( \bar{X}^{\prime },\bar{X}\right) \left\vert 
\bar{\Psi}\left( \bar{K}^{\prime },\bar{X}^{\prime }\right) \right\vert ^{2}%
}{1+\underline{\bar{k}}\left( \bar{X}^{\prime }\right) }d\bar{K}^{\prime }=%
\frac{\bar{K}_{\bar{X}^{\prime }}\bar{k}_{\eta }\left( \bar{X}^{\prime },%
\bar{X}\right) \left\vert \bar{\Psi}\left( \bar{X}^{\prime }\right)
\right\vert ^{2}}{1+\underline{\bar{k}}\left( \bar{X}^{\prime }\right) }
\end{eqnarray*}%
\begin{eqnarray*}
\hat{S}_{1}^{B}\left( \hat{X}^{\prime },\hat{K}^{\prime },\bar{X}\right) &=&%
\frac{\hat{K}^{\prime }\underline{\hat{k}}_{1}^{B}\left( \hat{X}^{\prime },%
\bar{X}\right) \left\vert \hat{\Psi}\left( \hat{K}^{\prime },\hat{X}^{\prime
}\right) \right\vert ^{2}}{1+\underline{\hat{k}}\left( \hat{X}^{\prime
}\right) +\underline{\hat{k}}_{1}^{B}\left( \bar{X}^{\prime }\right) +\kappa %
\left[ \frac{\underline{\hat{k}}_{2}^{B}}{1+\bar{k}}\right] \left( \hat{X}%
^{\prime }\right) } \\
\hat{S}_{1}^{B}\left( \hat{X}^{\prime },\bar{X}\right) &=&\frac{\hat{K}_{%
\hat{X}^{\prime }}\underline{\hat{k}}_{1}^{B}\left( \hat{X}^{\prime },\bar{X}%
\right) \left\vert \hat{\Psi}\left( \hat{X}^{\prime }\right) \right\vert ^{2}%
}{1+\underline{\hat{k}}\left( \hat{X}^{\prime }\right) +\underline{\hat{k}}%
_{1}^{B}\left( \bar{X}^{\prime }\right) +\kappa \left[ \frac{\underline{\hat{%
k}}_{2}^{B}}{1+\bar{k}}\right] \left( \hat{X}^{\prime }\right) }
\end{eqnarray*}%
\begin{eqnarray*}
\hat{S}_{2}^{B}\left( \hat{X}^{\prime },\hat{K}^{\prime },\bar{X}\right) &=&%
\frac{\hat{K}^{\prime }\frac{\kappa \underline{\hat{k}}_{2}^{B}\left( \hat{X}%
^{\prime },\bar{X}\right) }{1+\underline{\bar{k}}\left( \bar{X}\right) }%
\left\vert \hat{\Psi}\left( \hat{K}^{\prime },\hat{X}^{\prime }\right)
\right\vert ^{2}}{1+\underline{\hat{k}}\left( \hat{X}^{\prime }\right) +%
\underline{\hat{k}}_{1}^{B}\left( \bar{X}^{\prime }\right) +\kappa \left[ 
\frac{\underline{\hat{k}}_{2}^{B}}{1+\bar{k}}\right] \left( \hat{X}^{\prime
}\right) } \\
\hat{S}_{2}^{B}\left( \hat{X}^{\prime },\bar{X}\right) &=&\frac{\hat{K}_{%
\hat{X}^{\prime }}\frac{\kappa \underline{\hat{k}}_{2}^{B}\left( \hat{X}%
^{\prime },\bar{X}\right) }{1+\underline{\bar{k}}\left( \bar{X}\right) }%
\left\vert \hat{\Psi}\left( \hat{X}^{\prime }\right) \right\vert ^{2}}{1+%
\underline{\hat{k}}\left( \hat{X}^{\prime }\right) +\underline{\hat{k}}%
_{1}^{B}\left( \bar{X}^{\prime }\right) +\kappa \frac{\underline{\hat{k}}%
_{2}^{B}\left( \bar{X}^{\prime }\right) }{1+\bar{k}}}
\end{eqnarray*}%
\begin{equation*}
\left[ \frac{\underline{k}_{2}^{B}}{1+\bar{k}}\right] \left( \hat{X}^{\prime
}\right) =\int \frac{\underline{k}_{2}^{B}\left( \hat{X}^{\prime },\bar{X}%
\right) }{1+\bar{k}\left( \bar{X}\right) }\bar{K}\left\vert \bar{\Psi}\left( 
\bar{K},\bar{X}\right) \right\vert ^{2}d\bar{K}d\bar{X}
\end{equation*}%
\begin{eqnarray}
S_{1}^{B}\left( X^{\prime },K^{\prime },\hat{X}\right) &=&\frac{K^{\prime }%
\underline{k}_{1}^{\left( B\right) }\left( X^{\prime },\bar{X}\right)
\left\vert \Psi \left( K^{\prime },X^{\prime }\right) \right\vert ^{2}}{1+%
\underline{k}\left( X^{\prime }\right) +\underline{k}_{1}^{\left( B\right)
}\left( X^{\prime }\right) +\kappa \left[ \frac{\underline{k}_{2}^{\left(
B\right) }}{1+\underline{\bar{k}}}\right] \left( X^{\prime }\right) }
\label{DFr} \\
S_{2}^{B}\left( X^{\prime },K^{\prime },\hat{X}\right) &=&\frac{\frac{%
\underline{k}_{2}^{\left( B\right) }\left( X^{\prime },\hat{X}\right) }{1+%
\underline{\bar{k}}\left( \bar{X}\right) }K^{\prime }\left\vert \Psi \left(
K^{\prime },X^{\prime }\right) \right\vert ^{2}}{1+\underline{k}\left(
X^{\prime }\right) +\underline{k}_{1}^{\left( B\right) }\left( X^{\prime
}\right) +\kappa \left[ \frac{\underline{k}_{2}^{\left( B\right) }}{1+%
\underline{\bar{k}}}\right] \left( X^{\prime }\right) }  \notag \\
S_{1}^{B}\left( X^{\prime },\hat{X}\right) &=&\frac{K_{X^{\prime }}%
\underline{k}_{1}^{\left( B\right) }\left( X^{\prime },\bar{X}\right)
\left\vert \Psi \left( K^{\prime },X^{\prime }\right) \right\vert ^{2}}{1+%
\underline{k}\left( X^{\prime }\right) +\underline{k}_{1}^{\left( B\right)
}\left( X^{\prime }\right) +\kappa \left[ \frac{\underline{k}_{2}^{\left(
B\right) }}{1+\underline{\bar{k}}}\right] \left( X^{\prime }\right) }%
\left\vert \Psi \left( X^{\prime }\right) \right\vert ^{2}  \notag \\
S_{2}^{B}\left( X^{\prime },\hat{X}\right) &=&\frac{\frac{\underline{k}%
_{2}^{\left( B\right) }\left( X^{\prime },\hat{X}\right) }{1+\underline{\bar{%
k}}\left( \bar{X}\right) }K_{X^{\prime }}\left\vert \Psi \left( X^{\prime
}\right) \right\vert ^{2}}{1+\underline{k}\left( X^{\prime }\right) +%
\underline{k}_{1}^{\left( B\right) }\left( X^{\prime }\right) +\kappa \left[ 
\frac{\underline{k}_{2}^{\left( B\right) }}{1+\underline{\bar{k}}}\right]
\left( X^{\prime }\right) }  \notag
\end{eqnarray}%
and the equation writes with this parametrization: 
\begin{eqnarray}
0 &=&\left( 1-\bar{S}_{1}\left( \bar{X}^{\prime },\bar{X}\right) \right) 
\frac{\bar{f}\left( \bar{X}^{\prime }\right) -\bar{r}}{1+\underline{%
\overline{\bar{k}}}_{2}\left( \bar{X}^{\prime }\right) }-\hat{S}%
_{1}^{B}\left( \hat{X}^{\prime },\bar{X}\right) \left( \frac{\hat{f}\left( 
\hat{X}^{\prime }\right) -\bar{r}}{1+\underline{\hat{k}}_{2}\left( \bar{X}%
^{\prime }\right) +\kappa \frac{\underline{\hat{k}}_{2}^{B}\left( \bar{X}%
^{\prime }\right) }{1+\bar{k}\left( \bar{X}\right) }}\right) \\
&&-\frac{\left( 1+\bar{f}\left( \bar{X}^{\prime }\right) \right) H\left(
-\left( 1+\bar{f}\left( \bar{X}^{\prime }\right) \right) \right) }{%
\underline{\overline{\bar{k}}}_{2}\left( \hat{X}^{\prime }\right) }\bar{S}%
_{2}\left( \bar{X}^{\prime },\bar{X}\right) -\frac{\left( 1+\hat{f}\left( 
\hat{X}^{\prime }\right) \right) H\left( -\left( 1+\hat{f}\left( \hat{X}%
^{\prime }\right) \right) \right) }{\underline{\hat{k}}_{2}\left( \bar{X}%
^{\prime }\right) +\kappa \left[ \frac{\underline{\hat{k}}_{2}^{B}}{1+\bar{k}%
}\right] \left( \hat{X}^{\prime }\right) }\hat{S}_{2}^{B}\left( \hat{X}%
^{\prime },\bar{X}\right)  \notag \\
&&-\frac{\left( 1+f_{1}^{\prime }\left( X^{\prime }\right) \right) H\left(
1+f_{1}^{\prime }\left( K^{\prime },X^{\prime }\right) \right) }{\underline{k%
}_{2}\left( X^{\prime }\right) +\kappa \left[ \frac{\underline{k}%
_{2}^{\left( B\right) }}{1+\underline{\bar{k}}}\right] \left( X^{\prime
}\right) }S_{1}^{B}\left( \hat{X}^{\prime },\bar{X}\right) -S_{1}^{B}\left( 
\hat{X}^{\prime },\bar{X}\right) \left( \frac{\left( f_{1}^{\prime }\left( 
\bar{K},\bar{X}\right) -\bar{r}\right) }{1+\underline{k}_{2}\left( \hat{X}%
^{\prime }\right) +\kappa \frac{\underline{k}_{2}^{\left( B\right) }\left( 
\bar{X}^{\prime }\right) }{1+\underline{\bar{k}}}}+\Delta F_{\tau }\left( 
\bar{R}\left( K,X\right) \right) \right)  \notag
\end{eqnarray}%
with constrant:%
\begin{equation*}
\int \left( \hat{S}_{1}\left( \hat{X}^{\prime },\hat{X}\right) +\hat{S}%
_{2}\left( \hat{X}^{\prime },\hat{X}\right) \right) d\hat{X}^{\prime }+\int
\left( S_{1}\left( X^{\prime },\hat{X}\right) +S_{2}\left( X^{\prime },\hat{X%
}\right) \right) dX^{\prime }=1
\end{equation*}

\begin{equation*}
\int \bar{S}\left( \bar{X}^{\prime },\bar{X}\right) \frac{\bar{K}_{\bar{X}%
}\left\vert \bar{\Psi}\left( \bar{X}\right) \right\vert ^{2}}{\bar{K}_{\bar{X%
}^{\prime }}\left\vert \bar{\Psi}\left( \bar{X}^{\prime }\right) \right\vert
^{2}}d\bar{X}=\frac{\underline{\bar{k}}\left( \hat{X}^{\prime }\right) }{1+%
\underline{\bar{k}}\left( \hat{X}^{\prime }\right) }
\end{equation*}%
\begin{equation*}
\bar{S}\left( \bar{X}^{\prime },\bar{X}\right) =\bar{S}_{1}\left( \bar{X}%
^{\prime },\bar{X}\right) +\bar{S}_{2}\left( \bar{X}^{\prime },\bar{X}\right)
\end{equation*}%
We can replace all coefficients in function of the new parametrization.
Writing:%
\begin{equation*}
\frac{1}{1+\underline{\bar{k}}\left( \hat{X}^{\prime }\right) }=1-\int \bar{S%
}\left( \bar{X}^{\prime },\bar{X}\right) \frac{\bar{K}_{\bar{X}}\left\vert 
\bar{\Psi}\left( \bar{X}\right) \right\vert ^{2}}{\bar{K}_{\bar{X}^{\prime
}}\left\vert \bar{\Psi}\left( \bar{X}^{\prime }\right) \right\vert ^{2}}d%
\bar{X}
\end{equation*}%
we find:%
\begin{eqnarray*}
\underline{\bar{k}}\left( \hat{X}^{\prime }\right) &=&\int \bar{S}\left( 
\bar{X}^{\prime },\bar{X}\right) \frac{\bar{K}_{\bar{X}}\left\vert \bar{\Psi}%
\left( \bar{X}\right) \right\vert ^{2}}{\bar{K}_{\bar{X}^{\prime
}}\left\vert \bar{\Psi}\left( \bar{X}^{\prime }\right) \right\vert ^{2}}d%
\bar{X} \\
\underline{\bar{k}}_{\eta }\left( \hat{X}^{\prime }\right) &=&\frac{\int 
\bar{S}_{\eta }\left( \bar{X}^{\prime },\bar{X}\right) \frac{\bar{K}_{\bar{X}%
}\left\vert \bar{\Psi}\left( \bar{X}\right) \right\vert ^{2}}{\bar{K}_{\bar{X%
}^{\prime }}\left\vert \bar{\Psi}\left( \bar{X}^{\prime }\right) \right\vert
^{2}}d\bar{X}}{1-\int \bar{S}\left( \bar{X}^{\prime },\bar{X}\right) \frac{%
\bar{K}_{\bar{X}}\left\vert \bar{\Psi}\left( \bar{X}\right) \right\vert ^{2}%
}{\bar{K}_{\bar{X}^{\prime }}\left\vert \bar{\Psi}\left( \bar{X}^{\prime
}\right) \right\vert ^{2}}d\bar{X}}
\end{eqnarray*}%
\begin{equation*}
1+\underline{\bar{k}}_{2}\left( \hat{X}^{\prime }\right) =\frac{1-\int \bar{S%
}_{1}\left( \bar{X}^{\prime },\bar{X}\right) \frac{\bar{K}_{\bar{X}%
}\left\vert \bar{\Psi}\left( \bar{X}\right) \right\vert ^{2}}{\bar{K}_{\bar{X%
}^{\prime }}\left\vert \bar{\Psi}\left( \bar{X}^{\prime }\right) \right\vert
^{2}}d\bar{X}}{1-\int \bar{S}\left( \bar{X}^{\prime },\bar{X}\right) \frac{%
\bar{K}_{\bar{X}}\left\vert \bar{\Psi}\left( \bar{X}\right) \right\vert ^{2}%
}{\bar{K}_{\bar{X}^{\prime }}\left\vert \bar{\Psi}\left( \bar{X}^{\prime
}\right) \right\vert ^{2}}d\bar{X}}
\end{equation*}%
\bigskip

Ultimately, using:%
\begin{equation*}
\hat{S}_{\eta }\left( \hat{X}^{\prime },\hat{X}\right) =\frac{\hat{K}_{\hat{X%
}^{\prime }}\hat{k}_{\eta }\left( \hat{X}^{\prime },\hat{X}\right)
\left\vert \hat{\Psi}\left( \hat{X}^{\prime }\right) \right\vert ^{2}}{1+%
\underline{\hat{k}}\left( \hat{X}^{\prime }\right) +\underline{\hat{k}}%
_{1}^{B}\left( \bar{X}^{\prime }\right) +\kappa \left[ \frac{\underline{\hat{%
k}}_{2}^{B}}{1+\bar{k}}\right] \left( \hat{X}^{\prime }\right) }
\end{equation*}%
\begin{equation*}
S_{\eta }\left( X^{\prime },\hat{X}\right) =\frac{k_{\eta }\left( X^{\prime
},\hat{X}\right) K_{X^{\prime }}\left\vert \Psi \left( X^{\prime }\right)
\right\vert ^{2}}{1+\underline{k}\left( \hat{X}^{\prime }\right) +\underline{%
k}_{1}^{\left( B\right) }\left( \bar{X}^{\prime }\right) +\kappa \left[ 
\frac{\underline{k}_{2}^{B}}{1+\bar{k}}\right] \left( X^{\prime }\right) }
\end{equation*}%
and the following relations:%
\begin{equation*}
\int \hat{S}\left( \hat{X}^{\prime },\hat{X}\right) \frac{\hat{K}_{\hat{X}%
}\left\vert \hat{\Psi}\left( \hat{X}\right) \right\vert ^{2}}{\hat{K}_{\hat{X%
}^{\prime }}\left\vert \hat{\Psi}\left( \hat{X}^{\prime }\right) \right\vert
^{2}}d\hat{X}=\frac{\underline{\hat{k}}\left( \hat{X}^{\prime }\right) }{1+%
\underline{\hat{k}}\left( \hat{X}^{\prime }\right) +\underline{\hat{k}}%
_{1}^{B}\left( \bar{X}^{\prime }\right) +\kappa \left[ \frac{\underline{\hat{%
k}}_{2}^{B}}{1+\bar{k}}\right] \left( \hat{X}^{\prime }\right) }
\end{equation*}%
\begin{equation*}
\int \hat{S}_{1}^{B}\left( \hat{X}^{\prime },\bar{X}\right) \frac{\bar{K}_{%
\bar{X}}\left\vert \bar{\Psi}\left( \bar{X}\right) \right\vert ^{2}}{\hat{K}%
_{\hat{X}^{\prime }}\left\vert \hat{\Psi}\left( \hat{X}^{\prime }\right)
\right\vert ^{2}}d\bar{X}=\frac{\underline{\bar{k}}_{1}\left( \hat{X}%
^{\prime }\right) }{1+\underline{\hat{k}}\left( \hat{X}^{\prime }\right) +%
\underline{\hat{k}}_{1}^{B}\left( \bar{X}^{\prime }\right) +\kappa \left[ 
\frac{\underline{\hat{k}}_{2}^{B}}{1+\bar{k}}\right] \left( \hat{X}^{\prime
}\right) }
\end{equation*}%
\begin{equation*}
\int \hat{S}_{2}^{B}\left( \hat{X}^{\prime },\bar{X}\right) \frac{\bar{K}_{%
\bar{X}}\left\vert \bar{\Psi}\left( \bar{X}\right) \right\vert ^{2}}{\hat{K}%
_{\hat{X}^{\prime }}\left\vert \hat{\Psi}\left( \hat{X}^{\prime }\right)
\right\vert ^{2}}=\frac{\left[ \frac{\underline{\hat{k}}_{2}^{B}}{1+\bar{k}}%
\right] \left( \hat{X}^{\prime }\right) }{1+\underline{\hat{k}}\left( \hat{X}%
^{\prime }\right) +\underline{\hat{k}}_{1}^{B}\left( \bar{X}^{\prime
}\right) +\kappa \left[ \frac{\underline{\hat{k}}_{2}^{B}}{1+\bar{k}}\right]
\left( \hat{X}^{\prime }\right) }
\end{equation*}%
writing:%
\begin{equation*}
\left[ \frac{\underline{\hat{k}}_{2}^{B}}{1+\bar{k}}\right] \left( \hat{X}%
^{\prime }\right) =\int \frac{\underline{\hat{k}}_{2}^{B}\left( \hat{X}%
^{\prime },\bar{X}\right) }{1+\bar{k}\left( \bar{X}\right) }\bar{K}%
\left\vert \bar{\Psi}\left( \bar{K},\bar{X}\right) \right\vert ^{2}d\bar{K}d%
\bar{X}
\end{equation*}%
leads to:%
\begin{eqnarray*}
&&\int \frac{\hat{S}\left( \hat{X}^{\prime },\hat{X}\right) \hat{K}_{\hat{X}%
}\left\vert \hat{\Psi}\left( \hat{X}\right) \right\vert ^{2}}{\hat{K}_{\hat{X%
}^{\prime }}\left\vert \hat{\Psi}\left( \hat{X}^{\prime }\right) \right\vert
^{2}}d\hat{X}+\int \frac{\left( \hat{S}_{1}^{B}\left( \hat{X}^{\prime },\bar{%
X}\right) +\hat{S}_{2}^{B}\left( \hat{X}^{\prime },\bar{X}\right) \right) 
\bar{K}_{\bar{X}}\left\vert \bar{\Psi}\left( \bar{X}\right) \right\vert ^{2}%
}{\hat{K}_{\hat{X}^{\prime }}\left\vert \hat{\Psi}\left( \hat{X}^{\prime
}\right) \right\vert ^{2}}d\bar{X} \\
&=&\frac{\underline{\hat{k}}\left( \hat{X}^{\prime }\right) +\underline{\hat{%
k}}_{1}^{B}\left( \bar{X}^{\prime }\right) +\left[ \frac{\underline{\hat{k}}%
_{2}^{B}}{1+\bar{k}}\right] \left( \hat{X}^{\prime }\right) }{1+\underline{%
\hat{k}}\left( \hat{X}^{\prime }\right) +\underline{\hat{k}}_{1}^{B}\left( 
\bar{X}^{\prime }\right) +\kappa \left[ \frac{\underline{\hat{k}}_{2}^{B}}{1+%
\bar{k}}\right] \left( \hat{X}^{\prime }\right) }
\end{eqnarray*}%
This allows to rewrite the parameters in terms of rates. First for the
investors:%
\begin{eqnarray*}
&&\frac{1}{1+\underline{\hat{k}}\left( \hat{X}^{\prime }\right) +\underline{%
\hat{k}}_{1}^{B}\left( \bar{X}^{\prime }\right) +\kappa \left[ \frac{%
\underline{\hat{k}}_{2}^{B}}{1+\bar{k}}\right] \left( \hat{X}^{\prime
}\right) } \\
&=&1-\left( \int \frac{\hat{S}\left( \hat{X}^{\prime },\hat{X}\right) \hat{K}%
_{\hat{X}}\left\vert \hat{\Psi}\left( \hat{X}\right) \right\vert ^{2}}{\hat{K%
}_{\hat{X}^{\prime }}\left\vert \hat{\Psi}\left( \hat{X}^{\prime }\right)
\right\vert ^{2}}d\hat{X}+\int \frac{\left( \hat{S}_{1}^{B}\left( \hat{X}%
^{\prime },\bar{X}\right) +\hat{S}_{2}^{B}\left( \hat{X}^{\prime },\bar{X}%
\right) \right) \bar{K}_{\bar{X}}\left\vert \bar{\Psi}\left( \bar{X}\right)
\right\vert ^{2}}{\hat{K}_{\hat{X}^{\prime }}\left\vert \hat{\Psi}\left( 
\hat{X}^{\prime }\right) \right\vert ^{2}}d\bar{X}\right)
\end{eqnarray*}%
\begin{eqnarray*}
\underline{\hat{k}}\left( \hat{X}^{\prime }\right) &=&\frac{\int \hat{S}%
\left( \hat{X}^{\prime },\hat{X}\right) \frac{\hat{K}_{\hat{X}}\left\vert 
\hat{\Psi}\left( \hat{X}\right) \right\vert ^{2}}{\hat{K}_{\hat{X}^{\prime
}}\left\vert \hat{\Psi}\left( \hat{X}^{\prime }\right) \right\vert ^{2}}d%
\hat{X}}{1-\left( \int \frac{\hat{S}\left( \hat{X}^{\prime },\hat{X}\right) 
\hat{K}_{\hat{X}}\left\vert \hat{\Psi}\left( \hat{X}\right) \right\vert ^{2}%
}{\hat{K}_{\hat{X}^{\prime }}\left\vert \hat{\Psi}\left( \hat{X}^{\prime
}\right) \right\vert ^{2}}d\hat{X}+\int \frac{\left( \hat{S}_{1}^{B}\left( 
\hat{X}^{\prime },\bar{X}\right) +\hat{S}_{2}^{B}\left( \hat{X}^{\prime },%
\bar{X}\right) \right) \bar{K}_{\bar{X}}\left\vert \bar{\Psi}\left( \bar{X}%
\right) \right\vert ^{2}}{\hat{K}_{\hat{X}^{\prime }}\left\vert \hat{\Psi}%
\left( \hat{X}^{\prime }\right) \right\vert ^{2}}d\bar{X}\right) } \\
\underline{\hat{k}}_{1}^{B}\left( \bar{X}^{\prime }\right) &=&\frac{\int 
\hat{S}_{1}^{B}\left( \hat{X}^{\prime },\bar{X}\right) \frac{\bar{K}_{\bar{X}%
}\left\vert \bar{\Psi}\left( \bar{X}\right) \right\vert ^{2}}{\hat{K}_{\hat{X%
}^{\prime }}\left\vert \hat{\Psi}\left( \hat{X}^{\prime }\right) \right\vert
^{2}}d\bar{X}}{1-\left( \int \frac{\hat{S}\left( \hat{X}^{\prime },\hat{X}%
\right) \hat{K}_{\hat{X}}\left\vert \hat{\Psi}\left( \hat{X}\right)
\right\vert ^{2}}{\hat{K}_{\hat{X}^{\prime }}\left\vert \hat{\Psi}\left( 
\hat{X}^{\prime }\right) \right\vert ^{2}}d\hat{X}+\int \frac{\left( \hat{S}%
_{1}^{B}\left( \hat{X}^{\prime },\bar{X}\right) +\hat{S}_{2}^{B}\left( \hat{X%
}^{\prime },\bar{X}\right) \right) \bar{K}_{\bar{X}}\left\vert \bar{\Psi}%
\left( \bar{X}\right) \right\vert ^{2}}{\hat{K}_{\hat{X}^{\prime
}}\left\vert \hat{\Psi}\left( \hat{X}^{\prime }\right) \right\vert ^{2}}d%
\bar{X}\right) } \\
\kappa \left[ \frac{\underline{\hat{k}}_{2}^{B}}{1+\bar{k}}\right] \left( 
\hat{X}^{\prime }\right) &=&\frac{\int \hat{S}_{2}^{B}\left( \hat{X}^{\prime
},\bar{X}\right) \frac{\bar{K}_{\bar{X}}\left\vert \bar{\Psi}\left( \bar{X}%
\right) \right\vert ^{2}}{\hat{K}_{\hat{X}^{\prime }}\left\vert \hat{\Psi}%
\left( \hat{X}^{\prime }\right) \right\vert ^{2}}d\bar{X}}{1-\left( \int 
\frac{\hat{S}\left( \hat{X}^{\prime },\hat{X}\right) \hat{K}_{\hat{X}%
}\left\vert \hat{\Psi}\left( \hat{X}\right) \right\vert ^{2}}{\hat{K}_{\hat{X%
}^{\prime }}\left\vert \hat{\Psi}\left( \hat{X}^{\prime }\right) \right\vert
^{2}}d\hat{X}+\int \frac{\left( \hat{S}_{1}^{B}\left( \hat{X}^{\prime },\bar{%
X}\right) +\hat{S}_{2}^{B}\left( \hat{X}^{\prime },\bar{X}\right) \right) 
\bar{K}_{\bar{X}}\left\vert \bar{\Psi}\left( \bar{X}\right) \right\vert ^{2}%
}{\hat{K}_{\hat{X}^{\prime }}\left\vert \hat{\Psi}\left( \hat{X}^{\prime
}\right) \right\vert ^{2}}d\bar{X}\right) }
\end{eqnarray*}

and then for the frms bu using:

\begin{equation*}
\int S\left( X^{\prime },\hat{X}\right) \frac{K_{X}\left\vert \Psi \left(
X\right) \right\vert ^{2}}{K_{X^{\prime }}\left\vert \Psi \left( X^{\prime
}\right) \right\vert ^{2}}dX=\frac{\underline{\bar{k}}\left( X^{\prime
}\right) }{1+\underline{k}\left( \hat{X}^{\prime }\right) +\underline{k}%
_{1}^{\left( B\right) }\left( \bar{X}^{\prime }\right) +\kappa \left[ \frac{%
\underline{k}_{2}^{B}}{1+\bar{k}}\right] \left( X^{\prime }\right) }
\end{equation*}

\begin{equation*}
\left[ \frac{\underline{k}_{2}^{B}}{1+\bar{k}}\right] \left( \hat{X}^{\prime
}\right) =\int \frac{\underline{k}_{2}^{B}\left( \hat{X}^{\prime },\bar{X}%
\right) }{1+\bar{k}\left( \bar{X}\right) }\bar{K}\left\vert \bar{\Psi}\left( 
\bar{K},\bar{X}\right) \right\vert ^{2}d\bar{K}d\bar{X}
\end{equation*}%
and (\ref{DFr}), we find:%
\begin{eqnarray*}
\underline{k}\left( X^{\prime }\right) &=&\frac{\int \frac{S\left( X^{\prime
},\hat{X}\right) \hat{K}_{\hat{X}}\left\vert \hat{\Psi}\left( \hat{X}\right)
\right\vert ^{2}}{K_{X^{\prime }}\left\vert \Psi \left( X^{\prime }\right)
\right\vert ^{2}}d\hat{X}}{1-\left( \int \frac{S\left( \hat{X}^{\prime },%
\hat{X}\right) \hat{K}_{\hat{X}}\left\vert \hat{\Psi}\left( \hat{X}\right)
\right\vert ^{2}}{K_{X^{\prime }}\left\vert \Psi \left( X^{\prime }\right)
\right\vert ^{2}}d\hat{X}+\int \frac{\left( S_{1}^{B}\left( \hat{X}^{\prime
},\bar{X}\right) +S_{2}^{B}\left( \hat{X}^{\prime },\bar{X}\right) \right) 
\bar{K}_{\bar{X}}\left\vert \bar{\Psi}\left( \bar{X}\right) \right\vert ^{2}%
}{K_{X^{\prime }}\left\vert \Psi \left( X^{\prime }\right) \right\vert ^{2}}d%
\bar{X}\right) } \\
\underline{k}_{1}^{\left( B\right) }\left( X^{\prime }\right) &=&\frac{\int
S_{1}^{B}\left( \hat{X}^{\prime },\bar{X}\right) \frac{\bar{K}_{\bar{X}%
}\left\vert \bar{\Psi}\left( \bar{X}\right) \right\vert ^{2}}{K_{X^{\prime
}}\left\vert \Psi \left( X^{\prime }\right) \right\vert ^{2}}d\bar{X}}{%
1-\left( \int \frac{S\left( \hat{X}^{\prime },\hat{X}\right) \hat{K}_{\hat{X}%
}\left\vert \hat{\Psi}\left( \hat{X}\right) \right\vert ^{2}}{K_{X^{\prime
}}\left\vert \Psi \left( X^{\prime }\right) \right\vert ^{2}}d\hat{X}+\int 
\frac{\left( S_{1}^{B}\left( \hat{X}^{\prime },\bar{X}\right)
+S_{2}^{B}\left( \hat{X}^{\prime },\bar{X}\right) \right) \bar{K}_{\bar{X}%
}\left\vert \bar{\Psi}\left( \bar{X}\right) \right\vert ^{2}}{K_{X^{\prime
}}\left\vert \Psi \left( X^{\prime }\right) \right\vert ^{2}}d\bar{X}\right) 
} \\
\kappa \left[ \frac{\underline{k}_{2}^{B}}{1+\bar{k}}\right] \left( \hat{X}%
^{\prime }\right) &=&\frac{\int S_{2}^{B}\left( \hat{X}^{\prime },\bar{X}%
\right) \frac{\bar{K}_{\bar{X}}\left\vert \bar{\Psi}\left( \bar{X}\right)
\right\vert ^{2}}{K_{X^{\prime }}\left\vert \Psi \left( X^{\prime }\right)
\right\vert ^{2}}d\bar{X}}{1-\left( \int \frac{S\left( \hat{X}^{\prime },%
\hat{X}\right) \hat{K}_{\hat{X}}\left\vert \hat{\Psi}\left( \hat{X}\right)
\right\vert ^{2}}{K_{X^{\prime }}\left\vert \Psi \left( X^{\prime }\right)
\right\vert ^{2}}d\hat{X}+\int \frac{\left( S_{1}^{B}\left( \hat{X}^{\prime
},\bar{X}\right) +S_{2}^{B}\left( \hat{X}^{\prime },\bar{X}\right) \right) 
\bar{K}_{\bar{X}}\left\vert \bar{\Psi}\left( \bar{X}\right) \right\vert ^{2}%
}{K_{X^{\prime }}\left\vert \Psi \left( X^{\prime }\right) \right\vert ^{2}}d%
\bar{X}\right) }
\end{eqnarray*}%
\bigskip

leading to:

\begin{eqnarray}
0 &=&\left( 1-\bar{S}_{1}\left( \bar{X}^{\prime },\bar{X}\right) \right)
\left( \bar{f}\left( \bar{X}^{\prime }\right) -\bar{r}\right) \frac{1-\int 
\bar{S}\left( \bar{X}^{\prime },\bar{X}\right) \frac{\bar{K}_{\bar{X}%
}\left\vert \bar{\Psi}\left( \bar{X}\right) \right\vert ^{2}}{\bar{K}_{\bar{X%
}^{\prime }}\left\vert \bar{\Psi}\left( \bar{X}^{\prime }\right) \right\vert
^{2}}d\bar{X}}{1-\int \bar{S}_{1}\left( \bar{X}^{\prime },\bar{X}\right) 
\frac{\bar{K}_{\bar{X}}\left\vert \bar{\Psi}\left( \bar{X}\right)
\right\vert ^{2}}{\bar{K}_{\bar{X}^{\prime }}\left\vert \bar{\Psi}\left( 
\bar{X}^{\prime }\right) \right\vert ^{2}}d\bar{X}} \\
&&-\hat{S}_{1}^{B}\left( \hat{X}^{\prime },\bar{X}\right) \left( \hat{f}%
\left( \hat{X}^{\prime }\right) -\bar{r}\right)  \notag \\
&&\times \frac{1-\left( \int \frac{\hat{S}\left( \hat{X}^{\prime },\hat{X}%
\right) \hat{K}_{\hat{X}}\left\vert \hat{\Psi}\left( \hat{X}\right)
\right\vert ^{2}}{\hat{K}_{\hat{X}^{\prime }}\left\vert \hat{\Psi}\left( 
\hat{X}^{\prime }\right) \right\vert ^{2}}d\hat{X}+\int \frac{\left( \hat{S}%
_{1}^{B}\left( \hat{X}^{\prime },\bar{X}\right) +\hat{S}_{2}^{B}\left( \hat{X%
}^{\prime },\bar{X}\right) \right) \bar{K}_{\bar{X}}\left\vert \bar{\Psi}%
\left( \bar{X}\right) \right\vert ^{2}}{\hat{K}_{\hat{X}^{\prime
}}\left\vert \hat{\Psi}\left( \hat{X}^{\prime }\right) \right\vert ^{2}}d%
\bar{X}\right) }{1-\left( \int \frac{\hat{S}_{1}\left( \hat{X}^{\prime },%
\hat{X}\right) \hat{K}_{\hat{X}}\left\vert \hat{\Psi}\left( \hat{X}\right)
\right\vert ^{2}}{\hat{K}_{\hat{X}^{\prime }}\left\vert \hat{\Psi}\left( 
\hat{X}^{\prime }\right) \right\vert ^{2}}d\hat{X}+\int \frac{\hat{S}%
_{1}^{B}\left( \hat{X}^{\prime },\bar{X}\right) \bar{K}_{\bar{X}}\left\vert 
\bar{\Psi}\left( \bar{X}\right) \right\vert ^{2}}{\hat{K}_{\hat{X}^{\prime
}}\left\vert \hat{\Psi}\left( \hat{X}^{\prime }\right) \right\vert ^{2}}d%
\bar{X}\right) }  \notag \\
&&-\left( 1+\bar{f}\left( \bar{X}^{\prime }\right) \right) H\left( -\left( 1+%
\bar{f}\left( \bar{X}^{\prime }\right) \right) \right) \bar{S}_{2}\left( 
\bar{X}^{\prime },\bar{X}\right) \frac{\left( 1-\int \bar{S}\left( \bar{X}%
^{\prime },\bar{X}\right) \frac{\bar{K}_{\bar{X}}\left\vert \bar{\Psi}\left( 
\bar{X}\right) \right\vert ^{2}}{\bar{K}_{\bar{X}^{\prime }}\left\vert \bar{%
\Psi}\left( \bar{X}^{\prime }\right) \right\vert ^{2}}d\bar{X}\right) }{\int 
\bar{S}_{2}\left( \bar{X}^{\prime },\bar{X}\right) \frac{\bar{K}_{\bar{X}%
}\left\vert \bar{\Psi}\left( \bar{X}\right) \right\vert ^{2}}{\bar{K}_{\bar{X%
}^{\prime }}\left\vert \bar{\Psi}\left( \bar{X}^{\prime }\right) \right\vert
^{2}}d\bar{X}}  \notag \\
&&-\left( 1+\hat{f}\left( \hat{X}^{\prime }\right) \right) H\left( -\left( 1+%
\hat{f}\left( \hat{X}^{\prime }\right) \right) \right) \hat{S}_{2}^{B}\left( 
\hat{X}^{\prime },\bar{X}\right)  \notag \\
&&\times \frac{1-\left( \int \frac{S\left( \hat{X}^{\prime },\hat{X}\right) 
\hat{K}_{\hat{X}}\left\vert \hat{\Psi}\left( \hat{X}\right) \right\vert ^{2}%
}{K_{X^{\prime }}\left\vert \Psi \left( X^{\prime }\right) \right\vert ^{2}}d%
\hat{X}+\int \frac{\left( S_{1}^{B}\left( \hat{X}^{\prime },\bar{X}\right)
+S_{2}^{B}\left( \hat{X}^{\prime },\bar{X}\right) \right) \bar{K}_{\bar{X}%
}\left\vert \bar{\Psi}\left( \bar{X}\right) \right\vert ^{2}}{K_{X^{\prime
}}\left\vert \Psi \left( X^{\prime }\right) \right\vert ^{2}}d\bar{X}\right) 
}{\int \frac{S_{2}\left( \hat{X}^{\prime },\hat{X}\right) \hat{K}_{\hat{X}%
}\left\vert \hat{\Psi}\left( \hat{X}\right) \right\vert ^{2}}{K_{X^{\prime
}}\left\vert \Psi \left( X^{\prime }\right) \right\vert ^{2}}d\hat{X}+\int
S_{2}^{B}\left( \hat{X}^{\prime },\bar{X}\right) \frac{\bar{K}_{\bar{X}%
}\left\vert \bar{\Psi}\left( \bar{X}\right) \right\vert ^{2}}{K_{X^{\prime
}}\left\vert \Psi \left( X^{\prime }\right) \right\vert ^{2}}d\bar{X}} 
\notag \\
&&-\left( 1+f_{1}^{\prime }\left( X^{\prime }\right) \right) H\left( -\left(
1+f_{1}^{\prime }\left( K^{\prime },X^{\prime }\right) \right) \right)
S_{2}^{B}\left( \hat{X}^{\prime },\bar{X}\right)  \notag \\
&&\times \frac{1-\left( \int \frac{\hat{S}\left( \hat{X}^{\prime },\hat{X}%
\right) \hat{K}_{\hat{X}}\left\vert \hat{\Psi}\left( \hat{X}\right)
\right\vert ^{2}}{\hat{K}_{\hat{X}^{\prime }}\left\vert \hat{\Psi}\left( 
\hat{X}^{\prime }\right) \right\vert ^{2}}d\hat{X}+\int \frac{\left( \hat{S}%
_{1}^{B}\left( \hat{X}^{\prime },\bar{X}\right) +\hat{S}_{2}^{B}\left( \hat{X%
}^{\prime },\bar{X}\right) \right) \bar{K}_{\bar{X}}\left\vert \bar{\Psi}%
\left( \bar{X}\right) \right\vert ^{2}}{\hat{K}_{\hat{X}^{\prime
}}\left\vert \hat{\Psi}\left( \hat{X}^{\prime }\right) \right\vert ^{2}}d%
\bar{X}\right) }{\int \hat{S}_{2}\left( \hat{X}^{\prime },\hat{X}\right) 
\frac{\hat{K}_{\hat{X}}\left\vert \hat{\Psi}\left( \hat{X}\right)
\right\vert ^{2}}{\hat{K}_{\hat{X}^{\prime }}\left\vert \hat{\Psi}\left( 
\hat{X}^{\prime }\right) \right\vert ^{2}}d\hat{X}+\int \hat{S}%
_{2}^{B}\left( \hat{X}^{\prime },\bar{X}\right) \frac{\bar{K}_{\bar{X}%
}\left\vert \bar{\Psi}\left( \bar{X}\right) \right\vert ^{2}}{\hat{K}_{\hat{X%
}^{\prime }}\left\vert \hat{\Psi}\left( \hat{X}^{\prime }\right) \right\vert
^{2}}d\bar{X}} \\
&&-S_{1}^{B}\left( X^{\prime },\bar{X}\right) \left( f_{1}^{\prime }\left(
X^{\prime }\right) -\bar{r}+\Delta F_{\tau }\left( \bar{R}\left( K,X\right)
\right) \right)  \notag \\
&&\times \frac{1-\left( \int \frac{S\left( X^{\prime },\hat{X}\right) \hat{K}%
_{\hat{X}}\left\vert \hat{\Psi}\left( \hat{X}\right) \right\vert ^{2}}{%
K_{X^{\prime }}\left\vert \Psi \left( X^{\prime }\right) \right\vert ^{2}}d%
\hat{X}+\int \frac{\left( S_{1}^{B}\left( X^{\prime },\bar{X}\right)
+S_{2}^{B}\left( X^{\prime },\bar{X}\right) \right) \bar{K}_{\bar{X}%
}\left\vert \bar{\Psi}\left( \bar{X}\right) \right\vert ^{2}}{K_{X^{\prime
}}\left\vert \Psi \left( X^{\prime }\right) \right\vert ^{2}}d\bar{X}\right) 
}{1-\int \frac{S_{1}\left( X^{\prime },\hat{X}\right) \hat{K}_{\hat{X}%
}\left\vert \hat{\Psi}\left( \hat{X}\right) \right\vert ^{2}}{K_{X^{\prime
}}\left\vert \Psi \left( X^{\prime }\right) \right\vert ^{2}}d\hat{X}-\int
S_{1}^{B}\left( X^{\prime },\bar{X}\right) \frac{\bar{K}_{\bar{X}}\left\vert 
\bar{\Psi}\left( \bar{X}\right) \right\vert ^{2}}{K_{X^{\prime }}\left\vert
\Psi \left( X^{\prime }\right) \right\vert ^{2}}d\bar{X}}  \notag
\end{eqnarray}%
\begin{equation*}
\int \left( \hat{S}_{1}\left( \hat{X}^{\prime },\hat{X}\right) +\hat{S}%
_{2}\left( \hat{X}^{\prime },\hat{X}\right) \right) d\hat{X}^{\prime }+\int
\left( S_{1}\left( X^{\prime },\hat{X}\right) +S_{2}\left( X^{\prime },\hat{X%
}\right) \right) dX^{\prime }=1
\end{equation*}%
\begin{equation*}
\int \left( \bar{S}_{1}\left( \bar{X}^{\prime },\bar{X}\right) +\bar{S}%
_{2}\left( \bar{X}^{\prime },\bar{X}\right) \right) d\bar{X}^{\prime }+\int 
\hat{S}_{1}^{B}\left( \hat{X}^{\prime },\bar{X}\right) d\hat{X}^{\prime
}+\int S_{1}^{B}\left( X^{\prime },\bar{X}\right) dX^{\prime }=1
\end{equation*}%
\begin{equation*}
\int \hat{S}_{2}^{B}\left( \hat{X}^{\prime },\bar{X}\right) d\hat{X}^{\prime
}+\int S_{2}^{B}\left( X^{\prime },\bar{X}\right) dX^{\prime }=1
\end{equation*}%
\bigskip

\begin{eqnarray*}
\hat{S}_{\eta }\left( \hat{X}^{\prime }\right) &=&\int \hat{S}_{\eta }\left( 
\hat{X}^{\prime },\hat{X}\right) \frac{\hat{K}_{\hat{X}}\left\vert \hat{\Psi}%
\left( \hat{X}\right) \right\vert ^{2}}{\hat{K}_{\hat{X}^{\prime
}}\left\vert \hat{\Psi}\left( \hat{X}^{\prime }\right) \right\vert ^{2}}d%
\hat{X} \\
\hat{S}\left( \hat{X}^{\prime }\right) &=&\hat{S}_{1}\left( \hat{X}^{\prime
}\right) +\hat{S}_{2}\left( \hat{X}^{\prime }\right)
\end{eqnarray*}%
\begin{eqnarray*}
\hat{S}_{\eta }^{B}\left( \hat{X}^{\prime }\right) &=&\int \frac{\hat{S}%
_{\eta }^{B}\left( \hat{X}^{\prime },\bar{X}\right) \bar{K}_{\bar{X}%
}\left\vert \bar{\Psi}\left( \bar{X}\right) \right\vert ^{2}}{\hat{K}_{\hat{X%
}^{\prime }}\left\vert \hat{\Psi}\left( \hat{X}^{\prime }\right) \right\vert
^{2}}d\bar{X} \\
\hat{S}^{B}\left( \hat{X}^{\prime }\right) &=&\hat{S}_{1}^{B}\left( \hat{X}%
^{\prime }\right) +\hat{S}_{2}^{B}\left( \hat{X}^{\prime }\right)
\end{eqnarray*}

\begin{eqnarray*}
\bar{S}_{\eta }\left( \bar{X}^{\prime }\right) &=&\int \bar{S}_{\eta }\left( 
\bar{X}^{\prime },\bar{X}\right) \frac{\bar{K}_{\bar{X}}\left\vert \bar{\Psi}%
\left( \bar{X}\right) \right\vert ^{2}}{\bar{K}_{\bar{X}^{\prime
}}\left\vert \bar{\Psi}\left( \bar{X}^{\prime }\right) \right\vert ^{2}}d%
\bar{X} \\
\bar{S}\left( \bar{X}^{\prime }\right) &=&\bar{S}_{1}\left( \bar{X}^{\prime
}\right) +\bar{S}_{2}\left( \bar{X}^{\prime }\right)
\end{eqnarray*}%
\begin{eqnarray*}
S_{\eta }\left( X^{\prime }\right) &=&\int \frac{S_{\eta }\left( X^{\prime },%
\hat{X}\right) \hat{K}_{\hat{X}}\left\vert \hat{\Psi}\left( \hat{X}\right)
\right\vert ^{2}}{K_{X^{\prime }}\left\vert \Psi \left( X^{\prime }\right)
\right\vert ^{2}}d\hat{X} \\
S\left( X^{\prime }\right) &=&S_{1}\left( X^{\prime }\right) +S_{2}\left(
X^{\prime }\right)
\end{eqnarray*}%
\begin{eqnarray*}
S_{\eta }^{B}\left( X^{\prime }\right) &=&\int \frac{S_{\eta }^{B}\left(
X^{\prime },\bar{X}\right) \bar{K}_{\bar{X}}\left\vert \bar{\Psi}\left( \bar{%
X}\right) \right\vert ^{2}}{K_{X^{\prime }}\left\vert \Psi \left( X^{\prime
}\right) \right\vert ^{2}}d\bar{X} \\
S^{B}\left( X^{\prime }\right) &=&S_{1}^{B}\left( X^{\prime }\right)
+S_{2}^{B}\left( X^{\prime }\right)
\end{eqnarray*}%
This lds:

\begin{eqnarray*}
0 &=&\int \left( \Delta \left( \hat{X},\hat{X}^{\prime }\right) -\hat{S}%
_{1}\left( \hat{X}^{\prime },\hat{X}\right) \right) \frac{1-\hat{S}\left( 
\hat{X}^{\prime }\right) }{1-\hat{S}_{1}\left( \hat{X}^{\prime }\right) }%
\left( f\left( \hat{X}^{\prime }\right) -\bar{r}\right) d\hat{X}^{\prime } \\
&&-\int S_{1}\left( X^{\prime },\hat{X}\right) \frac{1-S\left( X^{\prime
}\right) }{1-S_{1}\left( X^{\prime }\right) }\left( \left( f_{1}^{\prime
}\left( \hat{X}^{\prime }\right) -\bar{r}\right) +\Delta F_{\tau }\left( 
\bar{R}\left( K,X\right) \right) \right) dX^{\prime } \\
&&-\int \frac{1-\left( \hat{S}\left( \hat{X}^{\prime }\right) +\hat{S}%
_{1}^{B}\left( \hat{X}^{\prime }\right) +\hat{S}_{2}^{B}\left( \hat{X}%
^{\prime }\right) \right) }{\hat{S}_{2}\left( \hat{X}^{\prime }\right) }%
\left( 1+f\left( \hat{X}^{\prime }\right) \right) H\left( -\left( 1+f\left( 
\hat{X}^{\prime }\right) \right) \right) \hat{S}_{2}\left( \hat{X}^{\prime },%
\hat{X}\right) d\hat{X}^{\prime } \\
&&-\int \frac{1-\left( S\left( X^{\prime }\right) +S_{1}^{B}\left( X^{\prime
}\right) +S_{2}^{B}\left( X^{\prime }\right) \right) }{S_{2}\left( X^{\prime
}\right) }\left( 1+f_{1}^{\prime }\left( X^{\prime }\right) \right) H\left(
-\left( 1+f_{1}^{\prime }\left( X^{\prime }\right) \right) \right)
S_{2}\left( X^{\prime },\hat{X}\right) dX^{\prime }
\end{eqnarray*}%
for nvstrs%
\begin{eqnarray}
0 &=&\left( 1-\bar{S}_{1}\left( \bar{X}^{\prime },\bar{X}\right) \right)
\left( \bar{f}\left( \bar{X}^{\prime }\right) -\bar{r}\right) \frac{1-\bar{S}%
\left( \bar{X}^{\prime }\right) }{1-\bar{S}_{1}\left( \bar{X}^{\prime
}\right) } \\
&&-\hat{S}_{1}^{B}\left( \hat{X}^{\prime },\bar{X}\right) \left( \hat{f}%
\left( \hat{X}^{\prime }\right) -\bar{r}\right) \frac{1-\left( \hat{S}\left( 
\hat{X}^{\prime },\hat{X}\right) +\hat{S}_{1}^{B}\left( \hat{X}^{\prime
}\right) +\hat{S}_{2}^{B}\left( \hat{X}^{\prime }\right) \right) }{1-\left( 
\hat{S}_{1}\left( \hat{X}^{\prime }\right) +\hat{S}_{1}^{B}\left( \hat{X}%
^{\prime }\right) \right) }  \notag \\
&&-\left( 1+\bar{f}\left( \bar{X}^{\prime }\right) \right) H\left( -\left( 1+%
\bar{f}\left( \bar{X}^{\prime }\right) \right) \right) \bar{S}_{2}\left( 
\bar{X}^{\prime },\bar{X}\right) \frac{\left( 1-\bar{S}\left( \bar{X}%
^{\prime }\right) \right) }{\bar{S}_{2}\left( \bar{X}^{\prime }\right) } 
\notag \\
&&-\left( 1+\hat{f}\left( \hat{X}^{\prime }\right) \right) H\left( -\left( 1+%
\hat{f}\left( \hat{X}^{\prime }\right) \right) \right) \hat{S}_{2}^{B}\left( 
\hat{X}^{\prime },\bar{X}\right) \frac{1-\left( S\left( \hat{X}^{\prime
}\right) +\left( S_{1}^{B}\left( \hat{X}^{\prime }\right) +S_{2}^{B}\left( 
\hat{X}^{\prime }\right) \right) \right) }{S_{2}\left( \hat{X}^{\prime
}\right) +S_{2}^{B}\left( \hat{X}^{\prime }\right) }  \notag \\
&&-\left( 1+f_{1}^{\prime }\left( X^{\prime }\right) \right) H\left( -\left(
1+f_{1}^{\prime }\left( X^{\prime }\right) \right) \right) S_{2}^{B}\left( 
\hat{X}^{\prime },\bar{X}\right) \frac{1-\left( \hat{S}\left( \hat{X}%
^{\prime },\hat{X}\right) +\left( \hat{S}_{1}^{B}\left( \hat{X}^{\prime
}\right) +\hat{S}_{2}^{B}\left( \hat{X}^{\prime }\right) \right) \right) }{%
\hat{S}_{2}\left( \hat{X}^{\prime }\right) +\hat{S}_{2}^{B}\left( \hat{X}%
^{\prime }\right) }  \notag \\
&&-S_{1}^{B}\left( X^{\prime },\bar{X}\right) \left\{ \left( f_{1}^{\prime
}\left( X^{\prime }\right) -\bar{r}\right) \frac{1-\left( S\left( X^{\prime
}\right) +\left( S_{1}^{B}\left( X^{\prime }\right) +S_{2}^{B}\left(
X^{\prime }\right) \right) \right) }{1-S_{1}\left( X^{\prime }\right)
-S_{1}^{B}\left( X^{\prime }\right) }+\Delta F_{\tau }\left( \bar{R}\left(
K,X\right) \right) \right\}  \notag
\end{eqnarray}%
fr bnk.

\section*{Appendix 20 minimization equations}

To compute the saddle point equation we need to derive the functional
derivative of the action functional of the full system for investors and
banks with respect to field of investors and field of bank:%
\begin{eqnarray*}
&&-\hat{\Psi}^{\dag }\left( \hat{K},\hat{X}\right) \nabla ^{2}\hat{\Psi}%
\left( \hat{K},\hat{X}\right) +\left( \frac{\hat{g}^{2}\left( \hat{K},\hat{X}%
\right) }{2\sigma _{\hat{K}}^{2}}+\frac{\hat{g}\left( \hat{K},\hat{X}\right) 
}{2\hat{K}}\right) \left\vert \hat{\Psi}\left( \hat{K},\hat{X}\right)
\right\vert ^{2}+\frac{1}{2\hat{\mu}}\left( \left\vert \hat{\Psi}\left( \hat{%
K},\hat{X}\right) \right\vert ^{2}-\left\vert \hat{\Psi}_{0}\left( \hat{X}%
\right) \right\vert ^{2}\right) ^{2} \\
&&-\bar{\Psi}^{\dag }\left( \bar{K},\bar{X}\right) \nabla ^{2}\bar{\Psi}%
\left( \bar{K},\bar{X}\right) +\left( \frac{\bar{g}^{2}\left( \bar{K},\bar{X}%
\right) }{2\sigma _{\bar{K}}^{2}}+\frac{\bar{g}\left( \bar{K},\bar{X}\right) 
}{2\bar{K}}\right) \left\vert \bar{\Psi}\left( \bar{K},\bar{X}\right)
\right\vert ^{2}+\frac{1}{2\hat{\mu}}\left( \left\vert \bar{\Psi}\left( \bar{%
K}_{1},\bar{X}_{1}\right) \right\vert ^{2}-\left\vert \bar{\Psi}_{0}\left( 
\bar{X}_{1}\right) \right\vert ^{2}\right) ^{2}
\end{eqnarray*}

\subsection*{A20.1 Investors minimization equations}

The derivative of the investors field functional with respect to $\hat{\Psi}%
\left( \hat{K}_{1},\hat{X}_{1}\right) $ are:%
\begin{eqnarray*}
0 &=&\frac{\hat{K}_{1}^{2}\hat{g}^{2}\left( \hat{K}_{1},\hat{X}_{1}\right) }{%
2\sigma _{\hat{K}}^{2}}+\frac{\hat{g}\left( \hat{K}_{1},\hat{X}_{1}\right) }{%
2} \\
&&+\int \left\vert \hat{\Psi}\left( \hat{K},\hat{X}\right) \right\vert
^{2}\left( \frac{\hat{K}^{2}\hat{g}\left( \hat{K},\hat{X},\Psi ,\hat{\Psi}%
\right) }{\sigma _{\hat{K}}^{2}}+\frac{1}{2}\right) \frac{\delta \hat{g}%
\left( \hat{K},\hat{X}\right) }{\delta \left\vert \hat{\Psi}\left( \hat{K}%
_{1},\hat{X}_{1}\right) \right\vert ^{2}} \\
&&+\int \left\vert \bar{\Psi}\left( \bar{K},\bar{X}\right) \right\vert
^{2}\left( \frac{\hat{K}^{2}\bar{g}\left( \bar{K},\bar{X}\right) }{\sigma _{%
\hat{K}}^{2}}+\frac{1}{2}\right) \frac{\delta \bar{g}\left( \bar{K},\bar{X}%
\right) }{\delta \left\vert \hat{\Psi}\left( \hat{K}_{1},\hat{X}_{1}\right)
\right\vert ^{2}}+\frac{1}{\hat{\mu}}\left( \left\vert \hat{\Psi}\left( \hat{%
K},\hat{X}\right) \right\vert ^{2}-\left\vert \hat{\Psi}_{0}\left( \hat{X}%
\right) \right\vert ^{2}\right)
\end{eqnarray*}%
The derivatives are estimatd in the next appendx and w fnd:%
\begin{eqnarray*}
&&\frac{\delta }{\delta \left\vert \hat{\Psi}\left( \hat{K},\hat{X}\right)
\right\vert ^{2}}\bar{g}\left( \hat{K}^{\prime },\hat{X}^{\prime }\right) \\
&\simeq &\frac{\left( \left( \hat{k}_{1}^{B}\left( \hat{X},\left\langle \bar{%
X}\right\rangle \right) +\left\langle \underline{\hat{k}}_{1}^{B}\right%
\rangle \frac{\hat{k}_{1}\left( \hat{X}_{1},\left\langle \hat{X}%
\right\rangle \right) }{1-\left\langle \hat{k}_{1}\right\rangle }\right)
A-\left\langle \hat{k}_{1}^{B}\left( \hat{X},\left\langle \bar{X}%
\right\rangle \right) +\left\langle \underline{\hat{k}}_{1}^{B}\right\rangle 
\frac{\hat{k}_{1}\left( \hat{X}_{1},\left\langle \hat{X}\right\rangle
\right) }{1-\left\langle \hat{k}_{1}\right\rangle }\right\rangle
\left\langle A\right\rangle \right) }{\left( 1-\left\langle \bar{k}\left( 
\bar{X}^{\prime },\bar{X}\right) \right\rangle \right) ^{2}}\frac{\bar{K}}{%
\left\Vert \hat{\Psi}\right\Vert ^{2}\left\langle \hat{K}\right\rangle } \\
&=&\Delta \left( \hat{k}^{B}\left( \hat{X},\left\langle \bar{X}\right\rangle
\right) A\right) \frac{\bar{K}}{\left\Vert \hat{\Psi}\right\Vert
^{2}\left\langle \hat{K}\right\rangle }
\end{eqnarray*}%
so that:%
\begin{equation*}
\left\langle \frac{\delta }{\delta \left\vert \hat{\Psi}\left( \hat{K},\hat{X%
}\right) \right\vert ^{2}}\bar{g}\left( \hat{K}^{\prime },\hat{X}^{\prime
}\right) \right\rangle =0
\end{equation*}%
nd:%
\begin{eqnarray*}
&&\frac{\delta }{\delta \left\vert \hat{\Psi}\left( \hat{K},\hat{X}\right)
\right\vert ^{2}}\hat{g}\left( \hat{K},\hat{X}\right) \\
&\simeq &-\frac{\left( \kappa \left\langle \left[ \frac{\underline{\hat{k}}%
_{2}^{B}}{1+\bar{k}}\right] \right\rangle \left( 1-\left\langle \hat{k}%
\right\rangle \right) +\left\langle \hat{k}_{1}^{B}\right\rangle
\left\langle \hat{k}_{2}\right\rangle \right) \left( \left\langle \hat{g}%
\left( \hat{K},\hat{X}\right) \right\rangle +\frac{1}{1-\left\langle 
\underline{\hat{k}}\right\rangle }\bar{N}\left\langle \bar{g}\left( \hat{K}%
^{\prime },\hat{X}^{\prime }\right) \right\rangle \right) \frac{\left\Vert 
\bar{\Psi}\right\Vert ^{2}\left\langle \bar{K}\right\rangle }{\left\Vert 
\hat{\Psi}\right\Vert ^{2}\left\langle \hat{K}\right\rangle }\frac{\hat{K}%
_{1}}{\left\langle \hat{K}\right\rangle }}{\left( 1-\left( \left\langle \hat{%
k}_{1}\right\rangle +\left\langle \hat{k}_{1}^{B}\right\rangle \frac{%
\left\Vert \bar{\Psi}\right\Vert ^{2}\left\langle \bar{K}\right\rangle }{%
\left\Vert \hat{\Psi}\right\Vert ^{2}\left\langle \hat{K}\right\rangle }%
\right) \right) \left( 1-\left( \left\langle \underline{\hat{k}}%
\right\rangle +\left( \left\langle \underline{\hat{k}}_{1}^{B}\right\rangle
+\kappa \left\langle \left[ \frac{\underline{\hat{k}}_{2}^{B}}{1+\bar{k}}%
\right] \right\rangle \right) \frac{\left\Vert \bar{\Psi}\right\Vert
^{2}\left\langle \bar{K}\right\rangle }{\left\Vert \hat{\Psi}\right\Vert
^{2}\left\langle \hat{K}\right\rangle }\right) \right) \left\Vert \hat{\Psi}%
\right\Vert ^{2}}
\end{eqnarray*}%
wth:%
\begin{equation*}
\bar{N}\rightarrow \left\langle \hat{k}_{1}^{B}\right\rangle +\kappa \frac{%
\left\langle \hat{k}_{2}^{B}\right\rangle }{1+\left\langle \bar{k}%
\right\rangle }\left( 1-\frac{\left\langle \bar{k}\right\rangle }{\left(
1+\left\langle \bar{k}\right\rangle \right) ^{2}}\right)
\end{equation*}%
and the equation reduces to:%
\begin{eqnarray*}
0 &=&\frac{\hat{K}_{1}^{2}\hat{g}^{2}\left( \hat{K}_{1},\hat{X}_{1}\right) }{%
2\sigma _{\hat{K}}^{2}}+\frac{\hat{g}\left( \hat{K}_{1},\hat{X}_{1}\right) }{%
2} \\
&&+\int \left\vert \hat{\Psi}\left( \hat{K},\hat{X}\right) \right\vert
^{2}\left( \frac{\hat{K}^{2}\hat{g}\left( \hat{K},\hat{X},\Psi ,\hat{\Psi}%
\right) }{\sigma _{\hat{K}}^{2}}+\frac{1}{2}\right) \frac{\delta \hat{g}%
\left( \hat{K},\hat{X}\right) }{\delta \left\vert \hat{\Psi}\left( \hat{K}%
_{1},\hat{X}_{1}\right) \right\vert ^{2}}+\frac{1}{\hat{\mu}}\left(
\left\vert \hat{\Psi}\left( \hat{K},\hat{X}\right) \right\vert
^{2}-\left\vert \hat{\Psi}_{0}\left( \hat{X}\right) \right\vert ^{2}\right)
\end{eqnarray*}

\subsection*{A20.2 Banks minimization equations}

The minimzation with respect to:%
\begin{eqnarray*}
0 &=&\left( \frac{\bar{K}_{1}^{2}\bar{g}^{2}\left( \bar{K}_{1},\bar{X}%
_{1}\right) }{\sigma _{\hat{K}}^{2}}+\frac{\bar{g}\left( \bar{K}_{1},\bar{X}%
_{1}\right) }{2}\right) \\
&&+\int \left\vert \bar{\Psi}\left( \bar{K},\bar{X}\right) \right\vert
^{2}\left( \frac{\hat{K}^{2}\bar{g}\left( \bar{K},\bar{X}\right) }{\sigma _{%
\hat{K}}^{2}}+\frac{1}{2}\right) \frac{\delta \bar{g}\left( \bar{K},\bar{X}%
\right) }{\delta \left\vert \bar{\Psi}\left( \bar{K}_{1},\bar{X}_{1}\right)
\right\vert ^{2}} \\
&&+\int \left\vert \hat{\Psi}\left( \hat{K},\hat{X}\right) \right\vert
^{2}\left( \frac{\hat{K}^{2}\hat{g}\left( \hat{K},\hat{X}\right) }{\sigma _{%
\hat{K}}^{2}}+\frac{1}{2}\right) \frac{\delta \hat{g}\left( \hat{K},\hat{X}%
\right) }{\delta \left\vert \bar{\Psi}\left( \bar{K}_{1},\bar{X}_{1}\right)
\right\vert ^{2}}+\frac{1}{\hat{\mu}}\left( \left\vert \bar{\Psi}\left( \bar{%
K}_{1},\bar{X}_{1}\right) \right\vert ^{2}-\left\vert \bar{\Psi}_{0}\left( 
\bar{X}_{1}\right) \right\vert ^{2}\right)
\end{eqnarray*}%
with:%
\begin{eqnarray*}
&&\frac{\delta }{\delta \left\vert \bar{\Psi}\left( \bar{K},\bar{X}\right)
\right\vert ^{2}}\hat{g}\left( \hat{K},\hat{X}\right) \\
&\rightarrow &-\frac{\kappa \left\langle \left[ \frac{\underline{\hat{k}}%
_{2}^{B}}{1+\bar{k}}\right] \right\rangle \left( 1-\left\langle \hat{k}%
\left( \hat{X}^{\prime },\hat{X}\right) \right\rangle \right) +\left\langle 
\hat{k}_{1}^{B}\right\rangle \left\langle \hat{k}_{2}\right\rangle }{\left(
1-\left( \left\langle \hat{k}\right\rangle +\left( \left\langle \hat{k}%
_{1}^{B}\right\rangle +\kappa \left\langle \left[ \frac{\underline{\hat{k}}%
_{2}^{B}}{1+\bar{k}}\right] \right\rangle \right) \frac{\left\Vert \bar{\Psi}%
\right\Vert ^{2}\left\langle \bar{K}\right\rangle }{\left\Vert \hat{\Psi}%
\right\Vert ^{2}\left\langle \hat{K}\right\rangle }\right) \right) \left(
1-\left( \left\langle \hat{k}_{1}\right\rangle +\left\langle \hat{k}%
_{1}^{B}\right\rangle \frac{\left\Vert \bar{\Psi}\right\Vert
^{2}\left\langle \bar{K}\right\rangle }{\left\Vert \hat{\Psi}\right\Vert
^{2}\left\langle \hat{K}\right\rangle }\right) \right) \left\Vert \hat{\Psi}%
\right\Vert ^{2}}\frac{\bar{K}}{\left\langle \hat{K}\right\rangle } \\
&&\times \left( \left\langle \hat{g}\left( \hat{K},\hat{X}\right)
\right\rangle +\left( 1-\hat{M}\right) ^{-1}\bar{N}\left\langle \bar{g}%
\left( \hat{K}^{\prime },\hat{X}^{\prime }\right) \right\rangle \right)
\end{eqnarray*}%
\begin{eqnarray}
&&\frac{\delta }{\delta \left\vert \bar{\Psi}\left( \bar{K},\bar{X}\right)
\right\vert ^{2}}\bar{g}\left( \hat{K}^{\prime },\hat{X}^{\prime }\right) \\
&\simeq &\left( \frac{\left( 1-\left\langle \bar{k}\right\rangle \right) ^{2}%
}{\left\Vert \bar{\Psi}\right\Vert ^{2}\left\langle \bar{K}\right\rangle }%
\right) ^{-1}\left\{ \left( \left\langle \underline{\hat{k}}%
_{1}^{B}\right\rangle +\frac{\left\langle \underline{\hat{k}}%
_{1}^{B}\right\rangle }{\left( 1-\left\langle \hat{k}_{1}\right\rangle
\right) }\right) \right.  \notag \\
&&\left. \times \left( \left( \hat{k}_{1}^{B}\left( \hat{X}^{\prime },\bar{X}%
\right) -\left\langle \hat{k}_{1}^{B}\right\rangle \right) +\kappa \frac{%
\left( \hat{k}_{2}^{B}\left( \hat{X}^{\prime },\bar{X}\right) -\left\langle 
\hat{k}_{2}^{B}\right\rangle \right) }{1+\left\langle \bar{k}\left( \bar{X}%
^{\prime },\bar{X}^{\prime \prime }\right) \right\rangle }\right)
\left\langle A\right\rangle +\frac{\left( 1-\left\langle \bar{k}%
\right\rangle \right) ^{2}}{1-\left\langle \bar{k}_{1}\right\rangle }\frac{%
\left( \bar{k}_{2}\left( \left\langle \bar{X}\right\rangle ,\bar{X}\right)
-\left\langle \bar{k}_{2}\right\rangle \right) }{\left\Vert \bar{\Psi}%
\right\Vert ^{2}\left\langle \bar{K}\right\rangle }\left\langle \bar{g}%
\right\rangle \right\} \frac{\bar{K}}{\left\Vert \hat{\Psi}\right\Vert
^{2}\left\langle \hat{K}\right\rangle }  \notag \\
&=&\left( \left( \left\langle \underline{\hat{k}}_{1}^{B}\right\rangle +%
\frac{\left\langle \underline{\hat{k}}_{1}^{B}\right\rangle }{\left(
1-\left\langle \hat{k}_{1}\right\rangle \right) }\right) \Delta \left( \hat{k%
}^{B}\left( \hat{X},\left\langle \bar{X}\right\rangle \right) A\right) +%
\frac{\Delta \bar{k}_{2}\left( \left\langle \bar{X}\right\rangle ,\bar{X}%
\right) }{\left( 1-\left\langle \bar{k}_{1}\right\rangle \right) \left\Vert 
\bar{\Psi}\right\Vert ^{2}\left\langle \bar{K}\right\rangle }\left\langle 
\bar{g}\right\rangle \right) \frac{\bar{K}}{\left\Vert \hat{\Psi}\right\Vert
^{2}\left\langle \hat{K}\right\rangle }
\end{eqnarray}%
whr:%
\begin{eqnarray*}
\left\langle A\right\rangle &=&\left\langle \frac{\left( 1-\hat{M}\right) 
\hat{g}\left( \hat{K}^{\prime },\hat{X}^{\prime }\right) +\bar{N}\bar{g}%
\left( \hat{K}^{\prime },\hat{X}^{\prime }\right) }{1+\underline{\hat{k}}%
_{2}\left( \bar{X}^{\prime }\right) +\kappa \left[ \frac{\underline{\hat{k}}%
_{2}^{B}}{1+\bar{k}}\right] \left( \hat{X}^{\prime }\right) }\right\rangle \\
&=&\frac{1-\left( \left\langle \underline{\hat{k}}\right\rangle
+\left\langle \underline{\hat{k}}_{1}^{B}\right\rangle \frac{\left\Vert \bar{%
\Psi}\right\Vert ^{2}\left\langle \bar{K}\right\rangle }{\left\Vert \hat{\Psi%
}\right\Vert ^{2}\left\langle \hat{K}\right\rangle }+\kappa \left\langle %
\left[ \frac{\underline{\hat{k}}_{2}^{B}}{1+\bar{k}}\right] \right\rangle 
\frac{\left\Vert \bar{\Psi}\right\Vert ^{2}\left\langle \bar{K}\right\rangle 
}{\left\Vert \hat{\Psi}\right\Vert ^{2}\left\langle \hat{K}\right\rangle }%
\right) }{\left( 1-\left( \left\langle \hat{k}_{1}\right\rangle
+\left\langle \hat{k}_{1}^{B}\right\rangle \frac{\left\Vert \bar{\Psi}%
\right\Vert ^{2}\left\langle \bar{K}\right\rangle }{\left\Vert \hat{\Psi}%
\right\Vert ^{2}\left\langle \hat{K}\right\rangle }\right) \right)
\left\Vert \hat{\Psi}\right\Vert ^{2}\left\langle \hat{K}\right\rangle } \\
&&\times \left( \left( 1-\left\langle \hat{k}\right\rangle \right)
\left\langle \hat{g}\right\rangle +\left( \left\langle \hat{k}%
_{1}^{B}\right\rangle +\kappa \frac{\left\langle \hat{k}_{2}^{B}\right%
\rangle }{1+\left\langle \bar{k}\right\rangle }\left( 1-\frac{\left\langle 
\bar{k}\right\rangle }{\left( 1+\left\langle \bar{k}\right\rangle \right)
^{2}}\right) \right) \left\langle \bar{g}\right\rangle \right)
\end{eqnarray*}%
as a consequence, in averag:%
\begin{equation*}
\left\langle \frac{\delta }{\delta \left\vert \bar{\Psi}\left( \bar{K},\bar{X%
}\right) \right\vert ^{2}}\bar{g}\left( \hat{K}^{\prime },\hat{X}^{\prime
}\right) \right\rangle =0
\end{equation*}%
so tht the equation reduces to:%
\begin{eqnarray*}
0 &=&\left( \frac{\bar{K}_{1}^{2}\bar{g}^{2}\left( \bar{K}_{1},\bar{X}%
_{1}\right) }{\sigma _{\hat{K}}^{2}}+\frac{\bar{g}\left( \bar{K}_{1},\bar{X}%
_{1}\right) }{2}\right) \\
&&+\int \left\vert \hat{\Psi}\left( \hat{K},\hat{X}\right) \right\vert
^{2}\left( \frac{\hat{K}^{2}\hat{g}\left( \hat{K},\hat{X}\right) }{\sigma _{%
\hat{K}}^{2}}+\frac{1}{2}\right) \frac{\delta \hat{g}\left( \hat{K},\hat{X}%
\right) }{\delta \left\vert \bar{\Psi}\left( \bar{K}_{1},\bar{X}_{1}\right)
\right\vert ^{2}}+\frac{1}{\hat{\mu}}\left( \left\vert \bar{\Psi}\left( \bar{%
K}_{1},\bar{X}_{1}\right) \right\vert ^{2}-\left\vert \bar{\Psi}_{0}\left( 
\bar{X}_{1}\right) \right\vert ^{2}\right)
\end{eqnarray*}

\subsection*{A20.3 Solving saddle point for investors and banks}

\subsubsection*{A20.3.1 Expression for the field}

Solving for $\left\vert \hat{\Psi}\left( \hat{K}_{1},\hat{X}_{1}\right)
\right\vert ^{2}$ and $\left\vert \bar{\Psi}\left( \bar{K}_{1},\bar{X}%
_{1}\right) \right\vert ^{2}$ yields directly:%
\begin{equation}
\left\vert \hat{\Psi}\left( \hat{K}_{1},\hat{X}_{1}\right) \right\vert
^{2}=\left\Vert \hat{\Psi}_{0}\left( \hat{X}_{1}\right) \right\Vert ^{2}-%
\hat{\mu}\left\{ \left( \frac{\hat{K}_{1}^{2}\hat{g}^{2}\left( \hat{X}%
_{1}\right) }{2\sigma _{\hat{K}}^{2}}+\frac{\hat{g}\left( \hat{X}_{1}\right) 
}{2}\right) -\left( \frac{\left\langle \hat{K}\right\rangle ^{2}\left\langle 
\hat{g}\right\rangle }{\sigma _{\hat{K}}^{2}}+\frac{1}{2}\right)
\left\langle \hat{g}^{ef}\right\rangle \frac{\hat{K}_{1}}{\left\langle \hat{K%
}\right\rangle }\right\}  \label{FDN}
\end{equation}%
\begin{equation}
\left\vert \bar{\Psi}\left( \bar{K}_{1},\bar{X}_{1}\right) \right\vert
^{2}=\left\vert \bar{\Psi}_{0}\left( \bar{X}_{1}\right) \right\vert ^{2}-%
\hat{\mu}\left\{ \left( \frac{\bar{K}_{1}^{2}\bar{g}^{2}\left( \bar{X}%
_{1}\right) }{\sigma _{\hat{K}}^{2}}+\frac{\bar{g}\left( \bar{X}_{1}\right) 
}{2}\right) -\left( \frac{\left\langle \hat{K}\right\rangle ^{2}\left\langle 
\hat{g}\right\rangle }{\sigma _{\hat{K}}^{2}}+\frac{1}{2}\right)
\left\langle \hat{g}^{Bef}\right\rangle \frac{\bar{K}_{1}}{\left\langle \hat{%
K}\right\rangle }\right\}  \label{FDT}
\end{equation}%
and these formula will be used to compute average capital per sector.

\subsubsection*{A20.3.2 Finding the maximal capital}

the maximal value for $\hat{K}_{0}$ and $\hat{K}_{0}$ is found by setting $%
\left\vert \hat{\Psi}\left( \hat{K}_{1},\hat{X}_{1}\right) \right\vert
^{2}=0 $.%
\begin{equation}
0=\left\Vert \hat{\Psi}_{0}\left( \hat{X}_{1}\right) \right\Vert ^{2}-\hat{%
\mu}\left\{ \left( \frac{\hat{K}_{0}^{2}\hat{g}^{2}\left( \hat{X}_{1}\right) 
}{2\sigma _{\hat{K}}^{2}}+\frac{\hat{g}\left( \hat{X}_{1}\right) }{2}\right)
-\left( \frac{\left\langle \hat{K}\right\rangle ^{2}\left\langle \hat{g}%
\right\rangle }{\sigma _{\hat{K}}^{2}}+\frac{1}{2}\right) \left\langle \hat{g%
}^{ef}\right\rangle \frac{\hat{K}_{0}}{\left\langle \hat{K}\right\rangle }%
\right\}
\end{equation}%
\begin{equation*}
0=\left\Vert \bar{\Psi}_{0}\left( \bar{X}_{1}\right) \right\Vert ^{2}-\hat{%
\mu}\left\{ \left( \frac{\bar{K}_{0}^{2}\bar{g}^{2}\left( \bar{X}_{1}\right) 
}{\sigma _{\hat{K}}^{2}}+\frac{\bar{g}\left( \bar{X}_{1}\right) }{2}\right)
-\left( \frac{\left\langle \hat{K}\right\rangle ^{2}\left\langle \hat{g}%
\right\rangle }{\sigma _{\hat{K}}^{2}}+\frac{1}{2}\right) \left\langle \hat{g%
}^{Bef}\right\rangle \frac{\bar{K}_{0}}{\left\langle \hat{K}\right\rangle }%
\right\}
\end{equation*}%
and we find:%
\begin{eqnarray}
\hat{K}_{0}^{2} &\simeq &\frac{2\sigma _{\hat{K}}^{2}}{\hat{g}^{2}\left( 
\hat{X}_{1}\right) }\left( \frac{\left\Vert \hat{\Psi}_{0}\left( \hat{X}%
_{1}\right) \right\Vert ^{2}}{\hat{\mu}}-\left( \frac{\hat{g}\left( \hat{X}%
_{1}\right) }{2}+\left( \frac{\left\langle \hat{K}\right\rangle
^{2}\left\langle \hat{g}\right\rangle }{\sigma _{\hat{K}}^{2}}+\frac{1}{2}%
\right) \left\langle \hat{g}^{ef}\right\rangle \frac{\left\langle \hat{K}%
_{0}\right\rangle }{\left\langle \hat{K}\right\rangle }\right) \right) \\
&\simeq &2\frac{\sigma _{\hat{K}}^{2}}{\hat{g}^{2}\left( \hat{X}_{1}\right) }%
\left( \frac{\left\Vert \hat{\Psi}_{0}\left( \hat{X}_{1}\right) \right\Vert
^{2}}{\hat{\mu}}-\left( \frac{\left\langle \hat{K}\right\rangle
^{2}\left\langle \hat{g}\right\rangle }{\sigma _{\hat{K}}^{2}}+\frac{1}{2}%
\right) \left\langle \hat{g}^{ef}\right\rangle \frac{\left\langle \hat{K}%
_{0}\right\rangle }{\left\langle \hat{K}\right\rangle }\right)  \notag
\end{eqnarray}%
and:%
\begin{equation*}
\bar{K}_{0}^{2}\simeq 2\frac{\sigma _{\hat{K}}^{2}}{\bar{g}^{2}\left( \hat{X}%
_{1}\right) }\left( \frac{\left\Vert \bar{\Psi}_{0}\left( \bar{X}_{1}\right)
\right\Vert ^{2}}{\hat{\mu}}-\left( \frac{\left\langle \hat{K}\right\rangle
^{2}\left\langle \hat{g}\right\rangle }{\sigma _{\hat{K}}^{2}}+\frac{1}{2}%
\right) \left\langle \hat{g}^{Bef}\right\rangle \frac{\left\langle \bar{K}%
_{0}\right\rangle }{\left\langle \hat{K}\right\rangle }\right)
\end{equation*}%
In first approximation, we obtain:%
\begin{equation}
\left\langle \hat{K}_{0}\right\rangle ^{2}\simeq 2\frac{\sigma _{\hat{K}}^{2}%
}{\left\langle \hat{g}\right\rangle ^{2}}\left( \frac{\left\Vert \hat{\Psi}%
_{0}\right\Vert ^{2}}{\hat{\mu}}-\left( \frac{\left\langle \hat{K}%
\right\rangle ^{2}\left\langle \hat{g}\right\rangle }{\sigma _{\hat{K}}^{2}}+%
\frac{1}{2}\right) \left\langle \hat{g}^{ef}\right\rangle \frac{\left\langle 
\hat{K}_{0}\right\rangle }{\left\langle \hat{K}\right\rangle }\right)
\end{equation}%
or:%
\begin{equation*}
\left\langle \hat{K}_{0}\right\rangle ^{2}\simeq 2\frac{\sigma _{\hat{K}}^{2}%
}{\left\langle \hat{g}\right\rangle ^{2}}\left( \frac{\left\Vert \hat{\Psi}%
_{0}\right\Vert ^{2}}{\hat{\mu}}-\left( \frac{\left\langle \hat{K}%
\right\rangle ^{2}\left\langle \hat{g}\right\rangle }{\sigma _{\hat{K}}^{2}}%
\right) \left\langle \hat{g}^{ef}\right\rangle \frac{\left\langle \hat{K}%
_{0}\right\rangle }{\left\langle \hat{K}\right\rangle }\right)
\end{equation*}%
and:%
\begin{equation*}
\left\langle \bar{K}_{0}\right\rangle ^{2}\simeq 2\frac{\sigma _{\hat{K}}^{2}%
}{\left\langle \bar{g}\right\rangle ^{2}}\left( \frac{\left\Vert \bar{\Psi}%
_{0}\right\Vert ^{2}}{\hat{\mu}}-\left( \frac{\left\langle \hat{K}%
\right\rangle ^{2}\left\langle \hat{g}\right\rangle }{\sigma _{\hat{K}}^{2}}%
\right) \left\langle \hat{g}^{Bef}\right\rangle \frac{\left\langle \bar{K}%
_{0}\right\rangle }{\left\langle \hat{K}\right\rangle }\right)
\end{equation*}

\subsection*{A20.4 Fields per sector and average capital per sector}

\subsubsection*{A20.4.1 Fields expression}

Integrating the expressions (\ref{FDN}) and (\ref{FDT}) for the field yields
the number of investors in one sector:%
\begin{eqnarray}
\left\Vert \hat{\Psi}\left( \hat{X}_{1}\right) \right\Vert ^{2} &=&\hat{K}%
_{0}\left\Vert \hat{\Psi}_{0}\left( \hat{X}_{1}\right) \right\Vert ^{2}-\hat{%
\mu}\left\{ \left( \frac{\hat{K}_{0}^{3}\hat{g}^{2}\left( \hat{X}_{1}\right) 
}{6\sigma _{\hat{K}}^{2}}+\frac{\hat{K}_{0}\hat{g}\left( \hat{X}_{1}\right) 
}{2}\right) \right.  \label{PSN} \\
&&-\left. \hat{K}_{0}\left( \frac{\left\langle \hat{K}\right\rangle
^{2}\left\langle \hat{g}\right\rangle }{\sigma _{\hat{K}}^{2}}+\frac{1}{2}%
\right) \left\langle \hat{g}^{ef}\right\rangle \frac{\hat{K}_{0}}{%
2\left\langle \hat{K}\right\rangle }\right\}  \notag
\end{eqnarray}%
using that $\hat{K}_{0}^{2}$ satisfies (\ref{KR}) we obtain:%
\begin{eqnarray}
\left\Vert \hat{\Psi}\left( \hat{X}_{1}\right) \right\Vert ^{2} &=&\hat{\mu}%
\hat{K}_{0}^{3}\frac{\hat{g}^{2}\left( \hat{X}_{1}\right) }{3\sigma _{\hat{K}%
}^{2}}-\hat{\mu}\frac{\hat{K}_{0}^{2}}{2\left\langle \hat{K}\right\rangle }%
\left( \frac{\left\langle \hat{K}\right\rangle ^{2}\left\langle \hat{g}%
\right\rangle }{\sigma _{\hat{K}}^{2}}+\frac{1}{2}\right) \left\langle \hat{g%
}^{ef}\right\rangle  \label{PSD} \\
&\simeq &\hat{\mu}\frac{\hat{K}_{0}^{3}}{\sigma _{\hat{K}}^{2}}\left( \frac{%
\hat{g}^{2}\left( \hat{X}_{1}\right) }{3}-\frac{\left\langle \hat{K}%
\right\rangle }{2\hat{K}_{0}}\left\langle \hat{g}\right\rangle \left\langle 
\hat{g}^{ef}\right\rangle \right)  \notag
\end{eqnarray}%
Similarly, the value of $\left\Vert \bar{\Psi}\left( \bar{X}_{1}\right)
\right\Vert ^{2}$ is given by 
\begin{equation*}
\left\Vert \bar{\Psi}\left( \bar{X}_{1}\right) \right\Vert ^{2}\simeq \hat{%
\mu}\frac{\bar{K}_{0}^{3}}{\sigma _{\hat{K}}^{2}}\left( \frac{\bar{g}%
^{2}\left( \bar{X}_{1}\right) }{3}-\frac{\left\langle \hat{K}\right\rangle }{%
2\bar{K}_{0}}\left\langle \hat{g}\right\rangle \left\langle \hat{g}%
^{Bef}\right\rangle \right)
\end{equation*}

\subsubsection*{A20.4.2 Capital per sector}

Multiplying (\ref{FDN}) and (\ref{FDT}) by $\hat{K}$ and $\bar{K}$
respectively and integrating leads:%
\begin{eqnarray*}
\hat{K}_{\hat{X}}\left\Vert \hat{\Psi}\left( \hat{X}_{1}\right) \right\Vert
^{2} &=&\frac{\hat{K}_{0}^{2}}{2}\left\Vert \hat{\Psi}_{0}\left( \hat{X}%
_{1}\right) \right\Vert ^{2}-\hat{\mu}\hat{K}_{0}\left\{ \left( \frac{\hat{K}%
_{0}^{3}\hat{g}\left( \hat{X}_{1}\right) }{8\sigma _{\hat{K}}^{2}}+\frac{%
\hat{K}_{0}\hat{g}\left( \hat{X}_{1}\right) }{4}\right) \right. \\
&&-\left. \frac{\hat{K}_{0}^{2}}{2}\left( \frac{\left\langle \hat{K}%
\right\rangle ^{2}\left\langle \hat{g}\right\rangle }{\sigma _{\hat{K}}^{2}}+%
\frac{1}{2}\right) \left( \frac{2\hat{K}_{0}}{3\left\langle \hat{K}%
\right\rangle }\left\langle \hat{g}^{ef}\right\rangle \right) \right\}
\end{eqnarray*}%
and since $\hat{K}_{0}$ satitsfies (\ref{KR}), we find:%
\begin{eqnarray}
\hat{K}_{\hat{X}}\left\Vert \hat{\Psi}\left( \hat{X}_{1}\right) \right\Vert
^{2} &=&\hat{\mu}\frac{\hat{K}_{0}^{4}\hat{g}^{2}\left( \hat{X}_{1}\right) }{%
8\sigma _{\hat{K}}^{2}}-\hat{\mu}\frac{\hat{K}_{0}^{2}}{6\left\langle \hat{K}%
\right\rangle }\left( \frac{\left\langle \hat{K}\right\rangle
^{2}\left\langle \hat{g}\right\rangle }{\sigma _{\hat{K}}^{2}}+\frac{1}{2}%
\right) \left\langle \hat{g}^{ef}\right\rangle \left\langle \hat{g}%
\right\rangle  \label{KM} \\
&\simeq &\hat{\mu}\frac{\hat{K}_{0}^{4}}{2\sigma _{\hat{K}}^{2}}\left( \frac{%
\hat{g}^{2}\left( \hat{X}_{1}\right) }{4}-\frac{\left\langle \hat{K}%
\right\rangle \left\langle \hat{g}\right\rangle }{3\left\langle \hat{K}%
_{0}\right\rangle }\left\langle \hat{g}^{ef}\right\rangle \right)  \notag
\end{eqnarray}%
we also have:%
\begin{equation}
\bar{K}_{\bar{X}}\left\Vert \bar{\Psi}\left( \bar{X}_{1}\right) \right\Vert
^{2}\simeq \hat{\mu}\frac{\bar{K}_{0}^{4}}{2\sigma _{\hat{K}}^{2}}\left( 
\frac{\bar{g}^{2}\left( \hat{X}_{1}\right) }{4}-\frac{\left\langle \hat{K}%
\right\rangle \left\langle \hat{g}\right\rangle }{3\left\langle \bar{K}%
_{0}\right\rangle }\left\langle \hat{g}^{Bef}\right\rangle \right)
\label{KB}
\end{equation}

\subsection*{A20.5 Average field and capital}

\subsubsection*{A20.5.1 Averages and ratio $\frac{\left\langle \hat{K}%
\right\rangle }{\left\langle \hat{K}_{0}\right\rangle }$}

avergs for (\ref{PST}) and (\ref{PK}) become:%
\begin{equation}
\left\Vert \hat{\Psi}\right\Vert ^{2}\simeq \hat{\mu}V\frac{\left\langle 
\hat{K}_{0}\right\rangle ^{3}}{\sigma _{\hat{K}}^{2}}\left( \frac{1}{3}-%
\frac{\left\langle \hat{K}\right\rangle }{2\left\langle \hat{K}%
_{0}\right\rangle }\frac{\left\langle \hat{g}^{ef}\right\rangle }{%
\left\langle \hat{g}\right\rangle }\right) \left\langle \hat{g}\right\rangle
^{2}
\end{equation}%
\begin{equation*}
\left\langle \hat{K}\right\rangle \left\Vert \hat{\Psi}\right\Vert ^{2}=\hat{%
\mu}V\frac{\left\langle \hat{K}_{0}\right\rangle ^{4}}{2\sigma _{\hat{K}}^{2}%
}\left( \frac{1}{4}-\frac{\left\langle \hat{K}\right\rangle }{3\left\langle 
\hat{K}_{0}\right\rangle }\frac{\left\langle \hat{g}^{ef}\right\rangle }{%
\left\langle \hat{g}\right\rangle }\right) \left\langle \hat{g}\right\rangle
^{2}
\end{equation*}%
wh $V$ the volume of sectors space. The ratio average capital over maximal
capital is given by:%
\begin{equation*}
\frac{\left\langle \hat{K}\right\rangle }{\left\langle \hat{K}%
_{0}\right\rangle }=\frac{1}{2}\frac{\frac{1}{4}-\frac{\left\langle \hat{K}%
\right\rangle }{3\left\langle \hat{K}_{0}\right\rangle }\frac{\left\langle 
\hat{g}^{ef}\right\rangle }{\left\langle \hat{g}\right\rangle }}{\frac{1}{3}-%
\frac{\left\langle \hat{K}\right\rangle }{2\left\langle \hat{K}%
_{0}\right\rangle }\frac{\left\langle \hat{g}^{ef}\right\rangle }{%
\left\langle \hat{g}\right\rangle }}
\end{equation*}%
with solution%
\begin{equation}
\frac{\left\langle \hat{K}\right\rangle }{\left\langle \hat{K}%
_{0}\right\rangle }=\frac{1}{6\frac{\left\langle \hat{g}^{ef}\right\rangle }{%
\left\langle \hat{g}\right\rangle }}\left( 2+\frac{\left\langle \hat{g}%
^{ef}\right\rangle }{\left\langle \hat{g}\right\rangle }-\sqrt{\left( 1-%
\frac{\left\langle \hat{g}^{ef}\right\rangle }{\left\langle \hat{g}%
\right\rangle }\right) \left( 4-\frac{\left\langle \hat{g}^{ef}\right\rangle 
}{\left\langle \hat{g}\right\rangle }\right) }\right)  \label{RTY}
\end{equation}%
Coming back to the equation for $\left\langle \hat{K}_{0}\right\rangle ^{2}$ 
\begin{equation}
\left\langle \hat{K}_{0}\right\rangle ^{2}\simeq 2\frac{\sigma _{\hat{K}}^{2}%
}{\left\langle \hat{g}\right\rangle ^{2}}\left( \frac{\left\Vert \hat{\Psi}%
_{0}\right\Vert ^{2}}{\hat{\mu}}-\left( \frac{\left\langle \hat{K}%
\right\rangle ^{2}\left\langle \hat{g}\right\rangle }{\sigma _{\hat{K}}^{2}}%
\right) \left\langle \hat{g}^{ef}\right\rangle \frac{\left\langle \hat{K}%
_{0}\right\rangle }{\left\langle \hat{K}\right\rangle }\right)  \label{RTV}
\end{equation}%
allows to find the exprssion for $\left\langle \hat{K}_{0}\right\rangle $.
Actually (\ref{RTV}) wrts: 
\begin{equation*}
1\simeq 2\frac{\sigma _{\hat{K}}^{2}}{\left\langle \hat{g}\right\rangle ^{2}}%
\left( \frac{\left\Vert \hat{\Psi}_{0}\right\Vert ^{2}}{\hat{\mu}%
\left\langle \hat{K}_{0}\right\rangle ^{2}}-\left( \frac{\left\langle \hat{K}%
\right\rangle \left\langle \hat{g}\right\rangle }{\sigma _{\hat{K}%
}^{2}\left\langle \hat{K}_{0}\right\rangle }\right) \left\langle \hat{g}%
^{ef}\right\rangle \right)
\end{equation*}%
that becomes:%
\begin{equation*}
2\frac{\left\langle \hat{K}\right\rangle \left\langle \hat{g}%
^{ef}\right\rangle }{\left\langle \hat{K}_{0}\right\rangle \left\langle \hat{%
g}\right\rangle }=2\frac{\sigma _{\hat{K}}^{2}}{\left\langle \hat{g}%
\right\rangle ^{2}}\frac{\left\Vert \hat{\Psi}_{0}\right\Vert ^{2}}{\hat{\mu}%
\left\langle \hat{K}_{0}\right\rangle ^{2}}-1
\end{equation*}%
and yields an expression for $\frac{\left\langle \hat{K}\right\rangle }{%
\left\langle \hat{K}_{0}\right\rangle }$ that completes (\ref{RTY}): 
\begin{equation}
\frac{\left\langle \hat{K}\right\rangle }{\left\langle \hat{K}%
_{0}\right\rangle }=\frac{1}{2}\frac{\left\langle \hat{g}\right\rangle }{%
\left\langle \hat{g}^{ef}\right\rangle }\left( 2\frac{\sigma _{\hat{K}}^{2}}{%
\left\langle \hat{g}\right\rangle ^{2}}\frac{\left\Vert \hat{\Psi}%
_{0}\right\Vert ^{2}}{\hat{\mu}\left\langle \hat{K}_{0}\right\rangle ^{2}}%
-1\right)  \label{RTZ}
\end{equation}%
Equating (\ref{RTZ}) and (\ref{RTY}) yields:%
\begin{equation*}
\frac{1}{2}\frac{\left\langle \hat{g}\right\rangle }{\left\langle \hat{g}%
^{ef}\right\rangle }\left( 2\frac{\sigma _{\hat{K}}^{2}}{\left\langle \hat{g}%
\right\rangle ^{2}}\frac{\left\Vert \hat{\Psi}_{0}\right\Vert ^{2}}{\hat{\mu}%
\left\langle \hat{K}_{0}\right\rangle ^{2}}-1\right) =\frac{1}{6\frac{%
\left\langle \hat{g}^{ef}\right\rangle }{\left\langle \hat{g}\right\rangle }}%
\left( 2+\frac{\left\langle \hat{g}^{ef}\right\rangle }{\left\langle \hat{g}%
\right\rangle }-\sqrt{\left( 1-\frac{\left\langle \hat{g}^{ef}\right\rangle 
}{\left\langle \hat{g}\right\rangle }\right) \left( 4-\frac{\left\langle 
\hat{g}^{ef}\right\rangle }{\left\langle \hat{g}\right\rangle }\right) }%
\right)
\end{equation*}%
leading ultimately to the expressions for $\left\langle \hat{K}%
_{0}\right\rangle ^{2}$ and $\left\langle \hat{K}\right\rangle ^{2}$: 
\begin{equation}
\left\langle \hat{K}_{0}\right\rangle ^{2}=6\frac{\sigma _{\hat{K}}^{2}}{%
\hat{\mu}}\frac{\left\Vert \hat{\Psi}_{0}\right\Vert ^{2}}{\left\langle \hat{%
g}\right\rangle ^{2}\left( 5+\frac{\left\langle \hat{g}^{ef}\right\rangle }{%
\left\langle \hat{g}\right\rangle }-\sqrt{\left( 1-\frac{\left\langle \hat{g}%
^{ef}\right\rangle }{\left\langle \hat{g}\right\rangle }\right) \left( 4-%
\frac{\left\langle \hat{g}^{ef}\right\rangle }{\left\langle \hat{g}%
\right\rangle }\right) }\right) }  \label{KCM}
\end{equation}%
\begin{eqnarray}
\left\langle \hat{K}\right\rangle ^{2} &=&\left( \frac{\left\langle \hat{g}%
\right\rangle }{6\left\langle \hat{g}^{ef}\right\rangle }\left( 2+\frac{%
\left\langle \hat{g}^{ef}\right\rangle }{\left\langle \hat{g}\right\rangle }-%
\sqrt{\left( 1-\frac{\left\langle \hat{g}^{ef}\right\rangle }{\left\langle 
\hat{g}\right\rangle }\right) \left( 4-\frac{\left\langle \hat{g}%
^{ef}\right\rangle }{\left\langle \hat{g}\right\rangle }\right) }\right)
\right) ^{2}\left\langle \hat{K}_{0}\right\rangle ^{2}  \label{KCR} \\
&\rightarrow &\frac{1}{6}\frac{\sigma _{\hat{K}}^{2}}{\hat{\mu}}\frac{\left(
\left( 2+\frac{\left\langle \hat{g}^{ef}\right\rangle }{\left\langle \hat{g}%
\right\rangle }-\sqrt{\left( 1-\frac{\left\langle \hat{g}^{ef}\right\rangle 
}{\left\langle \hat{g}\right\rangle }\right) \left( 4-\frac{\left\langle 
\hat{g}^{ef}\right\rangle }{\left\langle \hat{g}\right\rangle }\right) }%
\right) \right) ^{2}\left\Vert \hat{\Psi}_{0}\right\Vert ^{2}}{\left\langle 
\hat{g}^{ef}\right\rangle ^{2}\left( 5+\frac{\left\langle \hat{g}%
^{ef}\right\rangle }{\left\langle \hat{g}\right\rangle }-\sqrt{\left( 1-%
\frac{\left\langle \hat{g}^{ef}\right\rangle }{\left\langle \hat{g}%
\right\rangle }\right) \left( 4-\frac{\left\langle \hat{g}^{ef}\right\rangle 
}{\left\langle \hat{g}\right\rangle }\right) }\right) }  \notag
\end{eqnarray}

\subsubsection*{A20.5.2 Computation of $\frac{\left\langle \bar{K}%
_{0}\right\rangle }{\left\langle \hat{K}_{0}\right\rangle }$}

We proceed similarly for fild and average capital for banks:%
\begin{equation*}
\left\Vert \bar{\Psi}\right\Vert ^{2}=\hat{\mu}V\frac{\left\langle \bar{K}%
_{0}\right\rangle ^{3}}{\sigma _{\hat{K}}^{2}}\left( \frac{1}{3}-\frac{%
\left\langle \hat{K}\right\rangle }{2\left\langle \bar{K}_{0}\right\rangle }%
\frac{\left\langle \hat{g}^{Bef}\right\rangle }{\left\langle \bar{g}%
\right\rangle }\right) \left\langle \bar{g}\right\rangle ^{2}
\end{equation*}%
\begin{equation*}
\left\langle \bar{K}\right\rangle \left\Vert \bar{\Psi}\right\Vert ^{2}=\hat{%
\mu}V\frac{\left\langle \bar{K}_{0}\right\rangle ^{4}}{2\sigma _{\hat{K}}^{2}%
}\left( \frac{1}{4}-\frac{\left\langle \hat{K}\right\rangle }{3\left\langle 
\bar{K}_{0}\right\rangle }\frac{\left\langle \hat{g}^{Bef}\right\rangle }{%
\left\langle \bar{g}\right\rangle }\right) \left\langle \bar{g}\right\rangle
^{2}
\end{equation*}%
along with the value of average maximal capital:%
\begin{equation*}
\left\langle \bar{K}_{0}\right\rangle ^{2}\simeq 2\frac{\sigma _{\hat{K}}^{2}%
}{\left\langle \bar{g}\right\rangle ^{2}}\left( \frac{\left\Vert \bar{\Psi}%
_{0}\right\Vert ^{2}}{\hat{\mu}}-\left( \frac{\left\langle \hat{K}%
\right\rangle ^{2}\left\langle \hat{g}\right\rangle }{\sigma _{\hat{K}}^{2}}%
\right) \left\langle \hat{g}^{Bef}\right\rangle \frac{\left\langle \bar{K}%
_{0}\right\rangle }{\left\langle \hat{K}\right\rangle }\right)
\end{equation*}%
again this leads to write:%
\begin{equation*}
1\simeq 2\frac{\sigma _{\hat{K}}^{2}}{\left\langle \bar{g}\right\rangle ^{2}}%
\left( \frac{\left\Vert \bar{\Psi}_{0}\right\Vert ^{2}}{\hat{\mu}%
\left\langle \bar{K}_{0}\right\rangle ^{2}}-\frac{\left\langle \hat{K}%
\right\rangle \left\langle \hat{g}\right\rangle }{\left\langle \bar{K}%
_{0}\right\rangle \sigma _{\hat{K}}^{2}}\left\langle \hat{g}%
^{Bef}\right\rangle \right)
\end{equation*}%
and the ratio of averages maximal capital is:%
\begin{equation*}
\frac{\left\langle \bar{K}_{0}\right\rangle ^{2}}{\left\langle \hat{K}%
_{0}\right\rangle ^{2}}\simeq 2\frac{\sigma _{\hat{K}}^{2}}{\left\langle 
\bar{g}\right\rangle ^{2}}\left( \frac{\left\Vert \bar{\Psi}_{0}\right\Vert
^{2}}{\hat{\mu}\left\langle \hat{K}_{0}\right\rangle ^{2}}-\frac{%
\left\langle \hat{K}\right\rangle \left\langle \hat{g}\right\rangle }{%
\left\langle \hat{K}_{0}\right\rangle \sigma _{\hat{K}}^{2}}\left\langle 
\hat{g}^{Bef}\right\rangle \frac{\left\langle \bar{K}_{0}\right\rangle }{%
\left\langle \hat{K}_{0}\right\rangle }\right)
\end{equation*}%
using (\ref{RTY}):%
\begin{equation*}
\frac{\left\langle \hat{K}\right\rangle }{\left\langle \hat{K}%
_{0}\right\rangle }\simeq \frac{\left\langle \hat{g}\right\rangle }{%
6\left\langle \hat{g}^{ef}\right\rangle }\left( 2+\frac{\left\langle \hat{g}%
^{ef}\right\rangle }{\left\langle \hat{g}\right\rangle }-\sqrt{\left( 1-%
\frac{\left\langle \hat{g}^{ef}\right\rangle }{\left\langle \hat{g}%
\right\rangle }\right) \left( 4-\frac{\left\langle \hat{g}^{ef}\right\rangle 
}{\left\langle \hat{g}\right\rangle }\right) }\right)
\end{equation*}%
the ratio $\frac{\left\langle \bar{K}_{0}\right\rangle ^{2}}{\left\langle 
\hat{K}_{0}\right\rangle ^{2}}$ rewrts: 
\begin{eqnarray*}
\frac{\left\langle \bar{K}_{0}\right\rangle ^{2}}{\left\langle \hat{K}%
_{0}\right\rangle ^{2}} &\simeq &2\frac{\sigma _{\hat{K}}^{2}}{\left\langle 
\bar{g}\right\rangle ^{2}}\left( \frac{\left\Vert \bar{\Psi}_{0}\right\Vert
^{2}}{\hat{\mu}\left\langle \hat{K}_{0}\right\rangle ^{2}}-\frac{%
\left\langle \hat{K}\right\rangle \left\langle \hat{g}\right\rangle }{%
\left\langle \hat{K}_{0}\right\rangle \sigma _{\hat{K}}^{2}}\left\langle 
\hat{g}^{Bef}\right\rangle \frac{\left\langle \bar{K}_{0}\right\rangle }{%
\left\langle \hat{K}_{0}\right\rangle }\right) \\
&\simeq &2\frac{\sigma _{\hat{K}}^{2}}{\left\langle \bar{g}\right\rangle ^{2}%
}\left( \frac{\left\Vert \bar{\Psi}_{0}\right\Vert ^{2}}{\hat{\mu}%
\left\langle \hat{K}_{0}\right\rangle ^{2}}-\frac{1}{6}\frac{\left\langle 
\hat{g}\right\rangle ^{2}}{\sigma _{\hat{K}}^{2}}\frac{\left\langle \hat{g}%
^{Bef}\right\rangle }{\left\langle \hat{g}^{ef}\right\rangle }\left( 2+\frac{%
\left\langle \hat{g}^{ef}\right\rangle }{\left\langle \hat{g}\right\rangle }-%
\sqrt{\left( 1-\frac{\left\langle \hat{g}^{ef}\right\rangle }{\left\langle 
\hat{g}\right\rangle }\right) \left( 4-\frac{\left\langle \hat{g}%
^{ef}\right\rangle }{\left\langle \hat{g}\right\rangle }\right) }\right) 
\frac{\left\langle \bar{K}_{0}\right\rangle }{\left\langle \hat{K}%
_{0}\right\rangle }\right)
\end{eqnarray*}%
with solution:%
\begin{equation*}
\frac{\left\langle \bar{K}_{0}\right\rangle }{\left\langle \hat{K}%
_{0}\right\rangle }=-\frac{\left\langle \hat{K}\right\rangle \left\langle 
\hat{g}\right\rangle \left\langle \hat{g}^{Bef}\right\rangle }{\left\langle 
\bar{g}\right\rangle ^{2}\left\langle \hat{K}_{0}\right\rangle }+\sqrt{%
\left( \frac{\left\langle \hat{K}\right\rangle \left\langle \hat{g}%
\right\rangle \left\langle \hat{g}^{Bef}\right\rangle }{\left\langle \bar{g}%
\right\rangle ^{2}\left\langle \hat{K}_{0}\right\rangle }\right) ^{2}+2\frac{%
\sigma _{\hat{K}}^{2}}{\left\langle \bar{g}\right\rangle ^{2}}\frac{%
\left\Vert \bar{\Psi}_{0}\right\Vert ^{2}}{\hat{\mu}\left\langle \hat{K}%
_{0}\right\rangle ^{2}}}
\end{equation*}%
or in first approximation:%
\begin{equation}
\frac{\left\langle \bar{K}_{0}\right\rangle }{\left\langle \hat{K}%
_{0}\right\rangle }\rightarrow -\frac{\left\langle \hat{K}\right\rangle
\left\langle \hat{g}\right\rangle \left\langle \hat{g}^{Bef}\right\rangle }{%
\left\langle \bar{g}\right\rangle ^{2}\left\langle \hat{K}_{0}\right\rangle }%
+\sqrt{2\frac{\sigma _{\hat{K}}^{2}}{\left\langle \bar{g}\right\rangle ^{2}}%
\frac{\left\Vert \bar{\Psi}_{0}\right\Vert ^{2}}{\hat{\mu}\left\langle \hat{K%
}_{0}\right\rangle ^{2}}}  \label{RTW}
\end{equation}%
replacing for $\left\langle \hat{K}_{0}\right\rangle $ (\ref{KCM}) and ug (%
\ref{KCR}) leads to the following expression:%
\begin{eqnarray*}
\frac{\left\langle \bar{K}_{0}\right\rangle }{\left\langle \hat{K}%
_{0}\right\rangle } &\rightarrow &-\frac{\left\langle \hat{K}\right\rangle
\left\langle \hat{g}\right\rangle \left\langle \hat{g}^{Bef}\right\rangle }{%
\left\langle \hat{K}_{0}\right\rangle \left\langle \bar{g}\right\rangle ^{2}}%
+\sqrt{\left( \frac{\left\langle \hat{K}\right\rangle \left\langle \hat{g}%
\right\rangle \left\langle \hat{g}^{Bef}\right\rangle }{\left\langle \hat{K}%
_{0}\right\rangle \left\langle \bar{g}\right\rangle ^{2}}\right) ^{2}+2\frac{%
\sigma _{\hat{K}}^{2}}{\left\langle \bar{g}\right\rangle ^{2}}\frac{%
\left\Vert \bar{\Psi}_{0}\right\Vert ^{2}}{\hat{\mu}}} \\
&=&-\frac{6\left\langle \hat{g}\right\rangle ^{2}\left\langle \hat{g}%
^{Bef}\right\rangle }{\left\langle \bar{g}\right\rangle ^{2}\left\langle 
\hat{g}^{ef}\right\rangle }\left( 2+\frac{\left\langle \hat{g}%
^{ef}\right\rangle }{\left\langle \hat{g}\right\rangle }-\sqrt{\left( 1-%
\frac{\left\langle \hat{g}^{ef}\right\rangle }{\left\langle \hat{g}%
\right\rangle }\right) \left( 4-\frac{\left\langle \hat{g}^{ef}\right\rangle 
}{\left\langle \hat{g}\right\rangle }\right) }\right) \\
&&+\sqrt{36\left( \frac{\left\langle \hat{g}\right\rangle ^{2}\left\langle 
\hat{g}^{Bef}\right\rangle }{\left\langle \bar{g}\right\rangle
^{2}\left\langle \hat{g}^{ef}\right\rangle }\left( 2+\frac{\left\langle \hat{%
g}^{ef}\right\rangle }{\left\langle \hat{g}\right\rangle }-\sqrt{\left( 1-%
\frac{\left\langle \hat{g}^{ef}\right\rangle }{\left\langle \hat{g}%
\right\rangle }\right) \left( 4-\frac{\left\langle \hat{g}^{ef}\right\rangle 
}{\left\langle \hat{g}\right\rangle }\right) }\right) \right) ^{2}+2\frac{%
\sigma _{\hat{K}}^{2}}{\left\langle \bar{g}\right\rangle ^{2}}\frac{%
\left\Vert \bar{\Psi}_{0}\right\Vert ^{2}}{\hat{\mu}}}
\end{eqnarray*}%
**%
\begin{eqnarray*}
&&\frac{\left\langle \bar{K}_{0}\right\rangle }{\left\langle \hat{K}%
_{0}\right\rangle }\rightarrow -\frac{3}{8}\frac{\left\langle \hat{g}%
\right\rangle \left\langle \hat{g}^{Bef}\right\rangle }{\left\langle \bar{g}%
\right\rangle ^{2}}+\sqrt{\frac{\left\langle \hat{g}\right\rangle ^{2}}{%
3\left\langle \bar{g}\right\rangle ^{2}}\frac{\left\Vert \bar{\Psi}%
_{0}\right\Vert ^{2}\left( 5+\frac{\left\langle \hat{g}^{ef}\right\rangle }{%
\left\langle \hat{g}\right\rangle }-\sqrt{\left( 1-\frac{\left\langle \hat{g}%
^{ef}\right\rangle }{\left\langle \hat{g}\right\rangle }\right) \left( 4-%
\frac{\left\langle \hat{g}^{ef}\right\rangle }{\left\langle \hat{g}%
\right\rangle }\right) }\right) }{\left\Vert \hat{\Psi}_{0}\right\Vert ^{2}}}
\\
&\rightarrow &-\frac{3}{8}\frac{\left\langle \hat{g}\right\rangle
\left\langle \hat{g}^{Bef}\right\rangle }{\left\langle \bar{g}\right\rangle
^{2}}+\sqrt{\frac{\left\langle \hat{g}\right\rangle ^{2}}{\left\langle \bar{g%
}\right\rangle ^{2}}\frac{\left\Vert \bar{\Psi}_{0}\right\Vert ^{2}\left( 1+%
\frac{3}{4}\frac{\left\langle \hat{g}^{ef}\right\rangle }{\left\langle \hat{g%
}\right\rangle }\right) }{\left\Vert \hat{\Psi}_{0}\right\Vert ^{2}}}
\end{eqnarray*}

\subsubsection*{A20.5.3 Computation of $\frac{\left\langle \bar{K}%
_{0}\right\rangle }{\left\langle \hat{K}\right\rangle }$}

Multiply (\ref{RTW}) by $\frac{\left\langle \hat{K}_{0}\right\rangle }{%
\left\langle \hat{K}\right\rangle }$: 
\begin{equation*}
\frac{\left\langle \bar{K}_{0}\right\rangle }{\left\langle \hat{K}%
\right\rangle }=-\frac{\left\langle \hat{g}\right\rangle \left\langle \hat{g}%
^{Bef}\right\rangle }{\left\langle \bar{g}\right\rangle ^{2}}+\sqrt{\left( 
\frac{\left\langle \hat{g}\right\rangle \left\langle \hat{g}%
^{Bef}\right\rangle }{\left\langle \bar{g}\right\rangle ^{2}}\right) ^{2}+2%
\frac{\sigma _{\hat{K}}^{2}}{\left\langle \bar{g}\right\rangle ^{2}}\frac{%
\left\Vert \bar{\Psi}_{0}\right\Vert ^{2}}{\hat{\mu}\left\langle \hat{K}%
\right\rangle ^{2}}}
\end{equation*}%
and replacing $\left\langle \hat{K}\right\rangle ^{2}$ by its expressn (\ref%
{KCR}) we find:%
\begin{equation}
\frac{\left\langle \bar{K}_{0}\right\rangle }{\left\langle \hat{K}%
\right\rangle }=-\frac{\left\langle \hat{g}\right\rangle \left\langle \hat{g}%
^{Bef}\right\rangle }{\left\langle \bar{g}\right\rangle ^{2}}+\sqrt{\left( 
\frac{\left\langle \hat{g}\right\rangle \left\langle \hat{g}%
^{Bef}\right\rangle }{\left\langle \bar{g}\right\rangle ^{2}}\right) ^{2}+2%
\frac{\sigma _{\hat{K}}^{2}}{\left\langle \bar{g}\right\rangle ^{2}}\frac{%
\left\Vert \bar{\Psi}_{0}\right\Vert ^{2}}{\hat{\mu}\left\langle \hat{K}%
\right\rangle ^{2}}}\simeq -\frac{\left\langle \hat{g}\right\rangle
\left\langle \hat{g}^{Bef}\right\rangle }{\left\langle \bar{g}\right\rangle
^{2}}+\sqrt{2\frac{\sigma _{\hat{K}}^{2}}{\left\langle \bar{g}\right\rangle
^{2}}\frac{\left\Vert \bar{\Psi}_{0}\right\Vert ^{2}}{\hat{\mu}\left\langle 
\hat{K}\right\rangle ^{2}}}  \label{RTM}
\end{equation}

and (\ref{RTW}) bcms:

\subsubsection*{A20.5.4 Computation of $\frac{\left\langle \bar{K}%
\right\rangle \left\Vert \bar{\Psi}\right\Vert ^{2}}{\left\langle \hat{K}%
\right\rangle \left\Vert \hat{\Psi}\right\Vert ^{2}}$}

We define the ratio:%
\begin{equation*}
Z=\frac{\left\langle \bar{K}\right\rangle \left\Vert \bar{\Psi}\right\Vert
^{2}}{\left\langle \hat{K}\right\rangle \left\Vert \hat{\Psi}\right\Vert ^{2}%
}
\end{equation*}%
which satisfies:%
\begin{equation*}
Z=\frac{\left\langle \bar{K}_{0}\right\rangle ^{4}\left( \frac{1}{4}-\frac{%
\left\langle \hat{K}\right\rangle }{3\left\langle \bar{K}_{0}\right\rangle }%
\frac{\left\langle \hat{g}^{Bef}\right\rangle }{\left\langle \bar{g}%
\right\rangle }\right) \left\langle \bar{g}\right\rangle ^{2}}{\left\langle 
\hat{K}_{0}\right\rangle ^{4}\left( \frac{1}{4}-\frac{\left\langle \hat{K}%
\right\rangle }{3\left\langle \hat{K}_{0}\right\rangle }\frac{\left\langle 
\hat{g}^{ef}\right\rangle }{\left\langle \hat{g}\right\rangle }\right)
\left\langle \hat{g}\right\rangle ^{2}}
\end{equation*}%
\begin{equation*}
\frac{\left\langle \hat{K}\right\rangle }{\left\langle \hat{K}%
_{0}\right\rangle }=\frac{1}{6\frac{\left\langle \hat{g}^{ef}\right\rangle }{%
\left\langle \hat{g}\right\rangle }}\left( 2+\frac{\left\langle \hat{g}%
^{ef}\right\rangle }{\left\langle \hat{g}\right\rangle }-\sqrt{\left( 1-%
\frac{\left\langle \hat{g}^{ef}\right\rangle }{\left\langle \hat{g}%
\right\rangle }\right) \left( 4-\frac{\left\langle \hat{g}^{ef}\right\rangle 
}{\left\langle \hat{g}\right\rangle }\right) }\right)
\end{equation*}%
Leading to:%
\begin{equation*}
\frac{\left\langle \hat{K}\right\rangle }{3\left\langle \hat{K}%
_{0}\right\rangle }\frac{\left\langle \hat{g}^{ef}\right\rangle }{%
\left\langle \hat{g}\right\rangle }=\frac{1}{18}\left( 2+\frac{\left\langle 
\hat{g}^{ef}\right\rangle }{\left\langle \hat{g}\right\rangle }-\sqrt{\left(
1-\frac{\left\langle \hat{g}^{ef}\right\rangle }{\left\langle \hat{g}%
\right\rangle }\right) \left( 4-\frac{\left\langle \hat{g}^{ef}\right\rangle 
}{\left\langle \hat{g}\right\rangle }\right) }\right)
\end{equation*}%
The formula for $\left\langle \bar{K}_{0}\right\rangle ^{2}$: 
\begin{equation*}
\left\langle \bar{K}_{0}\right\rangle ^{2}\simeq 2\frac{\sigma _{\hat{K}}^{2}%
}{\left\langle \bar{g}\right\rangle ^{2}}\left( \frac{\left\Vert \bar{\Psi}%
_{0}\right\Vert ^{2}}{\hat{\mu}}-\left( \frac{\left\langle \hat{K}%
\right\rangle ^{2}\left\langle \hat{g}\right\rangle }{\sigma _{\hat{K}}^{2}}%
\right) \left\langle \hat{g}^{Bef}\right\rangle \frac{\left\langle \bar{K}%
_{0}\right\rangle }{\left\langle \hat{K}\right\rangle }\right)
\end{equation*}%
allows to find $2\frac{\sigma _{\hat{K}}^{2}}{\left\langle \bar{g}%
\right\rangle ^{2}}\frac{\left\Vert \bar{\Psi}_{0}\right\Vert ^{2}}{\hat{\mu}%
\left\langle \hat{K}\right\rangle ^{2}}$:

\begin{eqnarray*}
&&2\frac{\sigma _{\hat{K}}^{2}}{\left\langle \bar{g}\right\rangle ^{2}}\frac{%
\left\Vert \bar{\Psi}_{0}\right\Vert ^{2}}{\hat{\mu}\left\langle \hat{K}%
\right\rangle ^{2}} \\
&\rightarrow &\frac{12\left\langle \hat{g}^{ef}\right\rangle ^{2}\left( 5+%
\frac{\left\langle \hat{g}^{ef}\right\rangle }{\left\langle \hat{g}%
\right\rangle }-\sqrt{\left( 1-\frac{\left\langle \hat{g}^{ef}\right\rangle 
}{\left\langle \hat{g}\right\rangle }\right) \left( 4-\frac{\left\langle 
\hat{g}^{ef}\right\rangle }{\left\langle \hat{g}\right\rangle }\right) }%
\right) \left\Vert \bar{\Psi}_{0}\right\Vert ^{2}}{\left\langle \bar{g}%
\right\rangle ^{2}\left( \left( 2+\frac{\left\langle \hat{g}%
^{ef}\right\rangle }{\left\langle \hat{g}\right\rangle }-\sqrt{\left( 1-%
\frac{\left\langle \hat{g}^{ef}\right\rangle }{\left\langle \hat{g}%
\right\rangle }\right) \left( 4-\frac{\left\langle \hat{g}^{ef}\right\rangle 
}{\left\langle \hat{g}\right\rangle }\right) }\right) \right) ^{2}\left\Vert 
\hat{\Psi}_{0}\right\Vert ^{2}}\simeq \frac{196\left\Vert \bar{\Psi}%
_{0}\right\Vert ^{2}\left\langle \hat{g}\right\rangle ^{2}}{27\left\Vert 
\hat{\Psi}_{0}\right\Vert ^{2}\left\langle \bar{g}\right\rangle ^{2}}
\end{eqnarray*}%
and rewrite (\ref{RTM}) as:%
\begin{eqnarray*}
\frac{\left\langle \bar{K}_{0}\right\rangle }{\left\langle \hat{K}%
\right\rangle } &=&-\frac{\left\langle \hat{g}\right\rangle \left\langle 
\hat{g}^{Bef}\right\rangle }{\left\langle \bar{g}\right\rangle ^{2}}+\sqrt{%
\left( \frac{\left\langle \hat{g}\right\rangle \left\langle \hat{g}%
^{Bef}\right\rangle }{\left\langle \bar{g}\right\rangle ^{2}}\right) ^{2}+2%
\frac{\sigma _{\hat{K}}^{2}}{\left\langle \bar{g}\right\rangle ^{2}}\frac{%
\left\Vert \bar{\Psi}_{0}\right\Vert ^{2}}{\hat{\mu}\left\langle \hat{K}%
\right\rangle ^{2}}} \\
&\rightarrow &-\frac{\left\langle \hat{g}\right\rangle \left\langle \hat{g}%
^{Bef}\right\rangle }{\left\langle \bar{g}\right\rangle ^{2}}+\sqrt{\left( 
\frac{\left\langle \hat{g}\right\rangle \left\langle \hat{g}%
^{Bef}\right\rangle }{\left\langle \bar{g}\right\rangle ^{2}}\right) ^{2}+2%
\frac{\sigma _{\hat{K}}^{2}}{\left\langle \bar{g}\right\rangle ^{2}}\frac{%
\left\Vert \bar{\Psi}_{0}\right\Vert ^{2}}{\hat{\mu}\left\langle \hat{K}%
\right\rangle ^{2}}}\simeq \sqrt{\frac{196\left\Vert \bar{\Psi}%
_{0}\right\Vert ^{2}}{27\left\Vert \hat{\Psi}_{0}\right\Vert ^{2}}}\frac{%
\left\langle \hat{g}\right\rangle }{\left\langle \bar{g}\right\rangle }-%
\frac{\left\langle \hat{g}\right\rangle \left\langle \hat{g}%
^{Bef}\right\rangle }{\left\langle \bar{g}\right\rangle ^{2}}
\end{eqnarray*}%
\begin{eqnarray*}
&&\frac{\left\langle \bar{K}_{0}\right\rangle }{\left\langle \hat{K}%
_{0}\right\rangle }=-\frac{3}{8}\frac{\left\langle \hat{g}\right\rangle
\left\langle \hat{g}^{Bef}\right\rangle }{\left\langle \bar{g}\right\rangle
^{2}}+\sqrt{\frac{\left\langle \hat{g}\right\rangle ^{2}}{3\left\langle \bar{%
g}\right\rangle ^{2}}\frac{\left\Vert \bar{\Psi}_{0}\right\Vert ^{2}\left( 5+%
\frac{\left\langle \hat{g}^{ef}\right\rangle }{\left\langle \hat{g}%
\right\rangle }-\sqrt{\left( 1-\frac{\left\langle \hat{g}^{ef}\right\rangle 
}{\left\langle \hat{g}\right\rangle }\right) \left( 4-\frac{\left\langle 
\hat{g}^{ef}\right\rangle }{\left\langle \hat{g}\right\rangle }\right) }%
\right) }{\left\Vert \hat{\Psi}_{0}\right\Vert ^{2}}} \\
&\simeq &-\frac{3}{8}\frac{\left\langle \hat{g}\right\rangle \left\langle 
\hat{g}^{Bef}\right\rangle }{\left\langle \bar{g}\right\rangle ^{2}}+\sqrt{%
\frac{\left\langle \hat{g}\right\rangle ^{2}}{\left\langle \bar{g}%
\right\rangle ^{2}}\frac{\left\Vert \bar{\Psi}_{0}\right\Vert ^{2}\left( 1+%
\frac{3}{4}\frac{\left\langle \hat{g}^{ef}\right\rangle }{\left\langle \hat{g%
}\right\rangle }\right) }{\left\Vert \hat{\Psi}_{0}\right\Vert ^{2}}}
\end{eqnarray*}%
and the ratio $Z$ becomes: 
\begin{eqnarray*}
Z &=&\frac{\left\langle \bar{K}_{0}\right\rangle ^{4}\left( \frac{1}{4}-%
\frac{\left\langle \hat{K}\right\rangle }{3\left\langle \bar{K}%
_{0}\right\rangle }\frac{\left\langle \hat{g}^{Bef}\right\rangle }{%
\left\langle \bar{g}\right\rangle }\right) \left\langle \bar{g}\right\rangle
^{2}}{\left\langle \hat{K}_{0}\right\rangle ^{4}\left( \frac{1}{4}-\frac{%
\left\langle \hat{K}\right\rangle }{3\left\langle \hat{K}_{0}\right\rangle }%
\frac{\left\langle \hat{g}^{ef}\right\rangle }{\left\langle \hat{g}%
\right\rangle }\right) \left\langle \hat{g}\right\rangle ^{2}} \\
&\rightarrow &\frac{\left\langle \bar{K}_{0}\right\rangle ^{4}\left( \frac{1%
}{4}-\frac{1}{3\left( \sqrt{\frac{196\left\Vert \bar{\Psi}_{0}\right\Vert
^{2}}{27\left\Vert \hat{\Psi}_{0}\right\Vert ^{2}}}\frac{\left\langle \hat{g}%
\right\rangle }{\left\langle \bar{g}\right\rangle }-\frac{\left\langle \hat{g%
}\right\rangle \left\langle \hat{g}^{Bef}\right\rangle }{\left\langle \bar{g}%
\right\rangle ^{2}}\right) }\frac{\left\langle \hat{g}^{Bef}\right\rangle }{%
\left\langle \bar{g}\right\rangle }\right) \left\langle \bar{g}\right\rangle
^{2}}{\left\langle \hat{K}_{0}\right\rangle ^{4}\left( \frac{1}{4}-\frac{1}{%
18}\left( 2+\frac{\left\langle \hat{g}^{ef}\right\rangle }{\left\langle \hat{%
g}\right\rangle }-\sqrt{\left( 1-\frac{\left\langle \hat{g}%
^{ef}\right\rangle }{\left\langle \hat{g}\right\rangle }\right) \left( 4-%
\frac{\left\langle \hat{g}^{ef}\right\rangle }{\left\langle \hat{g}%
\right\rangle }\right) }\right) \right) \left\langle \hat{g}\right\rangle
^{2}}\simeq \frac{\left\langle \bar{K}_{0}\right\rangle ^{4}\left( \frac{1}{4%
}-\sqrt{\frac{3\left\Vert \hat{\Psi}_{0}\right\Vert ^{2}}{196\left\Vert \bar{%
\Psi}_{0}\right\Vert ^{2}}}\frac{\left\langle \hat{g}^{Bef}\right\rangle }{%
\left\langle \hat{g}\right\rangle }\right) \left\langle \bar{g}\right\rangle
^{2}}{\left\langle \hat{K}_{0}\right\rangle ^{4}\left( \frac{1}{4}-\frac{1}{%
18}\frac{5}{4}\frac{\left\langle \hat{g}^{ef}\right\rangle }{\left\langle 
\hat{g}\right\rangle }\right) \left\langle \hat{g}\right\rangle ^{2}}
\end{eqnarray*}%
that is:%
\begin{equation*}
Z\simeq \frac{\left\langle \bar{K}_{0}\right\rangle ^{4}\left( 1-\frac{1}{2}%
\sqrt{\frac{\left\Vert \hat{\Psi}_{0}\right\Vert ^{2}}{\left\Vert \bar{\Psi}%
_{0}\right\Vert ^{2}}}\frac{\left\langle \hat{g}^{Bef}\right\rangle }{%
\left\langle \hat{g}\right\rangle }\right) \left\langle \bar{g}\right\rangle
^{2}}{\left\langle \hat{K}_{0}\right\rangle ^{4}\left\langle \hat{g}%
\right\rangle ^{2}}
\end{equation*}%
with expanded form:%
\begin{equation*}
Z\simeq \left( -\frac{3}{8}\frac{\left\langle \hat{g}^{Bef}\right\rangle }{%
\left\langle \bar{g}\right\rangle }+\sqrt{\frac{\left\Vert \bar{\Psi}%
_{0}\right\Vert ^{2}\left( 1+\frac{3}{4}\frac{\left\langle \hat{g}%
^{ef}\right\rangle }{\left\langle \hat{g}\right\rangle }\right) }{\left\Vert 
\hat{\Psi}_{0}\right\Vert ^{2}}}\right) ^{2}\left( 1-\frac{1}{2}\sqrt{\frac{%
\left\Vert \hat{\Psi}_{0}\right\Vert ^{2}}{\left\Vert \bar{\Psi}%
_{0}\right\Vert ^{2}}}\frac{\left\langle \hat{g}^{Bef}\right\rangle }{%
\left\langle \hat{g}\right\rangle }\right)
\end{equation*}%
\begin{equation*}
Z=\left( 1+\frac{3}{4}\frac{\left\langle \hat{g}^{ef}\right\rangle }{%
\left\langle \hat{g}\right\rangle }\right) \frac{\left\Vert \bar{\Psi}%
_{0}\right\Vert ^{2}}{\left\Vert \hat{\Psi}_{0}\right\Vert ^{2}}-\left( 
\frac{3}{4}\frac{\left\langle \hat{g}^{Bef}\right\rangle }{\left\langle \bar{%
g}\right\rangle }+\frac{1}{2}\frac{\left\langle \hat{g}^{Bef}\right\rangle }{%
\left\langle \hat{g}\right\rangle }\right) \sqrt{\frac{\left\Vert \bar{\Psi}%
_{0}\right\Vert ^{2}}{\left\Vert \hat{\Psi}_{0}\right\Vert ^{2}}}
\end{equation*}%
where:%
\begin{equation*}
\left\langle \hat{g}^{ef}\right\rangle =-\frac{\left( \kappa \left\langle %
\left[ \frac{\underline{\hat{k}}_{2}^{B}}{1+\bar{k}}\right] \right\rangle
\left( 1-\left\langle \underline{\hat{k}}\right\rangle \right) +\left\langle 
\hat{k}_{1}^{B}\right\rangle \left\langle \hat{k}_{2}\right\rangle \right)
\left( \left\langle \hat{g}\right\rangle +\frac{1}{1-\left\langle \underline{%
\hat{k}}\right\rangle }\bar{N}\left\langle \bar{g}\right\rangle \right) 
\frac{\left\Vert \bar{\Psi}_{0}\right\Vert ^{2}}{\left\Vert \hat{\Psi}%
_{0}\right\Vert ^{2}}}{\left( 1-\left( \left\langle \hat{k}_{1}\right\rangle
+\left\langle \hat{k}_{1}^{B}\right\rangle \frac{\left\Vert \bar{\Psi}%
_{0}\right\Vert ^{2}}{\left\Vert \hat{\Psi}_{0}\right\Vert ^{2}}\right)
\right) \left( 1-\left( \left\langle \underline{\hat{k}}\right\rangle
+\left( \left\langle \underline{\hat{k}}_{1}^{B}\right\rangle +\kappa
\left\langle \left[ \frac{\underline{\hat{k}}_{2}^{B}}{1+\bar{k}}\right]
\right\rangle \right) \frac{\left\Vert \bar{\Psi}_{0}\right\Vert ^{2}}{%
\left\Vert \hat{\Psi}_{0}\right\Vert ^{2}}\right) \right) }
\end{equation*}%
and:%
\begin{eqnarray*}
\left\langle \hat{g}^{Bef}\right\rangle &=&-\frac{\kappa \left\langle \left[ 
\frac{\underline{\hat{k}}_{2}^{B}}{1+\bar{k}}\right] \right\rangle \left(
1-\left\langle \hat{k}\right\rangle \right) +\left\langle \hat{k}%
_{1}^{B}\right\rangle \left\langle \hat{k}_{2}\right\rangle }{\left(
1-\left( \left\langle \hat{k}\right\rangle +\left( \left\langle \hat{k}%
_{1}^{B}\right\rangle +\kappa \left\langle \left[ \frac{\underline{\hat{k}}%
_{2}^{B}}{1+\bar{k}}\right] \right\rangle \right) \frac{\left\Vert \bar{\Psi}%
_{0}\right\Vert ^{2}}{\left\Vert \hat{\Psi}_{0}\right\Vert ^{2}}\right)
\right) \left( 1-\left( \left\langle \hat{k}_{1}\right\rangle +\left\langle 
\hat{k}_{1}^{B}\right\rangle \frac{\left\Vert \bar{\Psi}_{0}\right\Vert ^{2}%
}{\left\Vert \hat{\Psi}_{0}\right\Vert ^{2}}\right) \right) } \\
&&\times \left( \left\langle \hat{g}\right\rangle +\left( 1-\hat{M}\right)
^{-1}\bar{N}\left\langle \bar{g}\right\rangle \right)
\end{eqnarray*}%
We thus obtain:%
\begin{equation*}
\left\langle \bar{K}_{0}\right\rangle ^{2}\simeq 2\frac{\sigma _{\hat{K}}^{2}%
}{\left\langle \bar{g}\right\rangle ^{2}}\left( \frac{\left\Vert \bar{\Psi}%
_{0}\right\Vert ^{2}}{\hat{\mu}}-\left( \frac{\left\langle \hat{K}%
\right\rangle ^{2}\left\langle \hat{g}\right\rangle }{\sigma _{\hat{K}}^{2}}%
\right) \left\langle \hat{g}^{Bef}\right\rangle \frac{\left\langle \bar{K}%
_{0}\right\rangle }{\left\langle \hat{K}\right\rangle }\right)
\end{equation*}%
and:%
\begin{eqnarray*}
\left\langle \hat{K}\right\rangle \left\Vert \hat{\Psi}\right\Vert ^{2} &=&%
\hat{\mu}V\frac{\left\langle \hat{K}_{0}\right\rangle ^{4}}{2\sigma _{\hat{K}%
}^{2}}\left( \frac{1}{4}-\frac{\left\langle \hat{K}\right\rangle }{%
3\left\langle \hat{K}_{0}\right\rangle }\frac{\left\langle \hat{g}%
^{ef}\right\rangle }{\left\langle \hat{g}\right\rangle }\right) \left\langle 
\hat{g}\right\rangle ^{2} \\
&=&\frac{18\sigma _{\hat{K}}^{2}}{\hat{\mu}}V\left( \frac{\left\Vert \hat{%
\Psi}_{0}\right\Vert ^{2}}{\left\langle \hat{g}\right\rangle ^{2}\left( 5+%
\frac{\left\langle \hat{g}^{ef}\right\rangle }{\left\langle \hat{g}%
\right\rangle }-\sqrt{\left( 1-\frac{\left\langle \hat{g}^{ef}\right\rangle 
}{\left\langle \hat{g}\right\rangle }\right) \left( 4-\frac{\left\langle 
\hat{g}^{ef}\right\rangle }{\left\langle \hat{g}\right\rangle }\right) }%
\right) }\right) ^{2} \\
&&\times \left( \frac{1}{4}-\frac{1}{18}\left( 2+\frac{\left\langle \hat{g}%
^{ef}\right\rangle }{\left\langle \hat{g}\right\rangle }-\sqrt{\left( 1-%
\frac{\left\langle \hat{g}^{ef}\right\rangle }{\left\langle \hat{g}%
\right\rangle }\right) \left( 4-\frac{\left\langle \hat{g}^{ef}\right\rangle 
}{\left\langle \hat{g}\right\rangle }\right) }\right) \right) \left\langle 
\hat{g}\right\rangle ^{2}
\end{eqnarray*}%
\bigskip

\subsection*{A20.6 Expression for field and capital per sector}

We use the formula (\ref{PSN}) and (\ref{PSD}):%
\begin{equation}
\left\Vert \hat{\Psi}\left( \hat{X}_{1}\right) \right\Vert ^{2}\simeq \hat{%
\mu}\frac{\hat{K}_{0}^{3}}{\sigma _{\hat{K}}^{2}}\left( \frac{\hat{g}%
^{2}\left( \hat{X}_{1}\right) }{3}-\frac{\left\langle \hat{K}\right\rangle }{%
2\hat{K}_{0}}\left\langle \hat{g}\right\rangle \left\langle \hat{g}%
^{ef}\right\rangle \right)
\end{equation}%
and:%
\begin{equation*}
\left\Vert \bar{\Psi}_{0}\left( \bar{X}_{1}\right) \right\Vert ^{2}\simeq 
\hat{\mu}\frac{\bar{K}_{0}^{3}}{\sigma _{\hat{K}}^{2}}\left( \frac{\bar{g}%
^{2}\left( \bar{X}_{1}\right) }{3}-\frac{\left\langle \hat{K}\right\rangle }{%
2\bar{K}_{0}}\left\langle \hat{g}\right\rangle \left\langle \hat{g}%
^{Bef}\right\rangle \right)
\end{equation*}

We also use our previous formula for capital (\ref{KM}) and (\ref{KB}):%
\begin{equation}
\hat{K}_{\hat{X}}\left\Vert \hat{\Psi}\left( \hat{X}_{1}\right) \right\Vert
^{2}\simeq \hat{\mu}\frac{\hat{K}_{0}^{4}}{2\sigma _{\hat{K}}^{2}}\left( 
\frac{\hat{g}^{2}\left( \hat{X}_{1}\right) }{4}-\frac{\left\langle \hat{K}%
\right\rangle \left\langle \hat{g}\right\rangle }{3\left\langle \hat{K}%
_{0}\right\rangle }\left\langle \hat{g}^{ef}\right\rangle \right)
\end{equation}%
and:%
\begin{equation}
\bar{K}_{\bar{X}}\left\Vert \bar{\Psi}\left( \bar{X}_{1}\right) \right\Vert
^{2}\simeq \hat{\mu}\frac{\bar{K}_{0}^{4}}{2\sigma _{\hat{K}}^{2}}\left( 
\frac{\hat{g}^{2}\left( \hat{X}_{1}\right) }{4}-\frac{\left\langle \hat{K}%
\right\rangle \left\langle \hat{g}\right\rangle }{3\left\langle \bar{K}%
_{0}\right\rangle }\left\langle \hat{g}^{Bef}\right\rangle \right)
\end{equation}

\subsection*{A20.7 Formula for $\hat{K}_{0}^{2}$ and $\bar{K}_{0}^{2}$}

\begin{equation}
\hat{K}_{0}^{2}\simeq 2\frac{\sigma _{\hat{K}}^{2}}{\hat{g}^{2}\left( \hat{X}%
_{1}\right) }\left( \frac{\left\Vert \hat{\Psi}_{0}\left( \hat{X}_{1}\right)
\right\Vert ^{2}}{\hat{\mu}}-\left( \frac{\left\langle \hat{K}\right\rangle
^{2}\left\langle \hat{g}\right\rangle }{\sigma _{\hat{K}}^{2}}+\frac{1}{2}%
\right) \left\langle \hat{g}^{ef}\right\rangle \frac{\left\langle \hat{K}%
_{0}\right\rangle }{\left\langle \hat{K}\right\rangle }\right)
\end{equation}%
\begin{equation*}
\bar{K}_{0}^{2}\simeq 2\frac{\sigma _{\hat{K}}^{2}}{\bar{g}^{2}\left( \hat{X}%
_{1}\right) }\left( \frac{\left\Vert \bar{\Psi}_{0}\left( \bar{X}_{1}\right)
\right\Vert ^{2}}{\hat{\mu}}-\left( \frac{\left\langle \hat{K}\right\rangle
^{2}\left\langle \hat{g}\right\rangle }{\sigma _{\hat{K}}^{2}}+\frac{1}{2}%
\right) \left\langle \hat{g}^{Bef}\right\rangle \frac{\left\langle \bar{K}%
_{0}\right\rangle }{\left\langle \hat{K}\right\rangle }\right)
\end{equation*}

\subsection*{A20.8 Including deviation from averages in functional derivative%
}

Using the previous results for averages (\ref{KCM}) and ug (\ref{KCR}) and
including the deviation to averages for the functional derivatives:

\begin{equation*}
\frac{\delta }{\delta \left\vert \hat{\Psi}\left( \hat{K},\hat{X}\right)
\right\vert ^{2}}\bar{g}\left( \hat{K}^{\prime },\hat{X}^{\prime }\right)
=\Delta \left( \hat{k}^{B}\left( \hat{X},\left\langle \bar{X}\right\rangle
\right) A\right) \frac{\hat{K}}{\left\Vert \hat{\Psi}\right\Vert
^{2}\left\langle \hat{K}\right\rangle }
\end{equation*}%
\begin{equation*}
\frac{\delta }{\delta \left\vert \bar{\Psi}\left( \bar{K},\bar{X}\right)
\right\vert ^{2}}\bar{g}\left( \hat{K}^{\prime },\hat{X}^{\prime }\right)
=\left( \left\langle \underline{\hat{k}}^{B}\right\rangle ^{ef}\Delta \left( 
\hat{k}^{B}\left( \hat{X},\left\langle \bar{X}\right\rangle \right) A\right)
+\frac{\Delta \bar{k}_{2}\left( \left\langle \bar{X}\right\rangle ,\bar{X}%
\right) }{\left( 1-\left\langle \bar{k}_{1}\right\rangle \right) \left\Vert 
\bar{\Psi}\right\Vert ^{2}\left\langle \bar{K}\right\rangle }\left\langle 
\bar{g}\right\rangle \right) \frac{\bar{K}}{\left\Vert \hat{\Psi}\right\Vert
^{2}\left\langle \hat{K}\right\rangle }
\end{equation*}%
and using, including normalizations:%
\begin{equation*}
\left\langle \underline{\hat{k}}^{B}\right\rangle ^{ef}=\left\langle 
\underline{\hat{k}}_{1}^{B}\right\rangle +\frac{\left\langle \underline{\hat{%
k}}_{1}^{B}\right\rangle }{\left( 1-\left\langle \hat{k}_{1}\right\rangle
\right) }
\end{equation*}

\subsection*{A20.9 Expressions for fields}

The solutions for the fields including the deviations becomes:%
\begin{eqnarray*}
\left\vert \hat{\Psi}\left( \hat{K}_{1},\hat{X}_{1}\right) \right\vert ^{2}
&=&\left\Vert \hat{\Psi}_{0}\left( \hat{X}_{1}\right) \right\Vert ^{2}-\hat{%
\mu}\left\{ \left( \frac{\hat{K}_{1}^{2}\hat{g}^{2}\left( \hat{X}_{1}\right) 
}{2\sigma _{\hat{K}}^{2}}+\frac{\hat{g}\left( \hat{X}_{1}\right) }{2}\right)
\right. \\
&&\left. +\left( \frac{\left\langle \bar{K}\right\rangle ^{2}\left\langle 
\bar{g}\right\rangle }{\sigma _{\hat{K}}^{2}}+\frac{1}{2}\right) \Delta
\left( \hat{k}^{B}\left( \hat{X}_{1},\left\langle \bar{X}\right\rangle
\right) A\right) \frac{\left\Vert \bar{\Psi}\right\Vert ^{2}\hat{K}_{1}}{%
\left\Vert \hat{\Psi}\right\Vert ^{2}\left\langle \hat{K}\right\rangle }%
-\left( \frac{\left\langle \hat{K}\right\rangle ^{2}\left\langle \hat{g}%
\right\rangle }{\sigma _{\hat{K}}^{2}}+\frac{1}{2}\right) \left\langle \hat{g%
}^{ef}\right\rangle \frac{\hat{K}_{1}}{\left\langle \hat{K}\right\rangle }%
\right\}
\end{eqnarray*}%
\begin{eqnarray*}
\left\vert \bar{\Psi}\left( \bar{K}_{1},\bar{X}_{1}\right) \right\vert ^{2}
&=&\left\vert \bar{\Psi}_{0}\left( \bar{X}_{1}\right) \right\vert ^{2}-\hat{%
\mu}\left\{ \left( \frac{\bar{K}_{1}^{2}\bar{g}^{2}\left( \bar{X}_{1}\right) 
}{\sigma _{\hat{K}}^{2}}+\frac{\bar{g}\left( \bar{X}_{1}\right) }{2}\right)
\right. \\
&&+\left( \frac{\left\langle \bar{K}\right\rangle ^{2}\left\langle \bar{g}%
\right\rangle }{\sigma _{\hat{K}}^{2}}+\frac{1}{2}\right) \left(
\left\langle \underline{\hat{k}}^{B}\right\rangle ^{ef}\Delta \left( \hat{k}%
^{B}\left( \hat{X}_{1},\left\langle \bar{X}\right\rangle \right) A\right) +%
\frac{\Delta \bar{k}_{2}\left( \left\langle \bar{X}\right\rangle ,\bar{X}%
\right) }{\left( 1-\left\langle \bar{k}_{1}\right\rangle \right) \left\Vert 
\bar{\Psi}\right\Vert ^{2}\left\langle \bar{K}\right\rangle }\left\langle 
\bar{g}\right\rangle \right) \frac{\left\Vert \bar{\Psi}\right\Vert ^{2}\bar{%
K}_{1}}{\left\Vert \hat{\Psi}\right\Vert ^{2}\left\langle \hat{K}%
\right\rangle } \\
&&\left. -\left( \frac{\left\langle \hat{K}\right\rangle ^{2}\left\langle 
\hat{g}\right\rangle }{\sigma _{\hat{K}}^{2}}+\frac{1}{2}\right)
\left\langle \hat{g}^{Bef}\right\rangle \frac{\bar{K}_{1}}{\left\langle \hat{%
K}\right\rangle }\right\}
\end{eqnarray*}%
and integratng:%
\begin{eqnarray}
\left\Vert \hat{\Psi}\left( \hat{X}_{1}\right) \right\Vert ^{2} &\simeq &%
\hat{\mu}\frac{\hat{K}_{0}^{3}}{\sigma _{\hat{K}}^{2}}\left( \frac{\hat{g}%
^{2}\left( \hat{X}_{1}\right) }{3}+\Delta \left( \hat{k}^{B}\left( \hat{X}%
_{1},\left\langle \bar{X}\right\rangle \right) A\right) Z\frac{\left\langle 
\bar{K}\right\rangle }{2\hat{K}_{0}}\left\langle \bar{g}\right\rangle -\frac{%
\left\langle \hat{K}\right\rangle }{2\hat{K}_{0}}\left\langle \hat{g}%
\right\rangle \left\langle \hat{g}^{ef}\right\rangle \right) \\
&\simeq &\hat{\mu}\frac{\hat{K}_{0}^{3}}{\sigma _{\hat{K}}^{2}}\left( \frac{%
\hat{g}^{2}\left( \hat{X}_{1}\right) }{3}-\frac{\left\langle \hat{K}%
\right\rangle }{2\hat{K}_{0}}\left\langle \hat{g}\right\rangle \left\langle 
\hat{g}^{ef}\right\rangle \right)  \notag
\end{eqnarray}%
similarly:%
\begin{eqnarray*}
&&\left\Vert \bar{\Psi}\left( \bar{X}_{1}\right) \right\Vert ^{2} \\
&\simeq &\hat{\mu}\frac{\bar{K}_{0}^{3}}{\sigma _{\hat{K}}^{2}}\left( \frac{%
\bar{g}^{2}\left( \bar{X}_{1}\right) }{3}+\left( \left\langle \underline{%
\hat{k}}^{B}\right\rangle ^{ef}\Delta \left( \hat{k}^{B}\left( \hat{X}%
_{1},\left\langle \bar{X}\right\rangle \right) A\right) +\frac{\Delta \bar{k}%
_{2}\left( \left\langle \bar{X}\right\rangle ,\bar{X}_{1}\right) }{\left(
1-\left\langle \bar{k}_{1}\right\rangle \right) \left\Vert \bar{\Psi}%
\right\Vert ^{2}\left\langle \bar{K}\right\rangle }\left\langle \bar{g}%
\right\rangle \right) \frac{\left\langle \bar{K}\right\rangle }{2\bar{K}_{0}}%
\left\langle \bar{g}\right\rangle -\frac{\left\langle \hat{K}\right\rangle }{%
2\bar{K}_{0}}\left\langle \hat{g}\right\rangle \left\langle \hat{g}%
^{Bef}\right\rangle \right) \\
&\simeq &\hat{\mu}\frac{\bar{K}_{0}^{3}}{\sigma _{\hat{K}}^{2}}\left( \frac{%
\bar{g}^{2}\left( \bar{X}_{1}\right) }{3}-\frac{\left\langle \hat{K}%
\right\rangle }{2\bar{K}_{0}}\left\langle \hat{g}\right\rangle \hat{g}%
^{Bef}\left( \bar{X}_{1}\right) \right)
\end{eqnarray*}

\subsection*{A20.10 Exprssion for capital per sector}

multiply $\left\vert \bar{\Psi}\left( \bar{K}_{1},\bar{X}_{1}\right)
\right\vert ^{2}$ by $\bar{K}_{1}$ and integrate:%
\begin{eqnarray}
\hat{K}_{\hat{X}}\left\Vert \hat{\Psi}\left( \hat{X}_{1}\right) \right\Vert
^{2} &\simeq &\hat{\mu}\frac{\hat{K}_{0}^{4}}{2\sigma _{\hat{K}}^{2}}\left( 
\frac{\hat{g}^{2}\left( \hat{X}_{1}\right) }{4}+\Delta \left( \hat{k}%
^{B}\left( \hat{X}_{1},\left\langle \bar{X}\right\rangle \right) A\right) Z%
\frac{\left\langle \bar{K}\right\rangle }{3\hat{K}_{0}}\left\langle \bar{g}%
\right\rangle -\frac{\left\langle \hat{K}\right\rangle \left\langle \hat{g}%
\right\rangle }{3\left\langle \hat{K}_{0}\right\rangle }\left\langle \hat{g}%
^{ef}\right\rangle \right) \\
&\simeq &\hat{\mu}\frac{\hat{K}_{0}^{4}}{2\sigma _{\hat{K}}^{2}}\left( \frac{%
\hat{g}^{2}\left( \hat{X}_{1}\right) }{4}-\frac{\left\langle \hat{K}%
\right\rangle }{3\bar{K}_{0}}\left\langle \hat{g}\right\rangle \hat{g}%
^{Bef}\left( \hat{X}_{1}\right) \right)  \notag
\end{eqnarray}%
we also have:%
\begin{eqnarray}
&&\bar{K}_{\bar{X}}\left\Vert \bar{\Psi}\left( \bar{X}_{1}\right)
\right\Vert ^{2} \\
&\simeq &\hat{\mu}\frac{\bar{K}_{0}^{4}}{2\sigma _{\hat{K}}^{2}}\left( \frac{%
\bar{g}^{2}\left( \hat{X}_{1}\right) }{4}+\left( \left\langle \underline{%
\hat{k}}^{B}\right\rangle ^{ef}\Delta \left( \hat{k}^{B}\left( \hat{X}%
_{1},\left\langle \bar{X}\right\rangle \right) A\right) +\frac{\Delta \bar{k}%
_{2}\left( \left\langle \bar{X}\right\rangle ,\bar{X}_{1}\right) }{\left(
1-\left\langle \bar{k}_{1}\right\rangle \right) \left\Vert \bar{\Psi}%
\right\Vert ^{2}\left\langle \bar{K}\right\rangle }\left\langle \bar{g}%
\right\rangle \right) \frac{\left\langle \bar{K}\right\rangle }{3\bar{K}_{0}}%
\left\langle \bar{g}\right\rangle -\frac{\left\langle \hat{K}\right\rangle
\left\langle \hat{g}\right\rangle }{3\left\langle \bar{K}_{0}\right\rangle }%
\left\langle \hat{g}^{Bef}\right\rangle \right)  \notag \\
&\simeq &\hat{\mu}\frac{\bar{K}_{0}^{4}}{2\sigma _{\hat{K}}^{2}}\left( \frac{%
\bar{g}^{2}\left( \hat{X}_{1}\right) }{4}-\frac{\left\langle \hat{K}%
\right\rangle \left\langle \bar{g}\right\rangle }{3\left\langle \bar{K}%
_{0}\right\rangle }\hat{g}^{Bef}\left( \bar{X}_{1}\right) \right)  \notag
\end{eqnarray}%
where:%
\begin{equation}
\hat{K}_{0}^{2}\simeq 2\frac{\sigma _{\hat{K}}^{2}}{\hat{g}^{2}\left( \hat{X}%
_{1}\right) }\left( \frac{\left\Vert \hat{\Psi}_{0}\left( \hat{X}_{1}\right)
\right\Vert ^{2}}{\hat{\mu}}+\left( \frac{\left\langle \bar{K}\right\rangle
^{2}\left\langle \bar{g}\right\rangle }{\sigma _{\hat{K}}^{2}}+\frac{1}{2}%
\right) \Delta \left( \hat{k}^{B}\left( \hat{X}_{1},\left\langle \bar{X}%
\right\rangle \right) A\right) Z\left\langle \bar{g}\right\rangle -\left( 
\frac{\left\langle \hat{K}\right\rangle ^{2}\left\langle \hat{g}%
\right\rangle }{\sigma _{\hat{K}}^{2}}+\frac{1}{2}\right) \left\langle \hat{g%
}^{ef}\right\rangle \frac{\left\langle \hat{K}_{0}\right\rangle }{%
\left\langle \hat{K}\right\rangle }\right)
\end{equation}%
and:%
\begin{eqnarray*}
\bar{K}_{0}^{2} &\simeq &2\frac{\sigma _{\hat{K}}^{2}}{\bar{g}^{2}\left( 
\hat{X}_{1}\right) }\left( \frac{\left\Vert \bar{\Psi}_{0}\left( \bar{X}%
_{1}\right) \right\Vert ^{2}}{\hat{\mu}}+\left( \frac{\left\langle \bar{K}%
\right\rangle ^{2}\left\langle \bar{g}\right\rangle }{\sigma _{\hat{K}}^{2}}+%
\frac{1}{2}\right) \left( \left\langle \underline{\hat{k}}^{B}\right\rangle
^{ef}\Delta \left( \hat{k}^{B}\left( \hat{X}_{1},\left\langle \bar{X}%
\right\rangle \right) A\right) +\frac{\Delta \bar{k}_{2}\left( \left\langle 
\bar{X}\right\rangle ,\bar{X}_{1}\right) }{\left( 1-\left\langle \bar{k}%
_{1}\right\rangle \right) \left\Vert \bar{\Psi}\right\Vert ^{2}\left\langle 
\bar{K}\right\rangle }\left\langle \bar{g}\right\rangle \right)
Z\left\langle \bar{g}\right\rangle \right. \\
&&\left. -\left( \frac{\left\langle \hat{K}\right\rangle ^{2}\left\langle 
\hat{g}\right\rangle }{\sigma _{\hat{K}}^{2}}+\frac{1}{2}\right)
\left\langle \hat{g}^{Bef}\right\rangle \frac{\left\langle \bar{K}%
_{0}\right\rangle }{\left\langle \hat{K}\right\rangle }\right)
\end{eqnarray*}%
These formula can be written depending on some parameters:%
\begin{eqnarray*}
\hat{K}\left[ \hat{X}_{1}\right] &=&\hat{K}_{\hat{X}}\left\Vert \hat{\Psi}%
\left( \hat{X}_{1}\right) \right\Vert ^{2}\simeq \hat{\mu}\frac{\hat{K}%
_{0}^{4}}{2\sigma _{\hat{K}}^{2}}\left( \frac{\hat{g}^{2}\left( \hat{X}%
_{1}\right) }{4}+\Delta \left( \hat{k}^{B}\left( \hat{X}_{1},\left\langle 
\bar{X}\right\rangle \right) A\right) Z\frac{\left\langle \bar{K}%
\right\rangle }{3\hat{K}_{0}}\left\langle \bar{g}\right\rangle -\frac{%
\left\langle \hat{K}\right\rangle \left\langle \hat{g}\right\rangle }{3\hat{K%
}_{0}}\left\langle \hat{g}^{ef}\right\rangle \right) \\
&=&\frac{\hat{\mu}}{2\sigma _{\hat{K}}^{2}}\left( 2\frac{\sigma _{\hat{K}%
}^{2}}{\hat{g}^{2}\left( \hat{X}_{1}\right) }\left( \frac{\left\Vert \hat{%
\Psi}_{0}\left( \hat{X}_{1}\right) \right\Vert ^{2}}{\hat{\mu}}-\hat{D}%
\left( \hat{X}_{1}\right) \right) \right) ^{2}\left( \frac{\hat{g}^{2}\left( 
\hat{X}_{1}\right) }{4}-\frac{\left\langle \hat{g}\right\rangle \hat{g}%
^{ef}\left( \hat{X}_{1}\right) }{3}\right)
\end{eqnarray*}%
and $\hat{g}^{ef}\left( \hat{X}_{1}\right) $ and $\hat{D}\left( \hat{X}%
_{1}\right) $ defined b:%
\begin{equation*}
\hat{g}^{ef}\left( \hat{X}_{1}\right) =4\left( \frac{\left\langle \hat{K}%
\right\rangle }{3\left\langle \hat{K}_{0}\right\rangle }\left\langle \hat{g}%
^{ef}\right\rangle -\Delta \left( \hat{k}^{B}\left( \hat{X}_{1},\left\langle 
\bar{X}\right\rangle \right) A\right) Z\frac{\left\langle \bar{K}%
\right\rangle }{3\hat{K}_{0}}\frac{\left\langle \bar{g}\right\rangle }{%
\left\langle \hat{g}\right\rangle }\right)
\end{equation*}%
\begin{equation*}
\hat{D}\left( \hat{X}_{1}\right) =\left( \frac{\left\langle \hat{K}%
\right\rangle ^{2}\left\langle \hat{g}\right\rangle }{\sigma _{\hat{K}}^{2}}+%
\frac{1}{2}\right) \frac{\hat{K}_{0}}{\left\langle \hat{K}\right\rangle }%
\left\langle \hat{g}^{ef}\right\rangle -\left( \frac{\left\langle \bar{K}%
\right\rangle ^{2}\left\langle \bar{g}\right\rangle }{\sigma _{\hat{K}}^{2}}+%
\frac{1}{2}\right) \Delta \left( \hat{k}^{B}\left( \hat{X}_{1},\left\langle 
\bar{X}\right\rangle \right) A\right) Z\left\langle \bar{g}\right\rangle
\end{equation*}%
Similarly:%
\begin{eqnarray*}
&&\bar{K}_{\bar{X}_{1}}\left\Vert \bar{\Psi}\left( \bar{X}_{1}\right)
\right\Vert ^{2} \\
&\simeq &\hat{\mu}\frac{\bar{K}_{0}^{4}}{2\sigma _{\hat{K}}^{2}}\left( \frac{%
\bar{g}^{2}\left( \hat{X}_{1}\right) }{4}+\left( \left\langle \underline{%
\hat{k}}^{B}\right\rangle ^{ef}\Delta \left( \hat{k}^{B}\left( \hat{X}%
_{1},\left\langle \bar{X}\right\rangle \right) A\right) +\frac{\Delta \bar{k}%
_{2}\left( \left\langle \bar{X}\right\rangle ,\bar{X}_{1}\right) }{\left(
1-\left\langle \bar{k}_{1}\right\rangle \right) \left\Vert \bar{\Psi}%
\right\Vert ^{2}\left\langle \bar{K}\right\rangle }\left\langle \bar{g}%
\right\rangle \right) \frac{\left\langle \bar{K}\right\rangle }{3\bar{K}_{0}}%
\left\langle \bar{g}\right\rangle -\frac{\left\langle \hat{K}\right\rangle
\left\langle \hat{g}\right\rangle }{3\bar{K}_{0}}\left\langle \hat{g}%
^{Bef}\right\rangle \right) \\
&=&\frac{\hat{\mu}}{2\sigma _{\hat{K}}^{2}}\left( 2\frac{\sigma _{\hat{K}%
}^{2}}{\bar{g}^{2}\left( \hat{X}_{1}\right) }\left( \frac{\left\Vert \bar{%
\Psi}_{0}\left( \bar{X}_{1}\right) \right\Vert ^{2}}{\hat{\mu}}-\bar{D}%
\left( \bar{X}_{1}\right) \right) \right) ^{2} \\
&&\times \left( \frac{\hat{g}^{2}\left( \hat{X}_{1}\right) }{4}-\frac{%
\left\langle \hat{K}\right\rangle \left\langle \hat{g}\right\rangle }{3\bar{K%
}_{0}}\left\langle \hat{g}^{Bef}\right\rangle +\left( \left\langle 
\underline{\hat{k}}^{B}\right\rangle ^{ef}\Delta \left( \hat{k}^{B}\left( 
\hat{X}_{1},\left\langle \bar{X}\right\rangle \right) A\right) +\frac{\Delta 
\bar{k}_{2}\left( \left\langle \bar{X}\right\rangle ,\bar{X}_{1}\right) }{%
\left( 1-\left\langle \bar{k}_{1}\right\rangle \right) \left\Vert \bar{\Psi}%
\right\Vert ^{2}\left\langle \bar{K}\right\rangle }\left\langle \bar{g}%
\right\rangle \right) \frac{\left\langle \bar{K}\right\rangle }{3\bar{K}_{0}}%
\left\langle \bar{g}\right\rangle \right)
\end{eqnarray*}%
where:%
\begin{eqnarray*}
\bar{D}\left( \bar{X}_{1}\right) &=&\left( \frac{\left\langle \hat{K}%
\right\rangle ^{2}\left\langle \hat{g}\right\rangle }{\sigma _{\hat{K}}^{2}}+%
\frac{1}{2}\right) \left\langle \hat{g}^{Bef}\right\rangle \frac{\bar{K}_{0}%
}{\left\langle \hat{K}\right\rangle } \\
&&-\left( \frac{\left\langle \bar{K}\right\rangle ^{2}\left\langle \bar{g}%
\right\rangle }{\sigma _{\hat{K}}^{2}}+\frac{1}{2}\right) \left(
\left\langle \underline{\hat{k}}^{B}\right\rangle ^{ef}\Delta \left( \hat{k}%
^{B}\left( \hat{X}_{1},\left\langle \bar{X}\right\rangle \right) A\right) +%
\frac{\Delta \bar{k}_{2}\left( \left\langle \bar{X}\right\rangle ,\bar{X}%
_{1}\right) }{\left( 1-\left\langle \bar{k}_{1}\right\rangle \right)
\left\Vert \bar{\Psi}\right\Vert ^{2}\left\langle \bar{K}\right\rangle }%
\left\langle \bar{g}\right\rangle \right) Z\left\langle \bar{g}\right\rangle
\end{eqnarray*}

\subsection*{A20.11 Compact expressions for field and capital}

\subsubsection*{A20.11.1 Investors}

We use the previous formula for $\hat{K}_{0}^{2}$, $\left\langle \hat{K}%
\right\rangle ^{2}$ to compute $\hat{K}\left[ \hat{X}_{1}\right] $ the
amount of capital in sector $\hat{X}_{1}$. We found: 
\begin{equation*}
\hat{K}_{0}^{2}=6\frac{\sigma _{\hat{K}}^{2}}{\hat{\mu}}\frac{\left\Vert 
\hat{\Psi}_{0}\left( \hat{X}_{1}\right) \right\Vert ^{2}}{\hat{g}^{2}\left( 
\bar{X}_{1}\right) \left( 5+\frac{\left\langle \hat{g}^{ef}\right\rangle }{%
\hat{g}\left( \bar{X}_{1}\right) }-\sqrt{\left( 1-\frac{\hat{g}^{ef}\left( 
\bar{X}_{1}\right) }{\hat{g}\left( \bar{X}_{1}\right) }\right) \left( 4-%
\frac{\hat{g}^{ef}\left( \bar{X}_{1}\right) }{\hat{g}\left( \bar{X}%
_{1}\right) }\right) }\right) }
\end{equation*}%
\begin{eqnarray*}
\left\langle \hat{K}\right\rangle ^{2} &=&\left( \frac{\left\langle \hat{g}%
\right\rangle }{6\left\langle \hat{g}^{ef}\right\rangle }\left( 2+\frac{%
\left\langle \hat{g}^{ef}\right\rangle }{\left\langle \hat{g}\right\rangle }-%
\sqrt{\left( 1-\frac{\left\langle \hat{g}^{ef}\right\rangle }{\left\langle 
\hat{g}\right\rangle }\right) \left( 4-\frac{\left\langle \hat{g}%
^{ef}\right\rangle }{\left\langle \hat{g}\right\rangle }\right) }\right)
\right) ^{2}\left\langle \hat{K}_{0}\right\rangle ^{2} \\
&\rightarrow &\frac{1}{6}\frac{\sigma _{\hat{K}}^{2}}{\hat{\mu}}\frac{\left(
\left( 2+\frac{\left\langle \hat{g}^{ef}\right\rangle }{\left\langle \hat{g}%
\right\rangle }-\sqrt{\left( 1-\frac{\left\langle \hat{g}^{ef}\right\rangle 
}{\left\langle \hat{g}\right\rangle }\right) \left( 4-\frac{\left\langle 
\hat{g}^{ef}\right\rangle }{\left\langle \hat{g}\right\rangle }\right) }%
\right) \right) ^{2}\left\Vert \hat{\Psi}_{0}\right\Vert ^{2}}{\left\langle 
\hat{g}^{ef}\right\rangle ^{2}\left( 5+\frac{\left\langle \hat{g}%
^{ef}\right\rangle }{\left\langle \hat{g}\right\rangle }-\sqrt{\left( 1-%
\frac{\left\langle \hat{g}^{ef}\right\rangle }{\left\langle \hat{g}%
\right\rangle }\right) \left( 4-\frac{\left\langle \hat{g}^{ef}\right\rangle 
}{\left\langle \hat{g}\right\rangle }\right) }\right) }
\end{eqnarray*}%
so that the formula for $\hat{K}\left[ \hat{X}_{1}\right] $ becomes: 
\begin{eqnarray}
\hat{K}\left[ \hat{X}_{1}\right] &=&\hat{K}_{\hat{X}}\left\Vert \hat{\Psi}%
\left( \hat{X}_{1}\right) \right\Vert ^{2}\simeq \hat{\mu}\frac{\hat{K}%
_{0}^{4}}{2\sigma _{\hat{K}}^{2}}\left( \frac{\hat{g}^{2}\left( \hat{X}%
_{1}\right) }{4}+\Delta \left( \hat{k}^{B}\left( \hat{X}_{1},\left\langle 
\bar{X}\right\rangle \right) A\right) Z\frac{\left\langle \bar{K}%
\right\rangle }{3\hat{K}_{0}}\left\langle \bar{g}\right\rangle -\frac{%
\left\langle \hat{K}\right\rangle \left\langle \hat{g}\right\rangle }{3\hat{K%
}_{0}}\left\langle \hat{g}^{ef}\right\rangle \right) \\
&=&\frac{\hat{\mu}}{2\sigma _{\hat{K}}^{2}}\left( 6\frac{\sigma _{\hat{K}%
}^{2}}{\hat{\mu}}\frac{\left\Vert \hat{\Psi}_{0}\left( \hat{X}_{1}\right)
\right\Vert ^{2}}{\hat{g}^{2}\left( \bar{X}_{1}\right) \left( 5+\frac{\hat{g}%
^{ef}\left( \bar{X}_{1}\right) }{\hat{g}\left( \bar{X}_{1}\right) }-\sqrt{%
\left( 1-\frac{\hat{g}^{ef}\left( \bar{X}_{1}\right) }{\hat{g}\left( \bar{X}%
_{1}\right) }\right) \left( 4-\frac{\hat{g}^{ef}\left( \bar{X}_{1}\right) }{%
\hat{g}\left( \bar{X}_{1}\right) }\right) }\right) }\right) ^{2}\left( \frac{%
\hat{g}^{2}\left( \hat{X}_{1}\right) }{4}-\frac{\left\langle \hat{g}%
\right\rangle \hat{g}^{ef}\left( \hat{X}_{1}\right) }{3}\right)  \notag \\
&\simeq &\frac{\hat{\mu}}{2\sigma _{\hat{K}}^{2}}\left( 6\frac{\sigma _{\hat{%
K}}^{2}}{\hat{\mu}}\frac{\left\Vert \hat{\Psi}_{0}\left( \hat{X}_{1}\right)
\right\Vert ^{2}}{\left( 5+\frac{\left\langle \hat{g}^{ef}\right\rangle }{%
\left\langle \hat{g}\right\rangle }-\sqrt{\left( 1-\frac{\left\langle \hat{g}%
^{ef}\right\rangle }{\left\langle \hat{g}\right\rangle }\right) \left( 4-%
\frac{\left\langle \hat{g}^{ef}\right\rangle }{\left\langle \hat{g}%
\right\rangle }\right) }\right) }\right) ^{2}\left( \frac{1}{4\hat{g}%
^{2}\left( \bar{X}_{1}\right) }-\frac{\left\langle \hat{g}\right\rangle \hat{%
g}^{ef}\left( \hat{X}_{1}\right) }{3\hat{g}^{4}\left( \bar{X}_{1}\right) }%
\right)
\end{eqnarray}%
and the associated field is:%
\begin{equation*}
\left\Vert \hat{\Psi}\left( \hat{X}_{1}\right) \right\Vert ^{2}=\frac{\hat{%
\mu}}{2\sigma _{\hat{K}}^{2}}\left( 6\frac{\sigma _{\hat{K}}^{2}}{\hat{\mu}}%
\frac{\left\Vert \hat{\Psi}_{0}\left( \hat{X}_{1}\right) \right\Vert ^{2}}{%
\left( 5+\frac{\left\langle \hat{g}^{ef}\right\rangle }{\left\langle \hat{g}%
\right\rangle }-\sqrt{\left( 1-\frac{\left\langle \hat{g}^{ef}\right\rangle 
}{\left\langle \hat{g}\right\rangle }\right) \left( 4-\frac{\left\langle 
\hat{g}^{ef}\right\rangle }{\left\langle \hat{g}\right\rangle }\right) }%
\right) }\right) ^{\frac{3}{2}}\left( \frac{1}{3\hat{g}^{2}\left( \bar{X}%
_{1}\right) }-\frac{\left\langle \hat{g}\right\rangle \hat{g}^{ef}\left( 
\hat{X}_{1}\right) }{2\hat{g}^{4}\left( \bar{X}_{1}\right) }\right)
\end{equation*}%
with:%
\begin{eqnarray*}
\hat{g}^{ef}\left( \hat{X}_{1}\right) &=&\frac{\left\langle \hat{K}%
\right\rangle }{\left\langle \hat{K}_{0}\right\rangle }\hat{g}^{ef}\left( 
\hat{X}_{1}\right) \\
&=&\frac{\left\langle \hat{g}\right\rangle }{6}\left( 2+\frac{\hat{g}%
^{ef}\left( \hat{X}_{1}\right) }{\left\langle \hat{g}\right\rangle }-\sqrt{%
\left( 1-\frac{\hat{g}^{ef}\left( \hat{X}_{1}\right) }{\left\langle \hat{g}%
\right\rangle }\right) \left( 4-\frac{\hat{g}^{ef}\left( \hat{X}_{1}\right) 
}{\left\langle \hat{g}\right\rangle }\right) }\right) \simeq \frac{3}{8}\hat{%
g}^{ef}\left( \hat{X}_{1}\right)
\end{eqnarray*}%
Then, replacing:%
\begin{equation*}
\frac{\hat{\mu}}{18\sigma _{\hat{K}}^{2}}\frac{\hat{K}\left[ \hat{X}_{1}%
\right] }{\left( \frac{\left\Vert \hat{\Psi}_{0}\left( \hat{X}_{1}\right)
\right\Vert ^{2}}{\left( 5+\frac{\left\langle \hat{g}^{ef}\right\rangle }{%
\left\langle \hat{g}\right\rangle }-\sqrt{\left( 1-\frac{\left\langle \hat{g}%
^{ef}\right\rangle }{\left\langle \hat{g}\right\rangle }\right) \left( 4-%
\frac{\left\langle \hat{g}^{ef}\right\rangle }{\left\langle \hat{g}%
\right\rangle }\right) }\right) }\right) ^{2}}=\left( \frac{1}{4\hat{g}%
^{2}\left( \bar{X}_{1}\right) }-\frac{\left\langle \hat{g}\right\rangle \hat{%
g}^{ef}\left( \hat{X}_{1}\right) }{3\hat{g}^{4}\left( \bar{X}_{1}\right) }%
\right)
\end{equation*}%
leads to:%
\begin{equation}
\hat{g}\left( \hat{X}_{1}\right) \simeq \frac{\left\Vert \hat{\Psi}%
_{0}\left( \hat{X}_{1}\right) \right\Vert ^{2}}{\left( 5+\frac{\left\langle 
\hat{g}^{ef}\right\rangle }{\left\langle \hat{g}\right\rangle }-\sqrt{\left(
1-\frac{\left\langle \hat{g}^{ef}\right\rangle }{\left\langle \hat{g}%
\right\rangle }\right) \left( 4-\frac{\left\langle \hat{g}^{ef}\right\rangle 
}{\left\langle \hat{g}\right\rangle }\right) }\right) \sqrt{\frac{2\hat{\mu}%
}{9\sigma _{\hat{K}}^{2}}\hat{K}\left[ \hat{X}_{1}\right] +\frac{4\hat{g}%
^{ef}\left( \hat{X}_{1}\right) }{3\left\langle \hat{g}\right\rangle ^{3}}%
\left( \frac{\left\Vert \hat{\Psi}_{0}\left( \hat{X}_{1}\right) \right\Vert
^{2}}{\left( 5+\frac{\left\langle \hat{g}^{ef}\right\rangle }{\left\langle 
\hat{g}\right\rangle }-\sqrt{\left( 1-\frac{\left\langle \hat{g}%
^{ef}\right\rangle }{\left\langle \hat{g}\right\rangle }\right) \left( 4-%
\frac{\left\langle \hat{g}^{ef}\right\rangle }{\left\langle \hat{g}%
\right\rangle }\right) }\right) }\right) ^{2}}}  \label{GH}
\end{equation}%
\bigskip

\subsubsection*{A20.11.2 Banks}

Similarly the value of $\bar{K}\left[ \bar{X}_{1}\right] $:%
\begin{equation*}
\bar{K}\left[ \bar{X}_{1}\right] =\bar{K}_{\bar{X}_{1}}\left\Vert \bar{\Psi}%
\left( \bar{X}_{1}\right) \right\Vert ^{2}\simeq \hat{\mu}\frac{\bar{K}%
_{0}^{4}}{2\sigma _{\hat{K}}^{2}}\left( \frac{\bar{g}^{2}\left( \bar{X}%
_{1}\right) }{4}-\frac{\left\langle \hat{K}\right\rangle \left\langle \hat{g}%
\right\rangle }{3\left\langle \bar{K}_{0}\right\rangle }\hat{g}^{Bef}\left( 
\bar{X}_{1}\right) \right)
\end{equation*}%
is writen by taking into account:%
\begin{eqnarray*}
\frac{\bar{K}_{0}}{\left\langle \hat{K}\right\rangle } &=&-\frac{%
\left\langle \hat{g}\right\rangle \hat{g}^{Bef}\left( \bar{X}_{1}\right) }{%
\bar{g}^{2}\left( \bar{X}_{1}\right) }+\sqrt{\left( \frac{\left\langle \hat{g%
}\right\rangle \hat{g}^{Bef}\left( \bar{X}_{1}\right) }{\bar{g}^{2}\left( 
\bar{X}_{1}\right) }\right) ^{2}+2\frac{\sigma _{\hat{K}}^{2}}{\bar{g}%
^{2}\left( \bar{X}_{1}\right) }\frac{\left\Vert \bar{\Psi}_{0}\right\Vert
^{2}}{\hat{\mu}\left\langle \hat{K}\right\rangle ^{2}}} \\
&\rightarrow &-\frac{\left\langle \hat{g}\right\rangle \hat{g}^{Bef}\left( 
\bar{X}_{1}\right) }{\bar{g}^{2}\left( \bar{X}_{1}\right) }+\sqrt{\left( 
\frac{\left\langle \hat{g}\right\rangle \hat{g}^{Bef}\left( \bar{X}%
_{1}\right) }{\bar{g}^{2}\left( \bar{X}_{1}\right) }\right) ^{2}+2\frac{%
\sigma _{\hat{K}}^{2}}{\bar{g}^{2}\left( \bar{X}_{1}\right) }\frac{%
\left\Vert \bar{\Psi}_{0}\right\Vert ^{2}}{\hat{\mu}\left\langle \hat{K}%
\right\rangle ^{2}}}\simeq \sqrt{\frac{196\left\Vert \bar{\Psi}%
_{0}\right\Vert ^{2}}{27\left\Vert \hat{\Psi}_{0}\right\Vert ^{2}}}\frac{%
\left\langle \hat{g}\right\rangle }{\bar{g}^{2}\left( \bar{X}_{1}\right) }-%
\frac{\left\langle \hat{g}\right\rangle \hat{g}^{Bef}\left( \bar{X}%
_{1}\right) }{\bar{g}^{2}\left( \bar{X}_{1}\right) }
\end{eqnarray*}%
and:%
\begin{equation*}
\hat{g}^{Bef}\left( \bar{X}_{1}\right) \rightarrow \frac{\left\langle \hat{K}%
\right\rangle }{\left\langle \bar{K}_{0}\right\rangle }\hat{g}^{Bef}\left( 
\bar{X}_{1}\right) =\frac{\hat{g}^{Bef}\left( \bar{X}_{1}\right) }{\sqrt{%
\frac{196\left\Vert \bar{\Psi}_{0}\right\Vert ^{2}}{27\left\Vert \hat{\Psi}%
_{0}\right\Vert ^{2}}}\frac{\left\langle \hat{g}\right\rangle }{\bar{g}%
^{2}\left( \bar{X}_{1}\right) }-\frac{\left\langle \hat{g}\right\rangle \hat{%
g}^{Bef}\left( \bar{X}_{1}\right) }{\bar{g}^{2}\left( \bar{X}_{1}\right) }}
\end{equation*}%
with:%
\begin{eqnarray*}
&&\frac{\bar{K}_{0}}{\left\langle \hat{K}_{0}\right\rangle }\rightarrow -%
\frac{3}{8}\frac{\left\langle \hat{g}\right\rangle \hat{g}^{Bef}\left( \bar{X%
}_{1}\right) }{\bar{g}^{2}\left( \bar{X}_{1}\right) }+\sqrt{\frac{%
\left\langle \hat{g}\right\rangle ^{2}}{3\bar{g}^{2}\left( \bar{X}%
_{1}\right) }\frac{\left\Vert \bar{\Psi}_{0}\right\Vert ^{2}\left( 5+\frac{%
\left\langle \hat{g}^{ef}\right\rangle }{\left\langle \hat{g}\right\rangle }-%
\sqrt{\left( 1-\frac{\left\langle \hat{g}^{ef}\right\rangle }{\left\langle 
\hat{g}\right\rangle }\right) \left( 4-\frac{\left\langle \hat{g}%
^{ef}\right\rangle }{\left\langle \hat{g}\right\rangle }\right) }\right) }{%
\left\Vert \hat{\Psi}_{0}\right\Vert ^{2}}} \\
&\rightarrow &-\frac{3}{8}\frac{\left\langle \hat{g}\right\rangle \hat{g}%
^{Bef}\left( \bar{X}_{1}\right) }{\bar{g}^{2}\left( \bar{X}_{1}\right) }+%
\sqrt{\frac{\left\langle \hat{g}\right\rangle ^{2}}{\bar{g}^{2}\left( \bar{X}%
_{1}\right) }\frac{\left\Vert \bar{\Psi}_{0}\right\Vert ^{2}\left( 1+\frac{3%
}{4}\frac{\left\langle \hat{g}^{ef}\right\rangle }{\left\langle \hat{g}%
\right\rangle }\right) }{\left\Vert \hat{\Psi}_{0}\right\Vert ^{2}}}
\end{eqnarray*}

Consequently, we find ultimately:

\begin{eqnarray}
\bar{K}_{\bar{X}_{1}}\left\Vert \bar{\Psi}\left( \bar{X}_{1}\right)
\right\Vert ^{2} &\simeq &\frac{\hat{\mu}}{2\sigma _{\hat{K}}^{2}}\left( 
\sqrt{\frac{\left\langle \hat{g}\right\rangle ^{2}}{\bar{g}^{2}\left( \bar{X}%
_{1}\right) }\frac{\left\Vert \bar{\Psi}_{0}\left( \hat{X}_{1}\right)
\right\Vert ^{2}\left( 1+\frac{3}{4}\frac{\left\langle \hat{g}%
^{ef}\right\rangle }{\left\langle \hat{g}\right\rangle }\right) }{\left\Vert 
\hat{\Psi}_{0}\left( \bar{X}_{1}\right) \right\Vert ^{2}}}-\frac{3}{8}\frac{%
\left\langle \hat{g}\right\rangle \hat{g}^{Bef}\left( \bar{X}_{1}\right) }{%
\bar{g}^{2}\left( \bar{X}_{1}\right) }\right) ^{4}  \label{KBP} \\
&&\times \left( 6\frac{\sigma _{\hat{K}}^{2}}{\hat{\mu}}\frac{\left\Vert 
\hat{\Psi}_{0}\left( \hat{X}_{1}\right) \right\Vert ^{2}}{\left( 5+\frac{%
\left\langle \hat{g}^{ef}\right\rangle }{\left\langle \hat{g}\right\rangle }-%
\sqrt{\left( 1-\frac{\left\langle \hat{g}^{ef}\right\rangle }{\left\langle 
\hat{g}\right\rangle }\right) \left( 4-\frac{\left\langle \hat{g}%
^{ef}\right\rangle }{\left\langle \hat{g}\right\rangle }\right) }\right) }%
\right) ^{2}\left( \frac{\bar{g}^{2}\left( \bar{X}_{1}\right) }{4}-\frac{%
\left\langle \hat{g}\right\rangle }{3}\hat{g}^{Bef}\left( \bar{X}_{1}\right)
\right)  \notag \\
&=&18\frac{\sigma _{\hat{K}}^{2}}{\hat{\mu}}\left( \frac{\sqrt{\left\langle 
\hat{g}\right\rangle ^{2}\left( 1+\frac{3}{4}\frac{\left\langle \hat{g}%
^{ef}\right\rangle }{\left\langle \hat{g}\right\rangle }\right) }\left\Vert 
\bar{\Psi}_{0}\left( \hat{X}_{1}\right) \right\Vert -\frac{3}{8}\left\langle 
\hat{g}\right\rangle \frac{\hat{g}^{Bef}\left( \bar{X}_{1}\right) }{\bar{g}%
\left( \bar{X}_{1}\right) }\left\Vert \hat{\Psi}_{0}\left( \bar{X}%
_{1}\right) \right\Vert }{\sqrt{5+\frac{\left\langle \hat{g}%
^{ef}\right\rangle }{\left\langle \hat{g}\right\rangle }-\sqrt{\left( 1-%
\frac{\left\langle \hat{g}^{ef}\right\rangle }{\left\langle \hat{g}%
\right\rangle }\right) \left( 4-\frac{\left\langle \hat{g}^{ef}\right\rangle 
}{\left\langle \hat{g}\right\rangle }\right) }}}\right) ^{4}  \notag \\
&&\times \left( \frac{1}{4\bar{g}^{2}\left( \bar{X}_{1}\right) }-\frac{%
\left\langle \hat{g}\right\rangle \hat{g}^{Bef}\left( \bar{X}_{1}\right) }{3%
\bar{g}^{4}\left( \bar{X}_{1}\right) }\right)  \notag
\end{eqnarray}%
and the field:%
\begin{equation*}
\left\Vert \bar{\Psi}\left( \bar{X}_{1}\right) \right\Vert ^{2}\simeq 18%
\frac{\sigma _{\hat{K}}^{2}}{\hat{\mu}}\left( \frac{\sqrt{\left\langle \hat{g%
}\right\rangle ^{2}\left( 1+\frac{3}{4}\frac{\left\langle \hat{g}%
^{ef}\right\rangle }{\left\langle \hat{g}\right\rangle }\right) }\left\Vert 
\bar{\Psi}_{0}\left( \hat{X}_{1}\right) \right\Vert -\frac{3}{8}\left\langle 
\hat{g}\right\rangle \frac{\hat{g}^{Bef}\left( \bar{X}_{1}\right) }{\bar{g}%
\left( \bar{X}_{1}\right) }\left\Vert \hat{\Psi}_{0}\left( \bar{X}%
_{1}\right) \right\Vert }{\sqrt{5+\frac{\left\langle \hat{g}%
^{ef}\right\rangle }{\left\langle \hat{g}\right\rangle }-\sqrt{\left( 1-%
\frac{\left\langle \hat{g}^{ef}\right\rangle }{\left\langle \hat{g}%
\right\rangle }\right) \left( 4-\frac{\left\langle \hat{g}^{ef}\right\rangle 
}{\left\langle \hat{g}\right\rangle }\right) }}}\right) ^{4}\left( \frac{1}{3%
\bar{g}^{2}\left( \bar{X}_{1}\right) }-\frac{\left\langle \hat{g}%
\right\rangle \hat{g}^{Bef}\left( \bar{X}_{1}\right) }{2\bar{g}^{4}\left( 
\bar{X}_{1}\right) }\right)
\end{equation*}%
allowing to express the return:%
\begin{equation}
\bar{g}\left( \bar{X}_{1}\right) \simeq \frac{\left( \sqrt{\left\langle \hat{%
g}\right\rangle ^{2}\left( 1+\frac{3}{4}\frac{\left\langle \hat{g}%
^{ef}\right\rangle }{\left\langle \hat{g}\right\rangle }\right) }\left\Vert 
\bar{\Psi}_{0}\left( \hat{X}_{1}\right) \right\Vert -\frac{3}{8}\left\langle 
\hat{g}\right\rangle \frac{\hat{g}^{Bef}\left( \bar{X}_{1}\right) }{%
\left\langle \bar{g}\right\rangle }\left\Vert \hat{\Psi}_{0}\left( \bar{X}%
_{1}\right) \right\Vert \right) ^{2}}{\left( 5+\frac{\left\langle \hat{g}%
^{ef}\right\rangle }{\left\langle \hat{g}\right\rangle }-\sqrt{\left( 1-%
\frac{\left\langle \hat{g}^{ef}\right\rangle }{\left\langle \hat{g}%
\right\rangle }\right) \left( 4-\frac{\left\langle \hat{g}^{ef}\right\rangle 
}{\left\langle \hat{g}\right\rangle }\right) }\right) ^{2}\sqrt{\frac{2\hat{%
\mu}}{9\sigma _{\hat{K}}^{2}}\hat{K}\left[ \hat{X}_{1}\right] +\frac{%
4\left\langle \hat{g}\right\rangle \hat{g}^{ef}\left( \hat{X}_{1}\right) }{%
3\left\langle \hat{g}\right\rangle ^{4}}\left( \frac{\left\Vert \hat{\Psi}%
_{0}\left( \hat{X}_{1}\right) \right\Vert ^{2}}{\left( 5+\frac{\left\langle 
\hat{g}^{ef}\right\rangle }{\left\langle \hat{g}\right\rangle }-\sqrt{\left(
1-\frac{\left\langle \hat{g}^{ef}\right\rangle }{\left\langle \hat{g}%
\right\rangle }\right) \left( 4-\frac{\left\langle \hat{g}^{ef}\right\rangle 
}{\left\langle \hat{g}\right\rangle }\right) }\right) }\right) ^{2}}}
\label{GB}
\end{equation}%
Finally, we also need the ratio $Z\left( \bar{X}_{1}\right) $:

\begin{equation*}
Z\left( \bar{X}_{1}\right) =\frac{\bar{K}_{\bar{X}}\left\Vert \bar{\Psi}%
\left( \bar{X}_{1}\right) \right\Vert ^{2}}{\hat{K}_{\hat{X}}\left\Vert \hat{%
\Psi}\left( \hat{X}_{1}\right) \right\Vert ^{2}}\simeq \frac{\left( \frac{1}{%
\bar{g}^{2}\left( \hat{X}_{1}\right) }\left( \frac{\left\Vert \bar{\Psi}%
_{0}\left( \bar{X}_{1}\right) \right\Vert ^{2}}{\hat{\mu}}-\bar{D}\left( 
\bar{X}_{1}\right) \right) \right) ^{2}\left( \frac{\hat{g}^{2}\left( \hat{X}%
_{1}\right) }{4}-\frac{\left\langle \hat{K}\right\rangle \left\langle \hat{g}%
\right\rangle }{3\left\langle \bar{K}_{0}\right\rangle }\left\langle \hat{g}%
^{Bef}\right\rangle \right) }{\left( \frac{1}{\hat{g}^{2}\left( \hat{X}%
_{1}\right) }\left( \frac{\left\Vert \hat{\Psi}_{0}\left( \hat{X}_{1}\right)
\right\Vert ^{2}}{\hat{\mu}}-\hat{D}\left( \hat{X}_{1}\right) \right)
\right) ^{2}\left( \frac{\hat{g}^{2}\left( \hat{X}_{1}\right) }{4}-\frac{%
\left\langle \hat{g}\right\rangle \hat{g}^{ef}\left( \hat{X}_{1}\right) }{3}%
\right) }
\end{equation*}%
that can be detailed using again:%
\begin{eqnarray*}
&&\frac{\bar{K}_{0}}{\hat{K}_{0}} \\
&\rightarrow &-\frac{3}{8}\frac{\hat{g}\left( \hat{X}_{1}\right)
\left\langle \hat{g}^{Bef}\right\rangle }{\bar{g}^{2}\left( \bar{X}%
_{1}\right) }+\sqrt{\frac{\hat{g}^{2}\left( \hat{X}_{1}\right) }{3\bar{g}%
^{2}\left( \bar{X}_{1}\right) }\frac{\left\Vert \bar{\Psi}_{0}\right\Vert
^{2}\left( 5+\frac{\left\langle \hat{g}^{ef}\right\rangle }{\hat{g}\left( 
\hat{X}_{1}\right) }-\sqrt{\left( 1-\frac{\left\langle \hat{g}%
^{ef}\right\rangle }{\hat{g}\left( \hat{X}_{1}\right) }\right) \left( 4-%
\frac{\left\langle \hat{g}^{ef}\right\rangle }{\hat{g}\left( \hat{X}%
_{1}\right) }\right) }\right) }{\left\Vert \hat{\Psi}_{0}\right\Vert ^{2}}}
\\
&\rightarrow &-\frac{3}{8}\frac{\hat{g}\left( \hat{X}_{1}\right)
\left\langle \hat{g}^{Bef}\right\rangle }{\bar{g}^{2}\left( \bar{X}%
_{1}\right) }+\sqrt{\frac{\hat{g}^{2}\left( \hat{X}_{1}\right) }{\bar{g}%
^{2}\left( \bar{X}_{1}\right) }\frac{\left\Vert \bar{\Psi}_{0}\right\Vert
^{2}\left( 1+\frac{3}{4}\frac{\left\langle \hat{g}^{ef}\right\rangle }{\hat{g%
}\left( \hat{X}_{1}\right) }\right) }{\left\Vert \hat{\Psi}_{0}\right\Vert
^{2}}}
\end{eqnarray*}%
\begin{eqnarray*}
Z &=&\frac{\left\langle \bar{K}_{0}\right\rangle ^{4}\left( \frac{1}{4}-%
\frac{\left\langle \hat{K}\right\rangle }{3\bar{K}_{0}}\frac{\left\langle 
\hat{g}^{Bef}\right\rangle }{\bar{g}\left( \bar{X}_{1}\right) }\right)
\left\langle \bar{g}\right\rangle ^{2}}{\left\langle \hat{K}%
_{0}\right\rangle ^{4}\left( \frac{1}{4}-\frac{\left\langle \hat{K}%
\right\rangle }{3\hat{K}_{0}}\frac{\left\langle \hat{g}^{ef}\right\rangle }{%
\hat{g}\left( \hat{X}_{1}\right) }\right) \left\langle \hat{g}\right\rangle
^{2}} \\
&\rightarrow &\frac{\bar{K}_{0}^{4}\left( \frac{1}{4}-\frac{1}{3\left( \sqrt{%
\frac{196\left\Vert \bar{\Psi}_{0}\left( \bar{X}_{1}\right) \right\Vert ^{2}%
}{27\left\Vert \hat{\Psi}_{0}\left( \hat{X}_{1}\right) \right\Vert ^{2}}}%
\frac{\hat{g}\left( \hat{X}_{1}\right) }{\bar{g}\left( \bar{X}_{1}\right) }-%
\frac{\hat{g}\left( \hat{X}_{1}\right) \left\langle \hat{g}%
^{Bef}\right\rangle }{\bar{g}^{2}\left( \bar{X}_{1}\right) }\right) }\frac{%
\left\langle \hat{g}^{Bef}\right\rangle }{\bar{g}\left( \bar{X}_{1}\right) }%
\right) \bar{g}^{2}\left( \bar{X}_{1}\right) }{\hat{K}_{0}^{4}\left( \frac{1%
}{4}-\frac{1}{18}\left( 2+\frac{\left\langle \hat{g}^{ef}\right\rangle }{%
\hat{g}\left( \hat{X}_{1}\right) }-\sqrt{\left( 1-\frac{\left\langle \hat{g}%
^{ef}\right\rangle }{\hat{g}\left( \hat{X}_{1}\right) }\right) \left( 4-%
\frac{\left\langle \hat{g}^{ef}\right\rangle }{\hat{g}\left( \hat{X}%
_{1}\right) }\right) }\right) \right) \hat{g}^{2}\left( \hat{X}_{1}\right) }%
\simeq \frac{\bar{K}_{0}^{4}\left( \frac{1}{4}-\sqrt{\frac{3\left\Vert \hat{%
\Psi}_{0}\right\Vert ^{2}}{196\left\Vert \bar{\Psi}_{0}\right\Vert ^{2}}}%
\frac{\left\langle \hat{g}^{Bef}\right\rangle }{\hat{g}\left( \hat{X}%
_{1}\right) }\right) \bar{g}^{2}\left( \bar{X}_{1}\right) }{\hat{K}%
_{0}^{4}\left( \frac{1}{4}-\frac{1}{18}\frac{5}{4}\frac{\left\langle \hat{g}%
^{ef}\right\rangle }{\hat{g}\left( \hat{X}_{1}\right) }\right) \hat{g}%
^{2}\left( \hat{X}_{1}\right) }
\end{eqnarray*}%
This simplifies in frst apprpximation:%
\begin{equation*}
Z\left( \bar{X}_{1}\right) \simeq \left( 1+\frac{3}{4}\frac{\left\langle 
\hat{g}^{ef}\right\rangle }{\hat{g}\left( \hat{X}_{1}\right) }\right) \frac{%
\left\Vert \bar{\Psi}_{0}\left( \bar{X}_{1}\right) \right\Vert ^{2}}{%
\left\Vert \hat{\Psi}_{0}\left( \hat{X}_{1}\right) \right\Vert ^{2}}-\left( 
\frac{3}{4}\frac{\left\langle \hat{g}^{Bef}\right\rangle }{\bar{g}\left( 
\bar{X}_{1}\right) }+\frac{1}{2}\frac{\left\langle \hat{g}%
^{Bef}\right\rangle }{\hat{g}\left( \hat{X}_{1}\right) }\right) \sqrt{\frac{%
\left\Vert \bar{\Psi}_{0}\left( \bar{X}_{1}\right) \right\Vert ^{2}}{%
\left\Vert \hat{\Psi}_{0}\right\Vert ^{2}}}
\end{equation*}

\subsubsection*{A20.11.3 Alternate expression for field and capital}

We present the forms that will be useful to derive the capital equation:%
\begin{eqnarray}
\left\Vert \hat{\Psi}\left( \hat{X}_{1}\right) \right\Vert ^{2} &\simeq &%
\hat{\mu}\frac{\hat{K}_{0}^{3}}{\sigma _{\hat{K}}^{2}}\left( \frac{\hat{g}%
^{2}\left( \hat{X}_{1}\right) }{3}+\Delta \left( \hat{k}^{B}\left( \hat{X}%
_{1},\left\langle \bar{X}\right\rangle \right) A\right) Z\frac{\left\langle 
\bar{K}\right\rangle }{2\hat{K}_{0}}\left\langle \bar{g}\right\rangle -\frac{%
\left\langle \hat{K}\right\rangle }{2\hat{K}_{0}}\left\langle \hat{g}%
\right\rangle \left\langle \hat{g}^{ef}\right\rangle \right) \\
&=&\frac{\hat{\mu}}{\sigma _{\hat{K}}^{2}}\left( 2\frac{\sigma _{\hat{K}}^{2}%
}{\hat{g}^{2}\left( \hat{X}_{1}\right) }\left( \frac{\left\Vert \hat{\Psi}%
_{0}\left( \hat{X}_{1}\right) \right\Vert ^{2}}{\hat{\mu}}-D\left( \hat{X}%
_{1}\right) \right) \right) ^{\frac{3}{2}}\left( \frac{\hat{g}^{2}\left( 
\hat{X}_{1}\right) }{4}-\frac{\left\langle \hat{g}\right\rangle \hat{g}%
^{ef}\left( \hat{X}_{1}\right) }{4}\right)  \notag
\end{eqnarray}%
\begin{eqnarray}
\hat{K}\left[ \hat{X}_{1}\right] &=&\hat{K}_{\hat{X}}\left\Vert \hat{\Psi}%
\left( \hat{X}_{1}\right) \right\Vert ^{2}\simeq \hat{\mu}\frac{\hat{K}%
_{0}^{4}}{2\sigma _{\hat{K}}^{2}}\left( \frac{\hat{g}^{2}\left( \hat{X}%
_{1}\right) }{4}-\frac{\left\langle \hat{g}\right\rangle \hat{g}^{ef}\left( 
\hat{X}_{1}\right) }{4}\right)  \label{KP} \\
&=&\frac{\hat{\mu}}{2\sigma _{\hat{K}}^{2}}\left( 2\frac{\sigma _{\hat{K}%
}^{2}}{\hat{g}^{2}\left( \hat{X}_{1}\right) }\left( \frac{\left\Vert \hat{%
\Psi}_{0}\left( \hat{X}_{1}\right) \right\Vert ^{2}}{\hat{\mu}}-D\left( \hat{%
X}_{1}\right) \right) \right) ^{2}\left( \frac{\hat{g}^{2}\left( \hat{X}%
_{1}\right) }{4}-\frac{\left\langle \hat{g}\right\rangle \hat{g}^{ef}\left( 
\hat{X}_{1}\right) }{4}\right)  \notag
\end{eqnarray}%
\begin{equation}
\hat{K}_{\hat{X}_{1}}=\frac{\hat{K}\left[ \hat{X}_{1}\right] }{\left\Vert 
\hat{\Psi}\left( \hat{X}_{1}\right) \right\Vert ^{2}}=\frac{\sqrt{2\frac{%
\sigma _{\hat{K}}^{2}}{\hat{g}^{2}\left( \hat{X}_{1}\right) }\left( \frac{%
\left\Vert \hat{\Psi}_{0}\left( \hat{X}_{1}\right) \right\Vert ^{2}}{\hat{\mu%
}}-D\left( \hat{X}_{1}\right) \right) }\left( \frac{\hat{g}^{2}\left( \hat{X}%
_{1}\right) }{4}-\frac{\left\langle \hat{g}\right\rangle \hat{g}^{ef}\left( 
\hat{X}_{1}\right) }{4}\right) }{2\left( \frac{\hat{g}^{2}\left( \hat{X}%
_{1}\right) }{4}-\frac{3\left\langle \hat{g}\right\rangle \hat{g}^{ef}\left( 
\hat{X}_{1}\right) }{8}\right) }  \label{KPS}
\end{equation}%
\bigskip

\subsection*{A20.12 Computation of $\hat{g}$}

While deriving the capital expression we found the return (\ref{GH}) and (%
\ref{GB}). We present an alternative formulation that will be usefl. Gvn (%
\ref{KP}):%
\begin{equation}
\hat{K}\left[ \hat{X}_{1}\right] =\frac{2\sigma _{\hat{K}}^{2}}{\hat{\mu}}%
\left( \frac{\left\Vert \hat{\Psi}_{0}\left( \hat{X}_{1}\right) \right\Vert
^{2}-\hat{\mu}D\left( \hat{X}_{1}\right) }{\hat{g}^{2}\left( \hat{X}%
_{1}\right) }\right) ^{2}\left( \frac{\hat{g}^{2}\left( \hat{X}_{1}\right) }{%
4}-\frac{\left\langle \hat{g}\right\rangle \hat{g}^{ef}\left( \hat{X}%
_{1}\right) }{4}\right)
\end{equation}%
\begin{equation*}
D\left( \hat{X}_{1}\right) =\left( \frac{\left\langle \hat{K}\right\rangle
^{2}\left\langle \hat{g}\right\rangle ^{2}}{\sigma _{\hat{K}}^{2}}+\frac{%
\left\langle \hat{g}\right\rangle }{2}\right) \left( \frac{\underline{\hat{k}%
}^{ef}\left( \hat{X}_{1}\right) }{\underline{\hat{k}}_{2}^{Bef}\left(
\left\langle \hat{X}\right\rangle \right) }-\frac{6\underline{\hat{k}}%
_{2}^{Bef}}{2+\underline{\hat{k}}_{2}^{Bef}-\sqrt{\left( 2+\underline{\hat{k}%
}\right) ^{2}-\underline{\hat{k}}_{2}^{Bef}}}\right) \underline{\hat{k}}%
_{2}^{Bef}
\end{equation*}%
we can write the followng equation for $\hat{g}\left( \hat{X}_{1}\right) $: 
\begin{equation*}
0=\frac{\hat{\mu}\hat{K}\left[ \hat{X}_{1}\right] }{2\sigma _{\hat{K}}^{2}}%
\left( \hat{g}^{2}\left( \hat{X}_{1}\right) \right) ^{2}-\left( \left\Vert 
\hat{\Psi}_{0}\left( \hat{X}_{1}\right) \right\Vert ^{2}-\hat{\mu}D\left( 
\hat{X}_{1}\right) \right) ^{2}\frac{\hat{g}^{2}\left( \hat{X}_{1}\right) }{4%
}+\left( \left\Vert \hat{\Psi}_{0}\left( \hat{X}_{1}\right) \right\Vert ^{2}-%
\hat{\mu}D\left( \hat{X}_{1}\right) \right) ^{2}\frac{\left\langle \hat{g}%
\right\rangle \hat{g}^{ef}\left( \hat{X}_{1}\right) }{4}
\end{equation*}

with solution:%
\begin{equation*}
\frac{\left( \left\Vert \hat{\Psi}_{0}\left( \hat{X}_{1}\right) \right\Vert
^{2}-\hat{\mu}D\left( \hat{X}_{1}\right) \right) ^{2}-\sqrt{\left(
\left\Vert \hat{\Psi}_{0}\left( \hat{X}_{1}\right) \right\Vert ^{2}-\hat{\mu}%
D\left( \hat{X}_{1}\right) \right) ^{4}-\frac{\hat{\mu}\hat{K}\left[ \hat{X}%
_{1}\right] }{2\sigma _{\hat{K}}^{2}}\left( \left\Vert \hat{\Psi}_{0}\left( 
\hat{X}_{1}\right) \right\Vert ^{2}-\hat{\mu}D\left( \hat{X}_{1}\right)
\right) ^{2}\left\langle \hat{g}\right\rangle \hat{g}^{ef}\left( \hat{X}%
_{1}\right) }}{\frac{\hat{\mu}\hat{K}\left[ \hat{X}_{1}\right] }{\sigma _{%
\hat{K}}^{2}}}
\end{equation*}%
and in first approximation:%
\begin{eqnarray*}
\hat{K}\left[ \hat{X}_{1}\right] &=&\frac{2\sigma _{\hat{K}}^{2}}{\hat{\mu}}%
\left( \frac{\left\Vert \hat{\Psi}_{0}\left( \hat{X}_{1}\right) \right\Vert
^{2}-\hat{\mu}D\left( \hat{X}_{1}\right) }{\hat{g}^{2}\left( \hat{X}%
_{1}\right) }\right) ^{2}\left( \frac{\hat{g}^{2}\left( \hat{X}_{1}\right) }{%
4}-\frac{\left\langle \hat{g}\right\rangle \hat{g}^{ef}\left( \hat{X}%
_{1}\right) }{4}\right) \\
&\simeq &\frac{\sigma _{\hat{K}}^{2}}{\hat{\mu}2\hat{g}^{2}\left( \hat{X}%
_{1}\right) }\left( \left\Vert \hat{\Psi}_{0}\left( \hat{X}_{1}\right)
\right\Vert ^{2}-\hat{\mu}D\left( \hat{X}_{1}\right) \right) ^{2}\left( 1-%
\frac{\hat{g}^{ef}\left( \hat{X}_{1}\right) }{\left\langle \hat{g}%
\right\rangle }\right)
\end{eqnarray*}%
equivalently:%
\begin{equation}
\hat{g}\left( \hat{X}_{1}\right) \simeq \frac{\left( \left\Vert \hat{\Psi}%
_{0}\left( \hat{X}_{1}\right) \right\Vert ^{2}-\hat{\mu}D\left( \hat{X}%
_{1}\right) \right) \sqrt{1-\frac{\hat{g}^{ef}\left( \hat{X}_{1}\right) }{%
\left\langle \hat{g}\right\rangle }}}{\sqrt{\frac{2\hat{\mu}\hat{K}\left[ 
\hat{X}_{1}\right] }{\sigma _{\hat{K}}^{2}}}}
\end{equation}

\section*{Appendix 21 Solving for Investors' returns}

Recall the formula for investors returns equation:%
\begin{eqnarray*}
\hat{f}\left( \hat{X}^{\prime }\right) &=&\left( 1+\underline{\hat{k}}%
_{2}\left( \hat{X}^{\prime }\right) +\kappa \left[ \frac{\underline{\hat{k}}%
_{2}^{B}}{1+\bar{k}}\right] \left( \hat{X}^{\prime }\right) \right) \left(
\Delta \left( \hat{X},\hat{X}^{\prime }\right) -\frac{\hat{K}^{\prime }\hat{k%
}_{1}\left( \hat{X}^{\prime },\hat{X}\right) \left\vert \hat{\Psi}\left( 
\hat{K}^{\prime },\hat{X}^{\prime }\right) \right\vert ^{2}}{1+\underline{%
\hat{k}}\left( \hat{X}^{\prime }\right) +\underline{\hat{k}}_{1}^{B}\left( 
\bar{X}^{\prime }\right) +\kappa \left[ \frac{\underline{\hat{k}}_{2}^{B}}{1+%
\bar{k}}\right] \left( \hat{X}^{\prime }\right) }\right) ^{-1} \\
&&\times \frac{k_{1}\left( \hat{X}^{\prime },X\right) \hat{K}^{\prime }}{1+%
\underline{k}\left( \hat{X}^{\prime }\right) +\underline{k}_{1}^{\left(
B\right) }\left( \bar{X}^{\prime }\right) +\kappa \left[ \frac{\underline{k}%
_{2}^{B}}{1+\bar{k}}\right] \left( X^{\prime }\right) }\frac{f_{1}^{\prime
}\left( \hat{K},\hat{X},\Psi ,\hat{\Psi}\right) }{1+\underline{k}_{2}\left( 
\hat{X}^{\prime }\right) +\kappa \left[ \frac{\underline{k}_{2}^{B}}{1+\bar{k%
}}\right] \left( X^{\prime }\right) }
\end{eqnarray*}%
and its equivalent in term of $\hat{g}\left( \hat{X}\right) $:%
\begin{eqnarray*}
&&\left( \Delta \left( \hat{X},\hat{X}^{\prime }\right) -\frac{\hat{K}%
^{\prime }\hat{k}_{1}\left( \hat{X}^{\prime },\hat{X}\right) \left\vert \hat{%
\Psi}\left( \hat{K}^{\prime },\hat{X}^{\prime }\right) \right\vert ^{2}}{1+%
\underline{\hat{k}}\left( \hat{X}^{\prime }\right) +\underline{\hat{k}}%
_{1}^{B}\left( \bar{X}^{\prime }\right) +\kappa \left[ \frac{\underline{\hat{%
k}}_{2}^{B}}{1+\bar{k}}\right] \left( \hat{X}^{\prime }\right) }\right) 
\frac{1}{1+\underline{\hat{k}}_{2}\left( \hat{X}^{\prime }\right) +\kappa %
\left[ \frac{\underline{\hat{k}}_{2}^{B}}{1+\bar{k}}\right] \left( \hat{X}%
^{\prime }\right) }\left( 1-\hat{M}\right) \hat{g}\left( \hat{X}^{\prime
}\right) \\
&=&\frac{k_{1}\left( \hat{X}^{\prime },X\right) \hat{K}^{\prime }}{1+%
\underline{k}\left( \hat{X}^{\prime }\right) +\underline{k}_{1}^{\left(
B\right) }\left( \bar{X}^{\prime }\right) +\kappa \left[ \frac{\underline{k}%
_{2}^{B}}{1+\bar{k}}\right] \left( X^{\prime }\right) }\frac{f_{1}^{\prime
}\left( \hat{K},\hat{X},\Psi ,\hat{\Psi}\right) }{1+\underline{k}_{2}\left( 
\hat{X}^{\prime }\right) +\kappa \left[ \frac{\underline{k}_{2}^{B}}{1+\bar{k%
}}\right] \left( X^{\prime }\right) }
\end{eqnarray*}

\subsection*{A21.1 Intermediate quantities}

\subsubsection{A21.1.1 Ratios}

We need to define the various ratios:%
\begin{equation}
\delta =\frac{k\left( X\right) \hat{K}_{X}\left\vert \hat{\Psi}\left( \hat{X}%
\right) \right\vert ^{2}}{\left( k_{1}^{B}\left( X\right) +\frac{\kappa
k_{2}^{B}\left( X\right) }{1+\bar{k}\left( \bar{X}\right) \frac{\left\Vert 
\bar{\Psi}\right\Vert ^{2}}{\left\vert \bar{\Psi}_{0}\left( \bar{X}\right)
\right\vert ^{2}}}\right) \bar{K}_{X}\left\vert \bar{\Psi}\left( \bar{X}%
\right) \right\vert ^{2}}=\frac{k\left( X\right) \hat{K}\left[ \hat{X}\right]
}{\left( k_{1}^{B}\left( X\right) +\frac{\kappa k_{2}^{B}\left( X\right) }{1+%
\bar{k}\frac{\left\Vert \bar{\Psi}\right\Vert ^{2}}{\left\vert \bar{\Psi}%
_{0}\left( \hat{X}\right) \right\vert ^{2}}}\right) \bar{K}_{X}\left[ \bar{X}%
\right] }  \label{DT}
\end{equation}%
is the ratio of investors invested capital versus banks invested capital in
one sector.

We estimatr the following coefficients:

\begin{equation*}
\underline{k}\left( X\right) =\int \frac{k\left( X\right) }{\left\langle
K\right\rangle }\frac{\hat{K}_{X^{\prime }}\left\vert \hat{\Psi}\left( \hat{X%
}^{\prime }\right) \right\vert ^{2}}{\left\Vert \Psi _{0}\right\Vert ^{2}}%
\simeq \frac{k\left( X\right) }{\left\langle K\right\rangle }\hat{K}_{X}%
\frac{\left\vert \hat{\Psi}\left( \hat{X}\right) \right\vert ^{2}}{%
\left\vert \Psi _{0}\left( X\right) \right\vert ^{2}}
\end{equation*}%
\begin{equation*}
\underline{k}_{1}^{\left( B\right) }\left( X^{\prime }\right) =\int \frac{%
k_{1}^{B}\left( X^{\prime }\right) }{\left\langle K\right\rangle }\bar{K}_{%
\bar{X}}\frac{\left\vert \bar{\Psi}\left( \bar{X}\right) \right\vert ^{2}}{%
\left\vert \Psi _{0}\left( X^{\prime }\right) \right\vert ^{2}}\simeq \frac{%
k_{1}^{B}\left( X^{\prime }\right) }{\left\langle K\right\rangle }\bar{K}%
_{X^{\prime }}\frac{\left\vert \bar{\Psi}\left( X^{\prime }\right)
\right\vert ^{2}}{\left\vert \Psi _{0}\left( X\right) \right\vert ^{2}}
\end{equation*}%
\begin{equation*}
\kappa \underline{k}_{2}^{\left( B\right) }\left( X^{\prime }\right) =\int
\kappa \frac{k_{2}^{B}\left( X^{\prime }\right) }{\left\langle
K\right\rangle }\bar{K}_{\bar{X}}\frac{\left\vert \bar{\Psi}\left( \bar{X}%
\right) \right\vert ^{2}}{\left\vert \Psi _{0}\left( X^{\prime }\right)
\right\vert ^{2}}\simeq \kappa k_{2}^{B}\left( X^{\prime }\right) \bar{K}%
_{X^{\prime }}\frac{\left\vert \bar{\Psi}\left( X^{\prime }\right)
\right\vert ^{2}}{\left\vert \Psi _{0}\left( X\right) \right\vert ^{2}}
\end{equation*}%
\begin{equation*}
\kappa \left[ \frac{\underline{k}_{2}^{B}}{1+\bar{k}}\right] \left(
X^{\prime }\right) \simeq \frac{\kappa k_{2}^{B}\left( X^{\prime }\right) 
\bar{K}_{X^{\prime }}\frac{\left\vert \bar{\Psi}\left( X^{\prime }\right)
\right\vert ^{2}}{\left\vert \Psi _{0}\left( X\right) \right\vert ^{2}}}{1+%
\frac{\bar{k}}{\left\langle \bar{K}\right\rangle }\frac{\left\langle \bar{K}%
\right\rangle \left\Vert \bar{\Psi}\right\Vert ^{2}}{\left\Vert \bar{\Psi}%
\right\Vert ^{2}}}\simeq \frac{\kappa k_{2}^{B}\left( X^{\prime }\right) 
\bar{K}_{X^{\prime }}\frac{\left\vert \bar{\Psi}\left( X^{\prime }\right)
\right\vert ^{2}}{\left\vert \Psi _{0}\left( X\right) \right\vert ^{2}}}{1+%
\underline{\bar{k}}}
\end{equation*}%
We also define the averages%
\begin{equation*}
\underline{\bar{k}}=\left\langle \bar{k}\left( \bar{X}\right) \right\rangle
\end{equation*}%
and the coefficients $k\left( X\right) $ by: 
\begin{equation*}
\underline{k}\left( X\right) =k\left( X\right) \frac{\hat{K}\left[ \hat{X}%
\right] }{\left\langle K\right\rangle \left\vert \Psi _{0}\left( X\right)
\right\vert ^{2}}
\end{equation*}%
so that we can replace $\hat{K}\left[ \hat{X}\right] $ as a function of $%
\underline{k}\left( X\right) $ 
\begin{equation*}
\hat{K}\left[ \hat{X}\right] =\underline{k}\left( X\right) \frac{%
\left\langle K\right\rangle \left\vert \Psi _{0}\left( X\right) \right\vert
^{2}}{k\left( X\right) }
\end{equation*}%
with the factor:%
\begin{equation*}
k=\frac{\left\langle K\right\rangle \left\vert \Psi _{0}\left( X\right)
\right\vert ^{2}}{k\left( X\right) }
\end{equation*}%
So that%
\begin{equation*}
\hat{K}\left[ \hat{X}\right] =\frac{\underline{k}\left( X\right) }{k\left(
X\right) }\left\langle K\right\rangle \left\vert \Psi _{0}\left( X\right)
\right\vert ^{2}\rightarrow \frac{\underline{k}\left( X\right) }{k}
\end{equation*}%
Moreover, $k\left( X\right) $ can be expressed as a function of the averages
capital: Note that given the formula (\ref{DT}) 
\begin{equation*}
\underline{k}\left( X\right) =\delta \underline{k}^{B}\left( X\right)
\end{equation*}%
we have:%
\begin{eqnarray*}
\underline{k}\left( X\right) &=&\frac{k\left( X\right) }{\left\langle
K\right\rangle }\hat{K}_{X}\frac{\left\vert \hat{\Psi}\left( \hat{X}\right)
\right\vert ^{2}}{\left\vert \Psi _{0}\left( X\right) \right\vert ^{2}} \\
&=&\delta \left( \frac{k_{1}^{B}\left( X\right) }{\left\langle
K\right\rangle }\bar{K}_{X}\frac{\left\vert \bar{\Psi}\left( \hat{X}\right)
\right\vert ^{2}}{\left\vert \Psi _{0}\left( X\right) \right\vert ^{2}}+%
\frac{\kappa \frac{k_{2}^{B}\left( X\right) }{\left\langle K\right\rangle }%
\bar{K}_{X}\frac{\left\vert \bar{\Psi}\left( \hat{X}\right) \right\vert ^{2}%
}{\left\vert \Psi _{0}\left( X\right) \right\vert ^{2}}}{1+\bar{k}\left( 
\bar{X}\right) \frac{\left\Vert \bar{\Psi}\right\Vert ^{2}}{\left\vert \bar{%
\Psi}\left( \bar{X}\right) \right\vert ^{2}}}\right) \\
&=&\delta \left( k_{1}^{B}\left( X\right) +\frac{\kappa k_{2}^{B}\left(
X\right) }{1+\bar{k}\left( \bar{X}\right) \frac{\left\Vert \bar{\Psi}%
\right\Vert ^{2}}{\left\vert \bar{\Psi}\left( \bar{X}\right) \right\vert ^{2}%
}}\right) \frac{\bar{K}_{X}}{\left\langle K\right\rangle }\frac{\left\vert 
\bar{\Psi}\left( \bar{X}\right) \right\vert ^{2}}{\left\vert \Psi _{0}\left(
X\right) \right\vert ^{2}} \\
&\simeq &\delta \left( k_{1}^{B}\left( X\right) +\frac{\kappa
k_{2}^{B}\left( X\right) }{1+\bar{k}\left( \bar{X}\right) \frac{\left\Vert 
\bar{\Psi}\right\Vert ^{2}}{\left\vert \bar{\Psi}_{0}\left( \bar{X}\right)
\right\vert ^{2}}}\right) \frac{\bar{K}_{X}}{\left\langle K\right\rangle }%
\frac{\left\vert \bar{\Psi}\left( \bar{X}\right) \right\vert ^{2}}{%
\left\vert \Psi _{0}\left( X\right) \right\vert ^{2}}
\end{eqnarray*}%
Ultimately, we can write a relation between $\delta $ and $Z\left( \bar{X}%
\right) =\frac{\bar{K}\left[ \bar{X}\right] }{\hat{K}\left[ \hat{X}\right] }$%
: 
\begin{eqnarray*}
\delta &=&\frac{k_{1}^{B}\left( X\right) +\frac{\kappa k_{2}^{B}\left(
X\right) }{1+\bar{k}\left( \bar{X}\right) \frac{\left\Vert \bar{\Psi}%
\right\Vert ^{2}}{\left\vert \bar{\Psi}_{0}\left( \bar{X}\right) \right\vert
^{2}}}}{k\left( X\right) }\frac{\bar{K}_{X}}{\hat{K}_{X}}\frac{\left\vert 
\bar{\Psi}\left( \bar{X}\right) \right\vert ^{2}}{\left\vert \hat{\Psi}%
\left( \hat{X}\right) \right\vert ^{2}} \\
&=&\frac{k_{1}^{B}\left( X\right) +\frac{\kappa k_{2}^{B}\left( X\right) }{1+%
\bar{k}\left( \bar{X}\right) \frac{\left\Vert \bar{\Psi}\right\Vert ^{2}}{%
\left\vert \bar{\Psi}_{0}\left( \bar{X}\right) \right\vert ^{2}}}}{k\left(
X\right) }\frac{\bar{K}\left[ \bar{X}\right] }{\hat{K}\left[ \hat{X}\right] }
\\
&=&\frac{k_{1}^{B}\left( X\right) +\frac{\kappa k_{2}^{B}\left( X\right) }{1+%
\bar{k}\left( \bar{X}\right) \frac{\left\Vert \bar{\Psi}\right\Vert ^{2}}{%
\left\vert \bar{\Psi}_{0}\left( \bar{X}\right) \right\vert ^{2}}}}{k\left(
X\right) }Z\left( \bar{X}\right)
\end{eqnarray*}%
with averges:%
\begin{equation*}
\left\langle \delta \right\rangle =\frac{\left\langle k_{1}^{B}\right\rangle
+\frac{\kappa \left\langle k_{2}^{B}\right\rangle }{1+\left\langle \bar{k}%
\right\rangle }}{\left\langle k\right\rangle }\frac{\left\langle \bar{K}%
\right\rangle \left\Vert \bar{\Psi}\right\Vert ^{2}}{\left\langle \hat{K}%
\right\rangle \left\Vert \hat{\Psi}\right\Vert ^{2}}
\end{equation*}%
where $Z\left( \bar{X}_{1}\right) $ was computed befr:%
\begin{equation*}
Z\left( \bar{X}_{1}\right) \simeq \left( 1+\frac{3}{4}\frac{\left\langle 
\hat{g}^{ef}\right\rangle }{\hat{g}\left( \hat{X}_{1}\right) }\right) \frac{%
\left\Vert \bar{\Psi}_{0}\left( \bar{X}_{1}\right) \right\Vert ^{2}}{%
\left\Vert \hat{\Psi}_{0}\left( \hat{X}_{1}\right) \right\Vert ^{2}}-\left( 
\frac{3}{4}\frac{\left\langle \hat{g}^{Bef}\right\rangle }{\bar{g}\left( 
\bar{X}_{1}\right) }+\frac{1}{2}\frac{\left\langle \hat{g}%
^{Bef}\right\rangle }{\hat{g}\left( \hat{X}_{1}\right) }\right) \sqrt{\frac{%
\left\Vert \bar{\Psi}_{0}\left( \bar{X}_{1}\right) \right\Vert ^{2}}{%
\left\Vert \hat{\Psi}_{0}\right\Vert ^{2}}}
\end{equation*}%
or, in first approximation%
\begin{equation*}
Z\left( \bar{X}\right) \simeq \left( 1+\frac{3}{4}\frac{\left\langle \hat{g}%
^{ef}\right\rangle }{\left\langle \hat{g}\right\rangle }\right) \frac{%
\left\Vert \bar{\Psi}_{0}\left( \bar{X}\right) \right\Vert ^{2}}{\left\Vert 
\hat{\Psi}_{0}\left( \hat{X}\right) \right\Vert ^{2}}
\end{equation*}%
\begin{equation*}
\underline{k}^{B}\left( X\right) =\frac{k^{B}\left( X\right) }{\left\langle 
\bar{K}_{X}\right\rangle }\bar{K}_{X}\left[ \bar{X}\right]
\end{equation*}

\subsubsection*{A21.1.2 Dedinition and estimation of several coefficints}

We first define:%
\begin{equation*}
\frac{1}{1+\underline{\hat{k}}_{2}\left( \hat{X}^{\prime }\right) +\kappa %
\left[ \frac{\underline{\hat{k}}_{2}^{B}}{1+\bar{k}}\right] \left( \hat{X}%
^{\prime }\right) }\rightarrow \frac{1}{1+\left[ \underline{\hat{k}}%
_{2}\left( \hat{X}^{\prime }\right) \right] _{\kappa }}
\end{equation*}%
Given the normalizatn of coefficients, this quantity rewrits:%
\begin{eqnarray*}
&&\frac{1}{1+\underline{\hat{k}}_{2}\left( \hat{X}^{\prime }\right) +\kappa %
\left[ \frac{\underline{\hat{k}}_{2}^{B}}{1+\bar{k}}\right] \left( \hat{X}%
^{\prime }\right) } \\
&\simeq &\frac{1-\left( \left\langle \hat{k}\right\rangle +\left(
\left\langle \hat{k}_{1}^{B}\right\rangle +\kappa \left\langle \left[ \frac{%
\underline{\hat{k}}_{2}^{B}}{1+\bar{k}}\right] \right\rangle \right) \frac{%
\left\Vert \bar{\Psi}\right\Vert ^{2}\left\langle \bar{K}\right\rangle }{%
\left\Vert \hat{\Psi}\right\Vert ^{2}\left\langle \hat{K}\right\rangle }%
\right) }{\left( 1-\left( \left\langle \hat{k}_{1}\right\rangle
+\left\langle \hat{k}_{1}^{B}\right\rangle \frac{\left\Vert \bar{\Psi}%
\right\Vert ^{2}\left\langle \bar{K}\right\rangle }{\left\Vert \hat{\Psi}%
\right\Vert ^{2}\left\langle \hat{K}\right\rangle }\right) \right) \left( 1+%
\frac{\underline{\hat{k}}_{2}\left( \hat{X}^{\prime }\right) +\kappa \left[ 
\frac{\underline{\hat{k}}_{2}^{B}}{1+\bar{k}}\right] \left( \hat{X}^{\prime
}\right) }{1-\left( \left\langle \hat{k}_{1}\right\rangle +\left\langle \hat{%
k}_{1}^{B}\right\rangle \frac{\left\Vert \bar{\Psi}\right\Vert
^{2}\left\langle \bar{K}\right\rangle }{\left\Vert \hat{\Psi}\right\Vert
^{2}\left\langle \hat{K}\right\rangle }\right) }\right) } \\
&\simeq &\frac{1-\left\langle \hat{K}\right\rangle }{\left( 1-\left\langle 
\hat{K}_{1}\right\rangle \right) }\frac{1}{1+\frac{\underline{\hat{k}}%
_{2}\left( \hat{X}^{\prime }\right) +\kappa \left[ \frac{\underline{\hat{k}}%
_{2}^{B}}{1+\bar{k}}\right] \left( \hat{X}^{\prime }\right) }{1-\left\langle 
\hat{K}_{1}\right\rangle }}
\end{eqnarray*}%
We also define average connnectivity for banks plus firms:%
\begin{eqnarray*}
\left\langle \hat{k}^{\Sigma }\right\rangle &=&\left( \left\langle \hat{k}%
\right\rangle +\left( \left\langle \hat{k}_{1}^{B}\right\rangle +\kappa
\left\langle \left[ \frac{\underline{\hat{k}}_{2}^{B}}{1+\bar{k}}\right]
\right\rangle \right) \frac{\left\Vert \bar{\Psi}\right\Vert
^{2}\left\langle \bar{K}\right\rangle }{\left\Vert \hat{\Psi}\right\Vert
^{2}\left\langle \hat{K}\right\rangle }\right) \\
\left\langle \hat{k}_{1}^{\Sigma }\right\rangle &=&\left( \left\langle \hat{k%
}_{1}\right\rangle +\left\langle \hat{k}_{1}^{B}\right\rangle \frac{%
\left\Vert \bar{\Psi}\right\Vert ^{2}\left\langle \bar{K}\right\rangle }{%
\left\Vert \hat{\Psi}\right\Vert ^{2}\left\langle \hat{K}\right\rangle }%
\right)
\end{eqnarray*}%
so that we can compute:%
\begin{eqnarray*}
&&\frac{1}{1+\underline{\hat{k}}_{2}\left( \hat{X}^{\prime }\right) +\kappa %
\left[ \frac{\underline{\hat{k}}_{2}^{B}}{1+\bar{k}}\right] \left( \hat{X}%
^{\prime }\right) } \\
&\simeq &\frac{1-\left\langle \hat{k}^{\Sigma }\right\rangle }{%
1-\left\langle \hat{k}_{1}^{\Sigma }\right\rangle }\frac{1}{1+\frac{%
\underline{\hat{k}}_{2}\left( \hat{X}^{\prime }\right) +\kappa \left[ \frac{%
\underline{\hat{k}}_{2}^{B}}{1+\bar{k}}\right] \left( \hat{X}^{\prime
}\right) }{1-\left\langle \hat{K}_{1}\right\rangle }} \\
&=&\frac{1-\left\langle \hat{k}^{\Sigma }\right\rangle }{1-\left\langle \hat{%
k}_{1}^{\Sigma }\right\rangle +\underline{\hat{k}}_{2}\left( \hat{X}^{\prime
}\right) +\kappa \left[ \frac{\underline{\hat{k}}_{2}^{B}}{1+\bar{k}}\right]
\left( \hat{X}^{\prime }\right) }\simeq \frac{1-\left\langle \hat{k}^{\Sigma
}\right\rangle }{1-\left\langle \hat{k}^{\Sigma }\right\rangle +\hat{k}%
_{2}\left( \hat{X}^{\prime },\left\langle \hat{X}\right\rangle \right)
+\kappa \left[ \frac{\hat{k}_{2}^{B}\left( \hat{X}^{\prime },\left\langle 
\bar{X}\right\rangle \right) }{1+\bar{k}}\right] \frac{\left\Vert \bar{\Psi}%
\right\Vert ^{2}\left\langle \bar{K}\right\rangle }{\left\Vert \hat{\Psi}%
\right\Vert ^{2}\left\langle \hat{K}\right\rangle }} \\
&=&\frac{1}{1+\frac{\hat{k}_{2}\left( \hat{X}^{\prime },\left\langle \hat{X}%
\right\rangle \right) +\kappa \left[ \frac{\hat{k}_{2}^{B}\left( \hat{X}%
^{\prime },\left\langle \bar{X}\right\rangle \right) }{1+\bar{k}}\right] 
\frac{\left\Vert \bar{\Psi}\right\Vert ^{2}\left\langle \bar{K}\right\rangle 
}{\left\Vert \hat{\Psi}\right\Vert ^{2}\left\langle \hat{K}\right\rangle }}{%
1-\left\langle \hat{k}^{\Sigma }\right\rangle }}\simeq \frac{1}{1+\frac{\hat{%
k}_{2}\left( \hat{X}^{\prime },\left\langle \hat{X}\right\rangle \right)
+\kappa \left[ \frac{\hat{k}_{2}^{B}\left( \hat{X}^{\prime },\left\langle 
\bar{X}\right\rangle \right) }{1+\bar{k}}\right] \left( \left( 1+\frac{3}{4}%
\frac{\left\langle \hat{g}^{ef}\right\rangle }{\left\langle \hat{g}%
\right\rangle }\right) \frac{\left\Vert \bar{\Psi}_{0}\right\Vert ^{2}}{%
\left\Vert \hat{\Psi}_{0}\right\Vert ^{2}}\right) }{1-\left\langle \hat{k}%
^{\Sigma }\right\rangle }} \\
&\rightarrow &\frac{1}{1+\hat{k}_{2}^{\left( n\right) }\left( \hat{X}%
^{\prime },\left\langle \hat{X}\right\rangle \right) +\kappa \left[ \frac{%
\underline{\hat{k}}_{2}^{B\left( n\right) }\left( \hat{X}^{\prime
},\left\langle \bar{X}\right\rangle \right) }{1+\bar{k}}\right] \left(
\left( 1+\frac{3}{4}\frac{\left\langle \hat{g}^{ef}\right\rangle }{%
\left\langle \hat{g}\right\rangle }\right) \frac{\left\Vert \bar{\Psi}%
_{0}\right\Vert ^{2}}{\left\Vert \hat{\Psi}_{0}\right\Vert ^{2}}\right) } \\
&\equiv &\frac{1}{1+\left[ \underline{\hat{k}}_{2}^{n}\left( \hat{X}\right) %
\right] _{\kappa }}
\end{eqnarray*}

\subsection*{A21.2 Development of (\protect\ref{RTn})}

\paragraph*{A21.2.1 Computation of the left hand side}

The left hand side:%
\begin{equation}
\left( \Delta \left( \hat{X},\hat{X}^{\prime }\right) -\frac{\hat{K}^{\prime
}\hat{k}_{1}\left( \hat{X}^{\prime },\hat{X}\right) \left\vert \hat{\Psi}%
\left( \hat{K}^{\prime },\hat{X}^{\prime }\right) \right\vert ^{2}}{1+%
\underline{\hat{k}}\left( \hat{X}^{\prime }\right) +\underline{\hat{k}}%
_{1}^{B}\left( \bar{X}^{\prime }\right) +\kappa \left[ \frac{\underline{\hat{%
k}}_{2}^{B}}{1+\bar{k}}\right] \left( \hat{X}^{\prime }\right) }\right) 
\frac{1}{1+\underline{\hat{k}}_{2}\left( \hat{X}^{\prime }\right) +\kappa %
\left[ \frac{\underline{\hat{k}}_{2}^{B}}{1+\bar{k}}\right] \left( \hat{X}%
^{\prime }\right) }\left( 1-\hat{M}\right)  \label{LS}
\end{equation}%
writes:%
\begin{eqnarray*}
&&\frac{1}{1+\left[ \underline{\hat{k}}_{2}\left( \hat{X}\right) \right]
_{\kappa }}\left( 1-\frac{\hat{k}\left( \hat{X},\hat{X}^{\prime }\right) 
\hat{K}_{X}}{1+\underline{\hat{k}}\left( \hat{X}^{\prime }\right) +%
\underline{\hat{k}}_{1}^{B}\left( \hat{X}^{\prime }\right) +\kappa \left[ 
\frac{\underline{\hat{k}}_{2}^{B}}{1+\bar{k}}\right] \left( \hat{X}^{\prime
}\right) }\left\Vert \hat{\Psi}\left( \hat{X}^{\prime }\right) \right\Vert
^{2}\right) \\
&&-\frac{\hat{K}^{\prime }\hat{k}_{1}\left( \hat{X}^{\prime },\hat{X}\right)
\left\vert \hat{\Psi}\left( \hat{K}^{\prime },\hat{X}^{\prime }\right)
\right\vert ^{2}}{1+\underline{\hat{k}}\left( \hat{X}^{\prime }\right) +%
\underline{\hat{k}}_{1}^{B}\left( \bar{X}^{\prime }\right) +\kappa \left[ 
\frac{\underline{\hat{k}}_{2}^{B}}{1+\bar{k}}\right] \left( \hat{X}^{\prime
}\right) }\frac{1}{1+\left[ \underline{\hat{k}}_{2}\left( \hat{X}^{\prime
}\right) \right] _{\kappa }} \\
&&+\frac{\hat{K}^{\prime }\hat{k}_{1}\left( \hat{X}^{\prime },\hat{X}\right)
\left\vert \hat{\Psi}\left( \hat{K}^{\prime },\hat{X}^{\prime }\right)
\right\vert ^{2}}{1+\underline{\hat{k}}\left( \hat{X}^{\prime }\right) +%
\underline{\hat{k}}_{1}^{B}\left( \bar{X}^{\prime }\right) +\kappa \left[ 
\frac{\underline{\hat{k}}_{2}^{B}}{1+\bar{k}}\right] \left( \hat{X}^{\prime
}\right) }\frac{1}{1+\left[ \underline{\hat{k}}_{2}\left( \hat{X}^{\prime
}\right) \right] _{\kappa }}\frac{k\left( X^{\prime },X^{\prime \prime
}\right) K_{X^{\prime }}}{1+k\left( X^{\prime },X^{\prime \prime }\right)
K_{X^{\prime \prime }}\left\Vert \hat{\Psi}\left( \hat{X}^{\prime \prime
}\right) \right\Vert ^{2}}\left\Vert \hat{\Psi}\left( \hat{X}^{\prime \prime
}\right) \right\Vert ^{2}
\end{eqnarray*}%
with the notation:%
\begin{equation*}
\underline{\hat{k}}\left( \hat{X}^{\prime }\right) +\underline{\hat{k}}%
_{1}^{B}\left( \bar{X}^{\prime }\right) +\kappa \left[ \frac{\underline{\hat{%
k}}_{2}^{B}}{1+\bar{k}}\right] \left( \hat{X}^{\prime }\right) \rightarrow %
\left[ \underline{\hat{k}}\left( \hat{X}^{\prime }\right) \right] _{\kappa }
\end{equation*}%
we can replace matrices elments with their averages:

\begin{eqnarray*}
&&\frac{1}{1+\left[ \underline{\hat{k}}_{2}\left( \hat{X}\right) \right]
_{\kappa }}\left( 1-\frac{\hat{k}\left( \hat{X},\hat{X}^{\prime }\right) 
\hat{K}_{X}}{1+\left[ \underline{\hat{k}}\left( \hat{X}^{\prime }\right) %
\right] _{\kappa }}\left\Vert \hat{\Psi}\left( \hat{X}^{\prime }\right)
\right\Vert ^{2}\right) \\
&&-\frac{\hat{k}_{1}\left( \hat{X}^{\prime },\hat{X}\right) \hat{K}%
_{X}\left\Vert \hat{\Psi}\left( \hat{X}^{\prime }\right) \right\Vert ^{2}}{1+%
\left[ \underline{\hat{k}}\left( \hat{X}\right) \right] _{\kappa }}\frac{1}{%
1+\left[ \underline{\hat{k}}_{2}\left( \hat{X}^{\prime }\right) \right]
_{\kappa }} \\
&&+\frac{\hat{k}_{1}\left( \hat{X}^{\prime },\hat{X}\right) \hat{K}%
_{X}\left\Vert \hat{\Psi}\left( \hat{X}^{\prime }\right) \right\Vert ^{2}}{1+%
\left[ \underline{\hat{k}}_{2}\left( \hat{X}^{\prime }\right) \right]
_{\kappa }}\frac{1}{1+\left[ \underline{\hat{k}}_{2}\left( \hat{X}^{\prime
}\right) \right] _{\kappa }}\frac{k\left( X^{\prime },X^{\prime \prime
}\right) K_{X^{\prime }}}{1+\left[ \underline{\hat{k}}\left( \hat{X}^{\prime
}\right) \right] _{\kappa }}\left\Vert \hat{\Psi}\left( \hat{X}^{\prime
\prime }\right) \right\Vert ^{2}
\end{eqnarray*}%
so that the terms:%
\begin{eqnarray*}
&&\frac{\hat{k}_{1}\left( \hat{X}^{\prime },\hat{X}\right) \hat{K}%
_{X}\left\Vert \hat{\Psi}\left( \hat{X}^{\prime }\right) \right\Vert ^{2}}{1+%
\hat{k}\left( \hat{X},\hat{X}^{\prime }\right) \hat{K}_{X^{\prime }}}\frac{1%
}{1+\underline{\hat{k}}_{2}\left( \hat{X}^{\prime }\right) }\frac{k\left(
X^{\prime },X^{\prime \prime }\right) K_{X^{\prime }}}{1+\underline{\hat{k}}%
\left( \hat{X}^{\prime }\right) }\left\Vert \hat{\Psi}\left( \hat{X}^{\prime
\prime }\right) \right\Vert ^{2} \\
&&+\frac{\hat{k}_{1}\left( \left\langle X\right\rangle ,\hat{X}\right) \hat{K%
}_{X}\left\Vert \hat{\Psi}\left( \left\langle \hat{X}\right\rangle \right)
\right\Vert ^{2}}{1+\underline{\hat{k}}\left( \hat{X}\right) }\frac{1}{1+%
\underline{\hat{k}}_{2}\left( \left\langle X\right\rangle \right) }\frac{%
k\left( \left\langle X\right\rangle ,X^{\prime }\right) \left\langle
K\right\rangle }{1+\underline{\hat{k}}\left( \left\langle \hat{X}%
\right\rangle \right) }\left\Vert \hat{\Psi}\left( \hat{X}^{\prime }\right)
\right\Vert ^{2}
\end{eqnarray*}%
yield the contribution:

\begin{eqnarray*}
&&-\left( \frac{\hat{k}\left( \hat{X},\hat{X}^{\prime }\right) }{1+\left[ 
\underline{\hat{k}}_{2}\left( \hat{X}\right) \right] _{\kappa }}+\frac{\hat{k%
}_{1}\left( \hat{X}^{\prime },\hat{X}\right) }{1+\left[ \underline{\hat{k}}%
\left( \hat{X}^{\prime }\right) \right] _{\kappa }}\right) \frac{\hat{K}%
_{X}\left\Vert \hat{\Psi}\left( \hat{X}^{\prime }\right) \right\Vert ^{2}}{1+%
\left[ \underline{\hat{k}}\left( \hat{X}\right) \right] _{\kappa }} \\
&&+\frac{\hat{k}_{1}\left( \left\langle X\right\rangle ,\hat{X}\right) \hat{K%
}_{X}\left\Vert \hat{\Psi}\left( \left\langle \hat{X}\right\rangle \right)
\right\Vert ^{2}}{1+\left[ \underline{\hat{k}}\left( \hat{X}\right) \right]
_{\kappa }}\frac{1}{1+\left[ \underline{\hat{k}}_{2}\left( \left\langle \hat{%
X}\right\rangle \right) \right] _{\kappa }}\frac{k\left( \left\langle
X\right\rangle ,X^{\prime }\right) \left\langle K\right\rangle }{1+\left[ 
\underline{\hat{k}}\left( \left\langle \hat{X}\right\rangle \right) \right]
_{\kappa }}\left\Vert \hat{\Psi}\left( \hat{X}^{\prime }\right) \right\Vert
^{2} \\
&=&-\left( \left( \frac{\hat{k}\left( \hat{X},\hat{X}^{\prime }\right) }{1+%
\left[ \underline{\hat{k}}_{2}\left( \hat{X}\right) \right] _{\kappa }}+%
\frac{\hat{k}_{1}\left( \hat{X}^{\prime },\hat{X}\right) }{1+\left[ 
\underline{\hat{k}}_{2}\left( \hat{X}^{\prime }\right) \right] _{\kappa }}%
\right) -\frac{\hat{k}_{1}\left( \left\langle X\right\rangle ,\hat{X}\right)
\left\Vert \hat{\Psi}\left( \left\langle \hat{X}\right\rangle \right)
\right\Vert ^{2}}{1+\left[ \underline{\hat{k}}\left( \left\langle \hat{X}%
\right\rangle \right) \right] _{\kappa }}\frac{k\left( \left\langle
X\right\rangle ,X^{\prime }\right) \left\langle K\right\rangle }{1+\left[ 
\underline{\hat{k}}_{2}\left( \left\langle \hat{X}\right\rangle \right) %
\right] _{\kappa }}\right) \frac{\hat{K}_{X}\left\Vert \hat{\Psi}\left( \hat{%
X}^{\prime }\right) \right\Vert ^{2}}{1+\left[ \underline{\hat{k}}\left( 
\hat{X}^{\prime }\right) \right] _{\kappa }} \\
&=&-\left( \frac{\hat{k}\left( \hat{X},\hat{X}^{\prime }\right) -\frac{%
\underline{\hat{k}}_{1}\left( \left\langle X\right\rangle ,\hat{X}\right)
\left\Vert \hat{\Psi}\left( \left\langle \hat{X}\right\rangle \right)
\right\Vert ^{2}}{1+\left[ \underline{\hat{k}}\left( \left\langle \hat{X}%
\right\rangle \right) \right] _{\kappa }}k\left( \left\langle X\right\rangle
,X^{\prime }\right) }{1+\left[ \underline{\hat{k}}_{2}\left( \hat{X}\right) %
\right] _{\kappa }}+\frac{\hat{k}_{1}\left( \hat{X}^{\prime },\hat{X}\right) 
}{1+\left[ \underline{\hat{k}}_{2}\left( \hat{X}^{\prime }\right) \right]
_{\kappa }}\right) \frac{\hat{K}_{X}\left\Vert \hat{\Psi}\left( \hat{X}%
^{\prime }\right) \right\Vert ^{2}}{1+\left[ \underline{\hat{k}}\left( \hat{X%
}\right) \right] _{\kappa }}
\end{eqnarray*}%
and we can approximate:%
\begin{eqnarray*}
&&-\left( \frac{\hat{k}\left( \hat{X},\hat{X}^{\prime }\right) -\frac{%
\underline{\hat{k}}_{1}\left( \left\langle X\right\rangle ,\hat{X}\right)
\left\Vert \hat{\Psi}\left( \left\langle \hat{X}\right\rangle \right)
\right\Vert ^{2}}{1+\left[ \underline{\hat{k}}\left( \left\langle \hat{X}%
\right\rangle \right) \right] _{\kappa }}k\left( \left\langle X\right\rangle
,X^{\prime }\right) }{1+\left[ \underline{\hat{k}}_{2}^{\left( n\right)
}\left( \hat{X}\right) \right] _{\kappa }}+\frac{\hat{k}_{1}\left( \hat{X}%
^{\prime },\hat{X}\right) }{1+\left[ \underline{\hat{k}}_{2}^{\left(
n\right) }\left( \hat{X}^{\prime }\right) \right] _{\kappa }}\right) \frac{%
\hat{K}_{X}\left\Vert \hat{\Psi}\left( \hat{X}^{\prime }\right) \right\Vert
^{2}}{1+\left[ \underline{\hat{k}}\left( \hat{X}\right) \right] _{\kappa }}
\\
&\simeq &-\left( \frac{\hat{k}\left( \hat{X},\hat{X}^{\prime }\right) -%
\underline{\hat{k}}_{1}\left( \left\langle X\right\rangle ,\hat{X}\right)
k\left( \left\langle X\right\rangle ,X^{\prime }\right) }{1+\left[ 
\underline{\hat{k}}_{2}^{\left( n\right) }\left( \hat{X}\right) \right]
_{\kappa }}+\frac{\hat{k}_{1}\left( \hat{X}^{\prime },\hat{X}\right) }{1+%
\left[ \underline{\hat{k}}_{2}^{\left( n\right) }\left( \hat{X}^{\prime
}\right) \right] _{\kappa }}\right) \frac{\hat{K}_{X}\left\Vert \hat{\Psi}%
\left( \hat{X}^{\prime }\right) \right\Vert ^{2}}{1+\left[ \underline{\hat{k}%
}\left( \hat{X}\right) \right] _{\kappa }}
\end{eqnarray*}%
where:%
\begin{equation*}
\underline{\hat{k}}_{1}\left( \left\langle X\right\rangle ,\hat{X}\right) =%
\hat{k}_{1}\left( \left\langle X\right\rangle ,\hat{X}\right) \left\Vert 
\hat{\Psi}\left( \left\langle \hat{X}\right\rangle \right) \right\Vert
^{2}\left\langle K\right\rangle
\end{equation*}%
Ultimately, we write (\ref{LS}) as:%
\begin{equation}
\frac{\Delta \left( \hat{X},\hat{X}^{\prime }\right) }{1+\left[ \underline{%
\hat{k}}_{2}\left( \hat{X}\right) \right] _{\kappa }}-\hat{S}_{1}^{E}\left( 
\hat{X}^{\prime },\hat{X}\right)  \label{Lst}
\end{equation}

\subsubsection*{A21.2.2 Computation of the right hand side}

We introduce the notations:%
\begin{eqnarray*}
\underline{k}_{2}\left( \hat{X}\right) &=&\beta \underline{k}\left( \hat{X}%
\right) \\
\underline{k}_{2}\left( \hat{X}\right) &=&\left( 1-\beta \right) \underline{k%
}\left( \hat{X}\right)
\end{eqnarray*}%
\begin{equation*}
C^{\left( e\right) }=\frac{1+\underline{k}_{2}\left( X\right) +\kappa \left[ 
\frac{\underline{k}_{2}^{B}}{1+\bar{k}}\right] \left( X\right) }{1+%
\underline{k}\left( X\right) +\underline{k}_{1}^{B}\left( X\right) +\kappa %
\left[ \frac{\underline{k}_{2}^{B}}{1+\bar{k}}\right] \left( X\right) }C
\end{equation*}%
\begin{equation*}
\kappa \left[ \frac{\underline{k}_{2}^{B}}{1+\bar{k}}\right] \left( X\right)
>>1
\end{equation*}%
\begin{equation*}
\kappa \left[ \frac{\underline{k}_{2}^{B}}{1+\bar{k}}\right] \left( X\right)
=\beta ^{B}\left( \underline{k}_{1}^{B}\left( X\right) +\kappa \left[ \frac{%
\underline{k}_{2}^{B}}{1+\bar{k}}\right] \left( X\right) \right) =\beta ^{B}%
\underline{k}^{B}\left( X\right)
\end{equation*}%
note that:%
\begin{equation*}
1-\beta ^{B}<<1
\end{equation*}%
\begin{equation*}
\underline{k}\left( X\right) =\delta \underline{k}^{B}\left( X\right)
\end{equation*}%
which allows to write:%
\begin{equation*}
\frac{\underline{k}_{2}\left( X\right) +\kappa \left[ \frac{\underline{k}%
_{2}^{B}}{1+\bar{k}}\right] \left( X\right) }{\underline{k}\left( X\right) +%
\underline{k}^{B}\left( X\right) }C=\frac{\beta \delta +\beta ^{B}}{1+\delta 
}
\end{equation*}

\begin{equation*}
\frac{\underline{k}_{1}\left( X^{\prime },\hat{X}\right) }{\left( 1+%
\underline{k}^{B}\left( X^{\prime }\right) \right) }\rightarrow \left(
1-\beta \right) \delta
\end{equation*}%
and we compute the return as:%
\begin{eqnarray*}
f_{1}^{\left( e\right) }\left( X\right) &=&\left( 1+\underline{k}_{2}\left(
X\right) +\kappa \left[ \frac{\underline{k}_{2}^{B}}{1+\bar{k}}\right]
\left( X\right) \right) f_{1}\left( X\right) -\left( \underline{k}_{2}\left(
X\right) +\kappa \left[ \frac{\underline{k}_{2}^{B}}{1+\bar{k}}\right]
\left( X\right) \right) \bar{r} \\
&=&f_{1}\left( X\right) +\left( \beta \delta +\beta ^{B}\right) \underline{k}%
^{B}\left( X\right) \left( f_{1}\left( X\right) -R\right)
\end{eqnarray*}%
We use formula (\ref{GF}) for returns%
\begin{equation*}
\hat{g}\left( \hat{X}_{1}\right) \simeq \left( \left\Vert \hat{\Psi}%
_{0}\left( \hat{X}_{1}\right) \right\Vert ^{2}-\hat{\mu}D\left( \hat{X}%
_{1}\right) \right) \sqrt{\frac{\sigma _{\hat{K}}^{2}\left( 1-\frac{\hat{g}%
^{ef}\left( \hat{X}_{1}\right) }{\left\langle \hat{g}\right\rangle }\right) 
}{2\hat{\mu}\hat{K}\left[ \hat{X}_{1}\right] }}
\end{equation*}%
altrnately, we can use the full expanded forms (\ref{GH}) and (\ref{GB}),
leading to same results.

\subsubsection*{A21.2.3 Rewrting (\protect\ref{RTN})}

The previous expression allow to obtain a detailed form of (\ref{RTN}):%
\begin{eqnarray}
&&\left( \frac{1}{1+\left[ \underline{\hat{k}}_{2}^{n}\left( \hat{X}\right) %
\right] _{\kappa }}-\hat{S}_{1}^{E}\left( \hat{X}^{\prime },\hat{X}\right)
\right) \left( \left( \left\Vert \hat{\Psi}_{0}\left( \hat{X}^{\prime
}\right) \right\Vert ^{2}-\hat{\mu}D\left( \hat{X}^{\prime }\right) \right) 
\sqrt{\frac{\sigma _{\hat{K}}^{2}\left( 1-\frac{\hat{g}^{ef}\left( \hat{X}%
^{\prime }\right) }{\left\langle \hat{g}\right\rangle }\right) }{2\hat{\mu}%
\hat{K}\left[ \hat{X}^{\prime }\right] }}-\bar{r}^{\prime }\right) \\
&=&\left( \frac{\left( \left\Vert \hat{\Psi}_{0}\left( \hat{X}_{1}\right)
\right\Vert ^{2}-\hat{\mu}D\left( \hat{X}_{1}\right) \right) \sqrt{\frac{%
\sigma _{\hat{K}}^{2}\left( 1-\frac{\hat{g}^{ef}\left( \hat{X}_{1}\right) }{%
\left\langle \hat{g}\right\rangle }\right) }{2\hat{\mu}\hat{K}\left[ \hat{X}%
_{1}\right] }}-\bar{r}^{\prime }}{1+\left[ \underline{\hat{k}}_{2}^{n}\left( 
\hat{X}\right) \right] _{\kappa }}\right)  \notag \\
&&-\int \hat{S}_{1}^{E}\left( \hat{X}^{\prime },\hat{X}_{1}\right) \left(
\left( \left\Vert \hat{\Psi}_{0}\left( \hat{X}^{\prime }\right) \right\Vert
^{2}-\hat{\mu}D\left( \hat{X}^{\prime }\right) \right) \sqrt{\frac{\sigma _{%
\hat{K}}^{2}\left( 1-\frac{\hat{g}^{ef}\left( \hat{X}^{\prime }\right) }{%
\left\langle \hat{g}\right\rangle }\right) }{2\hat{\mu}\hat{K}\left[ \hat{X}%
^{\prime }\right] }}-\bar{r}^{\prime }\right)  \notag
\end{eqnarray}

which leads to:

\begin{eqnarray}
&&\frac{\left( 1-\beta \right) \delta }{3\left( 1+\delta \right) }\left( 
\frac{\left( X+C\frac{\beta \delta +\beta ^{B}}{1+\delta }\right)
^{2}\epsilon \left( \frac{\left( R+\Delta F_{\tau }\left( \bar{R}\left(
K,X\right) \right) \right) \left( 3X-\frac{\beta \delta +\beta ^{B}}{%
1+\delta }C\right) \left( \frac{\beta \delta +\beta ^{B}}{1+\delta }%
C+X\right) }{4\left( f_{1}\left( X\right) +\left( \beta \delta +\beta
^{B}\right) \underline{k}^{B}\left( X\right) R\right) }-\frac{C\left( 2X-C%
\frac{\beta \delta +\beta ^{B}}{1+\delta }\right) }{1+\left( 1+\delta
\right) \underline{k}^{B}\left( X\right) }\right) }{\sigma _{\hat{K}%
}^{2}\left( f_{1}\left( X\right) +\left( \beta \delta +\beta ^{B}\right) 
\underline{k}^{B}\left( X\right) R\right) }\right)  \label{RTWvz} \\
&=&\frac{\left( \left\Vert \hat{\Psi}_{0}\left( \hat{X}_{1}\right)
\right\Vert ^{2}-\hat{\mu}D\left( \hat{X}_{1}\right) \right) \sqrt{\frac{%
\sigma _{\hat{K}}^{2}\left( 1-\frac{\hat{g}^{ef}\left( \hat{X}_{1}\right) }{%
\left\langle \hat{g}\right\rangle }\right) }{2\hat{\mu}\hat{K}\left[ \hat{X}%
_{1}\right] }}-\bar{r}^{\prime }}{1+\left[ \underline{\hat{k}}_{2}^{n}\left( 
\hat{X}\right) \right] _{\kappa }}  \notag \\
&&-\int \hat{S}_{1}^{E}\left( \hat{X}^{\prime },\hat{X}_{1}\right) \left(
\left( \left\Vert \hat{\Psi}_{0}\left( \hat{X}^{\prime }\right) \right\Vert
^{2}-\hat{\mu}D\left( \hat{X}^{\prime }\right) \right) \sqrt{\frac{\sigma _{%
\hat{K}}^{2}\left( 1-\frac{\hat{g}^{ef}\left( \hat{X}^{\prime }\right) }{%
\left\langle \hat{g}\right\rangle }\right) }{2\hat{\mu}\hat{K}\left[ \hat{X}%
^{\prime }\right] }}-\bar{r}^{\prime }\right)  \notag
\end{eqnarray}

where:%
\begin{equation*}
\underline{\hat{k}}_{2}\left( \hat{X}_{1}\right) =\hat{\beta}\underline{\hat{%
k}}\left( \hat{X}\right)
\end{equation*}%
\begin{equation*}
\underline{\hat{k}}^{B}\left( \hat{X}_{1}\right) =\underline{\hat{k}}%
_{1}^{B}\left( \hat{X}_{1}\right) +\kappa \left[ \frac{\underline{\hat{k}}%
_{2}^{B}}{1+\bar{k}}\right] \left( \hat{X}_{1}\right)
\end{equation*}%
\begin{equation*}
\underline{\hat{k}}^{B}\rightarrow \frac{\left( \hat{k}_{1}^{B}\left(
X\right) +\frac{\kappa \hat{k}_{2}^{B}\left( X\right) }{1+\bar{k}\left( \bar{%
X}\right) \frac{\left\Vert \bar{\Psi}\right\Vert ^{2}}{\left\vert \bar{\Psi}%
_{0}\left( \bar{X}\right) \right\vert ^{2}}}\right) }{\left\langle \hat{K}%
\right\rangle \left\Vert \hat{\Psi}\right\Vert ^{2}}\bar{K}_{X}\left\vert 
\bar{\Psi}\left( \bar{X}\right) \right\vert ^{2}
\end{equation*}%
\begin{equation*}
\kappa \left[ \frac{\underline{k}_{2}^{B}}{1+\bar{k}}\right] =\hat{\beta}^{B}%
\underline{\hat{k}}^{B}
\end{equation*}%
\begin{eqnarray*}
\hat{\delta} &=&\frac{\hat{k}\left( X\right) \hat{K}_{X}\left\vert \hat{\Psi}%
\left( \hat{X}\right) \right\vert ^{2}}{\left( \hat{k}_{1}^{B}\left(
X\right) +\frac{\kappa \hat{k}_{2}^{B}\left( X\right) }{1+\bar{k}\left( \bar{%
X}\right) \frac{\left\Vert \bar{\Psi}\right\Vert ^{2}}{\left\vert \bar{\Psi}%
_{0}\left( \bar{X}\right) \right\vert ^{2}}}\right) \bar{K}_{X}\left\vert 
\bar{\Psi}\left( \bar{X}\right) \right\vert ^{2}} \\
&=&\frac{\hat{k}\left( X\right) \hat{K}\left[ \hat{X}\right] }{\left( \hat{k}%
_{1}^{B}\left( X\right) +\frac{\kappa \hat{k}_{2}^{B}\left( X\right) }{1+%
\bar{k}\left( \bar{X}\right) \frac{\left\Vert \bar{\Psi}\right\Vert ^{2}}{%
\left\vert \bar{\Psi}_{0}\left( \bar{X}\right) \right\vert ^{2}}}\right) 
\bar{K}_{X}\left( \bar{X}\right) }\rightarrow \frac{\hat{k}\left( X\right)
\left\langle \hat{K}\right\rangle \left\Vert \hat{\Psi}\right\Vert ^{2}}{%
\left( \hat{k}_{1}^{B}\left( X\right) +\frac{\kappa \hat{k}_{2}^{B}\left(
X\right) }{1+\bar{k}\left( \bar{X}\right) \frac{\left\Vert \bar{\Psi}%
\right\Vert ^{2}}{\left\vert \bar{\Psi}_{0}\left( \bar{X}\right) \right\vert
^{2}}}\right) \left\langle \hat{K}\right\rangle \left\Vert \bar{\Psi}%
\right\Vert ^{2}}
\end{eqnarray*}%
so that the following relation holds%
\begin{equation*}
\underline{\hat{k}}\left( \hat{X}_{1}\right) =\hat{\delta}\underline{\hat{k}}%
^{B}\left( \hat{X}_{1}\right)
\end{equation*}

\subsubsection*{A21.3 Solving (\protect\ref{RTWvz})}

Recall that:%
\begin{equation*}
R=f_{1}\left( X\right) -\bar{r}
\end{equation*}

\begin{equation*}
\underline{k}\left( X\right) =\frac{k\left( X\right) }{\left\langle
K\right\rangle }\hat{K}_{X}\frac{\left\vert \hat{\Psi}\left( \hat{X}\right)
\right\vert ^{2}}{\left\vert \Psi _{0}\left( X\right) \right\vert ^{2}}
\end{equation*}%
\begin{equation*}
\hat{K}\left[ \hat{X}\right] =\frac{\underline{k}\left( X\right) }{k\left(
X\right) }\left\langle K\right\rangle \left\vert \Psi _{0}\left( X\right)
\right\vert ^{2}\rightarrow \frac{\underline{k}\left( X\right) }{k}
\end{equation*}%
Define:%
\begin{equation*}
H=\int \hat{S}_{1}^{E}\left( \hat{X}^{\prime },\hat{X}_{1}\right) \left(
\left( \left\Vert \hat{\Psi}_{0}\left( \hat{X}^{\prime }\right) \right\Vert
^{2}-\hat{\mu}D\left( \hat{X}^{\prime }\right) \right) \sqrt{\frac{\sigma _{%
\hat{K}}^{2}\left( 1-\frac{\hat{g}^{ef}\left( \hat{X}^{\prime }\right) }{%
\left\langle \hat{g}\right\rangle }\right) }{2\hat{\mu}\hat{K}\left[ \hat{X}%
^{\prime }\right] }}-\bar{r}^{\prime }\right)
\end{equation*}%
and (\ref{RTWvz}) becomes:%
\begin{eqnarray*}
&&-H-\frac{\bar{r}^{\prime }}{\left( 1+\left( \hat{\beta}\hat{\delta}+\hat{%
\beta}^{B}\right) \underline{\hat{k}}^{B}\left( \hat{X}_{1}\right) \right) }
\\
&=&\frac{1}{3}\frac{\left( 1-\beta \right) \delta }{1+\delta }\left( \frac{%
\left( X+C\frac{\beta \delta +\beta ^{B}}{1+\delta }\right) ^{2}\epsilon
\left( \frac{\left( R+\Delta F_{\tau }\left( \bar{R}\left( K,X\right)
\right) \right) \left( 3X-\frac{\beta \delta +\beta ^{B}}{1+\delta }C\right)
\left( \frac{\beta \delta +\beta ^{B}}{1+\delta }C+X\right) }{4\left(
f_{1}\left( X\right) +\left( \beta \delta +\beta ^{B}\right) \underline{k}%
^{B}\left( X\right) R\right) }-\frac{C\left( 2X-C\frac{\beta \delta +\beta
^{B}}{1+\delta }\right) }{1+\left( 1+\delta \right) \underline{k}^{B}\left(
X\right) }\right) }{\sigma _{\hat{K}}^{2}\left( f_{1}\left( X\right) +\left(
\beta \delta +\beta ^{B}\right) \underline{k}^{B}\left( X\right) R\right) }%
\right) \\
&&-\frac{\left( \left\Vert \hat{\Psi}_{0}\left( \hat{X}_{1}\right)
\right\Vert ^{2}-\hat{\mu}D\left( \hat{X}_{1}\right) \right) \sqrt{\frac{%
\sigma _{\hat{K}}^{2}\left( 1-\frac{\hat{g}^{ef}\left( \hat{X}_{1}\right) }{%
\left\langle \hat{g}\right\rangle }\right) }{2\hat{\mu}\hat{K}\left[ \hat{X}%
_{1}\right] }}-\bar{r}^{\prime }}{1+\left[ \underline{\hat{k}}_{2}^{n}\left( 
\hat{X}\right) \right] _{\kappa }}
\end{eqnarray*}%
\begin{eqnarray*}
&&\frac{\left( 1-\beta \right) \delta }{3\left( 1+\delta \right) }\frac{%
\left( X+C\frac{\beta \delta +\beta ^{B}}{1+\delta }\right) ^{2}\epsilon }{%
\sigma _{\hat{K}}^{2}\left( f_{1}\left( X\right) +\left( \beta \delta +\beta
^{B}\right) \underline{k}^{B}\left( X\right) R\right) }\frac{C\left( 2X-C%
\frac{\beta \delta +\beta ^{B}}{1+\delta }\right) }{1+\left( 1+\delta
\right) \underline{k}^{B}\left( X\right) } \\
&&+\left( \frac{\left( \left\Vert \hat{\Psi}_{0}\left( \hat{X}_{1}\right)
\right\Vert ^{2}-\hat{\mu}D\left( \hat{X}_{1}\right) \right) \sqrt{k\frac{%
\sigma _{\hat{K}}^{2}}{2\hat{\mu}}\left( 1-\frac{\hat{g}^{ef}\left( \hat{X}%
_{1}\right) }{\left\langle \hat{g}\right\rangle }\right) }}{\left( 1+\left[ 
\underline{\hat{k}}_{2}^{n}\left( \hat{X}\right) \right] _{\kappa }\right) 
\sqrt{\underline{k}\left( \hat{X}_{1}\right) }}\right) \\
&=&H+\frac{\bar{r}^{\prime }}{1+\left[ \underline{\hat{k}}_{2}^{n}\left( 
\hat{X}\right) \right] _{\kappa }}+\frac{1}{3}\frac{\left( 1-\beta \right)
\delta }{1+\delta }\left( \frac{\left( X+C\frac{\beta \delta +\beta ^{B}}{%
1+\delta }\right) ^{2}\epsilon }{\sigma _{\hat{K}}^{2}f_{1}\left( X\right) }%
\frac{R\left( 3X-\frac{\beta \delta +\beta ^{B}}{1+\delta }C\right) \left( 
\frac{\beta \delta +\beta ^{B}}{1+\delta }C+X\right) }{4f_{1}\left( X\right) 
}+\Delta F_{\tau }\left( \bar{R}\left( K,X\right) \right) \right)
\end{eqnarray*}%
a first order expansion in $R$ leads to:%
\begin{eqnarray*}
&&\frac{1}{1+\left( 1+\delta \right) \underline{k}^{B}\left( X\right) }%
\left( 1-\frac{\left( \beta \delta +\beta ^{B}\right) \underline{k}%
^{B}\left( X\right) \left( R+\Delta F_{\tau }\left( \bar{R}\left( K,X\right)
\right) \right) }{f_{1}^{2}\left( X\right) }\right) \\
&&+\frac{3\left( 1+\delta \right) \sigma _{\hat{K}}^{2}f_{1}\left( X\right) 
}{\left( 1-\beta \right) \delta C\left( X+C\frac{\beta \delta +\beta ^{B}}{%
1+\delta }\right) ^{2}\left( 2X-C\frac{\beta \delta +\beta ^{B}}{1+\delta }%
\right) \epsilon }\left( \frac{\left( \left\Vert \hat{\Psi}_{0}\left( \hat{X}%
_{1}\right) \right\Vert ^{2}-\hat{\mu}D\left( \hat{X}_{1}\right) \right) 
\sqrt{k\sigma _{\hat{K}}^{2}\left( 1-\frac{\hat{g}^{ef}\left( \hat{X}%
_{1}\right) }{\left\langle \hat{g}\right\rangle }\right) }}{\left( 1+\left[ 
\underline{\hat{k}}_{2}^{n}\left( \hat{X}\right) \right] _{\kappa }\right) 
\sqrt{\underline{k}\left( \hat{X}_{1}\right) }}\right) \\
&=&\frac{3\left( 1+\delta \right) \sigma _{\hat{K}}^{2}f_{1}\left( X\right) 
}{\left( 1-\beta \right) \delta C\left( X+C\frac{\beta \delta +\beta ^{B}}{%
1+\delta }\right) ^{2}\left( 2X-C\frac{\beta \delta +\beta ^{B}}{1+\delta }%
\right) \epsilon }\left( H+\frac{\bar{r}^{\prime }}{1+\left\langle \hat{K}%
_{2}\right\rangle }\right) \\
&&+\frac{\left( 3X-\frac{\beta \delta +\beta ^{B}}{1+\delta }C\right) \left( 
\frac{\beta \delta +\beta ^{B}}{1+\delta }C+X\right) \left( R+\Delta F_{\tau
}\left( \bar{R}\left( K,X\right) \right) \right) }{4f_{1}\left( X\right)
C\left( 2X-C\frac{\beta \delta +\beta ^{B}}{1+\delta }\right) }
\end{eqnarray*}%
which becomes:%
\begin{eqnarray*}
&&\frac{1}{1+\left( 1+\delta \right) \underline{k}^{B}\left( X\right) }%
\left( 1+\frac{\left( \beta \delta +\beta ^{B}\right) \left( R+\Delta
F_{\tau }\left( \bar{R}\left( K,X\right) \right) \right) }{\left( 1+\delta
\right) f_{1}^{2}\left( X\right) }\right) \\
&&+\frac{3\left( 1+\delta \right) \sigma _{\hat{K}}^{2}f_{1}\left( X\right) 
}{\left( 1-\beta \right) \delta C\left( X+C\frac{\beta \delta +\beta ^{B}}{%
1+\delta }\right) ^{2}\left( 2X-C\frac{\beta \delta +\beta ^{B}}{1+\delta }%
\right) \epsilon }\left( \frac{\left( \left\Vert \hat{\Psi}_{0}\left( \hat{X}%
_{1}\right) \right\Vert ^{2}-\hat{\mu}D\left( \hat{X}_{1}\right) \right) 
\sqrt{k\sigma _{\hat{K}}^{2}\left( 1-\frac{\hat{g}^{ef}\left( \hat{X}%
_{1}\right) }{\left\langle \hat{g}\right\rangle }\right) }}{\left( 1+\left[ 
\underline{\hat{k}}_{2}^{n}\left( \hat{X}\right) \right] _{\kappa }\right) 
\sqrt{\underline{k}\left( \hat{X}_{1}\right) }}\right) \\
&=&\frac{3\left( 1+\delta \right) \sigma _{\hat{K}}^{2}f_{1}\left( X\right) 
}{\left( 1-\beta \right) \delta C\left( X+C\frac{\beta \delta +\beta ^{B}}{%
1+\delta }\right) ^{2}\left( 2X-C\frac{\beta \delta +\beta ^{B}}{1+\delta }%
\right) \epsilon }\left( H+\frac{\bar{r}^{\prime }}{\left( 1+\left( \hat{%
\beta}\hat{\delta}+\hat{\beta}^{B}\right) \underline{\hat{k}}^{B}\left( \hat{%
X}_{1}\right) \right) }\right) \\
&&+\frac{R\left( 3X-\frac{\beta \delta +\beta ^{B}}{1+\delta }C\right)
\left( \frac{\beta \delta +\beta ^{B}}{1+\delta }C+X\right) }{4f_{1}\left(
X\right) C\left( 2X-C\frac{\beta \delta +\beta ^{B}}{1+\delta }\right) }+%
\frac{\left( \beta \delta +\beta ^{B}\right) \left( R+\Delta F_{\tau }\left( 
\bar{R}\left( K,X\right) \right) \right) }{\left( 1+\delta \right)
f_{1}^{2}\left( X\right) }
\end{eqnarray*}%
This is an equation for $\underline{k}\left( X\right) $. Defining: 
\begin{equation*}
a=\frac{3\left( 1+\delta \right) \sigma _{\hat{K}}^{2}f_{1}\left( X\right) }{%
\left( 1-\beta \right) \delta C\left( X+C\frac{\beta \delta +\beta ^{B}}{%
1+\delta }\right) ^{2}\left( 2X-C\frac{\beta \delta +\beta ^{B}}{1+\delta }%
\right) \epsilon }\left( \frac{\left( \left\Vert \hat{\Psi}_{0}\left( \hat{X}%
_{1}\right) \right\Vert ^{2}-\hat{\mu}D\left( \hat{X}_{1}\right) \right) 
\sqrt{k\sigma _{\hat{K}}^{2}\left( 1-\frac{\hat{g}^{ef}\left( \hat{X}%
_{1}\right) }{\left\langle \hat{g}\right\rangle }\right) }}{1+\left[ 
\underline{\hat{k}}_{2}^{n}\left( \hat{X}\right) \right] _{\kappa }}\right)
\end{equation*}%
and:%
\begin{eqnarray*}
c &=&\frac{3\left( 1+\delta \right) \sigma _{\hat{K}}^{2}f_{1}\left(
X\right) }{\left( 1-\beta \right) \delta C\left( X+C\frac{\beta \delta
+\beta ^{B}}{1+\delta }\right) ^{2}\left( 2X-C\frac{\beta \delta +\beta ^{B}%
}{1+\delta }\right) \epsilon }\left( H+\frac{\bar{r}^{\prime }}{1+\left[ 
\underline{\hat{k}}_{2}^{n}\left( \hat{X}\right) \right] _{\kappa }}\right)
\\
&&+\frac{R\left( 3X-\frac{\beta \delta +\beta ^{B}}{1+\delta }C\right)
\left( \frac{\beta \delta +\beta ^{B}}{1+\delta }C+X\right) }{4f_{1}\left(
X\right) C\left( 2X-C\frac{\beta \delta +\beta ^{B}}{1+\delta }\right) }+%
\frac{\left( \beta \delta +\beta ^{B}\right) R}{\left( 1+\delta \right)
f_{1}^{2}\left( X\right) }+\Delta F_{\tau }\left( \bar{R}\left( K,X\right)
\right)
\end{eqnarray*}%
fr $\underline{k}\left( X\right) >>1$ nd $\epsilon <\sigma _{\hat{K}%
}^{2}f_{1}\left( X\right) $ equation simplifies as:%
\begin{equation*}
\frac{a}{\sqrt{\underline{k}\left( X\right) }}-c=0
\end{equation*}%
with solution:%
\begin{equation*}
\sqrt{\underline{k}\left( X\right) }=\frac{a}{c}
\end{equation*}%
\begin{equation*}
\underline{k}\left( X\right) =\left( \frac{\left( \left\Vert \hat{\Psi}%
_{0}\left( \hat{X}_{1}\right) \right\Vert ^{2}-\hat{\mu}D\left( \hat{X}%
_{1}\right) \right) \sqrt{k\sigma _{\hat{K}}^{2}\left( 1-\frac{\hat{g}%
^{ef}\left( \hat{X}_{1}\right) }{\left\langle \hat{g}\right\rangle }\right) }%
}{\left( 1+\left[ \underline{\hat{k}}_{2}^{n}\left( \hat{X}\right) \right]
_{\kappa }\right) \left[ H+\frac{\bar{r}^{\prime }}{\left( 1+\left[ 
\underline{\hat{k}}_{2}^{n}\left( \hat{X}_{1}\right) \right] _{\kappa
}\right) }+\left( \frac{A\left( X\right) }{f_{1}^{2}\left( X\right) }+\frac{%
B\left( X\right) }{f_{1}^{3}\left( X\right) }\right) \left( R+\Delta F_{\tau
}\left( \bar{R}\left( K,X\right) \right) \right) \right] }-\sqrt{\frac{%
\sigma _{\hat{K}}^{2}\hat{\mu}k}{2}}D\left( \hat{X}_{1}\right) \right) ^{2}
\end{equation*}%
$\allowbreak $ $\allowbreak $with:%
\begin{eqnarray*}
A\left( X\right) &=&\frac{\epsilon \left( 1-\beta \right) \delta \left( X+C%
\frac{\beta \delta +\beta ^{B}}{1+\delta }\right) ^{2}}{3\left( 1+\delta
\right) \sigma _{\hat{K}}^{2}}\frac{\left( 3X-\frac{\beta \delta +\beta ^{B}%
}{1+\delta }C\right) \left( \frac{\beta \delta +\beta ^{B}}{1+\delta }%
C+X\right) }{4C\left( 2X-C\frac{\beta \delta +\beta ^{B}}{1+\delta }\right) }
\\
B\left( X\right) &=&\frac{\epsilon \left( 1-\beta \right) \delta \left( X+C%
\frac{\beta \delta +\beta ^{B}}{1+\delta }\right) ^{2}}{3\left( 1+\delta
\right) \sigma _{\hat{K}}^{2}}\frac{\left( \beta \delta +\beta ^{B}\right)
C\left( 2X-C\frac{\beta \delta +\beta ^{B}}{1+\delta }\right) }{\left(
1+\delta \right) }
\end{eqnarray*}%
and the equation for return writes:%
\begin{eqnarray*}
\frac{\hat{g}\left( \hat{X}_{1}\right) }{1+\left[ \underline{\hat{k}}%
_{2}^{n}\left( \hat{X}_{1}\right) \right] _{\kappa }} &=&\frac{\left(
\left\Vert \hat{\Psi}_{0}\left( \hat{X}_{1}\right) \right\Vert ^{2}-\hat{\mu}%
D\left( \hat{X}_{1}\right) \right) \sqrt{k\sigma _{\hat{K}}^{2}\left( 1-%
\frac{\hat{g}^{ef}\left( \hat{X}_{1}\right) }{\left\langle \hat{g}%
\right\rangle }\right) }}{\left( 1+\left[ \underline{\hat{k}}_{2}^{n}\left( 
\hat{X}\right) \right] _{\kappa }\right) \sqrt{\underline{k}\left( \hat{X}%
_{1}\right) }} \\
&\simeq &H+\frac{\bar{r}^{\prime }}{1+\left[ \underline{\hat{k}}%
_{2}^{n}\left( \hat{X}\right) \right] _{\kappa }}+\frac{\frac{R\left( 3X-%
\frac{\beta \delta +\beta ^{B}}{1+\delta }C\right) \left( \frac{\beta \delta
+\beta ^{B}}{1+\delta }C+X\right) }{4f_{1}\left( X\right) C\left( 2X-C\frac{%
\beta \delta +\beta ^{B}}{1+\delta }\right) }+\frac{\left( \beta \delta
+\beta ^{B}\right) R}{\left( 1+\delta \right) f_{1}^{2}\left( X\right) }}{%
\frac{3\left( 1+\delta \right) \sigma _{\hat{K}}^{2}f_{1}\left( X\right) }{%
\left( 1-\beta \right) \delta C\left( X+C\frac{\beta \delta +\beta ^{B}}{%
1+\delta }\right) ^{2}\left( 2X-C\frac{\beta \delta +\beta ^{B}}{1+\delta }%
\right) \epsilon }} \\
&=&H+\frac{\bar{r}^{\prime }}{1+\underline{\hat{k}}_{2}\left( \hat{X}%
_{1}\right) }+\left( \frac{A\left( X\right) }{f_{1}^{2}\left( X\right) }+%
\frac{B\left( X\right) }{f_{1}^{3}\left( X\right) }\right) \left( R+\Delta
F_{\tau }\left( \bar{R}\left( K,X\right) \right) \right)
\end{eqnarray*}

\begin{eqnarray*}
\hat{g}\left( \hat{X}_{1}\right) -\bar{r}^{\prime } &=&\left( 1+\left[ 
\underline{\hat{k}}_{2}^{n}\left( \hat{X}\right) \right] _{\kappa }\right)
\int \hat{S}_{1}^{E}\left( \hat{X}^{\prime },\hat{X}_{1}\right) \left( \frac{%
\sqrt{k\sigma _{\hat{K}}^{2}\left( 1-\frac{\hat{g}^{ef}\left( \hat{X}%
_{1}\right) }{\left\langle \hat{g}\right\rangle }\right) }\left( \left\Vert 
\hat{\Psi}_{0}\left( \hat{X}^{\prime }\right) \right\Vert ^{2}-\hat{\mu}%
D\left( \hat{X}^{\prime }\right) \right) }{\sqrt{\underline{k}\left( \hat{X}%
^{\prime }\right) }}-\bar{r}^{\prime }\right) \\
&&+\left( 1+\left[ \underline{\hat{k}}_{2}^{n}\left( \hat{X}\right) \right]
_{\kappa }\right) \left( \frac{A\left( \hat{X}^{\prime }\right) }{%
f_{1}^{2}\left( \hat{X}^{\prime }\right) }+\frac{B\left( \hat{X}^{\prime
}\right) }{f_{1}^{3}\left( \hat{X}^{\prime }\right) }\right) \left( R+\Delta
F_{\tau }\left( \bar{R}\left( K,\hat{X}^{\prime }\right) \right) \right)
\end{eqnarray*}

with solution:%
\begin{eqnarray*}
\hat{g}\left( \hat{X}_{1}\right) -\bar{r}^{\prime } &=&\int \left( 1-\left(
1+\left[ \underline{\hat{k}}_{2}^{n}\left( \hat{X}_{1}\right) \right]
_{\kappa }\right) \hat{S}_{1}^{E}\left( \hat{X}^{\prime },\hat{X}_{1}\right)
\right) ^{-1} \\
&&\times \left( 1+\left[ \underline{\hat{k}}_{2}^{n}\left( \hat{X}%
_{1}\right) \right] _{\kappa }\right) \left( \frac{A\left( \hat{X}^{\prime
}\right) }{f_{1}^{2}\left( \hat{X}^{\prime }\right) }+\frac{B\left( \hat{X}%
^{\prime }\right) }{f_{1}^{3}\left( \hat{X}^{\prime }\right) }\right) \left(
R+\Delta F_{\tau }\left( \bar{R}\left( K,\hat{X}^{\prime }\right) \right)
\right)
\end{eqnarray*}%
Remark that given:%
\begin{equation*}
\hat{g}\left( \hat{X}_{1}\right) \left( \hat{X}_{1}\right) -\bar{r}^{\prime
}=\left( 1-M\right) ^{-1}\left( \hat{f}\left( \hat{X}_{1}\right) -\bar{r}%
\right)
\end{equation*}%
the solution for $\left( \hat{f}\left( \hat{X}_{1}\right) -\bar{r}\right) $
is obtained directly by multiplying by $\left( 1-M\right) $.

\subsection*{A21.3 Estimation of $\hat{S}_{1}^{E}\left( \hat{X}^{\prime },%
\hat{X}_{1}\right) $}

We estimate:%
\begin{eqnarray*}
&&\left( 1+\left[ \underline{\hat{k}}_{2}^{n}\left( \hat{X}_{1}\right) %
\right] _{\kappa }\right) \hat{S}_{1}^{E}\left( \hat{X}^{\prime },\hat{X}%
_{1}\right) \\
&\rightarrow &\frac{\left( 1+\left[ \underline{\hat{k}}_{2}^{n}\left( \hat{X}%
_{1}\right) \right] _{\kappa }\right) }{\left\langle \hat{K}\right\rangle
\left\Vert \hat{\Psi}\right\Vert ^{2}}\left( \frac{\hat{k}\left( \hat{X},%
\hat{X}^{\prime }\right) -\frac{\underline{\hat{k}}_{1}\left( \left\langle
X\right\rangle ,\hat{X}\right) \left\Vert \hat{\Psi}\left( \left\langle \hat{%
X}\right\rangle \right) \right\Vert ^{2}}{1+\left[ \underline{\hat{k}}\left(
\left\langle \hat{X}\right\rangle \right) \right] _{\kappa }}k\left(
\left\langle X\right\rangle ,X^{\prime }\right) }{1+\left[ \underline{\hat{k}%
}_{2}\left( \hat{X}\right) \right] _{\kappa }}+\frac{\hat{k}_{1}\left( \hat{X%
}^{\prime },\hat{X}\right) }{1+\left[ \underline{\hat{k}}_{2}\left( \hat{X}%
^{\prime }\right) \right] _{\kappa }}\right) \frac{\hat{K}_{X}\left\Vert 
\hat{\Psi}\left( \hat{X}^{\prime }\right) \right\Vert ^{2}}{1+\left[ 
\underline{\hat{k}}\left( \hat{X}\right) \right] _{\kappa }}
\end{eqnarray*}%
\bigskip

We use (\ref{KD}), (\ref{PK}) and (\ref{KPS}). and replace at the lowest
order $\hat{g}\left( \left\langle \hat{X}\right\rangle \right) \rightarrow 
\bar{r}^{\prime }$%
\begin{equation*}
\hat{D}\left( \hat{X}_{1}\right) =\left( \frac{\left\langle \hat{K}%
\right\rangle ^{2}\left\langle \hat{g}\right\rangle }{\sigma _{\hat{K}}^{2}}+%
\frac{1}{2}\right) \frac{\left\langle \hat{K}_{0}\right\rangle }{%
\left\langle \hat{K}\right\rangle }\left\langle \hat{g}^{ef}\right\rangle
-\left( \frac{\left\langle \bar{K}\right\rangle ^{2}\left\langle \bar{g}%
\right\rangle }{\sigma _{\hat{K}}^{2}}+\frac{1}{2}\right) \Delta \left( \hat{%
k}^{B}\left( \hat{X}_{1},\left\langle \bar{X}\right\rangle \right) A\right)
Z\left\langle \bar{g}\right\rangle
\end{equation*}%
\begin{eqnarray}
\hat{K}\left[ \hat{X}_{1}\right] &=&\hat{K}_{\hat{X}}\left\Vert \hat{\Psi}%
\left( \hat{X}_{1}\right) \right\Vert ^{2}\simeq \hat{\mu}\frac{\hat{K}%
_{0}^{4}}{2\sigma _{\hat{K}}^{2}}\left( \frac{\hat{g}^{2}\left( \hat{X}%
_{1}\right) }{4}-\frac{\left\langle \hat{g}\right\rangle \hat{g}^{ef}\left( 
\hat{X}_{1}\right) }{3}\right) \\
&=&\frac{\hat{\mu}}{2\sigma _{\hat{K}}^{2}}\left( 2\frac{\sigma _{\hat{K}%
}^{2}}{\hat{g}^{2}\left( \hat{X}_{1}\right) }\left( \frac{\left\Vert \hat{%
\Psi}_{0}\left( \hat{X}_{1}\right) \right\Vert ^{2}}{\hat{\mu}}-D\left( \hat{%
X}_{1}\right) \right) \right) ^{2}\left( \frac{\hat{g}^{2}\left( \hat{X}%
_{1}\right) }{4}-\frac{\left\langle \hat{g}\right\rangle \hat{g}^{ef}\left( 
\hat{X}_{1}\right) }{3}\right)  \notag
\end{eqnarray}%
\begin{eqnarray}
\left\Vert \hat{\Psi}\left( \hat{X}_{1}\right) \right\Vert ^{2} &\simeq &%
\hat{\mu}\frac{\hat{K}_{0}^{3}}{\sigma _{\hat{K}}^{2}}\left( \frac{\hat{g}%
^{2}\left( \hat{X}_{1}\right) }{3}-\frac{\left\langle \hat{K}\right\rangle }{%
2\hat{K}_{0}}\left\langle \hat{g}\right\rangle \left\langle \hat{g}%
^{ef}\right\rangle \right) \\
&=&\frac{\hat{\mu}}{\sigma _{\hat{K}}^{2}}\left( 2\frac{\sigma _{\hat{K}}^{2}%
}{\hat{g}^{2}\left( \hat{X}_{1}\right) }\left( \frac{\left\Vert \hat{\Psi}%
_{0}\left( \hat{X}_{1}\right) \right\Vert ^{2}}{\hat{\mu}}-D\left( \hat{X}%
_{1}\right) \right) \right) ^{\frac{3}{2}}\left( \frac{\hat{g}^{2}\left( 
\hat{X}_{1}\right) }{3}-\frac{\left\langle \hat{g}\right\rangle \hat{g}%
^{ef}\left( \hat{X}_{1}\right) }{2}\right)  \notag
\end{eqnarray}%
\begin{equation}
\hat{K}_{\hat{X}_{1}}=\frac{\hat{K}\left[ \hat{X}_{1}\right] }{\left\Vert 
\hat{\Psi}\left( \hat{X}_{1}\right) \right\Vert ^{2}}=\frac{\sqrt{2\frac{%
\sigma _{\hat{K}}^{2}}{\hat{g}^{2}\left( \hat{X}_{1}\right) }\left( \frac{%
\left\Vert \hat{\Psi}_{0}\left( \hat{X}_{1}\right) \right\Vert ^{2}}{\hat{\mu%
}}-D\left( \hat{X}_{1}\right) \right) }\left( \frac{\hat{g}^{2}\left( \hat{X}%
_{1}\right) }{4}-\frac{\left\langle \hat{g}\right\rangle \hat{g}^{ef}\left( 
\hat{X}_{1}\right) }{3}\right) }{2\left( \frac{\hat{g}^{2}\left( \hat{X}%
_{1}\right) }{3}-\frac{\left\langle \hat{g}\right\rangle \hat{g}^{ef}\left( 
\hat{X}_{1}\right) }{2}\right) }
\end{equation}%
\begin{equation}
\hat{g}\left( \hat{X}_{1}\right) \rightarrow \frac{\left( \left\Vert \hat{%
\Psi}_{0}\left( \hat{X}_{1}\right) \right\Vert ^{2}-\hat{\mu}D\left( \hat{X}%
_{1}\right) \right) \sqrt{\left( 1-\frac{\hat{g}^{ef}\left( \hat{X}^{\prime
}\right) }{\left\langle \hat{g}\right\rangle }\right) }}{\sqrt{\frac{\hat{\mu%
}\hat{K}\left[ \hat{X}_{1}\right] }{2\sigma _{\hat{K}}^{2}}}}\rightarrow 
\bar{r}^{\prime }
\end{equation}

\begin{equation}
\hat{K}_{\hat{X}}=\frac{\sqrt{\frac{\sigma _{\hat{K}}^{2}}{2}\left( \frac{%
\left\Vert \hat{\Psi}_{0}\left( \hat{X}\right) \right\Vert ^{2}}{\hat{\mu}}%
-D\left( \hat{X}\right) \right) }\left( \frac{\left( \bar{r}^{\prime
}\right) ^{2}}{4}-\frac{\bar{r}^{\prime }\hat{g}^{ef}}{3}\right) }{\left( 
\frac{\left( \bar{r}^{\prime }\right) ^{2}}{3}-\frac{\bar{r}^{\prime }\hat{g}%
^{ef}}{2}\right) \bar{r}^{\prime }}
\end{equation}%
\begin{equation}
\left\Vert \hat{\Psi}\left( \hat{X}^{\prime }\right) \right\Vert ^{2}\simeq 
\frac{2\sqrt{2\sigma _{\hat{K}}^{2}}\hat{\mu}}{\bar{r}^{\prime }}\left( 
\frac{\left\Vert \hat{\Psi}_{0}\left( \hat{X}^{\prime }\right) \right\Vert
^{2}}{\hat{\mu}}-D\left( \hat{X}^{\prime }\right) \right) ^{\frac{3}{2}%
}\left( \frac{1}{3}-\frac{\left\langle \hat{K}\right\rangle }{2\left\langle 
\hat{K}_{0}\right\rangle }\frac{\hat{g}^{ef}}{\bar{r}^{\prime }}\right)
\end{equation}%
\begin{equation*}
D\left( \hat{X}_{1}\right) \rightarrow \left( \frac{\left\langle \hat{K}%
\right\rangle ^{2}\left\langle \hat{g}\right\rangle }{\sigma _{\hat{K}}^{2}}+%
\frac{1}{2}\right) \frac{\left\langle \hat{K}_{0}\right\rangle }{%
\left\langle \hat{K}\right\rangle }\left\langle \hat{g}^{ef}\right\rangle
\end{equation*}%
\bigskip 
\begin{equation*}
\hat{K}_{\hat{X}}\left\Vert \hat{\Psi}\left( \hat{X}^{\prime }\right)
\right\Vert ^{2}\rightarrow \frac{2\sqrt{2}\sigma _{\hat{K}}^{2}}{\hat{\mu}%
\bar{r}^{\prime 2}}\sqrt{\left\Vert \hat{\Psi}_{0}\left( \hat{X}\right)
\right\Vert ^{2}-\hat{\mu}D\left( \hat{X}\right) }\left( \left\Vert \hat{\Psi%
}_{0}\left( \hat{X}^{\prime }\right) \right\Vert ^{2}-\hat{\mu}D\left( \hat{X%
}^{\prime }\right) \right) ^{\frac{3}{2}}
\end{equation*}

\bigskip 
\begin{equation*}
\frac{\left( 1+\left[ \underline{\hat{k}}_{2}^{n}\left( \hat{X}_{1}\right) %
\right] _{\kappa }\right) }{\left\langle \hat{K}\right\rangle \left\Vert 
\hat{\Psi}\right\Vert ^{2}}\left( \frac{\hat{k}\left( \hat{X},\hat{X}%
^{\prime }\right) -\frac{\underline{\hat{k}}_{1}\left( \left\langle
X\right\rangle ,\hat{X}\right) }{1+\left[ \underline{\hat{k}}\left(
\left\langle \hat{X}\right\rangle \right) \right] _{\kappa }}k\left(
\left\langle X\right\rangle ,X^{\prime }\right) }{1+\left[ \underline{\hat{k}%
}_{2}\left( \hat{X}\right) \right] _{\kappa }}+\frac{\hat{k}_{1}\left( \hat{X%
}^{\prime },\hat{X}\right) }{1+\left[ \underline{\hat{k}}_{2}\left( \hat{X}%
^{\prime }\right) \right] _{\kappa }}\right) \frac{\hat{K}_{X}\left\Vert 
\hat{\Psi}\left( \hat{X}^{\prime }\right) \right\Vert ^{2}}{1+\left[ 
\underline{\hat{k}}\left( \hat{X}\right) \right] _{\kappa }}
\end{equation*}%
\begin{eqnarray*}
&&\left( 1+\left[ \underline{\hat{k}}_{2}^{n}\left( \hat{X}_{1}\right) %
\right] _{\kappa }\right) \hat{S}_{1}^{E}\left( \hat{X}^{\prime },\hat{X}%
_{1}\right) \\
&\rightarrow &\frac{\left( 1+\left[ \underline{\hat{k}}_{2}^{n}\left( \hat{X}%
_{1}\right) \right] _{\kappa }\right) }{1+\left[ \underline{\hat{k}}\left( 
\hat{X}_{1}\right) \right] _{\kappa }}\left( \frac{\hat{k}\left( \hat{X},%
\hat{X}^{\prime }\right) -\frac{\underline{\hat{k}}_{1}\left( \left\langle
X\right\rangle ,\hat{X}\right) }{1+\left[ \underline{\hat{k}}\left(
\left\langle \hat{X}\right\rangle \right) \right] _{\kappa }}k\left(
\left\langle X\right\rangle ,X^{\prime }\right) }{1+\left[ \underline{\hat{k}%
}_{2}^{n}\left( \hat{X}_{1}\right) \right] _{\kappa }}+\frac{\hat{k}%
_{1}\left( \hat{X}^{\prime },\hat{X}\right) }{1+\left[ \underline{\hat{k}}%
_{2}^{n}\left( \hat{X}^{\prime }\right) \right] _{\kappa }}\right) \frac{2%
\sqrt{2}\sigma _{\hat{K}}^{2}}{\hat{\mu}\bar{r}^{\prime 2}} \\
&&\times \frac{\sqrt{\left\Vert \hat{\Psi}_{0}\left( \hat{X}\right)
\right\Vert ^{2}-\hat{\mu}D\left( \hat{X}\right) }\left( \left\Vert \hat{\Psi%
}_{0}\left( \hat{X}^{\prime }\right) \right\Vert ^{2}-\hat{\mu}D\left( \hat{X%
}^{\prime }\right) \right) ^{\frac{3}{2}}}{\left\langle \hat{K}\right\rangle
\left\Vert \hat{\Psi}\right\Vert ^{2}} \\
&\rightarrow &\frac{\left( 1+\left[ \underline{\hat{k}}_{2}^{n}\left( \hat{X}%
_{1}\right) \right] _{\kappa }\right) }{1+\left[ \underline{\hat{k}}\left( 
\hat{X}_{1}\right) \right] _{\kappa }}\left( \frac{\hat{k}\left( \hat{X},%
\hat{X}^{\prime }\right) -\underline{\hat{k}}_{1}\left( \left\langle
X\right\rangle ,\hat{X}\right) k\left( \left\langle X\right\rangle
,X^{\prime }\right) }{1+\left[ \underline{\hat{k}}_{2}^{n}\left( \hat{X}%
_{1}\right) \right] _{\kappa }}+\frac{\hat{k}_{1}\left( \hat{X}^{\prime },%
\hat{X}\right) }{1+\left[ \underline{\hat{k}}_{2}^{n}\left( \hat{X}^{\prime
}\right) \right] _{\kappa }}\right) \frac{\left( \left\Vert \hat{\Psi}%
_{0}\left( \hat{X}^{\prime }\right) \right\Vert ^{2}-\hat{\mu}D\left( \hat{X}%
^{\prime }\right) \right) ^{2}}{\left( \left\Vert \hat{\Psi}_{0}\right\Vert
^{2}-\hat{\mu}\left\langle D\right\rangle \right) ^{2}}
\end{eqnarray*}%
and usng vr frml fr:%
\begin{equation*}
\frac{\left( \left\Vert \hat{\Psi}_{0}\left( \hat{X}^{\prime }\right)
\right\Vert ^{2}-\hat{\mu}D\left( \hat{X}^{\prime }\right) \right) ^{2}}{%
\left( \left\Vert \hat{\Psi}_{0}\right\Vert ^{2}-\hat{\mu}\left\langle
D\right\rangle \right) ^{2}}
\end{equation*}%
lds t:%
\begin{eqnarray*}
&&\frac{\left( \left\Vert \hat{\Psi}_{0}\left( \hat{X}^{\prime }\right)
\right\Vert ^{2}-\hat{\mu}D\left( \hat{X}^{\prime }\right) \right) ^{2}}{%
\left( \left\Vert \hat{\Psi}_{0}\right\Vert ^{2}-\hat{\mu}\left\langle
D\right\rangle \right) ^{2}} \\
&\simeq &\left( \frac{\sqrt{\left\langle \hat{g}\right\rangle ^{2}\left( 1+%
\frac{3}{4}\frac{\left\langle \hat{g}^{ef}\right\rangle }{\left\langle \hat{g%
}\right\rangle }\right) }\left\Vert \bar{\Psi}_{0}\left( \hat{X}_{1}\right)
\right\Vert -\frac{3}{8}\left\langle \hat{g}\right\rangle \frac{\hat{g}%
^{Bef}\left( \bar{X}_{1}\right) }{\bar{g}\left( \bar{X}_{1}\right) }%
\left\Vert \hat{\Psi}_{0}\left( \bar{X}_{1}\right) \right\Vert }{\sqrt{%
\left\langle \hat{g}\right\rangle ^{2}\left( 1+\frac{3}{4}\frac{\left\langle 
\hat{g}^{ef}\right\rangle }{\left\langle \hat{g}\right\rangle }\right) }%
\left\Vert \bar{\Psi}_{0}\left( \hat{X}_{1}\right) \right\Vert -\frac{3}{8}%
\left\langle \hat{g}\right\rangle \frac{\left\langle \hat{g}%
^{Bef}\right\rangle }{\left\langle \bar{g}\right\rangle }\left\Vert \hat{\Psi%
}_{0}\left( \bar{X}_{1}\right) \right\Vert }\right) ^{4}\frac{\frac{1}{4\bar{%
g}^{2}\left( \bar{X}_{1}\right) }-\frac{\left\langle \hat{g}\right\rangle 
\hat{g}^{Bef}\left( \bar{X}_{1}\right) }{3\bar{g}^{4}\left( \bar{X}%
_{1}\right) }}{\frac{1}{4\left\langle \bar{g}\right\rangle ^{2}}-\frac{%
\left\langle \hat{g}\right\rangle \left\langle \hat{g}^{Bef}\right\rangle }{%
3\left\langle \bar{g}\right\rangle ^{2}}}
\end{eqnarray*}

\section*{Appendix 22 Banks returns}

The equation for bank returns expressed in terms of $\bar{g}$ is:

\begin{eqnarray}
&&\left( \Delta \left( \bar{X}^{\prime },\bar{X}\right) -\frac{\bar{K}%
^{\prime }\bar{k}_{1}\left( \bar{X}^{\prime },\bar{X}\right) \left\vert \bar{%
\Psi}\left( \bar{K}^{\prime },\bar{X}^{\prime }\right) \right\vert ^{2}}{1+%
\underline{\hat{k}}^{B}\left( \bar{X}^{\prime }\right) }\right) \frac{\left(
1-\bar{M}\right) \bar{g}\left( \hat{K}^{\prime },\hat{X}^{\prime }\right) }{%
1+\underline{\overline{\bar{k}}}_{2}\left( \bar{X}^{\prime }\right) }
\label{GBR} \\
&&-\frac{\hat{K}^{\prime }\underline{\hat{k}}_{1}^{B}\left( \hat{X}^{\prime
},\bar{X}\right) }{1+\underline{\hat{k}}\left( \hat{X}^{\prime }\right) +%
\underline{\hat{k}}_{1}^{B}\left( \bar{X}^{\prime }\right) +\kappa \left[ 
\frac{\underline{\hat{k}}_{2}^{B}}{1+\bar{k}}\right] \left( \hat{X}^{\prime
}\right) }\frac{\left( 1-\hat{M}\right) \hat{g}\left( \hat{K}^{\prime },\hat{%
X}^{\prime }\right) +\bar{N}\bar{g}\left( \hat{K}^{\prime },\hat{X}^{\prime
}\right) }{1+\underline{\hat{k}}_{2}\left( \bar{X}^{\prime }\right) +\kappa 
\frac{\underline{\hat{k}}_{2}^{B}\left( \bar{X}^{\prime }\right) }{1+\bar{k}%
\left( \bar{X}\right) }}  \notag \\
&=&\frac{\underline{k}_{1}^{\left( B\right) }\left( X^{\prime },\bar{X}%
\right) }{1+\underline{k}\left( \hat{X}^{\prime }\right) +\underline{k}%
_{1}^{\left( B\right) }\left( \bar{X}^{\prime }\right) +\kappa \frac{%
\underline{k}_{2}^{\left( B\right) }\left( \bar{X}^{\prime }\right) }{1+%
\underline{\bar{k}}}}\frac{\left( f_{1}^{\prime }\left( X^{\prime }\right)
K^{\prime }-\bar{C}\left( X^{\prime }\right) \right) }{1+\underline{k}%
_{2}\left( \hat{X}^{\prime }\right) +\kappa \frac{\underline{k}_{2}^{\left(
B\right) }\left( \bar{X}^{\prime }\right) }{1+\underline{\bar{k}}}}  \notag
\end{eqnarray}

\subsection*{A22.1 Use of normalizations and definition of coefficients}

Using the normalizations, we replace:%
\begin{eqnarray*}
&&1+\underline{\hat{k}}\left( \hat{X}^{\prime }\right) +\underline{\hat{k}}%
_{1}^{B}\left( \bar{X}^{\prime }\right) +\kappa \frac{\underline{\hat{k}}%
_{2}^{\left( B\right) }\left( \bar{X}^{\prime }\right) }{1+\underline{\bar{k}%
}} \\
&\rightarrow &1+\left( \hat{k}\left( \hat{X}^{\prime },\left\langle \hat{X}%
\right\rangle \right) -\left\langle \hat{k}\right\rangle \right) +\left( 
\hat{k}_{1}^{B}\left( \bar{X}^{\prime },\left\langle \bar{X}\right\rangle
\right) +\kappa \frac{\hat{k}_{2}^{\left( B\right) }\left( \bar{X}^{\prime
},\left\langle \bar{X}\right\rangle \right) }{1+\underline{\bar{k}}}-\left(
\left\langle \hat{k}_{1}^{B}\right\rangle +\kappa \frac{\left\langle \hat{k}%
_{2}^{\left( B\right) }\right\rangle }{1+\underline{\bar{k}}}\right) \right)
Z \\
&=&1+\left[ \underline{\hat{k}}\left( \hat{X}^{\prime }\right) \right]
_{\kappa }
\end{eqnarray*}%
and:%
\begin{eqnarray*}
1+\underline{\bar{k}}_{2}\left( \bar{X}_{1}\right) &=&\frac{\left(
1-\left\langle \bar{k}_{1}\right\rangle \right) \left( 1+\frac{\underline{%
\bar{k}}_{2}\left( \bar{X}_{1}\right) }{1-\left\langle \bar{k}%
_{1}\right\rangle }\right) }{1-\left\langle \bar{k}\right\rangle } \\
&=&\frac{1-\left\langle \bar{k}\right\rangle +\bar{k}_{2}\left( \bar{X}%
_{1},\left\langle \bar{X}\right\rangle \right) -\left\langle \bar{k}%
_{2}\right\rangle }{1-\left\langle \bar{k}\right\rangle } \\
&=&1+\frac{\bar{k}_{2}\left( \bar{X}_{1},\left\langle \bar{X}\right\rangle
\right) }{1-\left\langle \bar{k}\right\rangle }=1+\bar{k}_{2}^{n}\left( \bar{%
X}_{1}\right)
\end{eqnarray*}%
We will write:%
\begin{equation*}
\frac{1}{1+\underline{\bar{k}}_{2}\left( \bar{X}_{1}\right) }\rightarrow 
\frac{1}{1+\bar{k}_{2}^{n}\left( \bar{X}_{1}\right) }
\end{equation*}%
and:%
\begin{equation*}
1+\underline{\bar{k}}\left( \bar{X}^{\prime }\right) \simeq 1+\left( \bar{k}%
\left( \bar{X}^{\prime },\left\langle \bar{X}\right\rangle \right)
-\left\langle \bar{k}\right\rangle \right)
\end{equation*}

\subsection*{A22.2 Definition of diffusion functions}

We consider the terms of (\ref{GBR}) proportional to $\bar{g}\left( \bar{K}%
^{\prime },\bar{X}^{\prime }\right) $, this defines the functns $\bar{S}%
_{1}^{E}\left( \bar{X}^{\prime },\hat{X}_{1}\right) $and $\bar{S}%
_{1}^{B}\left( \hat{X}^{\prime },\hat{X}_{1}\right) $: 
\begin{eqnarray*}
&&\left( 1-\frac{\bar{K}^{\prime }\bar{k}_{1}\left( \bar{X}^{\prime },\bar{X}%
_{1}\right) \left\vert \bar{\Psi}\left( \bar{K}^{\prime },\bar{X}^{\prime
}\right) \right\vert ^{2}}{1+\underline{\bar{k}}\left( \bar{X}^{\prime
}\right) }\right) \frac{\left( 1-\bar{M}\right) \bar{g}\left( \bar{K}%
^{\prime },\bar{X}^{\prime }\right) }{1+\bar{k}_{2}^{n}\left( \bar{X}%
^{\prime }\right) }-\frac{\bar{K}^{\prime }\underline{\hat{k}}_{1}^{B}\left( 
\hat{X}^{\prime },\bar{X}_{1}\right) }{1+\left[ \underline{\hat{k}}%
^{n}\left( \hat{X}^{\prime }\right) \right] _{\kappa }}\frac{\bar{N}\bar{g}%
\left( \bar{K}^{\prime },\bar{X}^{\prime }\right) }{1+\left[ \underline{\hat{%
k}}_{2}^{n}\left( \hat{X}^{\prime }\right) \right] _{\kappa }} \\
&\rightarrow &1-\bar{S}_{1}^{E}\left( \bar{X}^{\prime },\bar{X}_{1}\right) -%
\bar{S}_{1}^{B}\left( \hat{X}^{\prime },\bar{X}_{1}\right)
\end{eqnarray*}

with:%
\begin{equation*}
\bar{S}_{1}^{E}\left( \bar{X}^{\prime },\bar{X}_{1}\right) \rightarrow
\left( \frac{\bar{k}\left( \bar{X}^{\prime },\bar{X}_{1}\right) -\frac{%
\underline{\bar{k}}_{1}\left( \left\langle X\right\rangle ,\bar{X}%
_{1}\right) \bar{k}\left( \left\langle X\right\rangle ,\bar{X}_{1}\right) }{%
1+\underline{\bar{k}}\left( \left\langle \bar{X}\right\rangle \right) }}{1+%
\bar{k}_{2}^{n}\left( \bar{X}^{\prime }\right) }+\frac{\bar{k}_{1}\left( 
\bar{X}^{\prime },\bar{X}_{1}\right) }{1+\bar{k}_{2}^{n}\left( \bar{X}%
^{\prime }\right) }\right) \frac{1}{1+\underline{\bar{k}}\left( \bar{X}%
^{\prime }\right) }
\end{equation*}%
and:%
\begin{eqnarray*}
\bar{S}_{1}^{B}\left( \bar{X}^{\prime },\bar{X}_{1}\right) &\rightarrow &%
\frac{\bar{K}^{\prime }\bar{k}_{1}\left( \bar{X}^{\prime },\bar{X}%
_{1}\right) }{1+\underline{\hat{k}}\left( \hat{X}^{\prime }\right) +%
\underline{\hat{k}}_{1}^{B}\left( \bar{X}^{\prime }\right) +\kappa \left[ 
\frac{\underline{\hat{k}}_{2}^{B}}{1+\bar{k}}\right] \left( \hat{X}^{\prime
}\right) }\frac{\bar{N}\left( \bar{X}^{\prime },\bar{X}_{1}\right) }{1+%
\underline{\hat{k}}_{2}\left( \bar{X}^{\prime }\right) +\kappa \frac{%
\underline{\hat{k}}_{2}^{B}\left( \bar{X}^{\prime }\right) }{1+\bar{k}\left( 
\bar{X}\right) }} \\
&=&\frac{\bar{K}^{\prime }\bar{k}_{1}\left( \bar{X}^{\prime },\bar{X}%
_{1}\right) }{1+\left[ \underline{\hat{k}}^{n}\left( \hat{X}\right) \right]
_{\kappa }}\frac{\bar{N}\left( \bar{X}^{\prime },\bar{X}_{1}\right) }{1+%
\left[ \underline{\hat{k}}_{2}^{n}\left( \hat{X}^{\prime }\right) \right]
_{\kappa }}
\end{eqnarray*}%
The coefficient $\bar{N}\left( \bar{X}^{\prime },\bar{X}_{1}\right) $ is
defined by: 
\begin{eqnarray*}
&&\bar{N}\left( \bar{X}^{\prime },\bar{X}_{1}\right) \\
&=&\frac{\left( \hat{k}_{1}^{B}\left( \hat{X},\bar{X}^{\prime }\right)
+\kappa \frac{\hat{k}_{2}^{B}\left( \hat{X},\bar{X}^{\prime }\right) }{%
1+\int \overline{\bar{k}}\left( \bar{X}^{\prime },\bar{X}^{\prime \prime
}\right) \bar{K}_{0}^{\prime \prime }\left\vert \bar{\Psi}\left( \bar{K}%
_{0}^{\prime \prime },\bar{X}^{\prime \prime }\right) \right\vert ^{2}}%
-\kappa \int \frac{\hat{k}_{2}^{B}\left( \bar{X},\bar{X}^{\prime \prime
}\right) \bar{K}_{0}^{\prime \prime }\overline{\bar{k}}\left( \bar{X}%
^{\prime \prime },\bar{X}^{\prime }\right) }{\left( 1+\int \overline{\bar{k}}%
\left( \bar{X}^{\prime \prime },\bar{Y}\right) \bar{K}_{0}^{Y}\left\vert 
\bar{\Psi}\left( \bar{K}_{0}^{Y},\bar{Y}\right) \right\vert ^{2}\right) ^{2}}%
\right) \hat{K}}{1+\int \hat{k}\left( \hat{X},\hat{X}^{\prime }\right)
\left\vert \hat{\Psi}\left( \hat{K}^{\prime },\hat{X}^{\prime }\right)
\right\vert ^{2}+\int \hat{k}_{1}^{B}\left( \hat{X},\bar{X}^{\prime }\right) 
\bar{K}_{0}^{\prime }\left\vert \bar{\Psi}\left( \bar{K}_{0}^{\prime },\bar{X%
}^{\prime }\right) \right\vert ^{2}+\kappa \int \hat{k}_{2}^{B}\left( \hat{X}%
,\bar{X}^{\prime }\right) \frac{\bar{K}_{0}^{\prime }\left\vert \bar{\Psi}%
\left( \bar{K}_{0}^{\prime },\bar{X}^{\prime }\right) \right\vert ^{2}}{%
1+\int \overline{\bar{k}}\left( \bar{X}^{\prime },\bar{X}^{\prime \prime
}\right) \bar{K}_{0}^{\prime \prime }\left\vert \bar{\Psi}\left( \bar{K}%
_{0}^{\prime \prime },\bar{X}^{\prime \prime }\right) \right\vert ^{2}}} \\
&\rightarrow &\frac{\left( \hat{k}_{1}^{B}\left( \hat{X},\bar{X}^{\prime
}\right) +\kappa \frac{\underline{\hat{k}}_{2}^{B}\left( \hat{X}\right) }{1+%
\underline{\overline{\bar{k}}}}\left( 1-\frac{\kappa }{1+\underline{%
\overline{\bar{k}}}}\right) \right) \hat{K}}{1+\left[ \underline{\hat{k}}%
_{2}^{n}\left( \hat{X}\right) \right] _{\kappa }}\rightarrow \frac{\left( 
\hat{k}_{1}^{B}\left( \hat{X}\right) +\kappa \frac{\underline{\hat{k}}%
_{2}^{B}\left( \hat{X}\right) }{1+\underline{\overline{\bar{k}}}}\left( 1-%
\frac{\kappa }{1+\underline{\overline{\bar{k}}}}\right) \right) }{1+\left[ 
\underline{\hat{k}}^{n}\left( \hat{X}\right) \right] _{\kappa }}
\end{eqnarray*}

The terms proportional to $\hat{g}\left( \hat{K}^{\prime },\hat{X}^{\prime
}\right) $ define the diffusion function $\hat{S}_{1}^{B}\left( \bar{X}%
^{\prime },\bar{X}_{1}\right) $:%
\begin{eqnarray*}
&&\frac{\hat{K}^{\prime }\underline{\hat{k}}_{1}^{B}\left( \hat{X}^{\prime },%
\bar{X}\right) }{1+\underline{\hat{k}}\left( \hat{X}^{\prime }\right) +%
\underline{\hat{k}}_{1}^{B}\left( \bar{X}^{\prime }\right) +\kappa \left[ 
\frac{\underline{\hat{k}}_{2}^{B}}{1+\bar{k}}\right] \left( \hat{X}^{\prime
}\right) }\frac{\left( 1-\hat{M}\right) \hat{g}\left( \hat{K}^{\prime },\hat{%
X}^{\prime }\right) }{1+\underline{\hat{k}}_{2}\left( \bar{X}^{\prime
}\right) +\kappa \frac{\underline{\hat{k}}_{2}^{B}\left( \bar{X}^{\prime
}\right) }{1+\bar{k}\left( \bar{X}\right) }} \\
&\rightarrow &\frac{\hat{K}^{\prime }\underline{\hat{k}}_{1}^{B}\left( \hat{X%
}^{\prime },\bar{X}\right) }{1+\left[ \underline{\hat{k}}\left( \hat{X}%
^{\prime }\right) \right] _{\kappa }}\frac{1}{1+\left[ \underline{\hat{k}}%
_{2}\left( \hat{X}\right) \right] _{\kappa }}\left( 1-\frac{\hat{k}\left( 
\hat{X},\hat{X}^{\prime }\right) \hat{K}_{X}}{1+\left[ \underline{\hat{k}}%
\left( \hat{X}^{\prime }\right) \right] _{\kappa }}\left\Vert \hat{\Psi}%
\left( \hat{X}^{\prime }\right) \right\Vert ^{2}\right) \hat{g}\left( \hat{K}%
^{\prime },\hat{X}^{\prime }\right) \\
&=&\hat{S}_{1}^{B}\left( \bar{X}^{\prime },\bar{X}_{1}\right) \hat{g}\left( 
\hat{K}^{\prime },\hat{X}^{\prime }\right)
\end{eqnarray*}%
and we write:%
\begin{equation*}
\hat{S}_{1}^{B}\left( \bar{X}^{\prime },\bar{X}_{1}\right) \simeq \frac{\hat{%
K}^{\prime }\underline{\hat{k}}_{1}^{B}\left( \hat{X}^{\prime },\bar{X}%
\right) }{1+\left[ \underline{\hat{k}}\left( \hat{X}^{\prime }\right) \right]
_{\kappa }}\frac{1}{1+\left[ \underline{\hat{k}}_{2}\left( \hat{X}\right) %
\right] _{\kappa }}\left( 1-\frac{\hat{k}\left( \hat{X},\hat{X}^{\prime
}\right) \hat{K}_{X}}{1+\left[ \underline{\hat{k}}\left( \hat{X}^{\prime
}\right) \right] _{\kappa }}\left\Vert \hat{\Psi}\left( \hat{X}^{\prime
}\right) \right\Vert ^{2}\right)
\end{equation*}%
Ultimately, we write:%
\begin{equation*}
\frac{\underline{k}_{1}^{\left( B\right) }\left( X^{\prime },\bar{X}\right) 
}{1+\underline{k}\left( \hat{X}^{\prime }\right) +\underline{k}_{1}^{\left(
B\right) }\left( \bar{X}^{\prime }\right) +\kappa \frac{\underline{k}%
_{2}^{\left( B\right) }\left( \bar{X}^{\prime }\right) }{1+\underline{\bar{k}%
}}}\rightarrow \frac{1-\beta ^{B}}{1+\delta }
\end{equation*}%
\bigskip

\subsection*{A22.3 Writing (\protect\ref{GBR})}

\begin{eqnarray}
&&\frac{1}{3}\frac{1-\beta ^{B}}{1+\delta }\left( \left( \frac{A\left( \bar{X%
}_{1}\right) }{f_{1}^{2}\left( \bar{X}_{1}\right) }+\frac{B\left( \bar{X}%
_{1}\right) }{f_{1}^{3}\left( \bar{X}_{1}\right) }\right) \left( R+\Delta
F_{\tau }\left( \bar{R}\left( K,\bar{X}_{1}\right) \right) \right) \right)
\label{GBr} \\
&&-\left( \frac{\left( \left\Vert \bar{\Psi}_{0}\left( \bar{X}_{1}\right)
\right\Vert ^{2}-\hat{\mu}D\left( \bar{X}_{1}\right) \right) \sqrt{k\frac{%
\sigma _{\hat{K}}^{2}}{2\hat{\mu}}\left( 1-\frac{\hat{g}^{Bef}\left( \bar{X}%
_{1}\right) }{\left\langle \bar{g}\right\rangle }\right) }}{\left( 1+\bar{k}%
_{2}^{n}\left( \bar{X}_{1}\right) \right) \sqrt{\underline{k}\left( \bar{X}%
_{1}\right) }}-\frac{\bar{r}^{\prime }}{1+\bar{k}_{2}^{n}\left( \bar{X}%
_{1}\right) }\right)  \notag \\
&=&-\left( \bar{S}_{1}^{E}\left( \bar{X}^{\prime },\hat{X}_{1}\right) +\bar{S%
}_{1}^{B}\left( \bar{X}^{\prime },\bar{X}_{1}\right) \right) \left( \frac{%
\left( \left\Vert \bar{\Psi}_{0}\left( \bar{X}^{\prime }\right) \right\Vert
^{2}-\hat{\mu}D\left( \bar{X}^{\prime }\right) \right) \sqrt{k\frac{\sigma _{%
\hat{K}}^{2}}{2\hat{\mu}}\left( 1-\frac{\hat{g}^{Bef}\left( \bar{X}^{\prime
}\right) }{\left\langle \bar{g}\right\rangle }\right) }}{\sqrt{\underline{k}%
\left( \bar{X}^{\prime }\right) }}-\bar{r}^{\prime }\right)  \notag \\
&&-\hat{S}_{1}^{B}\left( \hat{X}^{\prime },\bar{X}_{1}\right) \left( \frac{%
\left( \left\Vert \hat{\Psi}_{0}\left( \hat{X}^{\prime }\right) \right\Vert
^{2}-\hat{\mu}D\left( \hat{X}^{\prime }\right) \right) \sqrt{k\frac{\sigma _{%
\hat{K}}^{2}}{2\hat{\mu}}\left( 1-\frac{\hat{g}^{ef}\left( \hat{X}^{\prime
}\right) }{\left\langle \hat{g}\right\rangle }\right) }}{\sqrt{\underline{k}%
\left( \hat{X}^{\prime }\right) }}-\bar{r}^{\prime }\right)  \notag
\end{eqnarray}%
Using (\ref{RTn}): 
\begin{eqnarray*}
&&\frac{1}{3}\frac{1-\beta ^{B}}{1+\delta }\left( \left( \frac{A\left( \bar{X%
}_{1}\right) }{f_{1}^{2}\left( \bar{X}_{1}\right) }+\frac{B\left( \bar{X}%
_{1}\right) }{f_{1}^{3}\left( \bar{X}_{1}\right) }\right) \left( R+\Delta
F_{\tau }\left( \bar{R}\left( K,\bar{X}_{1}\right) \right) \right) \right) \\
&=&\frac{1-\beta ^{B}}{\left( 1-\beta \right) \delta }\left( \frac{\Delta
\left( \hat{X},\hat{X}^{\prime }\right) }{1+\left[ \underline{\hat{k}}%
_{2}^{n}\left( \hat{X}\right) \right] _{\kappa }}-\hat{S}_{1}^{E}\left( \hat{%
X}^{\prime },\hat{X}_{1}\right) \right) \left( \left( \left\Vert \hat{\Psi}%
_{0}\left( \hat{X}^{\prime }\right) \right\Vert ^{2}-\hat{\mu}D\left( \hat{X}%
^{\prime }\right) \right) \sqrt{\frac{\sigma _{\hat{K}}^{2}\left( 1-\frac{%
\hat{g}^{ef}\left( \hat{X}^{\prime }\right) }{\left\langle \hat{g}%
\right\rangle }\right) }{2\hat{\mu}\hat{K}\left[ \hat{X}^{\prime }\right] }}-%
\bar{r}^{\prime }\right)
\end{eqnarray*}%
we rewrite (\ref{GBr}):where:%
\begin{eqnarray*}
&&\frac{1-\beta ^{B}}{\left( 1-\beta \right) \delta }\left( \frac{\Delta
\left( \hat{X},\hat{X}^{\prime }\right) }{1+\left[ \underline{\hat{k}}%
_{2}^{n}\left( \hat{X}\right) \right] _{\kappa }}-\hat{S}_{1}^{E}\left( \hat{%
X}^{\prime },\hat{X}_{1}\right) \right) \left( \frac{\left( \left\Vert \bar{%
\Psi}_{0}\left( \bar{X}^{\prime }\right) \right\Vert ^{2}-\hat{\mu}D\left( 
\bar{X}^{\prime }\right) \right) \sqrt{k\frac{\sigma _{\hat{K}}^{2}}{2\hat{%
\mu}}\left( 1-\frac{\hat{g}^{Bef}\left( \bar{X}^{\prime }\right) }{%
\left\langle \bar{g}\right\rangle }\right) }}{\sqrt{\underline{k}\left( \bar{%
X}^{\prime }\right) }}-\bar{r}^{\prime }\right) \\
&&-\left( \frac{\left( \left\Vert \bar{\Psi}_{0}\left( \bar{X}_{1}\right)
\right\Vert ^{2}-\hat{\mu}D\left( \bar{X}_{1}\right) \right) \sqrt{k\frac{%
\sigma _{\hat{K}}^{2}}{2\hat{\mu}}\left( 1-\frac{\hat{g}^{Bef}\left( \bar{X}%
_{1}\right) }{\left\langle \bar{g}\right\rangle }\right) }}{\left( 1+\bar{k}%
_{2}^{n}\left( \bar{X}_{1}\right) \right) \sqrt{\underline{k}\left( \bar{X}%
_{1}\right) }}-\frac{\bar{r}^{\prime }}{1+\bar{k}_{2}^{n}\left( \bar{X}%
_{1}\right) }\right) \\
&=&-\left( \bar{S}_{1}^{E}\left( \bar{X}^{\prime },\hat{X}_{1}\right) +\bar{S%
}_{1}^{B}\left( \bar{X}^{\prime },\bar{X}_{1}\right) \right) \left( \frac{%
\left( \left\Vert \bar{\Psi}_{0}\left( \bar{X}^{\prime }\right) \right\Vert
^{2}-\hat{\mu}D\left( \bar{X}^{\prime }\right) \right) \sqrt{k\frac{\sigma _{%
\hat{K}}^{2}}{2\hat{\mu}}\left( 1-\frac{\hat{g}^{Bef}\left( \bar{X}^{\prime
}\right) }{\left\langle \bar{g}\right\rangle }\right) }}{\sqrt{\underline{k}%
\left( \bar{X}^{\prime }\right) }}-\bar{r}^{\prime }\right) \\
&&-\hat{S}_{1}^{B}\left( \hat{X}^{\prime },\bar{X}_{1}\right) \left( \frac{%
\left( \left\Vert \hat{\Psi}_{0}\left( \hat{X}^{\prime }\right) \right\Vert
^{2}-\hat{\mu}D\left( \hat{X}^{\prime }\right) \right) \sqrt{k\frac{\sigma _{%
\hat{K}}^{2}}{2\hat{\mu}}\left( 1-\frac{\hat{g}^{ef}\left( \hat{X}^{\prime
}\right) }{\left\langle \hat{g}\right\rangle }\right) }}{\sqrt{\underline{k}%
\left( \hat{X}^{\prime }\right) }}-\bar{r}^{\prime }\right)
\end{eqnarray*}%
and we find the followng equation:%
\begin{eqnarray*}
&&\left( 1+\bar{k}_{2}^{n}\left( \bar{X}_{1}\right) \right) \left( \frac{%
1-\beta ^{B}}{\left( 1-\beta \right) \delta }\left( \frac{\Delta \left( \hat{%
X},\hat{X}^{\prime }\right) }{1+\left[ \underline{\hat{k}}_{2}^{n}\left( 
\hat{X}\right) \right] _{\kappa }}-\hat{S}_{1}^{E}\left( \hat{X}^{\prime },%
\hat{X}_{1}\right) \right) +\hat{S}_{1}^{B}\left( \hat{X}^{\prime },\bar{X}%
_{1}\right) \right) \\
&&\times \left( \frac{\left( \left\Vert \bar{\Psi}_{0}\left( \bar{X}^{\prime
}\right) \right\Vert ^{2}-\hat{\mu}D\left( \bar{X}^{\prime }\right) \right) 
\sqrt{k\frac{\sigma _{\hat{K}}^{2}}{2\hat{\mu}}\left( 1-\frac{\hat{g}%
^{Bef}\left( \bar{X}^{\prime }\right) }{\left\langle \bar{g}\right\rangle }%
\right) }}{\sqrt{\underline{k}\left( \bar{X}^{\prime }\right) }}-\bar{r}%
^{\prime }\right) \\
&=&\left( 1-\left( 1+\bar{k}_{2}^{n}\left( \bar{X}_{1}\right) \right) \left( 
\bar{S}_{1}^{E}\left( \bar{X}^{\prime },\hat{X}_{1}\right) +\bar{S}%
_{1}^{B}\left( \bar{X}^{\prime },\bar{X}_{1}\right) \right) \right) \left( 
\frac{\left( \left\Vert \bar{\Psi}_{0}\left( \bar{X}^{\prime }\right)
\right\Vert ^{2}-\hat{\mu}D\left( \bar{X}^{\prime }\right) \right) \sqrt{k%
\frac{\sigma _{\hat{K}}^{2}}{2\hat{\mu}}\left( 1-\frac{\hat{g}^{Bef}\left( 
\bar{X}^{\prime }\right) }{\left\langle \bar{g}\right\rangle }\right) }}{%
\sqrt{\underline{k}\left( \bar{X}^{\prime }\right) }}-\bar{r}^{\prime
}\right)
\end{eqnarray*}%
and we find the following expression:%
\begin{eqnarray*}
&&\frac{\left( \left\Vert \bar{\Psi}_{0}\left( \bar{X}^{\prime }\right)
\right\Vert ^{2}-\hat{\mu}D\left( \bar{X}^{\prime }\right) \right) \sqrt{k%
\frac{\sigma _{\hat{K}}^{2}}{2\hat{\mu}}\left( 1-\frac{\hat{g}^{Bef}\left( 
\bar{X}^{\prime }\right) }{\left\langle \bar{g}\right\rangle }\right) }}{%
\sqrt{\underline{k}\left( \bar{X}^{\prime }\right) }}-\bar{r}^{\prime } \\
&=&\left( 1-\left( 1+\bar{k}_{2}^{n}\left( \bar{X}_{1}\right) \right) \left( 
\bar{S}_{1}^{E}\left( \bar{X}^{\prime },\hat{X}_{1}\right) +\bar{S}%
_{1}^{B}\left( \bar{X}^{\prime },\bar{X}_{1}\right) \right) \right) ^{-1} \\
&&\times \left( 1+\bar{k}_{2}^{n}\left( \bar{X}_{1}\right) \right) \left( 
\frac{1-\beta ^{B}}{\left( 1-\beta \right) \delta }\left( \frac{\Delta
\left( \hat{X},\hat{X}^{\prime }\right) }{1+\left[ \underline{\hat{k}}%
_{2}^{n}\left( \hat{X}\right) \right] _{\kappa }}-\hat{S}_{1}^{E}\left( \hat{%
X}^{\prime },\hat{X}_{1}\right) \right) +\hat{S}_{1}^{B}\left( \hat{X}%
^{\prime },\bar{X}_{1}\right) \right) \\
&&\times \left( \frac{\left( \left\Vert \bar{\Psi}_{0}\left( \bar{X}^{\prime
}\right) \right\Vert ^{2}-\hat{\mu}D\left( \bar{X}^{\prime }\right) \right) 
\sqrt{k\frac{\sigma _{\hat{K}}^{2}}{2\hat{\mu}}\left( 1-\frac{\hat{g}%
^{Bef}\left( \bar{X}^{\prime }\right) }{\left\langle \bar{g}\right\rangle }%
\right) }}{\sqrt{\underline{k}\left( \bar{X}^{\prime }\right) }}-\bar{r}%
^{\prime }\right)
\end{eqnarray*}%
or the equivalent formulation:%
\begin{eqnarray*}
&&\frac{\left( \left\Vert \bar{\Psi}_{0}\left( \bar{X}^{\prime }\right)
\right\Vert ^{2}-\hat{\mu}D\left( \bar{X}^{\prime }\right) \right) \sqrt{k%
\frac{\sigma _{\hat{K}}^{2}}{2\hat{\mu}}\left( 1-\frac{\hat{g}^{Bef}\left( 
\bar{X}^{\prime }\right) }{\left\langle \bar{g}\right\rangle }\right) }}{%
\sqrt{\underline{k}\left( \bar{X}^{\prime }\right) }}-\bar{r}^{\prime } \\
&=&\left( 1-\left( 1+\bar{k}_{2}^{n}\left( \bar{X}_{1}\right) \right) \left( 
\bar{S}_{1}^{E}\left( \bar{X}^{\prime },\hat{X}_{1}\right) +\bar{S}%
_{1}^{B}\left( \bar{X}^{\prime },\bar{X}_{1}\right) \right) \right) ^{-1} \\
&&\times \left( 1+\bar{k}_{2}^{n}\left( \bar{X}_{1}\right) \right) \left( 
\frac{1-\beta ^{B}}{\left( 1-\beta \right) \delta }\left( \frac{\Delta
\left( \hat{X},\hat{X}^{\prime }\right) }{1+\left[ \underline{\hat{k}}%
_{2}^{n}\left( \hat{X}\right) \right] _{\kappa }}-\hat{S}_{1}^{E}\left( \hat{%
X}^{\prime },\hat{X}_{1}\right) \right) +\hat{S}_{1}^{B}\left( \hat{X}%
^{\prime },\bar{X}_{1}\right) \right) \\
&&\times \left( 1-\left( 1+\left[ \underline{\hat{k}}_{2}^{n}\left( \hat{X}%
_{1}\right) \right] _{\kappa }\right) \hat{S}_{1}^{E}\left( \hat{X}^{\prime
},\hat{X}_{1}\right) \right) ^{-1} \\
&&\times \left( 1+\left[ \underline{\hat{k}}_{2}^{n}\left( \hat{X}%
_{1}\right) \right] _{\kappa }\right) \left( \left( \frac{A\left( \hat{X}%
^{\prime }\right) }{f_{1}^{2}\left( \hat{X}^{\prime }\right) }+\frac{B\left( 
\hat{X}^{\prime }\right) }{f_{1}^{3}\left( \hat{X}^{\prime }\right) }\right)
\left( R+\Delta F_{\tau }\left( \bar{R}\left( K,\hat{X}^{\prime }\right)
\right) \right) \right)
\end{eqnarray*}%
where: 
\begin{equation*}
\delta \rightarrow \frac{\frac{k\hat{K}\left[ \hat{X}_{1}\right] }{\sigma _{%
\hat{K}}^{2}}+k^{B}D\left( \bar{X}_{1}\right) A+\sqrt{\left( \frac{k\hat{K}%
\left[ \hat{X}_{1}\right] }{\sigma _{\hat{K}}^{2}}+k^{B}D\left( \bar{X}%
_{1}\right) A\right) ^{2}+4\left( \left\Vert \bar{\Psi}_{0}\left( \bar{X}%
_{1}\right) \right\Vert ^{4}-k^{B}D\left( \bar{X}_{1}\right) \bar{r}^{\prime
}\right) \frac{k\hat{K}\left[ \hat{X}_{1}\right] }{\sigma _{\hat{K}}^{2}}A}}{%
2\left( \left\Vert \bar{\Psi}_{0}\left( \bar{X}_{1}\right) \right\Vert
^{4}-k^{B}D\left( \bar{X}_{1}\right) \bar{r}^{\prime }\right) }
\end{equation*}%
\begin{equation*}
A=\frac{\left( 1-\beta ^{B}\right) \left( 1+\bar{\beta}\bar{k}\left( \bar{X}%
_{1}\right) \right) \left( \left\Vert \hat{\Psi}_{0}\left( \hat{X}%
_{1}\right) \right\Vert ^{4}-\bar{r}^{\prime }\left( \frac{\hat{K}\left[ 
\hat{X}_{1}\right] }{\sigma _{\hat{K}}^{2}}+D\left( \hat{X}_{1}\right)
\right) \right) }{\left( \frac{\hat{K}\left[ \hat{X}_{1}\right] }{\sigma _{%
\hat{K}}^{2}}+D\left( \hat{X}_{1}\right) \right) \left( \frac{\hat{K}\left[ 
\hat{X}_{1}\right] }{\sigma _{\hat{K}}^{2}}+D\left( \hat{X}_{1}\right)
\right) \left( 1-\beta \right) \left( 1+\left[ \underline{\hat{k}}%
_{2}^{n}\left( \hat{X}_{1}\right) \right] _{\kappa }\right) }
\end{equation*}

\begin{equation*}
\bar{K}\left[ \bar{X}\right] =\frac{\underline{k}^{B}\left( \bar{X}\right) }{%
k^{B}\left( \bar{X}\right) }=\frac{k\left( X\right) \hat{K}\left[ \hat{X}%
\right] }{\delta k^{B}\left( \bar{X}\right) }
\end{equation*}

\begin{equation*}
\delta \left( \hat{X}_{1}\right) =\delta \left( \left\langle \hat{X}%
\right\rangle ,\hat{X}_{1}\right)
\end{equation*}

\section*{Appendix 23 Averages quantities, investors and bank returns}

We compute the remaining average quantities: firm capital and return,
investors and banks average returns.

\subsection{A23.1 Firms average capital}

\subsubsection{A23.1.1 Average coefficients}

We define the following averages coefficients arising in the formula for
average capital.

The coefficient $k$ was defined by: 
\begin{equation*}
k\simeq k\left( X\right) \frac{\hat{K}\left[ \hat{X}\right] }{%
K_{X}\left\Vert \Psi _{0}\right\Vert ^{2}}
\end{equation*}%
so that in average:%
\begin{equation*}
\left\langle k\right\rangle \simeq k\frac{\left\langle \hat{K}\right\rangle
\left\Vert \hat{\Psi}\right\Vert ^{2}}{\left\langle K\right\rangle
\left\Vert \Psi _{0}\right\Vert ^{2}}
\end{equation*}%
\begin{eqnarray*}
\left\langle \underline{\hat{k}}\left( X\right) \right\rangle
&=&\left\langle \hat{k}\left( X,X^{\prime }\right) \frac{\hat{K}_{X^{\prime
}}\left\Vert \hat{\Psi}\left( \hat{X}^{\prime }\right) \right\Vert ^{2}}{%
\left\langle \hat{K}\right\rangle \left\Vert \hat{\Psi}\left( \hat{X}\right)
\right\Vert ^{2}}\right\rangle \simeq \hat{k}\left( \left\langle
X\right\rangle ,\left\langle X\right\rangle \right) =\left\langle \hat{k}%
\right\rangle =\hat{k} \\
\left\langle \underline{k}\left( X\right) \right\rangle &=&\left\langle
k\left( X,X^{\prime }\right) \frac{\hat{K}_{X^{\prime }}\left\Vert \hat{\Psi}%
\left( \hat{X}^{\prime }\right) \right\Vert ^{2}}{\left\langle
K\right\rangle \left\Vert \Psi \left( \hat{X}\right) \right\Vert ^{2}}%
\right\rangle \\
&\simeq &k\left( \left\langle X\right\rangle ,\left\langle \hat{X}%
\right\rangle \right) \frac{\left\langle \hat{K}\right\rangle }{\left\langle
K\right\rangle }\frac{\left\Vert \hat{\Psi}\right\Vert ^{2}}{\left\Vert \Psi
\right\Vert ^{2}}\simeq \left\langle k\right\rangle \frac{\left\langle \hat{K%
}\right\rangle }{\left\langle K\right\rangle }\frac{\left\Vert \hat{\Psi}%
_{0}\right\Vert ^{2}}{\left\Vert \Psi _{0}\right\Vert ^{2}}
\end{eqnarray*}%
\begin{eqnarray*}
\left\langle X^{\left( e\right) }\right\rangle &\rightarrow &\sqrt{\frac{%
\left\vert \Psi _{0}\right\vert ^{2}}{\epsilon }-\frac{1}{2}\left\langle
f_{1}\right\rangle -\frac{1}{2}\left( \beta \underline{k}\left( \left\langle
f_{1}\right\rangle -\bar{r}\right) \right) } \\
&\simeq &\sqrt{\frac{\left\vert \Psi _{0}\right\vert ^{2}}{\epsilon }-\frac{1%
}{2}\left\langle f_{1}\right\rangle }-\frac{\beta \underline{k}\left(
\left\langle f_{1}\right\rangle -\bar{r}\right) }{4\sqrt{\frac{\left\vert
\Psi _{0}\right\vert ^{2}}{\epsilon }-\frac{1}{2}\left\langle
f_{1}\right\rangle }}=\sqrt{\frac{\left\vert \Psi _{0}\right\vert ^{2}}{%
\epsilon }-\frac{1}{2}\left\langle f_{1}\right\rangle }-\frac{\beta
\left\langle \underline{k}\right\rangle \frac{\left\langle \hat{K}%
\right\rangle }{\left\langle K\right\rangle }\frac{\left\Vert \hat{\Psi}%
_{0}\right\Vert ^{2}}{\left\Vert \Psi _{0}\right\Vert ^{2}}\left(
\left\langle f_{1}\right\rangle -\bar{r}\right) }{4\sqrt{\frac{\left\vert
\Psi _{0}\right\vert ^{2}}{\epsilon }-\frac{1}{2}\left\langle
f_{1}\right\rangle }}
\end{eqnarray*}%
\begin{equation*}
\left\langle C^{\left( e\right) }\right\rangle \simeq \frac{\beta \delta
+\beta ^{B}}{1+\delta }C
\end{equation*}%
The effective productivity becomes in average:%
\begin{equation*}
\left\langle f_{1}^{\left( e\right) }\right\rangle =\left\langle
f_{1}\right\rangle +\beta \underline{k}\left( \left\langle
f_{1}\right\rangle -\bar{r}\right) =\left\langle f_{1}\right\rangle +\beta
\left\langle \underline{k}\right\rangle \frac{\left\langle \hat{K}%
\right\rangle }{\left\langle K\right\rangle }\frac{\left\Vert \hat{\Psi}%
_{0}\right\Vert ^{2}}{\left\Vert \Psi _{0}\right\Vert ^{2}}\left(
\left\langle f_{1}\right\rangle -\bar{r}\right)
\end{equation*}%
The ratio of bank capital invested in firms over investors capital invested
in firms is:%
\begin{equation*}
\left\langle \delta \right\rangle \simeq \frac{\left\langle
k_{1}^{B}\right\rangle +\frac{\kappa \left\langle k_{2}^{B}\right\rangle }{%
1+\left\langle \bar{k}\right\rangle }}{\left\langle k\right\rangle }Z\simeq
\left( \frac{\left\langle k_{1}^{B}\right\rangle }{\left\langle
k\right\rangle }+\frac{\kappa \left\langle k_{2}^{B}\right\rangle }{%
\left\langle k\right\rangle \left( 1+\left\langle \bar{k}\right\rangle
\right) }\right) \frac{\left\langle \bar{K}\right\rangle \left\Vert \bar{\Psi%
}\right\Vert ^{2}}{\left\langle \hat{K}\right\rangle \left\Vert \hat{\Psi}%
\right\Vert ^{2}}
\end{equation*}

\subsubsection*{A23.1.2 Formula for average capital}

\paragraph*{A23.1.2.1 Firms average capital}

We use that:%
\begin{equation*}
\left\langle K\right\rangle =\frac{1}{4\left\langle f_{1}^{\left( e\right)
}\right\rangle }\left\langle \frac{\left( 3X^{\left( e\right) }-C^{\left(
e\right) }\right) \left( C^{\left( e\right) }+X^{\left( e\right) }\right) }{%
2X^{\left( e\right) }-C^{\left( e\right) }}\right\rangle
\end{equation*}%
To rewrite average capital as:%
\begin{eqnarray*}
\left\langle K\right\rangle &\simeq &\left( 1-\frac{4\left( \beta
\left\langle \underline{k}\right\rangle \frac{\left\Vert \hat{\Psi}%
_{0}\right\Vert ^{2}}{\left\Vert \Psi _{0}\right\Vert ^{2}}\left\langle
K\right\rangle \left( \left\langle f_{1}\right\rangle -\bar{r}\right)
\right) \left( 2\left\langle X^{\left( e\right) }\right\rangle -\frac{\beta
\delta +\beta ^{B}}{1+\delta }C\right) }{\left( 3\left\langle X^{\left(
e\right) }\right\rangle -\frac{\beta \delta +\beta ^{B}}{1+\delta }C\right)
\left( \left\langle X^{\left( e\right) }\right\rangle +\frac{\beta \delta
+\beta ^{B}}{1+\delta }C\right) }\right) \\
&&\times \left( 3\left\langle X^{\left( e\right) }\right\rangle -\frac{\beta
\delta +\beta ^{B}}{1+\delta }C\right) \left( \left\langle X^{\left(
e\right) }\right\rangle +\frac{\beta \delta +\beta ^{B}}{1+\delta }C\right)
\\
&\simeq &\frac{\left( 3\sqrt{\frac{\left\vert \Psi _{0}\right\vert ^{2}}{%
\epsilon }-\frac{1}{2}\left\langle f_{1}\right\rangle }-\frac{\beta \delta
+\beta ^{B}}{1+\delta }C\right) \left( \sqrt{\frac{\left\vert \Psi
_{0}\right\vert ^{2}}{\epsilon }-\frac{1}{2}\left\langle f_{1}\right\rangle }%
+\frac{\beta \delta +\beta ^{B}}{1+\delta }C\right) }{4\left\langle
f_{1}^{\left( e\right) }\right\rangle \left( 2\sqrt{\frac{\left\vert \Psi
_{0}\right\vert ^{2}}{\epsilon }-\frac{1}{2}\left\langle f_{1}\right\rangle }%
-\frac{\beta \delta +\beta ^{B}}{1+\delta }C\right) }
\end{eqnarray*}

\paragraph*{A23.1.2.2 Investors average amount of capital}

\begin{eqnarray*}
\left\langle \hat{K}\right\rangle \left\Vert \hat{\Psi}\right\Vert ^{2} &=&%
\hat{\mu}V\frac{\left\langle \hat{K}_{0}\right\rangle ^{4}}{2\sigma _{\hat{K}%
}^{2}}\left( \frac{1}{4}-\frac{\left\langle \hat{K}\right\rangle }{%
3\left\langle \hat{K}_{0}\right\rangle }\frac{\left\langle \hat{g}%
^{ef}\right\rangle }{\left\langle \hat{g}\right\rangle }\right) \left\langle 
\hat{g}\right\rangle ^{2} \\
&=&\frac{18\sigma _{\hat{K}}^{2}}{\hat{\mu}}V\left( \frac{\left\Vert \hat{%
\Psi}_{0}\right\Vert ^{2}}{\left\langle \hat{g}\right\rangle ^{2}\left( 5+%
\frac{\left\langle \hat{g}^{ef}\right\rangle }{\left\langle \hat{g}%
\right\rangle }-\sqrt{\left( 1-\frac{\left\langle \hat{g}^{ef}\right\rangle 
}{\left\langle \hat{g}\right\rangle }\right) \left( 4-\frac{\left\langle 
\hat{g}^{ef}\right\rangle }{\left\langle \hat{g}\right\rangle }\right) }%
\right) }\right) ^{2} \\
&&\times \left( \frac{1}{4}-\frac{1}{18}\left( 2+\frac{\left\langle \hat{g}%
^{ef}\right\rangle }{\left\langle \hat{g}\right\rangle }-\sqrt{\left( 1-%
\frac{\left\langle \hat{g}^{ef}\right\rangle }{\left\langle \hat{g}%
\right\rangle }\right) \left( 4-\frac{\left\langle \hat{g}^{ef}\right\rangle 
}{\left\langle \hat{g}\right\rangle }\right) }\right) \right) \left\langle 
\hat{g}\right\rangle ^{2}
\end{eqnarray*}%
\begin{eqnarray*}
\left\langle k\right\rangle &\simeq &k\frac{\sigma _{\hat{K}}^{2}}{\hat{\mu}}%
V\left( \frac{\left\Vert \bar{\Psi}_{0}\right\Vert ^{2}-\frac{\hat{\mu}}{6}%
\left( \frac{\left\langle \hat{K}\right\rangle ^{2}\left\langle \hat{g}%
\right\rangle }{\sigma _{\hat{K}}^{2}}\right) \left\langle \hat{g}%
^{Bef}\right\rangle \hat{l}}{\left\langle \bar{g}\right\rangle ^{2}}\right)
^{2}\left( \frac{1}{2}-\frac{2+\frac{\left\langle \hat{g}^{ef}\right\rangle 
}{\left\langle \hat{g}\right\rangle }-\sqrt{\left( 1-\frac{\left\langle \hat{%
g}^{ef}\right\rangle }{\left\langle \hat{g}\right\rangle }\right) \left( 4-%
\frac{\left\langle \hat{g}^{ef}\right\rangle }{\left\langle \hat{g}%
\right\rangle }\right) }}{9}\right) \left\langle \hat{g}\right\rangle ^{2} \\
&&\times \frac{4\left\langle f_{1}^{\left( e\right) }\right\rangle \left( 2%
\sqrt{\frac{\left\vert \Psi _{0}\right\vert ^{2}}{\epsilon }-\frac{1}{2}%
\left\langle f_{1}\right\rangle }-\frac{\beta \delta +\beta ^{B}}{1+\delta }%
C\right) }{\left( 3\sqrt{\frac{\left\vert \Psi _{0}\right\vert ^{2}}{%
\epsilon }-\frac{1}{2}\left\langle f_{1}\right\rangle }-\frac{\beta \delta
+\beta ^{B}}{1+\delta }C\right) \left( \sqrt{\frac{\left\vert \Psi
_{0}\right\vert ^{2}}{\epsilon }-\frac{1}{2}\left\langle f_{1}\right\rangle }%
+\frac{\beta \delta +\beta ^{B}}{1+\delta }C\right) }
\end{eqnarray*}%
where:%
\begin{equation*}
\left\langle \underline{k}^{B}\right\rangle =\left\langle
k_{1}^{B}\right\rangle \frac{\left\langle \bar{K}_{0}\right\rangle }{%
\left\langle K\right\rangle }\frac{\left\Vert \bar{\Psi}_{0}\right\Vert ^{2}%
}{\left\Vert \Psi _{0}\right\Vert ^{2}}+\left\langle k_{2}^{B}\right\rangle 
\frac{\kappa \left\langle \bar{K}_{0}\right\rangle }{\left\langle
K\right\rangle }\frac{\left\Vert \bar{\Psi}_{0}\right\Vert ^{2}}{\left\Vert
\Psi _{0}\right\Vert ^{2}}
\end{equation*}%
\begin{equation*}
\left\langle \delta \right\rangle \simeq \frac{\left\langle
k_{1}^{B}\right\rangle +\frac{\kappa \left\langle k_{2}^{B}\right\rangle }{%
1+\left\langle \bar{k}\right\rangle }}{\left\langle k\right\rangle }Z\simeq
\left( \frac{\left\langle k_{1}^{B}\right\rangle }{\left\langle
k\right\rangle }+\frac{\kappa \left\langle k_{2}^{B}\right\rangle }{%
\left\langle k\right\rangle \left( 1+\left\langle \bar{k}\right\rangle
\right) }\right) \frac{\left\langle \bar{K}\right\rangle \left\Vert \bar{\Psi%
}\right\Vert ^{2}}{\left\langle \hat{K}\right\rangle \left\Vert \hat{\Psi}%
\right\Vert ^{2}}
\end{equation*}%
\begin{equation*}
\left\langle \hat{g}^{ef}\right\rangle =-\frac{\left( \kappa \left\langle %
\left[ \frac{\underline{\hat{k}}_{2}^{B}}{1+\bar{k}}\right] \right\rangle
\left( 1-\left\langle \underline{\hat{k}}\right\rangle \right) +\left\langle 
\hat{k}_{1}^{B}\right\rangle \left\langle \hat{k}_{2}\right\rangle \right)
\left( \left\langle \hat{g}\right\rangle +\frac{1}{1-\left\langle \underline{%
\hat{k}}\right\rangle }\bar{N}\left\langle \bar{g}\right\rangle \right) 
\frac{\left\langle \bar{K}\right\rangle \left\Vert \bar{\Psi}\right\Vert ^{2}%
}{\left\langle \hat{K}\right\rangle \left\Vert \hat{\Psi}\right\Vert ^{2}}}{%
\left( 1-\left( \left\langle \hat{k}_{1}\right\rangle +\left\langle \hat{k}%
_{1}^{B}\right\rangle \frac{\left\langle \bar{K}\right\rangle \left\Vert 
\bar{\Psi}\right\Vert ^{2}}{\left\langle \hat{K}\right\rangle \left\Vert 
\hat{\Psi}\right\Vert ^{2}}\right) \right) \left( 1-\left( \left\langle 
\underline{\hat{k}}\right\rangle +\left( \left\langle \underline{\hat{k}}%
_{1}^{B}\right\rangle +\kappa \left\langle \left[ \frac{\underline{\hat{k}}%
_{2}^{B}}{1+\bar{k}}\right] \right\rangle \right) \frac{\left\langle \bar{K}%
\right\rangle \left\Vert \bar{\Psi}\right\Vert ^{2}}{\left\langle \hat{K}%
\right\rangle \left\Vert \hat{\Psi}\right\Vert ^{2}}\right) \right) }
\end{equation*}%
and:%
\begin{eqnarray*}
\left\langle \hat{g}^{Bef}\right\rangle &=&-\frac{\kappa \left\langle \left[ 
\frac{\underline{\hat{k}}_{2}^{B}}{1+\bar{k}}\right] \right\rangle \left(
1-\left\langle \hat{k}\right\rangle \right) +\left\langle \hat{k}%
_{1}^{B}\right\rangle \left\langle \hat{k}_{2}\right\rangle }{\left(
1-\left( \left\langle \hat{k}\right\rangle +\left( \left\langle \hat{k}%
_{1}^{B}\right\rangle +\kappa \left\langle \left[ \frac{\underline{\hat{k}}%
_{2}^{B}}{1+\bar{k}}\right] \right\rangle \right) \frac{\left\langle \bar{K}%
\right\rangle \left\Vert \bar{\Psi}\right\Vert ^{2}}{\left\langle \hat{K}%
\right\rangle \left\Vert \hat{\Psi}\right\Vert ^{2}}\right) \right) \left(
1-\left( \left\langle \hat{k}_{1}\right\rangle +\left\langle \hat{k}%
_{1}^{B}\right\rangle \frac{\left\langle \bar{K}\right\rangle \left\Vert 
\bar{\Psi}\right\Vert ^{2}}{\left\langle \hat{K}\right\rangle \left\Vert 
\hat{\Psi}\right\Vert ^{2}}\right) \right) } \\
&&\times \left( \left\langle \hat{g}\right\rangle +\left( 1-\hat{M}\right)
^{-1}\bar{N}\left\langle \bar{g}\right\rangle \right)
\end{eqnarray*}%
with the relation:%
\begin{equation*}
\left\langle \hat{g}^{Bef}\right\rangle \simeq \frac{\left\langle \hat{g}%
^{ef}\right\rangle }{\frac{\left\langle \bar{K}\right\rangle \left\Vert \bar{%
\Psi}\right\Vert ^{2}}{\left\langle \hat{K}\right\rangle \left\Vert \hat{\Psi%
}\right\Vert ^{2}}}
\end{equation*}

\paragraph*{A23.1.2.3 Banks average amount of capital}

Using (\ref{KBP}):%
\begin{equation}
\left\langle \bar{K}\right\rangle \left\Vert \bar{\Psi}\right\Vert
^{2}\simeq 18\frac{\sigma _{\hat{K}}^{2}}{\left\langle \bar{g}\right\rangle
^{2}\hat{\mu}}\left( \frac{\sqrt{\left\langle \hat{g}\right\rangle
^{2}\left( 1+\frac{3}{4}\frac{\left\langle \hat{g}^{ef}\right\rangle }{%
\left\langle \hat{g}\right\rangle }\right) }\left\Vert \bar{\Psi}%
_{0}\right\Vert -\frac{3}{8}\left\langle \hat{g}\right\rangle \frac{%
\left\langle \hat{g}^{Bef}\right\rangle }{\left\langle \bar{g}\right\rangle }%
\left\Vert \hat{\Psi}_{0}\right\Vert }{\sqrt{5+\frac{\left\langle \hat{g}%
^{ef}\right\rangle }{\left\langle \hat{g}\right\rangle }-\sqrt{\left( 1-%
\frac{\left\langle \hat{g}^{ef}\right\rangle }{\left\langle \hat{g}%
\right\rangle }\right) \left( 4-\frac{\left\langle \hat{g}^{ef}\right\rangle 
}{\left\langle \hat{g}\right\rangle }\right) }}}\right) ^{4}\left( \frac{1}{4%
}-\frac{\left\langle \hat{g}\right\rangle \left\langle \hat{g}%
^{Bef}\right\rangle }{3\left\langle \bar{g}\right\rangle ^{2}}\right)
\end{equation}

\subsection*{A23.2 Investors average returns}

\subsubsection*{A23.2.1 Average coefficients}

Using normalizations:%
\begin{eqnarray*}
\left\langle \frac{1}{1+\left[ \underline{\hat{k}}_{2}^{n}\left( \hat{X}%
\right) \right] _{\kappa }}\right\rangle &\simeq &\frac{1}{1+\left\langle 
\hat{k}_{2}^{\left( n\right) }\right\rangle +\kappa \left\langle \frac{%
\underline{\hat{k}}_{2}^{B\left( n\right) }}{1+\bar{k}}\right\rangle \left(
\left( 1+\frac{3}{4}\frac{\left\langle \hat{g}^{ef}\right\rangle }{%
\left\langle \hat{g}\right\rangle }\right) \frac{\left\Vert \bar{\Psi}%
_{0}\right\Vert ^{2}}{\left\Vert \hat{\Psi}_{0}\right\Vert ^{2}}\right) } \\
&\simeq &\frac{1}{1+\left\langle \hat{k}_{2}^{\left( n\right) }\right\rangle
+\kappa \left\langle \frac{\underline{\hat{k}}_{2}^{B\left( n\right) }}{1+%
\bar{k}}\right\rangle \frac{\left\Vert \bar{\Psi}_{0}\right\Vert ^{2}}{%
\left\Vert \hat{\Psi}_{0}\right\Vert ^{2}}}
\end{eqnarray*}%
\begin{eqnarray*}
&&1+\left[ \underline{\hat{k}}\left( \hat{X}^{\prime }\right) \right]
_{\kappa } \\
&=&1+\left( \hat{k}\left( \hat{X}^{\prime },\left\langle \hat{X}%
\right\rangle \right) -\left\langle \hat{k}\right\rangle \right) +\left( 
\hat{k}_{1}^{B}\left( \bar{X}^{\prime },\left\langle \bar{X}\right\rangle
\right) +\kappa \frac{\hat{k}_{2}^{\left( B\right) }\left( \bar{X}^{\prime
},\left\langle \bar{X}\right\rangle \right) }{1+\underline{\bar{k}}}-\left(
\left\langle \hat{k}_{1}^{B}\right\rangle +\kappa \frac{\left\langle \hat{k}%
_{2}^{\left( B\right) }\right\rangle }{1+\underline{\bar{k}}}\right) \right)
Z
\end{eqnarray*}%
in averages:%
\begin{equation*}
\left\langle 1+\left[ \underline{\hat{k}}\left( \hat{X}_{1}\right) \right]
_{\kappa }\right\rangle \simeq 1
\end{equation*}%
\begin{equation*}
\left\langle 1+\left[ \underline{\hat{k}}_{2}^{n}\left( \hat{X}\right) %
\right] _{\kappa }\right\rangle =1+\left\langle \hat{k}_{2}^{n}\right\rangle
=1+\frac{\left\langle \hat{k}_{2}^{\left( n\right) }\right\rangle +\kappa
\left\langle \frac{\underline{\hat{k}}_{2}^{B\left( n\right) }}{1+\bar{k}}%
\right\rangle \frac{\left\Vert \bar{\Psi}_{0}\right\Vert ^{2}}{\left\Vert 
\hat{\Psi}_{0}\right\Vert ^{2}}}{1-\left\langle \hat{k}^{\Sigma
}\right\rangle }
\end{equation*}%
\begin{equation*}
\left\langle \frac{\bar{k}_{1}\left( \bar{X}^{\prime },\bar{X}_{1}\right) }{%
1+\left[ \underline{\hat{k}}^{n}\left( \hat{X}\right) \right] _{\kappa }}%
\right\rangle \left\langle \frac{\bar{N}\left( \bar{X}^{\prime },\bar{X}%
_{1}\right) }{1+\left[ \underline{\hat{k}}_{2}^{n}\left( \hat{X}^{\prime
}\right) \right] _{\kappa }}\right\rangle \simeq \frac{\left\langle \bar{k}%
_{1}\right\rangle }{1+\left\langle \hat{k}_{2}^{n}\right\rangle }%
\left\langle \bar{N}\left( \bar{X}^{\prime },\bar{X}_{1}\right) \right\rangle
\end{equation*}

\subsubsection*{A23.2.2 Average return equation}

In return equation for investors:%
\begin{eqnarray*}
\hat{g}\left( \hat{X}_{1}\right) -\bar{r}^{\prime } &=&\int \left( 1-\left(
1+\left[ \underline{\hat{k}}_{2}^{n}\left( \hat{X}_{1}\right) \right]
_{\kappa }\right) \hat{S}_{1}^{E}\left( \hat{X}^{\prime },\hat{X}_{1}\right)
\right) ^{-1} \\
&&\times \left( 1+\left[ \underline{\hat{k}}_{2}^{n}\left( \hat{X}%
_{1}\right) \right] _{\kappa }\right) \left( \frac{A\left( \hat{X}^{\prime
}\right) }{f_{1}^{2}\left( \hat{X}^{\prime }\right) }+\frac{B\left( \hat{X}%
^{\prime }\right) }{f_{1}^{3}\left( \hat{X}^{\prime }\right) }\right) \left(
R+\Delta F_{\tau }\left( \bar{R}\left( K,\hat{X}^{\prime }\right) \right)
\right)
\end{eqnarray*}%
we estimate the average inverse diffusion matrix:%
\begin{equation*}
\left( 1+\left[ \underline{\hat{k}}_{2}^{n}\left( \hat{X}_{1}\right) \right]
_{\kappa }\right) \hat{S}_{1}^{E}\left( \hat{X}^{\prime },\hat{X}_{1}\right)
\simeq \left( 1+\left\langle \hat{k}_{2}^{n}\right\rangle \right) \left(
\left\langle \hat{k}\right\rangle -\left\langle \hat{k}_{1}\right\rangle
\left\langle k\right\rangle +\left\langle \hat{k}_{1}\right\rangle \right)
=\left( 1+\left\langle \hat{k}_{2}^{n}\right\rangle \right) \left\langle 
\hat{k}\right\rangle \left( 1-\left\langle \hat{k}_{1}\right\rangle \right)
+\left\langle \hat{k}_{1}\right\rangle
\end{equation*}%
Using that in average:%
\begin{equation*}
\left\langle \Delta F_{\tau }\left( \bar{R}\left( K,X\right) \right)
\right\rangle =0
\end{equation*}%
we find:%
\begin{equation*}
\left( 1-\left( 1+\underline{\hat{k}}_{2}\left( \hat{X}_{1}\right) \right) 
\hat{S}_{1}^{E}\left( \hat{X}^{\prime },\hat{X}_{1}\right) \right)
^{-1}\simeq \left( 1-\left( 1+\left\langle \hat{k}_{2}^{n}\right\rangle
\right) \left( \left\langle \hat{k}\right\rangle \left( 1-\left\langle \hat{k%
}_{1}\right\rangle \right) +\left\langle \hat{k}_{1}\right\rangle \right)
\right) ^{-1}
\end{equation*}%
\begin{eqnarray*}
\left\langle \hat{g}\right\rangle &=&\bar{r}^{\prime }+\left( 1+\left\langle 
\hat{k}_{2}^{n}\right\rangle \right) \left( 1-\left( 1+\left\langle \hat{k}%
_{2}^{n}\right\rangle \right) \left( \left\langle \hat{k}\right\rangle
\left( 1-\left\langle \hat{k}_{1}\right\rangle \right) +\left\langle \hat{k}%
_{1}\right\rangle \right) \right) ^{-1} \\
&&\times \left\langle \left( \frac{A\left( \hat{X}^{\prime }\right) }{%
f_{1}^{2}\left( \hat{X}^{\prime }\right) }+\frac{B\left( \hat{X}^{\prime
}\right) }{f_{1}^{3}\left( \hat{X}^{\prime }\right) }\right) \left( R+\Delta
F_{\tau }\left( \bar{R}\left( K,\hat{X}^{\prime }\right) \right) \right)
\right\rangle
\end{eqnarray*}%
that is:%
\begin{eqnarray*}
\left\langle \hat{g}\right\rangle &\simeq &\bar{r}^{\prime }+\left( 1-\left(
1+\left\langle \hat{k}_{2}^{n}\right\rangle \right) \left( \left\langle \hat{%
k}\right\rangle \left( 1-\left\langle \hat{k}_{1}\right\rangle \right)
+\left\langle \hat{k}_{1}\right\rangle \right) \right) ^{-1} \\
&&\times \left( 1+\left\langle \hat{k}_{2}^{n}\right\rangle \right)
\left\langle \left( \frac{A\left( \hat{X}^{\prime }\right) }{f_{1}^{2}\left( 
\hat{X}^{\prime }\right) }+\frac{B\left( \hat{X}^{\prime }\right) }{%
f_{1}^{3}\left( \hat{X}^{\prime }\right) }\right) \left( R+\Delta F_{\tau
}\left( \bar{R}\left( K,\hat{X}^{\prime }\right) \right) \right)
\right\rangle
\end{eqnarray*}

\subsection*{A23.3 Banks average returns}

\subsubsection*{A23.3.1 Average returns}

\begin{eqnarray*}
\left\langle \bar{g}\right\rangle &=&\bar{r}^{\prime }+\left( 1+\bar{\beta}%
\left\langle \underline{\bar{k}}\right\rangle \right) \frac{\left( 1-\beta
^{B}\right) }{\left( 1-\beta \right) \left\langle \delta \right\rangle } \\
&&\times \left[ \left( 1-\left( 1+\bar{\beta}\left\langle \underline{\bar{k}}%
\right\rangle \right) \left( \left\langle \bar{S}_{1}^{E}\right\rangle
+\left\langle \bar{S}_{1}^{B}\right\rangle \right) \right) \right]
^{-1}\left( \frac{1}{\left( \frac{1}{1+\underline{\bar{k}}_{2}^{n}\left( 
\bar{X}^{\prime }\right) }\right) \left\langle \hat{k}^{B}\right\rangle }%
-\left( \left\langle \hat{S}_{1}^{E}\right\rangle -\frac{\left( 1-\beta
\right) \left\langle \delta \right\rangle }{1-\beta ^{B}}\left\langle \hat{S}%
_{1}^{B}\right\rangle \right) \right) \left( \left\langle \hat{g}%
\right\rangle -\bar{r}^{\prime }\right)
\end{eqnarray*}%
Using;%
\begin{eqnarray*}
\left\langle \frac{1}{1+\underline{\bar{k}}\left( \bar{X}^{\prime }\right) }%
\right\rangle &\simeq &1 \\
\left\langle \frac{1}{1+\underline{\bar{k}}_{2}^{n}\left( \bar{X}^{\prime
}\right) }\right\rangle &\simeq &\frac{1}{1+\left\langle \underline{\bar{k}}%
_{2}^{n}\right\rangle }=\frac{1}{1+\frac{\left\langle \bar{k}%
_{2}\right\rangle }{1-\left\langle \bar{k}\right\rangle }}
\end{eqnarray*}%
\begin{equation*}
\left\langle \bar{g}\right\rangle =\bar{r}^{\prime }+\frac{\left( 1-\beta
^{B}\right) }{\left( 1-\beta \right) \left\langle \delta \right\rangle }%
\left[ \left( 1-\left( \left\langle \bar{S}_{1}^{E}\right\rangle
+\left\langle \bar{S}_{1}^{B}\right\rangle \right) \right) \right]
^{-1}\left( \frac{1}{\left\langle \hat{k}^{B}\right\rangle }-\left(
\left\langle \hat{S}_{1}^{E}\right\rangle -\frac{\left( 1-\beta \right)
\left\langle \delta \right\rangle }{1-\beta ^{B}}\left\langle \hat{S}%
_{1}^{B}\right\rangle \right) \right) \left( \left\langle \hat{g}%
\right\rangle -\bar{r}^{\prime }\right)
\end{equation*}%
The terms arising in this expression can be rewritten:%
\begin{equation*}
\left\langle \bar{S}_{1}^{E}\right\rangle =\left\langle \bar{k}\right\rangle
\left( 1-\left\langle \bar{k}_{1}\right\rangle \right) +\left\langle \bar{k}%
_{1}\right\rangle
\end{equation*}%
\begin{equation*}
\bar{S}_{1}^{E}\left( \bar{X}^{\prime },\bar{X}_{1}\right) \rightarrow
\left( 1-\left\langle \bar{k}_{1}\right\rangle \right) \left\langle \bar{k}%
\right\rangle +\left\langle \bar{k}_{1}\right\rangle
\end{equation*}%
\begin{equation*}
\bar{S}_{1}^{B}\left( \bar{X}^{\prime },\bar{X}_{1}\right) \rightarrow \frac{%
\underline{\hat{k}}_{1}^{B}\left( \bar{X}^{\prime },\bar{X}_{1}\right) }{1+%
\underline{\hat{k}}\left( \hat{X}^{\prime }\right) +\underline{\hat{k}}%
_{1}^{B}\left( \bar{X}^{\prime }\right) +\kappa \left[ \frac{\underline{\hat{%
k}}_{2}^{B}}{1+\bar{k}}\right] \left( \hat{X}^{\prime }\right) }\frac{\bar{N}%
\left( \bar{X}^{\prime },\bar{X}_{1}\right) }{1+\underline{\hat{k}}%
_{2}\left( \bar{X}^{\prime }\right) +\kappa \frac{\underline{\hat{k}}%
_{2}^{B}\left( \bar{X}^{\prime }\right) }{1+\bar{k}\left( \bar{X}\right) }}
\end{equation*}%
Given that;%
\begin{equation*}
\bar{N}\left( \bar{X}^{\prime },\bar{X}_{1}\right) \rightarrow \frac{\left( 
\hat{k}_{1}^{B}\left( \hat{X}\right) +\kappa \frac{\underline{\hat{k}}%
_{2}^{B}\left( \hat{X}\right) }{1+\underline{\overline{\bar{k}}}}\left( 1-%
\frac{\kappa }{1+\underline{\overline{\bar{k}}}}\right) \right) }{1+%
\underline{\hat{k}}\left( \hat{X}\right) +\underline{\hat{k}_{1}^{B}}\left( 
\hat{X}\right) +\kappa \frac{\underline{\hat{k}}_{2}^{B}\left( \hat{X}%
\right) }{1+\underline{\overline{\bar{k}}}}}
\end{equation*}%
the average $\left\langle \bar{N}\left( \bar{X}^{\prime },\bar{X}_{1}\right)
\right\rangle $ is given by:%
\begin{eqnarray*}
\left\langle \bar{N}\left( \bar{X}^{\prime },\bar{X}_{1}\right)
\right\rangle &=&\left\langle \underline{\hat{k}}_{1}^{B}\right\rangle
+\kappa \frac{\left\langle \underline{\hat{k}}_{2}^{B}\right\rangle }{1+%
\underline{\bar{k}}}\left( 1-\frac{\kappa }{1+\underline{\bar{k}}}\right) \\
&\rightarrow &\left\langle \underline{\hat{k}}_{1}^{B}\right\rangle +\kappa
\left\langle \underline{\hat{k}}_{2}^{B}\right\rangle \left( 1-\kappa
\right) \simeq -\kappa ^{2}\left\langle \underline{\hat{k}}%
_{2}^{B}\right\rangle
\end{eqnarray*}%
and:%
\begin{equation*}
\left\langle \bar{S}_{1}^{B}\left( \bar{X}^{\prime },\bar{X}_{1}\right)
\right\rangle =\left\langle \underline{\hat{k}}_{1}^{B}\right\rangle \left(
\left\langle \underline{\hat{k}}_{1}^{B}\right\rangle +\kappa \frac{%
\left\langle \underline{\hat{k}}_{2}^{B}\right\rangle }{1+\underline{\bar{k}}%
}\left( 1-\frac{\kappa }{1+\underline{\bar{k}}}\right) \right)
\end{equation*}%
\begin{equation*}
\left( 1+\bar{\beta}\left\langle \underline{\bar{k}}\right\rangle \right)
\left\langle \bar{S}_{1}^{E}\right\rangle \rightarrow 1
\end{equation*}%
so that:%
\begin{equation*}
\left( 1+\bar{\beta}\left\langle \underline{\bar{k}}\right\rangle \right) 
\bar{S}_{1}^{B}\left( \bar{X}^{\prime },\bar{X}_{1}\right) \rightarrow
\left\langle \underline{\hat{k}}_{1}^{B}\right\rangle \left( \left\langle 
\underline{\hat{k}}_{1}^{B}\right\rangle +\kappa \frac{\left\langle 
\underline{\hat{k}}_{2}^{B}\right\rangle }{1+\underline{\bar{k}}}\left( 1-%
\frac{\kappa }{1+\underline{\bar{k}}}\right) \right)
\end{equation*}%
To compute $\left\langle \hat{S}_{1}^{B}\left( \bar{X}^{\prime },\bar{X}%
_{1}\right) \right\rangle $, we use: 
\begin{equation*}
\hat{S}_{1}^{B}\left( \bar{X}^{\prime },\bar{X}_{1}\right) \simeq \frac{\hat{%
K}^{\prime }\underline{\hat{k}}_{1}^{B}\left( \hat{X}^{\prime },\bar{X}%
\right) }{1+\left[ \underline{\hat{k}}\left( \hat{X}^{\prime }\right) \right]
_{\kappa }}\frac{1}{1+\left[ \underline{\hat{k}}_{2}\left( \hat{X}\right) %
\right] _{\kappa }}\left( 1-\frac{\hat{k}\left( \hat{X},\hat{X}^{\prime
}\right) \hat{K}_{X}}{1+\left[ \underline{\hat{k}}\left( \hat{X}^{\prime
}\right) \right] _{\kappa }}\left\Vert \hat{\Psi}\left( \hat{X}^{\prime
}\right) \right\Vert ^{2}\right)
\end{equation*}%
and in average, this is equal to:%
\begin{equation*}
\left\langle \hat{S}_{1}^{B}\left( \bar{X}^{\prime },\bar{X}_{1}\right)
\right\rangle \rightarrow \left\langle \hat{k}_{1}^{B}\right\rangle \left(
1-\left\langle \hat{k}\right\rangle \right)
\end{equation*}%
Ultimately, the average return becomes:%
\begin{eqnarray*}
\left\langle \bar{g}\right\rangle &=&\bar{r}^{\prime }+\frac{\left( 1-\beta
^{B}\right) }{\left( 1-\beta \right) \left\langle \delta \right\rangle }%
\left[ \left( 1-\left( \left( \left( 1-\left\langle \bar{k}_{1}\right\rangle
\right) \left\langle \bar{k}\right\rangle +\left\langle \bar{k}%
_{1}\right\rangle \right) +\left\langle \bar{S}_{1}^{B}\right\rangle \right)
\right) \right] ^{-1} \\
&&\times \left( \frac{1}{\left\langle \hat{k}^{B}\right\rangle }-\left(
\left( \left\langle \hat{k}\right\rangle \left( 1-\left\langle \hat{k}%
_{1}\right\rangle \right) +\left\langle \hat{k}_{1}\right\rangle \right) -%
\frac{\left( 1-\beta \right) \left\langle \delta \right\rangle }{1-\beta ^{B}%
}\left( \left\langle \hat{k}_{1}^{B}\right\rangle \left( 1-\left\langle \hat{%
k}\right\rangle \right) \right) \right) \right) \left( \left\langle \hat{g}%
\right\rangle -\bar{r}^{\prime }\right)
\end{eqnarray*}%
Given that$1-\beta ^{B}<<1$, in first approximation we can cnsdr tht $%
\left\langle \bar{g}\right\rangle =\bar{r}^{\prime }$. 
\begin{equation*}
\left\langle \delta \right\rangle \simeq \frac{\left\langle
k_{1}^{B}\right\rangle +\frac{\kappa \left\langle k_{2}^{B}\right\rangle }{%
1+\left\langle \bar{k}\right\rangle }}{\left\langle k\right\rangle }Z\simeq
\left( \frac{\left\langle k_{1}^{B}\right\rangle }{\left\langle
k\right\rangle }+\frac{\kappa \left\langle k_{2}^{B}\right\rangle }{%
\left\langle k\right\rangle \left( 1+\left\langle \bar{k}\right\rangle
\right) }\right) \frac{\left\Vert \bar{\Psi}_{0}\right\Vert ^{2}}{\left\Vert 
\hat{\Psi}_{0}\right\Vert ^{2}}
\end{equation*}

\subsection*{A23.4 Approximate solutions for average global capital}

\subsubsection*{A23.4.1 Expressions for coefficients}

To find solutions for $\left\langle \hat{g}\right\rangle $ and use some
approximtns. First, given that $1-\beta ^{B}<<1$, in first approximation we
can cnsdr tht 
\begin{equation*}
\left\langle \bar{g}\right\rangle =\bar{r}^{\prime }
\end{equation*}%
Recall tht: 
\begin{equation*}
\left\langle \delta \right\rangle \simeq \frac{\left\langle
k_{1}^{B}\right\rangle +\frac{\kappa \left\langle k_{2}^{B}\right\rangle }{%
1+\left\langle \bar{k}\right\rangle }}{\left\langle k\right\rangle }Z\simeq
\left( \frac{\left\langle k_{1}^{B}\right\rangle }{\left\langle
k\right\rangle }+\frac{\kappa \left\langle k_{2}^{B}\right\rangle }{%
\left\langle k\right\rangle \left( 1+\left\langle \bar{k}\right\rangle
\right) }\right) \frac{\left\Vert \bar{\Psi}_{0}\right\Vert ^{2}}{\left\Vert 
\hat{\Psi}_{0}\right\Vert ^{2}}>>1
\end{equation*}%
for $\kappa >>1$.

Moreover, in first pprxmtn:%
\begin{equation*}
\bar{N}=-\kappa ^{2}\left\langle \underline{\hat{k}}_{2}^{B}\right\rangle
\end{equation*}%
\begin{eqnarray*}
\left\langle \hat{g}^{ef}\right\rangle &=&-\frac{\left( \kappa \left\langle %
\left[ \frac{\underline{\hat{k}}_{2}^{B}}{1+\bar{k}}\right] \right\rangle
\left( 1-\left\langle \underline{\hat{k}}\right\rangle \right) +\left\langle 
\hat{k}_{1}^{B}\right\rangle \left\langle \hat{k}_{2}\right\rangle \right)
\left( \left\langle \hat{g}\right\rangle +\frac{1}{1-\left\langle \underline{%
\hat{k}}\right\rangle }\bar{N}\bar{r}^{\prime }\right) \frac{\left\langle 
\bar{K}\right\rangle \left\Vert \bar{\Psi}\right\Vert ^{2}}{\left\langle 
\hat{K}\right\rangle \left\Vert \hat{\Psi}\right\Vert ^{2}}}{\left( 1-\left(
\left\langle \hat{k}_{1}\right\rangle +\left\langle \hat{k}%
_{1}^{B}\right\rangle \frac{\left\langle \bar{K}\right\rangle \left\Vert 
\bar{\Psi}\right\Vert ^{2}}{\left\langle \hat{K}\right\rangle \left\Vert 
\hat{\Psi}\right\Vert ^{2}}\right) \right) \left( 1-\left( \left\langle 
\underline{\hat{k}}\right\rangle +\left( \left\langle \underline{\hat{k}}%
_{1}^{B}\right\rangle +\kappa \left\langle \left[ \frac{\underline{\hat{k}}%
_{2}^{B}}{1+\bar{k}}\right] \right\rangle \right) \frac{\left\langle \bar{K}%
\right\rangle \left\Vert \bar{\Psi}\right\Vert ^{2}}{\left\langle \hat{K}%
\right\rangle \left\Vert \hat{\Psi}\right\Vert ^{2}}\right) \right) } \\
&\simeq &\frac{\kappa ^{2}\left\langle \underline{\hat{k}}%
_{2}^{B}\right\rangle }{1-\left\langle \underline{\hat{k}}\right\rangle }%
\frac{\left( \kappa \left\langle \left[ \frac{\underline{\hat{k}}_{2}^{B}}{1+%
\bar{k}}\right] \right\rangle \left( 1-\left\langle \underline{\hat{k}}%
\right\rangle \right) +\left\langle \hat{k}_{1}^{B}\right\rangle
\left\langle \hat{k}_{2}\right\rangle \right) \frac{\left\langle \bar{K}%
\right\rangle \left\Vert \bar{\Psi}\right\Vert ^{2}}{\left\langle \hat{K}%
\right\rangle \left\Vert \hat{\Psi}\right\Vert ^{2}}}{\left( 1-\left(
\left\langle \hat{k}_{1}\right\rangle +\left\langle \hat{k}%
_{1}^{B}\right\rangle \frac{\left\langle \bar{K}\right\rangle \left\Vert 
\bar{\Psi}\right\Vert ^{2}}{\left\langle \hat{K}\right\rangle \left\Vert 
\hat{\Psi}\right\Vert ^{2}}\right) \right) \left( 1-\left( \left\langle 
\underline{\hat{k}}\right\rangle +\left( \left\langle \underline{\hat{k}}%
_{1}^{B}\right\rangle +\kappa \left\langle \left[ \frac{\underline{\hat{k}}%
_{2}^{B}}{1+\bar{k}}\right] \right\rangle \right) \frac{\left\langle \bar{K}%
\right\rangle \left\Vert \bar{\Psi}\right\Vert ^{2}}{\left\langle \hat{K}%
\right\rangle \left\Vert \hat{\Psi}\right\Vert ^{2}}\right) \right) }\bar{r}%
^{\prime } \\
&\simeq &\frac{\kappa ^{3}\left\langle \underline{\hat{k}}%
_{2}^{B}\right\rangle \left\langle \left[ \frac{\underline{\hat{k}}_{2}^{B}}{%
1+\bar{k}}\right] \right\rangle \frac{\left\langle \bar{K}\right\rangle
\left\Vert \bar{\Psi}\right\Vert ^{2}}{\left\langle \hat{K}\right\rangle
\left\Vert \hat{\Psi}\right\Vert ^{2}}}{\left( 1-\left( \left\langle \hat{k}%
_{1}\right\rangle +\left\langle \hat{k}_{1}^{B}\right\rangle \frac{%
\left\langle \bar{K}\right\rangle \left\Vert \bar{\Psi}\right\Vert ^{2}}{%
\left\langle \hat{K}\right\rangle \left\Vert \hat{\Psi}\right\Vert ^{2}}%
\right) \right) \left( 1-\left( \left\langle \underline{\hat{k}}%
\right\rangle +\left( \left\langle \underline{\hat{k}}_{1}^{B}\right\rangle
+\kappa \left\langle \left[ \frac{\underline{\hat{k}}_{2}^{B}}{1+\bar{k}}%
\right] \right\rangle \right) \frac{\left\langle \bar{K}\right\rangle
\left\Vert \bar{\Psi}\right\Vert ^{2}}{\left\langle \hat{K}\right\rangle
\left\Vert \hat{\Psi}\right\Vert ^{2}}\right) \right) }\bar{r}^{\prime }
\end{eqnarray*}%
and:%
\begin{eqnarray*}
\left\langle \hat{g}^{Bef}\right\rangle &=&-\frac{\kappa \left\langle \left[ 
\frac{\underline{\hat{k}}_{2}^{B}}{1+\bar{k}}\right] \right\rangle \left(
1-\left\langle \hat{k}\right\rangle \right) +\left\langle \hat{k}%
_{1}^{B}\right\rangle \left\langle \hat{k}_{2}\right\rangle \left(
\left\langle \hat{g}\right\rangle +\left( 1-\hat{M}\right) ^{-1}\bar{N}\bar{r%
}^{\prime }\right) }{\left( 1-\left( \left\langle \hat{k}\right\rangle
+\left( \left\langle \hat{k}_{1}^{B}\right\rangle +\kappa \left\langle \left[
\frac{\underline{\hat{k}}_{2}^{B}}{1+\bar{k}}\right] \right\rangle \right) 
\frac{\left\langle \bar{K}\right\rangle \left\Vert \bar{\Psi}\right\Vert ^{2}%
}{\left\langle \hat{K}\right\rangle \left\Vert \hat{\Psi}\right\Vert ^{2}}%
\right) \right) \left( 1-\left( \left\langle \hat{k}_{1}\right\rangle
+\left\langle \hat{k}_{1}^{B}\right\rangle \frac{\left\langle \bar{K}%
\right\rangle \left\Vert \bar{\Psi}\right\Vert ^{2}}{\left\langle \hat{K}%
\right\rangle \left\Vert \hat{\Psi}\right\Vert ^{2}}\right) \right) } \\
&\simeq &\frac{\kappa ^{3}\left\langle \underline{\hat{k}}%
_{2}^{B}\right\rangle \left\langle \left[ \frac{\underline{\hat{k}}_{2}^{B}}{%
1+\bar{k}}\right] \right\rangle }{\left( 1-\left( \left\langle \hat{k}%
_{1}\right\rangle +\left\langle \hat{k}_{1}^{B}\right\rangle \frac{%
\left\langle \bar{K}\right\rangle \left\Vert \bar{\Psi}\right\Vert ^{2}}{%
\left\langle \hat{K}\right\rangle \left\Vert \hat{\Psi}\right\Vert ^{2}}%
\right) \right) \left( 1-\left( \left\langle \underline{\hat{k}}%
\right\rangle +\left( \left\langle \underline{\hat{k}}_{1}^{B}\right\rangle
+\kappa \left\langle \left[ \frac{\underline{\hat{k}}_{2}^{B}}{1+\bar{k}}%
\right] \right\rangle \right) \frac{\left\langle \bar{K}\right\rangle
\left\Vert \bar{\Psi}\right\Vert ^{2}}{\left\langle \hat{K}\right\rangle
\left\Vert \hat{\Psi}\right\Vert ^{2}}\right) \right) }\bar{r}^{\prime }
\end{eqnarray*}

\subsubsection*{A23.4.2 Expression for capital}

Average amounts of capital are given by:%
\begin{eqnarray*}
\left\langle \hat{K}\right\rangle \left\Vert \hat{\Psi}\right\Vert ^{2} &=&%
\frac{18\sigma _{\hat{K}}^{2}}{\hat{\mu}\left\langle \hat{g}\right\rangle
^{2}}V\left( \frac{\left\Vert \hat{\Psi}_{0}\right\Vert ^{2}}{\left( 5+\frac{%
\left\langle \hat{g}^{ef}\right\rangle }{\left\langle \hat{g}\right\rangle }-%
\sqrt{\left( 1-\frac{\left\langle \hat{g}^{ef}\right\rangle }{\left\langle 
\hat{g}\right\rangle }\right) \left( 4-\frac{\left\langle \hat{g}%
^{ef}\right\rangle }{\left\langle \hat{g}\right\rangle }\right) }\right) }%
\right) ^{2} \\
&&\times \left( \frac{1}{4}-\frac{1}{18}\left( 2+\frac{\left\langle \hat{g}%
^{ef}\right\rangle }{\left\langle \hat{g}\right\rangle }-\sqrt{\left( 1-%
\frac{\left\langle \hat{g}^{ef}\right\rangle }{\left\langle \hat{g}%
\right\rangle }\right) \left( 4-\frac{\left\langle \hat{g}^{ef}\right\rangle 
}{\left\langle \hat{g}\right\rangle }\right) }\right) \right) \\
&\simeq &\frac{9\sigma _{\hat{K}}^{2}}{2\hat{\mu}\left\langle \hat{g}%
\right\rangle ^{2}}V\left\Vert \hat{\Psi}_{0}\right\Vert ^{4}
\end{eqnarray*}%
so that:%
\begin{equation}
\left\langle \hat{g}\right\rangle \simeq \frac{\sqrt{\frac{9\sigma _{\hat{K}%
}^{2}}{2\hat{\mu}}V}\left\Vert \hat{\Psi}_{0}\right\Vert ^{2}}{\sqrt{%
\left\langle \hat{K}\right\rangle \left\Vert \hat{\Psi}\right\Vert ^{2}}}
\label{Gh}
\end{equation}%
We have ssmd tht $\frac{\left\langle \hat{g}^{ef}\right\rangle }{%
\left\langle \hat{g}\right\rangle }<<1$ but correction at the first order
can be easily derived.

For banks capital: 
\begin{eqnarray}
\left\langle \bar{K}\right\rangle \left\Vert \bar{\Psi}\right\Vert ^{2}
&\simeq &18\frac{\sigma _{\hat{K}}^{2}V}{\left\langle \bar{g}\right\rangle
^{2}\hat{\mu}}\left( \frac{\sqrt{\left\langle \hat{g}\right\rangle
^{2}\left( 1+\frac{3}{4}\frac{\left\langle \hat{g}^{ef}\right\rangle }{%
\left\langle \hat{g}\right\rangle }\right) }\left\Vert \bar{\Psi}%
_{0}\right\Vert -\frac{3}{8}\left\langle \hat{g}\right\rangle \frac{%
\left\langle \hat{g}^{Bef}\right\rangle }{\left\langle \bar{g}\right\rangle }%
\left\Vert \hat{\Psi}_{0}\right\Vert }{\sqrt{5+\frac{\left\langle \hat{g}%
^{ef}\right\rangle }{\left\langle \hat{g}\right\rangle }-\sqrt{\left( 1-%
\frac{\left\langle \hat{g}^{ef}\right\rangle }{\left\langle \hat{g}%
\right\rangle }\right) \left( 4-\frac{\left\langle \hat{g}^{ef}\right\rangle 
}{\left\langle \hat{g}\right\rangle }\right) }}}\right) ^{4}\left( \frac{1}{4%
}-\frac{\left\langle \hat{g}\right\rangle \left\langle \hat{g}%
^{Bef}\right\rangle }{3\left\langle \bar{g}\right\rangle ^{2}}\right) \\
&\simeq &18\frac{\sigma _{\hat{K}}^{2}V\left\langle \hat{g}\right\rangle ^{4}%
}{\left\langle \bar{g}\right\rangle ^{2}\hat{\mu}}\left( \left\Vert \bar{\Psi%
}_{0}\right\Vert \right) ^{4}  \notag
\end{eqnarray}%
Using (\ref{Gh}) enables to rewrite at the lowest order:%
\begin{equation}
\left\langle \bar{K}\right\rangle \left\Vert \bar{\Psi}\right\Vert
^{2}\simeq 18\frac{\sigma _{\hat{K}}^{2}V\left( \frac{\sqrt{\frac{9\sigma _{%
\hat{K}}^{2}}{2\hat{\mu}}V}\left\Vert \hat{\Psi}_{0}\right\Vert ^{2}}{\sqrt{%
\left\langle \hat{K}\right\rangle \left\Vert \hat{\Psi}\right\Vert ^{2}}}%
\right) ^{4}}{\left\langle \bar{g}\right\rangle ^{2}\hat{\mu}}\left(
\left\Vert \bar{\Psi}_{0}\right\Vert \right) ^{4}\simeq 4\frac{\sigma _{\hat{%
K}}^{2}V\left( \frac{9\sigma _{\hat{K}}^{2}}{2\hat{\mu}}V\right)
^{3}\left\Vert \hat{\Psi}_{0}\right\Vert ^{8}\left\Vert \bar{\Psi}%
_{0}\right\Vert ^{4}}{\left( \bar{r}^{\prime }\right) ^{2}\hat{\mu}\left(
\left\langle \hat{K}\right\rangle \left\Vert \hat{\Psi}\right\Vert
^{2}\right) ^{2}\left\langle \hat{K}\right\rangle \left\Vert \hat{\Psi}%
\right\Vert ^{2}}  \label{KBp}
\end{equation}

\subsubsection*{A23.4.3 Equation for average capital}

We rewrite the equation for $\left\langle \hat{g}\right\rangle $ in terms of
our approximations: 
\begin{eqnarray*}
\left\langle \hat{g}\right\rangle &\simeq &\bar{r}^{\prime }+\left( 1-\left(
1+\left\langle \hat{k}_{2}^{n}\right\rangle \right) \left( \left\langle \hat{%
k}\right\rangle \left( 1-\left\langle \hat{k}_{1}\right\rangle \right)
+\left\langle \hat{k}_{1}\right\rangle \right) \right) ^{-1} \\
&&\times \left( 1+\left\langle \hat{k}_{2}^{n}\right\rangle \right)
\left\langle \left( \frac{A\left( \hat{X}^{\prime }\right) }{f_{1}^{2}\left( 
\hat{X}^{\prime }\right) }+\frac{B\left( \hat{X}^{\prime }\right) }{%
f_{1}^{3}\left( \hat{X}^{\prime }\right) }\right) \left( R+\Delta F_{\tau
}\left( \bar{R}\left( K,\hat{X}^{\prime }\right) \right) \right)
\right\rangle
\end{eqnarray*}%
Using that:%
\begin{equation*}
\delta >1
\end{equation*}%
this becoms%
\begin{eqnarray*}
\left\langle \hat{g}\right\rangle &\simeq &\bar{r}^{\prime }+\left( 1-\left(
1+\left\langle \hat{k}_{2}^{n}\right\rangle \right) \left( \left\langle \hat{%
k}\right\rangle \left( 1-\left\langle \hat{k}_{1}\right\rangle \right)
+\left\langle \hat{k}_{1}\right\rangle \right) \right) ^{-1} \\
&&\times \left( 1+\left\langle \hat{k}_{2}^{n}\right\rangle \right)
\left\langle \left( \frac{A\left( \hat{X}^{\prime }\right) }{f_{1}^{2}\left( 
\hat{X}^{\prime }\right) }+\frac{B\left( \hat{X}^{\prime }\right) }{%
f_{1}^{3}\left( \hat{X}^{\prime }\right) }\right) \left( R+\Delta F_{\tau
}\left( \bar{R}\left( K,\hat{X}^{\prime }\right) \right) \right)
\right\rangle
\end{eqnarray*}%
or reintroducing $\Delta F_{\tau }\left( \bar{R}\left( K,X\right) \right) $:%
\begin{eqnarray*}
\left\langle \hat{g}\right\rangle -\bar{r}^{\prime } &\simeq &\left(
1-\left( 1+\left\langle \hat{k}_{2}^{n}\right\rangle \right) \left(
\left\langle \hat{k}\right\rangle \left( 1-\left\langle \hat{k}%
_{1}\right\rangle \right) +\left\langle \hat{k}_{1}\right\rangle \right)
\right) ^{-1}\left( 1+\left\langle \hat{k}_{2}^{n}\right\rangle \right) \\
&&\times \left\langle \left( \frac{A}{\left( f_{1}\left( X,\hat{K}\left[ X%
\right] ,\bar{K}\left[ X\right] \right) \right) ^{2}}+\frac{B}{\left(
f_{1}\left( X,\hat{K}\left[ X\right] ,\bar{K}\left[ X\right] \right) \right)
^{3}}\right) \left( R+\Delta F_{\tau }\left( \bar{R}\left( K,X\right)
\right) \right) \right\rangle
\end{eqnarray*}

Here:%
\begin{equation*}
f_{1}\left( X,\hat{K}\left[ X\right] ,\bar{K}\left[ X\right] \right) =\frac{%
f_{1}\left( X\right) }{\left( 1+k\left( X\right) \hat{K}\left[ X\right]
+\left( k_{1}^{B}\left( X\right) +\kappa \left[ \frac{\underline{k}_{2}^{B}}{%
1+\bar{k}}\right] \right) \bar{K}\left[ X\right] \right) ^{r}}-C_{0}
\end{equation*}%
and:%
\begin{equation*}
\left\langle f_{1}\left( X,\hat{K}\left[ X\right] ,\bar{K}\left[ X\right]
\right) \right\rangle =\frac{\left\langle f_{1}\left( X\right) \right\rangle 
}{\left( 1+\left\langle k\right\rangle \left\langle \hat{K}\right\rangle
\left\Vert \hat{\Psi}\right\Vert ^{2}+\left( \left\langle
k_{1}^{B}\right\rangle +\kappa \left[ \frac{\underline{k}_{2}^{B}}{1+\bar{k}}%
\right] \right) \left\langle \bar{K}\right\rangle \left\Vert \bar{\Psi}%
\right\Vert ^{2}\right) ^{r}}-C_{0}
\end{equation*}%
Using (\ref{Gh}) and (\ref{KBp}) leads to write:%
\begin{equation}
\left\langle f_{1}\left( X,\hat{K}\left[ X\right] ,\bar{K}\left[ X\right]
\right) \right\rangle \simeq \frac{\left\langle f_{1}\left( X\right)
\right\rangle }{\left( 1+\left\langle k\right\rangle \left\langle \hat{K}%
\right\rangle \left\Vert \hat{\Psi}\right\Vert ^{2}+\left( \left\langle
k_{1}^{B}\right\rangle +\kappa \left[ \frac{\underline{k}_{2}^{B}}{1+\bar{k}}%
\right] \right) \frac{4\sigma _{\hat{K}}^{2}V\left( \frac{9\sigma _{\hat{K}%
}^{2}}{2\hat{\mu}}V\right) ^{3}\left\Vert \hat{\Psi}_{0}\right\Vert
^{8}\left\Vert \bar{\Psi}_{0}\right\Vert ^{4}}{\left( \bar{r}^{\prime
}\right) ^{2}\hat{\mu}\left( \left\langle \hat{K}\right\rangle \left\Vert 
\hat{\Psi}\right\Vert ^{2}\right) ^{2}}\right) ^{r}}-C_{0}  \label{FWN}
\end{equation}%
and return equation becomes an equation for $\left\langle \hat{K}%
\right\rangle \left\Vert \hat{\Psi}\right\Vert ^{2}$:%
\begin{eqnarray*}
\frac{\sqrt{\frac{9\sigma _{\hat{K}}^{2}}{2\hat{\mu}}V}\left\Vert \hat{\Psi}%
_{0}\right\Vert ^{2}}{\sqrt{\left\langle \hat{K}\right\rangle \left\Vert 
\hat{\Psi}\right\Vert ^{2}}}-\bar{r}^{\prime } &\simeq &\left( 1-\left(
1+\left\langle \hat{k}_{2}^{n}\right\rangle \right) \left( \left\langle \hat{%
k}\right\rangle \left( 1-\left\langle \hat{k}_{1}\right\rangle \right)
+\left\langle \hat{k}_{1}\right\rangle \right) \right) ^{-1}\left(
1+\left\langle \hat{k}_{2}^{n}\right\rangle \right) \\
&&\times \left\langle \left( \frac{A}{\left( f_{1}\left( X,\hat{K}\left[ X%
\right] ,\bar{K}\left[ X\right] \right) \right) ^{2}}+\frac{B}{\left(
f_{1}\left( X,\hat{K}\left[ X\right] ,\bar{K}\left[ X\right] \right) \right)
^{3}}\right) R\right\rangle
\end{eqnarray*}%
This equation is similar to the equation without banks, except that the
formula (\ref{FWN}), includes banks effects in the definition of disposable
ncomes of the firms. Numerical studies indicate that this stabilizes the
possibilities of multiple states and reduces the number of stats given some
average productivt.

Moreover, including the normaltn f cf:%
\begin{equation*}
\frac{\hat{k}_{\eta }\left( \hat{X}^{\prime },\hat{X}\right) }{\left\Vert 
\hat{\Psi}\right\Vert ^{2}\left\langle \hat{K}\right\rangle \left(
1-\left\langle \hat{k}\left( \hat{X}^{\prime },\hat{X}\right) \right\rangle
-\left( \left\langle \hat{k}_{1}^{B}\left( \hat{X}^{\prime },\bar{X}\right)
\right\rangle +\kappa \frac{\left\langle \hat{k}_{2}^{B}\left( \hat{X}%
^{\prime },\bar{X}\right) \right\rangle }{1+\left\langle \bar{k}%
\right\rangle \left\langle \bar{K}\right\rangle }\frac{\left\Vert \bar{\Psi}%
\right\Vert ^{2}\left\langle \bar{K}\right\rangle }{\left\Vert \hat{\Psi}%
\right\Vert ^{2}\left\langle \hat{K}\right\rangle }\right) \right) }
\end{equation*}%
The coeficients are higher in this case than in case with \ no banks where:%
\begin{equation*}
\frac{\hat{k}_{\eta }\left( \hat{X}^{\prime },\hat{X}\right) }{\left\Vert 
\hat{\Psi}\right\Vert ^{2}\left\langle \hat{K}\right\rangle \left(
1-\left\langle \hat{k}\left( \hat{X}^{\prime },\hat{X}\right) \right\rangle
\right) }
\end{equation*}%
As a consequence, the diffusion matrix is larger in this case than in part
1. Investors may now borrow from banks, incrreasing their disposable
capital, and invreasing their investment in other investors.

\section*{Appendix 24 Block interactions}

\subsection*{A24.1 Several groups}

To model groups dynamics, we use the alternate description for returns:%
\begin{eqnarray}
0 &=&\int \left( 1-\hat{S}_{1}\left( \hat{X}^{\prime },\hat{K}^{\prime },%
\hat{X}\right) \right) \frac{\hat{f}\left( \hat{X}^{\prime }\right) -\bar{r}%
}{1+\underline{\hat{k}}_{2}\left( \bar{X}^{\prime }\right) +\kappa \frac{%
\underline{\hat{k}}_{2}^{B}\left( \bar{X}^{\prime }\right) }{1+\bar{k}}}d%
\hat{X}^{\prime }d\hat{K}^{\prime } \\
&&-\int \left( \frac{1+\hat{f}\left( \hat{X}^{\prime }\right) }{\underline{%
\hat{k}}_{2}\left( \bar{X}^{\prime }\right) +\kappa \left[ \frac{\underline{%
\hat{k}}_{2}^{B}}{1+\bar{k}}\right] \left( \hat{X}^{\prime }\right) }H\left(
-\frac{1+\hat{f}\left( \hat{X}^{\prime }\right) }{\underline{\hat{k}}%
_{2}\left( \bar{X}^{\prime }\right) +\kappa \left[ \frac{\underline{\hat{k}}%
_{2}^{B}}{1+\bar{k}}\right] \left( \hat{X}^{\prime }\right) }\right) \right) 
\hat{S}_{2}\left( \hat{X}^{\prime },\hat{K}^{\prime },\hat{X}\right) d\hat{X}%
^{\prime }d\hat{K}^{\prime }  \notag \\
&&-\int \frac{1+f_{1}^{\prime }\left( X^{\prime }\right) }{\underline{k}%
_{2}\left( X^{\prime }\right) +\kappa \left[ \frac{\underline{k}_{2}^{B}}{1+%
\bar{k}}\right] \left( X^{\prime }\right) }H\left( -\frac{1+f_{1}^{\prime
}\left( X^{\prime }\right) }{\underline{k}_{2}\left( X^{\prime }\right)
+\kappa \left[ \frac{\underline{k}_{2}^{B}}{1+\bar{k}}\right] \left(
X^{\prime }\right) }\right) S_{2}\left( X^{\prime },K^{\prime },\hat{X}%
\right) -\int S_{1}\left( X^{\prime },K^{\prime },\hat{X}\right) \left( \hat{%
f}_{1}\left( \hat{K},\hat{X}\right) -\bar{r}\right)  \notag
\end{eqnarray}%
for investors, and:%
\begin{eqnarray}
0 &=&\left( 1-\bar{S}_{1}\left( \bar{X}^{\prime },\bar{X}\right) \right) 
\frac{\bar{f}\left( \bar{X}^{\prime }\right) -\bar{r}}{1+\underline{%
\overline{\bar{k}}}_{2}\left( \bar{X}^{\prime }\right) }-\hat{S}%
_{1}^{B}\left( \hat{X}^{\prime },\bar{X}\right) \left( \frac{\hat{f}\left( 
\hat{X}^{\prime }\right) -\bar{r}}{1+\underline{\hat{k}}_{2}\left( \bar{X}%
^{\prime }\right) +\kappa \frac{\underline{\hat{k}}_{2}^{B}\left( \bar{X}%
^{\prime }\right) }{1+\bar{k}\left( \bar{X}\right) }}\right) \\
&&-\frac{\left( 1+\bar{f}\left( \bar{X}^{\prime }\right) \right) H\left(
-\left( 1+\bar{f}\left( \bar{X}^{\prime }\right) \right) \right) }{%
\underline{\overline{\bar{k}}}_{2}\left( \hat{X}^{\prime }\right) }\bar{S}%
_{2}\left( \bar{X}^{\prime },\bar{X}\right) -\frac{\left( 1+\hat{f}\left( 
\hat{X}^{\prime }\right) \right) H\left( -\left( 1+\hat{f}\left( \hat{X}%
^{\prime }\right) \right) \right) }{\underline{\hat{k}}_{2}\left( \bar{X}%
^{\prime }\right) +\kappa \left[ \frac{\underline{\hat{k}}_{2}^{B}}{1+\bar{k}%
}\right] \left( \hat{X}^{\prime }\right) }\hat{S}_{2}^{B}\left( \hat{X}%
^{\prime },\bar{X}\right)  \notag \\
&&-\frac{\left( 1+f_{1}^{\prime }\left( X^{\prime }\right) \right) H\left(
1+f_{1}^{\prime }\left( K^{\prime },X^{\prime }\right) \right) }{\underline{k%
}_{2}\left( X^{\prime }\right) +\kappa \left[ \frac{\underline{k}%
_{2}^{\left( B\right) }}{1+\underline{\bar{k}}}\right] \left( X^{\prime
}\right) }S_{1}^{B}\left( \hat{X}^{\prime },\bar{X}\right) -S_{1}^{B}\left( 
\hat{X}^{\prime },\bar{X}\right) \left( \frac{\left( f_{1}^{\prime }\left( 
\bar{K},\bar{X}\right) -\bar{r}\right) }{1+\underline{k}_{2}\left( \hat{X}%
^{\prime }\right) +\kappa \frac{\underline{k}_{2}^{\left( B\right) }\left( 
\bar{X}^{\prime }\right) }{1+\underline{\bar{k}}}}+\Delta F_{\tau }\left( 
\bar{R}\left( K,X\right) \right) \right)  \notag
\end{eqnarray}%
for banks. The coefficients are defined in Appendix 23.4.

Applying these formula to groups averages, as in the first part, we are led
to define average intra and inter coefficiennts.

\subsection*{A24.2 Investor return}

\subsubsection*{A24.2.1 Mixed formulation}

The return equation for investors rewrites:%
\begin{eqnarray}
0 &=&\left( 
\begin{array}{cc}
1-\underline{\hat{S}}_{1}^{\left[ ii\right] } & -\underline{\hat{S}}_{1}^{%
\left[ ji\right] } \\ 
-\underline{\hat{S}}_{1}^{\left[ ij\right] } & 1-\underline{\hat{S}}_{1}^{%
\left[ jj\right] }%
\end{array}%
\right) \left( 
\begin{array}{c}
\frac{f^{\left[ i\right] }-\bar{r}}{1+\underline{\hat{k}}_{2}^{\left[ i%
\right] }+\kappa \left[ \frac{\underline{\hat{k}}_{2}^{B}}{1+\bar{k}}\right]
^{\left[ i\right] }} \\ 
\frac{f^{\left[ j\right] }-\bar{r}}{1+\underline{\hat{k}}_{2}^{\left[ j%
\right] }+\kappa \left[ \frac{\underline{\hat{k}}_{2}^{B}}{1+\bar{k}}\right]
^{\left[ j\right] }}%
\end{array}%
\right) -\left( 
\begin{array}{cc}
\underline{\hat{S}}_{2}^{\left[ ii\right] } & \underline{\hat{S}}_{2}^{\left[
ji\right] } \\ 
\underline{\hat{S}}_{2}^{\left[ ij\right] } & \underline{\hat{S}}_{2}^{\left[
jj\right] }%
\end{array}%
\right) \left( 
\begin{array}{c}
\frac{\left( 1+f^{\left[ i\right] }\right) H\left( -\left( 1+f^{\left[ i%
\right] }\right) \right) }{\underline{\hat{k}}_{2}^{\left[ i\right] }+\kappa %
\left[ \frac{\underline{\hat{k}}_{2}^{B}}{1+\bar{k}}\right] ^{\left[ i\right]
}} \\ 
\frac{\left( 1+f^{\left[ j\right] }\right) H\left( -\left( 1+f^{\left[ j%
\right] }\right) \right) }{\underline{\hat{k}}_{2}^{\left[ j\right] }+\kappa %
\left[ \frac{\underline{\hat{k}}_{2}^{B}}{1+\bar{k}}\right] ^{\left[ j\right]
}}%
\end{array}%
\right)  \notag \\
&&-\left( 
\begin{array}{cc}
\underline{S}_{2}^{\left[ ii\right] } & \underline{S}_{2}^{\left[ ji\right] }
\\ 
\underline{S}_{2}^{\left[ ij\right] } & \underline{S}_{2}^{\left[ jj\right] }%
\end{array}%
\right) \left( 
\begin{array}{c}
\frac{\left( 1+f_{1}^{\prime \left[ i\right] }\right) H\left( -\left(
1+f_{1}^{\prime \left[ i\right] }\right) \right) }{\underline{k}_{2}^{\left[
i\right] }+\left[ \frac{\underline{k}_{2}^{B}}{1+\bar{k}}\right] ^{\left[ i%
\right] }} \\ 
\frac{\left( 1+f_{1}^{\prime \left[ j\right] }\right) H\left( -\left(
1+f_{1}^{\prime \left[ j\right] }\right) \right) }{\underline{k}_{2}^{\left[
j\right] }+\left[ \frac{\underline{k}_{2}^{B}}{1+\bar{k}}\right] ^{\left[ j%
\right] }}%
\end{array}%
\right) -\left( 
\begin{array}{cc}
\underline{S}_{1}^{\left[ ii\right] } & \underline{S}_{1}^{\left[ ji\right] }
\\ 
\underline{S}_{1}^{\left[ ij\right] } & \underline{S}_{1}^{\left[ jj\right] }%
\end{array}%
\right) \left( 
\begin{array}{c}
\frac{f_{1}^{\prime \left[ i\right] }-\bar{r}}{1+\underline{k}_{2}^{\left[ i%
\right] }+\left[ \frac{\underline{k}_{2}^{B}}{1+\bar{k}}\right] ^{\left[ i%
\right] }} \\ 
\frac{f_{1}^{\prime \left[ j\right] }-\bar{r}}{1+\underline{k}_{2}^{\left[ j%
\right] }+\left[ \frac{\underline{k}_{2}^{B}}{1+\bar{k}}\right] ^{\left[ j%
\right] }}%
\end{array}%
\right)  \label{MF}
\end{eqnarray}%
where we defined:%
\begin{equation*}
\kappa \left( \frac{\underline{\hat{k}}_{\epsilon }^{B\left[ ii\right] }}{1+%
\bar{k}^{\left[ i\right] }}+\frac{\underline{\hat{k}}_{\epsilon }^{B\left[ ij%
\right] }}{1+\bar{k}^{\left[ j\right] }}\right) \rightarrow \kappa \left[ 
\frac{\underline{\hat{k}}_{\epsilon }^{B}}{1+\bar{k}}\right] ^{\left[ i%
\right] }
\end{equation*}%
\begin{equation*}
\kappa \left( \frac{\underline{\hat{k}}^{B\left[ ii\right] }}{1+\bar{k}^{%
\left[ i\right] }}+\frac{\underline{\hat{k}}^{B\left[ ij\right] }}{1+\bar{k}%
^{\left[ j\right] }}\right) \rightarrow \kappa \left[ \frac{\underline{\hat{k%
}}^{B}}{1+\bar{k}}\right] ^{\left[ i\right] }
\end{equation*}%
\begin{equation*}
\kappa \left( \frac{\underline{k}_{\epsilon }^{B\left[ ii\right] }}{1+\bar{k}%
^{\left[ i\right] }}+\frac{\underline{k}_{\epsilon }^{B\left[ ij\right] }}{1+%
\bar{k}^{\left[ j\right] }}\right) \rightarrow \kappa \left[ \frac{%
\underline{k}_{\epsilon }^{B}}{1+\bar{k}}\right] ^{\left[ i\right] }
\end{equation*}%
\begin{equation*}
\kappa \left( \frac{\underline{k}^{B\left[ ii\right] }}{1+\bar{k}^{\left[ i%
\right] }}+\frac{\underline{k}^{B\left[ ij\right] }}{1+\bar{k}^{\left[ j%
\right] }}\right) \rightarrow \kappa \left[ \frac{\underline{k}^{B}}{1+\bar{k%
}}\right] ^{\left[ i\right] }
\end{equation*}%
and the following rates:%
\begin{eqnarray*}
\underline{\hat{S}}_{\eta }^{\left[ ii\right] } &=&\frac{\underline{\hat{k}}%
_{\eta }^{\left[ ii\right] }}{\left( 1+\underline{\hat{k}}^{\left[ i\right]
}+\underline{\hat{k}}_{1}^{B\left[ i\right] }+\kappa \left[ \frac{\underline{%
\hat{k}}_{2}^{B}}{1+\bar{k}}\right] ^{\left[ i\right] }\right) } \\
\underline{\hat{S}}_{\eta }^{\left[ ij\right] } &=&\frac{\underline{\hat{k}}%
_{\eta }^{\left[ ij\right] }}{\left( 1+\underline{\hat{k}}^{\left[ i\right]
}+\underline{\hat{k}}_{1}^{B\left[ i\right] }+\kappa \left[ \frac{\underline{%
\hat{k}}_{2}^{B}}{1+\bar{k}}\right] ^{\left[ i\right] }\right) }
\end{eqnarray*}%
\begin{eqnarray*}
\underline{\hat{S}}^{\left[ ii\right] } &=&\underline{\hat{S}}_{1}^{\left[ ii%
\right] }+\underline{\hat{S}}_{2}^{\left[ ii\right] } \\
\underline{\hat{S}}^{\left[ ij\right] } &=&\underline{\hat{S}}_{1}^{\left[ ij%
\right] }+\underline{\hat{S}}_{2}^{\left[ ij\right] }
\end{eqnarray*}%
\begin{eqnarray*}
\underline{S}_{\eta }^{\left[ ii\right] } &=&\frac{k_{\eta }^{\left[ ii%
\right] }}{1+\underline{k}^{\left[ i\right] }+\underline{k}_{1}^{B\left[ i%
\right] }+\kappa \left[ \frac{\underline{k}_{2}^{B}}{1+\bar{k}}\right] ^{%
\left[ i\right] }} \\
\underline{S}_{\eta }^{\left[ ij\right] } &=&\frac{k_{\eta }^{\left[ ij%
\right] }}{1+\underline{k}^{\left[ i\right] }+\underline{k}_{1}^{B\left[ i%
\right] }+\kappa \left[ \frac{\underline{k}_{2}^{B}}{1+\bar{k}}\right] ^{%
\left[ i\right] }}
\end{eqnarray*}%
\begin{eqnarray*}
\underline{S}^{\left[ ii\right] } &=&\underline{S}_{1}^{\left[ ii\right]
}+S_{2}^{\left[ ii\right] } \\
\underline{S}^{\left[ ij\right] } &=&\underline{S}_{1}^{\left[ ij\right]
}+S_{2}^{\left[ ij\right] }
\end{eqnarray*}

\subsubsection*{A24.2.2 Constraints}

The follwing constraints apply on outgoing flux of capital:

\begin{eqnarray*}
1 &=&\frac{\underline{\hat{k}}^{\left[ ji\right] }}{1+\underline{\hat{k}}^{%
\left[ j\right] }+\underline{\hat{k}}_{1}^{B\left[ j\right] }+\kappa \left[ 
\frac{\underline{\hat{k}}_{2}^{B}}{1+\bar{k}}\right] ^{\left[ j\right] }}+%
\frac{\underline{\hat{k}}^{\left[ ii\right] }}{1+\underline{\hat{k}}^{\left[
i\right] }+\underline{\hat{k}}_{1}^{B\left[ i\right] }+\kappa \left[ \frac{%
\underline{\hat{k}}_{2}^{B}}{1+\bar{k}}\right] ^{\left[ i\right] }} \\
&&+\frac{\underline{k}^{\left[ ji\right] }}{1+\underline{k}^{\left[ j\right]
}+\underline{k}_{1}^{B\left[ j\right] }+\kappa \left[ \frac{\underline{k}%
_{2}^{B}}{1+\bar{k}}\right] ^{\left[ j\right] }}+\frac{\underline{k}^{\left[
ii\right] }}{1+\underline{k}^{\left[ i\right] }+\underline{k}_{1}^{B\left[ i%
\right] }+\kappa \left[ \frac{\underline{k}_{2}^{B}}{1+\bar{k}}\right] ^{%
\left[ i\right] }}
\end{eqnarray*}%
and:

\begin{eqnarray*}
1 &=&\frac{\underline{\hat{k}}_{2}^{\left[ ii\right] }}{1+\underline{\hat{k}}%
^{\left[ i\right] }+\underline{\hat{k}}_{1}^{B\left[ i\right] }+\kappa \left[
\frac{\underline{\hat{k}}_{2}^{B}}{1+\bar{k}}\right] ^{\left[ i\right] }}+%
\frac{\underline{\hat{k}}_{2}^{\left[ ji\right] }}{1+\underline{\hat{k}}^{%
\left[ j\right] }+\underline{\hat{k}}_{1}^{B\left[ j\right] }+\kappa \left[ 
\frac{\underline{\hat{k}}_{2}^{B}}{1+\bar{k}}\right] ^{\left[ j\right] }} \\
&&+\frac{k_{2}^{\left[ ii\right] }}{1+\underline{k}^{\left[ i\right] }+%
\underline{k}_{1}^{B\left[ i\right] }+\kappa \left[ \frac{\underline{k}%
_{2}^{B}}{1+\bar{k}}\right] ^{\left[ i\right] }}+\frac{k_{2}^{\left[ ji%
\right] }}{1+\underline{k}^{\left[ j\right] }+\underline{k}_{1}^{B\left[ j%
\right] }+\kappa \left[ \frac{\underline{k}_{2}^{B}}{1+\bar{k}}\right] ^{%
\left[ j\right] }} \\
&&+\frac{\underline{\hat{k}}_{1}^{\left[ ii\right] }}{1+\underline{\hat{k}}^{%
\left[ i\right] }+\underline{\hat{k}}_{1}^{B\left[ i\right] }+\kappa \left[ 
\frac{\underline{\hat{k}}_{2}^{B}}{1+\bar{k}}\right] ^{\left[ i\right] }}+%
\frac{\underline{\hat{k}}_{1}^{\left[ ji\right] }}{1+\underline{\hat{k}}^{%
\left[ j\right] }+\underline{\hat{k}}_{1}^{B\left[ j\right] }+\kappa \left[ 
\frac{\underline{\hat{k}}_{2}^{B}}{1+\bar{k}}\right] ^{\left[ j\right] }} \\
&&+\frac{k_{1}^{\left[ ii\right] }}{1+\underline{k}^{\left[ i\right] }+%
\underline{k}_{1}^{B\left[ i\right] }+\kappa \left[ \frac{\underline{k}%
_{2}^{B}}{1+\bar{k}}\right] ^{\left[ i\right] }}+\frac{k_{1}^{\left[ ji%
\right] }}{1+\underline{k}^{\left[ j\right] }+\underline{k}_{1}^{B\left[ j%
\right] }+\kappa \left[ \frac{\underline{k}_{2}^{B}}{1+\bar{k}}\right] ^{%
\left[ j\right] }}
\end{eqnarray*}%
\bigskip

that are translated in terms of shars:%
\begin{equation*}
\underline{\hat{S}}^{\left[ ii\right] }+\underline{\hat{S}}^{\left[ ji\right]
}+\underline{S}^{\left[ ii\right] }+\underline{S}^{\left[ ji\right] }=1
\end{equation*}

\subsubsection*{A24.2.3 Full alternate formulation}

The previous definitions allow to write all quantities as functions of the
new parameters. Actually we have for the coefficients related to investment
in firms:%
\begin{equation*}
\underline{\hat{S}}_{\eta }^{\left[ i\right] }=\underline{\hat{S}}_{\eta }^{%
\left[ ii\right] }+\underline{\hat{S}}_{\eta }^{\left[ ij\right] }=\frac{%
\underline{\hat{k}}_{\eta }^{\left[ i\right] }}{1+\underline{\hat{k}}^{\left[
i\right] }+\underline{\hat{k}}_{1}^{B\left[ i\right] }+\kappa \left[ \frac{%
\underline{\hat{k}}_{2}^{B}}{1+\bar{k}}\right] ^{\left[ i\right] }}
\end{equation*}%
\begin{equation*}
\underline{\hat{S}}^{\left[ i\right] }=\underline{\hat{S}}_{1}^{\left[ i%
\right] }+\underline{\hat{S}}_{2}^{\left[ i\right] }=\frac{\underline{\hat{k}%
}^{\left[ i\right] }}{1+\underline{\hat{k}}^{\left[ i\right] }+\underline{%
\hat{k}}_{1}^{B\left[ i\right] }+\kappa \left[ \frac{\underline{\hat{k}}%
_{2}^{B}}{1+\bar{k}}\right] ^{\left[ i\right] }}
\end{equation*}%
\begin{equation*}
\underline{S}^{\left[ i\right] }=\underline{S}_{1}^{\left[ i\right] }+%
\underline{S}_{2}^{\left[ i\right] }=\frac{k^{\left[ i\right] }}{1+%
\underline{k}^{\left[ i\right] }+\underline{k}_{1}^{B\left[ i\right]
}+\kappa \left[ \frac{\underline{k}_{2}^{B}}{1+\bar{k}}\right] ^{\left[ i%
\right] }}
\end{equation*}

\begin{eqnarray*}
\underline{\hat{S}}^{\left[ ii\right] } &=&\underline{\hat{S}}_{1}^{\left[ ii%
\right] }+\underline{\hat{S}}_{2}^{\left[ ii\right] } \\
\underline{\hat{S}}^{\left[ ij\right] } &=&\underline{\hat{S}}_{1}^{\left[ ij%
\right] }+\underline{\hat{S}}_{2}^{\left[ ij\right] }
\end{eqnarray*}%
\begin{equation*}
\underline{\hat{S}}^{\left[ i\right] }=\underline{\hat{S}}_{1}^{\left[ i%
\right] }+\underline{\hat{S}}_{2}^{\left[ i\right] }=\underline{\hat{S}}^{%
\left[ ii\right] }+\underline{\hat{S}}^{\left[ ij\right] }
\end{equation*}

\begin{equation*}
\underline{\hat{S}}^{\left[ ii\right] }+\underline{\hat{S}}^{\left[ ji\right]
}+\underline{S}^{\left[ ii\right] }+\underline{S}^{\left[ ji\right] }=1
\end{equation*}

While for the coefficients related to bank investment in investrs:%
\begin{eqnarray*}
\underline{\hat{S}}_{1}^{B\left[ ii\right] } &=&\frac{\underline{\hat{k}}%
_{1}^{B\left[ ii\right] }}{\left( 1+\underline{\hat{k}}^{\left[ i\right] }+%
\underline{\hat{k}}_{1}^{B\left[ i\right] }+\kappa \left[ \frac{\underline{%
\hat{k}}_{2}^{B}}{1+\bar{k}}\right] ^{\left[ i\right] }\right) } \\
\underline{\hat{S}}_{1}^{B\left[ ij\right] } &=&\frac{\underline{\hat{k}}%
_{1}^{B\left[ ij\right] }}{\left( 1+\underline{\hat{k}}^{\left[ i\right] }+%
\underline{\hat{k}}_{1}^{B\left[ i\right] }+\kappa \left[ \frac{\underline{%
\hat{k}}_{2}^{B}}{1+\bar{k}}\right] ^{\left[ i\right] }\right) }
\end{eqnarray*}%
\begin{eqnarray*}
\underline{\hat{S}}_{2}^{B\left[ ii\right] } &=&\frac{\kappa \frac{%
\underline{\hat{k}}_{2}^{B\left[ ii\right] }}{1+\bar{k}^{\left[ i\right] }}}{%
1+\underline{\hat{k}}^{\left[ i\right] }+\underline{\hat{k}}_{1}^{B\left[ i%
\right] }+\kappa \left[ \frac{\underline{\hat{k}}_{2}^{B}}{1+\bar{k}}\right]
^{\left[ i\right] }} \\
\underline{\hat{S}}_{2}^{B\left[ ij\right] } &=&\frac{\kappa \frac{%
\underline{\hat{k}}_{2}^{B\left[ ij\right] }}{1+\bar{k}^{\left[ j\right] }}}{%
1+\underline{\hat{k}}^{\left[ i\right] }+\underline{\hat{k}}_{1}^{B\left[ i%
\right] }+\kappa \left[ \frac{\underline{\hat{k}}_{2}^{B}}{1+\bar{k}}\right]
^{\left[ i\right] }}
\end{eqnarray*}%
\begin{eqnarray*}
\underline{\hat{S}}_{\eta }^{B\left[ i\right] } &=&\underline{\hat{S}}_{\eta
}^{B\left[ ii\right] }+\underline{\hat{S}}_{\eta }^{B\left[ ij\right] } \\
\underline{\hat{S}}^{B\left[ i\right] } &=&\underline{\hat{S}}_{1}^{B\left[ i%
\right] }+\underline{\hat{S}}_{2}^{B\left[ i\right] }=\underline{\hat{S}}^{B%
\left[ ii\right] }+\underline{\hat{S}}^{B\left[ ij\right] }
\end{eqnarray*}%
\begin{equation*}
\underline{\hat{S}}^{\left[ ii\right] }+\underline{\hat{S}}^{\left[ ij\right]
}+\underline{\hat{S}}^{B\left[ ii\right] }+\underline{\hat{S}}^{B\left[ ij%
\right] }=\frac{\underline{\hat{k}}^{\left[ i\right] }+\underline{\hat{k}}%
_{1}^{B\left[ i\right] }+\kappa \left[ \frac{\underline{\hat{k}}_{2}^{B}}{1+%
\bar{k}}\right] ^{\left[ i\right] }}{1+\underline{\hat{k}}^{\left[ i\right]
}+\underline{\hat{k}}_{1}^{B\left[ i\right] }+\kappa \left[ \frac{\underline{%
\hat{k}}_{2}^{B}}{1+\bar{k}}\right] ^{\left[ i\right] }}
\end{equation*}%
As a consequence, we can write the following coefficients:%
\begin{equation*}
\frac{1}{1+\underline{\hat{k}}^{\left[ i\right] }+\underline{\hat{k}}_{1}^{B%
\left[ i\right] }+\kappa \left[ \frac{\underline{\hat{k}}_{2}^{B}}{1+\bar{k}}%
\right] ^{\left[ i\right] }}=1-\left( \underline{\hat{S}}^{\left[ ii\right]
}+\underline{\hat{S}}^{\left[ ij\right] }+\underline{\hat{S}}^{B\left[ ii%
\right] }+\underline{\hat{S}}^{B\left[ ij\right] }\right)
\end{equation*}%
\begin{equation*}
\underline{\hat{k}}_{\eta }^{\left[ i\right] }\left( 1-\left( \underline{%
\hat{S}}^{\left[ ii\right] }+\underline{\hat{S}}^{\left[ ij\right] }+%
\underline{\hat{S}}^{B\left[ ii\right] }+\underline{\hat{S}}^{B\left[ ij%
\right] }\right) \right) =\underline{\hat{S}}_{\eta }^{\left[ ii\right] }+%
\underline{\hat{S}}_{\eta }^{\left[ ij\right] }
\end{equation*}%
and:%
\begin{equation*}
\underline{\hat{k}}_{\eta }^{\left[ i\right] }=\frac{\underline{\hat{S}}%
_{\eta }^{\left[ ii\right] }+\underline{\hat{S}}_{\eta }^{\left[ ij\right] }%
}{1-\left( \underline{\hat{S}}^{\left[ ii\right] }+\underline{\hat{S}}^{%
\left[ ij\right] }+\underline{\hat{S}}^{B\left[ ii\right] }+\underline{\hat{S%
}}^{B\left[ ij\right] }\right) }
\end{equation*}%
\begin{equation*}
\underline{\hat{k}}_{1}^{B\left[ i\right] }=\frac{\underline{\hat{S}}_{1}^{B%
\left[ ii\right] }+\underline{\hat{S}}_{1}^{B\left[ ij\right] }}{1-\left( 
\underline{\hat{S}}^{\left[ ii\right] }+\underline{\hat{S}}^{\left[ ij\right]
}+\underline{\hat{S}}^{B\left[ ii\right] }+\underline{\hat{S}}^{B\left[ ij%
\right] }\right) }
\end{equation*}%
\begin{equation*}
\kappa \left[ \frac{\underline{\hat{k}}_{2}^{B}}{1+\bar{k}}\right] ^{\left[ i%
\right] }=\frac{\underline{\hat{S}}_{2}^{B\left[ ii\right] }+\underline{\hat{%
S}}_{2}^{B\left[ ij\right] }}{1-\left( \underline{\hat{S}}^{\left[ ii\right]
}+\underline{\hat{S}}^{\left[ ij\right] }+\underline{\hat{S}}^{B\left[ ii%
\right] }+\underline{\hat{S}}^{B\left[ ij\right] }\right) }
\end{equation*}%
\begin{eqnarray*}
\underline{\hat{k}}_{2}^{\left[ i\right] }+\kappa \left[ \frac{\underline{%
\hat{k}}_{2}^{B}}{1+\bar{k}}\right] ^{\left[ i\right] } &=&\frac{\underline{%
\hat{S}}_{2}^{\left[ ii\right] }+\underline{\hat{S}}_{2}^{\left[ ij\right] }%
}{1-\left( \underline{\hat{S}}^{\left[ ii\right] }+\underline{\hat{S}}^{%
\left[ ij\right] }+\underline{\hat{S}}^{B\left[ ii\right] }+\underline{\hat{S%
}}^{B\left[ ij\right] }\right) }+\frac{\underline{\hat{S}}_{2}^{B\left[ ii%
\right] }+\underline{\hat{S}}_{2}^{B\left[ ij\right] }}{1-\left( \underline{%
\hat{S}}^{\left[ ii\right] }+\underline{\hat{S}}^{\left[ ij\right] }+%
\underline{\hat{S}}^{B\left[ ii\right] }+\underline{\hat{S}}^{B\left[ ij%
\right] }\right) } \\
&=&\frac{\underline{\hat{S}}_{2}^{\left[ ii\right] }+\underline{\hat{S}}%
_{2}^{\left[ ij\right] }+\underline{\hat{S}}_{2}^{B\left[ ii\right] }+%
\underline{\hat{S}}_{2}^{B\left[ ij\right] }}{1-\left( \underline{\hat{S}}^{%
\left[ ii\right] }+\underline{\hat{S}}^{\left[ ij\right] }+\underline{\hat{S}%
}^{B\left[ ii\right] }+\underline{\hat{S}}^{B\left[ ij\right] }\right) }
\end{eqnarray*}%
\begin{eqnarray*}
1+\underline{\hat{k}}_{2}^{\left[ i\right] }+\kappa \left[ \frac{\underline{%
\hat{k}}_{2}^{B}}{1+\bar{k}}\right] ^{\left[ i\right] } &=&1+\frac{%
\underline{\hat{S}}_{2}^{\left[ ii\right] }+\underline{\hat{S}}_{2}^{\left[
ij\right] }}{1-\left( \underline{\hat{S}}^{\left[ ii\right] }+\underline{%
\hat{S}}^{\left[ ij\right] }+\underline{\hat{S}}^{B\left[ ii\right] }+%
\underline{\hat{S}}^{B\left[ ij\right] }\right) }+\frac{\underline{\hat{S}}%
_{2}^{B\left[ ii\right] }+\underline{\hat{S}}_{2}^{B\left[ ij\right] }}{%
1-\left( \underline{\hat{S}}^{\left[ ii\right] }+\underline{\hat{S}}^{\left[
ij\right] }+\underline{\hat{S}}^{B\left[ ii\right] }+\underline{\hat{S}}^{B%
\left[ ij\right] }\right) } \\
&=&\frac{1-\left( \underline{\hat{S}}_{1}^{\left[ ii\right] }+\underline{%
\hat{S}}_{1}^{\left[ ij\right] }+\underline{\hat{S}}_{1}^{B\left[ ii\right]
}+\underline{\hat{S}}_{1}^{B\left[ ij\right] }\right) }{1-\left( \underline{%
\hat{S}}^{\left[ ii\right] }+\underline{\hat{S}}^{\left[ ij\right] }+%
\underline{\hat{S}}^{B\left[ ii\right] }+\underline{\hat{S}}^{B\left[ ij%
\right] }\right) }
\end{eqnarray*}

\bigskip 
\begin{eqnarray*}
\underline{S}_{2}^{B\left[ ii\right] } &=&\frac{\kappa \frac{\underline{k}%
_{2}^{\left( B\right) \left[ ii\right] }}{1+\underline{\bar{k}}^{\left[ i%
\right] }}}{\left( 1+\underline{k}^{\left[ i\right] }+\underline{k}%
_{1}^{\left( B\right) \left[ i\right] }+\kappa \left[ \frac{\underline{k}%
_{2}^{\left( B\right) }}{1+\underline{\bar{k}}}\right] ^{\left[ i\right]
}\right) } \\
\underline{S}_{2}^{B\left[ ij\right] } &=&\frac{\kappa \frac{\underline{k}%
_{2}^{\left( B\right) \left[ ij\right] }}{1+\underline{\bar{k}}^{\left[ j%
\right] }}}{\left( 1+\underline{k}^{\left[ i\right] }+\underline{k}%
_{1}^{\left( B\right) \left[ i\right] }+\kappa \left[ \frac{\underline{k}%
_{2}^{\left( B\right) }}{1+\underline{\bar{k}}}\right] ^{\left[ i\right]
}\right) }
\end{eqnarray*}

\begin{equation*}
\underline{k}_{\eta }^{\left[ i\right] }=\frac{\underline{S}_{\eta }^{\left[
ii\right] }+\underline{S}_{\eta }^{\left[ ij\right] }}{1-\left( \underline{S}%
^{\left[ ii\right] }+\underline{S}^{\left[ ij\right] }+\underline{S}^{B\left[
ii\right] }+\underline{S}^{B\left[ ij\right] }\right) }
\end{equation*}%
\begin{equation*}
\underline{k}_{1}^{\left( B\right) \left[ ij\right] }=\frac{\underline{S}%
_{1}^{B\left[ ii\right] }+\underline{S}_{1}^{B\left[ ij\right] }}{1-\left( 
\underline{S}^{\left[ ii\right] }+\underline{S}^{\left[ ij\right] }+%
\underline{S}^{B\left[ ii\right] }+\underline{S}^{B\left[ ij\right] }\right) 
}
\end{equation*}%
\begin{equation*}
\kappa \left[ \frac{\underline{k}_{2}^{\left( B\right) }}{1+\underline{\bar{k%
}}}\right] ^{\left[ i\right] }=\frac{\underline{S}_{2}^{B\left[ ii\right] }+%
\underline{S}_{2}^{B\left[ ij\right] }}{1-\left( \underline{S}^{\left[ ii%
\right] }+\underline{S}^{\left[ ij\right] }+\underline{S}^{B\left[ ii\right]
}+\underline{S}^{B\left[ ij\right] }\right) }
\end{equation*}%
\begin{equation*}
\underline{k}_{2}^{\left[ i\right] }+\left[ \frac{\underline{k}_{2}^{B}}{1+%
\bar{k}}\right] ^{\left[ i\right] }=\frac{\underline{S}_{2}^{\left[ ii\right]
}+\underline{S}_{2}^{\left[ ij\right] }+\underline{S}_{2}^{B\left[ ii\right]
}+\underline{S}_{2}^{B\left[ ij\right] }}{1-\left( \underline{S}^{\left[ ii%
\right] }+\underline{S}^{\left[ ij\right] }+\underline{S}^{B\left[ ii\right]
}+\underline{S}^{B\left[ ij\right] }\right) }
\end{equation*}%
\begin{equation*}
1+\underline{k}_{2}^{\left[ i\right] }+\left[ \frac{\underline{k}_{2}^{B}}{1+%
\bar{k}}\right] ^{\left[ i\right] }=\frac{1-\left( \underline{S}_{1}^{\left[
ii\right] }+\underline{S}_{1}^{\left[ ij\right] }+\underline{S}_{1}^{B\left[
ii\right] }+\underline{S}_{1}^{B\left[ ij\right] }\right) }{1-\left( 
\underline{S}^{\left[ ii\right] }+\underline{S}^{\left[ ij\right] }+%
\underline{S}^{B\left[ ii\right] }+\underline{S}^{B\left[ ij\right] }\right) 
}
\end{equation*}%
\begin{equation*}
f_{1}^{\prime \left[ i\right] }=\left( 1+\underline{k}_{2}^{\left[ i\right]
}+\left[ \frac{\underline{k}_{2}^{B}}{1+\bar{k}}\right] ^{\left[ i\right]
}\right) f_{1}^{\left[ i\right] }-\bar{r}\left( \underline{k}_{2}^{\left[ i%
\right] }+\left[ \frac{\underline{k}_{2}^{B}}{1+\bar{k}}\right] ^{\left[ i%
\right] }\right)
\end{equation*}

\begin{eqnarray*}
\underline{\hat{S}}_{1}^{B\left[ ii\right] } &=&\frac{\underline{\hat{k}}%
_{1}^{B\left[ ii\right] }}{\left( 1+\underline{\hat{k}}^{\left[ i\right] }+%
\underline{\hat{k}}_{1}^{B\left[ i\right] }+\kappa \left[ \frac{\underline{%
\hat{k}}_{2}^{B}}{1+\bar{k}}\right] ^{\left[ i\right] }\right) } \\
\underline{\hat{S}}_{1}^{B\left[ ij\right] } &=&\frac{\underline{\hat{k}}%
_{1}^{B\left[ ij\right] }}{\left( 1+\underline{\hat{k}}^{\left[ i\right] }+%
\underline{\hat{k}}_{1}^{B\left[ i\right] }+\kappa \left[ \frac{\underline{%
\hat{k}}_{2}^{B}}{1+\bar{k}}\right] ^{\left[ i\right] }\right) }
\end{eqnarray*}%
\begin{eqnarray*}
\underline{\hat{S}}_{2}^{B\left[ ii\right] } &=&\frac{\kappa \frac{%
\underline{\hat{k}}_{2}^{B\left[ ii\right] }}{1+\bar{k}^{\left[ i\right] }}}{%
1+\underline{\hat{k}}^{\left[ i\right] }+\underline{\hat{k}}_{1}^{B\left[ i%
\right] }+\kappa \left[ \frac{\underline{\hat{k}}_{2}^{B}}{1+\bar{k}}\right]
^{\left[ i\right] }} \\
\underline{\hat{S}}_{2}^{B\left[ ij\right] } &=&\frac{\kappa \frac{%
\underline{\hat{k}}_{2}^{B\left[ ij\right] }}{1+\bar{k}^{\left[ j\right] }}}{%
1+\underline{\hat{k}}^{\left[ i\right] }+\underline{\hat{k}}_{1}^{B\left[ i%
\right] }+\kappa \left[ \frac{\underline{\hat{k}}_{2}^{B}}{1+\bar{k}}\right]
^{\left[ i\right] }}
\end{eqnarray*}%
\begin{eqnarray*}
\underline{\hat{S}}_{\eta }^{B\left[ i\right] } &=&\underline{\hat{S}}_{\eta
}^{B\left[ ii\right] }+\underline{\hat{S}}_{\eta }^{B\left[ ij\right] } \\
\underline{\hat{S}}^{B\left[ i\right] } &=&\underline{\hat{S}}_{1}^{B\left[ i%
\right] }+\underline{\hat{S}}_{2}^{B\left[ i\right] }=\underline{\hat{S}}^{B%
\left[ ii\right] }+\underline{\hat{S}}^{B\left[ ij\right] }
\end{eqnarray*}%
\begin{equation*}
\frac{1}{1+\underline{\hat{k}}^{\left[ i\right] }+\underline{\hat{k}}_{1}^{B%
\left[ i\right] }+\kappa \left[ \frac{\underline{\hat{k}}_{2}^{B}}{1+\bar{k}}%
\right] ^{\left[ i\right] }}=1-\left( \underline{\hat{S}}^{B\left[ ii\right]
}+\underline{\hat{S}}^{B\left[ ij\right] }\right) =1-\underline{\hat{S}}^{B%
\left[ i\right] }
\end{equation*}%
\begin{equation*}
\underline{\hat{S}}_{\eta }^{B\left[ i\right] }=\frac{\underline{\hat{k}}%
_{\eta }^{B\left[ i\right] }}{1+\underline{\hat{k}}^{\left[ i\right] }+%
\underline{\hat{k}}_{1}^{B\left[ i\right] }+\kappa \left[ \frac{\underline{%
\hat{k}}_{2}^{B}}{1+\bar{k}}\right] ^{\left[ i\right] }}=\underline{\hat{k}}%
_{\eta }^{B\left[ i\right] }\left( 1-\left( \underline{\bar{S}}^{\left[ ii%
\right] }+\underline{\bar{S}}^{\left[ ij\right] }\right) \right)
\end{equation*}%
\begin{equation*}
\underline{\bar{k}}_{\eta }^{\left[ i\right] }=\frac{\underline{\bar{S}}%
_{\eta }^{\left[ i\right] }}{1-\underline{\bar{S}}^{\left[ i\right] }}=\frac{%
\underline{\bar{S}}_{\eta }^{\left[ ii\right] }+\underline{\bar{S}}_{\eta }^{%
\left[ ij\right] }}{1-\left( \underline{\bar{S}}^{\left[ ii\right] }+%
\underline{\bar{S}}^{\left[ ij\right] }\right) }
\end{equation*}%
\begin{eqnarray*}
\underline{S}_{2}^{B\left[ ii\right] } &=&\frac{\kappa \frac{\underline{k}%
_{2}^{\left( B\right) \left[ ii\right] }}{1+\underline{\bar{k}}^{\left[ i%
\right] }}}{\left( 1+\underline{k}^{\left[ i\right] }+\underline{k}%
_{1}^{\left( B\right) \left[ i\right] }+\kappa \left[ \frac{\underline{k}%
_{2}^{\left( B\right) }}{1+\underline{\bar{k}}}\right] ^{\left[ i\right]
}\right) } \\
\underline{S}_{2}^{B\left[ ij\right] } &=&\frac{\kappa \frac{\underline{k}%
_{2}^{\left( B\right) \left[ ij\right] }}{1+\underline{\bar{k}}^{\left[ j%
\right] }}}{\left( 1+\underline{k}^{\left[ i\right] }+\underline{k}%
_{1}^{\left( B\right) \left[ i\right] }+\kappa \left[ \frac{\underline{k}%
_{2}^{\left( B\right) }}{1+\underline{\bar{k}}}\right] ^{\left[ i\right]
}\right) }
\end{eqnarray*}

\begin{equation*}
\underline{\bar{S}}^{\left[ ii\right] }+\underline{\bar{S}}^{\left[ ji\right]
}+\underline{\hat{S}}^{B\left[ ii\right] }+\underline{\hat{S}}^{B\left[ ji%
\right] }+\underline{S}^{B\left[ ii\right] }+\underline{S}^{\left[ ji\right]
}=1
\end{equation*}%
\bigskip

\begin{eqnarray*}
\underline{\hat{k}}_{2}^{\left[ i\right] }+\kappa \left[ \frac{\underline{%
\hat{k}}_{2}^{B}}{1+\bar{k}}\right] ^{\left[ i\right] } &=&\frac{\underline{%
\hat{S}}_{2}^{\left[ ii\right] }+\underline{\hat{S}}_{2}^{\left[ ij\right] }%
}{1-\left( \underline{\hat{S}}^{\left[ ii\right] }+\underline{\hat{S}}^{%
\left[ ij\right] }+\underline{\hat{S}}^{B\left[ ii\right] }+\underline{\hat{S%
}}^{B\left[ ij\right] }\right) }+\frac{\underline{\hat{S}}_{2}^{B\left[ ii%
\right] }+\underline{\hat{S}}_{2}^{B\left[ ij\right] }}{1-\left( \underline{%
\hat{S}}^{\left[ ii\right] }+\underline{\hat{S}}^{\left[ ij\right] }+%
\underline{\hat{S}}^{B\left[ ii\right] }+\underline{\hat{S}}^{B\left[ ij%
\right] }\right) } \\
&=&\frac{\underline{\hat{S}}_{2}^{\left[ ii\right] }+\underline{\hat{S}}%
_{2}^{\left[ ij\right] }+\underline{\hat{S}}_{2}^{B\left[ ii\right] }+%
\underline{\hat{S}}_{2}^{B\left[ ij\right] }}{1-\left( \underline{\hat{S}}^{%
\left[ ii\right] }+\underline{\hat{S}}^{\left[ ij\right] }+\underline{\hat{S}%
}^{B\left[ ii\right] }+\underline{\hat{S}}^{B\left[ ij\right] }\right) }
\end{eqnarray*}%
\begin{eqnarray*}
1+\underline{\hat{k}}_{2}^{\left[ i\right] }+\kappa \left[ \frac{\underline{%
\hat{k}}_{2}^{B}}{1+\bar{k}}\right] ^{\left[ i\right] } &=&1+\frac{%
\underline{\hat{S}}_{2}^{\left[ ii\right] }+\underline{\hat{S}}_{2}^{\left[
ij\right] }}{1-\left( \underline{\hat{S}}^{\left[ ii\right] }+\underline{%
\hat{S}}^{\left[ ij\right] }+\underline{\hat{S}}^{B\left[ ii\right] }+%
\underline{\hat{S}}^{B\left[ ij\right] }\right) }+\frac{\underline{\hat{S}}%
_{2}^{B\left[ ii\right] }+\underline{\hat{S}}_{2}^{B\left[ ij\right] }}{%
1-\left( \underline{\hat{S}}^{\left[ ii\right] }+\underline{\hat{S}}^{\left[
ij\right] }+\underline{\hat{S}}^{B\left[ ii\right] }+\underline{\hat{S}}^{B%
\left[ ij\right] }\right) } \\
&=&\frac{1-\left( \underline{\hat{S}}_{1}^{\left[ ii\right] }+\underline{%
\hat{S}}_{1}^{\left[ ij\right] }+\underline{\hat{S}}_{1}^{B\left[ ii\right]
}+\underline{\hat{S}}_{1}^{B\left[ ij\right] }\right) }{1-\left( \underline{%
\hat{S}}^{\left[ ii\right] }+\underline{\hat{S}}^{\left[ ij\right] }+%
\underline{\hat{S}}^{B\left[ ii\right] }+\underline{\hat{S}}^{B\left[ ij%
\right] }\right) }
\end{eqnarray*}%
\begin{equation*}
\underline{k}_{2}^{\left[ i\right] }+\left[ \frac{\underline{k}_{2}^{B}}{1+%
\bar{k}}\right] ^{\left[ i\right] }=\frac{\underline{S}_{2}^{\left[ ii\right]
}+\underline{S}_{2}^{\left[ ij\right] }+\underline{S}_{2}^{B\left[ ii\right]
}+\underline{S}_{2}^{B\left[ ij\right] }}{1-\left( \underline{S}^{\left[ ii%
\right] }+\underline{S}^{\left[ ij\right] }+\underline{S}^{B\left[ ii\right]
}+\underline{S}^{B\left[ ij\right] }\right) }
\end{equation*}%
\begin{equation*}
\frac{\underline{\hat{k}}^{\left[ ji\right] }}{1+\underline{\hat{k}}^{\left[
j\right] }+\underline{\hat{k}}_{1}^{B\left[ j\right] }+\kappa \left[ \frac{%
\underline{\hat{k}}_{2}^{B}}{1+\bar{k}}\right] ^{\left[ j\right] }}+\frac{%
\underline{\hat{k}}^{\left[ ii\right] }}{1+\underline{\hat{k}}^{\left[ i%
\right] }+\underline{\hat{k}}_{1}^{B\left[ i\right] }+\kappa \left[ \frac{%
\underline{\hat{k}}_{2}^{B}}{1+\bar{k}}\right] ^{\left[ i\right] }}+\frac{%
\underline{k}^{\left[ ji\right] }}{1+\underline{k}^{\left[ j\right] }+%
\underline{k}_{1}^{B\left[ j\right] }+\kappa \left[ \frac{\underline{k}%
_{2}^{B}}{1+\bar{k}}\right] ^{\left[ j\right] }}+\frac{\underline{k}^{\left[
ii\right] }}{1+\underline{k}^{\left[ i\right] }+\underline{k}_{1}^{B\left[ i%
\right] }+\kappa \left[ \frac{\underline{k}_{2}^{B}}{1+\bar{k}}\right] ^{%
\left[ i\right] }}=1
\end{equation*}%
\begin{equation*}
\frac{\underline{\hat{k}}_{1}^{\left[ ii\right] }}{\left( 1+\underline{\hat{k%
}}^{\left[ i\right] }+\underline{\hat{k}}_{1}^{B\left[ i\right] }+\kappa %
\left[ \frac{\underline{\hat{k}}_{2}^{B}}{1+\bar{k}}\right] ^{\left[ i\right]
}\right) \left( 1+\underline{\hat{k}}_{2}^{\left[ i\right] }+\kappa \left[ 
\frac{\underline{\hat{k}}_{2}^{B}}{1+\bar{k}}\right] ^{\left[ i\right]
}\right) }=\frac{\underline{\hat{S}}_{1}^{\left[ ii\right] }}{1+\underline{%
\hat{k}}_{2}^{\left[ i\right] }+\kappa \left[ \frac{\underline{\hat{k}}%
_{2}^{B}}{1+\bar{k}}\right] ^{\left[ i\right] }}
\end{equation*}%
Replacing these coeficients in equation (\ref{MF}):

\begin{eqnarray*}
0 &=&\left( 
\begin{array}{cc}
1-\underline{\hat{S}}_{1}^{\left[ ii\right] } & -\underline{\hat{S}}_{1}^{%
\left[ ji\right] } \\ 
-\underline{\hat{S}}_{1}^{\left[ ij\right] } & 1-\underline{\hat{S}}_{1}^{%
\left[ jj\right] }%
\end{array}%
\right) \left( 
\begin{array}{c}
\left( f^{\left[ i\right] }-\bar{r}\right) \underline{\hat{s}}_{1}^{\left[ i%
\right] } \\ 
\left( f^{\left[ j\right] }-\bar{r}\right) \underline{\hat{s}}_{1}^{\left[ j%
\right] }%
\end{array}%
\right) -\left( 
\begin{array}{cc}
\underline{\hat{S}}_{2}^{\left[ ii\right] } & \underline{\hat{S}}_{2}^{\left[
ji\right] } \\ 
\underline{\hat{S}}_{2}^{\left[ ij\right] } & \underline{\hat{S}}_{2}^{\left[
jj\right] }%
\end{array}%
\right) \left( 
\begin{array}{c}
\left( 1+f^{\left[ i\right] }\right) \underline{\hat{s}}_{2}^{\left[ i\right]
}H\left( -\left( 1+f^{\left[ i\right] }\right) \right) \\ 
\left( 1+f^{\left[ j\right] }\right) \underline{\hat{s}}_{2}^{\left[ i\right]
}H\left( -\left( 1+f^{\left[ j\right] }\right) \right)%
\end{array}%
\right) \\
&&-\left( 
\begin{array}{cc}
\underline{S}_{2}^{\left[ ii\right] } & \underline{S}_{2}^{\left[ ji\right] }
\\ 
\underline{S}_{2}^{\left[ ij\right] } & \underline{S}_{2}^{\left[ jj\right] }%
\end{array}%
\right) \left( 
\begin{array}{c}
\left( 1+f_{1}^{\prime \left[ i\right] }\right) \underline{s}_{2}^{\left[ i%
\right] }H\left( -\left( 1+f_{1}^{\prime \left[ i\right] }\right) \right) \\ 
\left( 1+f_{1}^{\prime \left[ j\right] }\right) \underline{s}_{2}^{\left[ j%
\right] }H\left( -\left( 1+f_{1}^{\prime \left[ j\right] }\right) \right)%
\end{array}%
\right) -\left( 
\begin{array}{cc}
\underline{S}_{1}^{\left[ ii\right] } & \underline{S}_{1}^{\left[ ji\right] }
\\ 
\underline{S}_{1}^{\left[ ij\right] } & \underline{S}_{1}^{\left[ jj\right] }%
\end{array}%
\right) \left( 
\begin{array}{c}
\left( f_{1}^{\prime \left[ i\right] }-\bar{r}\right) \underline{s}_{1}^{%
\left[ i\right] } \\ 
\left( f_{1}^{\prime \left[ j\right] }-\bar{r}\right) \underline{s}_{1}^{%
\left[ j\right] }%
\end{array}%
\right)
\end{eqnarray*}

with:%
\begin{eqnarray*}
\underline{\hat{s}}_{1}^{\left[ i\right] } &=&\frac{1-\left( \underline{\hat{%
S}}^{\left[ ii\right] }+\underline{\hat{S}}^{\left[ ij\right] }+\underline{%
\hat{S}}^{B\left[ ii\right] }+\underline{\hat{S}}^{B\left[ ij\right]
}\right) }{1-\left( \underline{\hat{S}}_{1}^{\left[ ii\right] }+\underline{%
\hat{S}}_{1}^{\left[ ij\right] }+\underline{\hat{S}}_{1}^{B\left[ ii\right]
}+\underline{\hat{S}}_{1}^{B\left[ ij\right] }\right) }\text{, }\underline{%
\hat{s}}_{2}^{\left[ i\right] }=\frac{1-\left( \underline{\hat{S}}^{\left[ ii%
\right] }+\underline{\hat{S}}^{\left[ ij\right] }+\underline{\hat{S}}^{B%
\left[ ii\right] }+\underline{\hat{S}}^{B\left[ ij\right] }\right) }{%
\underline{\hat{S}}_{2}^{\left[ ii\right] }+\underline{\hat{S}}_{2}^{\left[
ij\right] }+\underline{\hat{S}}_{2}^{B\left[ ii\right] }+\underline{\hat{S}}%
_{2}^{B\left[ ij\right] }} \\
\underline{s}_{1}^{\left[ i\right] } &=&\frac{1-\left( \underline{S}^{\left[
ii\right] }+\underline{S}^{\left[ ij\right] }+\underline{S}^{B\left[ ii%
\right] }+\underline{S}^{B\left[ ij\right] }\right) }{1-\left( \underline{S}%
_{1}^{\left[ ii\right] }+\underline{S}_{1}^{\left[ ij\right] }+\underline{S}%
_{1}^{B\left[ ii\right] }+\underline{S}_{1}^{B\left[ ij\right] }\right) }%
\text{, }\underline{s}_{2}^{\left[ i\right] }=\frac{1-\left( \underline{S}^{%
\left[ ii\right] }+\underline{S}^{\left[ ij\right] }+\underline{S}^{B\left[
ii\right] }+\underline{S}^{B\left[ ij\right] }\right) }{\underline{S}_{2}^{%
\left[ ii\right] }+\underline{S}_{2}^{\left[ ij\right] }+\underline{S}_{2}^{B%
\left[ ii\right] }+\underline{S}_{2}^{B\left[ ij\right] }}
\end{eqnarray*}

\subsection*{A24.3 Bank return}

\subsubsection*{A24.3.1 Constraints}

The constraint for bank investments are:%
\begin{eqnarray*}
1 &=&\frac{\underline{k}_{1}^{\left( B\right) \left[ ii\right] }}{\left( 1+%
\underline{k}^{\left[ i\right] }+\underline{k}_{1}^{\left( B\right) \left[ i%
\right] }+\kappa \left[ \frac{\underline{k}_{2}^{\left( B\right) }}{1+%
\underline{\bar{k}}}\right] ^{\left[ i\right] }\right) }+\frac{\underline{k}%
_{1}^{\left( B\right) \left[ ji\right] }}{\left( 1+\underline{k}^{\left[ j%
\right] }+\underline{k}_{1}^{\left( B\right) \left[ j\right] }+\kappa \left[ 
\frac{\underline{k}_{2}^{\left( B\right) }}{1+\underline{\bar{k}}}\right] ^{%
\left[ j\right] }\right) } \\
&&+\frac{\underline{\hat{k}}_{1}^{B\left[ ii\right] }}{1+\underline{\hat{k}}%
^{\left[ i\right] }+\underline{\hat{k}}_{1}^{B\left[ i\right] }+\kappa \left[
\frac{\underline{\hat{k}}_{2}^{B}}{\left( 1+\underline{\bar{k}}\right) }%
\right] ^{\left[ i\right] }}+\frac{\underline{\hat{k}}_{1}^{B\left[ ji\right]
}}{1+\underline{\hat{k}}^{\left[ j\right] }+\underline{\hat{k}}_{1}^{B\left[
j\right] }+\kappa \left[ \frac{\underline{\hat{k}}_{2}^{B}}{\left( 1+%
\underline{\bar{k}}\right) }\right] ^{\left[ j\right] }}+\frac{\underline{%
\overline{\bar{k}}}^{\left[ ii\right] }}{1+\underline{\overline{\bar{k}}}^{%
\left[ i\right] }}+\frac{\underline{\overline{\bar{k}}}^{\left[ ji\right] }}{%
1+\underline{\overline{\bar{k}}}^{\left[ j\right] }}
\end{eqnarray*}%
which translates in terms of new set of paramters:%
\begin{equation*}
\underline{\bar{S}}^{\left[ ii\right] }+\underline{\bar{S}}^{\left[ ji\right]
}+\underline{\hat{S}}_{1}^{B\left[ ii\right] }+\underline{\hat{S}}_{1}^{B%
\left[ ji\right] }+\underline{S}_{1}^{B\left[ ii\right] }+\underline{S}%
_{1}^{B\left[ ji\right] }=1
\end{equation*}%
while the constraint for investors:%
\begin{eqnarray*}
1 &=&\frac{\kappa \frac{\underline{k}_{2}^{\left( B\right) \left[ ii\right] }%
}{1+\underline{\bar{k}}^{\left[ i\right] }}}{\left( 1+\underline{k}^{\left[ i%
\right] }+\underline{k}_{1}^{\left( B\right) \left[ i\right] }+\kappa \left[ 
\frac{\underline{k}_{2}^{\left( B\right) }}{1+\underline{\bar{k}}}\right] ^{%
\left[ i\right] }\right) }+\frac{\kappa \frac{\underline{k}_{2}^{\left(
B\right) \left[ ji\right] }}{1+\underline{\bar{k}}^{\left[ i\right] }}}{%
\left( 1+\underline{k}^{\left[ j\right] }+\underline{k}_{1}^{\left( B\right) %
\left[ j\right] }+\kappa \left[ \frac{\underline{k}_{2}^{\left( B\right) }}{%
1+\underline{\bar{k}}}\right] ^{\left[ j\right] }\right) } \\
&&+\frac{\kappa \frac{\underline{\hat{k}}_{2}^{B\left[ ii\right] }}{\left( 1+%
\underline{\bar{k}}^{\left[ i\right] }\right) }}{1+\underline{\hat{k}}^{%
\left[ i\right] }+\underline{\hat{k}}_{1}^{B\left[ i\right] }+\kappa \left[ 
\frac{\underline{\hat{k}}_{2}^{B}}{\left( 1+\underline{\bar{k}}\right) }%
\right] ^{\left[ i\right] }}+\frac{\kappa \frac{\underline{\hat{k}}_{2}^{B%
\left[ ji\right] }}{\left( 1+\underline{\bar{k}}^{\left[ j\right] }\right) }%
}{1+\underline{\hat{k}}^{\left[ j\right] }+\underline{\hat{k}}_{1}^{B\left[ j%
\right] }+\kappa \left[ \frac{\underline{\hat{k}}_{2}^{B}}{\left( 1+%
\underline{\bar{k}}\right) }\right] ^{\left[ j\right] }}
\end{eqnarray*}%
translates into:%
\begin{equation*}
\underline{\hat{S}}_{2}^{B\left[ ii\right] }+\underline{\hat{S}}_{2}^{B\left[
ji\right] }+\underline{S}_{2}^{B\left[ ii\right] }+\underline{S}_{2}^{B\left[
ji\right] }=1
\end{equation*}

\subsubsection*{A24.3.2 Bank equation}

The formula for average returns per group is:%
\begin{eqnarray*}
0 &=&\left( 
\begin{array}{cc}
1-\underline{\bar{S}}_{1}^{\left[ ii\right] } & -\underline{\bar{S}}_{1}^{%
\left[ ji\right] } \\ 
-\underline{\bar{S}}_{1}^{\left[ ij\right] } & 1-\underline{\bar{S}}_{1}^{%
\left[ jj\right] }%
\end{array}%
\right) \left( 
\begin{array}{c}
\frac{\bar{f}^{\left[ i\right] }-\bar{r}}{1+\underline{\bar{k}}^{\left[ i%
\right] }} \\ 
\frac{\bar{f}^{\left[ j\right] }-\bar{r}}{1+\underline{\bar{k}}^{\left[ j%
\right] }}%
\end{array}%
\right) -\left( 
\begin{array}{cc}
\underline{\hat{S}}_{2}^{\left[ ii\right] } & \underline{\hat{S}}_{\eta }^{%
\left[ ij\right] } \\ 
\underline{\hat{S}}_{\eta }^{\left[ ij\right] } & \underline{\hat{S}}_{2}^{%
\left[ jj\right] }%
\end{array}%
\right) \left( 
\begin{array}{c}
\frac{\hat{f}^{\left[ i\right] }-\bar{r}}{1+\underline{\hat{k}}_{2}^{\left[ i%
\right] }+\kappa \left[ \frac{\underline{\hat{k}}_{2}^{B}}{1+\bar{k}}\right]
^{\left[ i\right] }} \\ 
\frac{\hat{f}^{\left[ j\right] }-\bar{r}}{1+\underline{\hat{k}}_{2}^{\left[ j%
\right] }+\kappa \left[ \frac{\underline{\hat{k}}_{2}^{B}}{1+\bar{k}}\right]
^{\left[ j\right] }}%
\end{array}%
\right) \\
&&-\left( 
\begin{array}{cc}
\underline{\bar{S}}_{2}^{\left[ ii\right] } & \underline{\bar{S}}_{2}^{\left[
ji\right] } \\ 
\underline{\bar{S}}_{2}^{B\left[ ij\right] } & \underline{\bar{S}}_{2}^{B%
\left[ jj\right] }%
\end{array}%
\right) \left( 
\begin{array}{c}
\frac{1+\bar{f}^{\left[ i\right] }}{\underline{\overline{\bar{k}}}_{2}^{%
\left[ i\right] }}H\left( -\left( 1+\bar{f}^{\left[ i\right] }\right) \right)
\\ 
\frac{1+\bar{f}^{\left[ j\right] }}{\underline{\overline{\bar{k}}}_{2}^{%
\left[ j\right] }}H\left( -\left( 1+\bar{f}^{\left[ j\right] }\right) \right)%
\end{array}%
\right) -\left( 
\begin{array}{cc}
\underline{\hat{S}}_{2}^{B\left[ ii\right] } & \underline{\hat{S}}_{2}^{B%
\left[ ji\right] } \\ 
\underline{\hat{S}}_{2}^{B\left[ ij\right] } & \underline{\hat{S}}_{2}^{B%
\left[ jj\right] }%
\end{array}%
\right) \left( 
\begin{array}{c}
\frac{1+\hat{f}^{\left[ i\right] }}{\underline{\hat{k}}_{2}^{\left[ i\right]
}+\kappa \left[ \frac{\underline{\hat{k}}_{2}^{B}}{1+\bar{k}}\right] ^{\left[
i\right] }}H\left( -\left( 1+\hat{f}^{\left[ i\right] }\right) \right) \\ 
\frac{1+\hat{f}^{\left[ j\right] }}{\underline{\hat{k}}_{2}^{\left[ j\right]
}+\kappa \left[ \frac{\underline{\hat{k}}_{2}^{B}}{1+\bar{k}}\right] ^{\left[
j\right] }}H\left( -\left( 1+\hat{f}^{\left[ j\right] }\right) \right)%
\end{array}%
\right) \\
&&-\left( 
\begin{array}{cc}
\underline{S}_{2}^{B\left[ ii\right] } & \underline{S}_{2}^{B\left[ ji\right]
} \\ 
\underline{S}_{2}^{B\left[ ij\right] } & \underline{S}_{2}^{B\left[ jj\right]
}%
\end{array}%
\right) \left( 
\begin{array}{c}
\frac{1+f_{1}^{\prime \left[ i\right] }}{\underline{k}_{2}^{\left[ i\right]
}+\left[ \frac{\underline{k}_{2}^{B}}{1+\bar{k}}\right] ^{\left[ i\right] }}%
H\left( -\left( 1+f_{1}^{\prime \left[ i\right] }\right) \right) \\ 
\frac{1+f_{1}^{\prime \left[ j\right] }}{\underline{k}_{2}^{\left[ j\right]
}+\left[ \frac{\underline{k}_{2}^{B}}{1+\bar{k}}\right] ^{\left[ j\right] }}%
H\left( -\left( 1+f_{1}^{\prime \left[ j\right] }\right) \right)%
\end{array}%
\right) -\left( 
\begin{array}{cc}
\underline{S}_{1}^{B\left[ ii\right] } & \underline{S}_{1}^{B\left[ ji\right]
} \\ 
\underline{S}_{1}^{B\left[ ij\right] } & \underline{S}_{1}^{B\left[ jj\right]
}%
\end{array}%
\right) \left( 
\begin{array}{c}
f_{1}^{\left[ i\right] }-\bar{r} \\ 
f_{1}^{\left[ i\right] }-\bar{r}%
\end{array}%
\right)
\end{eqnarray*}%
where $f_{1}^{\prime \left[ i\right] }$ is defined by:%
\begin{equation*}
\frac{f_{1}^{\prime \left[ i\right] }-\bar{r}}{1+\underline{k}_{2}^{\left[ i%
\right] }+\left[ \frac{\underline{k}_{2}^{B}}{1+\bar{k}}\right] ^{\left[ i%
\right] }}=f_{1}^{\left[ i\right] }-\bar{r}
\end{equation*}%
where:%
\begin{eqnarray*}
\underline{\bar{S}}_{\eta }^{\left[ ii\right] } &=&\frac{\underline{\bar{k}}%
_{\eta }^{\left[ ii\right] }}{\left( 1+\underline{\bar{k}}^{\left[ i\right]
}\right) } \\
\underline{\bar{S}}_{\eta }^{\left[ ij\right] } &=&\frac{\underline{\bar{k}}%
_{\eta }^{\left[ ij\right] }}{\left( 1+\underline{\bar{k}}^{\left[ i\right]
}\right) }
\end{eqnarray*}%
with:%
\begin{eqnarray*}
\underline{\bar{S}}^{\left[ ii\right] } &=&\underline{\bar{S}}_{1}^{\left[ ii%
\right] }+\underline{\bar{S}}_{2}^{\left[ ii\right] } \\
\underline{\bar{S}}^{\left[ ij\right] } &=&\underline{\bar{S}}_{1}^{\left[ ij%
\right] }+\underline{\bar{S}}_{2}^{\left[ ij\right] }
\end{eqnarray*}

\subsubsection*{A24.3.3 Full alternate formulation of return equation}

From these formula we compute the coefficnts:%
\begin{equation*}
\underline{\bar{S}}^{\left[ i\right] }=\underline{\bar{S}}^{\left[ ii\right]
}+\underline{\bar{S}}^{\left[ ij\right] }=\frac{\underline{\bar{k}}^{\left[ i%
\right] }}{\left( 1+\underline{\bar{k}}^{\left[ i\right] }\right) }
\end{equation*}

\begin{equation*}
\frac{1}{1+\underline{\hat{k}}^{\left[ i\right] }}=1-\left( \underline{\bar{S%
}}^{\left[ ii\right] }+\underline{\bar{S}}^{\left[ ij\right] }\right) =1-%
\underline{\bar{S}}^{\left[ i\right] }
\end{equation*}%
\begin{equation*}
\underline{\bar{S}}_{\eta }^{\left[ i\right] }=\frac{\underline{\bar{k}}%
_{\eta }^{\left[ i\right] }}{\left( 1+\underline{\bar{k}}^{\left[ i\right]
}\right) }=\underline{\bar{k}}_{\eta }^{\left[ i\right] }\left( 1-\left( 
\underline{\bar{S}}^{\left[ ii\right] }+\underline{\bar{S}}^{\left[ ij\right]
}\right) \right)
\end{equation*}%
\begin{equation*}
\underline{\bar{k}}_{\eta }^{\left[ i\right] }=\frac{\underline{\bar{S}}%
_{\eta }^{\left[ i\right] }}{1-\underline{\bar{S}}^{\left[ i\right] }}=\frac{%
\underline{\bar{S}}_{\eta }^{\left[ ii\right] }+\underline{\bar{S}}_{\eta }^{%
\left[ ij\right] }}{1-\left( \underline{\bar{S}}^{\left[ ii\right] }+%
\underline{\bar{S}}^{\left[ ij\right] }\right) }
\end{equation*}%
\begin{equation*}
1+\underline{\bar{k}}_{\eta }^{\left[ i\right] }=\frac{1-\left( \underline{%
\bar{S}}_{3-\eta }^{\left[ ii\right] }+\underline{\bar{S}}_{3-\eta }^{\left[
ij\right] }\right) }{1-\left( \underline{\bar{S}}^{\left[ ii\right] }+%
\underline{\bar{S}}^{\left[ ij\right] }\right) }
\end{equation*}%
\begin{eqnarray*}
\underline{\bar{S}}_{\eta }^{\left[ ii\right] } &=&\frac{\underline{\bar{k}}%
_{\eta }^{\left[ ii\right] }}{\left( 1+\underline{\bar{k}}^{\left[ i\right]
}\right) } \\
\underline{\bar{S}}_{\eta }^{\left[ ij\right] } &=&\frac{\underline{\bar{k}}%
_{\eta }^{\left[ ij\right] }}{\left( 1+\underline{\bar{k}}^{\left[ i\right]
}\right) }
\end{eqnarray*}%
\begin{eqnarray*}
\underline{\bar{S}}^{\left[ ii\right] } &=&\underline{\bar{S}}_{1}^{\left[ ii%
\right] }+\underline{\bar{S}}_{2}^{\left[ ii\right] } \\
\underline{\bar{S}}^{\left[ ij\right] } &=&\underline{\bar{S}}_{1}^{\left[ ij%
\right] }+\underline{\bar{S}}_{2}^{\left[ ij\right] }
\end{eqnarray*}%
\begin{eqnarray*}
\underline{\hat{S}}_{\eta }^{B\left[ ii\right] } &=&\frac{\kappa \frac{%
\underline{\hat{k}}_{\eta }^{B\left[ ii\right] }}{1+\bar{k}^{\left[ i\right]
}}}{\left( 1+\underline{\hat{k}}^{\left[ i\right] }+\underline{\hat{k}}%
_{1}^{B\left[ i\right] }+\kappa \left[ \frac{\underline{\hat{k}}_{2}^{B}}{1+%
\bar{k}}\right] ^{\left[ i\right] }\right) } \\
\underline{\hat{S}}_{\eta }^{B\left[ ij\right] } &=&\frac{\kappa \frac{%
\underline{\hat{k}}_{\eta }^{B\left[ ij\right] }}{1+\bar{k}^{\left[ i\right]
}}}{\left( 1+\underline{\hat{k}}^{\left[ i\right] }+\underline{\hat{k}}%
_{1}^{B\left[ i\right] }+\kappa \left[ \frac{\underline{\hat{k}}_{2}^{B}}{1+%
\bar{k}}\right] ^{\left[ i\right] }\right) }
\end{eqnarray*}%
\begin{eqnarray*}
\underline{S}_{\eta }^{B\left[ ii\right] } &=&\frac{\kappa \frac{\underline{k%
}_{\eta }^{\left( B\right) \left[ ii\right] }}{1+\underline{\bar{k}}^{\left[
i\right] }}}{\left( 1+\underline{k}^{\left[ i\right] }+\underline{k}%
_{1}^{\left( B\right) \left[ i\right] }+\kappa \left[ \frac{\underline{k}%
_{2}^{\left( B\right) }}{1+\underline{\bar{k}}}\right] ^{\left[ i\right]
}\right) } \\
\underline{S}_{\eta }^{B\left[ ij\right] } &=&\frac{\kappa \frac{\underline{k%
}_{\eta }^{\left( B\right) \left[ ij\right] }}{1+\underline{\bar{k}}^{\left[
i\right] }}}{\left( 1+\underline{k}^{\left[ i\right] }+\underline{k}%
_{1}^{\left( B\right) \left[ i\right] }+\kappa \left[ \frac{\underline{k}%
_{2}^{\left( B\right) }}{1+\underline{\bar{k}}}\right] ^{\left[ i\right]
}\right) }
\end{eqnarray*}%
Finaly, the formulation with the new set of parameters becomes:%
\begin{eqnarray*}
0 &=&\left( 
\begin{array}{cc}
1-\underline{\bar{S}}_{1}^{\left[ ii\right] } & -\underline{\bar{S}}_{1}^{%
\left[ ji\right] } \\ 
-\underline{\bar{S}}_{1}^{\left[ ij\right] } & 1-\underline{\bar{S}}_{1}^{%
\left[ jj\right] }%
\end{array}%
\right) \left( 
\begin{array}{c}
\frac{\bar{f}^{\left[ i\right] }-\bar{r}}{1+\underline{\bar{k}}^{\left[ i%
\right] }} \\ 
\frac{\bar{f}^{\left[ j\right] }-\bar{r}}{1+\underline{\bar{k}}^{\left[ j%
\right] }}%
\end{array}%
\right) \\
&&-\left( 
\begin{array}{cc}
\underline{\bar{S}}_{2}^{\left[ ii\right] } & \underline{\bar{S}}_{2}^{\left[
ji\right] } \\ 
\underline{\bar{S}}_{2}^{\left[ ij\right] } & \underline{\bar{S}}_{2}^{\left[
jj\right] }%
\end{array}%
\right) \left( 
\begin{array}{c}
\frac{1+\bar{f}^{\left[ i\right] }}{\underline{\overline{\bar{k}}}_{2}^{%
\left[ i\right] }}H\left( -\left( 1+\bar{f}^{\left[ i\right] }\right) \right)
\\ 
\frac{1+\bar{f}^{\left[ j\right] }}{\underline{\overline{\bar{k}}}_{2}^{%
\left[ j\right] }}H\left( -\left( 1+\bar{f}^{\left[ j\right] }\right) \right)%
\end{array}%
\right) -\left( 
\begin{array}{cc}
\underline{\hat{S}}_{2}^{B^{\prime }\left[ ii\right] } & \underline{\hat{S}}%
_{2}^{B^{\prime }\left[ ji\right] } \\ 
\underline{\hat{S}}_{2}^{B^{\prime }\left[ ij\right] } & \underline{\hat{S}}%
_{2}^{B^{\prime }\left[ jj\right] }%
\end{array}%
\right) \left( 
\begin{array}{c}
\frac{1+\hat{f}^{\left[ i\right] }}{\underline{\hat{k}}_{2}^{\left[ i\right]
}+\kappa \left[ \frac{\underline{\hat{k}}_{2}^{B}}{1+\bar{k}}\right] ^{\left[
i\right] }}H\left( -\left( 1+\hat{f}^{\left[ i\right] }\right) \right) \\ 
\frac{1+\hat{f}^{\left[ j\right] }}{\underline{\hat{k}}_{2}^{\left[ j\right]
}+\kappa \left[ \frac{\underline{\hat{k}}_{2}^{B}}{1+\bar{k}}\right] ^{\left[
j\right] }}H\left( -\left( 1+\hat{f}^{\left[ j\right] }\right) \right)%
\end{array}%
\right) \\
&&-\left( 
\begin{array}{cc}
\underline{S}_{2}^{B^{\prime }\left[ ii\right] } & \underline{S}%
_{2}^{B^{\prime }\left[ ji\right] } \\ 
\underline{S}_{2}^{B^{\prime }\left[ ij\right] } & \underline{S}%
_{2}^{B^{\prime }\left[ jj\right] }%
\end{array}%
\right) \left( 
\begin{array}{c}
\frac{1+f_{1}^{\prime \left[ i\right] }}{\underline{k}_{2}^{\left[ i\right]
}+\left[ \frac{\underline{k}_{2}^{B}}{1+\bar{k}}\right] ^{\left[ i\right] }}%
H\left( -\left( 1+f_{1}^{\prime \left[ i\right] }\right) \right) \\ 
\frac{1+f_{1}^{\prime \left[ j\right] }}{\underline{k}_{2}^{\left[ j\right]
}+\left[ \frac{\underline{k}_{2}^{B}}{1+\bar{k}}\right] ^{\left[ j\right] }}%
H\left( -\left( 1+f_{1}^{\prime \left[ j\right] }\right) \right)%
\end{array}%
\right) -\left( 
\begin{array}{cc}
\underline{S}_{1}^{B^{\prime }\left[ ii\right] } & \underline{S}%
_{1}^{B^{\prime }\left[ ji\right] } \\ 
\underline{S}_{1}^{B^{\prime }\left[ ij\right] } & \underline{S}%
_{1}^{B^{\prime }\left[ jj\right] }%
\end{array}%
\right) \left( 
\begin{array}{c}
\frac{f_{1}^{\prime \left[ i\right] }-\bar{r}}{1+\underline{k}_{2}^{\left[ i%
\right] }+\left[ \frac{\underline{k}_{2}^{B}}{1+\bar{k}}\right] ^{\left[ i%
\right] }} \\ 
\frac{f_{1}^{\left[ i\right] }-\bar{r}}{\frac{f_{1}^{\prime \left[ j\right]
}-\bar{r}}{1+\underline{k}_{2}^{\left[ j\right] }+\left[ \frac{\underline{k}%
_{2}^{B}}{1+\bar{k}}\right] ^{\left[ j\right] }}}%
\end{array}%
\right)
\end{eqnarray*}

with:%
\begin{equation*}
\left( 
\begin{array}{cc}
\underline{\hat{S}}_{2}^{B^{\prime }\left[ ii\right] } & \underline{\hat{S}}%
_{2}^{B^{\prime }\left[ ji\right] } \\ 
\underline{\hat{S}}_{2}^{B^{\prime }\left[ ij\right] } & \underline{\hat{S}}%
_{2}^{B^{\prime }\left[ jj\right] }%
\end{array}%
\right) =\left( 
\begin{array}{cc}
\underline{\hat{S}}_{2}^{B\left[ ii\right] } & \underline{\hat{S}}_{2}^{B%
\left[ ji\right] } \\ 
\underline{\hat{S}}_{2}^{B\left[ ij\right] } & \underline{\hat{S}}_{2}^{B%
\left[ jj\right] }%
\end{array}%
\right) +\left( 
\begin{array}{cc}
\underline{\hat{S}}_{2}^{\left[ ii\right] } & \underline{\hat{S}}_{\eta }^{%
\left[ ij\right] } \\ 
\underline{\hat{S}}_{\eta }^{\left[ ij\right] } & \underline{\hat{S}}_{2}^{%
\left[ jj\right] }%
\end{array}%
\right) \left( 
\begin{array}{cc}
1-\underline{\hat{S}}_{1}^{\left[ ii\right] } & -\underline{\hat{S}}_{1}^{%
\left[ ji\right] } \\ 
-\underline{\hat{S}}_{1}^{\left[ ij\right] } & 1-\underline{\hat{S}}_{1}^{%
\left[ jj\right] }%
\end{array}%
\right) ^{-1}\left( 
\begin{array}{cc}
\underline{\hat{S}}_{2}^{\left[ ii\right] } & \underline{\hat{S}}_{2}^{\left[
ji\right] } \\ 
\underline{\hat{S}}_{2}^{\left[ ij\right] } & \underline{\hat{S}}_{2}^{\left[
jj\right] }%
\end{array}%
\right)
\end{equation*}%
\begin{equation*}
\left( 
\begin{array}{cc}
\underline{S}_{\eta }^{B^{\prime }\left[ ii\right] } & \underline{S}_{\eta
}^{B^{\prime }\left[ ji\right] } \\ 
\underline{S}_{\eta }^{B^{\prime }\left[ ij\right] } & \underline{S}_{\eta
}^{B^{\prime }\left[ jj\right] }%
\end{array}%
\right) =\left( 
\begin{array}{cc}
\underline{S}_{\eta }^{B\left[ ii\right] } & \underline{S}_{\eta }^{B\left[
ji\right] } \\ 
\underline{S}_{\eta }^{B\left[ ij\right] } & \underline{S}_{\eta }^{B\left[
jj\right] }%
\end{array}%
\right) +\left( 
\begin{array}{cc}
\underline{\hat{S}}_{2}^{\left[ ii\right] } & \underline{\hat{S}}_{\eta }^{%
\left[ ij\right] } \\ 
\underline{\hat{S}}_{\eta }^{\left[ ij\right] } & \underline{\hat{S}}_{2}^{%
\left[ jj\right] }%
\end{array}%
\right) \left( 
\begin{array}{cc}
1-\underline{\hat{S}}_{1}^{\left[ ii\right] } & -\underline{\hat{S}}_{1}^{%
\left[ ji\right] } \\ 
-\underline{\hat{S}}_{1}^{\left[ ij\right] } & 1-\underline{\hat{S}}_{1}^{%
\left[ jj\right] }%
\end{array}%
\right) ^{-1}\left( 
\begin{array}{cc}
\underline{S}_{\eta }^{\left[ ii\right] } & \underline{S}_{\eta }^{\left[ ji%
\right] } \\ 
\underline{S}_{\eta }^{\left[ ij\right] } & \underline{S}_{\eta }^{\left[ jj%
\right] }%
\end{array}%
\right)
\end{equation*}%
with:%
\begin{equation*}
\underline{\hat{S}}^{\left[ ii\right] }+\underline{\hat{S}}^{\left[ ji\right]
}+\underline{S}^{\left[ ii\right] }+\underline{S}^{\left[ ji\right] }=1
\end{equation*}%
\begin{equation*}
\underline{\bar{S}}^{\left[ ii\right] }+\underline{\bar{S}}^{\left[ ji\right]
}+\underline{\hat{S}}_{1}^{B\left[ ii\right] }+\underline{\hat{S}}_{1}^{B%
\left[ ji\right] }+\underline{S}_{1}^{B\left[ ii\right] }+\underline{S}%
_{1}^{B\left[ ji\right] }=1
\end{equation*}%
\begin{equation*}
\underline{\hat{S}}_{2}^{B\left[ ii\right] }+\underline{\hat{S}}_{2}^{B\left[
ji\right] }+\underline{S}_{2}^{B\left[ ii\right] }+\underline{S}_{2}^{B\left[
ji\right] }=1
\end{equation*}%
\begin{equation*}
\underline{\bar{k}}_{2}^{\left[ i\right] }=\frac{\underline{\bar{S}}_{2}^{%
\left[ i\right] }}{1-\underline{\bar{S}}^{\left[ i\right] }}=\frac{%
\underline{\bar{S}}_{2}^{\left[ ii\right] }+\underline{\bar{S}}_{2}^{\left[
ij\right] }}{1-\left( \underline{\bar{S}}^{\left[ ii\right] }+\underline{%
\bar{S}}^{\left[ ij\right] }\right) }
\end{equation*}%
\begin{equation*}
1+\underline{\bar{k}}_{2}^{\left[ i\right] }=\frac{1-\left( \underline{\bar{S%
}}_{1}^{\left[ ii\right] }+\underline{\bar{S}}_{1}^{\left[ ij\right]
}\right) }{1-\left( \underline{\bar{S}}^{\left[ ii\right] }+\underline{\bar{S%
}}^{\left[ ij\right] }\right) }
\end{equation*}%
\begin{equation*}
\frac{1}{1+\underline{\bar{k}}^{\left[ i\right] }}=\frac{1}{1-\left( 
\underline{\bar{S}}^{\left[ ii\right] }+\underline{\bar{S}}^{\left[ ij\right]
}\right) }
\end{equation*}%
\begin{eqnarray*}
0 &=&\left( 
\begin{array}{cc}
1-\underline{\bar{S}}_{1}^{\left[ ii\right] } & -\underline{\bar{S}}_{1}^{%
\left[ ji\right] } \\ 
-\underline{\bar{S}}_{1}^{\left[ ij\right] } & 1-\underline{\bar{S}}_{1}^{%
\left[ jj\right] }%
\end{array}%
\right) \left( 
\begin{array}{c}
\left( \bar{f}^{\left[ i\right] }-\bar{r}\right) \underline{\bar{s}}_{1}^{%
\left[ i\right] } \\ 
\left( \bar{f}^{\left[ j\right] }-\bar{r}\right) \underline{\bar{s}}_{1}^{%
\left[ j\right] }%
\end{array}%
\right) \\
&&-\left( 
\begin{array}{cc}
\underline{\bar{S}}_{2}^{\left[ ii\right] } & \underline{\bar{S}}_{2}^{\left[
ji\right] } \\ 
\underline{\bar{S}}_{2}^{\left[ ij\right] } & \underline{\bar{S}}_{2}^{\left[
jj\right] }%
\end{array}%
\right) \left( 
\begin{array}{c}
\left( 1+\bar{f}^{\left[ i\right] }\right) \underline{\bar{s}}_{2}^{\left[ i%
\right] }H\left( -\left( 1+\bar{f}^{\left[ i\right] }\right) \right) \\ 
\left( 1+\bar{f}^{\left[ j\right] }\right) \underline{\bar{s}}_{2}^{\left[ j%
\right] }H\left( -\left( 1+\bar{f}^{\left[ j\right] }\right) \right)%
\end{array}%
\right) -\left( 
\begin{array}{cc}
\underline{\hat{S}}_{2}^{B^{\prime }\left[ ii\right] } & \underline{\hat{S}}%
_{2}^{B^{\prime }\left[ ji\right] } \\ 
\underline{\hat{S}}_{2}^{B^{\prime }\left[ ij\right] } & \underline{\hat{S}}%
_{2}^{B^{\prime }\left[ jj\right] }%
\end{array}%
\right) \left( 
\begin{array}{c}
\left( 1+f^{\left[ i\right] }\right) \underline{\hat{s}}_{2}^{\left[ i\right]
}H\left( -\left( 1+f^{\left[ i\right] }\right) \right) \\ 
\left( 1+f^{\left[ j\right] }\right) \underline{\hat{s}}_{2}^{\left[ j\right]
}H\left( -\left( 1+f^{\left[ j\right] }\right) \right)%
\end{array}%
\right) \\
&&-\left( 
\begin{array}{cc}
\underline{S}_{2}^{B^{\prime }\left[ ii\right] } & \underline{S}%
_{2}^{B^{\prime }\left[ ji\right] } \\ 
\underline{S}_{2}^{B^{\prime }\left[ ij\right] } & \underline{S}%
_{2}^{B^{\prime }\left[ jj\right] }%
\end{array}%
\right) \left( 
\begin{array}{c}
\left( 1+f_{1}^{\prime \left[ i\right] }\right) \underline{s}_{2}^{\left[ i%
\right] }H\left( -\left( 1+f_{1}^{\prime \left[ i\right] }\right) \right) \\ 
\left( 1+f_{1}^{\prime \left[ j\right] }\right) \underline{s}_{2}^{\left[ j%
\right] }H\left( -\left( 1+f_{1}^{\prime \left[ j\right] }\right) \right)%
\end{array}%
\right) -\left( 
\begin{array}{cc}
\underline{S}_{1}^{B^{\prime }\left[ ii\right] } & \underline{S}%
_{1}^{B^{\prime }\left[ ji\right] } \\ 
\underline{S}_{1}^{B^{\prime }\left[ ij\right] } & \underline{S}%
_{1}^{B^{\prime }\left[ jj\right] }%
\end{array}%
\right) \left( 
\begin{array}{c}
\left( f_{1}^{\prime \left[ i\right] }-\bar{r}\right) \underline{s}_{1}^{%
\left[ i\right] } \\ 
\left( f_{1}^{\prime \left[ j\right] }-\bar{r}\right) \underline{s}_{1}^{%
\left[ j\right] }%
\end{array}%
\right)
\end{eqnarray*}%
\begin{equation*}
\underline{\bar{s}}_{1}^{\left[ i\right] }=\frac{1-\left( \underline{\bar{S}}%
_{1}^{\left[ ii\right] }+\underline{\bar{S}}_{1}^{\left[ ij\right] }\right) 
}{1-\left( \underline{\bar{S}}^{\left[ ii\right] }+\underline{\bar{S}}^{%
\left[ ij\right] }\right) }\text{, }\underline{\bar{s}}_{2}^{\left[ i\right]
}=\frac{1-\left( \underline{\bar{S}}^{\left[ ii\right] }+\underline{\bar{S}}%
^{\left[ ij\right] }\right) }{\underline{\bar{S}}_{2}^{\left[ ii\right] }+%
\underline{\bar{S}}_{2}^{\left[ ij\right] }}
\end{equation*}%
\begin{eqnarray*}
\underline{\hat{s}}_{1}^{\left[ i\right] } &=&\frac{1-\left( \underline{\hat{%
S}}^{\left[ ii\right] }+\underline{\hat{S}}^{\left[ ij\right] }+\underline{%
\hat{S}}^{B\left[ ii\right] }+\underline{\hat{S}}^{B\left[ ij\right]
}\right) }{1-\left( \underline{\hat{S}}_{1}^{\left[ ii\right] }+\underline{%
\hat{S}}_{1}^{\left[ ij\right] }+\underline{\hat{S}}_{1}^{B\left[ ii\right]
}+\underline{\hat{S}}_{1}^{B\left[ ij\right] }\right) }\text{, }\underline{%
\hat{s}}_{2}^{\left[ i\right] }=\frac{1-\left( \underline{\hat{S}}^{\left[ ii%
\right] }+\underline{\hat{S}}^{\left[ ij\right] }+\underline{\hat{S}}^{B%
\left[ ii\right] }+\underline{\hat{S}}^{B\left[ ij\right] }\right) }{%
\underline{\hat{S}}_{2}^{\left[ ii\right] }+\underline{\hat{S}}_{2}^{\left[
ij\right] }+\underline{\hat{S}}_{2}^{B\left[ ii\right] }+\underline{\hat{S}}%
_{2}^{B\left[ ij\right] }} \\
\underline{s}_{1}^{\left[ i\right] } &=&\frac{1-\left( \underline{S}^{\left[
ii\right] }+\underline{S}^{\left[ ij\right] }+\underline{S}^{B\left[ ii%
\right] }+\underline{S}^{B\left[ ij\right] }\right) }{1-\left( \underline{S}%
_{1}^{\left[ ii\right] }+\underline{S}_{1}^{\left[ ij\right] }+\underline{S}%
_{1}^{B\left[ ii\right] }+\underline{S}_{1}^{B\left[ ij\right] }\right) }%
\text{, }\underline{s}_{2}^{\left[ i\right] }=\frac{1-\left( \underline{S}^{%
\left[ ii\right] }+\underline{S}^{\left[ ij\right] }+\underline{S}^{B\left[
ii\right] }+\underline{S}^{B\left[ ij\right] }\right) }{\underline{S}_{2}^{%
\left[ ii\right] }+\underline{S}_{2}^{\left[ ij\right] }+\underline{S}_{2}^{B%
\left[ ii\right] }+\underline{S}_{2}^{B\left[ ij\right] }}
\end{eqnarray*}%
\bigskip

\section*{Appendix 25 Computation of the functionl derivatives}

We comput the returns derivativ with respect to the various flds.

\subsection*{A25.1 Derivatives with respect to $\left\vert \hat{\Psi}\left( 
\hat{K},\hat{X}\right) \right\vert ^{2}$}

Starting with investors nd banks return equation:%
\begin{eqnarray*}
&&\left( \Delta \left( \hat{X},\hat{X}^{\prime }\right) -\frac{\hat{K}%
^{\prime }\hat{k}_{1}\left( \hat{X}^{\prime },\hat{X}\right) \left\vert \hat{%
\Psi}\left( \hat{K}^{\prime },\hat{X}^{\prime }\right) \right\vert ^{2}}{1+%
\underline{\hat{k}}\left( \hat{X}^{\prime }\right) +\underline{\hat{k}}%
_{1}^{B}\left( \bar{X}^{\prime }\right) +\kappa \left[ \frac{\underline{\hat{%
k}}_{2}^{B}}{1+\bar{k}}\right] \left( \hat{X}^{\prime }\right) }\right) 
\frac{\left( 1-\hat{M}\right) \hat{g}\left( \hat{K}^{\prime },\hat{X}%
^{\prime }\right) +\bar{N}\bar{g}\left( \bar{K}^{\prime },\bar{X}^{\prime
}\right) }{1+\underline{\hat{k}}_{2}\left( \bar{X}^{\prime }\right) +\kappa %
\left[ \frac{\underline{\hat{k}}_{2}^{B}}{1+\bar{k}}\right] \left( \hat{X}%
^{\prime }\right) } \\
&=&\frac{k_{1}\left( X^{\prime },X^{\prime }\right) \left( f_{1}^{\prime
}\left( X^{\prime }\right) K^{\prime }-\bar{C}\left( X^{\prime }\right)
\right) \left\vert \Psi \left( K^{\prime },X^{\prime }\right) \right\vert
^{2}}{\left( 1+\underline{k}\left( \hat{X}^{\prime }\right) +\underline{k}%
_{1}^{\left( B\right) }\left( \bar{X}^{\prime }\right) +\kappa \left[ \frac{%
\underline{k}_{2}^{B}}{1+\bar{k}}\right] \left( X^{\prime }\right) \right)
\left( 1+\underline{k}_{2}\left( \hat{X}^{\prime }\right) +\kappa \left[ 
\frac{\underline{k}_{2}^{B}}{1+\bar{k}}\right] \left( X^{\prime }\right)
\right) }
\end{eqnarray*}%
\begin{eqnarray}
&&\left( \Delta \left( \bar{X}^{\prime },\bar{X}\right) -\frac{\bar{K}%
^{\prime }\bar{k}_{1}\left( \bar{X}^{\prime },\bar{X}\right) \left\vert \bar{%
\Psi}\left( \bar{K}^{\prime },\bar{X}^{\prime }\right) \right\vert ^{2}}{1+%
\underline{\bar{k}}\left( \bar{X}^{\prime }\right) }\right) \frac{\left( 1-%
\bar{M}\right) \bar{g}\left( \hat{K}^{\prime },\hat{X}^{\prime }\right) }{1+%
\underline{\overline{\bar{k}}}_{2}\left( \bar{X}^{\prime }\right) } \\
&&-\frac{\hat{K}^{\prime }\underline{\hat{k}}_{1}^{B}\left( \hat{X}^{\prime
},\bar{X}\right) \left\vert \Psi \left( K^{\prime },X^{\prime }\right)
\right\vert ^{2}}{\left\langle \hat{K}\right\rangle \left\Vert \hat{\Psi}%
\right\Vert ^{2}\left( 1+\underline{\hat{k}}\left( \hat{X}^{\prime }\right) +%
\underline{\hat{k}}_{1}^{B}\left( \bar{X}^{\prime }\right) +\kappa \left[ 
\frac{\underline{\hat{k}}_{2}^{B}}{1+\bar{k}}\right] \left( \hat{X}^{\prime
}\right) \right) }\frac{\left( 1-\hat{M}\right) \hat{g}\left( \hat{K}%
^{\prime },\hat{X}^{\prime }\right) +\bar{N}\bar{g}\left( \bar{K}^{\prime },%
\bar{X}^{\prime }\right) }{1+\underline{\hat{k}}_{2}\left( \bar{X}^{\prime
}\right) +\kappa \left[ \frac{\underline{\hat{k}}_{2}^{B}}{1+\bar{k}}\right]
\left( \hat{X}^{\prime }\right) }  \notag \\
&=&\frac{\underline{k}_{1}^{\left( B\right) }\left( X^{\prime },\bar{X}%
\right) }{1+\underline{k}\left( \hat{X}^{\prime }\right) +\underline{k}%
_{1}^{\left( B\right) }\left( \bar{X}^{\prime }\right) +\kappa \frac{%
\underline{k}_{2}^{\left( B\right) }\left( \bar{X}^{\prime }\right) }{1+%
\underline{\bar{k}}}}\frac{\left( f_{1}^{\prime }\left( X^{\prime }\right)
K^{\prime }-\bar{C}\left( X^{\prime }\right) \right) \left\vert \Psi \left(
K^{\prime },X^{\prime }\right) \right\vert ^{2}}{1+\underline{k}_{2}\left( 
\hat{X}^{\prime }\right) +\kappa \frac{\underline{k}_{2}^{\left( B\right)
}\left( \bar{X}^{\prime }\right) }{1+\underline{\bar{k}}}}  \notag
\end{eqnarray}%
and first consder:%
\begin{eqnarray*}
&&\frac{\delta }{\delta \left\vert \hat{\Psi}\left( \hat{K},\hat{X}\right)
\right\vert ^{2}}\frac{k_{1}\left( X^{\prime },X^{\prime }\right) \left(
f_{1}^{\prime }\left( X^{\prime }\right) K^{\prime }-\bar{C}\left( X^{\prime
}\right) \right) \left\vert \Psi \left( K^{\prime },X^{\prime }\right)
\right\vert ^{2}}{\left( 1+\underline{k}\left( \hat{X}^{\prime }\right) +%
\underline{k}_{1}^{\left( B\right) }\left( \bar{X}^{\prime }\right) +\kappa %
\left[ \frac{\underline{k}_{2}^{B}}{1+\bar{k}}\right] \left( X^{\prime
}\right) \right) \left( 1+\underline{k}_{2}\left( \hat{X}^{\prime }\right)
+\kappa \left[ \frac{\underline{k}_{2}^{B}}{1+\bar{k}}\right] \left(
X^{\prime }\right) \right) } \\
&\rightarrow &-\frac{k_{1}\left( X^{\prime },X^{\prime }\right) \left(
f_{1}^{\prime }\left( X^{\prime }\right) K^{\prime }-\bar{C}\left( X^{\prime
}\right) \right) \left\vert \Psi \left( K^{\prime },X^{\prime }\right)
\right\vert ^{2}}{\left( 1+\underline{k}\left( \hat{X}^{\prime }\right) +%
\underline{k}_{1}^{\left( B\right) }\left( \bar{X}^{\prime }\right) +\kappa %
\left[ \frac{\underline{k}_{2}^{B}}{1+\bar{k}}\right] \left( X^{\prime
}\right) \right) \left( 1+\underline{k}_{2}\left( \hat{X}^{\prime }\right)
+\kappa \left[ \frac{\underline{k}_{2}^{B}}{1+\bar{k}}\right] \left(
X^{\prime }\right) \right) ^{2}}\frac{\left\langle k_{2}\left( X^{\prime },%
\hat{X}\right) \right\rangle }{\left\langle K\right\rangle \left\Vert \Psi
\right\Vert ^{2}} \\
&&-\frac{k_{1}\left( X^{\prime },X^{\prime }\right) \left( f_{1}^{\prime
}\left( X^{\prime }\right) K^{\prime }-\bar{C}\left( X^{\prime }\right)
\right) \left\vert \Psi \left( K^{\prime },X^{\prime }\right) \right\vert
^{2}}{\left( 1+\underline{k}\left( \hat{X}^{\prime }\right) +\underline{k}%
_{1}^{\left( B\right) }\left( \bar{X}^{\prime }\right) +\kappa \left[ \frac{%
\underline{k}_{2}^{B}}{1+\bar{k}}\right] \left( X^{\prime }\right) \right)
^{2}\left( 1+\underline{k}_{2}\left( \hat{X}^{\prime }\right) +\kappa \left[ 
\frac{\underline{k}_{2}^{B}}{1+\bar{k}}\right] \left( X^{\prime }\right)
\right) }\frac{\left\langle k\left( X^{\prime },\hat{X}\right) \right\rangle 
}{\left\langle K\right\rangle \left\Vert \Psi \right\Vert ^{2}}
\end{eqnarray*}%
we use that:%
\begin{equation*}
\frac{\delta }{\delta \left\vert \hat{\Psi}\left( \hat{K},\hat{X}\right)
\right\vert ^{2}}\frac{k_{1}\left( X^{\prime },X^{\prime }\right) \left(
f_{1}^{\prime }\left( X^{\prime }\right) K^{\prime }-\bar{C}\left( X^{\prime
}\right) \right) \left\vert \Psi \left( K^{\prime },X^{\prime }\right)
\right\vert ^{2}}{\left( 1+\underline{k}\left( \hat{X}^{\prime }\right) +%
\underline{k}_{1}^{\left( B\right) }\left( \bar{X}^{\prime }\right) +\kappa %
\left[ \frac{\underline{k}_{2}^{B}}{1+\bar{k}}\right] \left( X^{\prime
}\right) \right) \left( 1+\underline{k}_{2}\left( \hat{X}^{\prime }\right)
+\kappa \left[ \frac{\underline{k}_{2}^{B}}{1+\bar{k}}\right] \left(
X^{\prime }\right) \right) }<<1
\end{equation*}%
similarly, we have:%
\begin{equation*}
\frac{\delta }{\delta \left\vert \hat{\Psi}\left( \hat{K},\hat{X}\right)
\right\vert ^{2}}\frac{\underline{k}_{1}^{\left( B\right) }\left( X^{\prime
},\bar{X}\right) }{1+\underline{k}\left( \hat{X}^{\prime }\right) +%
\underline{k}_{1}^{\left( B\right) }\left( \bar{X}^{\prime }\right) +\kappa 
\frac{\underline{k}_{2}^{\left( B\right) }\left( \bar{X}^{\prime }\right) }{%
1+\underline{\bar{k}}}}\frac{\left( f_{1}^{\prime }\left( X^{\prime }\right)
K^{\prime }-\bar{C}\left( X^{\prime }\right) \right) \left\vert \Psi \left(
K^{\prime },X^{\prime }\right) \right\vert ^{2}}{1+\underline{k}_{2}\left( 
\hat{X}^{\prime }\right) +\kappa \frac{\underline{k}_{2}^{\left( B\right)
}\left( \bar{X}^{\prime }\right) }{1+\underline{\bar{k}}}}<<1
\end{equation*}%
The derivativ for return becomes: 
\begin{equation*}
\frac{\delta }{\delta \left\vert \hat{\Psi}\left( \hat{K},\hat{X}\right)
\right\vert ^{2}}\left( \left( \Delta \left( \hat{X},\hat{X}^{\prime
}\right) -\frac{\hat{K}^{\prime }\hat{k}_{1}\left( \hat{X}^{\prime },\hat{X}%
\right) \left\vert \hat{\Psi}\left( \hat{K}^{\prime },\hat{X}^{\prime
}\right) \right\vert ^{2}}{1+\underline{\hat{k}}\left( \hat{X}^{\prime
}\right) +\underline{\hat{k}}_{1}^{B}\left( \bar{X}^{\prime }\right) +\kappa %
\left[ \frac{\underline{\hat{k}}_{2}^{B}}{1+\bar{k}}\right] \left( \hat{X}%
^{\prime }\right) }\right) A\right) \simeq 0
\end{equation*}%
and:%
\begin{equation}
\frac{\delta \left\{ \left( \Delta \left( \bar{X}^{\prime },\bar{X}\right) -%
\frac{\bar{K}^{\prime }\bar{k}_{1}\left( \bar{X}^{\prime },\bar{X}\right)
\left\vert \bar{\Psi}\left( \bar{K}^{\prime },\bar{X}^{\prime }\right)
\right\vert ^{2}}{1+\underline{\bar{k}}\left( \bar{X}^{\prime }\right) }%
\right) \frac{\left( 1-\bar{M}\right) \bar{g}\left( \hat{K}^{\prime },\hat{X}%
^{\prime }\right) }{1+\underline{\overline{\bar{k}}}_{2}\left( \bar{X}%
^{\prime }\right) }-\frac{\hat{K}^{\prime }\underline{\hat{k}}_{1}^{B}\left( 
\hat{X}^{\prime },\bar{X}\right) \left\vert \hat{\Psi}\left( \hat{K}^{\prime
},\hat{X}^{\prime }\right) \right\vert ^{2}}{\hat{D}\left( \hat{X}^{\prime
}\right) }A\right\} }{\delta \left\vert \hat{\Psi}\left( \hat{K},\hat{X}%
\right) \right\vert ^{2}}\simeq 0
\end{equation}%
where:%
\begin{eqnarray*}
A &=&\frac{\left( 1-\hat{M}\right) \hat{g}\left( \hat{K}^{\prime },\hat{X}%
^{\prime }\right) +\bar{N}\bar{g}\left( \hat{K}^{\prime },\hat{X}^{\prime
}\right) }{1+\underline{\hat{k}}_{2}\left( \bar{X}^{\prime }\right) +\kappa %
\left[ \frac{\underline{\hat{k}}_{2}^{B}}{1+\bar{k}}\right] \left( \hat{X}%
^{\prime }\right) } \\
B &=&\frac{\left( 1-\bar{M}\right) \bar{g}\left( \hat{K}^{\prime },\hat{X}%
^{\prime }\right) }{1+\underline{\overline{\bar{k}}}_{2}\left( \bar{X}%
^{\prime }\right) }
\end{eqnarray*}%
\begin{equation*}
\hat{D}\left( \hat{X}^{\prime }\right) =1+\underline{\hat{k}}\left( \hat{X}%
^{\prime }\right) +\underline{\hat{k}}_{1}^{B}\left( \bar{X}^{\prime
}\right) +\kappa \left[ \frac{\underline{\hat{k}}_{2}^{B}}{1+\bar{k}}\right]
\left( \hat{X}^{\prime }\right)
\end{equation*}%
and:%
\begin{equation*}
\bar{N}=\frac{\left( \hat{k}_{1}^{B}\left( \hat{X},\bar{X}^{\prime }\right)
+\kappa \frac{\hat{k}_{2}^{B}\left( \hat{X},\bar{X}^{\prime }\right) }{1+%
\overline{\bar{k}}\left( \bar{X}^{\prime },\left\langle \bar{X}^{\prime
\prime }\right\rangle \right) }-\kappa \frac{\hat{k}_{2}^{B}\left( \bar{X},%
\bar{X}^{\prime \prime }\right) \bar{K}_{0}^{\prime \prime }\overline{\bar{k}%
}\left( \bar{X}^{\prime \prime },\bar{X}^{\prime }\right) }{\left( 1+%
\overline{\bar{k}}\left( \bar{X}^{\prime },\left\langle \bar{X}^{\prime
\prime }\right\rangle \right) \right) ^{2}}\right) \hat{K}^{\prime
}\left\vert \hat{\Psi}\left( \hat{K}^{\prime },\hat{X}^{\prime }\right)
\right\vert ^{2}}{\hat{D}\left( \hat{X}^{\prime }\right) \left\langle \hat{K}%
\right\rangle \left\Vert \hat{\Psi}\right\Vert ^{2}}
\end{equation*}%
\begin{eqnarray}
0 &\simeq &\left( \Delta \left( \bar{X}^{\prime },\bar{X}\right) -\frac{\bar{%
K}^{\prime }\bar{k}_{1}\left( \bar{X}^{\prime },\bar{X}\right) \left\vert 
\bar{\Psi}\left( \bar{K}^{\prime },\bar{X}^{\prime }\right) \right\vert ^{2}%
}{1+\underline{\bar{k}}\left( \bar{X}^{\prime }\right) }\right) \frac{\left(
1-\bar{M}\right) }{1+\underline{\overline{\bar{k}}}_{2}\left( \bar{X}%
^{\prime }\right) }\frac{\delta }{\delta \left\vert \hat{\Psi}\left( \hat{K},%
\hat{X}\right) \right\vert ^{2}}\bar{g}\left( \hat{K}^{\prime },\hat{X}%
^{\prime }\right) \\
&&-\frac{\delta }{\delta \left\vert \hat{\Psi}\left( \hat{K},\hat{X}\right)
\right\vert ^{2}}\left\{ \frac{\hat{K}^{\prime }\underline{\hat{k}}%
_{1}^{B}\left( \hat{X}^{\prime },\bar{X}\right) \left\vert \hat{\Psi}\left( 
\hat{K}^{\prime },\hat{X}^{\prime }\right) \right\vert ^{2}}{1+\underline{%
\hat{k}}\left( \hat{X}^{\prime }\right) +\underline{\hat{k}}_{1}^{B}\left( 
\bar{X}^{\prime }\right) +\kappa \left[ \frac{\underline{\hat{k}}_{2}^{B}}{1+%
\bar{k}}\right] \left( \hat{X}^{\prime }\right) }A\right\}  \notag
\end{eqnarray}%
\begin{eqnarray}
&&\left( \Delta \left( \bar{X}^{\prime },\bar{X}\right) -\frac{\bar{K}%
^{\prime }\bar{k}_{1}\left( \bar{X}^{\prime },\bar{X}\right) \left\vert \bar{%
\Psi}\left( \bar{K}^{\prime },\bar{X}^{\prime }\right) \right\vert ^{2}}{1+%
\underline{\bar{k}}\left( \bar{X}^{\prime }\right) }\right) \frac{\left( 1-%
\bar{M}\right) }{1+\underline{\overline{\bar{k}}}_{2}\left( \bar{X}^{\prime
}\right) }\frac{\delta }{\delta \left\vert \hat{\Psi}\left( \hat{K},\hat{X}%
\right) \right\vert ^{2}}\bar{g}\left( \hat{K}^{\prime },\hat{X}^{\prime
}\right) \\
&\simeq &\left[ \frac{\delta }{\delta \left\vert \hat{\Psi}\left( \hat{K},%
\hat{X}\right) \right\vert ^{2}}\frac{\hat{K}^{\prime }\underline{\hat{k}}%
_{1}^{B}\left( \hat{X}^{\prime },\bar{X}\right) \left\vert \hat{\Psi}\left( 
\hat{K}^{\prime },\hat{X}^{\prime }\right) \right\vert ^{2}}{\left\langle 
\hat{K}\right\rangle \left\Vert \hat{\Psi}\right\Vert ^{2}\left( 1+%
\underline{\hat{k}}\left( \hat{X}^{\prime }\right) +\underline{\hat{k}}%
_{1}^{B}\left( \bar{X}^{\prime }\right) +\kappa \left[ \frac{\underline{\hat{%
k}}_{2}^{B}}{1+\bar{k}}\right] \left( \hat{X}^{\prime }\right) \right) }%
\right] A  \notag \\
&&+\frac{\hat{K}^{\prime }\underline{\hat{k}}_{1}^{B}\left( \hat{X}^{\prime
},\bar{X}\right) \left\vert \hat{\Psi}\left( \hat{K}^{\prime },\hat{X}%
^{\prime }\right) \right\vert ^{2}}{\left\langle \hat{K}\right\rangle
\left\Vert \hat{\Psi}\right\Vert ^{2}\left( 1+\underline{\hat{k}}\left( \hat{%
X}^{\prime }\right) +\underline{\hat{k}}_{1}^{B}\left( \bar{X}^{\prime
}\right) +\kappa \left[ \frac{\underline{\hat{k}}_{2}^{B}}{1+\bar{k}}\right]
\left( \hat{X}^{\prime }\right) \right) }\frac{\delta }{\delta \left\vert 
\hat{\Psi}\left( \hat{K},\hat{X}\right) \right\vert ^{2}}A  \notag
\end{eqnarray}%
so that:%
\begin{eqnarray}
&&\frac{\delta }{\delta \left\vert \hat{\Psi}\left( \hat{K},\hat{X}\right)
\right\vert ^{2}}\bar{g}\left( \hat{K}^{\prime },\hat{X}^{\prime }\right)
\label{Dg} \\
&\simeq &\left\{ \left( \Delta \left( \bar{X}^{\prime },\bar{X}\right) -%
\frac{\bar{K}^{\prime }\bar{k}_{1}\left( \bar{X}^{\prime },\bar{X}\right)
\left\vert \bar{\Psi}\left( \bar{K}^{\prime },\bar{X}^{\prime }\right)
\right\vert ^{2}}{1+\underline{\bar{k}}\left( \bar{X}^{\prime }\right) }%
\right) \frac{\left( 1-\bar{M}\right) }{1+\underline{\overline{\bar{k}}}%
_{2}\left( \bar{X}^{\prime }\right) }\right\} ^{-1}\times  \notag \\
&&\left[ \frac{\delta }{\delta \left\vert \hat{\Psi}\left( \hat{K},\hat{X}%
\right) \right\vert ^{2}}\frac{\hat{K}^{\prime }\underline{\hat{k}}%
_{1}^{B}\left( \hat{X}^{\prime },\bar{X}\right) }{\hat{D}\left( \hat{X}%
^{\prime }\right) }A+\frac{\hat{K}^{\prime }\underline{\hat{k}}%
_{1}^{B}\left( \hat{X}^{\prime },\bar{X}\right) \left\vert \hat{\Psi}\left( 
\hat{K}^{\prime },\hat{X}^{\prime }\right) \right\vert ^{2}}{\hat{D}\left( 
\hat{X}^{\prime }\right) }\frac{\delta }{\delta \left\vert \hat{\Psi}\left( 
\hat{K},\hat{X}\right) \right\vert ^{2}}A\right]  \notag
\end{eqnarray}

\subsubsection*{A25.1.1 Estimation of $\frac{\protect\delta }{\protect\delta %
\left\vert \hat{\Psi}\left( \hat{K},\hat{X}\right) \right\vert ^{2}}A$}

Recall that:%
\begin{equation*}
A=\frac{\left( 1-\hat{M}\right) \hat{g}\left( \hat{K}^{\prime },\hat{X}%
^{\prime }\right) +\bar{N}\bar{g}\left( \hat{K}^{\prime },\hat{X}^{\prime
}\right) }{1+\underline{\hat{k}}_{2}\left( \bar{X}^{\prime }\right) +\kappa %
\left[ \frac{\underline{\hat{k}}_{2}^{B}}{1+\bar{k}}\right] \left( \hat{X}%
^{\prime }\right) }
\end{equation*}

and:%
\begin{equation*}
\frac{\delta \left( \left( \Delta \left( \hat{X},\hat{X}^{\prime }\right) -%
\frac{\hat{K}^{\prime }\hat{k}_{1}\left( \hat{X}^{\prime },\hat{X}\right)
\left\vert \hat{\Psi}\left( \hat{K}^{\prime },\hat{X}^{\prime }\right)
\right\vert ^{2}}{1+\underline{\hat{k}}\left( \hat{X}^{\prime }\right) +%
\underline{\hat{k}}_{1}^{B}\left( \bar{X}^{\prime }\right) +\kappa \left[ 
\frac{\underline{\hat{k}}_{2}^{B}}{1+\bar{k}}\right] \left( \hat{X}^{\prime
}\right) }\right) \frac{\left( 1-\hat{M}\right) \hat{g}\left( \hat{K}%
^{\prime },\hat{X}^{\prime }\right) +\bar{N}\bar{g}\left( \hat{K}^{\prime },%
\hat{X}^{\prime }\right) }{1+\underline{\hat{k}}_{2}\left( \bar{X}^{\prime
}\right) +\kappa \left[ \frac{\underline{\hat{k}}_{2}^{B}}{1+\bar{k}}\right]
\left( \hat{X}^{\prime }\right) }\right) }{\delta \left\vert \hat{\Psi}%
\left( \hat{K},\hat{X}\right) \right\vert ^{2}}\simeq 0
\end{equation*}%
\begin{eqnarray}
&&\frac{\delta }{\delta \left\vert \hat{\Psi}\left( \hat{K},\hat{X}\right)
\right\vert ^{2}}\frac{\left( 1-\hat{M}\right) \hat{g}\left( \hat{K}^{\prime
},\hat{X}^{\prime }\right) +\bar{N}\bar{g}\left( \hat{K}^{\prime },\hat{X}%
^{\prime }\right) }{1+\underline{\hat{k}}_{2}\left( \bar{X}^{\prime }\right)
+\kappa \left[ \frac{\underline{\hat{k}}_{2}^{B}}{1+\bar{k}}\right] \left( 
\hat{X}^{\prime }\right) }  \label{FG} \\
&=&\left( \Delta \left( \hat{X},\hat{X}^{\prime }\right) -\frac{\hat{K}%
^{\prime }\hat{k}_{1}\left( \hat{X}^{\prime },\hat{X}\right) \left\vert \hat{%
\Psi}\left( \hat{K}^{\prime },\hat{X}^{\prime }\right) \right\vert ^{2}}{%
\hat{D}\left( \hat{X}^{\prime }\right) }\right) ^{-1}\frac{\delta }{\delta
\left\vert \hat{\Psi}\left( \hat{K},\hat{X}\right) \right\vert ^{2}}\frac{%
\hat{K}^{\prime }\hat{k}_{1}\left( \hat{X}^{\prime },\hat{X}\right)
\left\vert \hat{\Psi}\left( \hat{K}^{\prime },\hat{X}^{\prime }\right)
\right\vert ^{2}}{\hat{D}\left( \hat{X}^{\prime }\right) }A  \notag
\end{eqnarray}

\paragraph*{A25.1.1.1 Estimation of:}

\begin{equation*}
\frac{\delta }{\delta \left\vert \hat{\Psi}\left( \hat{K}_{1},\hat{X}%
_{1}\right) \right\vert ^{2}}\left( \frac{1}{\hat{D}\left( \hat{X}\right) }%
\right)
\end{equation*}%
given the normalizations:%
\begin{eqnarray*}
\hat{D}\left( \hat{X}\right) &=&1+\int \frac{\hat{k}\left( \hat{X},\hat{X}%
^{\prime }\right) -\left\langle \hat{k}\left( \hat{X},\hat{X}^{\prime
}\right) \right\rangle }{\left\langle \hat{K}\right\rangle \left\Vert \hat{%
\Psi}\right\Vert ^{2}}\hat{K}^{\prime }\left\vert \hat{\Psi}\left( \hat{K}%
^{\prime },\hat{X}^{\prime }\right) \right\vert ^{2}+\int \frac{\hat{k}%
_{1}^{B}\left( \hat{X},\bar{X}^{\prime }\right) -\left\langle \hat{k}%
_{1}^{B}\left( \hat{X},\bar{X}^{\prime }\right) \right\rangle }{\left\langle 
\hat{K}\right\rangle \left\Vert \hat{\Psi}\right\Vert ^{2}}\bar{K}%
_{0}^{\prime }\left\vert \bar{\Psi}\left( \bar{K}_{0}^{\prime },\bar{X}%
^{\prime }\right) \right\vert ^{2} \\
&&+\kappa \int \frac{\hat{k}_{2}^{B}\left( \hat{X},\bar{X}^{\prime }\right)
-\left\langle \hat{k}_{2}^{B}\left( \hat{X},\bar{X}^{\prime }\right)
\right\rangle }{\left\langle \hat{K}\right\rangle \left\Vert \hat{\Psi}%
\right\Vert ^{2}}\frac{\bar{K}_{0}^{\prime }\left\vert \bar{\Psi}\left( \bar{%
K}_{0}^{\prime },\bar{X}^{\prime }\right) \right\vert ^{2}}{1+\int \bar{k}%
\left( \bar{X}^{\prime },\bar{X}^{\prime \prime }\right) \bar{K}_{0}^{\prime
\prime }\left\vert \bar{\Psi}\left( \bar{K}_{0}^{\prime \prime },\bar{X}%
^{\prime \prime }\right) \right\vert ^{2}}
\end{eqnarray*}%
\begin{equation*}
\hat{k}\left( \hat{X},\hat{X}^{\prime }\right) \rightarrow \frac{\hat{k}%
\left( \hat{X},\hat{X}^{\prime }\right) }{\left\langle \hat{K}\right\rangle
\left\Vert \hat{\Psi}\right\Vert ^{2}}=\frac{\hat{k}\left( \hat{X},\hat{X}%
^{\prime }\right) }{\int \hat{K}\left\vert \hat{\Psi}\left( \hat{K},\hat{X}%
\right) \right\vert ^{2}}
\end{equation*}%
\begin{equation*}
\hat{k}_{\eta }^{B}\left( \hat{X},\bar{X}^{\prime }\right) \rightarrow \frac{%
\hat{k}_{\eta }^{B}\left( \hat{X},\bar{X}^{\prime }\right) }{\int \hat{K}%
\left\vert \hat{\Psi}\left( \hat{K},\hat{X}\right) \right\vert ^{2}}
\end{equation*}%
and:%
\begin{equation*}
\frac{\delta \left( \int \hat{k}\left( \hat{X},\hat{X}^{\prime }\right) \hat{%
K}^{\prime }\left\vert \hat{\Psi}\left( \hat{K}^{\prime },\hat{X}^{\prime
}\right) \right\vert ^{2}\right) }{\delta \left\vert \hat{\Psi}\left( \hat{K}%
_{1},\hat{X}_{1}\right) \right\vert ^{2}}<<1
\end{equation*}%
we find:%
\begin{eqnarray*}
&&\frac{\delta }{\delta \left\vert \hat{\Psi}\left( \hat{K}_{1},\hat{X}%
_{1}\right) \right\vert ^{2}}\left( \frac{1}{\hat{D}\left( \hat{X}\right) }%
\right) \\
&=&-\frac{\left( \hat{k}\left( \hat{X},\hat{X}_{1}\right) -\left\langle \hat{%
k}\left( \hat{X},\hat{X}^{\prime }\right) \right\rangle \right) \hat{K}_{1}}{%
\left\langle \hat{K}\right\rangle \left\Vert \hat{\Psi}\right\Vert
^{2}\left( \hat{D}\left( \hat{X}\right) \right) ^{2}} \\
&&+\left\{ \frac{\int \frac{\left( \hat{k}\left( \hat{X},\hat{X}^{\prime
}\right) -\left\langle \hat{k}\left( \hat{X},\hat{X}^{\prime }\right)
\right\rangle \right) \hat{K}^{\prime }\left\vert \hat{\Psi}\left( \hat{K}%
^{\prime },\hat{X}^{\prime }\right) \right\vert ^{2}}{\left\langle \hat{K}%
\right\rangle \left\Vert \hat{\Psi}\right\Vert ^{2}}+\int \frac{\left( \hat{k%
}_{1}^{B}\left( \hat{X},\bar{X}^{\prime }\right) -\left\langle \hat{k}%
_{1}^{B}\left( \hat{X},\bar{X}^{\prime }\right) \right\rangle \right) \bar{K}%
_{0}^{\prime }\left\vert \bar{\Psi}\left( \bar{K}_{0}^{\prime },\bar{X}%
^{\prime }\right) \right\vert ^{2}}{\int \hat{K}\left\vert \hat{\Psi}\left( 
\hat{K},\hat{X}\right) \right\vert ^{2}}}{\left( \hat{D}\left( \hat{X}%
\right) \right) ^{2}}\right. \\
&&+\left. \frac{\kappa }{\left( \hat{D}\left( \hat{X}\right) \right) ^{2}}%
\int \frac{\left( \hat{k}_{2}^{B}\left( \hat{X},\bar{X}^{\prime }\right)
-\left\langle \hat{k}_{2}^{B}\left( \hat{X},\bar{X}^{\prime }\right)
\right\rangle \right) \bar{K}_{0}^{\prime }\left\vert \bar{\Psi}\left( \bar{K%
}_{0}^{\prime },\bar{X}^{\prime }\right) \right\vert ^{2}}{\int \hat{K}%
\left\vert \hat{\Psi}\left( \hat{K},\hat{X}\right) \right\vert ^{2}\left(
1+\int \bar{k}\left( \bar{X}^{\prime },\bar{X}^{\prime \prime }\right) \bar{K%
}_{0}^{\prime \prime }\left\vert \bar{\Psi}\left( \bar{K}_{0}^{\prime \prime
},\bar{X}^{\prime \prime }\right) \right\vert ^{2}\right) }\right\} \\
&&\times \frac{\hat{K}_{1}}{\left( \int \hat{K}\left\vert \hat{\Psi}\left( 
\hat{K},\hat{X}\right) \right\vert \right) }
\end{eqnarray*}%
in averages this leads to:%
\begin{equation*}
\frac{\delta }{\delta \left\vert \hat{\Psi}\left( \hat{K}_{1},\hat{X}%
_{1}\right) \right\vert ^{2}}\left( \frac{1}{\hat{D}\left( \hat{X}\right) }%
\right) \rightarrow -\frac{\left( \hat{k}\left( \hat{X},\hat{X}_{1}\right)
-\left\langle \hat{k}\left( \hat{X},\hat{X}^{\prime }\right) \right\rangle
\right) \hat{K}_{1}}{\left\langle \hat{K}\right\rangle \left\Vert \hat{\Psi}%
\right\Vert ^{2}\left( \hat{D}\left( \hat{X}\right) \right) ^{2}}
\end{equation*}

\paragraph*{A25.1.1.2 Estimation of}

\begin{equation*}
\frac{\delta }{\delta \left\vert \hat{\Psi}\left( \hat{K},\hat{X}\right)
\right\vert ^{2}}\frac{\hat{K}^{\prime }\hat{k}_{1}\left( \hat{X}^{\prime },%
\hat{X}\right) \left\vert \hat{\Psi}\left( \hat{K}^{\prime },\hat{X}^{\prime
}\right) \right\vert ^{2}}{\hat{D}\left( \hat{X}^{\prime }\right) }A
\end{equation*}%
Using the previous result:%
\begin{eqnarray*}
&&\frac{\delta }{\delta \left\vert \hat{\Psi}\left( \hat{K}_{1},\hat{X}%
_{1}\right) \right\vert ^{2}}\left( \frac{\hat{K}^{\prime }\hat{k}_{1}\left( 
\hat{X}^{\prime },\hat{X}\right) \left\vert \hat{\Psi}\left( \hat{K}^{\prime
},\hat{X}^{\prime }\right) \right\vert ^{2}}{\left\langle \hat{K}%
\right\rangle \left\Vert \hat{\Psi}\right\Vert ^{2}\left( 1+\underline{\hat{k%
}}\left( \hat{X}^{\prime }\right) +\underline{\hat{k}}_{1}^{B}\left( \bar{X}%
^{\prime }\right) +\kappa \left[ \frac{\underline{\hat{k}}_{2}^{B}}{1+\bar{k}%
}\right] \left( \hat{X}^{\prime }\right) \right) }\right) \\
&\simeq &\frac{\hat{K}_{1}\hat{k}_{1}\left( \hat{X}_{1},\hat{X}\right) }{%
\left\langle \hat{K}\right\rangle \left\Vert \hat{\Psi}\right\Vert ^{2}\hat{D%
}}-\frac{\hat{K}^{\prime }\hat{k}_{1}\left( \hat{X}^{\prime },\hat{X}\right)
\left\vert \hat{\Psi}\left( \hat{K}^{\prime },\hat{X}^{\prime }\right)
\right\vert ^{2}}{\left\langle \hat{K}\right\rangle \left\Vert \hat{\Psi}%
\right\Vert ^{2}\hat{D}\left( \hat{X}^{\prime }\right) }\frac{\hat{K}_{1}}{%
\left\Vert \hat{\Psi}\right\Vert ^{2}\left\langle \hat{K}\right\rangle } \\
&&-\frac{\hat{K}^{\prime }\hat{k}_{1}\left( \hat{X}^{\prime },\hat{X}\right)
\left\vert \hat{\Psi}\left( \hat{K}^{\prime },\hat{X}^{\prime }\right)
\right\vert ^{2}}{\left\langle \hat{K}\right\rangle \left\Vert \hat{\Psi}%
\right\Vert ^{2}\hat{D}\left( \hat{X}^{\prime }\right) }\frac{\left( \hat{k}%
\left( \left\langle \hat{X}\right\rangle ,\hat{X}_{1}\right) -\left\langle 
\hat{k}\left( \hat{X},\hat{X}^{\prime }\right) \right\rangle \right) \hat{K}%
_{1}}{\left\langle \hat{K}\right\rangle \left\Vert \hat{\Psi}\right\Vert ^{2}%
\hat{D}\left( \hat{X}\right) }
\end{eqnarray*}%
as before, this is can be replaced in matricl elements by:%
\begin{eqnarray*}
&&\frac{\delta }{\delta \left\vert \hat{\Psi}\left( \hat{K}_{1},\hat{X}%
_{1}\right) \right\vert ^{2}}\frac{\hat{K}^{\prime }\hat{k}_{1}\left( \hat{X}%
^{\prime },\hat{X}\right) \left\vert \hat{\Psi}\left( \hat{K}^{\prime },\hat{%
X}^{\prime }\right) \right\vert ^{2}}{\hat{D}\left( \hat{X}^{\prime }\right) 
}A \\
&=&\frac{\hat{K}_{1}\hat{k}_{1}\left( \hat{X}_{1},\hat{X}\right) }{%
\left\langle \hat{K}\right\rangle \left\Vert \hat{\Psi}\right\Vert ^{2}}A-%
\frac{\left\langle \hat{k}_{1}\left( \hat{X}^{\prime },\hat{X}\right)
\right\rangle \left( 1+\left( \hat{k}\left( \left\langle \hat{X}%
\right\rangle ,\hat{X}_{1}\right) \right) -\left\langle \hat{k}\left( \hat{X}%
,\hat{X}^{\prime }\right) \right\rangle \right) \hat{K}_{1}}{\left\langle 
\hat{K}\right\rangle \left\Vert \hat{\Psi}\right\Vert ^{2}}\left\langle
A\right\rangle \\
&\simeq &\frac{\hat{K}_{1}\hat{k}_{1}\left( \hat{X}_{1},\hat{X}\right) }{%
\left\langle \hat{K}\right\rangle \left\Vert \hat{\Psi}\right\Vert ^{2}}A-%
\frac{\left\langle \hat{k}_{1}\left( \hat{X}^{\prime },\hat{X}\right)
\right\rangle \hat{K}_{1}}{\left\langle \hat{K}\right\rangle \left\Vert \hat{%
\Psi}\right\Vert ^{2}}\left\langle A\right\rangle
\end{eqnarray*}

\paragraph*{A25.1.1.3 formula for $\frac{\protect\delta }{\protect\delta %
\left\vert \hat{\Psi}\left( \hat{K},\hat{X}\right) \right\vert ^{2}}A$}

\begin{eqnarray}
&&\frac{\delta }{\delta \left\vert \hat{\Psi}\left( \hat{K},\hat{X}\right)
\right\vert ^{2}}\frac{\left( 1-\hat{M}\right) \hat{g}\left( \hat{K}^{\prime
},\hat{X}^{\prime }\right) +\bar{N}\bar{g}\left( \hat{K}^{\prime },\hat{X}%
^{\prime }\right) }{1+\underline{\hat{k}}_{2}\left( \bar{X}^{\prime }\right)
+\kappa \left[ \frac{\underline{\hat{k}}_{2}^{B}}{1+\bar{k}}\right] \left( 
\hat{X}^{\prime }\right) } \\
&=&\left( \Delta \left( \hat{X},\hat{X}^{\prime }\right) -\frac{\hat{K}%
^{\prime }\hat{k}_{1}\left( \hat{X}^{\prime },\hat{X}\right) \left\vert \hat{%
\Psi}\left( \hat{K}^{\prime },\hat{X}^{\prime }\right) \right\vert ^{2}}{%
\hat{D}\left( \hat{X}^{\prime }\right) }\right) ^{-1}\frac{\delta }{\delta
\left\vert \hat{\Psi}\left( \hat{K},\hat{X}\right) \right\vert ^{2}}\frac{%
\hat{K}^{\prime }\hat{k}_{1}\left( \hat{X}^{\prime },\hat{X}\right)
\left\vert \hat{\Psi}\left( \hat{K}^{\prime },\hat{X}^{\prime }\right)
\right\vert ^{2}}{\hat{D}\left( \hat{X}^{\prime }\right) }A  \notag \\
&=&\left( \Delta \left( \hat{X},\hat{X}^{\prime }\right) -\frac{\hat{K}%
^{\prime }\hat{k}_{1}\left( \hat{X}^{\prime },\hat{X}\right) \left\vert \hat{%
\Psi}\left( \hat{K}^{\prime },\hat{X}^{\prime }\right) \right\vert ^{2}}{%
\hat{D}\left( \hat{X}^{\prime }\right) }\right) ^{-1}\left( \frac{\hat{K}_{1}%
\hat{k}_{1}\left( \hat{X}_{1},\hat{X}\right) }{\left\langle \hat{K}%
\right\rangle \left\Vert \hat{\Psi}\right\Vert ^{2}}A-\frac{\left\langle 
\hat{k}_{1}\left( \hat{X}^{\prime },\hat{X}\right) \right\rangle \hat{K}_{1}%
}{\left\langle \hat{K}\right\rangle \left\Vert \hat{\Psi}\right\Vert ^{2}}%
\left\langle A\right\rangle \right)  \notag \\
&\simeq &\left( 1-\left\langle \hat{k}_{1}\left( \hat{X}^{\prime },\hat{X}%
\right) \right\rangle \right) ^{-1}\left( \frac{\hat{K}_{1}\hat{k}_{1}\left( 
\hat{X}_{1},\hat{X}\right) }{\left\langle \hat{K}\right\rangle \left\Vert 
\hat{\Psi}\right\Vert ^{2}}A-\frac{\left\langle \hat{k}_{1}\left( \hat{X}%
^{\prime },\hat{X}\right) \right\rangle \hat{K}_{1}}{\left\langle \hat{K}%
\right\rangle \left\Vert \hat{\Psi}\right\Vert ^{2}}\left\langle
A\right\rangle \right)  \notag
\end{eqnarray}

\subsubsection*{A25.1.2 Estimation of $\frac{\protect\delta }{\protect\delta %
\left\vert \hat{\Psi}\left( \hat{K},\hat{X}\right) \right\vert ^{2}}\bar{g}%
\left( \hat{K}^{\prime },\hat{X}^{\prime }\right) $}

We start with formula (\ref{Dg}):%
\begin{eqnarray}
&&\frac{\delta }{\delta \left\vert \hat{\Psi}\left( \hat{K},\hat{X}\right)
\right\vert ^{2}}\bar{g}\left( \hat{K}^{\prime },\hat{X}^{\prime }\right) \\
&\simeq &\left\{ \left( \Delta \left( \bar{X}^{\prime },\bar{X}\right) -%
\frac{\bar{K}^{\prime }\bar{k}_{1}\left( \bar{X}^{\prime },\bar{X}\right)
\left\vert \bar{\Psi}\left( \bar{K}^{\prime },\bar{X}^{\prime }\right)
\right\vert ^{2}}{1+\underline{\bar{k}}\left( \bar{X}^{\prime }\right) }%
\right) \frac{\left( 1-\bar{M}\right) }{1+\underline{\overline{\bar{k}}}%
_{2}\left( \bar{X}^{\prime }\right) }\right\} ^{-1}\times  \notag \\
&&\left[ \frac{\delta }{\delta \left\vert \hat{\Psi}\left( \hat{K},\hat{X}%
\right) \right\vert ^{2}}\frac{\hat{K}^{\prime }\underline{\hat{k}}%
_{1}^{B}\left( \hat{X}^{\prime },\bar{X}\right) \left\vert \hat{\Psi}\left( 
\hat{K}^{\prime },\hat{X}^{\prime }\right) \right\vert ^{2}}{\left\langle 
\hat{K}\right\rangle \left\Vert \hat{\Psi}\right\Vert ^{2}\hat{D}\left( \hat{%
X}^{\prime }\right) }A+\frac{\hat{K}^{\prime }\underline{\hat{k}}%
_{1}^{B}\left( \hat{X}^{\prime },\bar{X}\right) \left\vert \hat{\Psi}\left( 
\hat{K}^{\prime },\hat{X}^{\prime }\right) \right\vert ^{2}}{\left\langle 
\hat{K}\right\rangle \left\Vert \hat{\Psi}\right\Vert ^{2}\hat{D}\left( \hat{%
X}^{\prime }\right) }\frac{\delta }{\delta \left\vert \hat{\Psi}\left( \hat{K%
},\hat{X}\right) \right\vert ^{2}}A\right]  \notag
\end{eqnarray}%
and compute the various contributions:

\paragraph*{A25.1.2.1 Estimation of:}

\begin{equation*}
\frac{\delta }{\delta \left\vert \hat{\Psi}\left( \hat{K},\hat{X}\right)
\right\vert ^{2}}\frac{\hat{K}^{\prime }\underline{\hat{k}}_{1}^{B}\left( 
\hat{X}^{\prime },\bar{X}\right) }{\hat{D}\left( \hat{X}^{\prime }\right) }A
\end{equation*}%
\begin{eqnarray*}
&&\frac{\delta }{\delta \left\vert \hat{\Psi}\left( \hat{K},\hat{X}\right)
\right\vert ^{2}}\frac{\hat{K}^{\prime }\underline{\hat{k}}_{1}^{B}\left( 
\hat{X}^{\prime },\bar{X}\right) \left\vert \hat{\Psi}\left( \hat{K}^{\prime
},\hat{X}^{\prime }\right) \right\vert ^{2}}{\left\langle \hat{K}%
\right\rangle \left\Vert \hat{\Psi}\right\Vert ^{2}\left( 1+\underline{\hat{k%
}}\left( \hat{X}^{\prime }\right) +\underline{\hat{k}}_{1}^{B}\left( \bar{X}%
^{\prime }\right) +\kappa \left[ \frac{\underline{\hat{k}}_{2}^{B}}{1+\bar{k}%
}\right] \left( \hat{X}^{\prime }\right) \right) } \\
&\simeq &-\frac{\left( \hat{k}\left( \hat{X},\bar{X}\right) -\left\langle 
\hat{k}\left( \hat{X},\bar{X}\right) \right\rangle \right) \hat{K}}{%
\left\langle \hat{K}\right\rangle \left\Vert \hat{\Psi}\right\Vert
^{2}\left( \hat{D}\left( \hat{X}\right) \right) ^{2}}\underline{\hat{k}}%
_{1}^{B}\left( \left\langle \hat{X}\right\rangle ,\bar{X}\right) +\frac{\hat{%
K}\underline{\hat{k}}_{1}^{B}\left( \hat{X},\bar{X}\right) }{\left\langle 
\hat{K}\right\rangle \left\Vert \hat{\Psi}\right\Vert ^{2}\hat{D}\left( \hat{%
X}\right) }-\frac{\hat{K}^{\prime }\underline{\hat{k}}_{1}^{B}\left( \hat{X}%
^{\prime },\bar{X}\right) \left\vert \hat{\Psi}\left( \hat{K}^{\prime },\hat{%
X}^{\prime }\right) \right\vert ^{2}}{\left\langle \hat{K}\right\rangle
\left\Vert \hat{\Psi}\right\Vert ^{2}\hat{D}\left( \hat{X}\right) }\frac{%
\hat{K}}{\left\langle \hat{K}\right\rangle \left\Vert \hat{\Psi}\right\Vert
^{2}}
\end{eqnarray*}%
In average, this formula becomes:%
\begin{eqnarray*}
&&\frac{\delta }{\delta \left\vert \hat{\Psi}\left( \hat{K},\hat{X}\right)
\right\vert ^{2}}\frac{\hat{K}^{\prime }\underline{\hat{k}}_{1}^{B}\left( 
\hat{X}^{\prime },\bar{X}\right) \left\vert \hat{\Psi}\left( \hat{K}^{\prime
},\hat{X}^{\prime }\right) \right\vert ^{2}}{\left\langle \hat{K}%
\right\rangle \left\Vert \hat{\Psi}\right\Vert ^{2}\hat{D}\left( \hat{X}%
^{\prime }\right) }A \\
&\simeq &-\frac{\left( \hat{k}\left( \hat{X}^{\prime },\hat{X}\right)
-\left\langle \hat{k}\left( \hat{X}^{\prime },\hat{X}\right) \right\rangle
\right) \hat{K}}{\left\langle \hat{K}\right\rangle \left\Vert \hat{\Psi}%
\right\Vert ^{2}\left( \hat{D}\left( \hat{X}\right) \right) ^{2}}\left( 
\frac{\hat{K}^{\prime }\underline{\hat{k}}_{1}^{B}\left( \hat{X}^{\prime },%
\bar{X}\right) \left\vert \hat{\Psi}\left( \hat{K}^{\prime },\hat{X}^{\prime
}\right) \right\vert ^{2}}{\left\langle \hat{K}\right\rangle \left\Vert \hat{%
\Psi}\right\Vert ^{2}}\right) A \\
&&+\frac{\hat{K}\underline{\hat{k}}_{1}^{B}\left( \hat{X},\bar{X}\right) }{%
\left\langle \hat{K}\right\rangle \left\Vert \hat{\Psi}\right\Vert ^{2}\hat{D%
}\left( \hat{X}\right) }A-\frac{\hat{K}\underline{\hat{k}}_{1}^{B}\left(
\left\langle \hat{X}\right\rangle ,\bar{X}\right) }{\left\langle \hat{K}%
\right\rangle \left\Vert \hat{\Psi}\right\Vert ^{2}\hat{D}\left( \hat{X}%
\right) }\left\langle A\right\rangle
\end{eqnarray*}%
\begin{equation*}
\rightarrow -\frac{\underline{\hat{k}}_{1}^{B}\left( \left\langle \hat{X}%
\right\rangle ,\bar{X}\right) \left( \hat{k}\left( \hat{X}^{\prime },\hat{X}%
\right) -\left\langle \hat{k}\left( \hat{X}^{\prime },\hat{X}\right)
\right\rangle \right) \hat{K}}{\left\langle \hat{K}\right\rangle \left\Vert 
\hat{\Psi}\right\Vert ^{2}}\left\langle A\right\rangle +\frac{\hat{K}%
\underline{\hat{k}}_{1}^{B}\left( \hat{X},\bar{X}\right) }{\left\langle \hat{%
K}\right\rangle \left\Vert \hat{\Psi}\right\Vert ^{2}}A-\frac{\hat{K}%
\underline{\hat{k}}_{1}^{B}\left( \left\langle \hat{X}\right\rangle ,\bar{X}%
\right) }{\left\langle \hat{K}\right\rangle \left\Vert \hat{\Psi}\right\Vert
^{2}}\left\langle A\right\rangle
\end{equation*}

\paragraph*{A25.1.2.2 Summing terms}

\begin{eqnarray}
&&\frac{\delta }{\delta \left\vert \hat{\Psi}\left( \hat{K},\hat{X}\right)
\right\vert ^{2}}\bar{g}\left( \hat{K}^{\prime },\hat{X}^{\prime }\right) \\
&\simeq &\left\{ \left( \Delta \left( \bar{X}^{\prime },\bar{X}\right) -%
\frac{\bar{K}^{\prime }\bar{k}_{1}\left( \bar{X}^{\prime },\bar{X}\right)
\left\vert \bar{\Psi}\left( \bar{K}^{\prime },\bar{X}^{\prime }\right)
\right\vert ^{2}}{1+\underline{\bar{k}}\left( \bar{X}^{\prime }\right) }%
\right) \frac{\left( 1-\bar{M}\right) }{1+\underline{\overline{\bar{k}}}%
_{2}\left( \bar{X}^{\prime }\right) }\right\} ^{-1}  \notag \\
&&\times \left[ \frac{\underline{\hat{k}}_{1}^{B}\left( \hat{X},\bar{X}%
\right) \hat{K}A}{\left\langle \hat{K}\right\rangle \left\Vert \hat{\Psi}%
\right\Vert ^{2}}-\frac{\underline{\hat{k}}_{1}^{B}\left( \left\langle \hat{X%
}\right\rangle ,\bar{X}\right) \left( 1+\left( \hat{k}\left( \hat{X}^{\prime
},\hat{X}\right) -\left\langle \hat{k}\left( \hat{X}^{\prime },\hat{X}%
\right) \right\rangle \right) \right) \hat{K}\left\langle A\right\rangle }{%
\left\Vert \hat{\Psi}\right\Vert ^{2}\left\langle \hat{K}\right\rangle }%
\right.  \notag \\
&&+\frac{\hat{K}^{\prime }\underline{\hat{k}}_{1}^{B}\left( \hat{X}^{\prime
},\bar{X}\right) \left\vert \hat{\Psi}\left( \hat{K}^{\prime },\hat{X}%
^{\prime }\right) \right\vert ^{2}}{\hat{D}\left( \hat{X}^{\prime }\right) }%
\left( 1-\left\langle \hat{k}_{1}\left( \hat{X}^{\prime },\hat{X}\right)
\right\rangle \right) ^{-1}  \notag \\
&&\left. \times \left( \frac{\hat{K}_{1}\hat{k}_{1}\left( \hat{X}_{1},\hat{X}%
\right) }{\left\langle \hat{K}\right\rangle \left\Vert \hat{\Psi}\right\Vert
^{2}}A-\frac{\left\langle \hat{k}_{1}\left( \hat{X}^{\prime },\hat{X}\right)
\right\rangle \left( 1+\left( \hat{k}\left( \left\langle \hat{X}%
\right\rangle ,\hat{X}_{1}\right) \right) -\left\langle \hat{k}\left( \hat{X}%
,\hat{X}^{\prime }\right) \right\rangle \right) \hat{K}_{1}}{\left\langle 
\hat{K}\right\rangle \left\Vert \hat{\Psi}\right\Vert ^{2}}\left\langle
A\right\rangle \right) \right]  \notag
\end{eqnarray}%
\bigskip

Using the normalzations:%
\begin{equation*}
1+\underline{\bar{k}}_{2}\left( \bar{X}^{\prime }\right) \rightarrow \frac{%
1-\left\langle \bar{k}_{1}\left( \bar{X}^{\prime },\bar{X}\right)
\right\rangle }{1-\left\langle \bar{k}\left( \bar{X}^{\prime },\bar{X}%
\right) \right\rangle }\left( 1+\frac{\underline{\bar{k}}_{2}\left( \hat{X}%
^{\prime }\right) }{1-\left\langle \hat{k}_{1}\left( \hat{X}^{\prime },\hat{X%
}\right) \right\rangle }\right)
\end{equation*}%
\begin{equation*}
\frac{1}{1+\underline{\overline{\bar{k}}}_{2}\left( \bar{X}^{\prime }\right) 
}\rightarrow \frac{\left( 1-\left\langle \bar{k}\left( \hat{X}^{\prime },%
\hat{X}\right) \right\rangle \right) }{\left( 1-\left\langle \bar{k}%
_{1}\left( \hat{X}^{\prime },\hat{X}\right) \right\rangle \right) \left( 1+%
\frac{\underline{\bar{k}}_{2}\left( \hat{X}^{\prime }\right) }{%
1-\left\langle \bar{k}_{1}\left( \hat{X}^{\prime },\hat{X}\right)
\right\rangle }\right) }
\end{equation*}%
\begin{equation*}
\frac{1}{1+\underline{\overline{\bar{k}}}_{2}\left( \bar{X}^{\prime }\right) 
}\rightarrow \frac{\left( 1-\left\langle \bar{k}\right\rangle \right) }{%
\left( 1-\left\langle \bar{k}_{1}\right\rangle \right) }
\end{equation*}%
The derivative, when inserted in matrices in matrices elements writes:%
\begin{eqnarray}
&&\frac{\delta }{\delta \left\vert \hat{\Psi}\left( \hat{K},\hat{X}\right)
\right\vert ^{2}}\bar{g}\left( \hat{K}^{\prime },\hat{X}^{\prime }\right) \\
&\simeq &\left\{ \left( 1-\left\langle \bar{k}\left( \bar{X}^{\prime },\bar{X%
}\right) \right\rangle \right) ^{2}\right\} ^{-1}  \notag \\
&&\times \left[ \frac{\underline{\hat{k}}_{1}^{B}\left( \hat{X},\left\langle 
\bar{X}\right\rangle \right) \hat{K}}{\left\langle \hat{K}\right\rangle
\left\Vert \hat{\Psi}\right\Vert ^{2}}A-\frac{\left\langle \underline{\hat{k}%
}_{1}^{B}\right\rangle \left( 1+\left( \hat{k}\left( \hat{X}^{\prime },\hat{X%
}\right) -\left\langle \hat{k}\right\rangle \right) \right) \hat{K}}{%
\left\Vert \hat{\Psi}\right\Vert ^{2}\left\langle \hat{K}\right\rangle }%
\left\langle A\right\rangle \right.  \notag \\
&&\left. +\left\langle \underline{\hat{k}}_{1}^{B}\right\rangle \left\{
\left( 1-\left\langle \hat{k}_{1}\right\rangle \right) ^{-1}\left( \frac{%
\hat{K}_{1}\hat{k}_{1}\left( \hat{X}_{1},\left\langle \hat{X}\right\rangle
\right) }{\left\langle \hat{K}\right\rangle \left\Vert \hat{\Psi}\right\Vert
^{2}}A-\frac{\left\langle \hat{k}_{1}\right\rangle \left( 1+\left( \hat{k}%
\left( \hat{X}_{1},\hat{X}\right) -\left\langle \hat{k}\right\rangle \right)
\right) \hat{K}_{1}}{\left\langle \hat{K}\right\rangle \left\Vert \hat{\Psi}%
\right\Vert ^{2}}\left\langle A\right\rangle \right) \right\} \right]  \notag
\end{eqnarray}%
which is in averagzs:%
\begin{equation*}
\rightarrow \frac{\left( \left( \underline{\hat{k}}_{1}^{B}\left( \hat{X}%
,\left\langle \bar{X}\right\rangle \right) +\left\langle \underline{\hat{k}}%
_{1}^{B}\right\rangle \frac{\hat{k}_{1}\left( \hat{X}_{1},\left\langle \hat{X%
}\right\rangle \right) }{1-\left\langle \hat{k}_{1}\right\rangle }\right)
A-\left\langle \underline{\hat{k}}_{1}^{B}\left( \hat{X},\left\langle \bar{X}%
\right\rangle \right) +\left\langle \underline{\hat{k}}_{1}^{B}\right\rangle 
\frac{\hat{k}_{1}\left( \hat{X}_{1},\left\langle \hat{X}\right\rangle
\right) }{1-\left\langle \hat{k}_{1}\right\rangle }\right\rangle
\left\langle A\right\rangle \right) \hat{K}}{\left( 1-\left\langle \bar{k}%
\left( \bar{X}^{\prime },\bar{X}\right) \right\rangle \right)
^{2}\left\langle \hat{K}\right\rangle \left\Vert \hat{\Psi}\right\Vert ^{2}}-
\end{equation*}

\subsubsection*{A25.1.3 Estimation of $\frac{\protect\delta }{\protect\delta %
\left\vert \hat{\Psi}\left( \hat{K},\hat{X}\right) \right\vert ^{2}}\hat{g}%
\left( \hat{K}^{\prime },\hat{X}^{\prime }\right) $}

We start with:%
\begin{equation*}
A=\frac{\left( 1-\hat{M}\right) \hat{g}\left( \hat{K}^{\prime },\hat{X}%
^{\prime }\right) +\bar{N}\bar{g}\left( \hat{K}^{\prime },\hat{X}^{\prime
}\right) }{1+\underline{\hat{k}}_{2}\left( \bar{X}^{\prime }\right) +\kappa %
\left[ \frac{\underline{\hat{k}}_{2}^{B}}{1+\bar{k}}\right] \left( \hat{X}%
^{\prime }\right) }
\end{equation*}%
and:%
\begin{equation*}
\hat{g}\left( \hat{K},\hat{X}\right) =\left( 1-\hat{M}\right) ^{-1}\left(
\left( 1+\underline{\hat{k}}_{2}\left( \bar{X}^{\prime }\right) +\kappa %
\left[ \frac{\underline{\hat{k}}_{2}^{B}}{1+\bar{k}}\right] \left( \hat{X}%
^{\prime }\right) \right) A-\bar{N}\bar{g}\left( \hat{K}^{\prime },\hat{X}%
^{\prime }\right) \right)
\end{equation*}%
The derivative decomposes as:%
\begin{eqnarray*}
&&\frac{\delta }{\delta \left\vert \hat{\Psi}\left( \hat{K},\hat{X}\right)
\right\vert ^{2}}\hat{g}\left( \hat{K},\hat{X}\right) \\
&=&\left( 1-\hat{M}\right) ^{-1}\frac{\delta }{\delta \left\vert \hat{\Psi}%
\left( \hat{K},\hat{X}\right) \right\vert ^{2}}\hat{M}\hat{g}\left( \hat{K},%
\hat{X}\right) \\
&&+\left( 1-\hat{M}\right) ^{-1}\left( \frac{\delta }{\delta \left\vert \hat{%
\Psi}\left( \hat{K},\hat{X}\right) \right\vert ^{2}}\left( 1+\underline{\hat{%
k}}_{2}\left( \bar{X}^{\prime }\right) +\kappa \left[ \frac{\underline{\hat{k%
}}_{2}^{B}}{1+\bar{k}}\right] \left( \hat{X}^{\prime }\right) \right) A-%
\frac{\delta }{\delta \left\vert \hat{\Psi}\left( \hat{K},\hat{X}\right)
\right\vert ^{2}}\left( \bar{N}\bar{g}\left( \hat{K}^{\prime },\hat{X}%
^{\prime }\right) \right) \right)
\end{eqnarray*}%
and in avrage this is:%
\begin{eqnarray}
&&\left( 1-\hat{M}\right) ^{-1}\left( \frac{\delta }{\delta \left\vert \hat{%
\Psi}\left( \hat{K},\hat{X}\right) \right\vert ^{2}}\hat{M}\right) \hat{g}%
\left( \hat{K},\hat{X}\right)  \label{FRL} \\
&&+\frac{\frac{\delta }{\delta \left\vert \hat{\Psi}\left( \hat{K},\hat{X}%
\right) \right\vert ^{2}}\left( 1+\underline{\hat{k}}_{2}\left( \bar{X}%
^{\prime }\right) +\kappa \left[ \frac{\underline{\hat{k}}_{2}^{B}}{1+\bar{k}%
}\right] \left( \hat{X}^{\prime }\right) \right) }{\left( 1+\underline{\hat{k%
}}_{2}\left( \bar{X}^{\prime }\right) +\kappa \left[ \frac{\underline{\hat{k}%
}_{2}^{B}}{1+\bar{k}}\right] \left( \hat{X}^{\prime }\right) \right) }%
\left\langle \left( 1-\hat{M}\right) ^{-1}\left( 1+\underline{\hat{k}}%
_{2}\left( \bar{X}^{\prime }\right) +\kappa \left[ \frac{\underline{\hat{k}}%
_{2}^{B}}{1+\bar{k}}\right] \left( \hat{X}^{\prime }\right) \right)
A\right\rangle  \notag \\
&&+\left( 1-\hat{M}\right) ^{-1}\left( 1+\underline{\hat{k}}_{2}\left( \bar{X%
}^{\prime }\right) +\kappa \left[ \frac{\underline{\hat{k}}_{2}^{B}}{1+\bar{k%
}}\right] \left( \hat{X}^{\prime }\right) \right) \frac{\delta }{\delta
\left\vert \hat{\Psi}\left( \hat{K},\hat{X}\right) \right\vert ^{2}}A  \notag
\\
&&-\left( 1-\hat{M}\right) ^{-1}\frac{\delta }{\delta \left\vert \hat{\Psi}%
\left( \hat{K},\hat{X}\right) \right\vert ^{2}}\left( \bar{N}\bar{g}\left( 
\hat{K}^{\prime },\hat{X}^{\prime }\right) \right)  \notag
\end{eqnarray}%
The various contributions to (\ref{FRL}) are estimatd as:%
\begin{eqnarray*}
&&\frac{\frac{\delta }{\delta \left\vert \hat{\Psi}\left( \hat{K},\hat{X}%
\right) \right\vert ^{2}}\left( 1+\underline{\hat{k}}_{2}\left( \bar{X}%
^{\prime }\right) +\kappa \left[ \frac{\underline{\hat{k}}_{2}^{B}}{1+\bar{k}%
}\right] \left( \hat{X}^{\prime }\right) \right) }{\left( 1+\underline{\hat{k%
}}_{2}\left( \bar{X}^{\prime }\right) +\kappa \left[ \frac{\underline{\hat{k}%
}_{2}^{B}}{1+\bar{k}}\right] \left( \hat{X}^{\prime }\right) \right) }%
\left\langle \left( 1-\hat{M}\right) ^{-1}\left( 1+\underline{\hat{k}}%
_{2}\left( \bar{X}^{\prime }\right) +\kappa \left[ \frac{\underline{\hat{k}}%
_{2}^{B}}{1+\bar{k}}\right] \left( \hat{X}^{\prime }\right) \right)
A\right\rangle \\
&=&\frac{\frac{\delta }{\delta \left\vert \hat{\Psi}\left( \hat{K},\hat{X}%
\right) \right\vert ^{2}}\left( 1+\underline{\hat{k}}_{2}\left( \bar{X}%
^{\prime }\right) +\kappa \left[ \frac{\underline{\hat{k}}_{2}^{B}}{1+\bar{k}%
}\right] \left( \hat{X}^{\prime }\right) \right) }{\left( 1+\underline{\hat{k%
}}_{2}\left( \bar{X}^{\prime }\right) +\kappa \left[ \frac{\underline{\hat{k}%
}_{2}^{B}}{1+\bar{k}}\right] \left( \hat{X}^{\prime }\right) \right) }\left(
\left\langle \hat{g}\left( \hat{K},\hat{X}\right) \right\rangle +\left( 1-%
\hat{M}\right) ^{-1}\bar{N}\left\langle \bar{g}\left( \hat{K}^{\prime },\hat{%
X}^{\prime }\right) \right\rangle \right)
\end{eqnarray*}%
\begin{eqnarray*}
&&\left( 1-\hat{M}\right) ^{-1}\left( 1+\underline{\hat{k}}_{2}\left( \bar{X}%
^{\prime }\right) +\kappa \left[ \frac{\underline{\hat{k}}_{2}^{B}}{1+\bar{k}%
}\right] \left( \hat{X}^{\prime }\right) \right) \frac{\delta }{\delta
\left\vert \hat{\Psi}\left( \hat{K},\hat{X}\right) \right\vert ^{2}}A \\
&=&\left( 1-\left\langle \hat{k}\right\rangle \right) ^{-1}\left( 1+%
\underline{\hat{k}}_{2}\left( \bar{X}^{\prime }\right) +\kappa \left[ \frac{%
\underline{\hat{k}}_{2}^{B}}{1+\bar{k}}\right] \left( \hat{X}^{\prime
}\right) \right) \left( 1-\left\langle \hat{k}_{1}\right\rangle \right)
^{-1}\left( \frac{\hat{K}_{1}\hat{k}_{1}\left( \hat{X}_{1},\hat{X}\right) }{%
\left\langle \hat{K}\right\rangle \left\Vert \hat{\Psi}\right\Vert ^{2}}A-%
\frac{\left\langle \hat{k}_{1}\left( \hat{X}^{\prime },\hat{X}\right)
\right\rangle \hat{K}_{1}}{\left\langle \hat{K}\right\rangle \left\Vert \hat{%
\Psi}\right\Vert ^{2}}\left\langle A\right\rangle \right) \\
&\simeq &\left( \left( 1-\left\langle \hat{k}\right\rangle \right) \left(
1-\left\langle \hat{k}_{1}\right\rangle \right) \right) ^{-1}\left( \frac{%
\hat{K}_{1}\hat{k}_{1}\left( \hat{X}_{1},\hat{X}\right) }{\left\langle \hat{K%
}\right\rangle \left\Vert \hat{\Psi}\right\Vert ^{2}}-\frac{\left\langle 
\hat{k}_{1}\left( \hat{X}^{\prime },\hat{X}\right) \right\rangle \hat{K}_{1}%
}{\left\langle \hat{K}\right\rangle \left\Vert \hat{\Psi}\right\Vert ^{2}}%
\right) \\
&&\times \left( \left\langle \hat{g}\left( \hat{K},\hat{X}\right)
\right\rangle +\left( 1-\hat{M}\right) ^{-1}\bar{N}\left\langle \bar{g}%
\left( \hat{K}^{\prime },\hat{X}^{\prime }\right) \right\rangle \right)
\end{eqnarray*}%
and gathering the contributions leads to:%
\begin{eqnarray*}
&&\frac{\delta }{\delta \left\vert \hat{\Psi}\left( \hat{K},\hat{X}\right)
\right\vert ^{2}}\hat{g}\left( \hat{K},\hat{X}\right) \\
&=&\left( 1-\hat{M}\right) ^{-1}\frac{\delta }{\delta \left\vert \hat{\Psi}%
\left( \hat{K},\hat{X}\right) \right\vert ^{2}}\hat{M}\hat{g}\left( \hat{K},%
\hat{X}\right) \\
&&+\frac{\frac{\delta }{\delta \left\vert \hat{\Psi}\left( \hat{K},\hat{X}%
\right) \right\vert ^{2}}\left( 1+\underline{\hat{k}}_{2}\left( \bar{X}%
^{\prime }\right) +\kappa \left[ \frac{\underline{\hat{k}}_{2}^{B}}{1+\bar{k}%
}\right] \left( \hat{X}^{\prime }\right) \right) }{\left( 1+\underline{\hat{k%
}}_{2}\left( \bar{X}^{\prime }\right) +\kappa \left[ \frac{\underline{\hat{k}%
}_{2}^{B}}{1+\bar{k}}\right] \left( \hat{X}^{\prime }\right) \right) }\left(
\left\langle \hat{g}\left( \hat{K},\hat{X}\right) \right\rangle +\left( 1-%
\hat{M}\right) ^{-1}\bar{N}\left\langle \bar{g}\left( \hat{K}^{\prime },\hat{%
X}^{\prime }\right) \right\rangle \right) \\
&&+\left( \left( 1-\left\langle \hat{k}\right\rangle \right) \left(
1-\left\langle \hat{k}_{1}\right\rangle \right) \right) ^{-1}\left( \frac{%
\hat{K}_{1}\hat{k}_{1}\left( \hat{X}_{1},\hat{X}\right) }{\left\langle \hat{K%
}\right\rangle \left\Vert \hat{\Psi}\right\Vert ^{2}}-\frac{\left\langle 
\hat{k}_{1}\left( \hat{X}^{\prime },\hat{X}\right) \right\rangle \hat{K}_{1}%
}{\left\langle \hat{K}\right\rangle \left\Vert \hat{\Psi}\right\Vert ^{2}}%
\right) \left( \left\langle \hat{g}\left( \hat{K},\hat{X}\right)
\right\rangle +\left( 1-\hat{M}\right) ^{-1}\bar{N}\left\langle \bar{g}%
\left( \hat{K}^{\prime },\hat{X}^{\prime }\right) \right\rangle \right) \\
&&-\left( 1-\hat{M}\right) ^{-1}\left( \frac{\delta }{\delta \left\vert \hat{%
\Psi}\left( \hat{K},\hat{X}\right) \right\vert ^{2}}\left( \bar{N}\bar{g}%
\left( \hat{K}^{\prime },\hat{X}^{\prime }\right) \right) \right)
\end{eqnarray*}%
The successive contributions are computed below.

\paragraph*{A25.1.3.1 Estimtn of}

\begin{equation*}
\frac{\frac{\delta }{\delta \left\vert \hat{\Psi}\left( \hat{K},\hat{X}%
\right) \right\vert ^{2}}\left( 1+\underline{\hat{k}}_{2}\left( \bar{X}%
^{\prime }\right) +\kappa \left[ \frac{\underline{\hat{k}}_{2}^{B}}{1+\bar{k}%
}\right] \left( \hat{X}^{\prime }\right) \right) }{\left( 1+\underline{\hat{k%
}}_{2}\left( \bar{X}^{\prime }\right) +\kappa \left[ \frac{\underline{\hat{k}%
}_{2}^{B}}{1+\bar{k}}\right] \left( \hat{X}^{\prime }\right) \right) }
\end{equation*}%
This term is obtained by developping the contributions involved in its
definitn:%
\begin{eqnarray*}
&&\frac{\delta }{\delta \left\vert \hat{\Psi}\left( \hat{K},\hat{X}\right)
\right\vert ^{2}}\left( 1+\underline{\hat{k}}_{2}\left( \bar{X}^{\prime
}\right) +\kappa \left[ \frac{\underline{\hat{k}}_{2}^{B}}{1+\bar{k}}\right]
\left( \hat{X}^{\prime }\right) \right) \\
&\rightarrow &\frac{\delta }{\delta \left\vert \hat{\Psi}\left( \hat{K},\hat{%
X}\right) \right\vert ^{2}}\frac{\left( 1-\left( \left\langle \hat{k}%
_{1}\left( \hat{X}^{\prime },\hat{X}\right) \right\rangle +\left\langle \hat{%
k}_{1}^{B}\left( \hat{X}^{\prime },\hat{X}\right) \right\rangle \frac{%
\left\Vert \bar{\Psi}\right\Vert ^{2}\left\langle \bar{K}\right\rangle }{%
\left\Vert \hat{\Psi}\right\Vert ^{2}\left\langle \hat{K}\right\rangle }%
\right) \right) \left( 1+\frac{\underline{\hat{k}}_{2}\left( \hat{X}^{\prime
}\right) +\kappa \left[ \frac{\underline{\hat{k}}_{2}^{B}}{1+\bar{k}}\right]
\left( \hat{X}^{\prime }\right) }{1-\left( \left\langle \hat{k}_{1}\left( 
\hat{X}^{\prime },\hat{X}\right) \right\rangle +\left\langle \hat{k}%
_{1}^{B}\left( \hat{X}^{\prime },\hat{X}\right) \right\rangle \frac{%
\left\Vert \bar{\Psi}\right\Vert ^{2}\left\langle \bar{K}\right\rangle }{%
\left\Vert \hat{\Psi}\right\Vert ^{2}\left\langle \hat{K}\right\rangle }%
\right) }\right) }{\left( 1-\left( \left\langle \underline{\hat{k}}\left( 
\hat{X}^{\prime },\hat{X}\right) \right\rangle +\left\langle \underline{\hat{%
k}}_{1}^{B}\left( \hat{X}^{\prime },\bar{X}\right) \right\rangle \frac{%
\left\Vert \bar{\Psi}\right\Vert ^{2}\left\langle \bar{K}\right\rangle }{%
\left\Vert \hat{\Psi}\right\Vert ^{2}\left\langle \hat{K}\right\rangle }%
+\kappa \left\langle \left[ \frac{\underline{\hat{k}}_{2}^{B}\left( \hat{X}%
^{\prime },\bar{X}\right) }{1+\bar{k}}\right] \right\rangle \frac{\left\Vert 
\bar{\Psi}\right\Vert ^{2}\left\langle \bar{K}\right\rangle }{\left\Vert 
\hat{\Psi}\right\Vert ^{2}\left\langle \hat{K}\right\rangle }\right) \right) 
} \\
&=&-\left( \kappa \left\langle \left[ \frac{\underline{\hat{k}}%
_{2}^{B}\left( \hat{X}^{\prime },\bar{X}\right) }{1+\bar{k}}\right]
\right\rangle \left( 1-\left\langle \underline{\hat{k}}_{1}\left( \hat{X}%
^{\prime },\hat{X}\right) \right\rangle \right) +\left\langle \hat{k}%
_{1}^{B}\left( \hat{X}^{\prime },\hat{X}\right) \right\rangle \left\langle 
\hat{k}_{2}\left( \hat{X}^{\prime },\hat{X}\right) \right\rangle \right) \\
&&\times \frac{\left( 1+\frac{\underline{\hat{k}}_{2}\left( \hat{X}^{\prime
}\right) +\kappa \left[ \frac{\underline{\hat{k}}_{2}^{B}}{1+\bar{k}}\right]
\left( \hat{X}^{\prime }\right) }{1-\left( \left\langle \hat{k}_{1}\left( 
\hat{X}^{\prime },\hat{X}\right) \right\rangle +\left\langle \hat{k}%
_{1}^{B}\left( \hat{X}^{\prime },\hat{X}\right) \right\rangle \frac{%
\left\Vert \bar{\Psi}\right\Vert ^{2}\left\langle \bar{K}\right\rangle }{%
\left\Vert \hat{\Psi}\right\Vert ^{2}\left\langle \hat{K}\right\rangle }%
\right) }\right) \frac{\left\Vert \bar{\Psi}\right\Vert ^{2}\left\langle 
\bar{K}\right\rangle }{\left\Vert \hat{\Psi}\right\Vert ^{2}\left\langle 
\hat{K}\right\rangle }\hat{K}}{\left\Vert \hat{\Psi}\right\Vert
^{2}\left\langle \hat{K}\right\rangle \left( 1-\left( \left\langle 
\underline{\hat{k}}\left( \hat{X}^{\prime },\hat{X}\right) \right\rangle
+\left( \left\langle \underline{\hat{k}}_{1}^{B}\left( \hat{X}^{\prime },%
\bar{X}\right) \right\rangle +\kappa \left\langle \left[ \frac{\underline{%
\hat{k}}_{2}^{B}\left( \hat{X}^{\prime },\bar{X}\right) }{1+\bar{k}}\right]
\right\rangle \right) \frac{\left\Vert \bar{\Psi}\right\Vert
^{2}\left\langle \bar{K}\right\rangle }{\left\Vert \hat{\Psi}\right\Vert
^{2}\left\langle \hat{K}\right\rangle }\right) \right) ^{2}} \\
&&+\frac{\left( \hat{k}_{2}\left( \hat{X}^{\prime },\hat{X}\right)
-\left\langle \hat{k}_{2}\left( \hat{X}^{\prime },\hat{X}\right)
\right\rangle \right) }{\left\Vert \hat{\Psi}\right\Vert ^{2}\left\langle 
\hat{K}\right\rangle \left( 1-\left( \left\langle \underline{\hat{k}}\left( 
\hat{X}^{\prime },\hat{X}\right) \right\rangle +\left\langle \underline{\hat{%
k}}_{1}^{B}\left( \hat{X}^{\prime },\bar{X}\right) \right\rangle \frac{%
\left\Vert \bar{\Psi}\right\Vert ^{2}\left\langle \bar{K}\right\rangle }{%
\left\Vert \hat{\Psi}\right\Vert ^{2}\left\langle \hat{K}\right\rangle }%
+\kappa \left\langle \left[ \frac{\underline{\hat{k}}_{2}^{B}\left( \hat{X}%
^{\prime },\bar{X}\right) }{1+\bar{k}}\right] \right\rangle \frac{\left\Vert 
\bar{\Psi}\right\Vert ^{2}\left\langle \bar{K}\right\rangle }{\left\Vert 
\hat{\Psi}\right\Vert ^{2}\left\langle \hat{K}\right\rangle }\right) \right) 
}
\end{eqnarray*}%
and thus:%
\begin{eqnarray*}
&&\frac{\frac{\delta }{\delta \left\vert \hat{\Psi}\left( \hat{K},\hat{X}%
\right) \right\vert ^{2}}\left( 1+\underline{\hat{k}}_{2}\left( \bar{X}%
^{\prime }\right) +\kappa \left[ \frac{\underline{\hat{k}}_{2}^{B}}{1+\bar{k}%
}\right] \left( \hat{X}^{\prime }\right) \right) }{\left( 1+\underline{\hat{k%
}}_{2}\left( \bar{X}^{\prime }\right) +\kappa \left[ \frac{\underline{\hat{k}%
}_{2}^{B}}{1+\bar{k}}\right] \left( \hat{X}^{\prime }\right) \right) } \\
&\simeq &-\frac{\left( \kappa \left\langle \left[ \frac{\underline{\hat{k}}%
_{2}^{B}}{1+\bar{k}}\right] \right\rangle \left( 1-\left\langle \underline{%
\hat{k}}\right\rangle \right) +\left\langle \hat{k}_{1}^{B}\right\rangle
\left\langle \hat{k}_{2}\right\rangle \right) \frac{\left\Vert \bar{\Psi}%
\right\Vert ^{2}\left\langle \bar{K}\right\rangle }{\left\Vert \hat{\Psi}%
\right\Vert ^{2}\left\langle \hat{K}\right\rangle }\frac{\hat{K}}{%
\left\langle \hat{K}\right\rangle }}{\left( 1-\left( \left\langle \hat{k}%
_{1}\right\rangle +\left\langle \hat{k}_{1}^{B}\right\rangle \frac{%
\left\Vert \bar{\Psi}\right\Vert ^{2}\left\langle \bar{K}\right\rangle }{%
\left\Vert \hat{\Psi}\right\Vert ^{2}\left\langle \hat{K}\right\rangle }%
\right) \right) \left( 1-\left( \left\langle \underline{\hat{k}}%
\right\rangle +\left\langle \underline{\hat{k}}_{1}^{B}\right\rangle \frac{%
\left\Vert \bar{\Psi}\right\Vert ^{2}\left\langle \bar{K}\right\rangle }{%
\left\Vert \hat{\Psi}\right\Vert ^{2}\left\langle \hat{K}\right\rangle }%
+\kappa \left\langle \left[ \frac{\underline{\hat{k}}_{2}^{B}}{1+\bar{k}}%
\right] \right\rangle \frac{\left\Vert \bar{\Psi}\right\Vert
^{2}\left\langle \bar{K}\right\rangle }{\left\Vert \hat{\Psi}\right\Vert
^{2}\left\langle \hat{K}\right\rangle }\right) \right) \left\Vert \hat{\Psi}%
\right\Vert ^{2}} \\
&&+\frac{\left( \hat{k}_{2}\left( \hat{X}^{\prime },\hat{X}\right)
-\left\langle \hat{k}_{2}\right\rangle \right) \hat{K}}{\left\Vert \hat{\Psi}%
\right\Vert ^{2}\left\langle \hat{K}\right\rangle \left( 1-\left(
\left\langle \hat{k}_{1}\right\rangle +\left\langle \hat{k}%
_{1}^{B}\right\rangle \frac{\left\Vert \bar{\Psi}\right\Vert
^{2}\left\langle \bar{K}\right\rangle }{\left\Vert \hat{\Psi}\right\Vert
^{2}\left\langle \hat{K}\right\rangle }\right) \right) }
\end{eqnarray*}

\paragraph*{A25.1.3.2 Estimation of}

\begin{equation*}
\left( \frac{\delta }{\delta \left\vert \hat{\Psi}\left( \hat{K},\hat{X}%
\right) \right\vert ^{2}}\hat{M}\right) \hat{g}\left( \hat{K}^{\prime },\hat{%
X}^{\prime }\right)
\end{equation*}%
Defining:%
\begin{equation*}
\hat{M}\left( \left( \hat{K},\hat{X}\right) ,\left( \hat{K}^{\prime },\hat{X}%
^{\prime }\right) \right) =\frac{\hat{k}\left( \hat{X},\hat{X}^{\prime
}\right) \hat{K}\left\vert \hat{\Psi}\left( \hat{K}^{\prime },\hat{X}%
^{\prime }\right) \right\vert ^{2}}{\hat{D}\left( \hat{X}\right) }
\end{equation*}%
Given normalztions:%
\begin{eqnarray*}
\hat{D}\left( \hat{X}\right) &=&1+\int \frac{\hat{k}\left( \hat{X},\hat{X}%
^{\prime }\right) -\left\langle \hat{k}\left( \hat{X},\hat{X}^{\prime
}\right) \right\rangle }{\left\langle \hat{K}\right\rangle \left\Vert \hat{%
\Psi}\right\Vert ^{2}}\hat{K}^{\prime }\left\vert \hat{\Psi}\left( \hat{K}%
^{\prime },\hat{X}^{\prime }\right) \right\vert ^{2}+\int \frac{\hat{k}%
_{1}^{B}\left( \hat{X},\bar{X}^{\prime }\right) -\left\langle \hat{k}%
_{1}^{B}\left( \hat{X},\bar{X}^{\prime }\right) \right\rangle }{\left\langle 
\hat{K}\right\rangle \left\Vert \hat{\Psi}\right\Vert ^{2}}\bar{K}%
_{0}^{\prime }\left\vert \bar{\Psi}\left( \bar{K}_{0}^{\prime },\bar{X}%
^{\prime }\right) \right\vert ^{2} \\
&&+\kappa \int \frac{\hat{k}_{2}^{B}\left( \hat{X},\bar{X}^{\prime }\right)
-\left\langle \hat{k}_{2}^{B}\left( \hat{X},\bar{X}^{\prime }\right)
\right\rangle }{\left\langle \hat{K}\right\rangle \left\Vert \hat{\Psi}%
\right\Vert ^{2}}\frac{\bar{K}_{0}^{\prime }\left\vert \bar{\Psi}\left( \bar{%
K}_{0}^{\prime },\bar{X}^{\prime }\right) \right\vert ^{2}}{1+\int \bar{k}%
\left( \bar{X}^{\prime },\bar{X}^{\prime \prime }\right) \bar{K}_{0}^{\prime
\prime }\left\vert \bar{\Psi}\left( \bar{K}_{0}^{\prime \prime },\bar{X}%
^{\prime \prime }\right) \right\vert ^{2}}
\end{eqnarray*}%
and:%
\begin{equation*}
\hat{k}\left( \hat{X},\hat{X}^{\prime }\right) \rightarrow \frac{\hat{k}%
\left( \hat{X},\hat{X}^{\prime }\right) }{\left\langle \hat{K}\right\rangle
\left\Vert \hat{\Psi}\right\Vert ^{2}}=\frac{\hat{k}\left( \hat{X},\hat{X}%
^{\prime }\right) }{\int \hat{K}\left\vert \hat{\Psi}\left( \hat{K},\hat{X}%
\right) \right\vert ^{2}}
\end{equation*}%
In average, this leads to:%
\begin{equation*}
\frac{\delta }{\delta \left\vert \hat{\Psi}\left( \hat{K}_{1},\hat{X}%
_{1}\right) \right\vert ^{2}}\left( \frac{1}{\hat{D}\left( \hat{X}\right) }%
\right) \rightarrow -\frac{\left( \hat{k}\left( \hat{X},\hat{X}_{1}\right)
-\left\langle \hat{k}\left( \hat{X},\hat{X}^{\prime }\right) \right\rangle
\right) \hat{K}_{1}}{\left\langle \hat{K}\right\rangle \left\Vert \hat{\Psi}%
\right\Vert ^{2}\left( \hat{D}\left( \hat{X}\right) \right) ^{2}}
\end{equation*}

\begin{eqnarray*}
\frac{\delta \hat{M}}{\delta \left\vert \hat{\Psi}\left( \hat{K}_{1},\hat{X}%
_{1}\right) \right\vert ^{2}} &=&\frac{\hat{k}\left( \hat{X},\hat{X}%
_{1}\right) \hat{K}}{\hat{D}\left\langle \hat{K}\right\rangle \left\Vert 
\hat{\Psi}\right\Vert ^{2}}-\int \frac{\hat{k}\left( \hat{X},\hat{X}^{\prime
}\right) \hat{K}^{\prime }\left\vert \hat{\Psi}\left( \hat{K}^{\prime },\hat{%
X}^{\prime }\right) \right\vert ^{2}}{\hat{D}\left\langle \hat{K}%
\right\rangle \left\Vert \hat{\Psi}\right\Vert ^{2}}\frac{\hat{K}_{1}}{%
\left( \int \hat{K}\left\vert \hat{\Psi}\left( \hat{K},\hat{X}\right)
\right\vert \right) } \\
&&-\int \frac{\hat{k}\left( \hat{X},\hat{X}^{\prime }\right) \hat{K}^{\prime
}\left\vert \hat{\Psi}\left( \hat{K}^{\prime },\hat{X}^{\prime }\right)
\right\vert ^{2}}{\hat{D}\left\langle \hat{K}\right\rangle \left\Vert \hat{%
\Psi}\right\Vert ^{2}}\frac{\left( \hat{k}\left( \hat{X},\hat{X}_{1}\right)
-\left\langle \hat{k}\left( \hat{X},\hat{X}^{\prime }\right) \right\rangle
\right) \hat{K}_{1}}{\left\langle \hat{K}\right\rangle \left\Vert \hat{\Psi}%
\right\Vert ^{2}\hat{D}\left( \hat{X}\right) }
\end{eqnarray*}%
\begin{eqnarray*}
&&\frac{\hat{k}\left( \left\langle \hat{X}\right\rangle ,\hat{X}_{1}\right)
\left\langle \hat{K}\right\rangle }{\hat{D}\left\langle \hat{K}\right\rangle
\left\Vert \hat{\Psi}\right\Vert ^{2}}-\frac{\left\langle \hat{k}\left( \hat{%
X},\hat{X}^{\prime }\right) \right\rangle }{\hat{D}}\frac{\hat{K}_{1}}{%
\left\langle \hat{K}\right\rangle \left\Vert \hat{\Psi}\right\Vert ^{2}} \\
&&-\frac{\left\langle \hat{k}\left( \hat{X},\hat{X}^{\prime }\right)
\right\rangle }{\hat{D}}\frac{\left( \hat{k}\left( \left\langle \hat{X}%
\right\rangle ,\hat{X}_{1}\right) -\left\langle \hat{k}\left( \hat{X},\hat{X}%
^{\prime }\right) \right\rangle \right) \hat{K}_{1}}{\left\langle \hat{K}%
\right\rangle \left\Vert \hat{\Psi}\right\Vert ^{2}\hat{D}\left( \hat{X}%
\right) }
\end{eqnarray*}%
Given that in average:%
\begin{equation*}
\hat{D}\left( \hat{X}\right) \rightarrow \hat{D}\left( \left\langle \hat{X}%
\right\rangle \right) \simeq 1
\end{equation*}%
The contribution for:%
\begin{equation*}
\left( \frac{\delta }{\delta \left\vert \hat{\Psi}\left( \hat{K},\hat{X}%
\right) \right\vert ^{2}}\hat{M}\right) \hat{g}\left( \hat{K}^{\prime },\hat{%
X}^{\prime }\right)
\end{equation*}%
is given:%
\begin{eqnarray*}
&&\left( \frac{\hat{k}\left( \left\langle \hat{X}\right\rangle ,\hat{X}%
_{1}\right) }{\left\Vert \hat{\Psi}\right\Vert ^{2}}-\frac{\left\langle \hat{%
k}\left( \hat{X},\hat{X}^{\prime }\right) \right\rangle \hat{K}_{1}}{%
\left\langle \hat{K}\right\rangle \left\Vert \hat{\Psi}\right\Vert ^{2}}-%
\frac{\left\langle \hat{k}\left( \hat{X},\hat{X}^{\prime }\right)
\right\rangle \left( \hat{k}\left( \left\langle \hat{X}\right\rangle ,\hat{X}%
_{1}\right) -\left\langle \hat{k}\left( \hat{X},\hat{X}^{\prime }\right)
\right\rangle \right) \hat{K}_{1}}{\left\langle \hat{K}\right\rangle
\left\Vert \hat{\Psi}\right\Vert ^{2}}\right) \\
&=&\left( \frac{\hat{k}\left( \left\langle \hat{X}\right\rangle ,\hat{X}%
_{1}\right) }{\left\Vert \hat{\Psi}\right\Vert ^{2}}-\frac{\left\langle \hat{%
k}\left( \hat{X},\hat{X}^{\prime }\right) \right\rangle \left( 1+\left( \hat{%
k}\left( \left\langle \hat{X}\right\rangle ,\hat{X}_{1}\right) \right)
-\left\langle \hat{k}\left( \hat{X},\hat{X}^{\prime }\right) \right\rangle
\right) \hat{K}_{1}}{\left\langle \hat{K}\right\rangle \left\Vert \hat{\Psi}%
\right\Vert ^{2}}\right)
\end{eqnarray*}

\paragraph*{A25.1.3.3 Estimation of:}

\begin{equation*}
\frac{\delta }{\delta \left\vert \hat{\Psi}\left( \hat{K},\hat{X}\right)
\right\vert ^{2}}\left( \bar{N}\bar{g}\left( \hat{K}^{\prime },\hat{X}%
^{\prime }\right) \right)
\end{equation*}%
We use that:%
\begin{equation*}
\bar{N}\rightarrow \frac{\left( \hat{k}_{1}^{B}\left( \hat{X},\bar{X}%
^{\prime }\right) +\kappa \frac{\hat{k}_{2}^{B}\left( \hat{X},\bar{X}%
^{\prime }\right) }{1+\overline{\bar{k}}\left( \bar{X}^{\prime
},\left\langle \bar{X}^{\prime \prime }\right\rangle \right) }-\kappa \int 
\frac{\hat{k}_{2}^{B}\left( \bar{X},\bar{X}^{\prime \prime }\right) \bar{K}%
_{0}^{\prime \prime }\overline{\bar{k}}\left( \bar{X}^{\prime \prime },\bar{X%
}^{\prime }\right) }{\left( 1+\overline{\bar{k}}\left( \bar{X}^{\prime
},\left\langle \bar{X}^{\prime \prime }\right\rangle \right) \right) ^{2}}%
\right) \hat{K}}{1+\int \hat{k}\left( \hat{X},\hat{X}^{\prime }\right)
\left\vert \hat{\Psi}\left( \hat{K}^{\prime },\hat{X}^{\prime }\right)
\right\vert ^{2}+\int \hat{k}_{1}^{B}\left( \hat{X},\bar{X}^{\prime }\right) 
\bar{K}_{0}^{\prime }\left\vert \bar{\Psi}\left( \bar{K}_{0}^{\prime },\bar{X%
}^{\prime }\right) \right\vert ^{2}+\kappa \int \hat{k}_{2}^{B}\left( \hat{X}%
,\bar{X}^{\prime }\right) \frac{\bar{K}_{0}^{\prime }\left\vert \bar{\Psi}%
\left( \bar{K}_{0}^{\prime },\bar{X}^{\prime }\right) \right\vert ^{2}}{%
1+\int \overline{\bar{k}}\left( \bar{X}^{\prime },\bar{X}^{\prime \prime
}\right) \bar{K}_{0}^{\prime \prime }\left\vert \bar{\Psi}\left( \bar{K}%
_{0}^{\prime \prime },\bar{X}^{\prime \prime }\right) \right\vert ^{2}}}
\end{equation*}

Given the normalztn, this reduces to:%
\begin{equation*}
\bar{N}\rightarrow \left\langle \hat{k}_{1}^{B}\left( \hat{X},\bar{X}%
^{\prime }\right) \right\rangle +\kappa \frac{\left\langle \hat{k}%
_{2}^{B}\left( \hat{X},\bar{X}^{\prime }\right) \right\rangle }{%
1+\left\langle \overline{\bar{k}}\left( \bar{X}^{\prime },\left\langle \bar{X%
}^{\prime \prime }\right\rangle \right) \right\rangle }\left( 1-\frac{%
\left\langle \overline{\bar{k}}\left( \bar{X}^{\prime \prime },\bar{X}%
^{\prime }\right) \right\rangle }{\left( 1+\left\langle \overline{\bar{k}}%
\left( \bar{X}^{\prime },\left\langle \bar{X}^{\prime \prime }\right\rangle
\right) \right\rangle \right) ^{2}}\right)
\end{equation*}%
so that:%
\begin{eqnarray*}
&&\frac{\delta }{\delta \left\vert \hat{\Psi}\left( \hat{K},\hat{X}\right)
\right\vert ^{2}}\bar{N}\bar{g}\left( \hat{K}^{\prime },\hat{X}^{\prime
}\right) \\
&\rightarrow &\frac{\delta }{\delta \left\vert \hat{\Psi}\left( \hat{K},\hat{%
X}\right) \right\vert ^{2}}\frac{\left( \hat{k}_{1}^{B}\left( \hat{X}%
^{\prime },\bar{X}^{\prime }\right) +\kappa \frac{\hat{k}_{2}^{B}\left( \hat{%
X}^{\prime },\bar{X}^{\prime }\right) }{1+\overline{\bar{k}}\left( \bar{X}%
^{\prime },\left\langle \bar{X}^{\prime \prime }\right\rangle \right) }%
-\kappa \frac{\hat{k}_{2}^{B}\left( \bar{X}^{\prime },\bar{X}^{\prime \prime
}\right) \bar{K}_{0}^{\prime \prime }\overline{\bar{k}}\left( \bar{X}%
^{\prime \prime },\bar{X}^{\prime }\right) }{\left( 1+\overline{\bar{k}}%
\left( \bar{X}^{\prime },\left\langle \bar{X}^{\prime \prime }\right\rangle
\right) \right) ^{2}}\right) \hat{K}^{\prime }\left\vert \hat{\Psi}\left( 
\hat{K}^{\prime },\hat{X}^{\prime }\right) \right\vert ^{2}}{\hat{D}\left( 
\hat{X}^{\prime }\right) }\bar{g}\left( \hat{K}^{\prime },\hat{X}^{\prime
}\right) \\
&&+\bar{N}\frac{\delta }{\delta \left\vert \hat{\Psi}\left( \hat{K},\hat{X}%
\right) \right\vert ^{2}}\bar{g}\left( \hat{K}^{\prime },\hat{X}^{\prime
}\right)
\end{eqnarray*}%
and the corresponding contribution writes:%
\begin{eqnarray*}
&\rightarrow &-\frac{\left( \hat{k}\left( \hat{X},\hat{X}\right)
-\left\langle \hat{k}\left( \hat{X},\hat{X}^{\prime }\right) \right\rangle
\right) \hat{K}_{1}}{\left\langle \hat{K}\right\rangle \left\Vert \hat{\Psi}%
\right\Vert ^{2}}\left\langle \bar{g}\left( \hat{K}^{\prime },\hat{X}%
^{\prime }\right) \right\rangle \\
&&+\frac{\left( \left( \hat{k}_{1}^{B}\left( \hat{X},\bar{X}^{\prime
}\right) -\hat{k}_{1}^{B}\left( \left\langle \hat{X}\right\rangle ,\bar{X}%
^{\prime }\right) \right) +\kappa \frac{\hat{k}_{2}^{B}\left( \hat{X},\bar{X}%
^{\prime }\right) -\hat{k}_{2}^{B}\left( \left\langle \hat{X}\right\rangle ,%
\bar{X}^{\prime }\right) }{1+\overline{\bar{k}}\left( \bar{X}^{\prime
},\left\langle \bar{X}^{\prime \prime }\right\rangle \right) }\left( 1-\frac{%
1}{1+\overline{\bar{k}}\left( \bar{X}^{\prime },\left\langle \bar{X}^{\prime
\prime }\right\rangle \right) }\right) \right) \left\langle \bar{K}%
\right\rangle \left\Vert \bar{\Psi}\right\Vert ^{2}}{\left\langle \hat{K}%
\right\rangle \left\Vert \hat{\Psi}\right\Vert ^{2}}\left\langle \bar{N}%
\right\rangle \left\langle \bar{g}\left( \hat{K},\hat{X}\right) \right\rangle
\\
&&+\left\langle \bar{N}\right\rangle \frac{\delta }{\delta \left\vert \hat{%
\Psi}\left( \hat{K},\hat{X}\right) \right\vert ^{2}}\bar{g}\left( \hat{K}%
^{\prime },\hat{X}^{\prime }\right)
\end{eqnarray*}

\paragraph*{A25.1.3.4 Estimation of $\frac{\protect\delta }{\protect\delta %
\left\vert \hat{\Psi}\left( \hat{K},\hat{X}\right) \right\vert ^{2}}\hat{g}%
\left( \hat{K},\hat{X}\right) $}

\begin{equation*}
\frac{\delta }{\delta \left\vert \hat{\Psi}\left( \hat{K},\hat{X}\right)
\right\vert ^{2}}\hat{g}\left( \hat{K},\hat{X}\right)
\end{equation*}

We consider th sum of contributions:%
\begin{eqnarray*}
&&\left( 1-\hat{M}\right) ^{-1}\frac{\delta }{\delta \left\vert \hat{\Psi}%
\left( \hat{K},\hat{X}\right) \right\vert ^{2}}\hat{M}\hat{g}\left( \hat{K},%
\hat{X}\right) \\
&&+\frac{\frac{\delta }{\delta \left\vert \hat{\Psi}\left( \hat{K},\hat{X}%
\right) \right\vert ^{2}}\left( 1+\underline{\hat{k}}_{2}\left( \bar{X}%
^{\prime }\right) +\kappa \left[ \frac{\underline{\hat{k}}_{2}^{B}}{1+\bar{k}%
}\right] \left( \hat{X}^{\prime }\right) \right) }{\left( 1+\underline{\hat{k%
}}_{2}\left( \bar{X}^{\prime }\right) +\kappa \left[ \frac{\underline{\hat{k}%
}_{2}^{B}}{1+\bar{k}}\right] \left( \hat{X}^{\prime }\right) \right) }\left(
\left\langle \hat{g}\left( \hat{K},\hat{X}\right) \right\rangle +\left( 1-%
\hat{M}\right) ^{-1}\bar{N}\left\langle \bar{g}\left( \hat{K}^{\prime },\hat{%
X}^{\prime }\right) \right\rangle \right) \\
&&+\left( \left( 1-\left\langle \hat{k}\right\rangle \right) \left(
1-\left\langle \hat{k}_{1}\right\rangle \right) \right) ^{-1}\left( \frac{%
\hat{K}_{1}\hat{k}_{1}\left( \hat{X}_{1},\hat{X}\right) }{\left\langle \hat{K%
}\right\rangle \left\Vert \hat{\Psi}\right\Vert ^{2}}-\frac{\left\langle 
\hat{k}_{1}\left( \hat{X}^{\prime },\hat{X}\right) \right\rangle \hat{K}_{1}%
}{\left\langle \hat{K}\right\rangle \left\Vert \hat{\Psi}\right\Vert ^{2}}%
\right) \left( \left\langle \hat{g}\left( \hat{K},\hat{X}\right)
\right\rangle +\left( 1-\hat{M}\right) ^{-1}\bar{N}\left\langle \bar{g}%
\left( \hat{K}^{\prime },\hat{X}^{\prime }\right) \right\rangle \right) \\
&&-\left( 1-\hat{M}\right) ^{-1}\left( \frac{\delta }{\delta \left\vert \hat{%
\Psi}\left( \hat{K},\hat{X}\right) \right\vert ^{2}}\left( \bar{N}\bar{g}%
\left( \hat{K}^{\prime },\hat{X}^{\prime }\right) \right) \right)
\end{eqnarray*}%
leadng t:%
\begin{eqnarray*}
&\simeq &-\frac{\left( \kappa \left\langle \left[ \frac{\underline{\hat{k}}%
_{2}^{B}}{1+\bar{k}}\right] \right\rangle \left( 1-\left\langle \underline{%
\hat{k}}\right\rangle \right) +\left\langle \hat{k}_{1}^{B}\right\rangle
\left\langle \hat{k}_{2}\right\rangle \right) \frac{\left\Vert \bar{\Psi}%
\right\Vert ^{2}\left\langle \bar{K}\right\rangle }{\left\Vert \hat{\Psi}%
\right\Vert ^{2}\left\langle \hat{K}\right\rangle }\frac{\hat{K}}{%
\left\langle \hat{K}\right\rangle }}{\left( 1-\left( \left\langle \hat{k}%
_{1}\right\rangle +\left\langle \hat{k}_{1}^{B}\right\rangle \frac{%
\left\Vert \bar{\Psi}\right\Vert ^{2}\left\langle \bar{K}\right\rangle }{%
\left\Vert \hat{\Psi}\right\Vert ^{2}\left\langle \hat{K}\right\rangle }%
\right) \right) \left( 1-\left( \left\langle \underline{\hat{k}}%
\right\rangle +\left\langle \underline{\hat{k}}_{1}^{B}\right\rangle \frac{%
\left\Vert \bar{\Psi}\right\Vert ^{2}\left\langle \bar{K}\right\rangle }{%
\left\Vert \hat{\Psi}\right\Vert ^{2}\left\langle \hat{K}\right\rangle }%
+\kappa \left\langle \left[ \frac{\underline{\hat{k}}_{2}^{B}}{1+\bar{k}}%
\right] \right\rangle \frac{\left\Vert \bar{\Psi}\right\Vert
^{2}\left\langle \bar{K}\right\rangle }{\left\Vert \hat{\Psi}\right\Vert
^{2}\left\langle \hat{K}\right\rangle }\right) \right) \left\Vert \hat{\Psi}%
\right\Vert ^{2}} \\
&&+\frac{\left( \hat{k}_{2}\left( \hat{X}^{\prime },\hat{X}\right)
-\left\langle \hat{k}_{2}\right\rangle \right) \hat{K}}{\left\Vert \hat{\Psi}%
\right\Vert ^{2}\left\langle \hat{K}\right\rangle \left( 1-\left(
\left\langle \hat{k}_{1}\right\rangle +\left\langle \hat{k}%
_{1}^{B}\right\rangle \frac{\left\Vert \bar{\Psi}\right\Vert
^{2}\left\langle \bar{K}\right\rangle }{\left\Vert \hat{\Psi}\right\Vert
^{2}\left\langle \hat{K}\right\rangle }\right) \right) }
\end{eqnarray*}%
which is equal to:%
\begin{eqnarray*}
&=&\left( 1-\left\langle \hat{k}\left( \hat{X},\hat{X}^{\prime }\right)
\right\rangle \right) ^{-1}\left( \frac{\hat{k}\left( \left\langle \hat{X}%
\right\rangle ,\hat{X}_{1}\right) }{\left\Vert \hat{\Psi}\right\Vert ^{2}}-%
\frac{\left\langle \hat{k}\left( \hat{X},\hat{X}^{\prime }\right)
\right\rangle \left( 1+\left( \hat{k}\left( \left\langle \hat{X}%
\right\rangle ,\hat{X}_{1}\right) \right) -\left\langle \hat{k}\left( \hat{X}%
,\hat{X}^{\prime }\right) \right\rangle \right) \hat{K}_{1}}{\left\langle 
\hat{K}\right\rangle \left\Vert \hat{\Psi}\right\Vert ^{2}}\right) \hat{g}%
\left( \hat{K},\hat{X}\right) \\
&&+\frac{\left( \hat{k}_{2}\left( \hat{X}^{\prime },\hat{X}\right)
-\left\langle \hat{k}_{2}\right\rangle \right) \hat{K}_{1}\left(
\left\langle \hat{g}\left( \hat{K},\hat{X}\right) \right\rangle +\left( 1-%
\hat{M}\right) ^{-1}\bar{N}\left\langle \bar{g}\left( \hat{K}^{\prime },\hat{%
X}^{\prime }\right) \right\rangle \right) }{\left\Vert \hat{\Psi}\right\Vert
^{2}\left\langle \hat{K}\right\rangle \left( 1-\left( \left\langle \hat{k}%
_{1}\right\rangle +\left\langle \hat{k}_{1}^{B}\right\rangle \frac{%
\left\Vert \bar{\Psi}\right\Vert ^{2}\left\langle \bar{K}\right\rangle }{%
\left\Vert \hat{\Psi}\right\Vert ^{2}\left\langle \hat{K}\right\rangle }%
\right) \right) } \\
&&-\frac{\left( \kappa \left\langle \left[ \frac{\underline{\hat{k}}_{2}^{B}%
}{1+\bar{k}}\right] \right\rangle \left( 1-\left\langle \underline{\hat{k}}%
\right\rangle \right) +\left\langle \hat{k}_{1}^{B}\right\rangle
\left\langle \hat{k}_{2}\right\rangle \right) \frac{\left\Vert \bar{\Psi}%
\right\Vert ^{2}\left\langle \bar{K}\right\rangle }{\left\Vert \hat{\Psi}%
\right\Vert ^{2}\left\langle \hat{K}\right\rangle }\frac{\hat{K}_{1}}{%
\left\langle \hat{K}\right\rangle }\left( \left\langle \hat{g}\left( \hat{K},%
\hat{X}\right) \right\rangle +\left( 1-\hat{M}\right) ^{-1}\bar{N}%
\left\langle \bar{g}\left( \hat{K}^{\prime },\hat{X}^{\prime }\right)
\right\rangle \right) }{\left( 1-\left( \left\langle \hat{k}%
_{1}\right\rangle +\left\langle \hat{k}_{1}^{B}\right\rangle \frac{%
\left\Vert \bar{\Psi}\right\Vert ^{2}\left\langle \bar{K}\right\rangle }{%
\left\Vert \hat{\Psi}\right\Vert ^{2}\left\langle \hat{K}\right\rangle }%
\right) \right) \left( 1-\left( \left\langle \underline{\hat{k}}%
\right\rangle +\left\langle \underline{\hat{k}}_{1}^{B}\right\rangle \frac{%
\left\Vert \bar{\Psi}\right\Vert ^{2}\left\langle \bar{K}\right\rangle }{%
\left\Vert \hat{\Psi}\right\Vert ^{2}\left\langle \hat{K}\right\rangle }%
+\kappa \left\langle \left[ \frac{\underline{\hat{k}}_{2}^{B}}{1+\bar{k}}%
\right] \right\rangle \frac{\left\Vert \bar{\Psi}\right\Vert
^{2}\left\langle \bar{K}\right\rangle }{\left\Vert \hat{\Psi}\right\Vert
^{2}\left\langle \hat{K}\right\rangle }\right) \right) \left\Vert \hat{\Psi}%
\right\Vert ^{2}} \\
&&+\left( \left( 1-\left\langle \hat{k}\right\rangle \right) \left(
1-\left\langle \hat{k}_{1}\right\rangle \right) \right) ^{-1}\left( \frac{%
\hat{K}_{1}\hat{k}_{1}\left( \hat{X}_{1},\hat{X}\right) }{\left\langle \hat{K%
}\right\rangle \left\Vert \hat{\Psi}\right\Vert ^{2}}-\frac{\left\langle 
\hat{k}_{1}\left( \hat{X}^{\prime },\hat{X}\right) \right\rangle \hat{K}_{1}%
}{\left\langle \hat{K}\right\rangle \left\Vert \hat{\Psi}\right\Vert ^{2}}%
\right) \\
&&\times \left( \left\langle \hat{g}\left( \hat{K},\hat{X}\right)
\right\rangle +\left( 1-\hat{M}\right) ^{-1}\bar{N}\left\langle \bar{g}%
\left( \hat{K}^{\prime },\hat{X}^{\prime }\right) \right\rangle \right) \\
&&-\left( 1-\left\langle \hat{k}\left( \hat{X},\hat{X}^{\prime }\right)
\right\rangle \right) ^{-1}\left( \frac{\delta }{\delta \left\vert \hat{\Psi}%
\left( \hat{K},\hat{X}\right) \right\vert ^{2}}\left( \bar{N}\bar{g}\left( 
\hat{K}^{\prime },\hat{X}^{\prime }\right) \right) \right)
\end{eqnarray*}%
Ultimately, we use that:%
\begin{eqnarray*}
&&\frac{\delta }{\delta \left\vert \hat{\Psi}\left( \hat{K},\hat{X}\right)
\right\vert ^{2}}\left( \bar{N}\bar{g}\left( \hat{K}^{\prime },\hat{X}%
^{\prime }\right) \right) \\
&\rightarrow &-\frac{\left( \hat{k}\left( \hat{X},\hat{X}\right)
-\left\langle \hat{k}\left( \hat{X},\hat{X}^{\prime }\right) \right\rangle
\right) \hat{K}_{1}}{\left\langle \hat{K}\right\rangle \left\Vert \hat{\Psi}%
\right\Vert ^{2}}\left\langle \bar{g}\left( \hat{K}^{\prime },\hat{X}%
^{\prime }\right) \right\rangle \\
&&+\frac{\left( \left( \hat{k}_{1}^{B}\left( \hat{X},\bar{X}^{\prime
}\right) -\hat{k}_{1}^{B}\left( \left\langle \hat{X}\right\rangle ,\bar{X}%
^{\prime }\right) \right) +\kappa \frac{\hat{k}_{2}^{B}\left( \hat{X},\bar{X}%
^{\prime }\right) -\hat{k}_{2}^{B}\left( \left\langle \hat{X}\right\rangle ,%
\bar{X}^{\prime }\right) }{1+\overline{\bar{k}}\left( \bar{X}^{\prime
},\left\langle \bar{X}^{\prime \prime }\right\rangle \right) }\left( 1-\frac{%
1}{1+\overline{\bar{k}}\left( \bar{X}^{\prime },\left\langle \bar{X}^{\prime
\prime }\right\rangle \right) }\right) \right) \left\langle \bar{K}%
\right\rangle \left\Vert \bar{\Psi}\right\Vert ^{2}}{\left\langle \hat{K}%
\right\rangle \left\Vert \hat{\Psi}\right\Vert ^{2}}\left\langle \bar{N}%
\right\rangle \left\langle \bar{g}\left( \hat{K},\hat{X}\right) \right\rangle
\\
&&+\left\langle \bar{N}\right\rangle \frac{\delta }{\delta \left\vert \hat{%
\Psi}\left( \hat{K},\hat{X}\right) \right\vert ^{2}}\bar{g}\left( \hat{K}%
^{\prime },\hat{X}^{\prime }\right)
\end{eqnarray*}%
with:%
\begin{eqnarray*}
&&\frac{\delta }{\delta \left\vert \hat{\Psi}\left( \hat{K},\hat{X}\right)
\right\vert ^{2}}\bar{g}\left( \hat{K}^{\prime },\hat{X}^{\prime }\right) \\
&\rightarrow &\frac{\left( \left( \underline{\hat{k}}_{1}^{B}\left( \hat{X}%
,\left\langle \bar{X}\right\rangle \right) +\left\langle \underline{\hat{k}}%
_{1}^{B}\right\rangle \frac{\hat{k}_{1}\left( \hat{X}_{1},\left\langle \hat{X%
}\right\rangle \right) }{1-\left\langle \hat{k}_{1}\right\rangle }\right)
A-\left\langle \underline{\hat{k}}_{1}^{B}\left( \hat{X},\left\langle \bar{X}%
\right\rangle \right) +\left\langle \underline{\hat{k}}_{1}^{B}\right\rangle 
\frac{\hat{k}_{1}\left( \hat{X}_{1},\left\langle \hat{X}\right\rangle
\right) }{1-\left\langle \hat{k}_{1}\right\rangle }\right\rangle
\left\langle A\right\rangle \right) \hat{K}}{\left( 1-\left\langle \bar{k}%
\left( \bar{X}^{\prime },\bar{X}\right) \right\rangle \right)
^{2}\left\langle \hat{K}\right\rangle \left\Vert \hat{\Psi}\right\Vert ^{2}}
\end{eqnarray*}

\paragraph{A25.1.3.5 Dominant contribution}

Considering all contributions, they are all given by deviation around
averages, except:%
\begin{eqnarray*}
&&\frac{\delta }{\delta \left\vert \hat{\Psi}\left( \hat{K},\hat{X}\right)
\right\vert ^{2}}\hat{g}\left( \hat{K},\hat{X}\right) \\
&\simeq &-\frac{\left( \kappa \left\langle \left[ \frac{\underline{\hat{k}}%
_{2}^{B}}{1+\bar{k}}\right] \right\rangle \left( 1-\left\langle \underline{%
\hat{k}}\right\rangle \right) +\left\langle \hat{k}_{1}^{B}\right\rangle
\left\langle \hat{k}_{2}\right\rangle \right) \left( \left\langle \hat{g}%
\left( \hat{K},\hat{X}\right) \right\rangle +\frac{1}{1-\left\langle 
\underline{\hat{k}}\right\rangle }\bar{N}\left\langle \bar{g}\left( \hat{K}%
^{\prime },\hat{X}^{\prime }\right) \right\rangle \right) \frac{\left\Vert 
\bar{\Psi}\right\Vert ^{2}\left\langle \bar{K}\right\rangle }{\left\Vert 
\hat{\Psi}\right\Vert ^{2}\left\langle \hat{K}\right\rangle }\frac{\hat{K}%
_{1}}{\left\langle \hat{K}\right\rangle }}{\left( 1-\left( \left\langle \hat{%
k}_{1}\right\rangle +\left\langle \hat{k}_{1}^{B}\right\rangle \frac{%
\left\Vert \bar{\Psi}\right\Vert ^{2}\left\langle \bar{K}\right\rangle }{%
\left\Vert \hat{\Psi}\right\Vert ^{2}\left\langle \hat{K}\right\rangle }%
\right) \right) \left( 1-\left( \left\langle \underline{\hat{k}}%
\right\rangle +\left( \left\langle \underline{\hat{k}}_{1}^{B}\right\rangle
+\kappa \left\langle \left[ \frac{\underline{\hat{k}}_{2}^{B}}{1+\bar{k}}%
\right] \right\rangle \right) \frac{\left\Vert \bar{\Psi}\right\Vert
^{2}\left\langle \bar{K}\right\rangle }{\left\Vert \hat{\Psi}\right\Vert
^{2}\left\langle \hat{K}\right\rangle }\right) \right) \left\Vert \hat{\Psi}%
\right\Vert ^{2}}
\end{eqnarray*}

\subsection*{A25.2 Derivatives with respect to $\left\vert \bar{\Psi}\left( 
\bar{K},\bar{X}\right) \right\vert ^{2}$}

We use the followng conditions:%
\begin{equation*}
\frac{\delta }{\delta \left\vert \bar{\Psi}\left( \bar{K},\bar{X}\right)
\right\vert ^{2}}\frac{k_{1}\left( X^{\prime },X^{\prime }\right) \left(
f_{1}^{\prime }\left( X^{\prime }\right) K^{\prime }-\bar{C}\left( X^{\prime
}\right) \right) \left\vert \Psi \left( K^{\prime },X^{\prime }\right)
\right\vert ^{2}}{\left( 1+\underline{k}\left( \hat{X}^{\prime }\right) +%
\underline{k}_{1}^{\left( B\right) }\left( \bar{X}^{\prime }\right) +\kappa %
\left[ \frac{\underline{k}_{2}^{B}}{1+\bar{k}}\right] \left( X^{\prime
}\right) \right) \left( 1+\underline{k}_{2}\left( \hat{X}^{\prime }\right)
+\kappa \left[ \frac{\underline{k}_{2}^{B}}{1+\bar{k}}\right] \left(
X^{\prime }\right) \right) }<<1
\end{equation*}%
\begin{equation*}
\frac{\delta }{\delta \left\vert \bar{\Psi}\left( \bar{K},\bar{X}\right)
\right\vert ^{2}}\frac{\underline{k}_{1}^{\left( B\right) }\left( X^{\prime
},\bar{X}\right) }{1+\underline{k}\left( \hat{X}^{\prime }\right) +%
\underline{k}_{1}^{\left( B\right) }\left( \bar{X}^{\prime }\right) +\kappa 
\frac{\underline{k}_{2}^{\left( B\right) }\left( \bar{X}^{\prime }\right) }{%
1+\underline{\bar{k}}}}\frac{\left( f_{1}^{\prime }\left( X^{\prime }\right)
K^{\prime }-\bar{C}\left( X^{\prime }\right) \right) \left\vert \Psi \left(
K^{\prime },X^{\prime }\right) \right\vert ^{2}}{1+\underline{k}_{2}\left( 
\hat{X}^{\prime }\right) +\kappa \frac{\underline{k}_{2}^{\left( B\right)
}\left( \bar{X}^{\prime }\right) }{1+\underline{\bar{k}}}}<<1
\end{equation*}%
\begin{equation*}
\frac{\delta }{\delta \left\vert \bar{\Psi}\left( \bar{K},\bar{X}\right)
\right\vert ^{2}}\left( \Delta \left( \hat{X},\hat{X}^{\prime }\right) -%
\frac{\hat{K}^{\prime }\hat{k}_{1}\left( \hat{X}^{\prime },\hat{X}\right)
\left\vert \hat{\Psi}\left( \hat{K}^{\prime },\hat{X}^{\prime }\right)
\right\vert ^{2}}{1+\underline{\hat{k}}\left( \hat{X}^{\prime }\right) +%
\underline{\hat{k}}_{1}^{B}\left( \bar{X}^{\prime }\right) +\kappa \left[ 
\frac{\underline{\hat{k}}_{2}^{B}}{1+\bar{k}}\right] \left( \hat{X}^{\prime
}\right) }\right) \frac{\left( 1-\hat{M}\right) \hat{g}\left( \hat{K}%
^{\prime },\hat{X}^{\prime }\right) +\bar{N}\bar{g}\left( \hat{K}^{\prime },%
\hat{X}^{\prime }\right) }{1+\underline{\hat{k}}_{2}\left( \bar{X}^{\prime
}\right) +\kappa \left[ \frac{\underline{\hat{k}}_{2}^{B}}{1+\bar{k}}\right]
\left( \hat{X}^{\prime }\right) }\simeq 0
\end{equation*}%
\begin{eqnarray}
&&\frac{\delta }{\delta \left\vert \bar{\Psi}\left( \bar{K},\bar{X}\right)
\right\vert ^{2}}\left( \left( \Delta \left( \bar{X}^{\prime },\bar{X}%
\right) -\frac{\bar{K}^{\prime }\bar{k}_{1}\left( \bar{X}^{\prime },\bar{X}%
\right) \left\vert \bar{\Psi}\left( \bar{K}^{\prime },\bar{X}^{\prime
}\right) \right\vert ^{2}}{1+\underline{\bar{k}}\left( \bar{X}^{\prime
}\right) }\right) \frac{\left( 1-\bar{M}\right) \bar{g}\left( \hat{K}%
^{\prime },\hat{X}^{\prime }\right) }{1+\underline{\overline{\bar{k}}}%
_{2}\left( \bar{X}^{\prime }\right) }\right. \\
&&\left. -\frac{\hat{K}^{\prime }\underline{\hat{k}}_{1}^{B}\left( \hat{X}%
^{\prime },\bar{X}\right) }{1+\underline{\hat{k}}\left( \hat{X}^{\prime
}\right) +\underline{\hat{k}}_{1}^{B}\left( \bar{X}^{\prime }\right) +\kappa %
\left[ \frac{\underline{\hat{k}}_{2}^{B}}{1+\bar{k}}\right] \left( \hat{X}%
^{\prime }\right) }\frac{\left( 1-\hat{M}\right) \hat{g}\left( \hat{K}%
^{\prime },\hat{X}^{\prime }\right) +\bar{N}\bar{g}\left( \hat{K}^{\prime },%
\hat{X}^{\prime }\right) }{1+\underline{\hat{k}}_{2}\left( \bar{X}^{\prime
}\right) +\kappa \left[ \frac{\underline{\hat{k}}_{2}^{B}}{1+\bar{k}}\right]
\left( \hat{X}^{\prime }\right) }\right) \simeq 0  \notag
\end{eqnarray}

\subsubsection*{A25.2.1 Computation of $\frac{\protect\delta }{\protect%
\delta \left\vert \bar{\Psi}\left( \bar{K},\bar{X}\right) \right\vert ^{2}}%
\bar{g}\left( \hat{K}^{\prime },\hat{X}^{\prime }\right) $}

\begin{eqnarray*}
&&\frac{\delta }{\delta \left\vert \bar{\Psi}\left( \bar{K},\bar{X}\right)
\right\vert ^{2}}\left( \left( \Delta \left( \bar{X}^{\prime },\bar{X}%
\right) -\frac{\bar{K}^{\prime }\bar{k}_{1}\left( \bar{X}^{\prime },\bar{X}%
\right) \left\vert \bar{\Psi}\left( \bar{K}^{\prime },\bar{X}^{\prime
}\right) \right\vert ^{2}}{1+\underline{\bar{k}}\left( \bar{X}^{\prime
}\right) }\right) \frac{\left( 1-\bar{M}\right) }{1+\underline{\overline{%
\bar{k}}}_{2}\left( \bar{X}^{\prime }\right) }\right) \bar{g}\left( \hat{K}%
^{\prime },\hat{X}^{\prime }\right) \\
&&+\left( \left( \Delta \left( \bar{X}^{\prime },\bar{X}\right) -\frac{\bar{K%
}^{\prime }\bar{k}_{1}\left( \bar{X}^{\prime },\bar{X}\right) \left\vert 
\bar{\Psi}\left( \bar{K}^{\prime },\bar{X}^{\prime }\right) \right\vert ^{2}%
}{1+\underline{\bar{k}}\left( \bar{X}^{\prime }\right) }\right) \frac{\left(
1-\bar{M}\right) }{1+\underline{\overline{\bar{k}}}_{2}\left( \bar{X}%
^{\prime }\right) }\right) \frac{\delta }{\delta \left\vert \bar{\Psi}\left( 
\bar{K},\bar{X}\right) \right\vert ^{2}}\bar{g}\left( \hat{K}^{\prime },\hat{%
X}^{\prime }\right) \\
&=&-\frac{\delta }{\delta \left\vert \bar{\Psi}\left( \bar{K},\bar{X}\right)
\right\vert ^{2}}\frac{\hat{K}^{\prime }\underline{\hat{k}}_{1}^{B}\left( 
\hat{X}^{\prime },\bar{X}\right) }{1+\underline{\hat{k}}\left( \hat{X}%
^{\prime }\right) +\underline{\hat{k}}_{1}^{B}\left( \bar{X}^{\prime
}\right) +\kappa \left[ \frac{\underline{\hat{k}}_{2}^{B}}{1+\bar{k}}\right]
\left( \hat{X}^{\prime }\right) }\frac{\left( 1-\hat{M}\right) \hat{g}\left( 
\hat{K}^{\prime },\hat{X}^{\prime }\right) +\bar{N}\bar{g}\left( \hat{K}%
^{\prime },\hat{X}^{\prime }\right) }{1+\underline{\hat{k}}_{2}\left( \bar{X}%
^{\prime }\right) +\kappa \left[ \frac{\underline{\hat{k}}_{2}^{B}}{1+\bar{k}%
}\right] \left( \hat{X}^{\prime }\right) } \\
&&-\frac{\hat{K}^{\prime }\underline{\hat{k}}_{1}^{B}\left( \hat{X}^{\prime
},\bar{X}\right) }{1+\underline{\hat{k}}\left( \hat{X}^{\prime }\right) +%
\underline{\hat{k}}_{1}^{B}\left( \bar{X}^{\prime }\right) +\kappa \left[ 
\frac{\underline{\hat{k}}_{2}^{B}}{1+\bar{k}}\right] \left( \hat{X}^{\prime
}\right) }\frac{\delta }{\delta \left\vert \bar{\Psi}\left( \bar{K},\bar{X}%
\right) \right\vert ^{2}}\frac{\left( 1-\hat{M}\right) \hat{g}\left( \hat{K}%
^{\prime },\hat{X}^{\prime }\right) +\bar{N}\bar{g}\left( \hat{K}^{\prime },%
\hat{X}^{\prime }\right) }{1+\underline{\hat{k}}_{2}\left( \bar{X}^{\prime
}\right) +\kappa \left[ \frac{\underline{\hat{k}}_{2}^{B}}{1+\bar{k}}\right]
\left( \hat{X}^{\prime }\right) }
\end{eqnarray*}%
so that: 
\begin{eqnarray*}
&&\frac{\delta }{\delta \left\vert \bar{\Psi}\left( \bar{K},\bar{X}\right)
\right\vert ^{2}}\bar{g}\left( \hat{K}^{\prime },\hat{X}^{\prime }\right) \\
&=&\left( \left( \Delta \left( \bar{X}^{\prime },\bar{X}\right) -\frac{\bar{K%
}^{\prime }\bar{k}_{1}\left( \bar{X}^{\prime },\bar{X}\right) \left\vert 
\bar{\Psi}\left( \bar{K}^{\prime },\bar{X}^{\prime }\right) \right\vert ^{2}%
}{1+\underline{\bar{k}}\left( \bar{X}^{\prime }\right) }\right) \frac{\left(
1-\bar{M}\right) }{1+\underline{\overline{\bar{k}}}_{2}\left( \bar{X}%
^{\prime }\right) }\right) ^{-1} \\
&&\left\{ -\frac{\delta }{\delta \left\vert \bar{\Psi}\left( \bar{K},\bar{X}%
\right) \right\vert ^{2}}\frac{\hat{K}^{\prime }\underline{\hat{k}}%
_{1}^{B}\left( \hat{X}^{\prime },\bar{X}\right) \left\vert \hat{\Psi}\left( 
\hat{K}^{\prime },\hat{X}^{\prime }\right) \right\vert ^{2}}{\left\Vert \hat{%
\Psi}\right\Vert ^{2}\left\langle \hat{K}\right\rangle \left( 1+\underline{%
\hat{k}}\left( \hat{X}^{\prime }\right) +\underline{\hat{k}}_{1}^{B}\left( 
\bar{X}^{\prime }\right) +\kappa \left[ \frac{\underline{\hat{k}}_{2}^{B}}{1+%
\bar{k}}\right] \left( \hat{X}^{\prime }\right) \right) }\frac{\left( 1-\hat{%
M}\right) \hat{g}\left( \hat{K}^{\prime },\hat{X}^{\prime }\right) +\bar{N}%
\bar{g}\left( \hat{K}^{\prime },\hat{X}^{\prime }\right) }{1+\underline{\hat{%
k}}_{2}\left( \bar{X}^{\prime }\right) +\kappa \left[ \frac{\underline{\hat{k%
}}_{2}^{B}}{1+\bar{k}}\right] \left( \hat{X}^{\prime }\right) }\right. \\
&&-\frac{\hat{K}^{\prime }\underline{\hat{k}}_{1}^{B}\left( \hat{X}^{\prime
},\bar{X}\right) \left\vert \hat{\Psi}\left( \hat{K}^{\prime },\hat{X}%
^{\prime }\right) \right\vert ^{2}}{\left\Vert \hat{\Psi}\right\Vert
^{2}\left\langle \hat{K}\right\rangle \left( 1+\underline{\hat{k}}\left( 
\hat{X}^{\prime }\right) +\underline{\hat{k}}_{1}^{B}\left( \bar{X}^{\prime
}\right) +\kappa \left[ \frac{\underline{\hat{k}}_{2}^{B}}{1+\bar{k}}\right]
\left( \hat{X}^{\prime }\right) \right) }\frac{\delta }{\delta \left\vert 
\bar{\Psi}\left( \bar{K},\bar{X}\right) \right\vert ^{2}}\frac{\left( 1-\hat{%
M}\right) \hat{g}\left( \hat{K}^{\prime },\hat{X}^{\prime }\right) +\bar{N}%
\bar{g}\left( \hat{K}^{\prime },\hat{X}^{\prime }\right) }{1+\underline{\hat{%
k}}_{2}\left( \bar{X}^{\prime }\right) +\kappa \left[ \frac{\underline{\hat{k%
}}_{2}^{B}}{1+\bar{k}}\right] \left( \hat{X}^{\prime }\right) } \\
&&\left. -\frac{\delta }{\delta \left\vert \bar{\Psi}\left( \bar{K},\bar{X}%
\right) \right\vert ^{2}}\left( \left( \Delta \left( \bar{X}^{\prime },\bar{X%
}\right) -\frac{\bar{K}^{\prime }\bar{k}_{1}\left( \bar{X}^{\prime },\bar{X}%
\right) \left\vert \bar{\Psi}\left( \bar{K}^{\prime },\bar{X}^{\prime
}\right) \right\vert ^{2}}{1+\underline{\bar{k}}\left( \bar{X}^{\prime
}\right) }\right) \frac{\left( 1-\bar{M}\right) }{1+\underline{\overline{%
\bar{k}}}_{2}\left( \bar{X}^{\prime }\right) }\right) \bar{g}\left( \hat{K}%
^{\prime },\hat{X}^{\prime }\right) \right\}
\end{eqnarray*}%
Various contributions are computed below:

\paragraph*{A25.2.1.1 Estimation of}

\begin{equation*}
\frac{\delta }{\delta \left\vert \bar{\Psi}\left( \bar{K},\bar{X}\right)
\right\vert ^{2}}\hat{M}
\end{equation*}

\begin{eqnarray*}
\frac{\delta }{\delta \left\vert \bar{\Psi}\left( \bar{K},\bar{X}\right)
\right\vert ^{2}}\hat{M} &\simeq &-\frac{\hat{k}\left( \hat{X},\hat{X}%
^{\prime }\right) \hat{K}\left\vert \hat{\Psi}\left( \hat{K}^{\prime },\hat{X%
}^{\prime }\right) \right\vert ^{2}}{\hat{D}\left( \hat{X}\right) } \\
&&\times \frac{\left( \hat{k}_{1}^{B}\left( \hat{X},\bar{X}\right) +\kappa %
\left[ \frac{\underline{\hat{k}}_{2}^{B}\left( \hat{X},\bar{X}\right) }{1+%
\bar{k}}\right] -\left( \left\langle \hat{k}_{1}^{B}\left( \hat{X},\bar{X}%
\right) \right\rangle +\kappa \left\langle \left[ \frac{\underline{\hat{k}}%
_{2}^{B}\left( \hat{X},\bar{X}\right) }{1+\bar{k}}\right] \right\rangle
\right) \right) \bar{K}}{\hat{D}\left( \hat{X}\right) \left\Vert \hat{\Psi}%
\right\Vert ^{2}\left\langle \hat{K}\right\rangle } \\
&=&-\frac{\hat{K}^{\prime }\hat{k}\left( \hat{X}^{\prime },\hat{X}\right)
\left\vert \hat{\Psi}\left( \hat{K}^{\prime },\hat{X}^{\prime }\right)
\right\vert ^{2}}{\left( 1+\underline{\hat{k}}\left( \hat{X}^{\prime
}\right) +\underline{\hat{k}}_{1}^{B}\left( \bar{X}^{\prime }\right) +\kappa %
\left[ \frac{\underline{\hat{k}}_{2}^{B}}{1+\bar{k}}\right] \left( \hat{X}%
^{\prime }\right) \right) ^{2}} \\
&&\times \left( \left( \hat{k}_{1}^{B}\left( \hat{X}^{\prime },\bar{X}%
_{1}\right) -\left\langle \hat{k}_{1}^{B}\left( \hat{X}^{\prime },\bar{X}%
_{1}\right) \right\rangle \right) +\kappa \frac{\int \left( \hat{k}%
_{2}^{B}\left( \hat{X}^{\prime },\bar{X}_{1}\right) -\left\langle \hat{k}%
_{2}^{B}\left( \hat{X}^{\prime },\bar{X}_{1}\right) \right\rangle \right) }{%
1+\int \bar{k}\left( \bar{X}^{\prime },\bar{X}^{\prime \prime }\right) \bar{K%
}_{0}^{\prime \prime }\left\vert \bar{\Psi}\left( \bar{K}_{0}^{\prime \prime
},\bar{X}^{\prime \prime }\right) \right\vert ^{2}}\right) \bar{K}
\end{eqnarray*}%
In average:%
\begin{equation*}
\hat{D}\left( \hat{X}\right) =1
\end{equation*}%
and:%
\begin{eqnarray}
&&\frac{\delta }{\delta \left\vert \bar{\Psi}\left( \bar{K},\bar{X}\right)
\right\vert ^{2}}\hat{M}  \label{Dm} \\
&\rightarrow &-\frac{\hat{K}^{\prime }\hat{k}\left( \hat{X}^{\prime },\hat{X}%
\right) \left\vert \hat{\Psi}\left( \hat{K}^{\prime },\hat{X}^{\prime
}\right) \right\vert ^{2}}{\left\Vert \hat{\Psi}\right\Vert ^{2}\left\langle 
\hat{K}\right\rangle }\left( \left( \hat{k}_{1}^{B}\left( \hat{X}^{\prime },%
\bar{X}_{1}\right) -\left\langle \hat{k}_{1}^{B}\left( \hat{X}^{\prime },%
\bar{X}_{1}\right) \right\rangle \right) +\kappa \frac{\left( \hat{k}%
_{2}^{B}\left( \hat{X}^{\prime },\bar{X}_{1}\right) -\left\langle \hat{k}%
_{2}^{B}\left( \hat{X}^{\prime },\bar{X}_{1}\right) \right\rangle \right) }{%
1+\int \bar{k}\left( \bar{X}^{\prime },\bar{X}^{\prime \prime }\right) \bar{K%
}_{0}^{\prime \prime }\left\vert \bar{\Psi}\left( \bar{K}_{0}^{\prime \prime
},\bar{X}^{\prime \prime }\right) \right\vert ^{2}}\right) \bar{K}  \notag \\
&\simeq &-\left\langle \hat{k}\left( \hat{X}^{\prime },\hat{X}\right)
\right\rangle \left( \left( \hat{k}_{1}^{B}\left( \hat{X}^{\prime },\bar{X}%
_{1}\right) -\left\langle \hat{k}_{1}^{B}\left( \hat{X}^{\prime },\bar{X}%
_{1}\right) \right\rangle \right) +\kappa \frac{\left( \hat{k}_{2}^{B}\left( 
\hat{X}^{\prime },\bar{X}_{1}\right) -\left\langle \hat{k}_{2}^{B}\left( 
\hat{X}^{\prime },\bar{X}_{1}\right) \right\rangle \right) }{1+\left\langle 
\bar{k}\left( \bar{X}^{\prime },\bar{X}^{\prime \prime }\right)
\right\rangle }\right) \bar{K}  \notag
\end{eqnarray}

\paragraph*{A25.2.1.2 Estimation of}

\begin{eqnarray*}
&&\frac{\delta }{\delta \left\vert \bar{\Psi}\left( \bar{K},\bar{X}\right)
\right\vert ^{2}}\left( \Delta \left( \hat{X},\hat{X}^{\prime }\right) -%
\frac{\hat{K}^{\prime }\hat{k}_{1}\left( \hat{X}^{\prime },\hat{X}\right)
\left\vert \hat{\Psi}\left( \hat{K}^{\prime },\hat{X}^{\prime }\right)
\right\vert ^{2}}{1+\underline{\hat{k}}\left( \hat{X}^{\prime }\right) +%
\underline{\hat{k}}_{1}^{B}\left( \bar{X}^{\prime }\right) +\kappa \left[ 
\frac{\underline{\hat{k}}_{2}^{B}}{1+\bar{k}}\right] \left( \hat{X}^{\prime
}\right) }\right) \\
&\rightarrow &\frac{\hat{K}^{\prime }\hat{k}_{1}\left( \hat{X}^{\prime },%
\hat{X}\right) \left\vert \hat{\Psi}\left( \hat{K}^{\prime },\hat{X}^{\prime
}\right) \right\vert ^{2}}{\left( 1+\underline{\hat{k}}\left( \hat{X}%
^{\prime }\right) +\underline{\hat{k}}_{1}^{B}\left( \bar{X}^{\prime
}\right) +\kappa \left[ \frac{\underline{\hat{k}}_{2}^{B}}{1+\bar{k}}\right]
\left( \hat{X}^{\prime }\right) \right) ^{2}} \\
&&\times \left( \left( \hat{k}_{1}^{B}\left( \hat{X}^{\prime },\bar{X}%
\right) -\left\langle \hat{k}_{1}^{B}\left( \hat{X}^{\prime },\bar{X}\right)
\right\rangle \right) +\kappa \frac{\left( \hat{k}_{2}^{B}\left( \hat{X}%
^{\prime },\bar{X}\right) -\left\langle \hat{k}_{2}^{B}\left( \hat{X}%
^{\prime },\bar{X}\right) \right\rangle \right) }{1+\int \bar{k}\left( \bar{X%
}^{\prime },\bar{X}^{\prime \prime }\right) \frac{\bar{K}_{0}^{\prime \prime
}\left\vert \bar{\Psi}\left( \bar{K}_{0}^{\prime \prime },\bar{X}^{\prime
\prime }\right) \right\vert ^{2}}{\left\langle \bar{K}\right\rangle
\left\Vert \bar{\Psi}\right\Vert ^{2}}}\right) \bar{K} \\
&\rightarrow &\left( \left( \hat{k}_{1}^{B}\left( \hat{X}^{\prime },\bar{X}%
\right) -\left\langle \hat{k}_{1}^{B}\left( \hat{X}^{\prime },\bar{X}\right)
\right\rangle \right) +\kappa \frac{\left( \hat{k}_{2}^{B}\left( \hat{X}%
^{\prime },\bar{X}\right) -\left\langle \hat{k}_{2}^{B}\left( \hat{X}%
^{\prime },\bar{X}\right) \right\rangle \right) }{1+\left\langle \bar{k}%
\left( \bar{X}^{\prime },\bar{X}^{\prime \prime }\right) \right\rangle }%
\right) \bar{K}
\end{eqnarray*}

\paragraph*{A25.2.1.3 Estimation of}

\begin{equation*}
\frac{\delta }{\delta \left\vert \bar{\Psi}\left( \bar{K},\bar{X}\right)
\right\vert ^{2}}\frac{1}{1+\underline{\overline{\bar{k}}}_{2}\left( \bar{X}%
^{\prime }\right) }
\end{equation*}%
gv nrml 
\begin{equation*}
\frac{1}{1+\underline{\overline{\bar{k}}}_{2}\left( \bar{X}^{\prime }\right) 
}\rightarrow \frac{\left( 1-\left\langle \bar{k}\left( \hat{X}^{\prime },%
\hat{X}\right) \right\rangle \right) }{\left( 1-\left\langle \bar{k}%
_{1}\left( \hat{X}^{\prime },\hat{X}\right) \right\rangle \right) \left( 1+%
\frac{\underline{\bar{k}}_{2}\left( \hat{X}^{\prime }\right) }{%
1-\left\langle \bar{k}_{1}\left( \hat{X}^{\prime },\hat{X}\right)
\right\rangle }\right) }
\end{equation*}%
\begin{equation*}
=-\frac{\left( 1-\left\langle \bar{k}\left( \hat{X}^{\prime },\hat{X}\right)
\right\rangle \right) }{\left( \left( 1-\left\langle \bar{k}_{1}\left( \hat{X%
}^{\prime },\hat{X}\right) \right\rangle \right) \left( 1+\frac{\underline{%
\bar{k}}_{2}\left( \hat{X}^{\prime }\right) }{1-\left\langle \bar{k}%
_{1}\left( \hat{X}^{\prime },\hat{X}\right) \right\rangle }\right) \right)
^{2}}\times \frac{\left( \bar{k}_{2}\left( \left\langle \bar{X}\right\rangle
,\bar{X}_{1}\right) -\left\langle \bar{k}_{2}\left( \bar{X},\bar{X}^{\prime
}\right) \right\rangle \right) }{\left\Vert \bar{\Psi}\right\Vert
^{2}\left\langle \bar{K}\right\rangle }\bar{K}
\end{equation*}

In matrices elements, this becomes:%
\begin{equation*}
-\frac{\left( 1-\left\langle \bar{k}\left( \hat{X}^{\prime },\hat{X}\right)
\right\rangle \right) }{\left( \left( 1-\left\langle \bar{k}_{1}\left( \hat{X%
}^{\prime },\hat{X}\right) \right\rangle \right) \right) ^{2}}\times \frac{%
\left( \bar{k}_{2}\left( \left\langle \bar{X}\right\rangle ,\bar{X}\right)
-\left\langle \bar{k}_{2}\left( \bar{X},\bar{X}^{\prime }\right)
\right\rangle \right) }{\left\Vert \bar{\Psi}\right\Vert ^{2}\left\langle 
\bar{K}\right\rangle }\bar{K}
\end{equation*}

\paragraph*{A25.2.1.4 Estimation of}

\begin{equation*}
\frac{\delta }{\delta \left\vert \bar{\Psi}\left( \bar{K},\bar{X}\right)
\right\vert ^{2}}\left( \left( \Delta \left( \bar{X}^{\prime },\bar{X}%
\right) -\frac{\bar{K}^{\prime }\bar{k}_{1}\left( \bar{X}^{\prime },\bar{X}%
\right) \left\vert \bar{\Psi}\left( \bar{K}^{\prime },\bar{X}^{\prime
}\right) \right\vert ^{2}}{1+\underline{\bar{k}}\left( \bar{X}^{\prime
}\right) }\right) \frac{\left( 1-\bar{M}\right) }{1+\underline{\overline{%
\bar{k}}}_{2}\left( \bar{X}^{\prime }\right) }\right)
\end{equation*}%
Assuming contributions:%
\begin{eqnarray*}
&&\left( \left( \hat{k}_{1}^{B}\left( \hat{X}^{\prime },\bar{X}\right)
-\left\langle \hat{k}_{1}^{B}\left( \hat{X}^{\prime },\bar{X}\right)
\right\rangle \right) +\kappa \frac{\left( \hat{k}_{2}^{B}\left( \hat{X}%
^{\prime },\bar{X}\right) -\left\langle \hat{k}_{2}^{B}\left( \hat{X}%
^{\prime },\bar{X}\right) \right\rangle \right) }{1+\left\langle \bar{k}%
\left( \bar{X}^{\prime },\bar{X}^{\prime \prime }\right) \right\rangle }%
\right) \bar{K}\frac{\left( 1-\bar{M}\right) }{1+\underline{\overline{\bar{k}%
}}_{2}\left( \bar{X}^{\prime }\right) }\left\langle \bar{g}\left( \hat{K}%
^{\prime },\hat{X}^{\prime }\right) \right\rangle \\
&&+\left( 1-\left\langle \hat{k}_{1}\left( \hat{X}^{\prime },\hat{X}\right)
\right\rangle \right) \left\langle \hat{k}\left( \hat{X}^{\prime },\hat{X}%
\right) \right\rangle \\
&&\times \left( \left( \hat{k}_{1}^{B}\left( \hat{X}^{\prime },\bar{X}%
_{1}\right) -\left\langle \hat{k}_{1}^{B}\left( \hat{X}^{\prime },\bar{X}%
_{1}\right) \right\rangle \right) +\kappa \frac{\left( \hat{k}_{2}^{B}\left( 
\hat{X}^{\prime },\bar{X}_{1}\right) -\left\langle \hat{k}_{2}^{B}\left( 
\hat{X}^{\prime },\bar{X}_{1}\right) \right\rangle \right) }{1+\left\langle 
\bar{k}\left( \bar{X}^{\prime },\bar{X}^{\prime \prime }\right)
\right\rangle }\right) \bar{K}\frac{1}{1+\underline{\overline{\bar{k}}}%
_{2}\left( \bar{X}^{\prime }\right) }\left\langle \bar{g}\left( \hat{K}%
^{\prime },\hat{X}^{\prime }\right) \right\rangle \\
&&-\left( 1-\left\langle \hat{k}_{1}\left( \hat{X}^{\prime },\hat{X}\right)
\right\rangle \right) \left( 1-\left\langle \bar{k}\left( \hat{X}^{\prime },%
\hat{X}\right) \right\rangle \right) \frac{\left( 1-\left\langle \bar{k}%
\left( \hat{X}^{\prime },\hat{X}\right) \right\rangle \right) }{\left(
\left( 1-\left\langle \bar{k}_{1}\left( \hat{X}^{\prime },\hat{X}\right)
\right\rangle \right) \right) ^{2}}\times \frac{\left( \bar{k}_{2}\left(
\left\langle \bar{X}\right\rangle ,\bar{X}\right) -\left\langle \bar{k}%
_{2}\left( \bar{X},\bar{X}^{\prime }\right) \right\rangle \right) }{%
\left\Vert \bar{\Psi}\right\Vert ^{2}\left\langle \bar{K}\right\rangle }\bar{%
K}
\end{eqnarray*}%
with:%
\begin{eqnarray*}
&&\left( \left( \hat{k}_{1}^{B}\left( \hat{X}^{\prime },\bar{X}\right)
-\left\langle \hat{k}_{1}^{B}\left( \hat{X}^{\prime },\bar{X}\right)
\right\rangle \right) +\kappa \frac{\left( \hat{k}_{2}^{B}\left( \hat{X}%
^{\prime },\bar{X}\right) -\left\langle \hat{k}_{2}^{B}\left( \hat{X}%
^{\prime },\bar{X}\right) \right\rangle \right) }{1+\left\langle \bar{k}%
\left( \bar{X}^{\prime },\bar{X}^{\prime \prime }\right) \right\rangle }%
\right) \bar{K}\frac{\left( 1-\left\langle \bar{k}\left( \hat{X}^{\prime },%
\hat{X}\right) \right\rangle \right) ^{2}}{\left( 1-\left\langle \bar{k}%
_{1}\left( \hat{X}^{\prime },\hat{X}\right) \right\rangle \right) }%
\left\langle \bar{g}\left( \hat{K}^{\prime },\hat{X}^{\prime }\right)
\right\rangle \\
&&+\left\langle \hat{k}\left( \hat{X}^{\prime },\hat{X}\right) \right\rangle
\left( \left( \hat{k}_{1}^{B}\left( \hat{X}^{\prime },\bar{X}_{1}\right)
-\left\langle \hat{k}_{1}^{B}\left( \hat{X}^{\prime },\bar{X}_{1}\right)
\right\rangle \right) +\kappa \frac{\left( \hat{k}_{2}^{B}\left( \hat{X}%
^{\prime },\bar{X}_{1}\right) -\left\langle \hat{k}_{2}^{B}\left( \hat{X}%
^{\prime },\bar{X}_{1}\right) \right\rangle \right) }{1+\left\langle \bar{k}%
\left( \bar{X}^{\prime },\bar{X}^{\prime \prime }\right) \right\rangle }%
\right) \\
&&\times \bar{K}\left( 1-\left\langle \bar{k}\left( \hat{X}^{\prime },\hat{X}%
\right) \right\rangle \right) \left\langle \bar{g}\left( \hat{K}^{\prime },%
\hat{X}^{\prime }\right) \right\rangle \\
&&-\frac{\left( 1-\left\langle \bar{k}\left( \hat{X}^{\prime },\hat{X}%
\right) \right\rangle \right) ^{2}}{1-\left\langle \bar{k}_{1}\left( \hat{X}%
^{\prime },\hat{X}\right) \right\rangle }\times \frac{\left( \bar{k}%
_{2}\left( \left\langle \bar{X}\right\rangle ,\bar{X}\right) -\left\langle 
\bar{k}_{2}\left( \bar{X},\bar{X}^{\prime }\right) \right\rangle \right) }{%
\left\Vert \bar{\Psi}\right\Vert ^{2}\left\langle \bar{K}\right\rangle }\bar{%
K}\left\langle \bar{g}\left( \hat{K}^{\prime },\hat{X}^{\prime }\right)
\right\rangle
\end{eqnarray*}

and:%
\begin{eqnarray*}
&&\left( \frac{\left( 1-\left\langle \bar{k}\left( \hat{X}^{\prime },\hat{X}%
\right) \right\rangle \right) ^{2}}{\left( 1-\left\langle \bar{k}_{1}\left( 
\hat{X}^{\prime },\hat{X}\right) \right\rangle \right) }+\left\langle \hat{k}%
\left( \hat{X}^{\prime },\hat{X}\right) \right\rangle \left( 1-\left\langle 
\bar{k}\left( \hat{X}^{\prime },\hat{X}\right) \right\rangle \right) \right)
\\
&&\times \left( \left( \hat{k}_{1}^{B}\left( \hat{X}^{\prime },\bar{X}%
\right) -\left\langle \hat{k}_{1}^{B}\left( \hat{X}^{\prime },\bar{X}\right)
\right\rangle \right) +\kappa \frac{\left( \hat{k}_{2}^{B}\left( \hat{X}%
^{\prime },\bar{X}\right) -\left\langle \hat{k}_{2}^{B}\left( \hat{X}%
^{\prime },\bar{X}\right) \right\rangle \right) }{1+\left\langle \bar{k}%
\left( \bar{X}^{\prime },\bar{X}^{\prime \prime }\right) \right\rangle }%
\right) \bar{K}\left\langle \bar{g}\left( \hat{K}^{\prime },\hat{X}^{\prime
}\right) \right\rangle \\
&&-\frac{\left( 1-\left\langle \bar{k}\left( \hat{X}^{\prime },\hat{X}%
\right) \right\rangle \right) ^{2}}{1-\left\langle \bar{k}_{1}\left( \hat{X}%
^{\prime },\hat{X}\right) \right\rangle }\frac{\left( \bar{k}_{2}\left(
\left\langle \bar{X}\right\rangle ,\bar{X}\right) -\left\langle \bar{k}%
_{2}\left( \bar{X},\bar{X}^{\prime }\right) \right\rangle \right) }{%
\left\Vert \bar{\Psi}\right\Vert ^{2}\left\langle \bar{K}\right\rangle }\bar{%
K}\left\langle \bar{g}\left( \hat{K}^{\prime },\hat{X}^{\prime }\right)
\right\rangle
\end{eqnarray*}

\paragraph*{A25.2.1.5 Estimation of}

\bigskip 
\begin{eqnarray*}
&&S=-\frac{\delta }{\delta \left\vert \bar{\Psi}\left( \bar{K},\bar{X}%
\right) \right\vert ^{2}}\frac{\hat{K}^{\prime }\underline{\hat{k}}%
_{1}^{B}\left( \hat{X}^{\prime },\bar{X}\right) \left\vert \hat{\Psi}\left( 
\hat{K}^{\prime },\hat{X}^{\prime }\right) \right\vert ^{2}}{\left\Vert \hat{%
\Psi}\right\Vert ^{2}\left\langle \hat{K}\right\rangle \left( 1+\underline{%
\hat{k}}\left( \hat{X}^{\prime }\right) +\underline{\hat{k}}_{1}^{B}\left( 
\bar{X}^{\prime }\right) +\kappa \left[ \frac{\underline{\hat{k}}_{2}^{B}}{1+%
\bar{k}}\right] \left( \hat{X}^{\prime }\right) \right) }\frac{\left( 1-\hat{%
M}\right) \hat{g}\left( \hat{K}^{\prime },\hat{X}^{\prime }\right) +\bar{N}%
\bar{g}\left( \hat{K}^{\prime },\hat{X}^{\prime }\right) }{1+\underline{\hat{%
k}}_{2}\left( \bar{X}^{\prime }\right) +\kappa \left[ \frac{\underline{\hat{k%
}}_{2}^{B}}{1+\bar{k}}\right] \left( \hat{X}^{\prime }\right) } \\
&=&\frac{\underline{\hat{k}}_{1}^{B}\left( \hat{X}^{\prime },\bar{X}\right)
\left( \left( \hat{k}_{1}^{B}\left( \hat{X}^{\prime },\bar{X}\right)
-\left\langle \hat{k}_{1}^{B}\left( \hat{X}^{\prime },\bar{X}\right)
\right\rangle \right) +\kappa \frac{\left( \hat{k}_{2}^{B}\left( \hat{X}%
^{\prime },\bar{X}\right) -\left\langle \hat{k}_{2}^{B}\left( \hat{X}%
^{\prime },\bar{X}\right) \right\rangle \right) }{1+\left\langle \bar{k}%
\left( \bar{X}^{\prime },\bar{X}^{\prime \prime }\right) \right\rangle }%
\right) \bar{K}}{\left\Vert \hat{\Psi}\right\Vert ^{2}\left\langle \hat{K}%
\right\rangle \left( 1+\underline{\hat{k}}\left( \hat{X}^{\prime }\right) +%
\underline{\hat{k}}_{1}^{B}\left( \bar{X}^{\prime }\right) +\kappa \left[ 
\frac{\underline{\hat{k}}_{2}^{B}}{1+\bar{k}}\right] \left( \hat{X}^{\prime
}\right) \right) ^{2}} \\
&&\times \left\langle \frac{\left( 1-\hat{M}\right) \hat{g}\left( \hat{K}%
^{\prime },\hat{X}^{\prime }\right) +\bar{N}\bar{g}\left( \hat{K}^{\prime },%
\hat{X}^{\prime }\right) }{1+\underline{\hat{k}}_{2}\left( \bar{X}^{\prime
}\right) +\kappa \left[ \frac{\underline{\hat{k}}_{2}^{B}}{1+\bar{k}}\right]
\left( \hat{X}^{\prime }\right) }\right\rangle
\end{eqnarray*}%
which becomes in average:%
\begin{eqnarray*}
&&\frac{\left\langle \underline{\hat{k}}_{1}^{B}\left( \hat{X}^{\prime },%
\bar{X}\right) \right\rangle \left( \left( \hat{k}_{1}^{B}\left( \hat{X}%
^{\prime },\bar{X}\right) -\left\langle \hat{k}_{1}^{B}\left( \hat{X}%
^{\prime },\bar{X}\right) \right\rangle \right) +\kappa \frac{\left( \hat{k}%
_{2}^{B}\left( \hat{X}^{\prime },\bar{X}\right) -\left\langle \hat{k}%
_{2}^{B}\left( \hat{X}^{\prime },\bar{X}\right) \right\rangle \right) }{%
1+\left\langle \bar{k}\left( \bar{X}^{\prime },\bar{X}^{\prime \prime
}\right) \right\rangle }\right) \bar{K}}{\left\Vert \hat{\Psi}\right\Vert
^{2}\left\langle \hat{K}\right\rangle } \\
&&\times \left\langle \frac{\left( 1-\left\langle \hat{k}\left( \hat{X},\hat{%
X}^{\prime }\right) \right\rangle \right) \hat{g}\left( \hat{K}^{\prime },%
\hat{X}^{\prime }\right) +\bar{N}\bar{g}\left( \hat{K}^{\prime },\hat{X}%
^{\prime }\right) }{1+\underline{\hat{k}}_{2}\left( \bar{X}^{\prime }\right)
+\kappa \left[ \frac{\underline{\hat{k}}_{2}^{B}}{1+\bar{k}}\right] \left( 
\hat{X}^{\prime }\right) }\right\rangle
\end{eqnarray*}%
We can write:%
\begin{eqnarray*}
&&\frac{1}{1+\underline{\hat{k}}_{2}\left( \bar{X}^{\prime }\right) +\kappa %
\left[ \frac{\underline{\hat{k}}_{2}^{B}}{1+\bar{k}}\right] \left( \hat{X}%
^{\prime }\right) } \\
&=&\frac{1-\left( \left\langle \underline{\hat{k}}\left( \hat{X}^{\prime },%
\hat{X}\right) \right\rangle +\left( \left\langle \underline{\hat{k}}%
_{1}^{B}\left( \hat{X}^{\prime },\bar{X}\right) \right\rangle +\kappa
\left\langle \left[ \frac{\underline{\hat{k}}_{2}^{B}\left( \hat{X}^{\prime
},\bar{X}\right) }{1+\bar{k}}\right] \right\rangle \right) \frac{\left\Vert 
\bar{\Psi}\right\Vert ^{2}\left\langle \bar{K}\right\rangle }{\left\Vert 
\hat{\Psi}\right\Vert ^{2}\left\langle \hat{K}\right\rangle }\right) }{%
\left( 1-\left( \left\langle \hat{k}_{1}\left( \hat{X}^{\prime },\hat{X}%
\right) \right\rangle +\left\langle \hat{k}_{1}^{B}\left( \hat{X}^{\prime },%
\hat{X}\right) \right\rangle \frac{\left\Vert \bar{\Psi}\right\Vert
^{2}\left\langle \bar{K}\right\rangle }{\left\Vert \hat{\Psi}\right\Vert
^{2}\left\langle \hat{K}\right\rangle }\right) \right) \left( 1+\frac{%
\underline{\hat{k}}_{2}\left( \hat{X}^{\prime }\right) +\kappa \left[ \frac{%
\underline{\hat{k}}_{2}^{B}}{1+\bar{k}}\right] \left( \hat{X}^{\prime
}\right) }{1-\left( \left\langle \hat{k}_{1}\left( \hat{X}^{\prime },\hat{X}%
\right) \right\rangle +\left\langle \hat{k}_{1}^{B}\left( \hat{X}^{\prime },%
\hat{X}\right) \right\rangle \frac{\left\Vert \bar{\Psi}\right\Vert
^{2}\left\langle \bar{K}\right\rangle }{\left\Vert \hat{\Psi}\right\Vert
^{2}\left\langle \hat{K}\right\rangle }\right) }\right) }
\end{eqnarray*}%
In average: 
\begin{eqnarray*}
\frac{1}{1+\underline{\hat{k}}_{2}\left( \bar{X}^{\prime }\right) +\kappa %
\left[ \frac{\underline{\hat{k}}_{2}^{B}}{1+\bar{k}}\right] \left( \hat{X}%
^{\prime }\right) } &\rightarrow &\frac{1-\left( \left\langle \underline{%
\hat{k}}\left( \hat{X}^{\prime },\hat{X}\right) \right\rangle +\left\langle 
\underline{\hat{k}}_{1}^{B}\left( \hat{X}^{\prime },\bar{X}\right)
\right\rangle \frac{\left\Vert \bar{\Psi}\right\Vert ^{2}\left\langle \bar{K}%
\right\rangle }{\left\Vert \hat{\Psi}\right\Vert ^{2}\left\langle \hat{K}%
\right\rangle }+\kappa \left\langle \left[ \frac{\underline{\hat{k}}%
_{2}^{B}\left( \hat{X}^{\prime },\bar{X}\right) }{1+\bar{k}}\right]
\right\rangle \frac{\left\Vert \bar{\Psi}\right\Vert ^{2}\left\langle \bar{K}%
\right\rangle }{\left\Vert \hat{\Psi}\right\Vert ^{2}\left\langle \hat{K}%
\right\rangle }\right) }{\left( 1-\left( \left\langle \hat{k}_{1}\left( \hat{%
X}^{\prime },\hat{X}\right) \right\rangle +\left\langle \hat{k}%
_{1}^{B}\left( \hat{X}^{\prime },\hat{X}\right) \right\rangle \frac{%
\left\Vert \bar{\Psi}\right\Vert ^{2}\left\langle \bar{K}\right\rangle }{%
\left\Vert \hat{\Psi}\right\Vert ^{2}\left\langle \hat{K}\right\rangle }%
\right) \right) } \\
&\simeq &\frac{1-\left( \left\langle \underline{\hat{k}}\right\rangle
+\left\langle \underline{\hat{k}}_{1}^{B}\right\rangle \frac{\left\Vert \bar{%
\Psi}\right\Vert ^{2}\left\langle \bar{K}\right\rangle }{\left\Vert \hat{\Psi%
}\right\Vert ^{2}\left\langle \hat{K}\right\rangle }+\kappa \left\langle %
\left[ \frac{\underline{\hat{k}}_{2}^{B}}{1+\bar{k}}\right] \right\rangle 
\frac{\left\Vert \bar{\Psi}\right\Vert ^{2}\left\langle \bar{K}\right\rangle 
}{\left\Vert \hat{\Psi}\right\Vert ^{2}\left\langle \hat{K}\right\rangle }%
\right) }{\left( 1-\left( \left\langle \hat{k}_{1}\right\rangle
+\left\langle \hat{k}_{1}^{B}\right\rangle \frac{\left\Vert \bar{\Psi}%
\right\Vert ^{2}\left\langle \bar{K}\right\rangle }{\left\Vert \hat{\Psi}%
\right\Vert ^{2}\left\langle \hat{K}\right\rangle }\right) \right) }
\end{eqnarray*}%
\begin{equation*}
\bar{N}\rightarrow \left\langle \hat{k}_{1}^{B}\left( \hat{X},\bar{X}%
^{\prime }\right) \right\rangle +\kappa \frac{\left\langle \hat{k}%
_{2}^{B}\left( \hat{X},\bar{X}^{\prime }\right) \right\rangle }{%
1+\left\langle \bar{k}\left( \bar{X}^{\prime },\left\langle \bar{X}^{\prime
\prime }\right\rangle \right) \right\rangle }\left( 1-\frac{\left\langle 
\bar{k}\left( \bar{X}^{\prime \prime },\bar{X}^{\prime }\right)
\right\rangle }{\left( 1+\left\langle \bar{k}\left( \bar{X}^{\prime
},\left\langle \bar{X}^{\prime \prime }\right\rangle \right) \right\rangle
\right) ^{2}}\right)
\end{equation*}%
\begin{eqnarray*}
S &\rightarrow &\frac{\left\langle \underline{\hat{k}}_{1}^{B}\left(
\left\langle \hat{X}^{\prime }\right\rangle ,\bar{X}\right) \right\rangle }{%
\left\Vert \hat{\Psi}\right\Vert ^{2}\left\langle \hat{K}\right\rangle }%
\left( \left( \hat{k}_{1}^{B}\left( \hat{X}^{\prime },\bar{X}\right)
-\left\langle \hat{k}_{1}^{B}\left( \hat{X}^{\prime },\bar{X}\right)
\right\rangle \right) +\kappa \frac{\left( \hat{k}_{2}^{B}\left( \hat{X}%
^{\prime },\bar{X}\right) -\left\langle \hat{k}_{2}^{B}\left( \hat{X}%
^{\prime },\bar{X}\right) \right\rangle \right) }{1+\left\langle \bar{k}%
\left( \bar{X}^{\prime },\bar{X}^{\prime \prime }\right) \right\rangle }%
\right) \bar{K} \\
&&\times \frac{1-\left( \left\langle \underline{\hat{k}}\left( \hat{X}%
^{\prime },\hat{X}\right) \right\rangle +\left( \left\langle \underline{\hat{%
k}}_{1}^{B}\left( \hat{X}^{\prime },\bar{X}\right) \right\rangle +\kappa
\left\langle \left[ \frac{\underline{\hat{k}}_{2}^{B}\left( \hat{X}^{\prime
},\bar{X}\right) }{1+\bar{k}}\right] \right\rangle \right) \frac{\left\Vert 
\bar{\Psi}\right\Vert ^{2}\left\langle \bar{K}\right\rangle }{\left\Vert 
\hat{\Psi}\right\Vert ^{2}\left\langle \hat{K}\right\rangle }\right) }{%
\left( 1-\left( \left\langle \hat{k}_{1}\left( \hat{X}^{\prime },\hat{X}%
\right) \right\rangle +\left\langle \hat{k}_{1}^{B}\left( \hat{X}^{\prime },%
\hat{X}\right) \right\rangle \frac{\left\Vert \bar{\Psi}\right\Vert
^{2}\left\langle \bar{K}\right\rangle }{\left\Vert \hat{\Psi}\right\Vert
^{2}\left\langle \hat{K}\right\rangle }\right) \right) } \\
&&\times \left( \left( 1-\left\langle \hat{k}\left( \hat{X},\hat{X}^{\prime
}\right) \right\rangle \right) \left\langle \hat{g}\left( \hat{K}^{\prime },%
\hat{X}^{\prime }\right) \right\rangle \right. \\
&&\left. +\left( \left\langle \hat{k}_{1}^{B}\left( \hat{X},\bar{X}^{\prime
}\right) \right\rangle +\kappa \frac{\left\langle \hat{k}_{2}^{B}\left( \hat{%
X},\bar{X}^{\prime }\right) \right\rangle }{1+\left\langle \overline{\bar{k}}%
\left( \bar{X}^{\prime },\left\langle \bar{X}^{\prime \prime }\right\rangle
\right) \right\rangle }\left( 1-\frac{\left\langle \overline{\bar{k}}\left( 
\bar{X}^{\prime \prime },\bar{X}^{\prime }\right) \right\rangle }{\left(
1+\left\langle \overline{\bar{k}}\left( \bar{X}^{\prime },\left\langle \bar{X%
}^{\prime \prime }\right\rangle \right) \right\rangle \right) ^{2}}\right)
\right) \left\langle \bar{g}\left( \hat{K}^{\prime },\hat{X}^{\prime
}\right) \right\rangle \right)
\end{eqnarray*}

\paragraph*{A25.2.1.6 Estimation of}

\begin{equation*}
\frac{\delta }{\delta \left\vert \bar{\Psi}\left( \bar{K},\bar{X}\right)
\right\vert ^{2}}\frac{\left( 1-\hat{M}\right) \hat{g}\left( \hat{K}^{\prime
},\hat{X}^{\prime }\right) +\bar{N}\bar{g}\left( \hat{K}^{\prime },\hat{X}%
^{\prime }\right) }{1+\underline{\hat{k}}_{2}\left( \bar{X}^{\prime }\right)
+\kappa \left[ \frac{\underline{\hat{k}}_{2}^{B}}{1+\bar{k}}\right] \left( 
\hat{X}^{\prime }\right) }
\end{equation*}%
\begin{eqnarray*}
&&\frac{\delta }{\delta \left\vert \bar{\Psi}\left( \bar{K},\bar{X}\right)
\right\vert ^{2}}\frac{\left( 1-\hat{M}\right) \hat{g}\left( \hat{K}^{\prime
},\hat{X}^{\prime }\right) +\bar{N}\bar{g}\left( \hat{K}^{\prime },\hat{X}%
^{\prime }\right) }{1+\underline{\hat{k}}_{2}\left( \bar{X}^{\prime }\right)
+\kappa \left[ \frac{\underline{\hat{k}}_{2}^{B}}{1+\bar{k}}\right] \left( 
\hat{X}^{\prime }\right) } \\
&=&-\left( \Delta \left( \hat{X},\hat{X}^{\prime }\right) -\frac{\hat{K}%
^{\prime }\hat{k}_{1}\left( \hat{X}^{\prime },\hat{X}\right) \left\vert \hat{%
\Psi}\left( \hat{K}^{\prime },\hat{X}^{\prime }\right) \right\vert ^{2}}{1+%
\underline{\hat{k}}\left( \hat{X}^{\prime }\right) +\underline{\hat{k}}%
_{1}^{B}\left( \bar{X}^{\prime }\right) +\kappa \left[ \frac{\underline{\hat{%
k}}_{2}^{B}}{1+\bar{k}}\right] \left( \hat{X}^{\prime }\right) }\right) ^{-1}
\\
&&\times \frac{\delta }{\delta \left\vert \bar{\Psi}\left( \bar{K},\bar{X}%
\right) \right\vert ^{2}}\left( \Delta \left( \hat{X},\hat{X}^{\prime
}\right) -\frac{\hat{K}^{\prime }\hat{k}_{1}\left( \hat{X}^{\prime },\hat{X}%
\right) \left\vert \hat{\Psi}\left( \hat{K}^{\prime },\hat{X}^{\prime
}\right) \right\vert ^{2}}{1+\underline{\hat{k}}\left( \hat{X}^{\prime
}\right) +\underline{\hat{k}}_{1}^{B}\left( \bar{X}^{\prime }\right) +\kappa %
\left[ \frac{\underline{\hat{k}}_{2}^{B}}{1+\bar{k}}\right] \left( \hat{X}%
^{\prime }\right) }\right) \frac{\left( 1-\hat{M}\right) \hat{g}\left( \hat{K%
}^{\prime },\hat{X}^{\prime }\right) +\bar{N}\bar{g}\left( \hat{K}^{\prime },%
\hat{X}^{\prime }\right) }{1+\underline{\hat{k}}_{2}\left( \bar{X}^{\prime
}\right) +\kappa \left[ \frac{\underline{\hat{k}}_{2}^{B}}{1+\bar{k}}\right]
\left( \hat{X}^{\prime }\right) }
\end{eqnarray*}%
\begin{eqnarray*}
&&-\frac{\delta }{\delta \left\vert \bar{\Psi}\left( \bar{K},\bar{X}\right)
\right\vert ^{2}}\left( \Delta \left( \hat{X},\hat{X}^{\prime }\right) -%
\frac{\hat{K}^{\prime }\hat{k}_{1}\left( \hat{X}^{\prime },\hat{X}\right)
\left\vert \hat{\Psi}\left( \hat{K}^{\prime },\hat{X}^{\prime }\right)
\right\vert ^{2}}{1+\underline{\hat{k}}\left( \hat{X}^{\prime }\right) +%
\underline{\hat{k}}_{1}^{B}\left( \bar{X}^{\prime }\right) +\kappa \left[ 
\frac{\underline{\hat{k}}_{2}^{B}}{1+\bar{k}}\right] \left( \hat{X}^{\prime
}\right) }\right) \\
&\rightarrow &-\frac{\hat{K}^{\prime }\hat{k}_{1}\left( \hat{X}^{\prime },%
\hat{X}\right) \left\vert \hat{\Psi}\left( \hat{K}^{\prime },\hat{X}^{\prime
}\right) \right\vert ^{2}\left( \left( \hat{k}_{1}^{B}\left( \hat{X}^{\prime
},\bar{X}\right) -\left\langle \hat{k}_{1}^{B}\left( \hat{X}^{\prime },\bar{X%
}\right) \right\rangle \right) +\kappa \frac{\left( \hat{k}_{2}^{B}\left( 
\hat{X}^{\prime },\bar{X}\right) -\left\langle \hat{k}_{2}^{B}\left( \hat{X}%
^{\prime },\bar{X}\right) \right\rangle \right) }{1+\int \bar{k}\left( \bar{X%
}^{\prime },\bar{X}^{\prime \prime }\right) \frac{\bar{K}_{0}^{\prime \prime
}\left\vert \bar{\Psi}\left( \bar{K}_{0}^{\prime \prime },\bar{X}^{\prime
\prime }\right) \right\vert ^{2}}{\left\langle \bar{K}\right\rangle
\left\Vert \bar{\Psi}\right\Vert ^{2}}}\right) }{\left( 1+\underline{\hat{k}}%
\left( \hat{X}^{\prime }\right) +\underline{\hat{k}}_{1}^{B}\left( \bar{X}%
^{\prime }\right) +\kappa \left[ \frac{\underline{\hat{k}}_{2}^{B}}{1+\bar{k}%
}\right] \left( \hat{X}^{\prime }\right) \right) ^{2}}\bar{K}
\end{eqnarray*}%
\begin{eqnarray*}
&&\frac{\delta }{\delta \left\vert \bar{\Psi}\left( \bar{K},\bar{X}\right)
\right\vert ^{2}}\frac{\left( 1-\hat{M}\right) \hat{g}\left( \hat{K}^{\prime
},\hat{X}^{\prime }\right) +\bar{N}\bar{g}\left( \hat{K}^{\prime },\hat{X}%
^{\prime }\right) }{1+\underline{\hat{k}}_{2}\left( \bar{X}^{\prime }\right)
+\kappa \left[ \frac{\underline{\hat{k}}_{2}^{B}}{1+\bar{k}}\right] \left( 
\hat{X}^{\prime }\right) } \\
&\rightarrow &-\left( \Delta \left( \hat{X},\hat{X}^{\prime }\right) -\frac{%
\hat{K}^{\prime }\hat{k}_{1}\left( \hat{X}^{\prime },\hat{X}\right)
\left\vert \hat{\Psi}\left( \hat{K}^{\prime },\hat{X}^{\prime }\right)
\right\vert ^{2}}{1+\underline{\hat{k}}\left( \hat{X}^{\prime }\right) +%
\underline{\hat{k}}_{1}^{B}\left( \bar{X}^{\prime }\right) +\kappa \left[ 
\frac{\underline{\hat{k}}_{2}^{B}}{1+\bar{k}}\right] \left( \hat{X}^{\prime
}\right) }\right) ^{-1} \\
&&\times \frac{\hat{K}^{\prime }\hat{k}_{1}\left( \hat{X}^{\prime },\hat{X}%
\right) \left\vert \hat{\Psi}\left( \hat{K}^{\prime },\hat{X}^{\prime
}\right) \right\vert ^{2}\left( \left( \hat{k}_{1}^{B}\left( \hat{X}^{\prime
},\bar{X}\right) -\left\langle \hat{k}_{1}^{B}\left( \hat{X}^{\prime },\bar{X%
}\right) \right\rangle \right) +\kappa \frac{\left( \hat{k}_{2}^{B}\left( 
\hat{X}^{\prime },\bar{X}\right) -\left\langle \hat{k}_{2}^{B}\left( \hat{X}%
^{\prime },\bar{X}\right) \right\rangle \right) }{1+\int \bar{k}\left( \bar{X%
}^{\prime },\bar{X}^{\prime \prime }\right) \frac{\bar{K}_{0}^{\prime \prime
}\left\vert \bar{\Psi}\left( \bar{K}_{0}^{\prime \prime },\bar{X}^{\prime
\prime }\right) \right\vert ^{2}}{\left\langle \bar{K}\right\rangle
\left\Vert \bar{\Psi}\right\Vert ^{2}}}\right) }{\left\langle \hat{K}%
\right\rangle \left\Vert \hat{\Psi}\right\Vert ^{2}\left( 1+\underline{\hat{k%
}}\left( \hat{X}^{\prime }\right) +\underline{\hat{k}}_{1}^{B}\left( \bar{X}%
^{\prime }\right) +\kappa \left[ \frac{\underline{\hat{k}}_{2}^{B}}{1+\bar{k}%
}\right] \left( \hat{X}^{\prime }\right) \right) ^{2}}\bar{K} \\
&&\times \frac{\left( 1-\hat{M}\right) \hat{g}\left( \hat{K}^{\prime },\hat{X%
}^{\prime }\right) +\bar{N}\bar{g}\left( \hat{K}^{\prime },\hat{X}^{\prime
}\right) }{1+\underline{\hat{k}}_{2}\left( \bar{X}^{\prime }\right) +\kappa %
\left[ \frac{\underline{\hat{k}}_{2}^{B}}{1+\bar{k}}\right] \left( \hat{X}%
^{\prime }\right) }
\end{eqnarray*}%
In matrices element of 
\begin{equation*}
\frac{\delta }{\delta \left\vert \bar{\Psi}\left( \bar{K},\bar{X}\right)
\right\vert ^{2}}\frac{\left( 1-\hat{M}\right) \hat{g}\left( \hat{K}^{\prime
},\hat{X}^{\prime }\right) +\bar{N}\bar{g}\left( \hat{K}^{\prime },\hat{X}%
^{\prime }\right) }{1+\underline{\hat{k}}_{2}\left( \bar{X}^{\prime }\right)
+\kappa \left[ \frac{\underline{\hat{k}}_{2}^{B}}{1+\bar{k}}\right] \left( 
\hat{X}^{\prime }\right) }
\end{equation*}%
where:%
\begin{eqnarray*}
&&-\left( 1-\left\langle \hat{k}_{1}\left( \hat{X}^{\prime },\hat{X}\right)
\right\rangle \right) ^{-1}\left( \left( \hat{k}_{1}^{B}\left( \hat{X}%
^{\prime },\bar{X}\right) -\left\langle \hat{k}_{1}^{B}\left( \hat{X}%
^{\prime },\bar{X}\right) \right\rangle \right) +\kappa \frac{\left( \hat{k}%
_{2}^{B}\left( \hat{X}^{\prime },\bar{X}\right) -\left\langle \hat{k}%
_{2}^{B}\left( \hat{X}^{\prime },\bar{X}\right) \right\rangle \right) }{%
1+\left\langle \bar{k}\left( \bar{X}^{\prime },\bar{X}^{\prime \prime
}\right) \right\rangle \frac{\left\langle \bar{K}\right\rangle \left\vert
\left\Vert \bar{\Psi}\right\Vert \right\vert ^{2}}{\left\langle \bar{K}%
\right\rangle \left\Vert \bar{\Psi}\right\Vert ^{2}}}\right) \bar{K} \\
&&\times \frac{\left( 1-\hat{M}\right) \hat{g}\left( \hat{K}^{\prime },\hat{X%
}^{\prime }\right) +\bar{N}\bar{g}\left( \hat{K}^{\prime },\hat{X}^{\prime
}\right) }{1+\underline{\hat{k}}_{2}\left( \bar{X}^{\prime }\right) +\kappa %
\left[ \frac{\underline{\hat{k}}_{2}^{B}}{1+\bar{k}}\right] \left( \hat{X}%
^{\prime }\right) }
\end{eqnarray*}%
we have:%
\begin{eqnarray}
&&-\left( 1-\left\langle \hat{k}_{1}\left( \hat{X}^{\prime },\hat{X}\right)
\right\rangle \right) ^{-1}\left( \left( \hat{k}_{1}^{B}\left( \hat{X}%
^{\prime },\bar{X}\right) -\left\langle \hat{k}_{1}^{B}\left( \hat{X}%
^{\prime },\bar{X}\right) \right\rangle \right) +\kappa \frac{\left( \hat{k}%
_{2}^{B}\left( \hat{X}^{\prime },\bar{X}\right) -\left\langle \hat{k}%
_{2}^{B}\left( \hat{X}^{\prime },\bar{X}\right) \right\rangle \right) }{%
1+\left\langle \bar{k}\left( \bar{X}^{\prime },\bar{X}^{\prime \prime
}\right) \right\rangle \frac{\left\langle \bar{K}\right\rangle \left\vert
\left\Vert \bar{\Psi}\right\Vert \right\vert ^{2}}{\left\langle \bar{K}%
\right\rangle \left\Vert \bar{\Psi}\right\Vert ^{2}}}\right) \bar{K}  \notag
\\
&&\times \left\langle \frac{\left( 1-\hat{M}\right) \hat{g}\left( \hat{K}%
^{\prime },\hat{X}^{\prime }\right) +\bar{N}\bar{g}\left( \hat{K}^{\prime },%
\hat{X}^{\prime }\right) }{1+\underline{\hat{k}}_{2}\left( \bar{X}^{\prime
}\right) +\kappa \left[ \frac{\underline{\hat{k}}_{2}^{B}}{1+\bar{k}}\right]
\left( \hat{X}^{\prime }\right) }\right\rangle  \label{Da}
\end{eqnarray}

\paragraph*{A25.2.1.7 Gathering terms}

Using that in average:%
\begin{eqnarray*}
&&\frac{\left( 1-\bar{M}\right) }{1+\underline{\bar{k}}_{2}\left( \bar{X}%
^{\prime }\right) } \\
&\rightarrow &\frac{\left( 1-\left\langle \bar{k}\left( \hat{X}^{\prime },%
\hat{X}\right) \right\rangle \right) ^{2}}{\left\Vert \bar{\Psi}\right\Vert
^{2}\left\langle \bar{K}\right\rangle \left( 1-\left\langle \bar{k}%
_{1}\left( \hat{X}^{\prime },\hat{X}\right) \right\rangle \right) \left( 1+%
\frac{\underline{\bar{k}}_{2}\left( \hat{X}^{\prime }\right) }{%
1-\left\langle \bar{k}_{1}\left( \hat{X}^{\prime },\hat{X}\right)
\right\rangle }\right) }
\end{eqnarray*}%
and:%
\begin{eqnarray*}
&&\left( \Delta \left( \hat{X},\hat{X}^{\prime }\right) -\frac{\hat{K}%
^{\prime }\hat{k}_{1}\left( \hat{X}^{\prime },\hat{X}\right) \left\vert \hat{%
\Psi}\left( \hat{K}^{\prime },\hat{X}^{\prime }\right) \right\vert ^{2}}{1+%
\underline{\hat{k}}\left( \hat{X}^{\prime }\right) +\underline{\hat{k}}%
_{1}^{B}\left( \bar{X}^{\prime }\right) +\kappa \left[ \frac{\underline{\hat{%
k}}_{2}^{B}}{1+\bar{k}}\right] \left( \hat{X}^{\prime }\right) }\right) 
\frac{\left( 1-\bar{M}\right) }{1+\underline{\bar{k}}_{2}\left( \bar{X}%
^{\prime }\right) } \\
&\rightarrow &\left( 1-\left\langle \bar{k}\left( \hat{X}^{\prime },\hat{X}%
\right) \right\rangle \right) ^{2}\frac{1}{\left\Vert \bar{\Psi}\right\Vert
^{2}\left\langle \bar{K}\right\rangle }
\end{eqnarray*}%
We find: 
\begin{eqnarray*}
&&\frac{\delta }{\delta \left\vert \bar{\Psi}\left( \bar{K},\bar{X}\right)
\right\vert ^{2}}\bar{g}\left( \hat{K}^{\prime },\hat{X}^{\prime }\right) \\
&=&\left( \left( \Delta \left( \bar{X}^{\prime },\bar{X}\right) -\frac{\bar{K%
}^{\prime }\bar{k}_{1}\left( \bar{X}^{\prime },\bar{X}\right) \left\vert 
\bar{\Psi}\left( \bar{K}^{\prime },\bar{X}^{\prime }\right) \right\vert ^{2}%
}{1+\underline{\bar{k}}\left( \bar{X}^{\prime }\right) }\right) \frac{\left(
1-\bar{M}\right) }{1+\underline{\overline{\bar{k}}}_{2}\left( \bar{X}%
^{\prime }\right) }\right) ^{-1} \\
&&\left\{ -\frac{\delta }{\delta \left\vert \bar{\Psi}\left( \bar{K},\bar{X}%
\right) \right\vert ^{2}}\frac{\hat{K}^{\prime }\underline{\hat{k}}%
_{1}^{B}\left( \hat{X}^{\prime },\bar{X}\right) \left\vert \hat{\Psi}\left( 
\hat{K}^{\prime },\hat{X}^{\prime }\right) \right\vert ^{2}}{\left\Vert \hat{%
\Psi}\right\Vert ^{2}\left\langle \hat{K}\right\rangle \left( 1+\underline{%
\hat{k}}\left( \hat{X}^{\prime }\right) +\underline{\hat{k}}_{1}^{B}\left( 
\bar{X}^{\prime }\right) +\kappa \left[ \frac{\underline{\hat{k}}_{2}^{B}}{1+%
\bar{k}}\right] \left( \hat{X}^{\prime }\right) \right) }\frac{\left( 1-\hat{%
M}\right) \hat{g}\left( \hat{K}^{\prime },\hat{X}^{\prime }\right) +\bar{N}%
\bar{g}\left( \hat{K}^{\prime },\hat{X}^{\prime }\right) }{1+\underline{\hat{%
k}}_{2}\left( \bar{X}^{\prime }\right) +\kappa \left[ \frac{\underline{\hat{k%
}}_{2}^{B}}{1+\bar{k}}\right] \left( \hat{X}^{\prime }\right) }\right. \\
&&-\frac{\hat{K}^{\prime }\underline{\hat{k}}_{1}^{B}\left( \hat{X}^{\prime
},\bar{X}\right) \left\vert \hat{\Psi}\left( \hat{K}^{\prime },\hat{X}%
^{\prime }\right) \right\vert ^{2}}{\left\Vert \hat{\Psi}\right\Vert
^{2}\left\langle \hat{K}\right\rangle \left( 1+\underline{\hat{k}}\left( 
\hat{X}^{\prime }\right) +\underline{\hat{k}}_{1}^{B}\left( \bar{X}^{\prime
}\right) +\kappa \left[ \frac{\underline{\hat{k}}_{2}^{B}}{1+\bar{k}}\right]
\left( \hat{X}^{\prime }\right) \right) }\frac{\delta }{\delta \left\vert 
\bar{\Psi}\left( \bar{K},\bar{X}\right) \right\vert ^{2}}\frac{\left( 1-\hat{%
M}\right) \hat{g}\left( \hat{K}^{\prime },\hat{X}^{\prime }\right) +\bar{N}%
\bar{g}\left( \hat{K}^{\prime },\hat{X}^{\prime }\right) }{1+\underline{\hat{%
k}}_{2}\left( \bar{X}^{\prime }\right) +\kappa \left[ \frac{\underline{\hat{k%
}}_{2}^{B}}{1+\bar{k}}\right] \left( \hat{X}^{\prime }\right) } \\
&&\left. -\frac{\delta }{\delta \left\vert \bar{\Psi}\left( \bar{K},\bar{X}%
\right) \right\vert ^{2}}\left( \left( \Delta \left( \bar{X}^{\prime },\bar{X%
}\right) -\frac{\bar{K}^{\prime }\bar{k}_{1}\left( \bar{X}^{\prime },\bar{X}%
\right) \left\vert \bar{\Psi}\left( \bar{K}^{\prime },\bar{X}^{\prime
}\right) \right\vert ^{2}}{1+\underline{\bar{k}}\left( \bar{X}^{\prime
}\right) }\right) \frac{\left( 1-\bar{M}\right) }{1+\underline{\overline{%
\bar{k}}}_{2}\left( \bar{X}^{\prime }\right) }\right) \bar{g}\left( \hat{K}%
^{\prime },\hat{X}^{\prime }\right) \right\}
\end{eqnarray*}%
that is:%
\begin{eqnarray}
&&\frac{\delta }{\delta \left\vert \bar{\Psi}\left( \bar{K},\bar{X}\right)
\right\vert ^{2}}\bar{g}\left( \hat{K}^{\prime },\hat{X}^{\prime }\right)
\label{Dt} \\
&\simeq &\left( \frac{\left( 1-\left\langle \bar{k}\right\rangle \right) ^{2}%
}{\left\Vert \bar{\Psi}\right\Vert ^{2}\left\langle \bar{K}\right\rangle }%
\right) ^{-1}\left\{ \left( \left\langle \underline{\hat{k}}_{1}^{B}\left(
\left\langle \hat{X}^{\prime }\right\rangle ,\bar{X}\right) \right\rangle +%
\frac{\left\langle \underline{\hat{k}}_{1}^{B}\left( \hat{X}^{\prime },\bar{X%
}\right) \right\rangle }{\left( 1-\left\langle \hat{k}_{1}\left( \hat{X}%
^{\prime },\hat{X}\right) \right\rangle \right) }\right) \right.  \notag \\
&&\times \left( \left( \hat{k}_{1}^{B}\left( \hat{X}^{\prime },\bar{X}%
\right) -\left\langle \hat{k}_{1}^{B}\left( \hat{X}^{\prime },\bar{X}\right)
\right\rangle \right) +\kappa \frac{\left( \hat{k}_{2}^{B}\left( \hat{X}%
^{\prime },\bar{X}\right) -\left\langle \hat{k}_{2}^{B}\left( \hat{X}%
^{\prime },\bar{X}\right) \right\rangle \right) }{1+\left\langle \bar{k}%
\left( \bar{X}^{\prime },\bar{X}^{\prime \prime }\right) \right\rangle }%
\right) \left\langle A\right\rangle \frac{\bar{K}}{\left\Vert \hat{\Psi}%
\right\Vert ^{2}\left\langle \hat{K}\right\rangle }  \notag \\
&&\left. +\frac{\left( 1-\left\langle \bar{k}\left( \hat{X}^{\prime },\hat{X}%
\right) \right\rangle \right) ^{2}}{1-\left\langle \bar{k}_{1}\left( \hat{X}%
^{\prime },\hat{X}\right) \right\rangle }\frac{\left( \bar{k}_{2}\left(
\left\langle \bar{X}\right\rangle ,\bar{X}\right) -\left\langle \bar{k}%
_{2}\left( \bar{X},\bar{X}^{\prime }\right) \right\rangle \right) }{%
\left\Vert \bar{\Psi}\right\Vert ^{2}\left\langle \bar{K}\right\rangle }\bar{%
K}\left\langle \bar{g}\left( \hat{K}^{\prime },\hat{X}^{\prime }\right)
\right\rangle \right\}  \notag
\end{eqnarray}%
We can replace the following averages:%
\begin{equation*}
\left\langle \frac{1}{1+\underline{\hat{k}}_{2}\left( \bar{X}^{\prime
}\right) +\kappa \left[ \frac{\underline{\hat{k}}_{2}^{B}}{1+\bar{k}}\right]
\left( \hat{X}^{\prime }\right) }\right\rangle \rightarrow \frac{1-\left(
\left\langle \underline{\hat{k}}\right\rangle +\left\langle \underline{\hat{k%
}}_{1}^{B}\right\rangle \frac{\left\Vert \bar{\Psi}\right\Vert
^{2}\left\langle \bar{K}\right\rangle }{\left\Vert \hat{\Psi}\right\Vert
^{2}\left\langle \hat{K}\right\rangle }+\kappa \left\langle \left[ \frac{%
\underline{\hat{k}}_{2}^{B}}{1+\bar{k}}\right] \right\rangle \frac{%
\left\Vert \bar{\Psi}\right\Vert ^{2}\left\langle \bar{K}\right\rangle }{%
\left\Vert \hat{\Psi}\right\Vert ^{2}\left\langle \hat{K}\right\rangle }%
\right) }{\left( 1-\left( \left\langle \hat{k}_{1}\right\rangle
+\left\langle \hat{k}_{1}^{B}\right\rangle \frac{\left\Vert \bar{\Psi}%
\right\Vert ^{2}\left\langle \bar{K}\right\rangle }{\left\Vert \hat{\Psi}%
\right\Vert ^{2}\left\langle \hat{K}\right\rangle }\right) \right)
\left\Vert \hat{\Psi}\right\Vert ^{2}\left\langle \hat{K}\right\rangle }
\end{equation*}%
\begin{eqnarray*}
\left\langle A\right\rangle &=&\left\langle \frac{\left( 1-\hat{M}\right) 
\hat{g}\left( \hat{K}^{\prime },\hat{X}^{\prime }\right) +\bar{N}\bar{g}%
\left( \hat{K}^{\prime },\hat{X}^{\prime }\right) }{1+\underline{\hat{k}}%
_{2}\left( \bar{X}^{\prime }\right) +\kappa \left[ \frac{\underline{\hat{k}}%
_{2}^{B}}{1+\bar{k}}\right] \left( \hat{X}^{\prime }\right) }\right\rangle \\
&=&\frac{1-\left( \left\langle \underline{\hat{k}}\right\rangle
+\left\langle \underline{\hat{k}}_{1}^{B}\right\rangle \frac{\left\Vert \bar{%
\Psi}\right\Vert ^{2}\left\langle \bar{K}\right\rangle }{\left\Vert \hat{\Psi%
}\right\Vert ^{2}\left\langle \hat{K}\right\rangle }+\kappa \left\langle %
\left[ \frac{\underline{\hat{k}}_{2}^{B}}{1+\bar{k}}\right] \right\rangle 
\frac{\left\Vert \bar{\Psi}\right\Vert ^{2}\left\langle \bar{K}\right\rangle 
}{\left\Vert \hat{\Psi}\right\Vert ^{2}\left\langle \hat{K}\right\rangle }%
\right) }{\left( 1-\left( \left\langle \hat{k}_{1}\right\rangle
+\left\langle \hat{k}_{1}^{B}\right\rangle \frac{\left\Vert \bar{\Psi}%
\right\Vert ^{2}\left\langle \bar{K}\right\rangle }{\left\Vert \hat{\Psi}%
\right\Vert ^{2}\left\langle \hat{K}\right\rangle }\right) \right)
\left\Vert \hat{\Psi}\right\Vert ^{2}\left\langle \hat{K}\right\rangle } \\
&&\times \left( \left( 1-\left\langle \hat{k}\right\rangle \right)
\left\langle \hat{g}\right\rangle +\left( \left\langle \hat{k}%
_{1}^{B}\right\rangle +\kappa \frac{\left\langle \hat{k}_{2}^{B}\right%
\rangle }{1+\left\langle \bar{k}\right\rangle }\left( 1-\frac{\left\langle 
\bar{k}\right\rangle }{\left( 1+\left\langle \bar{k}\right\rangle \right)
^{2}}\right) \right) \left\langle \bar{g}\right\rangle \right)
\end{eqnarray*}%
and:%
\begin{equation*}
\bar{N}\rightarrow \left\langle \hat{k}_{1}^{B}\right\rangle +\kappa \frac{%
\left\langle \hat{k}_{2}^{B}\right\rangle }{1+\left\langle \bar{k}%
\right\rangle }\left( 1-\frac{\left\langle \bar{k}\right\rangle }{\left(
1+\left\langle \bar{k}\right\rangle \right) ^{2}}\right)
\end{equation*}%
We are thus led to:%
\begin{eqnarray}
&&\frac{\delta }{\delta \left\vert \bar{\Psi}\left( \bar{K},\bar{X}\right)
\right\vert ^{2}}\bar{g}\left( \hat{K}^{\prime },\hat{X}^{\prime }\right) \\
&\simeq &\left( \frac{\left( 1-\left\langle \bar{k}\right\rangle \right) ^{2}%
}{\left\Vert \bar{\Psi}\right\Vert ^{2}\left\langle \bar{K}\right\rangle }%
\right) ^{-1}\left\{ \left( \left\langle \underline{\hat{k}}%
_{1}^{B}\right\rangle +\frac{\left\langle \underline{\hat{k}}%
_{1}^{B}\right\rangle }{\left( 1-\left\langle \hat{k}_{1}\right\rangle
\right) }\right) \right.  \notag \\
&&\times \left( \left( \hat{k}_{1}^{B}\left( \hat{X}^{\prime },\bar{X}%
\right) -\left\langle \hat{k}_{1}^{B}\right\rangle \right) +\kappa \frac{%
\left( \hat{k}_{2}^{B}\left( \hat{X}^{\prime },\bar{X}\right) -\left\langle 
\hat{k}_{2}^{B}\right\rangle \right) }{1+\left\langle \bar{k}\left( \bar{X}%
^{\prime },\bar{X}^{\prime \prime }\right) \right\rangle }\right)
\left\langle A\right\rangle  \notag \\
&&\left. +\frac{\left( 1-\left\langle \bar{k}\right\rangle \right) ^{2}}{%
1-\left\langle \bar{k}_{1}\right\rangle }\frac{\left( \bar{k}_{2}\left(
\left\langle \bar{X}\right\rangle ,\bar{X}\right) -\left\langle \bar{k}%
_{2}\right\rangle \right) }{\left\Vert \bar{\Psi}\right\Vert
^{2}\left\langle \bar{K}\right\rangle }\left\langle \bar{g}\right\rangle
\right\} \frac{\bar{K}}{\left\Vert \hat{\Psi}\right\Vert ^{2}\left\langle 
\hat{K}\right\rangle }  \notag
\end{eqnarray}

\subsubsection*{A25.2.2 Computation of $\frac{\protect\delta \hat{g}\left( 
\hat{K}^{\prime },\hat{X}^{\prime }\right) }{\protect\delta \left\vert \bar{%
\Psi}\left( \bar{K},\bar{X}\right) \right\vert ^{2}}$}

We define, as before:

\begin{equation*}
A=\frac{\left( 1-\hat{M}\right) \hat{g}\left( \hat{K}^{\prime },\hat{X}%
^{\prime }\right) +\bar{N}\bar{g}\left( \hat{K}^{\prime },\hat{X}^{\prime
}\right) }{1+\underline{\hat{k}}_{2}\left( \bar{X}^{\prime }\right) +\kappa %
\left[ \frac{\underline{\hat{k}}_{2}^{B}}{1+\bar{k}}\right] \left( \hat{X}%
^{\prime }\right) }
\end{equation*}%
and $\hat{g}\left( \hat{K},\hat{X}\right) $ is given by: 
\begin{equation*}
\hat{g}\left( \hat{K},\hat{X}\right) =\left( 1-\hat{M}\right) ^{-1}\left(
\left( 1+\underline{\hat{k}}_{2}\left( \bar{X}^{\prime }\right) +\kappa %
\left[ \frac{\underline{\hat{k}}_{2}^{B}}{1+\bar{k}}\right] \left( \hat{X}%
^{\prime }\right) \right) A-\bar{N}\bar{g}\left( \hat{K}^{\prime },\hat{X}%
^{\prime }\right) \right)
\end{equation*}%
As a consequence, the derivative we are seeking for is given by:%
\begin{eqnarray*}
&&\frac{\delta }{\delta \left\vert \bar{\Psi}\left( \bar{K},\bar{X}\right)
\right\vert ^{2}}\hat{g}\left( \hat{K},\hat{X}\right) \\
&=&\left( 1-\hat{M}\right) ^{-1}\frac{\delta }{\delta \left\vert \bar{\Psi}%
\left( \bar{K},\bar{X}\right) \right\vert ^{2}}\hat{M}\hat{g}\left( \hat{K},%
\hat{X}\right) \\
&&+\left( 1-\hat{M}\right) ^{-1}\left( \frac{\delta }{\delta \left\vert \bar{%
\Psi}\left( \bar{K},\bar{X}\right) \right\vert ^{2}}\left( 1+\underline{\hat{%
k}}_{2}\left( \bar{X}^{\prime }\right) +\kappa \left[ \frac{\underline{\hat{k%
}}_{2}^{B}}{1+\bar{k}}\right] \left( \hat{X}^{\prime }\right) \right) A-%
\frac{\delta }{\delta \left\vert \bar{\Psi}\left( \bar{K},\bar{X}\right)
\right\vert ^{2}}\left( \bar{N}\bar{g}\left( \hat{K}^{\prime },\hat{X}%
^{\prime }\right) \right) \right)
\end{eqnarray*}%
In average, this becomes:%
\begin{eqnarray}
&&\left( 1-\hat{M}\right) ^{-1}\left( \frac{\delta }{\delta \left\vert \bar{%
\Psi}\left( \bar{K},\bar{X}\right) \right\vert ^{2}}\hat{M}\right) \hat{g}%
\left( \hat{K},\hat{X}\right)  \label{MCG} \\
&&+\frac{\frac{\delta }{\delta \left\vert \bar{\Psi}\left( \bar{K},\bar{X}%
\right) \right\vert ^{2}}\left( 1+\underline{\hat{k}}_{2}\left( \bar{X}%
^{\prime }\right) +\kappa \left[ \frac{\underline{\hat{k}}_{2}^{B}}{1+\bar{k}%
}\right] \left( \hat{X}^{\prime }\right) \right) }{\left( 1+\underline{\hat{k%
}}_{2}\left( \bar{X}^{\prime }\right) +\kappa \left[ \frac{\underline{\hat{k}%
}_{2}^{B}}{1+\bar{k}}\right] \left( \hat{X}^{\prime }\right) \right) }%
\left\langle \left( 1-\hat{M}\right) ^{-1}\left( 1+\underline{\hat{k}}%
_{2}\left( \bar{X}^{\prime }\right) +\kappa \left[ \frac{\underline{\hat{k}}%
_{2}^{B}}{1+\bar{k}}\right] \left( \hat{X}^{\prime }\right) \right)
A\right\rangle  \notag \\
&&+\left( 1-\hat{M}\right) ^{-1}\left( 1+\underline{\hat{k}}_{2}\left( \bar{X%
}^{\prime }\right) +\kappa \left[ \frac{\underline{\hat{k}}_{2}^{B}}{1+\bar{k%
}}\right] \left( \hat{X}^{\prime }\right) \right) \frac{\delta }{\delta
\left\vert \bar{\Psi}\left( \bar{K},\bar{X}\right) \right\vert ^{2}}A  \notag
\\
&&-\left( 1-\hat{M}\right) ^{-1}\frac{\delta }{\delta \left\vert \bar{\Psi}%
\left( \bar{K},\bar{X}\right) \right\vert ^{2}}\left( \bar{N}\bar{g}\left( 
\hat{K}^{\prime },\hat{X}^{\prime }\right) \right)  \notag
\end{eqnarray}%
where the terms in (\ref{MCG}) are:%
\begin{eqnarray*}
&&\frac{\frac{\delta }{\delta \left\vert \bar{\Psi}\left( \bar{K},\bar{X}%
\right) \right\vert ^{2}}\left( 1+\underline{\hat{k}}_{2}\left( \bar{X}%
^{\prime }\right) +\kappa \left[ \frac{\underline{\hat{k}}_{2}^{B}}{1+\bar{k}%
}\right] \left( \hat{X}^{\prime }\right) \right) }{\left( 1+\underline{\hat{k%
}}_{2}\left( \bar{X}^{\prime }\right) +\kappa \left[ \frac{\underline{\hat{k}%
}_{2}^{B}}{1+\bar{k}}\right] \left( \hat{X}^{\prime }\right) \right) }%
\left\langle \left( 1-\hat{M}\right) ^{-1}\left( 1+\underline{\hat{k}}%
_{2}\left( \bar{X}^{\prime }\right) +\kappa \left[ \frac{\underline{\hat{k}}%
_{2}^{B}}{1+\bar{k}}\right] \left( \hat{X}^{\prime }\right) \right)
A\right\rangle \\
&=&\frac{\frac{\delta }{\delta \left\vert \bar{\Psi}\left( \bar{K},\bar{X}%
\right) \right\vert ^{2}}\left( 1+\underline{\hat{k}}_{2}\left( \bar{X}%
^{\prime }\right) +\kappa \left[ \frac{\underline{\hat{k}}_{2}^{B}}{1+\bar{k}%
}\right] \left( \hat{X}^{\prime }\right) \right) }{\left( 1+\underline{\hat{k%
}}_{2}\left( \bar{X}^{\prime }\right) +\kappa \left[ \frac{\underline{\hat{k}%
}_{2}^{B}}{1+\bar{k}}\right] \left( \hat{X}^{\prime }\right) \right) }\left(
\left\langle \hat{g}\left( \hat{K},\hat{X}\right) \right\rangle +\left( 1-%
\hat{M}\right) ^{-1}\bar{N}\left\langle \bar{g}\left( \hat{K}^{\prime },\hat{%
X}^{\prime }\right) \right\rangle \right)
\end{eqnarray*}%
and, using (\ref{Da}) 
\begin{eqnarray*}
&&\left( 1-\hat{M}\right) ^{-1}\left( 1+\underline{\hat{k}}_{2}\left( \bar{X}%
^{\prime }\right) +\kappa \left[ \frac{\underline{\hat{k}}_{2}^{B}}{1+\bar{k}%
}\right] \left( \hat{X}^{\prime }\right) \right) \frac{\delta }{\delta
\left\vert \bar{\Psi}\left( \bar{K},\bar{X}\right) \right\vert ^{2}}A \\
&=&-\left( 1-\left\langle \hat{k}\right\rangle \right) ^{-1}\left( 1+%
\underline{\hat{k}}_{2}\left( \bar{X}^{\prime }\right) +\kappa \left[ \frac{%
\underline{\hat{k}}_{2}^{B}}{1+\bar{k}}\right] \left( \hat{X}^{\prime
}\right) \right) \left( 1-\left\langle \hat{k}_{1}\right\rangle \right) ^{-1}
\\
&&\times \left( \left( \hat{k}_{1}^{B}\left( \hat{X}^{\prime },\bar{X}%
\right) -\left\langle \hat{k}_{1}^{B}\right\rangle \right) +\kappa \frac{%
\left( \hat{k}_{2}^{B}\left( \hat{X}^{\prime },\bar{X}\right) -\left\langle 
\hat{k}_{2}^{B}\right\rangle \right) }{1+\left\langle \bar{k}\left( \bar{X}%
^{\prime },\bar{X}^{\prime \prime }\right) \right\rangle \frac{\left\langle 
\bar{K}\right\rangle \left\vert \left\Vert \bar{\Psi}\right\Vert \right\vert
^{2}}{\left\langle \bar{K}\right\rangle \left\Vert \bar{\Psi}\right\Vert ^{2}%
}}\right) \bar{K}\left\langle A\right\rangle \\
&\simeq &-\left( \left( 1-\left\langle \hat{k}\right\rangle \right) \left(
1-\left\langle \hat{k}_{1}\right\rangle \right) \right) ^{-1}\left( \left( 
\hat{k}_{1}^{B}\left( \hat{X}^{\prime },\bar{X}\right) -\left\langle \hat{k}%
_{1}^{B}\right\rangle \right) +\kappa \frac{\left( \hat{k}_{2}^{B}\left( 
\hat{X}^{\prime },\bar{X}\right) -\left\langle \hat{k}_{2}^{B}\right\rangle
\right) }{1+\left\langle \bar{k}\left( \bar{X}^{\prime },\bar{X}^{\prime
\prime }\right) \right\rangle \frac{\left\langle \bar{K}\right\rangle
\left\vert \left\Vert \bar{\Psi}\right\Vert \right\vert ^{2}}{\left\langle 
\bar{K}\right\rangle \left\Vert \bar{\Psi}\right\Vert ^{2}}}\right) \\
&&\times \left( \left\langle \hat{g}\left( \hat{K},\hat{X}\right)
\right\rangle +\left( 1-\hat{M}\right) ^{-1}\bar{N}\left\langle \bar{g}%
\left( \hat{K}^{\prime },\hat{X}^{\prime }\right) \right\rangle \right)
\end{eqnarray*}%
Consequently, we have:%
\begin{eqnarray*}
&&\frac{\delta }{\delta \left\vert \bar{\Psi}\left( \bar{K},\bar{X}\right)
\right\vert ^{2}}\hat{g}\left( \hat{K},\hat{X}\right) \\
&=&\left( 1-\hat{M}\right) ^{-1}\frac{\delta }{\delta \left\vert \bar{\Psi}%
\left( \bar{K},\bar{X}\right) \right\vert ^{2}}\hat{M}\hat{g}\left( \hat{K},%
\hat{X}\right) \\
&&+\frac{\frac{\delta }{\delta \left\vert \bar{\Psi}\left( \bar{K},\bar{X}%
\right) \right\vert ^{2}}\left( 1+\underline{\hat{k}}_{2}\left( \bar{X}%
^{\prime }\right) +\kappa \left[ \frac{\underline{\hat{k}}_{2}^{B}}{1+\bar{k}%
}\right] \left( \hat{X}^{\prime }\right) \right) }{\left( 1+\underline{\hat{k%
}}_{2}\left( \bar{X}^{\prime }\right) +\kappa \left[ \frac{\underline{\hat{k}%
}_{2}^{B}}{1+\bar{k}}\right] \left( \hat{X}^{\prime }\right) \right) }\left(
\left\langle \hat{g}\left( \hat{K},\hat{X}\right) \right\rangle +\left( 1-%
\hat{M}\right) ^{-1}\bar{N}\left\langle \bar{g}\left( \hat{K}^{\prime },\hat{%
X}^{\prime }\right) \right\rangle \right) \\
&&-\left( \left( 1-\left\langle \hat{k}\right\rangle \right) \left(
1-\left\langle \hat{k}_{1}\right\rangle \right) \right) ^{-1}\left( \left( 
\hat{k}_{1}^{B}\left( \hat{X}^{\prime },\bar{X}\right) -\left\langle \hat{k}%
_{1}^{B}\right\rangle \right) +\kappa \frac{\left( \hat{k}_{2}^{B}\left( 
\hat{X}^{\prime },\bar{X}\right) -\left\langle \hat{k}_{2}^{B}\right\rangle
\right) }{1+\left\langle \bar{k}\left( \bar{X}^{\prime },\bar{X}^{\prime
\prime }\right) \right\rangle \frac{\left\langle \bar{K}\right\rangle
\left\vert \left\Vert \bar{\Psi}\right\Vert \right\vert ^{2}}{\left\langle 
\bar{K}\right\rangle \left\Vert \bar{\Psi}\right\Vert ^{2}}}\right) \\
&&\times \left( \left\langle \hat{g}\left( \hat{K},\hat{X}\right)
\right\rangle +\left( 1-\hat{M}\right) ^{-1}\bar{N}\left\langle \bar{g}%
\left( \hat{K}^{\prime },\hat{X}^{\prime }\right) \right\rangle \right)
-\left( 1-\hat{M}\right) ^{-1}\left( \frac{\delta }{\delta \left\vert \bar{%
\Psi}\left( \bar{K},\bar{X}\right) \right\vert ^{2}}\left( \bar{N}\bar{g}%
\left( \hat{K}^{\prime },\hat{X}^{\prime }\right) \right) \right)
\end{eqnarray*}

\paragraph*{A25.2.2.1 Estimation of:}

\begin{equation*}
\frac{\frac{\delta }{\delta \left\vert \bar{\Psi}\left( \bar{K},\bar{X}%
\right) \right\vert ^{2}}\left( 1+\underline{\hat{k}}_{2}\left( \bar{X}%
^{\prime }\right) +\kappa \left[ \frac{\underline{\hat{k}}_{2}^{B}}{1+\bar{k}%
}\right] \left( \hat{X}^{\prime }\right) \right) }{\left( 1+\underline{\hat{k%
}}_{2}\left( \bar{X}^{\prime }\right) +\kappa \left[ \frac{\underline{\hat{k}%
}_{2}^{B}}{1+\bar{k}}\right] \left( \hat{X}^{\prime }\right) \right) }
\end{equation*}

The derivative has th xpanded form:

\begin{eqnarray*}
&&\frac{\delta }{\delta \left\vert \bar{\Psi}\left( \bar{K},\bar{X}\right)
\right\vert ^{2}}\left( 1+\underline{\hat{k}}_{2}^{B}\left( \bar{X}^{\prime
}\right) +\kappa \left[ \frac{\underline{\hat{k}}_{2}^{B}}{1+\bar{k}}\right]
\left( \hat{X}^{\prime }\right) \right) \\
&\rightarrow &\frac{\delta }{\delta \left\vert \bar{\Psi}\left( \bar{K},\bar{%
X}\right) \right\vert ^{2}}\frac{\left( 1-\left( \left\langle \hat{k}%
_{1}\left( \hat{X}^{\prime },\hat{X}\right) \right\rangle +\left\langle \hat{%
k}_{1}^{B}\left( \hat{X}^{\prime },\hat{X}\right) \right\rangle \frac{%
\left\Vert \bar{\Psi}\right\Vert ^{2}\left\langle \bar{K}\right\rangle }{%
\left\Vert \hat{\Psi}\right\Vert ^{2}\left\langle \hat{K}\right\rangle }%
\right) \right) \left( 1+\frac{\underline{\hat{k}}_{2}\left( \hat{X}^{\prime
}\right) +\kappa \left[ \frac{\underline{\hat{k}}_{2}^{B}}{1+\bar{k}}\right]
\left( \hat{X}^{\prime }\right) }{1-\left( \left\langle \hat{k}_{1}\left( 
\hat{X}^{\prime },\hat{X}\right) \right\rangle +\left\langle \hat{k}%
_{1}^{B}\left( \hat{X}^{\prime },\hat{X}\right) \right\rangle \frac{%
\left\Vert \bar{\Psi}\right\Vert ^{2}\left\langle \bar{K}\right\rangle }{%
\left\Vert \hat{\Psi}\right\Vert ^{2}\left\langle \hat{K}\right\rangle }%
\right) }\right) }{1-\left( \left\langle \underline{\hat{k}}\left( \hat{X}%
^{\prime },\hat{X}\right) \right\rangle +\left( \left\langle \underline{\hat{%
k}}_{1}^{B}\left( \hat{X}^{\prime },\bar{X}\right) \right\rangle +\kappa
\left\langle \left[ \frac{\underline{\hat{k}}_{2}^{B}\left( \hat{X}^{\prime
},\bar{X}\right) }{1+\bar{k}}\right] \right\rangle \right) \frac{\left\Vert 
\bar{\Psi}\right\Vert ^{2}\left\langle \bar{K}\right\rangle }{\left\Vert 
\hat{\Psi}\right\Vert ^{2}\left\langle \hat{K}\right\rangle }\right) } \\
&\rightarrow &\frac{1}{\left( 1+\frac{\underline{\hat{k}}_{2}\left( \hat{X}%
^{\prime }\right) +\kappa \left[ \frac{\underline{\hat{k}}_{2}^{B}}{1+\bar{k}%
}\right] \left( \hat{X}^{\prime }\right) }{1-\left( \left\langle \hat{k}%
_{1}\left( \hat{X}^{\prime },\hat{X}\right) \right\rangle +\left\langle \hat{%
k}_{1}^{B}\left( \hat{X}^{\prime },\hat{X}\right) \right\rangle \frac{%
\left\Vert \bar{\Psi}\right\Vert ^{2}\left\langle \bar{K}\right\rangle }{%
\left\Vert \hat{\Psi}\right\Vert ^{2}\left\langle \hat{K}\right\rangle }%
\right) }\right) ^{2}\left\Vert \hat{\Psi}\right\Vert ^{2}\left\langle \hat{K%
}\right\rangle } \\
&&\times \frac{\left( \kappa \left[ \frac{\underline{\hat{k}}_{2}^{B}}{1+%
\bar{k}}\right] \left( \hat{X}^{\prime },,\bar{X}\right) -\left\langle
\kappa \left[ \frac{\underline{\hat{k}}_{2}^{B}}{1+\bar{k}}\right] \left( 
\hat{X}^{\prime },,\bar{X}\right) \right\rangle \right) \bar{K}}{\left(
1-\left( \left\langle \underline{\hat{k}}\left( \hat{X}^{\prime },\hat{X}%
\right) \right\rangle +\left( \left\langle \underline{\hat{k}}_{1}^{B}\left( 
\hat{X}^{\prime },\bar{X}\right) \right\rangle +\kappa \left\langle \left[ 
\frac{\underline{\hat{k}}_{2}^{B}\left( \hat{X}^{\prime },\bar{X}\right) }{1+%
\bar{k}}\right] \right\rangle \right) \frac{\left\Vert \bar{\Psi}\right\Vert
^{2}\left\langle \bar{K}\right\rangle }{\left\Vert \hat{\Psi}\right\Vert
^{2}\left\langle \hat{K}\right\rangle }\right) \right) } \\
&&-\frac{1}{\left( 1+\frac{\underline{\hat{k}}_{2}\left( \hat{X}^{\prime
}\right) +\kappa \left[ \frac{\underline{\hat{k}}_{2}^{B}}{1+\bar{k}}\right]
\left( \hat{X}^{\prime }\right) }{1-\left( \left\langle \hat{k}_{1}\left( 
\hat{X}^{\prime },\hat{X}\right) \right\rangle +\left\langle \hat{k}%
_{1}^{B}\left( \hat{X}^{\prime },\hat{X}\right) \right\rangle \frac{%
\left\Vert \bar{\Psi}\right\Vert ^{2}\left\langle \bar{K}\right\rangle }{%
\left\Vert \hat{\Psi}\right\Vert ^{2}\left\langle \hat{K}\right\rangle }%
\right) }\right) \left\Vert \hat{\Psi}\right\Vert ^{2}\left\langle \hat{K}%
\right\rangle } \\
&&\times \frac{\left( \left( 1-\left\langle \underline{\hat{k}}\left( \hat{X}%
^{\prime },\hat{X}\right) \right\rangle \right) \kappa \left\langle \left[ 
\frac{\underline{\hat{k}}_{2}^{B}\left( \hat{X}^{\prime },\bar{X}\right) }{1+%
\bar{k}}\right] \right\rangle +\left\langle \underline{\hat{k}}%
_{1}^{B}\left( \hat{X}^{\prime },\bar{X}\right) \right\rangle \left\langle 
\underline{\hat{k}}_{2}\left( \hat{X}^{\prime },\hat{X}\right) \right\rangle
\right) \bar{K}}{\left( 1-\left( \left\langle \underline{\hat{k}}\left( \hat{%
X}^{\prime },\hat{X}\right) \right\rangle +\left( \left\langle \underline{%
\hat{k}}_{1}^{B}\left( \hat{X}^{\prime },\bar{X}\right) \right\rangle
+\kappa \left\langle \left[ \frac{\underline{\hat{k}}_{2}^{B}\left( \hat{X}%
^{\prime },\bar{X}\right) }{1+\bar{k}}\right] \right\rangle \right) \frac{%
\left\Vert \bar{\Psi}\right\Vert ^{2}\left\langle \bar{K}\right\rangle }{%
\left\Vert \hat{\Psi}\right\Vert ^{2}\left\langle \hat{K}\right\rangle }%
\right) \right) ^{2}}
\end{eqnarray*}%
and:%
\begin{eqnarray*}
&&-\left( 1-\left( \left\langle \hat{k}_{1}\left( \hat{X}^{\prime },\hat{X}%
\right) \right\rangle +\left\langle \hat{k}_{1}^{B}\left( \hat{X}^{\prime },%
\hat{X}\right) \right\rangle \frac{\left\Vert \bar{\Psi}\right\Vert
^{2}\left\langle \bar{K}\right\rangle }{\left\Vert \hat{\Psi}\right\Vert
^{2}\left\langle \hat{K}\right\rangle }\right) \right) ^{-1} \\
&&\times \left( \frac{\left( \kappa \left[ \frac{\underline{\hat{k}}_{2}^{B}%
}{1+\bar{k}}\right] \left( \hat{X}^{\prime },,\bar{X}\right) -\left\langle
\kappa \left[ \frac{\underline{\hat{k}}_{2}^{B}}{1+\bar{k}}\right] \left( 
\hat{X}^{\prime },,\bar{X}\right) \right\rangle \right) \bar{K}}{\left( 1+%
\frac{\underline{\hat{k}}_{2}\left( \hat{X}^{\prime }\right) +\kappa \left[ 
\frac{\underline{\hat{k}}_{2}^{B}}{1+\bar{k}}\right] \left( \hat{X}^{\prime
}\right) }{1-\left( \left\langle \hat{k}_{1}\left( \hat{X}^{\prime },\hat{X}%
\right) \right\rangle +\left\langle \hat{k}_{1}^{B}\left( \hat{X}^{\prime },%
\hat{X}\right) \right\rangle \frac{\left\Vert \bar{\Psi}\right\Vert
^{2}\left\langle \bar{K}\right\rangle }{\left\Vert \hat{\Psi}\right\Vert
^{2}\left\langle \hat{K}\right\rangle }\right) }\right) \left\Vert \hat{\Psi}%
\right\Vert ^{2}\left\langle \hat{K}\right\rangle }\right. \\
&&\left. +\frac{\left( \left( 1-\left\langle \underline{\hat{k}}\left( \hat{X%
}^{\prime },\hat{X}\right) \right\rangle \right) \kappa \left\langle \left[ 
\frac{\underline{\hat{k}}_{2}^{B}\left( \hat{X}^{\prime },\bar{X}\right) }{1+%
\bar{k}}\right] \right\rangle \bar{K}+\left\langle \underline{\hat{k}}%
_{1}^{B}\left( \hat{X}^{\prime },\bar{X}\right) \right\rangle \left\langle 
\underline{\hat{k}}_{2}\left( \hat{X}^{\prime },\hat{X}\right) \right\rangle
\right) \bar{K}}{\left( 1-\left( \left\langle \underline{\hat{k}}\left( \hat{%
X}^{\prime },\hat{X}\right) \right\rangle +\left( \left\langle \underline{%
\hat{k}}_{1}^{B}\left( \hat{X}^{\prime },\bar{X}\right) \right\rangle
+\kappa \left\langle \left[ \frac{\underline{\hat{k}}_{2}^{B}\left( \hat{X}%
^{\prime },\bar{X}\right) }{1+\bar{k}}\right] \right\rangle \right) \frac{%
\left\Vert \bar{\Psi}\right\Vert ^{2}\left\langle \bar{K}\right\rangle }{%
\left\Vert \hat{\Psi}\right\Vert ^{2}\left\langle \hat{K}\right\rangle }%
\right) \right) }\right)
\end{eqnarray*}%
which is, in average:%
\begin{eqnarray*}
&&\frac{\frac{\delta }{\delta \left\vert \bar{\Psi}\left( \bar{K},\bar{X}%
\right) \right\vert ^{2}}\left( 1+\underline{\hat{k}}_{2}\left( \bar{X}%
^{\prime }\right) +\kappa \left[ \frac{\underline{\hat{k}}_{2}^{B}}{1+\bar{k}%
}\right] \left( \hat{X}^{\prime }\right) \right) }{\left( 1+\underline{\hat{k%
}}_{2}\left( \bar{X}^{\prime }\right) +\kappa \left[ \frac{\underline{\hat{k}%
}_{2}^{B}}{1+\bar{k}}\right] \left( \hat{X}^{\prime }\right) \right) } \\
&\rightarrow &\frac{\left( \kappa \left[ \frac{\underline{\hat{k}}_{2}^{B}}{%
1+\bar{k}}\right] \left( \hat{X}^{\prime },,\bar{X}\right) -\left\langle
\kappa \left[ \frac{\underline{\hat{k}}_{2}^{B}}{1+\bar{k}}\right] \left( 
\hat{X}^{\prime },,\bar{X}\right) \right\rangle \right) -\frac{\kappa
\left\langle \left[ \frac{\underline{\hat{k}}_{2}^{B}}{1+\bar{k}}\right]
\right\rangle \left( 1-\left\langle \hat{k}\left( \hat{X}^{\prime },\hat{X}%
\right) \right\rangle \right) +\left\langle \hat{k}_{1}^{B}\right\rangle
\left\langle \hat{k}_{2}\right\rangle }{\left( 1-\left( \left\langle \hat{k}%
\right\rangle +\left( \left\langle \hat{k}_{1}^{B}\right\rangle +\kappa
\left\langle \left[ \frac{\underline{\hat{k}}_{2}^{B}}{1+\bar{k}}\right]
\right\rangle \right) \frac{\left\Vert \bar{\Psi}\right\Vert
^{2}\left\langle \bar{K}\right\rangle }{\left\Vert \hat{\Psi}\right\Vert
^{2}\left\langle \hat{K}\right\rangle }\right) \right) }}{\left( 1-\left(
\left\langle \hat{k}_{1}\right\rangle +\left\langle \hat{k}%
_{1}^{B}\right\rangle \frac{\left\Vert \bar{\Psi}\right\Vert
^{2}\left\langle \bar{K}\right\rangle }{\left\Vert \hat{\Psi}\right\Vert
^{2}\left\langle \hat{K}\right\rangle }\right) \right) \left\Vert \hat{\Psi}%
\right\Vert ^{2}}\frac{\bar{K}}{\left\langle \hat{K}\right\rangle }
\end{eqnarray*}

\paragraph{A25.2.2.2 Estimation of $\frac{\protect\delta }{\protect\delta %
\left\vert \bar{\Psi}\left( \bar{K},\bar{X}\right) \right\vert ^{2}}\hat{g}%
\left( \hat{K},\hat{X}\right) $}

Using (\ref{Dm}):%
\begin{eqnarray*}
&&\frac{\delta }{\delta \left\vert \bar{\Psi}\left( \bar{K},\bar{X}\right)
\right\vert ^{2}}\hat{g}\left( \hat{K},\hat{X}\right) \\
&=&-\left\langle \hat{k}\left( \hat{X}^{\prime },\hat{X}\right)
\right\rangle \left( 1-\left\langle \hat{k}\left( \hat{X}^{\prime },\hat{X}%
\right) \right\rangle \right) ^{-1} \\
&&\times \left( \left( \hat{k}_{1}^{B}\left( \hat{X}^{\prime },\bar{X}%
_{1}\right) -\left\langle \hat{k}_{1}^{B}\left( \hat{X}^{\prime },\bar{X}%
_{1}\right) \right\rangle \right) +\kappa \frac{\left( \hat{k}_{2}^{B}\left( 
\hat{X}^{\prime },\bar{X}_{1}\right) -\left\langle \hat{k}_{2}^{B}\left( 
\hat{X}^{\prime },\bar{X}_{1}\right) \right\rangle \right) }{1+\left\langle 
\bar{k}\left( \bar{X}^{\prime },\bar{X}^{\prime \prime }\right)
\right\rangle }\right) \frac{\bar{K}}{\left\Vert \hat{\Psi}\right\Vert
^{2}\left\langle \hat{K}\right\rangle }\hat{g}\left( \hat{K},\hat{X}\right)
\\
&&+\frac{\left( \kappa \left[ \frac{\underline{\hat{k}}_{2}^{B}}{1+\bar{k}}%
\right] \left( \hat{X}^{\prime },,\bar{X}\right) -\left\langle \kappa \left[ 
\frac{\underline{\hat{k}}_{2}^{B}}{1+\bar{k}}\right] \left( \hat{X}^{\prime
},,\bar{X}\right) \right\rangle \right) -\frac{\kappa \left\langle \left[ 
\frac{\underline{\hat{k}}_{2}^{B}}{1+\bar{k}}\right] \right\rangle \left(
1-\left\langle \hat{k}\left( \hat{X}^{\prime },\hat{X}\right) \right\rangle
\right) +\left\langle \hat{k}_{1}^{B}\right\rangle \left\langle \hat{k}%
_{2}\right\rangle }{\left( 1-\left( \left\langle \hat{k}\right\rangle
+\left( \left\langle \hat{k}_{1}^{B}\right\rangle +\kappa \left\langle \left[
\frac{\underline{\hat{k}}_{2}^{B}}{1+\bar{k}}\right] \right\rangle \right) 
\frac{\left\Vert \bar{\Psi}\right\Vert ^{2}\left\langle \bar{K}\right\rangle 
}{\left\Vert \hat{\Psi}\right\Vert ^{2}\left\langle \hat{K}\right\rangle }%
\right) \right) }}{\left( 1-\left( \left\langle \hat{k}_{1}\right\rangle
+\left\langle \hat{k}_{1}^{B}\right\rangle \frac{\left\Vert \bar{\Psi}%
\right\Vert ^{2}\left\langle \bar{K}\right\rangle }{\left\Vert \hat{\Psi}%
\right\Vert ^{2}\left\langle \hat{K}\right\rangle }\right) \right)
\left\Vert \hat{\Psi}\right\Vert ^{2}}\frac{\bar{K}}{\left\langle \hat{K}%
\right\rangle } \\
&&\times \left( \left\langle \hat{g}\left( \hat{K},\hat{X}\right)
\right\rangle +\left( 1-\hat{M}\right) ^{-1}\bar{N}\left\langle \bar{g}%
\left( \hat{K}^{\prime },\hat{X}^{\prime }\right) \right\rangle \right) \\
&&-\left( \left( 1-\left\langle \hat{k}\right\rangle \right) \left(
1-\left\langle \hat{k}_{1}\right\rangle \right) \right) ^{-1}\left( \left( 
\hat{k}_{1}^{B}\left( \hat{X}^{\prime },\bar{X}\right) -\left\langle \hat{k}%
_{1}^{B}\right\rangle \right) +\kappa \frac{\left( \hat{k}_{2}^{B}\left( 
\hat{X}^{\prime },\bar{X}\right) -\left\langle \hat{k}_{2}^{B}\right\rangle
\right) }{1+\left\langle \bar{k}\left( \bar{X}^{\prime },\bar{X}^{\prime
\prime }\right) \right\rangle \frac{\left\langle \bar{K}\right\rangle
\left\vert \left\Vert \bar{\Psi}\right\Vert \right\vert ^{2}}{\left\langle 
\bar{K}\right\rangle \left\Vert \bar{\Psi}\right\Vert ^{2}}}\right) \\
&&\times \left( \left\langle \hat{g}\left( \hat{K},\hat{X}\right)
\right\rangle +\left( 1-\hat{M}\right) ^{-1}\bar{N}\left\langle \bar{g}%
\left( \hat{K}^{\prime },\hat{X}^{\prime }\right) \right\rangle \right)
-\left( 1-\hat{M}\right) ^{-1}\left( \frac{\delta }{\delta \left\vert \bar{%
\Psi}\left( \bar{K},\bar{X}\right) \right\vert ^{2}}\left( \bar{N}\bar{g}%
\left( \hat{K}^{\prime },\hat{X}^{\prime }\right) \right) \right)
\end{eqnarray*}

\paragraph{A25.2.2.3 Estimation of $\frac{\protect\delta }{\protect\delta %
\left\vert \bar{\Psi}\left( \bar{K},\bar{X}\right) \right\vert ^{2}}\bar{N}%
\bar{g}\left( \hat{K}^{\prime },\hat{X}^{\prime }\right) $}

\begin{equation*}
\frac{\delta }{\delta \left\vert \bar{\Psi}\left( \bar{K},\bar{X}\right)
\right\vert ^{2}}\left( \bar{N}\bar{g}\left( \hat{K}^{\prime },\hat{X}%
^{\prime }\right) \right) =\left[ \frac{\delta }{\delta \left\vert \bar{\Psi}%
\left( \bar{K},\bar{X}\right) \right\vert ^{2}}\bar{N}\right] \bar{g}\left( 
\hat{K}^{\prime },\hat{X}^{\prime }\right) +\bar{N}\frac{\delta }{\delta
\left\vert \bar{\Psi}\left( \bar{K},\bar{X}\right) \right\vert ^{2}}\left( 
\bar{g}\left( \hat{K}^{\prime },\hat{X}^{\prime }\right) \right)
\end{equation*}%
The first term is obtained by wrtng:and this becomes:%
\begin{eqnarray*}
&&\frac{\delta }{\delta \left\vert \bar{\Psi}\left( \bar{K},\bar{X}\right)
\right\vert ^{2}}\bar{N} \\
&=&\frac{\delta }{\delta \left\vert \bar{\Psi}\left( \bar{K},\bar{X}\right)
\right\vert ^{2}}\frac{\left( \hat{k}_{1}^{B}\left( \hat{X},\bar{X}^{\prime
}\right) +\kappa \frac{\hat{k}_{2}^{B}\left( \hat{X},\bar{X}^{\prime
}\right) }{1+\overline{\bar{k}}\left( \bar{X}^{\prime },\left\langle \bar{X}%
^{\prime \prime }\right\rangle \right) }-\kappa \frac{\hat{k}_{2}^{B}\left( 
\bar{X},\bar{X}^{\prime \prime }\right) \bar{K}_{0}^{\prime \prime }%
\overline{\bar{k}}\left( \bar{X}^{\prime \prime },\bar{X}^{\prime }\right) }{%
\left( 1+\overline{\bar{k}}\left( \bar{X}^{\prime },\left\langle \bar{X}%
^{\prime \prime }\right\rangle \right) \right) ^{2}}\right) \hat{K}^{\prime
}\left\vert \hat{\Psi}\left( \hat{K}^{\prime },\hat{X}^{\prime }\right)
\right\vert ^{2}}{\hat{D}\left( \hat{X}^{\prime }\right) \left\langle \hat{K}%
\right\rangle \left\Vert \hat{\Psi}\right\Vert ^{2}} \\
&\rightarrow &-\frac{\left( \hat{k}_{1}^{B}\left( \hat{X},\bar{X}^{\prime
}\right) +\kappa \frac{\hat{k}_{2}^{B}\left( \hat{X},\bar{X}^{\prime
}\right) }{1+\overline{\bar{k}}\left( \bar{X}^{\prime },\left\langle \bar{X}%
^{\prime \prime }\right\rangle \right) }-\kappa \frac{\hat{k}_{2}^{B}\left( 
\bar{X},\bar{X}^{\prime \prime }\right) \bar{K}_{0}^{\prime \prime }%
\overline{\bar{k}}\left( \bar{X}^{\prime \prime },\bar{X}^{\prime }\right) }{%
\left( 1+\overline{\bar{k}}\left( \bar{X}^{\prime },\left\langle \bar{X}%
^{\prime \prime }\right\rangle \right) \right) ^{2}}\right) \hat{K}^{\prime
}\left\vert \hat{\Psi}\left( \hat{K}^{\prime },\hat{X}^{\prime }\right)
\right\vert ^{2}}{\hat{D}\left( \hat{X}^{\prime }\right) \left\langle \hat{K}%
\right\rangle \left\Vert \hat{\Psi}\right\Vert ^{2}} \\
&&\times \frac{\hat{k}_{1}^{B}\left( \hat{X}^{\prime },\bar{X}\right)
-\left\langle \hat{k}_{1}^{B}\left( \hat{X}^{\prime },\bar{X}\right)
\right\rangle +\kappa \left( \left[ \frac{\hat{k}_{2}^{B}}{1+\bar{k}}\right]
\left( \hat{X}^{\prime },\bar{X}\right) -\left\langle \left[ \frac{\hat{k}%
_{2}^{B}}{1+\bar{k}}\right] \left( \hat{X}^{\prime },\bar{X}\right)
\right\rangle \right) \bar{K}}{\hat{D}\left( \hat{X}^{\prime }\right)
\left\langle \hat{K}\right\rangle \left\Vert \hat{\Psi}\right\Vert ^{2}} \\
&\rightarrow &-\left( \left\langle \hat{k}_{1}^{B}\left( \hat{X},\bar{X}%
^{\prime }\right) \right\rangle +\kappa \frac{\left\langle \hat{k}%
_{2}^{B}\left( \hat{X},\bar{X}^{\prime }\right) \right\rangle }{1+\overline{%
\bar{k}}\left( \bar{X}^{\prime },\left\langle \bar{X}^{\prime \prime
}\right\rangle \right) }\left( 1-\kappa \frac{\left\langle \overline{\bar{k}}%
\left( \bar{X}^{\prime \prime },\bar{X}^{\prime }\right) \right\rangle }{1+%
\overline{\bar{k}}\left( \bar{X}^{\prime },\left\langle \bar{X}^{\prime
\prime }\right\rangle \right) }\right) \right) \\
&&\times \frac{\hat{k}_{1}^{B}\left( \hat{X}^{\prime },\bar{X}\right)
-\left\langle \hat{k}_{1}^{B}\left( \hat{X}^{\prime },\bar{X}\right)
\right\rangle +\kappa \left( \left[ \frac{\hat{k}_{2}^{B}}{1+\bar{k}}\right]
\left( \hat{X}^{\prime },\bar{X}\right) -\left\langle \left[ \frac{\hat{k}%
_{2}^{B}}{1+\bar{k}}\right] \left( \hat{X}^{\prime },\bar{X}\right)
\right\rangle \right) \bar{K}}{\left\langle \hat{K}\right\rangle \left\Vert 
\hat{\Psi}\right\Vert ^{2}}
\end{eqnarray*}%
and the second term $\frac{\delta }{\delta \left\vert \bar{\Psi}\left( \bar{K%
},\bar{X}\right) \right\vert ^{2}}\left( \bar{g}\left( \hat{K}^{\prime },%
\hat{X}^{\prime }\right) \right) $ is given by (\ref{Dt}).

\paragraph{A25.2.2.4 Dominant term}

All contributions, bt one, to the derivative $\frac{\delta }{\delta
\left\vert \bar{\Psi}\left( \bar{K},\bar{X}\right) \right\vert ^{2}}\hat{g}%
\left( \hat{K},\hat{X}\right) $ are deviations around the average. As a
consequence, in first approximation, we have: 
\begin{eqnarray*}
&&\frac{\delta }{\delta \left\vert \bar{\Psi}\left( \bar{K},\bar{X}\right)
\right\vert ^{2}}\hat{g}\left( \hat{K},\hat{X}\right) \\
&=&-\frac{\kappa \left\langle \left[ \frac{\underline{\hat{k}}_{2}^{B}}{1+%
\bar{k}}\right] \right\rangle \left( 1-\left\langle \hat{k}\left( \hat{X}%
^{\prime },\hat{X}\right) \right\rangle \right) +\left\langle \hat{k}%
_{1}^{B}\right\rangle \left\langle \hat{k}_{2}\right\rangle }{\left(
1-\left( \left\langle \hat{k}\right\rangle +\left( \left\langle \hat{k}%
_{1}^{B}\right\rangle +\kappa \left\langle \left[ \frac{\underline{\hat{k}}%
_{2}^{B}}{1+\bar{k}}\right] \right\rangle \right) \frac{\left\Vert \bar{\Psi}%
\right\Vert ^{2}\left\langle \bar{K}\right\rangle }{\left\Vert \hat{\Psi}%
\right\Vert ^{2}\left\langle \hat{K}\right\rangle }\right) \right) \left(
1-\left( \left\langle \hat{k}_{1}\right\rangle +\left\langle \hat{k}%
_{1}^{B}\right\rangle \frac{\left\Vert \bar{\Psi}\right\Vert
^{2}\left\langle \bar{K}\right\rangle }{\left\Vert \hat{\Psi}\right\Vert
^{2}\left\langle \hat{K}\right\rangle }\right) \right) \left\Vert \hat{\Psi}%
\right\Vert ^{2}}\frac{\bar{K}}{\left\langle \hat{K}\right\rangle } \\
&&\times \left( \left\langle \hat{g}\left( \hat{K},\hat{X}\right)
\right\rangle +\left( 1-\hat{M}\right) ^{-1}\bar{N}\left\langle \bar{g}%
\left( \hat{K}^{\prime },\hat{X}^{\prime }\right) \right\rangle \right)
\end{eqnarray*}

\end{document}